%% file: lhec_2020.tex
\pdfoutput=1
\documentclass[11pt, a4paper, english]{report} 

\providecommand{\main}{.}
\input{\main/main/lhec_2019.cls}
\input{\main/main/lhec_2019_commands.tex}

\newcommand{\ourauthor}[1]{\iffalse{#1}\fi}        

\begin{document}
%
   \def\biblio{}                       
   \def\subfilestableofcontents{}      
   \def\lhectitlepage{}                
   \def\lhecinstructions{}             
   \def\linenumbers{}                  
   %
   \input{\main/main/titlepage.tex}    
   %
   %
   \newpage
   ~
   \thispagestyle{empty}
   \newpage
   \subfile{\main/main/abstract}       
   %
   \input{\main/main/authors.tex}

\subfile{\main/main/foreword}
   \tableofcontents
   \newpage\null\thispagestyle{empty}\newpage 
   \subfile{\main/introduction/introduction}

\subfile{\main/characteristics/characteristics}

\subfile{\main/standardmodel/standardmodel.chapter}

\subfile{\main/nuclearphysics/nuclearphysics}

\subfile{\main/higgs/higgs}

\subfile{\main/bsm/bsm.chapter}

\subfile{\main/LHeC_and_HLLHC/LHeC_and_HLLHC}

\subfile{\main/accelerator/accelerator.chapter}

\subfile{\main/erl/erl.chapter}

\subfile{\main/detector/detector}

\subfile{\main/conclusion/conclusion}
   \newpage
   \appendix
   \subfile{\main/main/appendix}
   \footnotesize{
     \bibliography{\main/lhec}
   }

\end{document}

%% file: main/lhec_2019_commands.tex
\newcommand{\GeVsq}{\ensuremath{\mathrm{GeV}^2}\xspace}

\newcommand{\MeV}{\ensuremath{\mathrm{MeV}}\xspace}
\newcommand{\GeV}{\ensuremath{\mathrm{GeV}}\xspace}
\newcommand{\TeV}{\ensuremath{\mathrm{TeV}}\xspace}
\newcommand{\pt}{\ensuremath{p_\text{T}}\xspace}
\newcommand{\xbj}{\ensuremath{x_{\mathrm bj}}\xspace}

\newcommand{\Qsq}{\ensuremath{Q^{2}}\xspace}

\newcommand{\mw}{\ensuremath{M_{\mathrm W}}\xspace}
\newcommand{\mz}{\ensuremath{M_{\mathrm Z}}\xspace}
\newcommand{\as}{\ensuremath{\alpha_{\mathrm s}}\xspace}
\newcommand{\asmz}{\ensuremath{\alpha_{\mathrm s}(\mz)}\xspace}

\newcommand{\mur}{\ensuremath{\mu_{\mathrm R}}\xspace}

\newcommand{\lstar}{\ensuremath{L^{*}}\xspace}
\newcommand{\bstar}{\ensuremath{B^{*}}\xspace}
\newcommand{\degree}{\ensuremath{^{\circ}}\xspace}

\newcommand{\isotope}[3]{\ensuremath{^{#1}\mathrm{#2}^{#3}}}
\newcommand{\Pb}{\isotope{208}{Pb}{82+}} 

%% file: main/titlepage.tex

\begin{titlepage}

\noindent
CERN-ACC-Note-2020-0002 \\
Geneva,  July 28, 2020 \\ 

\begin{figure}[h]
\vspace{-2.2cm}
\hspace{13.5cm}
\includegraphics[clip=,width=.15\textwidth]{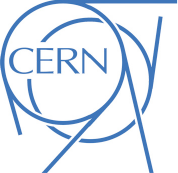}
\caption*{}
\end{figure}
\vspace{-1.5cm}
\begin{figure}[h]
\centering\includegraphics[clip=,width=0.45\textwidth]{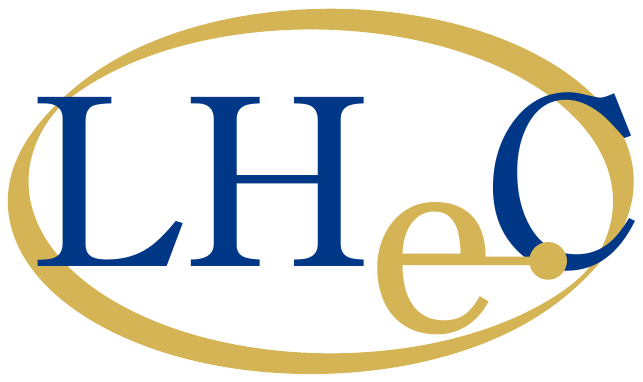}
\caption*{}
\end{figure}
\begin{center}
\vspace{-.5cm}
{\LARGE
\textbf{The Large Hadron-Electron Collider at the HL-LHC} \\
}
\vspace{2cm}
{\Large
\textbf{LHeC and FCC-he Study Group} \\
}
%
%
\vspace{-0.3cm}
\begin{figure}[h!]
\centering
\includegraphics[clip=,width=0.7\textwidth]{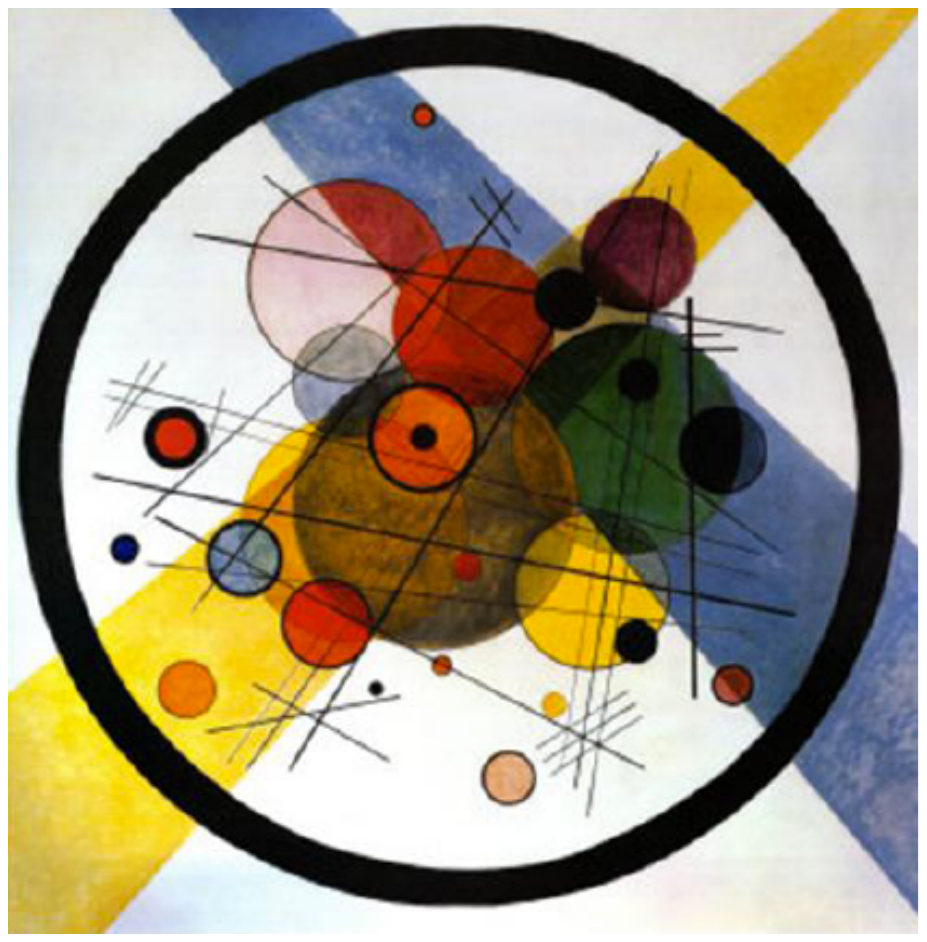}
\caption*{}
\end{figure}
\vspace{0.1cm}
To be submitted to J. Phys. G
\end{center}

\end{titlepage}

%% file: main/abstract.tex

  %
\thispagestyle{empty}
\noindent
CERN-ACC-Note-2020-0002 \\
Geneva,  July 28, 2020 \\ 
{
  \vspace{3.5cm}
  \begin{center}
    \textbf{\LARGE The Large Hadron-Electron Collider at the HL-LHC} \\
    \vspace{2cm}
    {\Large LHeC and FCC-he Study Group} \\
    \vspace{2cm}
    \textbf{Abstract}
  \end{center}
}

The Large Hadron electron Collider (LHeC) is designed to
move the field of deep inelastic scattering (DIS) to the energy and intensity frontier
of particle physics. Exploiting energy recovery technology, it collides a novel, intense electron beam 
with a proton or ion beam from the High Luminosity--Large Hadron Collider (HL-LHC).
The accelerator and interaction region are designed for concurrent electron-proton and proton-proton operation.
This report represents an update of the 
Conceptual Design Report (CDR) of the LHeC, published in 2012.
It comprises new results on
parton structure of the proton and heavier nuclei, QCD dynamics, electroweak and top-quark physics.
It is shown how the LHeC will open a new chapter of nuclear
particle physics in extending the accessible kinematic range in lepton-nucleus
scattering by several orders of magnitude. 
Due to enhanced luminosity, large energy and
the cleanliness of the hadronic final states, the LHeC has a strong
Higgs physics programme and its own discovery potential for new physics.
Building on the 2012 CDR, the report represents a detailed 
updated design of the energy recovery electron
linac (ERL) including new lattice, magnet, superconducting radio frequency technology and
further components. Challenges of energy recovery are described
and the lower energy, high current, 3-turn ERL facility,
PERLE at Orsay, is presented  which uses the LHeC characteristics
serving as a development facility for the design and operation of the LHeC.
An updated detector design is presented corresponding to the acceptance, resolution and
calibration goals which arise from the Higgs and parton density function physics programmes.
The paper also presents novel 
results on the Future Circular Collider in electron-hadron mode, FCC-eh, which 
 utilises the same ERL technology to further extend
the reach of DIS  to even higher centre-of-mass energies.
\clearpage

%

%% file: main/authors.tex

\input{\main/main/out_newcommands.tex}

\thispagestyle{empty}
\vspace{2cm}
{
  \centering
  \Large\textbf{LHeC Study Group}
} 
\\
\vspace{0.05cm}

{
  \begin{flushleft}
    \small
    \noindent
    \input{\main/main/out_authors.tex}
  
  \clearpage
  \small
  \input{\main/main/out_institutes.tex}
    \end{flushleft}
}

%% file: main/out_newcommands.tex
\newcommand{\SantiagodeCompostela}{1}
\newcommand{\UKahramanmaras}{2}

\newcommand{\UHamburg}{3}
\newcommand{\IProtvino}{4}
\newcommand{\UBirmingham}{5}
\newcommand{\UParisSaclay}{6}
\newcommand{\ULiverpool}{7}
\newcommand{\CERN}{8}
\newcommand{\DaresburyLaboratory}{9}
\newcommand{\CockroftInstitute}{10}
\newcommand{\UBasel}{11}
\newcommand{\ChineseAcademySciences}{12}
\newcommand{\LIP}{13}
\newcommand{\ULancaster}{14}
\newcommand{\InsYerevan}{15}
\newcommand{\UAnkara}{16}
\newcommand{\PrismaMainz}{17}
\newcommand{\JGUMainz}{18}
\newcommand{\UMontreal}{19}
\newcommand{\Montenegrin}{20}
\newcommand{\UOxford}{21}
\newcommand{\TechGuwahati}{22}
\newcommand{\DESY}{23}
\newcommand{\BrookhavenNL}{24}
\newcommand{\UStonyBrook}{25}
\newcommand{\RIKEN}{26}
\newcommand{\UBologna}{27}
\newcommand{\URavenshawCuttack}{28}
\newcommand{\JLab}{29}
\newcommand{\INFNRoma}{30}
\newcommand{\IRFUCEA}{31}
\newcommand{\URaziKermanshah}{32}
\newcommand{\CINVESTAVIPN}{33}
\newcommand{\URegensburg}{34}
\newcommand{\SLAC}{35}
\newcommand{\MPIMunich}{36}
\newcommand{\UGumushane}{37}
\newcommand{\IJCLab}{38}
\newcommand{\TechUPraque}{39}
\newcommand{\UKutahya}{40}
\newcommand{\UColumbia}{41}
\newcommand{\InstTechGandhinagar}{42}
\newcommand{\CELIABordeaux}{43}
\newcommand{\UJohannesburg}{44}
\newcommand{\IPPPDurham}{45}
\newcommand{\UToronto}{46}
\newcommand{\UOsaka}{47}
\newcommand{\UdelosAndes}{48}
\newcommand{\INFNFlorence}{49}
\newcommand{\UBoloAbantIzzet}{50}
\newcommand{\UMaryland}{51}
\newcommand{\URepublica}{52}
\newcommand{\UOrsay}{53}
\newcommand{\KhalsaColl}{54}
\newcommand{\Dubna}{55}
\newcommand{\UCornell}{56}

\newcommand{\UJyvaskyla}{57}

\newcommand{\MPIK}{58}
\newcommand{\UPuebla}{59}
\newcommand{\UMilano}{60}
\newcommand{\INFNMilano}{61}
\newcommand{\HefeiCUST}{62}
\newcommand{\Zurich}{63}
\newcommand{\InsTheoZur}{64}
\newcommand{\UNorthwestern}{65}
\newcommand{\RomaTorVergata}{66}
\newcommand{\InstSciBangalore}{67}
\newcommand{\UPelotas}{68}
\newcommand{\UDelhi}{69}
\newcommand{\Fermilab}{70}
\newcommand{\ULiaoningNormal}{71}
\newcommand{\PNPI}{72}
\newcommand{\UTokyo}{73}
\newcommand{\IPMU}{74}
\newcommand{\UMichiganState}{75}
\newcommand{\IPMTehran}{76}
\newcommand{\USouthernMethodist}{77}
\newcommand{\ISciWeizmann}{78}
\newcommand{\MathabhangaCol}{79}
\newcommand{\UPadua}{80}
\newcommand{\UStrasbourg}{81}
\newcommand{\UGiresun}{82}
\newcommand{\InsTechTokyo}{83}
\newcommand{\TOBBAnkara}{84}
\newcommand{\UAmsterdam}{85}
\newcommand{\UMazandaran}{86}
\newcommand{\UUludag}{87}
\newcommand{\UCumhuriyet}{88}
\newcommand{\UCatolicaSantaMaria}{89}
\newcommand{\UIstAydin}{90}
\newcommand{\InsBudker}{91}
\newcommand{\UWitwatersrand}{92}
\newcommand{\UTsinghua}{93}
\newcommand{\UTelAviv}{94}
\newcommand{\ZhejiangInstModPhys}{95}
\newcommand{\UZhejiang}{96}
\newcommand{\UXiamen}{97}
\newcommand{\UCollLondon}{98}
\newcommand{\InsSciTechHenan}{99}
\newcommand{\UVirginia}{100}
\newcommand{\TechDalian}{101}
\newcommand{\UdeRioGrandedoSul}{102}
\newcommand{\UValencia}{103}
\newcommand{\UHelsinki}{104}
\newcommand{\EcolePolytechnique}{105}
\newcommand{\UGenova}{106}
\newcommand{\INFNGenova}{107}
\newcommand{\InsResHarishChandra}{108}
\newcommand{\USouthampton}{109}
\newcommand{\UdeSaoPaulo}{110}
\newcommand{\UNigdeOmer}{111}
\newcommand{\UAndesBogota}{112}
\newcommand{\UAlabamaTuscaloosa}{113}
\newcommand{\InsTata}{114}
\newcommand{\UTechDarmstadt}{115}
\newcommand{\HomedayBerlin}{116}
\newcommand{\UniPraque}{117}
\newcommand{\INFNBologna}{118}
\newcommand{\URioGrandedoNorte}{119}
\newcommand{\KITmain}{120}
\newcommand{\IBMDeutschland}{121}
\newcommand{\IISER}{122}
\newcommand{\UPodgorica}{123}
\newcommand{\UPennsylvania}{124}
\newcommand{\UDelawareNewark}{125}
\newcommand{\ArgonneNL}{126}
\newcommand{\Oracle}{127}
\newcommand{\ACESanFran}{128}
\newcommand{\UUsak}{129}
\newcommand{\UMardelPlata}{130}
\newcommand{\UOklahomaState}{131}
\newcommand{\UPeking}{132}
\newcommand{\IMScChennai}{133}
\newcommand{\UJagiellonian}{134}
\newcommand{\UAnhui}{135}
\newcommand{\UPennState}{136}
\newcommand{\Sussex}{137}
\newcommand{\NCBJ}{138}
\newcommand{\UKansas}{139}
\newcommand{\KIAS}{140}
\newcommand{\UKastamonu}{141}
\newcommand{\NBI}{142}
\newcommand{\UBergen}{143}
\newcommand{\UWuhan}{144}
\newcommand{\APCTP}{145}
\newcommand{\UCRiverside}{146}
\newcommand{\UHebrew}{147}
\newcommand{\UPTCTunja}{148}
\newcommand{\UKobe}{149}
\newcommand{\LBN}{150}
\newcommand{\Auckland}{151}
\newcommand{\URostock}{152}
\newcommand{\NCTSHsinchu}{153}
\newcommand{\UNankai}{154}
\newcommand{\KITTTP}{155}
\newcommand{\KITIKP}{156}

%% file: main/out_authors.tex
P.~Agostini$^{\SantiagodeCompostela}$, 
H.~Aksakal$^{\UKahramanmaras}$, 
S.~Alekhin$^{\UHamburg,\IProtvino}$, 
P.~P.~Allport$^{\UBirmingham}$, 
N.~Andari$^{\UParisSaclay}$, 
K.~D.~J.~Andre$^{\ULiverpool,\CERN}$, 
D.~Angal-Kalinin$^{\DaresburyLaboratory,\CockroftInstitute}$, 
S.~Antusch$^{\UBasel}$, 
L.~Aperio~Bella$^{\ChineseAcademySciences}$, 
L.~Apolinario$^{\LIP}$
% ...

%% file: main/out_institutes.tex
$^{1}$ Universidade~de~Santiago~de~Compostela~(USC), Santiago~de~Compostela, Spain \\
$^{2}$ Kahramanmaras~Sutcu~Imam~University, Kahramanmaras, Turkey \\
$^{3}$ Universit\"at~Hamburg, Hamburg, Germany \\
$^{4}$ Institute~of~High~Energy~Physics~(IHEP), Protvino, Russia \\
$^{5}$ University~of~Birmingham, Birmingham, United~Kingdom \\
$^{6}$ Universit\'e~Paris-Saclay, Saint-Aubin, France \\
$^{7}$ University~of~Liverpool, Liverpool, United~Kingdom \\
$^{8}$ European~Organization~for~Nuclear~Research~(CERN), Gen\`eve, Switzerland \\
$^{9}$ Science~and~Technology~Facilities~Council~(STFC)~-~Daresbury Laboratory, Daresbury, United Kingdom \\
$^{10}$ Cockcroft~Institue~of~Accelerator~Science~and~Technology, Daresbury, United~Kingdom \\
$^{11}$ Universit\"at~Basel, Basel, Switzerland \\
$^{12}$ Chinese~Academy~of~Sciences~-~Institute~of~High~Energy~Physics~(IHEP), Beijing, China \\
$^{13}$ Laboratorio~de~Instrumentacao~e~Fisica~Experimental~de~Particulas~(LIP), Lisbon, Portugal \\
$^{14}$ University~of~Lancaster, Lancaster, United~Kingdom \\
$^{15}$ A.~Alikhanian~National~Laboratory~(AANL), Yerevan, Armenia \\
$^{16}$ Ankara~University, Ankara, Turkey \\
$^{17}$ Johannes~Gutenberg~University~Mainz~(JGU)~-~PRISMA~Cluster~of~Excellence, Mainz, Germany \\
$^{18}$ Johannes~Gutenberg-Universit\"at~Mainz~(JGU), Mainz, Germany \\
$^{19}$ Universit\'e~de~Montr\'eal, Montreal, Candada \\
$^{20}$ University~of~Montenegro, Podgorica, Montenegro \\
$^{21}$ University~of~Oxford, Oxford, United~Kingdom \\
$^{22}$ Department~of~Physics,~Indian~Institute~of~Technology, Guwahati, Assam,~India \\
$^{23}$ Deutsches~Elektronen-Synchrotron~(DESY), Hamburg, Germany \\
$^{24}$ Brookhaven~National~Laboratory~(BNL), Upton, USA \\
$^{25}$ Stony~Brook~University, Stony~Brook, USA \\
$^{26}$ BNL~Research~Center,~RIKEN, Upton,~NY, USA \\
$^{27}$ Universit\`a~di~Bologna, Bologna, Italy \\
$^{28}$ Ravenshaw~University, Cuttack, India \\
$^{29}$ Thomas~Jefferson~National~Accelerator~Facility~(Jefferson~Lab), Newport~News, USA \\
$^{30}$ Istituto~Nazionale~di~Fisica~Nucleare~(INFN)~-~Sezione~di~Roma, Rome, Italy \\
$^{31}$ Commissariat~\`a~l'Energie~Atomique~(CEA)~-~Institut~de~Recherche~sur~les Lois Fondamentales de l'Univers~(IRFU), Gif-sur-Yvette, France \\
$^{32}$ Razi~University, Kermanshah, Iran \\
$^{33}$  Centro de Investigaci\'{o}n y de Estudios Avanzados (CINVESTAV), M\'{e}rida, Mexico \\
$^{34}$ Universit\"at Regensburg, Regensburg, Germany \\
$^{35}$ SLAC~National~Accelerator~Laboratory, Menlo~Park, USA \\
$^{36}$ Max-Planck-Institut~f\"ur~Physik, Munich, Germany \\
$^{37}$ Gumushane~University, Gumushane, Turkey \\
$^{38}$ Universit\'e~Paris-Saclay,~CNRS/IN2P3,~IJCLab, Orsay, France \\
$^{39}$ Faculty~of~Nuclear~Sciences~and~Physical~Engineering,~Czech~Technical~University~in~Prague, Prague, Czech~Republic \\
$^{40}$ Kutahya~Dumlupinar~University, Kutahya, Turkey \\
$^{41}$ Columbia~University, New~York, USA \\
$^{42}$ Indian~Institute~of~Technology~(IIT), Gandhinagar, India \\
$^{43}$ Laboratoire Photonique, Numérique et Nanosciences (LP2N), IOGS-CNRS-Université Bordeaux, Talence, France  \\ 
$^{44}$ University~of~Johannesburg~(UJ), Johannesburg, South~Africa \\
$^{45}$ Institute~for~Particle~Physics~Phenomenology,~Durham~University, Durham, United~Kingdom \\
$^{46}$ University~of~Toronto, Toronto, Canada \\
$^{47}$ Osaka~University, Osaka, Japan \\
$^{48}$ Universidad~de~los~Andes, Santiago, Columbia \\
$^{49}$ Istituto~Nazionale~di~Fisica~Nucleare~(INFN)~-~Sezione~di~Firenze, Firenze, Italy \\
$^{50}$ Bolu~Abant~Izzet~Baysal~University, Bolu, Turkey \\
$^{51}$ University~of~Maryland, College~Park, USA \\
$^{52}$ Universidad~de~la~Republica~-~Instituto~de~Fisica~Facultad~de~Ciencias~(IFFC), Montevideo, Uruguay \\
$^{53}$ Universit\'e~Paris-Sud, Orsay, France \\
$^{54}$ Sri~Guru~Tegh~Badadur~Khalsa~College, Delhi, India \\
$^{55}$ Joint~Institute~for~Nuclear~Research~(JINR), Dubna, Russia \\
$^{56}$ Cornell~University, Ithaca, USA \\
$^{57}$ University~of~Jyv\"askyl\"a, Jyv\"askyl\"a, Finland \\
$^{58}$ Max-Planck-Institut~f\"ur~Kernphysik, Heidelberg, Germany \\
$^{59}$ Benemerita~Universidad~Autonoma~de~Puebla~(BUAP), Puebla, Mexico \\
$^{60}$ Universit\`a~degli~Studi~di~Milano, Milano, Italy \\
$^{61}$ Istituto~Nazionale~di~Fisica~Nucleare~(INFN)~-~Sezione~di~Milano, Milano, Italy \\
$^{62}$ University~of~Science~and~Technology~of~China~(USTC), Hefei, China \\
$^{63}$ Department~of~Physics,~Universit\"at~Z\"urich, Zurich, Switzerland \\
$^{64}$ Institute~for~Theoretical~Physics,~ETH, Zu{r}i{c}h, Switzerland \\
$^{65}$ Northwestern~University, Evanston, USA \\
$^{66}$ University~of~Rome~Tor~Vergata~and~INFN,~Sezione~di~Roma~2, Rome, Italy \\
$^{67}$ Indian~Institute~of~Science~(IISc), Bangalore, India \\
$^{68}$ Universidade~Federal~de~Pelotas~(UFPel), Pelotas, Brazil \\
$^{69}$ University~of~Delhi, Delhi, India \\
$^{70}$ Fermi~National~Accelerator~Laboratory~(FNAL), Batavia, USA \\
$^{71}$ Liaoning~Normal~University~(LNNU), Dalian, China \\
$^{72}$ Petersburg~Nuclear~Physics~Institute~(PNPI), Petersburg, Russia \\
$^{73}$ University~of~Tokyo, Tokyo, Japan \\
$^{74}$ Kavli~Institute~for~the~Physics~and~Mathematics~of~the~Universe~(KIPMU), Kashiwa, Japan \\
$^{75}$ Michigan~State~University, East~Lansing, USA \\
$^{76}$ Institute~for~Research~in~Fundamental~Sciences~(IPM), Tehran, Iran \\
$^{77}$ Southern~Methodist~University, Dallas, USA \\
$^{78}$ Weizmann~Institute~of~Science, Rehovot, Israel \\
$^{79}$ Department~of~Physics,~Mathabhanga~College, Cooch~Behar, West~Bengal,~India \\
$^{80}$ Universit\`a~degli~Studi~di~Padova, Padua, Italy \\
$^{81}$ Universit\'e~de~Strasbourg, Strasbourg, France \\
$^{82}$ Giresun~University, Giresun, Turkey \\
$^{83}$ Tokyo~Institute~of~Technology, Tokyo, Japan \\
$^{84}$ TOBB~University~of~Economic~and~Technology~(TOBB~ETU), Ankara, Turkey \\
$^{85}$ Vrije~University, Amsterdam, Netherlands \\
$^{86}$ University~of~Science~and~Technology~of~Mazandaran, Behshahr, Iran \\
$^{87}$ Uludag~University, Bursa, Turkey \\
$^{88}$ Sivas~Cumhuriyet~University, Sivas, Turkey \\
$^{89}$ Universidad~Tecnica~Federico~Santa~Maria, Valparaiso, Chile \\
$^{90}$ Istanbul~Aydin~University, Istanbul, Turkey \\
$^{91}$ Siberian~Branch~of~Russian~Academy~of~Science~-~Budker~Institute~of~Nuclear~Physics~(BINP), Novosibirsk, Russia \\
$^{92}$ University~of~the~Witwatersrand, Johannesburg, South~Africa \\
$^{93}$ Tsinghua~University, Beijing, China \\
$^{94}$ Tel-Aviv~University, Tel~Aviv, Israel \\
$^{95}$ Zhejiang~Institute~of~Modern~Physics~(ZIMP), Hangzhou, China \\
$^{96}$ Zhejiang~University~(ZJU), Hangzhou, China \\
$^{97}$ Xiamen~University~(XMU), Xiamen, China \\
$^{98}$ University~College~London, London, United~Kingdom \\
$^{99}$ Henan~Institute~of~Science~and~Technology~(HIST), Xinxiang, China \\
$^{100}$ University~of~Virginia, Charlottesville, USA \\
$^{101}$ Dalian~University~of~Technology~(DLUT), Dalian, China \\
$^{102}$ Universidade~Federal~do~Rio~Grande~do~Sul~(UFRGS), Porto~Alegre, Brazil \\
$^{103}$ Institut~de~F\'{i}sica~Corpuscular~--~CSIC/Universitat~de~Val\`{e}ncia, Paterna~(Valencia), Spain \\
$^{104}$ University~of~Helsinki, Helsinki, Finland \\
$^{105}$ CPHT,~CNRS,~Ecole~Polytechnique, I.~P.~Paris, France \\
$^{106}$ University~Genova, Genova, Italy \\
$^{107}$ Istituto~Nazionale~di~Fisica~Nucleare~(INFN)~-~Sezione~di~Genova, Genova, Italy \\
$^{108}$ Harish-Chandra~Research~Institute~(HRI), Allahabad, India \\
$^{109}$ University~of~Southampton, Southampton, United~Kingdom \\
$^{110}$ Universidade~de~Sao~Paulo~(USP), Sao, Paolo \\
$^{111}$ Nigde~Omer~Halisdemir~University, Nigde, Turkey \\
$^{112}$ Universidad~de~los~Andes, Carrera, Colombia \\
$^{113}$ The~University~of~Alabama, Tuscaloosa, USA \\
$^{114}$ Tata~Institute~of~Fundamental~Research~(TIFR), Mumbai, India \\
$^{115}$ Technische~Universit\"at~Darmstadt, Darmstadt, Germany \\
$^{116}$ Homeday~GmbH~Berlin, Berlin, Germany \\
$^{117}$ Charles~University, Praque, Czech~Republic \\
$^{118}$ Istituto~Nazionale~di~Fisica~Nucleare~(INFN)~-~Sezione~di~Bologna, Bologna, Italy \\
$^{119}$ Univ.~Federal~do~Rio~Grande~do~Norte, Natal, Brazil \\
$^{120}$ Karlsruher~Institut~f\"ur~Technologie~(KIT), Karlsruhe, Germany \\
$^{121}$ IBM~Deutschland~RnD,~GmbH, Urbar, Germany \\
$^{122}$ Indian~Institute~of~Science~Education~and~Research~(IISER), Kolkata, India \\
$^{123}$ Univ.~of~Montenegro, Podgorica, YUOGSLAVIA \\
$^{124}$ Shippensburg~University~of~Pennsylvania, Shippensburg,~Pennsylvania, USA \\
$^{125}$ University~of~Delaware, Newark, USA \\
$^{126}$ Argonne~National~Laboratory, Argonne, USA \\
$^{127}$ Oracle, San~Fransisco, USA \\
$^{128}$ Applied~AI~Center~of~Excellence, San~Francisco, USA \\
$^{129}$ Usak~University, Usak, Turkey \\
$^{130}$ National~University~of~Mar~del~Plata, Mar~del~Plata, Argentina \\
$^{131}$ Oklahoma~State~University~(OSU), Stillwater, USA \\
$^{132}$ Peking~University~(PKU), Beijing, China \\
$^{133}$ Institute~of~Mathematical~Sciences~(IMSc), Chennai, India \\
$^{134}$ Jagiellonian~University, Cracow, Poland \\
$^{135}$ Anhui~University~(AHU), Hefei, China \\
$^{136}$ Pennsylvania~State~University~(PSU), University~Park, USA \\
$^{137}$ University~of~Sussex, Sussex, United~Kingdom \\
$^{138}$ Narodowe~Centrum~Bada\'n~J\c{a}drowych~(NCBJ), Warsaw, Poland \\
$^{139}$ Kansas~State~University, Manhattan, USA \\
$^{140}$ Korea~Institute~for~Advanced~Study~(KIAS), Cheongryangri-dong, Korea \\
$^{141}$ Kastamonu~University, Kastamonu, Turkey \\
$^{142}$ K{\o}benhavns, Universitet~-~Niels~Bohr~Institutet~(NBI), Copenhagen \\
$^{143}$ University~of~Bergen, Bergen, Norway \\
$^{144}$ Wuhan~University~of~Technology, Wuhan, China \\
$^{145}$ Asia~Pacific~Center~for~Theoretical~Physics~(APCTP), Pohang, Korea \\
$^{146}$ University~of~California~(UC), Riverside, USA \\
$^{147}$ Hebrew~University~of~Jerusalem~-~Racah~Inst.~of~Physics, Jerusalem, Israel \\
$^{148}$ Universidad~Pedagogica~y~Tecnologica~de~Colombia, Tunja, Colombia \\
$^{149}$ Kobe~University, Kobe, Japan \\
$^{150}$ Lawrence~Berkeley~National~Laboratory~(LBNL), Berkeley, USA \\
$^{151}$ Fellow~Royal~Astronomical~Society~of~New~Zealand~(FRASNZ), Auckland, New~Zealand \\
$^{152}$ Universit\"at~Rostock, Rostock, Germany \\
$^{153}$ National~Center~for~Theoretical~Sciences~(NCTS), Hsinchu, Taiwan \\
$^{154}$ Nankai~University~(NKU), Tianjin, China \\
$^{155}$ Karlsruher~Institut~f\"ur~Technologie~(KIT)~-~Institut~f\"ur~Theoretische~Teilchenphysik~(TTP), Karlsruhe, Germany \\
$^{156}$ Karlsruher~Institut~f\"ur~Technologie~(KIT)~-~Institut~f\"ur~Kernphysik~(IKP), Karlsruhe, Germany \\

%% file: main/foreword.tex
\linenumbers

%
%
\chapter*{Preface}
\addcontentsline{toc}{chapter}{Preface}

%
%
\vspace{0.5cm}
This paper represents the updated design study of the Large Hadron-electron Collider, the LHeC,
a TeV energy scale electron-hadron ($eh$) collider which may come into operation during the
third decade of the lifetime of the Large Hadron Collider (LHC) at CERN.
It is an account, accompanied by numerous papers in the literature, for many
years of study and development, guided  by an International Advisory Committee (IAC)
which was charged by the CERN Directorate to advise  on the directions of energy frontier
electron-hadron physics at CERN. End of 2019 the IAC summarised its observations and
recommendations in a brief report to the Director General of CERN, which is here  reproduced as an Appendix.

\vspace{0.6cm}

The paper outlines a unique, far reaching physics programme on 
deep inelastic scattering (DIS), a design concept for a new
generation collider detector, together with a novel configuration of the intense, high energy electron beam.
This study builds on the previous, detailed LHeC Conceptual Design Report (CDR), which was published
eight years ago~\cite{AbelleiraFernandez:2012cc}. It surpasses
the initial study in essential characteristics: i)
the depth of the physics programme, owing to the insight obtained mainly with the LHC, 
and ii) the
luminosity prospect, for enabling a novel Higgs facility to be built and the  prospects to search
for and discover new physics 
to be strengthened. It builds on recent and forthcoming progress of modern technology, 
due to major advances especially of the superconducting RF technology and 
as well new detector techniques.

Unlike in 2012, there has now a decision been taken to configure the LHeC
as an electron linac-proton
 or nucleus
ring configuration, which leaves the ring-ring
option~\cite{Burkhardt:2012ta,AbelleiraFernandez:2012cc}
as a backup.
In $ep$,
the high instantaneous luminosity of about
$10^{34}$\,cm$^{-2}$s$^{-1}$ may be achieved with the electron accelerator
built as an energy recovery linac (ERL) and because the brightness of the
LHC exceeds early expectations by far, not least through the upgrade of the LHC to its
high luminosity version, the HL-LHC~\cite{ApollinariG.:2017ojx,Rossi:2019swj}.
For $e$Pb collisions, the corresponding per nucleon instantaneous luminosity would be about
$10^{33}$\,cm$^{-2}$s$^{-1}$.
The LHeC is designed to be compatible with concurrent operation with the LHC. It thus
 represents a unique opportunity to advance particle physics
by building on the singular investments which CERN and its global partners
have made into the LHC facility.

Since the 2012 document, significant experience with multi-turn ERL design, construction, and operation has been gained with the Cornell-BNL ERL Test Accelerator (CBETA), which has accelerated and energy recovered beam in all of its 4 turns~\cite{Hoffstaetter:2017jei,Bartnik:2020pos}.
Extending much beyond the CDR, a configuration has newly been designed for a low energy
ERL facility, termed PERLE~\cite{Angal-Kalinin:2017iup}, which is moving ahead to be built at Orsay by an international
collaboration.
%
The major parameters of PERLE have been taken
from the LHeC, such as the 3-turn configuration, source, the $802$\,MHz frequency and
cavity-cryomodule technology, in order to make PERLE a suitable 
facility for the development of LHeC ERL technology
and the accumulation of operating experience prior to and later in parallel with the LHeC.
In addition, the PERLE facility has a striking low energy physics programme,
industrial applications and will be an enabler for ERL technology as the
first facility to operate in the $10$\,MW power regime.

While the 2012 CDR  focussed the physics discussion on the genuine physics of
deep inelastic scattering leading much beyond HERA, 
a new focus arose through the challenges and opportunities posed
by the HL-LHC. It is demonstrated that DIS at the LHeC can play a crucial role 
in sustaining and enriching the LHC programme, a
consequence of the results obtained at the LHC, i.e.\ the discovery of the
Higgs boson, the non-observation of supersymmetry (SUSY) or
other non Standard Model (SM) exotic particles and, not least, the unexpected realisation of
the huge potential of the LHC for discovery through precision measurements in the
strong and electroweak sectors. Thus, it was felt time to summarise the recent seven years
of LHeC development, also in support of the current discussions on the future of
particle physics, especially at the energy frontier. Both for the
LHeC~\cite{Bruning:2652313,Bruning:2019scy,Bruning:2652335}
and PERLE~\cite{Klein:2652336}, documents were submitted for consideration to the European Strategy for Particle Physics Update.

The LHeC is a once in our lifetime opportunity for substantial progress in particle physics.
It comprises, with a linac shorter than the pioneering two-mile linac at SLAC, a most ambitious and exciting physics
programme, the introduction of novel accelerator technology and the complete exploitation of the unique values of
 and spendings into the LHC.
 %
 It requires probably less courage than that of Pief Panofsky 
and
%
%
colleagues half a century ago.  Finally, not least, one may realise
that the power LHeC needed without the energy recovery technique is beyond 1\,GW
while the electron beam is dumped at injection energy. It therefore is
a significant step towards green accelerator technology, a major general desire and requirement of our
 times. This paper aims at substantiating these statements in the various chapters following.
\\

\vspace{0.1cm}
\noindent
Oliver Br\"{u}ning (CERN) and Max Klein (University of Liverpool)

%

%% file: introduction/introduction.tex
\linenumbers
\lhectitlepage
\lhecinstructions
\subfilestableofcontents

%
%

%
\chapter{Introduction}\label{sec:intro}
\section{The Context}
\subsection{Particle Physics - at the Frontier of Fundamental Science}
Despite its striking success, the Standard Model (SM) has been recognised 
to have major deficiencies. These may be summarised in various ways. 
Some major questions can be condensed as follows:
\begin{itemize}
\item $\textbf{Higgs~boson}$~~Is the electroweak scale stabilised by new particles, interactions, symmetries? Is the Higgs boson discovered in 2012 the SM Higgs boson, what is its potential? Do more Higgs bosons exist as predicted, for example, in super-symmetric theories?
\item $\textbf{Elementary~Particles}$~~The SM has 61 identified particles: 12 leptons, 36 quarks and anti-quarks, 12 mediators, 1 Higgs boson. Are these too many or too few? Do right-handed neutrinos exist? Why are there three families? What makes leptons and quarks different? Do leptoquarks exist, is there a deeper substructure?
\item $\textbf{Strong~Interactions}$~~What is the true parton dynamics and structure inside the proton, inside other hadrons
 and inside nuclei -- at different levels of resolution? How is confinement explained and how do partons hadronise? How can the many body dynamics of the Quark Gluon Plasma (QGP) state be described in terms of the elementary fields of Quantum Chromodynamics? What is the meaning of the AdS/CFT relation and of supersymmetry in strong interactions? Do axions, odderons, instantons exist?
\item $\textbf{GUT}$~~Is there a genuine, grand unification of the interactions at high scales, would this include gravitation? What is the correct value of the strong coupling constant, is lattice theory correct in this respect? Is the proton stable?
\item $\textbf{Neutrinos}$~~Do Majorana or/and sterile neutrinos exist, is there CP violation in the neutrino sector?
\item $\textbf{Dark Matter}$ Is dark matter constituted of elementary particles or has it another origin? Do 
hidden or dark sectors of nature exist and would they be accessible to accelerator experiments?
\end{itemize}
These and other open problems are known, and they have been persistent questions to Particle Physics. They are intimately related and any future strategic programme should not be confined to only one or a few of these. The field of particle physics is far from being understood, despite the phenomenological success of the
SU$_L$(2)\,$\times$\,U(1)\,$\times$\,SU$_c$(3) gauge field theory termed the Standard Model.  Certain attempts to declare its end 
are in contradiction not only to the
 experience from a series of past revolutions in science but indeed contrary to the
 incomplete status of particle physics as sketched above.  
The question is not why to end particle physics but how to proceed. 
The answer is not hidden in philosophy but requires new, better, affordable experiments.
Indeed the situation is special as expressed by Guido Altarelli a few years ago: {\emph{It is now less 
unconceivable that no new physics will show up at the 
LHC$\dots$ We expected complexity and instead we have found a maximum of simplicity.
The possibility that the Standard model holds well beyond the electroweak 
scale must now be seriously considered~\cite{Altarelli:2014xxa}}}. This is reminiscent
of the time before 1969, prior to anything like a Standard Model, when gauge theory
was just for theorists, while a series of new accelerators, such as the 2\,mile electron linac
at Stanford or the SPS at CERN, were planned which resulted in a complete change of the 
paradigm of particle physics. 

Ingenious  theoretical hypotheses, such as on the existence of extra dimensions, 
on SUSY, of un-particles or the embedding in higher gauge groups, 
like E8, are a strong motivation to develop high energy physics 
rigorously further. In this endeavour, a substantial increase of 
precision, the conservation of diversity of projects and the extension of kinematic 
coverage are  a necessity, likely turning out to be of fundamental importance. 
The strategic question in this context, therefore, is not just which new collider
should be built next, as one often hears, but how we may challenge the current and incomplete
knowledge best. A realistic step to progress comprises a new $e^+e^-$ collider, built
perhaps in Asia, and  complementing the LHC  with an electron energy recovery linac
to synchronously operate $ep$ with $pp$ at the LHC, the topic of this paper.  

One may call these machines first technology generation colliders as their technology 
has been proven to principally work~\cite{jdhecfa}.  Beyond these times,
there is a long-term future reaching to the year 2050 and much beyond, 
of a second, further generation of hadron, lepton and electron-hadron colliders.
CERN has recently published a design study 
 of a future circular $hh,~eh$ and $e^+e^-$ collider (FCC)  
 complex~\cite{Abada:2019lih, Abada:2019zxq, Benedikt:2018csr},
  which would provide a corresponding base. For electron-hadron scattering this opens a new  horizon with the FCC-eh, an about $3$\,TeV centre-of-mass system (cms) energy collider which in this
  paper is also considered, mostly for comparison with the LHeC.
  A prospect similar to FCC is also being developed 
  in China~\cite{CEPC-SPPCStudyGroup:2015esa,CEPCStudyGroup:2018ghi}.

A new collider for CERN at the level of $\mathcal{O}(10^{10})$ CHF cost should have
the potential to change the paradigm of particle physics with 
direct, high energy discoveries in the $10$\,TeV mass range. This may
only be achieved with the FCC-hh including an $eh$ experiment. 
The FCC-hh/eh complex does access physics to several 
hundred TeV,  assisted by a qualitatively new level of QCD/DIS. 
A prime, very fundamental goal of the 
FCC-pp is the clarification of the Higgs vacuum potential which can not be
achieved in $e^+e^-$. This collider therefore has an overriding justification
beyond the unknown prospects of finding new physics nowadays termed ``exotics". 
It accesses rare Higgs boson decays, high scales and, when combined 
with $ep$, it measures the SM Higgs couplings to below percent precision. 
There is a huge, fundamental program on electroweak and strong 
interactions, flavour and heavy ions for FCC-hh to be explored. 
This represents CERN’s unique opportunity 
to build on the ongoing LHC program, for many decades ahead. 
The size of the FCC-hh requires
this to be established as a global enterprise. The HL-LHC and the LHeC can be understood as 
very important steps towards this major new facility, both in terms of physics and technology. 
The present report outlines a road towards realising a next generation, energy frontier
electron-hadron collider as part of this program, which would maximally exploit and support
the LHC.

\subsection{Deep Inelastic Scattering and HERA}
The field of deep inelastic lepton-hadron
 scattering (DIS)~\cite{Feynman:1973xc}  was born
with the discovery~\cite{Bloom:1969kc,Breidenbach:1969kd} 
of partons~\cite{Feynman:1969ej,Bjorken:1969ja} about $50$ years ago.
It readily contributed fundamental insights, for example on the development of QCD with the confirmation of fractional quark charges and of asymptotic freedom or 
with the spectacular  finding that the  weak isospin
charge of the right-handed electron was zero~\cite{Prescott:1978tm} which established the
Glashow-Weinberg-Salam ``Model of Leptons"~\cite{Weinberg:1967tq}  as the 
base of the united electroweak theory.  The quest to reach higher energies in accelerator based particle physics
led to generations of colliders, with HERA~\cite{Wiik:1985sb} as the so far only electron-proton one.

HERA collided electrons (and positrons) of $E_e=27.6$\,GeV energy off protons
of $E_p=920$\,GeV energy achieving a centre-of-mass energy, $\sqrt{s}=2 \sqrt{E_e E_p}$,
of about $0.3$\,TeV. It therefore
extended the kinematic range covered by fixed target experiments by two 
orders of magnitude in  Bjorken $x$ and in four-momentum transfer squared, 
$Q^2$, with its limit $Q^2_{max}=s$.
 HERA was built in less than a decade, and it operated for 16 years. 
Together with the Tevatron and LEP, HERA was pivotal to the development of the Standard Model.

HERA had a unique collider physics programme and success~\cite{Klein:2008di}.
It established QCD  as the correct description of proton
substructure and parton dynamics down to $10^{-19}$\,m. It demonstrated electroweak
theory to hold in the newly accessed range, especially with the measurement of neutral
and charged current $ep$ scattering cross sections beyond  $Q^2 \sim M_{W,Z}^2$ and
with the proof of electroweak interference at high scales through the measurement of 
the interference structure functions $F_2^{\gamma Z}$ and  $xF_3^{\gamma Z}$. The HERA
collider has provided the core base of the physics of parton distributions, not only in determining the
gluon, valence, light and heavy sea quark momentum 
distributions in a much extended range,
 but as well in  supporting the foundation of the theory of unintegrated, 
diffractive, photon, neutron PDFs through a series of corresponding measurements.
It discovered the rise of the parton distributions towards small momentum fractions, $x$,
supporting early QCD expectations on the asymptotic behaviour of the structure
functions~\cite{DeRujula:1974mnv}.   
Like the TeVatron and LEP/SLC colliders which  explored the Fermi scale of a few hundred GeV energy,
determined by the vacuum expectation value of the Higgs field, $v = 1/ \sqrt{\sqrt{2} G_F} =
2 M_W/g \simeq 246$\,GeV, HERA showed too that there was no supersymmetric
or other exotic particle with reasonable couplings existing at the Fermi energy scale.

HERA established electron-proton scattering as an integral part of modern high energy
particle physics. It demonstrated the 
richness of DIS physics, and the feasibility of constructing and operating energy frontier $ep$ colliders.
 What did we learn to take into
a next, higher energy $ep$ collider design? Perhaps there  arose three lessons about: 
\begin{itemize}
\item \emph{the~need~for~higher~energy}, for three reasons: i) 
to make  charged currents a real, precision part of $ep$ physics, for instance for the complete unfolding
of the flavour composition of the sea and valence quarks, ii) to produce heavier mass particles (Higgs, top,
exotics) with favourable cross sections, and iii) to discover or disproof the existence of 
gluon saturation for which one needs to measure at lower $x \propto Q^2/s$, i.e. higher
$s$ than HERA had available;
\item
\emph{the~need~for~much~higher~luminosity}: the first almost ten years of HERA provided 
just a hundred pb$^{-1}$. As a consequence, HERA could not accurately
 access the high $x$ region, and it was inefficient and short of statistics in resolving 
 puzzling event fluctuations; 
 \item  \emph{the~complexity~of~the~interaction~region} when
 a bent electron beam caused synchrotron radiation while the opposite proton beam
 generated quite some halo background through beam-gas and beam-wall proton-ion interactions.
 \end{itemize}
Based on these and further lessons a first LHeC paper was published in 
2006~\cite{Dainton:2006wd}. The LHeC design was then intensely worked on,
and a comprehensive CDR appeared in 2012~\cite{AbelleiraFernandez:2012cc}. 
This has now been pursued much further still recognising that 
the LHC is the only existing base to realise a TeV energy scale electron-hadron collider in the
accessible future. It offers highly energetic, intense hadron beams, a long time perspective
and a unique infrastructure and expertise, i.e. everything required for an energy frontier
DIS physics and innovative accelerator programme.
\section{The Paper}
\subsection{The LHeC Physics Programme}
\begin{figure}[!ht]
\centering
\includegraphics[width=0.7\textwidth,trim={0 70 0 70},clip]{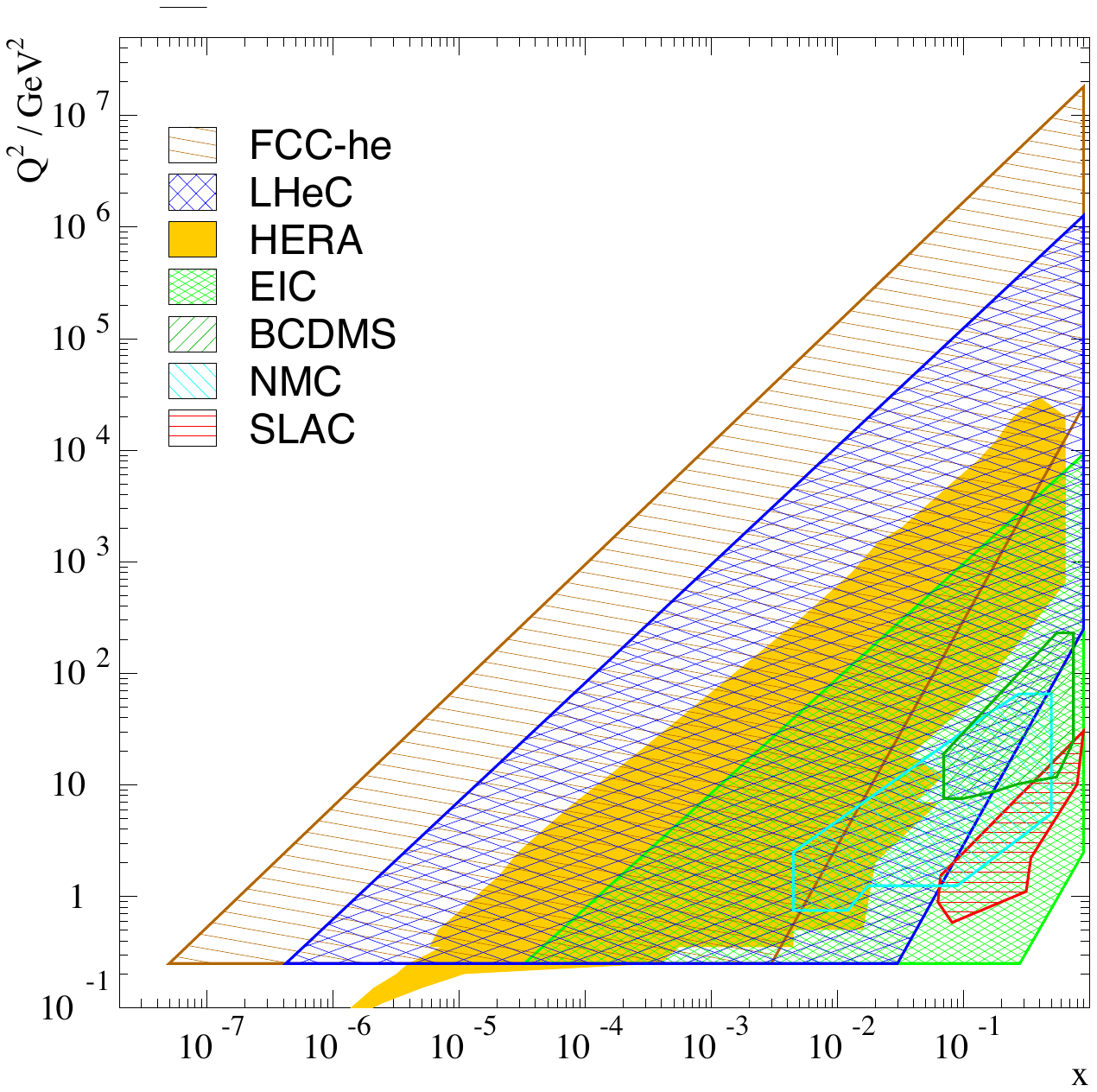}
\caption{
Coverage of the kinematic plane in deep inelastic lepton-proton scattering 
by some initial fixed target experiments, with electrons (SLAC) and
muons (NMS, BCDMS), and by the $ep$ colliders:  the EIC (green), HERA (yellow), the LHeC (blue) and 
the FCC-eh (brown). The low $Q^2$ region for the colliders is here limited to about $0.2$\,GeV$^2$, 
which is covered by the central detectors, roughly and perhaps using low electron
beam data. Electron taggers may extend this to even lower $Q^2$.
The high $Q^2$ limit at fixed $x$ is given by the line of inelasticity $y=1$.  Approximate limitations
of acceptance at medium $x$, low $Q^2$ are illustrated using polar angle limits of 
$\eta = - \ln \tan \theta /2$ of $4,~5,~6$ for the EIC, LHeC, and FCC-eh, respectively.
These lines are given by $x= \exp {\eta} \cdot \sqrt{Q^2}/(2 E_p)$, and can be moved to larger $x$
when $E_p$ is lowered below the nominal values.}.
\label{fig:kinplane}
\end{figure}

This paper presents a design concept of the LHeC, using a $50$\,GeV energy
electron beam to be scattered off the LHC hadron beams (proton and ion) in concurrent
operation\footnote{The CDR in 2012 used a 60\,GeV beam energy. Recent considerations of
cost, effort and synchrotron radiation effects led to preference of a small reduction of the
energy. Various physics studies presented here still use $60$\,GeV. While for BSM, top and Higgs physics the high energy is indeed important, the basic conclusions remain valid if eventually the energy was indeed
chosen somewhat smaller than previously considered. This is further discussed below.
A decision on the energy would come with the approval obviously.}.
Its main characteristics are presented in \textbf{Chapter\,2}.
The instantaneous luminosity is designed to be $10^{34}$\,cm$^{-2}$s$^{-1}$ 
exceeding that of HERA, which
achieved a few times $10^{31}$\,cm$^{-2}$s$^{-1}$, by a factor of several hundreds. The kinematic
range nominally is extended by a factor of about $15$, but in fact by a larger amount because 
of the hugely increased luminosity which is available for exploring the maximum $Q^2$ and 
large $x  \leq 1$ regions, which were major deficiencies at HERA. 
The coverage of the $Q^2,~x$ plane
by previous and future DIS experiments is illustrated in Fig.\,\ref{fig:kinplane}.

The LHeC would provide a major extension of the DIS kinematic range as is
required for the physics programme at the energy frontier. 
For the LHC, the $ep$/A detector would be a new major experiment.
  A number of major themes would be 
 explored with significant discovery potential. These are presented in quite some detail in 
 seven chapters of this paper dedicated to physics:    
\begin{itemize}
\item
Based on the unique hadron beams of the LHC and employing a point-like
probe, the LHeC would represent the world's cleanest, high resolution
microscope for exploring the substructure of and dynamics inside matter,
which  may be termed the Hubble telescope for the smallest dimensions. 
The first chapter on physics, \textbf{Chapter\,3}, is devoted to the measurement 
of parton distributions with the LHeC, and it also presents the potential to resolve
proton structure in 3D. 

\item 
\textbf{Chapter\,4} is devoted to the deep exploration of QCD. A key deliverable of the
LHeC is the clarification of the parton interaction dynamics at small Bjorken $x$,
in the new regime of very high parton densities but small coupling which HERA 
discovered but was unable to clarify for its energy was limited. It is first shown
that the LHeC can measure  $\alpha_s$ to per mille accuracy followed by 
various studies to illustrate the unique potential of the LHeC to pin down the 
dynamics at small $x$. The chapter also covers the seminal potential for diffractive DIS 
to be developed. It concludes with brief presentations on theoretical developments
on pQCD and of novel physics on the light cone.

\item
The maximum $Q^2$ exceeds the $Z,~W$ boson mass values (squared) by two orders of magnitude. The LHeC, supported by
variations of beam parameters and high luminosity, thus offers a unique potential to test the electroweak SM in the
spacelike region with unprecedented precision.
The high $ep$ cms energy leads to the copious production of
top quarks, of about $2 \cdot 10^6$ single top and $  5 \cdot 10^4$ $t \bar{t}$ events. Top
production could not be observed at HERA but will thus become a
central theme of precision and discovery physics with the LHeC.
In particular, the top momentum fraction, 
top couplings to the photon, the $W$ boson and possible flavour changing neutral currents
(FCNC) interactions can be studied in a uniquely clean environment (\textbf{Chapter\,5}).

\item
The LHeC 
 extends the kinematic range in lepton-nucleus scattering 
by nearly four orders of magnitude. It thus will transform 
nuclear particle physics completely, by resolving the hitherto hidden parton dynamics and substructure in nuclei and clarifying the QCD base for the collective dynamics 
observed in QGP phenomena (\textbf{Chapter\,6}).
\item
The clean DIS final state in neutral and charged current scattering  and the 
high integrated luminosity  enable a
high precision Higgs physics programme with the LHeC. The Higgs production
cross section is comparable to the one of Higgs-strahlung at $e^+e^-$. This opens
unexpected extra potential to independently test the Higgs sector of the SM,
with high precision 
insight especially into the $H - WW/ZZ$ and $H - bb/cc$ couplings (\textbf{Chapter\,7}).
\item
As  a new, unique, luminous TeV scale
collider, the LHeC has an outstanding opportunity to discover new
physics, such as in the exotic Higgs, dark matter, heavy neutrino and QCD
areas (\textbf{Chapter\,8}).
\item
With concurrent $ep$ and $pp$ operation, the LHeC would transform the
LHC into a 3-beam, twin collider of greatly improved potential 
which is sketched in \textbf{Chapter\,9}. 
Through ultra-precise strong and electroweak measurements,
the $ep$ experiment would make the HL-LHC complex  a much more powerful
search and measurement laboratory than current expectations, based on $pp$
only, do entail.
 The joint $pp/ep$ LHC facility together with a 
novel $e^+e^-$ collider will make a major step in the study of the 
SM Higgs Boson, leading far beyond the HL-LHC. Putting $pp$ and $ep$
results together, as is illustrated for PDFs, will lead to new insight, especially
when compared with its single $pp$ and $ep$ components.

\end{itemize}
The development of particle physics, 
the future of CERN, the exploitation of the singular LHC investments,
 the culture of accelerator art, 
 all make the  LHeC a unique project of great interest. It is challenging in terms of technology,
 affordable given  budget constraints and it may still be realised in the two decades of currently
 projected LHC lifetime.

\subsection{The Accelerator}
The LHeC  provides an intense, high energy electron beam 
to collide with the LHC.  It represents the highest energy application of
energy recovery linac (ERL) technology which is increasingly recognised as one of the
major pilot technologies for the development of particle physics because
 it utilises and stimulates superconducting RF technology progress, and it
 increases intensity while keeping the power consumption low.
 
The LHeC instantaneous luminosity is determined through the integrated luminosity 
goal of $\mathcal{O}(1)$\,ab$^{-1}$ caused by various physics reasons.
The electron beam energy is
chosen to achieve  TeV cms collision energy  and enable competitive searches 
and precision Higgs boson measurements. A cost-physics-energy evaluation
is presented here which points to choosing $E_e \simeq 50$\,GeV as a new 
default value, which was $60$\,GeV before~\cite{AbelleiraFernandez:2012cc}.
The wall-plug power has been constrained to $100$\,MW. 
Two super-conducting linacs of about $900$\,m length, which are placed opposite to each other,
accelerate the passing electrons by $8.3$\,GeV each. This
leads to a final electron beam energy of about $50$\,GeV in 
a 3-turn racetrack energy recovery linac  configuration.
 
 For measuring at very low $Q^2$ 
and for determining the longitudinal 
structure function $F_L$, see below, the electron beam energy may be reduced to 
a minimum of  about $10$\,GeV. For maximising the acceptance at large Bjorken
$x$, the proton beam energy, $E_p$, may be reduced to $1$\,TeV. 
This determines a minimum cms energy of $200$\,GeV, below HERA's $319$\,GeV.
 If the ERL may be combined in the further future with the double energy
 HE-LHC~\cite{Abada:2019ono}, the proton beam energy $E_p$
could reach $14$\,TeV and $\sqrt{s}$ be increased to $1.7$\,TeV.
This is extended to $3.5$\,TeV for the FCC-eh with a $50$\,TeV proton energy beam.
We thus have the unique, exciting prospect for  future DIS $ep$ scattering at CERN with an
energy range from below HERA to the few TeV region, at hugely increased luminosity
and based on much more sophisticated experimental techniques than had been available
at HERA times.

A spectacular extension of the
kinematic range will be expected for deep inelastic lepton-nucleus scattering which 
was not pursued at DESY. Currently,
highest energy $lN$ data are due to fixed target muon-nucleus experiments, such as 
NMC and COMPASS, with a maximum $\sqrt{s}$ of about $20$\,GeV which permits a maximum $Q^2$
of $400$\,GeV$^2$. This will be extended with the EIC at Brookhaven
to about $10^4$\,GeV$^2$.
 The corresponding numbers for $e$Pb scattering at LHeC (FCC-eh) are
$\sqrt{s} \simeq 0.74~(2.2)$\,TeV and $Q^2_{max} = 0.54~(4.6)~10^6$\,GeV$^2$. 
The kinematic range in $e$A scattering  will thus be extended through the LHeC (FCC-eh)  
by three (four) orders of magnitude as compared to the current status. This
will thoroughly alter the understanding of parton and collective dynamics inside nuclei.

The ERL beam configuration
is located inside the LHC ring but
outside its tunnel, which minimises any interference
with the main hadron beam infrastructure. The electron accelerator may thus
be built independently, to a considerable extent, of the status of operation
of the proton machine.  
The length of the  ERL has configuration to be a fraction
$1/n$ of the LHC circumference as is required 
for the $e$ and $p$ matching of bunch patterns. Here the
return arcs count as two single half rings. 
The chosen electron beam energy  of $50$\,GeV
leads, for $n=5$, to a circumference $U$ of $5.4$\,km for the electron
racetrack~\footnote{The circumference
may eventually be chosen to be $6.8$\,km, the length of the SPS,
which would relax certain parameters and ease an energy upgrade.}. 
 A 3-pass ERL  configuration had been adopted also for the FCC-eh
albeit maintaining the original $60$\,GeV as default which had a $9$\,km circumference. 

For the LHC, the ERL would be tangential to IP2. According to current plans, 
IP2 is given to the ALICE detector with a program extending to LS4,
 the first long shutdown following
the three year pause of the LHC operation for upgrading the luminosity performance
and detectors. There are plans for a new heavy ion detector to move into IP2. 
 The LS4 shutdown is currently scheduled to begin in 2031
with certain likelihood of being postponed to 2032 or later as recent events seem
to move LS3 forward and extend its duration to three years.

For FCC-eh
the preferred position is interaction point L, for geological reasons mainly, and the time of operation fully depending on the progress with FCC-hh, beginning at the earliest in the late 40ies if
CERN went for the hadron collider directly after the LHC.

The LHeC operation is transparent to the LHC collider experiments owing to
the low lepton bunch charge and resulting small beam-beam tune shift
experienced by the protons. The LHeC is thus designed to run simultaneously with
$pp$ (or $p$A or AA) collisions with a dedicated final operation of a few years.

The paper presents in considerable detail the design of the LHeC (\textbf{Chapter\,10)}, i.e.\ the
optics and lattice, components, magnets, as well as
designs of the linac and interaction region besides special topics such 
as the prospects for electron-ion scattering, positron-proton operation and a novel
study of beam-beam interaction effects. With the more ambitious luminosity goal,
with a new lattice adapted to $50$\,GeV, with progress on the IR design,
a novel analysis of the civil engineering works and,
especially, the production and successful test~\cite{Marhauser:2018jxn} of the first  SC cavity 
at the newly chosen default frequency of $801.58$\,MHz, this report 
considerably extends beyond the initial CDR.  This holds especially since several
LHeC institutes have recently embarked on the development of the 
ERL technology with a low energy facility, PERLE, to be built at IJC Laboratory at Orsay.

\subsection{PERLE}

Large progress has been made in the development
of superconducting, high gradient cavities with quality factors, $Q_0$, beyond $10^{10}$.
This will enable the exploitation of  ERLs in high-energy physics colliders, with the 
LHeC as a prime example, while considerations are also brought forward for future
$e^+e^-$ colliders~\cite{Litvinenko:2019txu} and for proton beam cooling with an ERL
tangential to eRHIC. 
The status and challenges of energy recovery linacs
are summarised in \textbf{Chapter~\,11}. This chapter
 also presents the design, status and prospects for
the ERL development facility PERLE. The major parameters of PERLE have been taken
from the LHeC, such as the 3-turn configuration, source,  frequency and 
cavity-cryomodule technology, in order to make 
PERLE a suitable facility for the development of LHeC ERL technology
and the accumulation of operating experience prior to and later in parallel with the LHeC.

An international collaboration has been established to build PERLE at Orsay. With the design goals
of $500$\,MeV electron energy, obtained in three passes through two cryo-modules and
of $20$\,mA, corresponding to $500$\,nC charge at $40$\,MHz bunch frequency, PERLE
is set to become the first ERL facility to operate at $10$\,MW power. Following its
CDR~\cite{Angal-Kalinin:2017iup} and a paper submitted to 
the European strategy~\cite{Klein:2652336}, work is directed 
to build a first dressed cavity and to release a TDR by 2021/22. Besides its value
for accelerator and ERL technology, PERLE is also of importance for pursuing
a low energy physics programme, see~\cite{Angal-Kalinin:2017iup}, and for
several possible industrial applications. It also serves as a local hub for the education
of accelerator physicists at a place, previously called Linear Accelerator Laboratory (LAL),
which has long been at the forefront of accelerator design and operation. 

There are a number of related ERL projects as are characterised in Chapter\,11.
The realisation of the ERL for the LHeC at CERN
represents a unique opportunity not only for physics and technology but as well
for a next and the current generation of accelerator physicists, engineers and technicians
to realise an ambitious collider project while the plans for very expensive next machines
may take shape. Similarly, this holds for a new generation of detector experts, as the
design of the upgrade of the general purpose detectors (GPDs) at the LHC is reaching completion, with the question
increasingly posed about opportunities for new collider detector construction to not loose
the expertise nor the infrastructure for building trackers, calorimeters and alike. The
LHeC offers the opportunity for a novel $4\pi$ particle physics detector design, construction
and operation. As a linac-ring collider, it may serve one detector of a size smaller than CMS and larger than H1 or ZEUS.

\subsection{The Detector}
\textbf{Chapter\,12} on the detector relies to a large extent on the very detailed write-up
on the kinematics, design considerations, and realisation of a detector for the LHeC
presented in the CDR~\cite{AbelleiraFernandez:2012cc}. In the previous report
one finds detailed studies not only on the central detector and its magnets, a central solenoid for
momentum measurements and an extended dipole for ensuring head-on $ep$ collisions,
 but as well on the forward ($p$ and $n$) and backward ($e$ and $\gamma$) tagging devices.
 The work on the detector as presented here was focussed on an optimisation of the
 performance and on the scaling of the design towards higher proton beam energies. It presents
 a new, consistent design and summaries of the essential characteristics in support of
 many physics analyses that this paper entails.
 
 The most demanding performance requirements arise from the $ep$ Higgs measurement programme,
 especially the large acceptance and high precision desirable for heavy flavour tagging and
 the requirement to resolve the hadronic final state. This has been influenced by both the rapidity
 acceptance extensions and the technology progress of the HL-LHC detector upgrades. A key example, also discussed, is the HV-CMOS Silicon technology, for which the LHeC 
 is an ideal application due to the much limited radiation level as compared to $pp$.
 
  Therefore we have now completed two studies of design: previously,
  of a rather conventional detector with limited cost and, here, of a more ambitious device. Both
  of these designs appear feasible. This regards also the installation. The paper
  presents a brief description of the installation of the LHeC detector at IP2 with the result
  that it may proceed within two years, including the dismantling of the 
  there residing detector. This
  calls for modularity and pre-mounting of detector elements on the surface, as was done
  for CMS too. It will be for the LHeC detector Collaboration, to be established with and for the
  approval of the project, to eventually design the detector according to its understanding and 
technical capabilities. 

\section{Outline}

The paper is organised as follows. For a brief overview, Chapter\,2 summarises the LHeC characteristics. 
Chapter\,3 presents the physics of the LHeC seen as a microscope for
measuring PDFs and exploring the 3D structure of the proton.
Chapter\,4 contains further means to explore QCD, especially low $x$ dynamics, together with two sections on QCD theory developments.
Chapter\,5 describes the electroweak and top physics potential of the LHeC.
Chapter\,6 presents the seminal nuclear particle physics
potential of the LHeC, through luminous electron-ion scattering exploring
an unexplored kinematic territory.
Chapter\,7 presents a detailed analysis of the opportunity for precision SM Higgs boson 
physics with charged and neutral current
  $ep$ scattering. Chapter\,8 is a description of the salient
opportunities to discover physics beyond the Standard Model with the LHeC, including non-SM
Higgs physics, right-handed neutrinos, physics of the dark sector, heavy resonances
and exotic substructure phenomena.
Chapter\,9 describes the interplay of $ep$ and $pp$ physics, i.e. the necessity to have
the LHeC for fully exploiting the potential of the LHC facility, e.g. through
the large increase of electroweak precision measurements, the considerable extension of
 search ranges and the joint $ep$ and $pp$ Higgs physics potential.
Chapter\,10 presents the update of the design on the electron accelerator with many novel results
such as on the lattice and interaction region, updated parameters for $ep$ and $eA$ scattering,
new specifications of components, updates on the electron source,$\dots$ The chapter
also presents the encouraging  results of the first LHeC $802$\,MHz cavity. Chapter\,11 is 
devoted, first, to the status and challenges of energy recovery based accelerators and, second,
to the description of the PERLE facility between its CDR and a forthcoming TDR. Chapter\,12 
describes the update of the detector studies towards an optimum configuration in terms
of acceptance and performance. Chapter\,13 presents a summary of the paper including a time line
for realising the LHeC to operate with the LHC. An Appendix 
presents the statement of the International Advisory Committee
on its evaluation of the project together with recommendations about how to proceed. It also contains an account for the membership
in the LHeC organisation, i.e. the Coordination Group and finally the
list of Physics Working Group convenors.


\biblio

%% file: characteristics/characteristics.tex
\linenumbers
\lhectitlepage
\lhecinstructions
\subfilestableofcontents

%
%

%
%
%
\chapter{LHeC Configuration and Parameters}
\section{Introduction}
The Conceptual Design Report (CDR) of the LHeC was published in 2012~\cite{AbelleiraFernandez:2012cc}. 
The CDR default configuration uses a $60$\,GeV energy electron 
beam derived from
a racetrack, three-turn, intense energy recovery linac (ERL)  achieving 
a cms energy of $\sqrt{s}=1.3$\,TeV, where $s=4E_pE_e$ is determined 
by the electron and proton beam energies, $E_e$ and $E_p$. In 2012, 
the Higgs boson, $H$, was discovered which has become a central topic 
of current and future high energy physics. The Higgs production cross 
section in charged current (CC) deep inelastic scattering (DIS) at the 
LHeC is roughly $100$\,fb. The Large Hadron Collider has so far not led 
to the discovery of any exotic phenomenon. This forces searches to be 
pursued, in $pp$ but as well in $ep$, with the highest achievable precision
 in order to access a maximum range of phase space and possibly rare 
channels. The DIS cross section at large $x$ roughly behaves like
 $(1-x)^3/Q^4$, demanding very high luminosities for exploiting the 
unknown regions of Bjorken $x$ near  $1$ and very high $Q^2$, the 
negative four-momentum transfer squared between the electron and the 
proton. For the current update of the design of the LHeC this has set
 a luminosity goal about an order of magnitude higher
than the $10^{33}$\,cm$^{-2}$s$^{-1}$ which
had been adopted for the CDR. There arises the potential, as 
described subsequently in this paper, to transform the LHC into
a high precision electroweak, Higgs and top quark physics facility. 

The $ep$ Higgs production cross section rises approximately with $E_e$. 
New physics may be related to the heaviest known elementary particle, 
the top quark, the $ep$ production cross section of which rises more strongly 
than linearly with $E_e$ in the LHeC kinematic range as that is not very far 
from the $t \bar{t}$ threshold. Searches for heavy neutrinos, SUSY particles, 
etc.\ are the more promising the higher the energy is. 
The region of deep inelastic scattering and pQCD
requires that $Q^2$ be larger than $ M_p^2 \simeq 1$\,GeV$^2$.
Access with DIS to very low Bjorken $x$ requires high energies because of
$x=Q^2/s$, for inelasticity $y=1$. In DIS, one needs $Q^2 > M_p^2 \simeq 1$\,GeV$^2$.
 Physics therefore requires a maximally large energy. However, 
cost and effort set realistic limits such that twice the HERA 
electron beam energy, of about $27$\,GeV, 
appeared as a reasonable and affordable target value. 
 
In the CDR~\cite{AbelleiraFernandez:2012cc}
 the default electron energy was chosen to be $60$\,GeV. 
This can be achieved with an ERL circumference of $1/3$ of that of 
the LHC. Recently, the cost was estimated in quite some
 detail~\cite{ocost}, comparing also with other accelerator projects. Aiming 
at a cost optimisation and providing an option for a staged 
installation, the cost estimate lead to defining a new default configuration 
of $E_e=50$\,GeV with the option of starting in an initial phase 
with a beam energy of $E_e=30$\,GeV and a
a circumference  of $5.4$\,km which is $1/5$ of the LHC length.
Lowering $E_e$ is also advantageous for mastering the synchrotron radiation
challenges in the interaction region. Naturally, 
the decision on $E_e$ is not taken now. This paper comprises studies with 
different energy configurations, mainly $E_e=50$ and $60$\,GeV, which are close
in their centre-of-mass energy values of $1.2$ and $1.3$\,TeV, respectively.

Up to beam energies of about $60$\,GeV, the ERL cost is dominated by 
the cost for the superconducting RF of the linacs. Up to this energy 
the ERL cost scales approximately  linearly with the beam energy. Above
 this energy the return arcs represent the main contribution to the cost 
and to the ERL cost scaling is no longer linear.  Given the non-linear 
dependence of the cost on $E_e$, for energies larger than about $60$\,GeV, 
significantly larger electron beam energy values may only be justified 
by overriding arguments, such as, for example, the existence of 
leptoquarks~\footnote{If these existed with a mass of say $M=1.5$\,TeV 
this would require, at the LHC with $E_p=7$\,TeV, to choose $E_e$ to be
 larger than $90$\,GeV, and to pay for it. Leptoquarks would be produced
 by $ep$ fusion and appear as resonances, much like the $Z$ boson in
 $e^+e^-$ and would therefore fix $E_e$ (given certain $E_p$ which at 
the FCC exceeds $7$\,TeV). The genuine DIS kinematics, however, is 
spacelike, the exchanged four-momentum squared $q^2=-Q^2$ being negative,
 which implies that the choice of the energies is less constrained than 
in an $e^+e^-$ collider aiming at the study of the $Z$ or $H$ bosons.}.
Higher values of $\sqrt{s}$ are also provided with enlarged proton beam energies 
by the High Energy LHC ($E_p=13.5$\,TeV)~\cite{Abada:2019ono}
 and the FCC-hh~\cite{Benedikt:2018csr} with $E_p$ between 
$20$ and possibly $75$\,TeV, depending on the dipole magnet technology. 

%
\section{Cost Estimate, Default Configuration and Staging}
\label{subsec:cost}
In 2018 a detailed cost estimate was carried out~\cite{ocost} following 
the guidance and practice of CERN accelerator studies. The assumptions were 
also compared with the DESY XFEL cost. The result was that for the $60$\,GeV 
configuration about half of the total cost was due to the two SC linacs.  The
 cost of the arcs decreases more strongly than linearly with decreasing energy,
about $\propto E^4$ for synchrotron radiation losses and $\propto E^3$ when 
emittance dilution is required to be avoided~\cite{alexBerlin}.  It was 
therefore considered to set a new default of $50$\,GeV with a circumference 
of $1/5$ of that of the LHC, see Sect.\,\ref{subsec:config}, compared to 1/3 for
 $60$\,GeV. Furthermore, an initial phase at $30$\,GeV was considered, within 
the $1/5$ configuration but with only partially 
equipped linacs. The  HERA electron beam
 energy was $27$\,GeV. The main results, taken from~\cite{ocost} are 
reproduced in Tab.\,\ref{tab:cost}.
\begin{table}[ht]
  \centering
  \small  
  \begin{tabular}{lccc}
    \toprule
    Component   & CDR 2012 & Stage 1 & Default  \\
                &  (60\,GeV)  &   (30\,GeV)  &  (50\,GeV)   \\
    \midrule
    SRF System   & 805 &   402  &  670 \\
    SRF R+D and Prototyping  &  31 & 31 & 31 \\
    Injector   &  40 & 40 &  40 \\
    Arc Magnets and Vacuum & 215 & 103 &  103   \\
    SC IR Magnets & 105  &  105  &   105  \\
    Source and Dump System & 5 & 5 &  5 \\
    Cryogenic Infrastructure & 100  & 41   &  69  \\
    General Infrastructure and Installation & 69 & 58  & 58  \\
    Civil Engineering &  386   & 289  &   289   \\
    \midrule
    Total Cost   & 1756  & 1075   &   1371   \\
    \bottomrule
  \end{tabular}
  \caption{Summary of cost estimates,
    in MCHF, from~\cite{ocost}. The $60$\,GeV configuration 
    is built with a $9$\,km triple racetrack configuration as
    was considered in the CDR~\cite{AbelleiraFernandez:2012cc}.
    It is taken as the default configuration for FCC-eh, with 
    an additional CE cost of $40$\,MCHF due to the 
    larger depth on point L (FCC) as compared to IP2 (LHC).
    Both the $30$ and the $50$\,GeV assume a $5.4$\,km configuration,
    i.e.\ the $30$\,GeV is assumed to be a first stage of LHeC upgradeable
    to  $50$\,GeV ERL. Whenever a choice was to be made on estimates, 
    in~\cite{ocost} the conservative number was chosen. 
  }
  \label{tab:cost}
\end{table}

The choice of a default of $50$\,GeV at 1/5 of the LHC circumference results, as displayed, in a total cost 
of $1,075$\,MCHF for the initial $30$\,GeV configuration and an additional,
 upgrade cost to $50$\,GeV of $296$\,MCHF. If one restricted the  
LHeC to a non-upgradeable $30$\,GeV only configuration one would,
 still in a triple racetrack configuration, come to roughly a $1$\,km
 long structure with two linacs of about $500$\,m length, probably in 
a single linac tunnel configuration. The cost of this version of the 
LHeC is roughly $800$\,MCHF, i.e.\ about half the $60$\,GeV estimated cost. However,
this would essentially reduce the LHeC to a QCD and electroweak machine, still
very powerful but accepting substantial losses in its Higgs, top and BSM programme. 

\begin{figure}[h]
\centering
\includegraphics*[width=0.7\textwidth]{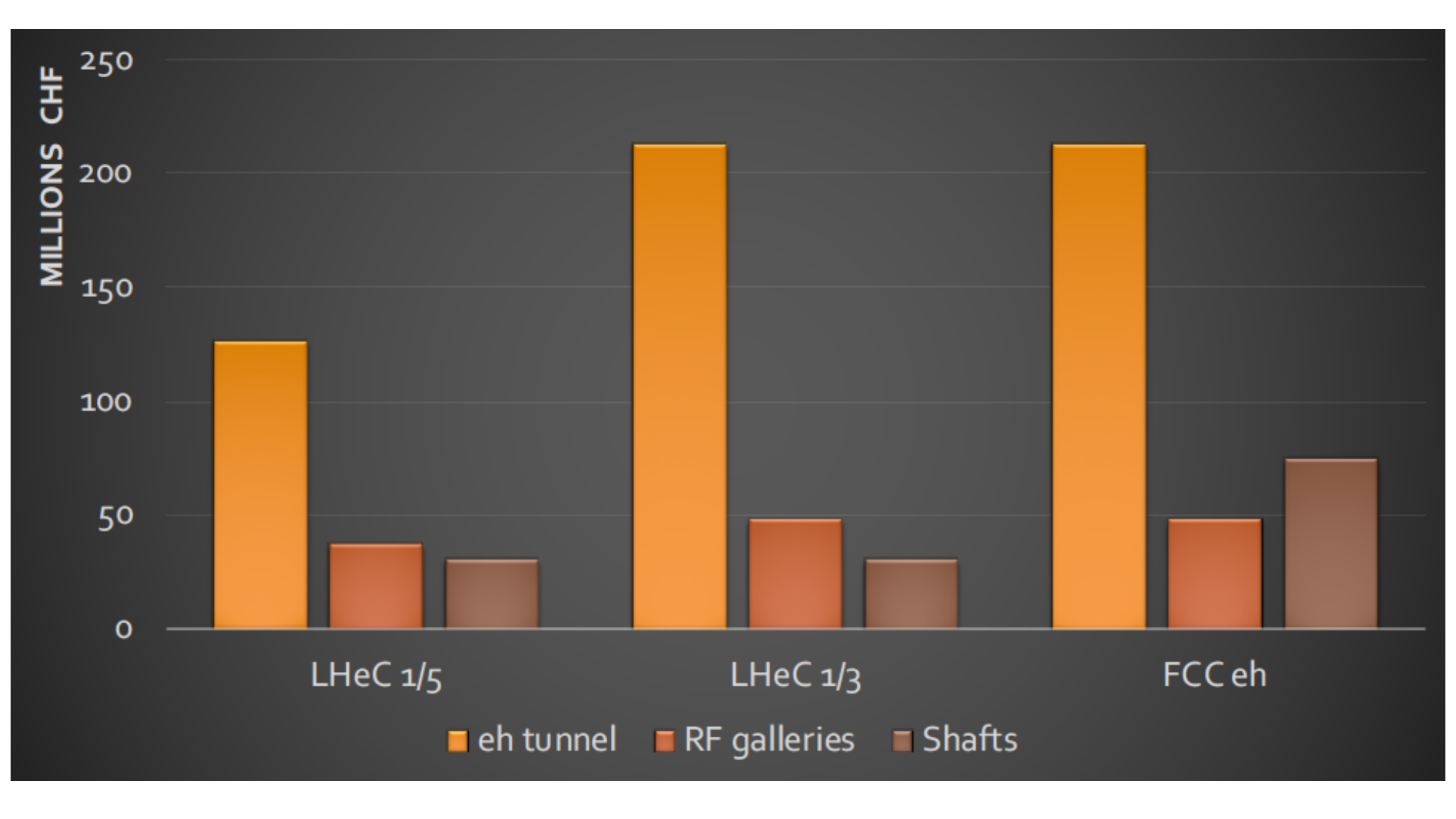}
\caption{
Cost estimate for the civil engineering work for the tunnel, rf galleries and shafts
for the LHeC at 1/5 of the LHC circumference (left), at 1/3 (middle) and the
FCC-eh (right). The unit costs and percentages are consistent
 with FCC and CLIC unit prices. The estimate is considered reliable to $30$\,\%.
The cost estimates include: Site investigations: $2$\,\%, Preliminary design, tender
documents and project changes: $12$\,\% and the Contractors profit: $3$\,\%. Surface
site work is not included, which for LHeC exists with IP2.}
\label{fig:cecost}
\end{figure}

A detailed study was made on the cost of the civil engineering, which is also
discussed subsequently. This concerned a comparison of the 1/3 vs the 1/5 LHC
circumference versions, and the FCC-eh. The result is illustrated in 
Fig.\,\ref{fig:cecost}. It shows that the CE cost for the 1/5 version is about 
a quarter of the total cost. The reduction from 1/3 to 1/5 economises 
about $100$\,MCHF.

Choices of the final energy will be made later. They  depend not only  on a budget
 but also on the future development of particle physics at large. For example, it
 may turn out that, for some years into the 
 future, the community   may not find the $\mathcal{O}$(10)\,GCHF 
required to build any of the $e^+e^-$ colliders currently considered. Then the only way 
to improve on the Higgs measurements beyond HL-LHC substantially is the high energy 
($50 - 60$\,GeV), high luminosity ($\int{L} = 1$\,ab$^{-1}$) LHeC. 
 Obviously, physics and cost are intimately related. 
Based on such considerations, but also taking into account technical constraints as
resulting from  
the amount of synchrotron radiation losses in the interaction region
 and the arcs, we have chosen $50$\,GeV in a $1/5$ of U(LHC) configuration
 as the new default. This economises about $400$\,MCHF as
compared to the CDR configuration.

If the LHeC ERL were built, it may later be transferred, with some 
reconfiguration and upgrades, to the FCC to serve as the FCC-eh. 
The  FCC-eh has its own location, L, for the ERL which requires a 
new accelerator tunnel. It has been decided to 
keep the $60$\,GeV configuration for the FCC, as described in the 
recently published CDR of the FCC~\cite{Benedikt:2018csr}. The LHeC
ERL configuration may also be used as a top-up injector for the $Z$ 
and possibly $WW$ phase of the FCC-e should the FCC-ee indeed 
precede the FCC-hh/eh phase. 

\section{Configuration Parameters}
\label{subsec:config}
\begin{figure}[th]
\centering
\includegraphics[width=0.7\textwidth]{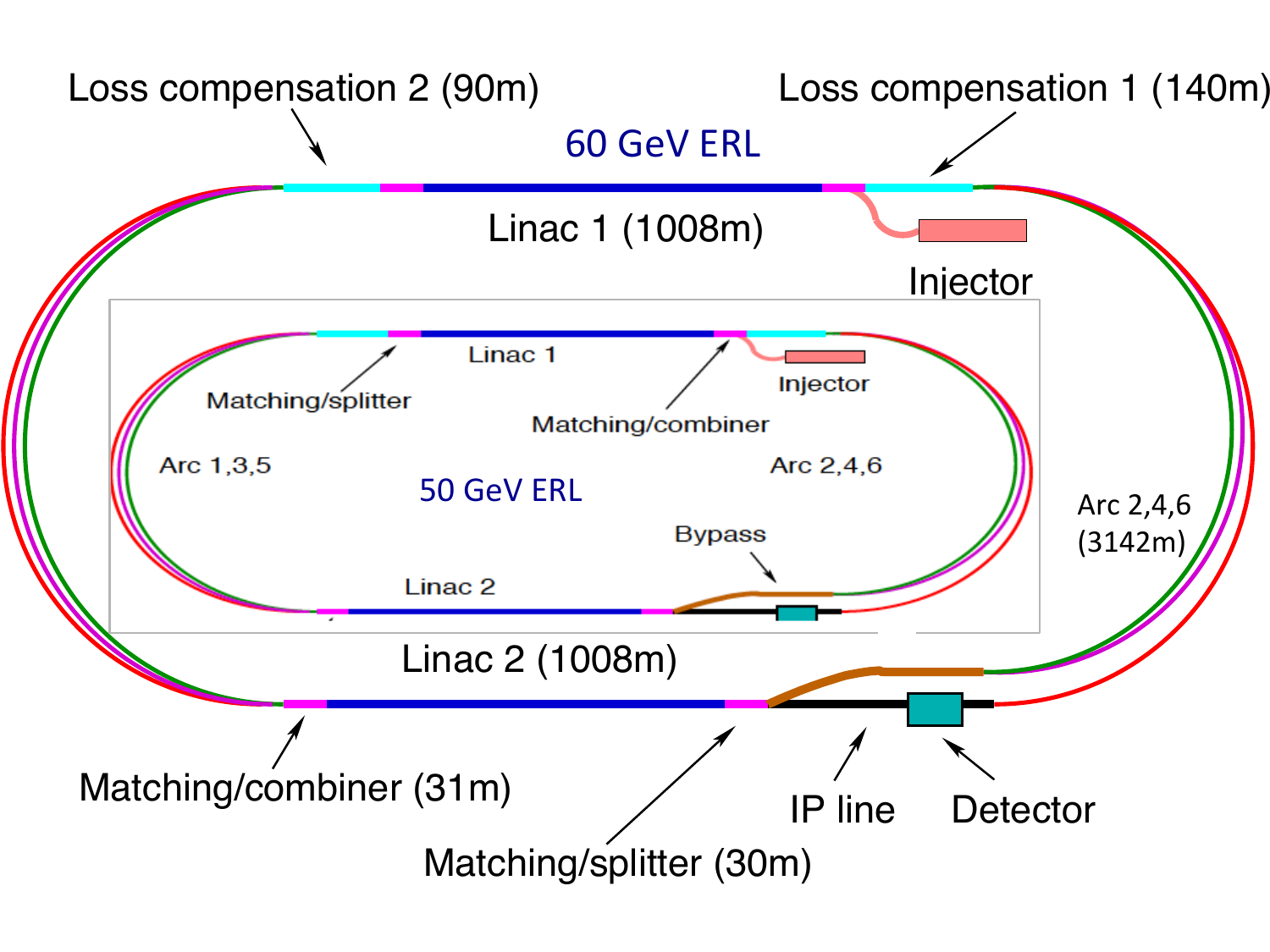}
\caption{
Schematic view of the three-turn LHeC configuration with two oppositely positioned 
electron linacs and three arcs housed in the same tunnel. Two configurations are 
shown: Outer: Default $E_e = 60$\,GeV with linacs of about $1$\,km length and $1$\,km arc radius
leading to an ERL circumference of about $9$\,km, or $1/3$ of the LHC length.
Inner: Sketch for $E_e = 50$\,GeV with linacs of about $0.8$\,km length and $0.55$\,km 
arc radius leading to an ERL circumference of $5.4$\,km, 
or $1/5$ of the LHC length, which is smaller than the size of the SPS.
The $1/5$ circumference configuration is flexible: it 
entails the possibility to stage the project as funds of physics dictate by using 
only partially equipped linacs, and it also permits upgrading to somewhat higher 
energies if one admits increased synchrotron power losses and operates at higher gradients.
}
\label{fig:lhecconf}
\end{figure}

A possible transition from the $60$\,GeV to the $50$\,GeV configuration
 of the LHeC was already envisaged  in 2018, as considered in the  
paper submitted to the European Strategy~\cite{Bruning:2019scy}. The 
machine layout shown in that paper is  reproduced in Fig.\,\ref{fig:lhecconf}.
 It is a rough sketch illustrating the reduction from a $60$\,GeV to a $50$\,GeV 
configuration, which results not only in a reduction of capital costs, as discussed 
above, but also of effort.

The ERL configuration has been recently revisited~\cite{alexBerlin} considering
its dependence on the electron beam energy. Applying a dimension scaling which
preserves the emittance dilution, the  results have been obtained as are 
summarised in Tab.\,\ref{tab:scaledim}.
\begin{table}[ht]
   \centering
   \small
   \begin{tabular}{lccccc}
     \toprule
     Parameter & Unit & \multicolumn{4}{c}{LHeC option} \\
     \cmidrule{3-6}
      & & 1/3 LHC  & 1/4 LHC  & 1/5 LHC & 1/6 LHC \\
     \midrule
     Circumference & m & 9000 & 6750 &  5332 & 4500 \\
     Arc radius &  m\,$\cdot$\, $2\pi$ & 1058 & 737 & 536 &  427 \\
     Linac length &  m\,$\cdot$\,2 & 1025 & 909 &  829 & 758 \\
     Spreader and recombiner length &  m\,$\cdot$\,4 & 76 & 76 & 76 & 76 \\
     Electron energy &  GeV & 61.1 & 54.2 & 49.1 & 45.2 \\  
     \bottomrule
   \end{tabular}
 \caption{Scaling of the electron beam energy, linac and further 
   accelerator element dimensions with the 
   choice of the total circumference in units $1/n$ of the LHC circumference. For
   comparison, the CERN SPS has a circumference of 6.9\,km, only 
   somewhat larger than $1/4$ of that of the LHC.}
 \label{tab:scaledim}
\end{table}%
The $1/5$ configuration is chosen as the new LHeC default while the CDR on
 the LHeC from 2012 and the recent CDR on FCC-eh have used the 1/3 configuration. 
 The energy and configuration may  be decided as physics, cost and effort dictate,
 once a decision is taken.

\section{Luminosity}
\label{subsec:parlu}
The luminosity $L$ for the LHeC in its linac-ring configuration
is determined as
\begin{equation}
\label{LLR}
L = \frac{N_e N_p n_p f_{rev} \gamma_p}{4 \pi \epsilon_p \beta^*} \cdot \prod_{i=1}^3{H_i},
\end{equation}
where $N_{e(p)}$ is the number of electrons (protons) per bunch, $n_{p}$ the number of proton bunches in the LHC, 
$f_{rev}$ the revolution frequency in the LHC [the bunch spacing in a batch 
is given by $\Delta$, equal to $25$\,ns for protons in the LHC] and 
$\gamma_p$ the relativistic factor $E_p/M_p$ of the proton beam.
Further,  $\epsilon_p$ denotes the normalised proton transverse beam emittance 
and $\beta^*$  the proton beta function at the IP, assumed to be equal in 
$x$ and $y$. The luminosity is moderated by the hourglass factor, $H_1=H_{geo} \simeq 0.9$, the
pinch or beam-beam correction factor, $H_2=H_{b-b} \simeq 1.3$, and the 
filling factor $H_3=H_{coll} \simeq 0.8$, should an ion clearing gap in the
electron beam be required. This justifies taking the product of these factors. As the product is close to
unity, the factors are not listed for simplicity in the subsequent tables.

The electron beam current is given as
\begin{equation}
  \label{Ie}
  I_e= e N_e  f\,,
\end{equation}
%
%
where $f$ is the bunch frequency $1/\Delta$. The current for the LHeC is limited by the charge delivery of the source. In the new default design
we have $I_e=20$\,mA which results from a charge of $500$\,pC for the bunch frequency
of $40$\,MHz. It is one of the tasks of the PERLE facility to investigate the stability of the
3-turn ERL configuration in view of the challenge for each cavity to hold the sixfold current due to the
simultaneous acceleration and deceleration of bunches at three different beam energies each.

\subsection{Electron-Proton Collisions}
The design parameters of the luminosity were recently provided in a note describing the 
FCC-eh configuration~\cite{LHeClumi}, including the LHeC.  
Tab.\,\ref{tab:lumip} represents an update comprising in addition the initial
 $30$\,GeV configuration and the lower energy version of the FCC-hh  based on the LHC 
magnets\footnote{
The low energy FCC-pp collider, as of today, uses a 6\,T LHC 
magnet in a 100\,km tunnel. If, sometime in the coming decades, high field magnets  become available 
based on HTS technology, then a $20$\,TeV proton beam energy may even be
achievable in the LHC tunnel. To this extent the  low energy FCC  considered here and 
an HTS based HE-LHC would be comparable options in terms of their energy reach.
}.  
For the LHeC, as noted above, we assume $E_e=50$\,GeV while for FCC-eh we retain 
$60$\,GeV. Since the source limits the electron current, the peak luminosity
 may be taken not to depend on $E_e$.
 Studies of the interaction region design, presented in this paper, show that one may be confident of reaching
a $\beta^*$ of $10$\,cm but it will be a challenge to reach even smaller values. Similarly,
it will be quite a challenge to operate with a current much beyond $20$\,mA.
That has nevertheless been considered~\cite{Bordry:2018gri} 
for a possible dedicated LHeC operation mode for a few years following the $pp$ operation program. 
%
\begin{table}[ht]
  \centering
  \small
  \begin{tabular}{lcccccccc}
    \toprule
    Parameter & Unit & \multicolumn{4}{c}{LHeC} & & \multicolumn{2}{c}{FCC-eh}  \\
    \cmidrule{3-6} \cmidrule{8-9}
     & & CDR  & Run 5 & Run 6 & Dedicated & & $E_p$=$20$\,\TeV & $E_p$=$50$\,\TeV \\
    \midrule
    $E_e$ & GeV & 60 & 30  & 50  &  50 & & 60  &  60 \\
    $N_p$ & $10^{11}$ & 1.7 & 2.2  & 2.2 & 2.2  && 1 & 1 \\
    $\epsilon_p$ & $\mu$m & 3.7 & 2.5 & 2.5 & 2.5  && 2.2 & 2.2 \\
    $I_e$ & mA & 6.4 & 15  & 20 & 50  && 20 & 20 \\
    $N_e$ & $10^9$ & 1 & 2.3  & 3.1 & 7.8  && 3.1 & 3.1 \\
    $\beta^*$ & cm & 10 & 10 & 7 & 7  && 12 & 15 \\
    Luminosity & $10^{33}$\,cm$^{-2}$s$^{-1}$ & 1 & 5 & 9 & 23& & 8 & 15 \\ 
    \bottomrule
  \end{tabular}
  \caption{Summary of luminosity parameter values for the LHeC and FCC-eh. Left: CDR 
    from 2012; Middle: LHeC in three stages, an initial run, possibly during Run\,5 of the LHC,
 the $50$\,GeV operation during Run\,6, both concurrently with the LHC, and a final,
 dedicated, stand-alone $ep$ phase; Right: FCC-eh with a 20
 and a 50\,TeV proton beam, in synchronous operation.}
  \label{tab:lumip}
\end{table}

The peak luminosity values exceed those at HERA by 2--3 orders of magnitude. 
The operation of HERA in its first, extended running period, 1992-2000, provided 
an integrated luminosity of about $0.1$\,fb$^{-1}$ for
the collider experiments H1 and ZEUS. This may now be expected to be taken in a day of initial LHeC operation. 
\subsection{Electron-Ion Collisions}

The design parameters and luminosity were also provided recently~\cite{LHeClumi}
for collisions of  electrons and lead nuclei (fully stripped 
$^{208}\mathrm{Pb}^{82+}$ ions). 
Tab.\,\ref{tab:lumeA} is an 
 update of the numbers presented there for consistency with the Run~6  
LHeC  configuration in Tab.~\ref{tab:lumip} and with the addition of 
 parameters corresponding to the 
$E_p=20\,\mathrm{TeV}$  FCC-hh configuration.  
Further discussion of this operating mode and motivations for the parameter choices 
in this table are provided in 
Section~\ref{sec:eAoperation}. 
 
\begin{table*}[ht] 
  \centering
  \small
  \begin{tabular}{lcccc}
    \toprule
    Parameter  & Unit  & LHeC  & FCC-eh & FCC-eh \\
               &       &       & ($E_p$=20\,\TeV) & ($E_p$=50\,\TeV) \\
    \midrule
    Ion  energy $E_\mathrm{Pb} $ & PeV &   0.574 &  1.64 &  4.1 \\ 
    Ion energy/nucleon $E_\mathrm{Pb}/A $ & TeV &   2.76 &  7.88 &  19.7 \\ 
    Electron beam energy $E_e$ & GeV  & 50 & 60 & 60 \\
    Electron-nucleon CMS  $\sqrt{s_{eN}}$  & TeV  & 0.74 & 1.4 & 2.2 \\
    Bunch spacing & ns  & 50 & 100 & 100 \\
    Number of bunches & & 1200 & 2072 & 2072 \\
    Ions per bunch & $10^{8}$  & 1.8 & 1.8 & 1.8 \\
    Normalised emittance $\epsilon_n$ & $\mu$m  & 1.5 & 1.5 & 1.5 \\
    Electrons per bunch & $10^9$ & 6.2 & 6.2 & 6.2 \\
    Electron current & mA  & 20 & 20 & 20 \\
    IP beta function $\beta^*_A$ & cm  & 10 & 10 & 15 \\
    e-N Luminosity   & $10^{32}$cm$^{-2}$s$^{-1}$  & 7 & 14 & 35 \\
    \bottomrule
  \end{tabular}
  \caption{
    Baseline parameters of future electron-ion collider
    configurations based on the electron ERL, in concurrent
    $eA$ and $AA$ operation mode with the LHC and the two versions of a future hadron 
    collider at CERN. 
    Following established convention in this field, the luminosity quoted, at the start of a fill, is the 
    \emph{electron-nucleon} luminosity which is a factor $A$ larger than the usual (i.e.\ electron-nucleus) luminosity.
    }
  \label{tab:lumeA}
\end{table*} 

One can expect the average luminosity during fills to be 
about 50\% of the peak in Tab.\,\ref{tab:lumeA} and we assume 
an overall operational efficiency of 50\%.   Then, a year of 
$eA$ operation, possibly composed 
by combining shorter periods of operation,
would have the potential to provide an integrated data set of about $5~(25)$\,fb$^{-1}$ for
the LHeC (FCC-eh), respectively. 
This exceeds the HERA electron-proton luminosity value by about tenfold for the LHeC and much more at FCC-eh
while the fixed target nuclear DIS experiment kinematics is extended by 3--4 orders of magnitude.
 These energy frontier  electron-ion configurations therefore have the unique
potential to radically modify our present view of nuclear structure and parton dynamics. 
This is discussed in Chapter~4.

\section{Linac Parameters}
\label{subsec:parlin}
The brief summary of the main LHeC characteristics here concludes with
a table of the main ERL parameters for the new default electron energy of $50$\,GeV,
Tab.\,\ref{tab:linpar}, which are discussed in detail in Chapter\,8.
\begin{table}[ht]
  \centering
  \small
  \begin{tabular}{lcc}
    \toprule
    Parameter  & Unit & Value \\
    \midrule
    Frequency  & MHz & 801.58  \\
    Bunch charge & pC & 499 \\
    Bunch spacing & ns & 24.95 \\ 
    Electron current & mA  & 20 \\
    Injector energy & MeV & 500 \\
    Gradient & MV/m & 19.73 \\
    Cavity length, active & m & 0.918 \\
    Cavity length, flange-to-flange & m & 1.5 \\
    Cavities per cryomodule &  &  4 \\
    Length of cryomodule & m & 7  \\
    Acceleration per cryomodule & MeV & 72.45  \\
    Total number of cryomodules &  & 112  \\
    Acceleration energy per pass & GeV & 8.1 \\
    \bottomrule
  \end{tabular}
  \caption{Basic LHeC ERL characteristics for the default configuration
    using two such linacs located opposite to each other in a racetrack of
    $5.4$\,km length. Each linac is passed three times for acceleration and three
    times for deceleration.
  }
  \label{tab:linpar}
\end{table}%
\section{Operation Schedule}
The LHeC parameters are determined to be compatible with a parasitic operation with the 
nominal HL-LHC proton-proton operation. This implies limiting the electron bunch current
to sufficiently small values so that the proton beam-beam parameter remains small enough
to be negligible for the proton beam dynamics.

Assuming a ten year construction period for the LHeC after approval of the project
 and a required installation window of two years for the LHeC detector, the earliest 
realistic operation period for the LHeC coincides with the LHC Run\,5 period in 2032 and 
with a detector installation during LS4 which is currently scheduled
during 2030 and would need to be extended by one year to 2031. The baseline HL-LHC operation
 mode assumes 160 days of proton operation, 20 days of ion operation and 20 days of machine
 development time for the Run 4 period, amounting to a total of 200 operation days per year. 
 After the Run 4 period the HL-LHC does not consider ion operation at present  and assumes 190 days for proton operation.
The HL-LHC project 
assumes an overall machine efficiency of $54$\,\% (e.g.\ fraction of scheduled operation time 
spent in physics production) and we assume that the ERL does not contribute to significant
 additional downtime for the operation. Assuming an initial $15$\,mA of electron beam current, a
 $\beta^{*}$ of $10$\,cm and HL-LHC proton beam parameters, the LHeC reaches a peak luminosity 
of $0.5\cdot 10^{34}$\,cm$^{-2}$s$^{-1}$. Assuming further a proton beam lifetime of 16.7 hours,
 a proton fill length of 11.7 hours and an average proton beam turnaround time of 4 hours,
 the LHeC can reach in this configuration an annual integrated luminosity of $20$\,fb$^{-1}$.

For the evaluation of the physics potential it is important to note that the Run\,5
initial $ep$ operation period may accumulate about $50$\,fb$^{-1}$ of integrated luminosity.
This is the hundredfold value which H1 (or ZEUS) took over a HERA lifetime of $15$ years.
As one may expect, for details see Chapter\,3, such a huge DIS luminosity is ample
for pursuing basically the complete QCD programme. In particular, the LHeC would deliver
on time for the HL-LHC precision analyses the external, precise PDFs and with just
a fraction of the $50$\,fb$^{-1}$ the secrets of low $x$ parton dynamics would unfold.
Higher $ep$ luminosity is necessary for ultimate precision and for
the top, BSM and the Higgs programme of the LHeC to be of competitive value.

For the Run\,6 period of the HL-LHC, the last of the HL-LHC operation periods, we assume
 that the number of machine development sessions for the LHC can be suppressed, providing
 an increase in the  operation time for physics production from 190 days to 200 days
 per year. Furthermore, we assume that the electron beam parameters can be slightly further 
pushed.
 Assuming a $\beta^{*}$ reduced to $7$\,cm, an electron beam current of up to $25$\,mA and still nominal 
HL-LHC proton beam parameters, the LHeC reaches a peak performance of $1.2 \cdot 10^{34}$\,cm$^{-2}$s$^{-1}$
 and an annual integrated luminosity of $50$\,fb$^{-1}$.
This would add up to an integrated luminosity of a few hundred fb$^{-1}$, 
a strong base for top, BSM and Higgs physics at the LHeC.

Beyond the HL-LHC exploitation period, the electron beam parameters
could be further pushed in dedicated $ep$ operation, when the requirement of a parasitic
 operation to the HL-LHC proton-proton operation may no longer be imposed. 
The proton beam lifetime without proton-proton collisions would be significantly 
larger than in the HL-LHC configuration. In the following we assume a proton beam
 lifetime of $100$ hours and a proton beam efficiency of $60$\,\% without proton-proton
 beam collisions. The electron beam current in this configuration would only be
 limited by the electron beam dynamics and the SRF beam current limit. Assuming
 that electron beam currents of up to $50$\,mA, the LHeC would reach a peak luminosity
 of $2.4\cdot 10^{34}$\,cm$^{-2}$s$^{-1}$ and an annual integrated luminosity of up to
 $180$\,fb$^{-1}$. Table\,\ref{op-schedule} summarises the LHeC configurations over these three 
periods of operation.


\begin{table}[ht]
  \centering
  \small
  \begin{tabular}{lcccc}
    \toprule
    Parameter  & Unit & Run\,5 Period  & Run\,6 Period & Dedicated\\
    \midrule
    Brightness $N_p/(\gamma\epsilon_p)$ & $10^{17}\text{m}^{-1}$ & 2.2/2.5 & 2.2/2.5 & 2.2/2.5  \\
    Electron beam current & mA & 15 & 25 & 50? \\
    Proton $\beta^*$ & m & 0.1 & 0.7 & 0.7 \\ 
    Peak luminosity & $10^{34}$\,\si{cm^{-2}s^{-1}}  & 0.5 & 1.2 & 2.4 \\
    Proton beam lifetime & h & 16.7 & 16.7 & 100 \\
    Fill duration & h & 11.7 & 11.7 & 21 \\
    Turnaround time & h & 4 & 4 & 3 \\
    Overall efficiency & \% & 54 & 54 & 60 \\
    Physics time / year & days &  160 & 180 & 185 \\
    Annual integrated lumi. & fb$^{-1}$ & 20 & 50 & 180  \\
    \bottomrule
  \end{tabular}
\caption{The LHeC performance levels during different operation modes.
}
\label{op-schedule}
\end{table}
Depending on the years available for a dedicated final operation (or through an
extension of the $pp$ LHC run, currently not planned but interesting for
collecting $4$ instead of $3$\,ab$^{-1}$ to, for example, observe di-Higgs production
at the LHC), a total luminosity of $1$\,ab$^{-1}$ could be available for the LHeC.
This would double the precision of Higgs couplings measured in $ep$ as compared to the
default HL-LHC run period with $ep$ added as described. It would also significantly
enlarge the potential to observe or/and quantify rare and new physics phenomena. Obviously
such considerations are subject to the grand developments at CERN. A period with most interesting
physics and on-site operation activity could be particularly welcome for narrowing a possible large time
gap between the LHC and its grand successor, the FCC-hh. One may, however, be interested
in ending LHC on time. It thus is important for the LHeC project to recognise its particular
value as an asset of the HL-LHC, and on its own, with even less than the ultimate luminosity,
albeit values which had been dreamt of at HERA.

%

\biblio

%% file: standardmodel/standardmodel.chapter.tex
\linenumbers
\lhectitlepage
\lhecinstructions
\subfilestableofcontents

\input{\main/standardmodel/standardmodel.tex}

\biblio

%% file: standardmodel/standardmodel.tex

\chapter{Parton Distributions - Resolving the Substructure of the Proton}
\label{chapter:pdf}

\section{Introduction}
\label{sect:pdfintro}
Since the discovery of quarks in the famous $ep \to eX$ scattering experiment at 
Stanford~\cite{Bloom:1969kc,Breidenbach:1969kd},
the deep inelastic scattering process has been established as the most reliable method
to resolve the substructure of protons, which was immediately recognised, not least by
Feynman~\cite{Feynman:1973xc}. Since that time, a series of electron, muon and neutrino 
DIS experiments validated the Quark-Parton Model and promoted the development of
Quantum Chromodynamics. A new quality of this physics was realised 
with HERA, the first electron-proton collider built, which extended the kinematic
range in momentum transfer squared to $Q^2_{max} =s \simeq 10^5$\,GeV$^2$, for
$s=4E_eE_p$. Seen from today's perspective, largely influenced by the LHC, 
it is necessary to advance to a further level in these investigations, with higher energy and 
much increased luminosity than HERA could achieve. This is a major motivation for building the LHeC,
with an extension of the $Q^2$ and $1/x$ range by more than an order of magnitude
and an increase of the luminosity by a factor of almost a thousand.
QCD may breakdown, be embedded in a higher gauge symmetry,
or unconfined colour might be observed;
These phenomena raise a number  a series of fundamental questions of the
QCD theory~\cite{Quigg:2013lya} and highlight the importance
of a precision DIS programme with the LHeC.

 The subsequent chapter
is mainly devoted to the exploration of the seminal potential of the LHeC to resolve the substructure
of the proton in an unprecedented range,  with the first ever complete and coherent 
measurement of the full set of parton distribution functions (PDFs) in one experiment. 
The precise determination of PDFs, consistently to 
high orders pQCD,   is crucial for the interpretation of LHC physics, i.e.
its precision electroweak and Higgs measurements as well as the exploration of the high mass region
where new physics may occur when the HL-LHC operates. 
Extra constraints on PDFs arise also from $pp$ scattering as is discussed in a later chapter. 
Conceptually, however, the LHeC provides the singular opportunity to completely separate
the PDF determination from proton-proton physics. This approach is not only more precise
for the PDFs, but it is theoretically more accurate and enables incisive tests of QCD, by confronting
independent predictions with LHC (and later FCC) measurements, as well as providing an indispensable base for reliable
interpretations of searches for new physics.

While the resolution of the longitudinal, collinear structure of the proton is key to the 
physics programme of the LHeC (and the LHC), the $ep$ collider
provides further fundamental insight in the structure
of the proton: semi-inclusive measurements of jets and vector mesons, and especially
Deeply Virtual Compton Scattering, a process established at HERA,
 will shed light on also the transverse structure of the 
proton in a new kinematic range. This is presented at the end of the current chapter.

\subsection{Partons in Deep Inelastic Scattering}

Parton Distribution Functions $xf(x,Q^2)$
represent a probabilistic view on hadron substructure  at
a given distance, $1/\sqrt{Q^2}$.  They depend on the parton
type $f=(q_i,~g)$, for quarks and gluons, and must be determined 
from experiment, most suitably DIS,
as perturbative QCD is not prescribing the parton density at a given momentum fraction Bjorken $x$.
PDFs are  important also for they determine Drell-Yan, hadron-hadron
scattering processes, supposedly universally  
through the QCD factorisation theorem~\cite{Collins:1989gx}~\footnote{In his referee report 
on the LHeC CDR, in 2012, Guido Altarelli noted on the factorisation
theorem in QCD for hadron colliders that: ``many 
people still advance doubts. Actually this question could be studied
experimentally, in that the LHeC, with its improved precision, could put
bounds on the allowed amount of possible factorisation violations (e.g.\ by
measuring in DIS the gluon at large $x$ and then comparing with jet 
production at large $p_T$ in hadron colliders).''
This question was addressed also in a previous LHeC
paper~\cite{AbelleiraFernandez:2012ty}.}.
%
The PDF programme of the  LHeC is of unprecedented reach for the following reasons: 
\begin{itemize}
\item For the first time it will resolve the partonic structure of the proton (and nuclei) 
completely, i.e. determine the $u_v$, $d_v$, $u$, $d$, $s$, $c$, $b$, and gluon momentum
distributions through neutral  and charged current  cross section as well as direct heavy quark PDF measurements,
performed in a huge kinematic range of DIS, from  $x=10^{-6}$ to $0.9$ and from $Q^2$  above $1$ to  $10^6$\,GeV$^2$. The LHeC explores the
 strange density and the momentum fraction carried by top 
 quarks~\cite{Boroun:2015yea} which was impossible at HERA.

\item Very high  luminosity and unprecedented precision, owing to both new detector technology and the redundant evaluation of the event kinematics from the leptonic and hadronic final states, will lead to extremely high PDF precision.
%
\item Because of the high LHeC energy, the weak probes ($W,~Z$) dominate the interaction at larger $Q^2$ which permits the  up and down sea and valence quark distributions to be resolved in the full range of $x$. Thus no additional data will be required~\footnote{The 
LHeC may be operated at basically HERA energies and collect a fb$^{-1}$ of luminosity for
cross checks and maximising the high $x$, medium $Q^2$ acceptance, see
Sect.\,\ref{sect:DISdata}.}: that is, there is no influence from higher twists nor nuclear uncertainties or data inconsistencies, which are main sources of uncertainty of current so-called global PDF determinations.
\end{itemize}

While PDFs are nowadays often seen as merely a tool for interpreting LHC data, in fact what really is involved is a new understanding of strong interaction dynamics and the deeper resolution of substructure extending into hitherto uncovered phase space regions, in particular the small $x$ region,
 by virtue of the very high energy $s$, and the  very small spatial dimension ($1/\sqrt{Q^2}$) and the $x \to 1$ region, owing to the high luminosity and energy. The QPM is not tested well enough, despite decades of DIS and other
experiments, and QCD is not developed fully either in these kinematic regimes. 

Examples of issues of fundamental interest for the LHeC to resolve are: i) the
long awaited resolution of the behaviour of $u/d$ near the kinematic limit ($x \to 1$);
 ii) the flavour democracy of the light quark sea (is $d \simeq u \simeq s$ ?);
  iii) the existence of quark-level charge-symmetry~\cite{Hobbs:2011vy}; 
iv) the behaviour of the ratio $\bar{d}/\bar{u}$ at small $x$;
 v) the turn-on and the values of heavy quark PDFs; vi) the value of
the strong coupling constant and vii) the question of the dynamics, linear or non-linear, at small $x$ where the gluon and quark densities rise.  

Of special further  interest is the gluon distribution, for
the gluon self-interaction prescribes all visible mass, the
gluon-gluon fusion process dominates Higgs production at hadron colliders
(the LHC and the FCC)
and because its large $x$ behaviour, essentially unknown today, affects predictions of BSM cross sections at the LHC.

The LHeC may be understood as an extension of HERA to a considerable extent.
It has the reach in $x \propto 1/s$ to resolve the question of new strong interaction dynamics
at small $x$ and it accesses 
 high $Q^2$, much larger than $M_{W,Z}^2$, with huge luminosity to make accurate
 use of weak NC and CC cross sections in DIS PDF physics for the first time. QCD analyses of HERA data are still
ongoing. For obvious reasons, there is no quantitative analysis of LHC related 
PDF physics possible without relying on the HERA data, and often on its QCD analyses.
These are introduced briefly next. Albeit with certain assumptions and limited luminosity,
HERA completely changed the field of PDF physics as compared to 
the times of solely fixed target data,
see Ref.~\cite{Klein:2010zzc}, and it opened the era of physics of high parton densities at small $x$.

\subsection{Fit Methodology and HERA PDFs}
\label{sec:hera}
The methodology of PDF determinations with HERA data has been developed
over decades by the H1 and ZEUS 
Collaborations~\cite{Klein:2008di,HERA:2009wt,Abramowicz:2015mha}, 
in close contact with many theorists. 
It has been essentially adopted with suitable modifications
 for the LHeC PDF prospect study as is detailed subsequently.

HERAPDF fits use information from both $e^{\pm} p$ neutral current  and charged current  
scattering  from exclusively the $ep$ collider experiments, H1 and
ZEUS, up to high $Q^2=30~000$\,GeV$^2$ and down to about $x=5 \cdot 10^{-5}$. 
The precision of the HERA combined data is 
below 1.5\,$\%$ over the $Q^2$ range of $3 < Q^2 < 500$\,GeV$^2$ 
and remains below 3$\%$ up to 
$Q^2= 3000$\,GeV$^2$. The precision for large $x > 0.5$ is rather poor
due to limited luminosity and high-$x$ acceptance limitations at medium $Q^2$.


The QCD analysis  is performed at LO, NLO and NNLO within the \emph{xFitter} framework~\cite{HERAFitter,HERA:2009wt,Aaron:2009kv},
 and the latest version is 
the HERAPDF2.0 family~\cite{Abramowicz:2015mha}.
The DGLAP evolution of the PDFs, as well as the light-quark
coefficient functions, are calculated using QCDNUM~\cite{Botje:2010ay,Botje:2016wbq}.
The contributions of heavy quarks are calculated in the general-mass
variable-flavour-number (GMVFN) scheme of Refs.~\cite{Thorne:2006qt,Thorne:2012az}.
Experimental uncertainties are determined using the Hessian method imposing 
a $\chi^2 + 1$ criterion. This is usually impossible in global fits over rather incoherent
data sets originating from different processes and experiments, but has been a major advantage of the solely HERA based QCD analyses.

In the HERAPDF analysis, as well as subsequently in the LHeC study, the starting scale is
chosen to be $Q^2_0 = 1.9$\,GeV$^2$ such that it is below the
charm mass, $m_c^2$.
The data is restricted to $Q^2_{min} \ge 3.5$\,GeV$^2$
in order to stay in the DIS kinematic range. The forward hadron final state acceptance
introduces a lower $W$ cut which removes the region which otherwise is potentially
sensitive to higher twist effects~\cite{Alekhin:2012ig}.
The strong coupling constant is set to
$\alpha_S(M_Z) =  0.118$~\footnote{
The strong coupling constant cannot be reliably determined from inclusive HERA data alone. 
DIS results, including fixed target data, have provided values which tend 
to be lower~\cite{Alekhin:2017kpj}
than the here chosen value, see for a discussion Ref.~\cite{Klein:2018rhq}. As is further presented in detail in
Sect.\,\ref{sec:alphas} the LHeC reaches a sensitivity to $\alpha_s$ at the per mille level based on inclusive and jet data as well as their combination.}.
All these assumptions are varied in the evaluation of model
uncertainties on the resulting fit. These variations will essentially have no significant effect with the LHeC as the sensitivity to the quark masses, for example, is hugely improved with respect to HERA, $\alpha_s$ known to 1--2 per mille, and the kinematic range of the data is much extended.

In HERAPDF fits, the quark distributions at the initial $Q^2_0$ are represented by the generic form
\begin{equation}
 xq_i(x) = A_i x^{B_i} (1-x)^{C_i} P_i(x),
\label{eqn:pdf}
\end{equation}
where $i$ specifies the flavour of the quark distribution 
and $P_i(x)= (1 + D_i x +E_i x^2)$. The inclusive NC and CC cross sections determine
four independent quark distributions, essentially the sums of the up and down quark and 
anti-quark densities. These may be decomposed into any four other distributions of up and 
down quarks with an ad-hoc assumption on the fraction of strange to anti-down quarks
which has minimal numeric effect on the PDFs, apart from that on $xs$ itself.
The parameterised quark distributions, $xq_i$, are chosen to be
the valence quark distributions ($xu_v,~xd_v$) and the
light anti-quark distributions
($x\bar{u},~x\bar{d}$). This has been adopted for the LHeC also.

The parameters $A_{u_v}$ and $A_{d_v}$ are fixed using
the quark counting rule.
The normalisation and slope parameters, $A$ and $B$,
of $\bar{u}$ and $\bar{d}$  are set equal such that
$x\bar{u} = x\bar{d}$ at  $x\to 0$, a crucial assumption
which the LHeC can validate. The strange quark PDF
$x\bar{s}$ is set as a fixed fraction $r_s=0.67$ of  $x\bar{d}$.
This fraction is varied in the determination of model uncertainties.
By default it is assumed that $xs=x\bar{s}$ and that $u$ and $d$ 
sea and anti-quarks have the  same distributions also. These assumptions
will be resolved by the LHeC and their uncertainties will essentially be eliminated, see Sect.\,\ref{sec:strange}.
The $D$ and $E$ parameters are used only if  required
by the data, following a $\chi^2$ saturation 
procedure~\cite{HERA:2009wt}. This leads for HERAPDF2.0 to
two non-zero parameters, $E_{u_v}$ and $D_{\bar{u}}$.

The gluon distribution is parameterised differently
\begin{equation}
xg(x) = A_g x^{B_g} (1-x)^{C_g}  - A'_g x^{B'_g} (1-x)^{C'_g}.
\label{eqn:gpdf}
\end{equation}
The normalisation parameter $A_g$ is calculated using the momentum sum rule.
Variations of the PDFs were also considered with $A'_g=0$ which for earlier
HERA data fits had been the default choice. The appearance of this
second term may be understood as coming from a not-well
constrained behaviour of $xg(x,Q^2)$ at small $x$. In fact, $xg$ is resembling a
 valence-quark distribution at $Q^2 \simeq Q_0^2$. The much extended
$Q^2$ range of the LHeC at a given small $x$ and the access to much smaller $x$ values
than probed at HERA will quite certainly enable 
this behaviour to be clarified.  Since also $C'_g$ had been set to just a large value, 
there is negligible effect of that second term in Eq.\,\eqref{eqn:gpdf} on the resulting
PDF uncertainties. Consequently $A'_g$ is set to zero in the LHeC study.

Alternative parameterisations are used in the evaluation of the 
parameterisation uncertainty. These variations include: 
introducing extra parameters $D$, $E$ for each quark distribution; the removal of primed
gluon parameters; and the relaxation of assumptions about the low-$x$ sea. These fits 
provide alternative extracted PDFs with similar fit $\chi^2$. The maximum deviation from the central PDF at each value of $x$ is taken as an envelope and added in quadrature with the experimental and model  uncertainties to give the total uncertainty. As for the model uncertainties, the extended
range and improved precision of the LHeC data may well be expected to 
render such variations negligible.


\begin{figure}[!th]
  \centering
  \includegraphics[width=0.4\textwidth]{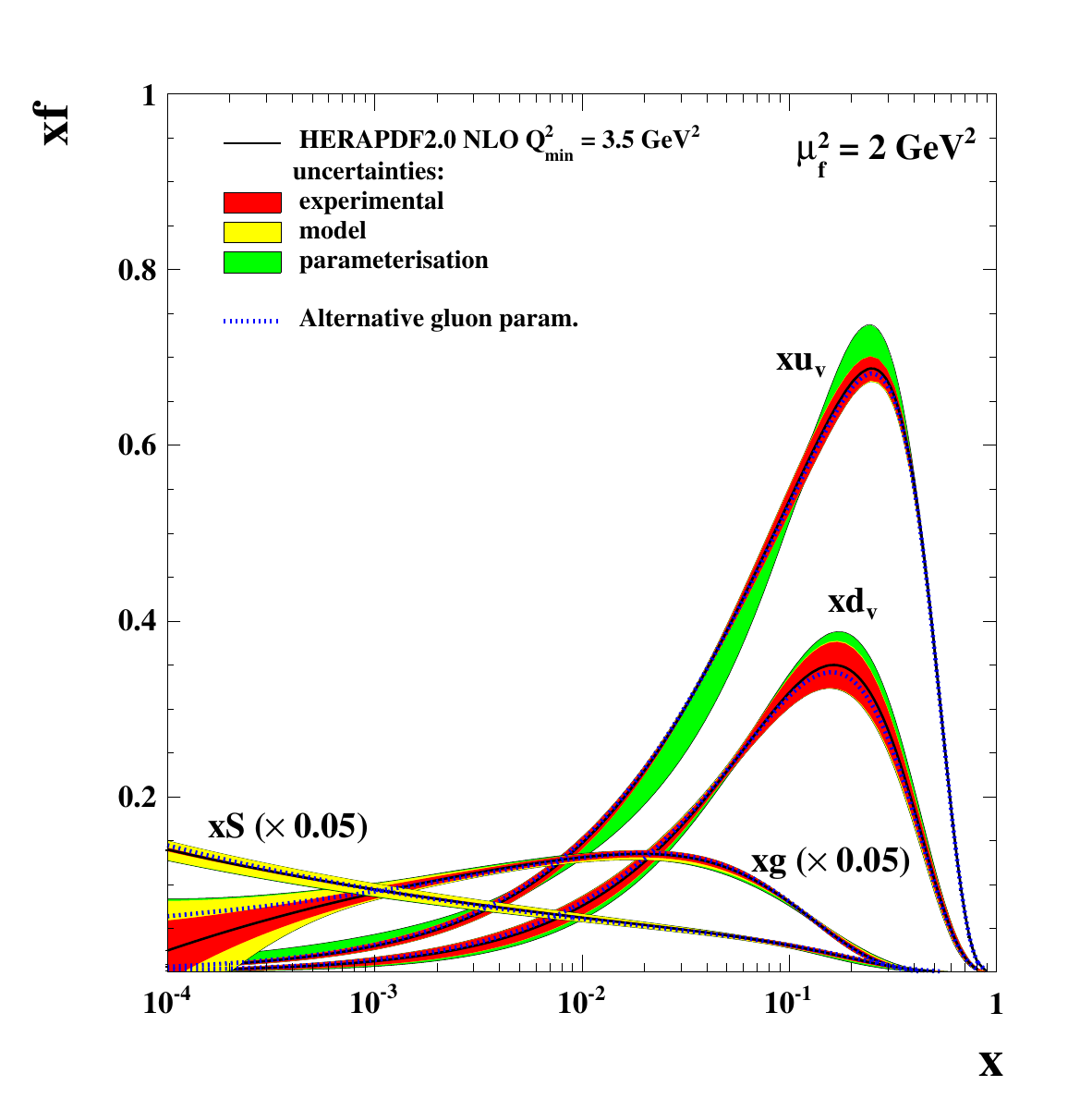} 
 \includegraphics[width=0.4\textwidth]{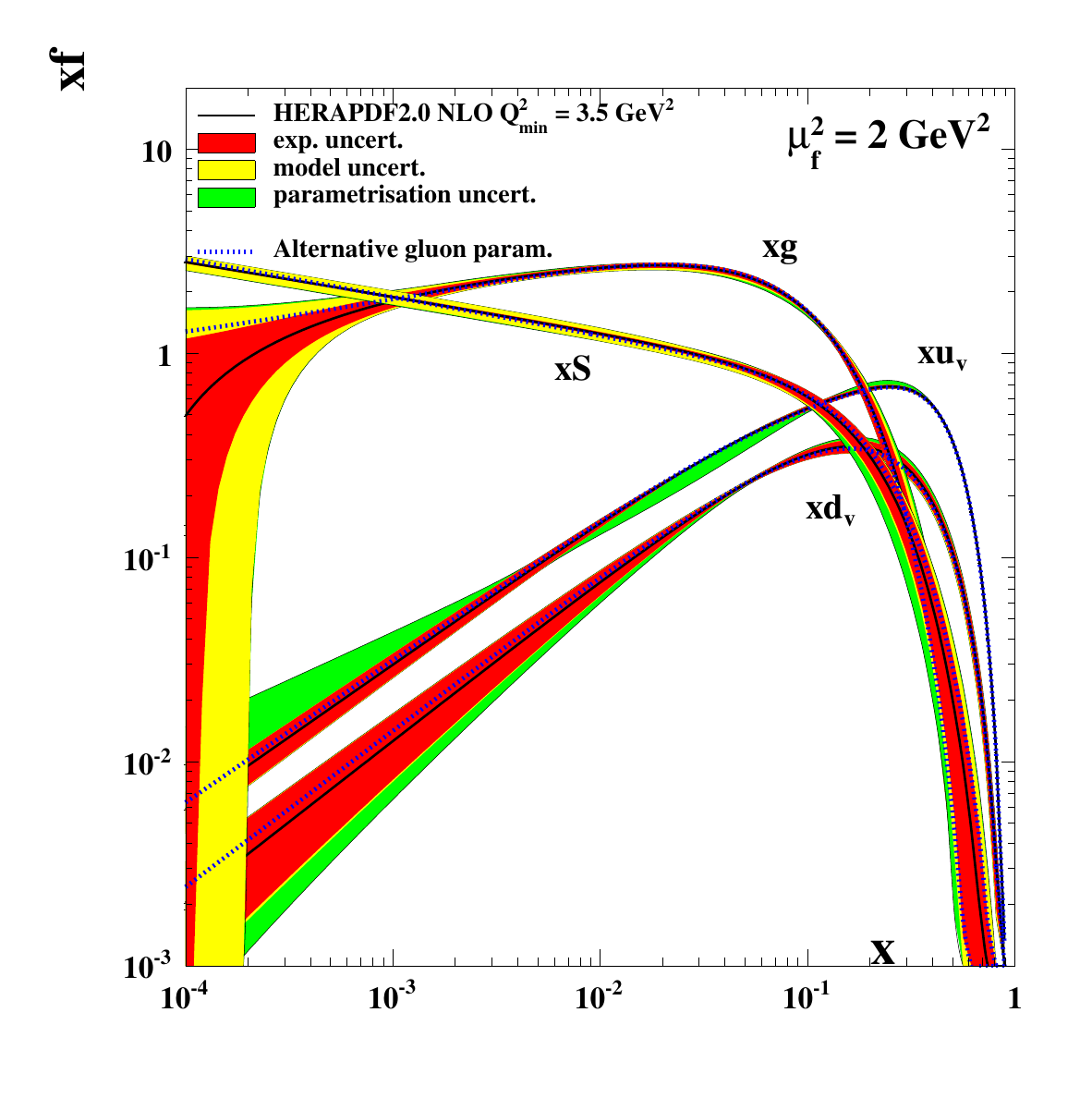} \\
\includegraphics[width=0.4\textwidth]{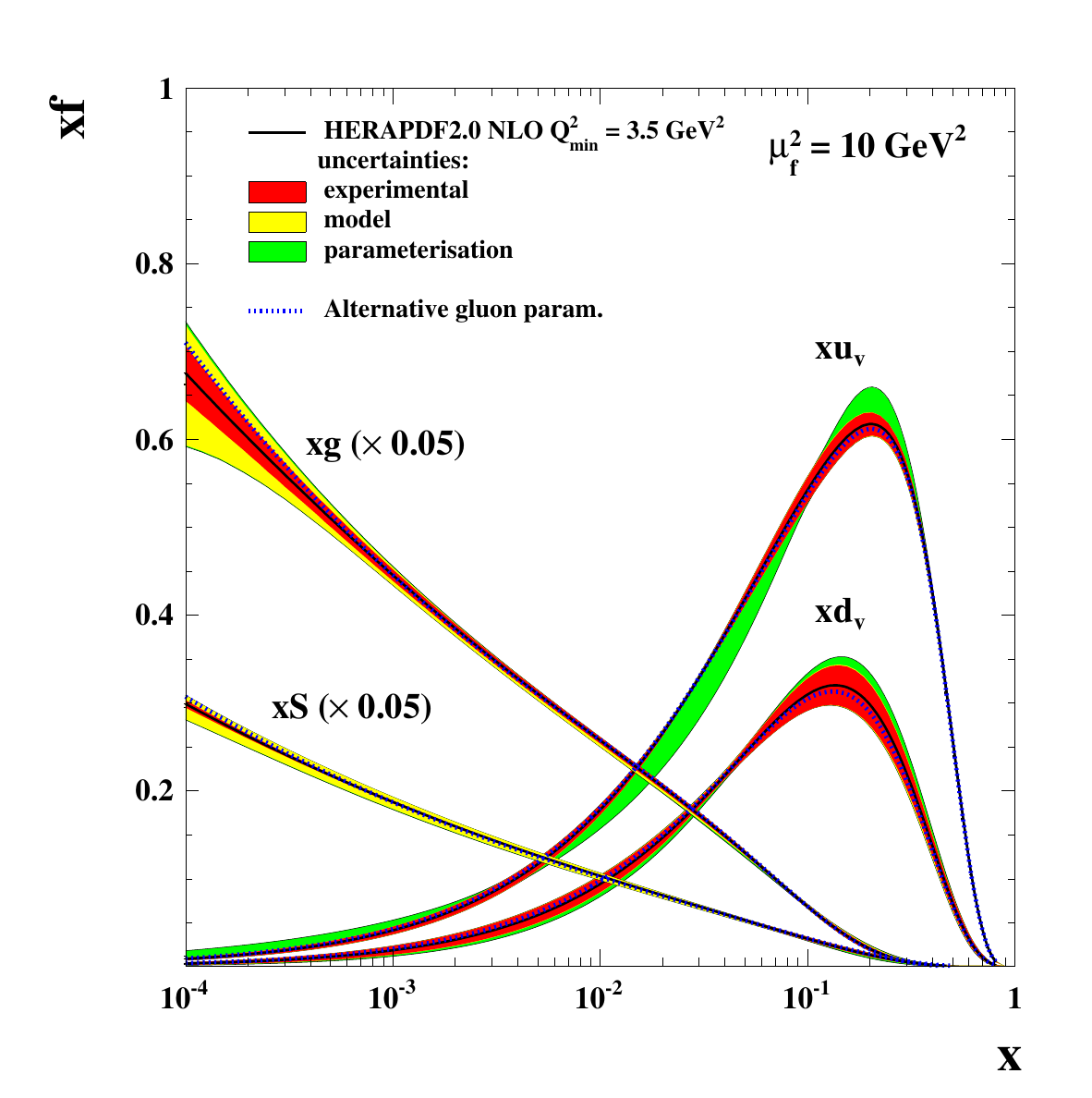} 
 \includegraphics[width=0.4\textwidth]{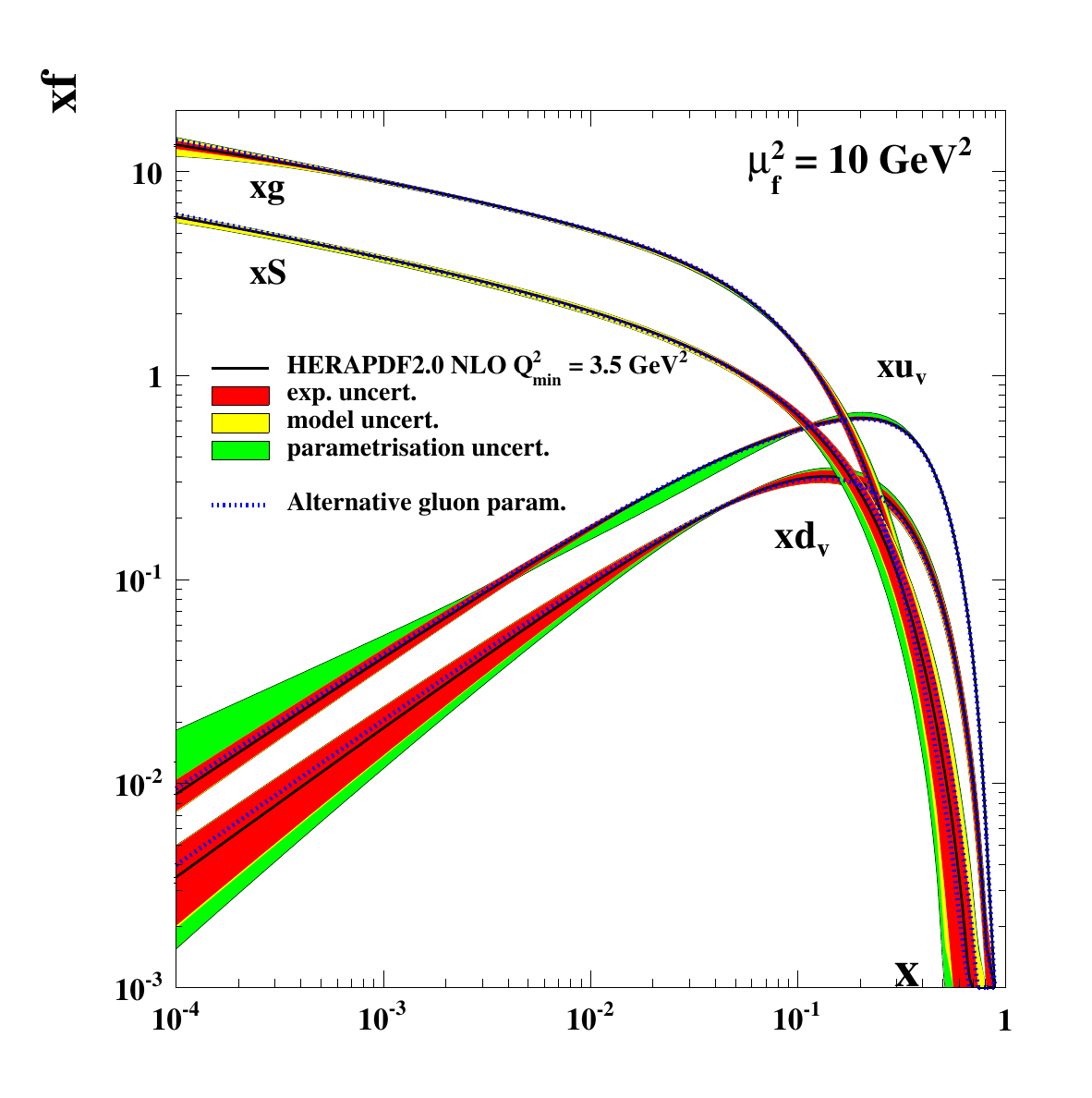} 
  \caption{
Parton distributions as determined by
the QCD fit to the combined H1 and ZEUS data
 at $Q^2 = 1.9$~GeV$^2$ (top) and at $Q^2 = 10$~GeV$^2$
 (bottom).  The color coding represents the experimental, model and
parameterisation uncertainties separately.
Here $xS = 2x(\overline{u} + \overline{d})$ 
denotes the total sea quark density. Note that $xg$ and $xS$ are scaled by $1/20$
in the left side plots with a linear $y$ scale. 
%
  }
  \label{fig:herapdf}
\end{figure}

The results of the HERA PDF analysis~\cite{Abramowicz:2015mha}
are shown in Fig.~\ref{fig:herapdf} for the HERAPDF2.0NNLO PDF set, 
displaying experimental, model and parameterisation uncertainties separately.
The structure of the proton is seen to depend on the resolution 
$ \propto 1/\sqrt{Q^2}$, with which
it is probed.  At $Q^2$ of  about $1-2$\,GeV$^2$,
corresponding to $0.2$\,fm, the parton contents may be decomposed
as is shown in Figure\,\ref{fig:herapdf} top.
The gluon distribution at $Q^2 \simeq 2$\,GeV$^2$ has a valence like shape, i.e.\ at very low $x$
the momentum is carried by sea quarks, see Fig.~\ref{fig:herapdf} (top). At medium $x \sim 0.05$ the
gluon density dominates over all quark densities. At largest $x$, above $0.3$, the proton 
structure is dominated by the up and down valence quarks. This picture
evolves such that below $10^{-16}$\,m, for $x \leq 0.1$, the gluon density
dominates also over the sea quark density, see Figure\,\ref{fig:herapdf} (bottom). 
The valence quark distributions are rather insensitive to the resolution which reflects
their non-singlet transformation behaviour in QCD.

The HERAPDF set differs 
from other PDF sets in that: i) it represents a fit to a consistent data set with small 
correlated systematic uncertainties;
ii) it uses data on solely a proton target such that no heavy target 
corrections are needed and the 
assumption of strong isospin invariance, $d_\text{proton}=u_\text{neutron}$,
 is not required; 
iii) a large $x,Q^2$ region is covered such that no regions where higher
twist effects are important are included in the analysis.

The limitations of HERA PDFs are known as well: i) the data is limited in statistics such that the region
$x > 0.5$ is poorly constrained; ii) the energy is limited such that the very low $x$
region, below $x \simeq 10^{-4}$, is not or not reliably accessed; iii) limits of luminosity and 
energy implied that the potential of the flavour resolution through weak interactions, in
NC and  CC, while remarkable, could not be utilised accurately and $\alpha_s$ not been determined
alongside PDFs in solely inclusive fits; iv) while the strange
quark density was not accessed by H1 and ZEUS, only initial measurements of $xc$ and $xb$
could be performed. The strong success with respect to the fixed target PDF situation $ante$ HERA
has yet been most remarkable. The thorough clarification of parton dynamics and the
establishment of a precision PDF base for the LHC and later hadron colliders, however, make a
next generation, high energy and luminosity $ep$ collider a necessity. 

The PDF potential
of the LHeC is presented next.   This study follows closely the first extended analysis, developed for the CDR and detailed subsequently in Ref.\,\cite{Klein:1564929}. The 
main differences compared to that analysis result from the choice of the
Linac-Ring LHeC configuration, with preferentially $e^-p$ of high
polarisation (and much less $e^+p$) combined with an order of magnitude enhanced 
luminosity and developments of the apparatus design.

%

%
\section{Simulated LHeC Data}
\label{sect:DISdata}
\subsection{Inclusive Neutral and Charged Current Cross Sections}
In order to estimate the uncertainties of PDFs from the LHeC, several
sets of LHeC inclusive NC/CC DIS data with a 
full set of uncertainties have been simulated and are described in the following.
The systematic uncertainties of the DIS cross  
sections have a number of sources, which can be classified as uncorrelated 
and correlated across bin boundaries.  For the NC case,
the uncorrelated sources, apart from event statistics, are a global efficiency uncertainty,
due for example to tracking or electron identification errors, as well as uncertainties
due to photo-production background, calorimeter noise and radiative corrections.
The correlated uncertainties result from imperfect electromagnetic and
hadronic energy scale and angle calibrations.
In the classic $ep$ kinematic reconstruction methods
used here, the scattered electron energy $E_e'$ and
polar electron angle $\theta_e$, complemented by the energy of the
hadronic final state $E_h$, can be employed to determine
$Q^2$ and $x$ in a redundant way.

Briefly, $Q^2$ is best determined with the electron kinematics 
and $x$ is calculated from $y=Q^2/sx$. At large $y$, the inelasticity is best
measured using the electron energy, $y_e \simeq 1 - E_e'/E_e$.
At low $y$, the relation $y_h = E_h \sin^2(\theta_h/2)/E_e$
can be used to provide a measurement of the inelasticity 
with the hadronic final state energy $E_h$ and
angle $\theta_h$. This results in
the uncertainty $\delta y_h /y_h \simeq \delta E_h/E_h$, which is determined
by the $E_h$ calibration uncertainty to good approximation.

There have been various refined methods
proposed to determine the DIS kinematics, such as the
double angle method~\cite{Bentvelsen:1992fu}, which is commonly used to calibrate the
electromagnetic energy scale, or the so-called $\Sigma$
method~\cite{Bassler:1994uq}, which exhibits reduced sensitivity to
QED radiative corrections, see a discussion in
Ref.~\cite{Bassler:1997tv}.
For the estimate of the cross section uncertainty the 
electron method ($Q^2_e, y_e$) is used at large $y$, while at low $y$ we use
$Q^2_e, y_h$, which is transparent and accurate
to better than a factor of two. In much of
the phase space, moreover, it is rather the uncorrelated
efficiency or further specific errors than the kinematic correlations,
which dominate the cross section measurement precision.

The assumptions used in the simulation of pseudodata are summarised in
Tab.\,\ref{tab:sys}. The procedure was gauged with full H1 Monte Carlo
simulations and the assumptions are corresponding to H1's 
achievements with an improvement by at most a factor of two. Using a numerical 
procedure developed in Ref.~\cite{Blumlein:1990dj}, the scale uncertainties are
transformed to kinematics-dependent correlated cross-section uncertainties caused
by imperfect measurements of $E_e'$, $\theta_e$ and $E_h$. 
\begin{table}[ht]
  \centering
  \small
  \begin{tabular}{lc}
    \toprule
    Source of uncertainty & Uncertainty \\
    \midrule
    Scattered electron energy scale $\Delta E_e' /E_e'$ & 0.1 \% \\
    Scattered electron polar angle  & 0.1\,mrad \\
    Hadronic energy scale $\Delta E_h /E_h$ & 0.5\,\% \\
    Radiative corrections & 0.3\,\% \\
    Photoproduction background (for $y > 0.5$) & 1\,\% \\
    Global efficiency error & 0.5\,\%  \\
    \bottomrule
 \end{tabular}
\caption{
Assumptions used in the simulation of the NC cross sections
on the size of uncertainties from various sources. The top three are
uncertainties on the calibrations which are transported to 
provide correlated systematic cross section errors. 
The lower three values are uncertainties of the cross section
caused by various sources. 
}
\label{tab:sys}
\end{table}
These data uncertainties were imposed for all data sets, NC and CC, as are subsequently listed and described.
\begin{figure}[!th]
  \centering
  \includegraphics[width=0.9\textwidth,trim={0 150 0 150 },clip]{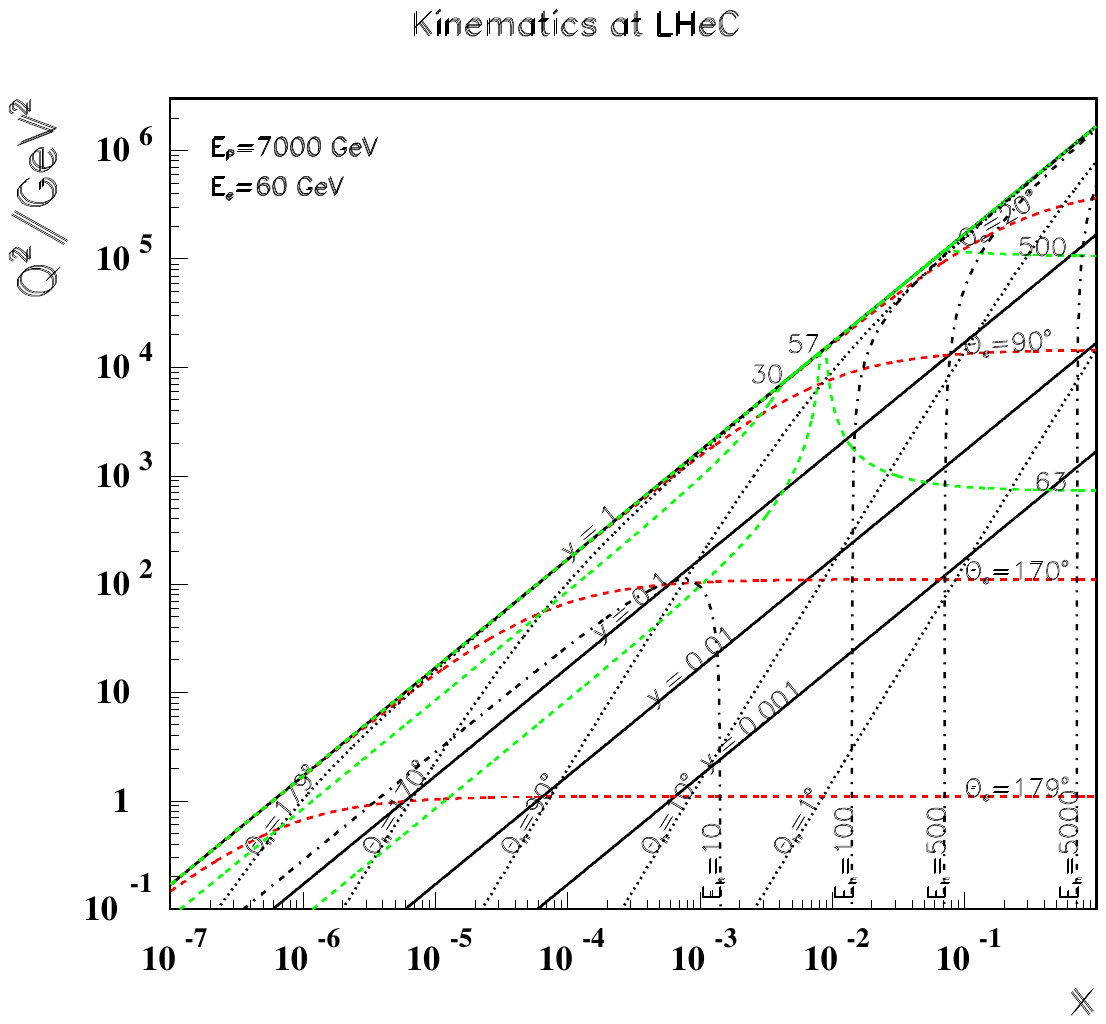}
\caption{
Kinematic plane covered with the maximum beam energies at the LHeC. Red dashed: Lines of constant scattered electron polar angle. Note that low $Q^2$ is measured with electrons scattered
into the backward region, highest $Q^2$ is reached with Rutherford backscattering;
Black dotted: lines of constant angle of the hadronic final state; Black solid: Lines of
constant inelasticity $y=Q^2/sx$; Green dashed: Lines of constant scattered
electron energy $E_e'$. Most of the central region is covered by what is termed
the kinematic peak, where $E_e' \simeq E_e$. The small $x$ region is accessed with 
small energies $E_e'$ below $E_e$ while the very forward, high $Q^2$ electrons 
carry TeV energies; Black dashed-dotted: lines of constant hadronic final state energy $E_h$. Note that the very forward, large $x$ region sees very high hadronic energy deposits too.}
\label{fig:kinisomax}
\end{figure}

The design of the LHeC assumes that it operates with the LHC in the high luminosity
phase, following LS4 at the earliest. As detailed in Chapter\,2, it is assumed there 
will be an initial phase, during which the LHeC may collect $50$\,fb$^{-1}$ of data. This may begin
with a sample of $5$\,fb$^{-1}$. Such values are very high when compared with HERA,
corresponding to the hundred(ten)-fold of luminosity which H1 collected in its lifetime
of about $15$ years. The total luminosity may come close to $1$\,ab$^{-1}$.

The bulk of the data is assumed to be taken with electrons, possibly at large negative
helicity $P_e$, because this configuration maximises the number of Higgs bosons that one
can produce at the LHeC: $e^-$ couples to $W^-$ which interacts primarily with
an up-quark and the CC cross section is proportional to $(1-P_e)$. However, for
electroweak physics there is a strong interest to vary the polarisation and 
charge~\footnote{With a linac source, the generation of an intense positron beam is very challenging
and will not be able to compete with the electron intensity. This is discussed in the accelerator chapter.}.
It was considered that the $e^+p$ luminosity may reach $1$\,fb$^{-1}$ while the tenfold has
been simulated  for sensitivity studies.
A dataset has also been produced with reduced proton beam energy as that 
enlarges the acceptance towards large $x$ at smaller $Q^2$. 
The full list of simulated sets is provided in Tab.\,\ref{tab:dsets}.
\begin{table}[ht]
  \centering
  \small
  \begin{tabular}{l@{\hspace{0.8em}}ccccccccccc}
\toprule
Parameter & Unit & & \multicolumn{9}{c}{Data set} \\
\cmidrule(lr){4-12}
 & & & D1 & D2 & D3 & D4 &D5 & D6 &D7 &D8 & D9 \\ 
\midrule
Proton beam energy       & TeV & &  7   & 7      & 7 &  7     &  1 &   7 &   7 &  7     & 7\\
Lepton charge       &     & & $-1$   & $-1$    & $-1$ & $-1$     & $-1$ & +1 & +1 & $-1$     & $-1$   \\
Longitudinal lepton polarisation &     & & $-0.8$ & $-0.8$ & 0  & $-0.8$  & 0  & 0   & 0   & +0.8 & +0.8   \\
Integrated luminosity & fb$^{-1}$ &  &  5   & 50   & 50& 1000 & 1 & 1   & 10 & 10    & 50 \\
\bottomrule
\end{tabular}
  \caption{Summary of characteristic parameters of data sets used to simulate neutral and charged current $e^\pm$ cross section data, for a lepton beam energy of $E_e=50$\,GeV. Sets D1-D4 are for $E_p = 7$\,TeV and $e^-p$ scattering, with
  varying assumptions on the integrated luminosity and the electron beam polarisation.
  The data set D1 corresponds to possibly the first year of LHeC data taking
  with the tenfold of luminosity which H1/ZEUS collected in their lifetime.
  Set D5 is a low $Ep$ energy run, essential to extend the acceptance at large $x$
  and medium $Q^2$. D6 and D7 are sets for smaller amounts of positron data.
  Finally, D8 and D9 are for high energy $e^-p$ scattering with positive helicity
  as is important for electroweak NC physics. These variations of data taking are
  subsequently  studied for their effect on PDF determinations.
}
\label{tab:dsets}
\end{table}

\begin{figure}[!th]
  \centering
  \includegraphics[width=0.9\textwidth,trim={0 150 0 150 },clip]{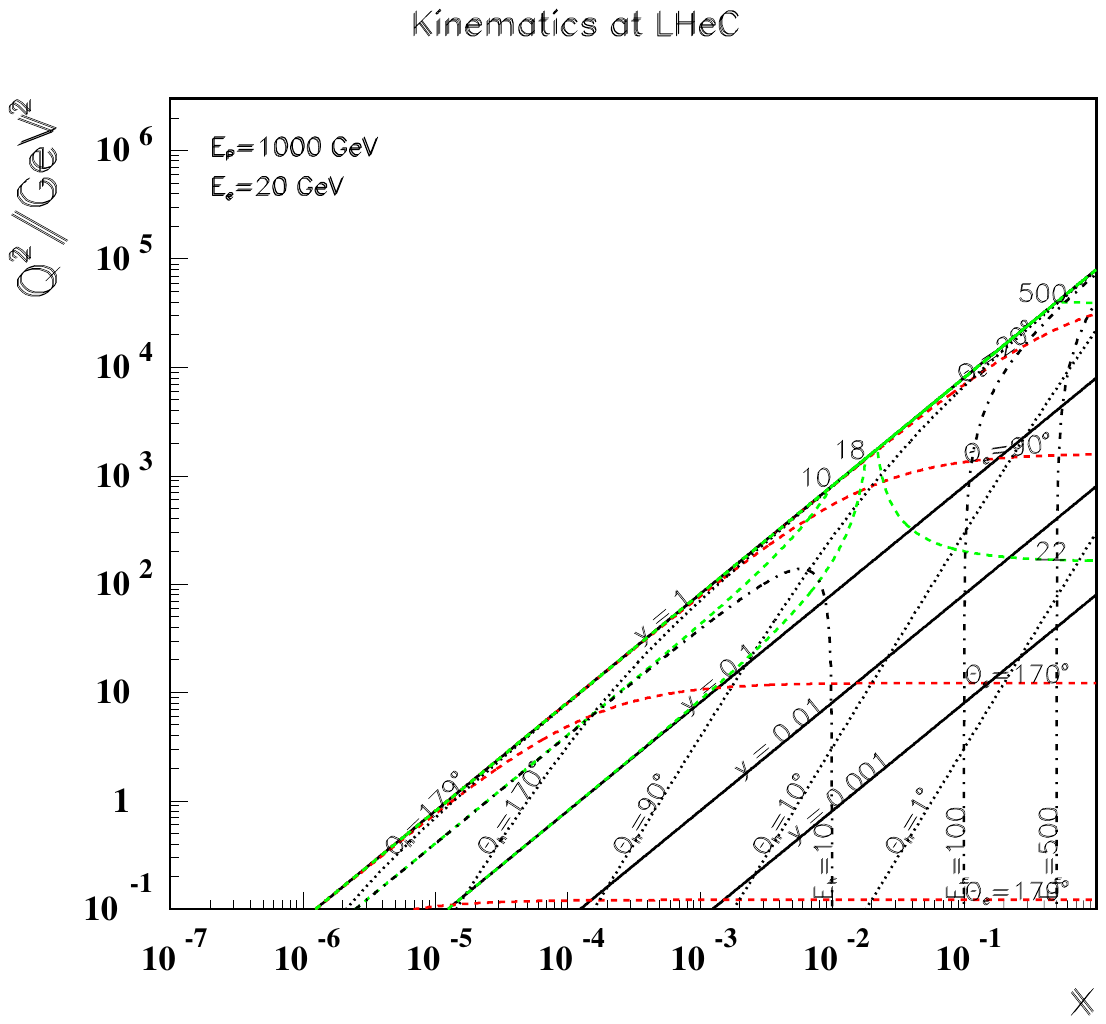}
\caption{
Kinematic plane covered with the minimum beam energies at LHeC. The meaning
of the curves is the same as in the previous figure. This coverage is very similar
to that by HERA as the energies are about the same.}
\label{fig:kinisomin}
\end{figure}

The highest energies obviously give access 
to the smallest $x$ at a given $Q^2$, and to the maximum $Q^2$ at fixed $x$. This
is illustrated with the kinematic plane and iso-energy and iso-angle lines, see
Fig.~\ref{fig:kinisomax}.
It is instructive to see how the variation of the proton beam energy changes the kinematics 
considerably and enables additional coverage of various regions.
 This is clear from
Fig.~\ref{fig:kinisomin} which shows the kinematic plane
choosing the approximate minimum energies the LHeC could operate with.
 There are striking changes one may note which are related to
kinematics (c.f.~Ref.~\cite{Blumlein:1990dj}). For example, one can see that the line of 
$\theta_e = 179^{\circ}$ now corresponds to $Q^2 \simeq 0.1$\,GeV$^2$ which 
is due to lowering $E_e$ as compared to $1$\,GeV$^2$ in the maximum energy case,
cf. Fig.~\ref{fig:kinisomax}. Similarly, comparing the two figures one finds that the
lower $Q^2$, larger $x$ region becomes more easily accessible with lower energies,
in this case solely owing to the reduction of $E_p$ from $7$ to $1$\,TeV.
It is worthwhile to note that the LHeC, when operating
at these low energies, would permit a complete repetition of the HERA programme, within
a short period of special data taking.

\begin{figure}[!th]
  \centering
  \includegraphics[width=0.90\textwidth]{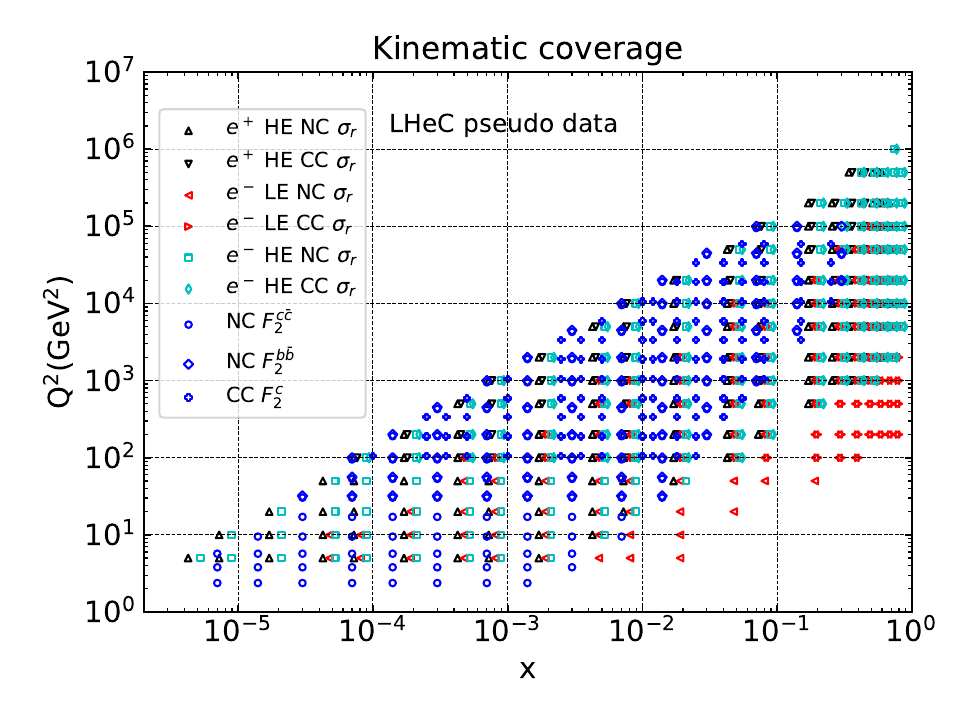}
  \caption{Illustration of the $x, Q^2$ values of simulated cross section
 and heavy quark density data used in LHeC studies. The red points illustrate
the gain in acceptance towards large $x$ at fixed $Q^2$ when $E_p$ is lowered,
see text.
  }
    \label{fig:pdfs_pseudodata}
\end{figure}

The coverage of the kinematic plane is illustrated in the plot of the 
$x,Q^2$ bin centers of data points used in simulations, see 
Fig.~\ref{fig:pdfs_pseudodata}~\cite{AbdulKhalek:2019mps}.
The full coverage at highest Bjorken-$x$, i.e.\ very close 
to $x=1$, is enabled by the high luminosity of the LHeC.
This was impossible to achieve for HERA as the NC/CC DIS
cross sections decrease proportional to some power of $(1-x)$ when $x$ approaches
$1$, as has long been established with Regge 
counting~\cite{Brodsky:1973kr,Brodsky:1974vy,Matveev:1973ra}.

It has been a prime goal, leading beyond previous PDF studies, to understand the
importance of these varying data taking conditions for measuring PDFs with the LHeC. This
holds especially for the question about what can be expected from an initial, lower luminosity
LHeC operation period, which is of highest interest for the LHC analyses during the
HL-LHC phase.
Some special data sets of lowered electron energy have also been
produced in order to evaluate the potential to measure $F_L$, see Sect.~\ref{sec:FL}.
These data sets have not been included in the bulk PDF analyses presented subsequently in this Chapter.


\subsection{Heavy Quark Structure Functions}
\label{sec:hquarks}
The LHeC is the ideal environment for a determination of the strange, charm and
bottom density distributions which is necessary for a comprehensive unfolding
of the parton contents and dynamics in protons and nuclei.
With charm-tagging techniques one can directly access $xs$ in CC or $xc$ in NC DIS,
while with bottom tagging one has access to $xb$ in NC DIS.
The inner Silicon detectors has a resolution of typically $10\,\mu$m, which is much
smaller than the typical decay lengths of charmed and bottomed of several hundreds $\mu$m.
Also, the transverse extension of the beam spot of only $(7$\,$\mu$m$)^2$ is comparably small.
The experimental challenges then are the beam pipe radius,
coping at the LHeC with strong synchrotron radiation effects, and the
forward tagging acceptance, similar to the HL-LHC challenges albeit much easier
through the absence of pile-up in $ep$ (see e.g.~\cite{Klein:2016uxu} for a brief discussion).
Very sophisticated  techniques are
being developed at the LHC in order to identify bottom production through 
jets~\cite{Aad:2019aic} which
are not touched upon here.

\begin{figure}[!th]
  \centering
  \includegraphics[width=0.80\textwidth]{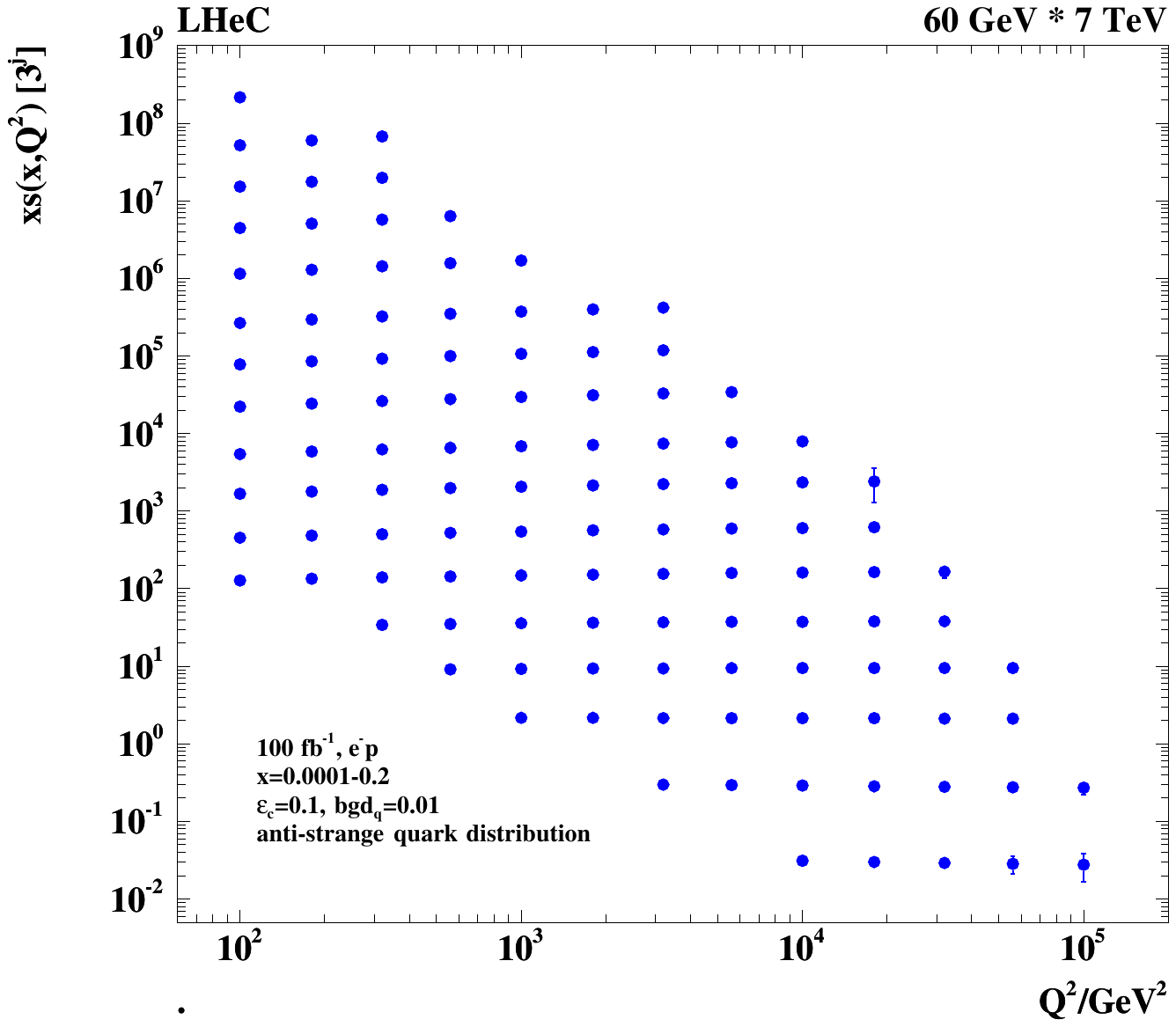}
  \caption{Simulation of the measurement of the (anti)-strange quark distribution,
$x\bar{s}(x,Q^2)$, in charged current $e^-p$ scattering through
the t-channel reaction  $W^- \bar{s} \rightarrow c$.
The vertical error bars indicate the full systematic
and statistical uncertainties added in quadrature, but are mostly smaller than the
marker size. The covered $x$ range extends from
$10^{-4}$ (top left bin), determined by the CC trigger threshold conservatively assumed to be at
$Q^2 = 100$\,GeV$^2$, to $x \simeq 0.2$ (bottom right) determined by the forward tagging acceptance
limits, which could be further extended by lowering $E_p$. 
  }
    \label{fig:xstrange}
\end{figure}

\begin{figure}[!th]
  \centering
  \includegraphics[width=0.80\textwidth]{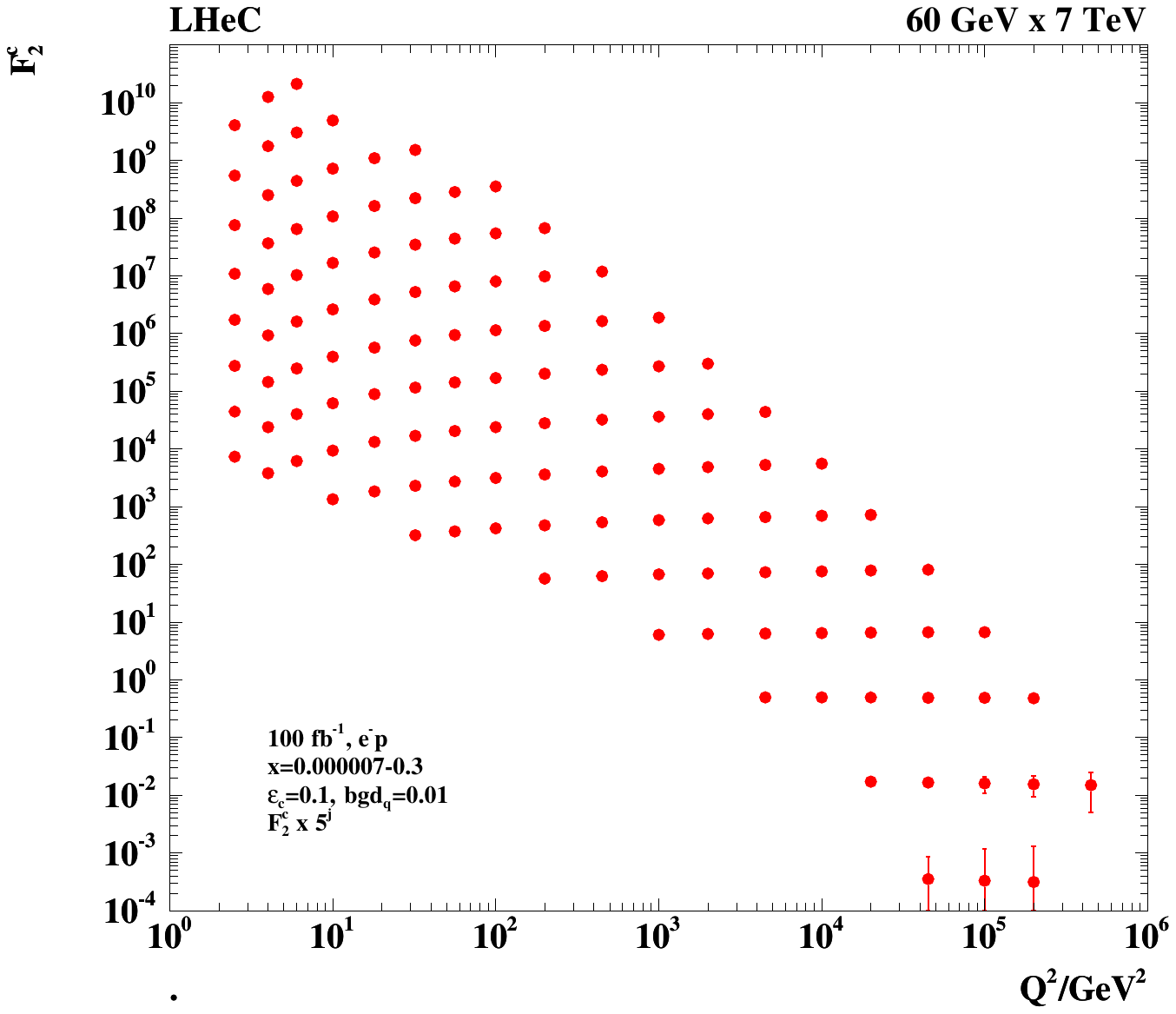}
  \caption{Simulation of the measurement of the charm quark distribution
expressed as $F_2^c=e_c^2x(c+\bar{c})$ in neutral current $e^-p$ scattering.
The vertical error bars indicate the full systematic
and statistical uncertainties added in quadrature, but are mostly smaller than the
marker size.
The minimum $x$ (left top bin) is at $7 \cdot 10^{-6}$, and the data extend to $x=0.3$ (right bottom bin).
The simulation uses a massless scheme and is only indicative near threshold albeit the uncertainties
entering the QCD PDF analysis are estimated consistently. 
  }
    \label{fig:xcharm}
\end{figure}

\begin{figure}[t]
  \centering
  \includegraphics[width=0.80\textwidth]{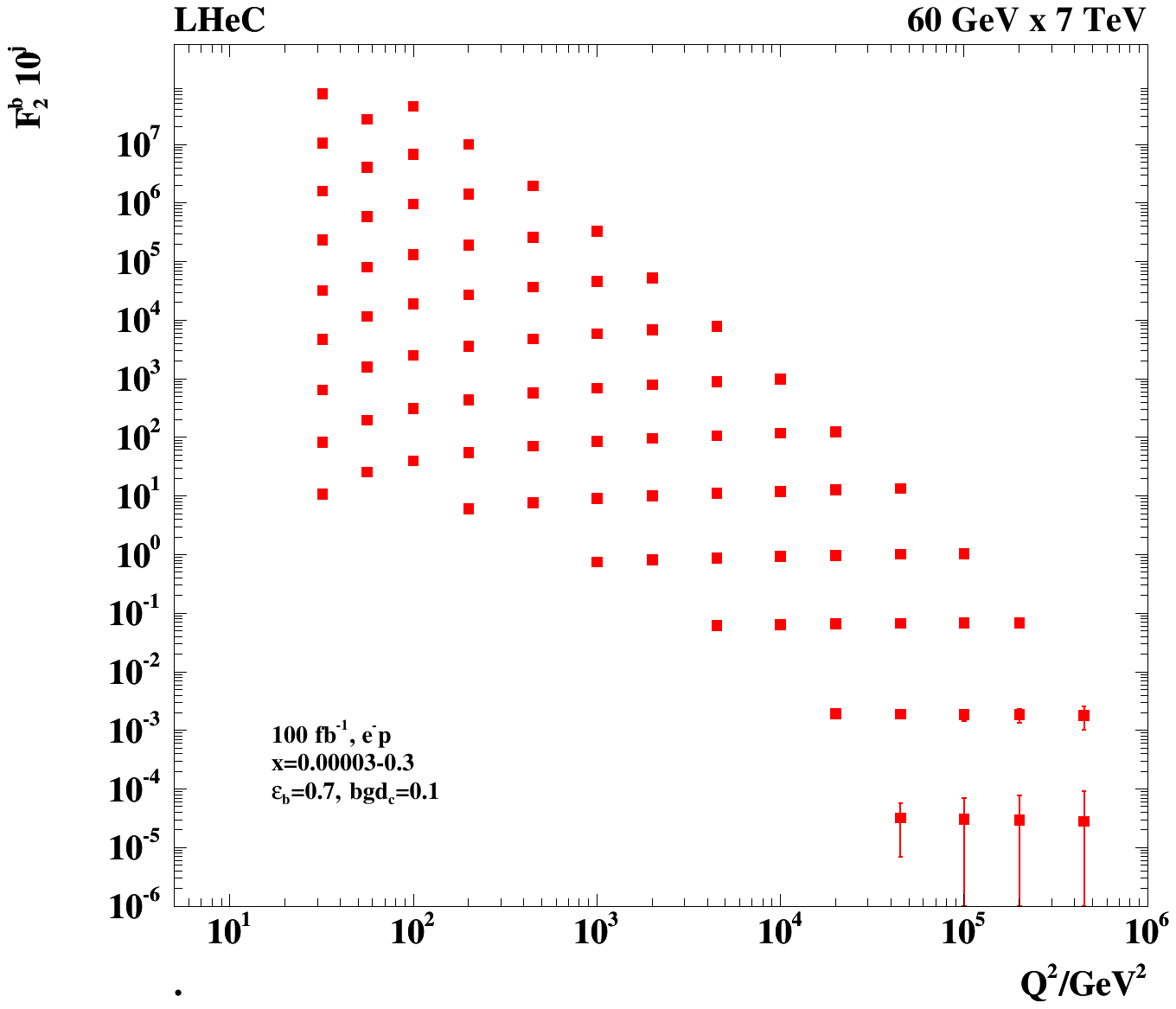}
  \caption{Simulation of the measurement of the bottom quark distribution
expressed as $F_2^b=e_b^2x(b+\bar{b})$  in neutral current $e^-p$ scattering.
The vertical error bars indicate the full systematic
and statistical uncertainties added in quadrature, but are mostly smaller than the
marker size.
The minimum $x$ (left top bin) is at $3 \cdot 10^{-5}$, and the data extend to $x=0.3$ (right bottom bin).
The simulation uses a massless scheme and is only indicative near threshold albeit the uncertainties
entering the QCD PDF analysis are estimated consistently. 
  }
    \label{fig:xbottom}
\end{figure}

A simulation of the measurement of the anti-strange density at the LHeC was performed
using impact parameter tagging in $ep$ CC scattering, see Fig.\,\ref{fig:xstrange}.
The measruement of the charm and beauty structure functions using $c$ and $b$ tagging
were simulated for NC DIS (see Figs.\,\ref{fig:xcharm}, \ref{fig:xbottom}).
The results served as input for the PDF study subsequently presented.

For this simulation, the charm and beauty tagging efficiencies are
assumed to be  $10$\,\% and $60$\,\%, respectively.
These values are derived from  heavy flavour tagging techniques at HERA and ATLAS.
Backgrounds arise from light-quark jets in the charm analysis, or
charm background in the beauty analysis.
The light-quark jet background is assumed to be reducible to the per cent level, and 
the charm-quark jet background is assumed to be 10\,\%.
The background contaminations, as well as the tagging efficiencies,
affect primarily the statistical uncertainty of the measurment, which
for the assumed $100$\,fb$^{-1}$ is relevant only in some edges of the
phase space, as the figures illustrate for all three distribution.

In addition, an uncorrelated systematic uncertainty is assumed 
in the simulated strange and beauty quark measurements of $3$\,\%
while for charm a $2$\,\% error is used. These errors determine
the measurement uncertainties in almost the full kinematic range. At higher
$Q^2$ and $x$, these increase, for example to $10,~5$ and $7$\,\% for
$xs,~xc$ and $xb$, respectively, at $x \simeq 0.1$ and $Q^2 \simeq 10^5$\,GeV$^2$.
As is specified in the figures, the $x$ and $Q^2$ ranges of these measurements 
extend over $3,~5$ and $4$ orders of magnitude for $s,~c$ and $b$. The coverage
of very high $Q^2$ values, much beyond $M_Z^2$, permits to determine
the $c$ and $b$ densities probed in $\gamma Z$ interference interactions for the
first time. At HERA, $xs$ was not directly accessible while pioneering
 measurements of $xc$ and $xb$ could be performed~\cite{H1:2018flt}, albeit
 in a smaller range and less precise than shall be achieved with the LHeC. 
 These measurements, as discussed below
and in much detail in the 2012 LHeC CDR~\cite{AbelleiraFernandez:2012cc}, 
are of vital importance for the development of QCD and for the interpretation of precision LHC data.

%
\section{Parton Distributions from the LHeC}
\label{sec:LHeCPDF}
\subsection{Procedure and Assumptions}
\label{sec:qcdfit}
In this section, PDF constraints from the simulation of LHeC
inclusive NC and CC cross section measurements and heavy quark 
densities are investigated. The analysis closely follows the one for HERA as presented above.

The expectations on PDFs for the ``LHeC  inclusive'' dataset,
corresponding to the combination of datasets D4+D5+D6+D9, are presented,
see Tab.\,\ref{tab:dsets}.
These datasets have the highest sensitivity to general aspects of PDF phenomenology.
Since the data are recorded concurrently to the HL-LHC
operation they will become available only after the end of the HL-LHC.
Therefore, these PDFs will be valuable for re-analysis or re-interpretation of (HL-)LHC data,
and for further future hadron colliders.

In order that LHeC will be useful already during the lifetime of the HL-LHC,
it is of high relevance that the
LHeC can deliver PDFs of transformative precision already on a short timescale.
Therefore, in the present study particular attention is paid to PDF constraints
that can be extracted from the first $50$\,$\text{fb}^{-1}$ of
electron-proton data, which corresponds to the first three years of LHeC operation.
The dataset is labelled D2 in Tab.\,\ref{tab:dsets} 
and also referred to as ``LHeC~1st~run'' in the following. 

Already the data recorded during the initial weeks of
data taking will be highly valuable and impose new PDF constraints.
This is because already the initial instantaneous luminosity will be
comparably high, and the kinematic range is largely extended in
comparison to the HERA data.
These initial analyses will provide the starting point for the LHeC PDF programme.
It may be recalled that the HERA I data period (1992-2000) provided just $0.1$\,fb$^{-1}$ of
data which was ample for discovering the rise of $F_2$ and of $xg$ towards small $x$
at low $Q^2$, and still today these data form the most important ingredient to
the combined legacy HERA data~\cite{Abramowicz:2015mha}.
The sets in Tab.\,\ref{tab:dsets} comprise D1, with $5$\,fb$^{-1}$, still
the tenfold of what H1 collected in 15 years, and D3, which resembles D2 but has the 
electron polarisation set to zero. 

Additional dedicated studies of the impact of $s, c, b$ data on the
PDFs are then also presented, based on $10$\,$\text{fb}^{-1}$ of $e^-p$ simulated data.
%
Further important PDF constraints that would be provided by  measurements of
$F_L$  and
jets  
are not considered in the present study. These remarks are significant in that 
they mean one has to be cautious when comparing the LHeC PDF potential 
with some global fits: $F_L$ will resolve the low $x$ non-linear parton interaction issue, 
see Sect.~\ref{sec:FL}, and jets 
are important to pin down the gluon density behaviour at large $x$
as well as providing a precision measurement of $\alpha_s$, Sect.\,\ref{sec:alphas}.

To assess the importance of different operating conditions, the impact of datasets with:
differing amounts of integrated luminosity (D1 vs.\ D4);
positrons (D6 vs.\ D7); and with different polarisation states for the
leptons (D3 vs.\ D8) are also considered.

In the following, PDF fits are presented, which make use of the
simulated data and NLO QCD predictions. 
Fits in NNLO have been performed as a cross check.
The analysis follows closely the HERAPDF procedure
(c.f.\ Sect.~\ref{sec:hera} and Ref.~\cite{Abramowicz:2015mha}).
The parametric functions in Eqs.\,\eqref{eqn:pdf} and \eqref{eqn:gpdf}
are used, and the parameterised PDFs are the valence distributions
$xu_v$ and  $xd_v$,  the gluon distribution $xg$, and the
$x\bar{U}$ and $x\bar{D}$ distributions, using $x\bar{U} = x\bar{u}$ and
$x\bar{D} = x\bar{d} +x\bar{s}$.
In total the following $14$ parameters are set free for the nominal
fits: $B_g$, $C_g$, $D_g$, $B_{uv}$, $C_{uv}$, $E_{uv}$, $B_{dv}$, $C_{dv}$,
$A_{\bar{U}}$, $B_{\bar{U}}$, $C_{\bar{U}}$, $A_{\bar{D}}$,
$B_{\bar{D}}$, $C_{\bar{D}}$.
These fit parameters are similar to HERAPDF2.0, albeit to
some extent more flexible due to the stronger constraints from the
LHeC.
Note, the $B$ parameters for $u_v$ and $d_v$, and the
$A$ and $B$ parameters for $\bar{U}$ and $\bar{D}$
are fitted independently, such that the up and down valence and sea
quark distributions are uncorrelated in the analysis, whereas for
HERAPDF2.0 $x\bar{u} \to x\bar{d}$ as $x \to 0$ is imposed.
The other main difference is that no negative gluon term has been
included, i.e.\ $A'_g =0$ but $D_g\neq0$.

This ansatz is natural to the
extent that the NC and CC inclusive cross sections determine
the sums of up and down quark distributions,
and their anti--quark distributions, as the four independent
sets of PDFs, which may be transformed to the ones chosen
if one assumes $u_v = U -\overline{U}$ and $d_v = D - \bar{D}$,
i.e. the equality of anti-- and sea--quark distributions of given
flavour.
For the majority of the QCD fits presented here,
the strange quark distribution  at $Q^2_0$ is assumed to be a constant fraction of $\bar{D}$,
$x\bar{s}= f_s  x\bar{D}$ with $f_s=0.4$ as for HERAPDF, while
this assumption is relaxed for the fits including simulated $s,c,b$ data.

Note, that the prospects presented here are illustrations for a different
era of PDF physics, which will be richer and deeper than one may be
able to simulate now.
For instance, without real data one cannot determine the actual
parameterisation needed for the PDFs.
In particular the low $x$
kinematic region was so far unexplored and the simulated data relies on a
simple extrapolation of current PDFs, and no reliable data or model is  available
that provides constraints on this region\,\footnote{ 
  It is expected that real LHeC data, and also the inclusion of
  further information such as $F_\text{L}$,
  will certainly lead to a quite different optimal parameterisation ansatz than
  was used in the present analysis. Though, it has been checked that
  with a more relaxed set of parameters,
  very similar results on the PDF uncertainties are obtained, which
  justifies the size of the prospected PDF uncertainties.
}.
The LHeC data explores new corners of phase space with high precision,
and therefore it will have a great potential, much larger than HERA had, to determine the
parameterisation. 
As another example, with LHeC data one can directly derive relations
for how the valence quarks are determined with a set of NC and CC
cross section data in a redundant way, since the gluon distribution at
small $x$ can be determined from the $Q^2$ derivative of $F_2$ and
from a measurement of $F_L$.
The question of the optimal gluon parameterisation may then be settled
by analysing these constraints and not by assuming some specific behaviour of a given fit.

Furthermore, the precise direct determinations of $s$, $c$ and $b$
densities with measurements of the impact parameter of their decays, 
will put the treatment of heavy flavours in PDF analyses on a new
level.
The need for the phenomenological introduction of the $f_s$ factor
will disappear and the debate on the value of fixed and variable heavy
flavour schemes will be settled.

\subsection{Valence Quarks}
\label{sec:udval}
Since the first measurements of DIS physics, 
it had been proposed to identify partons with
quarks and to consider the proton to consist of valence quarks together with
``an indefinite number of ($q\bar{q}$) pairs"~\cite{Kuti:1971ph}. 50 years later there
are still basic questions unanswered about the behaviour of valence quarks, such as
the $d_v/u_v$ ratio at large $x$, and PDF fits struggle to resolve the flavour 
composition and interaction dynamics of
the sea. The LHeC is the most suited machine to resolve these challenges.

\begin{figure}[!th]
  \centering
  \includegraphics[width=0.45\textwidth]{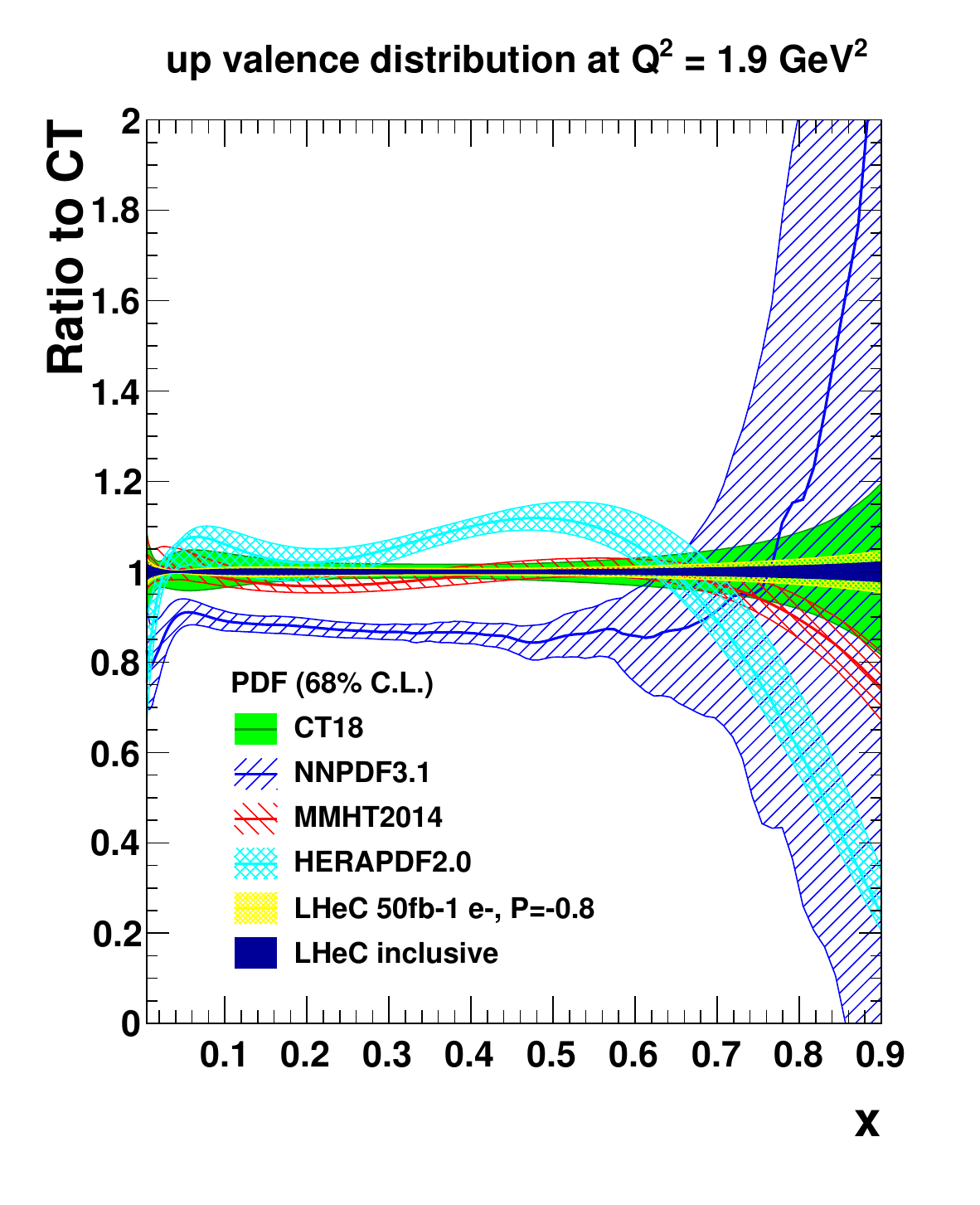}
  \includegraphics[width=0.45\textwidth]{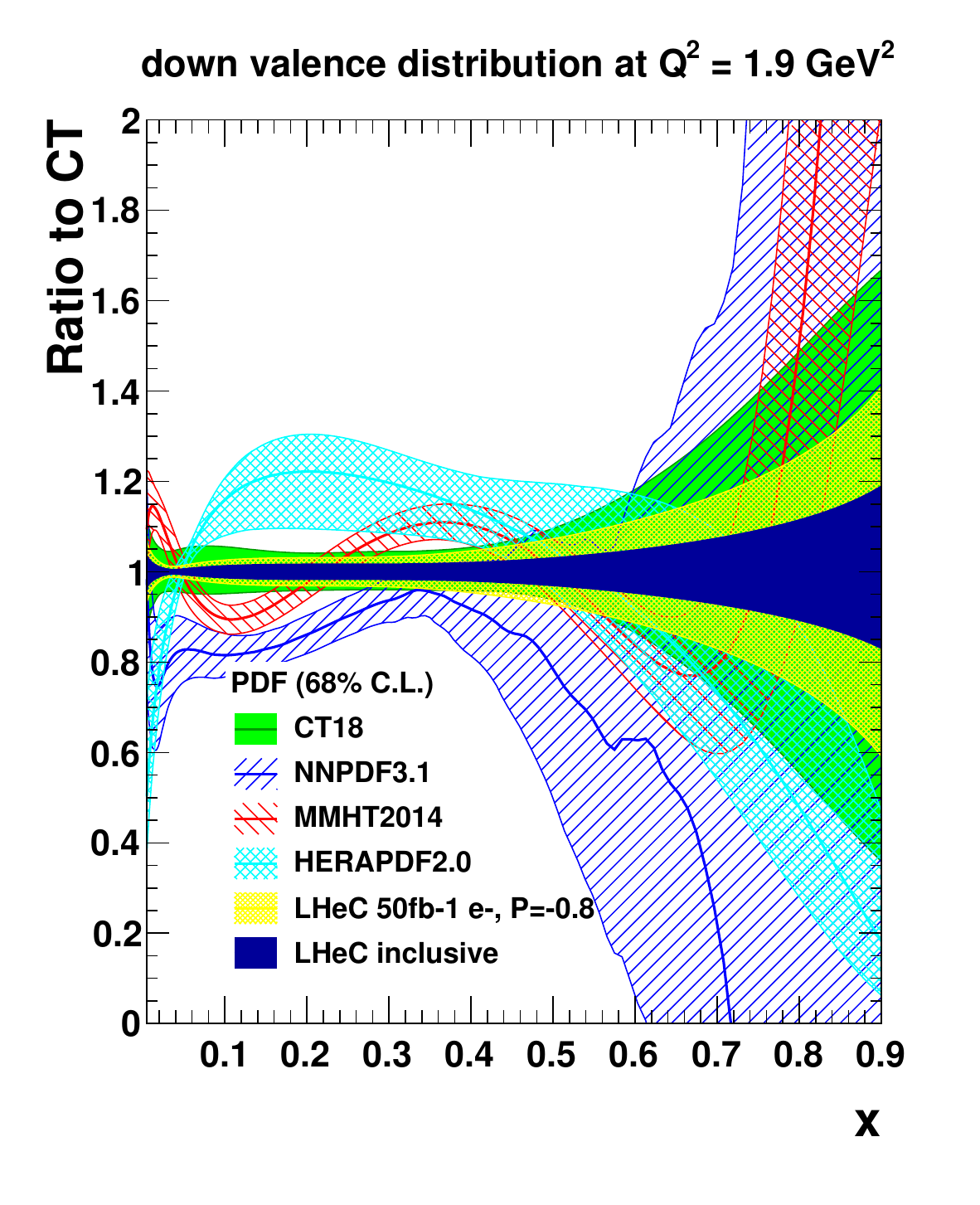}
  \vspace{-0.cm}
  \caption{
Valence quark distributions at $Q^2=1.9$\,GeV$^2$ as a function of $x$,
displayed as a ratio to the CT18~\cite{Hou:2019efy} PDF set. 
The yellow band corresponds to the ``LHeC 1st run" PDFs (D2),
while the dark blue shows the final ``LHeC inclusive" PDFs based on the data sets (D4+D5+D6+D9),
as described in Sect.~\ref{sec:qcdfit}.
For the purposes of illustrating the improvement to the uncertainties
more clearly, the central values of the LHeC PDFs have been scaled to the
CT18 PDF, which itself is displayed by the green band.
}
\label{fig:valence}
\end{figure}

The precision that can be expected for the valence quark distributions from the LHeC is illustrated
in Fig.~\ref{fig:valence}, and compared to a selection of recent PDF sets.
Today, the precision of the valence quark distributions, particularly
at large $x$, is fairly limited as it can be derived from the Figure.
This is due to the limited integrated luminosity of the HERA data, challenging
systematics that rise proportional to $1/(1-x)$, and to
uncertainties attributed to nuclear corrections.
At lower values of $x$ the valence quark distributions are very
small compared to the sea quarks and cannot be separated 
easily from these.

The $u$ valence distribution is nowadays known with higher precision than the $d$ valence, since it enters
the calculation of $F_2$ with a four-fold higher weight because of the different
electric charges of the quarks.
Nevertheless, a substantial improvement in $d_v$ by the LHeC is also visible, because
the relative weight of $d_v$ to $u_v$ is changing favourably towards 
the down quark due to the influence of weak NC and CC interactions at high $Q^2$
where the LHeC is providing very accurate data.
The strong constraints to the highest $x$ valence distributions
are due to the very high integrated luminosity.
Note, at the HL-LHC, albeit its high integrated luminosity, the highest $x$ are
there only accessible as convolutions with partons at lower $x$, and
those can therefore not be well constrained.

Note that the ``LHeC 1st run'' PDF, displayed by the yellow band in
Fig.~\ref{fig:valence}, includes only electron, i.e.\ no positron, data.
In fact, from the $e^{\pm}p$ cross section differences access to valence quarks at low $x$
can be obtained.
As has already been illustrated in the 
CDR from 2012~\cite{AbelleiraFernandez:2012cc} the sum of $2 u_v +d_v$
may be measured directly with the NC $\gamma Z$ interference structure
function $xF_3^{\gamma Z}$ down to $x \simeq 10^{-4}$ with very good precision.
Thus the LHeC will have a direct access to the valence quarks at small $x$.
This also tests the assumption of the equality of sea- and anti-quark densities
which  if different would cause  $xF_3^{\gamma Z}$ to rise towards small $x$.

As becomes evident in Fig.~\ref{fig:valence} there is a striking difference 
and even contradiction between the estimates of the uncertainties
of the parton distributions between the various fit groups. This is due to
different fit technologies but as well a result of different choices of data and
assumptions on the $d/u$ ratio. Such major uncertainties would be resolved
by the LHeC.


The precise determinations of the valence quark distributions at large $x$ have
strong implications for physics at the HL-LHC, in particular for BSM
searches.
%
\begin{figure}[!th]
  \centering
\includegraphics[width=0.8\textwidth]{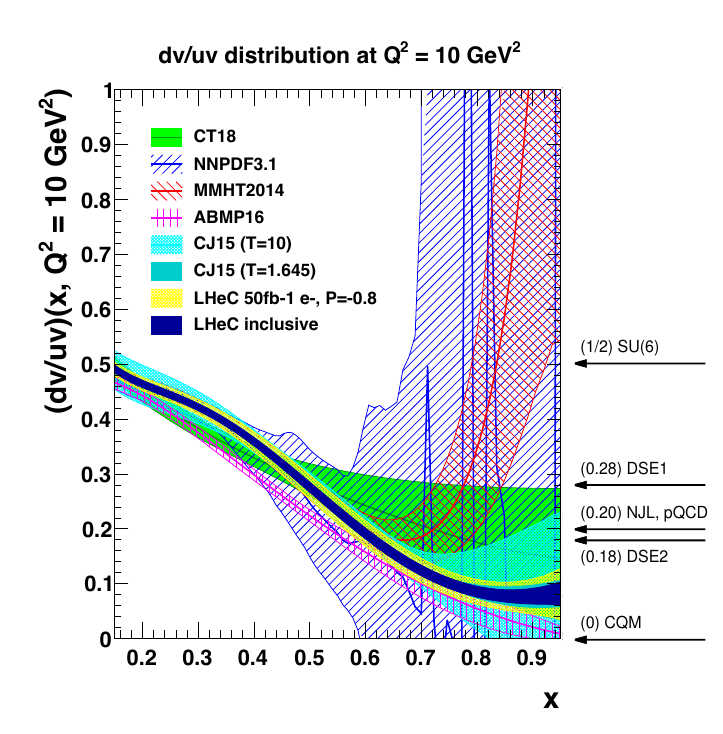}
  \vspace{-0.3cm}
  \caption{The $d_v/u_v$ distribution at $Q^2=10$\,GeV$^2$ as a function of $x$.
      The yellow band corresponds to the ``LHeC 1st run'' PDFs (D2),
      while the dark blue shows the final ``LHeC inclusive'' result.
      Both LHeC PDFs shown are scaled to the central value of CJ15. 
  }
\label{fig:dovu}
\end{figure}
The precise determinations of the valence quarks will resolve the
long standing mystery of the behaviour of the $d/u$ ratio at large $x$, see Fig.~\ref{fig:dovu}.
As exemplarily shown in Fig.~\ref{fig:dovu}, there are currently conflicting theoretical pictures for the central value of the $d/u$ ratio, albeit the large uncertainty bands of the different PDF sets mainly overlap.
As of today, the constraints from data are inconclusive statistically and also suffer from large
 uncertainties from the use of DIS data on nuclear targets, which therefore cause those large uncertainties.

%

\subsection{Light Sea Quarks}
\label{sec:lightsea}
Our knowledge today about the anti-quark distributions is fairly poor
and uncertainties are very large at smaller values of $x$, and also
at the highest $x$.
In particular, at low $x$ the size of the anti-quark PDFs are large and they
contribute significantly to precision SM measurements at the HL-LHC.
At high $x$, sea and valence need to be properly distinguished and accurately
be measured for reliable BSM searches at high mass.

Our knowledge about the anti-quark PDFs will be changed completely with
LHeC data. Precise constraints are obtained with inclusive NC/CC DIS data
despite the relaxation of any assumptions in the fit ansatz that would
force $\bar{u} \to \bar{d}$ as $x \to 0$, as it is present in other PDF
determinations today. At smaller $Q^2$ in DIS one measures
essentially $F_2 \propto 4 \bar{U} + \bar{D}$. Thus, at HERA, with limited precision at high $Q^2$,
one could not resolve
the two parts, and neither will that be possible at any other lower energy
$ep$ collider which cannot reach small $x$.
 At the LHeC, in contrast, the CC DIS cross sections are measured very
well down to $x$ values even below $10^{-4}$, and in addition there
are strong weak current contributions to the NC cross section which probe
the flavour composition differently than  the photon exchange does.
This enables this distinction of $\bar{U}$ and $\bar{D}$ at the LHeC.

\begin{figure}[!th]
  \centering
  \includegraphics[width=0.4\textwidth]{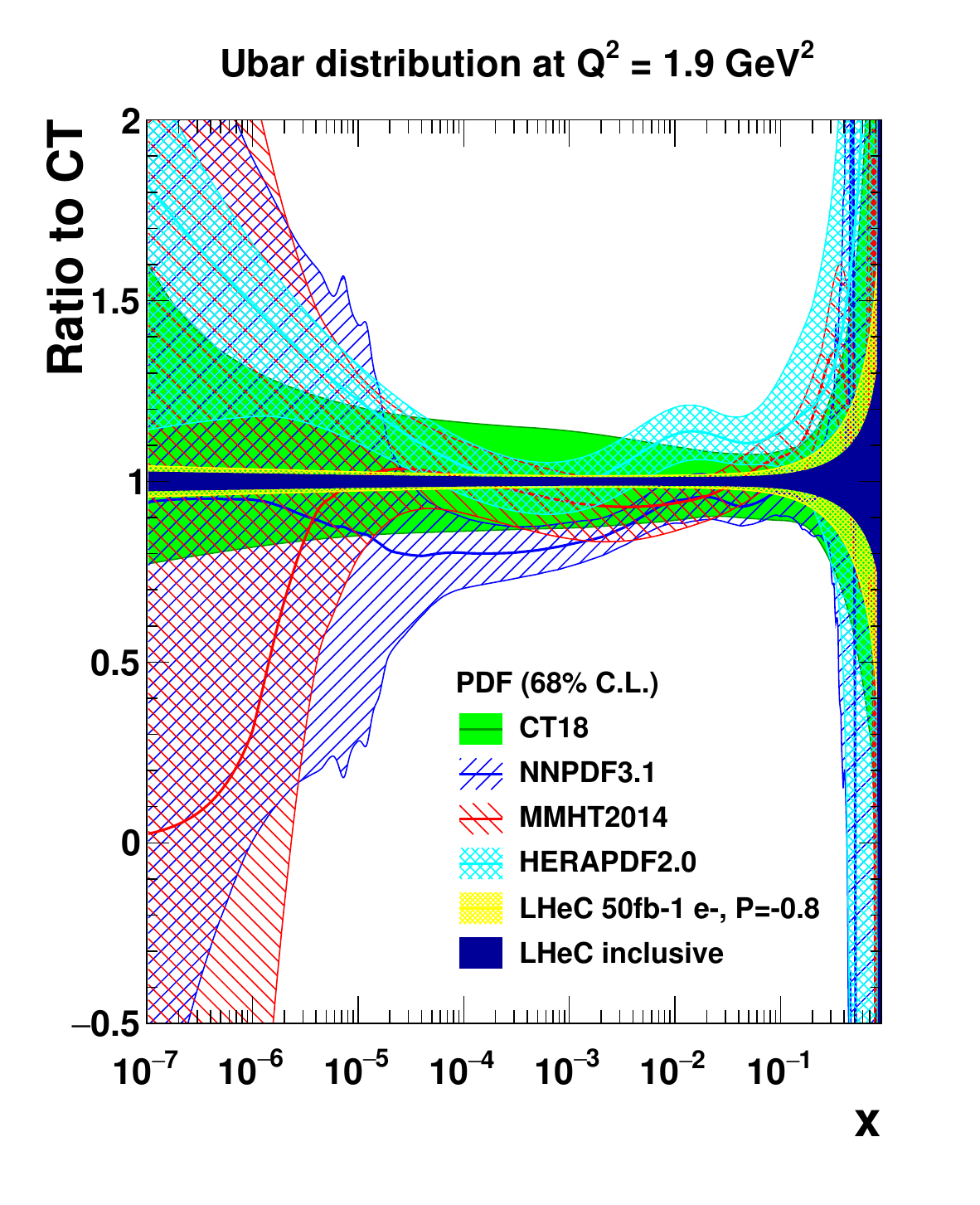}
  \includegraphics[width=0.4\textwidth]{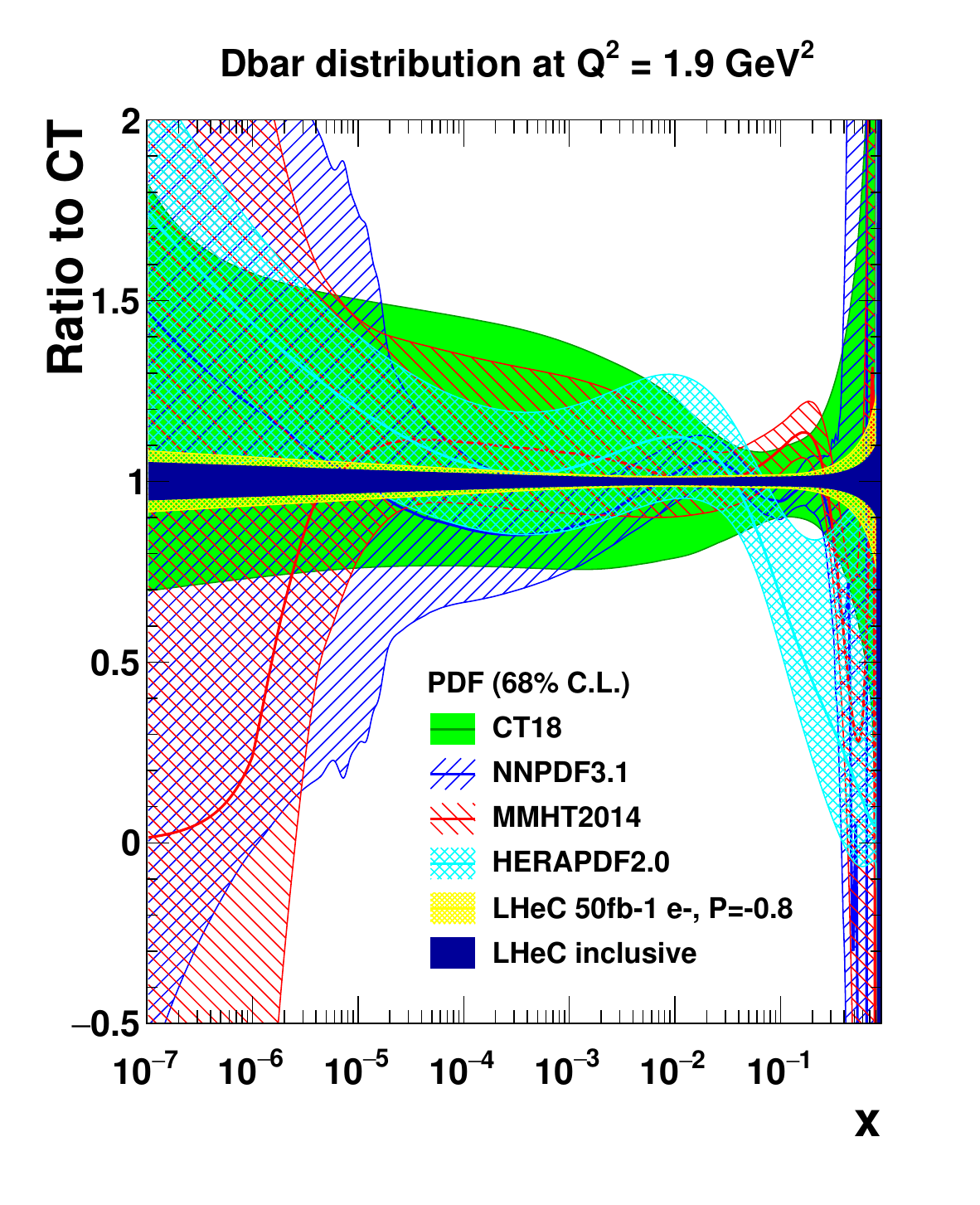}
  \vspace{-0.3cm}
  \caption{Sea quark distributions at $Q^2=1.9$\,GeV$^2$ as a function of $x$,
    displayed as the ratio to the CT18 PDF set.
    The yellow band corresponds to the ``LHeC 1st run" PDFs (D2),
    while the dark blue shows the final ``LHeC inclusive" PDFs (D4+D5+D6+D9),
    as described in the text.
    Both LHeC PDFs shown are scaled to the central value of CT18.
  }
\label{fig:sea}
\end{figure}

\begin{figure}[!th]
  \centering
  \includegraphics[width=0.4\textwidth]{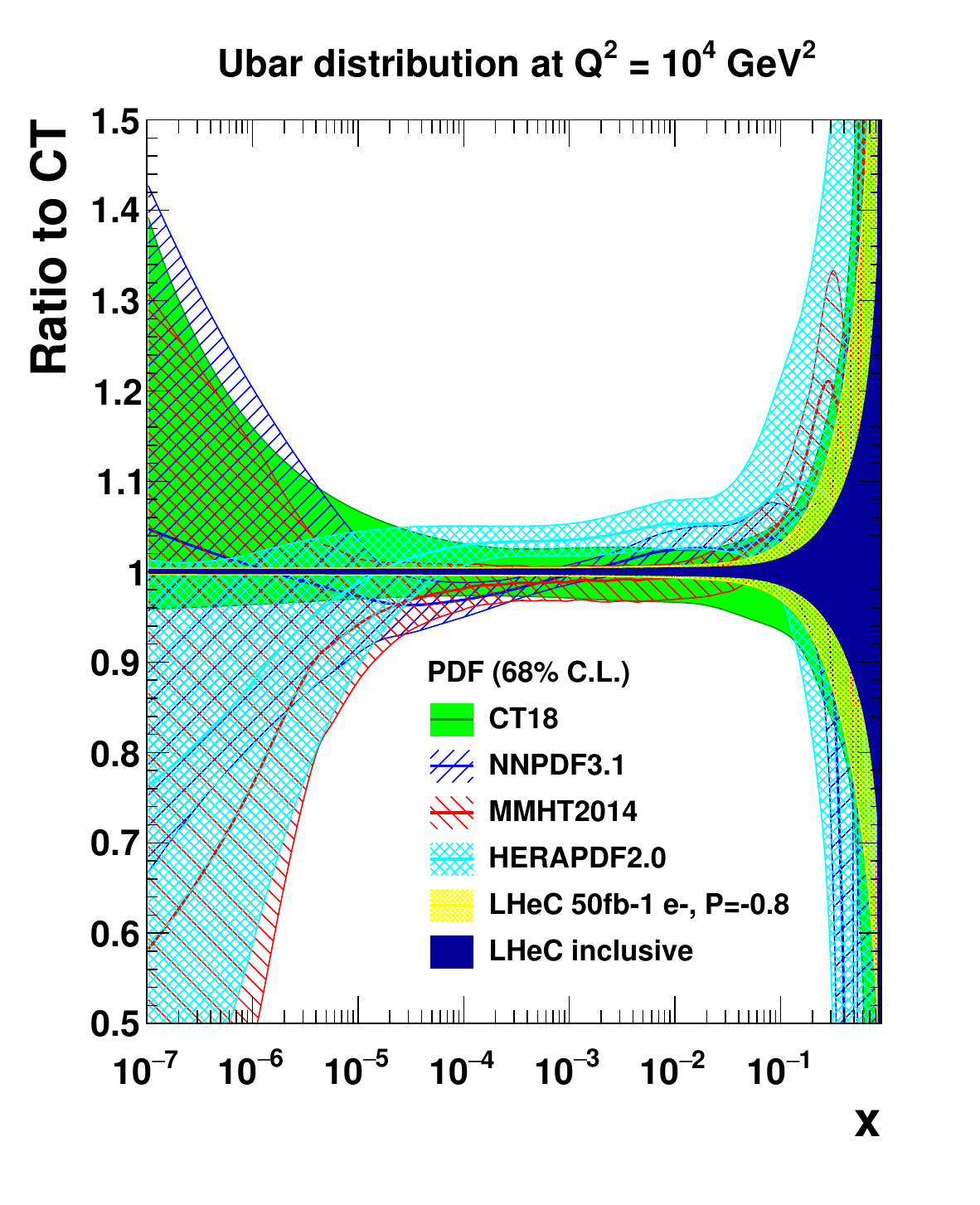}
  \includegraphics[width=0.4\textwidth]{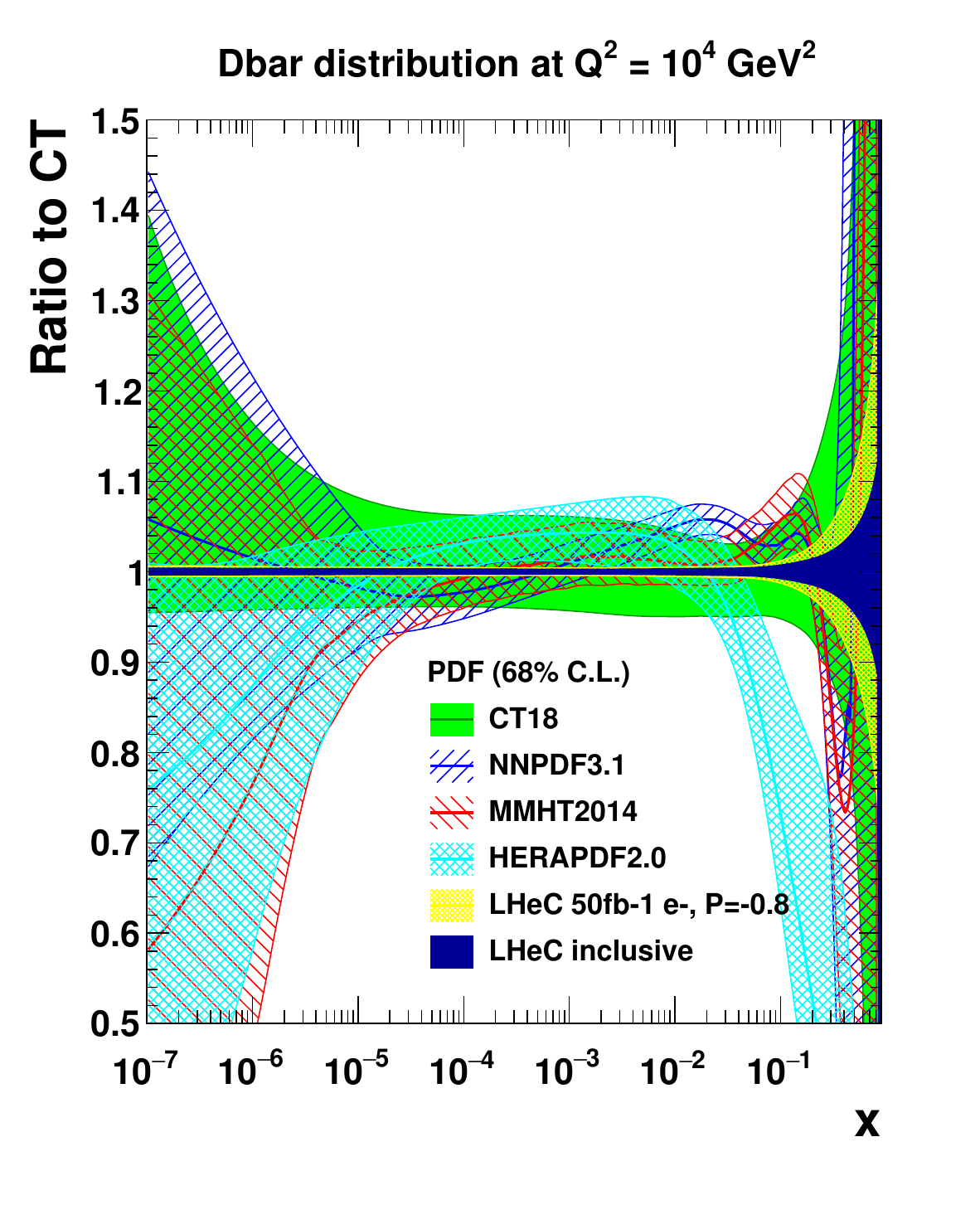}
  \vspace{-0.3cm}
  \caption{Sea quark distributions at $Q^2=10^4$\,GeV$^2$ as a function of $x$,
    displayed as the ratio to the CT18 PDF set.
    The yellow band corresponds to the ``LHeC 1st run" PDFs (D2),
    while the dark blue shows the final ``LHeC inclusive" PDFs (D4+D5+D6+D9),
    as described in the text.
    Both LHeC PDFs shown are scaled to the central value of CT18.
  }
\label{fig:sea10000}
\end{figure}

The distributions of $\bar{U}$ and $\bar{D}$ for the PDFs from the 1st
run and the ``LHeC inclusive data" are shown in Figs.~\ref{fig:sea} 
and \ref{fig:sea10000} for $Q^2=1.9$\,GeV$^2$ and $Q^2=10^4$\,GeV$^2$, 
respectively, and compared to present PDF analyses. One observes
a striking increase in precision for both $\bar{U}$ and $\bar{D}$ 
which persists from low to high scales. The relative uncertainty
is large at high $x \geq 0.5$. However, in that region the sea-quark contributions 
are already very tiny. In the high $x$ region one recognises the value of the 
full LHeC data sample fitted over the initial one while the uncertainties
below $x \simeq 0.1$ of both the small and the full data sets are 
of comparable, very small size.


\subsection{Strange Quark}
\label{sec:strange}
%
The determination of the strange PDF has generated significant controversy 
in the literature for more than a decade. 
Fixed-target neutrino DIS measurements \cite{Seligman:1997mc,Tzanov:2005kr,Onengut:2005kv,Berge:1989hr,Samoylov:2013xoa} typically prefer a strange PDF that is 
roughly half of the up and down sea distribution; $\kappa=(s+\bar{s})/(\bar{u}+\bar{d})\sim 0.5$.
The recent measurements from the
LHC~\cite{Aad:2012sb,Chatrchyan:2013uja,Aad:2014xca,Aaboud:2016btc} and related
studies~\cite{Alekhin:2017olj,Cooper-Sarkar:2018ufj} suggest  a
larger strange quark distribution, that may potentially even be larger than the up
and down sea quarks. The $x$ dependence of $xs$ is essentially unknown, and it may differ
from that of $x\bar{d}$, or $x(\bar{u}+\bar{d})$, by more than a normalisation factor. A recent paper ascribes the strange enhancement to a suppression of the anti-down distribution related
to suspected parameterisation effects and the behaviour of the ratio $d/u$ for $x \to 1$~\cite{Alekhin:2018dbs}. Apparently, a direct measurement
of $xs(x,Q^2)$ and the resolution of the complete light-quark
structure of the proton is required, which is a fundamental goal
of the LHeC.

The precise knowledge of the strange quark PDF is of high relevance,
since it provides a significant contribution to 
\emph{standard candle} measurements at the HL-LHC, such as $W$/$Z$ production, and it
imposes a significant uncertainty on the $W$ mass measurements at the LHC.
The question of light-sea flavour `democracy' is of principle relevance for QCD and 
the parton model. For the first time,  
as has been presented in Sect.\,\ref{sec:hquarks}, $x\bar{s}(x,Q^2)$
can be accurately measured, namely through the charm tagging $W s \to c$ reaction 
in CC $e^-p$ scattering at the LHeC. The inclusion of the CC charm data
in the PDF analysis will settle the question of how strange the strange quark 
distribution really is~\footnote{The provision of positron-proton data will enable very interesting
tests of charge symmetry, i.e. permit to search for a difference between the
strange and the anti-strange quark densities. This has not been studied in this paper.}.
This prospect has been analysed within the LHeC fit framework here introduced
and as well studied in detail in a profiling analysis using \emph{xFitter}. Both analyses yield rather
compatible results and are presented in the following.
  

In the standard LHeC fit studies, the parameterised PDFs are
the four quark distributions $xu_v$, $xd_v$, $x\bar{U}$, $x\bar{D}$ and $xg$
(constituting a 4+1 parameterisation), as the inclusive NC and CC data
determine only the sums of the up and down 
quark and anti-quark distribution, as discussed previously.
The strange quark PDF is then assumed to be a constant fraction of
$x\bar d$.

With the strange quark data available, the LHeC PDF fit parameterisations can be extended
to include $xs=x\bar{s}$, parameterised as $A_s x^{B_s} (1-x)^{C_s}$~\footnote{
It is worth mentioning that the $W,Z$ data~\cite{Aad:2012sb} essentially determine
only a moment of $xs$ at $x \sim 0.02$, not the $x$ dependence. Therefore, 
in analyses of HERA and ATLAS data such as Ref.\,\cite{Cooper-Sarkar:2018ufj},
there is no determination attempted of the relevant parameter, $B_s$, which instead is
set equal to $B_{\bar{d}}$. The kinematic dependence of $xs$ is basically not determined
by LHC data while the hint to the strange being unsuppressed has been persistent.}.
For the fits presented in the following, the $\bar{d}$ and $\bar{s}$
are treated now separately,
and therefore a total of five quark distributions are parameterised
($xu_v$, $xd_v$, $x\bar{U}$, $x\bar{d}$, $x\bar{s}$)
as well as $g$. This provides  a 5+1 parameterisation, and
the total number of free parameters of the PDF fit then becomes 17.

\begin{figure}[!th]
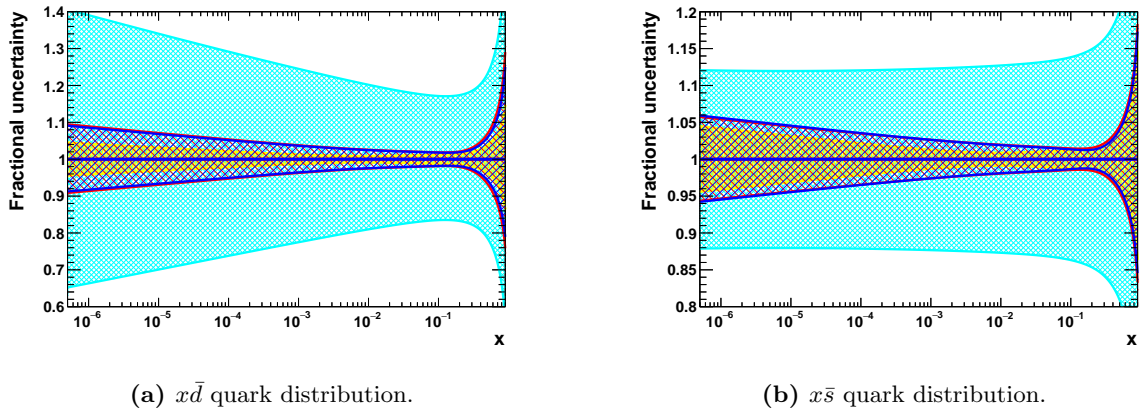

  \centering
  \begin{subfigure}{0.48\textwidth}
    \includegraphics[width=\linewidth]{claireHQ_xdbar_vs_x_log_restrictx.pdf}
    \caption{$x\bar{d}$ quark distribution.}
  \end{subfigure}
  \hspace*{\fill}
  \begin{subfigure}{0.48\textwidth}
    \includegraphics[width=\linewidth]{claireHQ_xsbar_vs_x_log_restrictx.pdf}
    \caption{$x \bar s$ quark distribution.}
  \end{subfigure}
  \caption{
    PDF uncertainties at $Q^2=1.9\,\GeVsq$ as a function of $x$ 
    for the $\bar{d}$ and $\bar{s}$ distributions. 
    The yellow band displays the uncertainties of the nominal ``LHeC
    inclusive'' PDF, which was obtained in a 4+1 PDF fit.
    From the same dataset, results of the more flexible 5+1 fit (see text) are
    displayed as a cyan band.
    The red band displays the results, when in addition an LHeC
    measurement of the $\bar{s}$ quark density is included.
    When even further including LHeC measurements of $F_2^c$ and
    $F_2^b$, the PDF fits yields uncertainties as displayed by the
    blue band.
  }
  \label{fig:hq}
\end{figure}

Results of the 5+1 PDF fits are shown in Fig.~\ref{fig:hq},
where fits to inclusive NC/CC DIS data are displayed as reference
(both for the 4+1 and 5+1 ansatz) and the fits where in
addition strange density measurements and even further measurements of
$F_2^{c,b}$ are considered. 
As expected, the uncertainties of the 5+1 fit to the inclusive DIS data, especially
on the $\bar{d}$ and $\bar{s}$ distributions (c.f.\ Fig.~\ref{fig:hq}), become substantially larger in comparison to the respective
4+1 fit, since the $\bar{d}$ and $\bar{s}$ distributions are treated now separately.
This demonstrates that the inclusive DIS data alone does not have the
flavour separating power to determine the individual distributions very precisely.

When including an LHeC measurement of the $\bar{s}$ quark density
based on $10\,\text{fb}^{-1}$ of $e^-p$ data,
the uncertainties on the $\bar{d}$ and $\bar{s}$ PDFs become
significantly smaller.
By chance, those uncertainties are then comparable to the 
4+1 fit in which $ x \bar s$ is linked to $x \bar d$ by a constant fraction.

The constraints from a measurement of charm quark production cross
sections in charged current DIS have also been studied in a profiling
analysis using \emph{xFitter}~\cite{Abdolmaleki:2019acd}. 
The treatment of  heavy quark production to higher orders in pQCD is 
discussed extensively in this paper.
At leading-order QCD,  the subprocess under consideration is $Ws\to c$,
where the $s$ represents an intrinsic strange quark. 
%
\begin{figure}[!thb]
  \centering
    \includegraphics[width=0.45\textwidth]{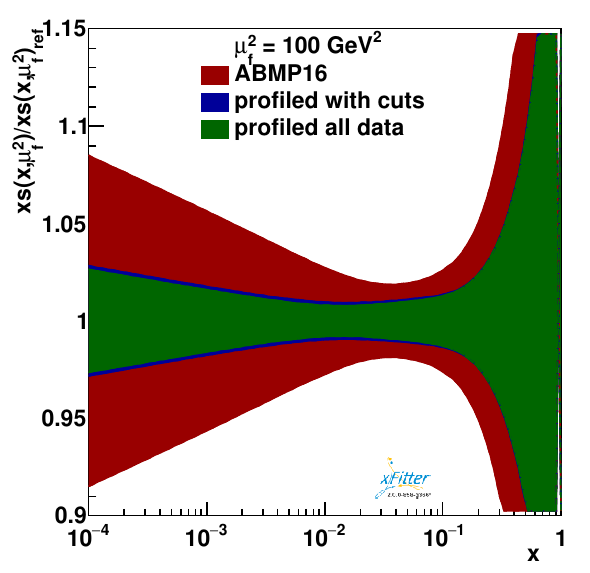}
  \includegraphics[width=0.45\textwidth]{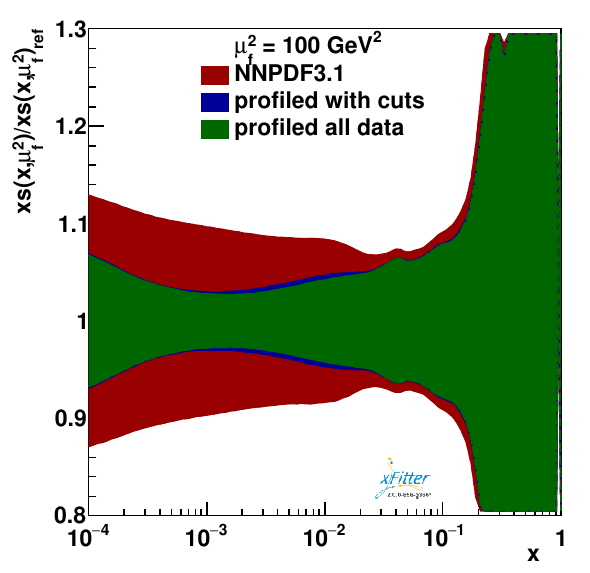}
  \caption{
    Constraints on the strange quark PDF $xs$
     using simulated data for charged-current
    production of charm quarks at the LHeC, from a profiling
    study~\cite{Abdolmaleki:2019acd}
    using the ABMP16 (left) and the NNPDF3.1 (right) PDF sets. 
    The red band displays the nominal PDF uncertainties,
    and the green and blue bands the improved uncertainties due to the
    LHeC strange quark data.
  }
  \label{fig:pdfs_sCombo}
\end{figure}
Fig.~\ref{fig:pdfs_sCombo} displays the tight constraints obtained
for the strange PDF when using the LHeC pseudo-data for the CC
charm production channel.
The results of this profiling analysis, both when based on the ABM16 and the NNPDF3.1 
PDF sets, and of the direct
fit presented above, are very similar, reaching about $3-5$\,\% precision for
$x$ below $\simeq 0.01$

In a variation of the study~\cite{Abdolmaleki:2019acd}, 
a large reduction of uncertainties is already observed when restricting 
the input data to the kinematic range where the differences between
the different heavy flavour schemes (VFNS and FFNS) are not larger than
the present PDF uncertainties.
This further indicates that the PDF constraints are stable and
independent of the particular heavy-flavour scheme.

It may thus be concluded that the LHeC, through high luminosity, energy and
precise kinematic reconstruction, will be able to solve a long standing question about
the role of the strange-quark density in the proton, and its integration
into a consistent QCD treatment of parton dynamics.

\subsection{Heavy Quarks}
\label{sec:HQ}
One of the unsolved mysteries of the Standard Model is the existence of three generations of quarks and leptons. The strongly interacting fermion sector contains altogether six quarks with masses differing by up to  five orders of magnitude.  This hierarchy of masses is on one hand a challenge to explain, on the other hand it offers a unique opportunity to explore dynamics at a variety of different scales and thus develop different facets of the strong interaction. While the light quarks at low scales are non-perturbative and couple strongly, the heavier quarks charm, bottom and top are separated from the soft sea by their masses and thus can serve as a suitable additional probe for the soft part of QCD. 

There are a number of deep and unresolved questions that can be posed in the context of the proton structure: what is the individual contribution of the different quark flavours to the structure functions?; are heavy quarks like charm and bottom radiatively generated or is there also an intrinsic heavy quark component in the proton?; to what extent do the universality and factorisation theorems work in the presence of heavy quarks?  It is therefore imperative to be able to perform precise measurements of each individual quark flavour and their contribution to the proton structure. The LHeC
is the ideal place for these investigations because it resolves the complete  composition
of the proton flavour by flavour. In particular, as shown in Sect.\,\ref{sec:hquarks}, the LHeC provides
data on $F_2^{c}$ and $F_2^{b}$ extending over nearly $5$ and $6$ orders of magnitude in $x,Q^2$, respectively. These are obtained through charm and beauty tagging with high precision in NC
$ep$ scattering. A thorough PDF analysis of the LHeC data thus can be based on the inclusive 
NC/CC cross sections and tagged $s,~c,~b$ data. In addition, one may use DIS jets, here used
for the $\alpha_s$ prospective study (Sect.\,\ref{sec:alphas}) and low energy data, here analysed
for resolving the low $x$ dynamics with a precision measurement of $F_L$ (Sect.\,\ref{sec:FL}).
The current studies in this chapter therefore must be understood as indicative only as we 
have not performed a comprehensive analysis using all these data as yet~\footnote{This is
to be considered when one compares the precision of the inclusive
PDF fits with so-called global analyses, for example regarding the behaviour of $xg$ at large $x$.}.


The production of heavy quarks at HERA (charm and bottom)
is an especially interesting process as the quark mass
introduces a new scale ($m=m_{c,b}$) which was neither
heavy or light (see e.g.\ reviews~\cite{Behnke:2015qja,Zenaiev:2016kfl}).
Actually, the treatment of heavy quark mass effects is essential in 
PDF fits which include data from fixed target to collider energies 
and thus require the computation of physical cross sections over a large
range of perturbative scales $\mu^2$.
With these scales passing through (or close to) the thresholds for charm,
bottom and, eventually, top, precise computations demand the incorporation of heavy quark mass effects close to threshold, $\mu^2 \sim m^2$, and the resummation of collinear logarithms $\ln (\mu^2/m^2)$ at scales far above the threshold, $\mu^2 \gg  m^2$.
The first problem can be dealt with through the use of massive matrix elements for the generation of heavy quark-antiquark pairs but keeping a fixed number of parton densities (fixed flavour number schemes, FFNS).
On the other hand, the proper  treatment of resummation is achieved through
the use of variable flavour number schemes (VFNS) which consider an increasing number of massless parton species, evolved through standard DGLAP, when the scale is increased above heavy quark mass thresholds.
At present, calculations involving heavy
quarks in DIS in different schemes (generalised mass VFNS) with different numbers of active flavours participating
to DGLAP evolution are combined to derive an expression for the coefficient functions
which is valid both close to threshold, and far above it.
Such multi-scale problems are
particularly difficult, and numerous techniques were
developed to cope with this challenging problem~\cite{Aivazis:1990pe, Aivazis:1993kh, Aivazis:1993pi, Thorne:2000zd,Alekhin:2012ig, Alekhin:2013nda, Alekhin:2009ni, Forte:2010ta,Martin:2010db, Ball:2011mu}.
Additional complications, see e.g.\ Ref.~\cite{Ball:2015dpa}, arise when the possibility of a non-perturbative origin of heavy quark distributions is allowed above the heavy quark mass threshold - intrinsic heavy flavour.
The ABMP16 analysis~\cite{Alekhin:2017kpj} underlines that the available DIS data are compatible with solely an FFNS treatment
assuming that the heavy quarks are generated in the final state. 

  %
\begin{figure}[!th]
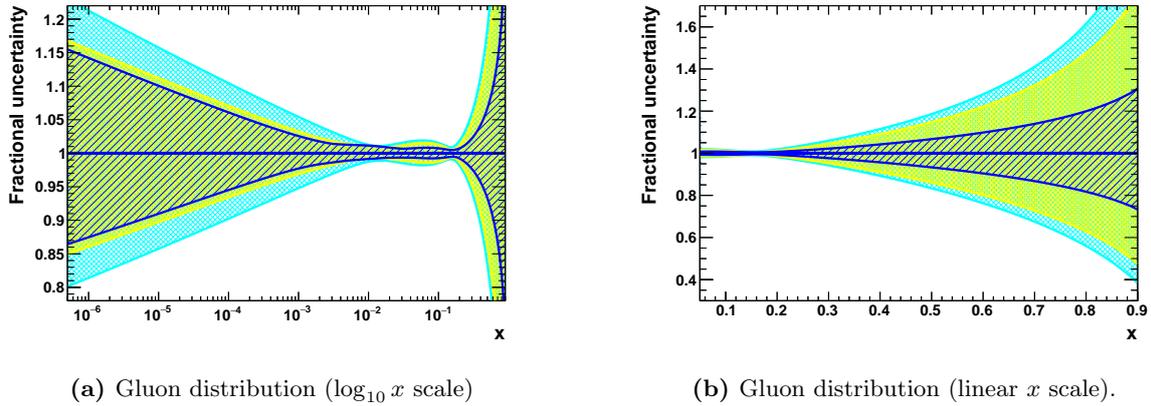

  \centering
  \begin{subfigure}{0.48\textwidth}
    \includegraphics[width=\linewidth]{claireHQ_xglu_vs_x_log_restrictx.pdf}
    \caption{Gluon distribution ($\log_{10}{x}$ scale)}
  \end{subfigure}
  \hspace*{\fill}
  \begin{subfigure}{0.48\textwidth}
    \includegraphics[width=\linewidth]{claireHQ_xglu_vs_x_lin.pdf}
    \caption{Gluon distribution (linear $x$ scale).}
  \end{subfigure}
  \caption{
    PDF uncertainties at $Q^2=1.9\,\GeVsq$ as a function of $x$ 
   to illustrate the constraints from additional heavy quark sensitive
    measurements at LHeC.
    Displayed is the gluon distribution on a logarithmic and linear
   scale.
    The yellow band illustrates the uncertainties of the nominal ``LHeC
    inclusive'' PDF, obtained in a 4+1 PDF fit.
    From the same dataset, results of the more flexible 5+1 fit (see text) are
    displayed as a cyan band.
    When further including LHeC measurements of $F_2^{c}$ and
    $F_2^{b}$, the PDF fits yields uncertainties as displayed by the
    blue band.
  }
  \label{fig:xghq}
\end{figure}

At the LHeC, as illustrated in Figs.\,\ref{fig:xcharm},\,\ref{fig:xbottom},
the large polar angle acceptance and the high centre-of-mass energy allow
heavy quark physics to be investigated from below threshold to almost $10^6$\,GeV$^2$.
The extended reach in comparison to HERA is dramatic.
This permits to comprehensively
explore the \emph{asymptotic} high energy limit where $m_{c,b}^2/Q^2\to  0$,
as well as the low energy \emph{decoupling} region $m_{c,b}^2/Q^2\sim 1$. 

For the PDF determination the obviously direct impact of the tagged charm 
and bottom data will be on the determination of $xc$ and $xb$, and the
clarification of their appropriate theoretical treatment. In addition, however,
there is a remarkable improvement achieved for the determination of the gluon density, see
Fig.\,\ref{fig:xghq}. The determination of $xg$ 
 will be discussed in much more detail in the following section.

These channels will also strongly improve the determination of the 
charm and bottom quark masses and  bring these uncertainties 
down to about 
$\delta m_{c(b)}\simeq 3(10)$\,MeV~\cite{AbelleiraFernandez:2012cc}~\footnote{
Such precision demands the availability of calculations with higher orders in pQCD, and those computations are already ongoing~\cite{Moch:2017uml,Herzog:2018kwj,Das:2019btv}.
Note than in PDF fits the heavy quark mass is an effective parameter that has to be related with the pole mass, see e.g.\ Ref.~\cite{Ball:2016qeg} and refs. therein.}.
These accuracies are crucial for eliminating the corresponding model uncertainties in the PDF fit.
Precision tagged charm and bottom data are
 also essential for the determination of the $W$-boson mass in $pp$, 
and the extraction of the Higgs $ \to c\bar c$ and $b \bar b$ couplings
 in $ep$, as is discussed further below.

\subsection{The Gluon PDF}
\label{sec:xgluon}
%
The LHeC, with hugely increased precision and extended kinematic range of DIS,
i.e.\ the most appropriate process to explore $xg(x,Q^2)$, can pin down the gluon
distribution much more accurately than it is known today.
This is primarily attributed to the huge kinematic range
and high precision of the measurement of $\partial F_2/\partial \ln Q^2$,
which at small $x$ is closely related to a direct measure of $xg$ .
The precision determination of the quark distributions, discussed previously,
also strongly constrains $xg$. Further sensitivity arises with the high-$y$ part
of the NC cross section which is controlled by the longitudinal structure function
as is discussed in Sect.\,\ref{sec:FL}.

\begin{figure}[!th]
  \centering
  \includegraphics[width=0.45\textwidth]{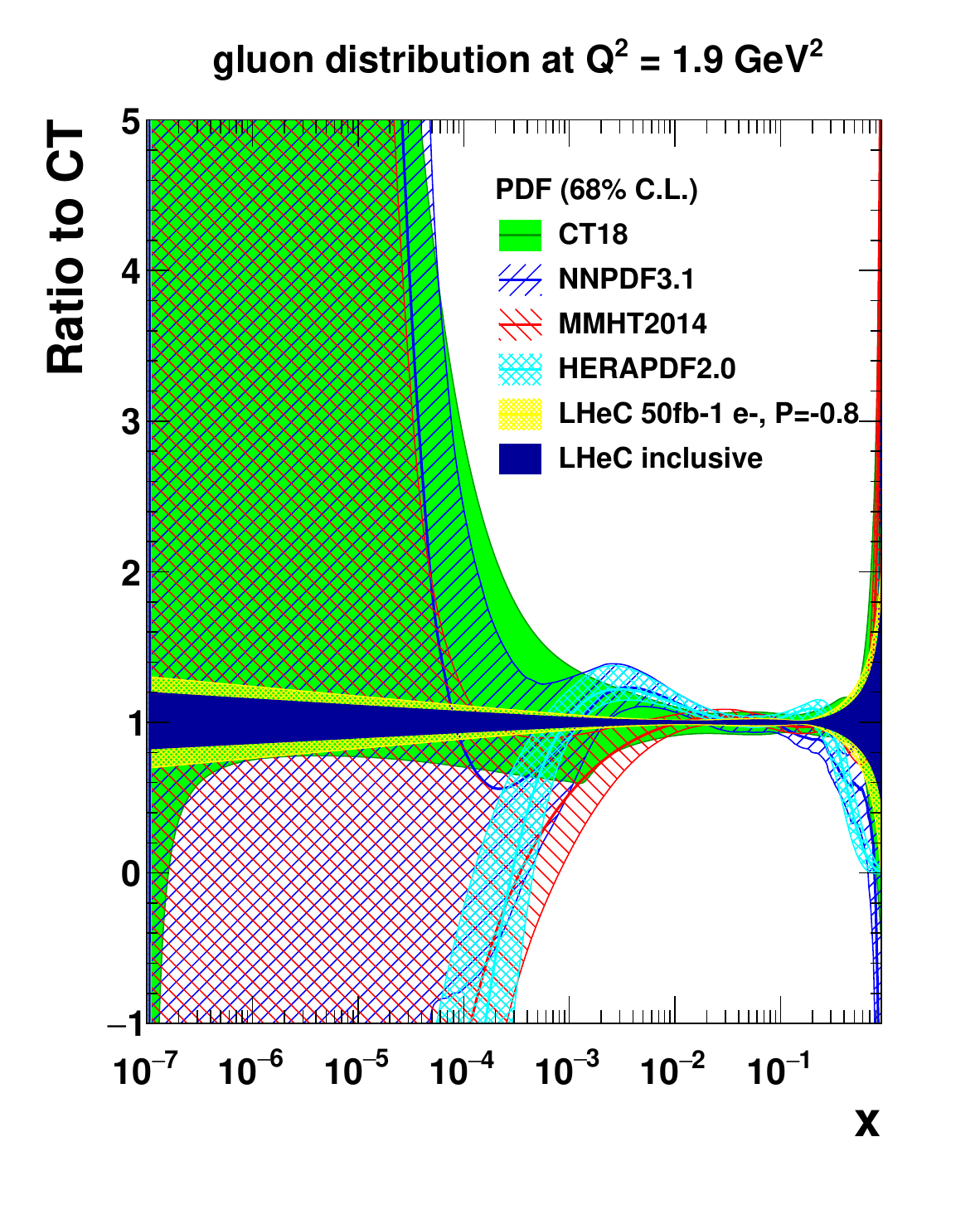}
  \includegraphics[width=0.45\textwidth]{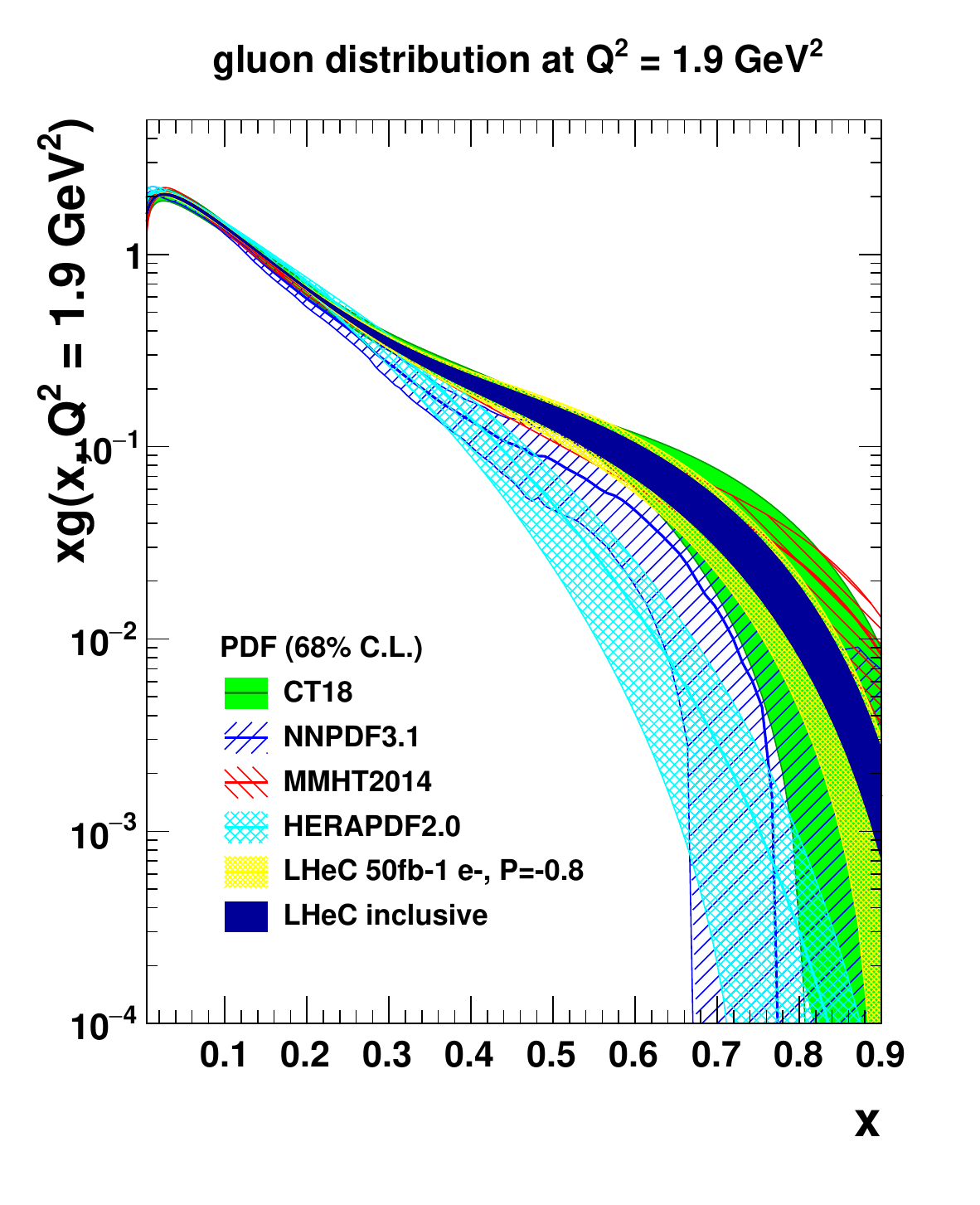}
  \vspace{-0.3cm}
\caption{Gluon distribution at $Q^2=1.9$\,GeV$^2$ as a function of
  $x$.
  Left: the distribution is displayed as the ratio to the CT18 PDF set
  and highlights the low $x$ region.
  Right: the distribution is shwon on a linear $x$ scale and
  highlights the high $x$ region.
The yellow band corresponds to the ``LHeC 1st run" PDFs (D2),
while the dark blue shows the ``LHeC inclusive" PDFs (D4+D5+D6+D9),
as described in the text. Both LHeC PDFs shown are scaled to the central value of CT18.
}
\label{fig:gluon}
\end{figure}

The gluon distribution, as it is obtained from the fit to the LHeC inclusive NC/CC data,
is shown in Fig.~\ref{fig:gluon}.
The determination of $xg$ will be radically improved
with the LHeC NC and CC precision data, which provide constraints
on $\partial F_2/\partial \ln Q^2$ down to very  low $x$ values, $ \geq 10^{-5}$,
and large $x \leq 0.8$.

Below $x \simeq 5 \cdot 10^{-4}$, the HERA data provide almost zero
constraints due to the kinematic limits and therefore  the gluon is
not well known today at lower $x$.
This can be seen in all modern PDF sets.
 With the LHeC, a precision of a few per cent at small $x$ is achieved down to 
 about $10^{-5}$. This should resolve the question of non-linear parton interactions
 at small $x$ (cf. Sect.\,\ref{sec:lowx}).
It also has direct implications for the LHC (and even stronger for the FCC): 
with the extension of the rapidity range 
to about $4$ at the HL-LHC by ATLAS and CMS, Higgs physics will become small
$x$ physics for which $xg$ must be known very accurately 
since $gg \to H$ is the dominant production mechanism.

At large $x$, i.e. at values greater than 0.3, the gluon distribution
becomes very small. In this region, the uncertainty on $xg$ is very
large, and the gluon distributions from several PDF groups differ
substantially. The limited experimental constraints are partially due
to the small luminosity at HERA, while uncertainties on jet
measurements are not negligible also. In addition, at high-$x$ the
valence quarks dominate, the non-singlet evolution of which is
insensitive to the gluon distribution. At the LHeC, the very large
luminosity provides NC and CC data to accurately access the highest
values of $x$, disentangling the sea from the dominant valence
part. The gluon distribution at high $x$ is then largely constrained
through the momentum sum-rule which at the LHeC (and FCC-eh) profits
from the seminal coverage from $x$ near $1$ down to very small values
of $x$. The resulting very small uncertainties on the high-$x$ quark
and gluon PDFs, as illustrated in the Figures, are of great importance
for BSM searches in hadron-hadron collisions at high scales as is
illustrated in this paper. If the LHeC established non-linear parton
interactions at small $x$, this will 
reflect also on the PDFs at high $x$. Furthermore, tests of the
factorisation theorem can be performed and electroweak effects be
measured to unprecedented precision, jointly with  PDFs
 (c.f.\ also Sect.~\ref{sec:EW}). 

The analysis presented here has not made use of
the additional information that is provided at the LHeC
in the measurements of $F_2^{c,b}$ (see Sect.~\ref{sec:HQ}) or $F_L$. 
The large $x$ situation can be expected to further improve by using
LHeC jet data, providing further, direct constraints
at large $x$ which, however, have not yet been studied in comparable detail.

The LHeC is the ideal laboratory to resolve all unknowns of the gluon density, which
is the origin for all visible mass in the universe, and one of the particular secrets
of particle physics for the gluon cannot directly be observed but is confined inside hadrons.
It is obvious that resolving this puzzle is an energy frontier DIS task and goal,
including electron-ion scattering since the gluon inside heavy matter is known even much less.
Therefore, the special importance of this part of high energy PDF physics is not primarily
related to the smallness of uncertainties: it is about a consistent understanding
and resolution of QCD at all regions of spatial and momentum dimensions which the LHeC
will explore.

\subsection{Luminosity and Beam Charge Dependence of LHeC PDFs}
\label{sec:setfits}
%
It is informative to study the transition of
the PDF uncertainties from the ``LHeC 1st run'' PDFs, which exploits only a
single electron-proton dataset, D2, through to the ``LHeC final
inclusive'' PDFs,
which makes use of the full datasets D4+D5+D6+D9 as listed in Tab.\,\ref{tab:dsets},
i.e. including high luminosity data (D4), small sets of low energy $E_p=1$\,TeV
and positron data (D5 and D6) together with $10$\,fb$^{-1}$ of opposite helicity
data.  Various intermediate PDF fits are performed using subsets of the data
in order to quantify the influence of the beam parameters on the precision of
the various PDFs. All fits use the same, standard 4+1 fit parameterisation and exclude
the use of $s,~c,~b$ data, the effect of which was evaluated before. 
The fits do neither include the low electron energy data sets generated for
the $F_L$ analysis, cf. Sect.\,\ref{sec:FL}, nor any jet $ep$ data. The
emphasis is on the development of the $u_v$, $d_v$, total sea and $xg$ uncertainty.
   
\begin{figure}[!htp]
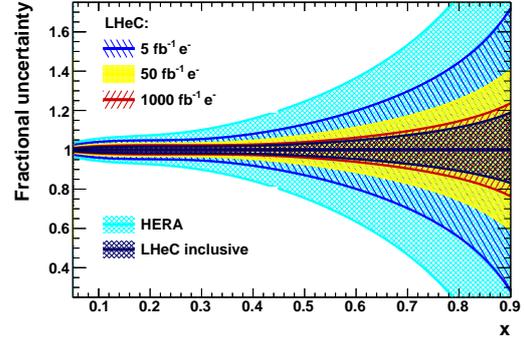
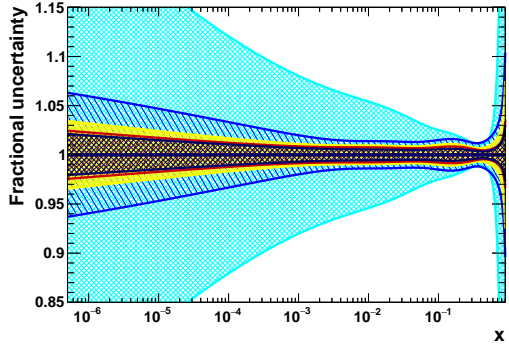
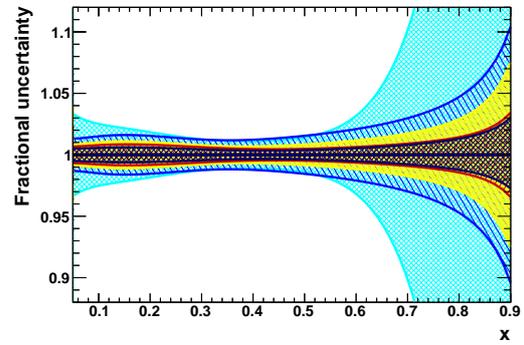
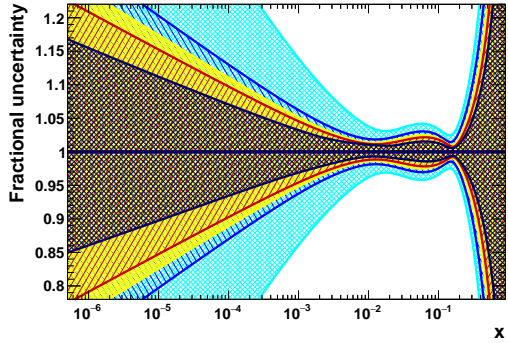
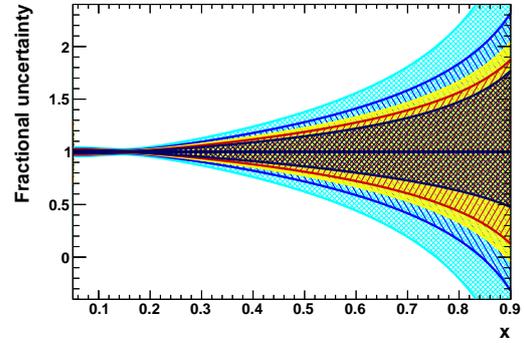

  \centering
  \begin{subfigure}{0.48\textwidth}
    \includegraphics[width=\linewidth]{claireLu_xuv_vs_x_lin.pdf}
    \caption{$u$-valence distribution.}
  \end{subfigure}
  \hspace*{\fill}
  \begin{subfigure}{0.48\textwidth}
    \includegraphics[width=\linewidth]{claireLu_xdv_vs_x_lin.pdf}
    \caption{$d$-valence distribution.}
  \end{subfigure}
  \begin{subfigure}{0.48\textwidth}
    \includegraphics[width=\linewidth]{claireLu_xsea_vs_x_log_restrictx.pdf}
    \caption{Sea quark distribution ($\log_{10}{x}$ scale).}
  \end{subfigure}
  \hspace*{\fill}
  \begin{subfigure}{0.48\textwidth}
    \includegraphics[width=\linewidth]{claireLu_xsea_vs_x_lin.pdf}
    \caption{Sea quark distribution (linear $x$ scale).}
  \end{subfigure}
    \begin{subfigure}{0.48\textwidth}
    \includegraphics[width=\linewidth]{claireLu_xglu_vs_x_log_restrictx.pdf}
    \caption{Gluon distribution ($\log_{10}{x}$ scale).}
  \end{subfigure}
  \hspace*{\fill}
  \begin{subfigure}{0.48\textwidth}
    \includegraphics[width=\linewidth]{claireLu_xglu_vs_x_lin.pdf}
    \caption{Gluon distribution (linear $x$ scale).}
  \end{subfigure}
  \caption{
    PDF distributions at $Q^2=1.9$\,${\GeV}^2$ as a function of $x$,
    illustrating the impact of different amounts of integrated luminosity.
    The blue, yellow and red bands correspond to LHeC PDFs
    using electron-only NC and CC inclusive measurements with 5, 50 and
    $1000$\,$\text{fb}^{-1}$ (datasets D1, D2 and D4), respectively.
    The yellow band is therefore equivalent to the ``LHeC 1st run'' PDF.
    For reference, the dark blue band shows the results of the final ``LHeC 
    inclusive'' PDF.
    For comparison, the cyan band represents an identical PDF fit using
    HERA combined inclusive NC and CC data~\cite{Abramowicz:2015mha},
    restricted to solely the experimental uncertainties.   
    Note that this, unlike the LHeC, extends everywhere beyond the narrow limits
    of the $y$ scale of the plots.
  }
  \label{fig:lumi}
\end{figure}

 A first study, Fig.\,\ref{fig:lumi}, shows the influence of the integrated luminosity.
 This compares four cases, three with evolving luminosity, from $5$ over $50$ to
 $1000$\,fb$^{-1}$. These assumptions, according to the luminosity scenarios presented
 elsewhere, correspond to year 1 (D1), the initial 3 years (D2) 
 and to the maximum attainable integrated luminosity (D4). The fourth case is represented
 by what is denoted the LHeC inclusive fit. One observes a number of peculiarities. For
 example, the initial $5$\,fb$^{-1}$ (yellow in  Fig.\,\ref{fig:lumi}), i.e.
 the tenfold of what H1 collected
 over its lifetime (albeit with different beam parameters), leads i) to an extension of 
 the HERA range to low and higher $x$, ii) to high precision at small $x$, for example
 of the sea quark density of $5$\,\% below $x=10^{-5}$ or iii) of also $5$\,\% for
 $u_v$ at very high $x=0.8$. With $50$\,fb$^{-1}$ the down valence distribution
 is measured to within $20$\,\% accuracy at $x=0.8$, an improvement by about a factor
 of two as compared to the $5$\,fb$^{-1}$ case, and a major improvement
 to what is currently known about $xd_v$ at large $x$, compare 
 with Fig.\,\ref{fig:valence}. The very high luminosity, here taken to be $1$\,ab$^{-1}$,
 leads to a next level of high precision, for example of $2$\,\% below $x=10^{-5}$
 for the total sea. The full data set further improves, especially the $xd_v$ and the
 gluon at high $x$. The valence quark improvement is mostly linked to the 
 positron data while the gluon improvement is related to the extension of
 the lever arm towards small values of $Q^2$ as the reduction of $E_p$
 extends the acceptance at large $x$. The visible improvement through the final
 inclusive fit is probably related to the increased precision at high $x$ for there
 exists a momentum sum rule correlation over the full $x$ range.
 In comparison to the analogous HERA fit, it becomes clear, that the
vast majority of the gain comes already from the first $5-50$\,$\text{fb}^{-1}$.

\begin{figure}[!htp]
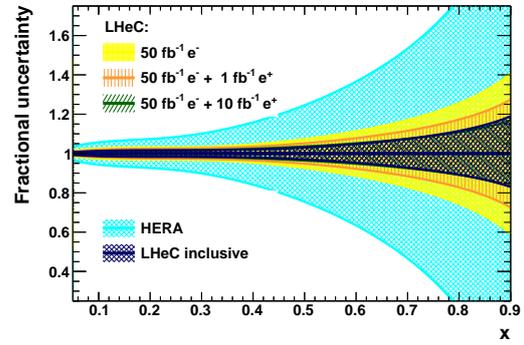
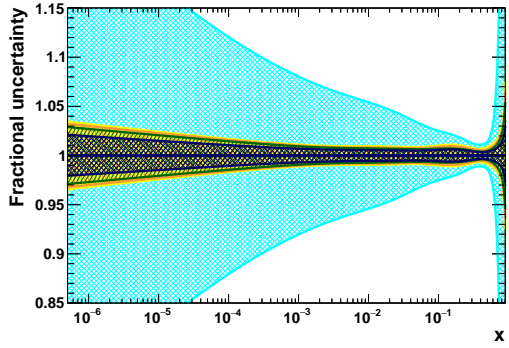
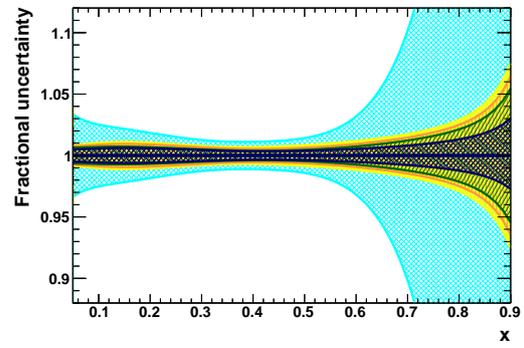
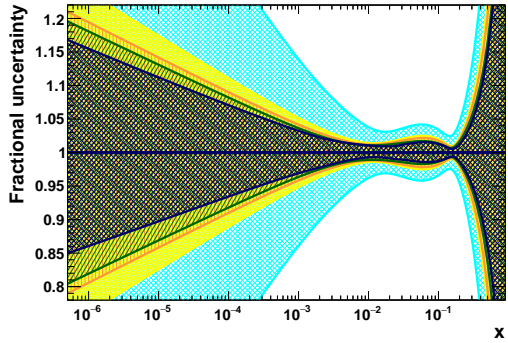
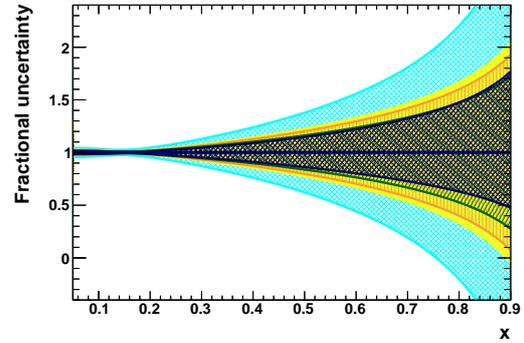

  \centering
  \begin{subfigure}{0.48\textwidth}
    \includegraphics[width=\linewidth]{clairePo_xuv_vs_x_lin.pdf}
    \caption{$u$-valence distribution.}
  \end{subfigure}
  \hspace*{\fill}
  \begin{subfigure}{0.48\textwidth}
    \includegraphics[width=\linewidth]{clairePo_xdv_vs_x_lin.pdf}
    \caption{$d$-valence distribution.}
  \end{subfigure}
  \begin{subfigure}{0.48\textwidth}
    \includegraphics[width=\linewidth]{clairePo_xsea_vs_x_log_restrictx.pdf}
    \caption{Sea quark distribution ($\log_{10}{x}$ scale).}
  \end{subfigure}
  \hspace*{\fill}
  \begin{subfigure}{0.48\textwidth}
    \includegraphics[width=\linewidth]{clairePo_xsea_vs_x_lin.pdf}
    \caption{Sea quark distribution (linear $x$ scale).}
  \end{subfigure}
    \begin{subfigure}{0.48\textwidth}
    \includegraphics[width=\linewidth]{clairePo_xglu_vs_x_log_restrictx.pdf}
    \caption{Gluon distribution ($\log_{10}{x}$ scale).}
  \end{subfigure}
  \hspace*{\fill}
  \begin{subfigure}{0.48\textwidth}
    \includegraphics[width=\linewidth]{clairePo_xglu_vs_x_lin.pdf}
    \caption{Gluon distribution (linear $x$ scale).}
  \end{subfigure}
  \caption{PDF distributions at $Q^2=1.9$\,${\GeV}^2$ as a function of $x$,
    illustrating the impact of including positron data.
    The yellow (``LHeC 1st run'') and dark blue (``LHeC final
    inclusive'')  and cyan bands (HERA data) are as in Fig.~\ref{fig:lumi}.
    The orange band corresponds to a fit with 1\,$\text{fb}^{-1}$ of inclusive NC and CC
    positron-proton data, in addition to 50\,$\text{fb}^{-1}$ of electron-proton data (D2+D6),
    while the green band is similar, but with 10\,$\text{fb}^{-1}$ of positron-proton data
    (D2+D7).
  }
  \label{fig:positron}
\end{figure}

The second study presented here regards
the impact on the PDF uncertainties when adding additionally positron
data of different luminosity to a baseline
fit on $50$\,fb$^{-1}$ of $e^-p$ data, the ``LHeC 1st run'' dataset.
The results are illustrated in Fig.\,\ref{fig:positron}.
It is observed, that the addition of positron data 
does bring benefits, which, however, are not striking in 
their effect on the here considered PDFs. 
A prominent improvement is obtained for the $d$-valence PDF,  
primarily due to the sensitivity gained via
the CC cross section of the positron data.
The benefit of the precise access to NC and CC weak interactions
by the LHeC is clearer when one studies the cross sections and their
impact on PDFs. This is illustrated in the subsequent section.

\subsection{Weak Interactions Probing Proton Structure}
\label{sec:weakpdfs}
%
%

It had long been considered to use the weak interactions to probe proton structure
in deep inelastic scattering~\cite{Klein:1983vs}. First important steps in this direction
could be pursued with HERA, especially with the measurements of the 
polarisation and  beam charge asymmetries in NC $ep$ scattering 
by H1 and ZEUS~\cite{Abramowicz:2015mha}. This area of research will become
a focus at the LHeC, because the $Q^2$ range 
extends by 2-3 orders of magnitude beyond the weak scale $Q^2 \simeq M_{W,Z}^2$,
with hugely increased luminosity.  In Sect.~\ref{sec:EW} below, 
the emphasis is on accessing the electroweak theory parameters at a new level of sensitivity.
Here we illustrate the importance of using the $Z$ and also $W$ exchange for pinning
down the parton contents of the proton. This has been implicite for the QCD fits
presented above, it yet emerges clearly only when one considers cross sections directly,
their asymmetries with respect to beam charge and polarisation,  and certain
kinematic limits. 

Parity violation is accessed in NC DIS through a variation of the lepton beam helicity, $P$,  as can be deduced from~\cite{Klein:1983vs}
\begin{equation} \label{eq:pasy}
\frac{ \sigma_{r,NC}^{\pm}(P_R) - \sigma_{r,NC}^{\pm}(P_L)}{P_R -P_L} 
 = \mp \kappa_Z  g_A^e     F_2^{\gamma Z} -  ( \kappa_Z  g_A^e)^2  \frac{Y_-}{Y_+} xF_3^{Z}
\end{equation}
where $\sigma_{r,NC}$ denotes the double differential NC scattering cross section
scaled by $Q^4 x / 2 \pi \alpha^2 Y_+$.
Here $\kappa_Z$ is of the order of $Q^2/M_Z^2$,  $F_2^{\gamma Z} = 2 x \sum {Q_q g_V^q (q - \bar q)}$
and the NC vector couplings are determined as $g_V^f = I_{3,L}^f - 2 Q_f \sin^2\theta_W$, where
$Q_f$ is the electric charge and
$I_{3,L}^f$ the left handed weak isospin charge of the fermion $f=e,q$, which also determines
the axial vector couplings $g_A^f$, with $g_A^e = -1/2$. 
At LHeC (unlike FCC-he) the second term in Eq.\,\ref{eq:pasy} is suppressed with respect
to the first one as it results from pure $Z$ exchange and because the $Y$ factor is small, $\propto y$
since $Y_{\mp} = (1 \mp (1-y)^2)$.

\begin{figure}[!th]
   \centering
   \includegraphics[width=0.98\textwidth]{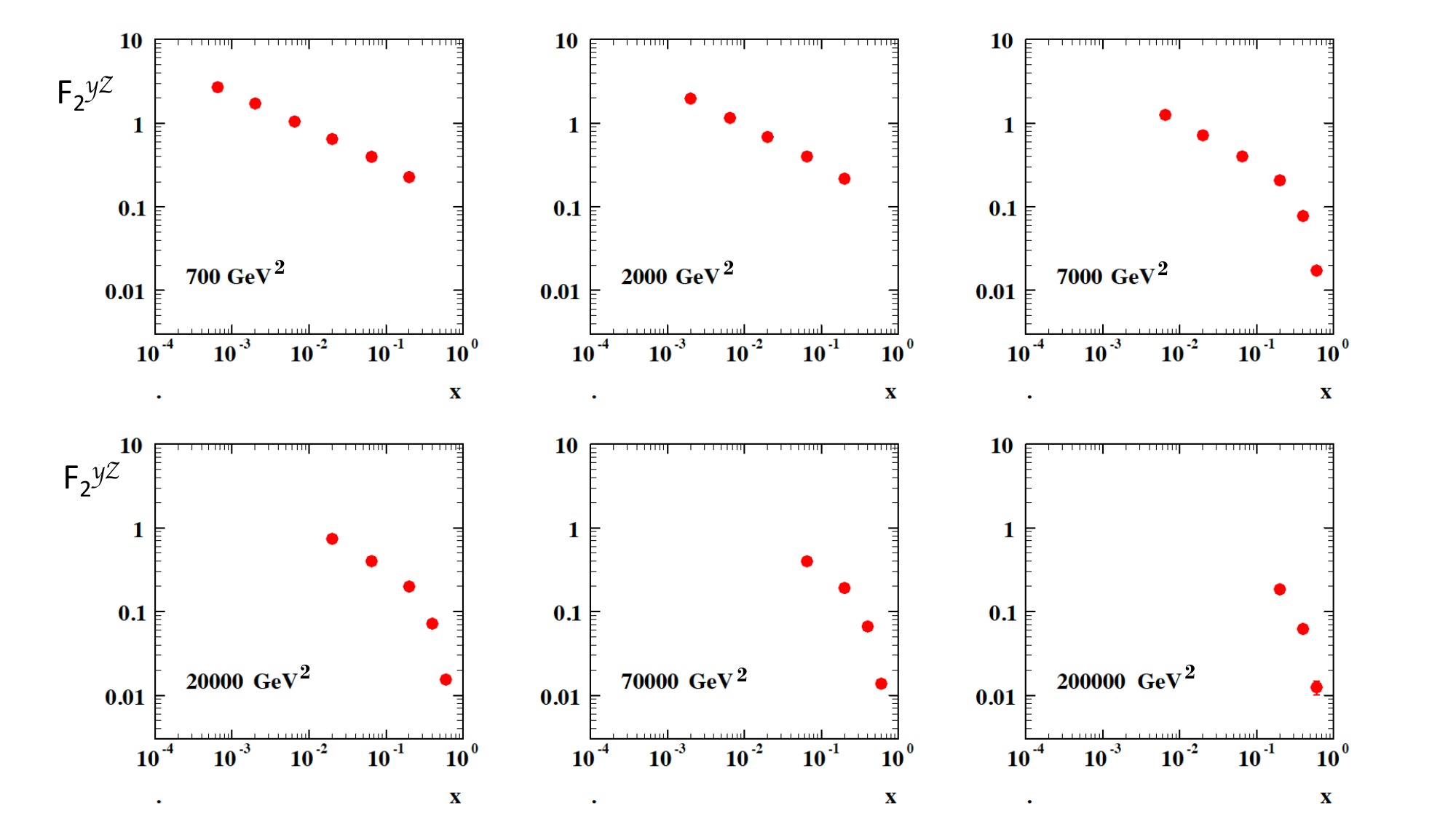}
 \caption{Prospective measurement of the photon-Z interference structure function
 $F_2^{\gamma Z}(x,Q^2)$ at the LHeC using polarised electron beams of helicity $\pm 0.8$
 and an integrated luminosity of $10$\,fb$^{-1}$ for each state. The uncertainties are
 only statistical.
  }
  \label{fig:g2}
\end{figure}

 For the approximate value
of the weak mixing angle $\sin^2\theta_W = 1/4$ one obtains $g_V^e=0,~ g_V^u=1/6$ and $g_V^d=-1/3$.
Consequently, one may write to good approximation 
\begin{equation} \label{g2}
 F_2^{\gamma Z}(x,Q^2)  = 2x \sum_q Q_q g_V^q (q - \bar q) \simeq  x \frac{2}{9} [ U + \bar U + D + \bar D]
 \end{equation}    
The beam helicity asymmetry therefore determines the total sea. A simulation is shown in 
Fig.\,\ref{fig:g2} for integrated luminosities of $10$\,fb$^{-1}$ and helicities of $P = \pm 0.8$.
Apparently, this asymmetry will provide a very precise measurement of the total sea.
 The combination of up and down quarks accessed with $F_2^{\gamma Z}$ (Eq.\,\ref{g2}) is different
from that provided by the known function 
\begin{equation} \label{f2}
 F_2(x,Q^2)  = 2x \sum_q Q_q^2 (q - \bar q) =  x \frac{1}{9} [ 4(U + \bar U) + D + \bar D]
 \end{equation} 
 because of the difference of the photon and $Z$ boson couplings to quarks. 
 Following Eq.\,\ref{eq:pasy}, the beam polarisation asymmetry
\begin{equation} \label{apm}
A^\pm = \frac{\sigma_{NC}^{\pm}(P_R) -\sigma_{NC}^{\pm}(P_L)}
                          {\sigma_{NC}^{\pm}(P_R) +\sigma_{NC}^{\pm}(P_L)}
             \simeq   \mp  (P_L - P_R)  \kappa_Z g_A^e \frac{F_2^{\gamma Z}}{F_2}.
\end{equation}
 measures  to a very good approximation the $F_2$ structure function ratio. The
 different composition of up and down quark contributions to $F_2^{\gamma Z}$ and $F_2$, see
 above, indicates that the weak neutral current interactions will assist to separate
 the up and down quark distributions which HERA had to link together by setting
 $B_d=B_u$.
 
Inserting $P_L = - P_R  = - P$  and considering the large $x$ limit,  one observes that
the asymmetry measures  the $d/u$ ratio of the valence quark distributions according to
\begin{equation} \label{doveru} 
           A^\pm   \simeq  \pm  \kappa_Z  P\frac{1+d_v/u_v}{4+d_v/u_v}.
\end{equation}
This quantity will be accessible with very high precision, as Fig.\,\ref{fig:g2}
illustrates, which is one reason, besides the CC cross sections, why the $d/u$ 
ratio comes out to be so highly constrained by the LHeC (see Fig.\,\ref{fig:dovu}).
 
 \begin{figure}[!th]
   \centering
   \includegraphics[width=0.65\textwidth]{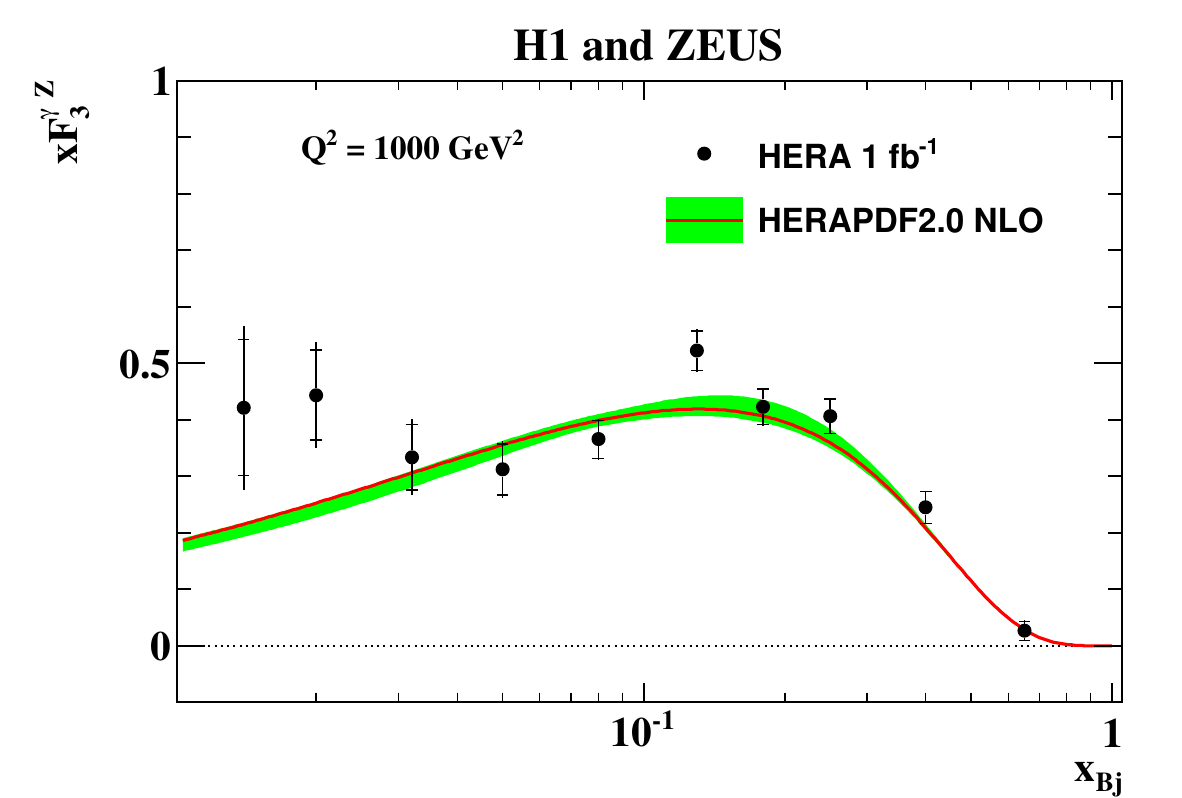}
 \caption{Combination of H1 and ZEUS measurement of the structure function $xF_3^{\gamma Z}(x,Q^2)$
 as a function of $x$ projected to a fixed $Q^2$ value of $2000$\,GeV$^2$, 
 from~\cite{Abramowicz:2015mha}. The inner error bar represents the statistical uncertainty.
  }
  \label{fig:xg3hera}
\end{figure}
A further interesting quantity is the the lepton beam charge asymmetry, which is given as
\begin{equation} \label{casy}
 \sigma_{r,NC}^+(P_1) - \sigma_{r,NC}^-(P_2) = \kappa_Z  a_e [ -(P_1 +P_2)  F_2^{\gamma Z} -   
 \frac{Y_-}{Y_+} ( 2 xF_3^{\gamma Z} + \kappa_Z  a_e (P_1 -P_2) xF_3^Z)] 
\end{equation}
neglecting terms $\propto g_V^e$. For zero polarisation this provides directly a parity conserving measurement of the structure function
\begin{equation} \label{eq:xg3}
  xF_3^{\gamma Z} (x,Q^2)  = 2 x \sum_q Q_q g_A^q ( q -\bar q) = 
   \frac{2}{3}  x (U - \bar U) +  \frac{1}{3}  x (D - \bar D). 
\end{equation}
The appearance of this function in weak NC DIS resembles that of $xW^3$ in CC, or fixed 
target neutrino-nucleon, scattering. It enables one to resolve the flavour contents of the proton.
The function $xF_3^{\gamma Z}$ was first measured by the BCDMS Collaboration in
$\mu^{\pm} C$ scattering~\cite{Argento:1983dj} at the SPS. 

The HERA result is shown in 
Fig.\,\ref{fig:xg3hera}. It covers the range from about $x=0.05$ to $x=0.6$ with typically $10$\,\%
statistical precision. Assuming that sea and anti-quark densities are equal, such as
$u_s = \bar u$ or $d_s = \bar d$,  $xF_3^{\gamma Z}$ is given as $x/3 (2 u_v + d_v)$. This function
therefore accesses valence quarks down to small values of $x$ where their densities
become much smaller than that of the sea quarks. Since the $Q^2$ evolution of the 
non-singlet valence quark distributions is very weak, it has been customary to project
the various charge asymmetry measurements to some lowish value of $Q^2$ and present
the measurement as the $x$ dependence of $xF_3^{\gamma Z}$.

 If, however, there would be 
differences between the sea and anti-quarks, if $s \ne \bar s$, 
 for example, one expected 
a rise of $xF_3^{\gamma Z}$ towards low $x$. This may be a cause for the undershoot
of the QCD fit below the HERA data near to $x \simeq 0.01$, see Fig.\,\ref{fig:xg3hera},
which yet are not precise enough. However, it is apparent 
that, besides providing constraints on the valence
quark densities, this measurement indeed has the the potential to discover a new
anti-symmetry in the quark sea. 

\begin{figure}[!th]
   \centering
   \includegraphics[width=0.9\textwidth]{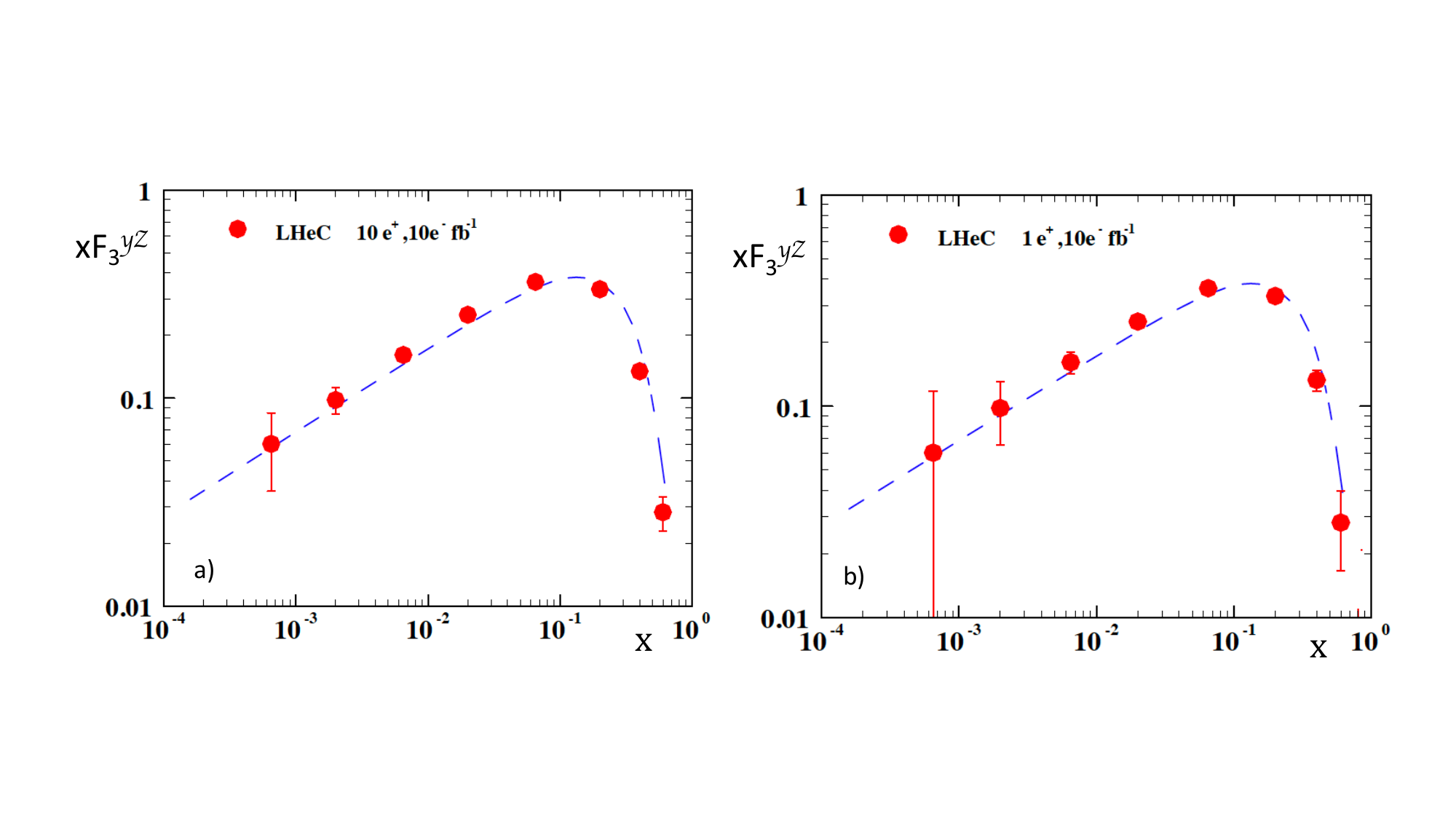}
   \vspace{-1.3cm}
 \caption{Prospective measurement of the photon-Z interference structure function
 $xF_3^{\gamma Z}(x,Q^2)$ at the LHeC projected to a fixed $Q^2$ value of $2000$\,GeV$^2$.
 The result corresponds to a cross section charge asymmetry for
 an unpolarised $e^-p$ beam with $10$\,fb$^{-1}$  luminosity combined
 with unpolarised $e^+p$ beams of a) $10$\,fb$^{-1}$ (left) and b) $1$\,fb$^{-1}$ (right).
The error bars represent the statistical uncertainty. The curve is drawn to guide the eye.
It is possible that the measurement would discover a rise of $xF_3^{\gamma Z}$ towards
low $x$ should there exits so far unknown differences between sea and anti-quark densities, see text. 
  }
  \label{fig:xg3lhec}
\end{figure}

Such a discovery  would be enabled by the LHeC as is illustrated in 
Fig.\,\ref{fig:xg3lhec} with an extension of the kinematic range by an order of magnitude
towards small $x$ and a much increased precision in the medium $x$ region.
 The simulation is performed for $10$ and for $1$\,fb$^{-1}$
of $e^+p$ luminosity. Obviously it would be very desirable to reach high values
 of integrated luminosity in positron-proton scattering too. 

It is finally of interest to consider the role of precisely measured cross sections
in CC scattering. The coupling of the $W$ boson to quarks is flavour dependent
resulting in the relations
\begin{eqnarray} \label{eq:sigcc}
 \sigma_{r,CC}^+ = (1+P) [x\bar U+ (1-y)^2xD], ~~\\
 \sigma_{r,CC}^-  = (1-P)  [xU +(1-y)^2 x\bar D] . ~~
\end{eqnarray}
Here $\sigma_{r,CC}$ is the double differential charged current DIS cross section
scaled by a factor $2 \pi x \cdot (M_W^2 +Q^2)^2 /( G_F M_W^2)^2$ with the
Fermi constant $G_F$ and the $W$ boson mass $M_W$. The positron beam at the LHeC
is most likely unpolarised, $P=0$. Maximum rate in $e^-p$ is achieved with large negative 
polarisation. In the valence-quark approximation,
the $e^+p$ CC cross section is proportional to $(1-y)^2 xd_v$ while  $ \sigma_{r,CC}^- \propto u_v$.
This provides direct, independent measurements of $d_v$ and $u_v$ as had been
illustrated already in the LHeC CDR~\cite{AbelleiraFernandez:2012cc}. 

Inclusive NC and CC DIS accesses four combinations of parton distributions, as is obvious from
Eq.\,\ref{eq:sigcc} for CC  above and from the NC relation
\begin{eqnarray} \label{f23ud}
\sigma^{\pm}_{r,NC} \simeq [c_u (U+\bar U) +c_d(D+\bar D)] + \kappa_Z [d_u(U-\bar U) + d_d (D-\bar D)] \nonumber \\
\mbox{with} \hspace{0.2cm} c_{u,d} = Q^2_{u,d} + \kappa_Z (-g_V^e \mp P g_A^e) Q_{u,d}g_V^{u,d} 
\hspace{0.2cm} \mbox{and} \hspace{0.2cm} d_{u,d} = \pm g_A^e g_A^{u,d} Q_{u,d},
\end{eqnarray}
restricted to photon and $\gamma Z$ interference contributions. These four PDF combinations are complemented by the $s,~c,~b$ measurements introduced before. The parton contents 
can therefore be completely resolved, which was impossible at HERA.

 It is the high energy and high luminosity access to DIS, the high precision NC/CC and
tagged heavy quark measurement programme, which makes the LHeC the uniquely suited
environment to uncover the secrets of parton structure and dynamics. This will establish
a new level with possible discoveries of strong interaction physics and also
provide the necessary base for precision electroweak and Higgs measurements at the LHC,
for massively extending the range of BSM searches and reliably interpreting 
new physics signals in hadron-hadron scattering at the LHC.

\subsection{Parton-Parton Luminosities}
\label{sec:partlumi}
The energy frontier in accelerator particle physics is the LHC, with a cms energy
of $\sqrt{s} = 2 E_p \simeq 14$\,TeV, with the horizon of a future circular hadron collider,
the FCC-hh, reaching energies up to $\sqrt{s} =100$\,TeV. Proton-proton collider reactions are characterised
by the Drell-Yan scattering~\cite{Drell:1970wh}. 
To leading order, the double differential
Drell-Yan scattering cross section~\cite{Kubar:1980zv} for the
neutral current
reaction $pp \rightarrow (\gamma,Z)X  \rightarrow e^+e^-X$
and the charged current (CC) reaction  $pp \rightarrow W^{\pm}X \rightarrow e\nu  X$,
can be written as
\begin{equation} \label{EqsigDY}
\frac{d^2 \sigma}{dM dy} = \frac{4 \pi \alpha^2(M)}{9} \cdot 2 M \cdot P(M) \cdot  \Phi(x_1,x_2,M^2) ~~~[\mathrm{nb\,GeV}^{-1}].
\end{equation}
Here  $M$ is the mass of the $e^+e^-$ and  $e^+\nu$ and $e^-\bar{\nu}$ systems for the NC and
CC process, respectively, and  $y$ is the boson rapidity.
The cross section implicitly depends on the Bjorken $x$ values
of the incoming quark $q$ and its anti-quark $\overline{q}$, which are related to the
rapidity $y$ as
\begin{equation}
x_1=\sqrt{\tau}e^y~~~~x_2=\sqrt{\tau}e^{-y}~~~~\tau=\frac{M^2}{s}.
\end{equation}
%
%
For the NC process, the cross section is a sum of a contribution
from  photon and $Z$ exchange as well as an interference term.
In the case of photon exchange, the propagator term $P(M)$
and the parton distribution term $\Phi$ are given by
\begin{eqnarray} \label{eq:fqqnc}
P_{\gamma}(M) = \frac{1}{M^4} ~~~~~~~~~~~~~  \Phi_{\gamma} =\sum_q{Q_q^2 F_{q\overline{q}}} ~~~~~~~~~~~~~~ \\
F_{q\overline{q}}=x_1x_2 \cdot [q(x_1,M^2)\overline{q}(x_2,M^2) +\overline{q}(x_1,M^2)q(x_2,M^2)].
\end{eqnarray}
Similar to DIS, the corresponding formulae for the $\gamma Z$ interference term read as
%
%
\begin{equation} \label{gammaZ}
P_{\gamma Z}=\frac{\kappa_Z g_V^e (M^2-M_Z^2)}{M^2[(M^2-M_Z^2)^2+(\Gamma_Z M_Z)^2]}  ~~~~~~~~~~
\Phi_{\gamma Z} =\sum_q 2 Q_q g_V^q F_{q\overline{q}}
\end{equation}
The interference contribution is small being proportional to the
vector coupling of the electron $g_V^e$. 
One also sees in Eq~\ref{gammaZ}
that the interference cross section contribution
changes sign from plus to minus as the mass increases and
passes $M_Z$. The expressions of $P$ and $\Phi$
for the pure $Z$ exchange part are
%
%
\begin{equation}
~~~~~~~~~~~P_{Z}=\frac{\kappa_Z^2 (g_V^{e^2}+g_A^{e^2})}{(M^2-M_Z^2)^2+(\Gamma_Z M_Z)^2}  ~~~~~~~~~~~~~~~\Phi_{Z} =\sum_q  ( g_V^{q^2} + g_A^{q^2}) F_{q\overline{q}}.~~~~~
\end{equation}
For the CC cross section the propagator term is
%
%
\begin{equation}
~~~~~~~~~~~P_{W}=\frac{\kappa_W^2 }{(M^2-M_W^2)^2+(\Gamma_W M_W)^2}
\end{equation}
and the charge dependent parton distribution forms are
\begin{equation} \label{eq:fqqcplu}
\Phi_{W^+} =x_1 x_2 [V_{ud}^2(u_1\overline{d}_2+u_2\overline{d}_1)+V_{cs}^2(c_1\overline{s}_2+c_2\overline{s}_1)+  \\
V_{us}^2(u_1\overline{s}_2+u_2\overline{s}_1)+V_{cd}^2(c_1\overline{d}_2+c_2\overline{d}_1) ]
\end{equation}
\begin{equation} \label{eq:fqqcmin}
\Phi_{W^-} =x_1 x_2 [V_{ud}^2(\overline{u}_1 d_2+\overline{u}_2 d_1)+V_{cs}^2(\overline{c}_1 s_2+ \overline{c}_2 s_1)+  \\
V_{us}^2(\overline{u}_1s_2+\overline{u}_2 s_1)+V_{cd}^2 (\overline{c}_1 d_2+\overline{c}_2 d_1) ],
\end{equation}
with $\kappa_W= 1 / (4 \sin^2 \Theta)$ and $q_i=q_i(x,M^2)$ and the CKM matrix elements $V_{ij}$.
The expressions given here are valid in the QPM. At higher order pQCD, Drell-Yan scattering 
comprises also quark-gluon and gluon-gluon contributions. Certain production channels are
sensitive to specific parton-parton reactions, Higgs production, for example, originating predominantly
from gluon-gluon fusion.  Based on the factorisation theorem~\cite{Collins:1989gx}
 one therefore opened a further testing 
ground for PDFs, and much of the current PDF analyses is about constraining parton distributions
by Drell-Yan scattering measurements and semi-inclusive production processes, such as top, jet and 
charm production, at the LHC. An account of this field is provided below, including a study as to
how LHeC would add to the ``global" PDF knowledge at the time of the HL-LHC. 

There are drawbacks to the use of Drell-Yan and other 
hadron collider data for the PDF determination and advantages
for $ep$ scattering:  i) DIS has the
ability to prescribe the reaction type and the kinematics ($x,~Q^2$) through the 
reconstruction of solely the leptonic vertex; ii) there are no colour reconnection and, for the
lepton vertex, no hadronisation effects disturbing the theoretical description; iii) the most
precise LHC data, on $W$ and $Z$ production, are located at a fixed equivalent $Q^2 = M_{W,Z}^2$
and represent a snapshot at a fixed scale which in DIS at the LHeC varies by more than $5$ orders of magnitude~\footnote{This is mitigated by measurements of Drell-Yan
 scattering at low masses, which are less
precise, however. At high masses, $M =\sqrt{s x_1 x_2}  >> M_{W,Z}$, one soon reaches the region where new physics may occur, i.e. there arises the difficulty to separate unknown physics from the
uncertainty of the quark and gluon densities at large $x$. High mass Drell Yan searches
often are performed at the edge of the data statistics, i.e. they can not really be guided 
by data but miss a reliable guidance for the behaviour of the SM background around and beyond a
(non-) resonant effect they would like to discover.}. 

There are further difficulties
inherent to the use of LHC data for PDF determinations, such as hadronisation corrections and 
incompatibility of data. For example, the most recent CT18~\cite{Hou:2019efy} global PDF analysis
had to arrange for a separate set (CT18A) because the standard fit would not respond well
to the most precise ATLAS $W,~Z$ data taken at $7$\,TeV cms. The intent to include
all data can only be realised with the introduction of so-called $\chi^2$ tolerance criteria
which fundamentally affect  the meaning of the quoted PDF uncertainties. 

\begin{figure}[!th]
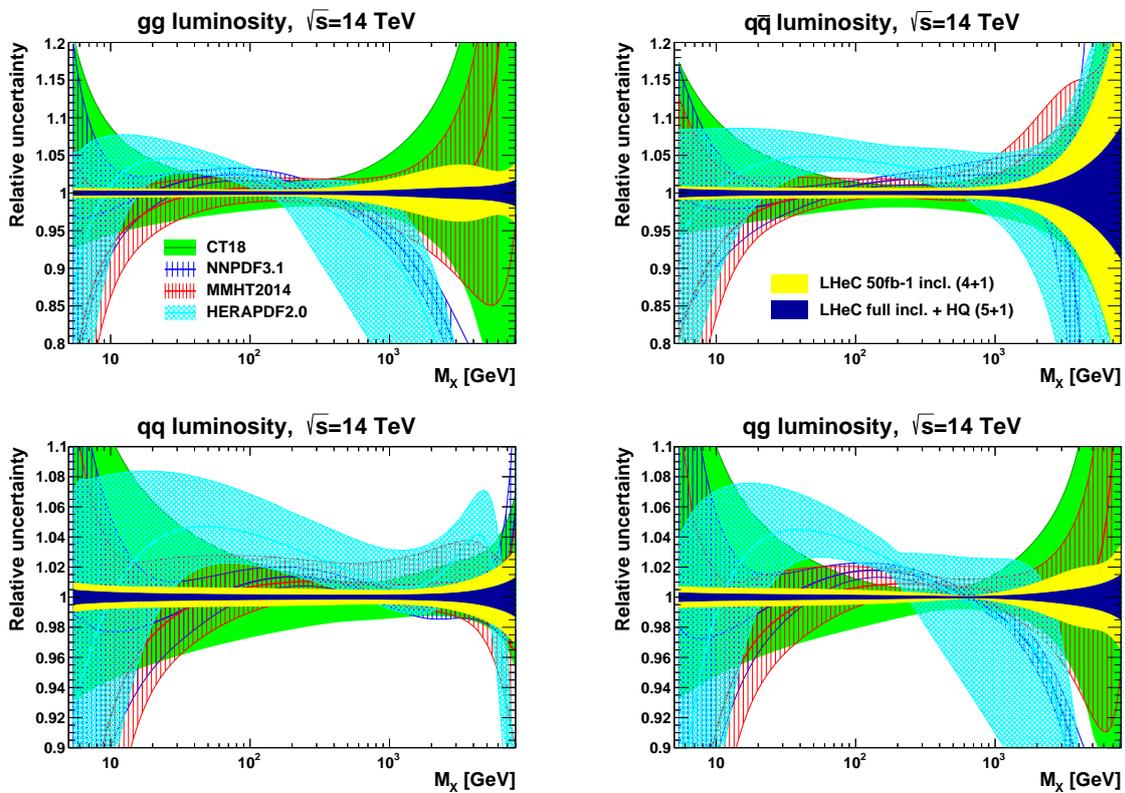

  \centering
    \includegraphics[width=0.49\textwidth]{claire_ratio_gg14_vs_mx_log.pdf}
    \includegraphics[width=0.49\textwidth]{claire_ratio_qqbar14_vs_mx_log.pdf} \\
    \includegraphics[width=0.49\textwidth]{claire_ratio_qq14_vs_mx_log.pdf}
   \includegraphics[width=0.49\textwidth]{claire_ratio_gq14_vs_mx_log.pdf}
  \caption{Uncertainty bands for parton luminosities as a function of the mass $M_X = \sqrt{s x_1 x_2}$
  for LHC energies. 
  The yellow band corresponds to the ``LHeC 1st run'' PDFs (D2),
  while the dark blue shows a fit to the LHeC inclusive data sets (D4+D5+D6+D9) 
  in Tab.\,\ref{tab:dsets} together with the simulated heavy flavour $s,~c,~b$ 
  data with a 5 quark distribution parameterisation as described in the text.
  Both LHeC PDFs shown are scaled to the central value of CT18.
  } 
  \label{fig:ablumis}
  \end{figure}

Conceptually, the LHeC enables us to change this approach completely. Instead of trying to use
all previous and current PDF sensitive data, to which nowadays one has no alternative,
it replaces these by pure $ep$ collider DIS data.
Then one will bring order back into the PDF field:
parton distributions, completely resolved, and from a single process, extending over nearly six
orders of magnitude and calculated from NLO pQCD up to probably even
N$^4$LO (see Sect.\,\ref{sec:hoqcd}).
These PDFs
will be applicable for i) identifying new dynamics and symmetries; ii) testing factorisation; 
iii) confronting other PDF analsyses at that time; iv)
performing high precision Higgs and electroweak analyses, and v) interpreting
any peculiar HL-LHC signal for BSM using that independent PDF.
higg
It has been customary, which is obvious from Eqs.\,\ref{eq:fqqnc}, \ref{eq:fqqcplu}
and \ref{eq:fqqcmin},
to express the usefulness of various PDF determinations and prospects for the LHC,
and similarly the FCC, with four so-called parton luminosities which are defined as
\begin{equation} \label{eq:partlum}
  L_{ab} (M_X) = \int dx_a dx_b \sum_q F_{ab} ~\delta (M_X^2 - s x_a x_b) 
\end{equation}
where $F_{ab}$ for $(a,b)=(q \bar q)$ is defined in Eq.\,\ref{eq:fqqnc} and 
$(a,b)$ could also be $(g,q)$, $(g,\bar q)$ and $(gg)$, without a sum over quarks in the latter case.  
The expectations for the quark and gluon related four parton luminosities are presented
in Fig.\,\ref{fig:ablumis}. The 
LHeC provides very precise parton luminosity predictions in the complete range of $M_X$ up to
the high mass edge of the search range at the LHC. This eliminates the 
currently sizeable PDF uncertainty of
precision electroweak measurements at the LHC, as for example for the anticipated
measurement of $M_W$ to within $10^{-4}$ uncertainty, see below.  One may also notice that the
gluon-gluon luminosity (left top in Fig.\,\ref{fig:ablumis}) is at a per cent level for
the Higgs mass  $M_X = M_H \simeq 125$\,GeV.  This is evaluated further in 
the chapter on Higgs physics with the LHeC.

 \section{The 3D Structure of the Proton}
 \label{sec:PSM_Disc_3D}

As is evident from the discussion in the previous Sections, the  LHeC machine will be able to measure the collinear parton distribution functions with unprecedented accuracy in its extended range of $x$ and $Q^2$.  Thus, it will provide a new insight into the details of the one-dimensional structure of the proton and nuclei, including  novel phenomena at low $x$. In addition to collinear dynamics, the LHeC opens  a new window into proton and nuclear structure by allowing a precise investigation of the partonic structure in more than just the one dimension of the longitudinal momentum. Precision DIS thus gives access  to multidimensional aspects of   hadron structure. This can be achieved by accurately measuring processes with more exclusive final states like production of   jets, semi-inclusive production of hadrons and exclusive processes, in particular the elastic diffractive production of vector mesons and  deeply virtual Compton (DVCS) scattering that were explored in the 2012 LHeC CDR~\cite{AbelleiraFernandez:2012cc}. These  processes have the potential to provide  information not only on the longitudinal distribution of partons in the proton or nucleus, but also on the dependence  of the parton distribution on  transverse momenta and momentum transfer.  Therefore, future, high precision DIS machines like the LHeC or the  Electron Ion Collider (EIC) in the US~\cite{Accardi:2012qut},  open a unique window into the details of the 3D structure of hadrons.
Note that the measurement of these processes requires a detector with large acceptance, $|\eta|<4$, see e.g.~\cite{AbelleiraFernandez:2012cc,Lomnitz:2018juf}. The current LHeC central detector design covers $|\eta|\lesssim 4.5$, see Chapter~\ref{chap:detector}.

The most general quantity that can be defined in QCD, that would contain very detailed information about the partonic content of the hadron,  is the Wigner distribution~\cite{Belitsky:2003nz}. This function $W(x,{\textbf k},{\textbf b})$  is a 1+4 dimensional function. One can think of it as the ``mother" or ``master" parton distribution, from which lower-dimensional distributions can be obtained. In the definition of the Wigner function, ${\textbf k}$ is the transverse momentum of the parton and ${\textbf b}$ is the 2-dimensional impact parameter, which can be defined as a Fourier conjugate to the momentum transfer of the process. The other, lower dimensional parton distributions can be obtained by integrating out different variables. Thus,  transverse momentum dependent  (TMD) parton distributions (or unintegrated parton distribution functions) $f_{\text{TMD}}(x,{\textbf k})$ can be obtained by integrating out the impact parameter ${\textbf b}$ in the Wigner function, while the generalised parton densities (GPD), $f_\text{GPD}(x,{\textbf b})$, can be obtained from the Wigner function through the integration over the transverse momentum ${\textbf k}$. In the regime of small $x$, or high energy, a suitable formalism is that of the dipole picture~\cite{Nikolaev:1990ja, Nikolaev:1991et, Nikolaev:1994kk, Nikolaev:1995xu, Mueller:1993rr, Mueller:1994jq}, where the fundamental quantity which contains the details of the partonic distribution is the dipole amplitude $N(x,{\textbf r},{\textbf b})$. This object contains the dependence on the impact parameter ${\textbf b}$ as well as another transverse size ${\textbf r}$, the dipole size, which can be related to the transverse momentum of the parton ${\textbf k}$ through a Fourier transform. The important  feature of the dipole amplitude is that it should obey the unitarity limit $N \le 1$. The dipole amplitude $N$ within this formalism can be roughly interpreted as a Wigner function in the high energy limit, as it contains information about the spatial distribution of the partons in addition to the dependence on the longitudinal momentum fraction $x$. 

\begin{figure}[!th]
\centering
     \includegraphics[width=0.35\linewidth,trim={0 160 0 160},clip]{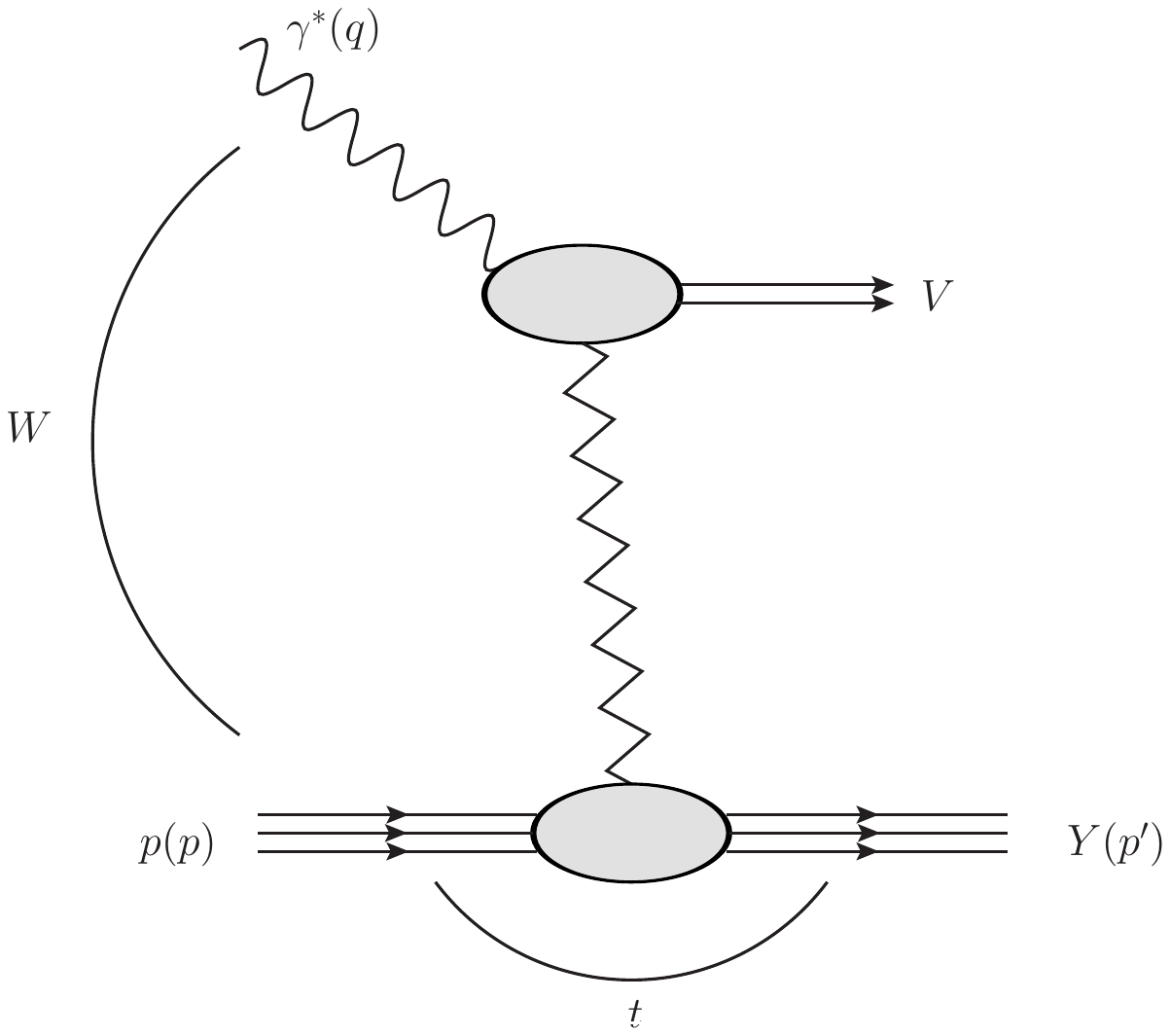}
     \hspace{0.05\textwidth}
     \includegraphics[width=0.35\linewidth,trim={0 160 0 160},clip]{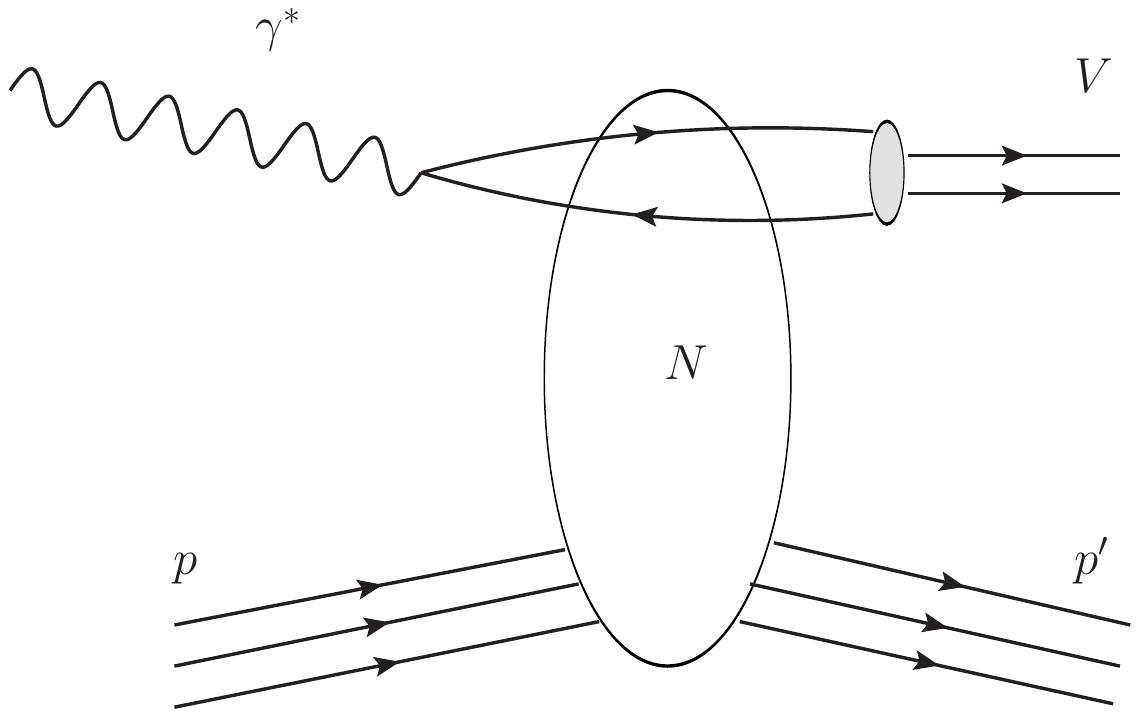}
    \caption{Left:  diagram for the quasi-elastic  production of the vector meson. Right: schematic illustration of the same process,  quasi-elastic  vector meson production, within  the framework of the  dipole picture (taken from~\cite{Berger:2012wx}). The initial virtual photon, fluctuates into a quark-antiquark pair which then scatters off the hadronic target and  forms the vector meson. The details of the hadronic interaction of the dipole with the target are encoded in the dipole amplitude $N$. 
       \label{fig:dipole_vm} }
\end{figure}

Detailed simulations of elastic $J/\psi$ vector meson production 
were performed for the LHeC kinematic region and beyond~\cite{AbelleiraFernandez:2012cc}, using the formalism of the dipole picture.  This particular process is shown in Fig.~\ref{fig:dipole_vm}, left plot. The proton is scattered elastically with  momentum transfer $t$, and the vector meson is produced, which is separated from the final state proton by a rapidity gap.  Of particular importance is the measurement of the $t$ slope of this process, since it can be related directly to the impact parameter distribution and is thus sensitive to the transverse variation of the partonic density in the target. The first type of analysis like this, in the context of  elastic scattering, was performed by Amaldi and Schubert~\cite{Amaldi:1979kd}, where it was demonstrated that the Fourier transform of the elastic cross section yields access to the impact parameter profile of the scattering amplitude. This method can be used in the context of vector meson scattering in DIS, where the transverse distribution of partons, in the perturbative regime, can be extracted through the appropriate Fourier transform~\cite{Munier:2001nr}. The additional advantage of studying  diffractive vector meson production is the fact that the partonic distributions can be studied as a function of the hard scale in this process given by the mass of the vector meson $M_V^2$ in the photoproduction case or $Q^2$ (or more precisely a combination of $Q^2$ and $M_V^2$) in the case of the diffractive DIS production of  vector mesons, as well as the energy $W$ of the photon-proton system available in the process which is closely related to $x$.

The differential cross section for  elastic vector meson production can be expressed in the following form:
\begin{equation}
    \frac{{d\sigma}^{\gamma^* p \rightarrow J/\psi p}}{dt} = \frac{1}{16 \pi }|{\mathcal A}(x,Q,\Delta)|^2 \; ,
    \label{eq:dipole_dt_xsection}
\end{equation}
where the amplitude for the process of elastic diffractive vector meson production in the high energy limit, in the dipole picture, is given by
\begin{equation}
{\mathcal A}(x,Q,\Delta) \; = \; \sum_{h\bar{h}} \int d^2 {\textbf r} \int dz \Psi^*_{h\bar{h}}(z,{\textbf r},Q) \, {\mathcal N}(x,{\textbf r},\Delta)\, \Psi^V_{h\bar{h}}(z,{\textbf r}) \; .
    \label{eq:dipole_vm_elastic}
\end{equation}
In the above formula, $\Psi^*_{h\bar{h}}(z,{\textbf r},Q)$ is the photon wave function which describes the splitting of the virtual photon $\gamma^*$ into a $q\bar{q}$ pair. This wave funtion  can be calculated in perturbative QCD. The function 
$\Psi^V_{h\bar{h}}(z,{\textbf r})$ is the wave function of the vector meson. Finally, ${\mathcal N}(x,{\textbf r},\Delta)$ is the dipole amplitude which contains all the information about the interaction of the quark-antiquark dipole with the target. The formula \eqref{eq:dipole_vm_elastic}
 can be interpreted as the process of  fluctuation of the virtual photon into a $q\bar{q}$ pair, which  subsequently interacts with the target through the dipole amplitude ${\mathcal N}$ and then forms the vector meson, given by the amplitude $\Psi^V$, see Fig.~\ref{fig:dipole_vm}, right plot. The two integrals in the definition Eq.~\eqref{eq:dipole_vm_elastic} are performed over the dipole size which is denoted by ${\textbf r}$, and $z$ which is the longitudinal momentum fraction of the photon carried by the quark. The scattering amplitude depends on the value of the momentum transfer $\Delta$, which is related to the Mandelstam variable $t=-\Delta^2$. The sum is performed over the helicity states of  the quark and antiquark.

The dipole amplitude ${\mathcal N}(x,{\textbf r},\Delta)$ can be related to the dipole amplitude in  coordinate space through the appropriate Fourier transform
\begin{equation}
   { N}(x,{\textbf r},{\textbf b}) =  \int d^2{\Delta} \, e^{i \Delta \cdot {\textbf b}} {\mathcal N}(x,{\textbf r},\Delta) \; .
   \label{eq:dipole_amplitude}
\end{equation}
We stress that ${\textbf r}$ and ${\textbf b}$ are two different transverse sizes here. The dipole size ${\textbf r}$ is conjugate to the transverse momentum of the partons ${\textbf k}$, whereas the impact parameter is roughly the distance between the centre of the scattering target to the centre-of-mass of the quark-antiquark dipole and is related to the Fourier conjugate variable, the momentum transfer $\Delta$.

The dipole amplitude ${ N}(x,{\textbf r},{\textbf b})$ contains rich information about the dynamics of the hadronic interaction. It is a 5-dimensional function and it depends on the longitudinal momentum fraction, and two two-dimensional coordinates. The dependence on the longitudinal momentum fraction is obviously related to the evolution with the centre-of-mass energy of the process, while the dependence on  ${\textbf b}$ provides information about the spatial distribution of the partons in the target. The dipole amplitude is related  to the distribution of gluons in impact parameter space. The dipole amplitude has a nice property that its value should be bounded from above by the unitarity requirement $N\le 1$. The complicated dependence on energy, dipole size and impact parameter of this amplitude can provide a unique insight into the dynamics of QCD, and on the approach to the dense partonic regime. Besides, from Eqs.~\eqref{eq:dipole_dt_xsection},\eqref{eq:dipole_vm_elastic} and \eqref{eq:dipole_amplitude} it is evident that the information about the spatial distribution in impact parameter ${\textbf b}$ is related through the Fourier transform to the dependence of the cross section on the momentum transfer $t=-\Delta^2$.

To see how the details of the distribution, and in particular the approach to unitarity, can be studied through the VM elastic production, calculations based on the dipole model were performed~\cite{Armesto:2014sma}, and extended to energies which can be reached at the LHeC as well as the FCC-eh.
The parameterisations used in the calculation were the so-called IP-Sat ~\cite{Kowalski:2003hm,Kowalski:2006hc} and b-CGC ~\cite{Watt:2007nr} models. In both cases the impact parameter dependence has to be modelled phenomenologically. In 
the IP-Sat model the dipole amplitude has the following form
\begin{equation}
N(x,{\textbf r},{\textbf b}) \; = \; 1- \exp\left[-\frac{\pi^2 r^2}{2 N_c} \alpha_s(\mu^2) xg(x,\mu^2) T_G(b)\right] \; ,
    \label{eq:amplitude_IPSat}
\end{equation}
where $xg(x,\mu^2)$ is the collinear gluon density, evolved using LO DGLAP (without  quarks), from  an initial scale $\mu_0^2$ up to the scale $\mu^2$ set by the dipole size $\mu^2 = \frac{4}{r^2}+\mu_0^2$. $\alpha_s(\mu^2)$ is the strong coupling.
The parameterisation of the gluon density at the initial scale $\mu_0^2$ is given by
\begin{equation}
xg(x,\mu_0^2) = A_g x^{-\lambda_g} (1-x)^{5.6} \; ,
    \label{eq:initial_gluon_density}
\end{equation}
and the impact parameter profile for the gluon by
\begin{equation}
T_G(b) \; = \; \frac{1}{2\pi B_G} \exp(-b^2 / 2B_G) \; .
    \label{eq:impact_parameter_profile}
\end{equation}

An alternative parameterisation is given by the b-CGC model~\cite{Watt:2007nr} which has the form
\begin{align}
    N(x,{\textbf r},{\textbf b}) \; = \;
    \begin{cases}
    N_0 \left(\frac{rQ_s}{2}\right)^{2\gamma_\text{eff}} \;\;\; \text{for} \;\;\; rQ_s \le 2\; , \\
    1-\exp(-{\mathcal A} \ln^2 ({\mathcal B }r  Q_s))\;\;\;\text{for} \;\;\; rQ_s > 2\; .
    \end{cases}
    \label{eq:ampliutde_bcgc}
\end{align}
Here the effective anomalous dimension $\gamma_\text{eff}$ and the saturation scale $Q_s$ of the proton explicitly depend on
the impact parameter and are defined as
\begin{eqnarray}
\gamma_\textrm{eff} & = & \gamma_s + \frac{1}{\kappa \lambda \ln 1/x} \ln \left( \frac{2}{rQ_s}\right)\; , \nonumber \\
Q_s(x,{ b}) & = & \left(\frac{x_0}{x}\right)^{\lambda/2}  \exp \left[-\frac{b^2}{4\gamma_s B_\text{CGC}}\right]\;\;\; {\GeV}\; ,
\label{eq:bcgc_parameters}
\end{eqnarray}
where  $\kappa = \chi''(\gamma_s)/\chi'(\gamma_s)$, with $\chi(\gamma)$ being the leading-logarithmic BFKL kernel eigenvalue function~\cite{Lipatov:1985uk}. The parameters
${\mathcal A}$ and ${\mathcal B}$ in Eq.\eqref{eq:ampliutde_bcgc} are determined uniquely from the matching of the dipole amplitude and its logarithmic
derivatives at the limiting value of $rq_s=2$. The b-CGC model is constructed by smoothly interpolating between two analytically
known limiting cases~\cite{Watt:2007nr}, namely the solution of the BFKL equation in the vicinity of the saturation line for
small dipole sizes $r < 2/Q_s$, and the solution of the BK equation deep inside the saturation region for large dipole sizes $r>2/Q_s$. 

The parameters $\mu_0,A_g,\lambda_g$  of the IP-Sat model and $N_0,\gamma_s,x_0\lambda$ of the b-CGC model were fitted to obtain the best description of the inclusive data for the structure function $F_2$ at HERA. The slope parameters $B_g$ and $B_\text{CGC}$, which control the $b$ -dependence in both models, were fitted to obtain the best description   of  elastic diffractive  $J/\psi$ production, in particular its $t$-dependence, at small values of $t$.

\begin{figure}[!th]
    \centering
    \includegraphics[width=0.42\textwidth]{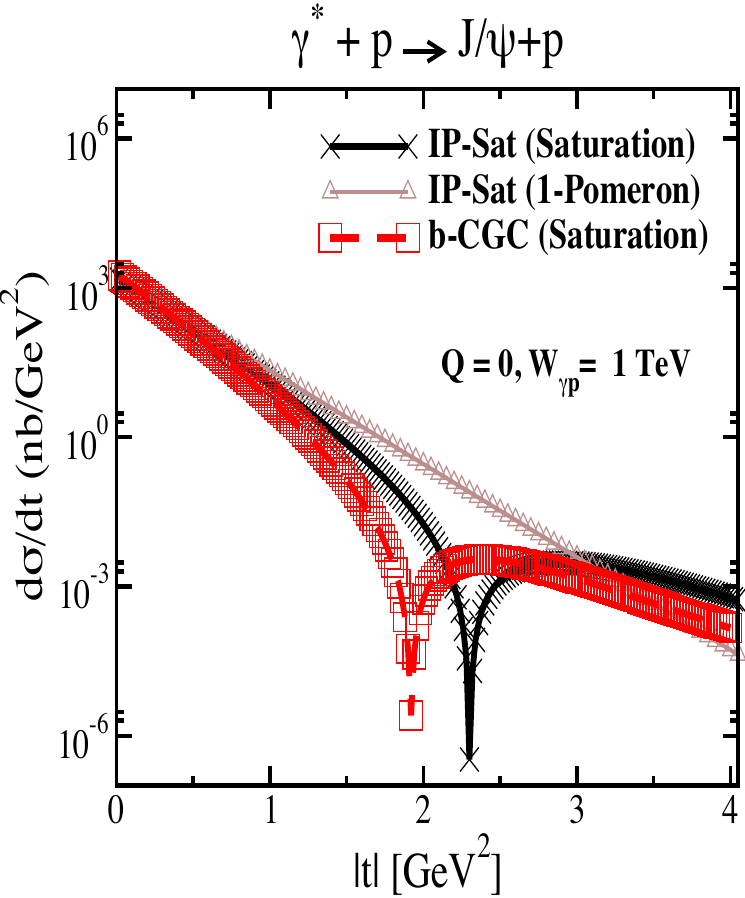}
    \hspace{0.05\textwidth}
    \includegraphics[width=0.42\textwidth]{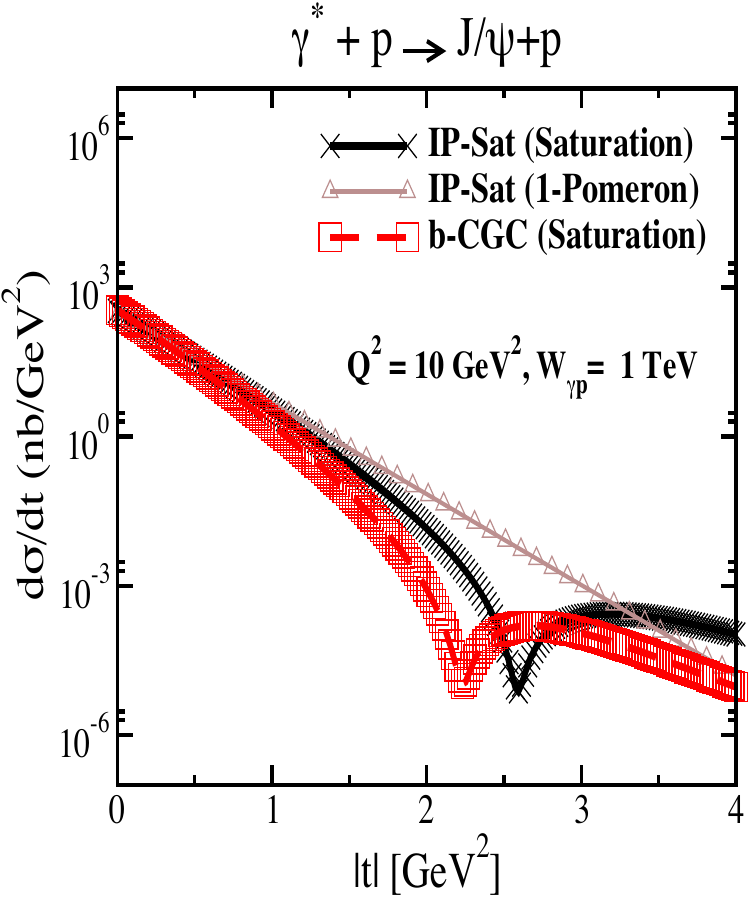}
    \caption{Differential cross section for the elastic $J/\psi$ production as a function of $|t|$ within the IP-Sat (saturation), b-CGC and 1-Pomeron models at a fixed $W\gamma p = 1 \, \text{TeV}$, which corresponds to the LHeC kinematics, and for two different values of photon virtuality $Q = 0$ and $Q^2=10 \; {\GeV}^2$. The thickness of points includes the uncertainties associated with the freedom to choose different values for the charm quark mass within the range $m_c=1.2-1.4\; {\GeV}$.}
    \label{fig:jpsi-t-1tev}
\end{figure}

\begin{figure}[!th]
    \centering
    \includegraphics[width=0.42\textwidth]{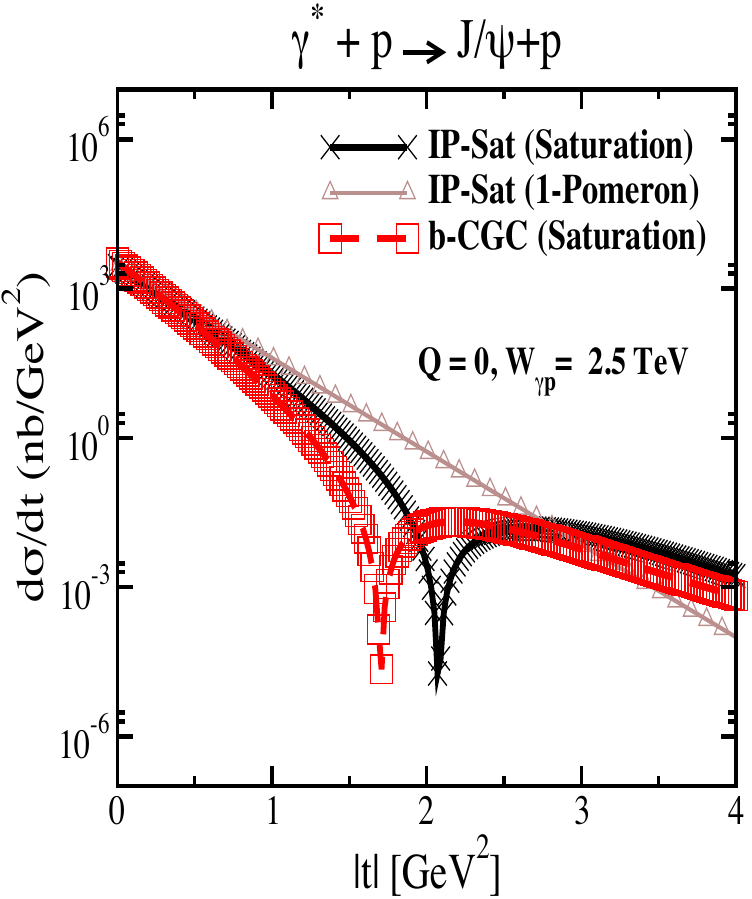}
    \hspace{0.05\textwidth}
    \includegraphics[width=0.42\textwidth]{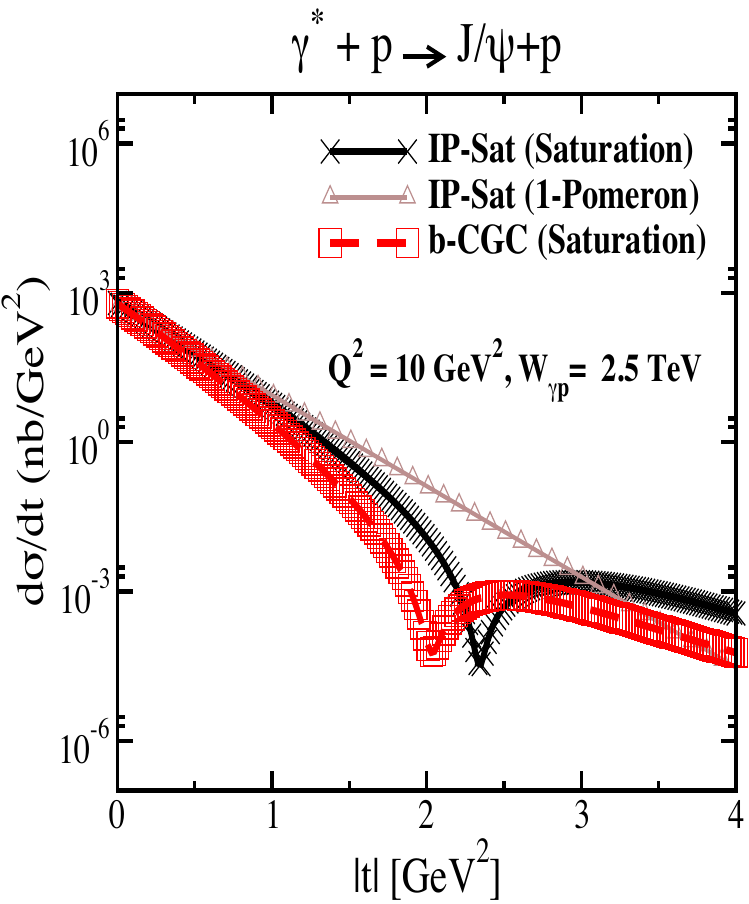}
 \caption{Differential cross section for elastic $J/\psi$ production as a function of $|t|$   within the IP-Sat (saturation), b-CGC and 1-Pomeron models at a fixed $W\gamma p = 2.5 \, \TeV$, which corresponds to the region that can be explored by FCC-eh, and for two different values of photon virtuality $Q = 0$ (left plot) and $Q^2=10 \; {\GeV^2}$ (right plot). The thickness of points includes the uncertainties associated with the freedom to choose different values for the charm quark mass within the range $m_c=1.2-1.4\; {\GeV}$ .}    \label{fig:jpsi-t-2tev}
\end{figure}

In Figs.~\ref{fig:jpsi-t-1tev} and~\ref{fig:jpsi-t-2tev} we show the simulated differential cross section $d\sigma/dt$  as a function of $|t|$  and study its variation with energy and virtuality, and its model dependence. First, in Fig.~\ref{fig:jpsi-t-1tev} we show the differential cross section as a  function of $t$ for fixed energy $W=1 \,{\TeV}$, in the case of the photoproduction of $J/\psi$ (left plot) and for the case of DIS with $Q^2=10 \, {\GeV^2}$ (right plot). The energy $W$ corresponds to the LHeC kinematics. There are three different calculations in each plot, using the IP-sat model, the b-CGC model and the 1-Pomeron approximation. The last  one is obtained by keeping just the first non-trivial term in the expansion of the eikonalised formula of the IP-Sat amplitude \eqref{eq:amplitude_IPSat}. First, let us observe that all three models coincide for very low values of $t$, where the dependence on $t$ is exponential.  This is because for low $|t|$, relatively large values of impact parameter are probed in Eq.~\eqref{eq:dipole_vm_elastic} where the amplitude is small, and therefore the tail in impact parameter is Gaussian in all three cases. Since the Fourier transform of the Gaussian in $b$ is an exponential in $t$, the result at low $t$ follows. On the other hand, the three scenarios differ significantly for large values of $|t|$. In the case of the 1-Pomeron approximation the dependence is still exponential, without any dips, which is easily understood since the impact parameter profile is perfectly Gaussian in this case. For the two other scenarios, dips in $d\sigma/dt$ as a function in $t$ emerge. They signal the departure from the Gaussian profile in $b$ for small values of $b$ where the system is dense. A similar pattern can be observed when performing the Fourier transform of the Wood-Saxon distribution, which is the typical distribution used for the description of the matter density in nuclei. When $Q^2$ is increased the pattern of dips also changes. This is illustrated in Fig.~\ref{fig:jpsi-t-1tev}. It is seen that the dips move to  higher values of $|t|$ for DIS than for photoproduction. This can be understood from the dipole  formula Eq.~\eqref{eq:dipole_vm_elastic} which contains the integral over the dipole size. Larger values of $Q^2$ select smaller values of dipole size $r$, where the amplitude is smaller and thus in the dilute regime, where the profile in $b$ is again Gaussian. On the other hand, small scales select large dipole sizes for which  the dipole amplitude is larger and thus the saturation effects more prominent, leading to the distortion of the impact parameter profile and therefore to the emergence of dips in the differential cross section $d\sigma/dt$ when studied as a function of $t$.

In Fig.~\ref{fig:jpsi-t-2tev} we show the same calculation but for even higher energy $W=2.5 \, {\TeV}$, which could be explored in the FCC-eh. In this case we see that the dips move to  lower values of $|t|$. This can be easily understood, as with  increasing energy the dipole scattering amplitude increases, and thus the dilute-dense boundary shifts to larger values of $b$, meaning that the deviation from the exponential fall off occurs for smaller values of $|t|$. Similar studies~\cite{Armesto:2014sma} show also the change of the position of the dips with the mass of the vector meson: for  lighter vector mesons like $\rho,\omega,\phi$ the dips occur at smaller $t$ than for the heavier vector mesons $J/\psi$ or $\Upsilon$. We note that, of course,  the positions of the dips depend crucially on the details of the models, which are currently not constrained by the existing HERA data. We also note the sizeable uncertainties due to the charm quark mass (the fits to inclusive HERA data from which parameters of the models have been extracted are performed at each fixed value of the charm mass that is then used to compute exclusive $J/\psi$ production).

We thus see that the precise measurement of the $t$-slope in the elastic  production of vector mesons at the LHeC, and its variation with $x$ and scales, provide a unique opportunity to explore the transition between the dilute and dense partonic regimes. 
As mentioned earlier, elastic diffractive production is one among several different measurements which can be performed to explore the 3D structure of the hadron. Another one is  Deeply Virtual Compton Scattering which is a process sensitive to the spatial distribution of quarks inside the hadron. Previous preliminary  analyses~\cite{AbelleiraFernandez:2012cc} indicate a  huge potential of LHeC for the measurement of DVCS. Another example of a process that could be studied at the LHeC, is  diffractive exclusive dijet production. It has been suggested~\cite{Hatta:2016dxp} that this process is sensitive to the Wigner function, and that the transverse momentum and spatial distribution of partons can be extracted by measuring this process. The transverse momentum of  jets would be sensitive to the transverse momentum of the participating partons, whereas the momentum transfer of the elastically scattered proton would give a handle on the impact parameter distribution of the partons in the target~\cite{Altinoluk:2015dpi,Mantysaari:2019csc,Salazar:2019ncp}, thus giving a possibility to extract  information about the Wigner distribution. 


So far we have referred to coherent diffraction, i.e.\ to a scenario in which the proton remains intact after the collision. There also exists incoherent diffraction, where the proton gets excited into some state with the quantum numbers of the proton and separated from the rest of the event by a large rapidity gap. In order to apply the dipole formalism to the incoherent case, see Sec.~\ref{sec:NPP_EVM} where the formulae applicable for both protons and nuclei are shown. Here one must consider a more involved structure of the proton (e.g.\ as composed by a fixed~\cite{Mantysaari:2016ykx,Mantysaari:2016jaz,Mantysaari:2017dwh,Mantysaari:2018zdd} or a growing number with $1/x$ of hot spots~\cite{Cepila:2016uku,Bendova:2018bbb,Krelina:2019gee}). As discussed in Sec.~\ref{sec:NPP_EVM}, coherent diffraction is sensitive to the gluon distribution in transverse space, while incoherent diffraction is particularly sensitive to fluctuations of the gluon distribution. A prediction of the model with a growing number  of hot spots, both in models where this increasing number is implemented by hand~\cite{Cepila:2016uku,Bendova:2018bbb,Krelina:2019gee} and in those where it is dynamically generated~\cite{Mantysaari:2018zdd} from a fixed number at larger $x$, is that the ratio of incoherent to coherent diffraction will decrease with $W$, and that this decrease is sensitive to the details of the distribution of hot spots, and thus, to the fluctuations of the gluon distribution in transverse space.
In order to check these ideas, both the experimental capability to separate coherent from incoherent diffraction and a large lever arm in $W$, as available at the LHeC, are required.

In conclusion, measurements at the LHeC (particularly exclusive diffractive production of vector mesons, photons and other final states like dijets) will offer unprecedented opportunities to unravel the three-dimensional structure of hadrons in a kinematic region complementary to that at the EIC. Note that, such structure varies with $x$ or energy, so its measurement at small enough $x$ is key as input for both analytic calculations and Monte Carlo simulators at high energy hadron colliders. And the large lever arms both in $x$ and $Q^2$, as those offered by the LHeC, are required to understand the perturbative evolution of such quantities, as much as it is required for collinear PDFs.
Ultraperipheral collisions at the LHC, see Refs.~\cite{Baltz:2007kq,Klein:2017vua} and references therein, offer an alternative albeit less precise and for photoproduction.



\chapter{Exploration of Quantum Chromodynamics}
\label{sec:QCD}
The gauge theory formalism of Quantum Chromodynamics
(QCD) provides a very successful description of strong interactions
between confined partons.
Despite the undoubted success of QCD, 
the strong force still remains one of the
least known fundamental sectors of (particle) physics
which needs to be explored much deeper.

For an improved understanding of strong interactions and to answer a
variety of those open questions additional measurements with highest precision have to be performed.
At the LHeC, deep-inelastic electron-proton and lepton-nucleus
reactions will extend tests of QCD phenomena to a new
and yet unexplored domain up to the TeV scale and to $x$ values as
low as $10^{-6}$, and QCD measurements can be performed with
very high experimental precision.
This is because the proton is a \emph{strongly} bound system and
in deep-inelastic scattering (DIS) the exchanged \emph{colourless} photon
(or $Z$) between the electron and the parton inside the proton acts as
a neutral observer with respect to the phenomena of the strong force.
In addition, the over-constrained kinematic system in DIS allows for
precise (\emph{in-situ}) calibrations of the detector to measure the
kinematics of the scattered lepton, and,
more importantly here, also the hadronic final state.
In DIS, in many cases, the virtuality of the exchanged $\gamma/Z$ boson often
provides a reasonable scale to stabilise theoretical predictions.

In this Chapter, selected topics of QCD studies at the LHeC are
discussed.

 \section{Determination of the strong coupling constant}
 \label{sec:alphas} 
Quantum Chromodynamics 
(QCD)~\cite{Zweig:1981pd,Fritzsch:1973pi} has been established as the theory of strong
interactions within the Standard Model of particle physics.
While there are manifold aspects both from the theoretical and
from the experimental point-of-view,  by far the most important parameter of
QCD is the coupling strength which is most commonly expressed at the
mass of the $Z$ boson, $M_Z$, as \asmz.
Its (renormalisation) scale dependence is given by the QCD gauge
group SU(3)~\cite{Gross:1973id,Politzer:1973fx}.
Predictions for numerous processes in $e^+e^-$, $pp$
or $ep$ collisions are then commonly performed in the framework of
perturbative QCD, and (the lack of) higher-order QCD corrections often
represent limiting aspects for precision physics.
Therefore, the determination of the strong coupling constant \asmz 
constitutes one of the most crucial tasks for future precision
physics, while at the same time the study of the scale dependence of
\as provides an inevitable test of the validity of QCD as the theory
of strong interactions and the portal for GUT theories.

Different processes and methodologies can be considered for a
determination of \asmz\
(see e.g.\ reviews~\cite{Dissertori:2015tfa,Tanabashi:2018oca,dEnterria:2019its}).
Since QCD is an asymptotically free theory, with free behaviour at
high scales but confinement at low scales, a high
sensitivity to the value of \asmz\ is naturally obtained from low-scale
measurements.
However, the high-scale behaviour must then be calculated by solving the
renormalisation group equation, which implies the strict validity
of the theory and an excellent understanding of all subleading
effects, such as the behaviour around quark-mass thresholds.

Precision measurements at the LHeC offer the unique opportunity to
exploit many of these aspects.
Measurements of jet production cross sections or
inclusive NC and CC DIS cross sections provide a high sensitivity to
the value of \asmz, since these measurements can be performed at
comparably low scales and with high experimental precision. At the
same time, the LHeC provides the opportunity to test the running of the
strong coupling constant over a large kinematic range.
In this Section, the prospects for a determination of the strong
coupling constant with inclusive jet cross sections and with inclusive
NC/CC DIS cross sections are studied.

\subsection{Strong coupling from inclusive jet cross sections}
\begin{figure}[!th]
  \centering
  \includegraphics[width=0.25\textwidth]{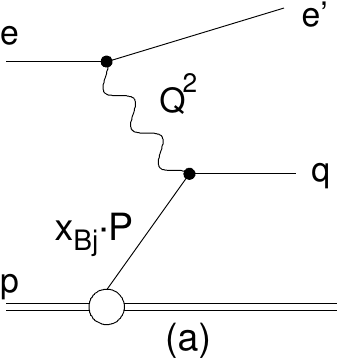}
  \hspace{0.05\textwidth}
  \includegraphics[width=0.25\textwidth]{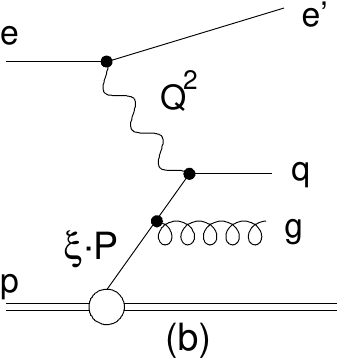}
  \hspace{0.05\textwidth}
  \includegraphics[width=0.25\textwidth]{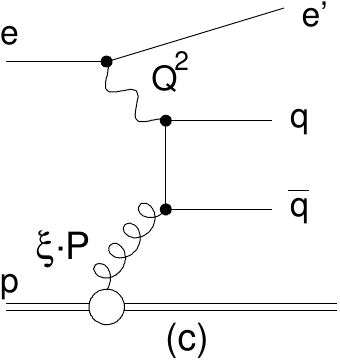}
  \caption{
    Leading order diagrams for inclusive DIS (a) and jet production (b,c) in
    the Breit frame (taken from Ref.~\cite{Aaron:2010ac}).
  }
  \label{fig:jetdiagrams}
\end{figure}

The measurement of inclusive jet or di-jet production cross sections in NC DIS provides a
high sensitivity to the strong coupling constant and to the gluon PDF
of the proton.
This is because jet cross sections in NC DIS are measured in the Breit reference frame~\cite{Streng:1979pv},
where the virtual boson $\gamma^\ast$ or $Z$ collides
head-on with the struck parton from the proton and the outgoing
jets are required to have a non-zero transverse momentum in that
reference frame.
The leading order QCD diagrams are QCD Compton and boson-gluon fusion
and are both $\mathcal{O}(\as)$, see Fig.~\ref{fig:jetdiagrams}.

At HERA, jets are most commonly defined by the
longitudinally invariant $k_t$ jet algorithm~\cite{Ellis:1993tq} with a distance
parameter $R=1.0$~\cite{Adloff:1998vc,Adloff:2000tq,Adloff:2002ew,Aktas:2003ja,Aktas:2004px,Aktas:2007aa,Aaron:2009vs,Aaron:2010ac,Andreev:2014wwa,Andreev:2016tgi,Breitweg:2000sv,Chekanov:2001fw,Chekanov:2002be,Chekanov:2004hz,Chekanov:2006xr,Chekanov:2006yc,Abramowicz:2010cka,Abramowicz:2010ke}.
This provides an infrared safe jet definition and the chosen distance
parameter guarantees a small dependence on non-perturbative effects,
such as hadronisation.
Differently than in $pp$ at the LHC~\cite{Khachatryan:2016mlc,Rabbertz:2017ssq,Aaboud:2017dvo,Aaboud:2017wsi}, jet algorithms at the LHeC do not
require any pile-up subtraction and any reduction of the dependence on
minimum bias or underlying event, due to the absence of such effects.
Therefore, for this study we adopt the choices made at HERA.

\begin{figure}[!th]
  \centering
  \includegraphics[width=0.55\textwidth]{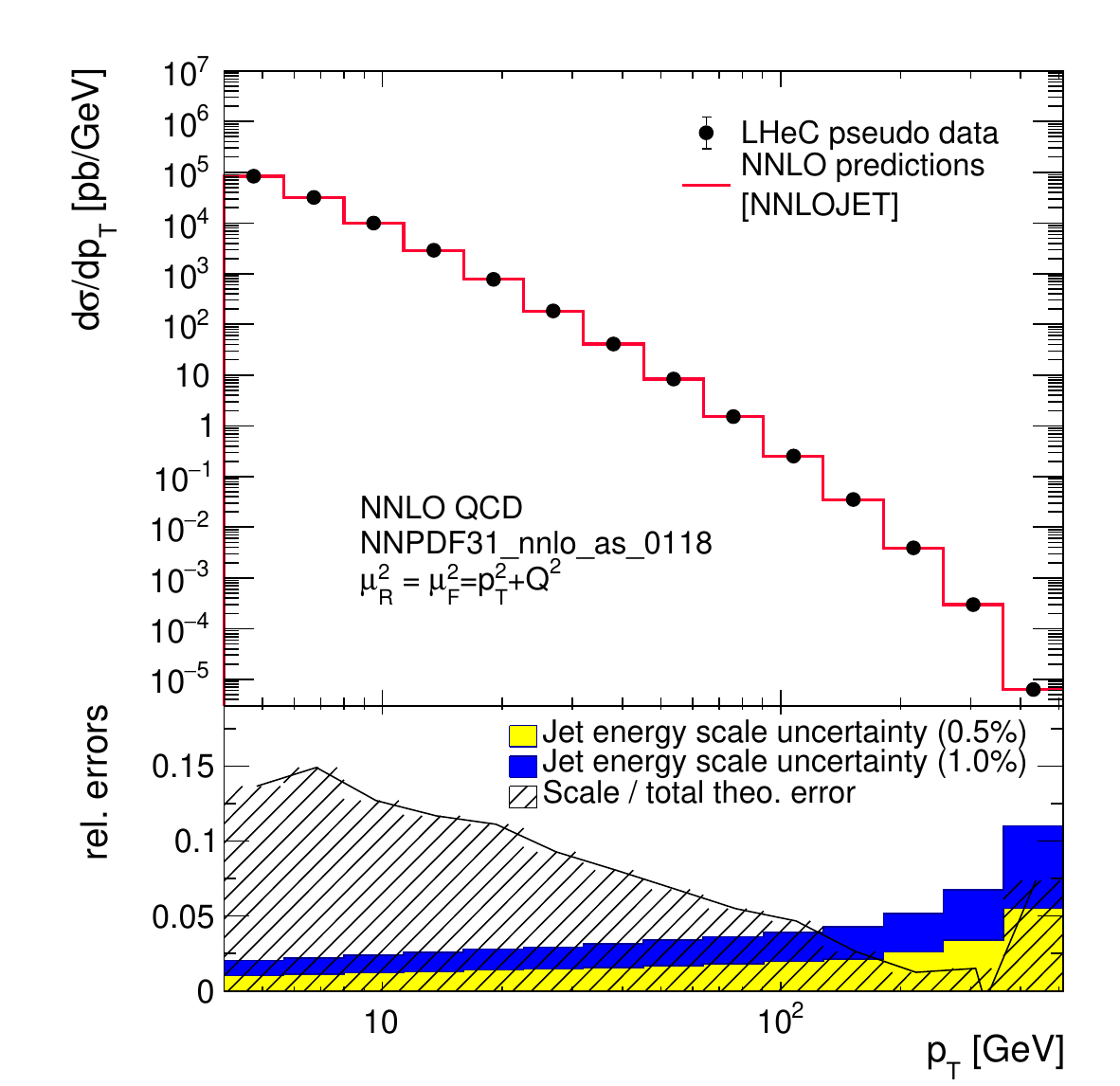}
  \caption{Inclusive jet cross sections calculated in NNLO QCD as a function of
    the jet transverse momentum in the Breit frame, $p_\text{T}$.
    The shaded area indicates NNLO scale uncertainties and the yellow 
    band shows the estimated experimental jet energy scale uncertainty (JES) of 0.5\,\%.
    The blue band shows a very conservative assumption on the JES of 1\,\%.
  }
  \label{fig:incljets}
\end{figure}
In Fig.~\ref{fig:incljets} the next-to-next-to-leading
order QCD (NNLO) predictions~\cite{Currie:2016ytq,Currie:2017tpe} for cross sections for inclusive jet
production in NC DIS as a function of the transverse 
momentum of the jets in the Breit frame are displayed.
The calculations are performed for an electron beam energy of $E_e=60\,\GeV$ and include
$\gamma/Z$ and $Z$ exchange terms and account for the electron
polarisation $P_e=-0.8$.
The NC DIS kinematic range is set to $\Qsq>4\,\GeVsq$.
The calculations are performed using the NNLOJET program~\cite{Gehrmann:2018szu} interfaced to the 
APPLfast library~\cite{Bendavid:2018nar,Amoroso:2020lgh,Britzger:2019kkb} which provides a generic 
interface to the APPLgrid~\cite{Carli:2005ji,Carli:2010rw} and fastNLO~\cite{Kluge:2006xs,Britzger:2012bs}
interpolation grid code.

The kinematically accessible range in jet-\pt ranges over two orders
of magnitude, $4<\pt\lesssim400\,\GeV$.
The size of the cross section extends over many orders
in magnitude, thus imposing challenging demands on LHeC experimental
conditions, triggers and DAQ bandwidth, calibration, and data processing
capabilities.
The scale uncertainty of the NNLO predictions is about 10\,\% at
low values of \pt and significantly decreases with increasing values
of \pt.
Future improved predictions will further reduce these theoretical
uncertainties.

For the purpose of estimating the uncertainty of \asmz\ in a
determination from inclusive jet cross sections at the LHeC,
double-differential cross sections as a function of \Qsq\ and \pt\ with
a full set of experimental uncertainties are generated.
Altogether 509 cross section values are calculated in the kinematic
range $8<\Qsq<500\,000\,\GeVsq$ and $4<\pt<512\,\GeV$, and the bin grid is
similar to the ones used by CMS, H1 or
ZEUS~\cite{Andreev:2017vxu,Abramowicz:2015mha,Khachatryan:2016mlc,Britzger:2019kkb}.
The various error sources considered are summarised in Tab.~\ref{tab:jetuncert}.
The uncertainties related to the reconstruction of the NC
DIS kinematic variables, \Qsq, $y$ and \xbj, are similar to the
estimates for the inclusive NC DIS cross sections (see section~\ref{sect:DISdata}).
For the reconstruction of hadronic final state particles which are the
input to the jet algorithm, jet energy scale uncertainty (JES),
calorimetric noise and the polar angle uncertainty are considered.
The size of the uncertainties is gauged with achieved values by H1,
ZEUS, ATLAS and CMS~\cite{Abramowicz:2010cka,Andreev:2014wwa,Khachatryan:2016kdb,Aaboud:2019ibw}.
The size of the dominant JES one is assumed to be 0.5\,\% for
reconstructed particles in the laboratory rest frame, yielding an
uncertainty of 0.2--4.4\,\% on the cross section after the
boost to the Breit frame.
A JES uncertainty of 0.5\,\% is well justified by improved calorimeters, since
already H1 and ZEUS reported uncertainties of 1\,\%~\cite{Abramowicz:2010cka,Kogler:2011zz,Andreev:2014wwa},
and ATLAS and CMS achieved
1\,\% over a wide range in \pt~\cite{Khachatryan:2016kdb,Aaboud:2019ibw}, albeit the presence of pile-up and
the considerably more complicated definition of a reference object for
the in-situ calibration.
The size of the JES uncertainty is also displayed in Fig.~\ref{fig:incljets}.
The calorimetric noise of $\pm 20\,\MeV$ on every calorimeter cluster,
as reported by H1, yields an uncertainty of up to 0.7\,\% on the jet
cross sections.
A minimum size of the statistical uncertainty of $0.15\,\%$ is imposed for each cross section bin.
An overall normalisation uncertainty of 1.0\,\% is
assumed, which will be mainly dominated by the luminosity uncertainty.
In addition, an uncorrelated uncertainty component of 0.6\,\% collects various smaller
error sources, such as for instance radiative corrections, unfolding
or model uncertainties.
Studies on the size and the correlation model of these uncertainties
are performed below.
\begin{table}[ht]
  \centering
  \small
  \begin{tabular}{lcc} 
    \toprule
    Exp. uncertainty & Shift & Size on $\sigma$ [\%] \\
    \midrule
    Statistics with $1\,\text{ab}^{-1}$ & min. 0.15\,\%         &  $0.15\,$--$5$ \\
    Electron energy       & 0.1\,\%               &  $0.02\,$--$0.62$ \\
    Polar angle           & 2\,mrad               &  $0.02\,$--$0.48$ \\
    Calorimeter noise     & $\pm20\,\text{MeV}$    &  $0.01\,$--$0.74$ \\
    Jet energy scale (JES)& 0.5\,\%               &  $0.2\,$--$4.4$ \\
    Uncorrelated uncert.  & $0.6\,\%$             &  $0.6$ \\
    Normalisation uncert. & $1.0\,\%$             &  $1.0$ \\
    \bottomrule
  \end{tabular}
  \caption{Anticipated uncertainties of inclusive jet cross section measurements at the LHeC.} 
  \label{tab:jetuncert}
\end{table}

The value and uncertainty of \asmz\ is obtained in a $\chi^2$-fit of NNLO
predictions~\cite{Currie:2016ytq,Currie:2017tpe}
to the simulated data with $\asmz$ being a free fit parameter.
The methodology follows closely analyses of HERA jet
data~\cite{Andreev:2017vxu,Britzger:2019kkb} and the $\chi^2$  quantity is
calculated from relative uncertainties, i.e.\ those of the right column of
Tab.~\ref{tab:jetuncert}.
The predictions for the cross section $\sigma$ account for both $\alpha_s$-dependent terms in the
NNLO calculations, i.e.\ in the DGLAP operator and the hard matrix
elements, by using 
\begin{equation}
  \sigma = f_{\mu_0} \otimes P_{\mu_0\rightarrow\mu_F}(\alpha_s(M_z))\otimes\hat{\sigma}(\alpha_s(M_z),\mu)\,,
\end{equation}
where $f_{\mu_0}$ are the PDFs at a scale of $\mu_0=30\,\GeV$, and $P_{\mu_0\rightarrow\mu_F}$ denotes the DGLAP operator, which is dependent on the value of \asmz.
The $\alpha_s$ uncertainty is obtained by linear error propagation and
is validated with a separate study of the $\Delta\chi^2=1$ criterion.   

In the fit of NNLO QCD predictions to the simulated
double-differential LHeC inclusive jet cross sections an uncertainty of
\begin{equation}
  \Delta\asmz \text{(jets)} = \pm0.00013_\text{(exp)} \pm 0.00010_\text{(PDF)}
\end{equation}
is found.
The PDF uncertainty is estimated from a PDF set obtained from LHeC
inclusive DIS data (see Sec.~\ref{sec:LHeCPDF}).
These uncertainties promise a determination of \asmz\ with the  highest
precision and would represent a considerable reduction of the current
world average value with a present uncertainty of $\pm0.00110$~\cite{Tanabashi:2018oca}.

\begin{figure}[!th]
  \centering
  \includegraphics[width=0.38\textwidth,trim={20 0 50 50},clip]{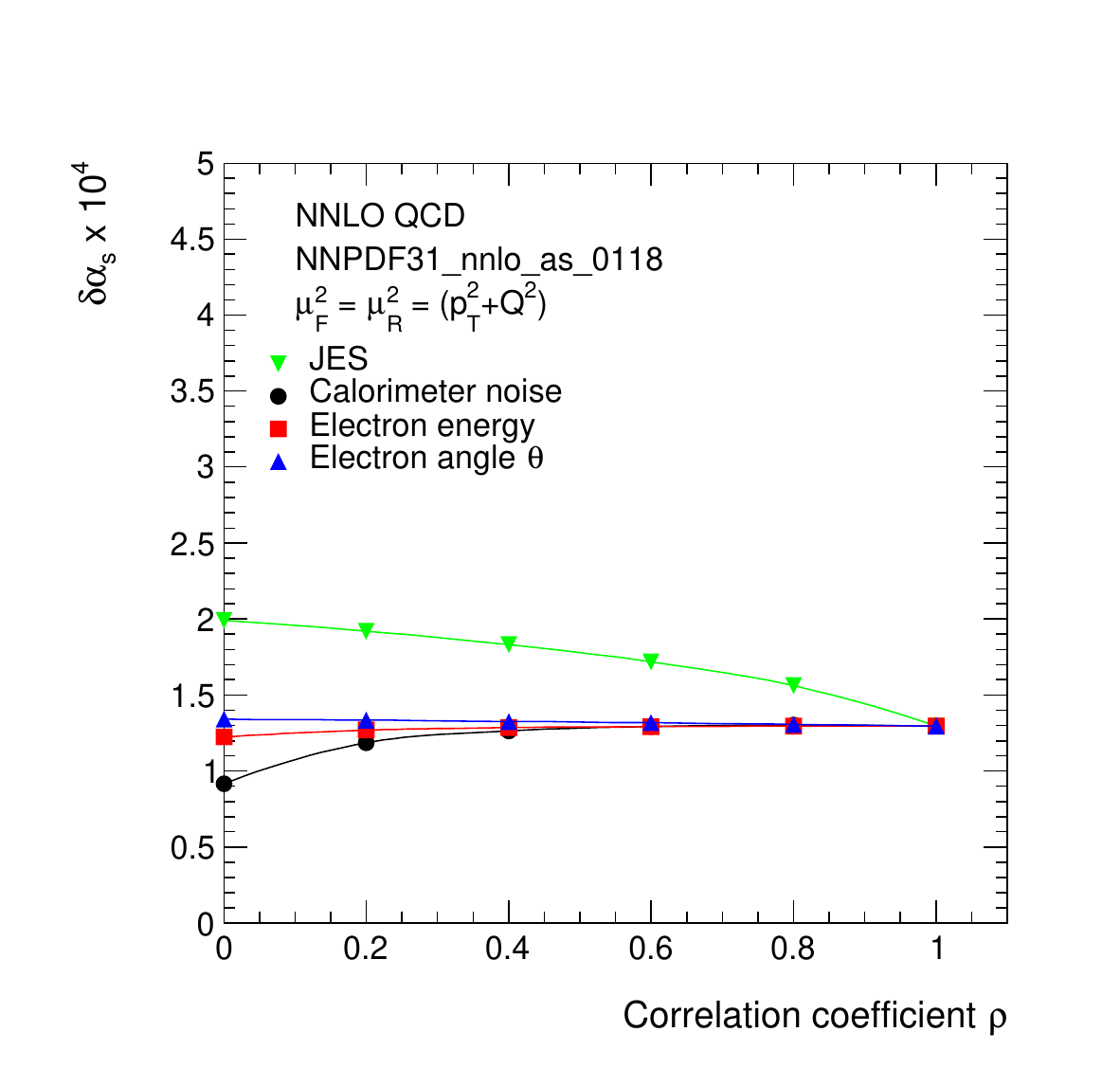}
  \hspace{0.05\textwidth}
  \includegraphics[width=0.38\textwidth,trim={20 0 50 50},clip]{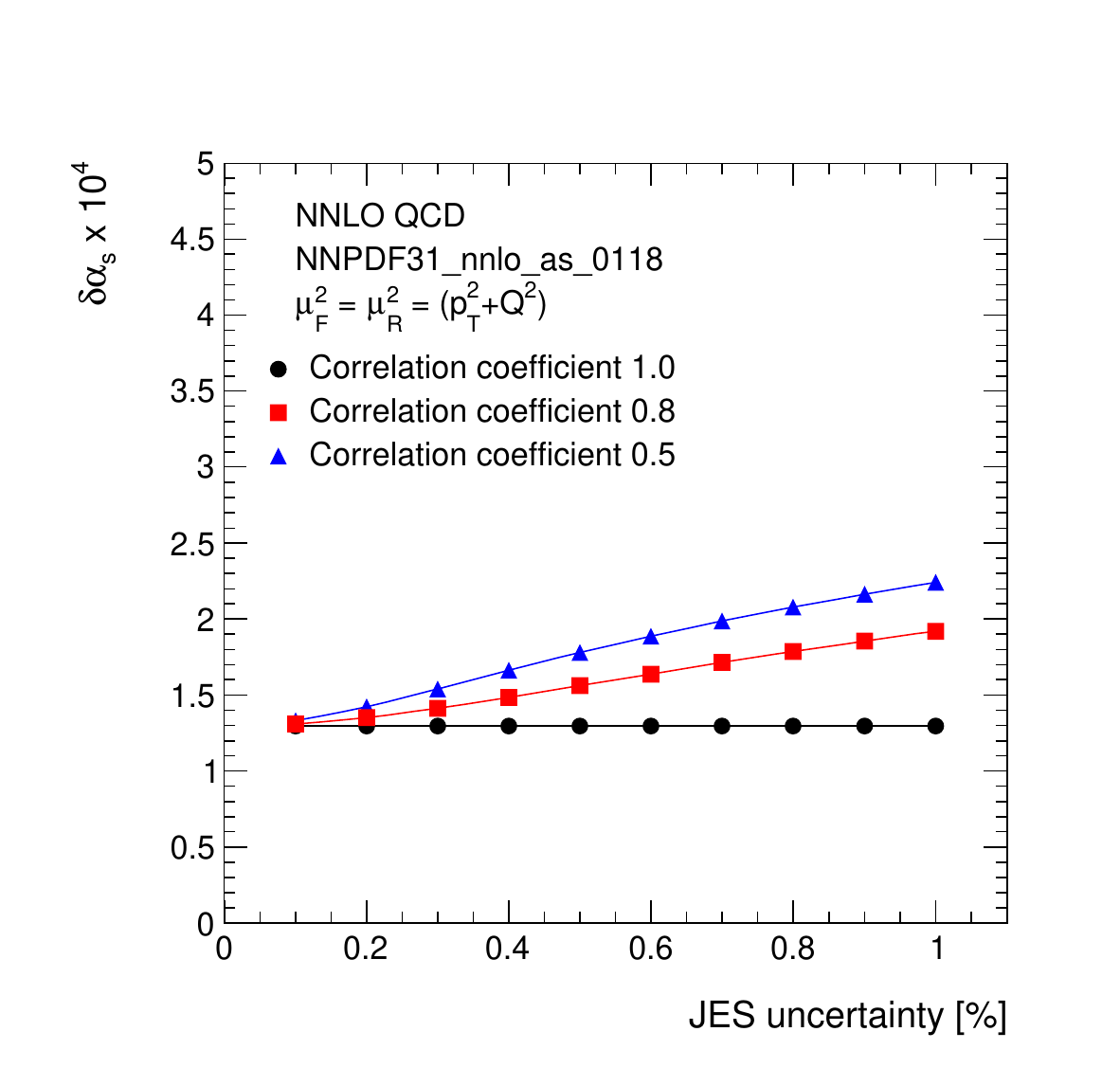} \\
  \includegraphics[width=0.38\textwidth,trim={20 0 50 50},clip]{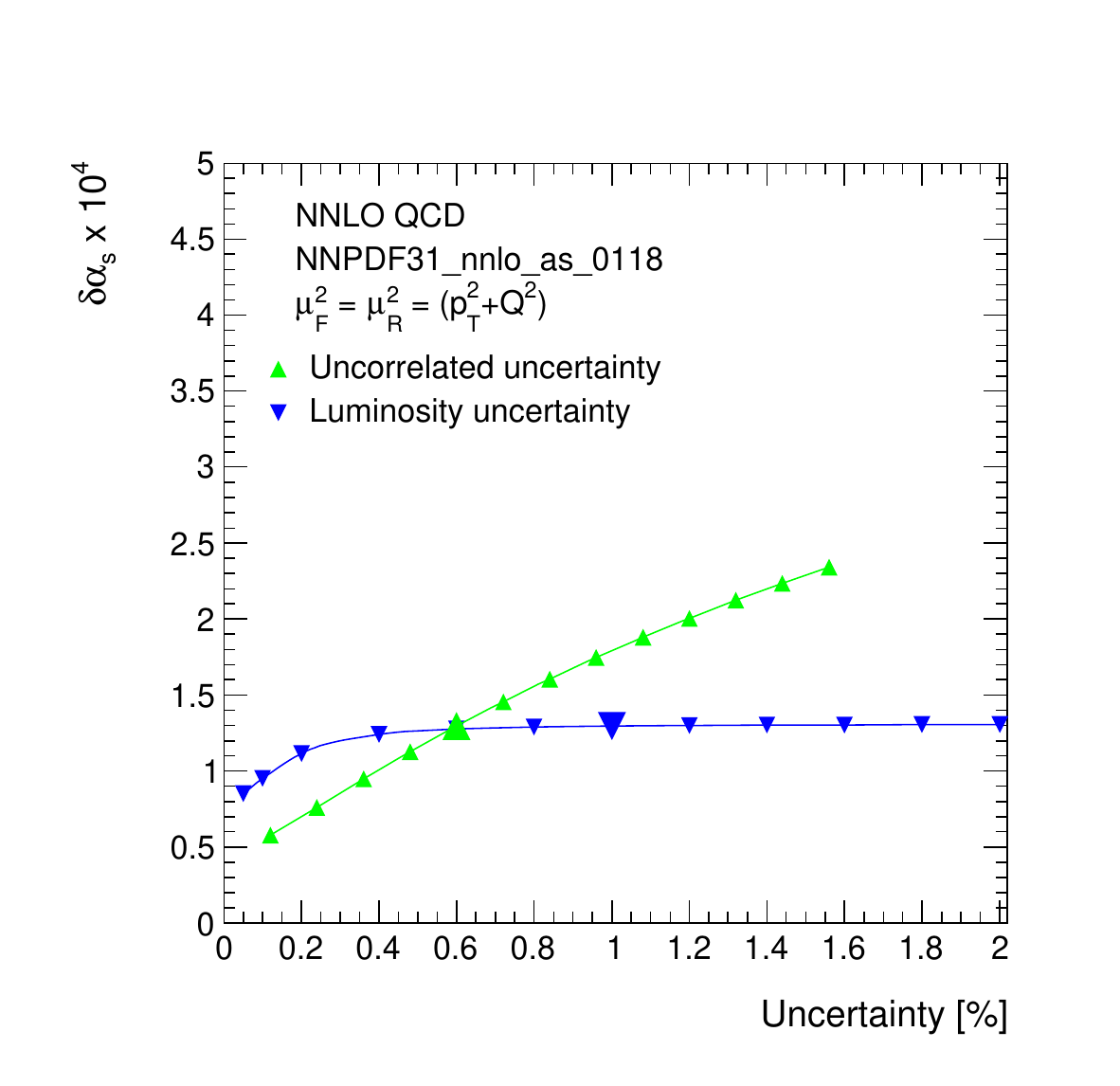} 
  \hspace{0.05\textwidth}
  \begin{minipage}{0.38\textwidth}
    \centering
    \vspace{-0.9\textwidth}
    \includegraphics[width=0.7\textwidth,trim={20 0 5 10},clip]{ErrorBudget}
  \end{minipage}
  \caption{
    Studies of the size and correlations of experimental uncertainties impacting the uncertainty of $\alpha_s(M_Z)$.
    Top left:
    Study of the value of the correlation coefficient $\rho$ for
    different systematic uncertainties. Common systematic
    uncertainties are considered as fully correlated, $\rho=1$.
    Top right:
    Size of the JES uncertainty for three different values of
    $\rho_\text{JES}$.
    Bottom left:
    Impact of the uncorrelated  and normalisation
    uncertainties on $\Delta\asmz$.
    Bottom right: Contribution of individual sources of experimental
    uncertainty to the total experimental uncertainty of \asmz.
  }
  \label{fig:aserrors}
\end{figure}
The uncertainty of $\alpha_s$ is studied for different values of the experimental uncertainties for the inclusive jet cross section measurement and
for different assumption on bin-to-bin correlations, expressed by the correlation
coefficient $\rho$, of
individual uncertainty sources, as shown in Fig.~\ref{fig:aserrors}.
It is observed that, even for quite conservative scenarios,
$\alpha_s(M_Z)$ will be determined with an uncertainty smaller than
2\,\textperthousand.
For this, it is important to keep the size of
the uncorrelated uncertainty or the uncorrelated components of other
systematic uncertainties under good control.
This is also visible from Fig.~\ref{fig:aserrors} (bottom right),
where the contributions of the individual uncertainty sources to the
total uncertainty of \asmz\ are displayed, and it is seen that the
uncorrelated and the normalisation uncertainty are the largest
individual uncertainty components.
It is further observed, that the size of the statistical uncertainty (stat.)
is non-negligible, which is, however, strongly dependent on the
\emph{ad hoc} assumption on the minimum size of 0.15\,\%.
The noise uncertainty contributes mainly to jets at low-\pt, and since
these have a particular high sensitivity to \asmz, due to their low
scale \mur.
It is of great importance to keep this experimental
uncertainty well under control, or make better use of track-based
information for the measurment of jets.

In the present formalism theoretical uncertainties from scale
variations of the NNLO predictions amount to about
$\Delta\asmz=0.0035\,{\text{(NNLO)}}$. 
These can be reduced with suitable cuts in \pt\ or \Qsq\ to about
$\Delta\asmz\approx0.0010$.
However, it is expected that improved predictions, e.g.\ with resummed
contributions or N$^3$LO predictions will significantly reduce these
uncertainties in the future.
Uncertainties on non-perturbative hadronisation effects will have to
be considered as well, but these will be under good control due to the
measurements of charged particle spectra at the LHeC and improved
phenomenological models.

\subsection[Pinning Down $\alpha_s$ with Inclusive and Jet LHeC Data]{\boldmath Pinning Down $\alpha_\textrm{s}$ with Inclusive and Jet LHeC Data}
\label{sec:alphasjets}
\begin{figure}[!th]
  \centering
  \includegraphics[width=0.60\textwidth]{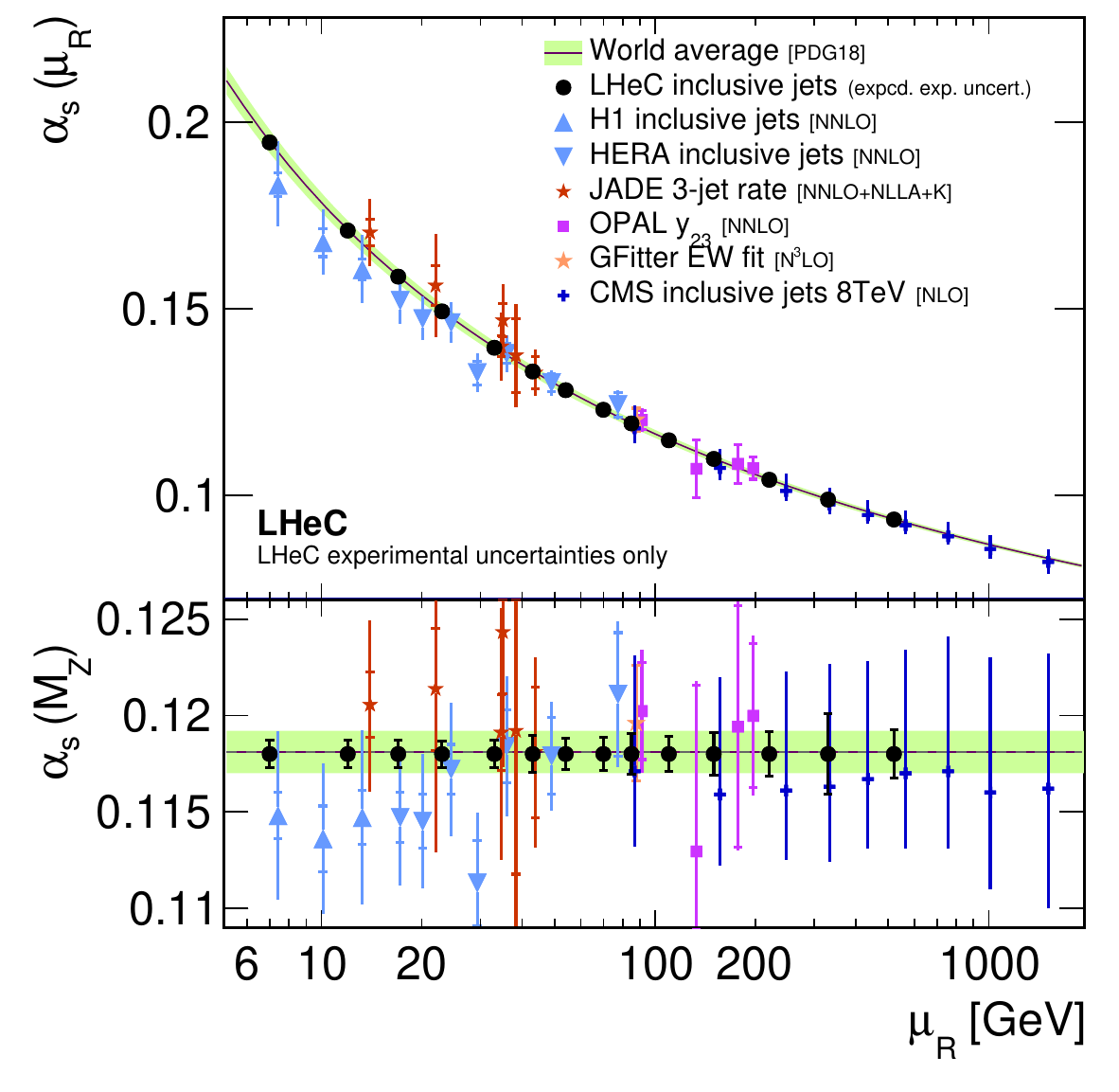}
  \caption{
    Uncertainties of $\alpha_s(M_Z)$ and corresponding
    $\alpha_s(\mu_R)$ in a determination of $\alpha_s$ using LHeC
    inclusive jet cross sections at different values of
    $\mu_R^2=Q^2+p_T^2$.
    Only experimental uncertainties are shown for
    LHeC and are compared with a number of presently available measurements
    and the world average value.
  }
    \label{fig:as_running}
\end{figure}

The dependence of the coupling strength as a function of the
renormalisation scale \mur\ is predicted by QCD, which
is often called the \emph{running} of the strong coupling.
Its study with experimental data represents an important consistency
and validity test of QCD.
Using inclusive jet cross sections the running of the
strong coupling can be tested by determining the value of \as\ at
different values of \mur\ by grouping data points with similar values
of \mur\ and determining the value of $\as(\mur)$ from these subsets
of data points.
The assumptions on the running of $\as(\mur)$ are then imposed only for
the limited range of the chosen interval, and not to the full measured interval as in the previous study.
Here we set $\mur^2=\Qsq+\pt^2$~\footnote{
  The choice of the scales follows a \emph{conventional} scale setting procedure and
  uncertainties for the scale choice and for unknown higher order terms are estimated by varying the scales.
  Such variations are sensitive only  to the terms which govern the behaviour of the running coupling,
  and may become unreliable due to renormalons~\cite{Ellis:1995jv}.
  An alternative  way to fix the scales is provided by the Principle of Maximum Conformality (PMC)~\cite{Brodsky:2011ta,Brodsky:2012rj,Brodsky:2011ig,Mojaza:2012mf,Brodsky:2013vpa}. 
  The PMC method was recently applied to predictions of event shape observables in $e^+ e^- \to $ hadrons~\cite{Wang:2019isi}.
  When applying the PMC method to observables in DIS, the alternative scale setting provides a
  profound alternative to verify the running of $\as(\mur)$.
  Such a procedure could be particularly relevant for DIS event shape observables,
  where the leading-order terms are insensitive to \as\ and conventional scale choices may not
  be adequately related to the $\alpha_s$-sensitive higher order QCD corrections.
}.
The experimental uncertainties from the fits to subsets of the inclusive
jet pseudodata are displayed in Fig.~\ref{fig:as_running}.
These results demonstrate a high sensitivity to $\alpha_s$
over two orders of magnitude in renormalisation scale
up to values of about $\mur\approx500\,\GeV$.
In the range $6<\mur\lesssim200\,\GeV$ the experimental
uncertainty is found to be smaller than the expectation from the world average
value~\cite{Tanabashi:2019}.
This region is of particular interest since it connects the
precision determinations from lattice calculations~\cite{Aoki:2019cca} or $\tau$ decay
measurements~\cite{Boito:2014sta}, which are at low scales $\mathcal{O}(\GeV)$,
to the measurements at the $Z$ pole~\cite{Baak:2014ora} and to the applications to scales which are relevant for the
LHC, e.g.\ for Higgs or top-quark physics or high-mass searches.
This kinematic region of scales $\mathcal{O}(10\,\GeV)$ cannot be
accessed by (HL-)LHC experiments because of limitations due to pile-up
and underlying event~\cite{Azzi:2019yne}.

Inclusive DIS cross sections are sensitive to \asmz\
through higher-order QCD corrections, contributions from the $F_L$
structure function and the scale dependence of the cross section at high $x$
(\emph{scaling violations}).
The value of \asmz\ can then be determined in a combined fit of the
PDFs and \asmz~\cite{Andreev:2017vxu}.
While a simultaneous determination of $\alpha_s(M_Z)$  and PDFs is not
possible with HERA inclusive DIS data
alone due to its limited precision and kinematic
coverage~\cite{Abramowicz:2015mha,Andreev:2017vxu}, the large
kinematic coverage, high precision and the integrated luminosity of
the LHeC data will allow for the first time such an \as\ analysis.

For the purpose of the determination of \asmz\ from inclusive NC/CC
DIS data, a combined PDF+\as fit to the simulated data is 
performed, similar to the studies presented above, in Chapter~\ref{chapter:pdf}.
Other technical details are outlined in Ref.~\cite{Andreev:2017vxu}.
In this fit, however, the numbers of free parameters of the gluon
parameterisation is increased, since the gluon PDF and \asmz\ are highly
correlated and LHeC data are sensitive to values down to $x<10^{-5}$,
which requires additional freedom for the gluon parameterisation.
The inclusive data are restricted to $\Qsq\geq5\,\GeVsq$ in order to
avoid a region where effects beyond fixed-order perturbation theory may
become sizeable~\cite{Abramowicz:2015mha,Abt:2017nkc}.


Exploiting the full LHeC inclusive NC/CC DIS
data with $E_e=50\,\GeV$, the value of \asmz\
can be determined with an uncertainty
$\Delta\asmz = \pm0.00038$.
With a more optimistic assumption on the dominant uncorrelated
uncertainty of $\delta\sigma_{\text{(uncor.)}}=0.25\,\%$, an uncertainty as small as
\begin{equation}
  \Delta\asmz \text{(incl. DIS)} = \pm0.00022_\text{(exp+PDF)}\,
  \label{eq:asDIS}
\end{equation}
is achieved.
This would represent a considerable improvement over the present world
average value.
Given these small uncertainties, theoretical uncertainties from missing higher
orders or heavy quark effects have to be considered in addition.
\begin{figure}[!th]
  \centering
  \includegraphics[width=0.50\textwidth]{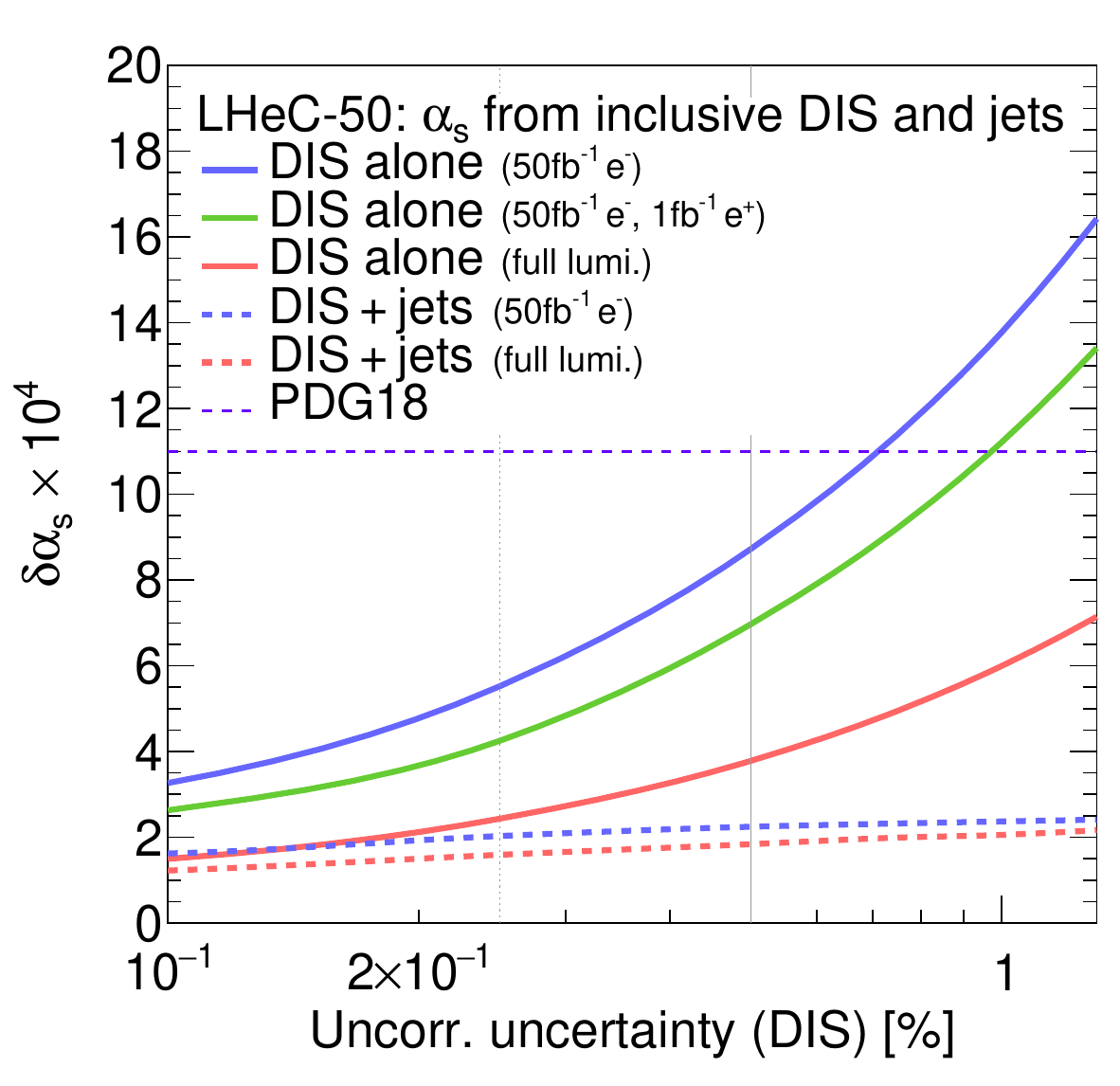}
  \caption{
    Uncertainties of $\alpha_s(M_Z)$ from simultaneous fits of
    \asmz\ and PDFs to inclusive NC/CC DIS data
    as a function of the size of the uncorrelated  uncertainty of the
    NC/CC DIS data.
    The full lines indicate the uncertainties obtained with different
    assumptions on the data taking scenario and integrated luminosity.
    The dashed lines indicate results where, additionally to the
    inclusive NC/CC DIS data, inclusive jet cross section data are considered.
  }
  \label{fig:asDIS}
\end{figure}
In a dedicated study, the fit is repeated with a reduced data set
which can be accumulated already during a single year of
operation~\footnote{Two different assumptions are made. One fit is
  performed with only electron data corresponding to
  $\mathcal{L}\sim50\,\text{fb}^{-1}$, and an alternative scenario
  considers further positron data corresponding to
  $\mathcal{L}\sim1\,\text{fb}^{-1}$.},
corresponding to about $\mathcal{L}\sim50\,\text{fb}^{-1}$.
Already these data will be able to improve the world average value.
These studies are displayed in Fig.~\ref{fig:asDIS}.

High sensitivity to \asmz\ and an optimal treatment of the PDFs
is obtained by using inclusive jet data together with inclusive NC/CC
DIS data in a combined determination of \asmz\ and the PDFs. 
The jet data will provide an enhanced sensitivity to \asmz, while inclusive
DIS data has the highest sensitivity to the determination of the
PDFs.
In such combined QCD analyses, also heavy quark data may be further
analysed to determine $m_c$ and $m_b$.
However, since jet cross sections have sufficiently high scale
($\pt\gg m_b$) these are fairly insensitive to the actual
value of the heavy quark masses.
Contrary, heavy quark data is predominantly sensitive to the quark
mass parameters rather than to \asmz, and their correlation is
commonly found to be small in such combined analyses, see
e.g.\ Ref~\cite{Alekhin:2017kpj}.
Infact, at LHeC the masses of charm and bottom quarks can be determined
with high precision and uncertainties of $3$\,MeV and 10\,MeV are
expected, respectively~\cite{AbelleiraFernandez:2012cc}. 
Therefore, for our sole purpose of estimating the uncertainty of
\asmz\ from LHeC data, we do not consider heavy quark data, nor free
values of $m_c$ or $m_b$ in the analysis, and we leave the outcome of
such a complete QCD analysis to the time when real data are available
and the actual value of the parameters are of interest.
At this time, also better theoretical predictions will be used,
including higher order corrections, heavy quark mass 
effects or higher-twist terms, as can be expected from steady
progress~\cite{Ablinger:2010ty,Ablinger:2014lka,Ablinger:2014nga,Behring:2014eya,Ablinger:2017xml,Ablinger:2018brx}.

For this study, the double-differential inclusive jet data as
described above, and additionally the inclusive NC/CC DIS data with
$E_e=50\,\GeV$ as introduced in Sec.~\ref{sect:DISdata}, are
employed.
Besides the normalisation uncertainty, all sources of systematic
uncertainties are considered as uncorrelated between the two
processes.
A fit of NNLO QCD predictions to these data sets is then performed, and
\asmz\ and the parameters of the PDFs are determined.
The methodology follows closely the methodology sketched in
Sect.~\ref{chapter:pdf}.
Using inclusive jet and inclusive DIS data in a single analysis,
the value of $\alpha_s(M_Z)$ is determined with an
uncertainty of
\begin{equation}
  \Delta\asmz \text{(incl. DIS \& jets)} = \pm0.00018_\text{(exp+PDF)}\,.
  \label{eq:asdisjets}
\end{equation}
This result will improve the world average value considerably.
However, theoretical uncertainties are not included and new mathematical tools and an improved understanding of
QCD will be needed in order to achieve  small values similar to the experimental ones.
The dominant sensitivity in this study arises from the jet data.
This can be seen from Fig.~\ref{fig:asDIS}, where $\Delta\asmz$ changes
only moderately with different assumptions imposed on the inclusive NC/CC
DIS data.
Assumptions made for the uncertainties of the inclusive jet data have
been studied above, and these results can be translated easily to this
PDF+\as fit.

\begin{figure}[!th]
  \centering
  \includegraphics[width=0.55\textwidth]{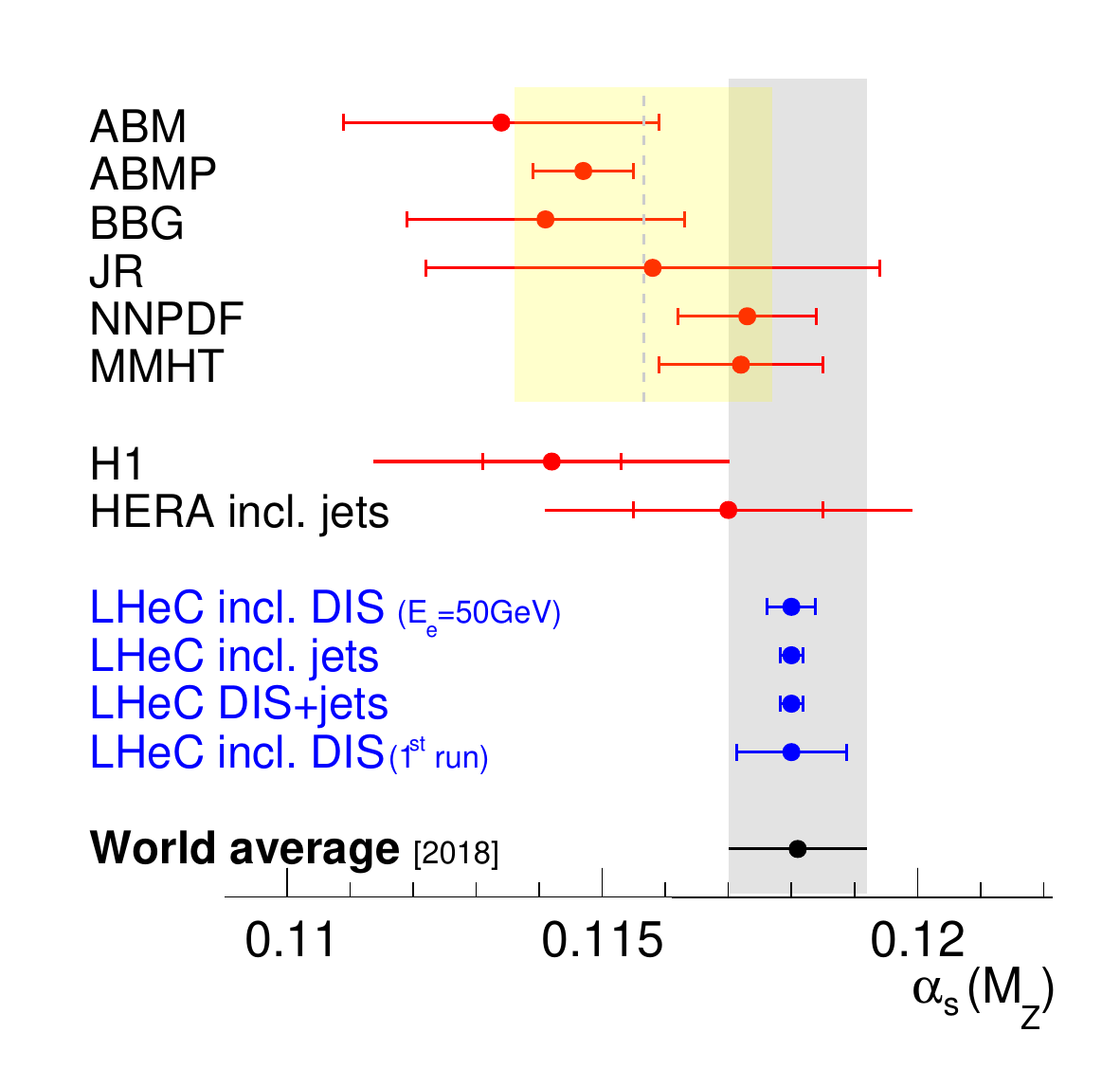}
  \caption{Summary of \asmz\ values in comparison with present values.
  }
  \label{fig:as_summary}
\end{figure}

The expected values for \asmz\ obtained from inclusive jets or from 
inclusive NC/CC DIS data are compared in
Fig.~\ref{fig:as_summary} with present determinations from global fits
based on DIS data (called \emph{PDF fits}) and the world average value~\cite{Tanabashi:2018oca}. 
It is observed that LHeC will have the potential to improve
considerably the world average value.  
Already after one year of data taking, the experimental uncertainties
of the NC/CC DIS data are competitive with the world average value.
The measurement of jet cross sections will further improve that value (not shown).

Furthermore, LHeC will be able to address a long standing puzzle. All
\as\ determinations from global fits based on NC/CC DIS data find a lower
value of \asmz\ than determinations in the lattice QCD framework, from
$\tau$ decays or in a global electroweak fit.
With the expected precision from LHeC this discrepancy will be resolved.


\subsection{Strong coupling from other processes}
A detailed study for the determination of \asmz\ from NC/CC DIS and from inclusive 
jet data was presented in the previous paragraphs.
However, a large number of additional processes and observables that are measured at the
LHeC can also be considered for a determination of \asmz.
Suitable observables or processes are  di-jet and multi-jet production,
heavy flavour production, jets in photoproduction or event shape observables.
These processes all exploit the \as\ dependence of the hard
interaction.
Using suitable predictions, also \emph{softer} processes
can be exploited for an \as\ determination. Examples could be
jet shapes or other substructure observables, or charged particle
multiplicities.

Since \asmz\ is a parameter of a phenomenological model, the total
uncertainty of \asmz\ is always a sum of experimental
 and theoretical uncertainties which are related to the
definition of the observable and to the applied model,
e.g.\ hadronisation uncertainties, diagram removal/subtraction
uncertainties or uncertainties from missing higher orders.
Therefore, credible prospects for the total uncertainty of \asmz\ from
other observables or processes altogether are difficult to predict, even more
since LHeC will explore a new kinematic regime that was
previously unmeasured.

In a first approximation, for any process the sensitivity to \asmz\
scales with the order $n$ of $\as$ in the leading-order diagram, $\as^n$.
The higher the power $n$ the higher the sensitivity to \asmz.
Consequently, the experimental uncertainty of an \as\ fit may reduce with
increasing power $n$.
Already at HERA three-jet cross section were proven to have a high
sensitivity to \asmz albeit their sizeable statistical
uncertainties~\cite{Aaron:2010ac,Andreev:2016tgi}.
At the LHeC, due to the higher $\sqrt{s}$ and huge integrated luminosity,
as well as the larger acceptance of the detector,
three-, four- or five-jet cross sections represent
highly sensitive observables for a precise determination of \asmz,
and high experimental precision can be achieved.
In these cases, fixed order pQCD predictions may become limiting factors, since they are
more complicated for large $n$.

Di-jet observables are expected to yield a fairly
similar experimental uncertainty than inclusive jet cross sections,
as studied in the previous paragraphs, since both have $n=1$ at LO.
However, their theoretical uncertainties may be smaller,
since di-jet observables are less sensitive to additional higher-order
radiation, in particular at lower scales where $\as(\mur)$ is larger.

Event shape observables in DIS exploit additional radiation in
DIS events (see e.g.\ review~\cite{Dasgupta:2003iq} or HERA measurements~\cite{Aktas:2005tz,Chekanov:2006hv}).
Consequently, once measured at the LHeC the experimental uncertainties of 
\asmz\ from these observables are expected to become very similar to that in
Eq.~\eqref{eq:asdisjets}, since both the event sample and the process is
similar to the inclusive jet cross sections~\footnote{It shall be
  noted, that event shape observables in NC DIS can be defined in the
  laboratory rest frame or the Breit frame.}.
However, different reconstruction techniques of the observables may
yield reduced experimental uncertainties,
and the calculation of event shape observables allow for the resummation
of large logarithms, and steady theoretical advances promise
small theoretical uncertainties~\cite{Kang:2013nha,Kang:2013lga,Kang:2015swk,Abelof:2016pby,Hoche:2018gti,Currie:2018fgr,Gehrmann:2019hwf}.

Jet production cross sections in photoproduction
represents a unique opportunity for another precision determination
of \asmz.
Such measurements have been performed at HERA~\cite{Adloff:2003nr,Aktas:2006qe,Chekanov:2007ac,Abramowicz:2012jz}.
The sizeable photoproduction cross section provides a huge event
sample, which is statistically independent from NC DIS events, and already the
leading-order predictions are sensitive to $\asmz$~\cite{Klasen:2002xb}.
Also its running can be largely
measured since the scale of the process is well estimated by the
transverse momentum of the jets $\mu_R \sim \pt^\text{jet}$.
Limiting theoretical aspects are due to the presence of a quasi-real
photon and the poorly known photon PDF~\cite{Gluck:1991jc,Sasaki:2018hud}.

A different class of observables represent heavy flavour (HF) cross
sections, which are discussed in Sec.~\ref{sec:HQ}.
Due to flavour conservation, these are commonly proportional
to $\mathcal{O}(\as^1)$ at leading-order.
However, when considering inclusive HF cross sections
above the heavy quark mass threshold heavy quarks can be
factorised into the PDFs, and the leading structure functions
$F_2^{c,b}$ are sensitive to \as\ only beyond the LO approximation
(see reviews~\cite{Behnke:2015qja,Zenaiev:2016kfl}, recent
HERA measurements~\cite{H1:2015dma,H1:2018flt} and references therein).
The presence of the heavy quark mass as an additional scale stabilises
perturbative calculations, and reduced theoretical uncertainties are expected.

At the LHeC the structure of jets and the formation of hadrons can be
studied with unprecedented precision.
This is so because of the presence of a single  hadron in the
initial state. Therefore, limiting effects like the underlying event or
pile-up are absent or greatly diminished.
Precise measurements of jet shape observables, or the study of jet
substructure observables~\cite{Larkoski:2017jix}, are highly sensitive to the value of \asmz,
because parton shower and hadronisation take place at lower
scales where the strong coupling becomes large and an increased
sensitivity to \asmz\ is attained~\cite{Bendavid:2018nar,Ringer:2019hdg}.

Finally, also the determination of \asmz\ from inclusive NC DIS cross
sections can be improved.
For NC DIS the dominant sensitivity to \as\ arises from the $F_L$
structure function and from scaling violations of $F_2$ at lower
values of \Qsq\ but at very high values of $x$.
Dedicated measurements of these kinematic regions will further improve
the experimental uncertainties from the estimated values in Eq.~\eqref{eq:asDIS}.

\section{Discovery of New Strong Interaction Dynamics at Small \emph{x}}
\label{sec:lowx}


The LHeC machine will offer access to a completely novel kinematic regime of DIS characterised by very small values of $x$. 
From the kinematical plane in $(x,Q^2)$ depicted in Fig.~\ref{fig:kinplane}, it is clear that the LHeC will be able to probe Bjorken-$x$ values as low as $10^{-6}$ for perturbative values of $Q^2$. At low values of $x$ various phenomena may occur which go  beyond the standard collinear perturbative description based on DGLAP evolution. Since the seminal works of Balitsky, Fadin, Kuraev and Lipatov~\cite{Balitsky:1978ic, Kuraev:1977fs, Lipatov:1985uk} it has been known that, at large values of centre-of-mass energy $\sqrt{s}$ or, to be more precise, in the Regge limit, there are large logarithms of energy which need to be resummed. Thus, even at low values of the strong coupling $\alpha_s$,  logarithms of energy $\ln s$ may be sufficiently large, such that terms like $(\alpha_s \ln s)^n$ will start to dominate the cross section.

In addition,   other novel effects may appear in the  low $x$ regime, which are related to the high gluon densities. At large parton densities the recombination of the gluons may become important in addition to the gluon splitting. This is known as the parton saturation phenomenon in QCD, and is deeply related to the restoration of the unitarity in QCD. As a result,  the linear evolution equations will need to be modified by the additional nonlinear terms in the gluon density.  In the next two subsections we shall explore the potential and sensitivity  of the LHeC to  these small $x$ phenomena in $ep$ collisions. Note also that, being a density effect, the non-linear phenomena leading to parton saturation are enhanced by increasing the mass number of the nucleus in $e$A. Chapter~\ref{chapter:nuclearphysics}, devoted to the physics opportunities with $e$A collisions at the LHeC, discusses this aspect, see also Ref.~\cite{AbelleiraFernandez:2012cc}.

\subsection{Resummation at small \emph{x}}
\label{sec:PSM_Disc_smallx}

The calculation of scattering amplitudes
in the high-energy limit
and the resummation of  $(\alpha_s \ln s)^n$ series in the leading logarithmic order was performed in Refs.~\cite{Balitsky:1978ic, Kuraev:1977fs, Lipatov:1985uk} and it resulted in the famous BFKL evolution equation.
This  small $x$  evolution equation, written for the so-called gluon Green's function or the unintegrated gluon density, is a differential equation in $\ln 1/x$. An important property of this equation is that it keeps the transverse momenta unordered along the gluon cascade. This has to be contrasted with DGLAP evolution which is differential in the hard scale $Q^2$ and relies on the strong ordering in the transverse momenta of the exchanged partons in the parton cascade.
The solution to the BFKL equation is a gluon density which grows sharply with decreasing $x$, as a power  i.e.\ $\sim x^{-\omega_{I\!P}}$, where $\omega_{I\!P}$
is the hard Pomeron intercept, and in the leading logarithmic approximation equals $\frac{N_c \alpha_s}{\pi} 4\ln 2 $, which gives a value of about $0.5$ for typical values of the strong coupling. The leading logarithmic (LLx) result yielded a growth of the gluon density which was too steep for the experimental data at HERA. The next-to-leading logarithmic (NLLx) calculation performed in the late 90s~\cite{Fadin:1998py,Ciafaloni:1998gs} resulted in large negative corrections to the LLx value of the hard Pomeron intercept and yielded some instabilities in the cross section~\cite{Ross:1998xw,Kovchegov:1998ae,Levin:1998pka,Armesto:1998gt} and it is important to account for subleading effects, since these are large~\cite{Blumlein:1995jp,Blumlein:1997em}.

The appearance of the large negative corrections at NLLx motivated the search for the appropriate resummation which would stabilize the result. It was understood very early that the large corrections which appear in BFKL at NLLx are mostly due to the kinematics~\cite{Ciafaloni:1987ur, Andersson:1995ju, Kwiecinski:1996td} as well as  DGLAP terms and the running of the strong coupling. First attempts at combining the BFKL and DGLAP dynamics together with the proper kinematics~\cite{Ellis:1993rb,Ellis:1995gv,Kwiecinski:1997ee} yielded encouraging  results, and allowed a description of HERA data on structure functions with good  accuracy. The complete resummation program was developed in a series of works~\cite{Catani:1993ww,Catani:1993rn,Catani:1994sq,Salam:1998tj,Ciafaloni:1999au,Ciafaloni:1999yw,Ciafaloni:2003ek,Ciafaloni:2003kd,Ciafaloni:2003rd,Ciafaloni:2007gf,Altarelli:1999vw,Altarelli:2000mh,Altarelli:2001ji,Altarelli:2003hk,Altarelli:2008aj,Thorne:2001nr,Vera:2005jt}.
 In these works the  resummation for the gluon Green's function and the splitting functions was developed. 

\begin{figure}[!th]
    \centering
\includegraphics[width=0.47\textwidth]{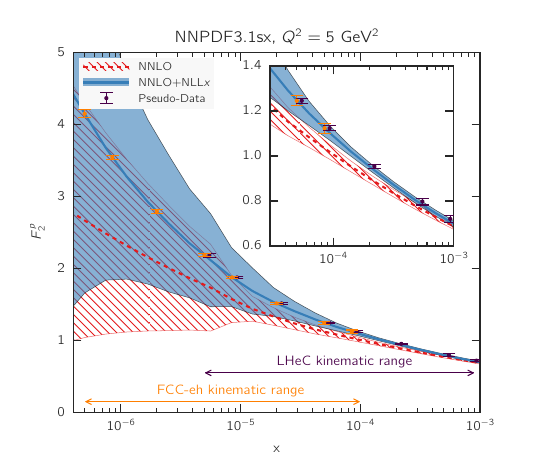}
\hspace{0.02\textwidth}
\includegraphics[width=0.47\textwidth]{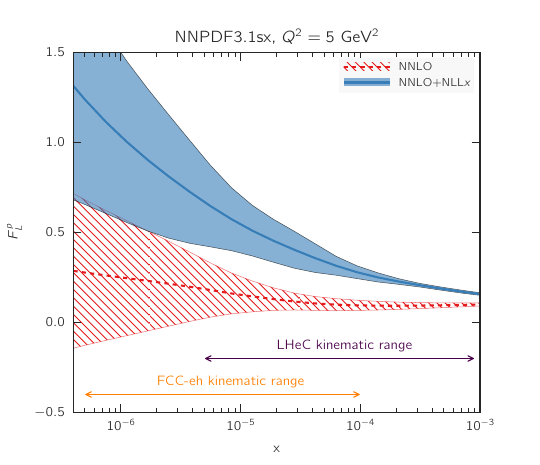}
\caption{Predictions for the $F_2$ and $F_L$ structure functions using the NNPDF3.1sx NNLO and NNLO+NLLx fits at $Q^2 = 5 \, \GeVsq$ for the kinematics of the LHeC and FCC-eh. In the case of $F_2$, we also show the expected total experimental uncertainties based on the simulated pseudodata, assuming the NNLO+NLLx values as the central prediction. A small offset has been applied to the LHeC pseudodata as some of the values of $x$ overlap with the FCC-eh pseudodata points. The inset in the left plot shows a magnified view in the kinematic region $x > 3 \times 10^{-5}$, corresponding to the reach of HERA data. Figure taken from Ref.~\cite{Bonvini:2016wki}.}
\label{fig:NNPDF_resum}
\end{figure}

The low-$x$ resummation  was recently applied to the description of  structure function data at HERA using the methodology of NNPDF~\cite{Bonvini:2016wki}. It was  demonstrated  that the resummed fits provide a better description of the structure function data than the pure DGLAP based fits at fixed NNL order. In particular, it was shown that the $\chi^2$ of the fits does not vary appreciably when more  small $x$ data are included in the case of the fits which include  the effects of the small-$x$ resummation. On the other hand, the fits based on NNLO DGLAP evolution exhibit a worsening of their quality in the region of low $x$ and low to moderate values of $Q^2$.  This indicates that there is some tension in the fixed order fits based on DGLAP, and that resummation alleviates it. In addition, it was shown that the description of the longitudinal structure function $F_L$ from HERA data is improved in the fits with the small $x$ resummation. This analysis  suggests that the small $x$ resummation effects are indeed visible in the HERA kinematic region. Such effects will  be strongly magnified at the LHeC, which probes values of $x$ more than one  order of magnitude lower than HERA. The NNPDF group also performed simulation of the structure functions $F_2$ and $F_L$  with and without resummation in the LHeC range as well as for the next generation electron-hadron collider FCC-eh~\cite{Bonvini:2016wki}. The predictions for the structure functions as a function of $x$ for fixed values of $Q^2$ are shown  in  Figs.~\ref{fig:NNPDF_resum}.


The simulations were done using APFEL~\cite{Bertone:2013vaa} together with the HELL package~\cite{Bonvini:2017ogt} which implements the small $x$ resummation.  From  Fig.~\ref{fig:NNPDF_resum} it is clear that LHeC will have much higher sensitivity to discriminate between fixed order and resummed scenarios  than the  HERA collider, with even better discrimination at the FCC-eh. The differences between the central values for the two predictions are of the order of $15\%$ for the case of $F_2$ and this is much larger than the projected error bar on the reduced cross section or structure function $F_2$ which could be measured at LHeC. For comparison, the simulated pseudodata for $F_2$ are shown together with the expected experimental uncertainties. The total uncertainties of the simulated pseudodata are at the few percent level at most, and are therefore  much smaller than the uncertainties coming from the PDFs in most of the kinematic range.

It is evident that  fits to the LHeC data will have  power to discriminate between the different frameworks. In the right plot in Fig.~\ref{fig:NNPDF_resum},  the predictions for the longitudinal structure function are shown. We see that in the case of  the $F_L$ structure function, the differences between the fixed order and resummed predictions are even larger, consistently over the entire range of $x$. This indicates the importance of the measurement of the longitudinal structure function $F_L$ which can provide further vital constraints on the QCD dynamics in the low $x$ region due to its sensitivity to the gluon density in the proton.

To further illustrate the power of a high energy DIS collider like the LHeC in exploring the dynamics at low $x$, fits which include the simulated data were performed.   The NNLO+NLLx resummed calculation was used to obtain the simulated pseudodata, both for the LHeC, in a scenario of a $60 \;\GeV$ electron beam on a $7 \;\TeV$ proton beam  as well  as in the case of the FCC-eh scenario with a $50\; \TeV$ proton beam. All the experimental uncertainties for the pseudodata have been added in quadrature. Next, fits were performed to the DIS HERA as well as LHeC and FCC-eh pseudodata using the theory with and without the resummation at low $x$. Hadronic data like jet, Drell-Yan or top, were not included for this analysis but, as demonstrated in~\cite{Bonvini:2016wki}, these data do not have much of the constraining  power at low $x$, and therefore the results of the analysis at low $x$ are independent of the additional non-DIS data sets. The quality of the fits characterised  by the $\chi^2$ was markedly worse when the NNLO DGLAP framework was used to fit the HERA data and the pseudodata from LHeC and/or FCC-eh than was the case with resummation. To be precise, the $\chi^2$ per degree of freedom for the HERA data set was equal to $1.22$ for the NNLO fit, and $1.07$ for the resummed fit. For the case of the LHeC/FCC-eh the $\chi^2$ per degree of freedom was equal to 1.71/2.72 and 1.22/1.34 for NNLO and NNLO+resummation fits, respectively. These results demonstrate the huge discriminatory power of the new DIS machines between the DGLAP and resummed  frameworks, and the large sensitivity to the low $x$ region while simultaneously probing low to moderate $Q^2$ values.

\begin{figure}[!th]
    \centering
    \includegraphics[width=0.47\textwidth]{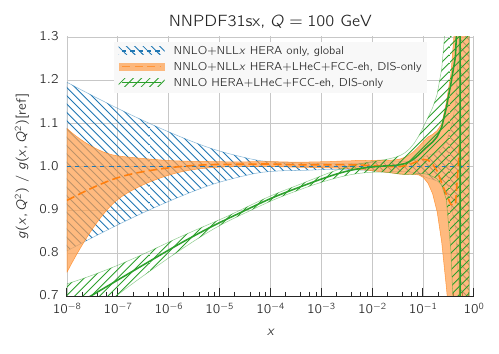}
    \hspace{0.02\textwidth}
    \includegraphics[width=0.47\textwidth]{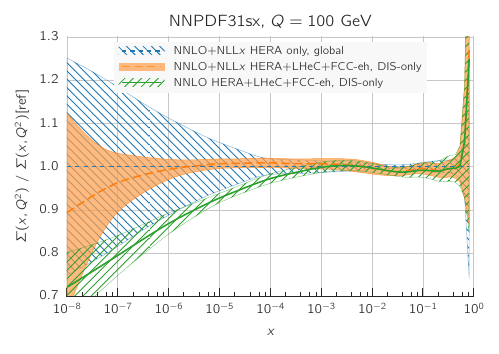}
    \caption{Comparison between the gluon (left plot) and the quark singlet (right plot) PDFs in the NNPDF3.1sx NNLO+NNLx fits without (blue hatched band) and with the LHeC+FCC-eh pseudodata (orange band) on inclusive structure functions. For completeness, we also show the results of the corresponding NNPDF3.1sx NNLO fit with LHeC+FCC-eh pseudodata (green hatched band). Figure taken from Ref.~\cite{Bonvini:2016wki}.}
    \label{fig:pdfs_resum}
\end{figure}

In Fig.~\ref{fig:pdfs_resum} the comparison of the gluon and quark distributions from the NNLO + NLLx fits is shown at $Q=100 \; \GeV$ as a function of $x$, with and without including the simulated pseudodata from LHeC as well as FCC-eh. The differences at large $x$ are due to the fact that only DIS data were included in the fits, and not the hadronic data. The central values of the extracted PDFs using only HERA or using HERA and the simulated pseudodata coincide with each other, but a large reduction in uncertainty is visible when the new data are included. The uncertainties from the fits based on the HERA data only increase sharply already at $x\sim 10^{-4}$. On the other hand, including the pseudodata from LHeC and/or FCC-eh can extend this regime by order(s) of magnitude down in $x$.   Furthermore, fits without resummation, based only on NNLO DGLAP, were performed to the HERA data and the pseudodata. We see that in this case the extracted gluon and singlet quark densities differ significantly from the fits using the NNLO+NLLx. Already at $x=10^{-4}$ the central values of the gluon differ by $10\%$ and at $x=10^{-5}$, which is the LHeC regime, the central values for the gluon differ by  $15\%$. This difference is  much larger than the precision with which the gluon can be extracted from the DIS data, which is of the order of $\sim 1\%$. 

The presented analysis demonstrates that the fixed order  prediction based on the DGLAP evolution would likely fail to describe accurately the structure function data in the new DIS machines and that  in that regime new dynamics including resummation are mandatory for quantitative predictions. Therefore, the LHeC machine  has an unprecedented potential to pin down the details of the QCD  dynamics at low values of Bjorken $x$. 

\def\gsim{\mathrel{\rlap{\lower4pt\hbox{\hskip1pt$\sim$}}
    \raise1pt\hbox{$>$}}}         
\def\lsim{\mathrel{\rlap{\lower4pt\hbox{\hskip1pt$\sim$}}
     \raise1pt\hbox{$<$}}}         
\subsection{Disentangling non-linear QCD dynamics at the LHeC}
  \label{sec:SM_nonlinearQCD}
%
%
As mentioned previously the kinematic extension of the LHeC
 will allow unprecedented tests of the strong interaction in the extremely low $x$ region, and allow for the tests of the novel QCD dynamics at low $x$. 
The second effect, in addition to resummation, that may be expected is the parton saturation phenomenon, which may manifest itself as the 
 deviation from the linear DGLAP evolution, and the emergence of the saturation scale.

In particular, it has been argued
that the strong growth of the gluon PDF at small-$x$ should eventually lead
to gluon recombination~\cite{Mueller:1989st} to avoid violating the unitary bounds.
The onset of such non-linear dynamics, also known as saturation, has been extensively
searched but so far there is no conclusive evidence of its presence, at least
within the HERA inclusive structure function measurements.
In this context, the extended kinematic range of the LHeC provides unique avenues
to explore the possible onset of non-linear QCD dynamics at small-$x$.
The discovery of saturation, a radically new regime of QCD, would then
represent an important milestone in our understanding of the strong
interactions.

The main challenge in disentangling saturation lies in the fact that non-linear corrections
are expected to be moderate even at the LHeC, since they
are small (if present at all) in the region covered by HERA.
Therefore, great care needs to be employed in order to separate such effects
from those of standard DGLAP linear evolution.
Indeed, it is well known that
HERA data at small-$x$ in the perturbative region can be equally well
described, at least
at the qualitative level, both
by PDF fits based on the DGLAP framework as well as by saturation-inspired models.
However, rapid progress both in theory calculations and methodological developments
have pushed QCD fits to a new level of sophistication, and recently it has been shown
that subtle but clear evidence of BFKL resummation at small-$x$ is present in HERA
data, both for inclusive and for heavy quark structure
functions~\cite{Ball:2017otu,Abdolmaleki:2018jln}.
Such studies highlight how it should be possible to tell apart non-linear from linear
dynamics using state-of-the-art fitting methods even if these are moderate,
 provided that they are within the LHeC reach.

Here we want to assess the sensitivity of the LHeC to detect the possible onset
of non-linear saturation dynamics.
This study will be carried out by generalising a recent analysis~\cite{AbdulKhalek:2019mps} that 
 quantified the impact of LHeC
 inclusive and semi-inclusive measurements on the PDF4LHC15
 PDFs~\cite{Butterworth:2015oua,Carrazza:2015aoa}
 by means
 of Hessian profiling~\cite{Paukkunen:2014zia}.
There, the LHeC pseudodata was generated assuming that linear DGLAP evolution was valid
in the entire LHeC kinematic range using the PDF4LHC15 set as input.
To ascertain the possibility of pinning down saturation at the LHeC, here
we have revisited this study but now generating the LHeC pseudodata
by means of a saturation-inspired calculation.
By monitoring the statistical significance of the tension that will
be introduced (by construction) between the saturation pseudodata
and the
DGLAP  theory assumed in the PDF fit, we aim to determine the likelihood of disentangling
non-linear from linear evolution effects at the LHeC.
See also~\cite{Rojo:2009ut} for previous related studies along the same direction.

\subsubsection{Analysis settings}
In this study we adopt the
settings of~\cite{Khalek:2018mdn,AbdulKhalek:2019mps},
to which we refer the interested reader for further details.
In Ref.~\cite{AbdulKhalek:2019mps} the impact on the proton PDFs
of inclusive and semi-inclusive neutral-current (NC) and charged
current (CC) DIS structure functions
from the LHeC was quantified.
These results were then compared with the corresponding projections
for the PDF sensitivity of the High-Luminosity upgrade of
the LHC (HL-LHC).
In Fig.~\ref{fig:pdfs_pseudodata}  the  kinematic range
in the $(x,Q^2)$ plane
of the LHeC pseudodata employed
in that analysis is displayed, which illustrated how the LHeC can provide unique constraints
on the behaviour of the quark and gluon PDFs in the very small-$x$ region.

\begin{figure}[!th]
\centering
    \includegraphics[width=0.7\linewidth]{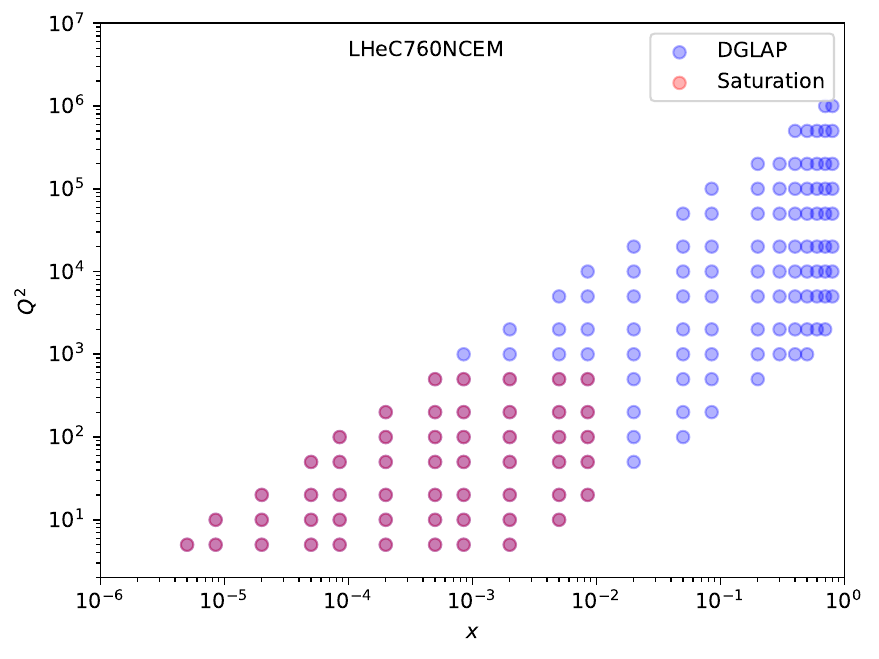}
    \caption{The kinematic coverage of the NC $e^-p$ scattering pseudodata
      at the LHeC, where the blue (red) points indicate those bins for which
      DGLAP (saturation) predictions are available.
       \label{fig:kin} }
\end{figure}

Since non-linear dynamics are known to become sizeable only at small-$x$,
for the present analysis it is
 sufficient to consider the  NC $e^-p$ inclusive scattering cross sections 
from proton beam energies of $E_p=7$\,TeV and $E_p=1$\,TeV.
In Fig.~\ref{fig:kin} we show the bins in $(x,Q^2)$
for which LHeC pseudodata for inclusive structure functions
has been generated according to a saturation-based calculation.
Specifically, we have adopted here
the DGLAP-improved saturation model of Ref.~\cite{Bartels:2002cj}, in which the scattering matrix is modelled through eikonal iteration of two gluon exchanges.
This model was further
extended to include heavy flavour in Ref.~\cite{GolecBiernat:2006ba}.
The specific parameters that we use
were taken from Fit 2 in Ref.~\cite{Golec-Biernat:2017lfv},
where parameterisations are provided that can be used for $x<0.01$ and $Q^2<700$\,GeV$^2$.
These parameters were extracted from a fit to the HERA legacy
inclusive structure function measurements~\cite{Abramowicz:2015mha}
restricted to $x<0.01$ and $0.045 < Q^2 < 650$\,GeV$^2$.
In contrast to other saturation models, the one we assume here~\cite{Golec-Biernat:2017lfv}
provides a reasonable description for large $Q^2$ in the small $x$ region,
where it ensure a smooth transition to standard fixed-order perturbative results.

%
%
%
%

Note that the above discussion refers only to the generated LHeC pseudodata:
all other aspects of the QCD analysis of Ref.~\cite{AbdulKhalek:2019mps} are left
unchanged.
In particular, the PDF profiling will be carried out using theory
calculations obtained by means of DGLAP evolution with the NNLO PDF4LHC15
set (see also Ref.~\cite{Gao:2017yyd}),
with heavy quark structure functions evaluated by means of the FONLL-B
general-mass variable flavour number scheme~\cite{Forte:2010ta}.
In order to ensure consistency with the PDF4LHC15 prior, here we will replace
the DGLAP pseudodata by the saturation calculation only in the kinematic
region for $x \lsim 10^{-4}$,
rather than for all the bins indicated in red in Fig.~\ref{fig:kin}.
The reason for this choice is that PDF4LHC15 already includes HERA data down to $x \simeq 10^{-4}$
which is successfully described via the DGLAP framework, and therefore if
we assume departures from DGLAP in the LHeC pseudodata this should only
be done for smaller values of $x$.

\subsubsection{Results and discussion}
Using the analysis settings described above,
we have  carried out the profiling of PDF4LHC15
with the LHeC inclusive
structure function pseudodata, which for $x \le 10^{-4}$ ($x > 10^{-4}$)
has been generated using the GBW
saturation (DGLAP) calculations, and compare them
with the results of the profiling where the
pseudodata follows the DGLAP prediction.
We have generated  $N_\text{exp}=500$ independent
sets LHeC pseudodata, each one characterised by different random
fluctuations (determined by the experimental uncertainties) around
the underlying central value.

\begin{figure}[!th]
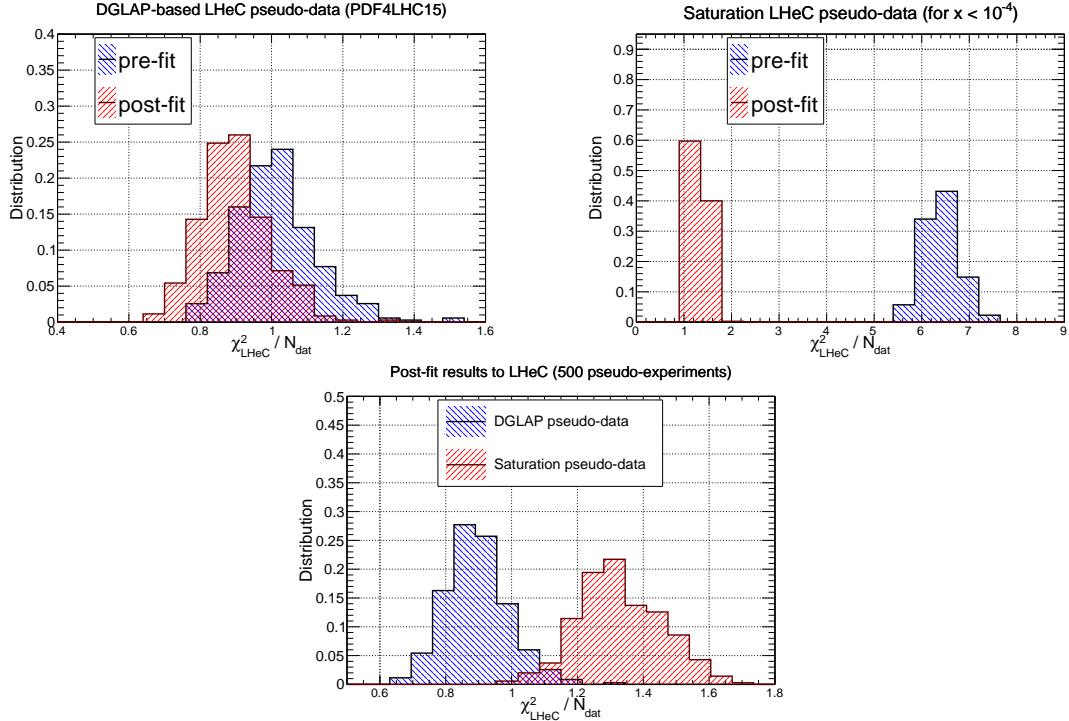

  \centering
    \includegraphics[width=0.44\linewidth]{plots/distribution-chi2.pdf}
    \hspace{0.02\textwidth}
    \includegraphics[width=0.44\linewidth]{plots/distribution-chi2-sat.pdf}
    \includegraphics[width=0.44\linewidth]{plots/distribution-chi2-postfit.pdf}
    \caption{Upper plots: the distribution of pre-fit
      and post-fit  values of $\chi^2/n_\text{dat}$ for the $N_\text{exp}=500$
      sets of generated LHeC pseudodata.
      We compare the results of the profiling of the LHeC
      pseudodata based on DGLAP calculations
      in the entire range of $x$ (left) with those where the pseudodata
      is based on the saturation
      model in the region $x< 10^{-4}$ (right plot).
      Bottom plot: comparison of the post-fit $\chi^2/n_\text{dat}$
      distributions between these two scenarios for the pseudodata generation.
      }
      \label{fig:chi2_distribution}
\end{figure}

To begin with, it is instructive to
compare the data versus theory agreement, $\chi^2/n_\text{dat}$,
between the pre-fit and post-fit calculations,
in order to assess the differences between the DGLAP and saturation
cases.
In the upper plots of Fig.~\ref{fig:chi2_distribution}
we show the distributions of pre-fit and post-fit values
of $\chi^2/n_\text{dat}$ for the $N_\text{exp}=500$
sets of generated LHeC pseudodata.
We compare the results of the profiling of the LHeC
pseudodata based on DGLAP calculations
in the entire range of $x$ with those where the pseudodata
is based on the saturation
model in the region $x< 10^{-4}$.
Then in the bottom plot
we compare of the post-fit $\chi^2$
distributions between the two scenarios.
Note that in these three plots the ranges in the $x$ axes are different.

From this comparison we can observe that
for the case where the pseudodata is generated using a consistent DGLAP framework (PDF4LHC15)
as the one adopted for the theory calculations used in the fit,
as expected the agreement is already good at the pre-fit level, and it is further improved
at the post-fit level.
However the situation is rather different in the case where a subset of the LHeC pseudodata is generated
using a saturation model: at the pre-fit level the agreement between theory and pseudodata is poor,
with $\chi^2/n_\text{dat}\simeq 7$.
The situation markedly improves at the post-fit level, where now the $\chi^2/n_\text{dat}$
distributions peaks around 1.3.
This result
implies that the DGLAP fit manages to absorb most of the differences in theory present
in the saturation pseudodata.
This said, the DGLAP fit cannot entirely \emph{fit away} the non-linear corrections: as shown
in the lower plot of Fig.~\ref{fig:chi2_distribution}, even at the post-fit level one can
still tell apart the $\chi^2/n_\text{dat}$ distributions between the two cases, with the DGLAP (saturation)
pseudodata peaking at around 0.9~(1.3).
This comparison highlights that it is not possible for the DGLAP fit to completely absorb
the saturation effects into a PDF redefinition.

\begin{figure}[!th]
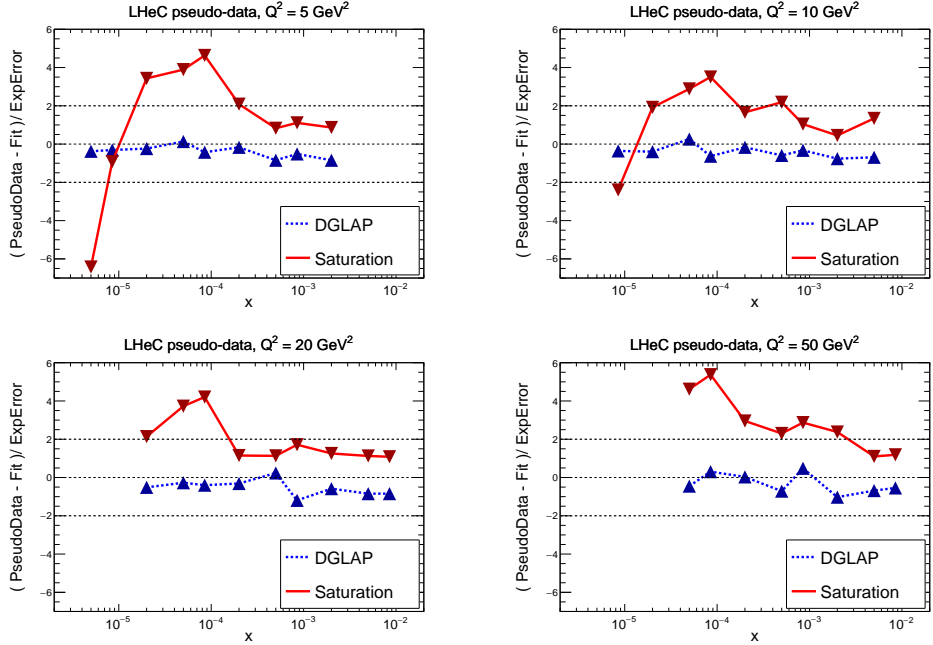

  \centering
    \includegraphics[width=0.38\textwidth,trim=0 -25 0 0,clip]{plots/pulls-5gev2.pdf}
    \hspace{0.02\textwidth}  
    \includegraphics[width=0.38\textwidth,trim=0 -25 0 0,clip]{plots/pulls-10gev2.pdf}
    \includegraphics[width=0.38\textwidth]{plots/pulls-20gev2.pdf}
    \hspace{0.02\textwidth}   
    \includegraphics[width=0.38\textwidth]{plots/pulls-50gev2.pdf}
    \caption{The pulls between  the
      central value of the LHeC pseudodata and post-fit prediction, Eq.~(\ref{eq:pulls}), for four different bins
      in $Q^2$.
      We compare the results of the profiling where the LHeC pseudodata has been generated
      using a consistent DGLAP theory with that partially
      based  on the saturation calculations.
     \label{fig:pulls} }
\end{figure}

In order to identify the origin of the worse agreement between theory predictions
and LHeC pseudodata
in the saturation case, it is illustrative to take a closer look at the pulls
defined as
\begin{equation}
\label{eq:pulls}
P(x,Q^2) = \frac{\mathcal{F}_\text{dat}(x,Q^2)-
  \mathcal{F}_\text{fit}(x,Q^2) }{\delta_\text{exp} \mathcal{F}(x,Q^2)} \, ,
\end{equation}
where $\mathcal{F}_\text{fit}$ is the central value of the profiled results for
the observable $\mathcal{F}$ (in this case the reduced neutral current DIS cross section),
$\mathcal{F}_\text{dat}$ is the corresponding central value of the pseudodata,
and $\delta_\text{exp} \mathcal{F}$ represents the associated total experimental uncertainty.
In Fig.~\ref{fig:pulls} we display
the pulls between the post-fit prediction and the
central value of the LHeC pseudodata for different bins
in $Q^2$.
We compare the cases where the pseudodata has been generated
using a consistent theory calculation (DGLAP) with that
based on the GBW saturation model.      

The comparisons in Fig.~\ref{fig:pulls} show first of all that in the DGLAP case
the pulls are $\mathcal{O}(1)$ in the entire kinematical range.
This is of course expected, given that the LHeC pseudodata is generated
using the same theory as the one subsequently used for the fit.
In the case where the pseudodata has been  partially
 generated with the saturation calculation, on the other hand, one finds a systematic tension
between the theory used for the fit (DGLAP) and the one used to generate
the pseudodata (saturation).
Indeed, we find that at the smallest values of $x$ the theory prediction
overshoots the data by a significant amount, while at higher $x$ the opposite
behaviour takes place.
One can also see that in the region $10^{-4} \lsim x \lsim 10^{-3}$ the fit undershoots
the pseudodata by a large amount.

These comparisons highlight how a QCD fit to the saturation pseudodata is obtained as a compromise
between opposite trends: the theory wants to overshoot the data at very small $x$ and undershoot it
at larger values of $x$.
These tensions result in a distorted fit, explaining the larger
$\chi^2/n_\text{dat}$ values as compared to the DGLAP case.
Such a behaviour can be partially traced back by the different scaling in $Q^2$ between DGLAP
and GBW:
while a different $x$ dependence could eventually
be absorbed into a change of the PDFs at the parameterisation
scale $Q_0$, this is not possible with a $Q^2$ dependence.

The pull analysis of Fig.~\ref{fig:pulls}
highlights how in order to tell apart linear from non-linear
QCD evolution effects at small-$x$ it would be crucial to ensure a lever arm in $Q^2$ as large
as possible in the perturbative region.
This way it becomes possible to disentangle the different scaling in $Q^2$ for the two cases.
The lack of a sufficiently large lever arm in $Q^2$ at HERA at small $x$ could 
explain in part why both frameworks are able to describe the same structure function measurements
at the qualitative level.
Furthermore, we find that amplifying the significance of these subtle effects
can be achieved
by monitoring the $\chi^2$ behaviour in the $Q^2$ bins more affected by the saturation
corrections.
The reason is that the total $\chi^2$, such as that reported in
Fig.~\ref{fig:chi2_distribution}, is somewhat less
informative since the deviations at small-$Q$ are washed out
by the good agreement between theory and pseudodata in the rest
of the kinematical range of the LHeC summarised in Figs.~\ref{fig:pdfs_pseudodata} and \ref{fig:kin}.

\begin{figure}[!th]
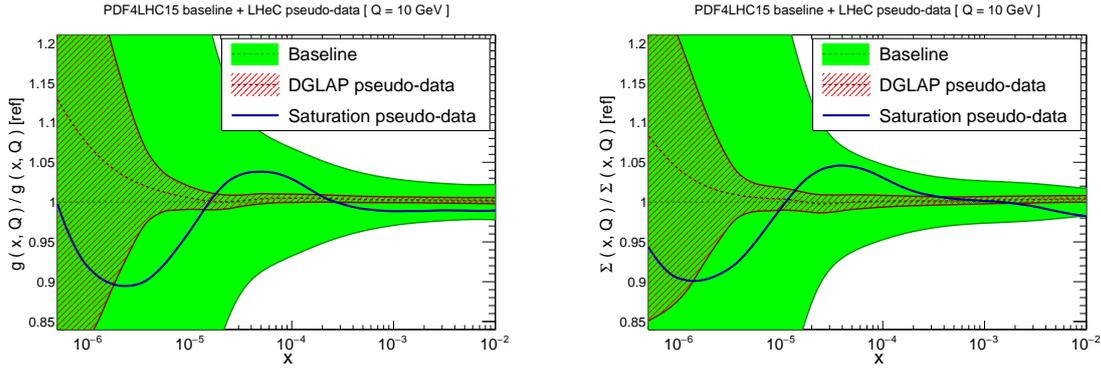

\centering
\includegraphics[width=0.45\textwidth]{plots/xg-pdf4lhc-saturation.pdf}
\hspace{0.02\textwidth}
\includegraphics[width=0.45\textwidth]{plots/xsinglet-pdf4lhc-saturation.pdf}
    \caption{Comparison between the PDF4LHC15 baseline (green band)
      with the results of the profiling of the LHeC pseudodata
      for the gluon (left) and quark singlet (right) for $Q=10$\,GeV.
      We show the cases where the pseudodata is generated using DGLAP
      calculations (red hatched band) and where it is partially
      based on the GBW saturation model (blue curve).
     \label{fig:pdfplots} }
\end{figure}

To conclude this analysis,
in Fig.~\ref{fig:pdfplots} we display the comparison between the
PDF4LHC15 baseline 
with the results of the PDF profiling of the LHeC pseudodata
for the gluon (left) and quark singlet (right) for $Q=10$\,GeV.
We show the cases where the pseudodata is generated using DGLAP
calculations  and where it is partially
based on the GBW saturation model (for $x\lsim 10^{-4}$).
We find that the distortion induced by the mismatch between theory and pseudodata
in the saturation case is typically larger than the PDF uncertainties
expected once the LHeC constraints are taken into account.
While of course in a realistic situation such a comparison would not be possible,
the results of Fig.~\ref{fig:pdfplots} show that saturation-induced effects
are expected to be larger than the typical PDF errors in the LHeC era,
and thus that it should be possible to tell them apart using
for example tools such as the pull analysis of Fig.~\ref{fig:pulls}
or other statistical methods.

\subsubsection{Summary}
Here we have assessed the feasibility of disentangling DGLAP evolution
from non-linear effects at the LHeC.
By means of a QCD analysis where LHeC pseudodata is generated using
a saturation model, we have demonstrated that the LHeC should be possible to identify
non-linear effects with large statistical significance, provided
their size is the one predicted by current calculations
such as the that of~\cite{Golec-Biernat:2017lfv} that have been tuned
to HERA data.
A more refined analysis would require to study whether or not small-$x$ BFKL resummation
effects can partially mask the impact of non-linear dynamics, though
this is unlikely since the main difference arises in their $Q^2$ scaling.
The discovery of non-linear dynamics would represent an important milestone
for the physics program of the LHeC, demonstrating the onset of a new gluon-dominated
regime of the strong interactions and paving the way for detailed studies
of the properties of this new state of matter.
Such discovery would have also implications outside nuclear and particle
physics, for instance it would affect the theory predictions
for the scattering of ultra-high energy neutrinos with matter~\cite{Bertone:2018dse}.


\subsection[Low ${x}$ and the Longitudinal Structure Function ${F_L}$]{\boldmath Low ${x}$ and the Longitudinal Structure Function ${F_L}$}
\label{sec:FL}
\subsubsection[DIS Cross Section and the Challenge to Access ${F_L}$]{\boldmath DIS Cross Section and the Challenge to Access ${F_L}$}
The inclusive, deep inelastic electron-proton scattering cross section at low $Q^2 \ll M_Z^2$,
\begin{equation}
 \frac{Q^4 x} {2\pi \alpha^2 Y_+} \cdot \frac{d^2\sigma}{dxdQ^2} =     \sigma_r \simeq
    F_2(x,Q^2) - f(y) \cdot F_L(x,Q^2) = F_2 \cdot \left(1 -f(y) \frac{R}{1+R} \right)
       \label{eq:sigr}
  \end{equation}  
is defined by two proton structure functions, $F_2$ and $F_L$, with
$y=Q^2/sx$, $Y_+ = 1+ (1-y)^2$ and $f(y)=y^2/Y_+$. 
The cross section may also be 
expressed~\cite{Hand:1963zz} as a 
sum of two contributions, $\sigma_r \propto (\sigma_T + \epsilon \sigma_L)$,
referring to the transverse and longitudinal polarisation state of the exchanged boson, 
with $\epsilon$ characterising the ratio of the longitudinal to the transverse 
polarisation. The ratio of the longitudinal to transverse cross sections is termed
\begin{equation}
R(x,Q^2) = \frac{\sigma_L}{\sigma_T} = \frac{F_L}{F_2-F_L},
\label{R}
\end{equation}
which is related to $F_2$ and $F_L$ as given above. Due to the positivity of the 
cross sections $\sigma_{L,T}$ one observes that $F_L \leq F_2$. The
reduced cross section $\sigma_r$, Eq.~\eqref{eq:sigr},
is therefore a direct measure of $F_2$, apart from a limited region of high $y$
where a contribution of $F_L$ may be sizeable. To leading order, for spin 1/2 particles,
one expected $R=0$. The initial measurements of $R$
 at SLAC~\cite{Miller:1971qb,Riordan:1974te}
showed that $R$ was indeed small, $R \simeq 0.18$,
which was taken as evidence for quarks to carry spin 1/2. 

The task to measure $F_L$ thus
requires to precisely measure the inclusive DIS cross section near to $y=1$ and
to then disentangle the two structure functions by exploiting the $f(y)=y^2/Y_+$ variation
which depends on $x$, $Q^2$ and $s$. By varying the centre-of-mass (cms) beam energy, $s$, one can disentangle $F_2$ and $F_L$ obtaining independent measurements at each common, fixed point of $x,Q^2$. This  is particularly challenging not only because  the $F_L$ part is 
small, calling for utmost precision, but also because 
it requires to measure at high $y$. The inelasticity
$y = 1 - E'/E_e$, however, is large only for scattered electron
energies $E_e'$ much smaller than the electron beam energy $E_e$, for example $E_e' = 2.7$\,GeV for $y=0.9$ at HERA~\footnote{The nominal electron beam energy  $E_e$ at the LHeC is doubled as compared to HERA. Ideally
one would like to vary the proton beam energy in an $F_L$ measurement at the
LHeC, which yet would affect the hadron collider operation. In the present study it was
therefore considered to lower $E_e$ which may be done independently of the HL-LHC.}. In the region where $E'$ is a few GeV only, the electron identification becomes a major problem and the electromagnetic ($\pi^0 \rightarrow \gamma \gamma$) and hadronic backgrounds, mainly from unrecognised photoproduction, rise strongly.

\begin{figure}[!th]
  \centering
  \includegraphics[width=0.65\textwidth]{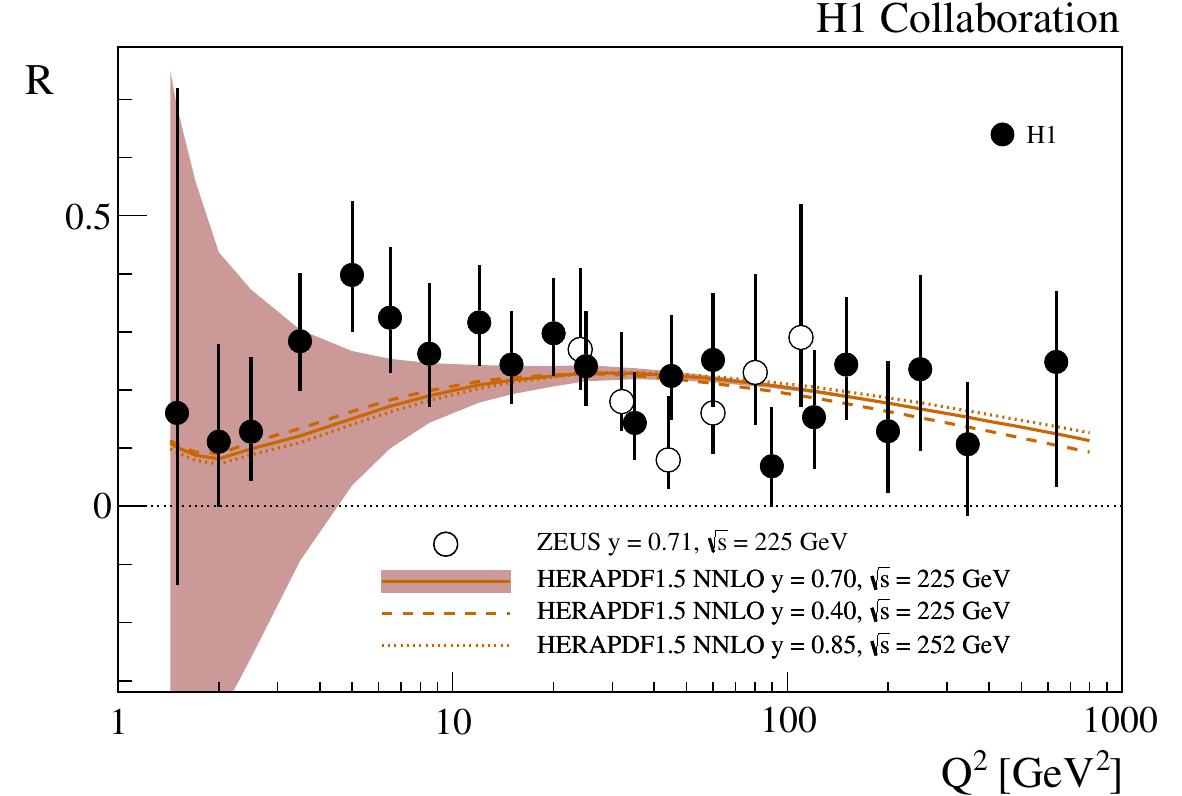}
    \caption{
      Measurement of the structure function ratio $R=F_L/(F_2-F_L)$
      by H1 (solid points) and ZEUS (open circles), from a variation of 
      proton beam energy in the final half year of HERA operation. The curve represents 
      an NNLO QCD fit analysis of the other HERA data. This becomes uncertain for $Q^2$ below
      $10$\,GeV$^2$ where the $Q^2$ dependence of $F_2$ at HERA does not permit
      an accurate determination of the gluon density which dominates the prediction on $F_L$.
    }
    \label{figRh1}
\end{figure}
The history and achievements on $F_L$, the role of HERA and the prospects as
sketched in the CDR of the LHeC, were summarised in detail in~\cite{Klein:2018rhq}.
The  measurement of $F_L$ at HERA~\cite{Collaboration:2010ry} was given very limited time
and it collected about $5.9$ and $12.2$\,pb$^{-1}$ of data at reduced beam energies which
were analysed together with about $100$\,pb$^{-1}$ at nominal HERA energies. The result 
may well be illustrated with the data obtained on the ratio $R(x,Q^2)$ shown in Fig.~\ref{figRh1}. To good approximation,
$R(x,Q^2)$ is a constant which was determined as $R=0.23 \pm 0.04$, in 
good agreement with the SLAC values of $R \simeq 0.18$ despite the hugely extended kinematic range. The rather small variation of $R$ towards small $x$, at fixed $y = Q^2/sx$, may appear
to be astonishing as one observed $F_2$ to strongly rise towards low $x$. A constant $R$ of e.g.\ $0.25$
means that $F_2=(1+R)F_L/R$ is five times larger than $F_L$, and that they rise together, as they
have a common origin, the rise of the gluon density. This can be understood in approximations
to the DGLAP expression of the $Q^2$ derivative of $F_2$ and the so-called 
Altarelli-Martinelli relation of $F_L$ to the parton densities~\cite{Altarelli:1978tq,Gluck:1980cp}, see the discussion in Ref.~\cite{Klein:2018rhq}. The resulting H1 value also obeyed
the condition $R \leq 0.37$, which had been obtained in a rigorous attempt to derive the 
dipole model for inelastic DIS~\cite{Ewerz:2006vd}.

\subsubsection[Parton Evolution at Low ${x}$]{\boldmath Parton Evolution at Low ${x}$}
Parton distributions are to be extracted from experiment as their
$x$ dependence and flavour sharing are not predicted in QCD. They
acquire a particular meaning through the theoretical prescription
of their kinematic evolution. PDFs, as they are frequently used
for LHC analyses, are predominantly defined through the now classic
DGLAP formalism, in which the $Q^2$ dependence of parton distributions is regulated  by splitting functions while the DIS cross
section, determined by the structure functions,  is calculable by folding the PDFs with coefficient functions.
 Deep inelastic scattering is known to be the most suited process to extract PDFs
from the experiment, for which the HERA collider has so far delivered
the most useful data.  Through factorisation theorems the PDFs are considered to be universal such that PDFs extracted in $ep$ DIS
shall be suited to describe for example Drell-Yan scattering cross sections in $pp$ at the LHC. This view has been formulated to
third order pQCD already and been quite successful in the
interpretation of LHC measurements, which by themselves also constrain PDFs
in  parton-parton scattering sub-processes. 

\begin{figure}[!th]
  \centering
  \includegraphics*[width=0.7\textwidth]{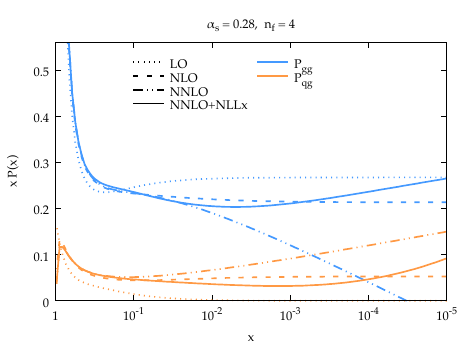}
  \caption{
      Calculation of splitting functions $P_{gg}$ (top, blue) and $P_{qg}$ (bottom, brown)
      in resummed NNLO (solid) as compared to non-resummed calculations at LO (dotted),
      NLO (dashed) and NNLO (dashed-dotted) as functions of $x$ for $n_f=4$ at a large
      value of $\alpha_s$ corresponding to a $Q^2$ of a few GeV$^2$, from Ref.~\cite{Abdolmaleki:2018jln}.
      The resummed calculation is seen to restore the dominance of $P_{gg}$ over $P_{qg}$ as
      $x$ becomes small (towards the right side), which is violated at NNLO. 
    }
    \label{fig:Presum}
\end{figure}

As commented in Sec.~\ref{sec:PSM_Disc_smallx}, the question has long been posed about the universal validity
of the DGLAP formalism, especially for the region of small Bjorken $x$ where logarithms $\propto \ln(1/x)$ become very sizeable. This feature of the perturbation expansion is expected to significantly modify the 
splitting functions.  This in turn changes the theory underlying the physics
of parton distributions, and predictions for the LHC and its successor will correspondingly have to be altered. 
 This mechanism,  for an equivalent $Q^2$ of a few GeV$^2$,  is illustrated in Fig.~\ref{fig:Presum}, taken from Ref.~\cite{Abdolmaleki:2018jln}. It shows the $x$ dependence 
of the gluon-gluon and the quark-gluon splitting functions, $P_{gg}$
and $P_{qg}$, calculated in DGLAP QCD.  It is observed that at NNLO
 $P_{gg}$ strongly decreases towards small $x$, becoming smaller 
than $P_{qg}$ for $x$ below $10^{-4}$. Resummation of the 
large $\ln (1/x)$ terms,  see Ref.~\cite{Abdolmaleki:2018jln}, 
here performed to next-to-leading log x, restores the
dominance of the $gg$ splitting over the $qg$ one. Consequently,
the gluon distribution in the resummed theory exceeds the one
derived in pure DGLAP. 
While this observation has been supported by the HERA data, it yet relies on limited
kinematic coverage and precision. The LHeC will examine this in detail, at a hugely extended range and  is
thus expected to resolve the long known question about the validity of the BFKL evolution and
the transition from DGLAP to BFKL as $x$ decreases while $Q^2$ remains large  enough for 
pQCD to apply.

\subsubsection{Kinematics of Higgs Production at the HL-LHC} 
The clarification of the evolution and the accurate and complete determination of the parton 
distributions is of direct importance for the LHC. This can be illustrated with the kinematics of Higgs
production at HL-LHC which is dominated by gluon-gluon fusion. 
With the luminosity upgrade, the detector acceptance is being extended into the 
forward region to pseudorapidity values of  $|\eta| = 4$, where $\eta = \ln \tan{\theta/2}$ is a very
good approximation of the rapidity. In Drell-Yan scattering of two partons with Bjorken $x$ values of 
$x_{1,2}$ these are related to the rapidity via the relation $x_{1,2}=\exp{(\pm \eta)} \cdot M/\sqrt{s}$
where $\sqrt{s}=2E_p$ is the cms energy and $M$ the mass of the produced particle. It is interesting 
to see that $\eta = \pm 4$ corresponds to $x_1 = 0.5$ and $x=0.00016$ for the SM Higgs boson of mass
$M=125$\,GeV. Consequently, Higgs physics at the HL-LHC will depend on understanding PDFs at high $x$, 
a challenge resolved by the LHeC too, and on
clarifying the evolution at small $x$.
At the FCC-hh, in 
its $100$\,TeV energy version, the small $x$ value for $\eta = 4$ will be as low as $2 \cdot 10^{-5}$.
Both the laws of QCD and the resulting phenomenology of particle production at the HL-LHC and its 
successor demand to clarify the evolution of the parton contents at small $x$ as a function
of the resolution scale $Q^2$~\cite{Hautmann:2002tu,Marzani:2008az,Bonvini:2018ixe}. 
This concerns in particular the 
unambiguous, accurate determination of the gluon distribution, which dominates the small-$x$ 
parton densities and as well the production of the Higgs boson in $pp$ scattering.

\subsubsection[Indications for Resummation in H1 ${F_L}$ Data]{\boldmath Indications for Resummation in H1 ${F_L}$ Data}

The simultaneous measurement of the two structure functions $F_2$ and $F_L$ is the cleanest way to establish new parton dynamics at low $x$. This holds because their independent constraints on the 
dominating gluon density at low $x$ ought to lead to consistent results. In other words, one may constrain all partons with a complete
PDF analysis of the inclusive cross section in the kinematic region where its $F_L$ part is negligible and confront the $F_L$ measurement with this result. A significant deviation from $F_L$
data signals the necessity to introduce new, non-DGLAP physics
in the theory of parton evolution, especially at small $x$. The 
salient value of the $F_L$ structure function results from its
inclusive character enabling a clean theoretical treatment as 
has early on been recognised~\cite{Altarelli:1978tq,Gluck:1980cp}.
This procedure has recently been illustrated~\cite{Abdolmaleki:2018jln}  using the 
H1 data on $F_L$~\cite{Andreev:2013vha} which are the %
\begin{figure}[!th]
  \centering
  \includegraphics[width=0.7\textwidth]{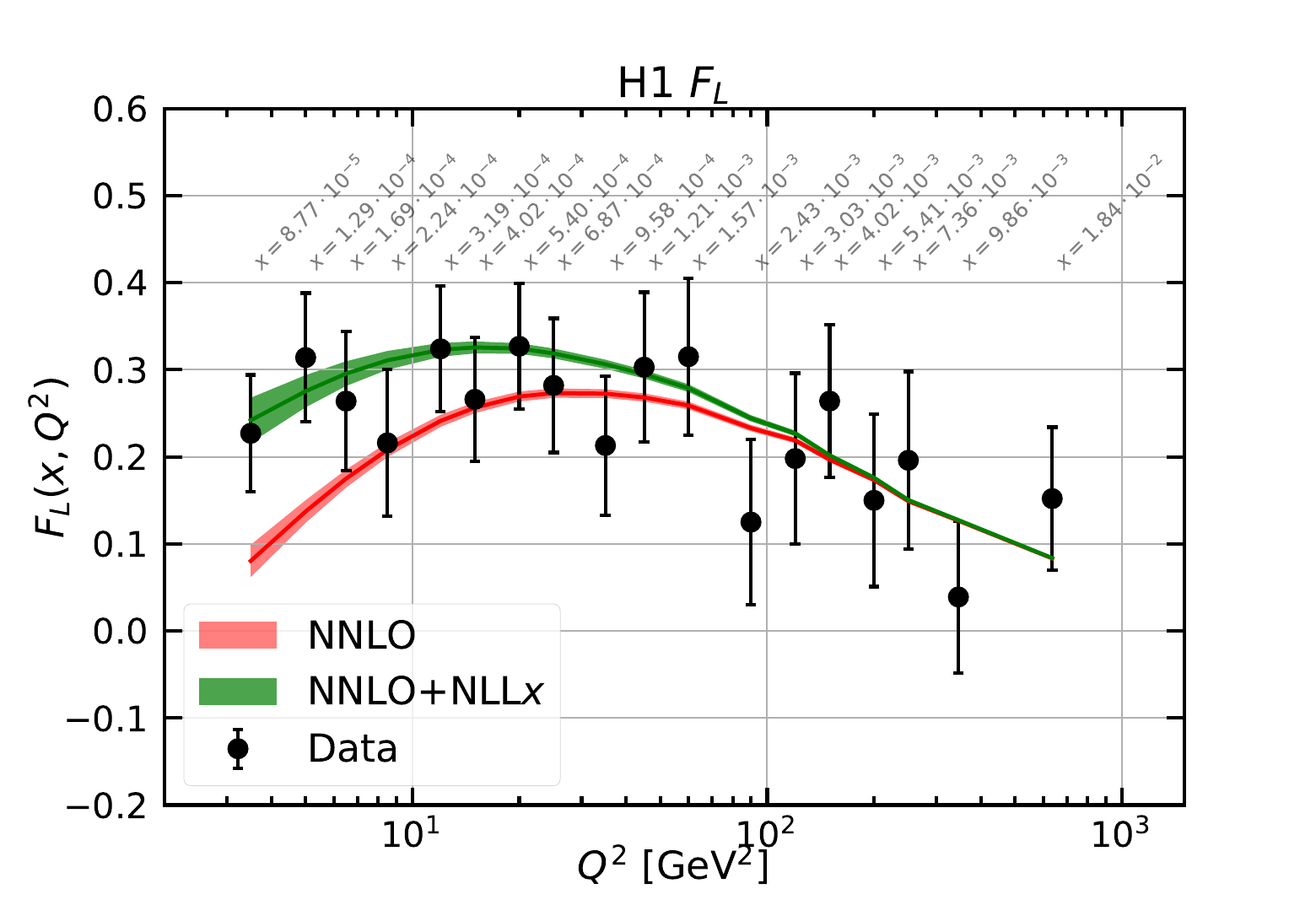}
  \caption{
    Measurement of the longitudinal structure function $F_L$, obtained
    as an average results over a number of $x$ dependent points at
    fixed $Q^2$, plotted vs $Q^2$ with the corresponding $x$ values
    indicated in grey. Red curve: NNLO fit to the H1 cross section data; green curve: NNLO fit including NLLx resummation,  from Ref.~\cite{Abdolmaleki:2018jln}. 
  }
  \label{fig:flh1resu}
\end{figure}
only accurate data from HERA at smallest $x$. The result is shown
in Fig.~\ref{fig:flh1resu}. One observes the trend described above: 
the resummed prediction is higher than the pure NNLO curve, and the
description at smallest $x$, below $5 \cdot 10^{-4}$, appears to be
improved. The difference between the two curves increases as $x$ 
decreases. However, due to the peculiarity of the DIS kinematics, which
relates $x$ to $Q^2/sy$, one faces the difficulty of $Q^2$ decreasing 
with $x$ at fixed $s$ for large $y \geq 0.6$, which is the region of
sensitivity to $F_L$. Thus one not only wishes to improve substantially
the precision of the $F_L$ data but also to increase substantially $s$
in order to avoid the region of non-perturbative behaviour while
testing theory at small $x$. This is the double and principal advantage
which the LHeC offers - a much increased precision and more than
a decade of extension of kinematic range.

\subsubsection{The Longitudinal Structure Function at the LHeC} 
Following the method described above, inclusive cross section data have been simulated
for $E_p = 7$\,TeV and three electron beam energies $E_e$ of $60$, $30$ and $20$\,GeV. The
assumed integrated luminosity values are $10$, $1$ and again $1$\,fb$^{-1}$, respectively. These are
about a factor of a hundred larger than the corresponding H1 luminosities. At large $y$,
the kinematics is best reconstructed using the scattered electron energy, $E_e'$, and polar
angle, $\theta_e$. The experimental methods to calibrate the angular and energy measurements are
described in~\cite{Collaboration:2010ry}. For the present study similar results are assumed:
for $E_e'$ a scale uncertainty of $0.5$\,\% at small $y$ (compared to $0.2$\,\% with H1) 
rising linearly to $1.2$\,\%, in the 
range of $y=0.4$ to $0.9$. For the polar angle, given the superior quality of the anticipated
LHeC Silicon tracker as compared to the H1 tracker, it is assumed that $\theta_e$ may be calibrated
to $0.2$\,mrad, as compared to $0.5$\,mrad at H1. The residual photo-production background
contamination is assumed to be $0.5$\,\% at largest $y$, twice better than with H1. There is
further an assumption made on the radiative corrections which are assumed to be uncertain
to $1$\,\% and treated as a correlated error. The main challenge is to reduce the uncorrelated
uncertainty, which here was varied between $0.2$ and $0.5$\,\%. This is about ten to three times more
accurate than  the H1 result which may be a reasonable assumption: the hundred fold increase
in statistics sets a totally different scale to the treatment of uncorrelated uncertainties, as from
imperfect simulations, trigger efficiency or Monte Carlo statistics. It is very difficult to transport
previous results to the modern and future conditions. It could, however, be an important fix
point if one knows that the most precise measurement of $Z$ boson production by ATLAS
at the LHC had a total systematic error of just $0.5$\,\%~\cite{Aad:2011dm}.  
\begin{figure}[!th]
  \centering
  \includegraphics[width=0.95\textwidth]{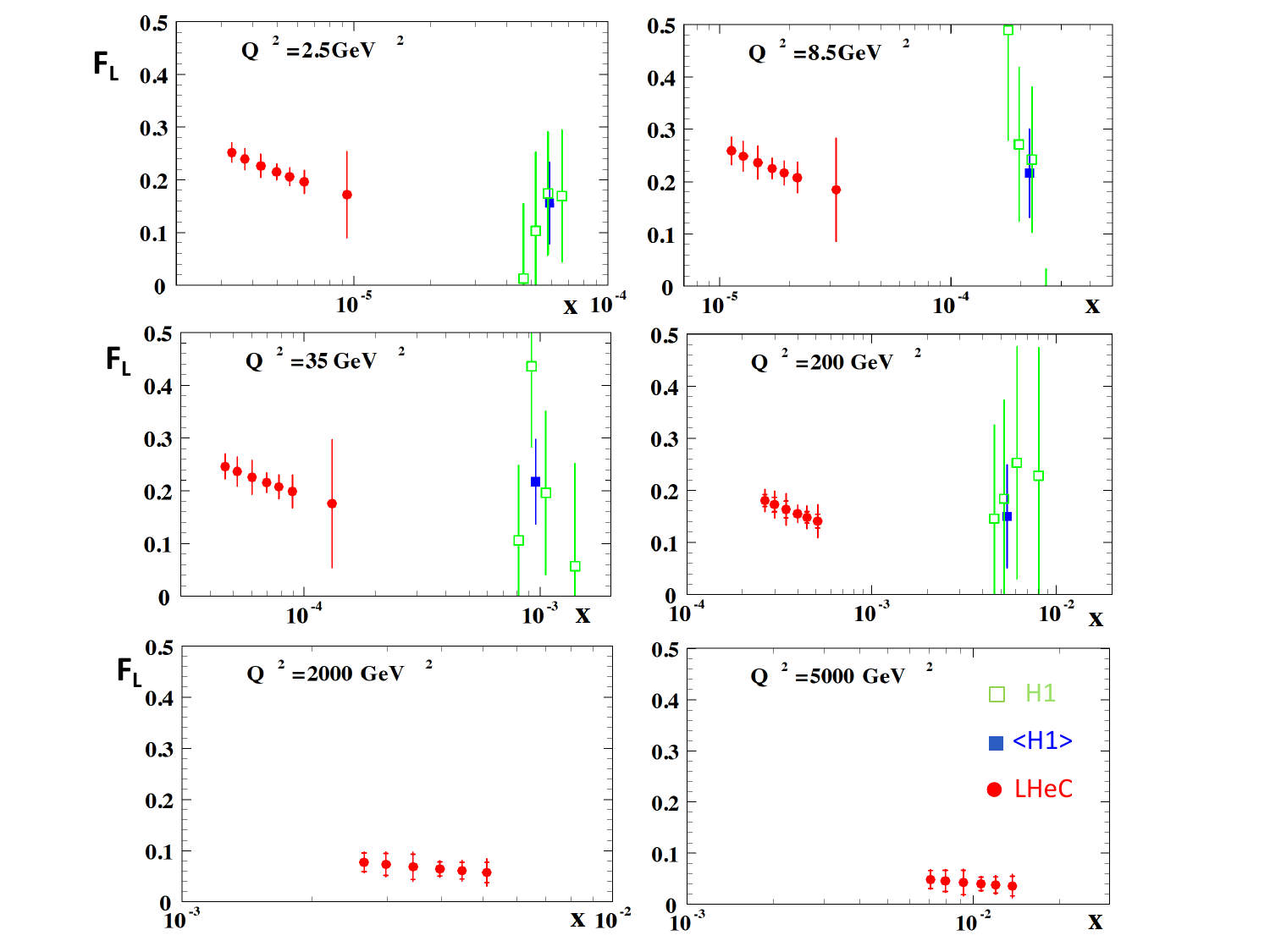}
  \caption{
    H1 measurement and LHeC simulation of data on the longitudinal
    structure function $F_L(x,Q^2)$. Green: Data by H1, for selected $Q^2$
    intervals from Ref.~\cite{Andreev:2013vha}; Blue: Weighted average of the (green) data points
    at fixed $Q^2$; Red: Simulated data from an $F_L$ measurement at the LHeC
    with varying beam energy, see text. The H1 error bars denote the total measurement uncertainty. The LHeC inner error bars represent the
    data statistics, visible only for $Q^2 \geq 200$\,GeV$^2$,  while the outer error  bars are the total uncertainty. Since the $F_L$ measurement is
    sensitive only at high values of inelasticity, $y=Q^2/sx$, each $Q^2$ 
    value is sensitive only to a certain limited interval of $x$ values which increase with $Q^2$. Thus each panel has a different $x$ axis. 
    The covered $x$ range similarly varies with $s$, i.e.\ H1 $x$ values are
    roughly twenty times larger at a given $Q^2$. There are no H1 data for 
    high $Q^2$, beyond $1000$\,GeV$^2$, see Ref.~\cite{Andreev:2013vha}.
  }
  \label{fig:fllhec}
\end{figure}

The method here used is that of a simple straight-line fit of $\sigma_r = F_2 - f(y) F_L$ (Eq.\,\eqref{eq:sigr}),
in which $F_L$ is obtained as the slope of the $f(y)$ dependence~\footnote{Better results were achieved by H1 using
a $\chi^2$ minimisation technique, see Ref.~\cite{Altarelli:2011zv}, which for the rough estimate
on the projected $F_L$ uncertainty at the LHeC has not been considered.}. The predictions for $F_2$ and 
$F_L$ were obtained using LO formulae for the PDF set of MSTW 2008. In this method any common
factor does not alter the absolute uncertainty of $F_L$.  This also implies that the estimated
absolute error on $F_L$ is independent of whether $F_L$ is larger or smaller than here assumed. 
For illustration, $F_L$ was scaled by a factor of two. Since $f(y) \propto y^2$, the
accuracy is optimised with a non-linear choice of lowered beam energies. The fit takes into account
cross section uncertainties and their correlations, calculated numerically 
following~\cite{Blumlein:1990dj,Blumlein:1992we}, by considering each source separately and adding the results of the 
various correlated sources to one correlated systematic error which is added quadratically to
the statistical and uncorrelated uncertainties to obtain one total error.

The result is illustrated 
in Fig.~\ref{fig:fllhec} presenting the $x$-dependent results, for some selected $Q^2$ values,
of both H1, with their average over $x$, and the prospect LHeC results. It reflects the huge extension of kinematic range, towards low $x$ and high $Q^2$ by the LHeC as compared to HERA.
 It also illustrates the striking improvement in 
precision which  the LHeC promises to provide. The $F_L$ measurement will cover an $x$ range
from $2 \cdot 10^{-6}$ to above $x=0.01$.  Surely, when comparing with Fig.~\ref{fig:flh1resu},
one can safely expect that any non-DGLAP parton evolution would be discovered with such data, in their 
combination with a very precise $F_2$ measurement.

A few comments are in order 
on the variation of the different error components with the kinematics, essentially $Q^2$
since the whole $F_L$ sensitivity is restricted to high $y$ which in turn for each $Q^2$ defines
a not wide interval of $x$ values covered. One observes in Fig.~\ref{fig:fllhec} that the
precision is spoiled towards large $x \propto 1/y$, see e.g.\ the result for $Q^2=8.5$\,GeV$^2$.
 The assumptions on the integrated luminosity
basically define a $Q^2$ range for the measurement. For example, the statistical uncertainty
for $Q^2=4.5$\,GeV$^2$ and $x= 10^{-5}$, a medium $x$ value at this $Q^2$ interval, is
only $0.6$\,\% (or $0.001$ in absolute for $F_L=0.22$). At $Q^2=2000$\,GeV$^2$ it rises to $21$\,\% (or
$0.012$ for $F_L=0.064$). One thus can perform the $F_L$ measurement at the LHeC, with 
a focus on only small $x$, with much less luminosity than the $1$\,fb$^{-1}$ here used. The relative
size of the various systematic error sources also varies considerably, which is due to the 
kinematic relations between angles and energies and their dependence on $x$ and $Q^2$. 
This is detailed in~\cite{Blumlein:1990dj}. It implies, for example, that the $0.2$\,mrad polar angle scale
uncertainty becomes the dominant error at small $Q^2$, which is the backward region where
the electron is scattered near the beam axis in the direction of the electron beam. For large $Q^2$, however,
the electron is more centrally scattered and the $\theta_e$ calibration requirement may be more relaxed.
The $E_e'$ scale uncertainty has a twice smaller effect than that due to the $\theta_e$ calibration
at lowest $Q^2$ but becomes the dominant correlated systematic error source at high $Q^2$. The here
used overall assumptions on scale uncertainties are therefore only rough first
approximations and would be replaced by kinematics and detector dependent requirements when
this measurement may be pursued. These could also exploit the cross calibration opportunities which 
result from the redundant determination of the inclusive DIS scattering kinematics through both
the electron and the hadronic final state. This had been noted 
very early at HERA times, see Ref.~\cite{Blumlein:1992we,Bentvelsen:1992fu,Bassler:1997tv} and was worked out in considerable detail by both H1 and ZEUS using independent and 
different methods. A feature used by H1 in their $F_L$ measurement
includes a number of decays such as $\pi^0 \to \gamma \gamma$ and
$J/\psi \to e^+e^-$ for calibrating the low energy measurement or $K_s^0 \to \pi^+\pi^-$ and $\Lambda \to
p \pi$ for the determination of tracker scales, see Ref.~\cite{Collaboration:2010ry}.

It is obvious that the prospect to measure $F_L$ as presented here is striking. 
For nearly a decade, Guido Altarelli was a chief theory advisor to the development of the LHeC. 
In 2011, he publishes an article~\cite{Altarelli:2011zv}, in honour 
of Mario Greco, about \emph{The Early Days of QCD (as seen from Rome)} 
in which he describes one of his main achievements~\cite{Altarelli:1978tq}, and 
persistent irritation, regarding 
the  longitudinal structure function, $F_L$,  and its measurement: \emph{$\dots$
  The present data, recently obtained by the H1 experiment at DESY, are in
agreement with our [!this] LO QCD prediction but the accuracy of the test is still far from
being satisfactory for such a basic quantity}. The LHeC developments had not been
rapid enough to let Guido see results of much higher quality on $F_L$ with which
the existence of departures from the DGLAP evolution, to high orders pQCD, may be 
expected to most safely  be discovered. 

%
%
%
\subsection{Associated jet final states at low $x$}

The dynamical effects from resummation or nonlinear corrections which we have 
discussed above can arise at the LHeC not only in the inclusive structure functions, as we have 
illustrated so far, but also in more exclusive observables describing the structure of the 
jet final states associated to low-$x$ DIS. 

Baseline predictions for jet final states in DIS are obtained  from perturbative finite-order 
calculations (see e.g.~\cite{Currie:2018fgr,Gehrmann:2018odt} for third-order 
calculations), supplemented by parton-shower Monte Carlo generators for realistic event 
simulation (as e.g.~in~\cite{Hoche:2018gti}). But owing to the large phase space opening up 
at LHeC energies and the complex kinematics possibly involving multiple hard scales, jet events 
are potentially sensitive to soft-gluon coherence effects of initial-state 
radiation~\cite{Ciafaloni:1987ur,Catani:1989sg,Marchesini:1992jw,Hautmann:2008vd}, 
which go beyond finite-order perturbative evaluations and collinear parton showers, and show up as 
logarithmic $x \to 0$ corrections to all orders of perturbation theory. These corrections can be 
resummed, and combined with large-$x$ contributions, via CCFM exclusive evolution 
equations~\cite{Ciafaloni:1987ur,Catani:1989sg}, and  affect the structure 
of jet multiplicities and angular jet correlations~\cite{Hautmann:2008vd} as well as heavy quark 
distributions~\cite{Marchesini:1992jw}. Observables based on forward jets,  transverse 
energy flow, angle and momentum correlations constitute probes of low-$x$ dynamics in DIS final 
states~\cite{Mueller:1990gy,Forshaw:1998yh,Hautmann:2009zzb}. Phenomenological 
studies started with HERA~\cite{Bartels:1996gr,Kwiecinski:1999wj,Andersson:1995jt}  
and will continue with the LHeC. 
 
\begin{figure}[htb]
\centering
\includegraphics[width=8.3cm]{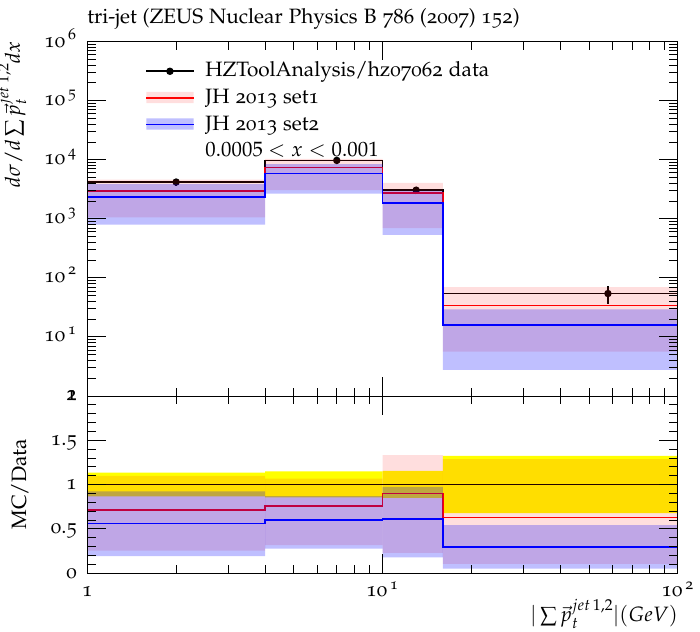}\includegraphics[width=8.3cm]{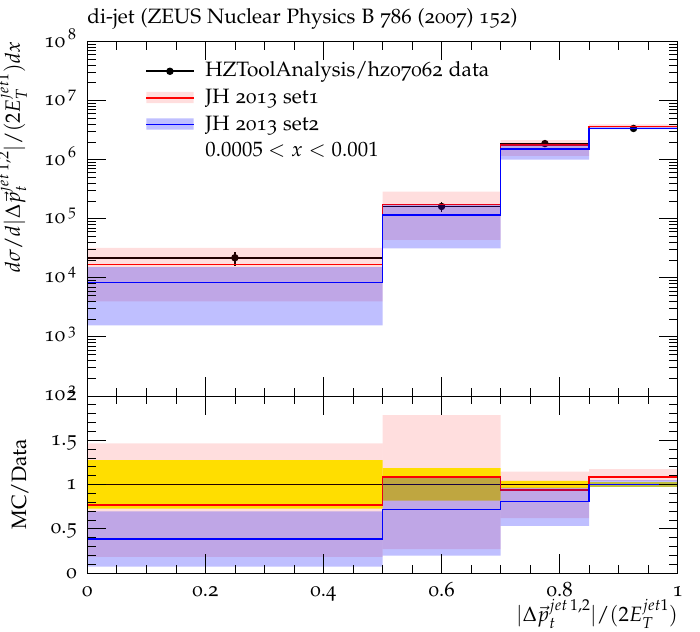}    
\caption{Momentum correlations in DIS multi-jet final states at low $x$ computed from the CCFM Monte Carlo~\protect\cite{Jung:2010si} with 
TMD parton densities JH2013~\protect\cite{Hautmann:2013tba}, compared with the measurements~\protect\cite{Chekanov:2007dx}: (left) tri-jets; (right) di-jets.}
\label{fig:hfsx-fig1}
\end{figure}

Computational tools are being developed to address the structure of multi-jet final states 
by including low-$x$ dynamical effects. These include CCFM Monte Carlo 
tools~\cite{Jung:2010si,Hautmann:2014uua}, off-shell matrix element parton-level 
generators~\cite{vanHameren:2016kkz,vanHameren:2019wzx}, BFKL 
Monte Carlo generators~\cite{Chachamis:2015zzp,Hoche:2007hg,Andersen:2017sht}. 
Fig.~\ref{fig:hfsx-fig1} gives an example of transverse momentum correlations in DIS 
di-jet and three-jet final states computed with the Monte Carlo~\cite{Jung:2010si}, compared with 
the measurements~\cite{Chekanov:2007dx}.  

\begin{figure}[htb]
\centering
\includegraphics[width=8.3cm]{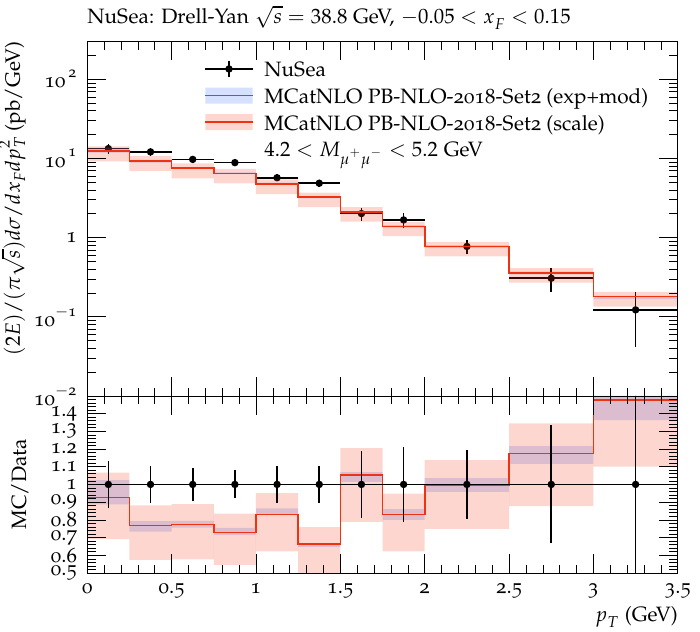}\includegraphics[width=8.3cm]{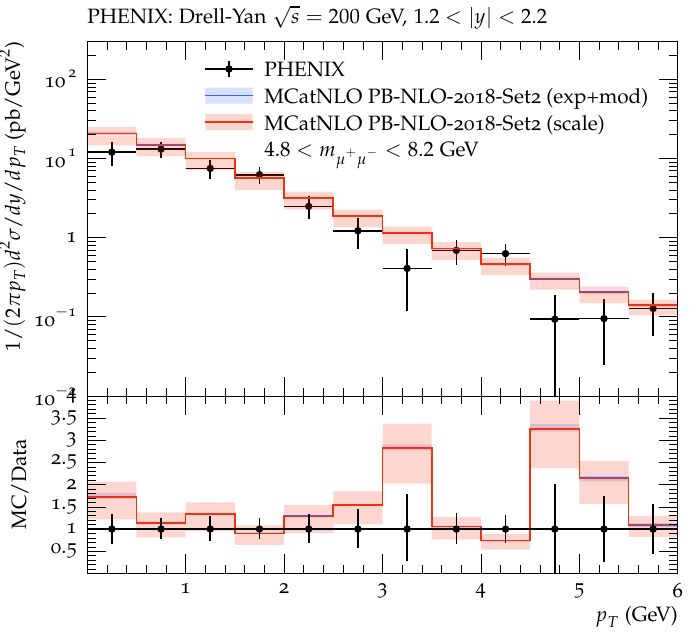}    
\caption{Transverse momentum spectra in low-mass Drell-Yan lepton pair production~\protect\cite{Martinez:2020fzs} from the 
parton branching (PB) method, compared with (left) NuSea measurements~\protect\cite{Webb:2003ps}  
and (right) PHENIX measurements~\protect\cite{Aidala:2018ajl}.}
\label{fig:hfsx-fig2}
\end{figure}

Furthermore, exclusive parton branching formalisms  are being proposed in which 
not only gluon distributions but also quark distributions are treated at unintegrated level in transverse 
momentum~\cite{Hautmann:2017xtx,Hautmann:2017fcj,Martinez:2018jxt}. 
This is instrumental in connecting low-$x$ approaches with DGLAP approaches to 
parton showers beyond leading order~\cite{Hoche:2017hno,Hoche:2017iem}. 
Applications of these new developments have so far been mostly  carried out 
for final states in hadron-hadron collisions, while extensions  to lepton-hadron collisions  are underway.
Fig.~\ref{fig:hfsx-fig2} gives examples of  transverse momentum spectra in low-mass Drell-Yan lepton pair production 
computed in~\cite{Martinez:2020fzs} by the parton branching (PB) method~\cite{Hautmann:2017fcj}, compared 
with the measurements~\cite{Webb:2003ps}  and~\cite{Aidala:2018ajl}.   

\subsection{Relation to Ultrahigh Energy Neutrino and Astroparticle physics}

The small-$x$ region probed by the LHeC is also very important in the context of  ultra-high energy neutrino physics and astroparticle physics. Highly energetic neutrinos provide a unique window into the Universe, due to their weak interaction with matter, for a review see for example~\cite{Gandhi:1998ri}. They can travel long distances from distant sources, undeflected by the magnetic fields inside and in between galaxies, and thus provide complementary information to  cosmic rays, gamma rays and gravitational wave signals. The IceCube observatory on  Antarctica~\cite{Aartsen:2016nxy} is sensitive to neutrinos with energies from $100 \, \GeV$ up (above $10 \, \GeV$ with the use of their Deep Core detector). Knowledge about low-$x$ physics becomes indispensable in two contexts: neutrino interactions and neutrino production. At  energies beyond the $\TeV$ scale the dominant part of the cross section is due to the neutrino DIS CC and NC interaction with the hadronic targets~\cite{Gandhi:1998ri}. 

\begin{figure}[!th]
    \centering
    \includegraphics[width=0.55\textwidth,trim={0 120 0 50},clip]{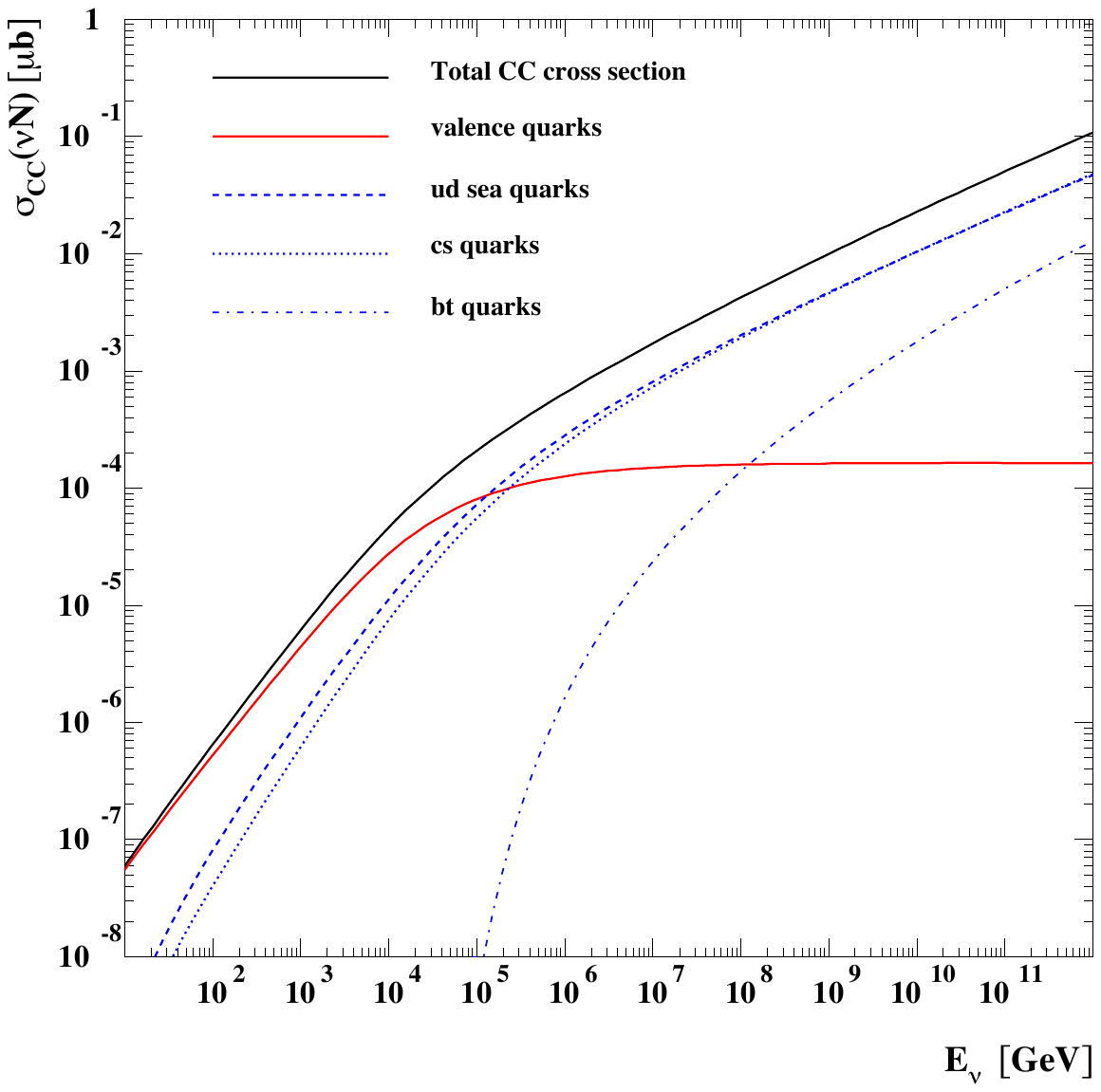}
    \caption{Charged current cross section for the neutrino - nucleon interaction on a isoscalar target as a function of  neutrino energy. The total CC cross section is broken down into several contributions due to valence, up-down,strange-charm and bottom-top quarks. The calculation was based on Ref.~\cite{Kwiecinski:1998yf}.}
    \label{fig:nucrosssection}
\end{figure}

\begin{figure}[!th]
    \centering
    \includegraphics[width=15cm]{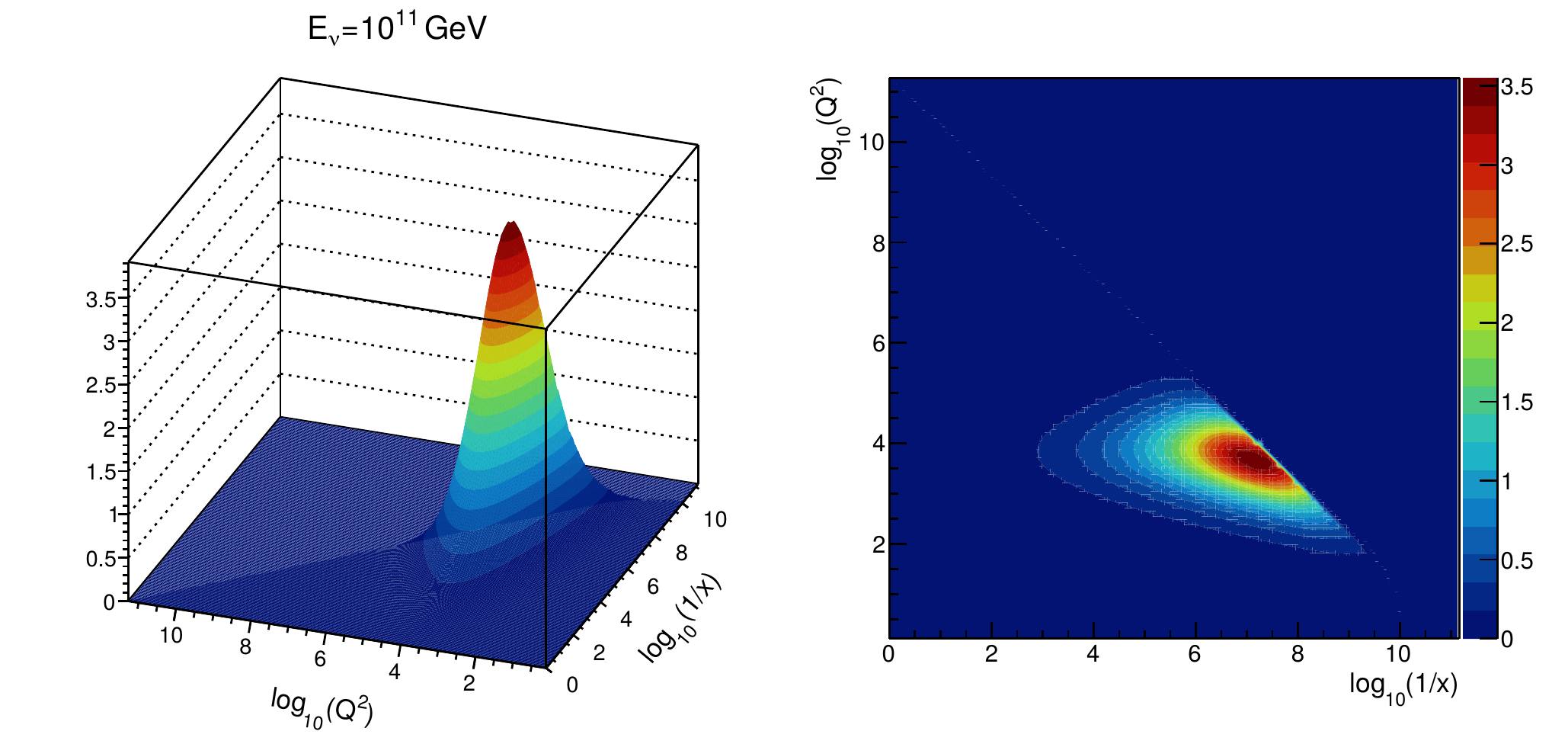}
    \caption{Differential charged current neutrino cross section $10^5 \cdot xQ^2d\sigma^{CC}/dxd Q^2\; [\text{nb}]$ as a function of $Q^2$ and $x$ for fixed neutrino energy $E_{\nu}=10^{11} \, \GeV$. Left:  surface plot; right:  contour plot. }
    \label{fig:disxq2}
\end{figure}

In Fig.~\ref{fig:nucrosssection} we show the charged current neutrino cross section as a function of the neutrino energy for an isoscalar target (in the laboratory frame where the target is at rest), using a calculation~\cite{Kwiecinski:1998yf} based on the resummed model in~\cite{Kwiecinski:1997ee}. We see that at energies below $\sim 50 \, \TeV$ the cross section grows roughly linearly with energy, and in this region it is dominated by contributions from the large-$x$ valence region. Beyond that energy the neutrino cross section grows slower, roughly as a power $\sim E_{\nu}^{\lambda}$ with $\lambda \simeq 0.3$. This high energy  behaviour is totally controlled by the small-$x$ behaviour of the parton distributions. The dominance of the sea contributions to the cross section is clearly seen in Fig.~\ref{fig:nucrosssection}.  To illustrate more precisely the contributing values of $x$ and $Q^2$, 
in Fig.~\ref{fig:disxq2} we show the differential cross section for the CC interaction $xQ^2d\sigma^{CC}/dxdQ^2$ for  a neutrino energy $E_{\nu}=10^{11} \; \GeV$ (in the frame where the hadronic target is at rest). We see a clear peak of the cross section at roughly a value of $Q^2 = M_W^2$ and an $x$ value
\begin{equation}
    x \simeq \frac{M_W^2}{2 M E_{\nu}} \; ,
\end{equation}
which in this case is about $3 \times 10^{-8}$. 
We note that IceCube extracted the DIS cross section from neutrino observations~\cite{Aartsen:2017kpd} in the region of neutrino energies $10-1000 \, \TeV$. The extraction is consistent, within the large error bands, with the predictions based on the QCD, like those illustrated in Fig.~\ref{fig:nucrosssection}. 
It is important to note that the IceCube extraction is limited to these  energies by the statistics due to the steeply falling flux of  neutrinos at high energy. We thus see that the neutrino interaction cross section at high energies is sensitive to a region which is currently completely unconstrained by existing precision DIS data.

Another instance where dynamics at low $x$ are crucial for neutrino physics is in understanding the mechanisms of ultra-high energy neutrino production. The neutrinos are produced in interactions which involve hadrons, either  in $\gamma p$ or in $pp$ interactions. They emerge as decay products of pions, kaons and charmed mesons, and possibly beauty mesons if the energy is high enough~\cite{Gaisser:1990vg}. For example, in the atmosphere  neutrinos are produced in the interactions of the highly energetic cosmic rays with nitrogen and oxygen nuclei. The lower energy part of the atmospheric neutrino spectrum, up to about $100 \, \TeV$ or so, is dominated by the decay of pions and kaons. This is called the conventional atmospheric neutrino flux. Above that energy the neutrino flux is dominated by the decay of the shorter-lived charmed mesons. Thus, this part of the neutrino flux is called the prompt-neutrino flux. The reason why the prompt-neutrino flux dominates at high energies is precisely related to the life-time of the intermediate mesons (and also baryons like $\Lambda_c$). The longer lived pions and kaons have a high probability of interacting before they decay, thus degrading their energy and leading to a steeply falling neutrino flux. The cross section for the production of charmed mesons is smaller than that for pions and kaons, but the charmed mesons $D^{\pm},D^0,D_s$ and baryon $\Lambda_c$ live shorter than  pions and kaons, and thus decay prior to any interaction. Thus, at energies about $100 \, \TeV$ the prompt neutrino flux will dominate over the conventional atmospheric neutrino flux. Therefore, the knowledge of this part of the spectrum is essential as it provides a background for the sought-after astrophysical neutrinos~\cite{Aartsen:2014gkd}. Charmed mesons in high energy hadron-hadron interactions are produced through gluon-gluon fusion into $c\bar{c}$ pairs, where one gluon carries rather large $x$ and the other one carries very small $x$. Since the scales are small, of the order of the charm masses, the values of the longitudinal momentum fractions involved are also very small  and thus the knowledge of the parton distributions in this region is essential~\cite{Gelmini:1999xq}. The predictions for the prompt neutrino flux become extremely sensitive to the behaviour of the gluon distribution at low $x$ (and low $Q^2$), where  novel QCD phenomena like resummation as well as  gluon saturation are likely to occur~\cite{Bhattacharya:2016jce}.

 In addition, the LHeC measurements could help pin down one enduring mystery - what is the composition of the most energetic cosmic rays?  The best measurements of composition at energies above ~ $10^{18}\, {\rm eV}$ are based on studies of how showers develop in the atmosphere.  The main observable is the depth (in the atmosphere) of shower maximum - so called  $X_{\rm max}$.  The absolute value of $X_{\rm max}$ and the elongation rate $dX_{\rm max}/dE$ of cosmic-rays depends on the assumed details of the  hadronic physics.  A change in the elongation rate, observed by the Auger observatory has often been interpreted as a signature for composition change (i.e. from mostly protons to mostly iron) with increasing energy \cite{Aab:2014aea,Aab:2014kda}.  However, new hadronic phenomena, such as a color glass condensate, might also lead to a change  in the elongation rate. Seeing saturation in a Large Hadron electron Collider would help select between these two options \cite{Drescher:2004sd,Klein:2019qfb}. 

Finally, the low-$x$ dynamics will become even more important at the HL-LHC and FCC hadron colliders, see Sect.~\ref{sec:smallx_impact}. With increasing centre-of-mass energy,  hadron colliders will probe values of $x$ previously unconstrained by HERA data. It is evident  that all the predictions in $pp$ interactions at high energy will heavily rely on the PDF extrapolations to the small $x$ region which carry large uncertainties. As discussed in detail in this Section, resummation will play an increasingly important role in the low $x$ region of PDFs. A precision DIS machine is thus an indispensable tool for constraining the QCD dynamics at low $x$ with great precision as well as for providing complementary information and independent measurements to hadronic colliders.


\section{Diffractive Deep Inelastic Scattering at the LHeC}
%
\label{sec:inclusive_diffraction}
\subsection{Introduction and Formalism}

The diffractive events in the Deep Inelastic Scattering  are characterized by the presence of the large, non-exponentially suppressed, rapidity gaps. By the large rapidity gap one means that there is a large region of the detector which is void of any particle activity between the proton (or a state with proton quantum numbers) and the rest of the produced particles. 
During the 90's both H1 and ZEUS experiments at HERA performed observations of diffraction in DIS ~\cite{Derrick:1993xh,Ahmed:1994nw,Adloff:1997sc,Breitweg:1997aa}, which constituted  large fraction, of about $10\%$, of all DIS events.

Diffractive events in DIS can be typically characterized by the presence of two scales: soft and hard. The soft scale is related to the size of the proton, which in the diffractive event remains typically intact\footnote{Or dissociates into a low mass excitation with quantum numbers of the proton.}, and the second scale, $Q^2$ which is perturbative. It is the presence of the latter, hard scale, which enabled to describe the diffraction in DIS in terms of the collinear factorization theorems ~\cite{Collins:1997sr,Berera:1995fj,Trentadue:1993ka}.  In a series of  ground-breaking papers \cite{Adloff:1997sc,Breitweg:1997aa,Chekanov:2005vv, Aktas:2006hx, Aktas:2006hy,Chekanov:2008fh,Chekanov:2009aa,Aaron:2010aa,Aaron:2012ad} (see  Ref.~\cite{Newman:2013ada} for a review) HERA experiments performed analysis and determined that the diffractive events could be described in terms of the diffractive parton densities.

The precise measurement of the diffractive DIS in wide kinematic range can provide unique insights into many facets of the strong interaction dynamics. Because of the presence of the large rapidity gap, it has been understood that the diffractive process proceeds through the exchange of the composite object which preserves color neutrality, and has quantum numbers of the vacuum. Thus the mechanism
  through which a composite  object interacts perturbatively ~\cite{Ingelman:1984ns,Buchmuller:1998jv,Hautmann:1998xn,Hautmann:1999ui,Brodsky:2004hi,Ingelman:2015qrt,Rasmussen:2015qgr} can offer information about the confinement and in general about emergent phenomena in strong interaction. 
  It has been established some time ago ~\cite{Bjorken:1992er,Bartels:1998ea,Hautmann:2000pw,Hautmann:2007cx} that the diffractive phenomena are closely related to the low-$x$ dynamics, and in particular to the partonic structure of the proton in this regime.
  Therefore an investigation of diffraction can offer unique insights into the role and importance of the higher twists and the non-linear parton evolution.
It has been also known \cite{Gribov:1968jf} that there is a relation between the diffraction in $ep$ and nuclear shadowing.   This relation has been used for example to successfully predict the amount of the shadowing in some processes at LHC ultra
peripheral collisions \cite{Frankfurt:2011cs,Guzey:2013qza}.
 Finally, precise measurements of the diffractive structure functions in the extended kinematic range of LHeC with respect to HERA will allow for the  accurate
extraction of diffractive parton distribution functions and provide more stringent constraints on the uncertainties. This in turn will  facilitate the tests of the  
range of validity of perturbative factorisation~\cite{Collins:1997sr,Berera:1995fj,Trentadue:1993ka} and potential importance of the higher twist effects.

In the following we will present the 
 studies of inclusive diffraction that will be possible at the LHeC. The detailed analysis was performed in Ref.~\cite{Armesto:2019gxy}, and we shall summarize these results in this and two following subsections. The LHeC will substantially extend the kinematic coverage of the HERA analyses, leading to much more detailed tests of theoretical ideas than have been possible hitherto. Although  the analysis done in \cite{Armesto:2019gxy} and summarized here was done at NLO of QCD, it is worth noting that similar analyses in the HERA context have recently extended to NNLO~\cite{Khanpour:2019pzq}.

\begin{figure}[!th]
  \centering
  \includegraphics[width=0.4\textwidth]{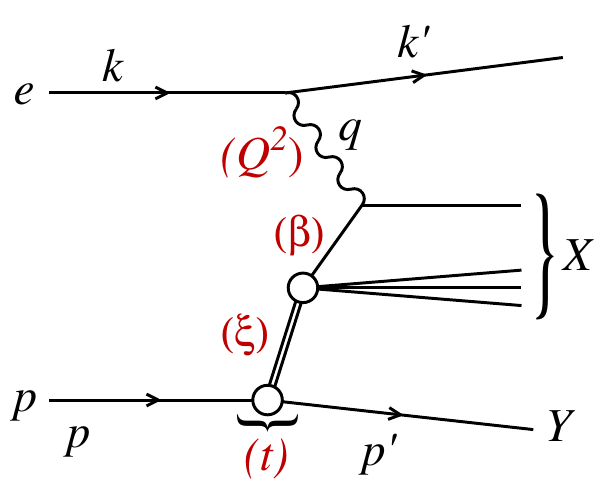}
\caption{A diagram of a diffractive NC event in DIS together with the corresponding variables, in the one-photon exchange approximation.  The incoming proton scatters elastically or is dissociated into a low mass excitation $Y$. The system of particles denoted by $X$ is then separated from the proton (or its excitation $Y$) by a large rapidity gap. Double line indicates a diffractive exchange in the $t$ channel.}
\label{fig:ddis}
\end{figure}

Diffractive deep inelastic event is schematically depicted in the diagram shown in Fig.~\ref{fig:ddis}. It is assumed that the process proceeds through the neutral current exchange.
 Charged currents could also be considered and  were measured at HERA~\cite{Aktas:2006hy} but with large statistical uncertainties and in a very restricted region of phase space. The LHeC and the FCC-eh will allow to measure charge currents in diffractive DIS with larger statistics and more extended kinematics than at HERA. However,  in the study \cite{Armesto:2019gxy} summarized here,   only neutral currents were considered, hence  we  shall also limit ourselves to that case.
 
The incoming electron or positron, with four momentum $k$, scatters off the proton, with  four momentum $p$.
Here, we only consider protons, though in principle one could also have nuclei. The inclusive diffraction in the nuclear case will be considered in Chapter 6.
The interaction proceeds through the exchange of a virtual photon with four-momentum $q$ and the diffractive exchange in the $t$ channel (indicated by the double line in Fig.~\ref{fig:ddis}). The kinematic variables   
for such an event include the standard deep inelastic variables
 \begin{equation}
 Q^2=-q^2\,, \qquad  x=\frac{-q^2}{2p\cdot q}\,,  \qquad y=\frac{p\cdot q}{p\cdot k}\,, 
 \end{equation} 
 where $Q^2$ is the (minus) photon virtuality, $x$ is the Bjorken variable and $y$ the inelasticity of the process. In addition, the variables
 \begin{equation}
 s=(k+p)^2\,, \qquad  W^2=(q+p)^2 \, ,
 \end{equation} 
 are  
 the electron-proton centre-of-mass energy squared and  the photon-proton centre-of-mass energy squared, respectively. A  diffractive event ${ep\rightarrow eXY}$ is uniquely characterized by   the presence of the large rapidity gap between the diffractive system, with the invariant mass 
$M_X$ and the final proton (or its low-mass excitation) $Y$
 with four momentum $p'$.  Therefore in order to fully describe the diffractive event in DIS, additional set of variables is necessary. They are defined as
\begin{equation}
t=(p-p')^2\,, \qquad \xi=\frac{Q^2+M_X^2-t}{Q^2+W^2}\,, \qquad \beta = \frac{Q^2}{Q^2+M_X^2-t}\, .
\end{equation}
In the above  $t$ is the squared
 four-momentum transfer   at the proton vertex, $\xi$ (alternatively denoted by $x_{I\!P}$)  can be interpreted as  the momentum fraction of the \emph{diffractive exchange}   with respect to the incoming hadron,  and  $\beta$ 
is the momentum fraction of the struck parton with respect to the diffractive exchange. 
 The two diffractive  momentum fractions are constrained to give  Bjorken-$x$, $x=\beta \xi$.

\begin{figure}[!th]
  \centering
  \includegraphics[width=0.6\textwidth]{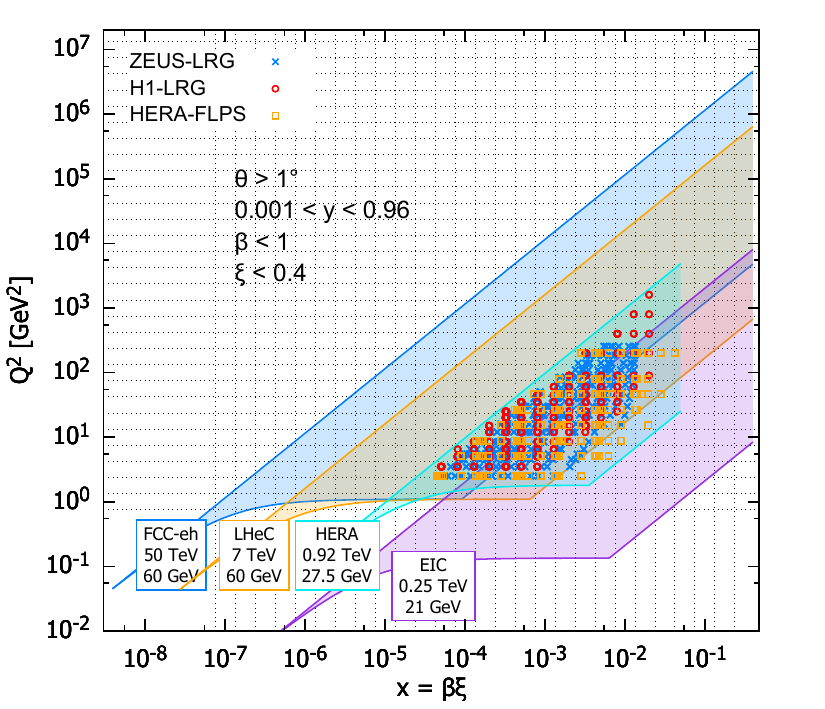}
\caption{Kinematic phase space for inclusive diffraction in $(x,Q^2)$ for  the Electron Ion Collider (EIC) (magenta region), the LHeC (orange region) and the FCC-eh (dark blue region) as compared with the HERA data (light blue region, ZEUS-LRG~\cite{Chekanov:2008fh}, H1-LRG~\cite{Aaron:2012ad}, HERA-FLPS~\cite{Aaron:2012hua}). The acceptance limit for the electron in the detector design has been assumed to be $ 1^{\circ}$, and we take $\xi<0.4$ (see text for details). Figure taken from \cite{Armesto:2019gxy}.}
\label{fig:phasespace_xQ}
\end{figure}

The  kinematic range in $(\beta,Q^2,\xi)$ that we consider at the LHeC is restricted by the following cuts:
\begin{itemize}
\item $Q^2 \ge 1.8\,\GeV^2$: due to the fact that the initial distribution for the DGLAP evolution is parameterised at $\mu_0^2=1.8 \,\GeV^2$. The renormalisation and factorisation scales are taken to be equal to $Q^2$.
\item $\xi<0.4$: constrained by physical and experimental limitations. This rather high $\xi$ value is an experimental challenge and physically enters the phase-space region where the Pomeron 
contribution should become negligible compared with sub-leading exchanges. 
Within the two-component model, see Eq.~\eqref{eq:param_2comp} below, at high $\xi$ the cross section is dominated by the secondary Reggeon contribution, which is poorly fixed by the HERA data. Nevertheless, we present this high $\xi$ ($> 0.1$) region for illustrative purpose and for the sake of discussion of the fit results below. It is also worth mentioning that with appropriate detector design it might be possible to reach this region of high $\xi$ which would be highly interesting and provide new constraints on the Pomeron and Reggeon contributions with respect to HERA.
\end{itemize}

In Fig.~\ref{fig:phasespace_xQ} the accessible kinematic range in $(x,Q^2)$ is 
shown for four machines: HERA, EIC, LHeC and FCC-eh \cite{Armesto:2019gxy}. For the LHeC design the 
range in $x$ is increased by a factor $\sim 20$ over HERA
and the maximum available $Q^2$ by a factor $\sim 100$. The FCC-eh machine would further increase this range with respect to LHeC by roughly one order of magnitude in both $x$ and $Q^2$. We also show the EIC kinematic region for comparison, which could cover high $x$ as well as low $Q^2$ range. The three different machines are clearly complementary in their kinematic coverage, with LHeC and EIC adding sensitivity at lower and higher $x$ than HERA, respectively. 

\begin{figure}[!th]
  \centering
  \includegraphics[width=0.92\textwidth,trim=0 0 10 50,clip]{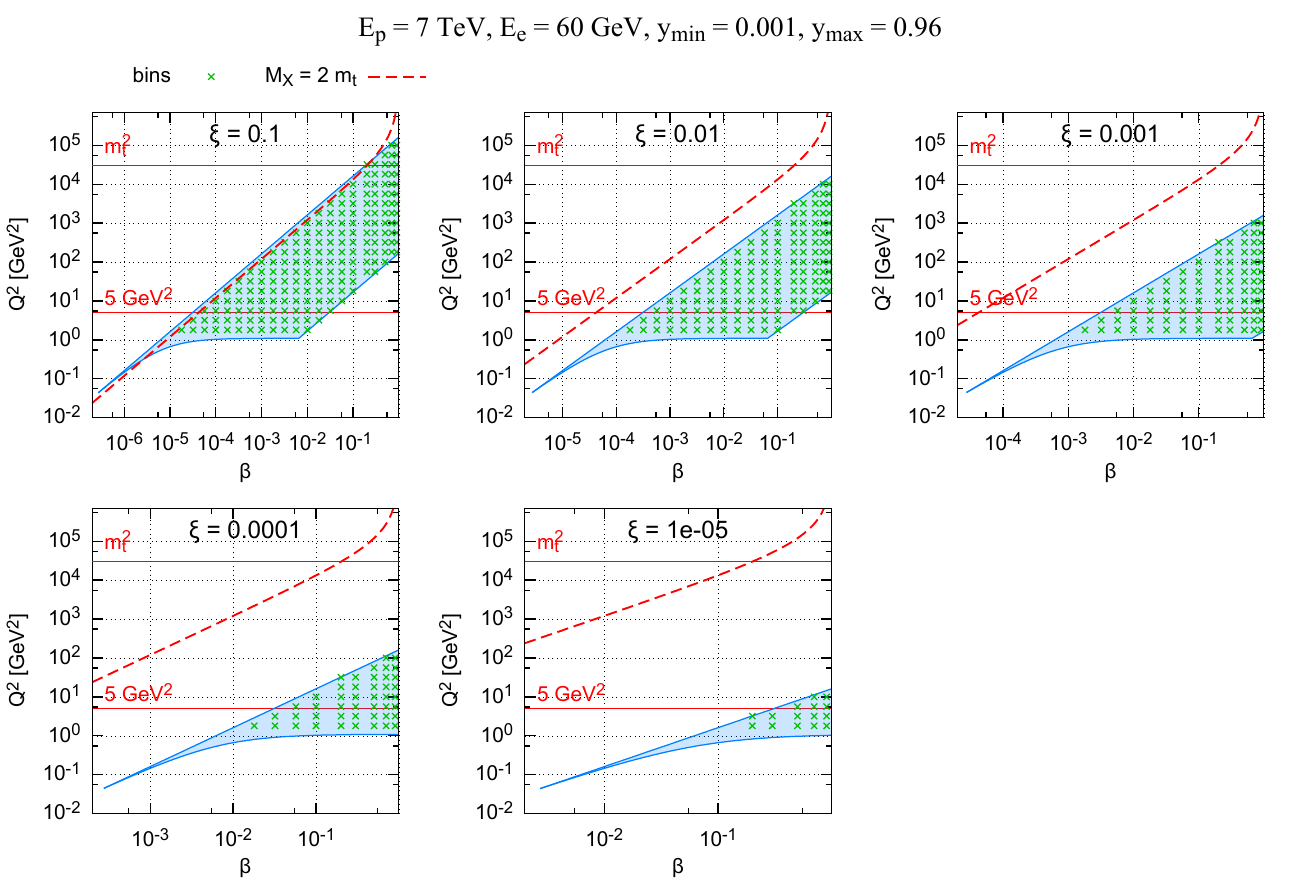}
\caption{Kinematic phase space for the inclusive diffraction in $(\beta,Q^2)$ for fixed values of $\xi$ for the LHeC design.
The horizontal lines indicate correspondingly, $Q^2=5 \; {\GeV}^2$, the lowest data value for the DGLAP fit performed in this study and $m_\text{t}^2$ the 6-flavour threshold. The dashed line marks the kinematic limit for $t\bar t$ production. Figure taken from \cite{Armesto:2019gxy}.
}
\label{fig:phasespace_bQ_lhec}
\end{figure}

In Fig.~\ref{fig:phasespace_bQ_lhec}  the phase space specific to diffractive processes in $(\beta,Q^2)$ is shown for fixed $\xi$ for the LHeC \cite{Armesto:2019gxy}.
Thanks to the high center of mass energy the LHeC machine probes very small values of $\xi$,  
reaching $10^{-4}$ with a wide range of $\beta$, and for the perturbative values of $Q^2$.  Of course, the ranges in $\beta$ and $\xi$ 
are correlated since $x=\beta\xi$. Therefore, for small values of $\xi$ only large values of $\beta$ are accessible while for large $\xi$ the range in $\beta$ extends to very small values.
The two horizontal lines denote $Q^2=5 \; \rm GeV^2 $ and $m_t^2$ threshold. The first value is the scale corresponding to the lowest data in the DGLAP fit  and DPDF extraction discussed later. The dashed line corresponds to the kinematic limit of the $t\bar{t}$ production.

In analogy to the inclusive case, the diffractive cross sections
 in the neutral current case can be represented in the form of the  reduced cross 
sections ~\cite{Aktas:2006hy}:
\begin{equation}
\label{eq:sigmared3}
\frac{d^3 \sigma^{{\mathrm{D}}}(3)}{d\xi d\beta dQ^2} = \frac{2\pi \alpha_\text{em}^2}{\beta Q^4} \, Y_+ \, \sigma_{\mathrm{red}}^{{\mathrm{D}}(3)}\, , 
\end{equation}
 where
 $Y_+= 1+(1-y)^2$ and the reduced cross sections can be expressed in terms of two diffractive structure functions
 $F_2^{{\mathrm{D}}}$ and $F_\mathrm{L}^{{\mathrm{D}}}$. 
 In the one-photon approximation, the relations are
\begin{equation}
\sigma_{\mathrm{red}}^{{\mathrm{D}}(3)} = F_2^{{\mathrm{D}}(3)}(\beta,\xi,Q^2) - \frac{y^2}{Y_+} F_\mathrm{L}^{{\mathrm{D}}(3)}(\beta,\xi,Q^2) \; .
\end{equation}
 In the above, both cross sections and structure functions are integrated over the momentum transfer $t$. This is indicated by $\sigma^{D(3)},F_2^{D(3)},F_L^{D(3)}$ notation where `3' means that the cross section or structure function depends on three variables, $(\beta,\xi,Q^2)$. Depending on the detector setup and luminosity it could also be possible to measure $\sigma^{D(4)},F_2^{D(4)},F_L^{D(4)}$ which depend on four variables $(\beta,\xi,Q^2,t)$. Also in principle, for the neutral current case, one needs to consider  $Z^0$ exchange in addition to photon, but in the analysis  \cite{Armesto:2019gxy} presented here it was neglected.

Both $\sigma_{\mathrm{red}}^{{\mathrm{D}}(3)}$ and $\sigma_{\mathrm{red}}^{{\mathrm{D}}(4)}$ have been measured at the HERA collider~\cite{Adloff:1997sc,Breitweg:1997aa,Chekanov:2005vv, Aktas:2006hx, Aktas:2006hy,Chekanov:2008fh,Chekanov:2009aa,Aaron:2010aa,Aaron:2012ad} and used to obtain QCD-inspired parameterisations.

The standard perturbative QCD approach to diffractive cross sections is based on the collinear factorisation~\cite{Collins:1997sr,Berera:1995fj,Trentadue:1993ka}. In these works it was demonstrated that, similarly to the inclusive DIS cross section, the diffractive cross section can be written, up to terms of  order  ${\mathcal O}(\Lambda^2/Q^2)$, where $\Lambda$ is the hadronic scale, in a factorised form
\begin{equation}
d\sigma^{ep\rightarrow eXY}(\beta,\xi,Q^2,t) \; = \; \sum_i \int_{\beta}^{1} dz \ d\hat{\sigma}^{ei}\left(\frac{\beta}{z},Q^2\right) \, f_i^\text{D}(z,\xi,Q^2,t) \; ,
\label{eq:collfac}
\end{equation}
where the sum is performed over all parton flavours (gluon, $d$-quark, $u$-quark, etc.).
The hard scattering partonic cross section $d\hat{\sigma}^{ei}$, corresponding to the short distance physics, can be computed order by order  in perturbative QCD and is the same as in the inclusive deep inelastic scattering case. The long distance 
part $f_i^\text{D}$ is the diffractive parton distribution function (DPDF). These functions
 can be interpreted as conditional probabilities for finding  partons  in the proton provided the latter is scattered into the final state system $Y$ with specified 4-momentum $p'$. 
They are evolved using the DGLAP evolution equations~\cite{Gribov:1972rt,Gribov:1972ri,Altarelli:1977zs,Dokshitzer:1977sg} similarly to the inclusive case.
The analogous formula to \ref{eq:collfac} for the $t$-integrated structure functions reads
\begin{equation}
\label{eq:FD3-fac}
F_{2/\text{L}}^{\text{D}(3)}(\beta,\xi, Q^2) =
\sum_i \int_{\beta}^1 \frac{dz}{z}\,
	 C_{2/\text{L},i}\Big(\frac{\beta}{z}\Big)\, f_i^{\text{D}(3)}(z,\xi,Q^2) \; ,
\end{equation}
where the coefficient functions $C_{2/\text{L},i}$ are the same as in  inclusive DIS, and can be computed perturbatively in QCD.

Fits to the diffractive structure functions usually
\cite{Aktas:2006hy,Chekanov:2009aa}
parameterise the diffractive PDFs in a two component 
model, which is a sum of two diffractive exchange contributions, ${I\!\!P}$ and ${I\!\!R}$:
\begin{equation}
f_i^{\text{D}(4)}(z,\xi,Q^2,t) =  f^p_{{I\!\!P}}(\xi,t) \, f_i^{{I\!\!P}}(z,Q^2)+f^p_{{I\!\!R}}(\xi,t) \, f_i^{{I\!\!R}}(z,Q^2) \;.
\label{eq:param_2comp}
\end{equation}
For both of these terms proton vertex factorisation is separately assumed, meaning that the diffractive exchange can be interpreted as colourless objects called a \emph{Pomeron} or a \emph{Reggeon} with  parton distributions $f_i^{{I\!\!P},{I\!\!R}}(\beta,Q^2)$.
Note that, this factorization is completely different than the collinear factorization for the structure functions mentioned above. It is an additional assumption motivated by the Regge theory and supported by the fits to the diffractive data. The flux factors  $f^p_{{I\!\!P},{I\!\!R}}(\xi,t)$ represent the probability that a Pomeron/Reggeon with given values $\xi,t$ couples to the proton.  They are parameterised using the form motivated by  Regge theory,
\begin{equation}
 f^p_{{I\!\!P},{I\!\!R}}(\xi,t) = A_{{I\!\!P},{I\!\!R}} \frac{e^{B_{{I\!\!P},{I\!\!R}}t}}{\xi^{2\alpha_{{I\!\!P},{I\!\!R}}(t)-1}} \; ,
\label{eq:flux}
\end{equation}
with a linear trajectory ${\alpha_{{I\!\!P},{I\!\!R}}(t)=\alpha_{{I\!\!P},{I\!\!R}}(0)+\alpha_{{I\!\!P},{I\!\!R}}'\,t}$, $B_{{I\!\!P},{I\!\!R}}$ being the $t$ slope and normalization factors $A_{{I\!\!P},{I\!\!R}}$.
One can also introduce the diffractive PDFs which correspond  to the $t$-integrated cross sections
\begin{equation}
f_i^{\text{D}(3)}(z,\xi,Q^2) =  \phi^{\;p}_{{I\!\!P}}(\xi) \, f_i^{{I\!\!P}}(z,Q^2) + \phi^{\;p}_{{I\!\!R}}(\xi) \, f_i^{{I\!\!R}}(z,Q^2) \; ,
\label{eq:fD3_2comp}
\end{equation}
with
\begin{equation}
	 \phi^{\;p}_{{I\!\!P},{I\!\!R}}(\xi) = \int \! dt\; f^p_{{I\!\!P},{I\!\!R}}(\xi,t) \;.
\end{equation}
Note that, the notions of \emph{Pomeron} and \emph{Reggeon} used here to model 
hard diffraction in DIS are, in principle, different from those describing the soft hadron-hadron interactions; in particular, the parameters of the fluxes may be different.

As is usual for the DGLAP evolution one needs to specify suitable initial conditions (see \cite{Armesto:2019gxy} for details). The diffractive parton distributions of the Pomeron at the initial scale $\mu_0^2 = 1.8\,\GeV^2$ are parameterised as
\begin{equation}
z f_i^{I\!\!P} (z,\mu_0^2)= A_i z^{B_i} (1-z)^{C_i} \; ,
\label{eq:initcond}
\end{equation}
as in ZEUS-SJ parametrisation, and
where $i$ is a gluon or a light quark and the momentum fraction $z=\beta$ in the case of quarks. In the diffractive parameterisations the contributions of all the 
light quarks (anti-quarks) are assumed to be equal. 
For the treatment of heavy flavours, a variable flavour number  scheme (VFNS) is adopted, where the charm and bottom quark DPDFs are generated radiatively via DGLAP evolution,  and no intrinsic heavy quark distributions are assumed.
The structure functions are calculated in a General-Mass Variable Flavour Number  scheme (GM-VFNS)~\cite{Collins:1986mp,Thorne:2008xf} which 
ensures a smooth transition of $F_\mathrm{2,L}$ across the flavour thresholds by including
$\mathcal{O}(m_h^2/Q^2)$ corrections.
The parton distributions for the Reggeon component are taken from a parameterisation which was obtained from fits to the pion structure function~\cite{Owens:1984zj,Gluck:1991ey}.

In Eq.~\eqref{eq:param_2comp} the normalisation factors of fluxes, $A_{{I\!\!P},{I\!\!R}}$ and of DPDFs, $A_i$ enter in the product. In order to resolve the ambiguity\footnote{Here, as in the HERA fits, $A_{{I\!\!P}}$ is fixed by normalizing $\phi^{\;p}_{{I\!\!P}}(0.003) = 1$.}  the normalization 
$A_{{I\!\!P}}$ was fixed
and  $f_i^{{I\!\!R}}(z,Q^2)$ was normalised to the pion structure function. This resulted in 
  $A_i$ and $A_{{I\!\!R}}$ being well defined free fit parameters. For full details of the parametrisations, see Ref.~\cite{Armesto:2019gxy}.

\subsection{Pseudodata for the  reduced cross section}
\label{sec:pseudo_data}
In order to generate the pseudodata for the LHeC one needs to use certain model which extrapolates the data from HERA. In the study \cite{Armesto:2019gxy} described here,
the reduced cross sections were extrapolated using the ZEUS-SJ DPDFs.
Following the scenario of the ZEUS fit~\cite{Chekanov:2009aa} we work
within the VFNS scheme at NLO accuracy. As mentioned before, there exists calculations at NNLO accuracy, however for the purposes of this analysis it is sufficient to work at NLO accuracy. The transition scales for 
DGLAP evolution are fixed by the heavy quark masses, $\mu^2 = m_h^2$
and the structure functions are calculated in the Thorne--Roberts GM-VFNS~\cite{Thorne:1997ga}.
The Reggeon PDFs are taken from the GRV pion set~\cite{Gluck:1991ey},
the numerical parameters are taken from Tables 1 and 3 of Ref.~\cite{Chekanov:2009aa},
the heavy quark masses are $m_c = 1.35\,\GeV, m_b = 4.3\,\GeV$,
and $\alpha_\mathrm{s}(M_Z^2) = 0.118$.

The pseudodata were generated \cite{Armesto:2019gxy} using the extrapolation of the fit to HERA data, which provides the central values,
amended with a random Gaussian smearing 
with  standard deviation corresponding to the relative 
error $\delta$. An uncorrelated $5\%$ systematic error was assumed giving a total uncertainty
\begin{equation}
\delta = \sqrt{\delta^2_\text{sys}+\delta^2_\text{stat}}\, .
\label{eq:uncertainty}
\end{equation}
The statistical error was computed assuming a very modest  integrated luminosity of $2 \, \text{fb}^{-1}$, see Ref.~\cite{Bordry:2018gri,LHeClumi}. For the binning adopted in the study \cite{Armesto:2019gxy}, the statistical uncertainties  have a very small effect on the uncertainties in the extracted DPDFs. Obviously, a much larger luminosity would allow a denser binning that would result in smaller DPDF uncertainties. 
An extended analysis in principle could be performed for the $\sigma^{D(4)}$ which would include $t$ dependence, provided the latter could be extracted by using proper forward instrumentation.  

\begin{figure}[!th]
  \centering
  \includegraphics[width=0.7\textwidth,trim=0 5 0 20,clip]{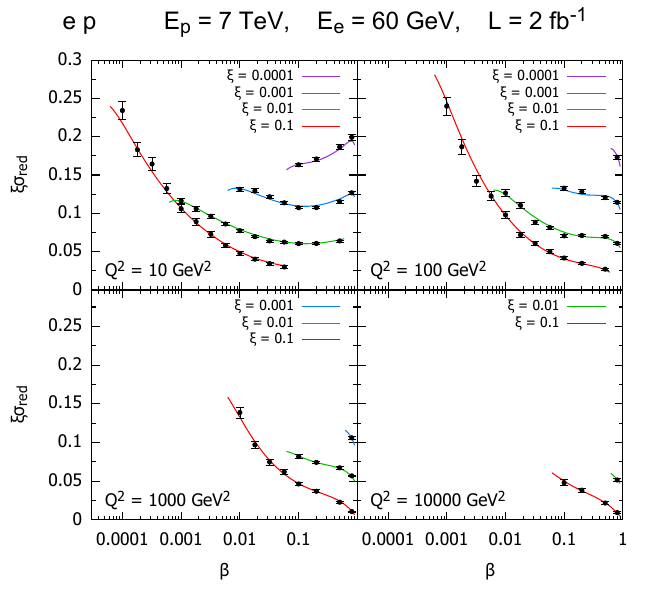}
\caption{Selected subset of the  simulated data for the diffractive reduced cross section as a function of $\beta$ in bins of $\xi$ and $Q^2$ for $ep$ collisions at the LHeC. Based on the extrapolation of the ZEUS-SJ fit to HERA data.
The curves for $\xi = 0.01, 0.001, 0.0001$ are shifted up by 0.04, 0.08, 0.12, respectively. Integrated luminosity is taken to be $2 \; \rm fb^{-1}$. Figure taken from \cite{Armesto:2019gxy}.}
\label{fig:sigred_ep_lhec}
\end{figure}

In Fig.~\ref{fig:sigred_ep_lhec} we show a 
subset of the simulated data for the diffractive reduced cross section 
$\xi\sigma_\text{red}$ as a function of $\beta$ in selected bins of $\xi$ 
and $Q^2$ for the LHeC \cite{Armesto:2019gxy}. For 
the most part the errors are very small, and are dominated by the systematics. The breaking of Regge factorisation, evident at large $\xi$, comes from the large Reggeon contribution in that region, whose validity could be further investigated at the LHeC. 

We see that for the LHeC parameters, the integrated luminosity is sufficient for the precise measurement of the diffractive reduced cross section. The study could be  further refined by implementing more information about the potential sources of the systematic errors including correlations.
In addition, by varying the centre-of-mass energy one could extract also longitudinal structure function $F_L^{D(3)}$.  Pioneering measurement of this quantity was performed at HERA \cite{Aaron:2012zz}, albeit with very limited precision. Longitudinal diffractive structure function could be extremely valuable information, as it is an independent   diffractive structure function and provides with an additional constraint on the diffractive PDFs. It also  may be more sensitive quantity to the higher twist contribution. More detailed analysis need to be however performed to determine the feasibility and precision of such measurement at the LHeC.

\subsection{Potential for constraining diffractive PDFs at the LHeC}

We next proceed to discuss the prospects for the extraction and constraining the diffractive PDFs from  the future experimental data obtained at the LHeC.
The strategy to assess the constraining potential was developed in \cite{Armesto:2019gxy} which we  summarize below. First,  the central values of the pseudodata using the central set of the ZEUS-SJ fit are generated, that are distributed according to a Gaussian with experimental width given by Eq.~\eqref{eq:uncertainty}, that also provides the uncertainty in the pseudodata. 
After that,   the pseudodata in a fit are included
alongside the existing HERA data using the same functional form of the initial parametrisation.
The quality of the resulting fit   was very good, as it was expected, and  $\chi^2/\mathrm{ndf} \sim 1$ was obtained, 
 which demonstrated the consistency of the approach.

To evaluate the experimental precision with which the DPDFs can be determined,
several pseudodata sets,
corresponding to independent random error samples, were generated \cite{Armesto:2019gxy}.
Each pseudodata set was fitted separately.
The minimal value of $Q^2$ for the data considered in the fits 
was set to  $Q^2_\text{min} = 5 \,\GeV^2$. The reason for this cut-off is  to show the feasibility of the fits including just the range in which standard twist-2 DGLAP evolution is expected to be trustable. At HERA, the $Q^2_\text{min}$ values giving acceptable 
DGLAP (twist-2) fits were $8\,\GeV^2$~\cite{Aktas:2006hy}
and $5\,\GeV^2$~\cite{Chekanov:2008fh} for  H1 and ZEUS, respectively. 
Below these values the fits were deteriorating.
The maximum value of $\xi$ was set 
by default to $\xi_\text{max} = 0.1$, above 
which the cross section starts to be dominated by the Reggeon exchange.
The binning adopted in the study \cite{Armesto:2019gxy} corresponds roughly to 4 bins per order of magnitude in each of $\xi, \beta, Q^2$.
For $Q^2_\text{min} = 5 \,\GeV^2$, $\xi_\text{max} = 0.1$ and below the top threshold
this results in 1229  pseudodata points for the LHeC.
The top-quark region adds 17 points for the LHeC. The LHeC offers window to study top quark contribution in diffraction in limited range of kinematics, going further to FCC-eh would expand that possibility greatly.
Lowering $Q^2_\text{min}$ down to $1.8\,\GeV^2$ we get 1589 pseudodata points,
while increasing $\xi$ up to 0.32 adds around 180 points for the LHeC machine. Of course, in the case of the lower value of $Q^2$ the collinear formalism with leading twist contribution may become questionable. Given the fact that the ZEUS and H1 fits based on DGLAP did not desrcibe well the data in the low $Q^2$ region may indicate that other effects may start playing important role. In Ref.~\cite{Motyka:2012ty} it was argued that this deviation from the leading twist DGLAP evolution might be an indication for higher twist effects. Larger lever arm in $x$, and high precision of the data from the LHeC  will be extremely helpful in mapping out the region of validity of the leading twist description and should help to constrain the higher twist effects in diffraction. Dedicated studies of the LHeC potential in this area will need to be performed,

The potential for determination of the gluon DPDF was investigated by fitting the inclusive diffractive DIS pseudodata with two  models with different numbers of parameters, named S and C (see Ref.~\cite{Armesto:2019gxy} for details)
with $\alpha_{I\!P,I\!R}(0)$ fixed, in order to focus on the shape of 
the Pomeron's PDFs.
At HERA, both S and C fits provide equally good 
descriptions of the data with $\chi^2/\mathrm{ndf} = 1.19$ and 1.18, respectively, 
despite different gluon DPDF shapes. 
The LHeC pseudodata are much more sensitive to gluons, resulting in
$\chi^2/\mathrm{ndf}$ values of 1.05 and 1.4 for the S and C fits, 
respectively.  It also shows clearly the potential of the LHeC to
better constrain the low-$x$ gluon and, therefore, unravel eventual departures from standard linear evolution.

\begin{figure}[!th]
\centering
\includegraphics[width=.48\textwidth,trim=0 0 8 34,clip]{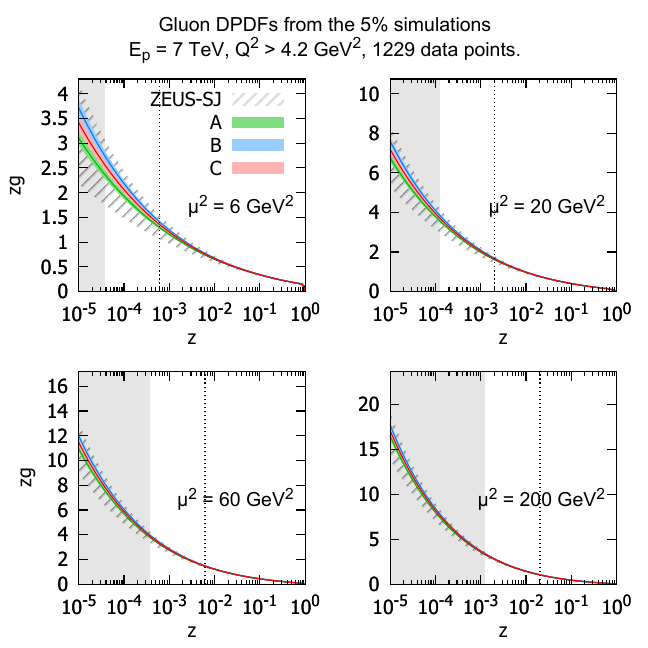}
\hspace{0.02\textwidth}
\includegraphics[width=.48\textwidth,trim=0 0 8 34,clip]{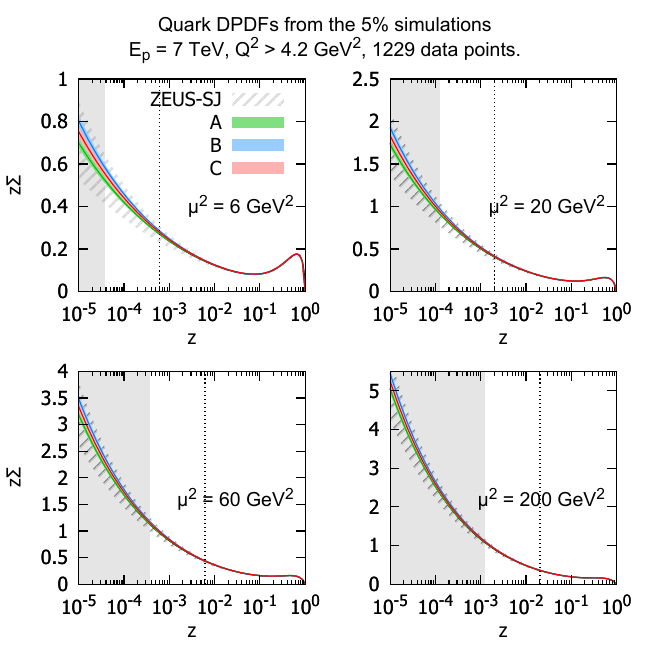}
\caption{Diffractive PDFs for gluon and quark in the LHeC kinematics as a function of momentum fraction $z$ for fixed values of scale
$\mu^2$. Results of fits to three (A,B,C) pseudodata 
replicas are shown together with the experimental error bands.
For comparison, the extrapolated 
ZEUS-SJ fit is also shown (black) with error bands marked with the hatched pattern.
The vertical dotted lines indicate the HERA kinematic limit. The bands indicate only the experimental uncertainties. The shaded band indicates region not accessible to LHeC. Figure taken from \cite{Armesto:2019gxy}.}
\label{fig:pdf_fits_lhec}
\end{figure}

In Fig.~\ref{fig:pdf_fits_lhec}  the diffractive gluon and quark 
distributions are shown for the LHeC, as a  function of longitudinal momentum fraction 
$z$ for fixed scales $\mu^2 = 6, 20, 60, 200\,\GeV^2$, see \cite{Armesto:2019gxy}.
The bands labelled
$A,B,C$ denote fits to three statistically independent pseudodata replicas, obtained from the same central values and statistical and systematic uncertainties. Hereafter the uncertainty bands
shown correspond to $\Delta\chi^2 = 2.7$ (90\,\% CL).
Also the extrapolated ZEUS-SJ DPDFs are shown with error bands marked by 
the `/' hatched area.
Note that the depicted uncertainty bands come solely from experimental errors, neglecting 
theoretical sources, such as fixed input parameters and parameterisation biases.
The extrapolation beyond the reach of LHeC is marked in grey and the HERA kinematic limit is marked with the vertical dotted line. 
The low $x$ DPDF determination accuracy improves with respect to 
HERA by a factor of 5--7 for the LHeC  and completely new kinematic regimes are accessed.

\begin{figure}[!th]
  \centering
  \includegraphics[width=0.9\textwidth,trim=0 0 0 48,clip]{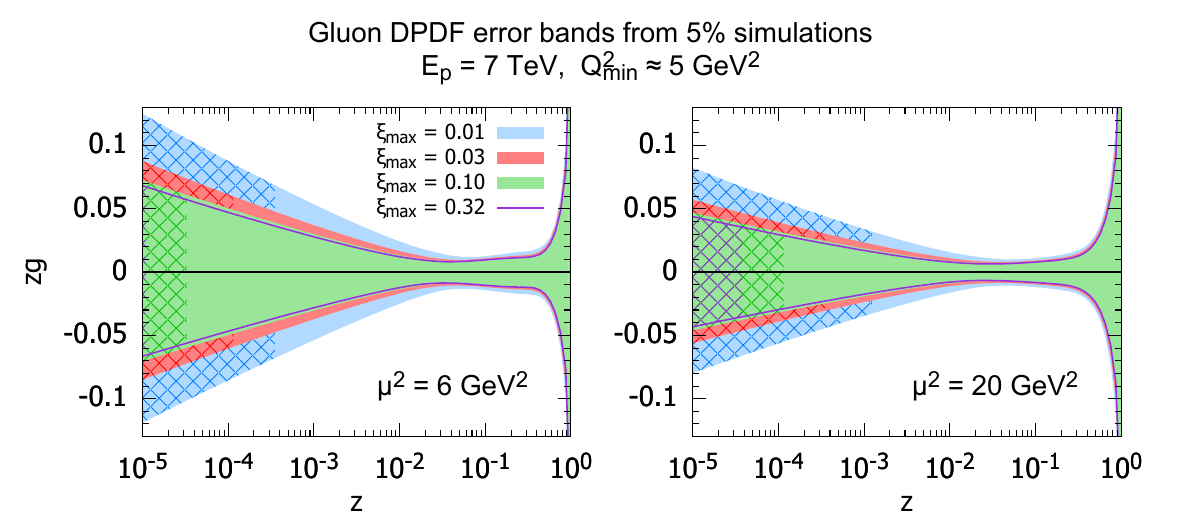}
\caption{Relative uncertainties on the diffractive gluon PDFs for the LHeC kinematics. Two different choices of scales are considered $\mu^2=6$ and $\mu^2=20$\,$\GeV^2$. The blue, red, green bands and magenta line correspond to different maximal values of $\xi = 0.01,0.03,0.1,0.32$, respectively.
The cross-hatched areas show kinematically excluded regions. The bands indicate only the experimental uncertainties, see the text. Figure taken from \cite{Armesto:2019gxy}.}
\label{fig:pdf_7_50_xi}
\end{figure}

For a better illustration of the precision, 
in Fig.~\ref{fig:pdf_7_50_xi}
the relative uncertainties  are shown for parton distributions at 
different scales, see \cite{Armesto:2019gxy}.
The different bands show the variation with the upper cut on the available 
$\xi$ range, from $0.01$ to $0.32$. In the best constrained region of $z \simeq 0.1$, the precision reaches the 1\% level. We observe only a modest 
improvement in the achievable accuracy of the extracted DPDFs with 
the change of $\xi$ by an order of magnitude from $0.01$ to $0.1$.  
An almost negligible effect is observed when further extending 
the $\xi$ range up to $0.32$.  This is very encouraging, 
since the measurement for the very large values of $\xi$ is challenging. It reflects
the dominance of the secondary Reggeon in this region.

We stress again that only experimental errors are included in our uncertainty bands. Neither theoretical uncertainties nor the parameterisation biases are considered. Of course such studies could be expanded  to obtain more precise estimates on the potential of the LHeC measurements to constrain and detect the deviations from the factorization of the importance of the higher twists, as an example.  For a detailed discussion of this and other aspects of the fits, see Ref.~\cite{Armesto:2019gxy}.

\subsection{Hadronic Final States in Diffraction and hard rapidity gap processes}
\label{sec:factest}

Various diffractive processes offer unique opportunity to investigate the factorisation properties and can help to disentangle DGLAP vs BFKL dynamics. 

The factorisation properties of diffractive DIS were a major topic of study at HERA~\cite{Newman:2013ada} and are highly relevant to the interpretation of diffractive processes at the LHC~\cite{Aad:2015xis}. A general theoretical framework is provided by the proof~\cite{Collins:1997sr} of a hard scattering collinear QCD factorisation theorem for semi-inclusive DIS scattering processes such as $ep \rightarrow epX$. This implies that the DPDFs extracted in fits to inclusive diffractive DIS may be used to predict perturbative cross sections for hadronic final state observables such as heavy flavour or jet production. Testing this factorisation pushes at the boundaries of applicability of perturbative QCD and will be a major topic of study at the LHeC.

Tests of diffractive factorisation at HERA are strongly limited by the kinematics. The mass of the dissociation system $X$ is limited to approximately $M_X < 30 \, {\GeV}$, which implies for example that jet transverse momenta cannot be larger than about $15 \, {\GeV}$ and more generally leaves very little phase space for any studies at perturbative scales. As well as restricting the kinematic range of studies, this restriction also implied large hadronisation and scale uncertainties in theoretical predictions, which in turn limit the precision with which tests can be made.  

The higher centre-of-mass energy of the LHeC opens up a completely new regime for diffractive hadronic final state observables in which masses and transverse momenta are larger and theoretical uncertainties are correspondingly reduced. For example, $M_X$ values in excess of $250 \, {\GeV}$ are accessible, whilst remaining in the region $\xi < 0.05$ where the leading diffractive (pomeron) exchange dominates. The precision of tests is also improved by the development of techniques for NNLO calculations for diffractive jets~\cite{Britzger:2018zvv}.

\begin{figure}[!th]
  \centering
  \includegraphics[width=0.65\textwidth]{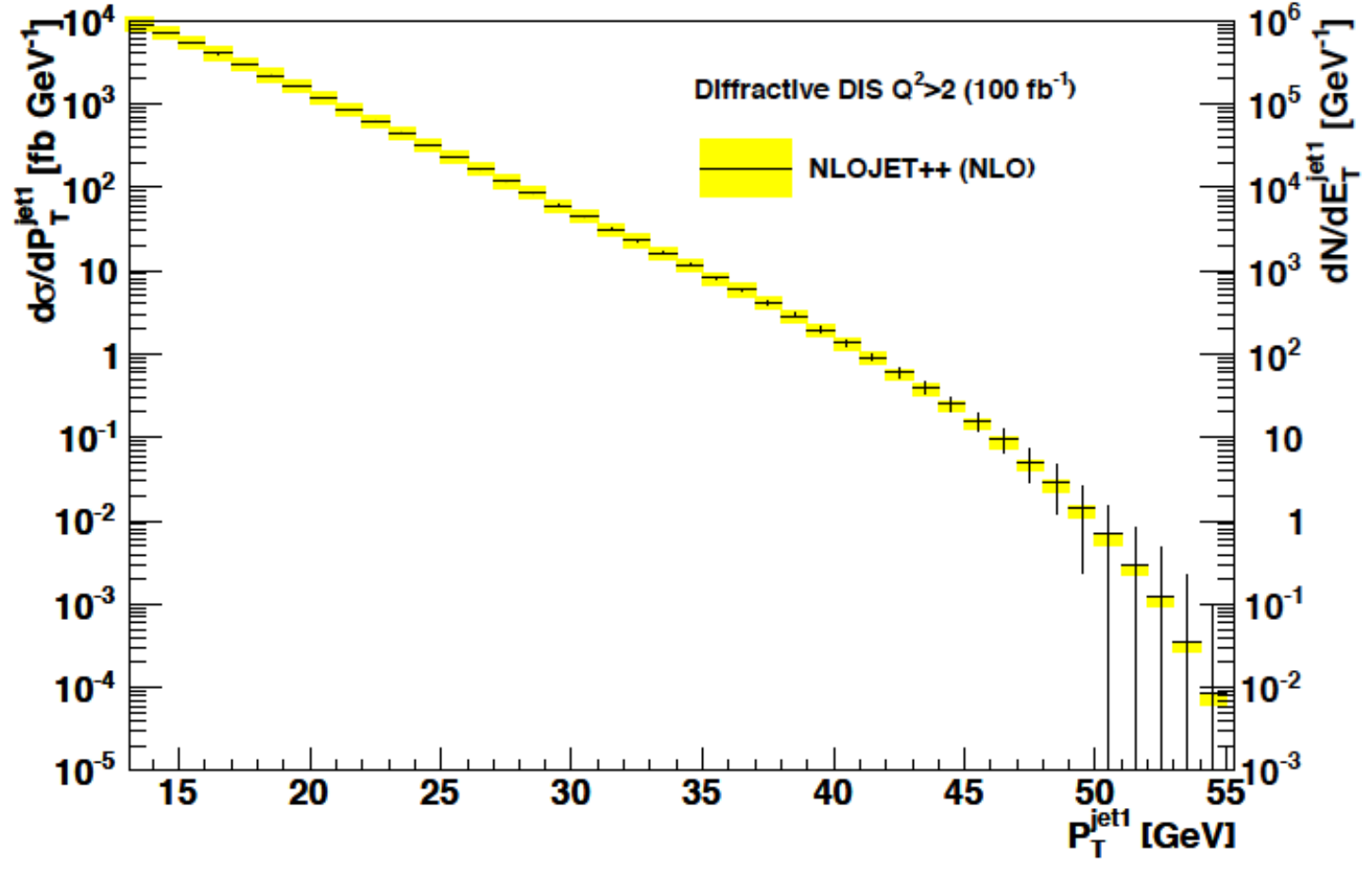}
\caption{Simulated diffractive dijet cross section as a function of leading jet transverse momentum in the kinematic range $Q^2 > 2 \, {\GeV^2}$ and $0.1 < y < 0.7$, with scattered electron angles in  excess of $1^\circ$. The error bars indicate predicted statistical uncertainties for a luminosity of $100 \, \text{fb}^{-1}$. The coloured bands correspond to theoretical uncertainties when varying the renormalisation and factorisation scales by factors of 2.}
\label{fig:difjets}
\end{figure}

Fig.~\ref{fig:difjets} shows a simulation of the expected diffractive jet cross section at the LHeC, assuming DPDFs extrapolated from H1 at HERA~\cite{Aktas:2006hy}, using the NLOJET++ framework~\cite{Nagy:2003tz}. 
An integrated luminosity of $100\,\text{fb}^{-1}$ is assumed and the kinematic range considered is $Q^2 > 2\,{\GeV^2}$, $0.1 < y < 0.7$ and scattered electron angles larger than $1^\circ$. Jets are reconstructed using the $k_T$ algorithm with $R = 1$. The statistical precision remains excellent up to jet transverse momenta of almost 50\,GeV and the theoretical scale uncertainties (shaded bands) are substantially reduced compared with HERA measurements. Comparing a measurement of this sort of quality with predictions refined using DPDFs from inclusive LHeC data would clearly provide an exacting test of diffractive factorisation. 

Further interesting hadronic final state observables that were studied at HERA and could be extended at the LHeC include open charm production, thrust and other event shapes, charged particle multiplicities and energy flows. In addition, the LHeC opens up completely new channels, notably diffractive beauty, $W$ and $Z$ production, the latter giving complementary sensitivity to the quark densities to that offered by inclusive diffraction.

Of  separate interest are hard rapidity gap processes, for example $\gamma^*p  \to J/\psi +  \mbox{rapidity gap}(\Delta y) + Y$ at large $-t\gg 1 \, {\rm GeV}^2$.
 In such processes 
DGLAP evolution is strongly suppressed and therefore this is  an ideal laboratory to investigate BFKL dynamics.   The dependence of the process on $\Delta y $ is expected to be given by
$\sigma \sim  \Delta y^{2\omega_P(t)}$. 
 Here, the effective pomeron trajectory is parameterized as $\alpha_P(t)= 1 + \omega(t)$. The current models give $\omega$ between 0.5 (LO BFKL) and $\omega =0.2 - 0.3$ for the 
resummed BFKL. With a proper large acceptance detector one would be able to study dependence on $\Delta y$  in a wide rapidity interval as well as the dependence on the momentum transfer $t$. Hence this process offers a powerful test for the theoretical predictions of the properties of the BFKL pomeron.

\section{Theoretical Developments}
\subsection{Prospects for Higher Order pQCD in DIS}
\label{sec:hoqcd}
With its large anticipated luminosity, the LHeC will be able to
perform highly precise measurements for a wide variety of final states
in deep inelastic scattering, often exploring novel kinematical
ranges, challenging the theory of QCD at an unprecedented level of
accuracy, and enabling precision determinations of QCD parameters and
of the proton’s parton structure. For this program to succeed, it will
be mandatory to be able to confront the LHeC precision data with
equally precise theoretical predictions.  

In the Standard Model, these predictions can be obtained through a
perturbative expansion to sufficiently high order. These calculations
are performed in the larger framework of QCD
factorisation~\cite{Collins:1989gx} and exploit the
process-independence of parton distributions, whose evolution is
controlled by the DGLAP equations. The DGLAP splitting functions are
known to NNLO level for already quite some
time~\cite{Moch:2004pa,Vogt:2004mw}, and important progress has been
made recently towards their N3LO
terms~\cite{Moch:2017uml,Herzog:2018kwj}. Moreover, mixed QCD/QED
corrections to them have been derived~\cite{deFlorian:2016gvk},
enabling a consistent combination of QCD and electroweak effects. 

The physics opportunities that are offered already by the HERA legacy
data set have motivated  substantial recent activity in precision QCD
calculations for deeply inelastic processes. At the inclusive level,
the QCD coefficients for the inclusive DIS structure functions are
known to three loops (N3LO) for some time~\cite{Vermaseren:2005qc},
they were improved upon recently by the computation of heavy quark
mass effects~\cite{Ablinger:2014vwa,Ablinger:2017err}. Fully
differential predictions for final states with jets, photons, heavy
quarks or hadrons are generally available to NLO in QCD, often dating
back to the HERA epoch. Technical developments that were made in the
context of fully differential higher-order QCD calculations for LHC
processes have enabled substantial advances in the theory precision of
DIS jet cross sections. Fully differential predictions for single jet
production are now available to NNLO~\cite{Abelof:2016pby} and
N3LO~\cite{Currie:2018fgr,Gehrmann:2018odt} for neutral-current and
charged current DIS, and  two-jet
production~\cite{Currie:2016ytq,Currie:2017tpe,Niehues:2018was} has
been computed to NNLO. The latter calculations are performed with
fully differential parton-level final state information, thereby
allowing their extension to jet production in diffractive
DIS~\cite{Britzger:2018zvv} and to DIS two-parton event
shapes~\cite{Gehrmann:2019hwf}. The newly derived NNLO jet cross
sections were partly used in the projections for LHeC precision jet
studies in Sections~\ref{sec:alphasjets} and~\ref{sec:factest} above.    

NLO calculations have been largely automated in
QCD~\cite{Hirschi:2011pa,Cascioli:2011va,Cullen:2014yla} and the
electroweak theory~\cite{Frederix:2018nkq,Buccioni:2019sur}, and are
now available in multi-purpose event generator
programs~\cite{Alwall:2014hca,Bellm:2015jjp,Bothmann:2019yzt} for
processes of arbitrary multiplicity. These can be combined with parton
shower approximations to provide NLO-accurate predictions for fully
exclusive final states.
Although most of the applications of these
tools were to hadron collider observables, they are also ready to be
used for DIS processes~\cite{Hoche:2018gti}, thereby offering novel
opportunities for precision studies for LHeC. In this context,
electroweak corrections may become particularly crucial for high-mass
final states at the LHeC, and have been largely unexplored up to now.
A similar level of automation has not yet been reached so far at NNLO,
where calculations are performed on a process-by-process basis. For
DIS processes, fully differential NNLO calculations for three-jet
final states or for heavy quark production could become feasible in
the near-term future.
Moreover, a whole set of calculations at this
order for specific final states (involving jets, vector bosons or
heavy quarks) in photoproduction could be readily taken over by
adapting the respective hadron collider results.  

The all-order resummation of large logarithmic corrections to hadron
collider processes has made very substantial advances in the recent
past, owing to the emergence of novel systematic frameworks from
soft-collinear effective theory, or in momentum space resummation. As
a result, threshold logarithms and transverse-momentum logarithms in
benchmark hadron collider processes can now be resummed up to the
third subleading logarithmic order (N3LL).
A similar accuracy has been reached for selected event shapes in
electron-position annihilation.
For DIS event shapes, currently available predictions
include only up to NLL resummation~\cite{Dasgupta:2002dc}. With the
newly available frameworks, they could be improved by two more
logarithmic orders, as demonstrated in exploratory work on the DIS
one-jettiness event shape~\cite{Kang:2013lga,Kang:2015swk}.
Applications of this framework to final states in DIS the small-$x$
limit (see Section~\ref{sec:lowx}) are 
largely unexplored, and may provide important novel insights into the
all-order dynamics in the high-energy  limit.  

The full exploitation of future LHeC data will require novel precision
calculations for a variety of benchmark processes, often combining
fixed-order, resummation and parton shower event generation to obtain
theory predictions of matching accuracy.
Recent advances in calculational techniques and an increasing degree
of automation will help to enable this progress. A close interplay
between experiment and theory will then be crucial to combine data and
predictions into precision measurements of physics parameters and into
probes of fundamental particle dynamics.

\subsection{Theoretical Concepts on the Light Cone}

\subsubsection{Intrinsic Heavy Quark Phenomena}
%
One of the most interesting nonperturbative quantum field theoretic aspects of hadron light front wavefunctions in QCD are the intrinsic heavy-quark 
Fock states~\cite{Brodsky:1980pb,Brodsky:2015fna,Brodsky:2015uwa}. 
Consider a heavy-quark loop insertion to the proton's self-energy.  The heavy-quark loop can be attached  by gluons to just one valence quark.  The cut of such diagrams yields the standard DGLAP gluon splitting contribution to the proton's heavy quark structure function.  In this case, the heavy quarks are produced at very small $x$. However, the heavy quark loop can also be attached to two or more valence quarks in the proton self-energy.  In the case of QED this corresponds to the  light-by-light lepton loop insertion in an atomic wavefunction. 
In the case of QCD, the heavy quark loop can be attached by three gluons to two or three valence  quarks in the proton self-energy.  This is  a non-Abelian  insertion to the hadron's self-energy. The cut of such diagrams gives the \emph{intrinsic} heavy-quark contribution to the proton's light-front wavefunction. In the case of QCD, the probability for an intrinsic  heavy $Q\bar Q$ pair scales as $\tfrac{1}{M^2_Q}$; this is in contrast to heavy $\ell \bar \ell $ lepton pairs in QED where the probability for heavy lepton pairs in an atomic wavefunction  scales as $\tfrac{1}{M^4_\ell}$.   This difference in heavy-particle scaling in mass distinguishes Abelian from non-Abelian theories. 

A basic property of hadronic light-front wavefunctions is that they have strong fall-off with the invariant mass of the Fock state. 
For example, the Light-Front Wave Functions (LFWFs) of the colour-confining AdS/QCD models~\cite{deTeramond:2008ht}
${\mathcal M}^2 = [\sum_i k^\mu_ i]^2$ of the Fock state constituents. This means that the probability is maximised when the constituents have equal true rapidity, 
i.e.\ $x_i \propto ({\vec k}_{\perp i}^2+ m^2_i)^{1/2}$.  
Thus the heavy quarks carry most of the momentum in an intrinsic heavy quark Fock state. For example, the charm quark in the intrinsic charm Fock state $|uudc\bar c\rangle$ of a proton carries about 40\,\% of the proton's momentum: $x_c \sim 0.4$.
After a high-energy collision, the co-moving constituents can then recombine to form the final state hadrons. along the proton.
Thus, in a $e p$ collision the comoving $udc $ quarks from the $ |uudc \bar c\rangle$ intrinsic 5-quark Fock state can recombine to a $\Lambda_c $, where $x_{\Lambda_c} = x_c + x_u + x_d \sim 0.5$.   
Similarly, the comoving $dcc $ in the $ |uudc \bar c c\bar c \rangle $ intrinsic 7-quark Fock state can recombine to a $\Xi(ccd)^{+}  $, with $x_{\Xi(ccd)}  = x_c +x_c + x_d \sim 0.9$.

Therefore, in the intrinsic heavy quark model the wavefunction of a hadron in QCD can be
represented as a superposition of Fock state fluctuations,
e.g.\ $\vert n_V
\rangle$, $\vert n_V g \rangle$, $\vert n_V Q \overline Q \rangle$,
\ldots components where $n_V \equiv dds$ for $\Sigma^-$, $uud$ for proton,
$\overline u d$ for $\pi^-$ and $u \overline d$ for $\pi^+$.  Charm hadrons
can be produced by coalescence in the wavefunctions of the moving hadron.
Doubly-charmed hadrons require fluctuations such as
$\vert n_V c \overline c c \overline c \rangle$. The probability for these
Fock state fluctuations to come on mass shell is inversely proportional to the
square of the quark mass, $\mathcal{O}(m_Q^{-2n})$ where $n$ is the number of
$Q \overline Q$ pairs in the hadron.
Thus the natural domain for heavy hadrons produced from heavy quark Fock states is ${\vec k_{\perp Q}}^2 \sim m^2_Q$ and high light-front momentum fraction $x_Q$~\cite{Brodsky:1980pb,Brodsky:2015fna,Brodsky:2015fna,Brodsky:2015uwa}.  For example, the rapidity regime for double-charm hadron production  $y_{ccd}\sim 3$ at low energies  is well  within the kinematic  experiment domain of a  fixed target experiment such as SELEX at the Tevatron~\cite{Ocherashvili:2004hi}. 
Note that the intrinsic heavy-quark mechanism can account for many previous observations of forward heavy hadron production single and double $J/\psi$ production by pions observed  at high $x_F>0.4$ in the low energy fixed target NA3 experiment, the high $x_F$ production of $p p \to \Lambda_c, +X$ and $p p \to  \Lambda_b +X$  observed at the ISR; single and double $\Upsilon(b \bar b)$ production, as well as \emph{quadra-bottom} tetraquark 
$ [bb\bar b\bar b]$ production observed recently by  the AnDY experiment at RHIC~\cite{Bland:2019aha}.  In addition the EMC collaboration observed that the charm quark distribution in the proton  at $x=0.42$ and $Q^2 = 75$\,GeV$^2$ is 30 times larger that expected from DGLAP evolution.  All of these experimental observations are naturally explained by the intrinsic heavy quark  mechanism.
The SELEX observation~\cite{Ocherashvili:2004hi} of double charm baryons at high $x_F$ reflects production from double intrinsic heavy quark Fock states of the baryon projectile.  Similarly, the high $x_F$ domain -- 
which would be accessible at forward high $x_F$ -- is the natural production domain for heavy hadron production at the LHeC.

The production of heavy hadrons based on intrinsic heavy quark Fock states is thus remarkable efficient and greatly extends the kinematic domain of the LHeC, e.g.\ for processes such as $\gamma^* b \to Z^0 b$. 
This is in contrast with the standard production cross sections based on gluon splitting, where only a small 
fraction of the incident momentum is effective in creating heavy hadrons.

\subsubsection{Light-Front Holography and Superconformal Algebra}
%
The LHeC has the potential of probing the  high mass spectrum of QCD, such as the spectroscopy and structure of hadrons consisting of heavy quarks.
Insights into this new domain of hadron physics can now be derived by new non-perturbative  colour-confining methods based on light-front (LF) holography.
A remarkable feature is universal Regge trajectories with universal slopes in both the principal quantum number $n$ and internal orbital angular momentum $L$.
A key feature is di-quark clustering and supersymmetric relations between the masses of meson, baryons, and tetraquarks. In addition the running coupling is determined at all scales, including the soft domain relevant to rescattering corrections to LHeC processes.
The  combination  of  lightfront holography with superconformal algebra leads to the novel prediction that hadron physics  has  supersymmetric  properties  in  both  spectroscopy  and  dynamics. 

\paragraph{Light-front holography and recent theoretical advances }
Five-dimensional AdS$_5$ space provides a geometrical representation of the conformal group.  
Remarkably,  AdS$_5$  is holographically dual to $3+1$  spacetime at fixed LF time $\tau$~\cite{Brodsky:2014yha}.  
A colour-confining LF  equation for mesons of arbitrary spin $J$ can be derived
from the holographic mapping of  the \emph{soft-wall model} modification of AdS$_5$ space for the specific dilaton 
profile $e^{+\kappa^2 z^2}$, where $z$ is the  fifth dimension variable of  the five-dimensional AdS$_5$ space. 
A holographic dictionary  maps the fifth dimension $z$  to the LF radial variable $\zeta$,  with $\zeta^2  =  b^2_\perp(1-x)$.   
The same physics transformation maps the AdS$_5$  and $(3+1)$ LF expressions for electromagnetic and  gravitational form factors to each other~\cite{deTeramond:2013it}.

A key tool is the remarkable dAFF principle~\cite{deAlfaro:1976vlx} 
 which shows how a mass scale can appear in a Hamiltonian and its equations of motion while retaining the conformal symmetry of the action. 
When applying it to LF holography, a mass scale $\kappa$ appears which determines universal Regge slopes, and the hadron masses.
The  resulting \emph{LF Schr\"odinger Equation}   incorporates colour confinement and other 
essential spectroscopic and dynamical features of hadron physics, including Regge theory, the Veneziano formula~\cite{Veneziano:1968yb}, a massless pion for zero quark mass and linear Regge trajectories with the universal slope  in the radial quantum number $n$   and the internal  orbital angular momentum $L$.  
The combination of LF dynamics, its holographic mapping to AdS$_5$ space, and the 
dAFF procedure provides new insight into the physics underlying colour confinement, the 
non-perturbative QCD coupling, and the QCD mass scale.  
The $q \bar q$ mesons and their valence LFWFs are the eigensolutions of the frame-independent a relativistic bound-state LF Schr\"odinger equation. 

The mesonic $q\bar  q$ bound-state eigenvalues for massless quarks are $M^2(n, L, S) = 4\kappa^2(n+L +S/2)$.
This equation predicts that the pion eigenstate  $n=L=S=0$ is massless for zero quark mass. 
When quark masses are included in the LF kinetic energy $\sum_i  \frac{k^2_{\perp i} + m^2}{x_i}$,  the  spectroscopy of  mesons  are predicted correctly, with equal slope in the principal quantum number $n$ and the internal orbital angular momentum $L$.
A comprehensive review is given in  Ref.~\cite{Brodsky:2014yha}. \\

\paragraph{The QCD Running Coupling at all Scales from Light-Front Holography}
 
The QCD running coupling $\alpha_s(Q^2)$ sets the strength of  the interactions of quarks and gluons as a function of the momentum transfer $Q$ (see Sec.~\ref{sec:alphas}).
The dependence of the coupling $Q^2$ is needed to describe hadronic interactions at 
both long and short distances~\cite{Deur:2016tte}. 
It can be defined~\cite{Grunberg:1980ja} at all momentum scales from a 
perturbatively calculable observable, such as the coupling $\alpha_s^{g_1}(Q^2)$, which is defined 
using the Bjorken sum rule~\cite{Bjorken:1966jh}, and determined from the sum rule prediction at high $Q^2$ and, below, from its measurements~\cite{Deur:2004ti,Deur:2014vea,Deur:2008ej}.  
At high $Q^2$, such \emph{effective charges}  satisfy asymptotic freedom, obey the usual pQCD renormalisation group equations, and can be related to each other without scale ambiguity by commensurate scale relations~\cite{Brodsky:1994eh}.  

\begin{figure}
\centering
\includegraphics[width=0.65\textwidth]{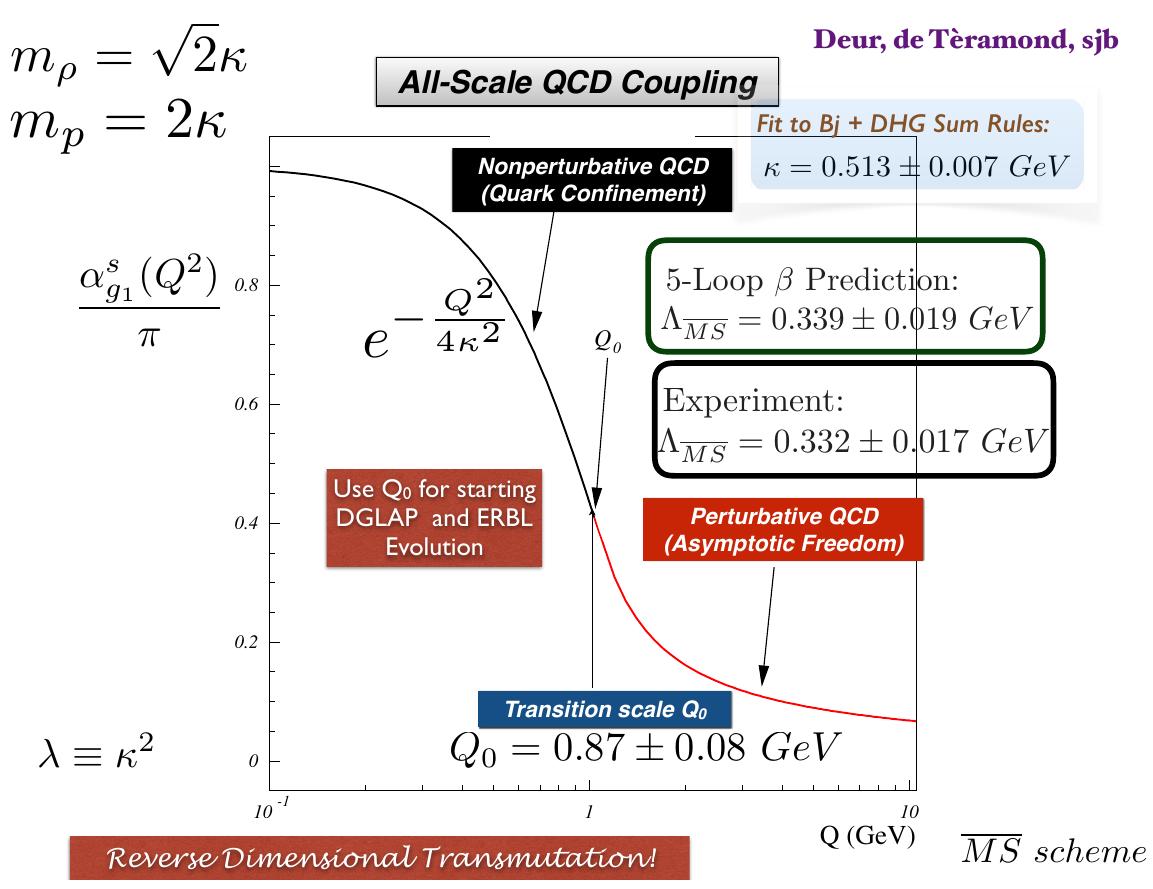}
\caption{
Prediction for the running coupling $\alpha_s^{g_1}(Q^2)$ at all scales. At lower \Qsq\ predictions are obtained from LF Holography and at higher \Qsq from perturbative QCD.   
The magnitude and derivative of the perturbative and non-perturbative coupling are matched at the scale $Q_0$.
This matching connects the perturbative scale 
$\Lambda_{\overline{MS}}$ to the non-perturbative scale $\kappa$ which underlies the hadron mass scale. 
\label{DeurCoupling}}
\end{figure}

The high $Q^2$ dependence of $\alpha_s^{g_1}(Q^2)$ is predicted by pQCD.
In the small $Q^2$ domain its functional behaviour can be predicted by the dilaton  $e^{+\kappa^2 z^2}$ soft-wall modification of the AdS$_5$ metric, together with LF holography~\cite{Brodsky:2010ur}, as ${\alpha_s^{g_1}(Q^2) = \pi e^{- Q^2 /4 \kappa^2 }}$. 
The parameter $\kappa$ determines the mass scale of  hadrons and Regge slopes in the zero quark mass limit, and it was shown that it can be connected to the  mass scale $\Lambda_s$, which controls the evolution of the pQCD coupling~\cite{Brodsky:2010ur,Deur:2014qfa,Brodsky:2014jia}.  
Measurements of  $\alpha_s^{g_1}(Q^2)$~\cite{Deur:2005cf,Deur:2008rf} are remarkably consistent  
with this predicted Gaussian form, and a fit gives $\kappa= 0.513 \pm 0.007$\,GeV, see Fig.~\ref{DeurCoupling}.

The matching of the high and low $Q^2$ regimes of $\alpha_s^{g_1}(Q^2)$
determines a scale $Q_0$, which
sets the interface between perturbative  and non-perturbative hadron dynamics.
This connection can be done for any choice of renormalisation scheme and one obtains an effective QCD coupling at all momenta.   
In the $\overline{\text{MS}}$ scheme one gets $Q_0 =0.87 \pm 0.08\,\GeV$~\cite{Brodsky:2019vqq}. 
The corresponding value of $\Lambda_{\overline{\text{MS}}}$ agrees well with the measured world average value and its value allows to compute hadron masses using the AdS/QCD superconformal predictions for hadron spectroscopy.
The value of $Q_0$ can further be used to set the factorisation scale for DGLAP evolution~\cite{Gribov:1972ri,Altarelli:1977zs,Dokshitzer:1977sg} or the ERBL evolution of distribution amplitudes~\cite{Lepage:1979zb,Efremov:1979qk}.
The use of  the scale $Q_0$  to  resolve  the factorisation scale uncertainty in structure functions and fragmentation  functions,  in combination with the scheme-independent \emph{principle of maximum conformality} 
(PMC)~\cite{Mojaza:2012mf} for  setting renormalisation scales,  can 
greatly improve the precision of pQCD predictions for collider phenomenology at LHeC and HL-LHC.

\paragraph{Superconformal Algebra and Hadron Physics with LHeC data}

If one generalises LF holography using \emph{superconformal algebra}
the resulting LF eigensolutions yield a unified Regge spectroscopy of mesons, baryons and tetraquarks, including remarkable supersymmetric relations between the masses of mesons and baryons of the same parity~\footnote{
QCD is not supersymmetrical in the usual sense, since the QCD Lagrangian is based on quark and gluonic fields, not squarks or gluinos. 
However, its hadronic eigensolutions conform to a representation of superconformal algebra, reflecting the underlying conformal symmetry of chiral QCD and its Pauli matrix representation.}~\cite{Brodsky:2013ar,Brodsky:2016nsn}.   
This generalisation further predicts hadron dynamics, including
 vector meson electroproduction,  hadronic LFWFs, distribution amplitudes, form factors, and valence structure functions~\cite{Sufian:2016hwn,deTeramond:2018ecg}.  
Applications to the deuteron elastic form factors and structure functions are given in Refs.~\cite{Gutsche:2015xva,Gutsche:2016lrz}

\begin{figure}
  \centering
  \includegraphics[width=0.65\textwidth]{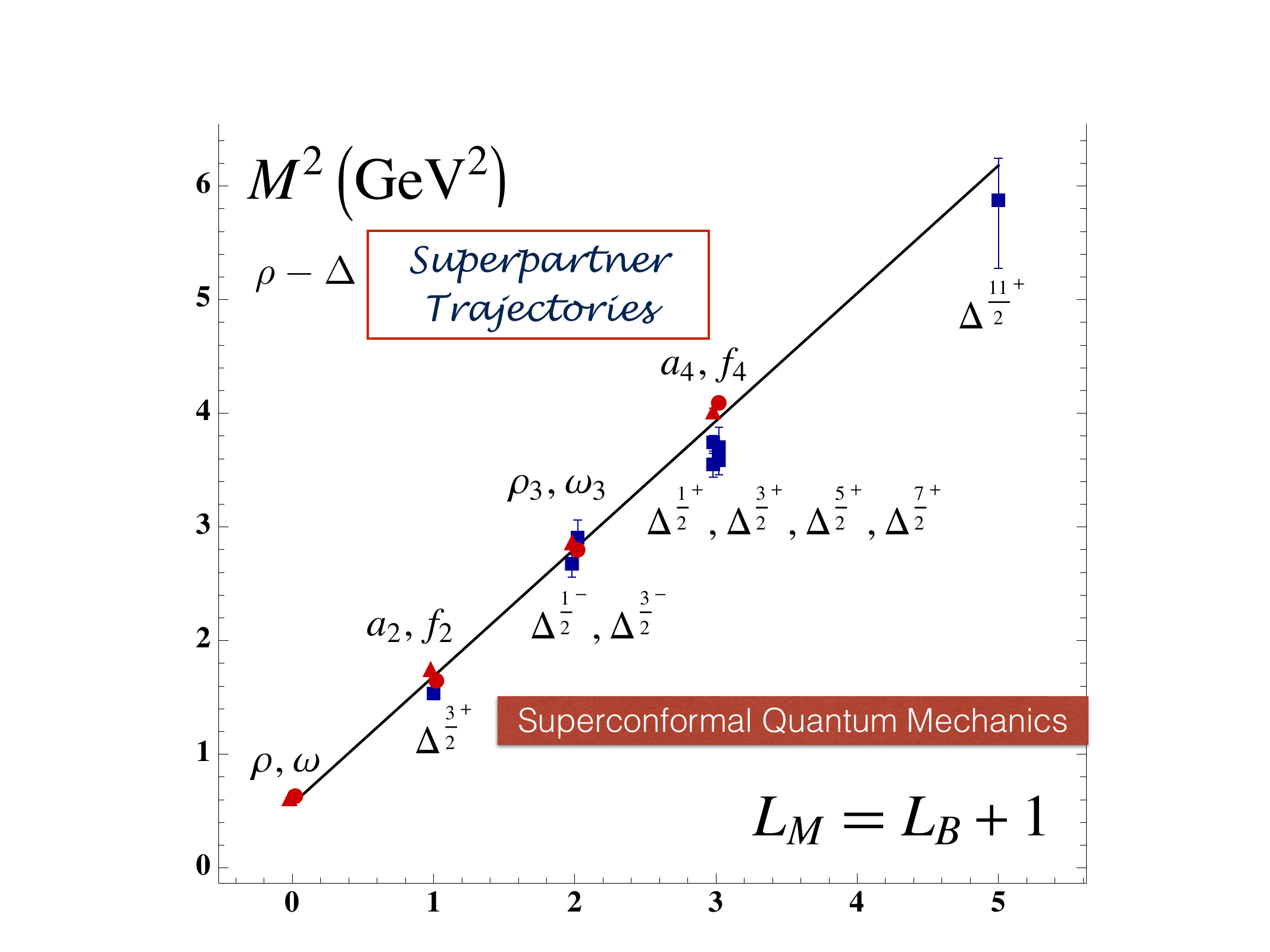}
  \caption{Comparison of the $\rho/\omega$ meson Regge trajectory with the $J=3/2$ $\Delta$  baryon trajectory. 
    Superconformal algebra  predicts the mass  degeneracy of the meson and baryon trajectories if one identifies a 
    meson with internal orbital angular momentum $L_M$ with its superpartner baryon with $L_M = L_B+1.$
    See Refs.~\cite{deTeramond:2014asa,Dosch:2015nwa}. 
    \label{NSTARFigB}}
\end{figure}

The eigensolutions of superconformal algebra predict the Regge spectroscopy of mesons, baryons, and tetraquarks of the same parity and twist as equal-mass members of the same 4-plet representation with  a universal Regge slope~\cite{Dosch:2015nwa, Brodsky:2016rvj, Nielsen:2018ytt}.
A comparison with experiment is shown in Fig.~\ref{NSTARFigB}.
The $q \bar q$ mesons with orbital angular momentum $L_M=L_B+1$ have the same mass as their baryonic partners with orbital angular momentum $L_B$~\cite{deTeramond:2014asa,Dosch:2015nwa}.

The predictions from LF holography and superconformal algebra can also be extended to mesons, baryons,  and tetraquarks with strange, charm and bottom quarks.  
Although conformal symmetry is strongly broken by the heavy quark masses, the basic underlying supersymmetric mechanism, which transforms mesons to baryons (and baryons to tetraquarks), still holds and gives remarkable mass degeneracy across the entire spectrum of light, heavy-light and double-heavy hadrons.

The 4-plet symmetry of quark-antiquark mesons, quark-diquark baryons, and diquark-antidiquark tetraquarks are important predictions by superconformal algebra~\cite{Brodsky:2013ar,Brodsky:2019vqq}.
Recently the AnDY experiment at RHIC  has reported the observation of a state at 18\,GeV which can be identified with the $[bb][\bar b \bar b]$ tetraquark~\cite{Bland:2019aha}.
The states with heavy quarks such as the $[bb][\bar b \bar b]$ tetraquark can be produced at the LHeC, especially at high $x_F$ along the proton beam direction.
New measurements at the LHeC are therefore inevitable to manifest the superconformal nature of hadronic bound states.

\chapter{Electroweak and Top Quark Physics}
        
\section{Electroweak Physics with Inclusive DIS data}
\label{sec:EW}
With the discovery of the Standard Model (SM) Higgs boson at the CERN
LHC experiments and subsequent measurements of its
properties, all fundamental parameters of the SM have now been measured
directly and with remarkable precision.
To further establish the validity of the theory of electroweak
interactions~\cite{Glashow:1961tr,Weinberg:1967tq,Weinberg:1971fb,Weinberg:1972tu,Salam:1964ry}, validate the mechanism of electroweak symmetry breaking
and the nature of the Higgs sector~\cite{Higgs:1964ia,Higgs:1964pj,Englert:1964et}, new
electroweak measurements have to be performed at highest precision.
Such high-precision measurements can be considered as a portal
to new physics, since non-SM contributions, as for instance
loop-insertions, may cause significant deviations for some precisely
measurable and calculable observables. At the LHeC, the greatly 
enlarged kinematic reach to higher mass scales in comparison to
HERA~\cite{Aktas:2005iv,Abramowicz:2016ztw,Spiesberger:2018vki} and the large targeted luminosity will enable electroweak 
measurements in $ep$ scattering with higher precision than ever 
before.

In this section the sensitivity of inclusive DIS cross
sections to electroweak parameters is discussed.
An extended analysis and a more comprehensive discussion is found in Ref.~\cite{Britzger:2020kgg}, and
some aspects are described in the following.
The direct production of $W$ and $Z$ bosons is discussed in the subsequent section.

\subsection{Electroweak effects in inclusive NC and CC DIS cross sections}
\newcommand{\ad} {\ensuremath{g_A^d}}
\newcommand{\vd} {\ensuremath{g_V^d}}
\newcommand{\au} {\ensuremath{g_A^u}}
\newcommand{\vu} {\ensuremath{g_V^u}}
\newcommand{\aq} {\ensuremath{g_A^q}}
\newcommand{\vq} {\ensuremath{g_V^q}}
\newcommand{\gae}{\ensuremath{g_A^e}}
\newcommand{\ve} {\ensuremath{g_V^e}}
\newcommand{\sw}{\ensuremath{\text{sin}^2\theta_\text{W}}}
\newcommand{\dr}{\ensuremath{\Delta r}}
\renewcommand{\mW}{\mw}
\newcommand{\mZ}{\mz}
Electroweak NC interactions in inclusive $e^\pm p$ DIS are mediated 
by exchange of a virtual photon $(\gamma)$ or a $Z$ boson in the
$t$-channel, while CC DIS is mediated exclusively
by $W$-boson exchange as a purely \emph{weak} process.  
Inclusive NC DIS cross sections are expressed in terms of
generalised structure functions $\tilde{F}_2^\pm$, $x\tilde{F}_3^\pm$
and $\tilde{F}_\text{L}^\pm$ at EW leading order (LO)
as
\begin{equation}
  \frac{d^2\sigma^\text{NC}(e^\pm p)}{dxd\Qsq} = \frac{2\pi\alpha^2}{xQ^4}\left[Y_+\tilde{F}_2^\pm(x,\Qsq) \mp Y_{-}  x\tilde{F}_3^\pm(x,\Qsq) - y^2 \tilde{F}_{\textrm\
 L}^\pm(x,\Qsq)\right]~,
  \label{eq:cs}
\end{equation}
where $\alpha$ denotes the fine structure constant.
The terms $Y_\pm = 1\pm(1-y)^2$, with $y=\Qsq/sx$, describe the helicity dependence of
the process.
The  generalised structure functions are separated into contributions
from pure $\gamma$- and $Z$-exchange and their interference~\cite{Tanabashi:2018oca,Klein:1983vs}:
\begin{align}
  \tilde{F}_2^\pm
  &= F_2
  -(\ve\pm P_e\gae)\kappa_ZF_2^{\gamma Z}
  +\left[(\ve\ve+\gae\gae)\pm2P_e\ve\gae\right]\kappa_Z^2F_2^Z~,
  \\
  \tilde{F}_3^\pm
  &=~~
  -(\gae\pm P_e\ve)\kappa_ZF_3^{\gamma Z}
  +\left[2\ve\gae\pm P_e(\ve\ve+\gae\gae)\right]\kappa_Z^2F_3^Z~.
  \label{eq:strfun}
\end{align}
Similar expressions hold for $\tilde{F}_L$.
In the naive quark-parton model, which corresponds to the LO QCD
approximation, the structure functions are calculated as
\begin{align}
  \left[F_2,F_2^{\gamma Z},F_2^Z\right]
  &= x\sum_q\left[Q_q^2,2Q_q\vq,\vq\vq+\aq\aq \right]\{q+\bar{q}\}~,
  \label{eq:last1}
  \\
  x\left[F_3^{\gamma Z},F_3^Z\right]
  &= x\sum_q\left[2Q_q\aq,2\vq\aq\right]\{q-\bar{q}\}~,
  \label{eq:last2}
\end{align}
representing two independent combinations of the quark and anti-quark
momentum distributions, $xq$ and $x\bar{q}$. 
In Eq.\,\eqref{eq:strfun}, the quantities $g_V^{f}$ and $g_A^{f}$ 
stand for the vector and axial-vector couplings of a fermion 
($f = e$ or $f = q$ for electron or quark) to the $Z$ boson, 
and the coefficient $\kappa_Z$ accounts for the $Z$-boson 
propagator including the normalisation of the weak couplings.
Both parameters are fully calculable from the electroweak theory.
The (effective) coupling parameters depend on the electric charge, 
$Q_{f}$ and the third component of the weak-isospin, 
$I^3_{\text{L},f}$. Using $\sw=1-\tfrac{\mw^2 }{\mz^2}$, one can 
write 
\begin{alignat}{2}
  g_V^{f} &= \sqrt{\rho_{\text{NC}, f}} \left(I^3_{\text{L},f} - 2
  Q_{f} \kappa_{\text{NC}, f}\ \sw \right)\,, ~~ && \text{and}\\
  g_A^{f} &= \sqrt{\rho_{\text{NC}, f}} \, I^3_{\text{L},f}\, && \text{with }f=(e,u,d)\,.
  \label{eq:gV-LO}
\end{alignat}
%
The parameters $\rho_{\text{NC}, f}$ and $\kappa_{\text{NC}, f}$ 
are calculated as real parts of complex form factors which include 
the higher-order loop
corrections~\cite{Bohm:1986na,Bardin:1988by,Hollik:1992bz}.
They contain non-leading flavour-specific components.

Predictions for CC DIS are written in terms of the CC structure
functions $W_2$, $xW_3$ and $W_L$ and higher-order electroweak effects
are collected in two form factors $\rho_{\text{CC},eq}$ and
$\rho_{\text{CC},e\bar{q}}$~\cite{Bohm:1987cg,Bardin:1989vz}.

In this study, the on-shell scheme is adopted for the calculation 
of higher-order corrections. This means that the independent 
parameters are chosen as the fine structure constant $\alpha$
and the masses of the weak bosons, the Higgs boson and the fermions.
The weak mixing angle is then fixed and \gf\ is a prediction,
whose higher-order corrections are included in the well-known 
correction factor
$\Delta r$~\cite{Sirlin:1980nh,Bohm:1986rj,Hollik:1988ii}
(see discussion of further contributions in Ref.~\cite{Tanabashi:2018oca}).

\begin{figure}[!th]
  \centering
  \includegraphics[width=0.55\textwidth]{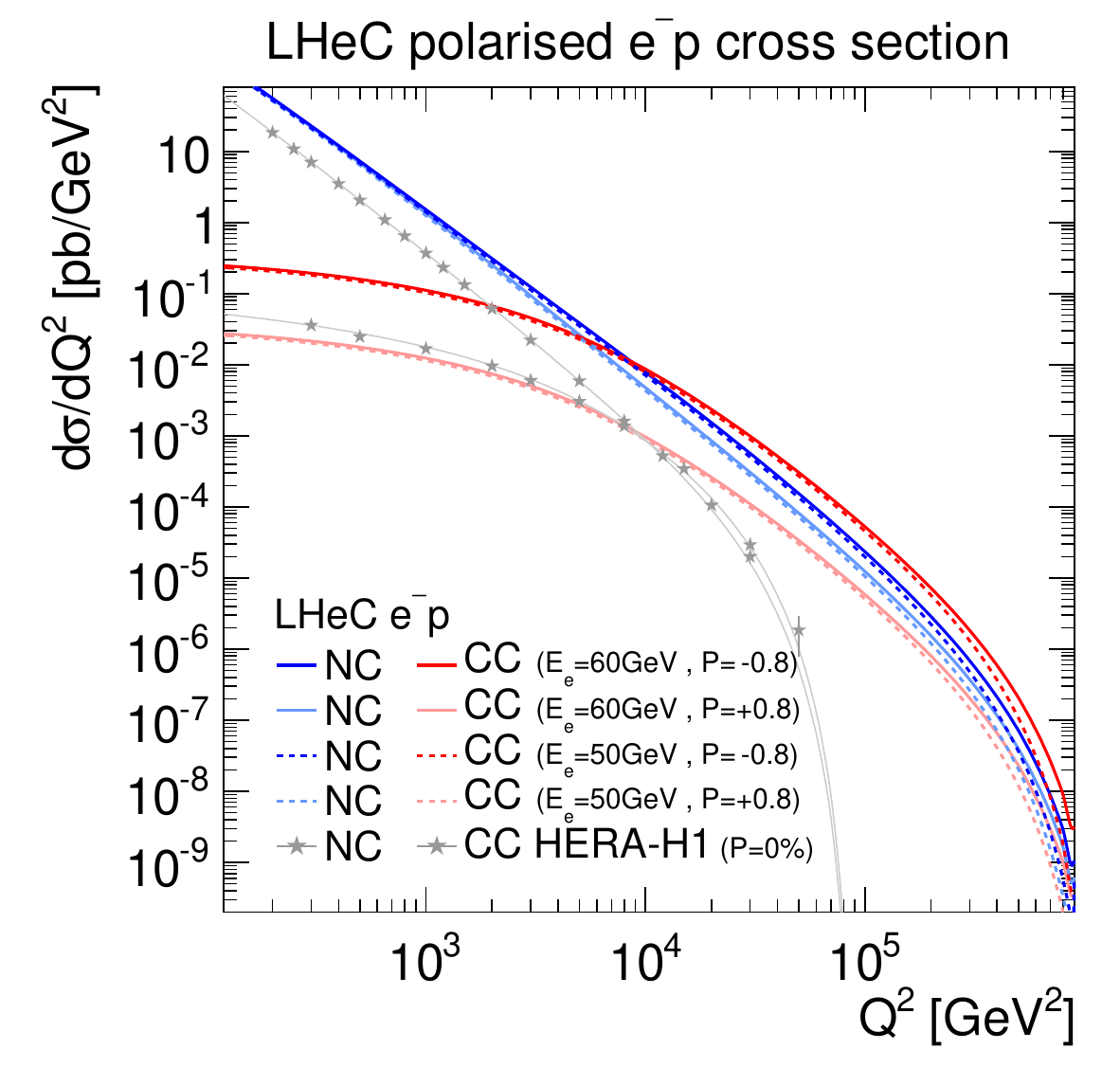}
  \caption{Single differential cross sections for polarised $e^-p$ NC
      and CC DIS at LHeC for two different electron beam energies
      ($E_e$). Cross sections for longitudinal electron beam
      polarisations of $P_e=-0.8$ and $+0.8$ are displayed.
      For comparison also measurements at centre-of-mass energies
      of $\sqrt{s}=920\,\GeV$ by H1 at HERA for unpolarised ($P_e=0\,\%$)
      electron beams are displayed~\cite{Aaron:2012qi}.
  }
  \label{fig:dSigma}
\end{figure}
The predicted single-differential inclusive NC and CC DIS cross sections
for polarised $e^-p$ scattering as a function of \Qsq are displayed in
Fig.~\ref{fig:dSigma}.
For NC DIS and at higher \Qsq, electroweak effects are important
through $\gamma Z$ interference and pure $Z$-exchange terms and the 
polarisation of the LHeC electron beam of $P_e=\pm0.8$ will 
considerably alter the cross sections.
For CC DIS, the cross section scales linearly with $P_e$.
Two different electron beam energies are displayed in
Fig.~\ref{fig:dSigma}, and albeit the impact of a reduction from
$E_e=60$ to $50\,\GeV$ appears to be small,
a larger electron beam energy would yield
higher precision for the measurement of electroweak parameters, 
since these are predominantly sensitive to the cross sections at 
highest scales, as will be shown in the following.

\subsection{Methodology of a combined EW and QCD fit}
A complete electroweak analysis of DIS data has to consider PDFs
together with electroweak parameters~\cite{Britzger:2017fuc}.
In this study, the uncertainties of electroweak parameters are
obtained in a combined fit of electroweak parameters and the PDFs,
and the inclusive NC and CC DIS pseudodata (see
Sec.~\ref{sec:pseudo_data}) are explored as input data.
The PDFs are parameterised with 13 parameters at a starting scale
$Q^2_0$ and NNLO DGLAP evolution is applied~\cite{Botje:2010ay,Botje:2016wbq}.
In this way, uncertainties from the PDFs are taken into
account, which is very reasonable, since the PDFs will predominantly
be determined from those LHeC data in the future.
The details of the PDF fit are altogether fairly similar to the PDF
fits outlined in Chapter.~\ref{chapter:pdf}.
Noteworthy differences are that additionally EW effects are
included into the calculation by considering the full set of 1-loop
electroweak corrections~\cite{Spiesberger:1995pr}, and
the $\chi^2$ quantity~\cite{Andreev:2014wwa}, which is input to the minimisation and
error propagation, is based on normal-distributed relative
uncertainties. In this way, a dependence on the actual size of 
the simulated cross sections is avoided.
The size of the pseudodata are therefore set equivalent to the
predictions~\cite{Cowan:2010js}. 

\subsection[Weak boson masses ${M_W}$ and $M_Z$]{\boldmath Weak boson masses ${M_W}$ and $M_Z$}
The expected uncertainties for a determination of the weak boson
masses, \mW\ and \mZ, are determined in the PDF+EW-fit, where one of
the masses is determined  together with the PDFs, while the other 
mass parameter is taken as external input. The expected 
uncertainties for \mW\ are 
\begin{alignat}{2}
  \Delta\mW(\text{LHeC-60})&=\pm8_{(\textrm{exp})}\pm5_{(\textrm{PDF})}\,\MeV  =\,&& 10_{\textrm{(tot)}}\,\MeV\textrm{~~and~~} \\
  \Delta\mW(\text{LHeC-50})&=\pm9_{(\textrm{exp})}\pm8_{(\textrm{PDF})}\,\MeV  =\,&& 12_{\textrm{(tot)}}\,\MeV
\nonumber
\end{alignat}
for LHeC with $E_e=60\,\GeV$ or 50\,\GeV, respectively.
The breakdown into experimental and PDF uncertainties is obtained by repeating the fit with PDF
parameters fixed.
\begin{figure}[!th]
    \centering
    \includegraphics[width=0.42\textwidth]{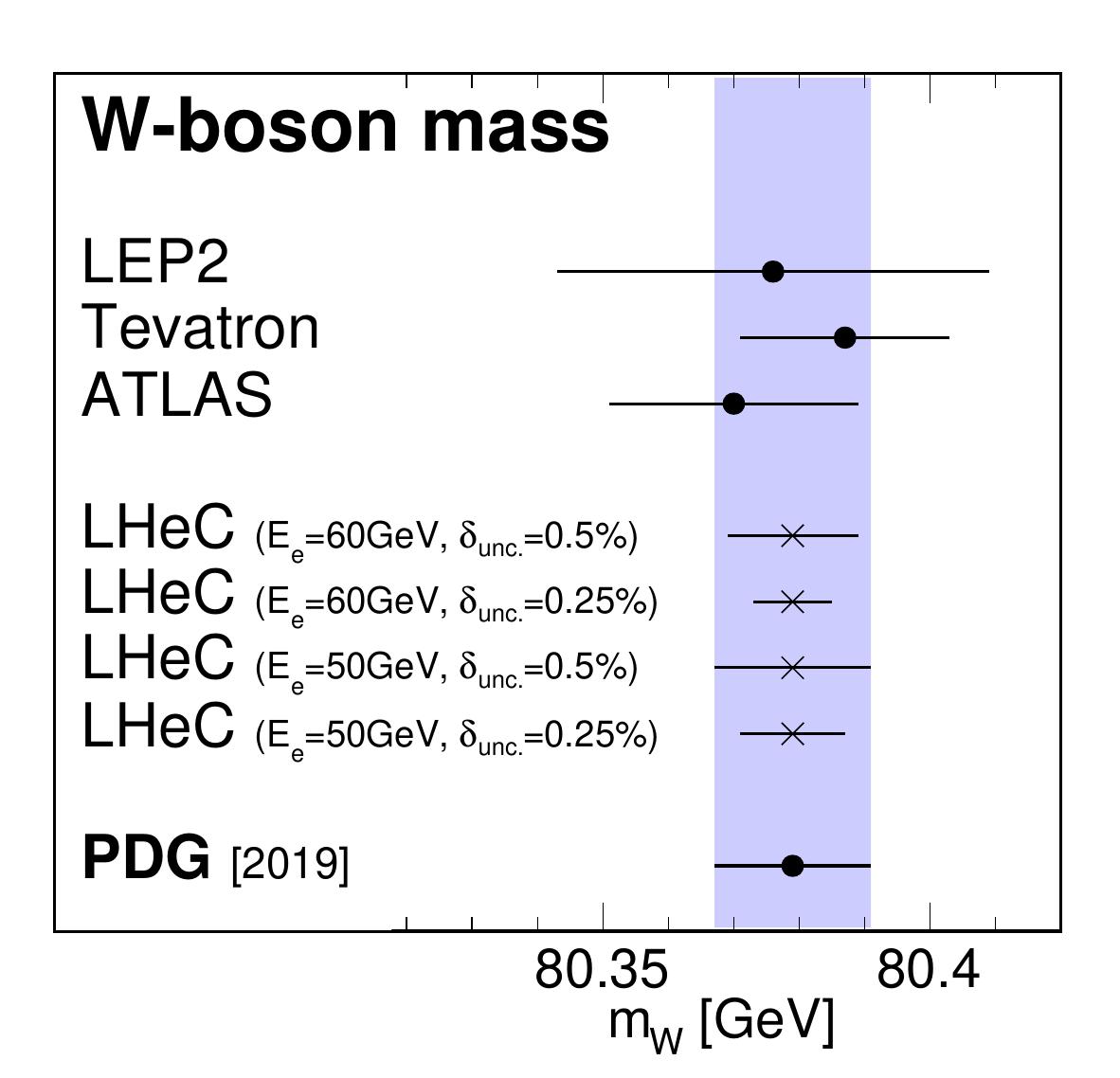}
    \hspace{0.02\textwidth}
    \includegraphics[width=0.42\textwidth]{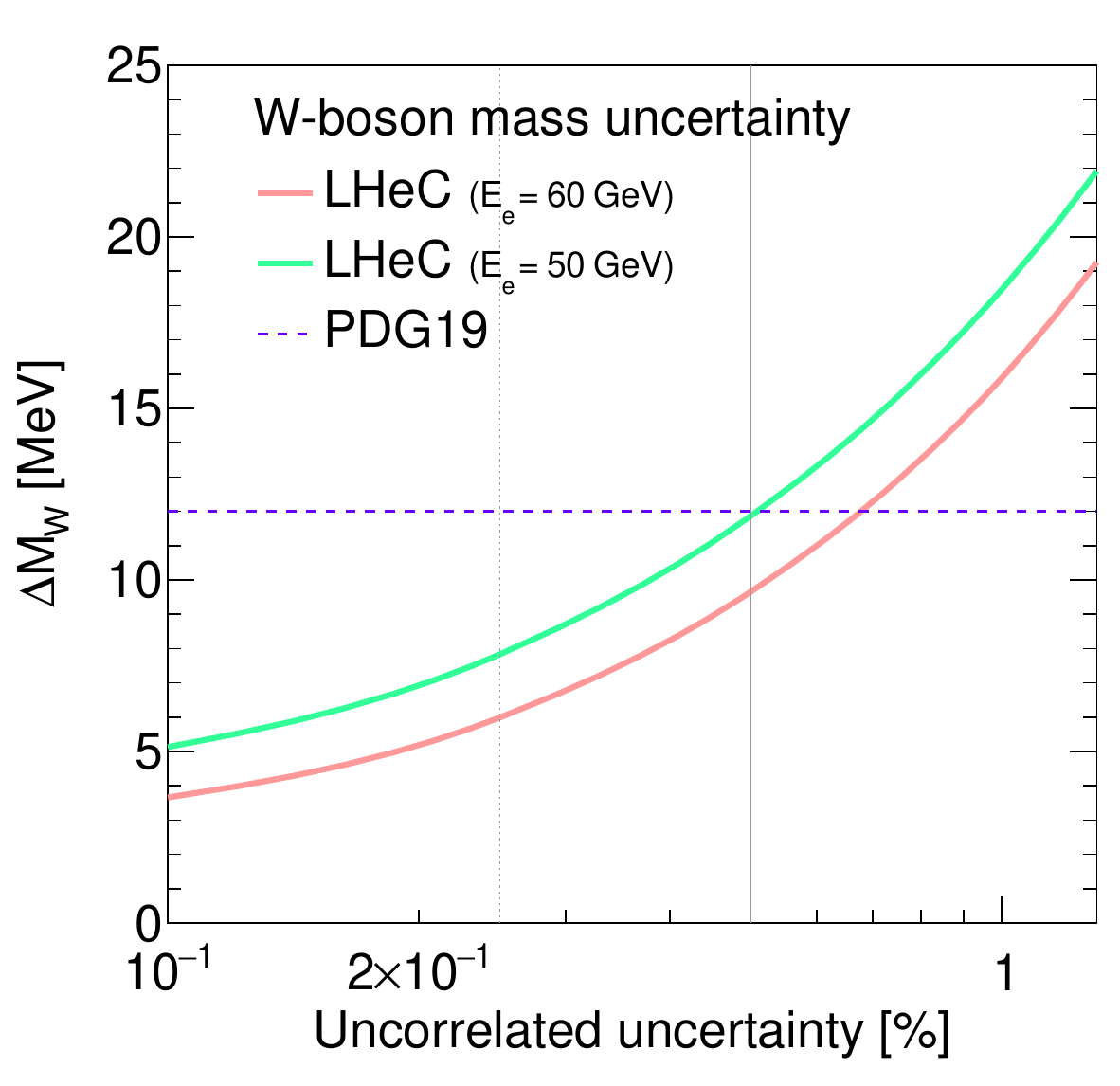}
    \caption{
      Left: Measurements of the $W$-boson mass assuming fixed 
      values for the top-quark
      and $Z$-boson masses at the LHeC for different 
      scenarios in comparison with today's
      measurements~\cite{Group:2012gb,Schael:2013ita,Aaboud:2017svj}
      and the  world average value (PDG19)~\cite{Tanabashi:2019}.
      For LHeC, prospects for $E_e=60\,\GeV$ and 50\,\GeV are
      displayed, as well as results for the two scenarios with 0.5\,\%
      or 0.25\,\% uncorrelated uncertainty (see text).
      Right: Comparison of the precision for $M_W$ for different assumptions of the uncorrelated
      uncertainty of the pseudodata.
      The uncertainty of the world average value is displayed as horizontal line.
      The nominal (and alternative) size of the uncorrelated uncertainty of the
      inclusive NC/CC DIS pseudodata is indicated by the vertical
      line (see text).
    }
    \label{fig:mWZ}
\end{figure}
These uncertainties are displayed in
Fig.~\ref{fig:mWZ} and compared to the values obtained by LEP2~\cite{Schael:2013ita},
Tevatron~\cite{Group:2012gb}, ATLAS~\cite{Aaboud:2017svj} and the PDG
value~\cite{Tanabashi:2019}. The LHeC measurement will become the most precise 
measurement from one single experiment and will greatly improve 
over the best measurement achieved
by H1, which was $\mW(\text{H1})=80.520\pm0.115\,\GeV$~\cite{Spiesberger:2018vki}.
%
If the dominating uncorrelated uncertainties can be reduced from
the prospected $0.5\,\%$ to $0.25\,\%$~\footnote{Due to performance
  reasons, the pseudodata are generated for a rather coarse grid. 
  With a
  binning which is closely related to the resolution of the LHeC
  detector, much finer grids in $x$ and \Qsq are feasible. Already
  such a change would alter the uncertainties of the fit parameters.
  However, such an effect can be reflected by a
  changed uncorrelated uncertainty, and a value of 0.25\,\% appears
  like an optimistic, but achievable, alternative scenario. },
a precision for \mW\ of up to
\begin{alignat}{2}
  \Delta\mW(\text{LHeC-60})&=\pm5_{(\text{exp})}\pm3_{(\text{PDF})}\,\MeV =\,&& 6_{\text{(tot)}}\,\MeV\text{~~and~~} \\
  \Delta\mW(\text{LHeC-50})&=\pm6_{(\text{exp})}\pm6_{(\text{PDF})}\,\MeV =\,&& 8_{\text{(tot)}}\,\MeV
\nonumber
\end{alignat}
for LHeC-60 and LHeC-50 may be achieved, respectively.
A complete dependence of the expected total experimental uncertainty 
$\Delta\mW$ on the size of the uncorrelated
uncertainty component 
is displayed in Fig.~\ref{fig:mWZ}, and with a more optimistic 
scenario an uncertainty of up to $\Delta\mW\approx5\,\MeV$ can 
be achieved. In view of such a high accuracy, it will be important 
to study carefully theoretical uncertainties. For instance the 
parameteric uncertainty due to the dependence on the top-quark mass 
of $0.5\,\GeV$ will yield an additional error 
of $\Delta\mW=2.5\,\MeV$. Also higher-order corrections, at least 
the dominating 2-loop corrections in DIS will have to be studied and 
kept under control. Then, the prospected determination of the 
$W$-boson mass from LHeC data will be among the most precise 
determinations and significantly improve the world average value 
of \mw. It will also become competitive with its prediction from 
global EW fits with present uncertainties of about
$\Delta\mw=7\,\MeV$~\cite{Tanabashi:2019,deBlas:2016ojx,Haller:2018nnx}.

While the determination of \mW\ from LHeC data is competitive 
with other measurements, the experimental uncertainties of a 
determination of \mZ\ are estimated to be about 11\,MeV and 
13\,MeV for LHeC-60 and LHeC-50, respectively.
Therefore, the precision of the determination of \mZ\ at LHeC cannot compete
with the precise measurements at the $Z$-pole by LEP+SLD and future
$e^+e^-$ colliders may even improve on that.

\begin{figure}
    \centering
    \includegraphics[width=0.4\textwidth,trim={0 20 0 25 },clip]{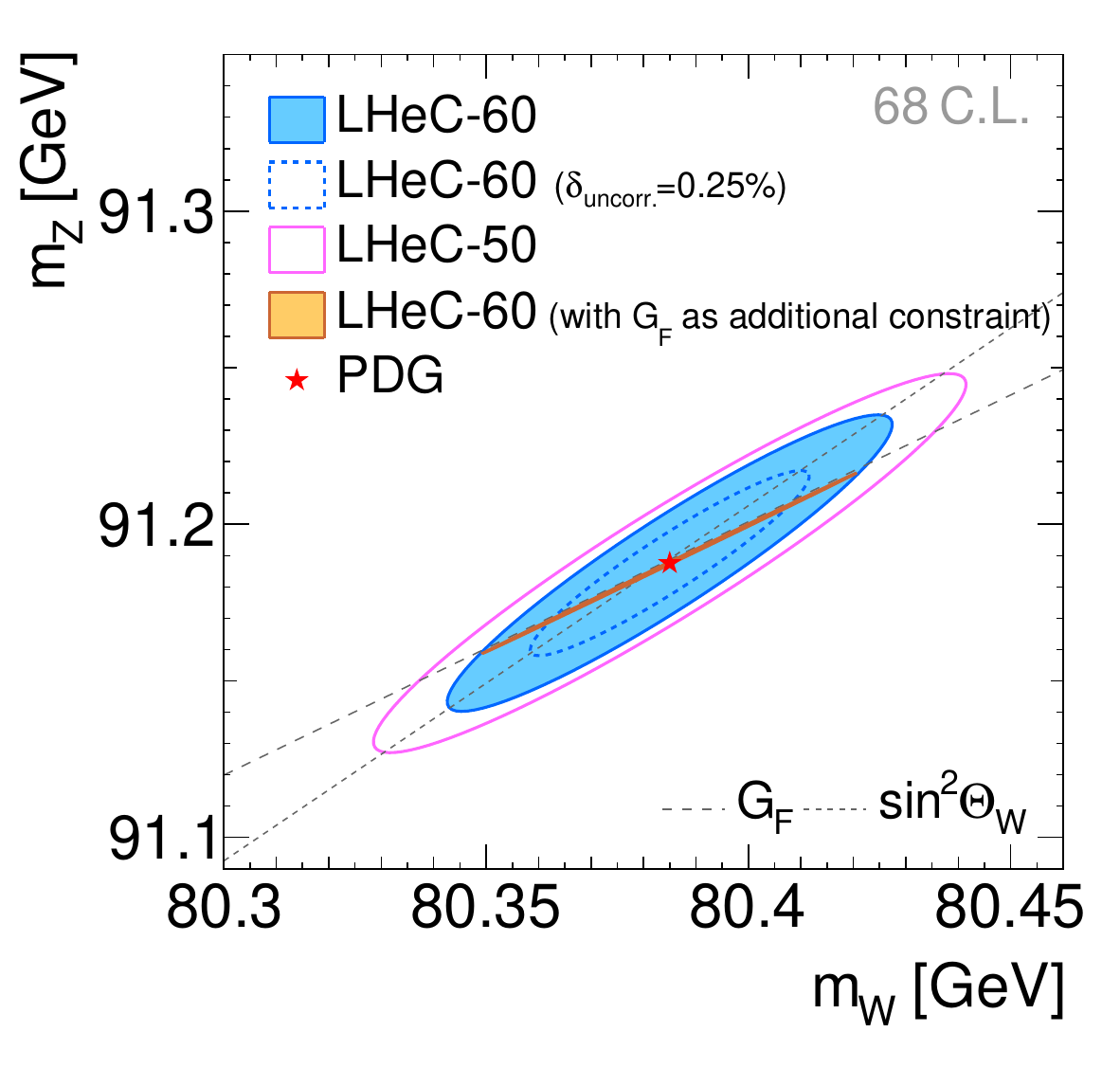}
    \hspace{0.02\textwidth}
    \includegraphics[width=0.4\textwidth,trim={0 20 0 25 },clip]{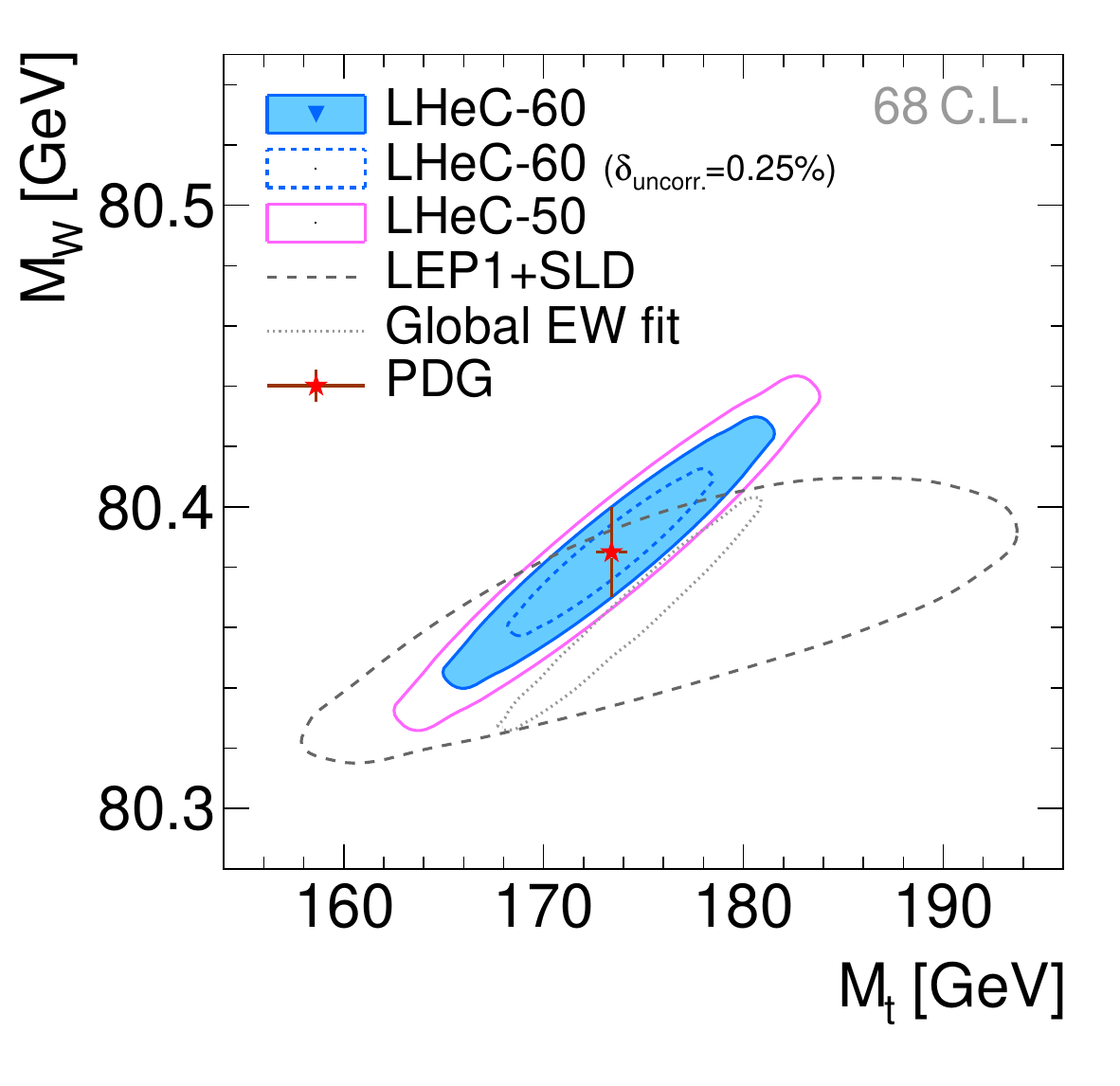}
    \caption{
      Simultaneous determination of the top-quark mass $M_t$ and
      $W$-boson mass \mw\ from LHeC-60 or LHeC-50 data (left).
      Simultaneous determination of the $W$-boson and $Z$-boson masses
      from LHeC-60 or LHeC-50 data (right).
    }
    \label{fig:mWmt}
\end{figure}

A simultaneous determination of \mW\ and \mZ\ is displayed in
Fig.~\ref{fig:mWmt} (left).
Although the precision of these two mass parameters is only moderate, 
a meaningful test of the high-energy behaviour of electroweak theory 
is obtained by using \gf\ as additional input:
The high precision of the \gf\ measurement~\cite{Tishchenko:2012ie} yields a very shallow
error ellipse and a precise test of the SM can be performed with only
NC and CC DIS cross sections alone.
Such a fit determines and simultaneously tests the
high-energy behaviour of electroweak theory, while using only low-energy parameters
$\alpha$ and \gf\ as input (plus values for masses like 
$M_t$ and $M_H$ needed for loop corrections).

\subsection{Further mass determinations}
Inclusive DIS data are sensitive to the top-quark mass $M_t$ 
indirectly through radiative corrections. $M_t$-dependent terms 
are dominantly due to corrections from the gauge boson self-energy 
corrections. They are contained in the $\rho$ and $\kappa$ parameters 
and in the correction factor $\Delta r$. The leading contributions 
are proportional to $M_t^2$. 
This allows for an indirect determination of the top-quark mass using
LHeC inclusive DIS data, and a determination of $M_t$ will yield an
uncertainty of $\Delta M_t = 1.8\,\GeV$ to 2.2\,\GeV. 
Assuming an uncorrelated uncertainty of the DIS data of
$0.25\,\%$ the uncertainty of $M_t$ becomes as small as
\begin{equation}
  \Delta M_t=1.1 \text{~~to~~} 1.4\,\GeV
\end{equation}
for 60 and 50\,\GeV electron beams, respectively.
This would
represent a very precise
indirect determination of the top-quark mass from purely electroweak
corrections and thus being fully complementary to measurements
based on real $t$-quark production, which often suffer from sizeable 
QCD corrections.
The precision achievable in this way will be competitive with 
indirect determinations from global EW fits after the 
HL-LHC~\cite{Schott:2019talk}.

More generally, and to some extent depending on the choice of the
renormalisation scheme, the leading self-energy corrections
are proportional to $\tfrac{M_t^2}{\mw^2}$ and thus a simultaneous
determination of $M_t$ and \mw is desirable.
The prospects for a simultaneous determination of $M_t$ and \mw is
displayed in Fig.~\ref{fig:mWmt} (right). It is remarkable that the 
precision of the LHeC is superior to that of the LEP+SLD 
combination~\cite{ALEPH:2005ab}. In an optimistic
scenario an uncertainty similar to the global electroweak
fit~\cite{Haller:2018nnx} can be achieved. In a fit without PDF 
parameters similar uncertainties are found (not shown),
which illustrates that the determination of EW parameters is
to a large extent independent of the QCD phenomenology and the PDFs. 

The subleading contributions to self-energy corrections
have a Higgs-boson mass dependence and are proportional to
$\log\tfrac{M^2_H}{\mw^2}$. When fixing all other EW parameters 
the Higgs boson mass could be constrained indirectly through 
these loop corrections with an experimental uncertainty of 
$\Delta m_H=^{+29}_{-23}$ to $^{+24}_{-20}\,\GeV$
for different LHeC scenarios, which is again similar to the indirect
constraints from a global electroweak fit~\cite{Haller:2018nnx},
but not competitive with direct measurements.

\subsection{Weak Neutral Current Couplings}
The vector and axial-vector couplings of up-type and
down-type quarks to the $Z$, $g_V^{q}$ and $g_A^{q}$, see
Eq.\,\eqref{eq:gV-LO}, are determined
in a fit of the four coupling parameters together with the PDFs.
\begin{table}[ht]
%
  \centering
  \small
  \begin{tabular}{cr@{\hskip4pt}lccc}
    \toprule
    Coupling  &  \multicolumn{2}{c}{PDG} & \multicolumn{3}{c}{Expected uncertainties} \\
    \cmidrule(lr){4-6}
    parameter &  \multicolumn{2}{c}{value}   & LHeC-60 & LHeC-60 ({\footnotesize$\delta_\text{uncor.}$=$0.25\,\%$})  & LHeC-50 \\
    \midrule
    $\au$  & $0.50$    & $^{+0.04}_{-0.05}$   & $0.0022$ & $0.0015$ & $0.0035$ \\
    $\ad$  & $-0.514$  & $^{+0.050}_{-0.029}$ & $0.0055$ & $0.0034$ & $0.0083$ \\
    $\vu$  & $0.18$    & $\pm0.05$            & $0.0015$ & $0.0010$ & $0.0028$ \\
    $\vd$  & $-0.35$   & $^{+0.05}_{-0.06}$   & $0.0046$ & $0.0027$ & $0.0067$ \\
    \bottomrule
  \end{tabular}
  \caption{
    Light-quark weak NC couplings ($\au$,$\ad$,$\vu$,$\vd$) 
    and their currently most precise values from the PDG~\cite{Tanabashi:2019} 
    compared with the prospected
    uncertainties for different LHeC scenarios.
    The LHeC prospects are obtained in a simultaneous fit of the PDF
    parameters and all four coupling parameters determined at a time.
  }
  \label{tab:couplings}
\end{table}

\begin{figure}
    \centering
    \includegraphics[width=0.42\textwidth,trim={0 10 0 10 },clip]{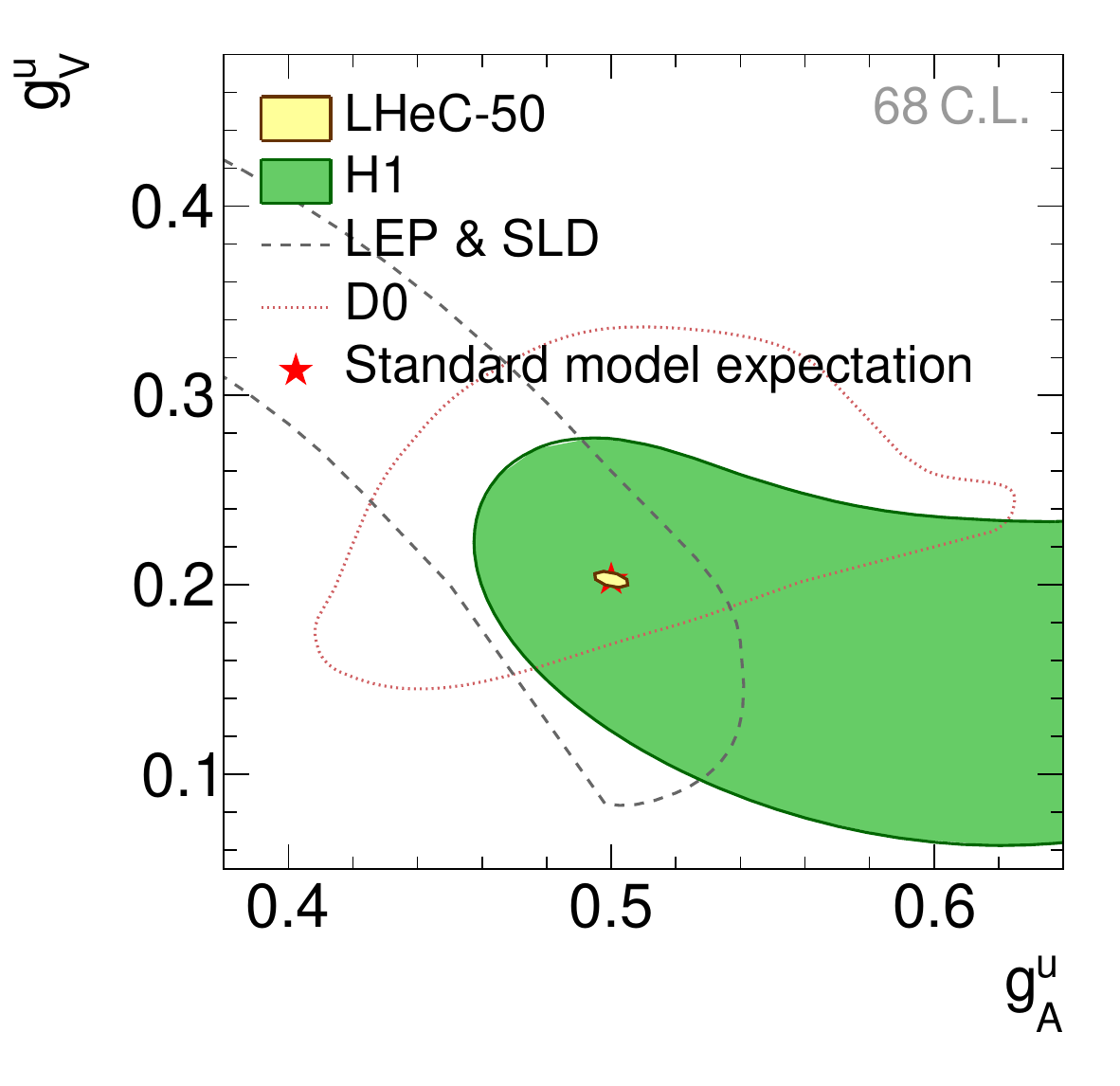}
    \hspace{0.02\textwidth}
    \includegraphics[width=0.42\textwidth,trim={0 10 0 10 },clip]{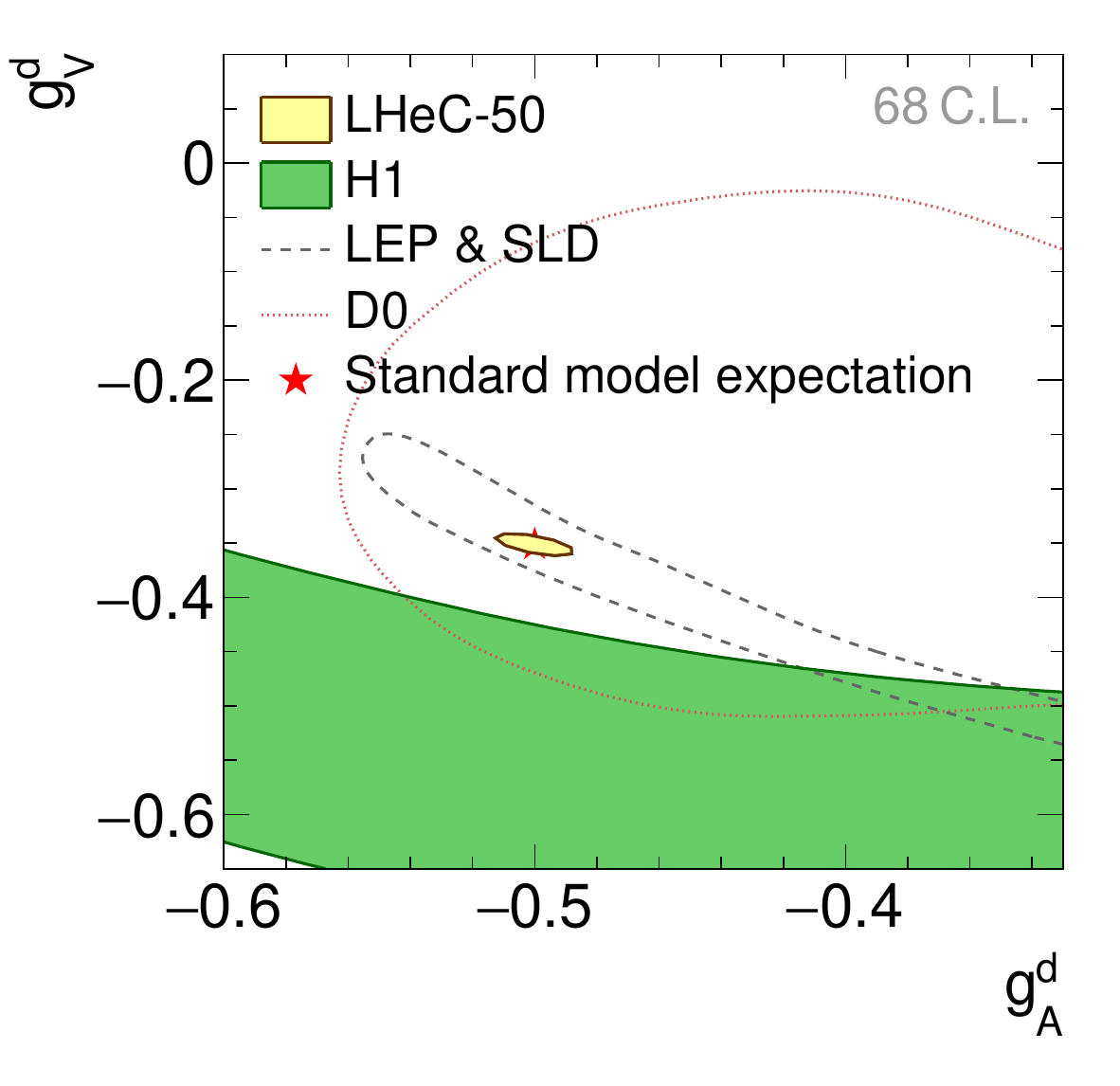}
  \caption{
    Weak NC vector and axial-vector couplings of $u$-type (left) 
    and $d$-type quarks (right) at 68\,\% confidence level~(C.L.) 
    for simulated LHeC data with $E_e=50\,\GeV$.
    The LHeC expectation is compared with results from the 
    combined LEP+SLD experiments~\cite{ALEPH:2005ab}, 
    a single measurement from D0~\cite{Abazov:2011ws} and one from  
    H1~\cite{Spiesberger:2018vki}.
    The standard model expectations are diplayed by a red star, partially hidden by the LHeC prospects.
  }
  \label{fig:couplings}
\end{figure}
The resulting uncertainties are collected in 
Tab.~\ref{tab:couplings}.
The two-dimensional uncertainty contours at 68\,\% confidence 
level obtained from LHeC data with $E_e=50\,\GeV$ are
displayed in Fig.~\ref{fig:couplings} for the two quark families and
compared with available measurements.
While all the current determinations from $e^+e^-$, $ep$ or $p\bar{p}$
data have a similar precision,
the future LHeC data will greatly improve the
precision of the weak neutral-current couplings and
expected uncertainties are an order of magnitude smaller than the
currently most precise ones~\cite{Tanabashi:2019}.
An increased electron beam energy of $E_e=60\,\GeV$ or improved
experimental uncertainties would further improve this measurement.

The determination of the couplings of the electron to the $Z$ boson,
\ve and \gae, can be determined at the LHeC with uncertainties of up to
$\Delta\ve=0.0013$ and $\Delta\gae=\pm0.0009$, which is similar to the
results of a single LEP experiment and about a factor three larger
than the LEP+SLD combination~\cite{ALEPH:2005ab}.

\subsection[The neutral current $\rho_\text{NC}$ and $\kappa_\text{NC}$ parameters]{\boldmath The neutral-current $\rho_\text{NC}$ and $\kappa_\text{NC}$ parameters}
Beyond Born approximation, the weak couplings are subject to 
higher-order loop corrections. These corrections are commonly 
parameterised by quantities called $\rho_\text{NC}$, 
$\kappa_\text{NC}$ and $\rho_\text{CC}$. They are sensitive to 
contributions beyond the SM and the structure of the Higgs sector.
It is important to keep in mind that these effective coupling 
parameters depend on the momentum transfer and are, indeed, 
form factors rather than constants. 
It is particularly interesting to investigate the so-called 
effective weak mixing angle defined as
$\sin^2\theta_{\text{W}}^{\text{eff}}=\kappa_{\text{NC}}\sw$. 
At the $Z$-pole it is well accessible through asymmetry measurements 
in $e^+e^-$ collisions. In DIS at the LHeC, the scale dependence 
of the effective weak mixing angle is not negligible. 
It can be determined only
together with the $\rho$ parameter due to the \Qsq\ dependence and the
presence of the photon exchange terms. Therefore, we introduce 
(multiplicative) anomalous contributions to these factors, denoted 
as $\rho_{\text{NC,CC}}^\prime$ and $\kappa^\prime_\text{NC}$, and test their
agreement with unity (for more details see
Ref.~\cite{Spiesberger:2018vki}), and uncertainties of these
parameters are obtained in a fit together with the PDFs.
\begin{figure}
    \centering
    \includegraphics[width=0.31\textwidth,trim={0 0 18 20 },clip]{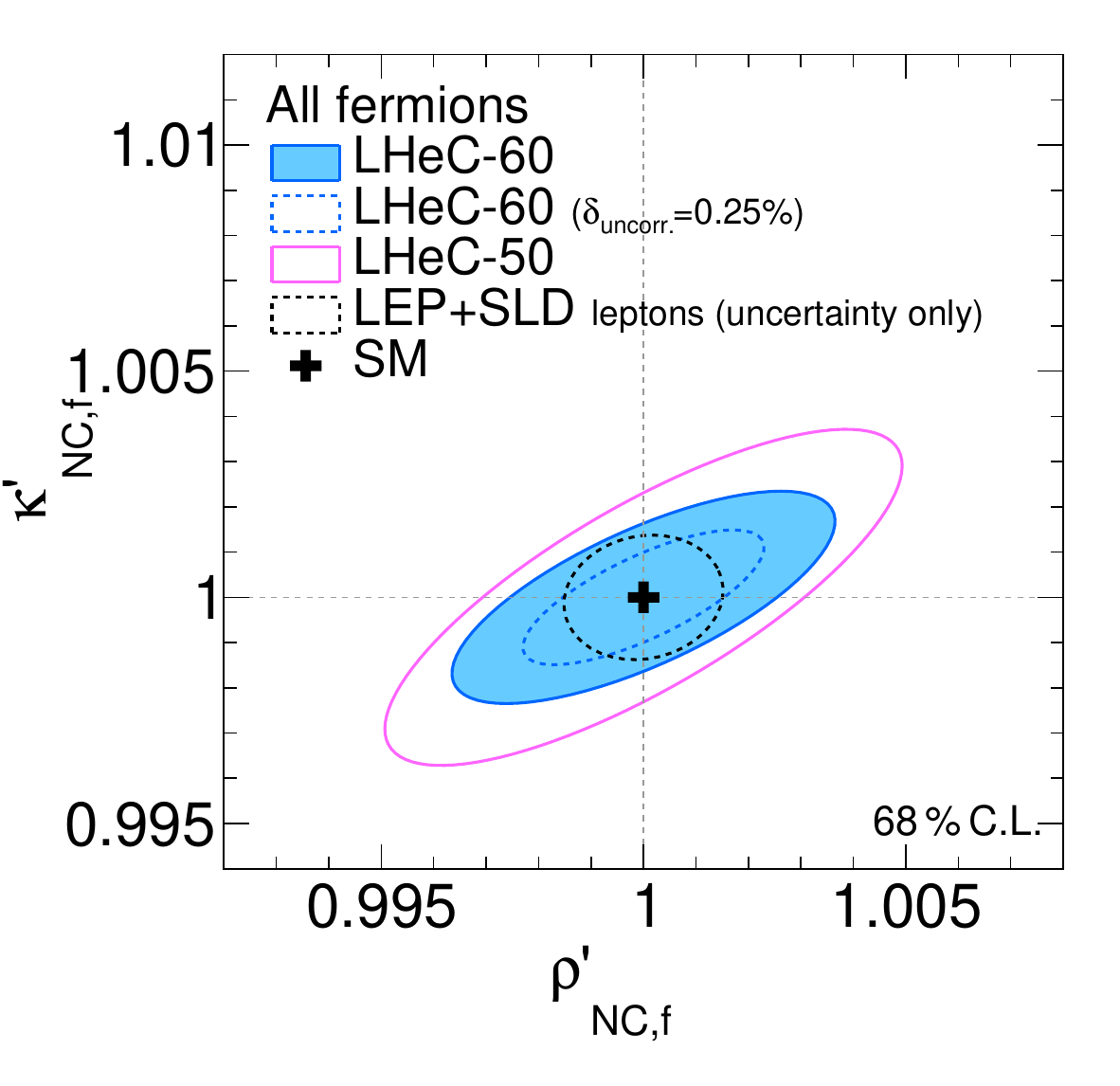}
    \hskip0.05\textwidth
    \includegraphics[width=0.31\textwidth,trim={0 0 18 20 },clip]{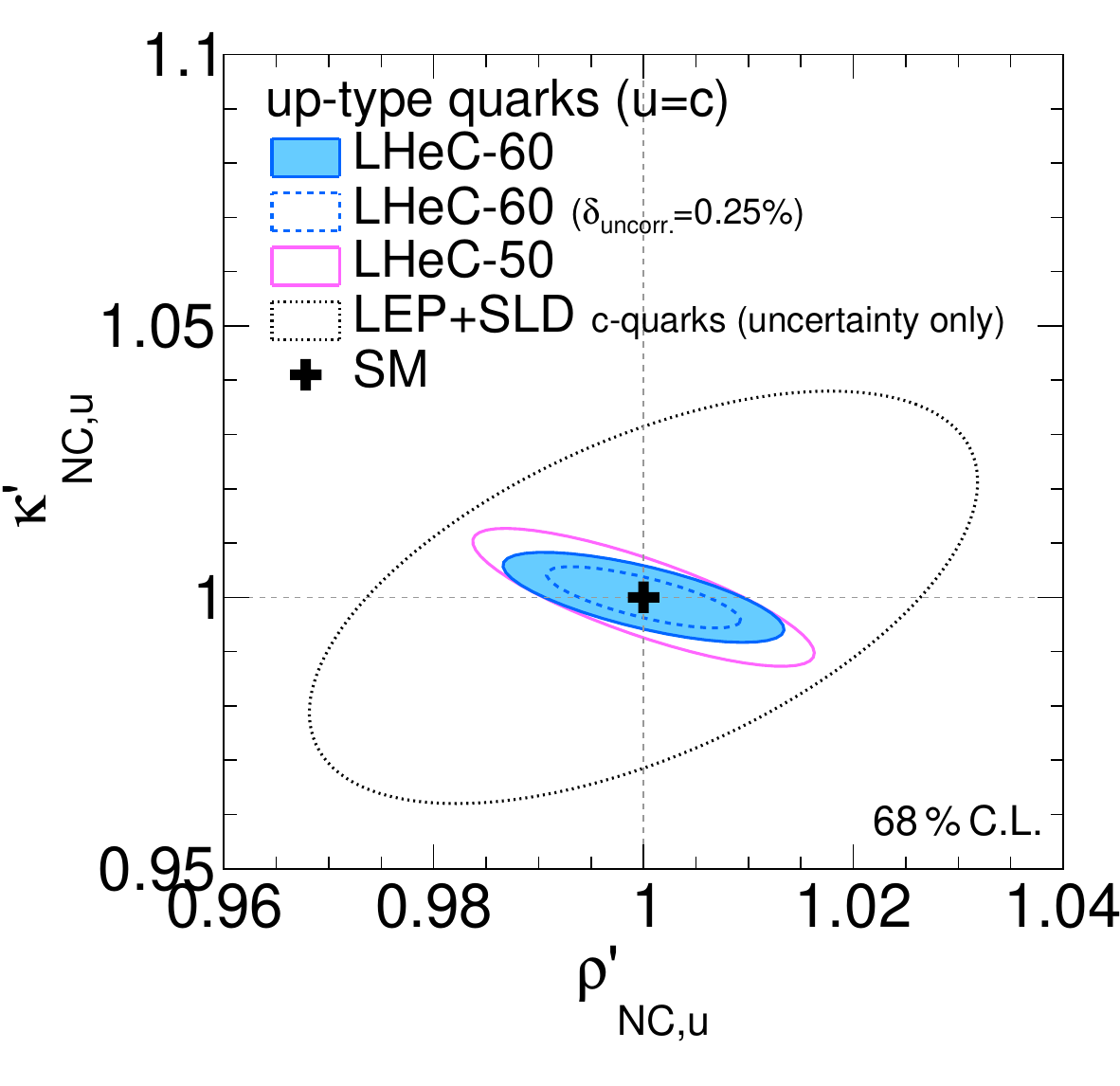}
    \includegraphics[width=0.31\textwidth,trim={0 0 18 20 },clip]{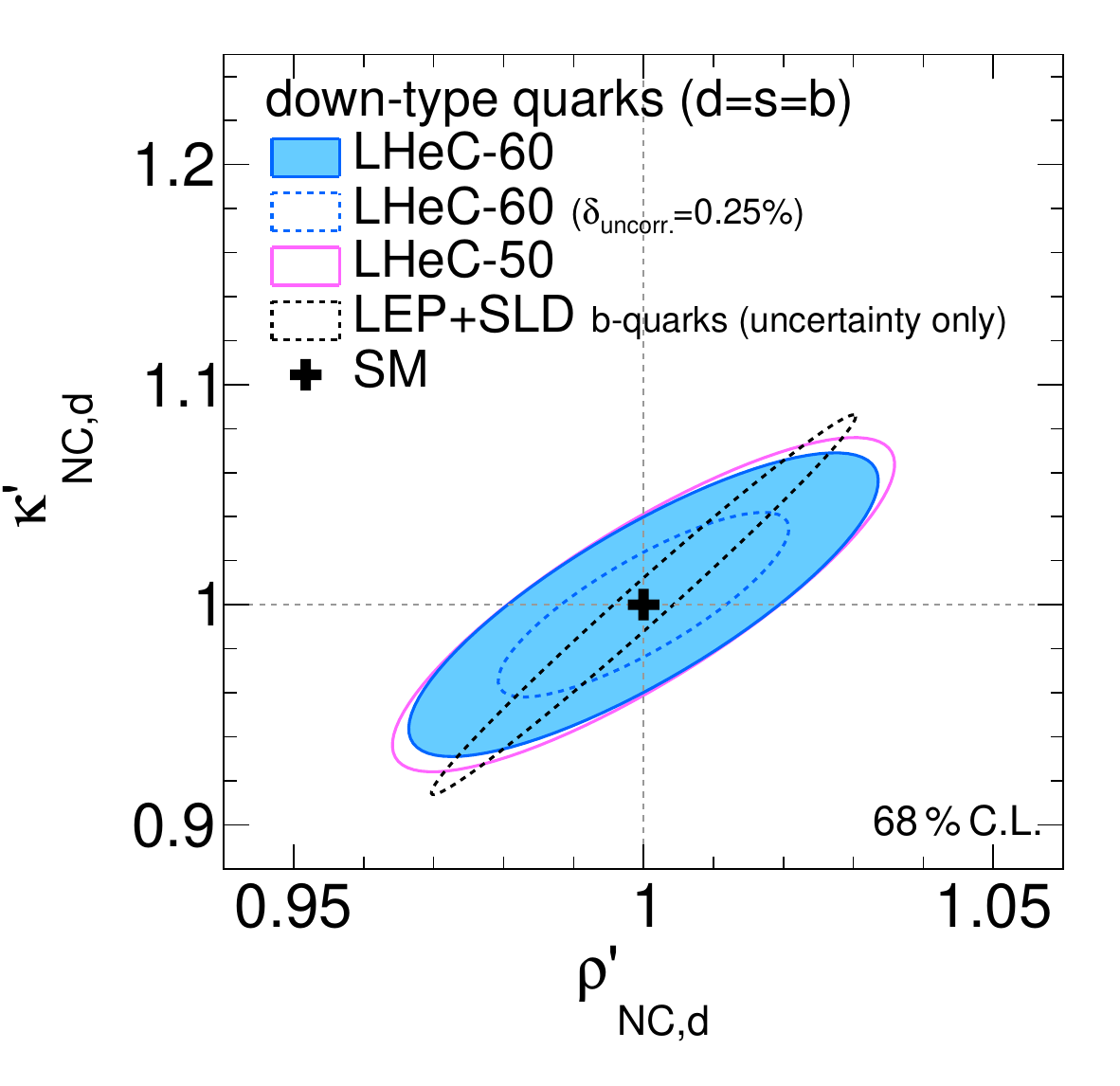}
  \caption{
    Expectations at 68\,\% confidence level for the determination
    of the $\rho_\text{NC}^\prime$ and $\kappa_\text{NC}^\prime$ parameters assuming 
    a single anomalous factor equal for all fermions (left). 
    The results for three different LHeC scenarios are
    compared with the achieved uncertainties from the LEP+SLD
    combination~\cite{ALEPH:2005ab} for the determination the
    respective leptonic quantities.
    Right: uncertainties for the simultaneous determination of the
    anomalous form factors for $u$ and $d$-type quarks, assuming known
    values for the electron parameters.
    The values are compared with uncertainties reported by LEP+SLD for
    the determination of the values $\rho_{\text{NC},(c,b)}$ and
    $\sin\theta_{\text{W}}^{\text{eff},(c,b)}$ for charm or bottom 
    quarks, respectively. 
  }
  \label{fig:rhokappa}
\end{figure}
The two-dimensional uncertainty contours of the anomalous form factors
$\rho^\prime_{\text{NC},f}$  and $\kappa^\prime_{\text{NC},f}$ are
displayed for three different 
LHeC scenarios in Fig.~\ref{fig:rhokappa} (left), and compared with
uncertainties from the LEP+SLD combination~\footnote{Since in the
  LEP+SLD analysis the values of $\rho_\text{NC}$ and $\kappa_\text{NC}\sw$ are
  determined, we compare only the size of the uncertainties in these
  figures. Furthermore it shall be noted, that LEP is mainly 
  sensitive to the parameters of leptons or heavy quarks, while 
  LHeC data is more sensitive to light quarks ($u$,$d$,$s$), and 
  thus the LHeC measurements are highly complementary.}~\cite{ALEPH:2005ab}.
It is found that these parameters can be determined with very high
experimental precision.

Assuming the couplings of the electron are given by the SM,
the anomalous form factors for the two quark families can be
determined and results are displayed in Fig.~\ref{fig:rhokappa} 
(right).
Since these measurements represent unique determinations of
parameters sensitive to the light-quark couplings,
we can compare only with nowadays measurements of the parameters for
heavy-quarks of the same charge and it is found that the
LHeC will provide high-precision determinations of the
$\rho^\prime_\text{NC}$ and $\kappa^\prime_\text{NC}$ parameters.

\begin{figure}
    \centering
    \includegraphics[width=0.44\textwidth]{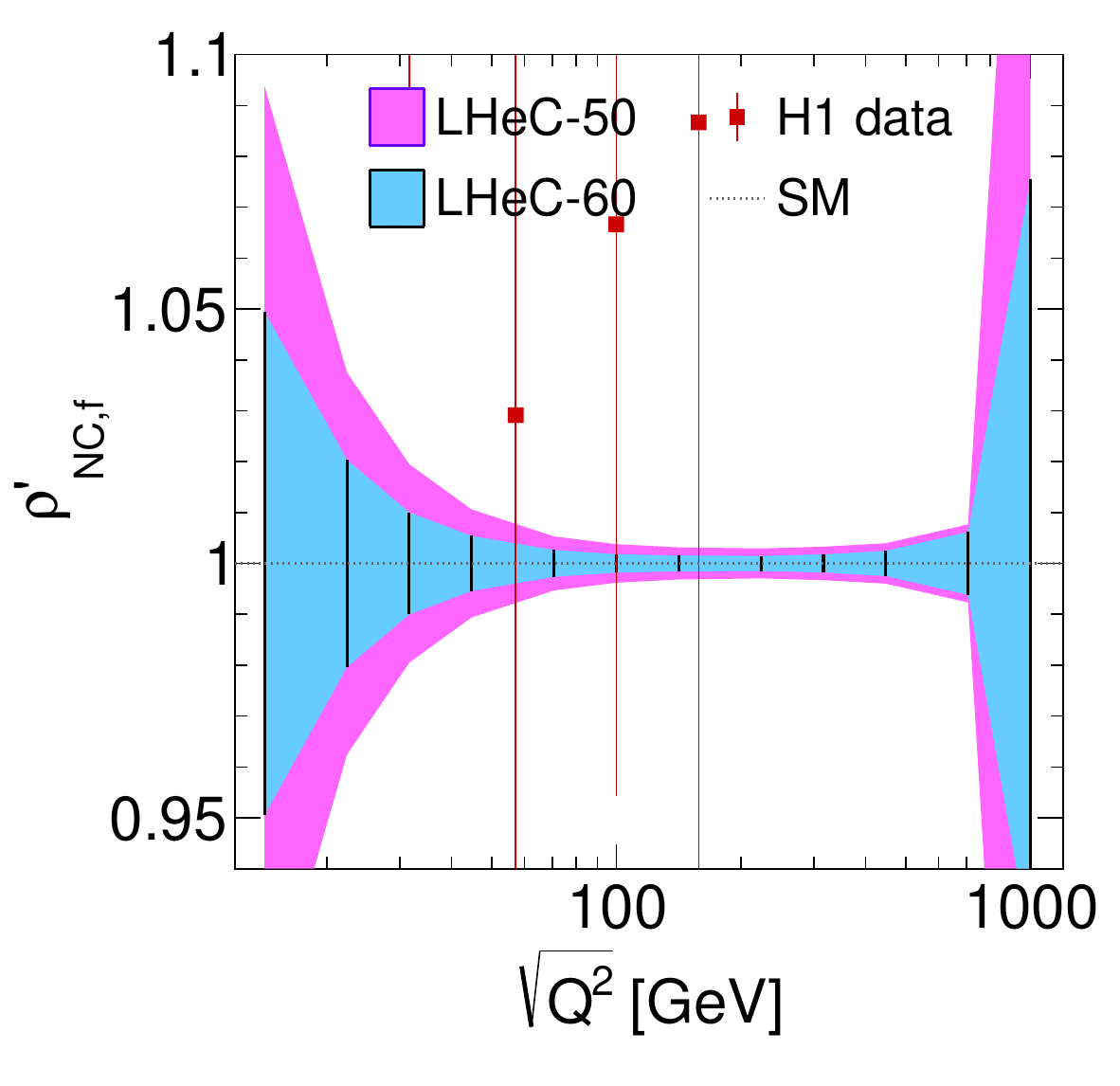}
    \hskip0.02\textwidth
    \includegraphics[width=0.44\textwidth]{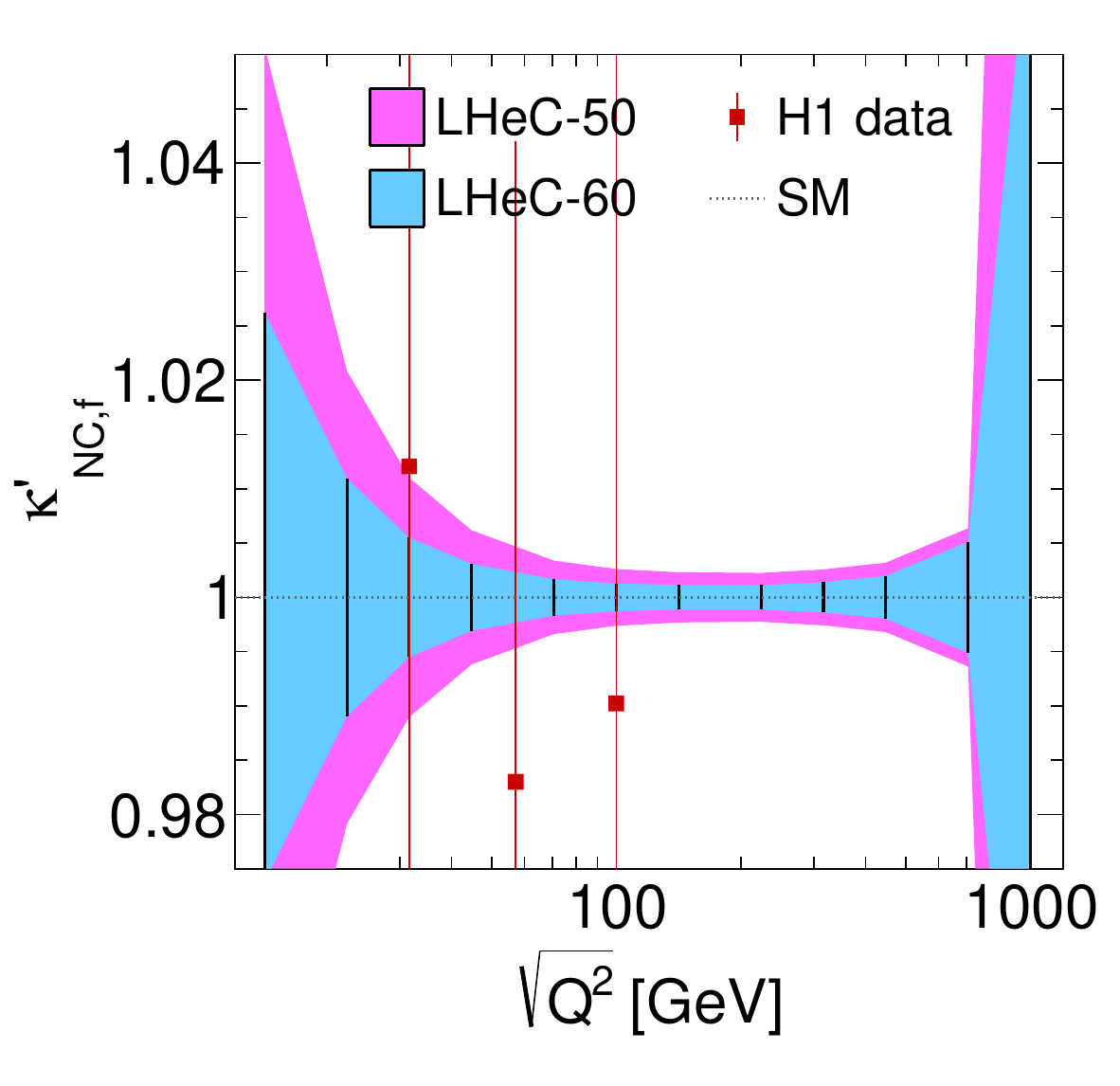}
    \caption{
      Test of the scale dependence of the anomalous $\rho$ and
      $\kappa$ parameters for two different LHeC scenarios.
      For the case of LHeC-60, i.e.\ $E_e=60\,\GeV$, we assume an uncorrelated 
      uncertainty of $0.25\,\%$.
      The uncertainties of the parameter $\kappa^\prime_\text{NC,f}$ can be interpreted 
      as sensitivity to the scale-dependence of the 
      weak mixing angle, 
      $\sin\theta_{\text{W}}^{\text{eff}}(\mu)$.
  }
  \label{fig:rhokappaQ2}
\end{figure}

A meaningful test of the SM can be performed by determining the 
effective coupling parameters as a function of the momentum 
transfer. In case of $\kappa_\text{NC}^\prime$, this is equivalent 
to measuring the running of the effective weak mixing angle,
$\sin\theta_{\text{W}}^{\text{eff}}(\mu)$ (see also Sec.~\ref{sec:sw2}).
However, DIS is quite complementary to other measurements 
since the process is mediated 
by space-like momentum transfer, i.e.\ $q^2=-\Qsq<0$ with
$q$ being the boson four-momentum. Prospects for a determination 
of $\rho_\text{NC}^\prime$ or $\kappa_\text{NC}^\prime$ at
different \Qsq\ values are displayed in 
Fig.~\ref{fig:rhokappaQ2} 
and compared to results obtaind by H1.
The value of 
$\kappa_\text{NC}^\prime(\mu)$ can be easily translated to a
measurement of $\sin\theta_{\text{W}}^{\text{eff}}(\mu)$. From 
Fig.~\ref{fig:rhokappaQ2} one can conclude that this
quantity can be determined with a precision of up to 0.1\,\% and
better than 1\,\% over a wide kinematic range of about 
$25<\sqrt{\Qsq}<700\,\GeV$.

\subsection[The effective weak mixing angle $\sin^2\theta_{\text{W}}^{\text{eff},\ell}$]{\boldmath The effective weak mixing angle $\sin^2\theta_{\text{W}}^{\text{eff},\ell}$}
\label{sec:sw2}
The leptonic effective weak mixing angle is defined as
$\sin^2\theta_{\text{W}}^{\text{eff},\ell}(\mu^2) = 
\kappa_{\text{NC},\ell}(\mu^2)\sw$. 
Due to its high sensitivity to loop corrections it represents an 
ideal quantity for precision tests of the Standard Model.
Its value is scheme dependent and it exhibits a scale dependence. 
Near the $Z$ pole, $\mu^2=\mz^2$, its value was precisely measured 
at LEP and at SLD. Those analyses were based on the measurement of 
asymmetries and their interpretation in terms of the leptonic 
weak mixing angle was simplified by the fact that many non-leptonic 
corrections and contributions from box graphs cancel or can be 
taken into account by subtracting their SM predictions. The highest 
sensitivity to $\sin^2\theta_{\text{W}}^{\text{eff},\ell}(M_Z)$ 
to date arises from a
measurement of $A_\text{fb}^{0,b}$~\cite{ALEPH:2005ab}, where the
non-universal flavour-specific corrections to the quark couplings are
taken from the SM and consequently these measurements are
interpreted to be sensitive only to the universal,
i.e.\ flavour-independent~\footnote{Flavour-specific tests have been
  discussed to some extent in the previous Section.},
non-SM contributions to $\kappa_{\text{NC}}$.
Applying this assumption also to the DIS cross sections, the
determination of $\kappa^\prime_{{\text{NC},f}}$ can directly be
interpreted as a sensitivity study of the leptonic effective weak
mixing angle $\sin^2\theta_{\text{W}}^{\text{eff},\ell}$.

\begin{table}
  \centering
  \small
  \begin{tabular}{lcccccc}
    \toprule
    Fit parameters &  Parameter & SM & \multicolumn{4}{c}{Expected uncertainties} \\
    \cmidrule(lr){4-7}
    &   of interest  &  value    & LHeC-50 & LHeC-60 & LHeC-50 & LHeC-60 \\
    &                &          &  \multicolumn{2}{c}{\footnotesize{($\delta_\text{uncor.}=0.50\,\%$)}} & \multicolumn{2}{c}{\footnotesize{($\delta_\text{uncor.}=0.25\,\%$)}} \\
    \midrule
    $\kappa^\prime_{\text{NC},f}$, PDFs
    & $\sin^2\theta_\text{W}^{\text{eff},\ell}(\mz^2)$ & 0.23154 & $0.00033$ & $0.00025$ & 0.00022 & $0.00015$ \\ 
    $\kappa^\prime_{\text{NC},f}$, $\rho^\prime_{\text{NC},f}$, PDFs 
    & $\sin^2\theta_\text{W}^{\text{eff},\ell}(\mz^2)$ & 0.23154 & 0.00071  & 0.00036 &  0.00056 & 0.00023 \\ 
    $\kappa^\prime_{\text{NC},e}$, PDFs   
    &  $\sin^2\theta_\text{W}^{\text{eff},e}(\mz^2)$   & 0.23154 & $0.00059$  & $0.00047$ & 0.00038 & $0.00028$ \\
    $\kappa^\prime_{\text{NC},e}$, $\kappa^\prime_{\text{NC},u}$, $\kappa^\prime_{\text{NC},d}$, PDFs
    &  $\sin^2\theta_\text{W}^{\text{eff},e}(\mz^2)$   & 0.23154 & 0.00111 & 0.00095 & 0.00069 & 0.00056\\ 
    $\kappa^\prime_{\text{NC},f}$
    & $\sin^2\theta_\text{W}^{\text{eff},\ell}(\mz^2)$ & 0.23154 & 0.00028 & 0.00023 & 0.00017 & 0.00014 \\ 
    \bottomrule
  \end{tabular}
  \caption{
    Determination of $\sin^2\theta_\text{W}^{\text{eff},\ell}(\mz^2)$
    with inclusive DIS data at the LHeC for different scenarios.
    Since the value of the effective weak mixing angle at the $Z$ pole
    cannot be determined directly in DIS, a fit of the
    $\kappa^\prime_{\text{NC},f}$ parameter is performed instead and
    its uncertainty is translated to
    $\sin^2\theta_\text{W}^{\text{eff},\ell}(\mz^2)$.
    Different assumptions on the fit parameters are studied, and
    results include uncertainties from the PDFs.
    Only the last line shows results where the PDF parameters are kept fixed.
    See text for more details.
  }
  \label{tab:sweff}
\end{table}

\begin{figure}
    \centering
    \includegraphics[width=0.50\textwidth]{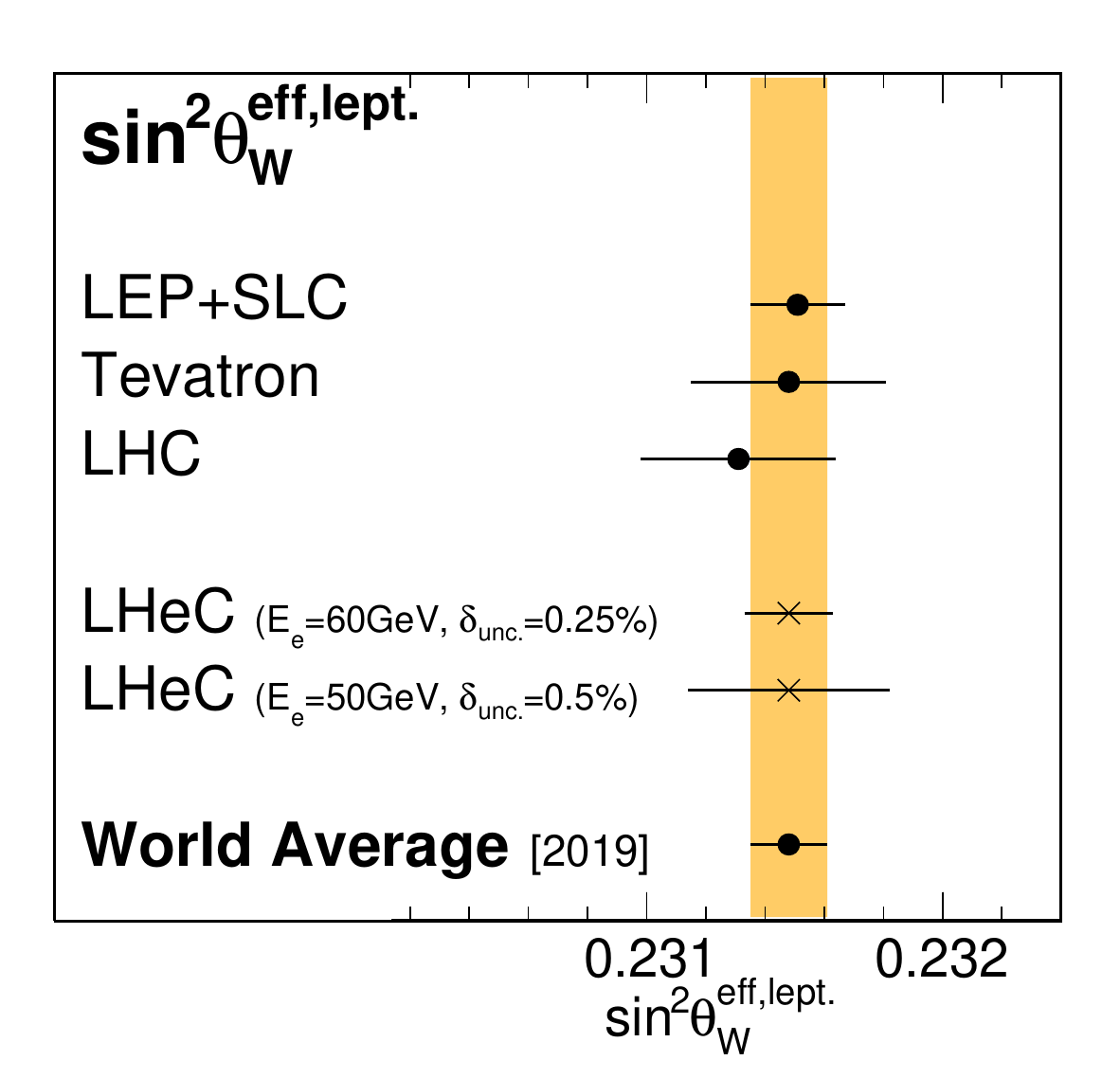}
    \caption{
      Comparison of the determination of 
      $\sin^2\theta_{\text{W}}^{\text{eff},\ell}(\mz^2)$
      from LHeC inclusive DIS data with recent averaged values.
      Results from LEP+SLC~\cite{ALEPH:2005ab}, Tevatron~\cite{Aaltonen:2018dxj},
      LHC~\cite{Aaij:2015lka,ATLAS:2018gqq,Sirunyan:2018swq,Erler:2019dcx} and the world average 
      value~\cite{Erler:2019dcx} are all obtained from a combination of various separate 
      measurements (not shown individually) (see also
      Ref.~\cite{Erler:2019hds} for additional discussions).
      For LHeC, the experimental and PDF uncertainties
      are displayed.
    }
  \label{fig:sw2eff}
\end{figure}

The prospects for a determination of 
$\sin^2\theta_{\text{W}}^{\text{eff},\ell}$
are listed in Tab.~\ref{tab:sweff}.
Two fits have been studied: one
with a fixed parameter $\rho^\prime_\text{NC}$ and one where
$\sin^2\theta_{\text{W}}^{\text{eff},\ell}$ is determined together 
with $\rho^\prime_\text{NC}$ (see Fig.~\ref{fig:rhokappa} (left)).
At the LHeC, it will be possible to determine the value of
$\sin^2\theta_{\text{W}}^{\text{eff},\ell}(\mz^2)$ with an 
experimental uncertainty of up to
\begin{equation}
  \Delta\sin^2\theta_{\text{W}}^{\text{eff},\ell} = \pm0.00015\,, 
\end{equation}
where PDF uncertainties are already included.
If the PDF parameters are artificially kept fixed, the uncertainties
are of very similar size, which demonstrates that these measurements
are fairly insensitive to the QCD effects and the PDFs.
The uncertainties are compared~\footnote{
  It shall be noted, that in order to compare the LHeC 
measurements with the $Z$-pole measurements at $\mu^2=\mZ^2$ in a conclusive 
way, one has to assume the validity of the SM framework. In 
particular the scale-dependence of $\kappa_{\text{NC},\ell}$ must be
known in addition to the flavour-specific corrections.
On the other hand, the 
scale dependence can be tested itself with the LHeC data 
which cover a large range of space-like $\Qsq$. 
In this aspect, DIS provides a unique opportunity for precision
measurements in the space-like regime ($\mu^2<0$) as has been discussed 
in the previous Section, see Fig.~\ref{fig:rhokappaQ2} (right).
} to recent average values in
Fig.~\ref{fig:sw2eff}. 
%
One can see that the LHeC measurement has 
the potential to become the most precise single measurement in 
the future with a significant impact to the world average value.
It is obvious that a conclusive interpretation of experimental 
results with such a high precision will require correspondingly 
precise theoretical predictions, and the investigation of two-loop 
corrections for DIS will become important.

This LHeC measurement will become competitive with measurements at the
HL-LHC~\cite{Azzi:2019yne}.
Since in $pp$ collisions one of the dominant uncertainties is from the
PDFs~\cite{ATLAS:2018gqq,Sirunyan:2018swq,Abdolmaleki:2019qmq,Accomando:2018nig,Accomando:2017scx}, future improvements can (only) be achieved with a
common analysis of LHeC and HL-LHC data.
Such a study will yield highest experimental precision and the
challenging theoretical and experimental aspects for a complete
understanding of such an analysis will deepen our
understanding of the electroweak sector.

It may be further of interest, to determine the value of the 
effective weak mixing angle of the electron separately in order 
to compare with measurements in $pp$ and test furthermore 
lepton-specific contributions to $\kappa_\text{NC,lept.}$. 
Such fits are summarised in Table~\ref{tab:sweff} and a 
reasonable precision is achieved with 
LHeC.

\begin{figure}
  \centering
  \includegraphics[width=0.55\textwidth]{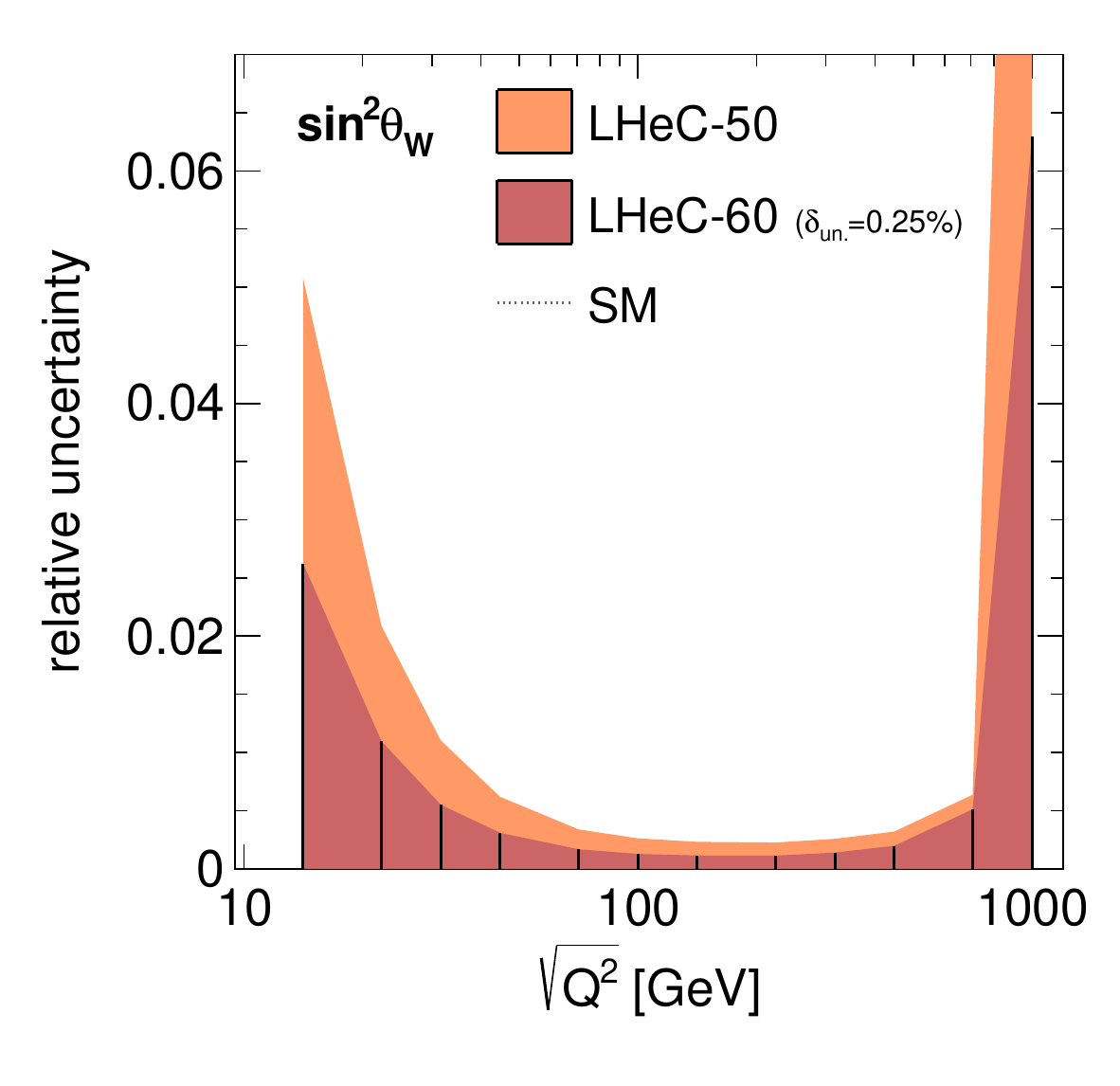}
  \caption{
    Expected uncertainties of the weak mixing angle determined
    in sub-regions of \Qsq. Two scenarios for the simulation
    of LHeC inclusive NC/CC DIS data are considered.
  }
\label{fig:sw2Q2}
\end{figure}
The measurement of
the weak mixing angle
 can be performed in 
sub-regions of \Qsq\ due to the wide kinematic range accessible at
the LHeC.
The relative uncertainties for the determination of the weak mixing
angle for different \Qsq\ intervals is displayed in Fig.~\ref{fig:sw2Q2}.
We find that the weak mixing angle
can be determined in the range of about $25 < \sqrt{\Qsq} < 700\,\GeV$
with a precision better than 0.1\,\%.
If a calculation of DIS cross sections including higher-order
EW corrections in the $\overline{\rm MS}$ scheme is available,
these relative uncertainties can be mapped
into a test of the running of the weak mixing angle.
Note, that in DIS the scattering process is mediated by boson
exchange with spacelike momenta and is therefore complementary to other
measurements since the scale is $\mu^2=-\Qsq$.

\subsection{Electroweak effects in charged-current scattering}
\begin{figure}[!thb]
    \centering
    \includegraphics[width=0.44\textwidth]{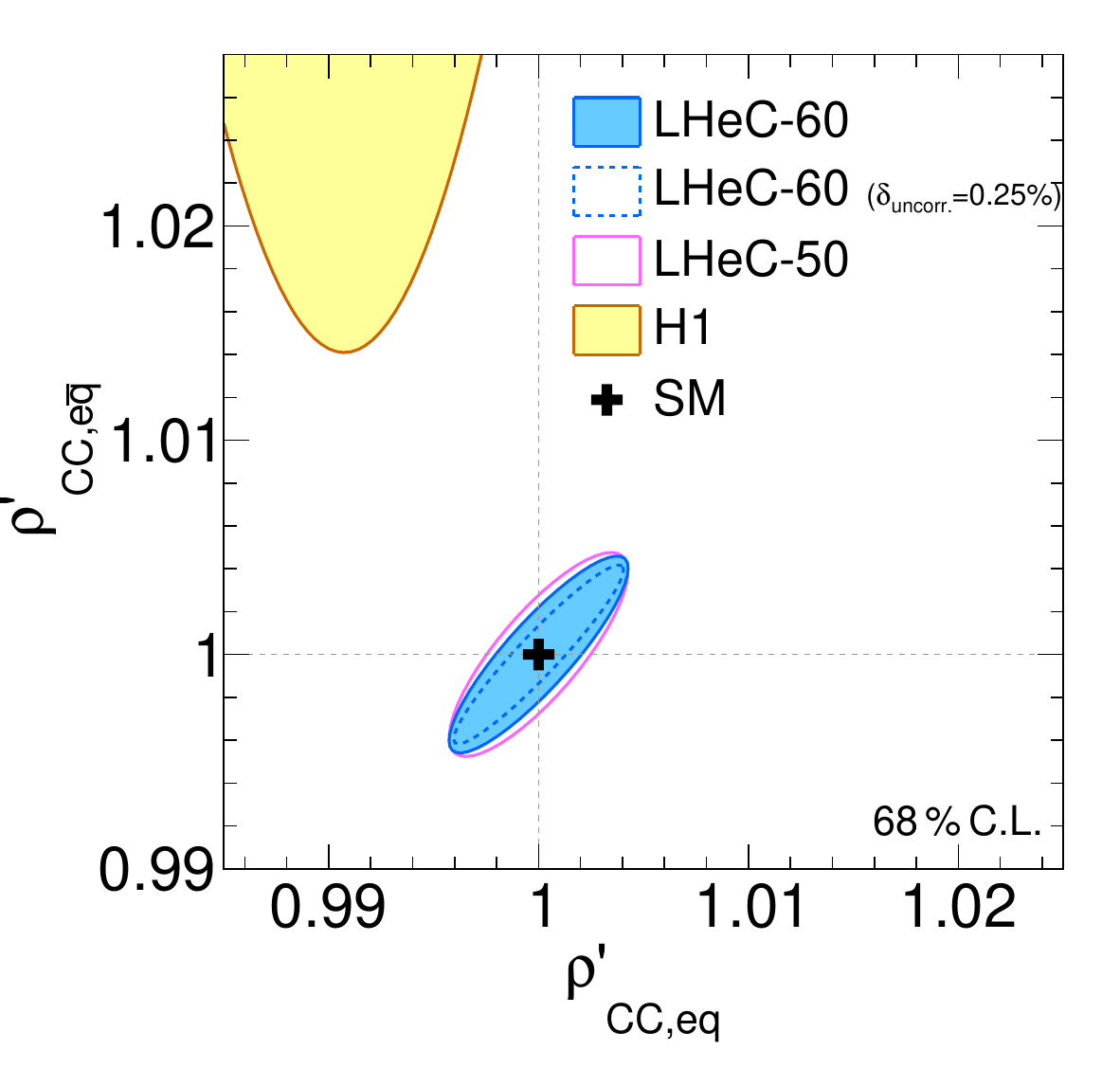}
    \hskip0.05\textwidth
    \includegraphics[width=0.44\textwidth]{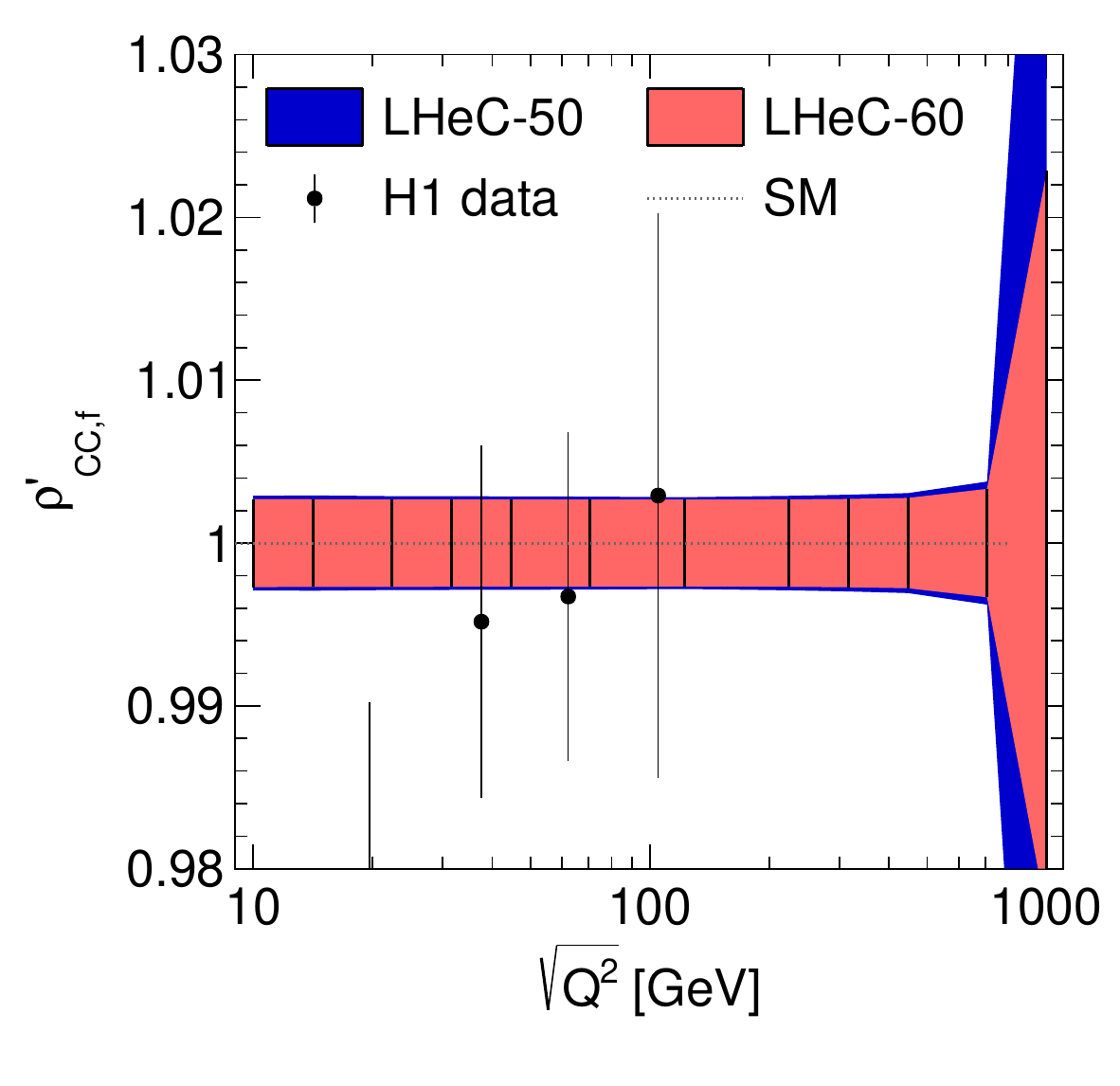}
  \caption{
    Left: anomalous modifications of the charged current form factors
    $\rho^\prime_{\text{CC},eq}$ and 
    $\rho^\prime_{\text{CC},e\bar{q}}$ for
    different LHeC scenarios in comparison with the H1
    measurement~\cite{Spiesberger:2018vki}.
    Right: scale dependent measurement of the anomalous modification
    of the charged current form factor 
    $\rho^\prime_{\text{CC}}(\Qsq)$, assuming 
    $\rho^\prime_{\text{CC},eq}=\rho^\prime_{\text{CC},e\bar{q}} 
    =\rho^\prime_{\text{CC}}$.
  }
  \label{fig:rhoCC}
\end{figure}

The charged-current sector of the SM can be uniquely measured at 
high scales over many orders of magnitude in $Q^2$ at the LHeC, 
due to the excellent tracking
detectors, calorimetry, and high-bandwidth triggers.
Similarly as in the NC case, the form factors of the effective
couplings of the fermions to the $W$ boson can be measured.
In the SM formalism, only two of these form factors are present,
$\rho_{\text{CC},eq}$ and $\rho_{\text{CC},e\bar{q}}$. We thus 
introduce two anomalous modifications to them,
$\rho_{\text{CC},(eq/e\bar{q})} \rightarrow 
\rho^\prime_{\text{CC},(eq/e\bar{q})} \rho_{\text{CC},(eq/e\bar{q})}$
(see Ref.~\cite{Spiesberger:2018vki}).
The prospects for the determination of these parameters are displayed
in Fig.~\ref{fig:rhoCC}, and it is found, that with the LHeC 
these parameters can be determined with a precision up to 
0.2--0.3\,\%. Also their 
\Qsq dependence can be uniquely studied with high precision up 
to $\sqrt{\Qsq}$ values of about 400\,\GeV.

\subsection{Conclusion}
With LHeC inclusive NC and CC DIS data, unique measurements  of
electroweak parameters can be performed with highest precision.
Since inclusive DIS is mediated through space-like momentum transfer
($t$-channel exchange) the results are often complementary to
other experiments, such as $pp$ or $e^+e^-$ collider experiments,
where measurements are performed in the time-like regime and most
often at the $Z$ peak.
Among many other quantities, measurements of the weak couplings
of the light quarks, $u$ and $d$, or their anomalous form factors
$\rho^\prime_{\text{NC},u/d}$  and $\kappa^\prime_{\text{NC},u/d}$,
can be performed uniquely due to the 
important contributions of valence quarks in the initial state.
Also scale dependent measurements of weak interactions can be
performed over a large range in $\sqrt{\Qsq}$, which provides an
interesting portal to BSM physics.
The $W$ boson mass can be determined with very small experimental
uncertainties, such that theoretical uncertainties are expected 
to become more important than experimental uncertainties. 
While the parameters of the PDFs are determined together with the EW
parameters in the present study, it is found that the PDFs do not
induce a limitation of the uncertainties.
Considering the dominating top-quark mass dependence of 
higher-order electroweak effects, one can realise that the 
LHeC will be competitive  with the global electroweak fit 
after the HL-LHC era~\cite{Azzi:2019yne,Schott:2019talk}.

Besides proving its own remarkable prospect on high-precision 
electroweak physics, the LHeC will further significantly improve 
the electroweak measurements in $pp$ collisions at the LHC by 
reducing the presently sizeable influence of PDF and $\alpha_s$ 
uncertainties. This is discussed in Sec.~\ref{chap:HLLHC}.

\section[Direct $W$ and $Z$ Production and Anomalous Triple Gauge Couplings]{\boldmath Direct $W$ and $Z$ Production and Anomalous Triple Gauge Couplings}
\subsection[Direct $W$ and $Z$ Production]{\boldmath Direct $W$ and $Z$ Production}
The direct production of single $W$ and $Z$ bosons as a crucial signal represents an important channel for EW precision measurements.
The production of $W$ bosons has been measured at $\sqrt{s}\simeq320\,\GeV$ at HERA~\cite{Chekanov:2008gn,Aaron:2009wp,Aaron:2009ab}.
With the full $e^\pm p$ data set collected by the H1 and ZEUS experiments together,
corresponding to an integrated luminosity of about $\mathcal{L}\sim1\,\text{fb}^{-1}$,
a few dozens of $W$ boson event candidates have been identified in the $e$, $\mu$ or $\tau$ decay channel.

Detailed studies of direct $W$/$Z$ production in $ep$ collisions at higher centre-of-mass energies have been presented in the past, see Refs.~\cite{Baur:1989gh,Baur:1990iw,Baur:1991pp}.
These theoretical studies were performed for a proton beam energy of $E_p=8\,\TeV$ and electron beam energies of $E_e=55\,\GeV$ or $100\,\GeV$, which correspond to a very similar centre-of-mass energy as the LHeC.
Measurements at the LHeC will benefit considerably from the large integrated luminosity, in comparison to earlier projections.

The $W$ or $Z$ direct production in $e^-p$ collisions can be classified into five processes
\begin{eqnarray}
e^{-}p \rightarrow e^{-}W^{+} j,&e^{-}p \rightarrow e^{-}W^{-} j, \nonumber\\ e^{-}p \rightarrow \nu_{e}^{-}W^{-} j,&e^{-}p \rightarrow \nu_{e}^{-}Z j
\label{eq:tgc-rel}
\end{eqnarray}
and
\begin{eqnarray}
e^{-}p \rightarrow {e}^{-}Z j,
\label{eq:tgc-irrel}
\end{eqnarray}
where $j$ denotes the hadronic the final state (i.e.\ the \emph{forward jet}).
According to the above classification, the four processes in Eq.\,\eqref{eq:tgc-rel} can be used to study Tripe Gauge Couplings (TGCs), e.g.\ $WW\gamma$ and $WWZ$ couplings, since some contributing diagrams represent Vector Boson Fusion~(VBF) processes.
The process shown in Eq.\,\eqref{eq:tgc-irrel} does not contain any TGC vertex.
The processes for positron-proton collisions can be easily derived from Eqs.~\eqref{eq:tgc-rel} and~\eqref{eq:tgc-irrel}, but are not discussed further here due to the small integrated luminosity of the LHeC $e^+p$ data.

The MadGraph5\_v2.4.2 program~\cite{Alwall:2014hca} is employed for matrix element calculation and event generation and
the PDF NNPDF23\_nlo\_as\_0119\_qed~\cite{Ball:2013hta} is used.
Technical cuts on the transverse momentum of the outgoing scattered lepton, $p_T^\ell$, of 10\,\GeV or alternatively 5\,\GeV, are imposed
and other basic cuts are $p_{T}^{j}> 20\,\text{GeV}$, $|\eta_{e,j}|<5$ and $\Delta R_{ej}<0.4$.
The resulting Standard Model total cross sections of the above processes are listed in Tab.~\ref{wbfXscetion}.
\begin{table}[ht]
  \centering
  \small
  \begin{tabular}{lccc}
    \toprule
    Process & $E_{e}=50\,\text{GeV}$,~$E_{p}=7\,\text{TeV}$ & $E_{e}=60\,\text{GeV}$,~$E_{p}=7\,\text{TeV}$ & $E_{e}=60\,\text{GeV}$,~$E_{p}=7\,\text{TeV}$ \\
    ~ & $p_{T}^{e}> 10\,\text{GeV}$ & $p_{T}^{e}> 10\,\text{GeV}$ & $p_{T}^{e}> 5\,\text{GeV}$  \\
    \midrule
    $e^{-}W^{+} j$ & 1.00\,\text{pb} & 1.18\,\text{pb} & 1.60\,\text{pb} \\
    $e^{-}W^{-} j$ & 0.930\,\text{pb} & 1.11\,\text{pb} & 1.41\,\text{pb}  \\
    $\nu_{e}^{-}W^{-} j$  & 0.796\,\text{pb} & 0.956\,\text{pb} & 0.956\,\text{pb} \\
    $\nu_{e}^{-}Z j$ & 0.412\,\text{pb} & 0.502\,\text{pb} & 0.502\,\text{pb} \\
    ${e}^{-}Z j$ & 0.177\,\text{pb} & 0.204\,\text{pb} & 0.242\,\text{pb} \\
    \bottomrule
  \end{tabular}
  \caption{The SM predictions of direct $W$ and $Z$ production cross sections in $e^-p$ collisions for different collider beam energy options, $E_e$, and final state forward electron transverse momentum cut, $p_{T}^{e}$.
  Two different electron beam energy options are considered, $E_e=50\,\GeV$ and $60\,\GeV$.
}
  \label{wbfXscetion}
\end{table}

The process with the largest production cross section in $e^-p$ scattering is the single $W^{+}$ boson production.
This will be the optimal channel of both the SM measurement and new physics probes in the EW sector.
Also, this channel is experimentally preferred since the $W^+$ is produced in NC scattering, so the beam electron is
measured in the detector, and the $W$-boson has opposite charge to the beam lepton and thus in a leptonic decay an opposite charge lepton and missing transverse momentum is observed.
Altogether, it is expected that a few million of direct $W$-boson events are measured at LHeC.

Several $10^5$ direct $Z$ events are measured, which corresponds approximately to the size of the event sample of the SLD experiment~\cite{ALEPH:2005ab}, but at the LHeC these $Z$ bosons are predominantly produced in VBF events.

All these total cross sections increase significantly with smaller transverse momentum of the outgoing scattered lepton.
Therefore it will become important to decrease that threshold with dedicated electron taggers, see Chapter\,\ref{chap:detector}.

\subsection{Anomalous Triple Gauge Couplings}
The measurement of gauge boson production processes provides a precise measurement of the triple gauge boson vertex.
The measurement is sensitive to new physics contributions in \emph{anomalous} Tripe Gauge Couplings~(aTGC).
The LHeC has advantages of a higher centre-of-mass energy and easier kinematic analysis in the measurement of aTGCs.

In the effective field theory language, aTGCs in the Lagrangian are generally parameterised as
\begin{eqnarray}
\mathcal{L}_{TGC}/g_{WWV}&=&ig_{1,V}(W_{\mu\nu}^{+}W_{\mu}^{-}V_{\nu}-W_{\mu\nu}^{-}W_{\mu}^{+}V_{\nu})+i\kappa_{V}W_{\mu}^{+}W_{\nu}^{-}V_{\mu\nu}+\frac{i\lambda_{V}}{M_{W}^{2}}W_{\mu\nu}^{+}W_{\nu\rho}^{-}V_{\rho\mu}\nonumber\\
&&+g_{5}^{V}\epsilon_{\mu\nu\rho\sigma}(W^{+}_{\mu}\overleftrightarrow{\partial}_{\rho}W^{-}_{\nu})V_{\sigma}-g_{4}^{V}W^{+}_{\mu}W^{-}_{\nu}(\partial_{\mu}V_{\nu}+\partial_{\nu}V_{\mu})\nonumber\\
&&+i\tilde{\kappa}_{V}W^{+}_{\mu}W^{-}_{\nu}\tilde{V}_{\mu\nu}+\frac{i\tilde{\lambda}_{V}}{M_{W}^{2}}W^{+}_{\lambda\mu}W^{-}_{\mu\nu}\tilde{V}_{\nu\lambda},
  \label{eq:lan-TGCs}
\end{eqnarray}
where $V=\gamma, Z$.
The gauge couplings $g_{WW\gamma}=-e$, $g_{WWZ}=-e\cot\theta_{W}$ and the weak mixing angle $\theta_{W}$ are from the SM. $\tilde{V}_{\mu\nu}$ and $A\overleftrightarrow{\partial}_{\mu}B$ are defined as $\tilde{V}_{\mu\nu}=\frac{1}{2}\epsilon_{\mu\nu\rho\sigma}V_{\rho\sigma}$, $A\overleftrightarrow{\partial}_{\mu}B=A(\partial_{\mu}B)-(\partial_{\mu}A)B$,~respectively. There are five aTGCs ($g_{1,Z}$, $\kappa_{V}$, and $\lambda_{V}$) conserving the $C$ and $CP$ condition with electromagnetic gauge symmetry requires $g_{1,\gamma}=1$.
Only three of them are independent because $\lambda_{Z}=\lambda_{\gamma}$ and $\Delta\kappa_{Z}=\Delta g_{1,Z}-\tan^{2}\theta_{W}\Delta\kappa_{\gamma}$~\cite{Hagiwara:1993ck, Hagiwara:1992eh, DeRujula:1991ufe}. The LHeC can set future constraints on $\Delta\kappa_{\gamma}$ and $\lambda_{\gamma}$.

In the direct $Z/\gamma$ production process, the anomalous $WWZ$ and $WW\gamma$ couplings can be separately measured without being influenced by their interference~\cite{Biswal:2014oaa,Cakir:2014swa}.
In the direct $W$ production process, both the deviation in signal cross section and the kinematic distributions can effectively constrain the $WW\gamma$ aTGC, while anomalous $WWZ$ contribution in this channel is insensitive as a result of the suppression from $Z$ boson mass~\cite{Li:2017kfk, Koksal:2019oqt, Gutierrez-Rodriguez:2019hek}. 

\begin{figure}[!th]
    \centering
    \includegraphics[width=0.42\textwidth]{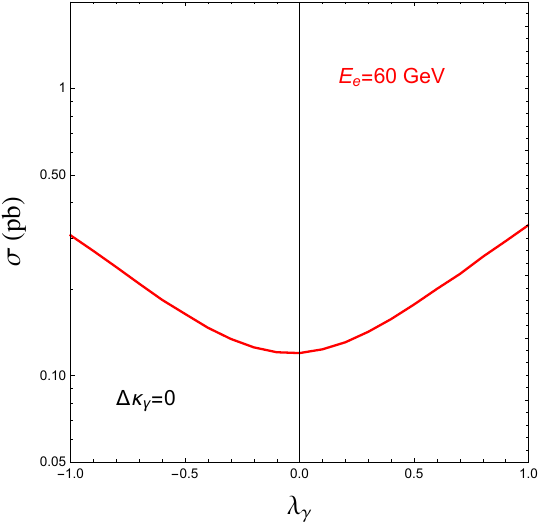}
    \hskip0.05\textwidth
     \includegraphics[width=0.42\textwidth]{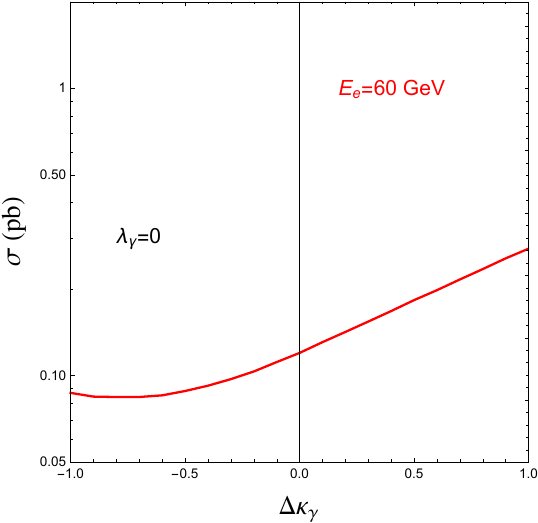}
    \caption{
Total cross sections of the $e^{-}p\rightarrow e^{-}\mu^{+}\nu_{\mu}j$ process with varying $\lambda_{\gamma}$~(left plot) and $\Delta\kappa_{\gamma}$~(right plot).
    }
    \label{fig:aTGC_Xsec}
\end{figure}

The $W$ decay into muon channel is the expected optimal measurement for the anomalous $WW\gamma$ coupling because of the discrimination of final states and mistagging efficiencies~\cite{Li:2017kfk}. Fig.~\ref{fig:aTGC_Xsec} shows the cross section of single $W^+$ production process followed by $W^+\to\mu^+\nu_{\mu}$ decay, with different $\lambda_{\gamma}$ and $\Delta\kappa_{\gamma}$ values.
Large anomalous coupling leads to measurable deviation to the SM prediction. The cross section increases monotonically with $\Delta\kappa_{\gamma}$ and the absolute value of $\lambda_{\gamma}$ within the region of $-1.0\leq\lambda_{\gamma}/\Delta\kappa_{\gamma}\leq1.0$. 

Kinematic analysis is necessary for the precise aTGC measurement.
At LHeC, the $e^{-}p\rightarrow e^-W^{\pm} j$ process with leptonic $W$ boson decay can be fully reconstructed because the undetected neutrino information is reconstructed either with energy-momentum conservation or the recoil mass method.
This allows to use angular correlation observables, which are sensitive to the $W$ boson polarisation.
Helicity amplitude calculation indicates that a non-SM value of $\lambda_{\gamma}$ leads to a significant enhancement in the transverse polarisation fraction of the $W$ boson in the $e^{-}p\rightarrow e^{-}W^{+}j$ process, while a non-SM value of $\Delta\kappa_{\gamma}$ leads to enhancement in the longitudinal component fraction~\cite{Baur:1989gh}.
The angle $\theta_{\ell W}$ is defined as the angle between the decay product lepton $\ell$ in the $W$ rest frame and $W$ moving direction in the collision rest frame. Making use of the energetic final states in the forward direction, a second useful angle $\Delta\phi_{ej}$ is defined as the separation of final state jet and electron on the azimuthal plane.
In an optimised analysis, assuming an integrated luminosity of $1\,\text{ab}^{-1}$, the observable $\Delta\phi_{ej}$ can impose stringent constraints on both $\lambda_{\gamma}$ and $\Delta\kappa_{\gamma}$, and uncertainties within [$-0.007$,~0.0056] and [$-0.0043$,~0.0054] are achieved, respectively.
The $\cos\theta_{\mu W}$ observable is also sensitive to $\Delta\kappa_{\gamma}$ at the same order, but fails to constrain $\lambda_{\gamma}$. The analysis is described in detail in Ref.~\cite{Li:2017kfk}.

\begin{figure}[!th]
    \centering
    \includegraphics[width=0.42\textwidth]{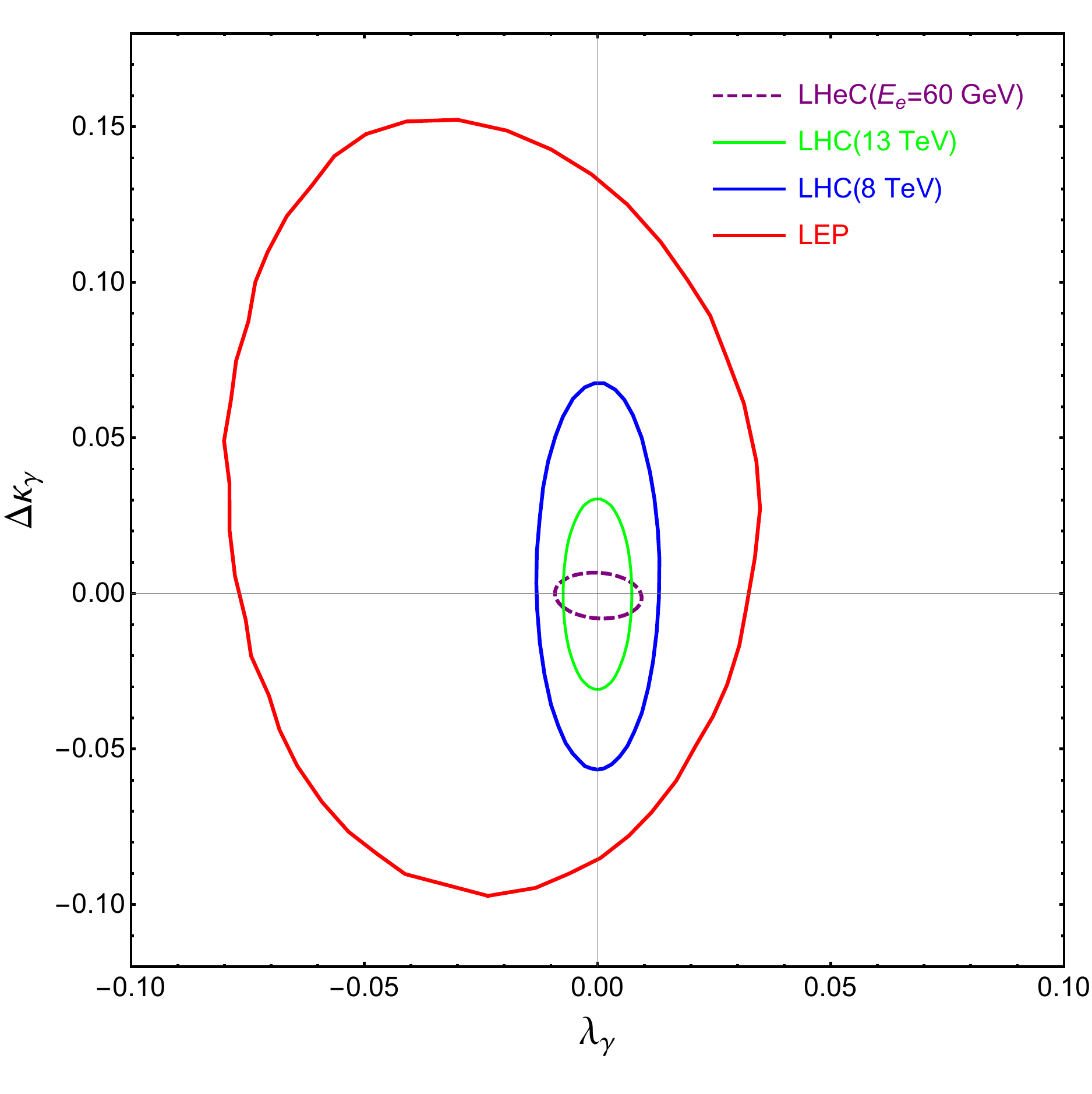}
    \caption{
      The $95\%~\text{C.L.}$ exclusion limit on the $\Delta\kappa_{\gamma}$-$\lambda_\gamma$ plane. The purple dashed contour is the projected LHeC exclusion limit with $1\,\text{ab}^{-1}$ integrated luminosity~\cite{Li:2017kfk}. The blue, green and red contours are current bounds from LHC~\cite{Sirunyan:2017bey, Sirunyan:2019gkh} and LEP~\cite{Villa:2004dx}. 
    }
    \label{fig:aTGC_contour}
\end{figure}

Fig.~\ref{fig:aTGC_contour} shows the two-parameter aTGC constraint on the $\lambda_{\gamma}$--$\Delta\kappa_{\gamma}$ plane based on a $\chi^2$ analysis of $\Delta\phi_{ej}$ at parton-level and assuming an electron beam energy of $E_{e}=60\,\text{GeV}$. 
When comparing with the current LHC~(blue and green) and LEP~(red) bounds, the LHeC has the potential to significantly improve the constraints, in particular on the $\Delta\kappa_{\gamma}$ parameter.
The polarised electron beam is found to improve the aTGC measurement~\cite{Cakir:2014swa, Gutierrez-Rodriguez:2019hek}.
In consideration of the \emph{realistic} analysis at detector level, one expects $2$-$3\,\text{ab}^{-1}$ integrated luminosity to achieve same results~\cite{Li:2017kfk}.

One uncertainty in the aTGC measurement at the (HL-)LHC comes from the PDF uncertainty. Future LHeC PDF measurement will improve the precision of aTGC measurement in the $x\simeq\mathcal{O}(10^{-2})$ region.

\section{Top Quark Physics}
\label{sec:top}
\newcommand{\ttbar} {\mbox{$t\bar{t}$}\xspace}

\begin{figure}[!th]
\centering
\includegraphics[width=0.75\textwidth]{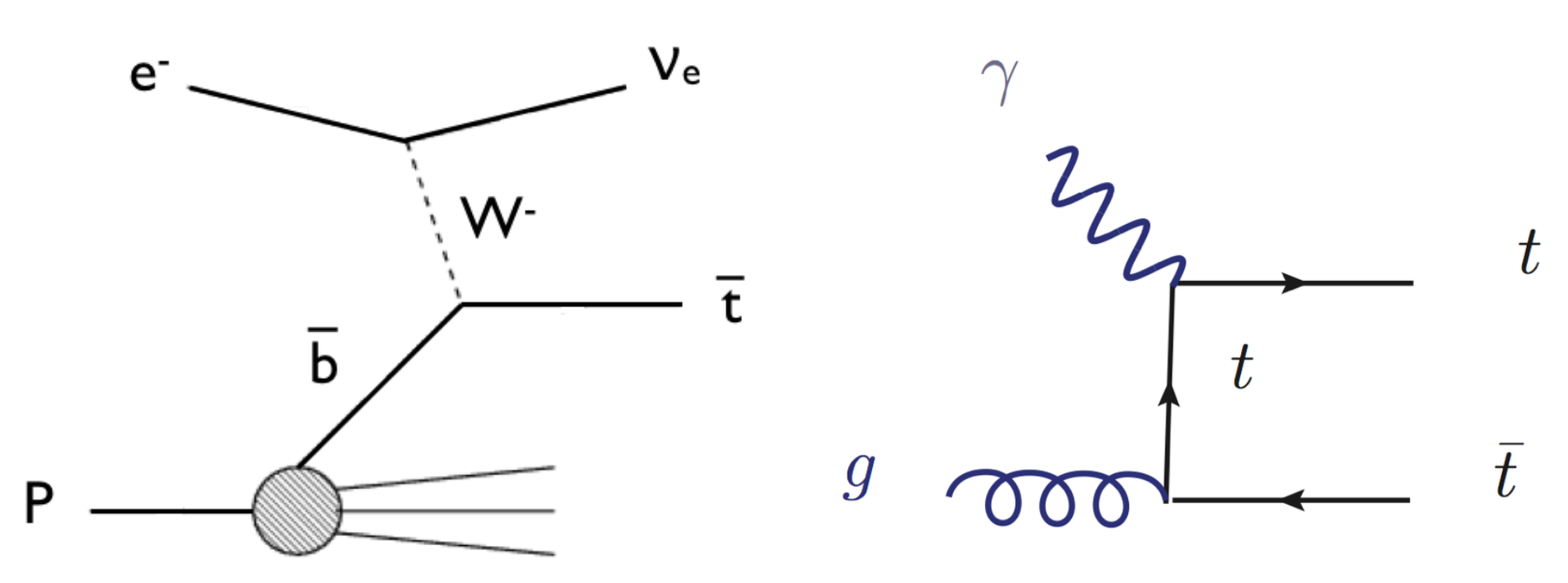}
\caption{
Example graphs for top quark production in CC DIS (left) and for $t\bar{t}$ photoproduction (right). 
}
\label{graphs_eh_top}
\end{figure}

SM top quark production at a future ep collider is dominated by single
top quark production, mainly via CC DIS production.
A leading-order Feynman diagram is displayed in Fig.~\ref{graphs_eh_top} (left). 
The total cross section for single top quark production is $1.89\,\text{pb}$ at the LHeC~\cite{Dutta:2013mva} and
a centre-of-mass energy of 1.3\,TeV, i.e.\ with an electron beam
energy of 60\,GeV and an LHC proton beam of 7\,TeV. 
The second important production
mode for top quarks at the LHeC is photoproduction
of top-antitop quark pairs (\ttbar), where a total cross section of
$0.05\,\text{pb}$ is expected at the LHeC~\cite{Bouzas:2013jha}.
Figure~\ref{graphs_eh_top} (right) shows an example Feynman diagram. 
This makes a future LHeC a top quark
factory and an ideal tool to study top quarks with a high precision,
and to analyse in particular their electroweak interaction. Selected
highlights in top quark physics are summarised here.

\subsection[$Wtq$ Couplings]{\boldmath $Wtq$ Couplings}

The top quark couplings to gauge bosons can be modified
significantly in models with new top (or third generation)
partners, such as in some extensions of the minimal
supersymmetric standard model, in little Higgs models,
top-color models, top seesaw, top compositeness, and
others. Testing them is therefore of utmost importance to find out
whether there are other sources of electroweak symmetry breaking that
are different from the standard Higgs mechanism. 

One highlight at the LHeC is the direct measurement of the CKM
matrix element $|V_{tb}|$.
Such a measurement can be done without making any model 
assumptions, like for instance on the unitarity of the CKM 
matrix or the number of quark generations.
An elaborate analysis of the single top quark CC DIS process at the
LHeC, which makes use of a detailed detector simulation using the DELPHES
package~\cite{Ovyn:2009tx}, shows that already at $100\,\text{fb}^{-1}$ of
integrated luminosity an uncertainty of 1\% can be expected. This compares
to a total uncertainty of 4.1\,\% of the currently most accurate result at the LHC
Run-I performed by the CMS experiment~\cite{Khachatryan:2014iya}. 

The same process is also highly sensitive to the
search for anomalous left- and right-handed $Wtb$ vector ($f_1^L$, $f_1^R$) and tensor ($f_2^L$, $f_2^R$)
couplings~\cite{Dutta:2013mva}.
These are given by an effective Lagrangian,
\begin{equation}
\mathcal{L}_{Wtb} = - \frac{g}{\sqrt{2}} \bar{b} \gamma^\mu V_{tb} ( f_1^L P_L - f_1^R
P_R) t W^-_\mu - \frac{g}{\sqrt{2}} \bar{b} \frac{ i \sigma^{\mu\nu} q_\nu}{M_W} ( f_2^L P_L - f_2^R
P_R) t W^-_\mu \,\, +h.c.
\label{eq:anom_Wtb_couplings}
\end{equation}
In the SM formalism $f_1^L=1$ and $f_1^R= f_2^L=f_2^R=0$.
The effect of
anomalous $Wtb$ couplings is consistently evaluated in the production
and the decay of the antitop quark, cf. Fig.~\ref{graphs_eh_top} (left).\footnote{Further studies of the top quark charged current coupling can be found in~\cite{Sarmiento-Alvarado:2014eha}. There, a more general framework is employed using the full basis of $SU(2)_L\times U(1)$ operators, including the relevant four-fermion ones.}

\begin{figure} [!hb]
\centering
\includegraphics[width=0.65\textwidth]{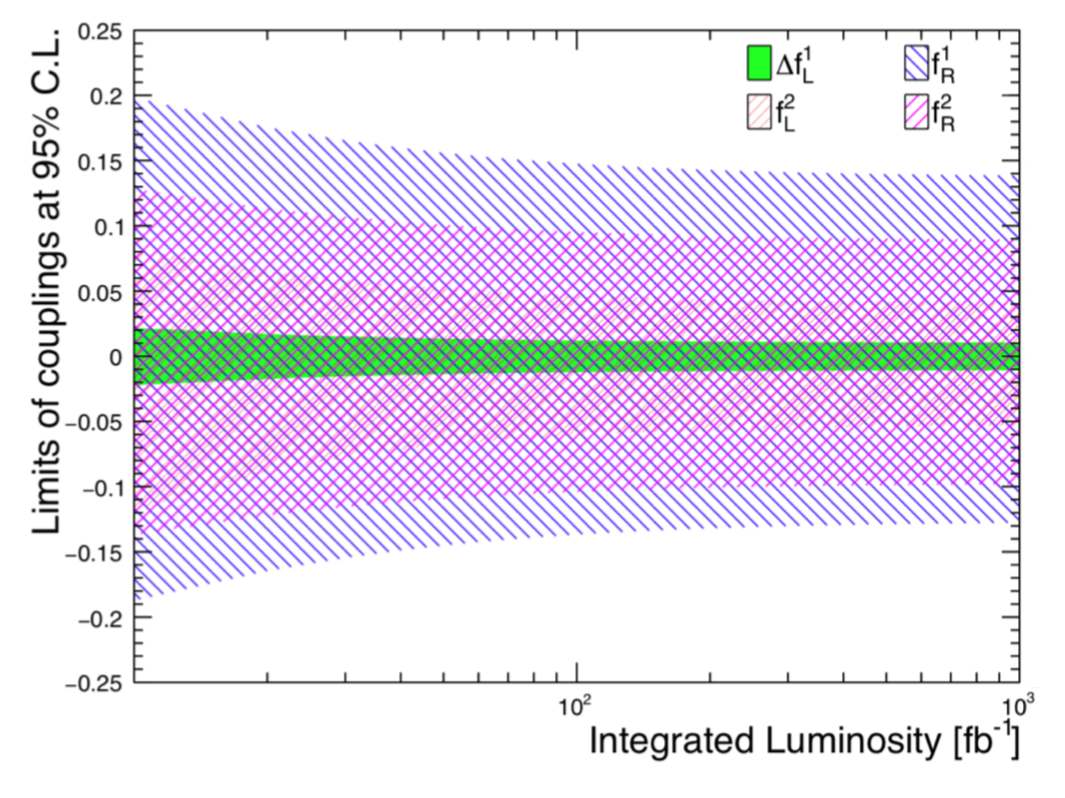}
\caption{
Expected sensitivities on
the SM and anomalous $Wtb$ couplings, as a function of the integrated luminosity~\cite{Dutta:2013mva}.
}
\label{fig_Wtb_couplings}
\end{figure}

The expected limits for the anomalous couplings at 95\,\% confidence level from a
measurement of single top quark production in CC DIS at the LHeC, is
displayed in Fig.~\ref{fig_Wtb_couplings}.
This analysis only exploits hadronic top quark decays~\cite{Dutta:2013mva}.
The coupling parameters are expected to be measured with accuracies of 1\,\% for the SM $f_1^L$ coupling
determining $|V_{tb}|$ (as discussed above) and of 4\,\% for $f_2^L$,
9\,\% for $f_2^R$, and 14\% for $f_1^R$ at $1\ \text{ab}^{-1}$.

\begin{figure} [!h]
\centering
\includegraphics[width=0.49\textwidth]{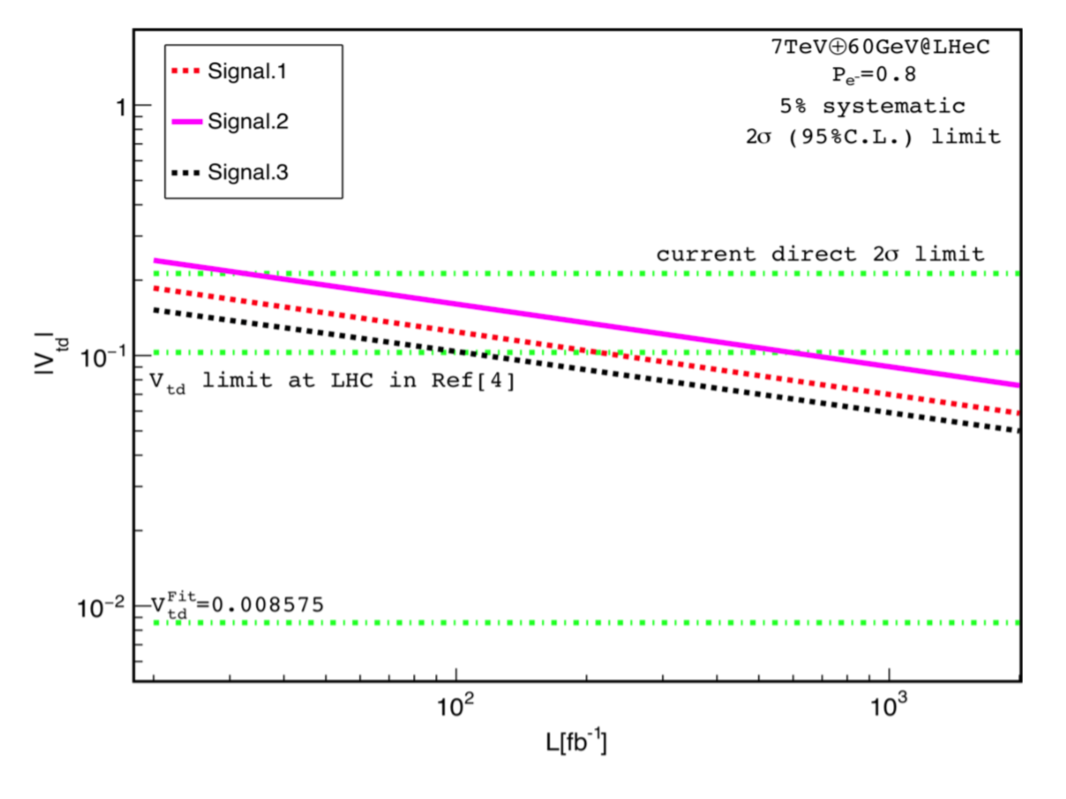}
\includegraphics[width=0.49\textwidth]{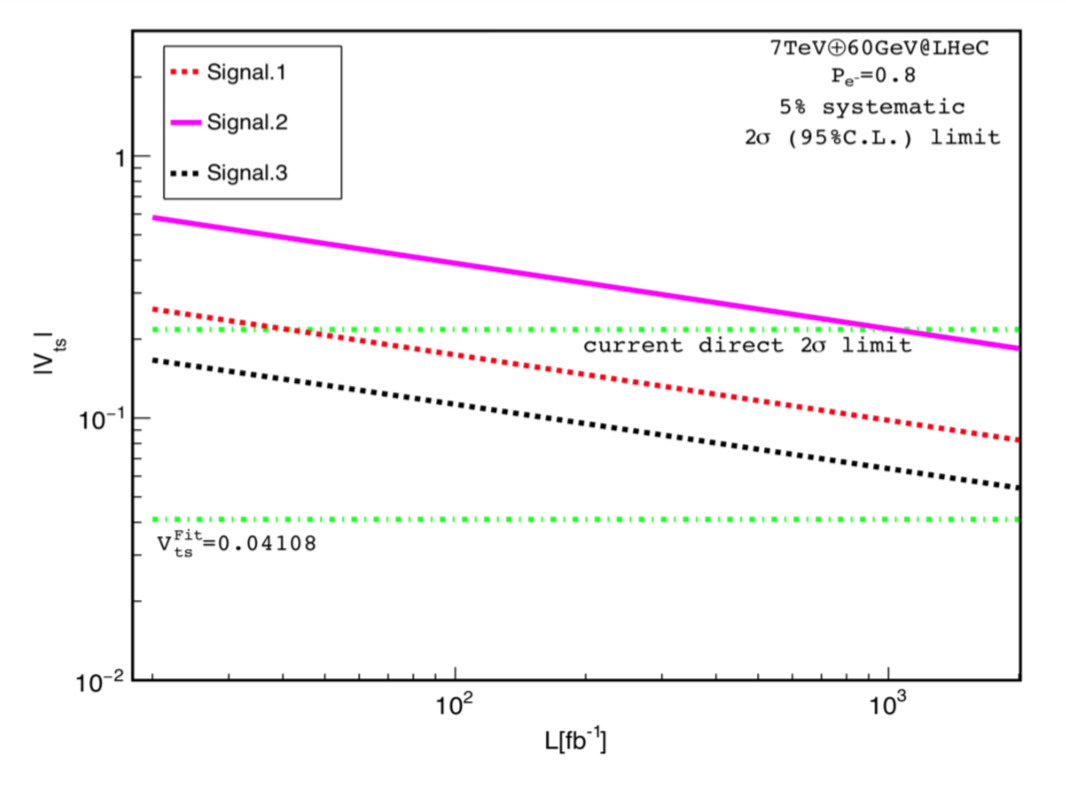}
\caption{
Expected sensitivities on $|V_{td}|$ (left) and $|V_{ts}|$ (right), as a function of the integrated luminosity~\cite{Sun:2018gqo}.
}
\label{fig_Wtx_couplings}
\end{figure}

In a similar way, through $W$ boson and bottom (light)
quark associated production, the CKM matrix elements $|V_{tx}|$ ($x=d,s$) can be extracted with very high precision utilizing a parameterisation of the deviations from the respective SM values. 
Here, the $W$ boson and the $b$-jet (light jet $j=d,s$) are produced via t-channel top quark exchange, or via s-channel single top
quark decay, as outlined in~\cite{Sun:2018gqo}. As an example, the processes 
\begin{itemize}
\item[ ] Signal 1: $p e^- \to \nu_e \bar{t} \to \nu_e W^- \bar{b} \to \nu_e \ell^-\nu_\ell \bar{b}$
\item[ ] Signal 2: $p e^- \to \nu_e W^- b \to \nu_e \ell^-\nu_\ell b$
\item[ ] Signal 3: $p e^- \to \nu_e \bar{t} \to \nu_e W^- j \to \nu_e \ell^-\nu_\ell j$
\end{itemize}
are analyzed in an elaborate study, where a detailed detector simulation is performed with the DELPHES
package~\cite{Ovyn:2009tx}. Fig.~\ref{fig_Wtx_couplings} shows the resulting accuracies at the $2\sigma$ confidence level (C.L.) for an expected measurement of 
$|V_{td}|$ and $|V_{ts}|$, respectively, as a function of the integrated luminosity.
At $1\,\text{ab}^{-1}$ of integrated luminosity and an electron
polarisation of $80\,\%$, the $2\sigma$ limits improve on existing limits from
the LHC~\cite{Khachatryan:2014nda} (interpreted by~\cite{AguilarSaavedra:2004wm})
by a factor of  $\approx 3.5$. Analyzing
Signal 3 alone, and even more when combining Signals 1, 2 and 3, will allow for the first time to achieve an accuracy of
the order of the actual SM value of $|V_{ts}^\text{SM}|=0.04108^{+0.0030}_{-0.0057}$ as
derived from an indirect global CKM matrix fit~\cite{Charles:2015gya}, and
will therefore represent a direct high precision measurement of this
important top quark property. In these studies, upper limits at the
$2\sigma$ level down to $|V_{ts}|<0.06$, and $|V_{td}|<0.06$ can be achieved.

\subsection{Top Quark Polarisation}

Single top quarks produced via the $e^+ p \rightarrow t \bar{\nu}$
processes possess a high degree of spin polarisation in terms of a basis
which decomposes the top quark spin in its rest frame along the
direction of the incoming $e$~beam~\cite{Atag:2006by}. It has been
investigated for $\sqrt{s} = 1.6\,\TeV$ in $e^+p$ scattering, that
the spin fraction defined as the ratio of the polarised cross section
to the unpolarised one, reaches $96\%$ allowing a detailed study of
the polarisation and the spin of the top quark. Exploring 
the angle
between the momentum direction of the charged lepton from top quark
decay and the spin quantisation axis in the top quark rest frame,
anomalous $Wtb$ couplings can be tested. Assuming a total systematic
uncertainty of $10\%$ the expected sensitivity for $\sqrt{s} = 1.6\,\TeV$
reaches $\pm 3\%$ for $f_2^L$, and $\pm 7\%$ for $f_2^R$ as defined in
Eq.~(\ref{eq:anom_Wtb_couplings}).

\subsection{Top-$\gamma$ and Top-$Z$ Couplings}

The LHeC is particularly well suited to measure the $t\bar{t}\gamma$ vertex,  
since in photoproducton of top quark pairs (see Fig.~\ref{graphs_eh_top}, right) the highly energetic incoming photon only couples to the top quark, and therefore the cross section directly depends on the $t\bar{t}\gamma$ vertex. This provides a direct measurement of the coupling between the top quark and the photon and therefore of another important top quark property, the top quark charge. In contrast, at the LHC the $t\bar{t}\gamma$ vertex vertex is probed in $t\bar{t}\gamma$ production, where the final state photon can also be produced from other vertices than the $t\bar{t}\gamma$ vertex, such as from initial state radiation or from radiation off charged top quark decay products. 

The LHeC also provides a high potential for measuring the $t\bar{t}\gamma$ magnetic and electric dipole moments (MDM and EDM, respectively) in $t\bar{t}$ production~\cite{Bouzas:2013jha}. In an effective Lagrangian framework, effective $t\bar{t}\gamma$ couplings can be written in terms of form factors:
\begin{equation}
\mathcal{L}_{Wtb} = e \, \bar{t} \left( Q_t \gamma^\mu A_\mu + \frac{1}{4\,m_t} \sigma^{\mu\nu}F_{\mu\nu} \, (\kappa+i\tilde{\kappa}\gamma_5) \right) t \,\, +h.c.
\label{eq:anom_ttgamma_couplings}
\end{equation}
with the anomalous MDM of the top quark, $\kappa$, and the EDM of the top quark, $\tilde{\kappa}$. The top quark charge is given by $e Q_t$. 
\begin{figure} [!b]
\centering
\includegraphics[width=0.65\textwidth]{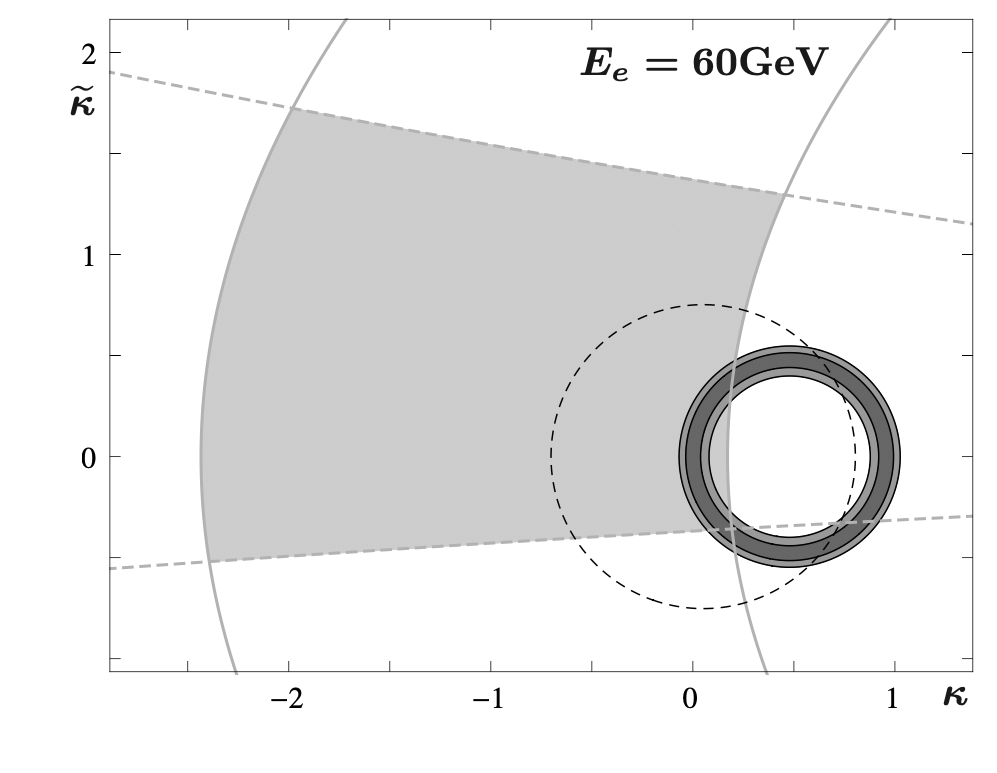}
\caption{
Allowed region of the magnetic dipole moment $\kappa$ and the electric dipole moment $\tilde{\kappa}$ of the top quark as expected in a measurement of the photoproduction cross section $\sigma (e(\gamma)p(g) \to t\bar{t})$ in semileptonic final states, assuming an experimental uncertainty of $8\%$ (dark grey), and of $16\%$ (dark+medium grey)~\cite{Bouzas:2013jha}. %
 Light gray area: region allowed by the measurements of the branching ratio (solid gray lines) and the CP asymmetry (dashed gray lines) of $B \to X_s \gamma$~\cite{Bouzas:2012av}. Black dashed line: region allowed by a hypothetical experimental result for $\sigma(pp \to t\bar{t}\gamma )$ utilizing semileptonic final states at the LHC at $\sqrt{s} = 14\,\TeV$ with phase-space cuts as defined in equations (5), (6) of Ref.~\cite{Bouzas:2012av} (including $E_T^\gamma > 10\,\GeV$), and assuming an experimental uncertainty of 5\%. 
 }
\label{fig_ttgamma_couplings}
\end{figure}

By solely measuring the $t\bar{t}$ production cross section, remarkably tight bounds can be derived on the MDM and the EDM of the top quark as presented in Fig.~\ref{fig_ttgamma_couplings}. In this parton level study, for the computation of the cross section a set of appropriate phase-space cuts are imposed on the final-state momenta. Applying further cuts to remove the background will result in a substantial reduction of the signal. It is therefore assumed that this would lead to a statistical uncertainty of about $8\%$, represented by the dark inner ring in Fig.~\ref{fig_ttgamma_couplings}. To include uncertainties due to mistagging and to allow for other unspecified sources of systematic uncertainty, it is assumed that the total uncertainty will be about $16\%$ corresponding to the full ring in Fig.~\ref{fig_ttgamma_couplings}. This would yield bounds of $-0.13 < \kappa < 0.18$, and $|\tilde{\kappa}| < 0.38$, respectively, at the $2\sigma$ C.L.
Figure~\ref{fig_ttgamma_couplings} shows that the LHeC could greatly improve the limits imposed by the indirect constraints from $b \to s\gamma$, and even the limits imposed by a future measurement of $t\bar{t}\gamma$ production at the LHC at $\sqrt{s} = 14\,\TeV$. 

Furthermore, the deep inelastic scattering (DIS) regime of $t\bar{t}$ production will allow to probe the $t\bar{t}Z$ coupling, albeit with less sensitivity~\cite{Bouzas:2013jha}.

\subsection{Top-Higgs Coupling}

The CP-nature of the top-Higgs coupling can be analysed
at the LHeC in $ep \to \bar{t}H$ production exploring the top quark
polarisation and other angular variables such as the difference in rapidity between the antitop quark and the Higgs boson. Measuring
just the fiducial inclusive production cross section gives already a powerful
probe of the CP properties of the $t\bar{t}H$
coupling~\cite{Coleppa:2017rgb}. 
Further details are given in Section~\ref{sec:topHinep}.

\subsection{Top Quark PDF and the Running of $\alpha_s$}

Parton distributions are usually released in a variable-flavor number
scheme, where the number of active flavors changes as the scale is
raised~\cite{Butterworth:2015oua}. However, $n_f=5$ is normally 
taken  by default as a maximum number of flavors, even though
in some PDF releases $n_f=6$ PDF sets are also made
available~\cite{Ball:2017nwa}. The top PDF is unlikely to be required
for precision phenomenology, even at very high scales, because the top
threshold is high enough that collinear resummation is not necessary
up to extremely large scales: indeed
$\frac{\alpha_s(M_t^2)}{\pi}\ln\frac{Q^2}{m_t^2}\sim \frac{1}{2}$  
only for  $Q\gtrsim 10^{6} \,m_t$. On the other hand, the use of $n_f=6$
active flavors in the running of $\alpha_s$ is important for precision
phenomenology, since the value of $\alpha_s$ with five and six active
flavors already differ by about 2\% at the TeV scale~\cite{Demartin:2010er}.
Investigations of the top quark structure inside the proton
are also discussed in Refs.~\cite{Boroun:2015yea,AbelleiraFernandez:2012cc}.

\subsection{FCNC Top Quark Couplings}
\label{SecFCNC}

Like all the Flavour Changing Neutral Currents (FCNCs) the top quark FCNC interactions are also extremely suppressed in the SM, which renders them a good test of new physics. The contributions from FCNC to top interactions can be parameterised via an effective theory and studied by analysing specific processes.

\begin{figure}
\centering
\includegraphics[width=0.9\textwidth]{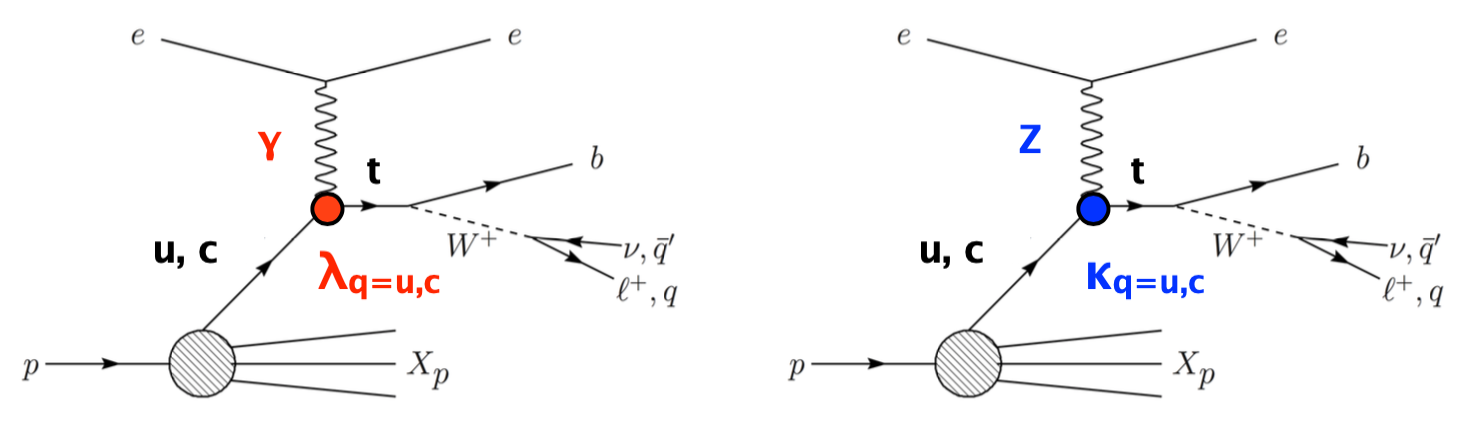}
\caption{
Example graphs for single top quark production via FCNC $tq\gamma$ (left) and $tuZ$ (right) couplings. 
}
\label{graphs_fcnc_top_quarks}
\end{figure}

NC DIS production of single top quarks can be explored to search for FCNC $tu\gamma$,  $tc\gamma$, $tuZ$,
and $tcZ$ couplings~\cite{TurkCakir:2017rvu,Cakir:2018ruj} as
represented by the Lagrangian
\begin{equation}
\mathcal{L}_{\textrm{FCNC}} = \sum_{q=u,c} \left( \frac{g_e}{2m_t} \bar{t} \sigma^{\mu\nu} 
      (\lambda^L_q P_L + \lambda^R_q P_R) q  A_{\mu\nu} 
+ \frac{g_W}{4c_Wm_Z} \bar{t} \sigma^{\mu\nu} 
      (\kappa^L_q P_L + \kappa^R_q P_R) q  Z_{\mu\nu} \right) 
 + h.c. \,\, .
 \label{eq:top_gamma_Z_fcnc}
\end{equation}
Here,  the electromagnetic (weak) coupling constant is denoted as $g_e$ ($g_W$), while
$c_W$ is the cosine of the weak mixing angle, $\lambda^{L,R}_q$ and
$\kappa^{L,R}_q$ are the anomalous top FCNC
coupling strengths. The values of these couplings vanish at the lowest order in
the SM. The study assumes that $\lambda^{L}_q = \lambda^{R}_q = \lambda_q$ and $\kappa^{L}_q = \kappa^{R}_q = \kappa_q$.
Top FCNC couplings as introduced in Eq.~(\ref{eq:top_gamma_Z_fcnc})
would lead to Feynman graphs as shown in Fig.~\ref{graphs_fcnc_top_quarks}.

In an elaborate analysis, events are selected that include at least one electron and three
jets from hadronic top quark decay, with high transverse momentum, and within the detector's pseudorapidity acceptance range. 
The invariant mass of two jets, reconstructing the $W$ boson mass, and an additional jet tagged as $b$-jet, are used to reconstruct the top quark mass. The respective distribution is used to further enhance signal over background events, mainly given by $W+\text{jets}$ production. Interference effects between signal and background are included. The DELPHES package~\cite{Ovyn:2009tx} is used to simulate the detector. 

\begin{figure} [!h]
\centering
\includegraphics[width=0.49\textwidth]{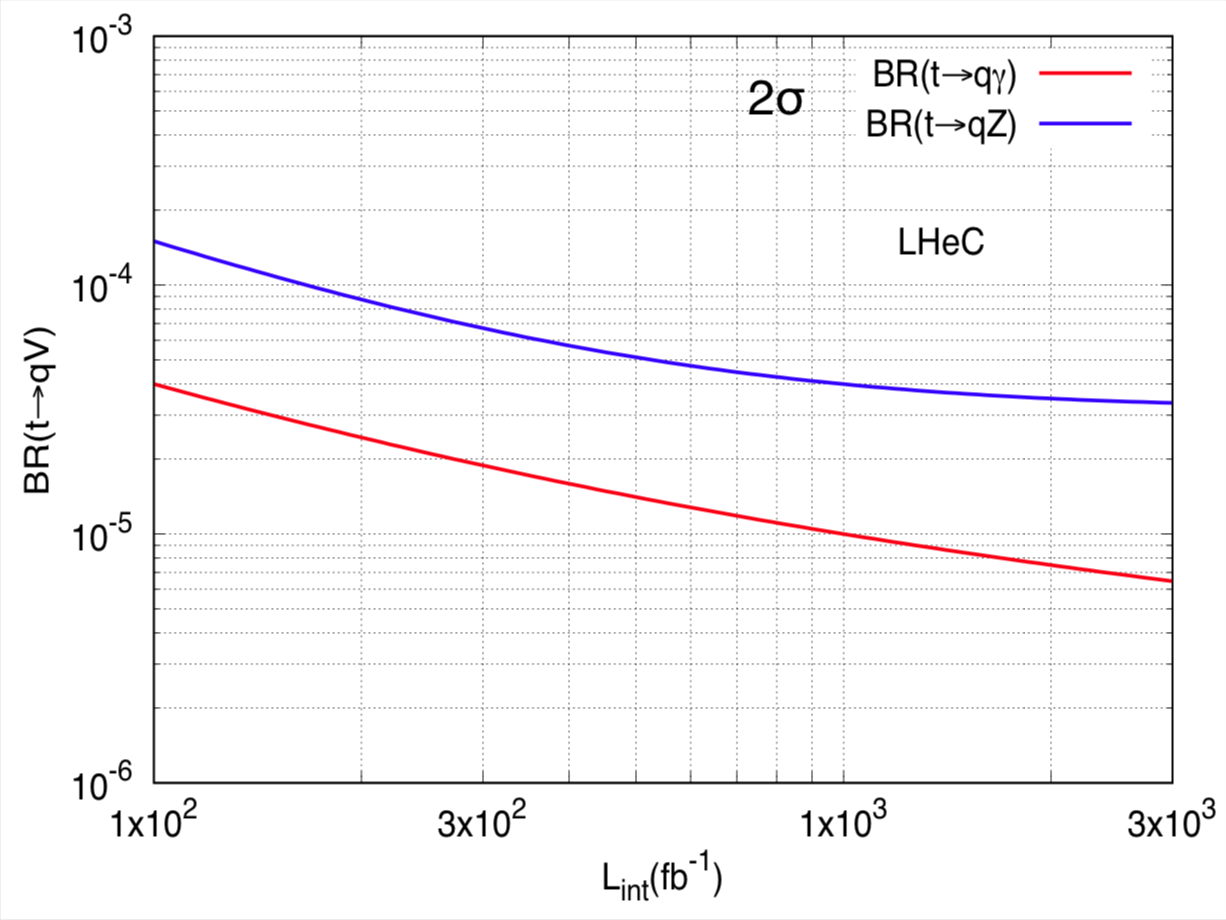}
\includegraphics[width=0.49\textwidth]{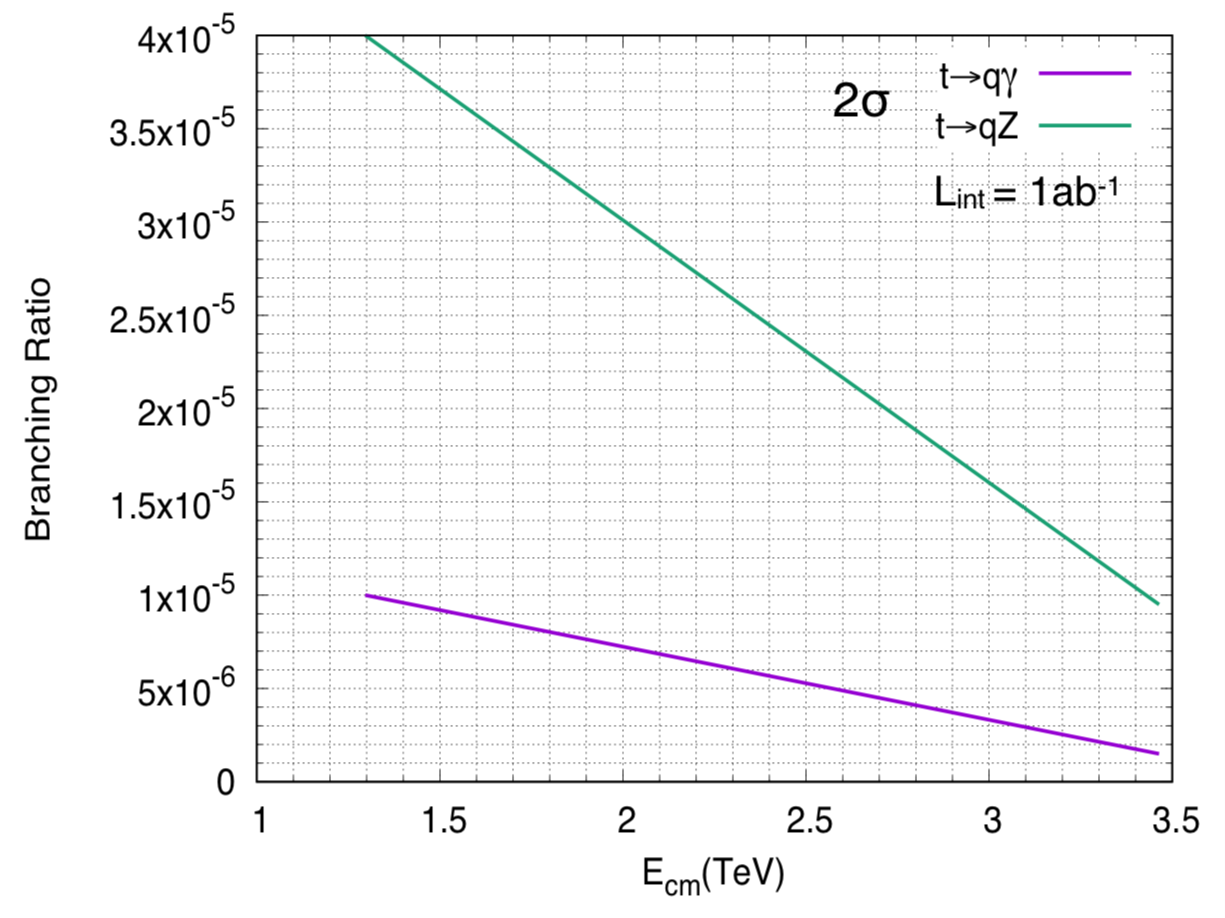}
\caption{
Expected sensitivities on
FCNC $t \to qV$ branching ratios (left),  as a function of the integrated luminosity~\cite{TurkCakir:2017rvu,Cakir:2018ruj}. Right: the expected upper limits on FCNC $t \to qV$ branching ratios are shown dependent on the centre-of-mass-energy.  
}
\label{fig_fcnc_top_quarks_couplings}
\end{figure}

Figure~\ref{fig_fcnc_top_quarks_couplings} (left) presents expected limits on the branching ratios $\text{BR}(t \to q \gamma)$
and $\text{BR}(t \to q Z)$ at the $2\sigma$ C.L., as a function of the integrated luminosity. Assuming an integrated luminosity of $1\,\textrm{ab}^{-1}$, 
limits of $\text{BR}(t \to q \gamma)<1 \cdot 10^{-5}$ and 
$\text{BR}(t \to q Z)< 4 \cdot 10^{-5}$ are expected.
This level of precision is close to concrete predictions by new phenomena models, that have the actual potential to produce FCNC top quark couplings, such as SUSY, little
Higgs, and technicolour.
These limits are expected to improve on existing limits from the LHC by one order of
magnitude~\cite{Abada:2019lih},  and will be similar to limits expected from the High
Luminosity-LHC (HL-LHC) with $3000\,\text{fb}^{-1}$~\cite{Azzi:2019yne}. They will also improve limits from the International Linear Collider
(ILC) with an integrated luminosity of $500\,\text{fb}^{-1}$ at a centre-of-mass energy of
$\sqrt{s}=250\,\GeV$~\cite{AguilarSaavedra:2001ab,Agashe:2013hma} by an order of magnitude (see also Fig.~\ref{fcncProj}).
The expected sensitivities on $\text{BR}(t \to q \gamma)$ and $\text{BR}(t \to q Z)$ are presented in Fig.~\ref{fig_fcnc_top_quarks_couplings} (right), as a
function of the centre-of-mass energy. 
At a future FCC-ep~\cite{Abada:2019lih} with a 60\,GeV electron
beam energy, and 
a 50\,TeV proton beam energy, leading to a centre-of-mass energy of
3.5\,TeV, the sensitivity to FCNC $tq\gamma$
couplings are expected to exceed sensitivities from the HL-LHC with $3000\,\text{fb}^{-1}$ at
$\sqrt{s}=14\,\TeV$~\cite{Azzi:2019yne}.

A further search for top quark FCNC $tuZ$ and $tcZ$ couplings has been performed~\cite{Behera:2018ryv} in a detailed analysis including a DELPHES~\cite{Ovyn:2009tx} detector simulation. The effective couplings investigated are of vector and tensor nature, the latter corresponding to those in Eq.~(\ref{eq:top_gamma_Z_fcnc}). The effect of these couplings is probed in single top quark production (see Fig.~\ref{graphs_fcnc_top_quarks} right). It is shown that the polar angle $\theta$ of the scattered initial state electron in association with top quark polarisation asymmetries constructed from angular distributions of the lepton from top quark decay, allow to distinguish the Lorentz structure of the couplings. From a multi-parameter analysis, a reach of the order ${\cal O} (10^{-2})$ in case of vector couplings and $0.1-0.5$ TeV$^{-1}$ in case of tensor couplings are obtained at the 95\% C.L. for an integrated luminosity of 2~ab$^{-1}$. This corresponds to respective limits on the branching ratio $\text{BR}(t \to u Z)$ of $9 \cdot 10^{-5}$ ($15 \cdot 10^{-5}$) for the left-(right-)handed vector coupling, and of $4 \cdot 10^{-5}$ ($6 \cdot 10^{-5}$) for the left-(right-)handed tensor coupling. 

\begin{figure} 
\centering
\includegraphics[width=0.4\textwidth]{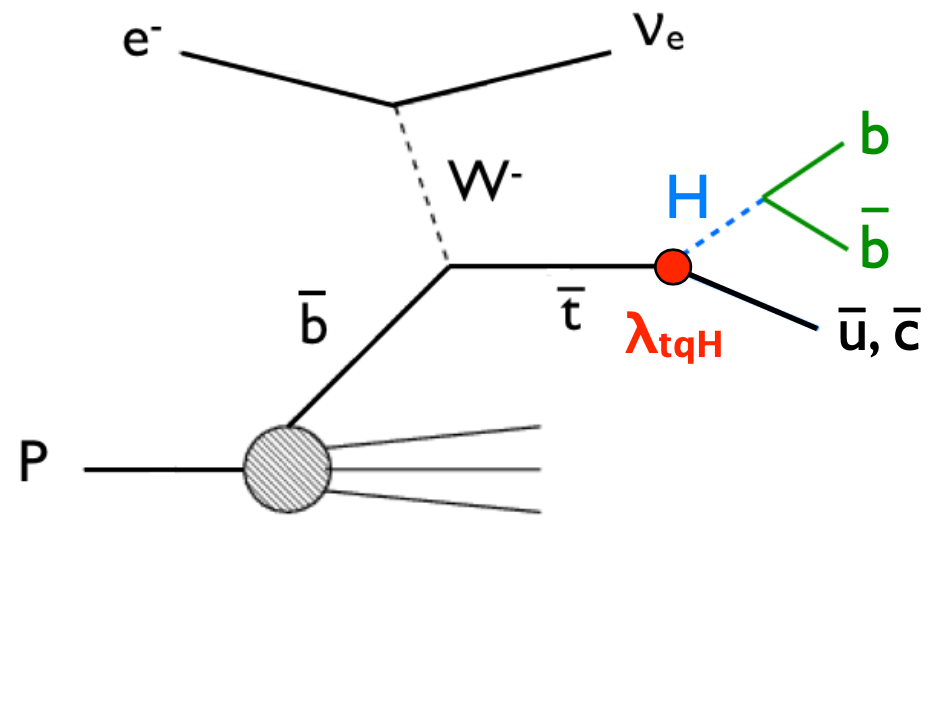}
\includegraphics[width=0.53\textwidth]{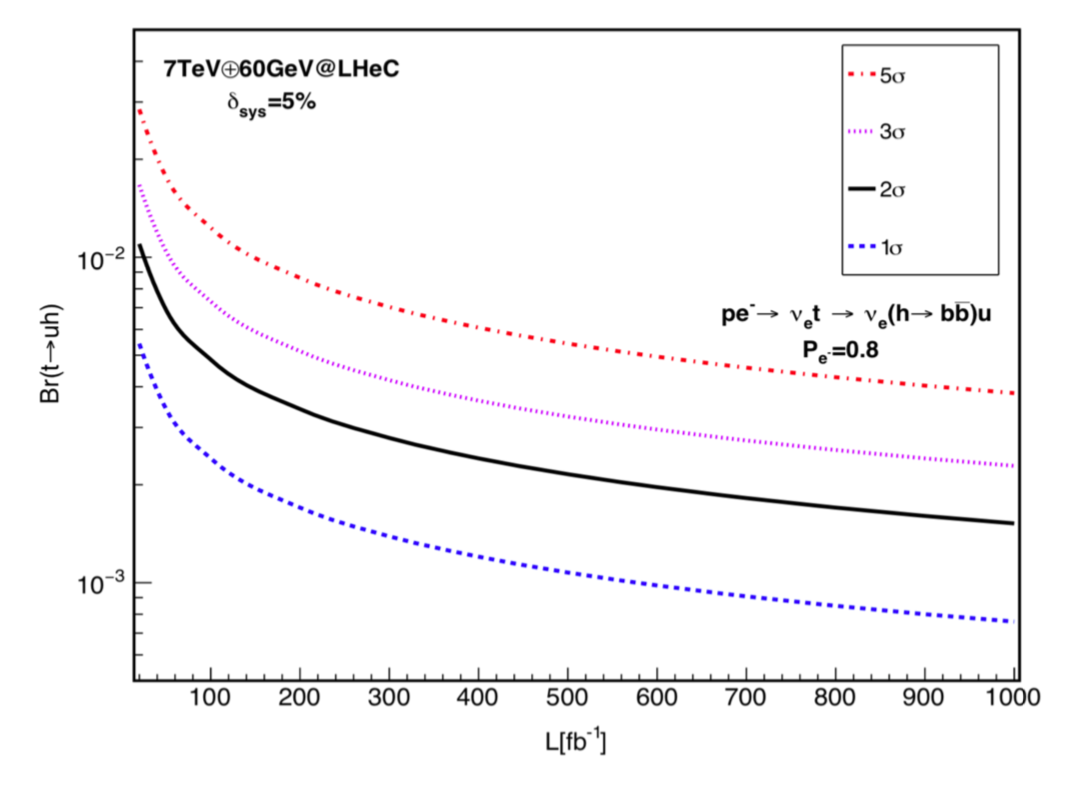}
\caption{
Example Feynman graph for associated single top quark and Higgs boson production via FCNC $tqH$ couplings (left). 
Expected sensitivities on
FCNC $t \to uH$ branching ratios are given as a function of the integrated luminosity~\cite{Sun:2016kek} (right). 
}
\label{fig_fcnc_top_Higgs_couplings}
\end{figure}

\begin{figure}[!th]
\centering
\includegraphics[width=0.85\textwidth]{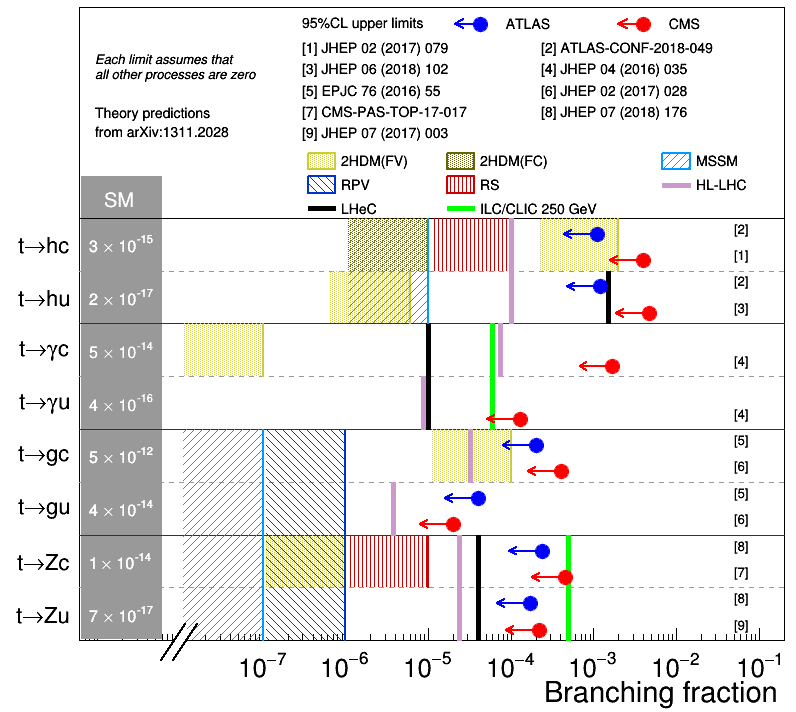}
\caption{
Summary of $95\%$ C.L. limits on top quark branching fractions in searches for FCNC in top quark production or decays. The LHeC results (black lines) are compared to current LHC limits (blue and red dots), to HL-LHC predictions with $3000\,\textrm{fb}^{-1}$ at $\sqrt{s}=14\,\TeV$~\cite{Azzi:2019yne} (magenta lines), and to predictions from a future ILC collider with $500\,\textrm{fb}^{-1}$ at $\sqrt{s}=250\,\GeV$~\cite{AguilarSaavedra:2001ab,Agashe:2013hma} (green lines). The results are also compared to various theory predictions (hached areas).
}
\label{fcncProj}
\end{figure}

Another sensitive search for FCNC $tqH$ couplings as defined in
\begin{equation}
\mathcal{L}_{\textrm{FCNC}}  = \kappa_{tuH} \, \bar{t}uH + \kappa_{tcH} \, \bar{t}cH + h.c.
\end{equation}
can be performed in CC DIS production as shown in Fig.~\ref{fig_fcnc_top_Higgs_couplings} (left). Here, singly produced top anti-quarks could decay via such couplings into a light up-type anti-quark and
a Higgs boson which further decays into a bottom quark-antiquark pair, $e^-p \to
\nu_e \bar{t} \to \nu_e H \bar{q} \to \nu_e b \bar{b}
\bar{q}$~\cite{Sun:2016kek}. 
Another signal process is given by the appearance of the FCNC $tqH$ coupling in the production vertex, involving a light quark from the proton interacting
via t-channel top quark exchange with a $W$ boson, which is radiated from the initial electron, and produces a $b$ 
quark and a Higgs boson decaying into a bottom quark-antiquark 
pair, $e^-p \to \nu_e H b \to \nu_e b \bar{b} b$~\cite{Sun:2016kek}. This
channel has a similar sensitivity as the previous one because of the clean experimental environment that can be achieved by requiring three jets to be identified as $b$-jets.
The most important backgrounds
are expected to be $Z \to b \bar{b}$, SM $H \to b \bar{b}$, and single top
quark production where the top quark decays hadronically. In order to account for the limited accuracy of the background yield calculations, $5\,\%$ of systematic uncertainty is added. In this analysis, parameterisations for the resolutions of electrons, photons, muons, jets and
unclustered energy are applied utilizing typical parameter values  as measured at the ATLAS experiment. 
Furthermore, it is assumed that the $b$-tag rate is 60\,\%, the $c$-jet fake rate is 10\%, and the
light-jet fake rate is 1\%. For the different signal contributions, separate selections are established and optimised. As a result, 
the expected upper limits on the branching ratio $\text{Br}(t \to Hu)$ with $1\sigma$, 
$2\sigma$, $3\sigma$, and $5\sigma$ C.L. are presented in 
Fig.~\ref{fig_fcnc_top_Higgs_couplings} (right), as a function of the integrated luminosity. The signal process $e^-p \to
\nu_e \bar{t} \to \nu_e H \bar{q} \to \nu_e b \bar{b}$ is presented. Upper limits of $\text{Br}(t \to Hu) < 1.5 \cdot
10^{-3}$  are expected at the $2\sigma$ C.L. for an integrated luminosity of $1\,\text{ab}^{-1}$.  
%

In Fig.~\ref{fcncProj} the different expected limits
on various FCNC top quark couplings
from the LHeC are summarised, and compared to results
from the LHC and the HL-LHC. This
documents the competitiveness of the LHeC results, and
clearly shows the complementarity of the results gained at different colliders. 

%

\subsection{Summary of Top Quark Physics}
%
Top quark physics at the LHeC represents a very rich and
diverse field of research involving high precision measurements of top
quark properties, and sensitive searches for new
physics. In particular the top couplings to the photon, the $W$ boson and possible FCNC interactions can be studied in a uniquely clean environment. 
One signature analysis
is the expected direct measurement of the CKM matrix element
$|V_{tb}|$ with a precision of less than $1\,\%$ in CC DIS. In top quark pair photoproduction the magnetic and electric dipole moments of the top quark can be probed directly with higher sensitivity than indirect limits from $b \to s\gamma$ and the potential limits from the LHC through $t\bar{t}\gamma$ production. Furthermore, FCNC top quark couplings can be studied with a precision high enough to explore those couplings in a regime that might be affected by actual new phenomena models, such as SUSY, little Higgs, and technicolour.

It has been shown~\cite{Abada:2019lih}, that results from future
$e^+e^-$-colliders, $eh$-colliders, and 
$hh$-colliders deliver complimentary information and will therefore
give us a more complete understanding of the properties of the
heaviest elementary particle known to date, and of the top quark
sector in general.


%


%% file: nuclearphysics/nuclearphysics.tex
\linenumbers
\lhectitlepage
\lhecinstructions
\subfilestableofcontents


\chapter{Nuclear Particle Physics with Electron-Ion  Scattering at the LHeC  \ourauthor{Nestor Armesto}}
\label{chapter:nuclearphysics}

\section{Introduction \ourauthor{Anna Stasto}}
\label{sec:NPP_intro}

\begin{figure}[!th]
  \centering
  \includegraphics[width=0.75\textwidth]{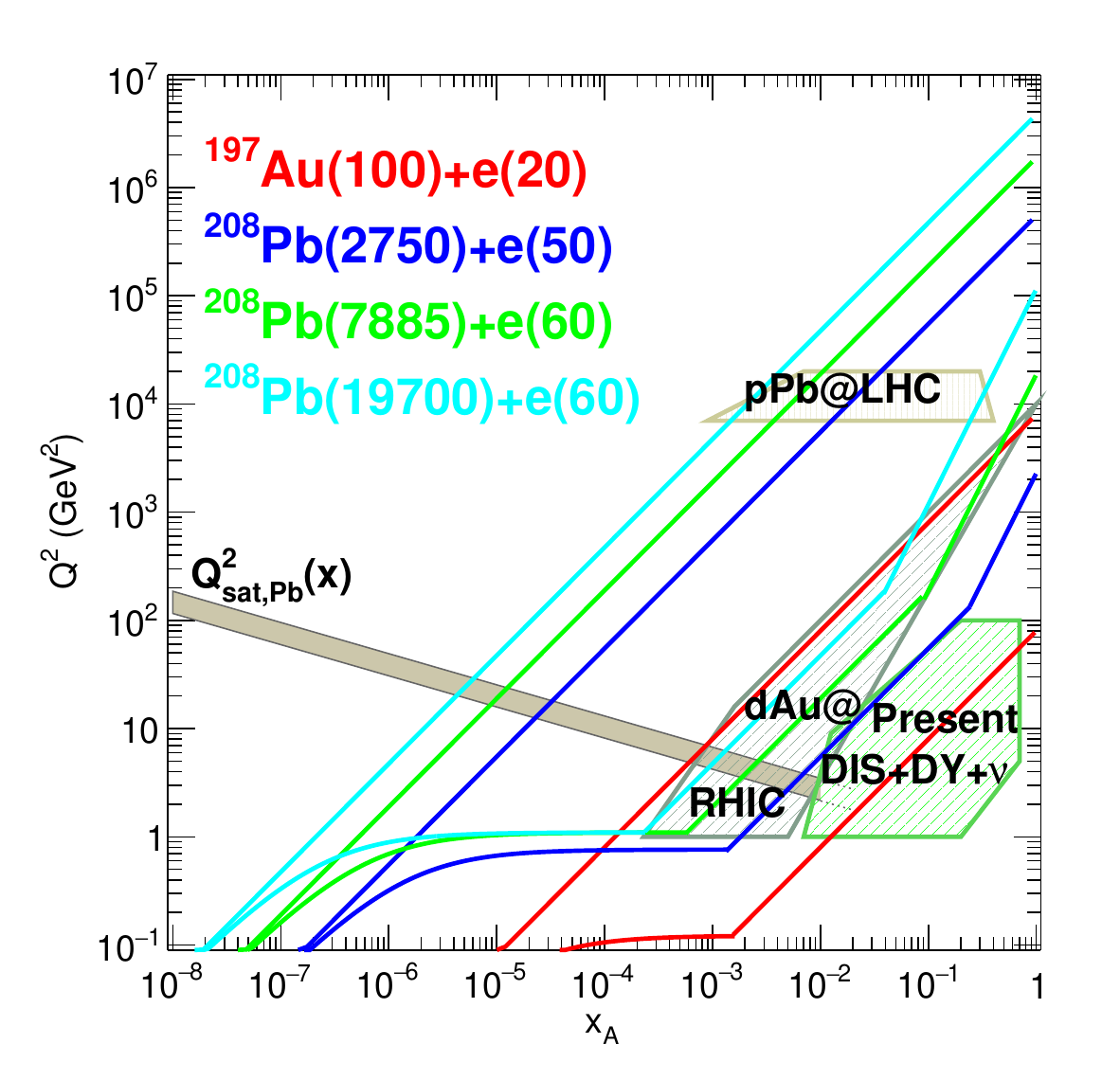}
\caption{Kinematic regions in the $x-Q^2$ plane explored by different data sets (charged lepton and neutrino DIS, DY, $d$Au at RHIC and $p$Pb at the LHC) used in present nPDF analyses~\cite{Eskola:2016oht}, compared to the ones achievable at the EIC (red), the LHeC (ERL against the HL-LHC beams, dark blue) and two FCC-eh versions (with Pb beams corresponding to proton energies  $E_p= 20$\,TeV - green and $E_p=50$\,TeV - light blue). Acceptance of the detector for the electrons is taken to be $1^\circ < \theta < 179 ^\circ$, and $0.01 (0.001)<y < 1$ for the EIC (all other colliders). The saturation scale $Q_{\textrm sat}$ shown here for indicative purposes only, see also~\cite{Salgado:2011wc}, has been drawn for a Pb nucleus considering an uncertainty $\sim 2$ and a behaviour with energy following the model in~\cite{GolecBiernat:1998js}. Note that it only indicates a region where saturation effects are expected to be important but there is no sharp transition between the linear and non-linear regimes.}
\label{fig:NPP_kinplot}
\end{figure}

The LHeC accelerator, in addition to being a powerful machine for exploring proton structure, will allow for the first time studies of DIS off nuclei in a collider mode at the energy frontier. The nuclear structure has been previously  studied in fixed target experiments with charged lepton and neutrino beams, see~\cite{Aubert:1983xm,Onengut:2005kv,Gomez:1993ri,Amaudruz:1995tq,Tzanov:2005kr,Arneodo:1995cs,Arneodo:1996rv,Ashman:1992kv,Arneodo:1996ru,Amaudruz:1991nw,Berge:1989hr,Arneodo:1992wf,Geesaman:1995yd} and references therein. Due to the energy limitations of the machines operating in this  mode, the kinematic range covered by these experiments is rather narrow,  mostly limited to relatively large values of $x\ge 0.01$ and low to moderate $Q^2$, in the range $Q^2 < 100$\,GeV$^2$. The precise kinematic range covered by  experiments is shown in Fig.~\ref{fig:NPP_kinplot}, where the DIS experiments overlap to a large degree with the data from hadronic collisions using the Drell-Yan (DY) process. These fixed target DIS and DY data dominate the data sets used in the fits for the nuclear parton distribution functions. In addition, in some analyses of nuclear PDFs, data on inclusive single hadron production in $d$Au collisions at RHIC and on EW bosons and dijets in $p$Pb collisions at the LHC are included. 

As is clear from Fig.~\ref{fig:NPP_kinplot}, the LHeC will be able to cover a very large range in $(x,Q^2)$ in $e$A, previously unexplored in experiments. It will extend the range in $x$ down to $\sim 10^{-6}$ and have a huge lever arm in $Q^2$ from very low values  up to $\sim 10^{6}$\,GeV$^2$.   It will also be complementary to the EIC~\cite{Accardi:2012qut} machine,  extending the range in $x$ and $Q^2$ by about two orders of magnitude with respect to it. The extension of these ranges will be even larger at the FCC-eh.

Due to large statistics and modern, specialised detectors, it will be possible to study nuclear structure at the LHeC with unprecedented precision in a kinematical range far wider than previously possible and with the controlled systematics of one single experiment. There are a large number of important physics topics that can be addressed in $e$A collisions at the LHeC:
\begin{itemize}
    \item A precise determination of nuclear parton densities for a single nucleus (lead, and eventually lighter ions) will be possible. In  particular, the current huge uncertainties in nuclear gluon and sea quark densities at low $x$ will be dramatically improved using the data from the LHeC.  In analogy to the proton PDF extraction described in previous sections, full flavour decomposition in the nuclear case could be achieved using both NC and CC data with heavy flavour identification.
    \item Precision measurement of semi-inclusive and exclusive processes will enable an exploration of new details of the nuclear structure. Similarly to the proton case,  DVCS and exclusive  vector-meson production will provide unique insight into 3D nuclear structure.
    \item The LHeC will offer unprecedented opportunities to extract diffractive parton densities in nuclei for the first time. A first detailed analysis~\cite{Armesto:2019gxy} indicates that the achievable precision on diffractive PDFs in nuclei will be comparable to that possible in the proton case.  The measurements of diffraction on protons and nuclei as well as the inclusive structure functions in the nuclear case will allow us to explore the very important  relation between nuclear shadowing and diffraction~\cite{Frankfurt:2011cs}.
    \item The LHeC will be able to test and  establish or exclude the phenomenon of parton saturation at low $x$ in protons and nuclei. According to the Color Glass Condensate framework~\cite{Gelis:2010nm,Kovchegov:2012mbw}, parton saturation is a density effect that can be achieved in two ways, either by decreasing  the value of $x$ or by increasing the size of the target by increasing $A$. The LHeC will be a unique machine to address both of their variations, such that the ideas of saturation could be precisely tested. It will be possible to search for parton saturation in a variety of ways which include, among others,  the search for tensions in DGLAP fits, the study of the diffraction, in particular the ratios of diffractive to inclusive cross sections, and  the study of  particle azimuthal de-correlations. 
    \item Finally, the LHeC machine in $e$A mode will have a huge impact onto physics explored in $p$A and AA collisions, see Sec.~\ref{sec:LHeConHLLHC_heavyions}, where it will provide vital input and constraints on the `baseline' initial state in nuclear collisions, measurements of the impact of a cold nuclear medium on hard probes and effects of hadronisation. It will also explore the initial state correlations on the final state observables relevant for understanding  collectivity in small systems explored in $pp$ or $p$A collisions.
\end{itemize}

As commented below, these aims will require an experimental apparatus with large rapidity coverage and associated forward and backward electron, photons, hadron and nuclear detectors. In addition the detector design should allow to precisely measure diffractive events in $e$A and allow the clean separation of radiative events, most important for the case of DVCS and exclusive diffraction.

Photonuclear interactions at high energies can also be studied through ultraperipheral collisions at RHIC and the LHC~\cite{Baltz:2007kq,Klein:2017vua,Citron:2018lsq,Klein:2019qfb} that offer an alternative although with less precision. This is briefly discussed in Chapter~\ref{chap:HLLHC} where the relation between the LHeC and the HL-LHC is presented.

In this Chapter we do not address issues on the nuclear modification on jet yields and fragmentation that are expected to show dramatic effects and to be of great importance for heavy-ion collisions. All these aspects were previously discussed in Ref.~\cite{AbelleiraFernandez:2012cc}. Besides, electron-deuteron collisions that offer additional possibilities for determining proton and neutron parton densities, and for studying weak interactions with neutron targets at high energies, are not considered here, see Ref.~\cite{AbelleiraFernandez:2012cc} where an analysis of parton densities in $e$D collisions can be found.


%
\section{Nuclear Parton Densities \ourauthor{Nestor Armesto}}
\label{sec:NPP_nPDFs}

PDFs are essential ingredients in our understanding of the dynamics of the strong interaction. First, they encode important information about  the structure of hadrons~\cite{Ioffe:2010zz,Collins:2011zzd}. Second, they are indispensable for  the description of hadronic collisions within standard collinear factorisation~\cite{Collins:1989gx}. Concerning nuclei, it has been known for more than 40 years that structure functions are strongly affected by the nuclear environment~\cite{Arneodo:1992wf,Geesaman:1995yd} so that they cannot be interpreted as a  simple superposition of structure functions of free nucleons. 
In the standard approach, within collinear factorization, the nuclear modification is included in the parametrisation of the parton densities. This means that the parton densities in a bound nucleon are different from those in a free nucleon, and the difference is encoded in the non-perturbative initial conditions of the parton densities at some low, initial scale $Q_0^2$.
 The present status of nuclear parton densities (nPDFs), see for example~\cite{Paukkunen:2017bbm,Paukkunen:2018kmm},  can be summarised as follows:
\begin{itemize}
\item Modern analyses~\cite{Eskola:2009uj,deFlorian:2011fp,Kovarik:2015cma,Eskola:2016oht} are performed at next-to-leading order (NLO) and next-to-next-to-leading order (NNLO)~\cite{Khanpour:2016pph,AbdulKhalek:2019mzd}. Differences between the different groups mainly arise from the different sets of data  included in the analyses~\footnote{The main difference lies in the use or not of neutrino-Pb cross sections (whose usage has been controversial~\cite{Paukkunen:2010hb,Kovarik:2010uv,Paukkunen:2013grz}, particularly the NuTeV data~\cite{Onengut:2005kv} from the Fe nucleus) from CHORUS and $\pi^{0,\pm}$ transverse momentum spectra from $d$Au collisions at the Relativistic Heavy Ion Collider (RHIC).} and from the different functional forms employed for the initial conditions. 
\item Many sets of data are presented as ratios of cross section for a given nucleus over that in deuterium, which is loosely bound and isoscalar. Therefore, it has become customary to work in terms of ratios of nPDFs:
\begin{equation}
R_i(x,Q^2)=\frac{f_i^{\textrm A}(x,Q^2)}{A f_i^{p}(x,Q^2)}\ ,\ \ i=u,d,s,c,b,g,\dots ,
\label{eq:ratio}
\end{equation}
with $f_i^{p\textrm{(A)}}(x,Q^2)$ the corresponding parton density in a free proton $p$ or in nucleus A. 
These nuclear modification factors are parametrised at  initial scale $Q_0^2$ (assuming isospin symmetry to hold). The nPDFs are then obtained multiplying the nuclear modification factors by some given set of free proton PDFs.
\item The available data come from a large variety of nuclei and the number of data points for any of them individually is very small compared to the proton analyses.  In particular, for the Pb nucleus there are less than 50 points coming from the fixed target DIS and DY experiments and from particle production data in $p$Pb collisions at the LHC. 
The fit for a single nucleus is therefore impossible and the modelling of the $A$-dependence of the parameters in the initial conditions becomes mandatory~\cite{Eskola:2016oht,Kovarik:2015cma}. The most up to date analyses include between 1000 and 2000 data points for 14 nuclei.
\item The kinematic coverage in $Q^2$ and $x$ with existing data is very small compared to that of present hadronic colliders. 
The ultimate precision and large coverage of the kinematic plane for nPDFs can only be provided by a high energy electron-ion collider. Meanwhile,
the only experimental collision system where nPDFs can be currently constrained are hadronic and ultraperipheral collisions (UPCs). It is important to stress that extracting PDFs from these collisions presents many theoretical challenges. These are related to the question of applicability of collinear factorization for nuclear collisions,  higher twist effects, scale choices and other theoretical uncertainties.

 \end{itemize}

All parton species are very weakly constrained at small~$x<10^{-2}$~\cite{Armesto:2006ph}, gluons are poorly known at large~$x>0.2$, and the flavour decomposition is largely unknown - a natural fact for $u$ and $d$ due to the approximate isospin symmetry in nuclei~\footnote{The $u$-$d$ difference is suppressed by a factor $2Z/A-1$.}. 
The impact of presently available LHC data, studied using reweighting~\cite{Paukkunen:2014zia,Eskola:2019dui} in~\cite{Armesto:2015lrg,Kusina:2016fxy} and included in the fit in~\cite{Eskola:2016oht}, is quite modest with some constrains on the gluon and the strange quark in the region $0.01<x<0.3$. 
On the other hand,  theoretical predictions for nuclear shadowing of quark and gluon PDFs  based on $s$-channel unitarity and diffractive nucleon PDFs are available down to $x \sim 10^{-4} -10^{-5}$~\cite{Frankfurt:2011cs,Armesto:2003fi,Armesto:2010kr,Krelina:2020ipn}. Predictions on the flavour dependence of nuclear effects in the antishadowing region~\cite{Brodsky:2004qa} cannot be confirmed with present data.

Future runs at the LHC will offer some further possibilities for improving our knowledge on nPDFs~\cite{Citron:2018lsq}. However, the ideal place to determine parton densities is DIS, either at the Electron Ion Collider (EIC)~\cite{Accardi:2012qut} in the USA or, in a much larger kinematic domain (see Fig.~\ref{fig:NPP_kinplot}), at the LHeC. DIS measurements in such configurations offer unprecedented possibilities to enlarge our knowledge of parton densities through a complete unfolding of all flavours. 

In the following, we show the possibilities for constraining the PDFs for a  Pb nucleus at the LHeC. In the next subsection, Subsec.~\ref{sec:NPP_nPDFs_pseudodata}, we discuss the corresponding pseudodata for the inclusive cross section in electron-nucleus scattering. Next, in Subsec.~\ref{sec:NPP_nPDFs_globalfit} we discuss how the pseudodata  will be introduced in a global nPDF fit. Finally, in Subsec.~\ref{sec:NPP_nPDFs_Pbonly} it is demonstrated how   the PDFs of Pb can be extracted with a very good precision from the LHeC data only, without requiring any other set of data.

\subsection{Pseudodata \ourauthor{Max Klein}}
\label{sec:NPP_nPDFs_pseudodata}

$e$A scattering at the LHeC provides measurements of inclusive neutral and charged
current cross sections in the deep inelastic
scattering region $1<Q^2<5 \cdot 10^5$\,GeV$^2$ and $x$ from a few times $10^{-6}$ 
to near $x=1$, see Ref.~\cite{Klein:2016uxu} which contains the material that is summarised in this Subsection. Achieving $Q^2$ much larger than the $W$-boson mass squared, CC measurements together with the NC contribution from photon and $Z$-boson exchange will be most important for flavour separation. In CC, charm tagging will determine the anti-strange quark contribution to  $10-20$\,\%
accuracy. In NC, charm and beauty tagging will precisely constrain nuclear $xc$ and $xb$. The use of data from a single experiment will allow nPDF uncertainties to follow from a straightforward $\Delta \chi^2 =1$
criterion. As often emphasised, the knowledge of the heavy quark densities is of key importance for our
understanding nuclear structure and for the development of QCD. 

The subsequent QCD analyses  of LHeC   cross section pseudodata employ sets of simulated NC and CC measurements. The corresponding assumptions on precision  are
summarised in Table\,\ref{tabsim}, see Ref.~\cite{Klein:2016uxu}. The cross section simulations were done
 employing derivative formulae from~\cite{Blumlein:1990dj}. They compare well to detailed Monte Carlo simulations for the conditions of the H1 experiment. The assumptions made, reasonable when compared to the H1 achievements, leave room for further improvements if new detector techniques and higher statistics would be considered. A special challenge is the
control of radiative corrections which in $e$A scattering grow $\propto Z^2$. Therefore, the LHeC detector requires to be equipped
with  photon detectors. The exploitation of energy-momentum
conservation, via  $E-p_z$ cuts, should further reduce the effect of photon
radiation to a few per cent level. Note that semi-inclusive
measurements of the $s$, $c$ and $b$ quark distributions contain further
uncertainties for tagging, acceptance and background influences.
\begin{table}[!htb]
  \centering
  \small
  \begin{tabular}{lc}
    \toprule
    Source of uncertainty  & Error on the source or cross section    \\
    \midrule
    Scattered electron energy scale  & 0.1\,\%  \\
    Scattered electron polar angle  & 0.1\,mrad  \\
    Hadronic energy scale  & 0.5\,\%  \\
    Calorimeter noise ($y < 0.01$)  & 1--3\,\%  \\
    Radiative corrections & 1--2\,\% \\
    Photoproduction background  & 1\,\%  \\
    Global efficiency error &  0.7\,\%  \\
    \bottomrule
  \end{tabular}
  \caption{Summary of assumed systematic uncertainties for future 
    inclusive cross section measurements at the LHeC. Taken from Ref.~\cite{Klein:2016uxu}.}
  \label{tabsim}       
\end{table}

\begin{figure}[!th]
  \centering
  \includegraphics[width=0.42\textwidth,trim={0 0 30 0},clip]{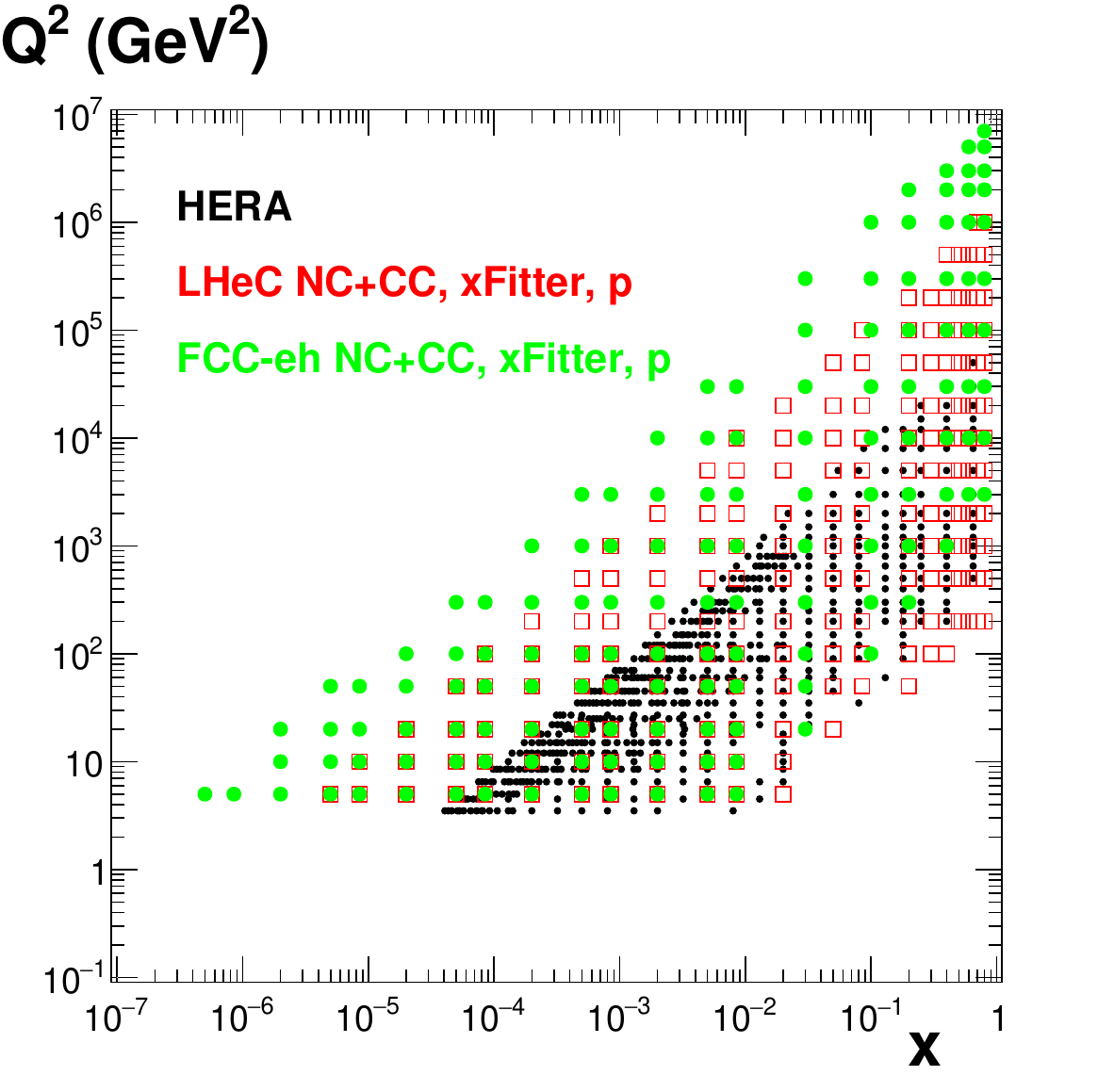}
  \hspace{0.02\textwidth}
  \includegraphics[width=0.42\textwidth,trim={0 0 30 0},clip]{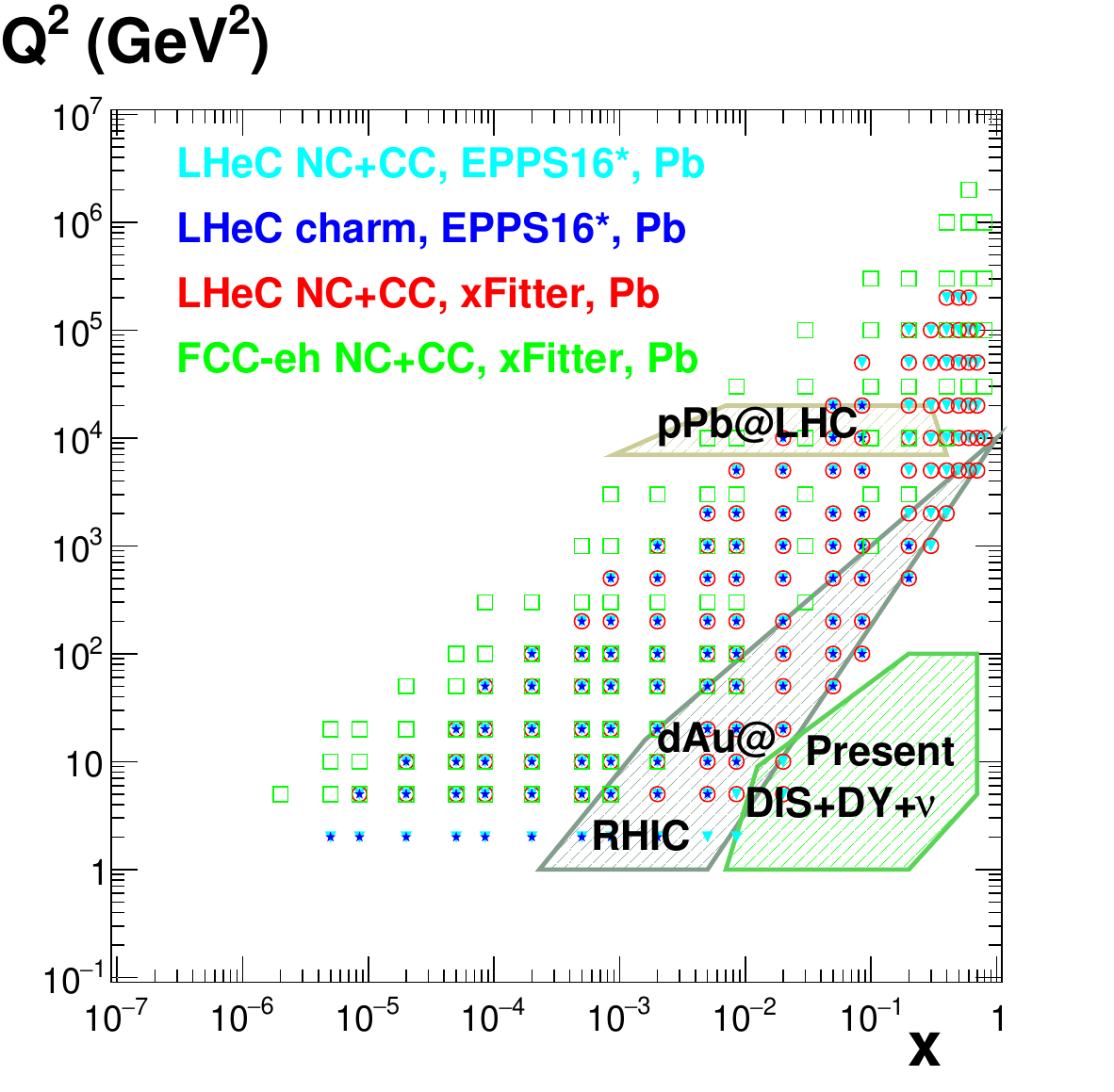}
\caption{Left: kinematic $x-Q^2$ plot of the NC+CC pseudodata on a proton at the LHeC (red symbols) and the FCC-eh (green symbols) used in the \emph{xFitter} analysis in Section~\ref{sec:NPP_nPDFs_Pbonly}; data used in analysis at HERA (black symbols) are shown for comparison. Right: kinematic $x-Q^2$ plot of the pseudodata on Pb used in the EPPS16 analysis at the LHeC (NC+CC, light blue symbols, and charm, dark blue symbols) in Section~\ref{sec:NPP_nPDFs_globalfit}, and in the \emph{xFitter} analysis in Subsec.~\ref{sec:NPP_nPDFs_Pbonly} (at the LHeC, red symbols, and the FCC-eh, green symbols); the regions explored by currently available data sets (charged lepton and neutrino DIS, DY, $d$Au at RHIC and $p$Pb at the LHC) used in present nPDF analyses~\cite{Eskola:2016oht} are shown for comparison.}
\label{fig:NPP_pseudodata}
\end{figure}

Fig.~\ref{fig:NPP_pseudodata} illustrates the kinematic reach of the NC+CC pseudodata at the LHeC and the FCC-eh, in $ep$ and $e$Pb collisions (for per nucleon integrated luminosities $\le 1$ and 10\,fb$^{-1}$ respectively).
In addition to inclusive data, semi-inclusive measurements with flavour sensitivity are also included. A determination of  
the strange, charm,
beauty and even top PDFs will thus become possible. The main technique required for flavour studies is charm  
(in CC for $xs$, in NC for $xc$) and beauty tagging (in NC for $xb$), for which the following consideration are in order, see Ref.~\cite{Klein:2016uxu}.
The transverse extension of the LHeC beam spot of the LHeC is about $(7$\,$\mu$m$)^2$.
Typical decay lengths of charm and beauty particles are of hundreds of
$\mu$m, to be compared with the resolution of a few microns for modern Si detectors. The experimental challenges are then the
forward tagging acceptance, similar to the situation at the HL-LHC, and the beam pipe radius,
coping at the LHeC with strong synchrotron radiation effects. 

\begin{figure}[!th]
  \centering
  \includegraphics[width=0.47\textwidth,trim={80 0 120 0},clip]{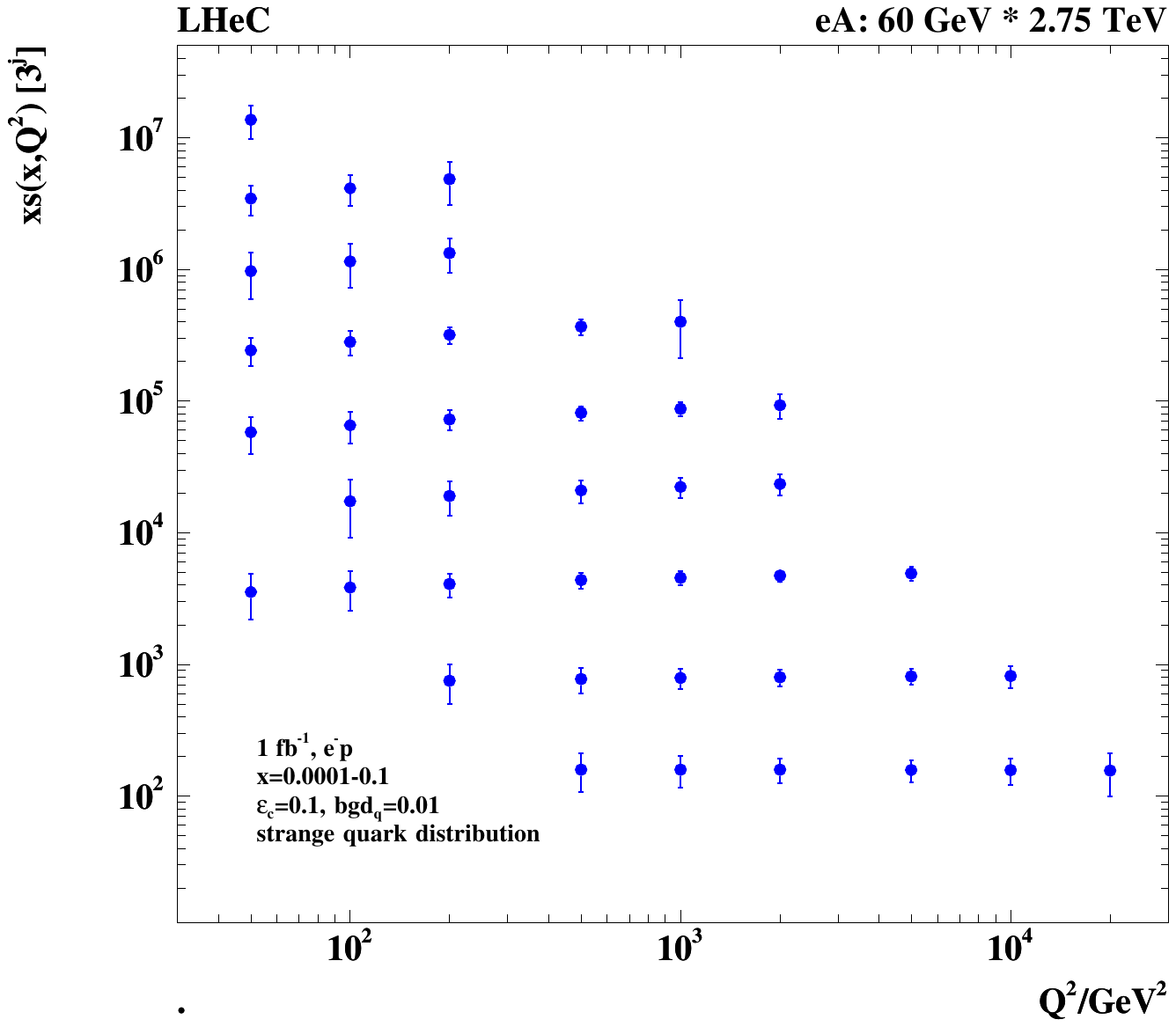}
  \caption{
    Simulation of the measurement of the (anti)-strange quark distribution
    $x\bar{s}(x,Q^2)$ in charged current $e$A scattering through
    the $t$-channel reaction  $W^- \bar{s} \rightarrow c$. The data are plotted with full systematic
    and statistical errors added in quadrature. Taken from Ref.~\cite{Klein:2016uxu}.
}
\label{fig:NPP_pseudodatabs1}       
\end{figure}

\begin{figure}[!th]
  \centering
  \includegraphics[width=0.47\textwidth,trim={80 0 120 0},clip]{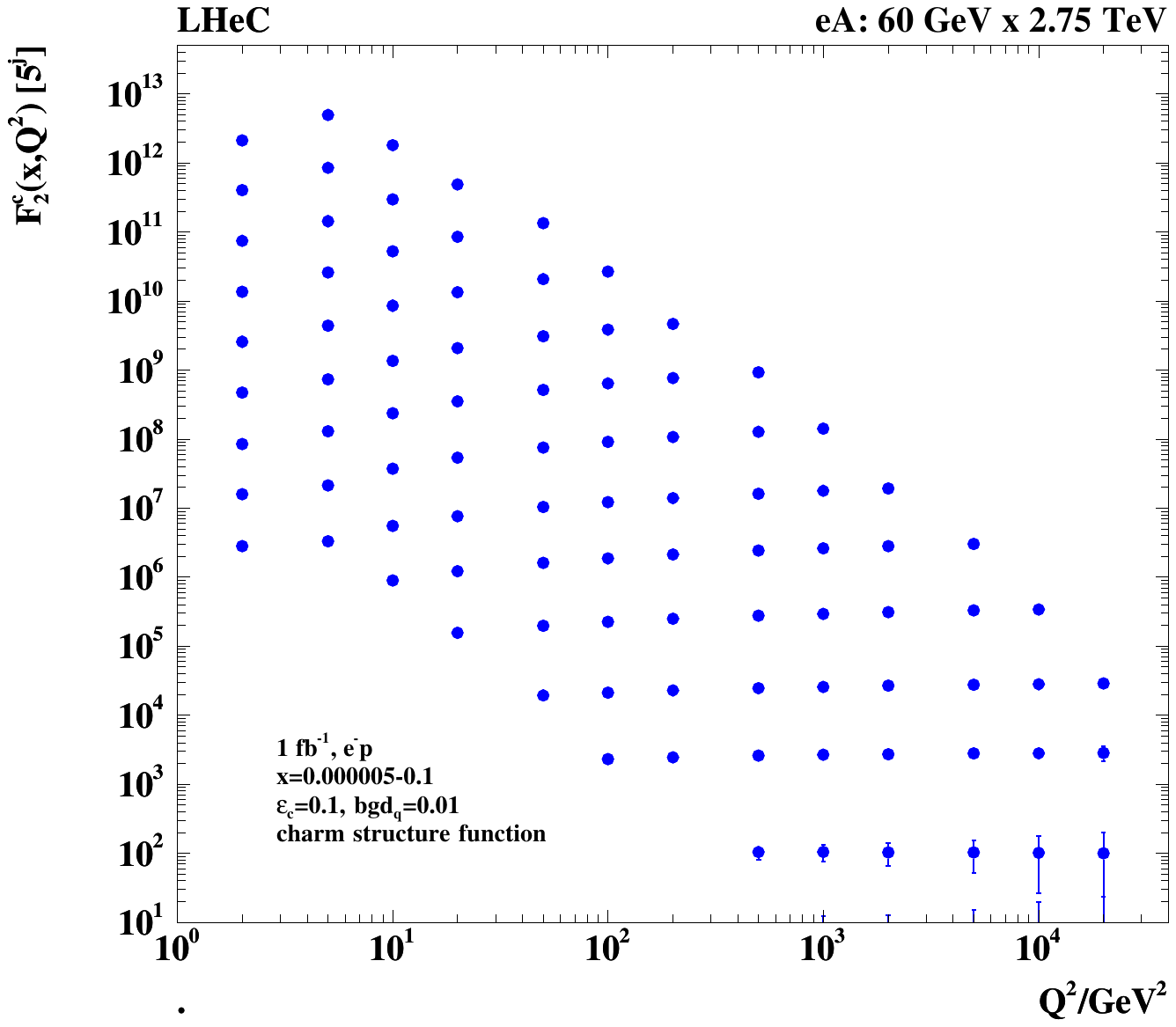}
  \hspace{0.02\textwidth}
  \includegraphics[width=0.47\textwidth,trim={80 0 120 0},clip]{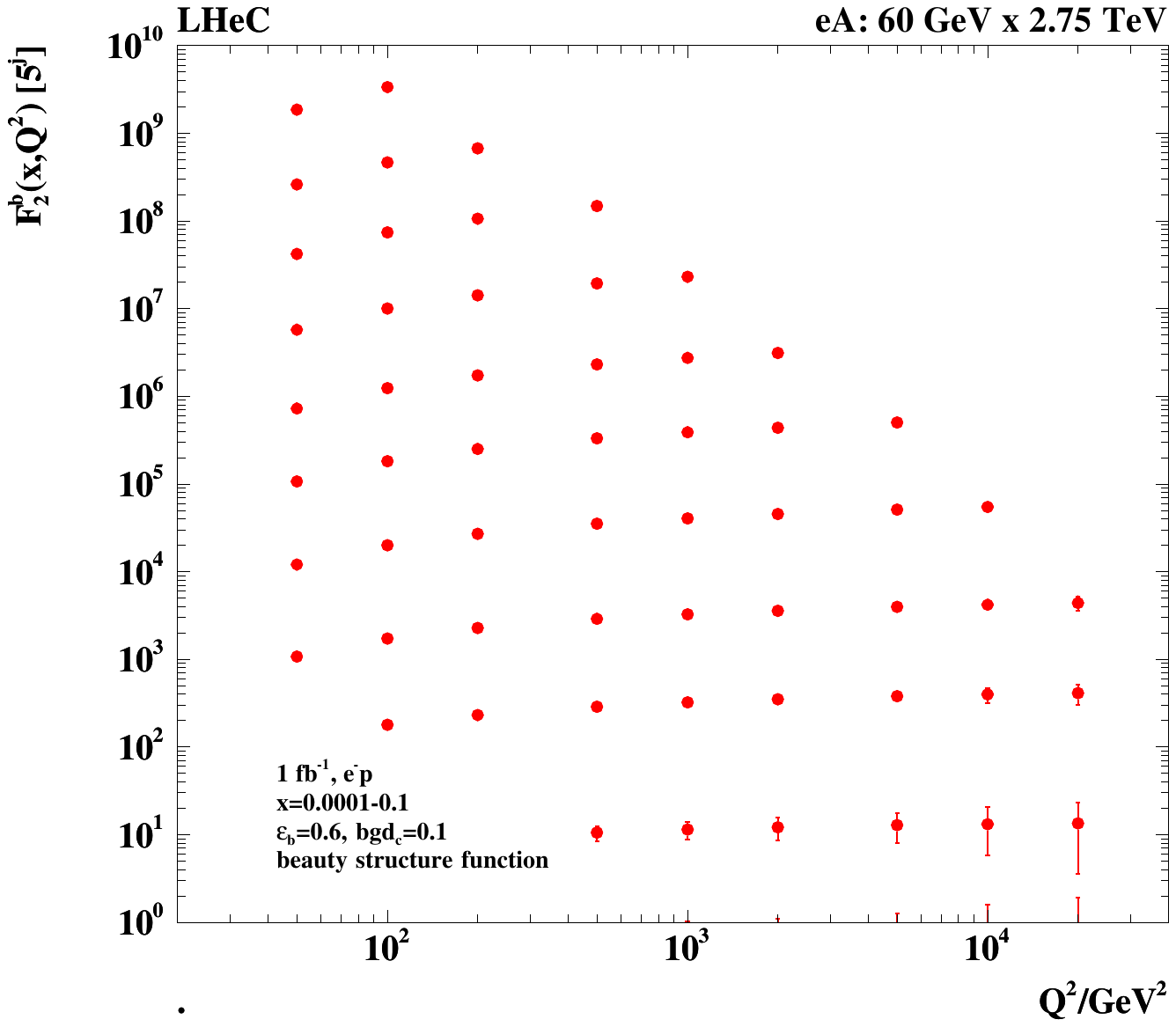}
  \caption{
    Left: Simulation of the measurement of the charm quark distribution
    expressed as $F_2^c=e_c^2x(c+\bar{c})$ in neutral current $e$A scattering;
    Right: Simulation of the measurement of the bottom quark distribution
    expressed as $F_2^b=e_b^2x(b+\bar{b})$  in neutral current $e$A scattering.
    The data are plotted with full systematic
    and statistical errors added in quadrature. Taken from Ref.~\cite{Klein:2016uxu}.
}
\label{fig:NPP_pseudodatabs2}       
\end{figure}

A study was made, Ref.~\cite{Klein:2016uxu}, of the possibilities for measurements of the nuclear anti-strange density 
(see Fig.~\ref{fig:NPP_pseudodatabs1}) through impact parameter tagging in $e$A CC scattering,
and of the charm and beauty cross sections in NC (see Fig.~\ref{fig:NPP_pseudodatabs2}).
Charm and beauty tagging efficiencies were assumed
to be $10$\,\% and $60$\,\%, respectively, following experience on heavy flavour tagging at HERA and ATLAS. Control of the light quark background
in the charm analysis and of  the charm background in the beauty tagging sample is assumed to be $1$ and $10$\,\%, respectively. Tagging efficiencies and background contaminations
affect the statistical error. Besides, an additional systematic
error of $3~(5)$\,\% is assumed in the simulated NC (CC) measurements.
These assumptions result in very promising measurements of  the heavier quark distributions,
to about $10-20$\,\% ($3-5$\,\%) total uncertainty on the strange (charm and beauty) measurements, for
$10^{-4}<x<0.1$ and $Q^2$ extending from below  threshold
$m_Q^2$ up to a few times $10^4$\,GeV$^2$.

\subsection{Nuclear gluon PDFs in a global-fit context \ourauthor{Hannu Paukkunen}}
\label{sec:NPP_nPDFs_globalfit}

\begin{figure}[!th]
  \centering
  \includegraphics[width=.46\textwidth,angle=0]{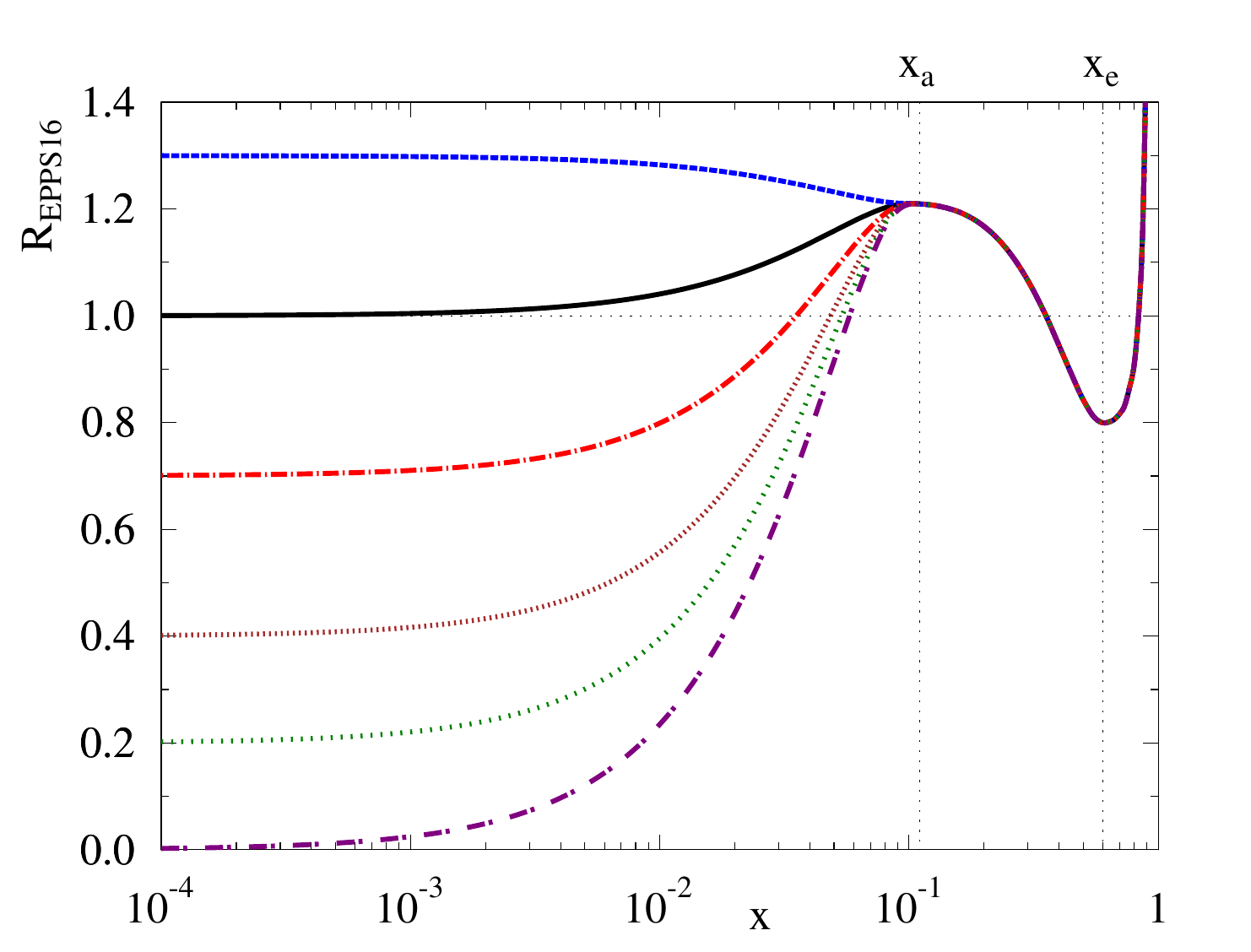}
  \includegraphics[width=.46\textwidth,angle=0]{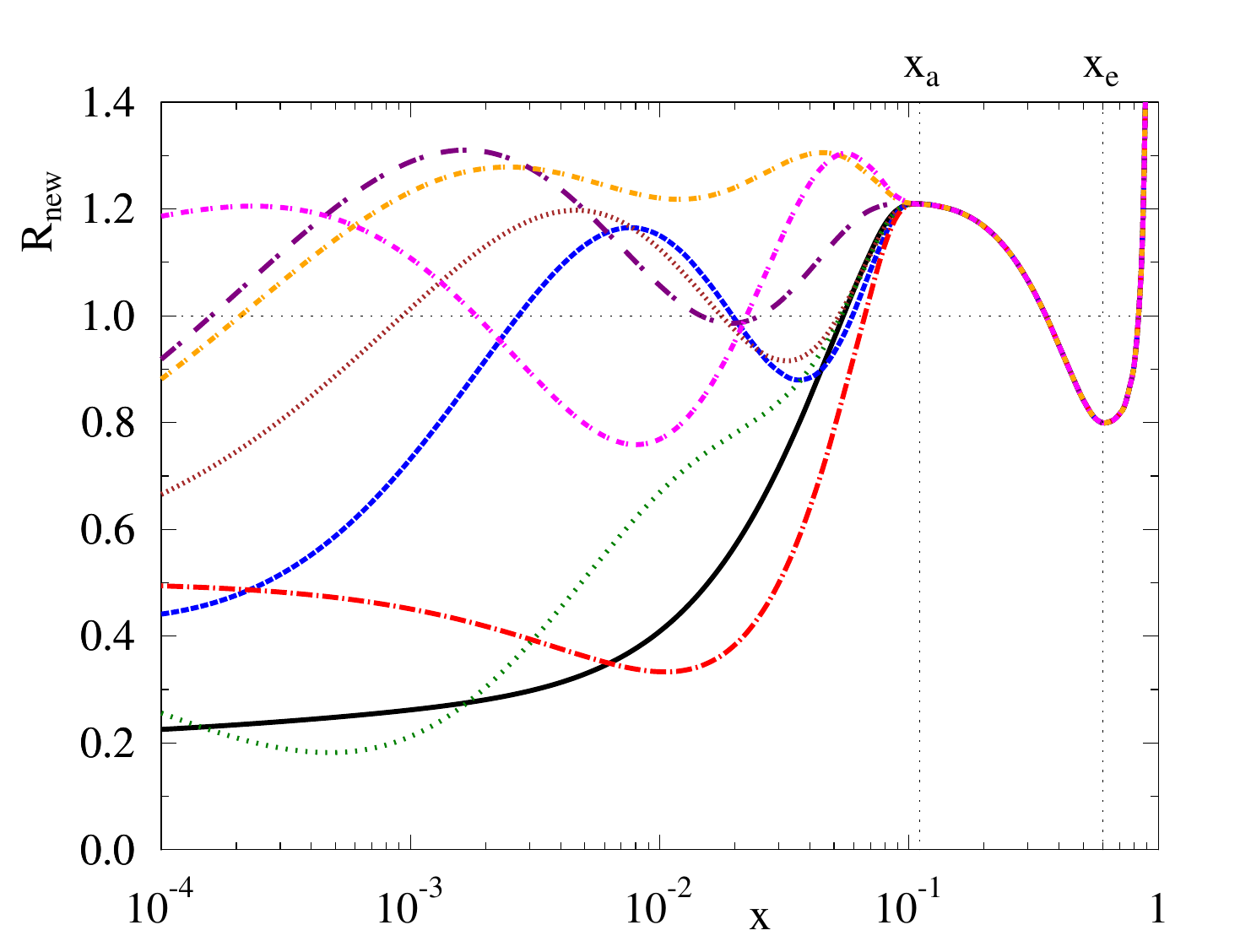}
  \caption{Left: Illustration of the functional behaviours allowed at small $x$ in the EPPS16 analysis. Right: Illustration of the possible functional variations at small $x$ in the extended parametrisation that we employ here.}
\label{fig:NPP_fitforms}
\end{figure}

\begin{figure}[!p]
  \centering
  \includegraphics[width=.75\textwidth,angle=0]{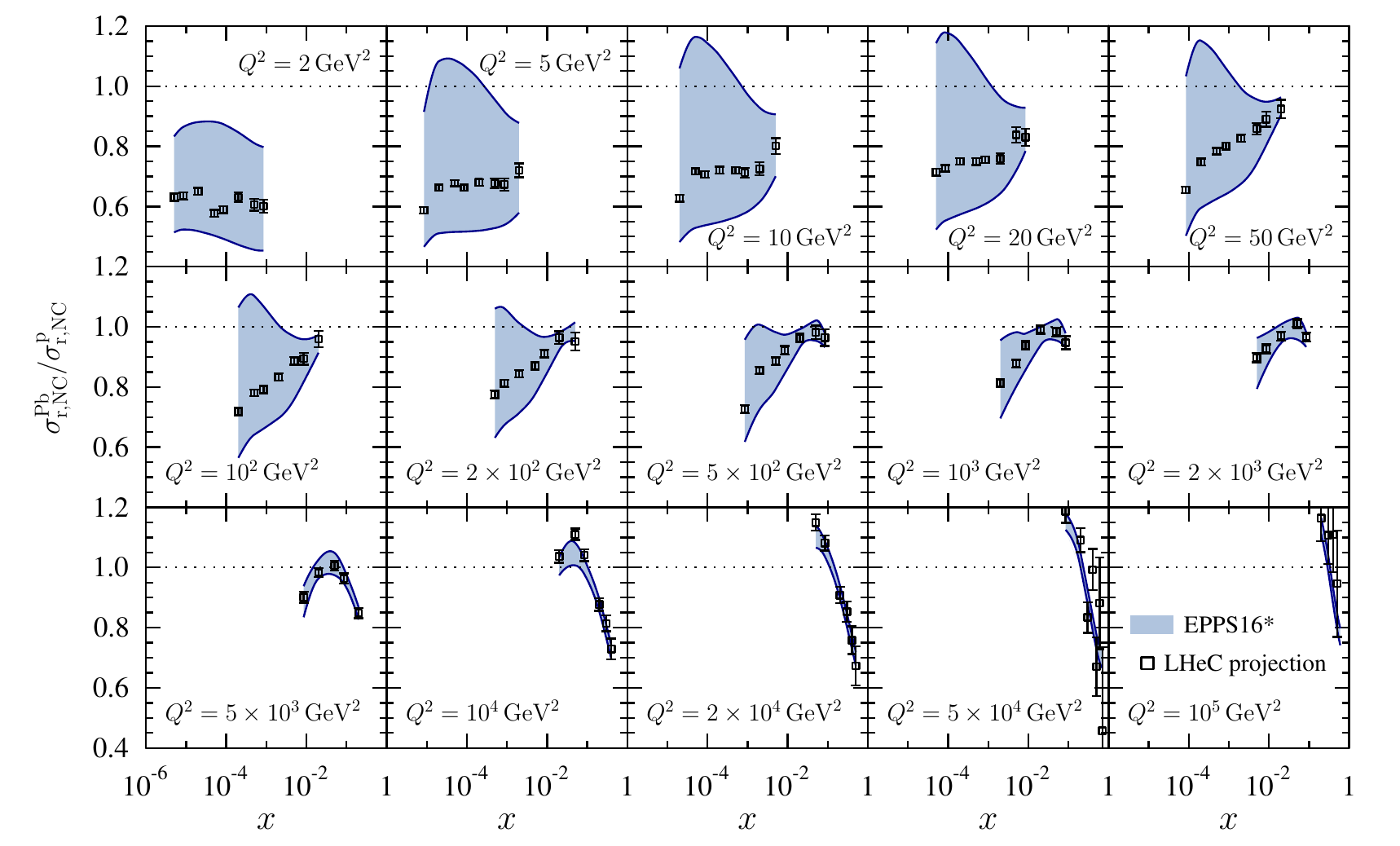}
  \includegraphics[width=.75\textwidth,angle=0]{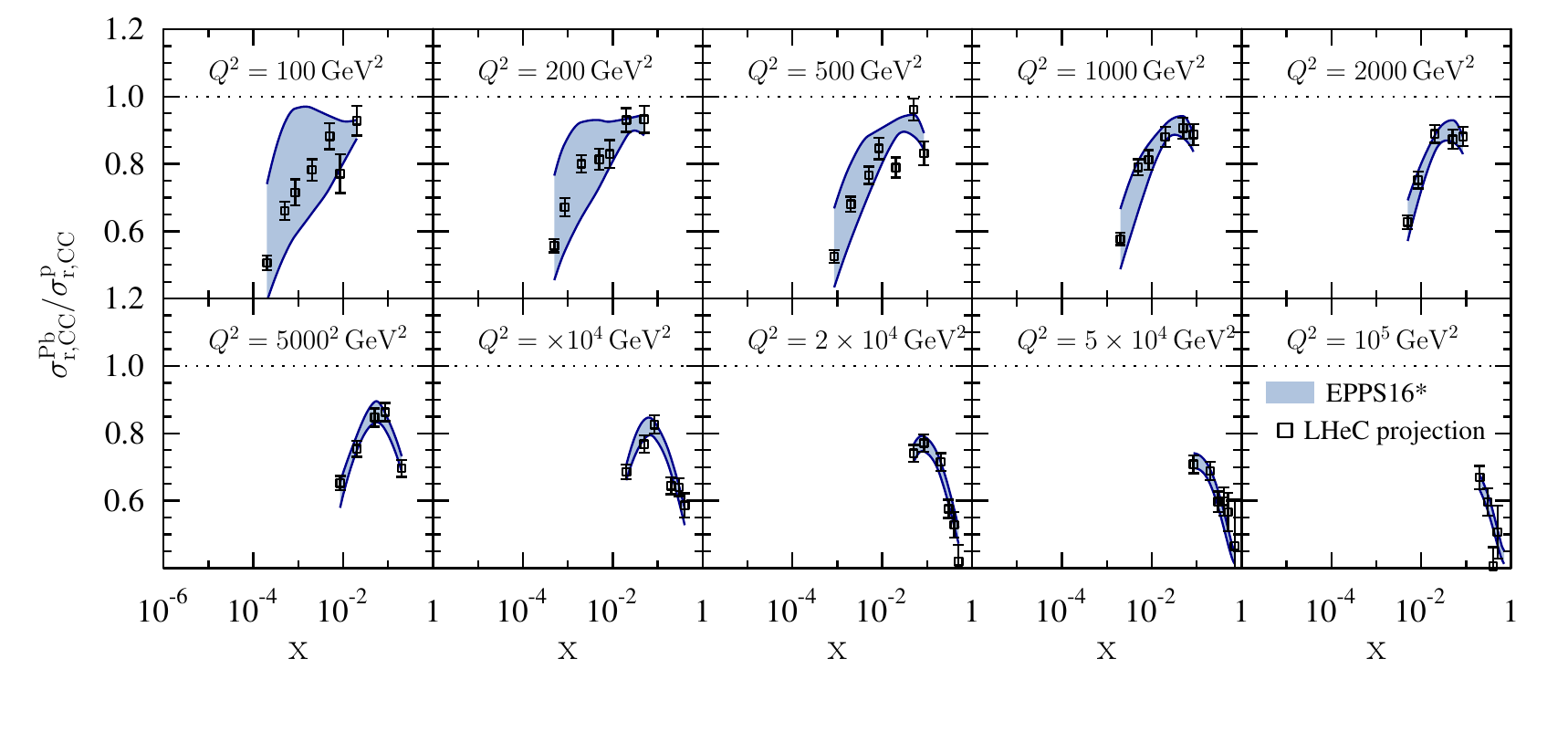}
  \includegraphics[width=.75\textwidth,angle=0]{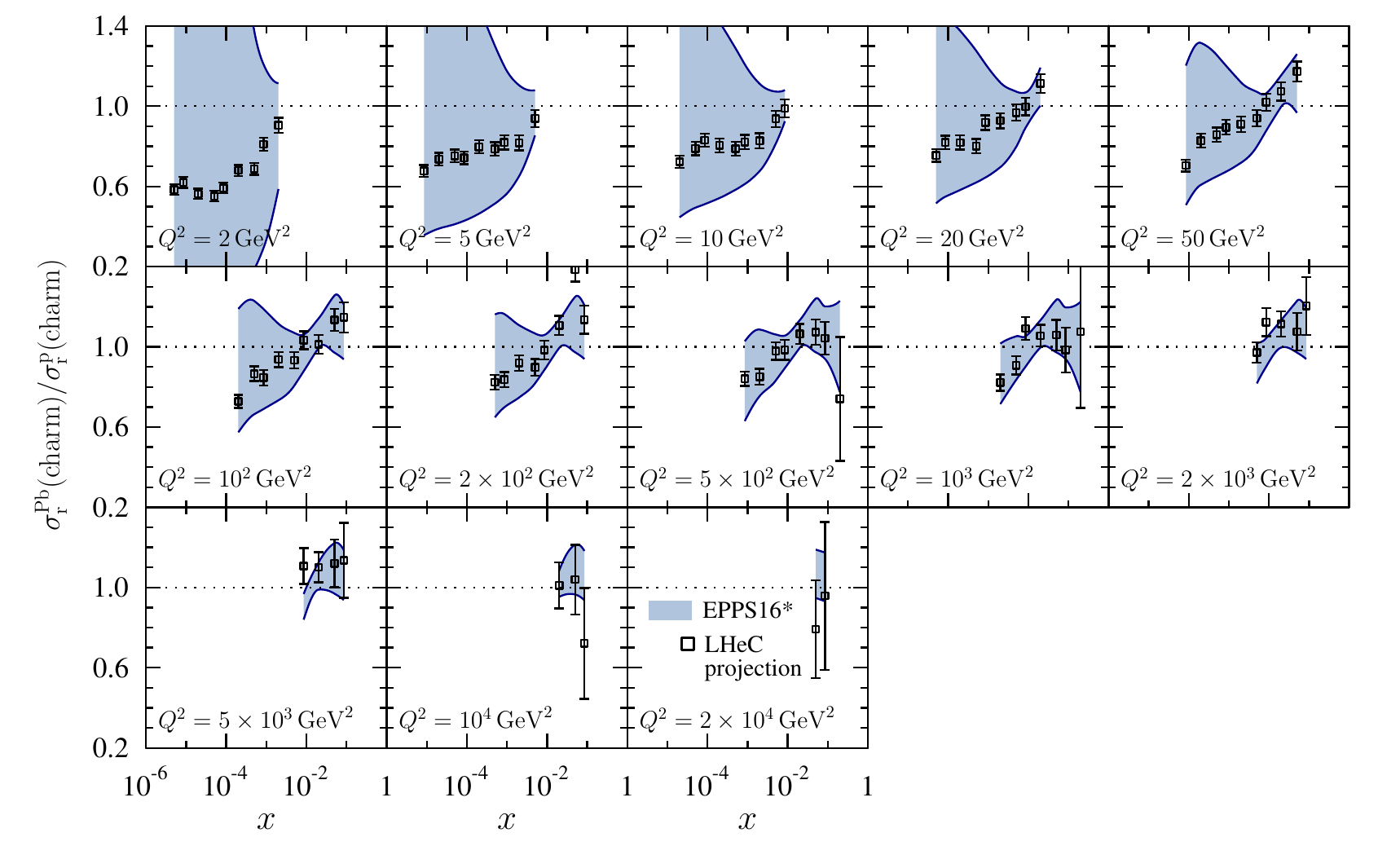}
\caption{Top: Simulated ratios of neutral-current reduced cross sections between $e$Pb and $ep$ collisions compared with the predictions from a EPPS16-type global fit of nuclear PDFs using an extended parametrisation for gluons. Middle: Charged-current cross section ratios. Bottom: Neutral-current charm-production cross section ratios.}
\label{fig:NPP_startingpoint}
\end{figure}

To illustrate the impact of the LHeC $e$Pb pseudodata in the global context, they have been added~\cite{HannuPaukkunenfortheLHeCstudygroup:2017ric} into the EPPS16 global analysis of nuclear PDFs~\cite{Eskola:2016oht}. The EPPS16 strategy is to parametrise the nuclear modification ratios $R_i(x,Q^2)$ between the bound-proton PDFs $f_i^{\textrm p/Pb}$ and proton PDFs $f_i^{\textrm p}$, 
\begin{equation}
R_i(x,Q^2) \equiv \frac{f_i^{\textrm p/Pb}(x,Q^2)}{f_i^{\textrm p}(x,Q^2)}\; ,
\end{equation}
at the charm mass threshold $Q^2=m_{\textrm charm}^2=(1.3 \,\textrm{GeV})^2$. At higher $Q^2$ the nuclear PDFs are obtained by solving the standard DGLAP evolution equations at next-to-leading order in QCD. As the LHeC pseudodata reach to significantly lower $x$ than the data that were used in the EPPS16 analysis, an extended small-$x$ parametrisation was used for gluons, see Figure~\ref{fig:NPP_fitforms}. The framework is almost identical to that in Ref.~\cite{Aschenauer:2017oxs}. The introduced functional form allows for rather wild -- arguably unphysical -- behaviour at small-$x$ where e.g.\ significant enhancement is allowed. This is contrary to the theoretical expectations from the saturation conjecture and looks also to be an improbable scenario given the recent LHCb D and B meson measurements~\cite{Aaij:2017gcy,Aaij:2019lkm} which impressively indicate~\cite{Eskola:2019bgf} gluon shadowing down to $x \sim 10^{-5}$ at interaction scales as low as $Q^2 \sim m_{\textrm charm}^2$. On the other hand, given that there are no prior DIS measurements in this kinematic range for  nuclei other than the proton, and that the D and B meson production in $p$Pb collisions could be affected by strong final-state effects (which could eventually be resolved by e.g.\ measurements of forward prompt photons ~\cite{Helenius:2014qla} in $p$Pb), we hypothesise that any kind of behaviour is  possible at this stage. 
Anyway, with the extended parametrisation -- called here EPPS16* -- the  uncertainties in the small-$x$ regime get significantly larger than in the standard  EPPS16 set. This is  reflected as significantly larger PDF error bands in comparison to the projected LHeC pseudodata. It is shown in Figure~\ref{fig:NPP_startingpoint} where EPPS16* predictions are compared with the LHeC pseudodata for inclusive NC and CC reactions, as well as charm production in neutral-current scattering. The uncertainties are estimated using the Hessian method~\cite{Pumplin:2001ct} and the same overall tolerance $\Delta\chi^2=52$ as in the EPPS16 analysis has been used when defining the error bands. Because there are no small-$x$ data constraints for gluons, the gluon uncertainty is enormous and the Hessian method used for estimating the uncertainties is not particularly accurate, i.e.\ the true $\Delta\chi^2=52$ error bands are likely to be even larger. At some point the downward uncertainty will be limited by positivity constraints e.g.\ for $F_{\textrm L}$, but will depend strongly on which $Q^2$ is used to set the positivity constraints (e.g.\ in the EPPS16 analysis $F_{\textrm L}$ is required to remain positive at $Q^2=m_{\textrm charm}^2$).

\begin{figure}[!p]
  \centering
  \includegraphics[width=.75\textwidth,angle=0]{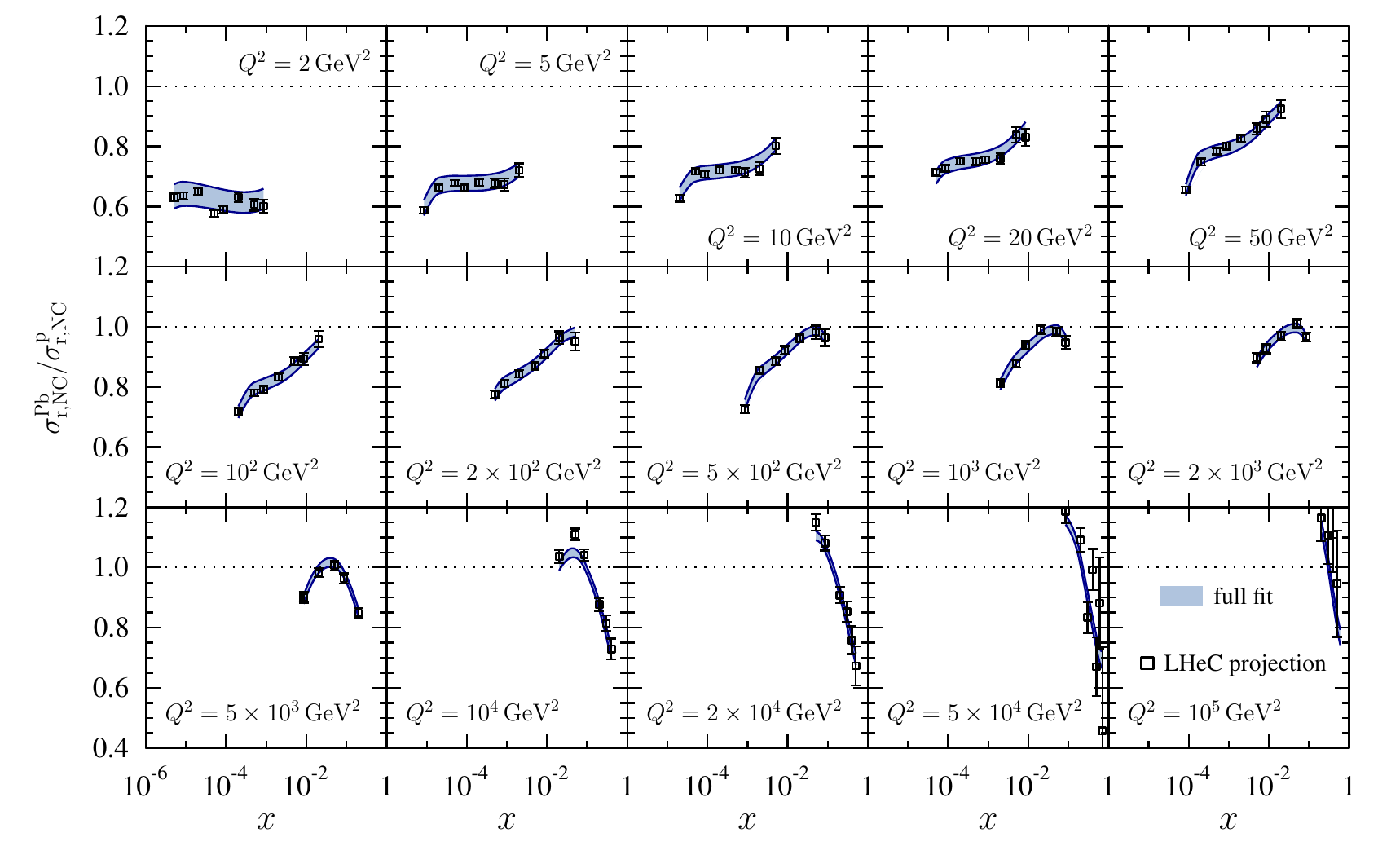}
  \includegraphics[width=.75\textwidth,angle=0]{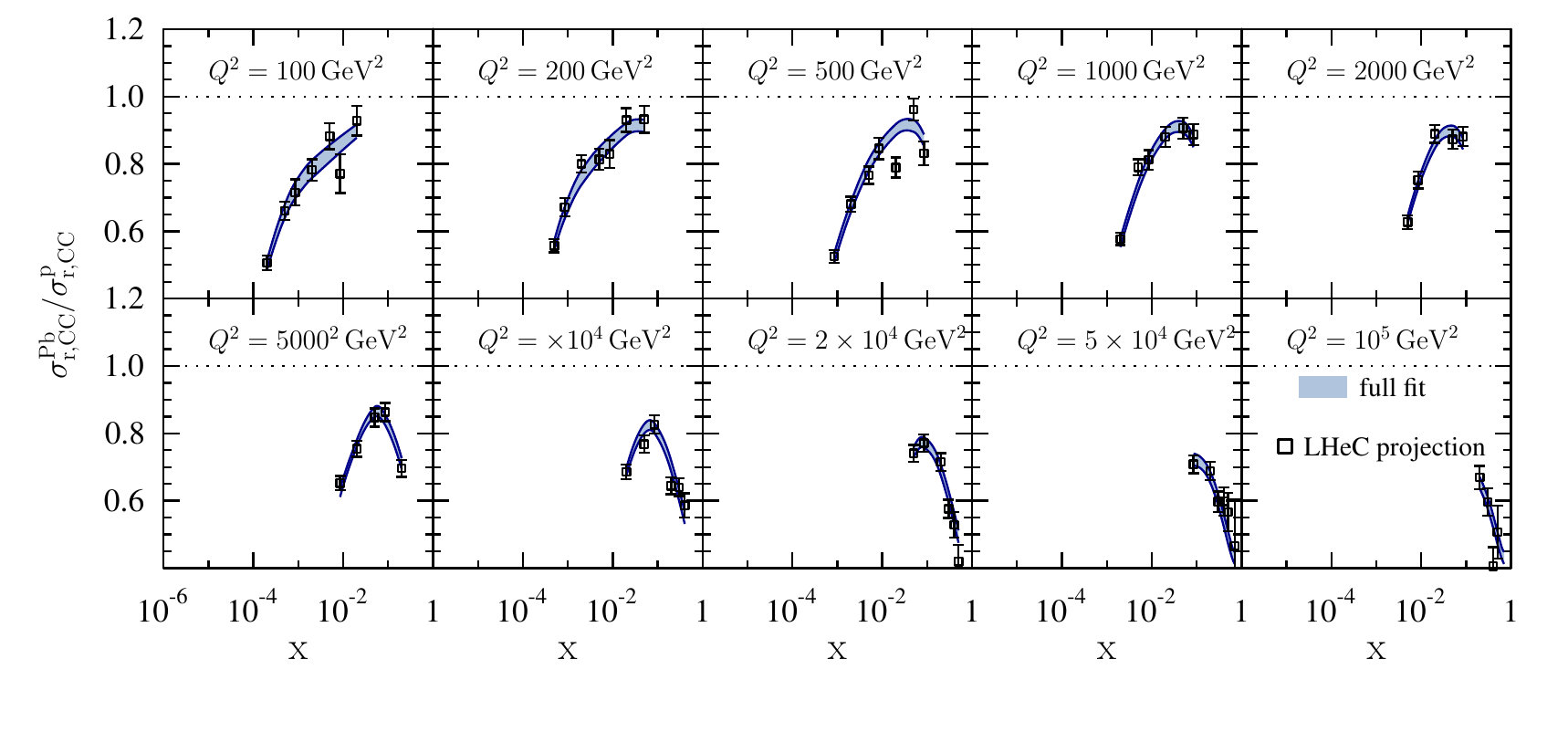}
  \includegraphics[width=.75\textwidth,angle=0]{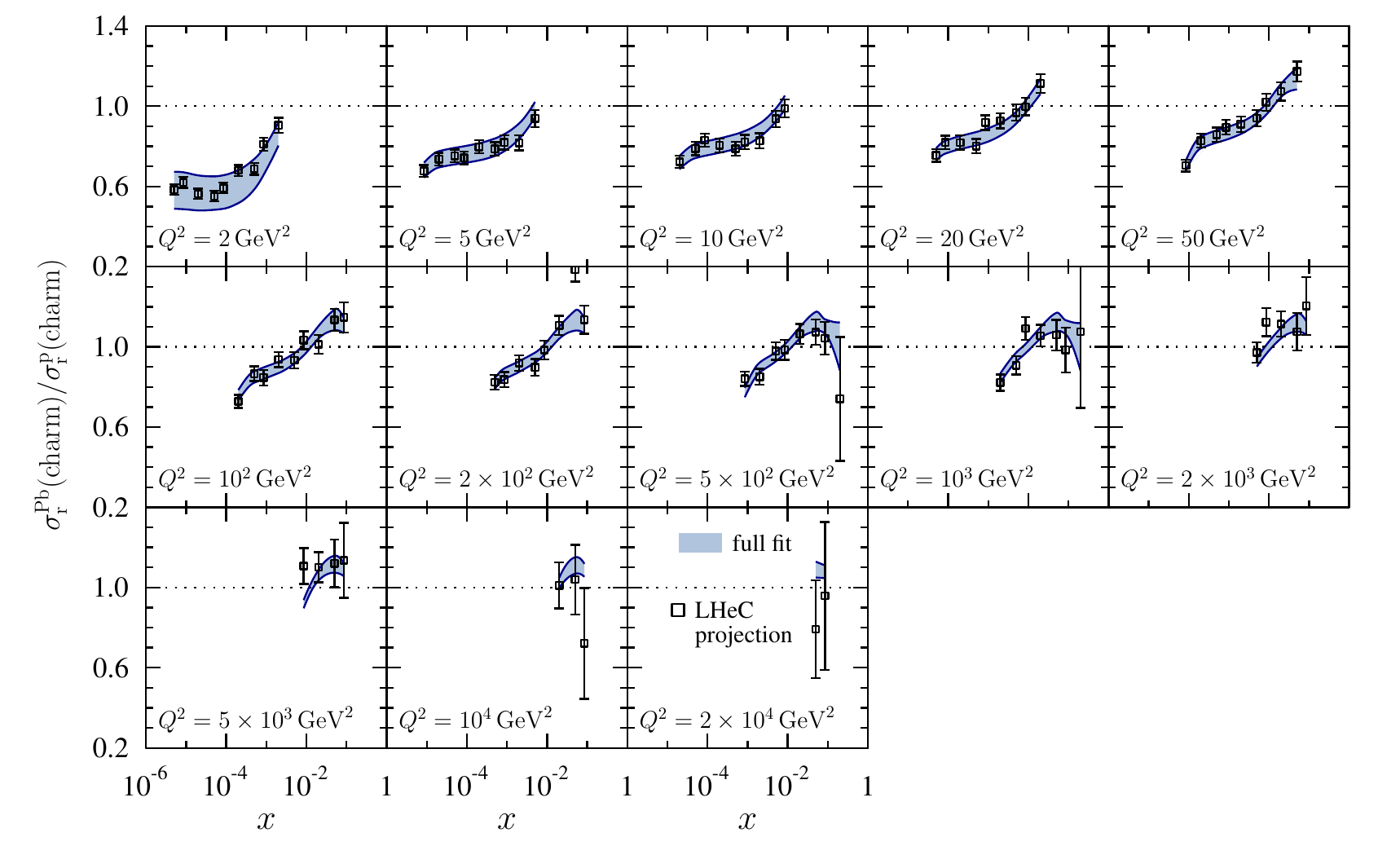}
\caption{As Figure~\ref{fig:NPP_startingpoint} but with fit results after including the LHeC pseudodata in the global analysis.}
\label{fig:NPP_thisiswhatweget}
\end{figure}

\begin{figure}[!th]
  \centering
  \includegraphics[width=.31\textwidth,angle=0]{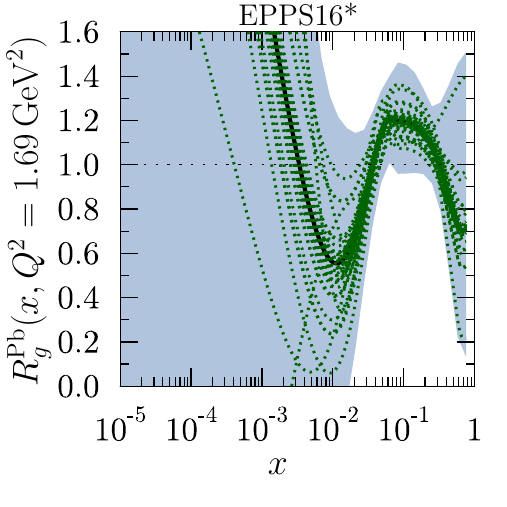}
  \includegraphics[width=.31\textwidth,angle=0]{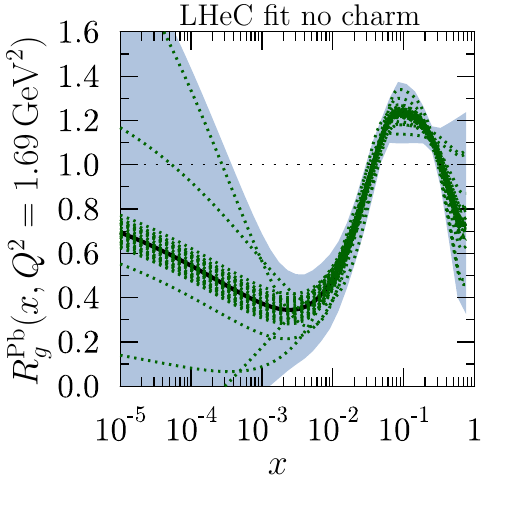}
  \includegraphics[width=.31\textwidth,angle=0]{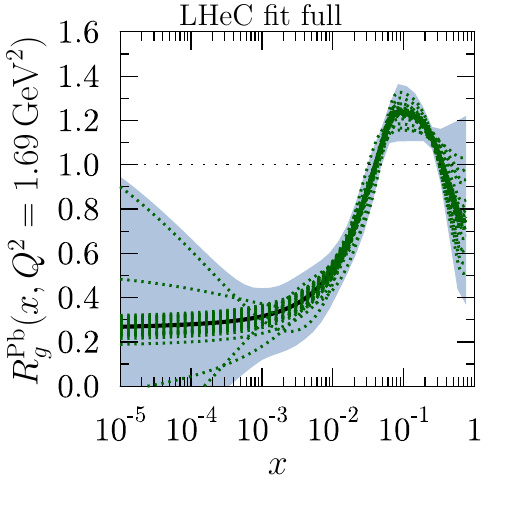}
  \includegraphics[width=.31\textwidth,angle=0]{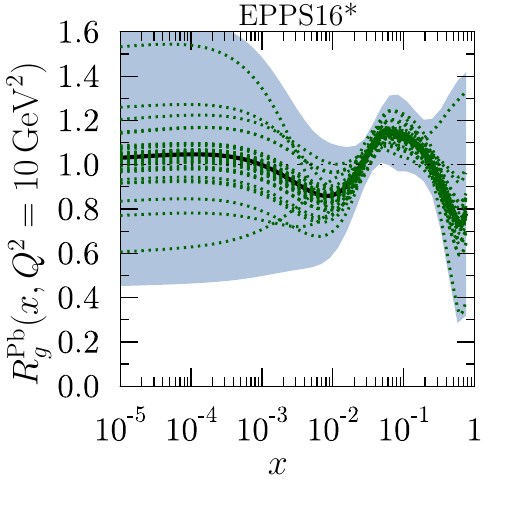}
  \includegraphics[width=.31\textwidth,angle=0]{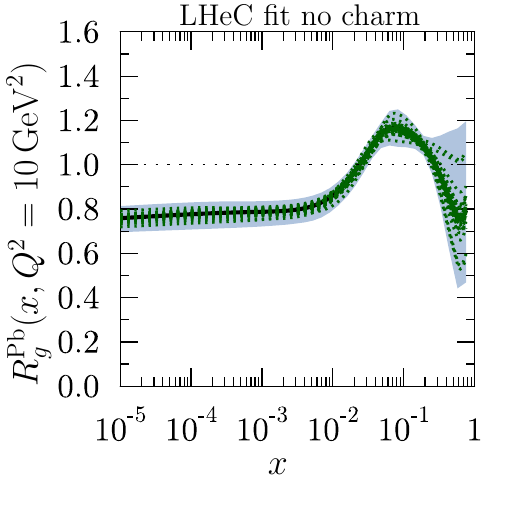}
  \includegraphics[width=.31\textwidth,angle=0]{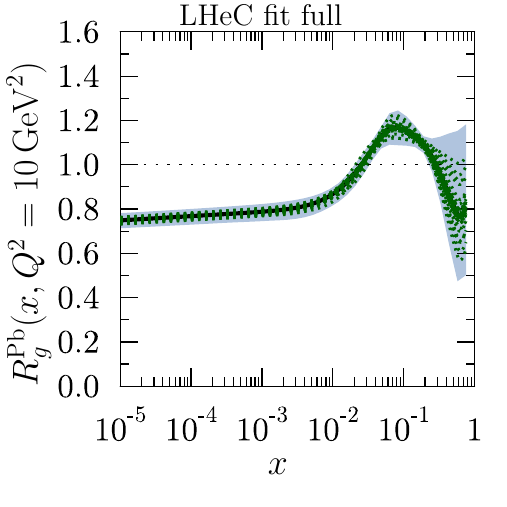}
\caption{
Upper panels: The gluon nuclear modification for the Pb nucleus at $Q^2=1.69\,\textrm{GeV}^2$ in EPPS16* (left), LHeC analysis without charm pseudodata (middle), and full LHeC analysis (right). The blue bands mark the total uncertainty and the green dotted curves correspond to individual Hessian error sets. Lower panels: As the upper panels but at $Q^2=10\,\textrm{GeV}^2$.
}
\label{fig:NPP_LHeCnuclearg}
\end{figure}

Upon including the LHeC $e$Pb pseudodata in the fit, the new nPDFs adapt to reproduce the pseudodata and their uncertainties are greatly reduced, as shown in Figure~\ref{fig:NPP_thisiswhatweget}. The overall tolerance has been kept fixed to the default value $\Delta\chi^2=52$. The impact on the nuclear modification of the gluon PDF is illustrated in Figure~\ref{fig:NPP_LHeCnuclearg} at two values of $Q^2$: $Q^2=1.69\,\textrm{GeV}^2$ (the parametrisation scale) and $Q^2=10\,\textrm{GeV}^2$.
Already the inclusive pseudodata are able to reduce the small-$x$ gluon uncertainty quite significantly, and the addition of the charm data promises an even more dramatic reduction in the errors. The analysis indicates that  the LHeC will nail the nuclear gluon PDF to a high precision down to $x$ of at least $10^{-5}$. 

\subsection{nPDFs from DIS on a single nucleus \ourauthor{Nestor Armesto}}
\label{sec:NPP_nPDFs_Pbonly}

Another approach that becomes possible with the large kinematic coverage and volume of data for a single nucleus, Pb, at the LHeC and FCC-eh, is to perform a fit to only Pb data in order to extract the Pb PDFs, removing the need to interpolate between different nuclei. Then the corresponding ratios or nuclear modification factors for each parton species
can be obtained using either a proton PDF set from a global fit or, as we do here (see~\cite {naFCCPW2018,Perez:2019hxm,Abada:2019lih}), from a fit to proton LHeC and FCC-eh pseudodata. In this way, there will be no need to introduce a nuclear size dependence in the parameters for the initial condition for DGLAP evolution. Such nPDFs can then be used for comparing to those obtained from global fits and for precision tests of collinear factorisation in nuclear collisions.

The fits are performed using \emph{xFitter}~\cite{Alekhin:2014irh}, where 484 (150) NC+CC Pb data points at the LHeC (FCC-eh) have been used in the fitted region $Q^2>3.5$\,GeV$^2$, see Fig.~\ref{fig:NPP_pseudodata}. A HERAPDF2.0-type parametrisation~\cite{Abramowicz:2015mha} has been employed to provide both the central values for the reduced cross sections (therefore, the extracted nuclear modification factors are centered at 1) and the fit functional form; in this way, neither theory uncertainties (treatment of heavy flavours, value of $\alpha_s$, order in the perturbative expansion) nor the uncertainty related to the functional form of the initial condition -- parametrisation bias -- are considered in our study, in agreement with our goal of estimating the {\em ultimate achievable experimental } precision in the extraction of nPDFs. We have worked at NNLO using the Roberts-Thorne improved heavy quark scheme, and $\alpha_s(m_Z^2)=0.118$. The treatment of systematics and the tolerance $\Delta \chi^2=1$ are identical to the approach in the HERAPDF2.0 fits, as achievable in a single experiment.

\begin{figure}
  \centering
  \includegraphics[width=.40\textwidth,angle=0,trim={0 0 40 0},clip]{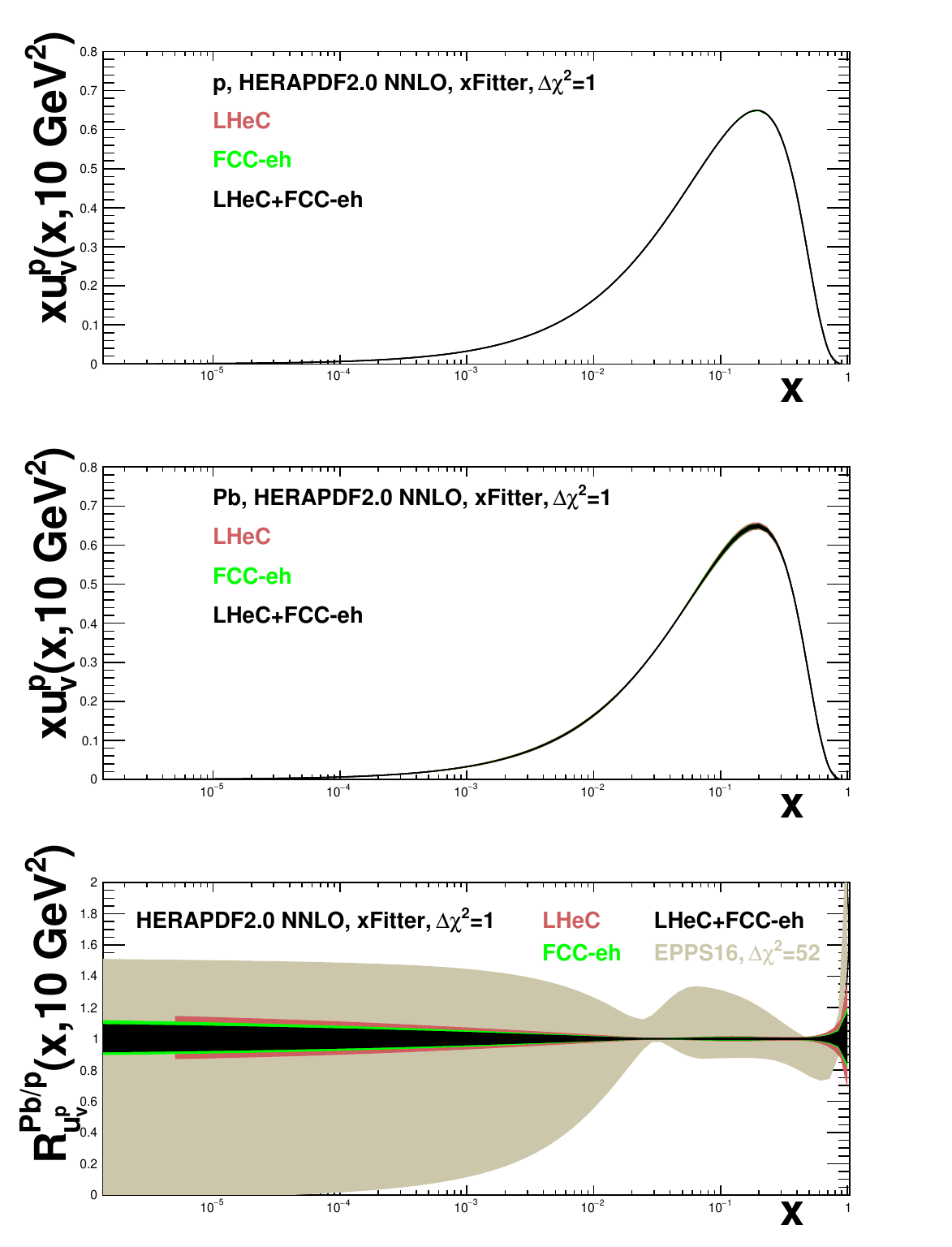}
  \includegraphics[width=.40\textwidth,angle=0,trim={0 0 40 0},clip]{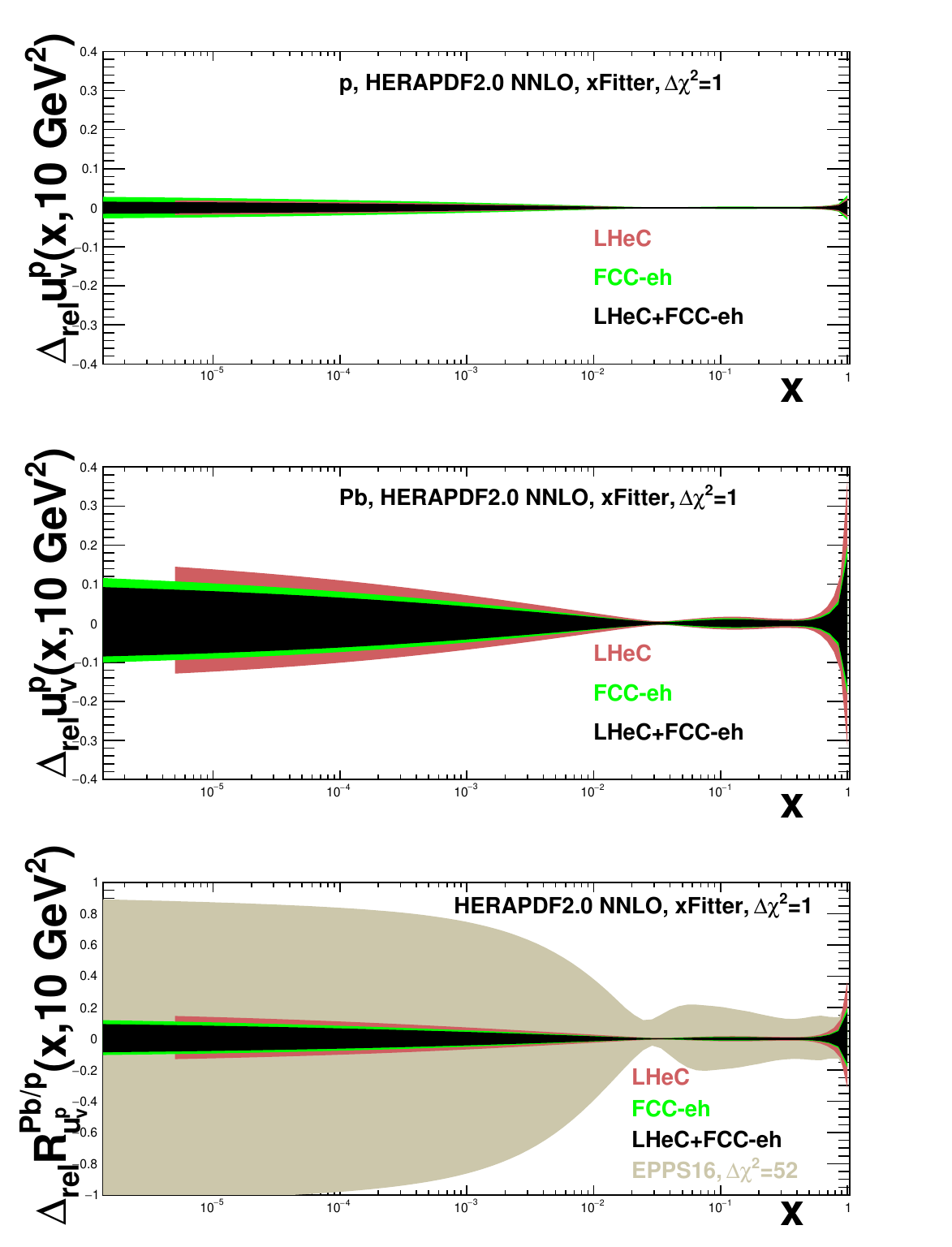}
\caption{Distributions (left) and their relative uncertainties (right) of the valence $u$-quark density in the proton (top), Pb (middle) and the corresponding nuclear modification factor (bottom) in an analysis of $ep$ and $e$Pb LHeC and FCC-eh NC plus CC pseudodata using \emph{xFitter} (both a single set of data and all combined), compared to the results of EPPS16~\cite{Eskola:2016oht}, see the text for details.}
\label{fig:NPP_xFval}
\end{figure}

\begin{figure}
  \centering
  \includegraphics[width=.40\textwidth,angle=0,trim={0 0 40 0},clip]{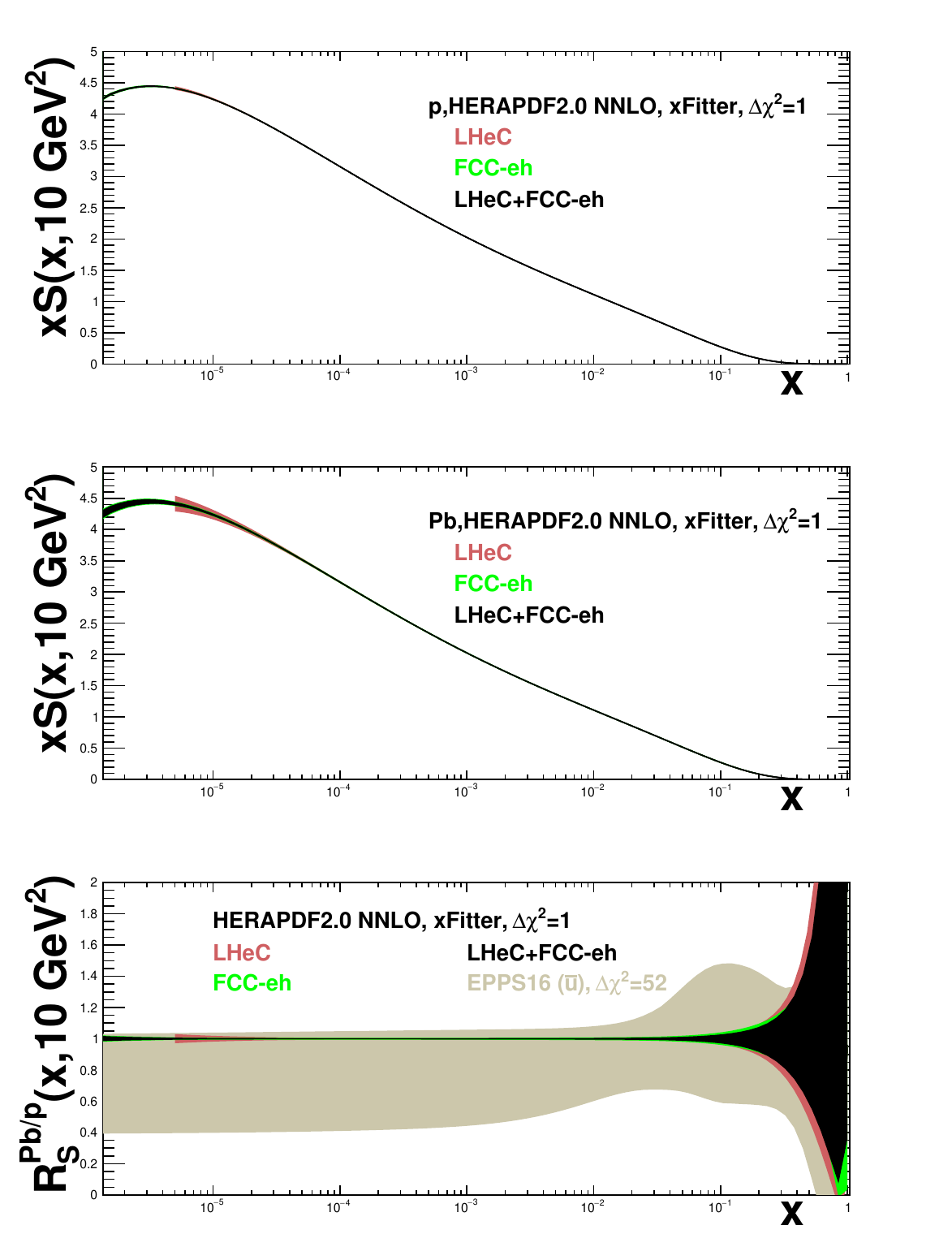}
  \includegraphics[width=.40\textwidth,angle=0,trim={0 0 40 0},clip]{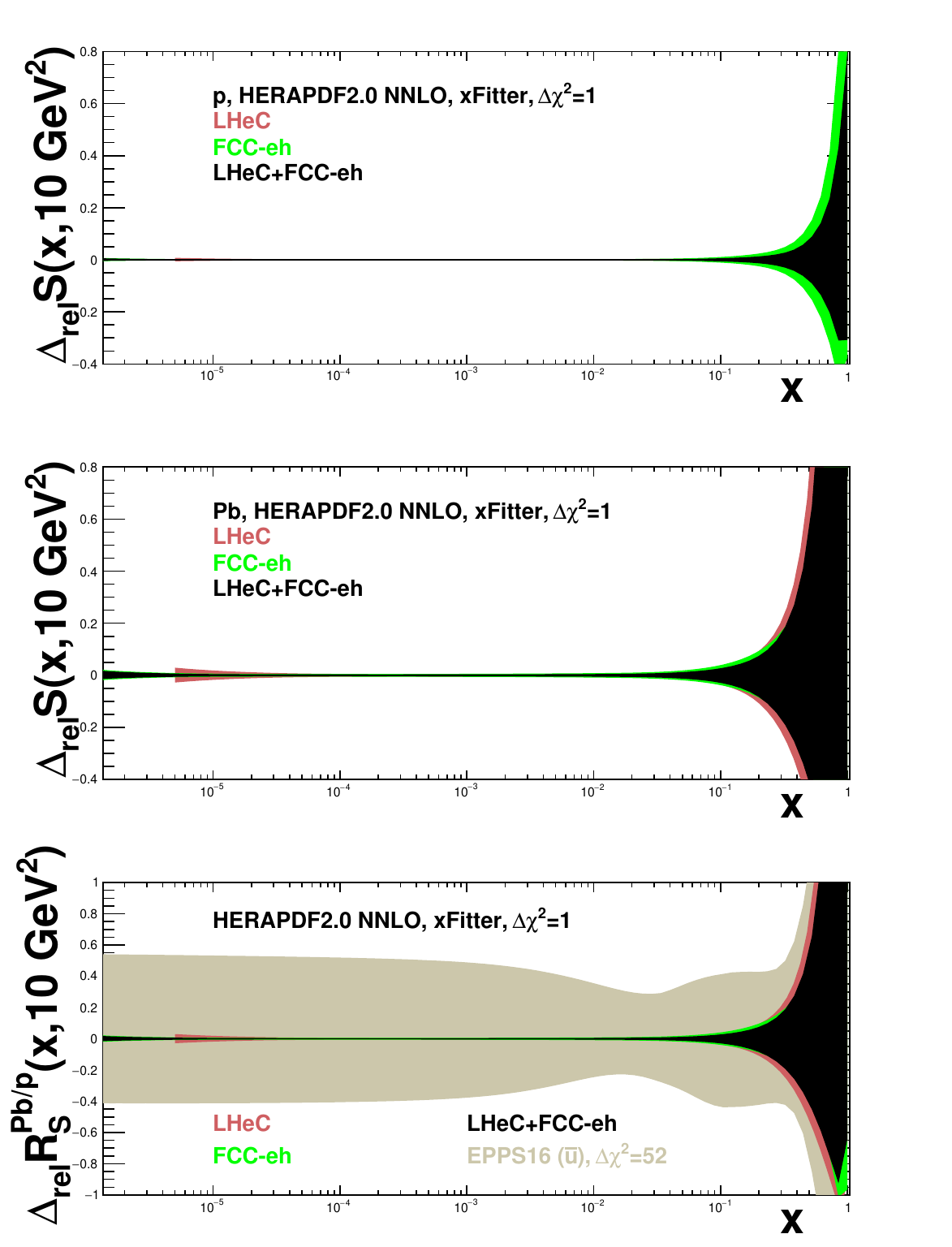}
\caption{Distributions (left) and their relative uncertainties (right) of the sea quark density in the proton (top), Pb (middle) and the corresponding nuclear modifications factor (bottom) in an analysis of $ep$ and $e$Pb LHeC and FCC-eh NC plus CC pseudodata using \emph{xFitter} (both a single set of data and all combined), compared to the results of EPPS16~\cite{Eskola:2016oht} for $\bar u$, see the text for details.}
\label{fig:NPP_xFsea}
\end{figure}

\begin{figure}
  \centering
  \includegraphics[width=.40\textwidth,angle=0,trim={0 0 40 0},clip]{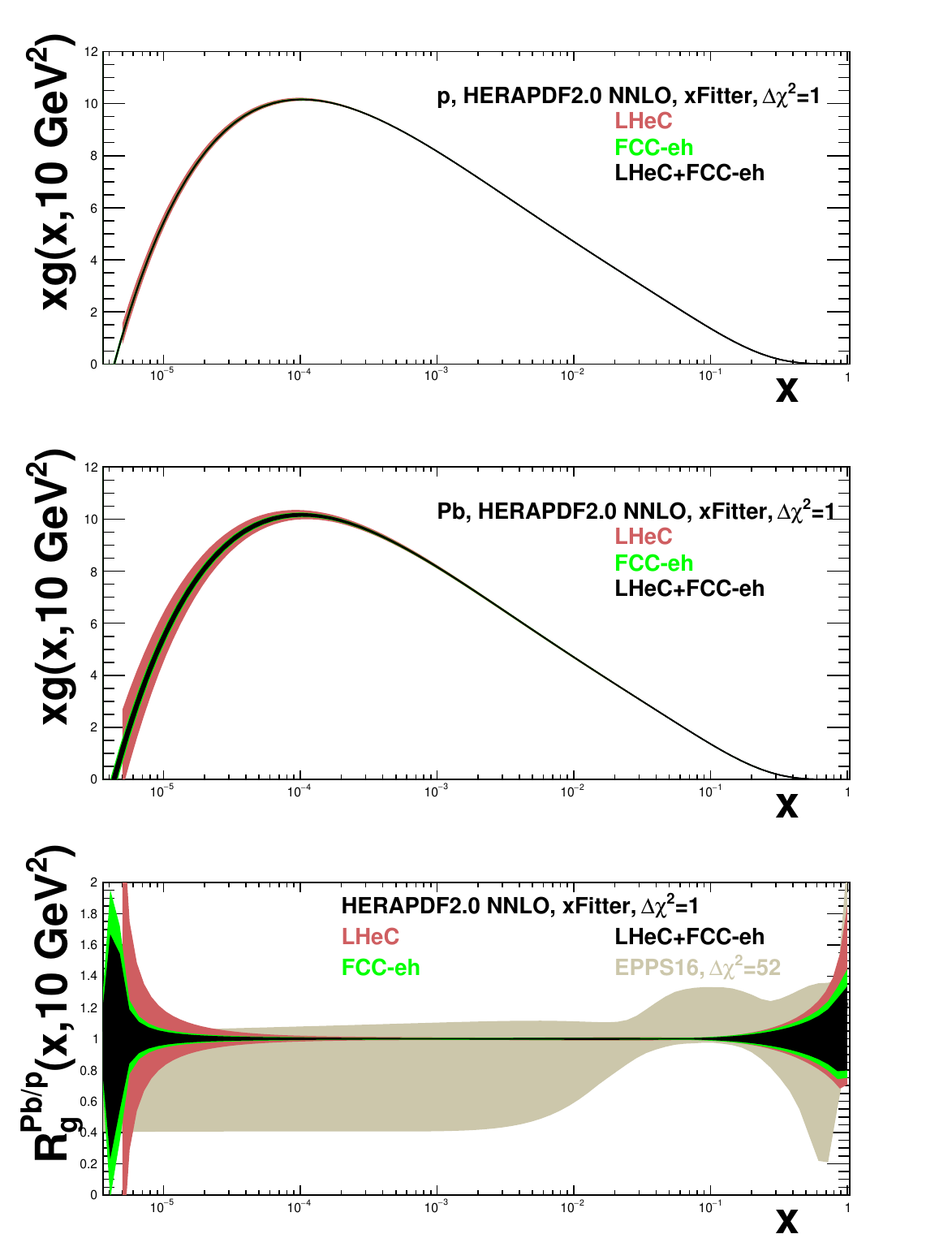}
  \includegraphics[width=.40\textwidth,angle=0,trim={0 0 40 0},clip]{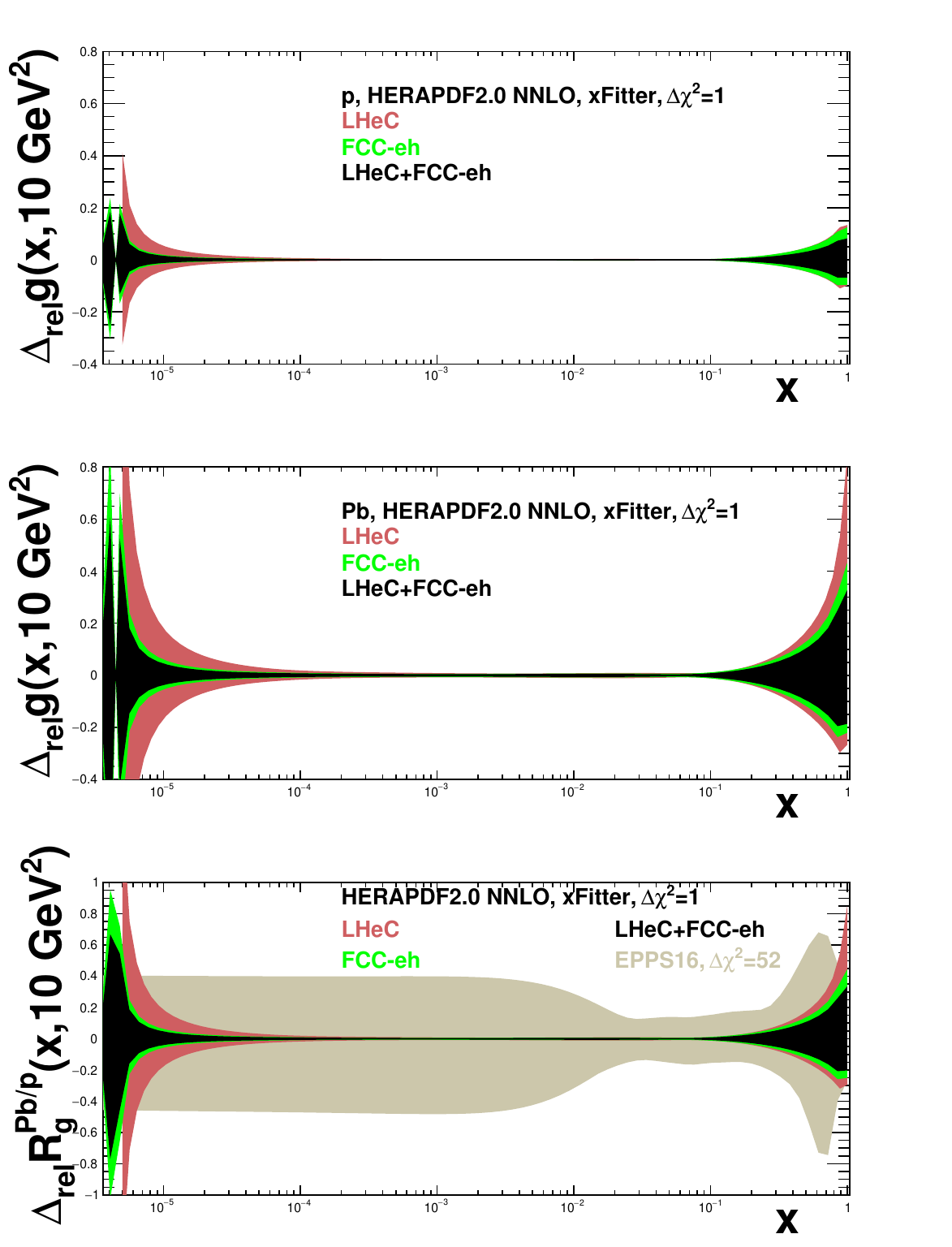}
\caption{Distributions (left) and their relative uncertainties (right) of the gluon density in the proton (top), Pb (middle) and the corresponding nuclear modifications factor (bottom) in an analysis of $ep$ and $e$Pb LHeC and FCC-eh NC plus CC pseudodata using \emph{xFitter} (both a single set of data and all combined), compared to the results of EPPS16~\cite{Eskola:2016oht}, see the text for details.}
\label{fig:NPP_xFglue}
\end{figure}

The results for the relative uncertainties in the nuclear modification factors are shown in Figs.~\ref{fig:NPP_xFval}, \ref{fig:NPP_xFsea} and \ref{fig:NPP_xFglue} for valence, sea and gluon, respectively.
The uncertainties in these plots reflect the assumed uncertainties in the pseudodata, both statistics (mainly at large $x$) and systematics from detector efficiencies, radiative corrections, etc., see Sec.~\ref{sec:NPP_nPDFs_pseudodata}. As expected, the uncertainty in the extraction of the valence at small $x$ is sizeably larger than that for the sea and gluon.

While a very high precision looks achievable at the LHeC and the FCC-eh, for the comparison with EPPS16 (or any other global fit) shown in the plots and with previous results including LHeC pseudodata in that setup, see Sect.~\ref{sec:NPP_nPDFs_globalfit} and~\cite{Aschenauer:2017oxs,HannuPaukkunenfortheLHeCstudygroup:2017ric}, some caution is required. First, the effective EPPS16 tolerance criterion $\Delta \chi^2\simeq 52$ implies that naively the uncertainty bands should be compared after rescaling by a factor $\sqrt{52}$. Second, the treatment of systematics is rather different, considering correlations in the \emph{xFitter} exercise and taking them as fully uncorrelated (and added quadratically to the statistical ones) in the EPPS16 approach. Finally, EPPS16 uses parametrisations for the nuclear modification factors for different parton species while in \emph{xFitter} just the (n)PDF combinations that enter the reduced cross sections are parametrised and employed for the fit~\footnote{In this respect let us note that, in analogy to proton PDFs, a full flavour decomposition can be achieved using both NC and CC with heavy flavour identification that will verify the existing ideas on flavour dependence of nuclear effects on parton densities~\cite{Brodsky:2004qa}.}. With all these considerations in mind, the results shown in this Section are fully compatible with those in the previous one.

\section{Nuclear diffraction \ourauthor{Anna Stasto, Paul Newman}}
\label{sec:NPP_nonconventional}

In Sec.~\ref{sec:PSM_Disc_3D} we have discussed  specific processes which will probe the details of the 3D structure of the proton. The same processes can be studied in the context of electron-ion scattering and used to learn about the partonic structure of nuclei. Inclusive diffraction on nuclei can provide important information about the nuclear diffractive parton distribution similarly to the diffraction on the proton, see Sec.~\ref{sec:inclusive_diffraction}.  Diffractive vector meson production can be studied in the nuclear case as well, e.g.\ within the framework of the dipole model suitable for high energy and including non-linear effects in density. In the nuclear case though, one needs to make a distinction between coherent and incoherent diffraction.  In the coherent process, the nucleus scatters elastically and stays intact after the collision. In incoherent diffraction, the nucleus breaks up, and individual nucleons can be set free. Still, there will be a large rapidity gap between the produced diffractive system and the dissociated nucleus. It is expected that this process will dominate the diffractive cross section for medium and large values of momentum transfer. It is only in the region of  small values of momentum transfer where elastic diffraction is the dominant contribution. 
Dedicated instrumentation in the forward region must be constructed in order to clearly distinguish between the two scenarios, see Chapter\,10.

\subsection{Exclusive vector meson diffraction}

\label{sec:NPP_EVM}

Calculations  for the case of  Pb for the coherent diffractive $J/\psi$ production were performed using the dipole model~\cite{Mantysaari:2017dwh}, see Sec.~\ref{sec:PSM_Disc_3D}. In order to apply the dipole model calculation to the nuclear case, one takes  the independent scattering approximation that is Glauber theory~\cite{Lappi:2013am}. The dipole amplitude can then be represented in the form
\begin{equation}
    N_A(x,{\textbf r},{\textbf b}) \; = \; 1- \prod_{i=1}^{A} \, [1-N(x,{\textbf r},{\textbf b}-{\textbf b}_i)] \; .
    \label{eq:dipole_nucleus}
\end{equation}
Here $N(x,{\textbf r},{\textbf b}-{\textbf b}_i)$ is the dipole amplitude for the nucleon (see Sec.~\ref{sec:PSM_Disc_3D}) and 
 ${\textbf b}_i$ denotes the transverse positions of the nucleons in
the nucleus. The interpretation of Eq.~\eqref{eq:dipole_nucleus} is that $1-N$ is the
probability not to scatter off an individual nucleon, and thus
$\prod_{i=1}^{A} \, [1-N({\textbf r},{\textbf b}-{\textbf b}_i,x)]$  is the probability not to scatter off the entire 
nucleus.

In addition,  the following simulation  includes the fluctuations of the density profile in the proton, following the prescription  given in~\cite{Mantysaari:2016ykx,Mantysaari:2016jaz,Mantysaari:2017dwh}.
To include these proton structure fluctuations one assumes that the gluonic density of the proton in the transverse plane is distributed around three constituent quarks (hot spots). These hot spots are assumed to be Gaussian. In practical terms one replaces the proton profile $T_p({\textbf b})$
\begin{equation}
    T_p({\textbf b}) = \frac{1}{2\pi B_p}e^{-b^2/(2B_p)} \; ,
\end{equation}
that appears in each individual nucleon scattering probability $N(x,{\textbf r},{\textbf b}-{\textbf b}_i)$
by the function
\begin{equation}
    T_p({\textbf b}) = \sum_{i=1}^3 T_q({\textbf b}-{\textbf b}_{q,i}) \; ,
\end{equation}
where the `quark' density profile is given by
\begin{equation}
    T_q({\textbf b}) = \frac{1}{2 \pi B_q} e^{-b^2 /(2 B_q)}\; .
\end{equation}
Here ${\textbf b}_{q,i}$ are the location of the hotspots that are sampled from a two dimensional Gaussian distribution whose width is given by parameter $B_{qc}$. The free parameters $B_q$ and $B_{qc}$ were obtained in~\cite{Mantysaari:2016jaz} by comparing with  HERA  data on  coherent and incoherent $J/\psi$ production at  a photon-proton centre-of-mass energy $W=75$\,{\textrm GeV}, corresponding to fractional hadronic target energy loss $x_{IP}=10^{-3}$. The proton fluctuation parameters obtained are $B_{qc} = 3.3\,\textrm{GeV}^{-2}$ and $B_q = 0.7\,\textrm{GeV}^{-2}$.

\begin{figure}[!th]
    \centering
    \includegraphics[width=0.44\textwidth]{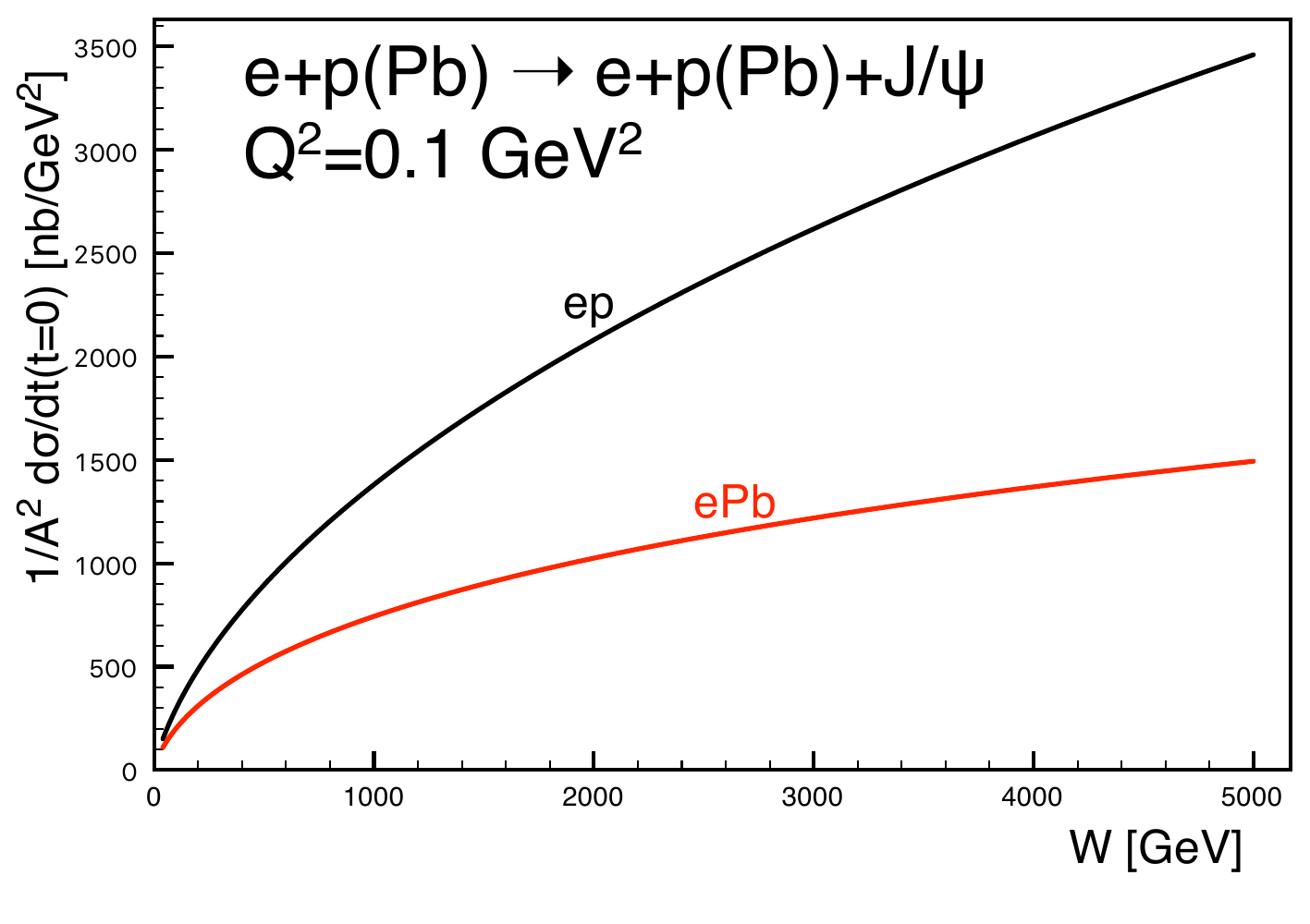}
    \hspace{0.02\textwidth}
    \includegraphics[width=0.44\textwidth]{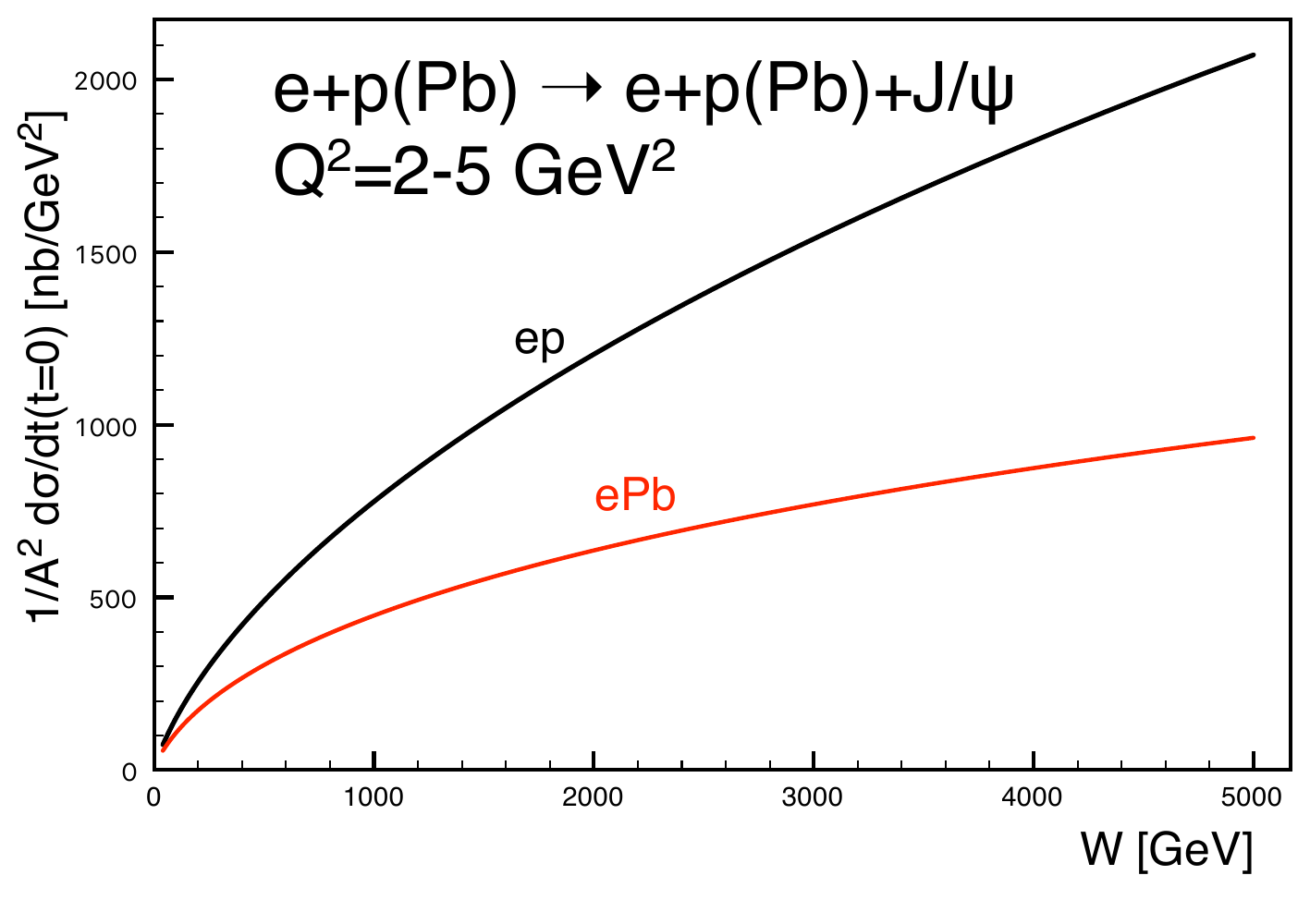}
    \caption{Cross section for the coherent diffractive production of the vector meson $J/\psi$ in $ePb$ (red solid curves) and $ep$ (black solid curves) collisions, as a function of the  energy $W$. Left: photoproduction case $Q^2 \simeq 0$, right: $Q^2=2-5 \;\textrm{GeV}^2$}
    \label{fig:sigvm_epeA_lowQ}
\end{figure}

\begin{figure}[!th]
    \centering
    \includegraphics[width=0.44\textwidth]{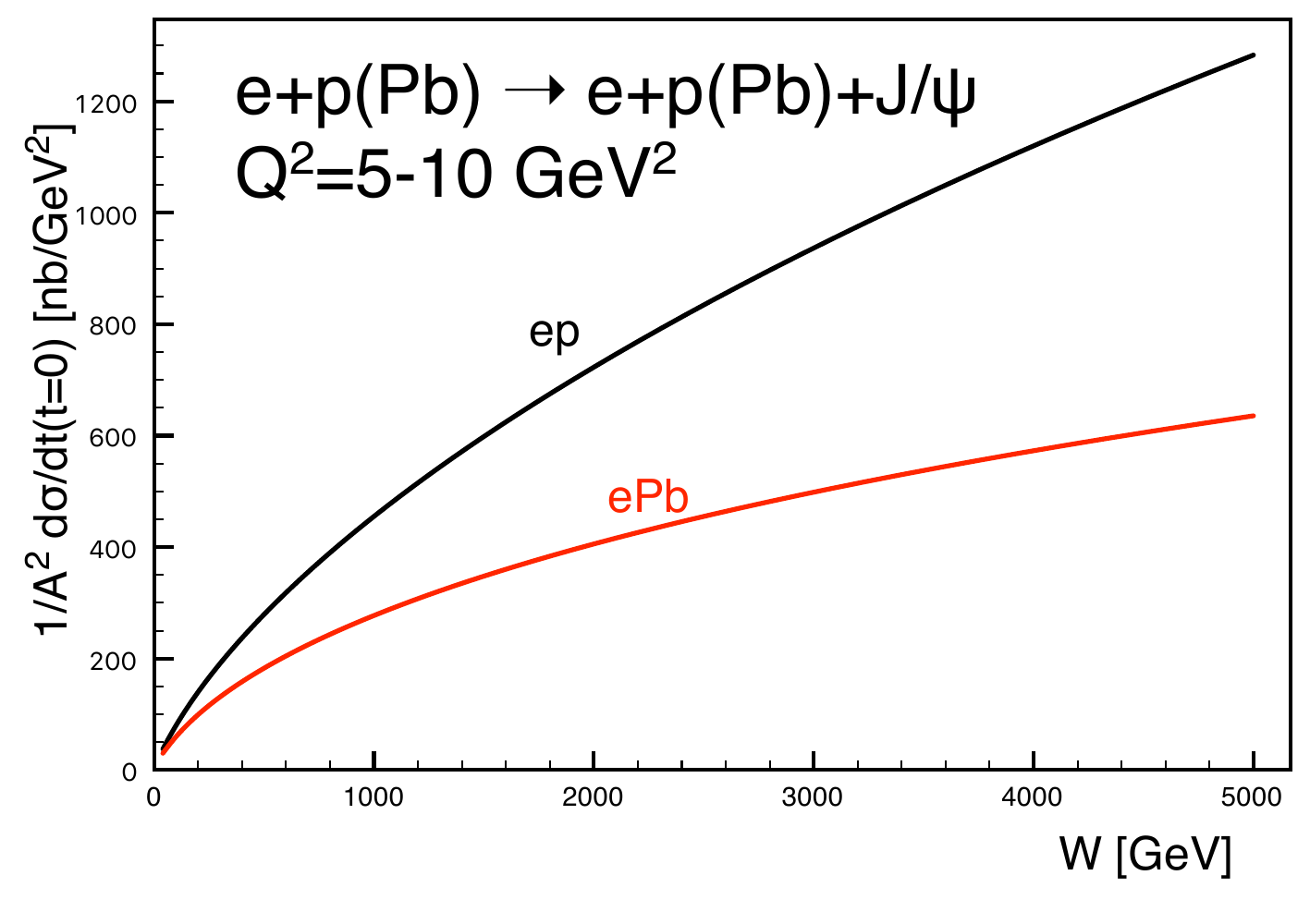}
    \hspace{0.02\textwidth}
    \includegraphics[width=0.44\textwidth]{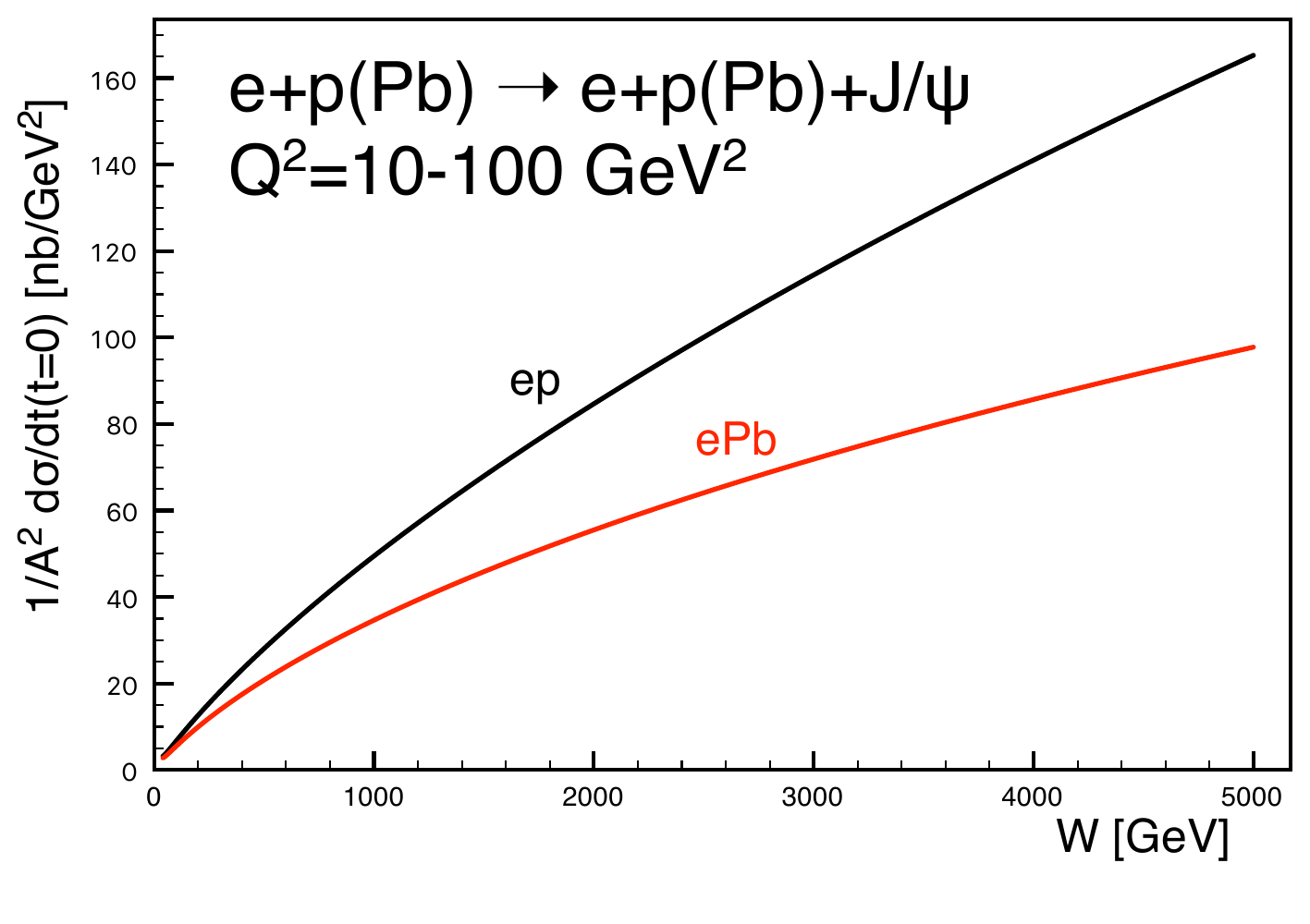}
   \caption{Cross section for the coherent diffractive production of the vector meson $J/\psi$ in $e$Pb (red solid curves) and $ep$ (black solid curves) collisions, as a function of the  energy $W$. Left:  $Q^2 = 5-10 \;\textrm{GeV}^2$, right: $Q^2=10-100 \;\textrm{GeV}^2$.}
    \label{fig:sigvm_epeA_hiQ}
\end{figure}

The results for the differential cross section at $t=0$  for  coherent production of $J/\psi$ as a function of (virtual) photon-proton energy $W$ for fixed values of $Q^2$ are shown in Figs.~\ref{fig:sigvm_epeA_lowQ} and Figs.~\ref{fig:sigvm_epeA_hiQ}.
The calculations for Pb are compared to those on the proton target. We see that the cross sections for the nuclear case increase with energy slower than for the proton case and are always smaller. Note that, we have already rescaled the diffractive cross section by a factor $A^2$, as appropriate for comparison of the diffractive cross section on the proton and nucleus. In the absence of nuclear corrections their ratio should be equal to $1$. The differences between the scattering off a nucleus and a proton are also a function of $Q^2$. They are larger for smaller values of $Q^2$ and for photoproduction. This is understood from the dipole formulae, see Eqs.~\eqref{eq:dipole_dt_xsection},
\eqref{eq:dipole_vm_elastic}, \eqref{eq:dipole_amplitude}. As explained previously, larger values of scale $Q^2$ select smaller size dipoles, for which the density effects are smaller. Similarly, the differences between the lead and proton cases are larger for higher energies. This is because the dipole amplitude grows with decreasing values of $x$ which are probed when the energy is increased, and thus the non-linear density effects are more prominent at low values of $x$ and $Q^2$.

\begin{figure}[!th]
    \centering
    \includegraphics[width=0.65\textwidth]{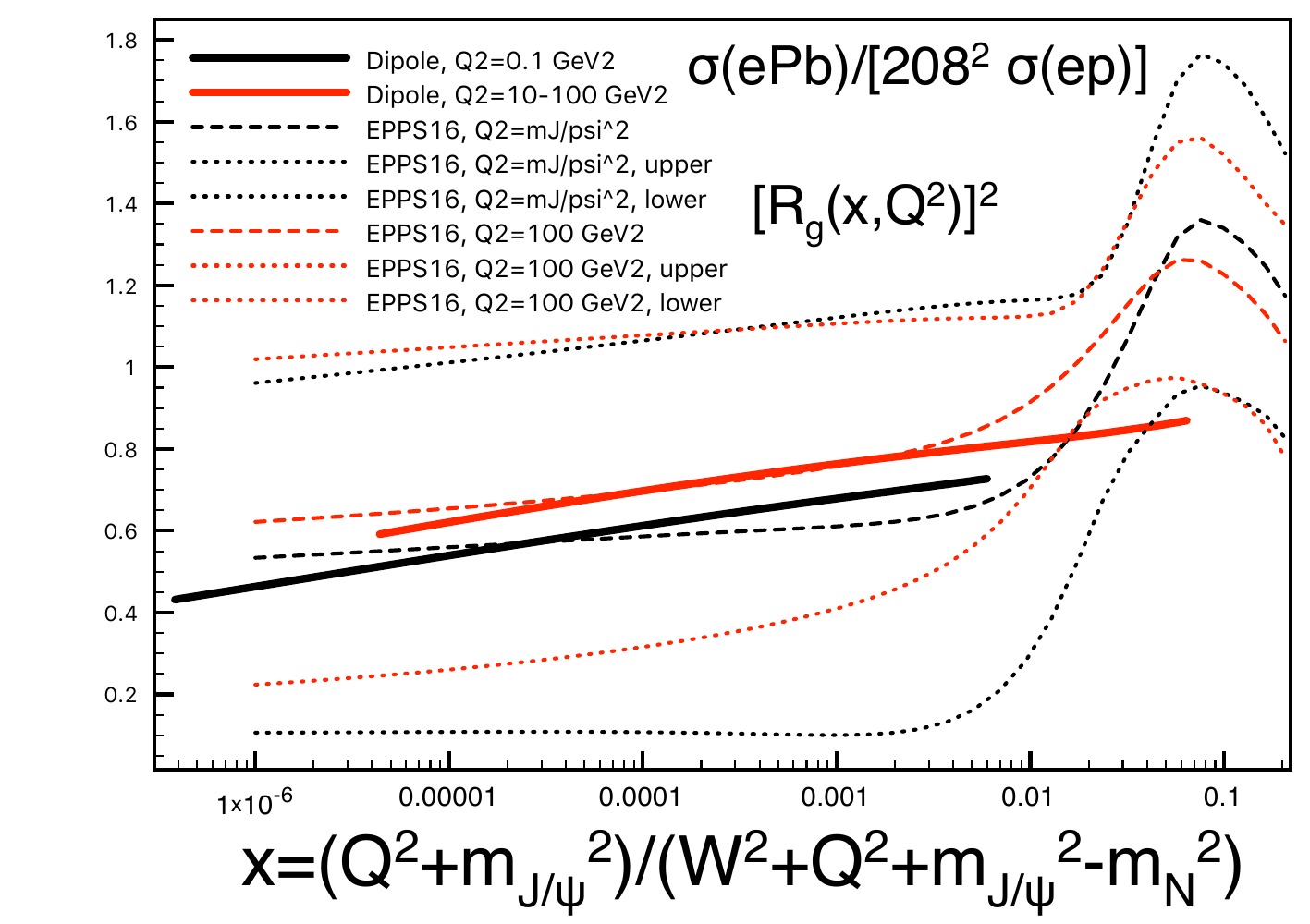}
    \caption{Ratio of coherent J/$\psi$ production diffractive cross sections for Pb and proton as a function of the variable $x$ (defined in Eq.~\eqref{eq:xvariable_jpsi} for the dipole model results). Solid lines: dipole model calculation, for $Q^2=0.1 \;\textrm{GeV}^2$ (black) and $Q^2=10-100 \;\textrm{GeV}^2$ (red). Dotted and dashed lines correspond to the nuclear ratio for the gluon density squared using the EPPS16 parametrisation~\cite{Eskola:2016oht} of the nuclear parton distribution functions. Black and red dashed lines are the central sets for $Q^2=M_{J/\psi}^2$ and $Q^2=100 \;\textrm{GeV}^2$. The dotted lines correspond to the low and high edges of the Hessian uncertainty in the EPPS16 parametrisation. The difference between the two dotted lines is thus indicative of the parametrisation uncertainty for the nuclear ratio. These ratios, that can also be measured in ultraperipheral collisions~\cite{Baltz:2007kq}, are larger that the values $0.2-0.4$ at $x\simeq 10^{-5}$ predicted by the relation between diffraction and nuclear shadowing~\cite{Frankfurt:2011cs}.} 
    \label{fig:ratio_dipole}
\end{figure}

These findings can be summarised by inspecting the ratio of the cross sections, presented as a function of $x$ defined as\footnote{This choice to translate $W$ and $Q^2$ into $x$ in the dipole model calculations differs from others in the literature but the difference is only significative at large $x$ where the dipole model is not applicable.}
\begin{equation}
    x =\frac{Q^2+m_{J/\psi}^2}{Q^2+W^2+m_{J/\psi}^2-m_N^2} 
\label{eq:xvariable_jpsi}
\end{equation}
which is shown in Fig.~\ref{fig:ratio_dipole}. We observe that the ratio is smaller for smaller values of  $Q^2$, and it decreases for decreasing values of $x$.
The results from the dipole model calculations are compared with the ratio of the gluon density squared (evaluated at $x$ and $Q^2$) obtained from the nuclear PDFs using  the EPPS16 set~\cite{Eskola:2016oht}. The reason why one can compare the diffractive cross section ratios with the ratios for the gluon density squared can be understood from Eqs.~\eqref{eq:dipole_dt_xsection} and \eqref{eq:dipole_vm_elastic}. The diffractive amplitude is proportional to the gluon density $xg(x,Q^2)$. On the other hand the diffractive cross section is proportional to the amplitude squared, thus having enhanced sensitivity to the gluon density. The nuclear PDFs have   large uncertainties, which is indicated by the region between the two sets of dotted lines. The EPPS16 parametrisation is practically unconstrained in the region below $x=0.01$. Nevertheless, the estimate based on the dipole model calculation and the central value of the EPPS16 parametrisation are consistent with each other. This strongly suggests that it will be hard to disentangle nuclear effects from saturation effects and that only through a detailed combined analysis of data on the proton and the nucleus  firm conclusions can be established on the existence of a new non-linear regime of QCD.

%
%

\begin{figure}[!th]
    \centering
    \includegraphics[width=0.65\textwidth,trim={0 0 20 0},clip]{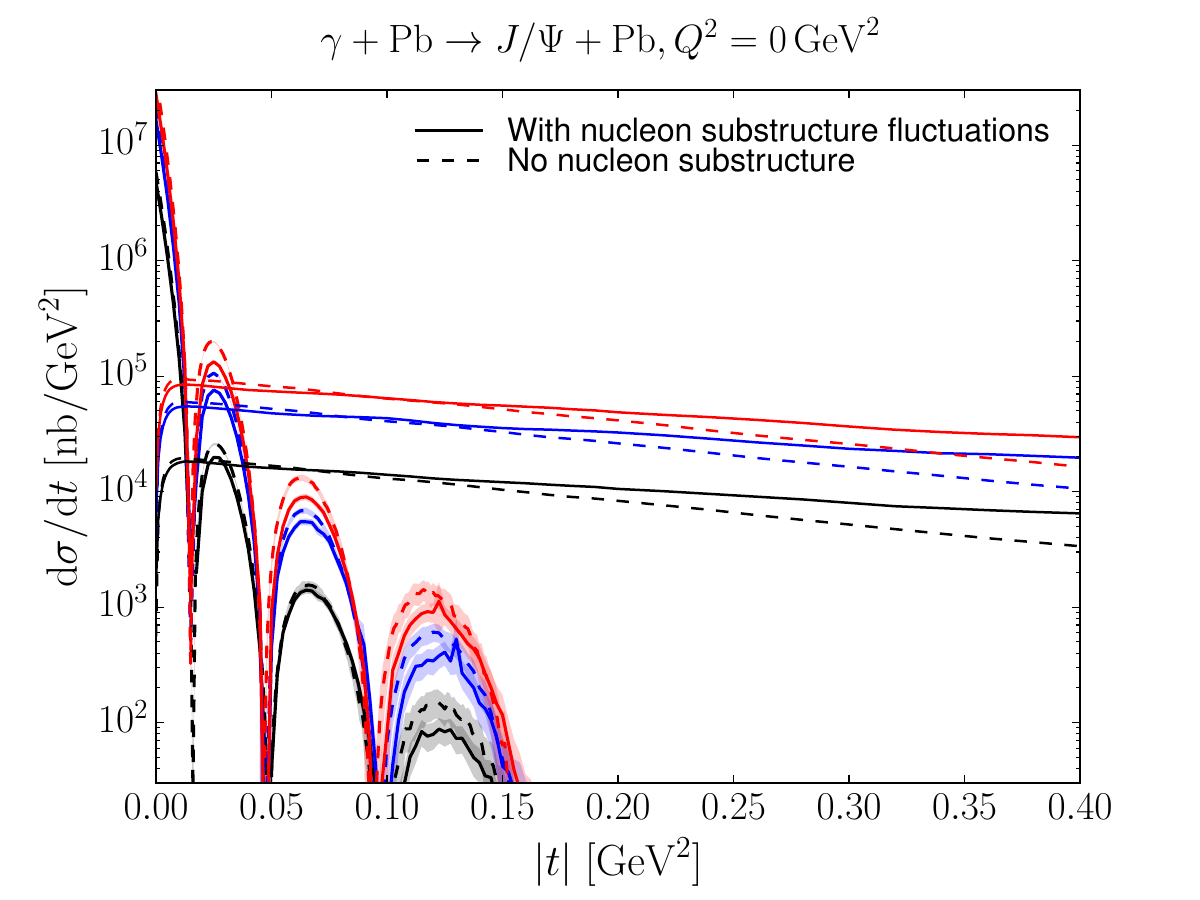}
    \caption{The differential cross sections for coherent and incoherent production of $J/\psi$ in $e$Pb as a function of the negative four momentum transfer squared $-t$, for photoproduction, $Q^2 = 0$. The lines showing dips are for  coherent production, and those extending to large $|t|$ are for incoherent. The solid (dashed) lines are the results with (without) nucleon substructure fluctuations. Black, blue, red are for $W=0.1, 0.813, 2.5$\,TeV, respectively.}
    \label{fig:dsigmadt_eA}
\end{figure}

The differential cross section $d\sigma/dt$ as a function of the negative four momentum transfer squared $-t$ for the case of coherent and incoherent production is shown in Fig.~\ref{fig:dsigmadt_eA}.
Coherent and incoherent diffraction cross sections are computed from the dipole model in the following way. The coherent diffractive cross section is obtained by averaging the diffractive scattering amplitude over the target configurations and taking the square
\begin{equation}
\frac{d\sigma}{dt} = \frac{1}{16 \pi} |\langle  {\mathcal A}(x,Q,\Delta)\rangle|^2 \; .
    \label{eq:coherent_eA}
\end{equation}
Here the brackets $\langle \dots \rangle$ refer to averages over different configurations of the target. The incoherent cross section is obtained by subtracting the coherent cross section from the total diffractive cross section. It is standardly assumed that it takes the form of a variance of the diffractive scattering amplitude
\begin{equation}
\frac{d\sigma}{dt} = \frac{1}{16 \pi}\bigg( \langle  |{\mathcal A}(x,Q,\Delta)|\rangle^2
- |\langle  {\mathcal A}(x,Q,\Delta)\rangle|^2\bigg)\; ,
    \label{eq:incoherent_eA}
\end{equation}
which should be valid for small $|t|$.
The $t$ dependence, and the relation between the impact parameter and $t$ through the Fourier transform, makes diffractive scattering a sensitive probe of the internal geometric structure of hadrons and nuclei, see Ref.~\cite{Adamczyk:2017vfu}  for an extraction of the transverse profile of the nucleus in ultraperipheral collisions at RHIC; also Ref.~\cite{Toll:2012mb} for an study for the EIC. In particular, because the incoherent cross section has the form of a variance of the amplitude, it is sensitive to the amount of fluctuations in  impact parameter space.

The results in Fig.~\ref{fig:dsigmadt_eA} (results for higher $Q^2$ are very similar)
indicate that the incoherent production is dominant for most values of $-t$, except for the very small momentum transfers,  about $|t|<0.02\,\textrm{GeV}^2$. Thus, dedicated instrumentation which will  allow us to distinguish between the two cases is essential if one wants to measure the coherent process in a reasonably wide range of $|t|$. As in the proton case, the coherent $t$ distribution exhibits characteristic dips. However, in the case of the nuclear targets the dips occur for much smaller values of $t$. This is related to the much larger value of the dipole  amplitude for a wide range of impact parameters in the case of nuclear targets  compared to the proton case.


Another interesting aspect, see Sec.~\ref{sec:PSM_Disc_3D}, is the effect of the transverse structure of the target in nuclear coherent and incoherent diffraction~\cite{Mantysaari:2020axf}. For example, in the  formulation shown above~\cite{Mantysaari:2017dwh} a fixed number of hot spots was considered, while in~\cite{Krelina:2019gee} (see also~\cite{Mantysaari:2018zdd} for a realisation using small-$x$ evolution) a growing number with $1/x$ is implemented. In both cases, the ratio of incoherent to coherent diffraction decreases with $W$, being smaller for larger nuclei. This decrease is sensitive to the details of the distribution of hot spots - thus, to the fluctuations of the gluon distribution in transverse space. It also shows  interesting dependencies on the mass of the produced vector meson and on $Q^2$, resulting in the ratio being  smaller for lighter vector mesons and for lower $Q^2$.
Besides, the hot spot treatment also has some effects on the distributions in momentum transfer, see Fig.~\ref{fig:dsigmadt_eA}.
In order to check these ideas, both the experimental capability to separate coherent form incoherent diffraction, and a large lever arm in $W$ and $Q^2$ as available at the LHeC, are required.


We thus conclude that by investigating  coherent and incoherent diffractive scattering on nuclei, one gets unique insight into the spatial structure of matter in nuclei. 
On the one hand, the coherent cross section, which is obtained by averaging the amplitude before squaring it,
 is sensitive to the average spatial density distribution of gluons in transverse space. On the other hand, 
the incoherent cross section, which is governed by the variance of the amplitude with respect to the initial
 nucleon configurations of the nucleus, measures fluctuations  of the gluon density inside the nucleus. In 
the case of a nucleus, the diffractive production rate is controlled by two different scales related to the
 proton and nucleus size. At momentum scales corresponding to the nucleon size $|t| \sim 1/R_p^2$  the 
diffractive cross section is almost purely incoherent. The $t$-distribution in coherent diffractive 
production off the nucleus gives rise to a dip-type structure for both saturation and non-saturation 
models, while in the case of incoherent production at small $|t|$, both saturation and non-saturation
 models do not lead to dips~\cite{Mantysaari:2017dwh}. This is in drastic contrast to the diffractive
 production off the proton where only saturation models lead to a dip-type structure in the $t$-distribution 
at values of $|t|$ that can be experimentally accessible. Therefore, diffractive production 
offers a unique opportunity to measure the spatial distribution of partons in the 
protons and nuclei. 
It is also an excellent tool to investigate the   approach to unitarity  in the high energy limit of QCD. 

While we have focused here on J/$\psi$ production, lighter vector mesons like $\rho,\omega,\phi$ could also be studied. They should show a different $Q^2$ dependence and their larger sizes would make them lie closer to the black disk regime. Also the dominance of two-jet events in photoproduction would provide sensitivity to the approach to the unitarity limit~\cite{Frankfurt:2011cs}.

\subsection{Inclusive diffraction on nuclei}

In Sec.~\ref{sec:inclusive_diffraction}, a study of the prospects for extracting diffractive parton densities in the proton was presented following~\cite{Armesto:2019gxy}.
Diffraction in $e$A is similar to that in $ep$, 
the main difference being a smaller contribution from 
incoherent $e+A \to e+X+A^*$ than from coherent $e+A \to e+X+A$ diffraction, with $A^*$ denoting a final state where the nucleus dissociates into at least two hadrons but the event still shows a rapidity gap. Incoherent diffraction dominates for $|t|$ larger than a few hundredths of a 
GeV$^2$. 
Forward detectors~\cite{AbelleiraFernandez:2012cc} will allow the separation of coherent diffraction, on which we focus in the following, summarising the study in Ref.~\cite{Armesto:2019gxy}.

Assuming that the same framework (collinear factorization for hard 
diffraction, Eq. \eqref{eq:collfac}, and Regge factorization,
Eq. \eqref{eq:param_2comp}) introduced in 
Sec.~\ref{sec:inclusive_diffraction} for $ep$ also
holds for $e$A, nuclear diffractive PDFs (nDPDFs) can be 
extracted from the diffractive reduced cross sections. Note that such nDPDFs have never been measured. For an electron energy $E_e=60$\,GeV and nuclear beams with $E_N=2.76$\,TeV/nucleon, the kinematic 
coverage at the LHeC is very similar to that shown in Fig.~\ref{fig:phasespace_xQ}. For details, see Ref.~\cite{Armesto:2019gxy}.

\begin{figure}
  \centering
  \includegraphics[width=0.65\textwidth,trim={0 0 0 11},clip]{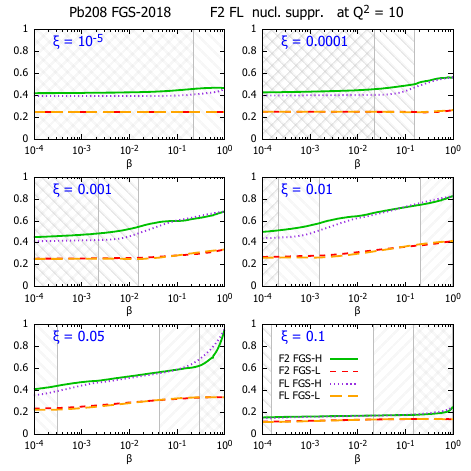}
\caption{Nuclear modification factor, Eq.~\eqref{eq:ndmf}, for $F_{2}^{D(3)}$ 
and $F_{L}^{D(3)}$ in $^{208}$Pb versus $\beta$, at $Q^2=10$\,GeV$^2$ and 
for different $\xi$, for the models H and L in~\cite{Frankfurt:2011cs}. 
The `\textbackslash' and `/' hatched areas show 
kinematically excluded regions for $E = 2.76$ and 19.7\,TeV/nucleon, respectively. Taken from Ref.~\cite{Armesto:2019gxy}.}
\label{fig:ratio_Pb-e60}
\end{figure}

\begin{figure}[!th]
\centering
	\includegraphics*[width=0.47\textwidth,trim={0 0 0 24},clip]{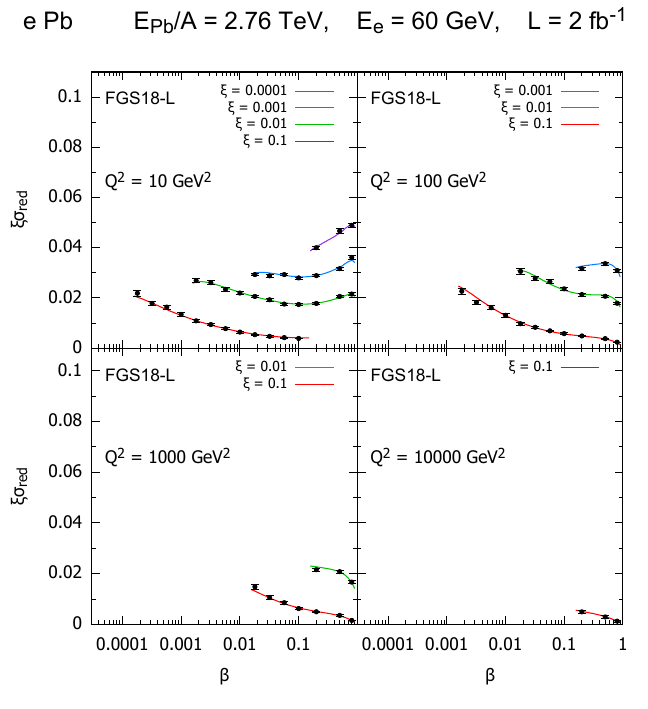}%
	\includegraphics*[width=0.47\textwidth,trim={0 0 0 24},clip]{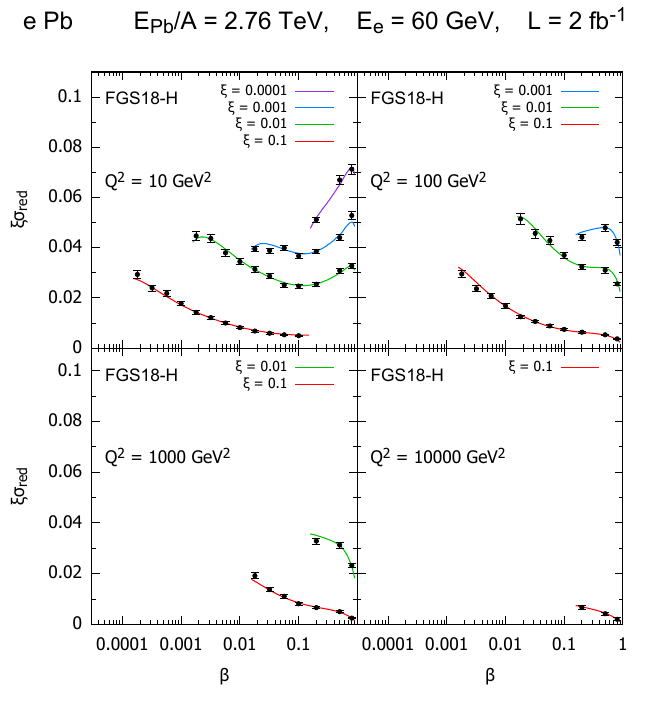}%
\caption{An indicative subset of simulated data for the diffractive reduced cross section as a function of $\beta$ in bins of $\xi$ and $Q^2$ for $e\,^{208}$Pb collisions at the LHeC, in the models in~\cite{Frankfurt:2011cs}.
The curves for $\xi = 0.01, 0.001, 0.0001$ are shifted up by 0.01, 0.02, 0.03, respectively. Taken from Ref.~\cite{Armesto:2019gxy}.}
\label{fig:sigred_Pb7e60}
\end{figure}

Defining, in analogy to Eq.~\eqref{eq:ratio},
the diffractive nuclear modification factor 
\begin{equation}
R_k^A(\beta,\xi,Q^2) = \frac{f_{k/A}^{D(3)}(\beta,\xi,Q^2)}{A\,f_{k/p}^{D(3)}(\beta,\xi,Q^2)}\;,
\label{eq:ndmf}
\end{equation}
we show in Fig.~\ref{fig:ratio_Pb-e60} the results for $F_{2}^{D(3)}$ 
and $F_{L}^{D(3)}$ from the FGS models~\cite{Frankfurt:2011cs}. These models are based on
Gribov inelastic shadowing~\cite{Gribov:1968jf} which relates diffraction in 
$ep$ to nuclear shadowing for total and diffractive 
$e$A cross sections. The nuclear wave function squared is approximated by the product of one-nucleon densities, the $t$-dependence of the diffractive $\gamma^*$-nucleon amplitude is neglected compared to the nuclear form factor, and a real part in the amplitudes~\cite{Gribov:1968uy} and  colour fluctuations for the inelastic intermediate nucleon states~\cite{Frankfurt:1994hf} are introduced. 
There are two models, named H and L, that correspond to different strengths of the colour fluctuations and result in larger and smaller probabilities for diffraction in nuclei with respect to that in proton, respectively.
In Figs.~\ref{fig:ratio_Pb-e60} and \ref{fig:sigred_Pb7e60} we show results~\cite{Armesto:2019gxy} for both models.

A  subset of the simulated pseudodata for the reduced cross sections is shown in Fig.~\ref{fig:sigred_Pb7e60}~\cite{Armesto:2019gxy}. They
are generated assuming 5\,\% systematic error and statistic errors calculated for an integrated luminosity of $2\,\mathrm{fb}^{-1}$.
Compared to Fig.~\ref{fig:sigred_ep_lhec}, the comparably large kinematic coverage and small (systematics-dominated) uncertainty illustrated in Fig.~\ref{fig:sigred_Pb7e60} show clearly that an extraction of nDPDFs in $^{208}$Pb  analogous to that shown in Figs.~\ref{fig:pdf_fits_lhec} and \ref{fig:pdf_7_50_xi} for the DPDFs, will be possible with similar accuracy in an extended kinematic region.


\section[New Dynamics at Small $x$ with Nuclear Targets]{\boldmath New Dynamics at Small $x$ with Nuclear Targets}
\label{sec:NPP_smallx}

As discussed in Sec.~\ref{sec:PSM_Disc_smallx}, theoretical expectations~\cite{Kovchegov:2012mbw} indicate that fixed-order perturbation theory leading to the DGLAP evolution equations should eventually fail. When $x$ decreases, $\alpha_s\ln 1/x$ becomes large and these large logarithms must be resummed, leading to the BFKL equation. Furthermore, when the parton density becomes large, the linear approximation that underlies both DGLAP and BFKL breaks, and non-linear processes must be taken into account to compute parton evolution. The CGC~\cite{Gelis:2010nm} offers a non-perturbative but weak coupling effective theory to treat dense parton systems
in a systematic and controlled way. One of the important predictions of the CGC is that  in a dense parton system  saturation occurs leading to the emergence of a new dynamical scale -- the saturation scale $Q_{\textrm sat}$, which increases with the energy.

The parton density in a hadron becomes high both through evolution -- when energy or $1/x$ becomes large, and/or when partons are accumulated by overlapping nucleons -- when mass number $A$ becomes large in a nucleus. In the nucleus rest frame, the virtual photon fluctuations at small $x<(2m_NR_A)^{-1}$, with $m_N$ the nucleon mass and $R_A$ the nuclear radius, acquire a lifetime larger than the time taken to traverse the nucleus and, thus, all partons within a transverse area $\sim 1/Q^2$ are simultaneously probed. Actually, the parameter determining the transition between linear and non-linear dynamics is the parton density and, therefore, the onset of this new regime of QCD and its explanation must be tested, as commented in~\cite{AbelleiraFernandez:2012cc}, exploring both decreasing values of $x$ and increasing values of $A$ in a kinematic $x-Q^2$ region where, in order to be sensitive to differences in evolution, enough lever arm in $Q^2\gg \Lambda_{\textrm{QCD}}^2$ at small $x$ is available. The saturation scale $Q_{\textrm sat}$ that characterises the typical gluon momentum in a saturated hadron wave function increases with nuclear size, $Q_{\textrm sat}^2 \propto A^{1/3}$. Therefore, in $e$A collisions the perturbatively saturated regime  is achieved at parametrically larger $x$ than in a proton -- a prediction not only of the CGC but of all multiple scattering models that anticipate an approach to the black disk, unitarity limit.

The opportunities to establish  the existence of saturation in lepton-nucleus collisions are numerous. They include inclusive observables, both total and diffractive cross sections, and less inclusive ones like correlations:
\begin{itemize}
\item \underline{Tension in DGLAP fits for inclusive observables}: As discussed in~\cite{AbelleiraFernandez:2012cc,Rojo:2009ut} and in Sec.~\ref{sec:SM_nonlinearQCD}, deviations from fixed-order perturbation theory can be tested by the tension that would appear in the description within a DGLAP fit of observables with different sensitivities to the sea and the glue, for example $F_2$ and $F_L$ (or reduced cross sections at different energies) or $F_2^{\textrm{inclusive}}$ and $F_2^{\textrm{heavy quarks}}$. In~\cite{Marquet:2017bga}, such an exercise was performed considering $F_2$ and $F_L$ pseudodata for $e$Au collisions at the EIC~\cite{Accardi:2012qut} using reweighting techniques. While the results for EIC energies are shown not to be conclusive due to the reduced lever arm in $Q^2>Q_{\textrm sat}^2\gg \Lambda_{\textrm{QCD}}^2$, the much larger centre-of-mass energies at the LHeC (and FCC-eh) should make possible a search for tensions between different observables.
\item \underline{Saturation effects in diffraction}: A longstanding prediction of saturation~\cite{Nikolaev:1995xu,Frankfurt:1991nx,Kowalski:2008sa} is a modification of the diffractive cross section in nuclei with respect to protons, with a suppression (enhancement) at small (large) $\beta$ due to the approach of the nucleus to the black disk limit, where elastic and diffractive scattering become maximal, and the behaviour of the different Fock components of the virtual photon wave function. Such effects can also be discussed in terms of a competition of nuclear shadowing with the probability that the event remains diffractive in the multiple scattering process~\cite{Frankfurt:2011cs}. This leads to the generic expectation of an enhancement of the ratio of the coherent diffractive cross section in nucleus over that in protons, in non-linear approaches with respect to linear ones~\cite{Accardi:2012qut}.
\item \underline{Correlations}: Correlations have been considered for a long time as sensitive probes of the underlying production dynamics. For example, the cross section for the production of two jets with the same hardness and widely separated in rapidity, called Mueller-Navelet jets~\cite{Mueller:1986ey}, was proposed as a test of BFKL versus DGLAP dynamics, but the effect of saturation has not been widely studied although it has the large potentiality of differentiating linear resummation from non-linear saturation where non-trivial nuclear effects could appear. Correlations between jets were analysed in ~\cite{AbelleiraFernandez:2012cc} for the LHeC kinematics, both in inclusive and diffractive events, see the formalism in~\cite{Deak:2011ga}.  On the other hand, the azimuthal decorrelation of particles and jets when saturation effects are at work -- at small $x$, studied by the difference between collisions involving proton and nuclei, was proposed long ago in $d$Au collisions at the Relativistic Hadron Collider~\cite {Albacete:2010pg,Lappi:2012nh}. It was studied in ~\cite{AbelleiraFernandez:2012cc} for the LHeC kinematics, see recent developments in~\cite{Stasto:2018rci} and the extension to forward dijet production in~\cite{vanHameren:2016ftb}. It could also be analysed in ultraperipheral collisions at the LHC, see Sec.~\ref{sec:LHeConHLLHC_heavyions}.
\end{itemize}

\section{Collective effects in dense environments -- the `ridge' \ourauthor{Nestor, Stan, Daniel}} 
\label{sect:ridge}
\begin{figure}[!th]
\centering
        \includegraphics[width=0.38\textwidth]{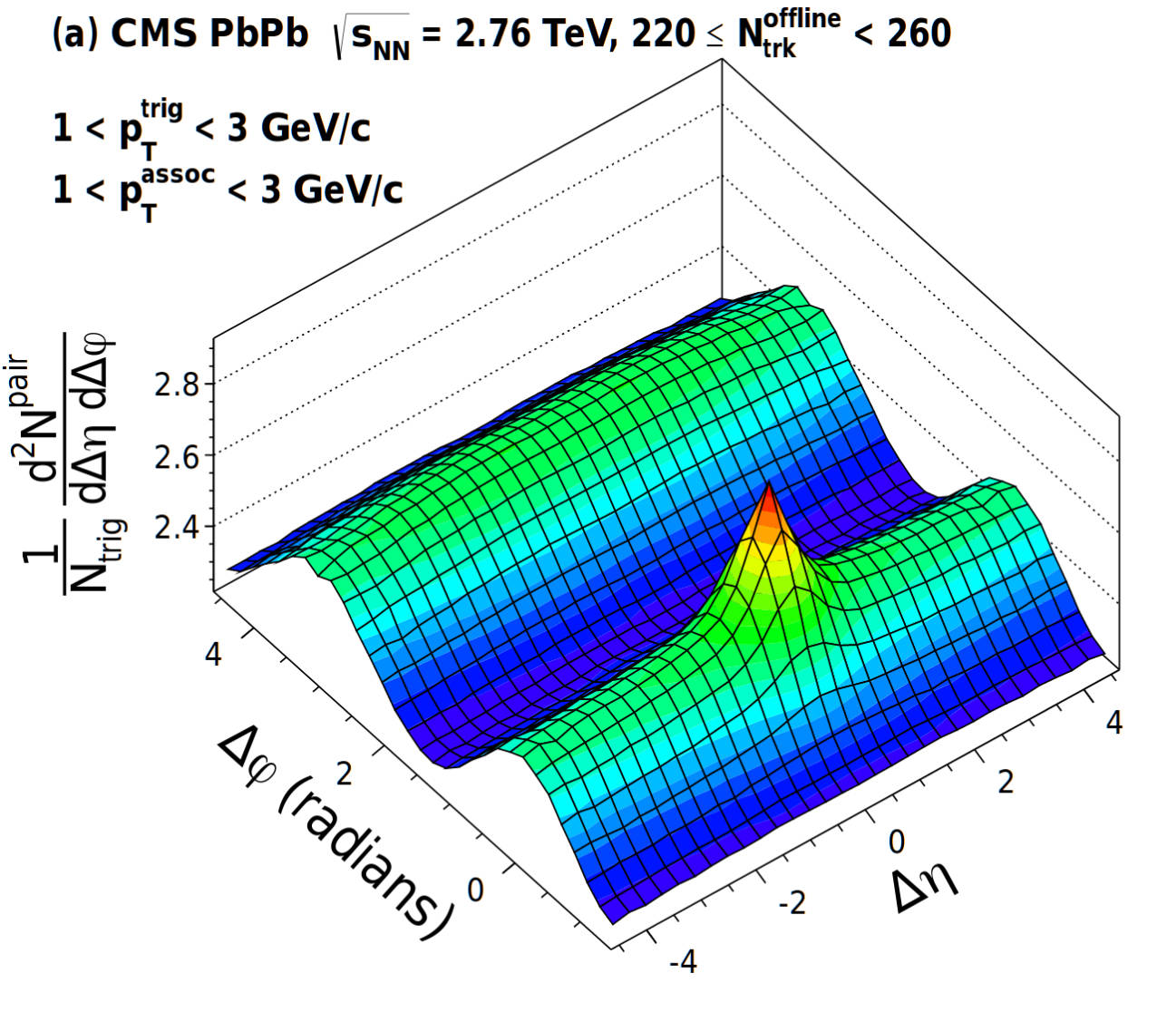}  \hspace{0.08\textwidth}%
        \includegraphics[width=0.38\textwidth]{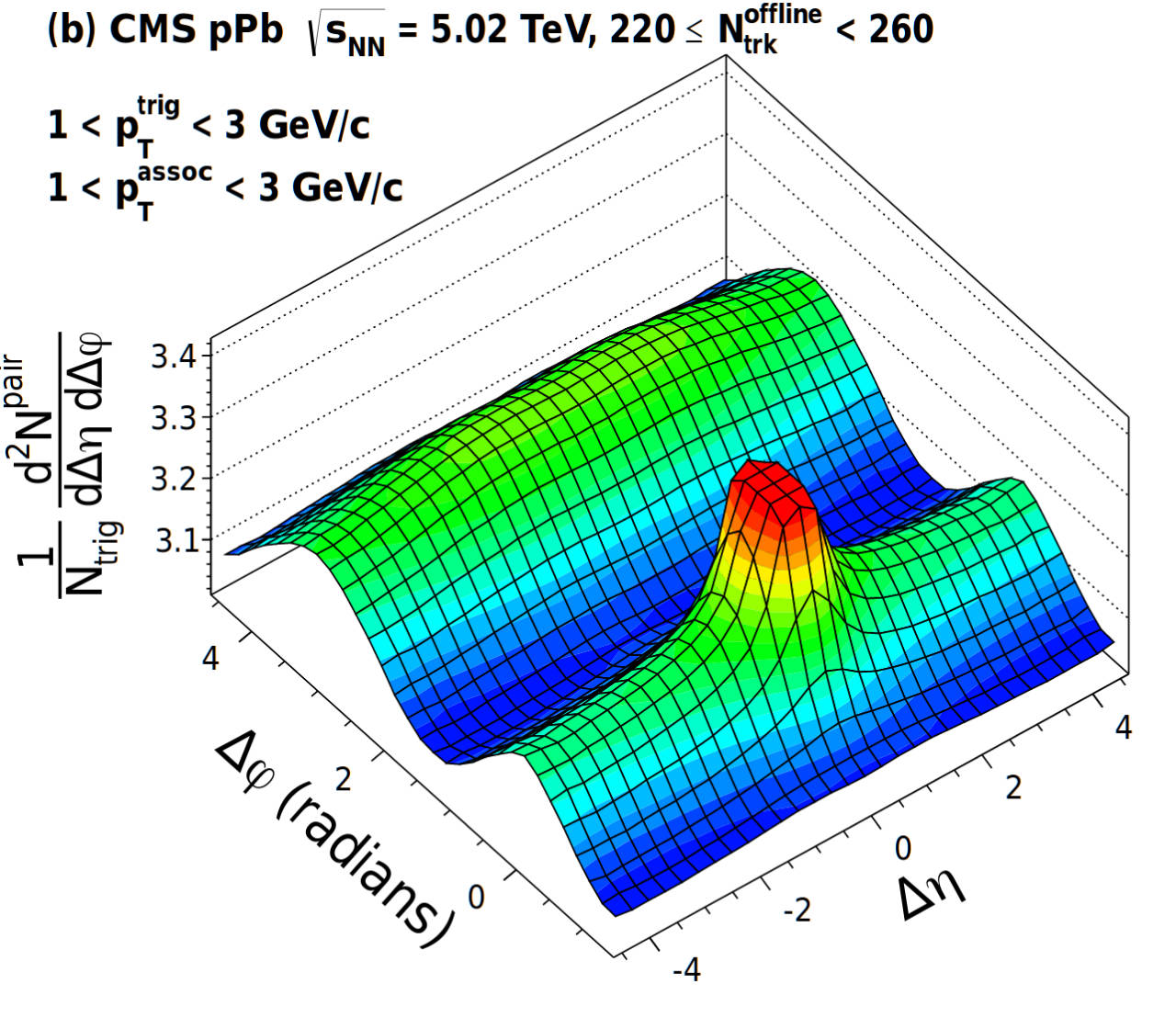}   \\ 
        \includegraphics[width=0.38\textwidth]{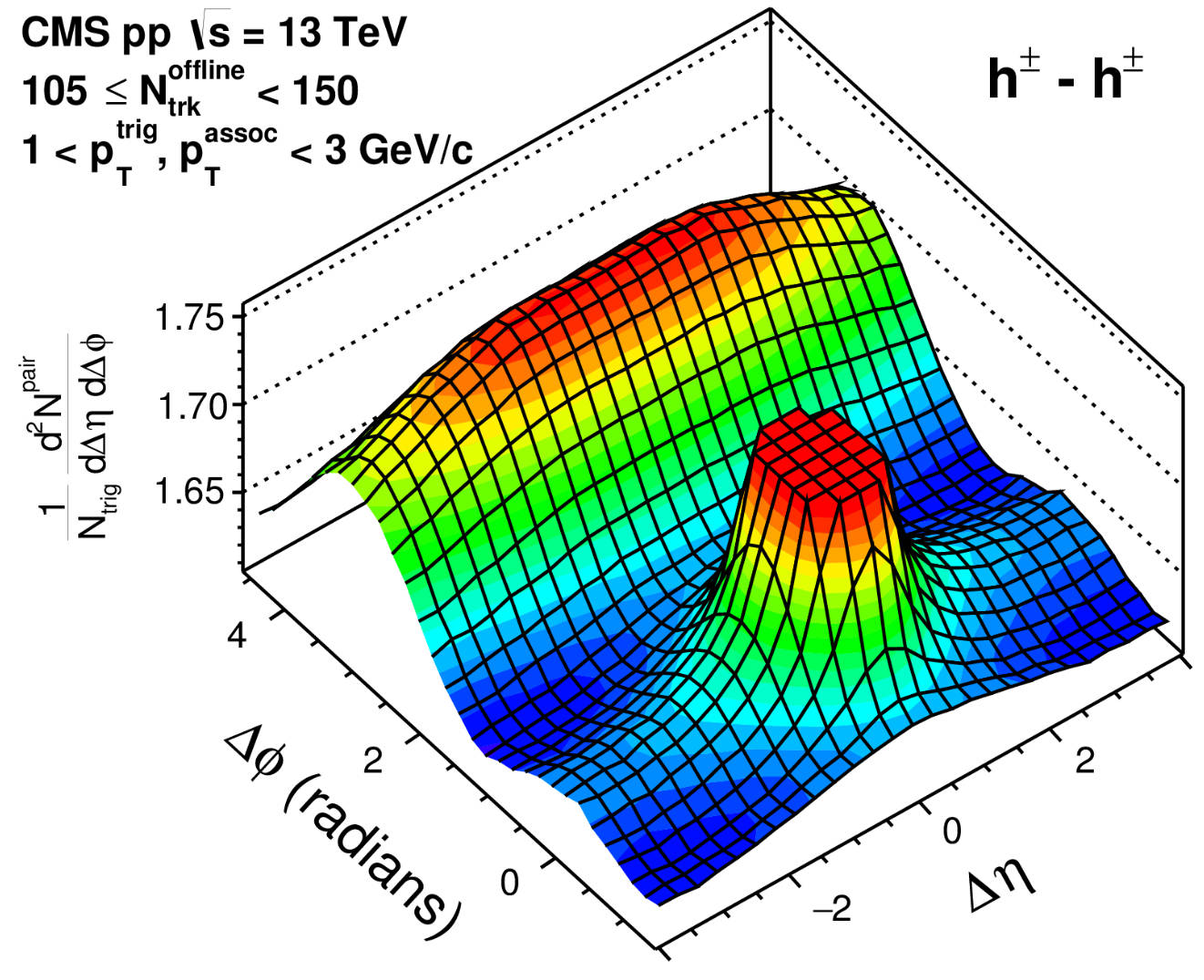}    \hspace{0.08\textwidth}%
        \includegraphics[width=0.38\textwidth,trim={0 -200 0 0},clip]{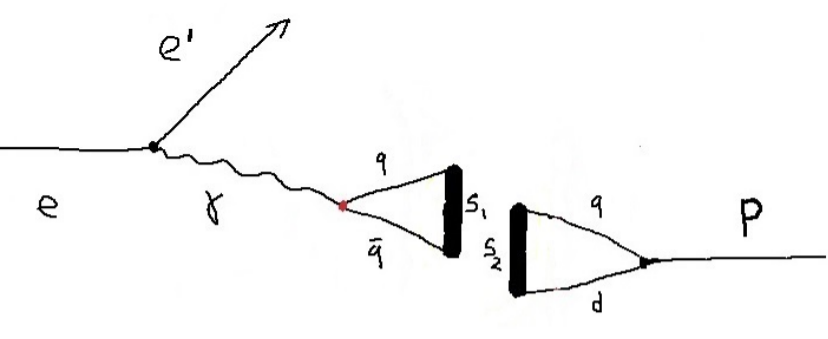}  
        \caption{ Left and top right: Collective effects seen in high-multiplicity two-particle azimuthal correlation, as observed
                  by CMS in PbPb, $p$Pb~\cite{Chatrchyan:2013nka}, and $pp$~\cite{Khachatryan:2016txc} collisions.
                 Bottom right: Schematic illustration for the production of \emph{ridge}-like effects in  $ep$ or $e$A scattering at the LHeC~\cite{Glazek:2018gqs}.
                 }
        \label{LHeC2019Ridge}
\end{figure}

%
One of the most striking discoveries~\cite{Khachatryan:2010gv} at the LHC is, that in all collision systems, from small ($pp$ and $p$A) to large (AA),  many of the features that are considered as indicative of the production of a dense hot partonic medium are observed (see e.g.\ reviews~\cite{Schlichting:2016sqo,Loizides:2016tew,Schenke:2017bog} and references therein).
The most celebrated of such features is the long rapidity range particle correlations collimated in azimuth, named the `ridge', shown in Fig.~\ref{LHeC2019Ridge}.
The dynamics underlying this phenomena, either the formation of QGP and the existence of strong final state interactions, or some initial state dynamics that leaves imprint on the final observables, is under discussion~\cite{Romatschke:2016hle}.
While observed in photoproduction on Pb in UPCs at the LHC~\cite{ATLAS:2019gsn}, its existence in smaller systems like $e^+e^-$~\cite{Badea:2019vey} at LEP and $ep$ at HERA~\cite{ZEUS:2019jya} has been scrutinised, but the results are not conclusive.

In this respect, measurements  in $ep$ and $e$A collisions at the LHeC at considerable centre-of-mass energies will offer crucial additional information.
For example, the collision of the virtual photon with the proton at the  LHeC can be considered as a high energy collision of two jets or `flux tubes',
as discussed in Refs.~\cite{Bjorken:2013boa,Glazek:2018gqs} and illustrated in Fig.~\ref{LHeC2019Ridge}.
This can lead to the production of `ridges'  and other novel configurations of gluons and quarks and will be measured uniquely at the LHeC.

%

\section{Novel QCD Nuclear Phenomena at the LHeC}

Beyond the topics discussed above there are many novel phenomena which can be explored in $e$A collisions at LHeC or FCC-eh, in a high energy regime and using dedicated instrumentation. We shall briefly review some of these phenomena, which can be understood utilizing the light-front framework of QCD, for a review see~\cite{Brodsky:1997de}.

One of the most important theoretical tools in high energy physics is Dirac's light-front (LF) time: $\tau =x^+ = t +z/c $,  the time along the light-front~\cite{Dirac:1949cp}, a concept which allows all of the tools and insights of 
Schr\"odinger's quantum mechanics and the Hamiltonian formalism to be applied to relativistic physics~\cite{Brodsky:1997de}.  
When one takes a photograph, the object is observed at a fixed LF time.  Similarly,  Compton  $\gamma p \to \gamma' p''$  and deep-inelastic lepton-proton scattering 
are measurements of proton structure at fixed LF time.    Unlike ordinary \emph{instant time} $t$,
physics at fixed $\tau$ is Poincar\'e invariant; i.e.\ independent of the observer's Lorentz frame.  Observations  at fixed $\tau$ are made within the causal horizon.  LF time  $\tau$ reduces to ordinary time $t$ in the nonrelativistic limit $c\to \infty.$

The LF wavefunctions (LFWF) of hadrons are superpositions of $\Psi^H_n(x_i, \vec k_{\perp i,} \lambda_i) = <\Psi_H|n>$, the Fock state projections of the eigensolution of the QCD
LF Hamiltonian $H_{QCD}|\Psi_H >= M^2_H \Psi_H >$. They encode the underlying structure of bound states in quantum field theory and underlie virtually every observable in hadron physics.
Hadronic LFWFs can also be measured directly by the Ashery method~\cite{Ashery:2005wa}, 
the coherent diffractive dissociation of high energy hadrons into jets~\cite{Bertsch:1981py,Frankfurt:2000jm}.
In the diffractive dissociation of a high energy hadron into quark and gluon jets by two-gluon exchange,  
the cross-section measures the  square of the second transverse derivative of the projectile LFWF. 
Similarly, the dissociation of a high energy atom such as positronium or \emph{true muonium} ($[\mu^+\mu^-]$) 
can be used to measure the transverse derivative of its LFWFs.

Hadronic LFWFs are defined at fixed $\tau =-x^+ = t+z/c$;
they are thus off-shell in the total $P^- =P^0 -P^z$, not energy $P^0$~\cite{Brodsky:1997de}.  Thus LFWFs are also off-shell in
${\mathcal M}^2=P^+ P^- -P^2_\perp =  [\sum_i k^\mu_i]^2 = \sum_i \tfrac{k^2_\perp+ m^2}{ x}_i,$   
the invariant mass squared of the constituents in the $n$-particle Fock state. LFWFs are thus functions of the invariant mass squared of the constituents in the Fock state. For a two-particle Fock state,
${\mathcal M}^2 = \tfrac{k^2_\perp+ m^2 }{ x(1-x)}$.   Thus, the constituent transverse momenta  $k^2_{\perp i} $ do appear alone as a separate factor in the LFWF; the transverse momenta are always  coupled to the longitudinal LF momentum fractions $x_i$.  This is
the light-front version of rotational invariance.  Only positive $ k^+_i  =k^0_i+k^z_i \ge 0$ and $0 \le x_i =\tfrac{k^+-}{ P^+}  \le 1$  appear,  where  
$\sum_i x_i = 1. $ 
In addition, $J^z = \sum_i { L^z_i+S^z_i}$,  as well as $P^+ =\sum_i k^+_i$ and $\vec P_\perp= \sum_i \vec k_{\perp i}$ are conserved at every vertex -- essential covariant kinematical constraints.  A remarkable property: the anomalous gravitomagnetic moment of every LF Fock state vanishes at $Q^2=0$.   The LFWFs of bound states are off-shell in $P^- = \sum_i k^-_i$, but they tend to be maximal at minimal off-shellness; i.e.\ minimal invariant mass. In fact, in the holographic LFWFs where colour is confined, the LFWFs of hadrons have fast Gaussian fall-off in invariant mass. This feature also underlie intrinsic heavy quark Fock states: the LFWFs have maximal support when all of the constituents
have the same rapidity $y_i$; i.e.\ $x_i \propto \sqrt{m^2_i + k^2_{\perp i}}$. Thus the heavy quarks have the highest momentum fractions $x_i$.

Conversely, light-front wavefunctions provide the boost-invariant transition amplitude which convert the free quark and gluons into the hadronic eigenstates of QCD. 
Thus, knowing the LFWFs allows one to compute \emph{hadronization at the amplitude level} -- 
how the coloured quarks and gluons produced in a deep inelastic scattering event  $e p \to e' X$ at the LHeC are confined and emerge as final-state hadrons.

The LF formalism leads to many novel nuclear phenomena, such as  \emph{hidden colour}~\cite{Brodsky:1983vf}
\emph{colour transparency}~\cite{Brodsky:1988xz},  \emph{nuclear-bound quarkonium}~\cite{Brodsky:1989jd}, \emph{nuclear shadowing and antishadowing} of nuclear structure functions, etc.  
For example, there are five distinct colour-singlet QCD Fock state representations of the 
six colour-triplet quarks of the deuteron.  These hidden-colour Fock states become manifest  when 
the deuteron fluctuates to a small transverse size,
as in measurements of the deuteron form factor at 
large momentum transfer. One can also probe the hidden-colour Fock states of the deuteron by 
studying the final state of the dissociation of the deuteron in deep inelastic lepton scattering at the LHeC $e D \to e' X$,
where $X$ can be $\Delta^{++} +\Delta^-$,  six quark jets, or other novel colour-singlet final states.

The LF wave functions provide the input for scattering experiments at the amplitude level, 
encoding the structure of a projectile at a single light-front time $\tau$~\cite{Brodsky:1997de}.  For example, consider photon-ion 
collisions. The incoming photon probes the finite size structure of  the incoming nucleus at fixed LF time, like 
a photograph -- not at a fixed instant time, which is acausal. 
Since the nuclear state is an eigenstate of the LF Hamiltonian, its structure is independent of its momentum, 
as required by Poincar\'e invariance.  One gets the same answer in the ion rest frame, the CM frame, or even if 
the incident particles move in the same direction, but collide transversely.  There are no colliding \emph{pancakes} using the LF formalism.

The resulting photon-ion cross-section is not point-like; it is shadowed: 
$\sigma(\gamma A \to X) = A^\alpha \sigma(\gamma N \to X)$, 
where  $A$ is the mass number of the ion, $N$ stands for a nucleon, and the power  $ \alpha \approx 0.8$  
reflects Glauber shadowing~\cite{Brodsky:1989qz}.  The shadowing stems from the destructive 
interference of two-step and one-step amplitudes, where the two-step processes involve diffractive reactions  
on a front-surface nucleon which shadows the interior nucleons. Thus the photon interacts primarily on the 
front surface. Similarly,  a high energy ion-ion collision  $A_1 + A_2 \to X$ involves the overlap of the incident
frame-independent LFWFs. The initial interaction on the front surface of the colliding ions can resemble a shock wave.

In the case of a deep inelastic lepton-nucleus collision $\gamma^* A \to X$, the two-step amplitude 
involves a leading-twist diffractive deep inelastic scattering (DDIS) 
$\gamma^* N_1\to V^* N_1$  on a front 
surface nucleon $N_1$ and then the on-shell propagation of the vector system $ V^*$ to a downstream 
nucleon $ N_2$ where it interacts inelastically: $V^* N_2 \to X$.  If the DDIS involves Pomeron exchange, 
the two-step amplitude  interferers destructively  with the one-step 
amplitude  $\gamma^* N_1 \to X$ thus producing shadowing of the nuclear parton distribution function at low $x < 0.1$.
On the other hand, if the DDIS process involves $I=1$ Reggeon exchange, the interference is constructive, 
producing \emph{flavour-dependent}  leading-twist antishadowing~\cite{Brodsky:1989qz}  in the domain $0.1 < x< 0.2$.

One can also show that  the Gribov-Glauber processes, which arise from leading-twist diffractive deep inelastic  scattering on nucleons 
and 
underly the shadowing and antishadowing of nuclear structure functions~\cite{Brodsky:1989qz}, prevent the application of the operator product expansion to the virtual Compton scattering amplitude $\gamma^* A \to\gamma^* A$  on nuclei and  thus negate the validity of the momentum sum rule for deep inelastic nuclear structure functions~\cite{Brodsky:2019jla}.

%
%
%
%
%
%
%

%

\biblio

%% file: higgs/higgs.tex
\linenumbers
\lhectitlepage
\lhecinstructions
\subfilestableofcontents

\chapter{Higgs Physics with LHeC}
\label{chapter:higgs}


\label{sec:smhiggsinep}
\section{Introduction}
The Higgs boson was discovered in 2012 by ATLAS~\cite{Aad:2012tfa} and 
CMS~\cite{Chatrchyan:2012xdj} at the Large Hadron Collider (LHC). It  is the most recently discovered and  
least explored part of the Standard Model.
The Higgs Boson (H) is of fundamental importance. It is related to the mechanism predicted by~\cite{Englert:1964et,Higgs:1964pj,Guralnik:1964eu} and independently by~\cite{Migdal:1966tq}], in which the intermediate vector bosons of the spontaneously broken electroweak symmetry acquire masses~\footnote{The mass of the $W$ boson, $M_W$, is generated through the vacuum expectation
value, $\eta$, of the Higgs field ($\Phi$) and given by the simple relation $M_W = g \eta /\sqrt{2}$ where
$g$ is the weak interaction coupling. Here $\eta=\sqrt{-\mu^2/2\lambda}$ with the
two parameters of the Higgs potential that is
predicted to be $V=-\mu^2 \Phi^+\Phi - \lambda (\Phi^+\Phi)^2$.
The Higgs mass is given as $M_H = 2\eta \sqrt{\lambda}$ while the mass of the $Z$ boson
is related to $M_W$ with the electroweak mixing angle, $M_Z=M_W/\cos\Theta_W$.} while the photon remain massless.
%
%
Fermions obtain a mass via the Yukawa couplings with the Higgs field. Following the 
discovery of the Higgs boson, its physics and thorough exploration has become a central
theme of the physics programme at the LHC. Any high-energy future collider project,
beginning with the high luminosity upgrade of the Large Hadron Collider, the HL-LHC,
underway to collect data in a decade hence, has put the potential to precisely
study the properties of the Higgs boson into its center of attention, for understanding
its characteristics and hoping to open a new window into physics extending beyond
the Standard Model, see for example~\cite{Gori:2017tvg,Cohen:2018mgv}.
In this section we present the potential to explore the SM Higgs physics at the 
LHeC and to certain extent at FCC-eh also.
 

A first challenge on the physics of the Higgs boson is to establish whether it
indeed satisfies the properties inherent to the Standard Model (SM) regarding 
its production and decay mechanisms. The SM neutral $H$ boson decays into pairs of fermions, $f\bar{f}$. The dominant decay
is $H \to b \bar{b}$ with a branching fraction of about $58$\,\%. The branching 
scales with the square of the fermion mass, $m_f^2$. The next prominent fermionic decay
therefore is $H \to \tau^+\tau^-$ with $6.3$\,\% followed by the charm decay with a predicted
branching fraction of $2.9$\,\%. The Higgs boson also decays into pairs of $W$ and $Z$ bosons
at a rate of $21.5$\,\% and $2.6$\,\%, respectively. Loop diagrams enable the decay into gluon 
and photon pairs with a branching of $8.2$ and $0.2$\,\%, respectively. The seven most frequent
 decay channels, ordered according to descending branching fractions, thus are into
 $b\bar{b}$, $W^+W^-$, $gg$, $\tau^+\tau^-$, $c\bar{c}$, $ZZ$ and $\gamma\gamma$. Together these 
are predicted to represent a total SM branching fraction of $99.9$\,\%. At the LHC these
 and rarer decays can be reconstructed, with the exception of the charm decay for reasons
 of prohibitive combinatorial  background. The main purpose of this paper is to evaluate
 the prospects for precisely measuring these channels in electron-proton scattering.

\section{Higgs Production in Deep Inelastic Scattering}
In deep inelastic electron-proton scattering, the 
Higgs boson is predominantly produced through
$WW$  fusion in charged  current DIS (CC) scattering, Fig.\,1.
The next large Higgs production mode in $ep$ is 
$ZZ \to H$ fusion 
in neutral current DIS (NC) scattering, Fig.\,1, 
which has a smaller but still sizable cross section.
\begin{figure}[!th]
\centering
    \includegraphics[width=0.35\textwidth]{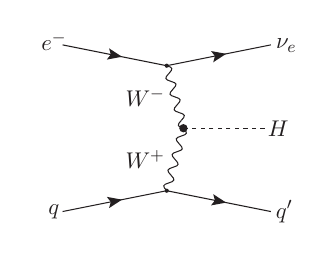}
    \hspace{0.05\textwidth}
    \includegraphics[width=0.35\textwidth]{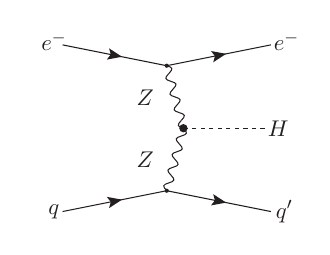}
   \caption{Higgs boson production in charged (left) and neutral (right) current deep inelastic electron-proton  scattering to leading order.}
    \label{fig:feynH}
\end{figure}
These $ep$ Higgs production processes are very clean for a number of reasons:
\begin{itemize}
    \item 
 even at the high luminosity of $10^{34}$\,cm$^{-2}$s$^{-1}$ the inclusive pileup is only 0.1 (1) for the LHeC (FCC-eh) and the final state signature therefore free from event overlap, in contrast to the HL-LHC where it will typically be 150; 
 \item in $ep$, contrary to $pp$, there is no initial nor final state colour (re)connection; 
 \item the higher-order corrections are small. For the total CC process they were estimated~\cite{Blumlein:1992eh} to be of the order of only $1$\,\% for the QCD part, subject to cut dependencies yielding shape changes up to 20\,\%, and $-5$\,\% for the QED part (with a weak dependence on the PDF choice).
 The smallness of the QCD corrections was attributed mainly to the absorption of gluon and quark radiation effects in the evolution of the parton distributions (PDFs)~\cite{Blumlein:1992eh}. The PDFs will be measured with very high precision at any of the $ep$ colliders here considered, see Chapter 3,
 thus allowing a unique self-consistency of Higgs cross section measurements.
\end{itemize}

 The NC reaction is even cleaner than the CC process as the scattered electron fixes the kinematics more accurately than the missing energy. While in $pp$ both 
$WW$ and $ZZ$ processes are hardly distinguishable, in $ep$ they uniquely are, which provides an important, precise  constraint on the $WWH$ and $ZZH$ couplings.


%
\subsection{Kinematics of Higgs Production}
\label{sec:kine}
\begin{figure}[!th]
\centering
\includegraphics[width=0.7\textwidth]{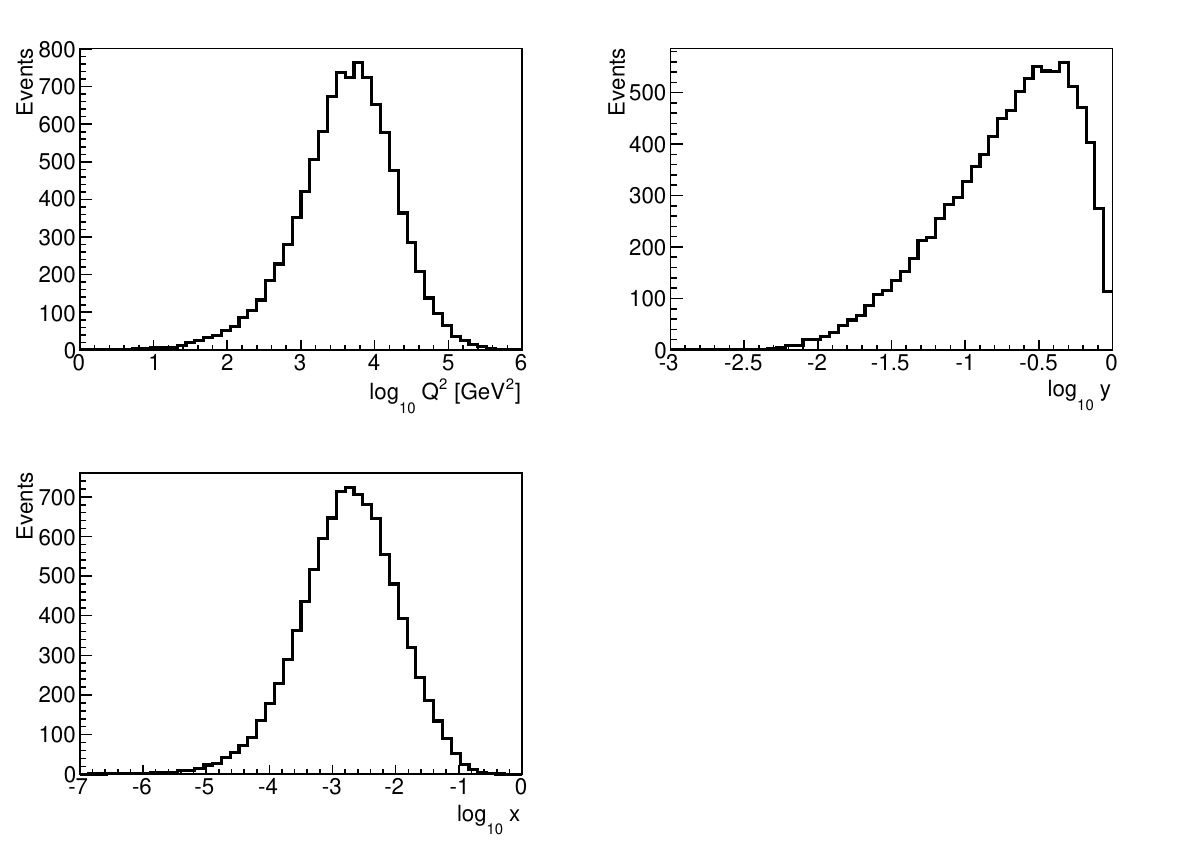}
\caption{Distributions for $ep \to \nu H X$ events, at parton-level, for the negative 4-momentum transfer squared, $Q^2$ (top left),
Bjorken $x$ (bottom left) and the inelasticity $y=Q^2/sx$ (top right) at $\sqrt{s}=3.5$\,TeV (FCC-eh). Events generated with MadGraph~\cite{Alwall:2014hca}, see Tab.~\ref{tab:Hcross}. }
\label{fig:qxy}
\end{figure}

%
 At HERA the kinematics was conveniently reconstructed through event-wise measurements of $Q^2$ and $y$. The reconstruction of the kinematics in charged currents uses the  inclusive hadronic final state measurements. Based on the energies $E_e'$ and $E_h$
and the polar angles $\Theta_e$ and $\Theta_h$ of the scattered electron and the hadronic final state, respectively, one obtains a redundant determination of the kinematics in neutral current scattering. This permits a cross calibration of calorimetric measurements, of the electromagnetic and hadronic parts and of different regions of the detector, which is a major means to achieve
superb, sub-percent  precision in $ep$ collider measurements.
Methods have been developed to optimise the kinematics reconstruction and maximise the acceptance by exploiting the redundant determination of the scattering kinematics, see for example~\cite{Bassler:1994uq}. 
The basic DIS kinematic distributions of $Q^2,~x$ and $y$ for Higgs production at $\sqrt{s}=3.5$~TeV are illustrated in Fig.\,\ref{fig:qxy}.
The average $Q^2$ and $x$ values probed 
are $Q^2 \approx 2000$\,GeV$^2$, $x\approx$0.02 at LHeC and $Q^2 \approx 6500$\,GeV$^2$, $x \approx$0.0016 at FCC-eh. 
%

\begin{figure}[!th]
  \centering
    \includegraphics*[width=0.65\textwidth]{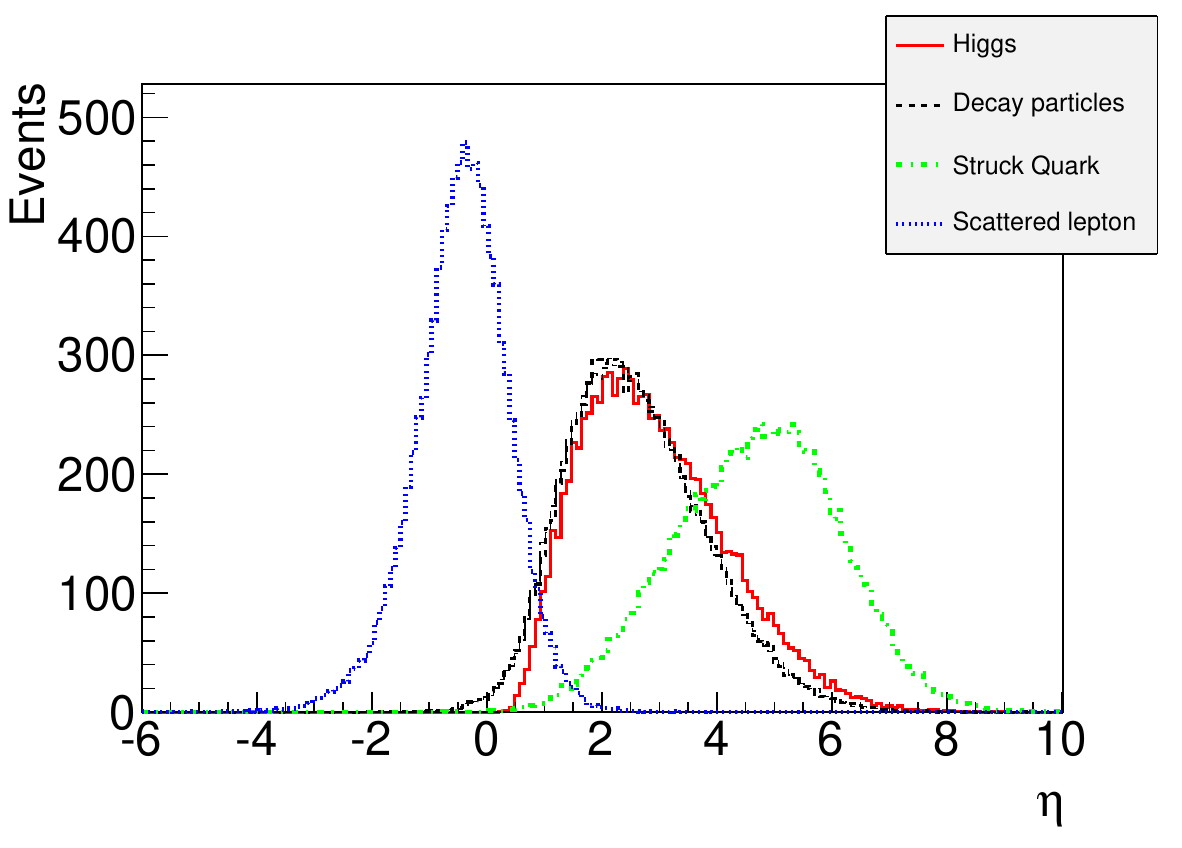}
\caption{Pseudorapidity ($\eta$) distributions, at parton-level, characterising the Vector-Boson-Fusion production and 
decay of the Higgs boson to $WW$ in 
DIS scattering at FCC-eh. The scattered lepton  (blue) in the NC case (or missing energy for CC) has an average $\eta$ of about $-0.5$, i.e. it is scattered somewhat backwards (in electron beam
direction). The pseudorapidity distributions of the generated 
Higgs boson (red) and its decay particles (black) are very similar and peak at $\eta \simeq 2$.
The struck quark, especially at the FCC-eh as compared to LHeC, generates a very forward jet requiring forward calorimetry up to $\eta \simeq 6$ as is foreseen in the FCC-eh detector 
design. Events are generated with MadGraph, 
see setup in Tab.~\ref{tab:Hcross}.
}
\label{fig:eta}
\end{figure}

As is described in this paper elsewhere,
constraints for a large pseudorapidity or polar angle, $\eta = \ln \tan \theta /2$, acceptance of the apparatus arise i) for the
backward region (the polar angle is defined w.r.t. the proton beam direction) from the need to reconstruct electrons at low $Q^2$ enabling low $x$ physics
and ii) for the forward region to cover a maximum region towards large $x$
at medium $Q^2$ with the reconstruction of the hadronic final state. The acceptance therefore
extends, for the LHeC, to pseudorapidities of $\eta = \pm 5$, which
for the FCC-eh case is extended to $\eta= \pm 6$. The large acceptance
is in particular suitable for the reconstruction of Vector-Boson-Fusion Higgs boson event signatures, see 
Fig.\,\ref{fig:eta} for the typical pseudorapidity distributions of Higgs boson event signature in DIS at the most asymmetric FCC-eh collider configuration.

\begin{figure}[htb!]
  \centering
  \includegraphics*[width=0.6\textwidth,trim={0 0 20 250},clip]{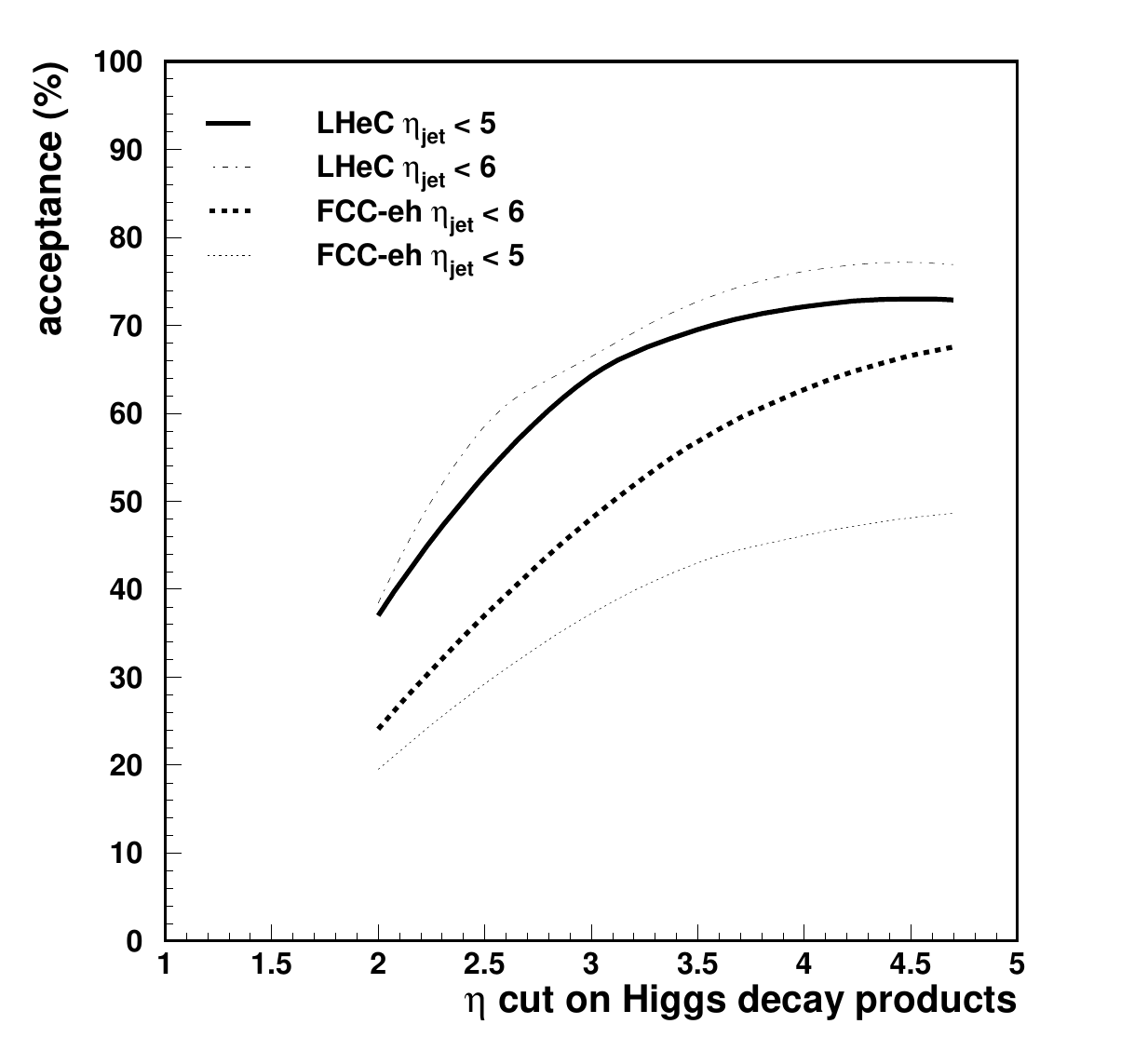}
\caption{Acceptance of DIS Higgs candidates ($y$ axis) in dependence on the pseudorapidity ($\eta$) cut requirement on the Higgs decay products ($x$ axis) for two scenarios of the coverage of the hadronic final states. All final states are selected with $p_T> 15$~GeV. The forward jet is accepted up to $\eta=5$ and $\eta=6$ for LHeC (full and dashed-dotted lines), and FCC-eh (dotted and dashed lines), respectively. Calculations are at parton-level using MadGraph. 
}
\label{fig:acc_eta}
\end{figure}

Geometric acceptances due to kinematic constraints in the pseudorapidity on the Higgs decay products for both LHeC and FCC-eh are further illustrated in Fig.~\ref{fig:acc_eta}. The acceptances are calculated for a basic selection of all final states with $p_T> 15$~GeV and a coverage of the forward jet up to $\eta=5$ and $\eta=6$, respectively,  for both colliders.
As seen from Fig.~\ref{fig:acc_eta}, the acceptances are higher for the less asymmetric LHeC beam configuration and about the same for hadronic calorimetry up to $\eta=5$ and $\eta=6$. Hence, the LHeC calorimeter is designed for $\eta=5$. 
The optimal hadronic calorimetry coverage for FCC-eh is clearly $\eta=6$ yielding significantly higher acceptances in comparison to an $\eta=5$ calorimetry.
From Fig.~\ref{fig:acc_eta}, it is apparent that for both collider configurations the Higgs decay products would require tagging capabilities up $\eta=3.5$, e.g. for heavy flavour and tau decays.
Suitably designed muon detectors covering $\eta=4$ appear feasible for both collider configurations, those would result in high $H \to \mu \mu$ acceptances of about 72\,\% (63\,\%) for LHeC (FCC-eh)  for selecting all final states with $p_T> 15$~GeV and a coverage of the forward jet up to $\eta=5$ ($\eta=6$).
A further extension to a 1$^{\circ}$ muon acceptance, would change the acceptances  marginally to 72.9\,\% (67.5\,\%) for LHeC (FCC-eh).

\subsection{Cross Sections and Rates}
\label{sec:cross}
The cross sections for Higgs production in CC and NC DIS $e^-$ scattering of a 60~GeV electron beam with protons at three different energies, for LHeC, HE-LHeC and FCC-eh, are summarised in Tab.\,\ref{tab:Hcross}. 
    The cross sections are calculated to leading order with MadGraph (MG5 v2.5.1) using the CTEQ6L1 proton PDF and $M_H=125$\,GeV. The CC $e^-p$ cross section is directly proportional to
    the beam polarisation, $P$, as $\sigma_{CC} \propto (1-P)$ while the NC cross section only weakly depends on the polarisation~\cite{Klein:1983vs}.
%
\begin{table}[th]
  \centering
  \small
\begin{tabular}{lccccc}
  \toprule
  Parameter & Unit & LHeC    & HE-LHeC  &  FCC-eh & FCC-eh \\
  \midrule
 $E_p$ & TeV &  7    &  13.5  & 20 & 50 \\
 $\sqrt{s}$ & TeV & 1.30 & 1.77 & 2.2 & 3.46 \\
 \addlinespace
 $\sigma_{CC}$ ($P=-0.8$) & fb & 197 & 372 & 516 & 1038 \\
  $\sigma_{NC}$ ($P=-0.8$) &fb  & 24 & 48 & 70 & 149 \\
 $\sigma_{CC}$ ($P=0$) & fb& 110 & 206 & 289 & 577 \\
  $\sigma_{NC}$ ($P=0$) & fb & 20 & 41 & 64 & 127 \\
 \addlinespace
 HH in CC & fb & 0.02 & 0.07 & 0.13 & 0.46 \\
 \bottomrule
\end{tabular}
\caption{Total cross sections, in fb, for inclusive Higgs production, $M_H=125$\,GeV, in charged and neutral current deep inelastic $e^-p$ scattering for an unpolarised ($P=0$) and polarised ($P=-0.8$) $E_e=60$\,GeV electron beam and
    four different proton beam energies, $E_p$,  for LHeC, HE-LHeC and two FCC-eh versions. The c.m.s. energy squared in $ep$ is $s=4E_eE_p$. The last row shows the double-Higgs CC production cross
    sections in fb.
     The calculations are at LO QCD using the CTEQ6L1 PDF~\cite{Pumplin:2002vw} and the default scale of MadGraph~\cite{Alwall:2014hca} with dependencies due to scale choices of  5-10\,\%.
 }
    \label{tab:Hcross}
\end{table}
It is observed that the CC Higgs production cross section at LHeC is comparable to that of a 250\,GeV $e^+e^-$ 
collider. One thus expects, roughly, results of comparable sensitivity. The difference being that $e^+e^-$
favours the $H$ to $ZZ$ couplings while $ep$ is dominantly sensitive to $WW \to H$ production. This provides a fundamental complementarity between $e^+e^-$ and $ep$ collider Higgs physics. 

The CC $e^-p$ cross section is enlarged with the (negative)
electron beam polarisation, $P_e$, while the NC cross section is less sensitive to $P_e$. The cross
section at FCC-eh reaches values of pb. Combined with long operation time one reaches sub-permille
precision of the Higgs couplings. Similarly, the $HH$ cross section approaches 0.5~fb values only with the
highest energy as expected for $\sqrt{s}>3$~TeV FCC-eh or CLIC-ee colliders. A first cut-based study to access the Higgs self-coupling at FCC-eh within 20\% is detailed in Ref.~\cite{Kumar:2015kca}. Further prospects are not discussed here since measuring the $HH$ coupling is one of the foremost tasks of HL-LHC and the FCC-hh~\cite{FCC_CDRv3}.

    The polarised $e^+p$ cross section is calculated to be significantly smaller than the $e^-p$ value,
by a factor of $197/58 \simeq 6$ at the LHeC, mainly because 
the $W^-u \rightarrow \bar{d}$ reaction is more frequent than $W^+d \rightarrow u$.  Furthermore, positron sources
are currently considered to be much less intense (by a factor of about ten or even a hundred)
than electron sources. It is desirable to take  $e^+p$ data at future $ep$ colliders
for electroweak physics but in the linac-ring version their amount will be limited and unlikely
suitable for precision Higgs physics. 

Table\,\ref{tab:Hehrates} provides an illustration of the statistics which is expected to be available in charged and neutral current scattering for nine decay channels ordered by their
branching ratios for the nominal LHeC and FCC-eh configurations. 
The statistics at LHeC would be about ten times lower than that at FCC-eh
since the cross section is diminished by $\simeq 1/5$ and due to a shorter expected running time, i.e. the integrated luminosity is assumed to be half of that at FCC-eh. Accessing rarer SM Higgs decay channels
is the particular strength of luminous
$pp$ scattering at highest energies rather than that of anticipated $ep$ or $e^+e^-$ colliders.
\begin{table}[ht!]
  \centering
  \small
\begin{tabular}{lcrrrr}
\toprule
 &  &  \multicolumn{4}{c}{Number of Events} \\
  \cmidrule(lr){3-6}
 &  & \multicolumn{2}{c}{Charged Current} & \multicolumn{2}{c}{Neutral Current} \\
 \cmidrule(lr){3-4} \cmidrule(lr){5-6}
 Channel & Fraction & LHeC & FCC-eh & LHeC & FCC-eh \\
 \midrule
 $b \overline{b} $ & 0.581  & 114\,500 & 1\,208\,000 & 14\,000 & 175\,000 \\
 $W^+W^- $ & 0.215  & 42\,300 &447\,000 &  5\,160 & 64\,000 \\
  $ g g $ & 0.082  &   16\,150 &171\,000 &  2000 & 25\,000 \\
    $ \tau^+ \tau^-$ & 0.063   & 12\,400 & 131\,000 &  1\,500 & 20\,000 \\
 $c \overline{c}  $ & 0.029  &  5700 & 60\,000 &  700 & 9\,000 \\
 $Z Z $ & 0.026  & 5\,100 & 54\,000 &  620 & 7\,900 \\
 \addlinespace
  $ \gamma \gamma $ & 0.0023  & 450 & 5\,000 &  55 & 700 \\
    $Z \gamma $ & 0.0015  &  300 & 3\,100 &  35 & 450 \\
  $ \mu^+ \mu^-$ & 0.0002  & 40 & 410 &  5 & 70 \\
  \midrule
 $\sigma$\,[pb]  &  & 0.197 & 1.04 & 0.024 & 0.15 \\
 \bottomrule
\end{tabular}
\caption{Total event rates, and cross sections, for SM Higgs decays in the charged 
($ep \to \nu HX$) and neutral ($ep \to e HX$) current
production in polarised ($P=-0.8$)
electron-proton deep inelastic scattering at LHeC ($\sqrt{s}=1.3$ TeV) and FCC-eh ($\sqrt{s}=3.5$ TeV),
for an integrated luminosity of $1$ and $2$\,ab$^{-1}$, respectively.  
The branching fractions are taken from~\cite{deFlorian:2016spz}. The estimates are at LO QCD using the CTEQ6L1 PDF and the default scale of MadGraph, 
see setup in Tab.~\ref{tab:Hcross}.
}
\label{tab:Hehrates}
\end{table}
The signal strength and coupling analyses subsequently presented deal with the seven most frequent decays representing $99.9$\,\% of the SM Higgs decays. In addition, there is a significant potential for a measurement of
the $H \to \mu \mu$ decay at the FCC-eh, which, as is seen in Tab.\,\ref{tab:Hehrates}, 
may provide about $500$ ($45$)\,events, from CC and NC DIS at FCC-eh (LHeC). Thus one may
be able to measure this process to about $6$\,\% precision at the FCC-eh and  $18$\,\% at LHeC. 

%
\section{Higgs Signal Strength Measurements }
%
Standard Model Higgs production in deep inelastic $ep$ scattering proceeds via Vector-Boson-Fusion in either charged
or neutral current scattering as it is illustrated in Fig.\,1. The scattering cross sections, including the decay of the Higgs boson into a pair  of particles $A_i \bar{A}_i$ can be written as
\begin{equation}
    \sigma_{CC}^i = \sigma_{CC} \cdot \frac{\Gamma^i}{\Gamma_H} ~~~\mathrm{and}~~~  \sigma_{NC}^i = \sigma_{NC} \cdot \frac{\Gamma^i}{\Gamma_H}.
    \label{eqsig1}
\end{equation}
Here the ratio of the partial to the total Higgs decay width defines the
branching ratio, $br_i$, for each decay  into $A_i\bar{A_i}$. The $ep$ Higgs production
cross section and the O(1)\,ab$^{-1}$ luminosity prospects enable to 
consider the seven most frequent SM Higgs decays, i.e. those into fermions ($b\bar{b},~c\bar{c},~\tau^+\tau^-$) and into gauge particles ($WW,~ZZ,~gg,~\gamma \gamma$) with high precision at the LHeC and its higher energy versions.

In $ep$ one 
obtains  constraints on the Higgs production characteristics from
CC and NC scattering, which probe uniquely either the HWW and the HZZ production, respectively. 
Event by event via the selection of the final state lepton which is either an electron (NC DIS) or missing energy (CC DIS) those production vertices can be uniquely distinguished, in contrast to $pp$.
 In $e^+e^-$, at the ILC, one has considered operation at $250$\,GeV and
separately at $500$\,GeV to optimise the HZZ versus the HWW sensitive production cross section measurements~\cite{Asner:2013psa}. For CLIC the c.m.s. energy may be set to $380$\,GeV as a compromise working point for  joint NC and CC measurements, including access to top production~\cite{Abramowicz:2016zbo}. The salient advantage of the $e^+e^-$ reaction, similarly considered for the more recent circular collider proposals, CEPC~\cite{CEPCStudyGroup:2018ghi} and FCC-ee~\cite{Abada:2019zxq}, stems from the kinematic constraint of the Higgs-strahlung, $e^+e^- \to Z^* \to Z H$, which determines the total Higgs production cross section independently of its decay.

The sum of the branching ratios 
for the seven Higgs decay channels here under study for $ep$ 
adds up to $99.87$\,\% of the 
total SM width~\cite{branchings}. As is 
discussed in Sect.\,\ref{sec:Hinv}, significant 
constraints of the $H \to invisible$
decay can be set with $ep$ also albeit not being 
able to exclude exotic, unnoticed Higgs decays. 
The accurate reconstruction of all decays considered here will  present a severe constraint on the total cross section and with that of the total decay width of the Higgs boson in the SM. For the evaluation of the measurement
accuracy, the cross section measurement prospects for a decay channel $i$ are 
presented here as relative signal strengths $\mu^i(NC,CC)$, obtained from division by the SM cross section.

Initially, detailed simulations and  Higgs extraction studies for LHeC were made for the 
dominant $H \to b\bar{b}$~\cite{Han:2009pe,tanaka,kay,klein_ichep14,uecha} and the 
challenging $H \to c\bar{c}$~\cite{uecha,izda1,izda2} channels. 
The focus on the $H \to b\bar{b}$ decay has been driven not only by its dominance but as well by the difficulty of its accurate reconstruction at the LHC. It has been natural to
extend this to the $H \to c\bar{c}$ which currently is considered to not be observable at 
the HL-LHC, for permutation and large background reasons. 
The results of the updated $b$ and $c$ decay studies, using cuts and boosted decision tree (BDT)
techniques, are presented below.

A next detailed analysis has been performed for the $H \rightarrow W^+W^-$ decay. The total of the $WW$ decays represents $21.5$\,\% of the Higgs branching into SM particles. There is a special interest in its reconstruction in the DIS charged current reaction as this channel uniquely determines the $HWW$ coupling to its
fourth power. A complete signal and background simulation and eventual BDT analysis 
of the $H \rightarrow W^+W^-$ decay in charged currents has been 
performed which is subsequently described. Unlike at LHC,
this uses the purely hadronic decays which in $pp$ are very difficult to exploit.

Finally, as summarised below, an analysis using acceptance, 
efficiency and signal-to-background scale factors has been 
established for the residual four of the seven dominant 
decay channels, Tab.\,\ref{tab:Hehrates}.  
This estimate could be successfully benchmarked with the
detailed simulations for heavy quark and $W$ decays.
The present study therefore covers more than $99$\,\% of the SM Higgs decays, which in $ep$
are redundantly measured, in both neutral and charged current reactions. This opens
interesting prospects for precision Higgs physics in $ep$, but as well
 in combination with $pp$, i.e. of LHeC with HL-LHC, and later of FCC-eh with FCC-hh.
%
%
%
\subsection{Higgs Decay into Bottom and Charm Quarks}
\label{sec:Hbc}
The  Higgs boson decays dominantly into $b\bar{b}$ with a $58$\,\% branching ratio
in the SM. Its reconstruction at the LHC has been complicated by large combinatorial background.
Recently this decay was established with signal strengths, relatively to the SM, of
$\mu_{bb} = 1.01 \pm 0.12 (stat) \pm^{0.16}_{0.15} (exp)$ by 
ATLAS~\cite{Aaboud:2018zhk}
with a luminosity of $79.8$\,fb$^{-1}$ and of  $\mu_{bb} = 1.01 \pm 0.22 $
by CMS~\cite{Sirunyan:2018kst}
with a luminosity of $41.3$\,fb$^{-1}$. This is a remarkable experimental LHC achievement since for long one expected to not be able to measure
this decay to better than about $10$\,\% at the future HL-LHC.
Meanwhile this expectation has become more optimistic with the updated HL-LHC prospects~\cite{Atlas:2019qfx},
however, the
most hopeful assumption for the $H \to c \bar{c}$ decay is a limit to two times the
SM expectation. 

Because of the special importance of determining
the frequent $b\bar{b}$ decay most accurately, and with it the full set of SM branchings, the prime attention of the LHeC Higgs prospect studies has been given to those  two channels. The first PGS detector-level study was published with the CDR~\cite{AbelleiraFernandez:2012cc} before the announcement of the
discovery of the Higgs boson and assuming $M_H=120$~GeV. 
This and subsequent analyses use samples generated by MadGraph5~\cite{Alwall:2014hca}, for both signal and background
events with fragmentation and hadronization via PYTHIA 6.4~\cite{Sj_strand_2006} in an $ep$ 
customised programme version\footnote{The hadronic showering is not expected to change the kinematics of the DIS scattered lepton. This has been shown, see page 11 of Ref.~\cite{Py_mod2014}, with the very good level of agreement of NC DIS electron kinematics with and without the ep-customized Pythia showering. Specifically, for 99.8\,\% of events the kinematics in the momentum vector components and for 98\,\% of the events the energy of the scattered electron remain unchanged. }.
Subsequent analyses have been updated to $M_H=125$~GeV and to state-of-the art
fast detector simulation with DELPHES 3~\cite{de_Favereau_2014} as testbed for $ep$ detector configurations. Both cut-based
and boosted decision tree (BDT) analyses were performed in independent evaluations.

\begin{figure}[htb!]
  \centering
      \includegraphics*[width=0.65\textwidth]{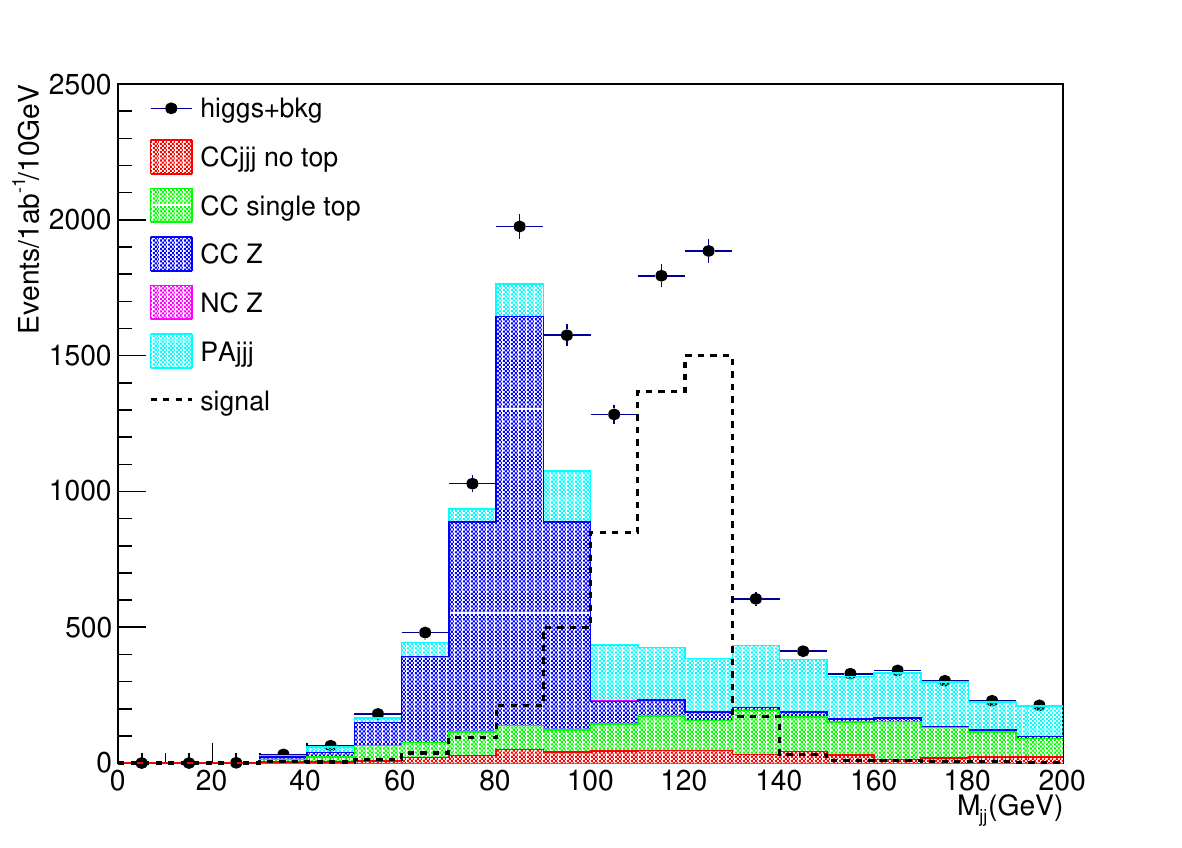}
\caption{Invariant dijet mass distribution at DELPHES detector-level expected for $1$~ab$^{-1}$ and -80\,\% electron polarisation at LHeC. The  $S/B$ is about $2.9$ for the events in the Higgs mass range of 100 to 130~GeV. Events are generated with MadGraph using $M_H=125$~GeV and showered with PYTHIA 6.4, 
 and subject to cut-based event selection criteria, see text for further details. Note that samples are generated with a minimum dijet mass cut of 60~GeV. 
}
\label{fig:cutHbb}
\end{figure}

As shown in the CDR, the $H \to b \bar{b}$ decay could be measured via applying classical kinematic selection requirements as follows:
\begin{itemize}
\item CC DIS kinematic cuts of $Q^2_h>500$ GeV$^2$, $y_h<0.9$, missing energy $E_T^\text{miss} > 30$~GeV, and no electrons in the final state to reject NC DIS;
 \item at least three anti-kt  $R=0.7$ jets with $p_T>20$~GeV which are subject to further b-tagging requirements;
 \item a Higgs candidate from two b-tagged jets  with b-tagging efficiencies of 60 to 75\,\%, charm (light quark) misidentification efficiencies of 10 to 5\,\% (1\,\%) ; 
 \item rejection of single-top events via requiring a dijet W candidate mass of greater than 130~GeV and a three-jet top candidate mass of larger than 250~GeV using a combination with one of the b-jets of the Higgs mass candidate;
 \item a forward scattered jet with $\eta>2$, and a large $\Delta \phi_{b,MET}>0.2$ between the b-tagged jet and the missing energy.
\end{itemize}
The dominant backgrounds are CC DIS multijet and single top production, while CC Z, W and NC Z contributions are small. The background due to multijets from photoproduction, where $Q^2 \sim 0$, can be reduced considerably due to the tagging of the small angle scattered electron with an electron tagger. The result of a cut-based analysis is shown in Fig.~\ref{fig:cutHbb} where clear $Z $ and $H \to b \bar{b}$ peaks are seen.
Assuming that the photoproduction background is vetoed with a 90\,\% efficiency, the resulting signal is shown in Fig.~\ref{fig:cutHbb} corresponding to a SM $H \to b \bar{b}$ signal strength $\delta \mu/\mu$ of 2\,\% 
for an integrated luminosity of 1000~fb$^{-1}$ and $P_e=-0.8$. This result is consistent with earlier analysis and 
robust w.r.t. the update of the Higgs mass from 120 to 125~GeV confirming the high $S/B>1$ (see also Ref.~\cite{klein_ichep14} where an alternative approach to estimate the multijet photoproduction background gives a similar signal strength uncertainty). The result illustrates that even with harsh kinematic requirements and already a  small luminosity of 100~fb$^{-1}$, this important decay channel could be measured to an uncertainty of about 6\,\%.

The stability of the cut-based results has been further shown for different hadronic calorimeter resolution setups 
\begin{eqnarray}
\frac{\sigma}{E} &= &\frac{a}{\sqrt{E} }\oplus b  \,\,\,\,\textrm{   for   } \,\,\,\,|\eta| <  |\eta_\text{min} |\,, \\
\frac{\sigma}{E} &= &\frac{c}{\sqrt{E} }\oplus d \,\,\,\,\textrm{   for   } \,\,\,\,|\eta_\text{min}| <  |\eta| < 5 \,,
\end{eqnarray}
where for $\eta_\text{min}=3$ the parameter $b$ ($d$) is varied within 1 (3) and 7 (9)\,\% for two resolution parameters $a$ ($c$) of either 30 (60) and 35 (45)\,\%. Alternatively, the central range was restricted to $\eta_\text{min}=2$ with parameter $b$ ($d$) of 3 (5)\,\% for resolution parameters $a$ ($c$) of 35 (45)\,\%.
While using the same analysis cuts, the signal yields varied within 34\,\%, it could be shown that with adjusted set of cuts (notably the choices of cuts for Higgs mass range, $\Delta\phi_{b,MET}$, and forward $\eta$) the SM $H \to b \bar{b}$ signal strength $\delta \mu/\mu$ varied with a fractional uncertainty of at most 7\,\%.

The cut-based $H \to b \bar{b}$ signal strength analyses are suffering from rather low acceptance times selection efficiencies in the range of 3 to 4\,\% only. 
Modern state-of-the-art analysis techniques, e.g. as performed for finding $H \to b \bar{b}$ at the LHC regardless of the overwhelming QCD jet background, are based on neural networks in the heavy flavour tagging as well as in the analysis.

\begin{figure}[ht!]
\centering 
\includegraphics[width=0.425\textwidth]{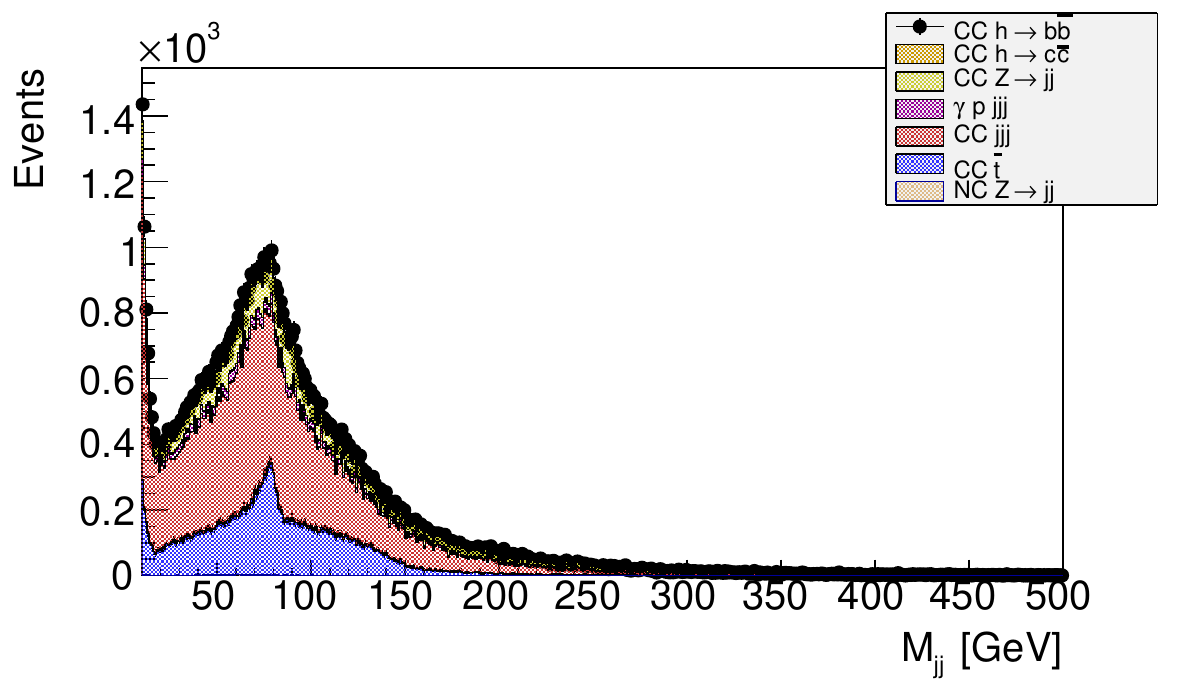}
\hspace{0.02\textwidth}
\includegraphics[width=0.425\textwidth]{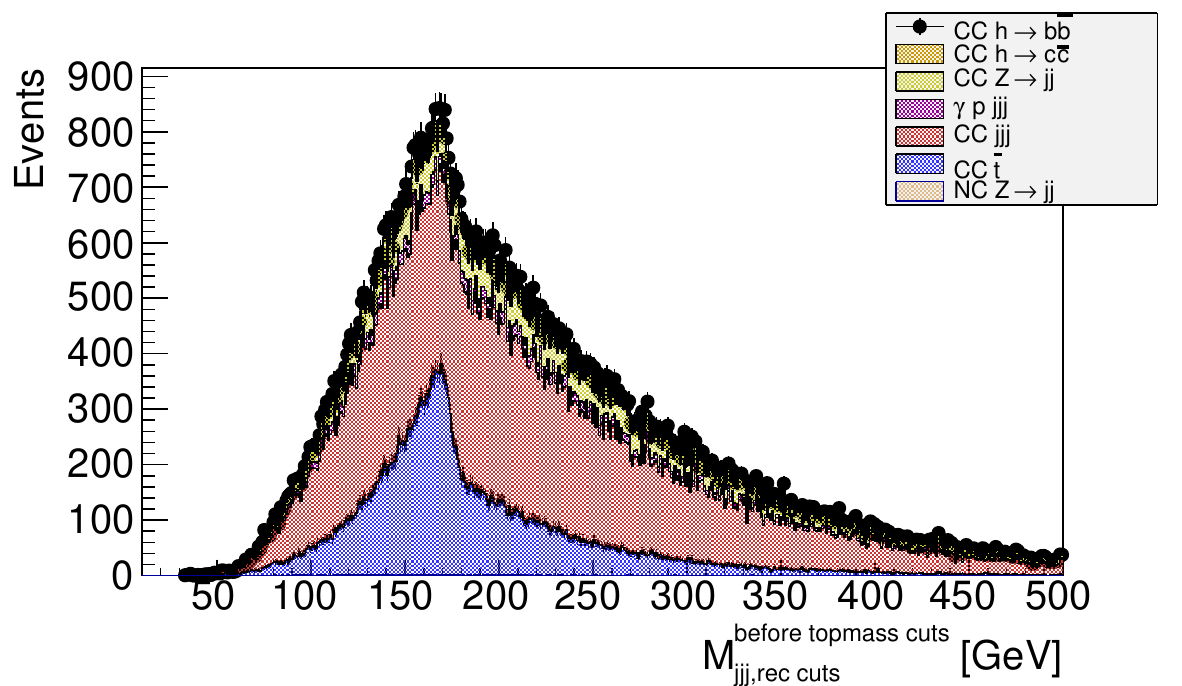}
\includegraphics[width=0.56\textwidth]{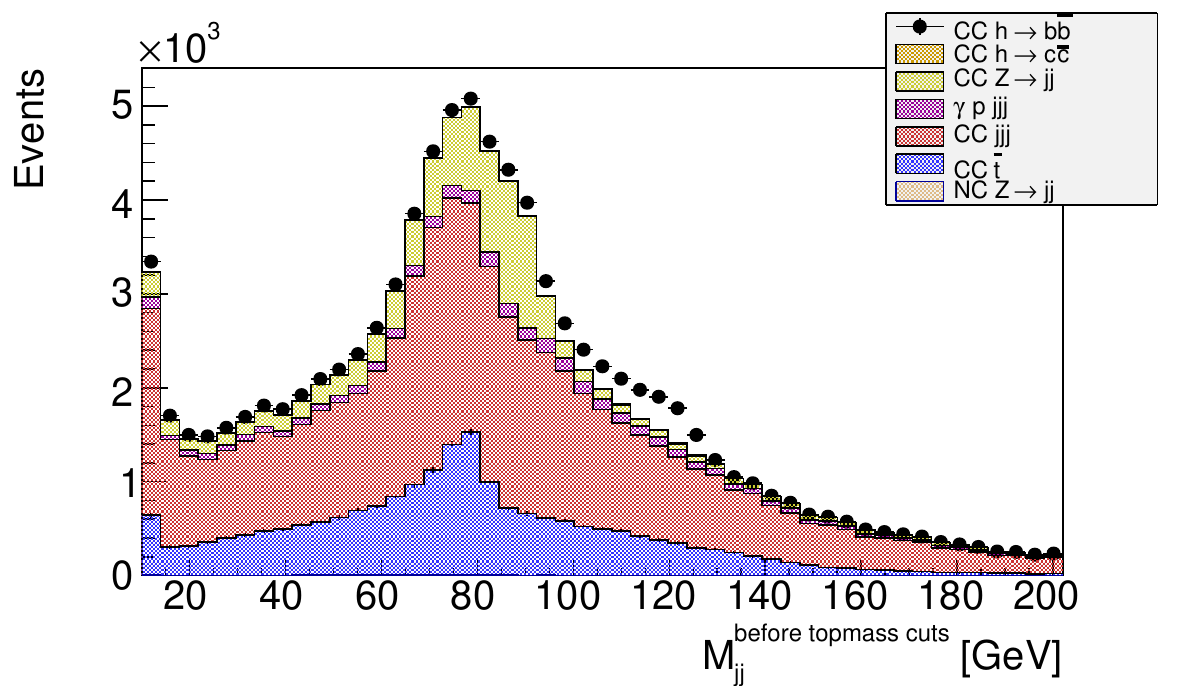}
\caption{Invariant mass distributions at DELPHES detector level for an integrated luminosity of 100~fb$^{-1}$ and $-80$\,\% electron polarisation. Events passed preselection cuts of $Q^2_h>400$ GeV$^2$, $y_h<0.9$, $E_T^\textrm{miss} > 20$~GeV and at least three, flavour-untagged anti-kt $R=0.5$ jets with $p_T>15$~GeV. The different colours show the contributions per process, the photoproduction background ($\gamma p$ jjj) is assumed to be vetoed with 90\,\%. Note that samples are generated with a minimum dijet mass cut of 60~GeV. Upper left: Invariant dijet mass, showing $W$ candidates from single top production (blue), based on combining jets with second and third lowest $|\eta|$ values per event. Upper right: Invariant mass distribution combining the three highest $p_T$ jets per event showing single top mass candidates (blue). Lower middle: Invariant dijet mass, showing Higgs candidates 
(black dots, including background), combining jets with the two lowest $|\eta|$ values per event.}
\label{fig:bbcc_presel}
\end{figure}

Boosted Decision Tree (BDT) $H \to b \bar{b}$ and $H \to c \bar{c}$ analyses using the Toolkit for Multivariate Data Analysis with ROOT (TMVA)~\cite{Hocker:2007ht} are performed using independently produced signal and background samples based on the same setup as for the cut-based analyses, see Fig.~\ref{fig:cutHbb}.
Those analyses start with loose preselections of at least three anti-kt jets with $p_T>15$~GeV  without any further heavy flavour tagging in addition to the CC DIS kinematic cuts of $Q^2_h>400$ GeV$^2$, $y_h<0.9$, and missing energy $E_T^\text{miss} > 20$~GeV. The invariant mass distributions using anti-kt $R=0.5$ jets are illustrated in Fig.\,\ref{fig:bbcc_presel}, where the mass distributions in the upper plots  illustrate in particular the single top contributions and the subsequent significant Higgs signal loss if simple anti-top cuts would be applied. 
In the lower plot  of Fig.\,\ref{fig:bbcc_presel} the invariant dijet mass distribution of untagged Higgs signal candidates is seen clearly above the background contributions in the expected mass range of 100 to 130~GeV.  
It is observed that the remaining background is dominated by CC multi-jets. 
The quantities represented in the three distributions of Fig.\,\ref{fig:bbcc_presel} are important inputs for the BDT neural network in addition to further variables describing e.g. the pseudorapidities of the Higgs and forward jet candidates including jet and track heavy flavour probabilities, see details below and further in Ref.~\cite{izda2}.

\begin{figure}[th!]
\centering
\includegraphics[width=0.65\textwidth,trim={30 250 50 0 },clip]{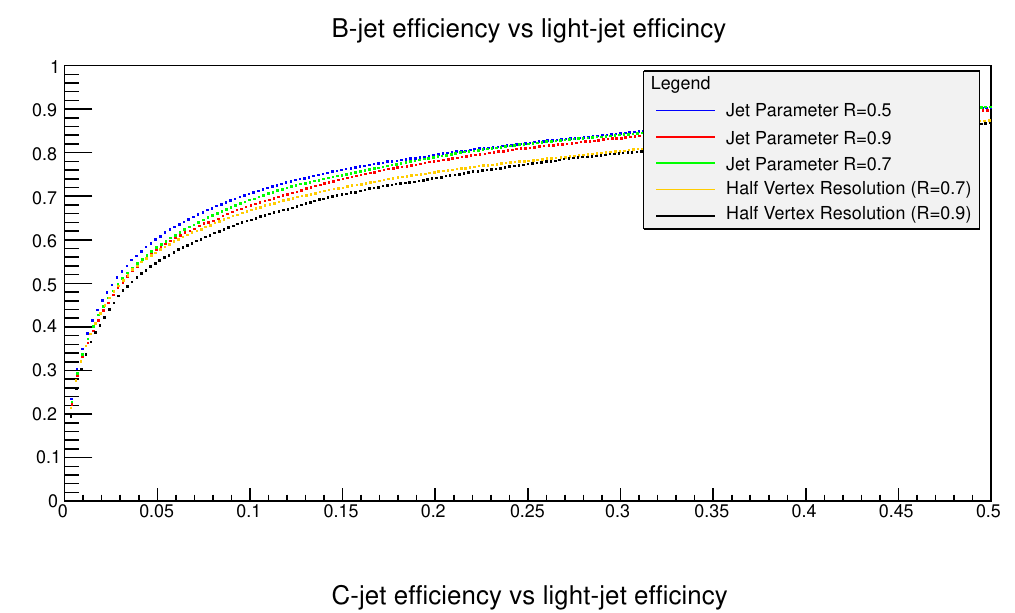}
\includegraphics[width=0.65\textwidth,trim={0 0 10 40 },clip]{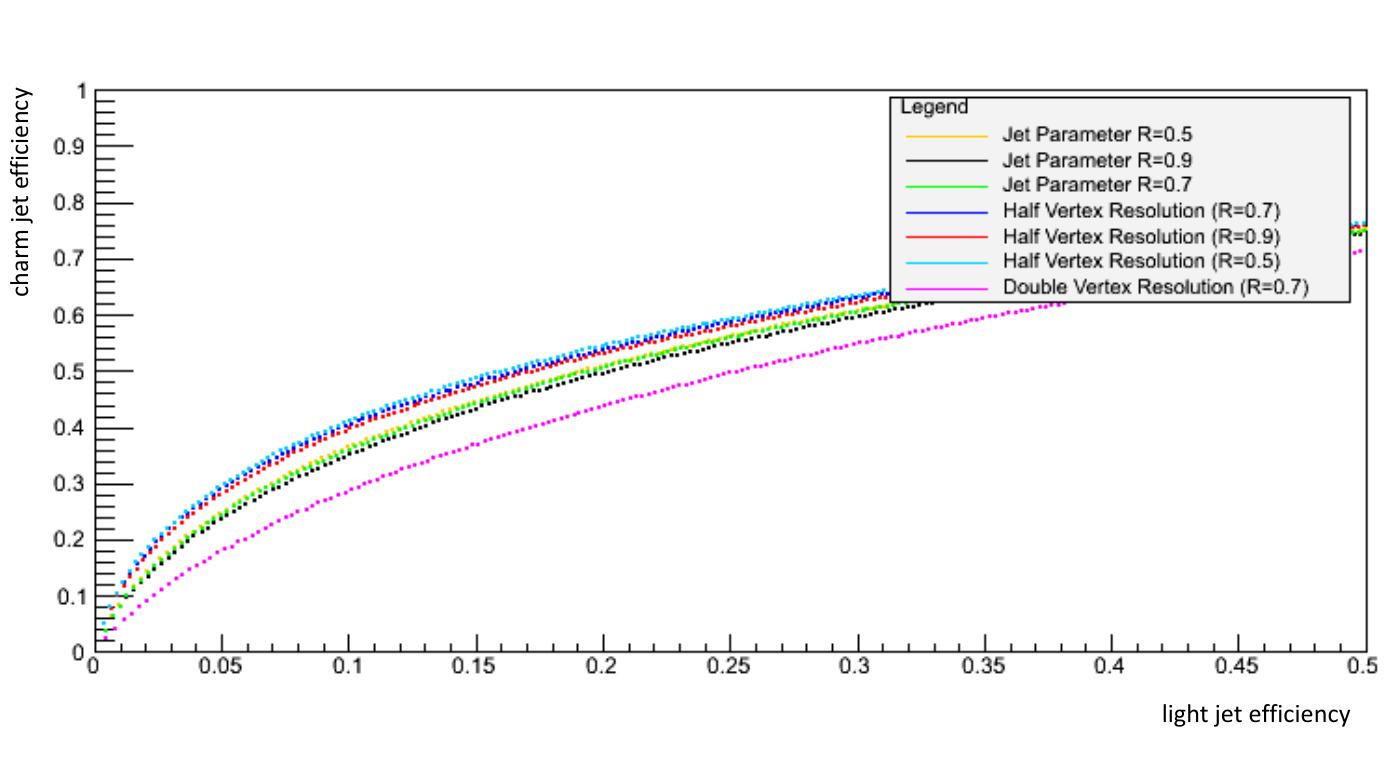}
\caption{Expected average efficiency to tag a $b$-jet (upper plot) and charm-jet (lower plot) versus the light-jet efficiency (x-axis) based on Tevatron-style jet tagging~\cite{Greder:2004sa}. Events are selected at DELPHES detector level  using a CC multi-jet sample and for an integrated luminosity of 100~fb$^{-1}$. The coloured lines correspond to  the choice of the anti-kt distance parameter $R$ and different assumptions in the impact parameter resolution of 10~(5)~$\mu$m (nominal, no text added in legend), 5~(2.5)~$\mu$m (Half Vertex Resolution), 20~(10)~$\mu$m (Double Vertex Resolution) for tracks with $0.5<p_T<5\,(>5)$~GeV within $|\eta|<3.5$. }
\label{fig:HFL_eff}
\end{figure}

As a novel element in these analyses, heavy flavour tagging based on track and jet probabilities has been implemented into the DELPHES detector analysis  following the Tevatron D0 experimental ansatz described e.g. in Ref.~\cite{Greder:2004sa}. The resulting $b$ and $c$-jet efficiency versus the light jet misidentification efficiencies are illustrated in Fig.~\ref{fig:HFL_eff} for  assumed nominal impact parameter resolution of 10~(5)~$\mu$m  for tracks with $0.5<p_T<5\,(>5)$~GeV and three choices of distance parameter $R=0.5,\,0.7,\,0.9$ for the anti-kt jets. In particular for the charm tagging, impact parameters are studied with resolutions of 5~(2.5)~$\mu$m (Half Vertex Resolution), 20~(10)~$\mu$m (Double Vertex Resolution) for tracks with $0.5<p_T<5\,(>5)$~GeV within $|\eta|<3.5$. For a conservative light jet efficiency of 5\,\%, the $b$-jet tagging efficiency is rather robust around 60\,\% for the considered nominal impact parameter performance and the three considered anti-kt distance parameters, in slight favour of the anti-kt $R=0.5$ choice. For the expected charm tagging, however, an excellent impact parameter resolution and $R=0.5$ jets give the best tagging efficiency  of around 30\,\%. This means a significant improvement e.g. w.r.t. a 23\,\% charm tagging efficiency for $R=0.9$ jets at a nominal impact parameter resolution. These tagging efficiencies can be considered as realistic but rather conservative in particular for the remaining light jet efficiency which is expected to be about 0.1\,\% at a $b$-jet efficiency of 60\,\% using LHC-style neural network based taggers.

\begin{figure}[ht!]
  \centering
  \includegraphics[height=5.3cm,trim={150 0 150 30},clip]{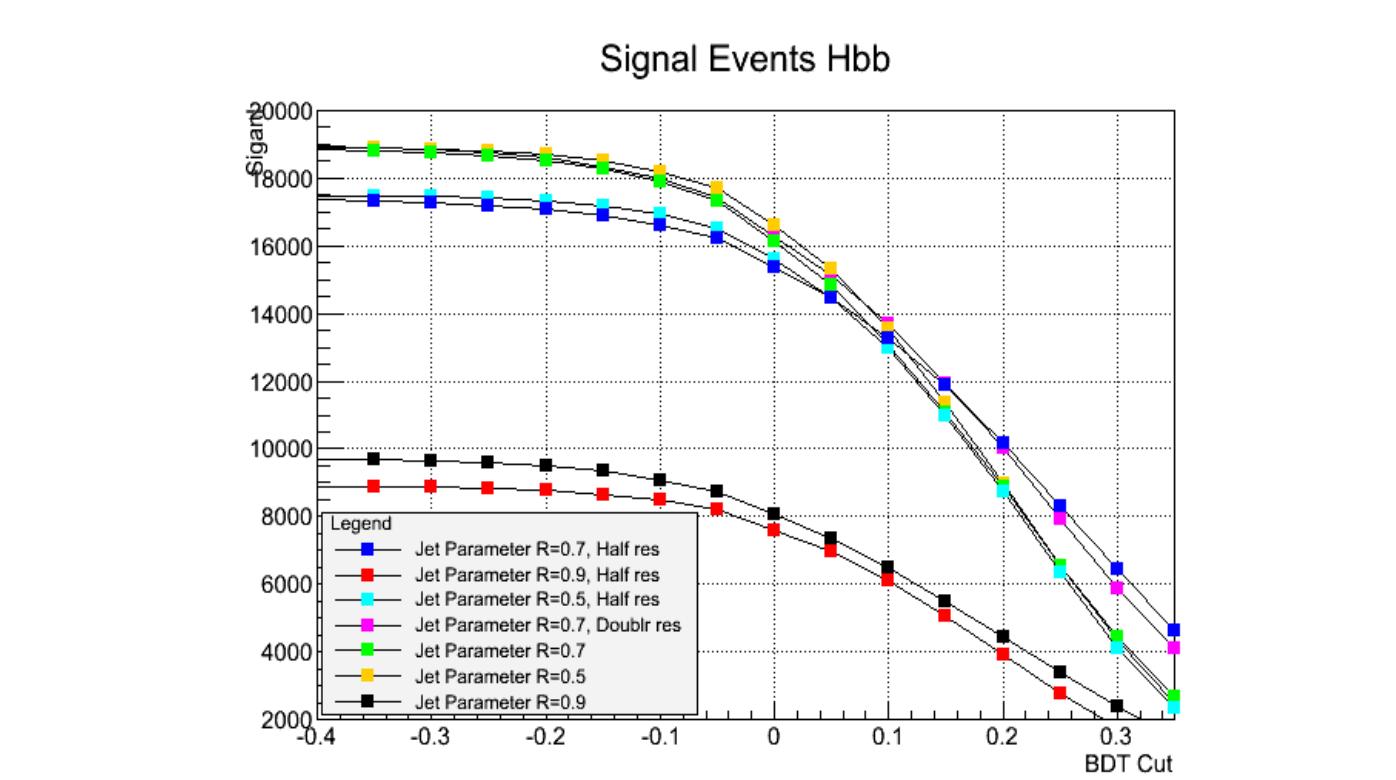}
  \hspace{0.02\textwidth}
  \includegraphics[height=5.3cm]{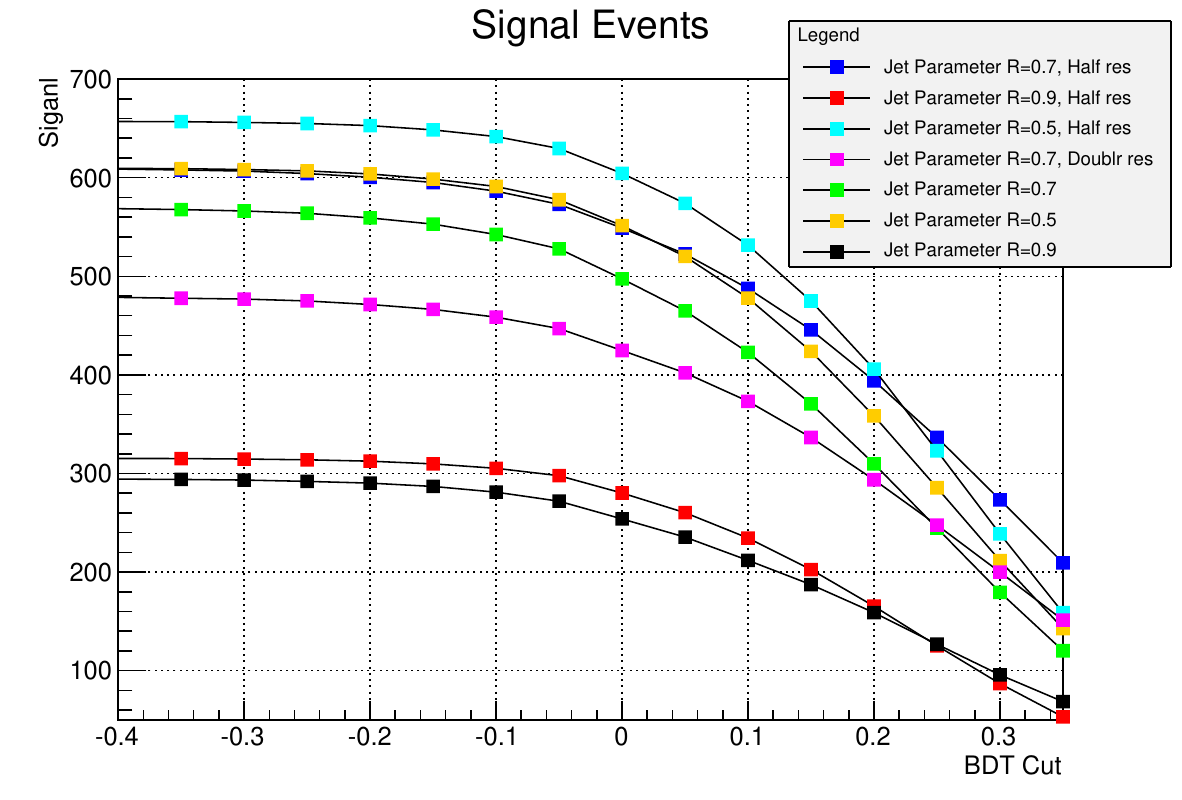}
\caption{Expected $H \to b \bar{b}$ (left) and $H \to c \bar{c}$ (right) signal events as a function of the BDT score. Events are selected at DELPHES detector level for an integrated luminosity of 1~ab$^{-1}$ and -80\,\% electron polarisation. 
The symbols correspond to  the choice of the anti-kt distance parameter $R$ and different assumptions in the impact parameter resolution of 10~(5)~$\mu$m (nominal, no further text in legend added), 20~(10)~$\mu$m (Doubl res), 5~(2.5)~$\mu$m (Half res) for tracks with $0.5<p_T<5\,(>5)$~GeV within $|\eta|<3.5$.}
\label{fig:ccH_bdt}
\end{figure}

A series of BDT score tests has been performed using the preselected signal samples and CC multi-jet as the main background sample to determine the optimal combination of the impact resolution parameters while resolving the two jets from the Higgs decay in dependence of $R$. The resulting number of $H \to b \bar{b} (c \bar{c})$ signal events versus the BDT score is illustrated in Fig.~\ref{fig:ccH_bdt}, which shows the evident interplay between detector performance and the choice of jet parameters $R$, where the $R=0.9$ anti-kt jets show the worst performance.   
At a score of BDT=0, the highest number of signal events are achieved for $R=0.5$ anti-kt jets for both charm and beauty decays, where the effect of the impact resolution is much more stringent for the charm than for the beauty tagging.
Following Fig.~\ref{fig:ccH_bdt}, the complete BDT-based $H \to b \bar{b} (c \bar{c})$ analyses are performed for anti-kt $R=0.5$ jets and impact parameter resolution of 5~(2.5)~$\mu$m (Half Vertex Resolution) for tracks with $0.5<p_T<5\,(>5)$~GeV within $|\eta|<3.5$. The acceptance times efficiency values are about 28\,\% for the $H \to b \bar{b} $  and about 11\,\% for the $H \to c \bar{c}$ channel at BDT=0.

\begin{figure}[htb!]
\centering
\includegraphics[width=0.8\textwidth,trim={15 10 15 15},clip]{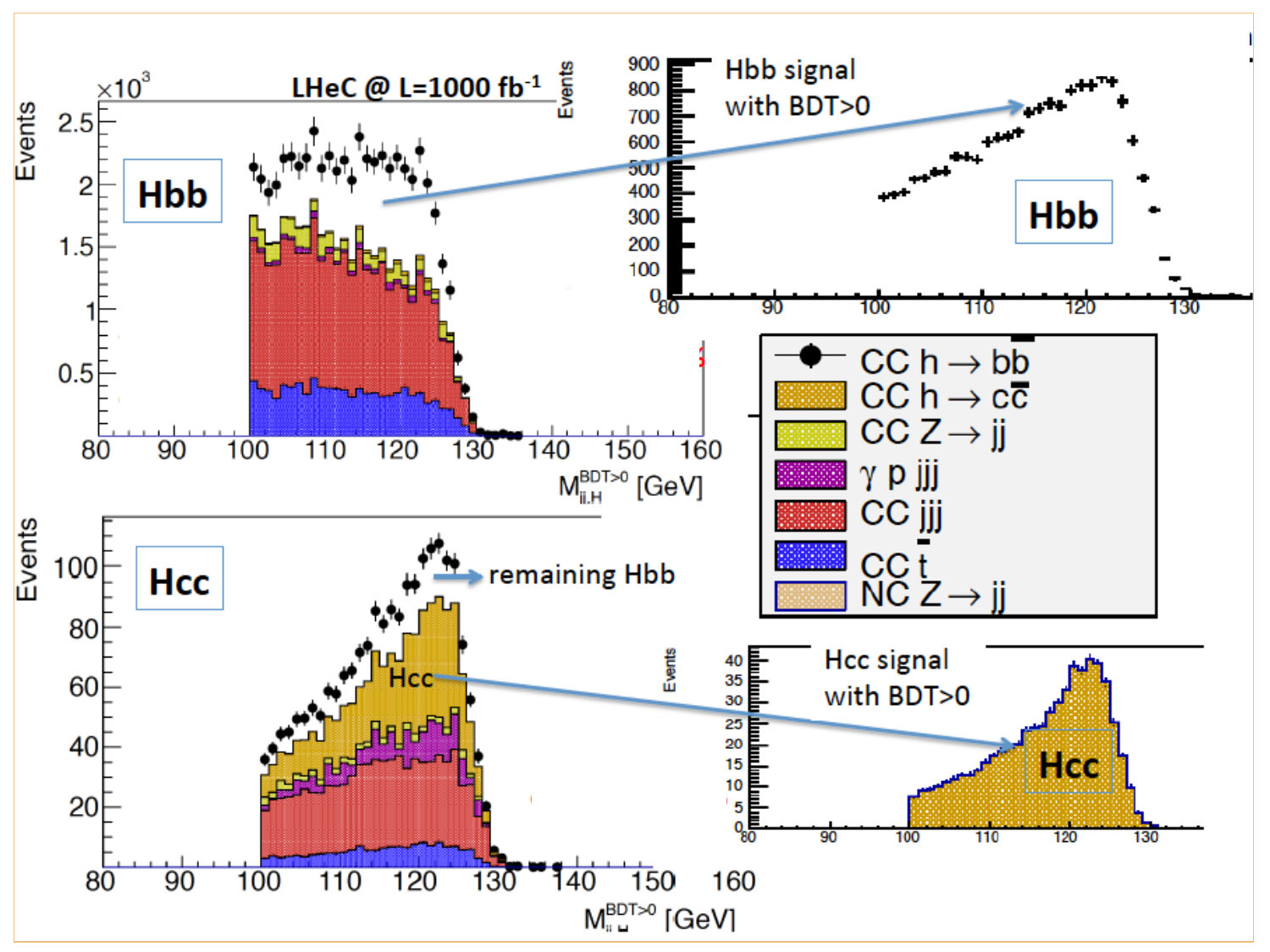}
\caption{Result of the joint $H \to b \bar{b}$ and $H \to c \bar{c}$ analysis for an integrated luminosity of 1~ab$^{-1}$ and -80\,\% electron polarisation at the LHeC. Left: Invariant mass distributions for the two channels with signal and background, see text. Right: Expected Higgs signal distributions
after background subtraction. The background is assumed to be at the 2\,\% level via control region measurements.}
\label{fig:bbccH}
\end{figure}

The results of the BDT $H \to b \bar{b}$ and $H \to c \bar{c}$ analyses, assuming that each background contribution is understood at the 2\,\% level via control regions and negligible statistical Monte Carlo uncertainties for the background predictions for the signal region,  are shown in Fig.\,\ref{fig:bbccH}. Using these assumptions, the resulting signal strengths are 0.8\,\% for the $H \to b \bar{b} $  and 7.4\,\% for the $H \to c \bar{c}$ channel.
For the latter, the SM Higgs decays, in particular $H \to b \bar{b}$, represent also
a part of the $cc$ background contribution but can be controlled by the high precision of the genuine $bb$ result. 
Advanced analysis strategies to distinguish $bb$ and $cc$ SM Higgs decays via several layers of neural networks are discussed e.g. in Ref.~\cite{b2009measurement} for an 250\,GeV ILC and $M_H=120$~GeV, where  the expected $H \to c \bar{c}$ cross section is 6.9\,fb for $M_H=120$~GeV yields a signal strength uncertainty of 8.8\,\% in the $ZH$ all hadronic channel ($Z \to q \bar{q}$) at an integrated luminosity of 250~fb$^{-1}$. The ILC charm cross section is quite similar to the 5.7\,fb cross section for $M_H=125$\,GeV at LHeC. The number of preselected charm events and SM Higgs contributions for the ILC analysis are at a similar level as in this analysis, while the non-Higgs background at ILC is by a factor 6.8 larger than for the LHeC preselected events. Comparing the two results gives confidence into the expected $H \to c \bar{c}$ signal strength results at LHeC using the before mentioned assumptions.

 In conclusion, Higgs to heavy flavour signal strength measurements require 
an excellent state-of-the-art calorimetry with high acceptance and 
excellent resolution as well as an impact parameter resolution as
 achieved e.g. with ATLAS inner b-layer. In addition, the details 
of the analysis strategy utilising neural network and advanced statistical methods
 (e.g. via RooStats/RooFit, see e.g. complex analysis methods using constraints
 via well measured control regions in signal 
fits~\cite{Aad:2019xxo}) will be important  to control a high signal at low background
 yields where the latter is expected to be constrained via control regions to better than a few\,\%. 

\subsection{Higgs Decay into WW}
\label{sec:WBDT}
\begin{figure}[ht!]
\centering  %
\includegraphics[angle=0.,width=0.75\textwidth]{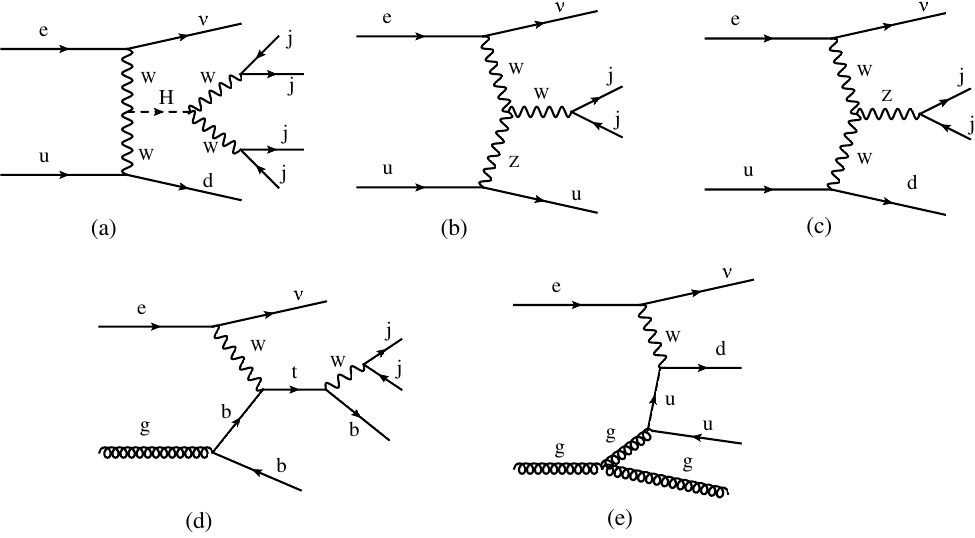}
\caption{Typical lepton-parton diagrams relevant to the $H \to W^*W$ analysis: a) signal: CC DIS with a Higgs produced in the t-channel and its decay into a pair of $W$ bosons which generates a four-jet final state, besides the forward jet. The other diagrams are examples to illustrate background channels which at higher orders, with extra emissions, may mimic the signal configuration: b) single $W$-boson production;  c) single $Z$-boson production; d) single top-quark production; e) QCD multi-jet production. 
}
\label{fig:WWdiag}
\end{figure}

Inclusive charged current scattering, the CC production of the Higgs boson with a $WW$ decay and the main backgrounds are illustrated in Fig.\,\ref{fig:WWdiag}. The $ep \to \nu H X \to \nu W^*W X$ process with hadronic $W$ decays, see  Fig.\,\ref{fig:WWdiag}\,a, causes a final state which to lowest order comprises $4 + 1$ jets and the escaping neutrino identified via missing energy (MET). The pure hadronic $WW$ Higgs decay has a branching ratio of about $45$\,\%. Using MadGraph (MG5) and a version of PYTHIA, customised for $ep$ DIS, events have been generated and analysed after passing a DELPHES
description of the FCC-eh detector. The present study has been performed for the most asymmetric beam configuration of $E_e=60$~GeV and $E_p=50$~TeV yielding $\sqrt{s}=3.5$~TeV.
 
 The analysis has been focussed on requiring four fully resolved jets from the Higgs decay and at least one forward jet, where the jets are reconstructed using the anti-$k_T$ algorithm with $R=0.7$. Further event categories where the jets from the Higgs decay products may  merge and yield either three or only two large-$R$ jets in the final state have been not considered yet. However, as shown from state-of-the art LHC-style studies, those event categories and the use of e.g. dedicated top and W-tagging based on large-$R$ jet substructures may give additional access to measure Higgs signal strengths.

\begin{figure}[!th]
\centering
\includegraphics[width=0.46\textwidth]{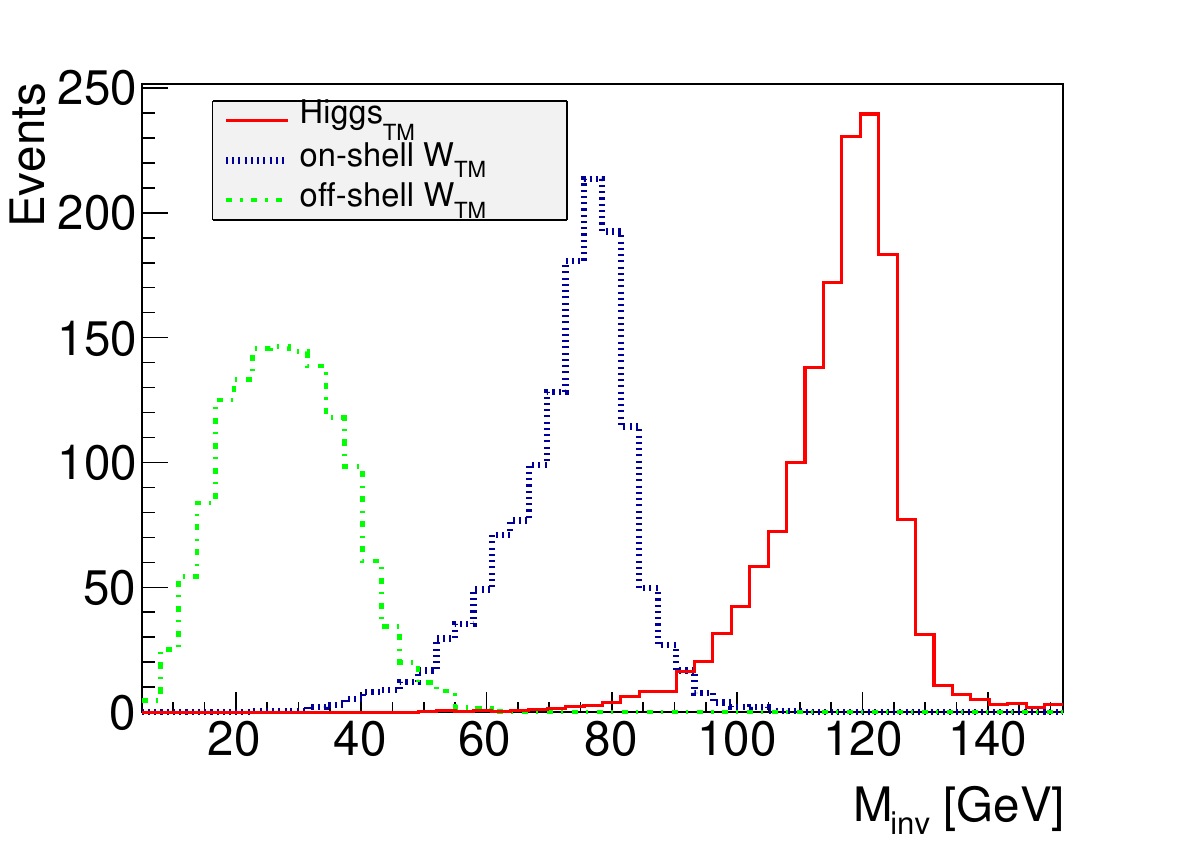}
\includegraphics[width=0.46\textwidth]{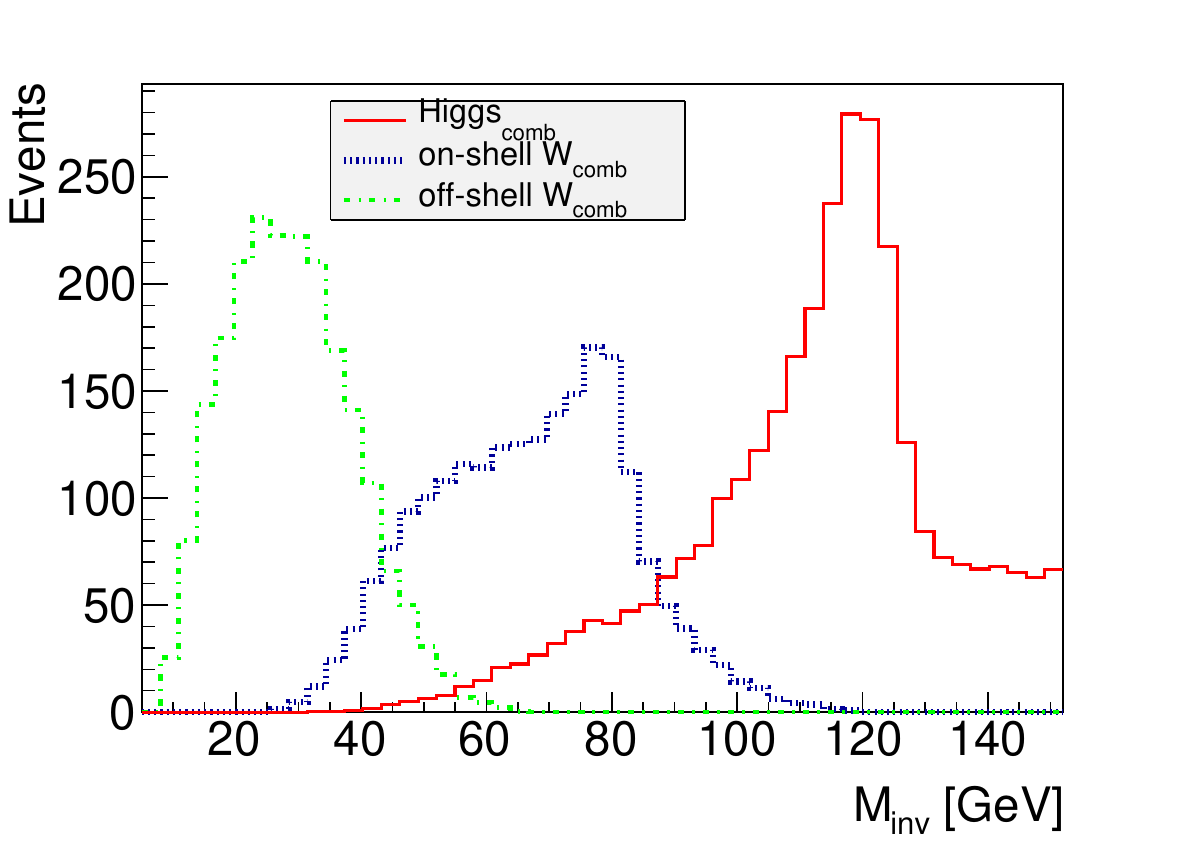}
\caption{Reconstructed signal mass distributions (at DELPHES detector level) of truth matched events (left)
and after the just combinatorial association of jets to the two $W$ bosons forming
Higgs candidates (right). Green: virtual
$W^*$ boson; blue: $W$ boson; red: Higgs signal from $W^*W$ reconstruction. It is observed
that the combination causes some background while the respective signal peaks are clearly
preserved with a purity of 68\,\% that the correct forward jet is identified.}
\label{fig:wwstar}
\end{figure}

The analysis requiring fully resolved jets from the $H\rightarrow W^*W \rightarrow 4j$ decay and at least one forward jet proceeds in the following steps:
\begin{itemize}
    \item Study of the reconstructed event configuration and recognition of its characteristics for
    defining a set of loose cuts. These are: the $p_T$ of any jet has to be larger than $6$\,GeV,
    the rapidity difference between the forward jet and the reconstructed 4-jet Higgs candidate
    to be larger than $1.5$, the azimuthal difference between that Higgs candidate and either
    the forward jet or the scattered lepton (MET) to be larger than $1$, two-jet masses of 
    the virtual and the real $W$ boson candidate to be larger than $12$\,GeV and below 90\,GeV ($Z$ mass). 
    \item Verification of truth matching to check that the combinatorial association 
    of jets reproduces the Higgs candidate (four jets) and its $W$ (two jets) decays (see Fig.\,\ref{fig:wwstar} and text).
    \item Application of this algorithm to the simulated background samples. The MadGraph single $W$, top and $Z$ production samples are turned to multi-jet background through PYTHIA. The
    cross sections are reliably calculated as there is a hard scale available. The initial cuts
    reduce this background to about $3$\,\% for single vector boson production and
    to $9$\,\% for top.
    \item Due to the size of the $H b \bar{b}$ decay and jet radiation, there occurs a residual
    background from the Higgs itself which is also reduced to $3$\,\%  
    through the cuts.
    \item The final background is due to multi-jets. The MadGraph cross section for a 
    4+1 jet CC configuration is considered much too large in view of the cross section
    measurement results as a function of the jet number, both at HERA and the LHC, see
     for example~\cite{Aaboud:2017hbk}. The sample was thus scaled using a
     conservative $\alpha_s$ renormalisation to the  inclusive cross section. The initial
     cuts reduce the multi-jet background to about $13$\,\%.
    \item Following a detailed training study, a BDT analysis was used. This determined
    a final event number of about 12k for to a signal-to-background ratio
    of 0.23. 
\end{itemize}    
    The result of this analysis translates to an estimated uncertainty
    on $\mu_{WW}$ of $1.9$\,\% at FCC-eh. The 4-jet mass distribution after the BDT
    requirement exhibits a clear $WW$ Higgs peak (see Fig.\,\ref{fig:wwstar}) which illustrates the suitability to
    use the electron-proton environment for Higgs measurements in indeed challenging 
    final state configurations.
\subsection{Accessing Further Decay Channels}
\label{sec:mus}
Following the detailed studies of the $b \bar{b}$ and $c \bar{c}$ decay channels,
presented above, a coarser analysis was established for other frequent decay channels
both in NC and CC. Here acceptances and backgrounds were 
estimated with MadGraph, and efficiencies, distinguishing leptonic and hadronic
decay channels for $W,~Z,$ and $\tau$, were taken from prospective studies
on Higgs coupling measurements at the LHC~\cite{Englert:2015hrx}.
This provided a systematic scale factor, $f$, on the pure statistical error $\delta_s$, which comprised the signal-to-background ratio, $S/B$,
and the product of acceptance, $A$, and extra reconstruction efficiency $\epsilon$, according to  
\begin{equation}
    f = \sqrt{\frac{1 + \frac{B}{S}}{A \cdot \epsilon}}
\label{eq:scalef}
\end{equation}
The error on the signal strength $\mu_i$
 for each of the Higgs decay channels $i$ is  determined 
 as $\delta \mu_i /\mu_i  = f_i \cdot \delta_s$.
%

\begin{table}[ht!]
  \centering
  \small
  \begin{tabular}{lccccccc}
    \toprule 
    Parameter & $b \bar{b}$ & WW  & gg& $\tau \tau$ & cc&  ZZ  &  $\gamma \gamma$ \\
    \midrule
    Branching fraction & 0.581 & 0.215 & 0.082 & 0.063 & 0.029 & 0.026 & 0.0023 \\
Statistical error ($\delta_s$) [\%] & 0.09 & 0.15 & 0.24 & 0.28 & 0.41 & 0.43 & 1.41 \\
\addlinespace
Acceptance ($A$) & 0.14 & 0.10 & 0.40 & 0.40 & 0.11 & 0.10&  0.40 \\
 Signal/background ($S/B$) & 9 & 0.2 & 0.1 & 0.2 & 0.43 & 0.33 & 0.5 \\
 Extra  efficiency ($\epsilon$)  & 1 & 0.3 & 0.5 & 0.43 & 1 & 0.5 & 0.7 \\
  Scale factor $f$ & 2.8 & 16 & 7.4 & 5.9 & 5.5 & 9.0 & 3.3 \\
\bottomrule
\end{tabular}
\caption{Statistical uncertainty for the seven most  abundant Higgs decay channels, for the  charged current Higgs measurement prospects with the FCC-eh, together with their systematic scale factor $f$, Eq.\,\ref{eq:scalef}, resulting from
acceptance, background and efficiency effects as given. Note that
the results for $b \bar{b}$ and $c \bar{c}$ are taken from the BDT analysis (Sect.\,\ref{sec:Hbc})
with efficiency 1. The $WW$ result is replaced by the BDT analysis (Sect.\,\ref{sec:WBDT})
for quoting the expected signal strength uncertainty.
}
\label{tab:muCCfcc}
\end{table}

\begin{figure}[!th]
\centering
\includegraphics[width=0.9\textwidth,trim={40 40 40 230},clip]{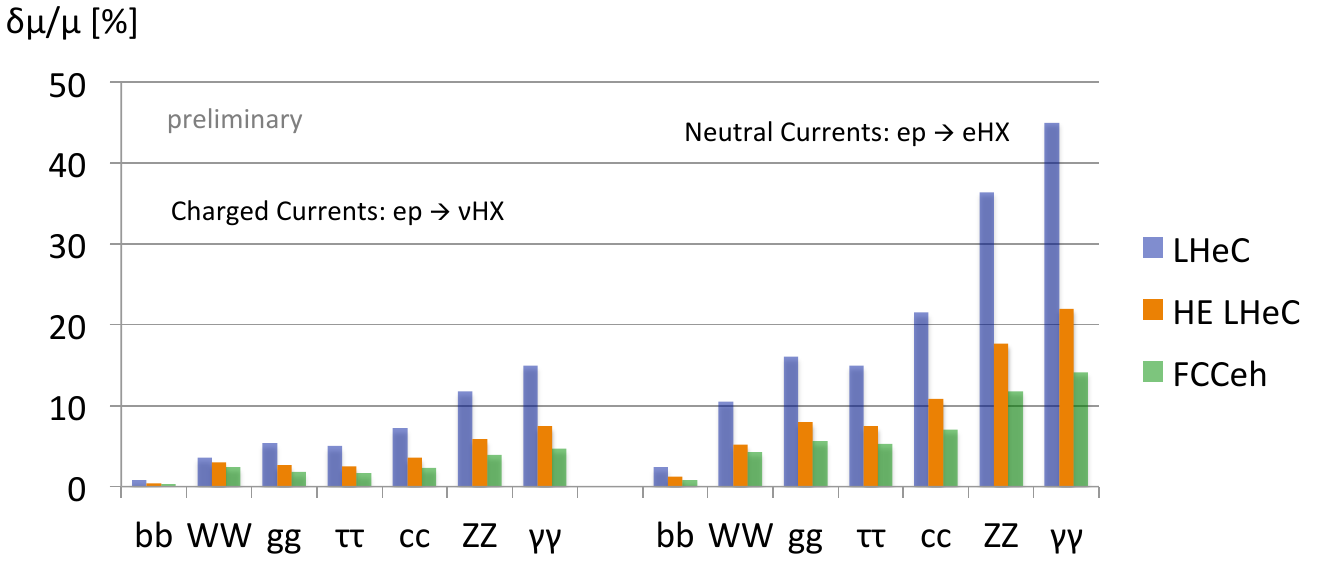}
\caption{
Uncertainties of signal strength determinations
 in the seven most
    abundant SM Higgs decay channels for the FCC-eh (green, $2$\,ab$^{-1}$),
    the HE LHeC (brown, $2$\,ab$^{-1}$) and LHeC (blue, $1$\,ab$^{-1}$),
    in  charged and neutral current DIS production.}
\label{fig:sigmus}
\end{figure}

To good approximation these factors apply to LHeC, HE-LHeC and FCC-eh
because the detector dimensions and acceptances scale with the proton
energy, conceptually using the same technology and very similar resolution assumptions. Therefore there is one main matrix used for the subsequent
experimental deterioration of the pure statistics precision, both for CC and NC. 
Future detailed analyses will lead to refining this expectation which for the 
current purpose was beyond the scope of the study.
The results of the 
 analysis of uncertainties are summarised in Tab.\,\ref{tab:muCCfcc} 
 for the CC channel at the FCC-eh. 
\begin{table}[h]
  \centering
  \small
  \begin{tabular}{lcccccccc}
    \toprule
     Setup & $b \bar{b}$ & $b \bar{b} \oplus \text{Thy}$ & WW  & gg& $\tau \tau$ & cc&  ZZ  &  $\gamma \gamma$ \\
   \midrule
 LHeC NC & 2.3 & 2.4  & 17  & 16  & 15  & 20  & 35 & 42 \\
 LHeC CC & 0.80 & 0.94  & 6.2 & 5.8 & 5.2 & 7.1 & 12 & 15 \\
 HE-LHeC NC & 1.15 & 1.25  & 8.9  & 8.3  & 7.5  & 10  & 17 & 21 \\
 HE-LHeC CC & 0.41 & 0.65 & 3.2 & 3.0 & 2.7 & 3.6 & 6.2 & 7.7 \\
 FCC-eh NC & 0.65 & 0.82 &  5.0 & 4.7 & 4.2 & 5.8 & 10 & 12 \\
 FCC-eh CC & 0.25 & 0.56 &  1.9 & 1.8 & 1.6 & 2.2 & 3.8 & 4.6 \\
 \bottomrule
\end{tabular}
\caption{Summary of estimates on the experimental uncertainty of the signal strength $\mu$, in per cent, for the seven most  abundant Higgs decay channels, in charged and neutral currents for the LHeC,
the HE-LHeC and the FCC-eh. The $b\bar{b}$ channel is the  one which is most sensitive to theoretical uncertainties  and for illustration is given two 
corresponding columns, see Sect.\,\ref{sec:syerr}.}
    \label{tab:musum}
\end{table}
The resulting signal strength uncertainty values are provided in Tab.\,\ref{tab:musum}. Note that for the beauty, charm and $WW$ channels the table contains the BDT analysis~\footnote{This is in very good agreement with the scale factor method: for example, the $WW$ result in Tab.\,\ref{tab:muCCfcc} leads to a value of $2.1$\,\% slightly worse than the BDT analysis.} results  of Sect.\,\ref{sec:Hbc} and Sect.\,\ref{sec:WBDT}, resp.  The beauty and charm CC results stem from
the BDT analysis for LHeC and are applied to FCC-eh with a factor of about $1/3$. The CC
$WW$ results are due to the FCC-eh BDT analysis and are used for LHeC, enlarged
by a factor of $3.2$, determined by the different cross sections and luminosities.
 For HE-LHC, the  values are about twice as precise as  the LHeC values because the cross section is enlarged by about a factor of two, see Tab.\,\ref{tab:Hcross}, and the integrated luminosity with $2$\,ab$^{-1}$ twice that of the LHeC. All signal strength uncertainties, in both CC and NC, for the  three collider configurations are shown in Fig.\,\ref{fig:sigmus}.
\subsection{Systematic and Theoretical Errors}
\label{sec:syerr}
The signal strength is expressed relatively to a theoretical calculation  of the charged current Higgs cross section, including its decay into a chosen channel, according to
\begin{equation}
    \mu = \frac{\sigma_{exp}}{\sigma_{thy}} = \frac{\sigma_{exp}}{\sigma_{Hty} \cdot br}.
\label{eq:muextybr}
\end{equation}
 Consequently one can decompose the (relative)  error of $\mu$ into the genuine 
 measurement error, denoted as $\delta \sigma_{exp}$, including a possible
 systematic error contribution, $E$, and two further components
\begin{equation}
    \frac{\delta \mu}{\mu} = \Biggl\{\biggl(\frac{\delta \sigma_{exp}}{\sigma_{exp}}\biggr)^2 
    \cdot (1 \oplus E) 
     + \biggl(\frac{\delta \sigma_{Hty}}{\sigma_{Hty}}\biggl)^2 +\biggl(\frac{\delta br}{br}\biggl)^2 \Biggr\}^{1/2},
     \label{eq:deltamu}
\end{equation}
%
which are due to
imperfections to theoretically model the 
Higgs production cross section, $\sigma_{Hty}$, and uncertainties 
on the branching ratio, $br$, in the channel under study.
Note, that  the experimental uncertainty takes into account possible variations of the backgrounds which are estimated conservatively and thus represent more than genuine statistics. 

The channel dependent signal strength uncertainties quoted in Tab.\,\ref{tab:musum}
are estimates of the first, experimental term in Eq.\,\ref{eq:deltamu} neglecting extra systematic error effects. They 
are derived as stated above from the purely statistical error ($\delta_s = 1/\sqrt{N}$), its increase due to 
acceptance ($A$) and efficiency ($\epsilon$) effects and, further, the modulation caused by the background-to-signal ratio ($B/S$). These factors are all involved in the BDT analysis but the scale factor equation, Eq.\,\ref{eq:scalef}, may be used to estimate further systematic effects for any channel. From the relation
\begin{equation}
    \frac{\delta \sigma_{exp}}{\sigma_{exp}} = \delta_s \cdot \sqrt{\frac{1+B/S}{A \cdot \epsilon}}
\end{equation}
the combined systematic error contribution, $E$, caused
by variations $\Delta$ of $A$, $\epsilon$ and the background $B$ 
can be estimated as
\begin{equation}
E= \frac{1}{2} \Biggl\{\biggl(\frac{ \Delta A}{A}\biggr)^2 +
    \biggl(\frac{ \Delta \epsilon}{\epsilon}\biggr)^2 + \biggl(\frac{ \Delta B}{B} \cdot \frac{B/S}{1+B/S}\biggr)^2\Biggr\}^{1/2}.
    \label{eq:deltaexp}
\end{equation}
The formula shows that if the background-to-signal ratio is very small, then
the background effect is suppressed  $\propto B/S$. If it is larger than $1$, the
relative uncertainty of the background enters as an additional component of the
signal strength error. 

Given the fact that the experimental $H \to b\bar{b}$ result in the CC reaction is especially precise, compare Tab.\,\ref{tab:musum}, an estimate was performed of the systematic 
error  in this channel. The following effects were included: a variation of the light-quark 
misidentification by a factor 3, a variation of the reduction of the photo-production
via tagging between $2$\,\% and $10$\,\%, a variation of the combined acceptance times
efficiency effect by $10$\,\% and a variation of the hadronic energy resolution,
studied in Ref.~\cite{tanaka}, leading to a $7$\,\% signal variation. The overall effect of these
contributions determines a systematic error of about $10$\,\% on $\mu_{bb}$, i.e.
$\delta \mu /\mu =0.80 \pm 0.09$ for  $H \to b\bar{b}$ at the LHeC in the CC channel. Similar
levels of uncertainty are expected to occur for other channels but have not been estimated 
to such detail as those channels are measured less precisely.

A separate effect arises from the measurement of the luminosity. While
that will be measured about as accurate as $0.5$\,\%, based on Bethe-Heitler
scattering and its accurate description to higher-order QEDC~\cite{AbelleiraFernandez:2012cc}, additionally it will be negligible
to a good approximation: the LHeC, and its successors, will provide a very precise,
determination of all parton distributions from the $ep$ data alone. Any systematic 
mistake in the normalisation will therefore affect both the measured and the 
calculated cross section and drop out in their ratio $\mu$.

A next uncertainty on the signal strength arises from the theoretical description of $\sigma_{CCH}$
to which the measured cross section is normalised. From a simulation of the systematic 
uncertainties due to imperfect calibrations and extra efficiencies one may expect
the cross section to be known to better than $1$\,\%. The prediction will be 
available to N$^3$LO, $\alpha_s$ be determined to $0.1-0.2$\,\% precision,  
and it can be gauged with the inclusive cross section measurement.
This uncertainty, following Eq.\,\ref{eq:deltaexp}, enters directly as a contribution
to the $\mu$ measurement result. A $0.5$\,\% uncertainty, as can be seen in Tab.\,\ref{tab:musum},
becomes noticeable in most of the $b \bar{b}$ results but is negligible for all other channels.
In the present analysis values of $0.5$\,\% and $1$\,\% uncertainty have been considered
and their effect on the $\kappa$ result been evaluated, see Sect.\,\ref{sec:kappaep}.

A final uncertainty is caused by the branching fractions and their uncertainty. A recent
uncertainty estimate~\cite{deFlorian:2016spz} quotes on the here most relevant 
$H \to b \bar{b}$ branching ratio
a theory contribution due to missing higher orders of $0.65$\,\%, 
a parameterisation uncertainty depending on the quark masses  of  $0.73$\,\%, and
an $\alpha_s$ induced part of $0.78$\,\%. The LHeC, or similarly the higher energy
$ep$ colliders, will determine the $b$ mass (in DIS) to about $10$\,MeV and $\alpha_s$ to per mille
precision~\cite{AbelleiraFernandez:2012cc}
which would render  corresponding uncertainty contributions to $br_{bb}$ negligible.
The genuine theoretical uncertainty would also be largely reduced with an extra order pQCD.
In the subsequent study the contribution from the branching fraction uncertainty
has  been neglected.
This may also be justified by the programme here sketched, and similarly for other future colliders: the $ep$ colliders will measure the 
couplings, especially of the $WW$, $bb$ and $ZZ$ very precisely, which will enable an iterative 
treatment of the branching ratio uncertainties. 

It may  be noticed~\cite{deFlorian:2016spz} that the $\alpha_s$ contribution to the $H \to gg$ branching fraction uncertainty
is about $3.7$\,\%, i.e. twice as large as the estimated signal strength measurement uncertainty
of this channel at the FCC-eh. There arises another important benefit of the future
$ep$ colliders and their high precision DIS programme for precision Higgs physics at the
combined $ep~\&~pp$ facilities.

\section{Higgs Coupling Analyses}
\label{sec:couplings}
In order to quantify possible deviations from the SM expectation  one may use the $\kappa$ parameterisation framework, introduced in Ref.~\cite{Dittmaier:2011ti}, 
which enables easy comparisons between different collider configurations independently of their ability to access the total Higgs decay width.
It should be noted that there are differences between the results in the EFT and in the $\kappa$ formalism~\cite{Barklow:2017suo}.
Therefore, it would be very interesting
to go beyond the $\kappa$ framework also for the $ep$ colliders here presented 
because out of the 2499 dimension-6 Wilson coefficients altogether $13 \cdot n_g^4 = 1053$
involve leptons and quarks~\cite{mtrott}, for $n_g=3$ generations. This, however,
has been beyond the scope of this study.
In the following results are presented
for the various $ep$ collider configurations (Sect.\,\ref{sec:cross}).
%

\label{sec:kappaep}

\begin{figure}[!th]
\centering
\includegraphics[width=0.9\textwidth]{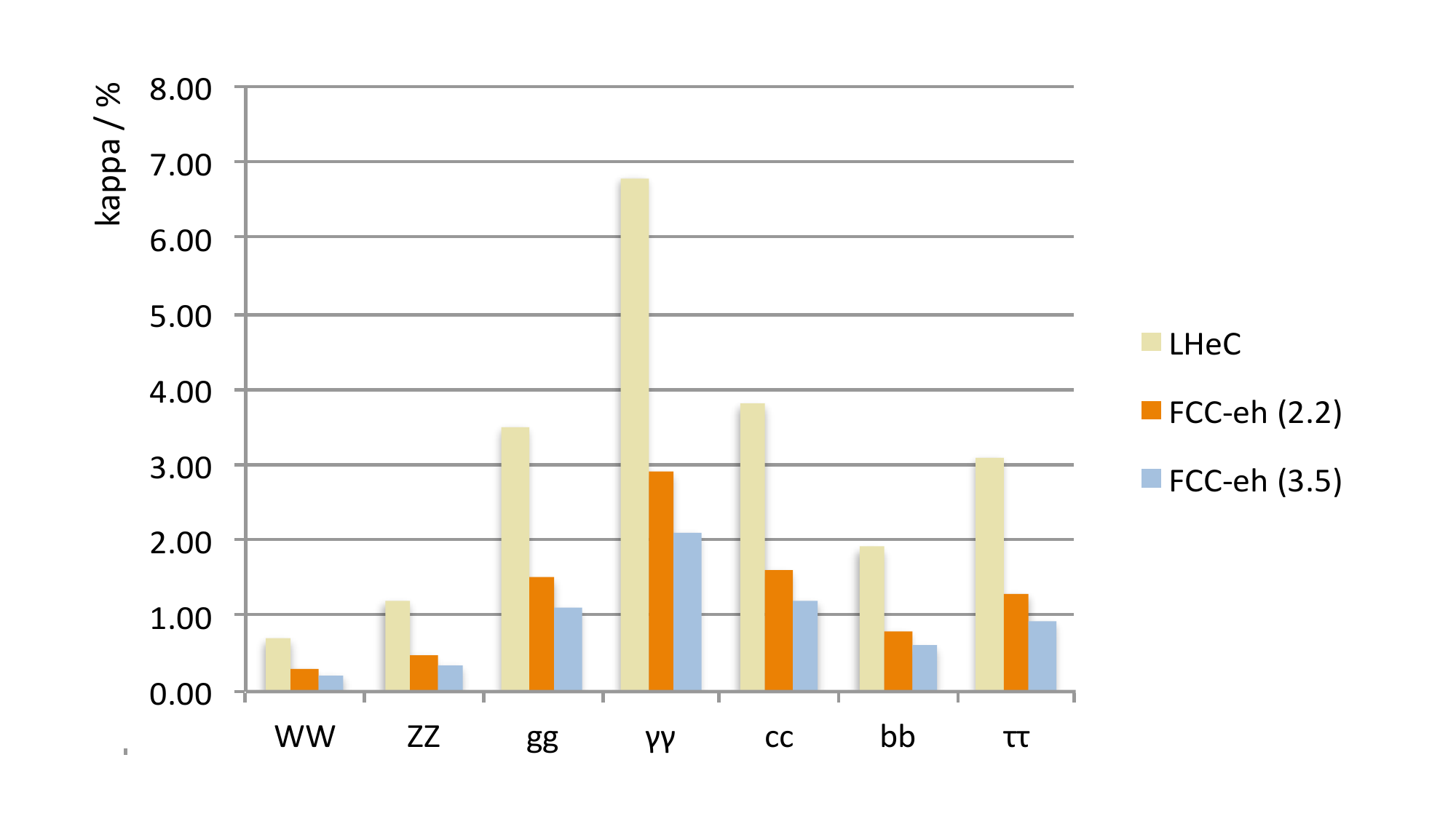}
\caption{
Summary of uncertainties of Higgs couplings from $ep$ for the seven most abundant decay channels,
for LHeC (gold), FCC-eh at 20\,TeV proton energy (brown) and for $E_p =50$\,TeV (blue). }
\label{fig:kappa3}
\end{figure}

The $\kappa$ parameters are factors to the various Higgs couplings, equal to one in the SM, which scale $\sigma_{NC/CC}$ with $\kappa^2_{Z/W}$, the width $\Gamma^i$ for a channel $i$ with $\kappa_i^2$ and lead to 
replacing $\Gamma_H$ by the sum $\Sigma_j \kappa_j \Gamma^j$, where we have assumed no non-SM H decays. This defines the following modifications of the cross sections (Eq.\,\ref{eqsig1})
\begin{equation}
    \sigma_{CC}^i = \sigma_{CC} ~ br_i  \cdot \kappa_W^2 \kappa_i^2  \frac{1}{\sum_j{\kappa_j^2 br_j}} 
   ~~\mathrm{and}~~ 
    \sigma_{NC}^i = \sigma_{NC} ~ br_i  \cdot \kappa_Z^2 \kappa_i^2  \frac{1}{\sum_j{\kappa_j^2 br_j}}. 
    \label{eqsig2}
\end{equation}
Dividing these expressions by the SM cross section predictions one obtains
the variations of the relative signal strengths, $\mu^i$, for charged and neutral currents and their $\kappa$ dependence
\begin{equation}
    \mu_{CC}^i =  \kappa_W^2 \kappa_i^2  \frac{1}{\sum_j{\kappa_j^2 br_j}} 
   ~~~~\mathrm{and}~~~~ 
    \mu_{NC}^i =  \kappa_Z^2 \kappa_i^2  \frac{1}{\sum_j{\kappa_j^2 br_j}}. 
    \label{eqmus}
\end{equation}
With seven decay channels considered in CC and NC, one finds that for each of the $ep$
collider configurations there exist eight constraints on $\kappa_W$ and $\kappa_Z$
and two on the other five $\kappa$ parameters.  Using the signal strength uncertainties
as listed in Tab.\,\ref{tab:musum} fits to all seven channels, in NC and CC, are
performed in a minimisation procedure to determine the resulting uncertainties
for the $\kappa$ parameters. These are done separately for each of the $ep$
collider configurations with results listed in Tab.\,\ref{tab:kappaep}. A naive 
expectation would have been that $\delta \kappa \simeq \delta \mu /2$. Comparing the results, for example for LHeC (top rows), of the signal strengths (Tab.\,\ref{tab:musum})
with the $\kappa$  fit results (Tab.\,\ref{tab:kappaep}) one observes that this
relation holds approximately for the $gg,~\tau \tau,~c\bar{c},~\gamma \gamma$ channels.
However, due to the dominance of $H\to b\bar{b}$ in the total H width and owing to the presence
of the $WWH$ and $ZZH$ couplings in the initial state, there occurs a reshuffling
of the precisions in the joint fit: $\kappa_{b}$ is relatively less precise than 
$\mu_{bb}$ while both $\kappa_{W}$ and $\kappa_{Z}$
 become more precise than naively estimated, even when 
one takes into account that the $H \to WW$ decay in CC measures
$\kappa_{W}^4$.
The seven channel results are displayed in Fig.\,\ref{fig:kappa3}.


\begin{table}[h]
  \centering
  \small
  \begin{tabular}{lccccccc}
    \toprule
      Setup & $b \bar{b}$ & WW  & gg& $\tau \tau$ & cc&  ZZ  &  $\gamma \gamma$ \\
   \midrule
 LHeC  & 1.9  & 0.70  & 3.5  & 3.1  & 3.8  & 1.2 & 6.8 \\
 HE-LHeC  & 1.0  & 0.38  & 1.8  & 1.6  & 1.9  & 0.6 & 3.5 \\
 FCC-eh  & 0.60 & 0.22 & 1.1 & 0.93 & 1.2 & 0.35 & 2.1 \\
 \bottomrule
\end{tabular}
\caption{Summary of $\kappa$ uncertainty values as obtained from separate fits to the
signal strength uncertainty estimates for the seven most abundant Higgs decay channels, in charged and neutral currents for the LHeC,
the HE-LHeC and the FCC-eh, see text.}
    \label{tab:kappaep}
\end{table}

In the electroweak theory there is an interesting relation between the ratio of the $W$ and $Z$
couplings and the  mixing angle,
\begin{equation}
    \frac{\sigma(WW \to H \to AA)}{\sigma(ZZ \to H \to AA)} = \frac{\kappa^2_W}{\kappa^2_Z}= (1-\sin^2\theta_W)^2
\end{equation}
This relation can be particularly well tested with the $ep$ colliders as they measure
both $WWH$ and $ZZH$ in one experiment and common theoretical environment. If one assumes
the $WW$ and $ZZ$ measurements to be independent, the resulting error on $\sin^2\theta_W \simeq 0.23$ is $0.003$ for the LHeC and $0.001$ for FCC-eh. However, this probably is smaller because
there exist correlations in the measurements which a genuine data based analysis would have to 
evaluate and take into account. 

The effect of the theory uncertainties has been studied for the FCC-eh where the experimental
precision is highest. Tab.\,\ref{tab:kappaepthy} presents the results of a $\kappa$ analysis
using the CC and NC FCC-eh signal strength input (Tab.\,\ref{tab:musum}) neglecting the
theoretical uncertainty and adding $0.5$\,\% or $1$\,\% in quadrature, to only $\mu_{bb}$ where
it matters. This results in an about linear increase of the uncertainty for $bb$ (by a factor of 1.5), $WW$ (by 1.7) and $ZZ$ (by 1.5), while all other $\kappa$ uncertainties only slightly deteriorate. The effect of such uncertainties for LHeC is much smaller as the $\mu$ uncertainties
are three times those of FCC-eh, see Tab.\,\ref{tab:musum}. Therefore, in the LHeC case, the theory uncertainties are neglected.
\begin{table}[!th]
  \centering
  \small
  \begin{tabular}{lccccccc}
    \toprule
    Setup & $b \bar{b}$ & WW  & gg& $\tau \tau$ & cc&  ZZ  &  $\gamma \gamma$ \\
   \midrule
 FCC-eh (no thy)       & 0.60 & 0.22 & 1.1 & 0.93 & 1.2 & 0.35 & 2.1 \\
 FCC-eh (0.5\,\% thy)  & 0.72 & 0.28 & 1.1 & 1.0 & 1.2 & 0.41 & 2.2 \\
 FCC-eh (1.0\,\% thy)  & 0.91 & 0.37 & 1.1 & 1.0 & 1.3 & 0.53 & 2.3 \\
 \bottomrule
\end{tabular}
\caption{Summary of $\kappa$ uncertainty values as obtained from separate fits to the
signal strength uncertainty estimates for the seven most abundant Higgs decay channels, in charged and neutral currents for the FCC-eh, with no
theoretical uncertainty, half a per cent and one per cent uncertainty added.}
    \label{tab:kappaepthy}
\end{table}

An interesting question regards the role of the electron beam polarisation. 
Assuming a maximum polarisation of $P=-0.8$, the CC (NC) Higgs cross section is
calculated to be $1.8~(1.09)$ times larger than that in unpolarised scattering. Therefore 
the signal CC and NC strength uncertainties scale like $1.34$ and $1.09$, respectively.
This is studied for the LHeC.
If the default fit is made, then the $\kappa$ uncertainties quoted in Tab.\,\ref{tab:kappaep}
for $bb,~WW,~gg,~\tau \tau$ and $cc$ are enhanced by a factor of $1.28$. This is due to
the combined effect of CC and NC which diminishes the deterioration a bit, from $1.34$ to
$1.28$. Thus, for example,
the $\kappa_{W}$ uncertainty moves from $0.7$ to $0.9$\,\% in the unpolarised case.
The uncertainty on $\kappa_{Z}$ is enhanced only by a factor of $1.14$, becoming
$1.38$ instead of $1.21$ because the NC channel has a particularly strong effect
on the $ZZH$ coupling. Since the prospect to detect the $\gamma \gamma$ channel in NC
is very poor, the $\kappa_{\gamma}$ uncertainty is enlarged by the full
CC factor of $1.34$. It is for maximum precision very desirable to have the
beam polarised. This, together with electroweak physics, represents an important
reason to continue to develop high current polarised electron sources.
\section{Measuring the Top-quark--Higgs Yukawa Coupling}
\label{sec:topHinep}
Electron-proton collisions at high energy are known to provide a unique window of opportunity to perform precision measurements in the 
top sector~\cite{Dutta:2013mva}. This is due to the large cross-sections of the production of single top, which amounts to about 2~pb for $E_e=60$~GeV 
and $E_p=7$~TeV, where clean signatures are provided without the challenges posed by pile-up. As a result, the cross-section of the SM in association 
with a single top in $e^-p$ collisions is large enough to perform competitive measurements. This includes the measurement of the absolute value of the 
top-Yukawa coupling and, most prominently, its CP-phase~\cite{Coleppa:2017rgb}.
\begin{figure}[th]
  \centering
  \includegraphics[width=0.25\textwidth]{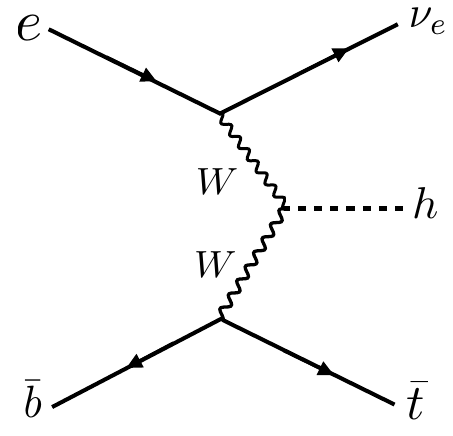}
  \hspace{0.05\textwidth}
  \includegraphics[clip,width=0.25\textwidth]{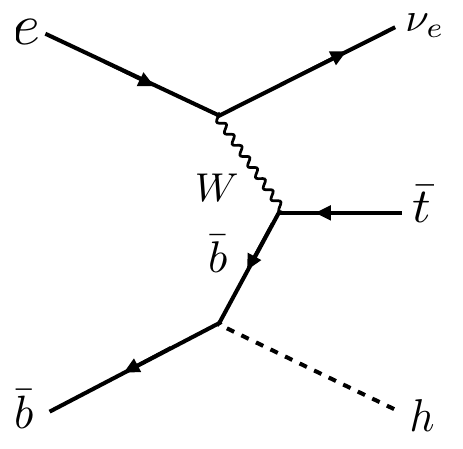}
  \hspace{0.05\textwidth}
  \includegraphics[width=0.25\textwidth]{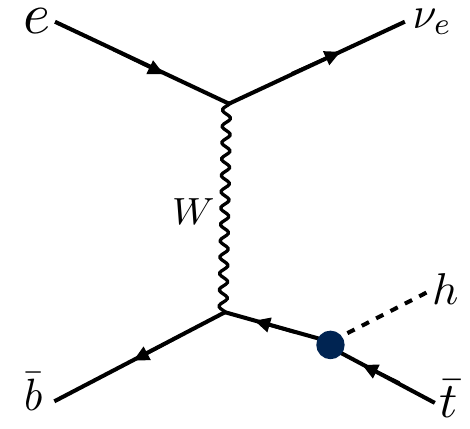}
  \caption{Leading order Feynman diagrams contributing to the process $p\, e^- \to \bar t\, h\, \nu_{e}$
    in high energy $e^-p$ collisions.
    The full circle in the right diagram shows the top-quark--Higgs coupling of interest in this section.}
\label{fig:figW}
\end{figure}

To investigate top Yukawa coupling, the SM interaction can be modified in terms of mixtures of CP-even and CP-odd states.  
In terms of a CP-phase ($\zeta_t$) generalised Lagrangian can be written as~\cite{Rindani:2016scj}:
\begin{align}
\mathcal{L}  =& -  \frac{m_t}{v} \bar t~[\kappa \cos\zeta_t+i\gamma_5\sin\zeta_t ]t\,h.
\label{eq:lphase}
\end{align}
Here, $\zeta_{t} = 0$ or $\zeta_{t} = \pi$ correspond to a pure scalar state while $\zeta_{t} =\frac{\pi}{2}$ to a pure pseudo scalar state. 
Therefore, the $\zeta_{t}$-ranges $0<\zeta_{t}<\pi/2$ or $\pi/2<\zeta_{t}<\pi$ represent a mixture of the different CP-states, and the case $\zeta_t = 0$ with $\kappa = 1$
corresponds to the SM.

In $e^-p$ collisions, the top-quark--Higgs couplings is accessed via associated production of the Higgs boson with an
anti-top quark through the process $e^-p \to \bar t \,h\,\nu_e$, where 5-flavour proton include the $b$-quark
parton distribution.
In Fig.~\ref{fig:figW}, the Feynman diagrams for the process of interest are displayed.
Interestingly this process involves three important couplings, namely $hWW$, $Wtb$ and the top-Higgs ($tth$). 
A detailed study of $hWW$ and $Wtb$ couplings at the $e^- p$ collider have been performed in 
Refs.~\cite{Biswal:2012mp,Dutta:2013mva}, respectively.

At the LHC~\cite{Rindani:2016scj}, quantitatively an interesting feature can be observed: in the pure SM case there is constructive interference 
between the diagrams shown in Fig.~\ref{fig:figW}, left and middle, for $\zeta_t >\pi/2$ resulting in an enhancement in the total production cross section of 
associated top-Higgs significantly. This is also true for $\zeta_t < \pi/2$ - however the degree of enhancement is much smaller owing to the flipped sign of 
the CP-even part of the coupling.
\begin{figure}[th]
\centering
  \includegraphics[width=0.70\textwidth,trim={0 10 0 0},clip]{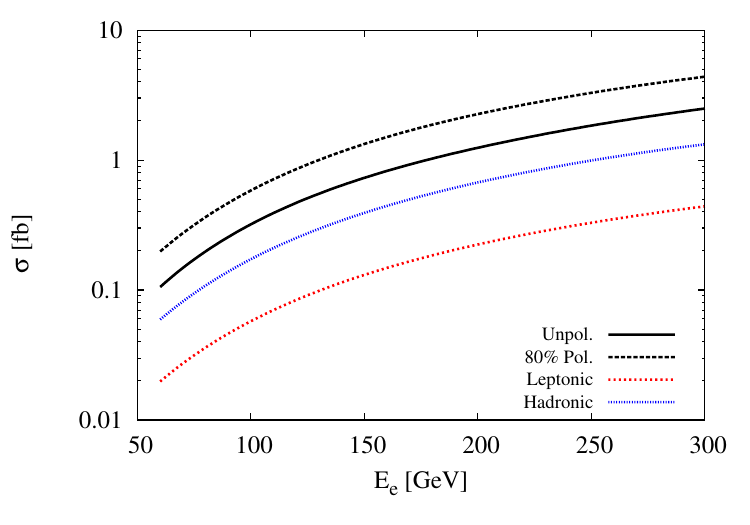}
  \caption{Cross-sections of the Higgs boson produced in association
    with a top quark in $e^-p$ collisions with  $E_p =$~7 TeV for
    different electron beam energies (taken from Ref~\cite{Coleppa:2017rgb}).
    The dotted and solid \emph{black} lines correspond to 
    $p~ e^- \rightarrow~ \bar t~ h~ \nu_{e}$ with and without longitudinal polarisation of the electron beam, respectively.
    The dotted \emph{red} and \emph{blue} lines correspond to $\sigma\times$BR for the leptonic and hadronic decay modes of
    $\bar{t}$  where for this estimation we use basic cuts (see text
    and Ref.~\cite{Coleppa:2017rgb}).}
\label{fig:cs_rs}
\end{figure}

A study of the sensitivity to the top-quark--Higgs couplings in terms
of $\zeta_t$ a model was presented in Ref.~\cite{Coleppa:2017rgb}.  
In the following the methodology and results are briefly described.
In order to assess the sensitivity to the top-quark--Higgs couplings,
a model file built in \texttt{FeynRules}~\cite{Alloul:2013bka}
which incorporates the Lagrangian, c.f.\ Eq.~\eqref{eq:lphase}. 
The associated top-Higgs production in the charged-current
channel $p\, e^- \to \bar t \, h\, \nu_e$ is then simulatend (c.f.\ Fig.~\ref{fig:figW}),  
where it is assumed that $h$ decays into a $b \bar b$ pair, and the
anti-top quark decays leptonically.
An electron-proton centre of mass energy of $\sqrt{s} \approx 1.3$~TeV is assumed.
In this study~\cite{Coleppa:2017rgb}, the analysis is performed at the
parton level.
For signal and background event generation the Monte Carlo event generator package 
\texttt{MadGraph5}~\cite{Alwall:2014hca} is employed together with
\texttt{NNPDF23\_lo\_as\_0130\_qed}~\cite{Ball:2012cx} parton
distribution functions.
The renormalisation and factorisation scales for the signal sample
are chosen to be $\mu_F = \mu_R = (m_t + m_h)/4$.
The background samples are generated using the default
\texttt{MadGraph5}~\cite{Alwall:2014hca} dynamic scales.
The longitudinal polarisation of the electron beam is asumed to be -0.8.

In Fig.~\ref{fig:cs_rs}  we present the variation of the total cross section against the electron beam energy for the signal process $ p\,e^- \to \bar t h \nu_e$, by considering 
un-polarised and polarised $e^-$ beam. Also, the effect of branchings of $h \to b \bar b$ and the $\bar{t}$ decay for both leptonic and hadronic modes are shown.
Possible background events typically arise from $W$+ multi-jet events, $Wb\bar b\bar b$ with missing energy which comes by considering only top-line, only Higgs-line, and 
without top- nor Higgs-line, in charged and neutral current deep-inelastic scattering and in photo-production by further decaying $W$ into leptonic mode.
To estimate the cross sections for signal and all possible backgrounds only basic cuts on rapidity $|\eta| \leq 10$ for light-jets, leptons and $b$-jets, the transverse momentum
cut $p_T \geq 10$\,GeV and $\Delta R_\text{min}$=0.4 for all particles considered.

\begin{figure}[th]
  \centering
  \includegraphics[width=0.7\textwidth,trim={0 10 0 0},clip]{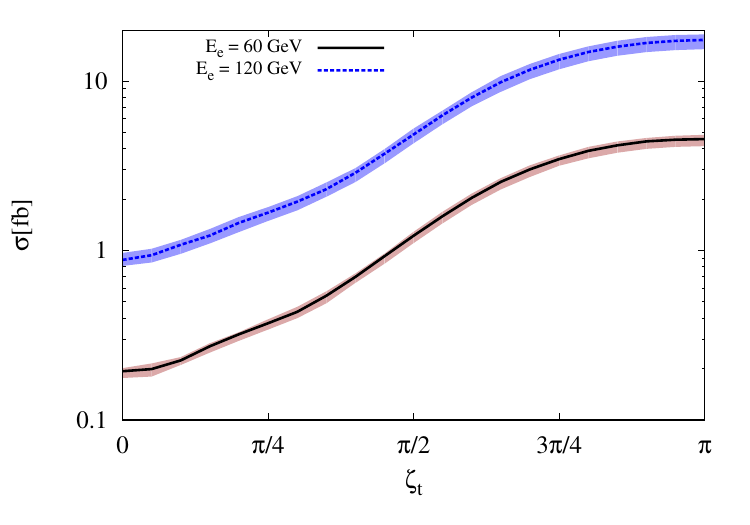}
    \caption{Total cross section of the Higgs boson produced in association with a single top as a function of $\zeta_t$, including scale uncertainties (taken from Ref.~\cite{Coleppa:2017rgb}). The \emph{black} solid and \emph{blue} dotted
      lines correspond to $E_e = 60$ and $120$\,GeV, respectively.
      These are obtained for fixed $E_p = 7$\,TeV and scales $\mu_F = \mu_R = (m_t + m_h)/4$.}
    \label{fig:mu}
\end{figure}

Estimation of the sensitivity of the associated top-Higgs production cross-section, $\sigma(\zeta_t)$, as a function of the CP phase of the $tth$-coupling is shown in 
Fig.~\ref{fig:mu}. In this study, the eletron and proton beams are
assumed to have an energy of 60\,GeV and 7\,TeV, respectively.
The scale uncertainties are obtained by varying the nominal scale,
$\mu_F = \mu_R \leq (m_t + m_h)/4$, by factors of 0.5 and 2.
It is observed that the size of the cross section is strongly dependent
on the value of $\zeta_t$, in particular in the region $\zeta_t > \frac{\pi}{2}$ where the 
interference between the diagrams becomes constructive.
At lower values of $\zeta_t$ the interference is still constructive,
but it decreases with decreasing $\zeta_t$.
Note, $\zeta_t=0$ represents the cross section in the strict SM formalism.

At $\zeta_t =\frac{\pi}{2}$, which corresponds to the pure CP-odd case,
the cross section is increased by about a factor of five in comparison to the SM expectation.
At $\zeta_t = \pi$, which corresponds to the pure CP-even case with opposite sign of $tth$-coupling,
the cross section can be enhanced by a factor of up to 24 times.
In the while range of $\zeta_t$, the scale uncertainty is found to be about 7\,\%.

To evaluate sensitivity to the measurement of the top-Yukawa coupling and its CP-phase, following criteria of fiducial selection taken~\cite{Coleppa:2017rgb}:
\begin{itemize}
 \item $p_T \ge 20$~GeV for $b$-tagged jets and light-jets, and $p_T \ge 10$~GeV for leptons.
 \item $b$-jets must be within $-2 \le \eta \le 5$
 \item light jets and the scattered lepton must be identified within $2 \le \eta \le 5$
 \item all final state particles must be separated by a distance $\Delta R$ greater than 0.4.
 \item the missing transverse energy must exceed 10\,GeV
 \item the invariant mass windows for the Higgs through $b$-tagged jets and the top
   are required to be $115 < m_{bb} < 130$~GeV and $160 < m_t < 177$~GeV.
\end{itemize}
In these selections the $b$-tagging efficiency is assumed to be 70\%,
with fake rates from $c$-initiated jets and light jets to the $b$-jets
to be 10\% and 1\%, respectively.  
The last requirement is important to reduce background processes.

\begin{figure}[th]
   \centering
   \includegraphics[width=0.7\textwidth]{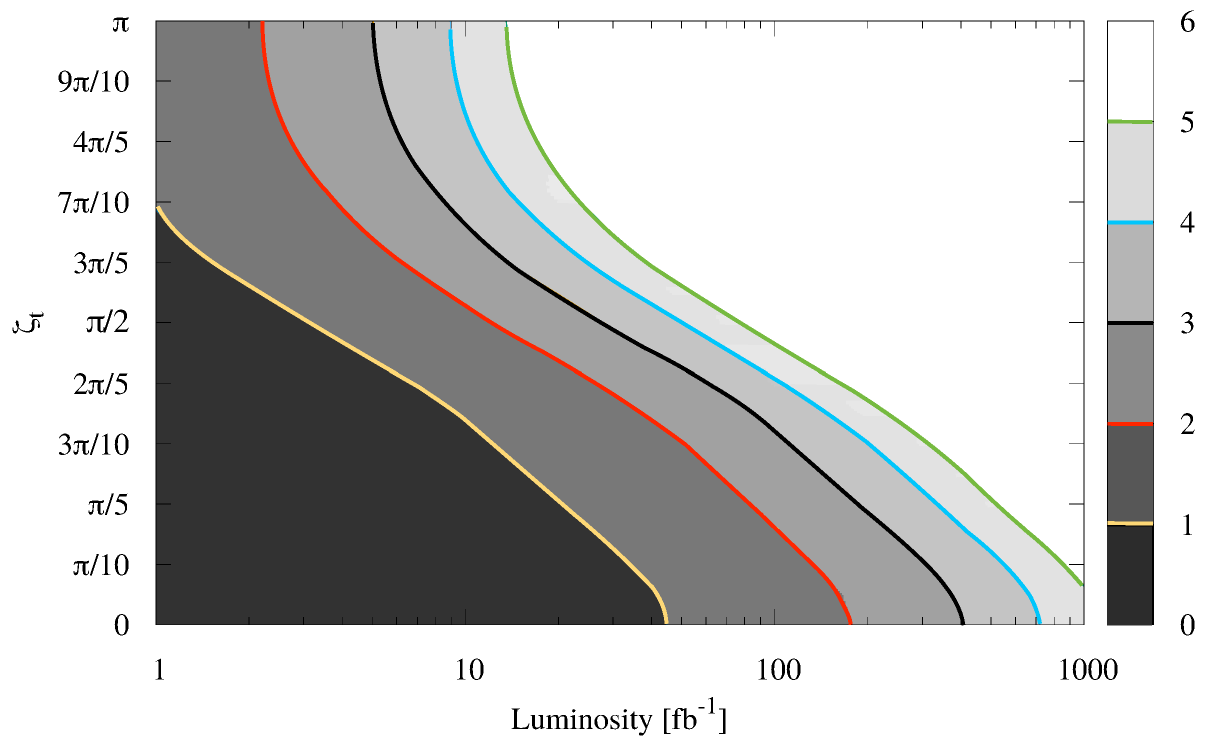}\\
   \caption{Exclusion contours for $\zeta_t$ as a function of
     integrated luminosity for $\sqrt{s}=1.3\,\TeV$ (from
     Ref.~\cite{Coleppa:2017rgb}).
     The shaded areas under a contour-line are excluded.
     The blue  and black lines represent the $4\sigma$ and
     $3\sigma$ regions.  Results are obtained based on fiducial
     cross-sections (see text).
   }
   \label{fig:Excl2}
 \end{figure}

Using the Poisson formula $S = \sqrt{2[(S+B)\log (1 + S/B) - S]}$,
where $S$ and $B$ are the number of expected signal and background
events at particular  
luminosity ($L$ in fb$^{-1}$), the exclusion regions regions of
$\zeta_t$ as a function of $L$ are estimated.
Here, 10\,\% systematic uncertainty is expected for the background yields. 
Fig.~\ref{fig:Excl2}, displays the exclusion contours at different
confidence levels.
It is observed that the shape of the exclusion limit changes at around
$\zeta_t=\pi/2$.
Therefore, in order to obtain significant exclusion limits in the
region $0<\zeta_t<\pi/2$ a larger integrated luminosity is required.
This is in keeping with the feature exhibited in 
As it is visible from Fig.~\ref{fig:mu}, for $\zeta_t$ values where
constructive interference between the signal diagrams enhances the
cross-section over the SM value, i.e.\ for $\zeta_t>\pi/2$,
less integrated luminosity is requried in order to obtain precise
exclusion limits.
For example, for $L = 100$~fb$^{-1}$, regions above $\pi/5 < \zeta_t
\leq \pi$ and $3\pi/10 < \zeta_t \leq \pi$ are excluded at a
confidence limit of 2$\sigma$ and 3$\sigma$, respectively.
With $L = 400$~fb$^{-1}$, regions above $\pi/6 < \zeta_t \leq \pi$ and
$\pi/4 < \zeta_t \leq \pi$ are excluded at 4$\sigma$ and
5$\sigma$~C.L., respectively.
The asymmetry studies at the HL-LHC~\cite{Rindani:2016scj} help probe
up to $\zeta_t=\pi/6$ for a total integrated luminosity of
3~ab$^{-1}$. However, LHeC provides a better environment to test the
CP nature of Higgs boson couplings.
For the targeted integrated luminosity of $L =1\,\text{ab}^{-1}$, almost all
values of $\zeta_t$ can be excluded with at a confidence level of at least 4$\sigma$. 

While investigating the overall sensitivity of $\zeta_t$ by applying these two observables, it is also important to measure the accuracy of SM $tth$ coupling $\kappa$ at the LHeC 
energies. By using the the formula $\sqrt{(S + B)}/{(2 S)}$ at a
selected luminosity of $L = 1\,\text{ab}^{-1}$, the value of $\kappa$
can be determined with an uncertainty of about $pm 0.17$.
In this estimate, 10\% systematic uncertainty is been taken for the background yields.

These results are obtained based on the evaluation of the fiducial cross-sections alone. As pointed out in Ref.~\cite{Coleppa:2017rgb},  a number of other observables 
carry sensitivity to the structure of the top-Higgs Yukawa coupling, such as the rapidity difference between the top quark and the Higgs boson and a number of angular 
variables. While the fiducial rate studied here is the single most sensitive observable, it is evident that a multi-variate approach will significantly enhance the sensitivity 
reported here.

\section{Higgs Decay into Invisible Particles}
\label{sec:Hinv}
The Higgs decay into invisible particles could be a key to BSM physics. 
The SM branching ratio of $H \to ZZ \to 4\nu$ is only 0.1\%. Any sizable
decay rate into invisible particles would thus indicate an exotic decay, 
for example to dark matter particles.
Its non-observation would give the SM
cross section measurement, reconstructing more than
$99$\,\% of the ordinary decays a higher meaning for 
constraining the total Higgs decay width. 
 
For the LHeC at a luminosity of $1$\,ab$^{-1}$, 
initial parton-level studies of this decay were presented in Ref.~\cite{Tang:2015uha}, with the estimate of a two $\sigma$ sensitivity to a branching fraction of $6$\,\%.
For this study, NC production via $ZZ$ fusion $eq \to eqZZ \to eqH$ was used,
 which has a cross section of about $25$\,fb at the LHeC.
The CC production via $WW$ fusion has a larger cross section, but entails a missing energy signal by itself which requires further study of potentially quite some gain in precision. This channel, when employed for the invisible decay study, results in a mono-jet signature which is hard to separate from the SM DIS CC background. 

\begin{figure}[th!]
\centering
\includegraphics[width=0.75\textwidth]{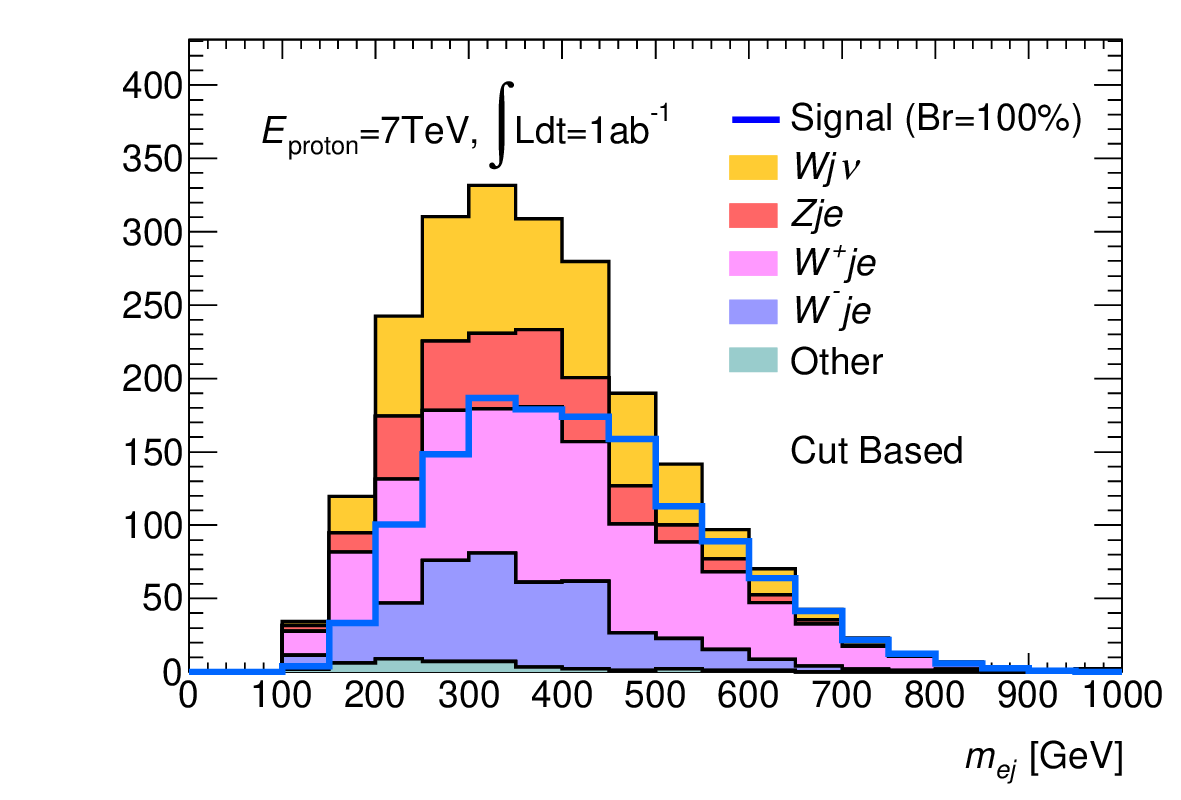}
\caption{
Electron-jet invariant mass distribution for the Higgs to
invisible decay signal (normalized to 100\%
branching ratio) and the stacked backgrounds for an integrated
 luminosity of 1~ab$^{-1}$ at the LHeC after all selection cuts.}
\label{fig:inv-fig1}
\end{figure}

The neutral current study has been repeated using the LHeC Higgs WG analysis tools, introduced above: MadGraph, Pythia and Delphes. Similar to~\cite{Tang:2015uha}, an electron beam of 60~GeV with a polarization of -$80\%$ is assumed.
The basic event topology contains the scattered electron, jet and missing transverse energy. Its main
background results from SM $W$ and $Z$ productions (followed
by $W \to \ell \nu$ and $Z \to \nu \bar{\nu}$). In the study
NC and CC $W$ production and NC $Z$ production are considered, while single-top, NC multijets and $W$ photoproduction  were all found to be negligible.
Requiring missing transverse energy of 60\,GeV, exactly one electron and one jet, and
 no other leptons (including $\tau$), as well as imposing several selection
criteria on the kinematics of electron, jet and missing transverse momentum,
 we get a two $\sigma$ sensitivity to a branching ratio of 7.2\,\%, which is similar  to the earlier result~\cite{Tang:2015uha}.
Fig.~\ref{fig:inv-fig1} shows the electron-jet invariant mass distribution
 after the selection for the signal (normalized to a 100\,\% branching ratio) and the background.

\begin{figure}[!htb]
  \centering
  \begin{minipage}[t]{0.48\textwidth}
    \centering\includegraphics[width=0.95\textwidth,trim={40 0 0 0},clip]{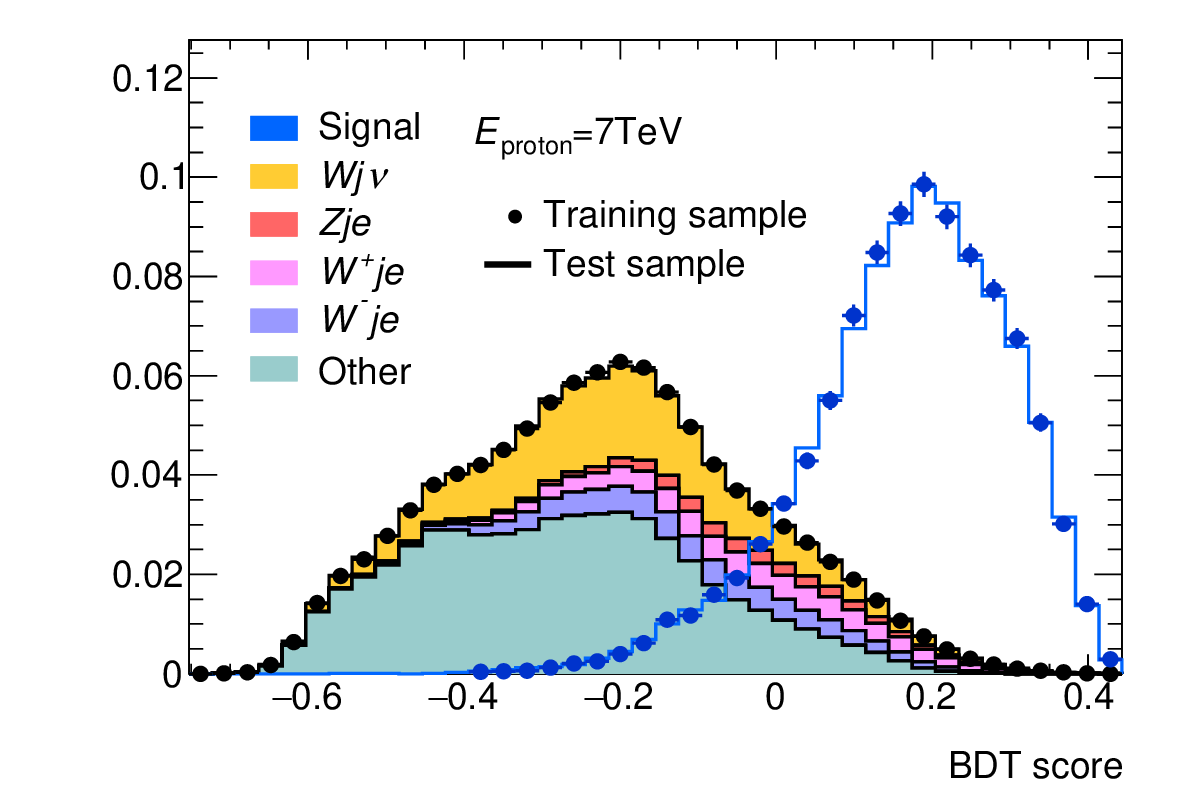}
    \caption{BDT output score distribution for the Higgs
to invisible decay signal and the stacked backgrounds (both area normalized) at the LHeC.}
    \label{fig:inv-fig2}
  \end{minipage}
  \hspace{0.02\textwidth}
  \begin{minipage}[t]{0.48\textwidth}
    \centering
    \includegraphics[width=0.95\textwidth,trim={40 0 0 0},clip]{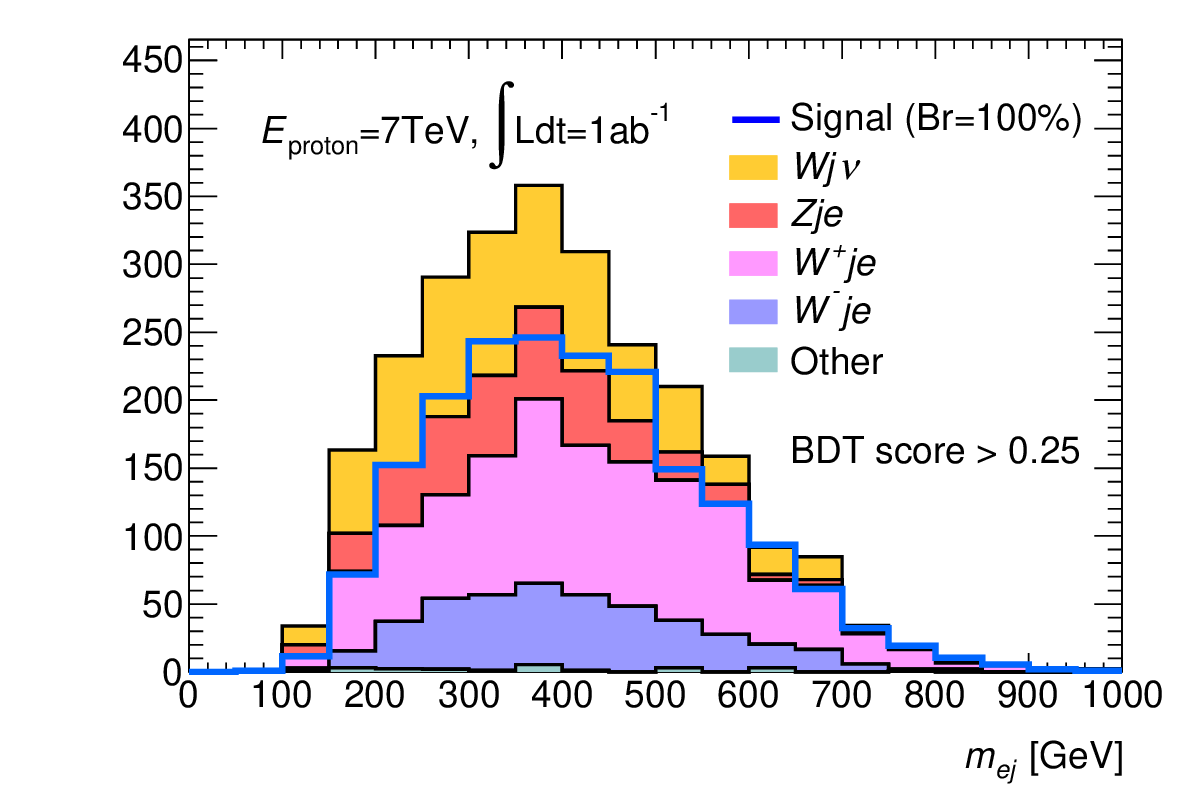}
    \caption{Electron-jet invariant mass distribution for the
Higgs to invisible decay signal (normalized to 100\%
branching ratio) and the stacked backgrounds for an integrated luminosity
 of 1~ab$^{-1}$ at the LHeC after the BDT score cut of 0.25.}
    \label{fig:inv-fig3}
  \end{minipage}
\end{figure}

The analysis has been further refined with a usage of multivariate analysis (Boosted Decision
Tree in TVMA package).
Basically the set of selection variables used in the cut-based analysis
above was used as inputs to the multivariate analysis, tuned to yield the
 best output score to discriminate the signal from backgrounds.
Fig.~\ref{fig:inv-fig2} shows the distribution of the discriminant variable
 for the signal and background (both area normalised).
An optimization on the statistical significance is found at the BDT
 score $>$ 0.25, and the resulting mass distribution is shown in
Fig.~\ref{fig:inv-fig3}. 
With $1$\,ab$^{-1}$ of integrated luminosity, a  two $\sigma$ sensitivity of 5.5\% is obtained consistent with the previous results. 
For a comparison, an estimate of 3.5\,\% is given for a HL-LHC
sensitivity study on this channel~\cite{Bernaciak:2014pna}.
The result on the LHeC may be further improved in the future with a refined BDT analysis when one introduces extra parameters, beyond those initially introduced with the cut based analysis.

In these initial studies no systematic uncertainties were considered. 
This may be justified with the very
a clean environment of electron-hadron collider, in which 
precise measurements of $W$ and $Z$ production will be made,
for example in their decays to muons, for accurately
controlling the systematics in the background prediction to 
a negligible level.

The BDT analysis was repeated for higher proton energies.  At the HE-LHeC ($E_p$=13.5~TeV)
the NC production cross section increases to 45\,fb and the branching ratio sensitivity 
improves to 3.4\,\% because the luminosity is doubled in the configurations here assumed. 
At the FCC-eh, the cross section rises to 120~fb and the  sensitivity of the branching ratio reaches about 1.7\,\%.
%
%

%

%
%

\biblio

%% file: bsm/bsm.chapter.tex
\linenumbers
\lhectitlepage
\lhecinstructions
\subfilestableofcontents

\input{\main/bsm/bsm.tex}

\biblio

%% file: bsm/bsm.tex
\chapter{Searches for Physics Beyond the Standard Model \ourauthor{ Georges Azuelos, Oliver Fischer, Monica D'Onofrio}}

\newcommand{\neutralino}[1]{\tilde{\chi}_{#1}^0}
\newcommand{\chargino}  [1]{\tilde{\chi}_{#1}^\pm}

\newcommand{\slepton} {\tilde{\ell}}
\newcommand{\sleptonL}{\tilde{\ell}_L}
\newcommand{\sleptonR}{\tilde{\ell}_R}
\newcommand{\sneutrino}{\tilde{\nu}}
\newcommand{\selectronL}{\tilde{e}_L}
\newcommand{\selectronR}{\tilde{e}_R}
\newcommand{\iab}{\ensuremath{\text{ab}^{-1}}}

\section{Introduction}
%



The LHC was originally envisioned as the ultimate machine to search for physics beyond the Standard Model at the TeV scale. 
Since electrons and quarks share only electroweak interactions, an electron-proton collider could allow to measure the same phenomena in a different environment with generally higher precision. It could add complementary search channels or lead to the discovery of a weak signal. The possibility of undiscovered New Physics (NP) below the TeV scale could thus be also addressed by the LHeC, which is projected to operate when the LHC will be in its high luminosity phase, in spite of the lower centre-of-mass energy.
Exotic phenomena that can be studied at $ep$ colliders have been reviewed, for example, in~\cite{Cashmore:1985xn}.
More recently, but when the LHC was only beginning to yield data in Run I, an overview of the potential of the LHeC for probing physics beyond the Standard Model has been given in the Conceptual Design Report ~\cite{AbelleiraFernandez:2012cc}. Since then, stringent constraints on NP phenomena have been obtained from the LHC and the
absence of hints from NP  to date is presently changing this paradigm to two alternative scenarios: NP may actually reside at an even larger energy scale; NP  may be at or below the TeV scale, but more weakly coupled, and thus hidden in the SM backgrounds~\cite{vonBuddenbrock:2019ajh}. 
 
A similar $pp$-$ep$ synergy could be envisaged with higher proton beam energies at the FCC 100\,km tunnel. 
With an electron beam of 60~GeV, the expected centre-of-mass energies for $ep$ could be $2.9$\,TeV for $E_p = 19$\,TeV (Low-Energy FCC) and $3.5$\,TeV for $E_p = 50$\,TeV (FCC). 
Below we list recent developments which discuss new physics opportunities at the LHeC and its potential future high-energy upgrades.

\section{Extensions of the SM Higgs Sector} 
Presently, given the precision of measurements in the Higgs sector, it appears that the discovered 125\,GeV scalar is indeed the SM Higgs boson. The question remains, however, if the scalar potential is truly that of the SM or if it should be extended, possibly with additional degrees of freedom. Several extensions of the Higgs sector have been proposed and can be studied at the $ep$ colliders with results often complementary to those of $pp$ colliders and other future  facilities.

\subsection{Modifications of the Top-Higgs interaction} 
In electron-proton collisions the heavy top-quarks can be produced in association with a Higgs boson, which allows us to study the sensitivity of the LHeC or the FCC-eh to the top-Higgs ($tH$) interaction.
In Ref.~\cite{Coleppa:2017rgb} the sensitivity of the process $p e^- \to {\bar t} H \nu_e$ to the CP nature of the $tH$ coupling is investigated by considering a CP phase $\zeta_t$ at the $ttH$ and $bbH$ vertices. The authors conclude, based on several observables and with appropriate error fitting methodology, that better limits on $\zeta_t$ are obtained at the LHeC than at the HL-LHC.  At the design luminosity of 1 ab$^{-1}$, almost all values of $\zeta_t$ are excluded up to 4$\sigma$ C.L. and the
SM top-Higgs coupling  could be measured relative to its SM value with a precision of $\kappa =1.00 \pm 0.17$. 

Flavour changing neutral currents (FCNC) are completely absent at tree-level in the SM and strongly constrained, especially by low energy experiments.
Anomalous flavour changing neutral current Yukawa interactions between the top quark, the Higgs boson, and either an up or charm quark are documented in Chapter~3, Sec.~\ref{SecFCNC}. Among other studies, in Ref.~\cite{Liu:2015kkp} the authors consider the Higgs decay modes $H \to \gamma\gamma, bb$ and $\tau\tau$ and $E_e=150$\,GeV.
The results are updated in Ref.~\cite{Sun:2016kek} for $E_e = 60$\,GeV, including estimates for lower electron beam energies, and the 2$\sigma$ sensitivity on the branching ratio Br$(t \to uh)$ is found to be $0.15 \times 10^{-2}$.
Making use of the polarisation of the electron beam and multivariate techniques, Ref.~\cite{Wang:2017pdg} shows that limits on the branching ratio Br$(t\to uh)$ of ${\mathcal O}(0.1)\,\%$ can be obtained, an improvement over present LHC limits of 0.19\,\% 
\cite{Aaboud:2018pob,Khachatryan:2016atv}. These results vary with $E_e$ and $E_p$.

\subsection{Charged scalars} 
The prospects to observe a light charged Higgs boson through the decay $H^+\to c\bar b$ are investigated within the framework of the Two Higgs Doublet Model (2HDM) Type III, assuming a four-zero texture in the Yukawa matrices and a general Higgs potential~\cite{Hernandez-Sanchez:2016vys}. 
The charged current production processes $e^- p \to \nu H^+ q$ are considered. The analysed signature stems from the subsequent decay $H^+ \to c \bar b$.
The parton level analysis accounts for irreducible SM backgrounds and considers scenarios up to a mass of 200\,GeV, consistent with present limits from Higgs and flavour physics. The authors show that for $L=100$\,fb$^{-1}$ a charged Higgs boson could be observed with about $3-4\sigma$ significance. This is to be compared with results from present LHC searches in which strong limits are set on the branching fraction $B(t\to H^+ b)$,  assuming $B(H^+ \to c \bar b)=1.0$ or $B(H^+ \to c \bar s)=1.0$ for the charged Higgs boson mass range $\sim 90-160$\,GeV \cite {Sirunyan:2018dvm,Khachatryan:2015uua}. 

A similar study, $H^\pm \to sc + su$, for the FCC-eh (with $\sqrt{s} \approx 3.5$\,TeV) is presented in Ref.~\cite{Das:2018vuk}, in the context of a  next-to-minimal supersymmetric model (NMSSM). 
Using dedicated optimisation techniques, the authors show that a light charged boson $H^\pm$ can be observed with maximal significance of 4.4 (2.2)$\sigma$ provided its mass is at most $m_{H^\pm} = 114 (121)$\,GeV, for the total luminosity of 1\,ab$^{-1}$. 

\begin{figure}
    \centering
    \includegraphics[width=0.45\textwidth]{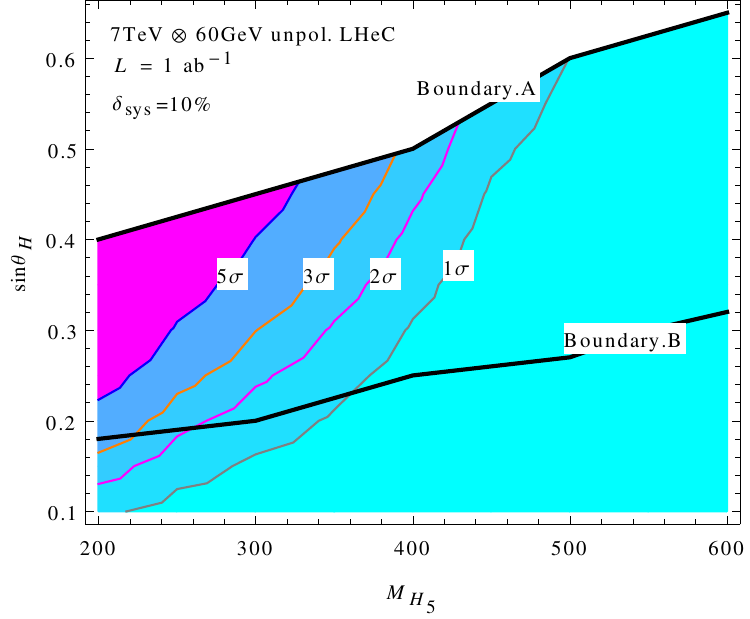}
    \includegraphics[width=0.51\textwidth]{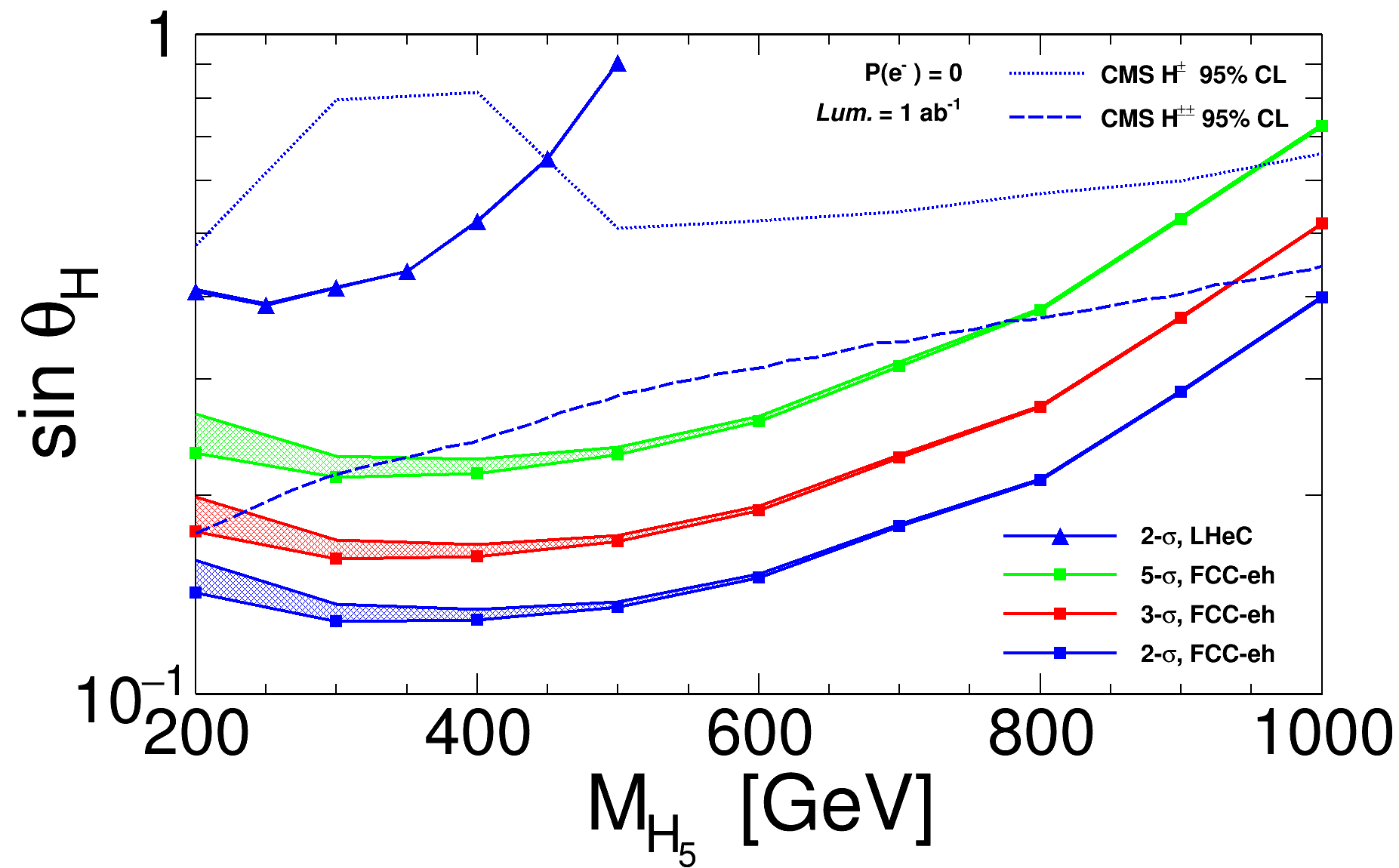}
    \caption{\emph{Left:} Discovery contour with respect to $\sin\theta_H$ and $M(H_5^{++/--})$ at LHeC with unpolarized beam; \emph{Right:} Limit Contours for the case of singly charged Higgs for FCC-eh and LHeC. The blue dotted curve and the blue dashed curves give the 95\% CL limit from CMS for $H_5^{+/-}$ and for $H^{++/--}$ respectively~\cite{Sirunyan:2017sbn,Sirunyan:2017ret}. An unpolarized beam of integrated luminosity of 1 ab$^{-1}$ and a 10\% systematic uncertainty for background yields is assumed in both plots. }
   \label{fig:H5_GM}
\end{figure}

 The Georgi-Machacek (GM) model extends the Higgs sector by including higher multiplet states while preserving custodial symmetry. The physical states include, besides the SM Higgs, a heavier singlet $H$,  a triplet $(H_3^+,H_3^0,H_3^-)$ and a quintuplet $(H_5^++,H_5^+,H_5^0,H_5^-,H_5^--)$. The $H_5$ scalars do not couple to fermions and can therefore only be produced by vector boson fusion.
 An analysis for the prospects to discover the doubly charged Higgs bosons in the GM model at the LHeC and the FCC-eh is presented in Ref.~\cite{Sun:2017mue}.
Therein the production of a doubly-charged member of five-plet Higgs-bosons ($H^{\pm\pm}_5$), produced from $W^\pm W^\pm$ fusion is studied.
The authors find 
that 2 to 3$\sigma$ limits can be obtained for mixings $\sin(\theta_H)$ as low as 0.2, for $M(H_5) < 300$ GeV.
The prospects can be improved at the FCC-eh collider, where doubly charged Higgs bosons can be tested for masses $M_{H_5} < 400$\,GeV, also for small scalar mixing angles (Fig.~\ref{fig:H5_GM} (left)). 

The discovery prospects for the singly charged Higgs, $H_5^\pm$ of  the Georgi-Machacek model, produced in $W^\pm Z$ fusion, are evaluated in Ref.~\cite{Azuelos:2017dqw}.
The authors perform a multivariate analysis, including a fast detector simulation, and consider the LHeC and the FCC-eh for a mass range from 200--1000\,GeV.
They find that the LHeC can improve over current LHC limits on $H_5^\pm$ for masses up to about 400\,GeV and scalar mixing angles $\sin \theta_H \sim 0.5$ (Fig.~\ref{fig:H5_GM} (right)).

\subsection{Neutral scalars}
Neutral scalar bosons generally appear in many extensions of the scalar sector.
They can be added directly, as $SU(1)$ singlets, or be part of higher representation $SU(2)$ multiplets. 
They generally mix with the SM Higgs boson, from which they inherit a Higgs-like phenomenology. 

The potential of testing the heavier CP-even scalar that is contained in the 2HDM Type-I is presented in Ref.~\cite{Mosomane:2017jcg}.
Therein, the lighter scalar particle is considered to be a SM-like Higgs boson and the properties of a heavy scalar, assumed to have the specific mass 270\,GeV, is discussed. The authors state that the final state $H\to Sh$, where $S$ is a scalar singlet with a mass around 100 GeV, is of particular interest, as it connects to the findings in Ref.~\cite{vonBuddenbrock:2019ajh}.

\begin{figure}
\centering
\includegraphics[width=0.75\textwidth]{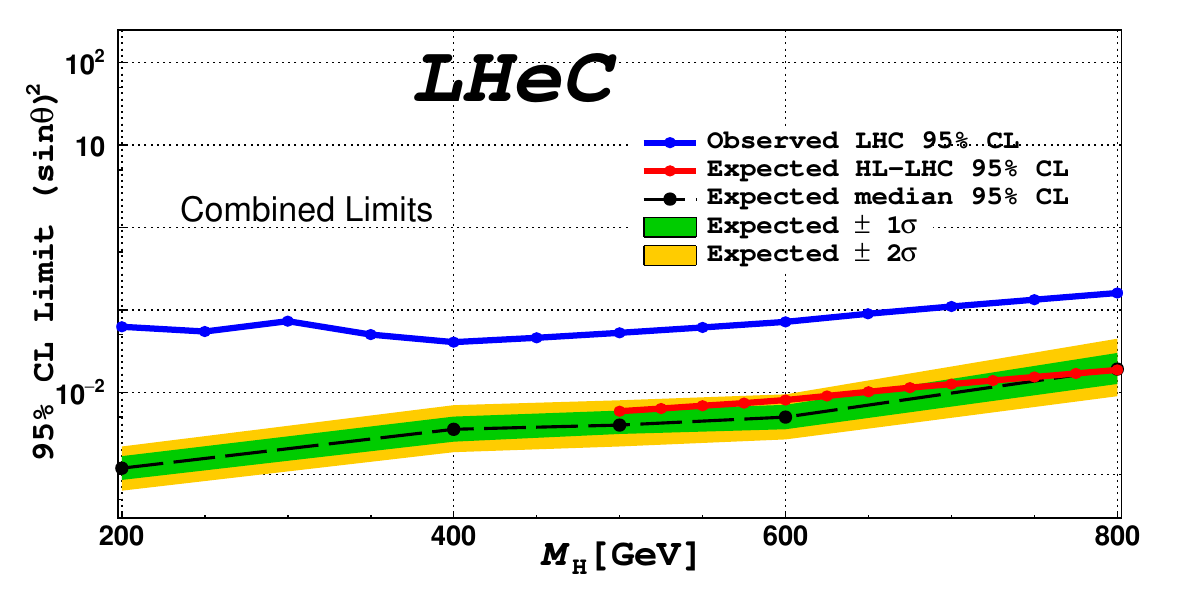}
\caption{Expected exclusion limits (green and yellow bnds) for a heavy scalar search at the LHeC, assuming a systematic uncertainty on the SM background of $2\,\%$ (from Ref.~\cite{DelleRose:2018ndz}). The blue line represents the current LHC limit at $95\,\%$ CL as extracted from \cite{Sirunyan:2018qlb}, the red line the forecast of the HL-LHC sensitivity via $h_2 \to ZZ$ searches from Ref.~\cite{CMS:2019qzn}. The LHeC results correspond to an integrated luminosity of 1\,ab$^{-1}$. } 
\label{fig:heavyhiggs}
\end{figure}

The prospects to search for a generic heavy neutral scalar particle are presented in detail Ref.~\cite{DelleRose:2018ndz}.
The model is a minimal extension of the SM with one additional complex scalar singlet that mixes with the SM Higgs doublet, which governs its production and decay mode.
The heavy scalar is produced via vector-boson fusion and decays into two vector bosons. A multivariate analysis is performed and 
detector simulation is taken into account. Masses between 200 and 800\,GeV and scalar mixings as small as $\sin^2 \alpha \sim 10^{-3}$ are considered. 
The resulting sensitivity for a total luminosity of 1\,ab$^{-1}$ is shown in Fig.~\ref{fig:heavyhiggs}, including existing bounds from the LHC and future HL-LHC projections. A significant improvement over existing LHC limits is found, with the LHeC probing scalar boson masses below $\sim 500$\,GeV, a region which remains difficult at the HL-LHC.

The scalar bosons from the 2HDM Type-III framework may give rise to flavour violating signatures, as discussed in Ref.~\cite{Das:2015kea}.
The prospects to observe the light and heavy CP-even neutral Higgs bosons via their decays into flavour violating $b\bar s$ channels were studied with specific Yukawa textures and a general Higgs potential. 
The signature consists of one jet originating from b-hadron fragmentation (b-tagged jets) and one light-flavour jet in the central rapidity region, with a remaining jet in the forward region.
Relevant SM backgrounds were considered and it is found that flavour violating decays of the SM-like Higgs boson would be accessible with $L=100$\,fb$^{-1}$ at $ep$ colliders. 

The prospects of observing the light CP-even neutral Higgs bosons of  the NMSSM framework, the MSSM with an additional singlet superfield, via their decays into b-quarks and in the neutral and charged current production processes, are studied in Ref.~\cite{Das:2016eob}.
In this work the following constraints are incorporated into the spectrum: neutralino relic density corresponding to the observed dark matter relic density; direct and indirect mass bounds from searches for specific sparticles; the SM-like Higgs boson has a mass around 126\,GeV and an invisible branching ratio below 0.25. 
The signal is given by three jets plus an electron or missing transverse momentum ($E^{miss}_{T}$) arising from the neutral (charged) current interaction, where two jets are required to be originating from a b-quark and the remaining jet is required to be in the forward region. 
For the cut-based analysis a number of reducible and irreducible SM backgrounds, generated with a fast detector simulation with an adaptation of the LHeC detector, are considered.
It is found that the boson $h_1$ could be observable for some of the NMSSM benchmark points, at up to 2.5$\sigma$ level in the $e+3j$ channel up to masses of 75\,GeV; in the $3j+E^{miss}_{T}$ channel $h_1$ could be discovered at  2.4$\sigma$ level up to masses of 88\,GeV with $L=100$\,fb$^{-1}$, and a 5$\sigma$ observation is possible with $\mathcal{L}=1$\,ab$^{-1}$ for masses up to $90$\,GeV.

\subsection{Modifications of Higgs self-couplings}
As in the chapter on Higgs physics above, the $e^- p$ collisions are a very convenient environment to study the property of the SM Higgs boson itself.
The latter is produced through vector-boson fusion processes and the precise measurement of its properties provides a unique opportunity to probe the interaction $HVV$, $(V = W^\pm, Z)$.
These interactions are in general sensitive to certain classes of beyond the SM physics, which can be parameterized, for instance, via higher dimensional operators and their coefficients, cf.\ Refs.~\cite{Biswal:2012mp,Senol:2012fc,Cakir:2013bxa,Kumar:2015kca,Hesari:2018ssq}. 

The prospects of inferring the strengths of the two couplings $HWW$ and $HZZ$ were studied in Refs.~\cite{Biswal:2012mp,Cakir:2013bxa} in the context of electron-proton collisions. The authors find that the higher-dimensional operator coefficients can be tested for values around ${\mathcal O}(10^{-1})$ at the LHeC. 
This sensitivity is improved at the FCC-eh due to larger centre-of-mass energies, which in general enhance the vector-boson fusion cross sections.

The Higgs self-coupling itself $HHH$ can be tested through the measurement of the di-Higgs production cross section as was shown in Ref.~\cite{Kumar:2015kca}. 
With appropriate error fitting methodology this study illustrates that the Higgs boson self-coupling could be measured with an accuracy of $g^{(1)}_{HHH} = 1.00^{+0.24(0.14)}_{-0.17(0.12)}$ of its expected SM value at $\sqrt{s} = 3.5(5.0)$\,TeV, considering an ultimate 10\,ab$^{-1}$ of integrated luminosity. 

An analysis presented in Ref.~\cite{Hesari:2018ssq} evaluates the LHeC sensitivity to dimension-six operators. The authors employ jet substructure techniques to reconstruct the boosted Higgs boson in the final state.
A shape analysis on the differential cross sections shows in some cases improvements with respect to the high-luminosity LHC forecasts.

\subsection{Exotic Higgs boson decays}
The LHeC sensitivity to an invisibly decaying Higgs boson was investigated in Ref.~\cite{Tang:2015uha}. 
Therein the focus is on the neutral current production channel due to the enhanced number of observables compared to the charged current counterpart. The signal contains one electron, one jet and large missing energy. A cut-based parton level analysis yields the estimated sensitivity of Br$(h \to $invisible)\,=\,6\,\% at 2$\sigma$ level.
%
Exotic decays of the Higgs boson into a pair of light spin-0 particles referred to as $\Phi$ was discussed in Ref.~\cite{Liu:2016ahc}.
The studied signature is a final state with 4 b-quarks, which is well motivated in models where the scalars can mix with the Higgs doublet, and suffers from multiple backgrounds at the LHC.
The analysis is carried out at the parton level, where simple selection requirements render the signature nearly free of SM  background and makes $\Phi$ with masses in the range [20, 60]\,GeV testable for a $hVV$ ($V = W, Z$) coupling strength relative to the SM at a few per-mille level and at 95\,\% confidence level.

\begin{figure}
\centering
\includegraphics[height=0.3\textwidth]{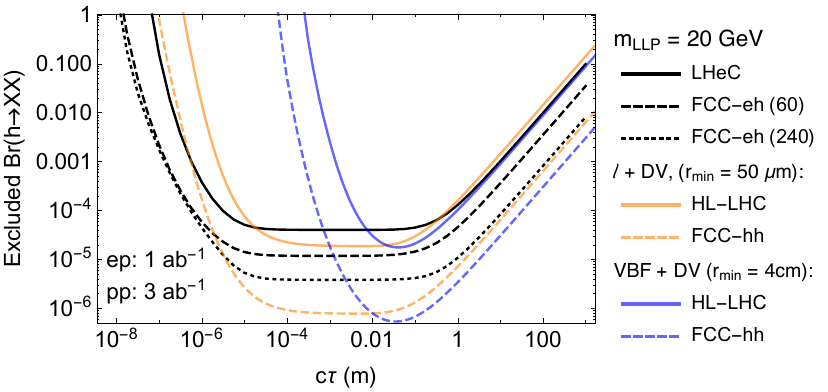}\qquad
\includegraphics[height=0.3\textwidth]{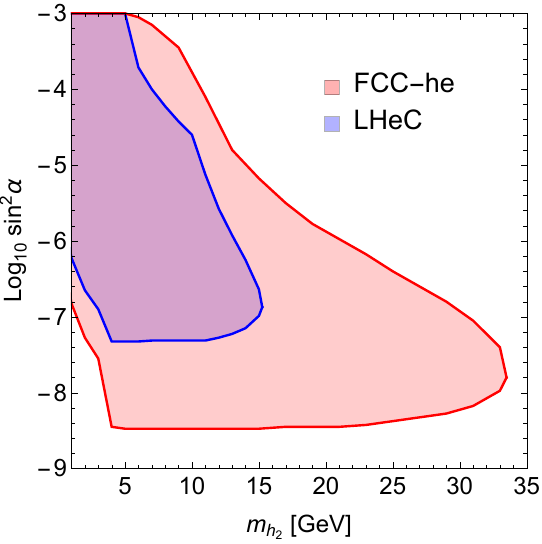}
\caption{Sensitivity contours for displaced vertex searches for Higgs decays into long-lived scalar particles (LLP), which are pair produced from decays of the Higgs boson, and which themselves decay via scalar mixing into fully visible final states.
Left: As a function of the LLP lifetime for a fixed mass from Ref.~\cite{Curtin:2017bxr}. Right: For a specific model, where lifetime and production rate of the LLP are governed by the scalar mixing angle. The contours are for 3 events and consider displacements larger than 50\,$\mu$m to be free of background.}
\label{fig:LLP}
\end{figure}

The prospects of testing exotic Higgs decays into pairs of light long-lived particles at the LHeC were studied in Ref.~\cite{Curtin:2017bxr} where it was shown that proper lifetimes as small as $\mu$m could be tested, which is significantly better compared to the LHC. This is shown in Fig.~\ref{fig:LLP} (left).
This information can be interpreted in a model where the long-lived particles are light scalars that mix with the Higgs doublet, where both, production and decay, are governed by this scalar mixing angle.
The area in the mass-mixing parameter space that give rise to at least 3 observable events with a displaced vertex are shown in Fig.~\ref{fig:LLP}. It is apparent that mixings as small as $\sin^2 \alpha \sim 10^{-7}$ can be tested at the LHeC for scalar masses between 5 and 15\,GeV~\cite{Curtin:2017bxr}. 

\section{Searches for supersymmetry}
Several SUSY scenarios might remain still elusive in searches performed at $pp$ colliders. While the null results from current searches by the LHC experiments have produced impressive constraints on the SUSY coloured sector (squarks and gluinos) because of their large production cross sections in strong interactions, less stringent constraints have been placed on weakly-produced SUSY particles, namely neutralinos  $\neutralino{}$, charginos $\chargino{}$, and sleptons $\slepton^{\pm}$. Some of these scenarios where $ep$ colliders might have discovery potential complementary to that of the HL-LHC are discussed below. These include R-parity conserving SUSY models, e.g.\ motivated by dark matter, or R-parity violating SUSY models, e.g.\ including single production of bottom and top squarks and low mass gluinos.

\subsection{Search for the SUSY Electroweak Sector: prompt signatures}
\begin{figure}[h]
\centering
\includegraphics[width=3.6cm,height=2.4cm]{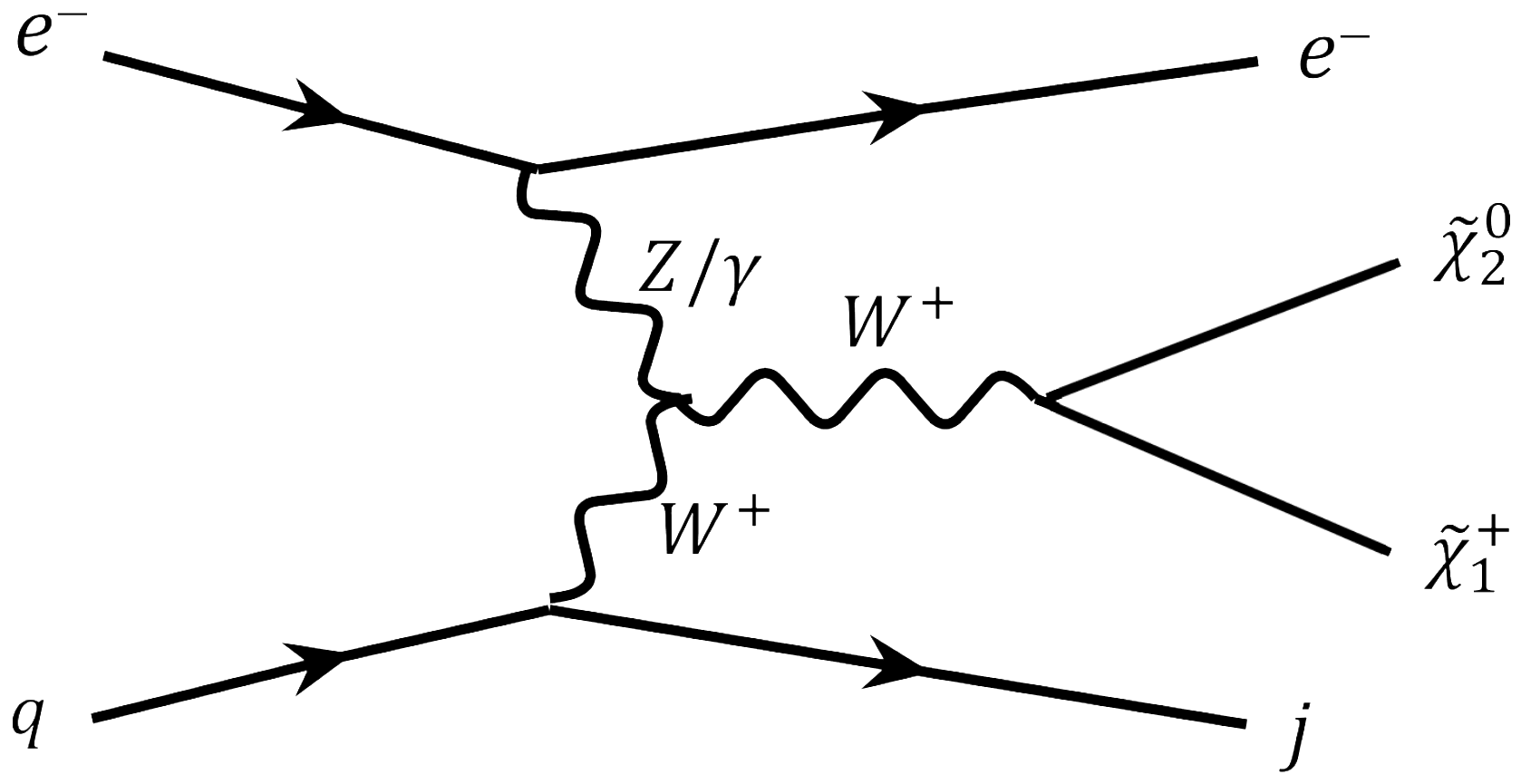}
\includegraphics[width=3.6cm,height=2.4cm]{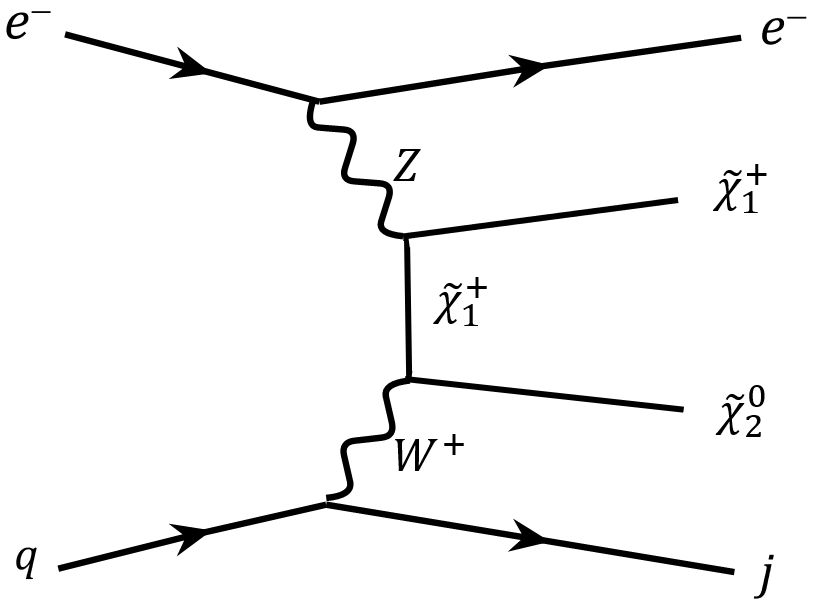}
\includegraphics[width=3.6cm,height=2.4cm]{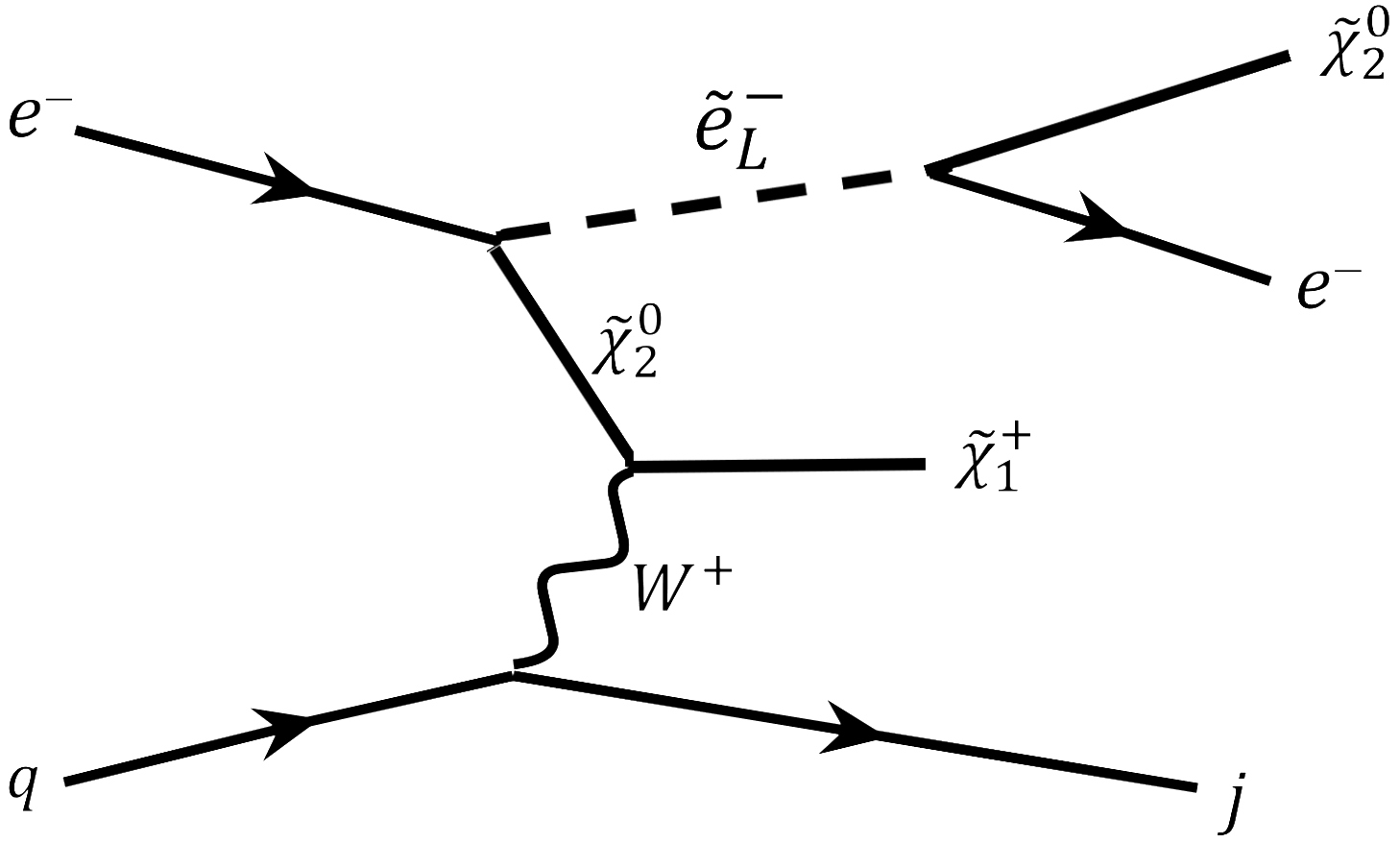}
\includegraphics[width=3.6cm,height=2.4cm]{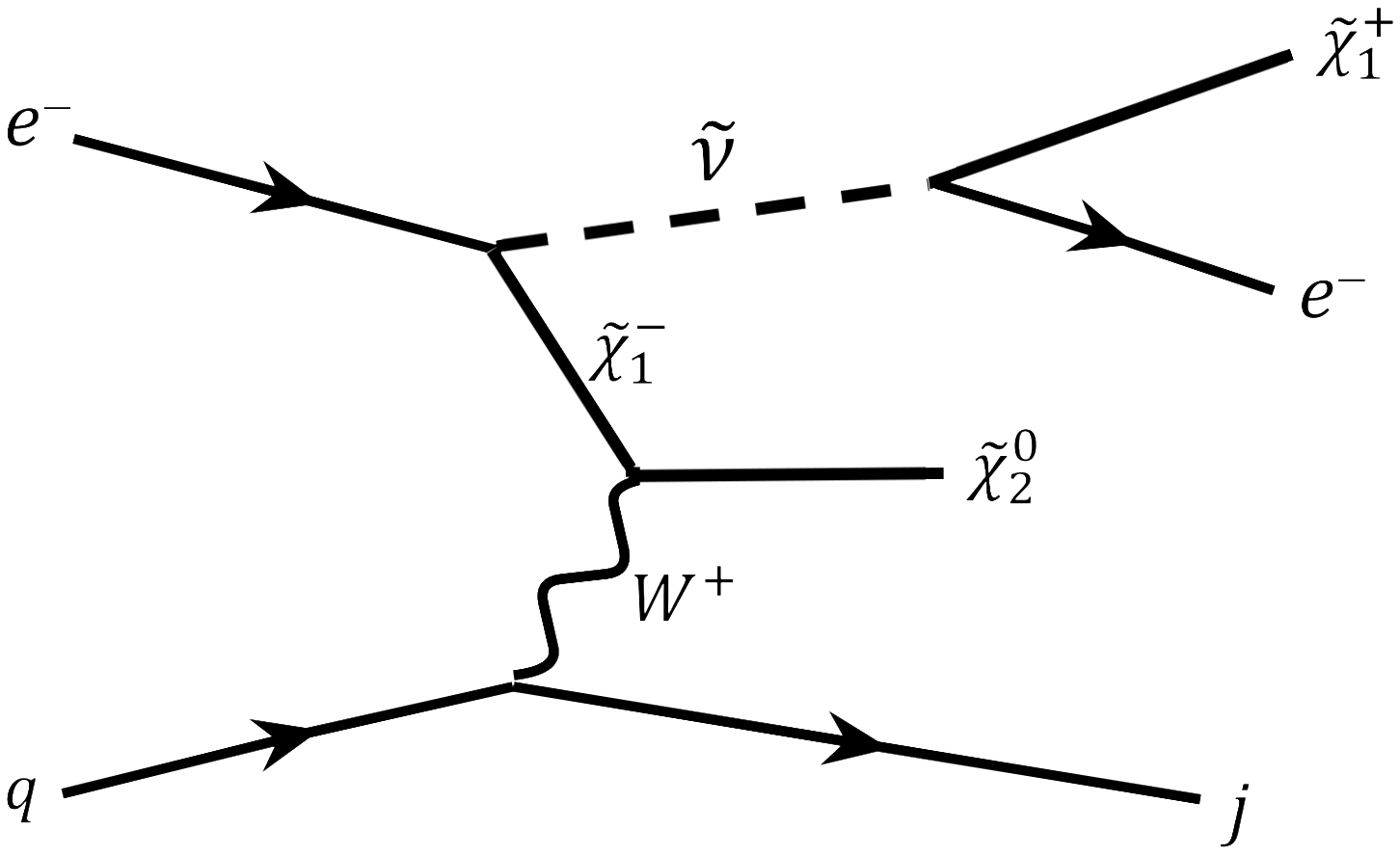}
\caption{
Representative production diagrams for the signal processes considered in Ref.~\cite{Azuelos:2019bwg}. The \emph{decoupled-slepton} scenario includes only the first two diagrams, while the \emph{compressed-slepton} scenario includes all four diagrams.
}
\label{fig:susyproduction}
\end{figure}

Electroweakino scenarios where charginos, neutralinos, and sleptons are close in mass can be characterised by the neutralino mass $m$ and the mass splitting between charginos and neutralinos $\Delta m$. 
We here refer to scenarios with $\Delta m < 50 \textrm{GeV}$ as
\textit{compressed}.
A subtlety arises for $\Delta m \leq 1$\,GeV, when the $\chargino1/\neutralino2$ becomes long lived and its decays are displaced.
For $\Delta m > 1$\,GeV the decays are prompt, the visible decay products from $\slepton$ and $\chargino1 / \neutralino2$ have very soft transverse momenta ($p_T$) and the SM backgrounds are kinematically similar to the signal.  
The analyses therefore become challenging and sensitivities decrease substantially. 
Two SUSY scenarios are considered in Ref.~\cite{Azuelos:2019bwg} and depicted in Fig.~\ref{fig:susyproduction} where the LSP $\neutralino1$ is Bino-like, $\chargino1$ and $\neutralino2$ are Wino-like with almost degenerate masses, and the mass difference between $\neutralino1$ and $\chargino1$ is small. 
The signal is produced via the process ``$p\, e^- \to j\, e^-\, \tilde{\chi} \tilde{\chi}$", where $\tilde{\chi}=\neutralino1$, $\chargino1$ or $\neutralino2$. Conservative leading order cross sections are considered for the SUSY signal models.  The kinematic observables are input to the TMVA package to perform a multivariate analysis at the detector level.

\begin{figure}
    \centering
    \includegraphics[width=0.45\textwidth]{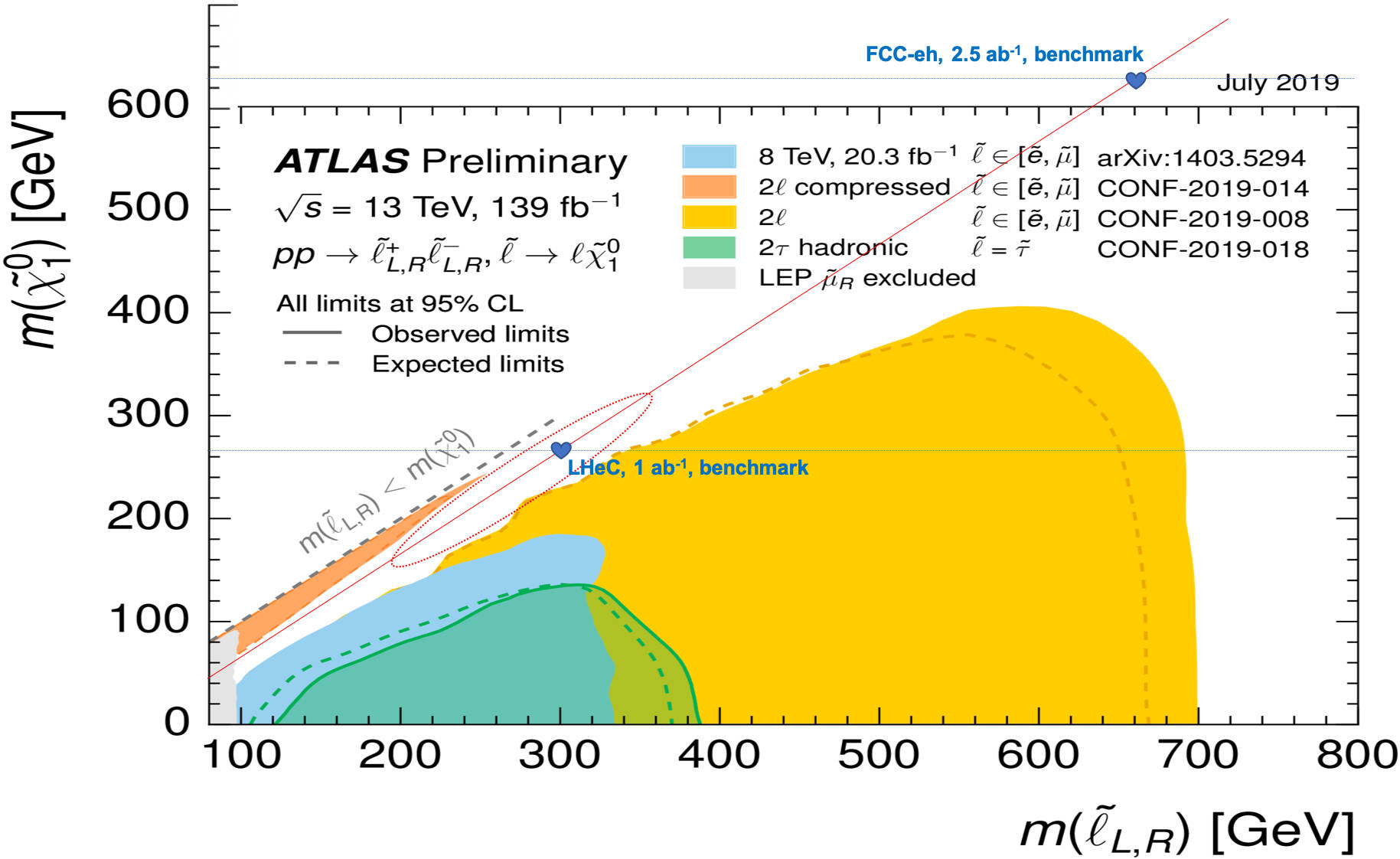}
    \includegraphics[width=0.51\textwidth]{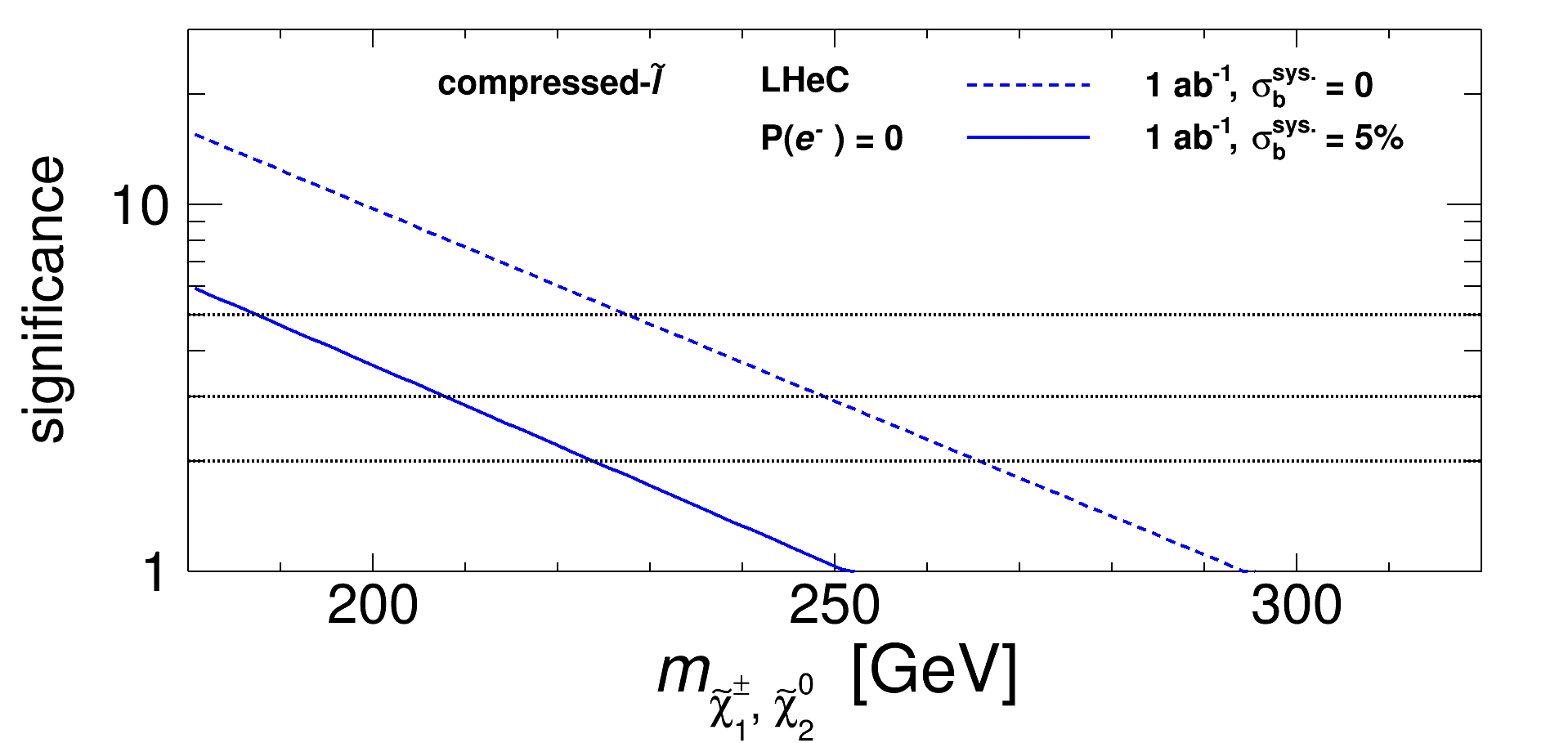}
    \caption{\emph{Left:} Benchmark assumption on slepton masses and 2019 reach of current ATLAS searches for sleptons (figure adapted from Ref.~\cite{ATLAS:2019nbu}). \emph{Right:} Significances as varying the masses of $\chargino1$ and $\neutralino2$ for the compressed-slepton scenario at the LHeC with unpolarised beams and 1 $\iab$ luminosity. For dashed (solid) curve, a systematic uncertainty of 0\,\% (5\,\%) on the background is considered. The figure is from Ref.~\cite{Azuelos:2019bwg}.}
    \label{fig:susy-complementarity}
\end{figure}

In the compressed-slepton scenario, the case where the left-handed slepton $\sleptonL$ and sneutrino $\sneutrino$ are slightly heavier than $\chargino1$ or $\neutralino2$ is considered.
When fixing the mass difference $\Delta m = m_{\slepton} - m_{\chargino1, \neutralino2} = $ 35 GeV and ignoring the systematic uncertainty on the background, the analysis indicates that the 2 (5)$\sigma$ limits on the $\chargino1$, $\neutralino2$ mass are 616 (517)\,GeV for 2.5\,$\iab$ luminosity at the FCC-eh, and 266 (227)\,GeV for 1\,$\iab$ luminosity at the LHeC, respectively. An illustration of the model assumptions in terms of sleptons and neutralino masses and the current constraints at the LHC is presented in Fig.~\ref{fig:susy-complementarity} (left). Results are illustrated in Fig.~\ref{fig:susy-complementarity} (right). The effects of varying $\Delta m$ are investigated: fixing $m_{\chargino1, \neutralino2}$ to be 400 GeV, it is found that at the FCC-eh the significance is maximal when $\Delta m$ is around 20\,GeV.

In the decoupled-slepton scenarios where only $\neutralino1$, $\chargino1$ and $\neutralino2$ are light and other SUSY particles are heavy and decoupled, 
the 2$\sigma$ limits obtained on the $\chargino1$, $\neutralino2$ mass are 230\,GeV for 2.5\,$\iab$ luminosity at the FCC-eh when neglecting the systematic uncertainty on the background. Large systematic uncertainties on the SM background processes can substantially affect the sensitivity, hence good control of experimental and theoretical sources of uncertainties is very important. 

Finally, it is also found that the possibility of having a negatively polarised electron beam ($P_{e} = 80$\,\%) could potentially extend the sensitivity to electroweakinos by up to 40\,\%. 

Overall, since the sensitivity to the electroweak SUSY sector depends on the mass hierarchy of $\chargino1$,  $\neutralino1$, $\neutralino2$ and sleptons, and given the difficulty to probe efficiently small $\Delta m$ regions at the current LHC and possibly at the HL-LHC, measurements at $ep$ colliders may prove to offer complementary or additional reaches, in particular for the compressed scenarios.

\subsection{Search for the SUSY Electroweak Sector: long-lived particles}
Studies on Higgsinos $(\chi)$ with masses ${\mathcal O}(100)$\,GeV are motivated by natural SUSY theories and help to avoid large fine-tuning on the Higgs boson mass. In these scenarios the low energy charginos $(\chi^+$)/neutralinos$(\chi^0)$ are all Higgsino-like and their masses are nearly degenerate, only slightly above the neutralino.

As mentioned above, a compressed spectrum with nearly degenerate masses results in a kinematic suppression of the heavier $\chi^+$ decays into $W^\pm \chi^0$, which has twofold consequences: it yields final states without hard leptons; it enhances the $\chi^+$ lifetime up to ${\mathcal O}(1)$\,mm.
At the LHC the absence of hard leptons with sizable transverse momentum makes this signature difficult to investigate. One possibility is to search for the tracks from $\chi^+$, which effectively disappear once it decays and are thus called \emph{disappearing tracks}.

The discovery prospects for prompt signatures of electroweakino decays in electron-proton collisions are presented in Ref.~\cite{Han:2018rkz}.
The light $\chi^+$ (and $\chi^0$) can be produced in pairs via in vector boson fusion of the charged or neutral currents.
A cut-based analysis of these processes at the LHeC, assuming prompt $\chi^+$ decays, yields $2\sigma$ discovery prospects for masses up to 120\,GeV.

Taking into account the finite lifetime of the charginos, two comments are in order: first, the lifetimes and boosts of the $\chi^+$ are in general too small to resolve a disappearing track; second, the soft final state is not a problem per se and can in principle be observed.

\begin{figure}
\centering
\includegraphics[width=0.65\textwidth]{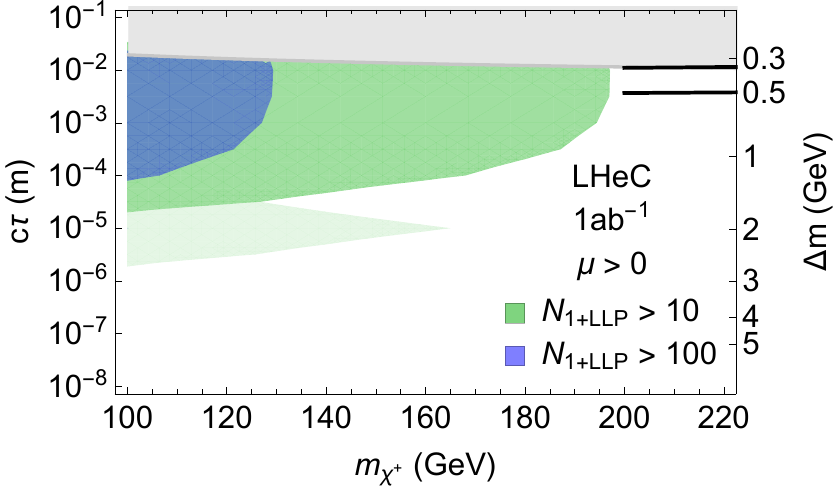}
\caption{Exclusion limits on Higgsino masses as a function of their lifetime from Ref.~\cite{Curtin:2017bxr}. Coloured regions denote where 10 or 100 events with at least one LLP decay are observed. Light shading indicates the uncertainty in the predicted number of events due to different hadronisation and LLP reconstruction assumptions. 
The black curves are the optimistic and pessimistic projected bounds from HL-LHC disappearing track searches.}
\label{fig:higgsino}
\end{figure}

Instead of searching for a disappearing track, the long lifetimes of the $\chi^+$ can be exploited via the measurement of the impact parameter of the soft hadronic final, as is discussed in Ref.~\cite{Curtin:2017bxr}. 
The crucial machine performance parameters are the tracking resolution, which is as good as ${\mathcal O}(10)\,\mu$m, and the absence of pile up, which allows to identify and measure a single soft pion's impact parameter.
In this way the LHeC can test $\chi$ with masses up to 200\,GeV. The corresponding sensitivity is shown in Fig.~\ref{fig:higgsino}, and the bounds on disappearing track searches at the HL-LHC are shown as black lines in the figure. 
By considering non-prompt decays of Higgsinos, the discovery prospects compared to the prompt analysis is thus significantly improved.
Further means of improving the prospects is an increased centre-of-mass energy, which enhances the production rate of the Higgsinos.

\subsection{R-parity violating signatures}
Supersymmetry typically evokes the so-called R-parity, which implies that each fundamental vertex contains an even number of sparticles and helps preventing rapid proton decays.
In general, R-parity need not be an exact symmetry of the theory, such that interactions can be present that allow for sparticles to decay into SM particles and include the possibility to violate lepton and/or baryon number.

R-partiy violating interactions are particularly interesting in electron-proton collisions, where single superpartners might be produced resonantly, and detected via the corresponding $2\to 2$ process.
This is discussed in Refs.~\cite{Kuday:2013cxa,Ren-You:2014pqa} for the case of the \emph{sbottom}, showing that a good level of precision could be achieved at LHeC compared with all the knowledge derived from indirect measurements. 

Single (anti-)top quark production associated with a lightest neutralino in the MSSM with R-parity breaking coupling is investigated in Ref.~\cite{Xiao-Peng:2013nha} for the LHeC.
The study, which includes calculations of QCD contributions at NLO, concludes that the available constraints would allow a notable production rate.

Certain SUSY scenarios might produce prompt signals of multiple soft jets, which generally resemble QCD backgrounds at the LHC and are thus notoriously difficult to test.
The largely QCD-free environment of electron-proton collisions allows to test this class of signatures. One example of this signal can come from gluinos, which are tested at the LHC via signatures that involve large amounts of missing energy.
If  the  gluino  has  an all-hadronic decay -- as in R-parity violating scenarios or Stealth SUSY models -- the current experimental searches have a gap in sensitivity for masses between about 50 to 70\,GeV \cite{Evans:2018scg}. Gluinos within this gap can be tested at the LHeC \cite{Curtin:2018xsc}, where a three sigma exclusion sensitivity was demonstrated with simple signal selection cuts.

\section{Feebly Interacting Particles }
New physics may interact with the SM via the so-called portal operators, including the vector, scalar, pseudoscalar, or neutrino portal.
In these scenarios, the SM is often extended by an entire sector of new physics, comprising new forces and several  particle species, which may be connected to the big open questions of Dark Matter or the origin of neutrino mass.

These hypothetical new sectors derive their typically very feeble interaction strength with the known particles from mass mixing with a SM particle that shares their quantum numbers.
Some examples are being discussed below.

\subsection{Searches for heavy neutrinos}
The observation of neutrino oscillations requires physics beyond the SM that gives rise to the light neutrino masses.
One well-motivated class of models for this purpose is the so-called symmetry protected type I seesaw scenario, which features heavy neutrinos with signatures that are in principle observable at colliders, cf.\ Ref.~\cite{Antusch:2015mia} and references therein.
A comprehensive overview over collider searches for the heavy and mostly sterile neutrinos can be found in Ref.~\cite{Antusch:2016ejd}, where the promising signatures for such searches at electron-proton colliders have been identified.

\begin{figure}
    \begin{minipage}{0.4\textwidth}
    \includegraphics[width=\textwidth]{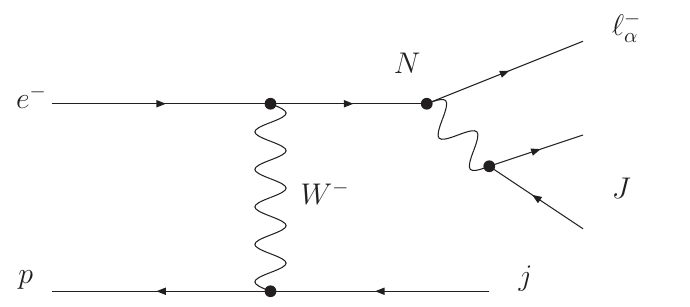}
    \end{minipage}
     \begin{minipage}{0.58\textwidth}
 \includegraphics[width=\textwidth]{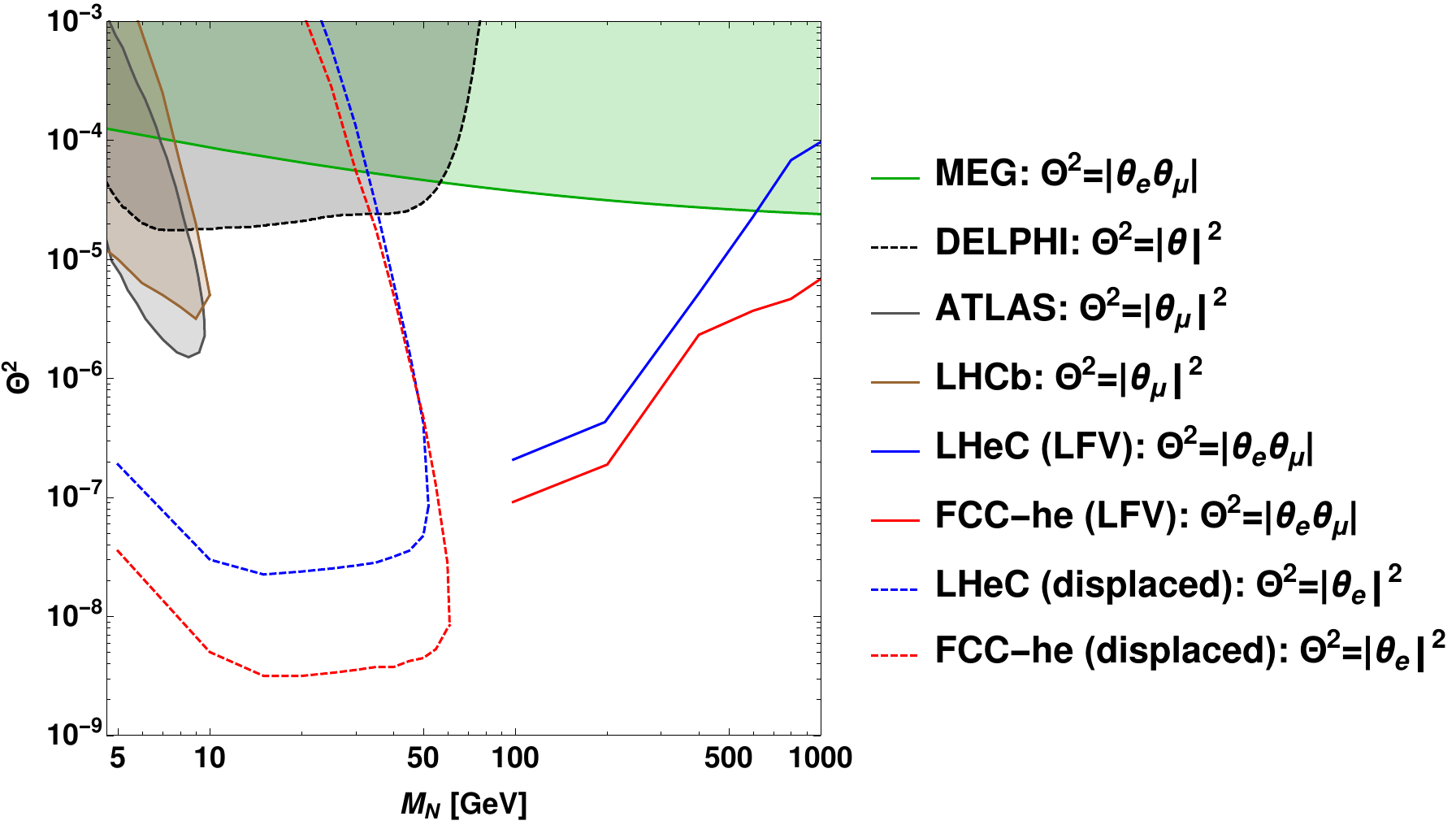}
    \end{minipage}

    \caption{Left: Dominant tree-level production mechanism for sterile neutrinos at the LHeC. The sterile neutrino decay via the charged current gives rise to the so-called lepton flavor violating lepton-trijet signature.
    Right: Sensitivity of the LFV lepton-trijet searches (at 95\,\% C.L.) and the displaced vertex searches (at 95\,\% C.L.) from Ref.~\cite{Antusch:2019eiz} compared to the current exclusion limits from ATLAS \cite{Aad:2019kiz}, LHCb \cite{Antusch:2017hhu}, LEP \cite{Abreu:1996pa}, and MEG \cite{Adam:2013mnn}.}
    \label{fig:bsm_sterileN}
\end{figure}

In electron-proton collisions heavy neutrinos can be produced via the charged current (see the left panel of Fig.~\ref{fig:bsm_sterileN}). The heavy neutrino production cross section is dependent on the active-sterile neutrino mixing with the electron flavour called $|\theta_e|^2$.
The most promising searches at the LHeC are given by processes with lepton flavour violating final states and displaced vertices, the prospects of which are evaluated in Ref.~\cite{Antusch:2019eiz} and are shown in the right panel of Fig.~\ref{fig:bsm_sterileN}.
It is remarkable, that the prospects to detect heavy neutrinos with masses above about 100~GeV are much better in electron-proton collisions compared to proton-proton or electron-positron, due to the much smaller reducible backgrounds.

The prospects of heavy neutrino detection can be further enhanced with jet substructure techniques when the $W$ boson in the decay $N \to e W, ~ W \to jj$ is highly boosted. Ref.~\cite{Das:2018usr} shows that these techniques can help to distinguish the heavy neutrino signal from the few SM backgrounds. A considerable improvement in the bounds of $|V_{eN}|^2$ over present limits from LHC, $0v2\beta$ experiments and from electroweak precision data is obtained with 1\,ab$^{-1}$ of integrated luminosity at the LHeC.

An alternative approach is employed in Ref.~\cite{Duarte:2014zea} where the dominant sterile neutrino interactions with the SM are taken to be higher dimension effective operators (parameterizing a wide variety of UV-complete new physics models) while contributions from neutrino mixing is neglected.
The study shows prospects of Majorana neutrino detection for masses lower than 700 and 1300\,GeV can be discovered at the LHeC with $E_e = 50$ and  $150$\,GeV, respectively, for $E_p = 7$\,TeV. Recently the influence of vector and scalar operators on the angular distribution of the final anti-lepton was investigated. The forward-backward asymmetry is studied in Ref.~\cite{Duarte:2018xst}, wherein, in particular, the feasibility of initial electron polarisation as a discriminator between different effective operators is studied.

Prospects of testing left-right symmetric models, featuring additional charged and neutral gauge bosons and heavy neutrinos, were studied in the context of electron-proton collisions in Refs.~\cite{Mondal:2015zba,Lindner:2016lxq}.
The authors show that the production of heavy right-handed neutrinos of mass ${\mathcal O}(10^2$-$10^3)$\,GeV at the LHeC, with a lepton number violating final state, can yield information on the parity breaking scale in left-right symmetric theories. 
Heavy neutrinos of sub-TeV mass in inverse see-saw model with Yukawa coupling of ${\mathcal O}(0.1)$ are investigated for the LHeC in Ref.~\cite{Mondal:2016kof}.

\subsection{Fermion triplets in type III seesaw}
Another technically natural way of generating the light neutrino masses is the so-called Type III seesaw mechanism, which extends the SM with a fermion $SU(2)$ triplet.
In minimal versions of these models the neutral and charged triplet fermions have almost degenerate masses around the TeV scale.

\begin{figure}
\centering
\includegraphics[width=0.38\textwidth]{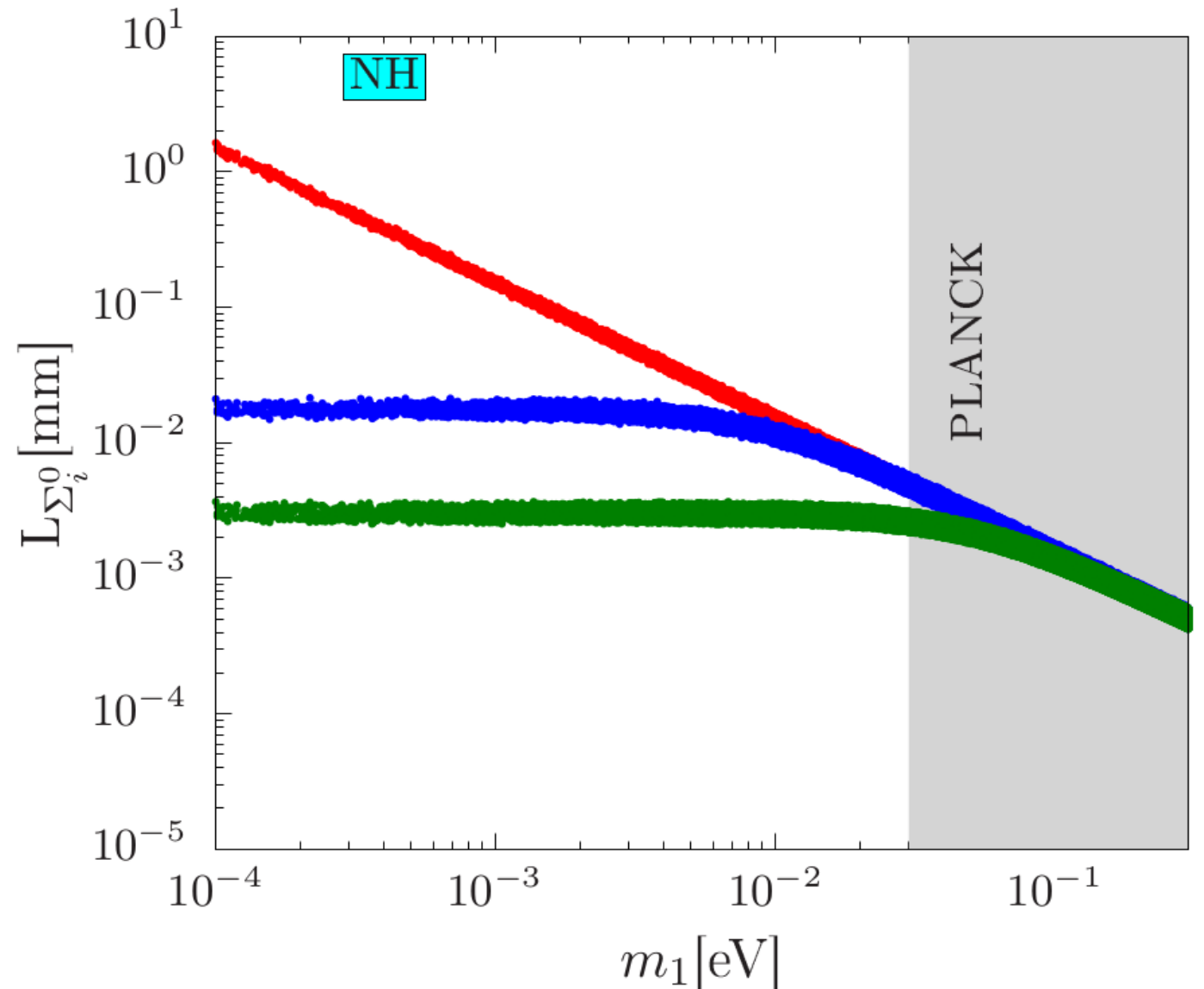}
\includegraphics[width=0.38\textwidth]{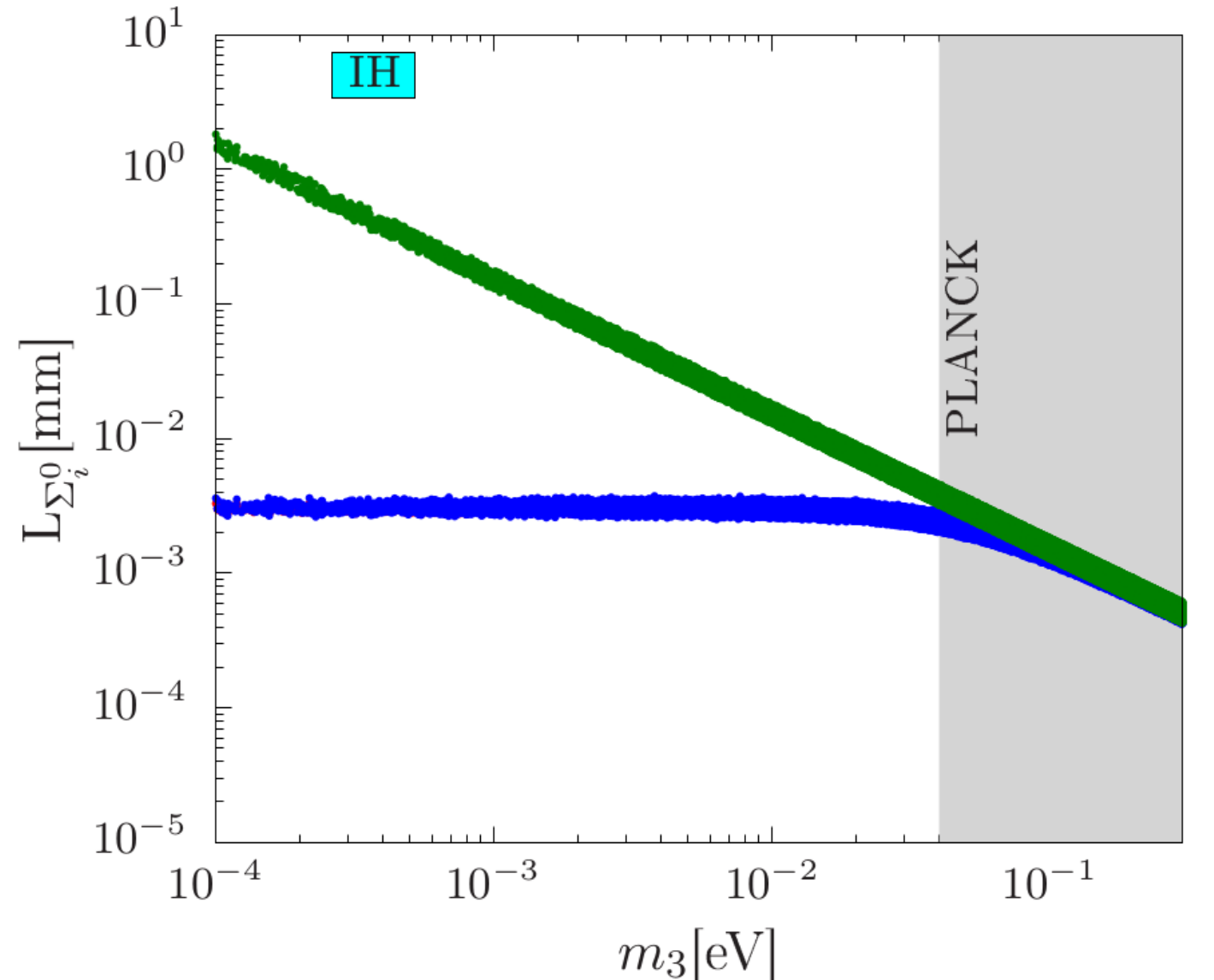}\\
\includegraphics[width=0.38\textwidth]{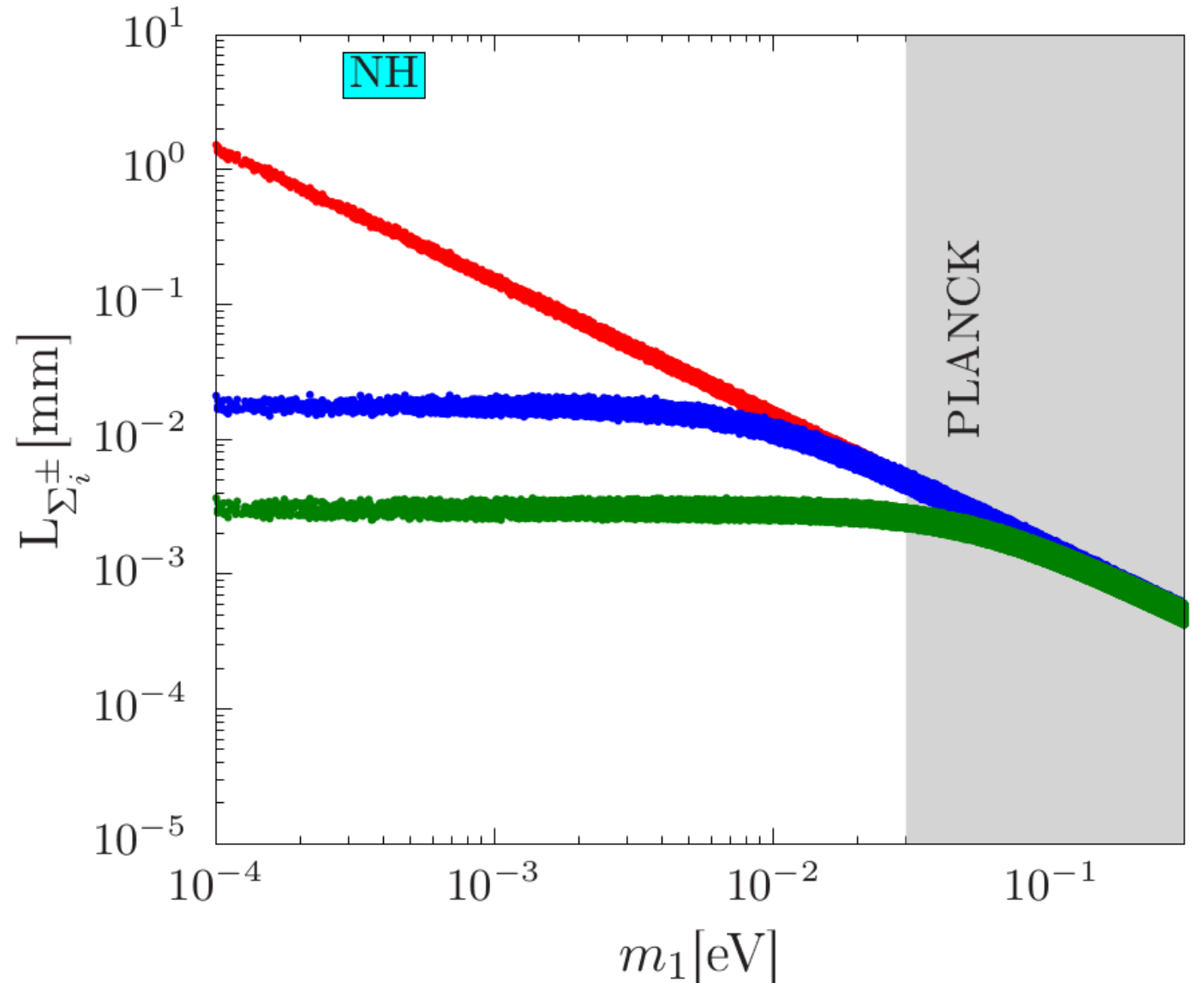}
\includegraphics[width=0.38\textwidth]{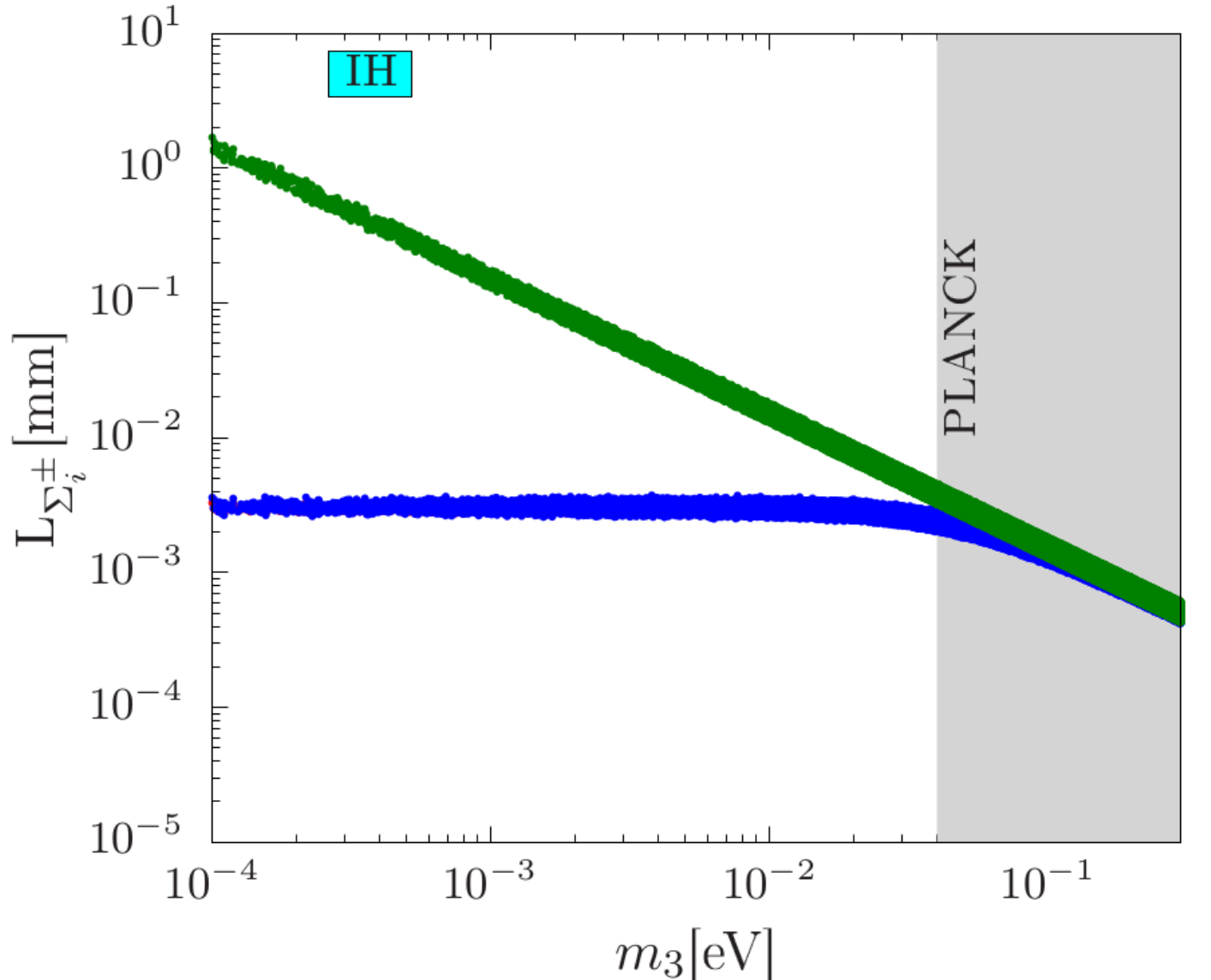}
\caption{Proper decay length of $\Sigma^0_i (\Sigma^{\pm}_i)$ with respect to the lightest light neutrino mass in the upper (lower) panel for $1$ TeV triplet.
The Normal (Inverted) hierarchy case is shown in the left (right) panel. The first generation triplet is represented by the red band, the second generation is represented by the blue band and the third generation is represented by the green band respectively. The shaded region is excluded by the PLANCK data \cite{Aghanim:2018eyx}.}
\label{fig1}
\end{figure}

In the three generation triplet extension of the type-III seesaw, the role of mixings between active neutrinos and neutral triplet fermions has been investigated in Ref.~\cite{Das:2020uer}. Depending upon the choices of Dirac Yukawa coupling, the mixing angles can take many possible values, from very small to large. With very small mixings, decay length of the triplet fermion can be very large. It can show a displaced decay inside the detector or outside the detector of the high energy colliders. The proper decay length as a function of the lightest light neutrino mass $m_1 (m_3)$ for the Normal (Inverted) hierarchy case are shown in Fig.~\ref{fig1}.

\begin{figure}
  \centering
    \begin{minipage}{0.3\textwidth}
    \includegraphics[width=\textwidth]{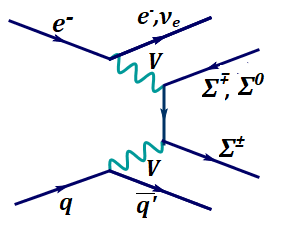}
    \hspace{0.05\textwidth}
    \end{minipage}
    \begin{minipage}{0.55\textwidth}
    \includegraphics[width=\textwidth]{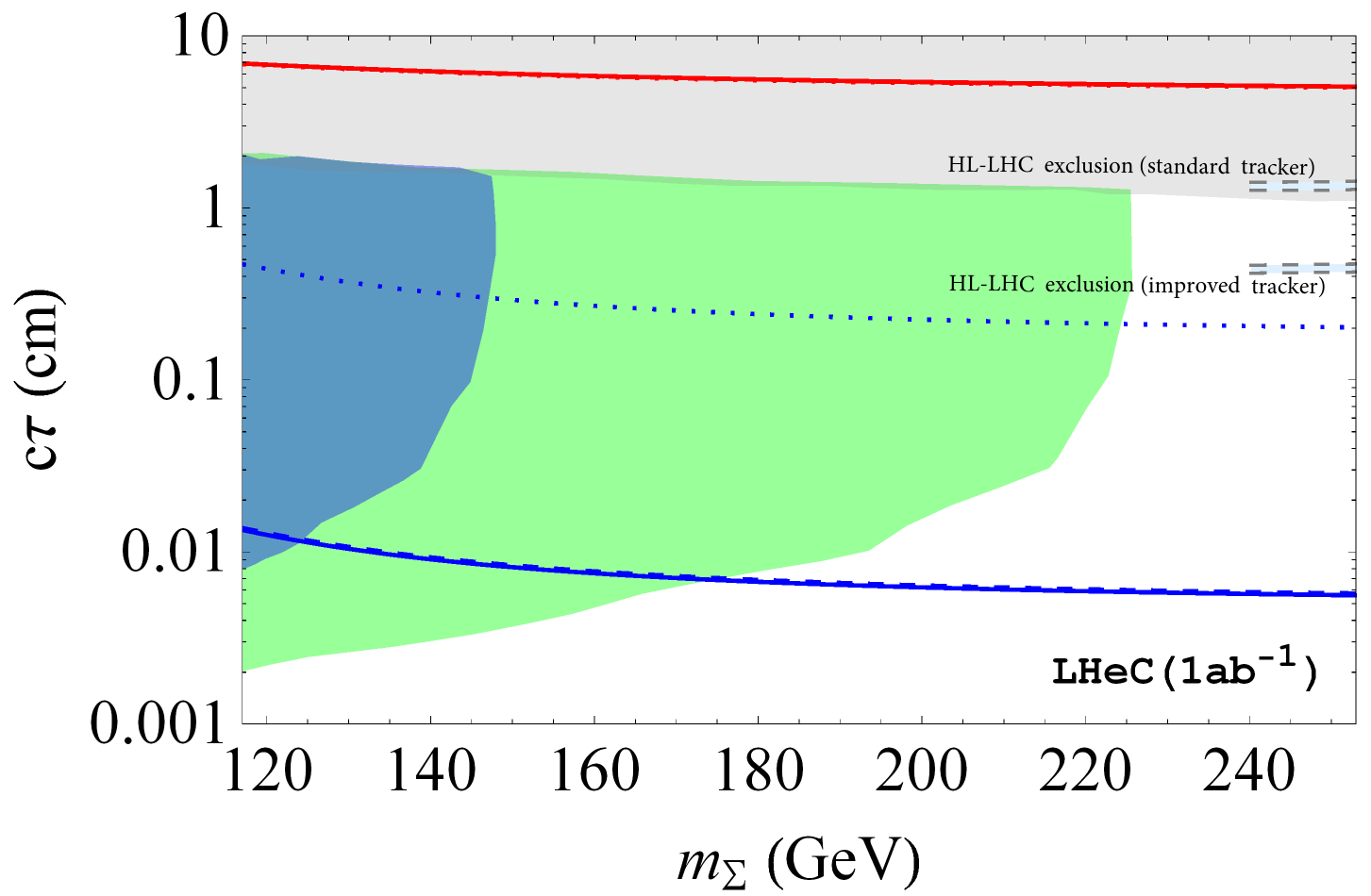}
    \end{minipage}
    \caption{Left: Dominant production diagram of triplet fermion pairs via their gauge interactions.
    Right: Prospects of displaced vertex searches from charged fermion triplet $\Sigma^\pm$. The blue and green shaded regions denote the expected observability of 10 (100) events, dashed lines denote HL-LHC exclusion sensitivity, and the red line is connected to the light neutrino properties. For details, see text and Ref.~\cite{Jana:2019tdm}.}
    \label{fig:bsm_hnl1}
\end{figure}

The prospects of probing this mechanism via searches for the new fermions are evaluated in Ref.~\cite{Jana:2019tdm}, wherein signatures from long-lived particles at various experiments were considered. The  triplet  fermions are primarily produced through their gauge interactions, as shown in the left panel of Fig.~\ref{fig:bsm_hnl1}, and can be observed via displaced vertices and disappearing track searches for masses of a few hundred GeV.
The authors find that the LHeC can observe displaced vertices from the decays of the charged fermion triplet components via the soft pion impact parameters for triplet masses up to about 220\,GeV and has a complementary sensitivity to the light neutrino mass scale, which governs the lifetime of the neutral fermion, compared the LHC and MATHUSLA.
The final results from Ref.~\cite{Jana:2019tdm} for the LHeC are shown in the right panel of Fig.~\ref{fig:bsm_hnl1}.

If the mixing becomes sufficiently large, and/or if the masses are ${\cal O}(100)$ GeV, the triplet fermions decay promptly. Also in this case, the heavy triplets can show a variety of interesting collider signatures including fat jets. The latter have been studied for FCC-he in Ref.~\cite{Das:2020gnt}.

\subsection{Dark photons }
Minimal extensions of the SM often involve additional gauge factors. In particular the $U(1)_X$ extensions are interesting, because they are often connected to a dark charge that can be associated with the dark matter.

An SM-extending $U(1)_X$ predicts an additional gauge boson that naturally mixes with the $U(1)_Y$ factor of the SM kinetically~\cite{Holdom:1985ag}. This kinetic mixing lets the SM photon couple to fermions that carry the dark charge $X$, and the other gauge boson to the electric charge. Both interactions are suppressed by the mixing parameter $\epsilon$.
In most models the additional gauge boson also receives a mass, possibly from spontaneous breaking of the $U(1)_X$, and the corresponding mass eigenstate is called a dark photon. 
Dark photons typically have masses around the GeV scale and their interactions are QED-like, scaled with the small mixing parameter $\epsilon$.
It can decay to pairs of leptons, hadrons, or quarks, which can give rise to a displaced vertex signal due to its long lifetime. 

\begin{figure}
		\centering
		\includegraphics[width=0.8\textwidth]{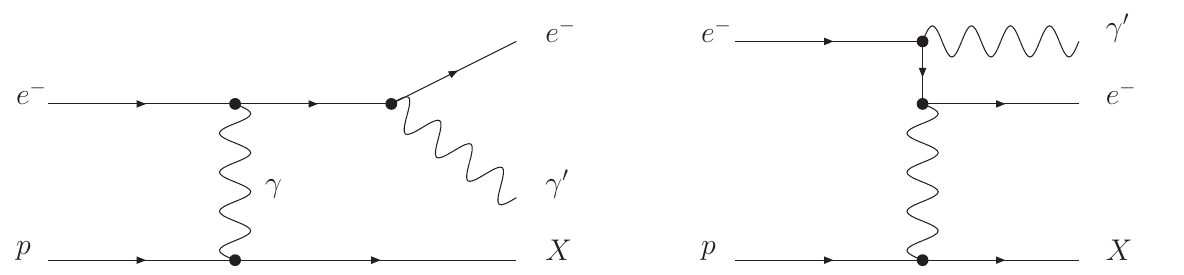}
		\caption{Feynman diagrams for the dark photon production processes in electron-proton collisions. Here $X$ denotes the final state hadrons after the scattering process.}
		\label{fig:production}
	\end{figure}

The prospects for the dark photon searches via their displaced decays in $ep$ collisions are presented in Ref.~\cite{DOnofrio:2019dcp}. The dark photon production process targeted in this search is depicted in Fig.\ \ref{fig:production}.
The signal is given by the process $e^- p \to e^- X \gamma'$, where $X$ denotes the final state hadrons, and the dark photon $\gamma'$ decays into two charged fermions or mesons. 

The most relevant performance characteristics of the LHeC are the very good tracking resolution and the very low level of background, which allow the detection of a secondary vertex with a displacement of ${\mathcal O}(0.1)$\,mm.

\begin{figure}
\centering
		\includegraphics[width=0.55\textheight]{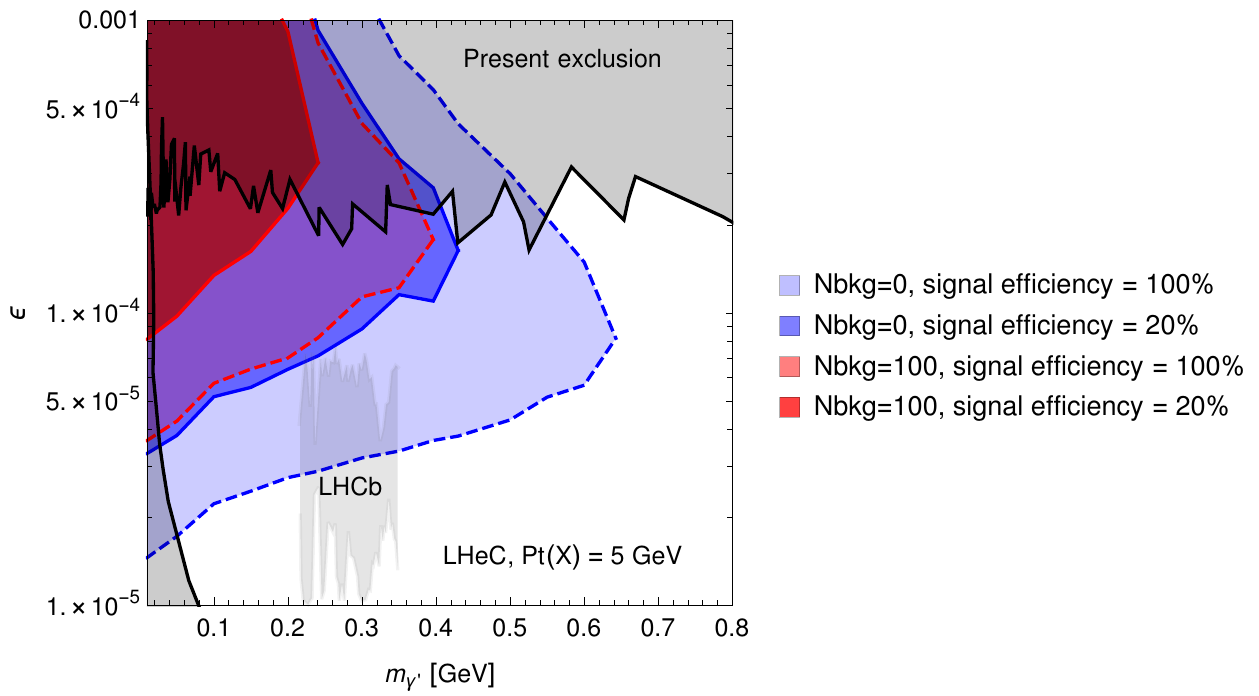}
		\caption{Projected sensitivity of dark photon searches at the LHeC via displaced dark photon decays from Ref.~\cite{DOnofrio:2019dcp}. The sensitivity contour lines are at the 90\,\% confidence level after a transverse momentum cut on the final state hadrons of 5\,GeV. The blue and red areas denote the assumption of zero and 100 background events, respectively, the solid and dashed lines correspond to a reconstruction efficiency of 100\,\% and 20\,\%, respectively. See Ref.~\cite{DOnofrio:2019dcp} for details.}
		\label{fig:dark_photon}
\end{figure}

The resulting sensitivity contours in the mass-mixing parameter space are shown in Fig.~\ref{fig:dark_photon}, where the different colours correspond to different assumptions on the irreducible background and the solid and dashed lines consider different signal reconstruction efficiencies. Also shown for comparison are existing exclusion limits from different experiments, and the region that is currently investigated by the LHCb collaboration \cite{Aaij:2017rft}.

The domain in parameter space tested in electron-proton collisions is complementary to other present and planned experiments.
In particular for masses below the di-muon threshold, searches at the LHC are practically impossible.
It is remarkable that dark photons in this mass range can be part of a dark sector that explains the observed Dark Matter in the Universe via a freeze-in mechanism, cf. e.g.\ Ref.~\cite{Heeba:2019jho}.

\subsection{Axion-like particles}
The axion is the Goldstone boson related to a global $U(1)$ symmetry, which is spontaneously broken at the so-called Peccei-Quinn scale, assumed to be around the GUT scale.
Its mass, being inversely proportional to the Peccei-Quinn scale, is therefore usually in the sub-eV regime and the axion provides a dynamical solution to the strong CP problem of the standard model.
Axions are a very attractive candidate for \emph{cold} dark matter, despite their tiny mass.

Axion-like particles (ALP) are motivated by the original idea of the QCD axion and similarly, they are good dark matter candidates.
ALPs are pseudoscalar particles that are usually assumed to be relatively light (i.e.\ with masses around and below one GeV) and couple to the QCD field strength.
In addition, they may have a number of further interactions, for instance they can interact with the other fields of the SM and also mix with the pion.
Particularly interesting is the possibility to produce ALPs via vector boson fusion processes.

\begin{figure}
    \begin{minipage}{0.38\textwidth}
    \includegraphics[width=\textwidth]{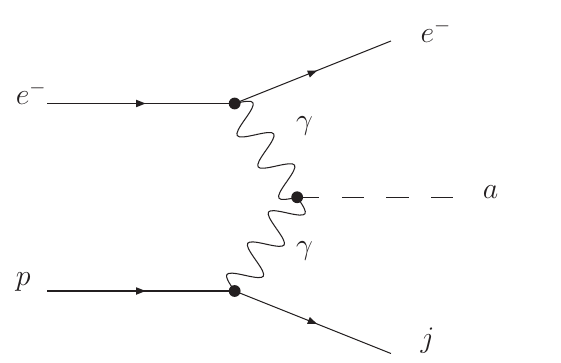}
    \end{minipage}
    \begin{minipage}{0.6\textwidth}
    \includegraphics[width=\textwidth]{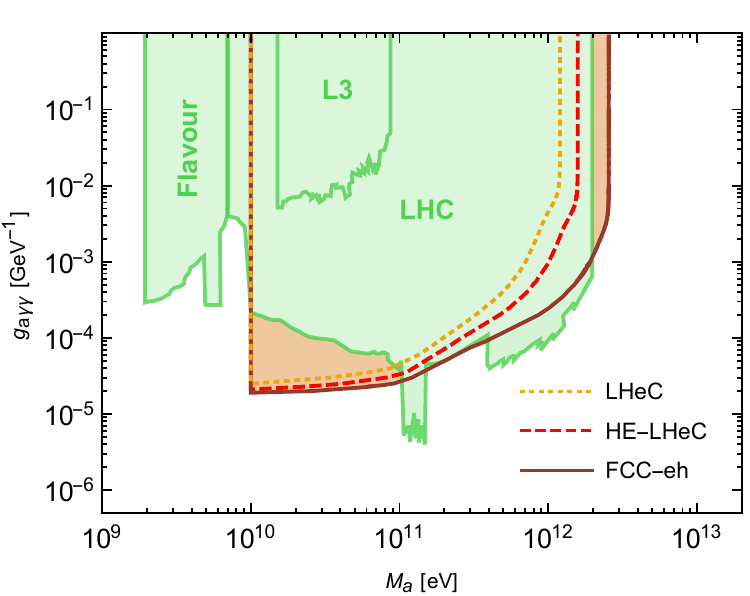}
    \end{minipage}
    \caption{Left: Production of axion-like particles (ALPs) via photon fusion.
    Right: Projected sensitivity of the LHeC to ALPs coupling with photons at 95\% CL. The existing exclusion limits are shown with the green regions. See Ref.~\cite{Yue:2019gbh} for details.}
    \label{fig:bsm_alp}
\end{figure}

A recent study~\cite{Yue:2019gbh} has evaluated the prospects of detecting ALPs at the LHeC via the process $e^- \gamma \to e^- a$, as shown in the left panel of Fig.~\ref{fig:bsm_alp}, in a model independent fashion.
The investigated signature is the decay $a\to \gamma \gamma$, which allows to test the effective ALP-photon coupling for ALPs with masses in the range of 10\,GeV $< m_a <$ 3\,TeV. 
It was found that sensitivities can improve current LHC bounds considerably, especially for ALP masses below 100\,GeV, as shown in the right panel of Fig.~\ref{fig:bsm_alp}. The authors state that ALP searches at $ep$ colliders might become an important handle on this class of new physics scenarios~\cite{Yue:2019gbh}.

\section{Anomalous Gauge Couplings }
New physics beyond the SM can modify SM interactions, for instance at the loop level.
Such contributions could either modify the interaction strength of SM particles or introduce additional interactions that are not present in the SM, like flavour changing neutral couplings. 

Searches for anomalous couplings of top quarks are summarised in Section~\ref{sec:top}. They are parametrised via an effective Lagrangian and are studied by analysing specific processes. For example, anomalous $Wtb$ couplings are studied in $e^- p \to \nu_e {\bar t}$, and anomalous $t\bar{t}\gamma $ and $t\bar{t}Z$ couplings are studied in top quark pair production. In addition FCNC $tu\gamma$ and $tuZ$ couplings are analysed in NC DIS single top quark production, and FCNC $tHu$ couplings are investigated in CC DIS single top quark production. Limits on the corresponding FCNC branching ratios are discussed in Section~\ref{SecFCNC} and summarised and compared to different colliders in Fig.~\ref{fcncProj}. 

%
Triple gauge boson couplings (TGC) $W^+W^-V$, $V = \gamma, Z$ are precisely defined in the SM and any significant deviation from the predicted values could indicate new physics.
Present constraints on anomalous triple vector boson couplings are dominated by LEP (but they are not free of assumptions) and the $WWZ$ and $WW\gamma$ vertices can be tested at LHeC in great detail.

The search for anomalous $WW\gamma$ and $WWZ$ couplings with polarised electron beam were studied in Ref.~\cite{Cakir:2014swa} via the processes $ep \to \nu q\gamma X$ and $ep \to \nu q Z X$. 
It was found that the LHeC sensitivity with $E_e=60$\,GeV and $L=100$/fb is comparable with existing experimental limits from lepton and hadron colliders, and that the sensitivity to anomalous $Z$ couplings might be better, reaching ($\Delta\kappa_{\gamma, Z}, \lambda_{\gamma, Z}$) as small as ${\mathcal O} (10^{-1}, 10^{-2})$.
In general, beam polarisation and larger electron beam energies improve the sensitivity, and the LHeC was found to give complementary information on the anomalous couplings
compared to the LHC.

The prospects of testing anomalous triple gauge couplings are also investigated in Ref.~\cite{Biswal:2014oaa}. Therein the authors study the kinematics of an isolated hard photon and a single jet with a substantial amount of missing transverse momentum. They show that the LHeC is sensitive to anomalous triple gauge couplings via the azimuthal angle differences in the considered final state. It is pointed out that, in such an analysis, it is possible to probe the $WW\gamma$ vertex separately with no contamination from possible BSM contributions to the $WWZ$ coupling. The estimations consider $E_e=100,\,140,\,200$\,GeV and it is claimed that, while higher energies yield better sensitivities, the differences are not very large. For an integrated luminosity of 200 fb$^{-1}$ and $E_e=140$\,GeV the exclusion power of the LHeC is superior to all existing bounds, including those from LEP.

The process $e^- p \to e^- \mu^+ \nu j$ is investigated in Ref.~\cite{Li:2017kfk}.
The analysis is carried out at the parton level and includes the cross section measurement and a shape analysis of angular variables, in particular of the distribution of the azimuthal angle between the final state forward electron and jet. 
It is shown that the full reconstruction of leptonic $W$ decay can be used for $W$ polarization which is another probe of anomalous triple gauge couplings. 
The results show that the LHeC could reach a sensitivity to $\lambda_\gamma$ and $\Delta k_\gamma$ as small as ${\mathcal O}(10^{-3})$ for $L=2-3$/ab.

\subsection{Radiation Amplitude Zero}
\begin{figure}
  \centering
  \includegraphics[width=0.65\textwidth]{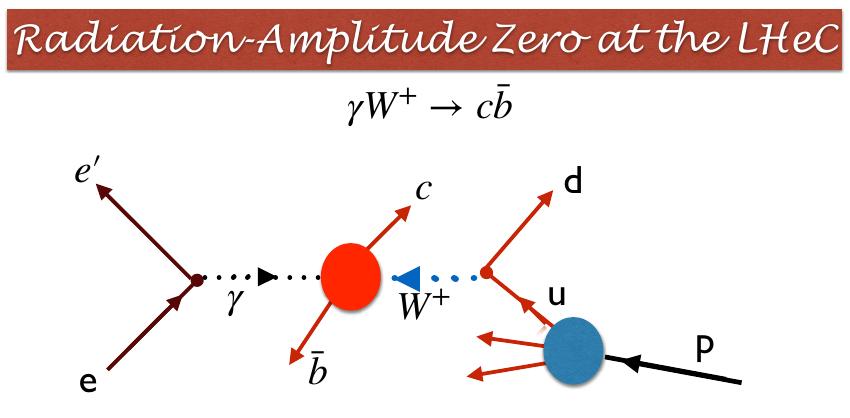} 
  \vskip 0.3cm
  \caption{The radiation amplitude zero of the Standard Model in $\gamma W^+ \to c \bar b$ and $\gamma u \to W^+ d$.
    A prediction for the angular distribution
  $\tfrac{d\sigma}{ d cos(\theta_{CM })} (\gamma u \to W^+ d)$ 
    is provided in Ref.~\cite{Samuel:1985gv}.}
  \label{RAZ}   
\end{figure}

The LHeC is ideal for testing a novel feature of the Standard Model: 
the \emph{radiation amplitude zero}~\cite{Mikaelian:1979nr,Brodsky:1982sh,Brown:1982xx,Samuel:1985gv}  of the amplitude $\gamma W^- \to c \bar b$  and related amplitudes, see Fig.~\ref{RAZ}.  
The Born amplitude is predicted to vanish and change sign at $\cos\theta_{CM} = \tfrac{e_{\bar b}}{ e_W^-}=-1/3$. 
This LHeC measurement tests  the compositeness of the $W$ boson and
its zero anomalous magnetic moment at leading order, where one has $g_W=2$, $\kappa_W=1$, as well as $g_q=2$ for the quarks.
More generally, one can also test the radiation amplitude zero for the
top quark from measurements of the process $\gamma b \to W^- t$.

\section{Theories with heavy resonances and contact interaction}
%
In many other BSM scenarios, new physics will manifest itself by the presence of new resonances. Although the high centre-of-mass energy of $pp$ colliders allow for a better reach in most of these scenarios, the LHeC and FCC-eh, thanks to the clean collision environment and the virtual absence of pileup, can complement the LHC in the search for these new phenomena. 
Deviations from Standard Model predictions could signal new physics even if it is at an energy scale beyond the centre-of-mass energy of the collider. In this case, the effective four-fermion contact interaction could be explained by the exchange of a virtual heavy particle, such as a leptoquark, a heavy boson or elementary constituents of quarks and leptons in composite models.  The effective contact interaction scale then represents the typical mass scale of the new particles.
Relevant studies on various topics including scalar and vector leptoquarks and excited leptons, are collected in this section.

\subsection{Leptoquarks}
\begin{figure}
    \centering
    \includegraphics[width=0.65\textwidth]{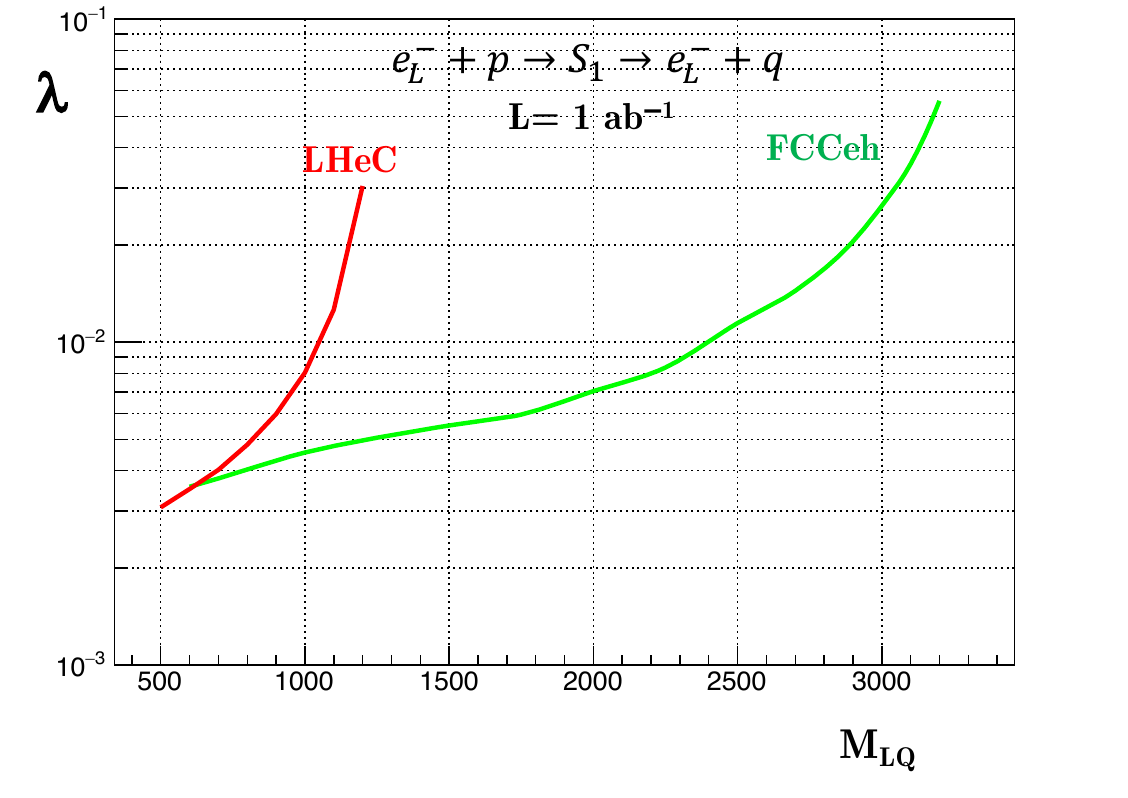}
    \caption{Estimated 2$\sigma$ significance for the coupling $\lambda$ at LHeC and FCC-eh for the scalar lepto-quark $S_1$ as a function of its mass, assuming 1\,ab$^{-1}$ luminosity  and no systematic uncertainty. }
    \label{fig:LQ}
\end{figure}
In recent years the experiments that study heavy flavoured mesons have revealed intriguing  hints for new physics in semileptonic decays of $B$ mesons. A violation of lepton flavour universality at the level of 3 to 5$\sigma$ is apparent in both the charged current and neutral current mediated processes~\cite{Amhis:2019ckw}. 
In this context BSM theories involving leptoquarks (LQs) have gained renewed interest as they can give rise to lepton universality violating decays of heavy mesons at tree level, provided they couple to the second and third generation of quarks. 
Leptoquarks first appeared in Ref.~\cite{Pati:1974yy} in Pati and Salam’s $SU(4)$ model, where lepton number was considered to be the fourth colour. They also appear in Grand Unified theories, extended technicolor models and compositeness models. The nomenclature and classification are based on their transformation properties under the SM gauge groups~\cite{Buchmuller:1986zs,Dorsner:2016wpm}.

In $ep$ collisions LQs can be produced in an s-channel resonance via their coupling to the first generation of quarks, the signature being a peak in the invariant mass of the outgoing $\ell q$ system.
Contrary to what is achievable in the LHC environment, it has been shown that at the LHeC many properties of the LQs can be measured with high precision~\cite{AbelleiraFernandez:2012cc}. 

The search for LQs at the LHC is essentially insensitive to the coupling LQ-e-q, characterized by the parameter $\lambda$, since the dominant process is pair production via the strong interaction. Recent searches have therefore been able to exclude LQs of the first generation of mass up to 1.4 TeV, assuming a branching ratio to charged leptons = 1.0. For other generations, the bounds are $\sim 1$\,TeV. (for the latest results, see, for example Ref.~\cite{ATLAS_bsm_summary_plots,CMS_bsm_summary_plots}). 
Under the assumption that the LQ has O(0.1) branching ratios to a number of tested final states, there remains some parameter space where the LHeC can make a significant contribution in the search for LQs.

For LQs with masses below the centre-of-mass energy of the collider, suitable searches promise a sensitivity to $\lambda$ as small as ${\mathcal O}(10^{-3})$. 
As shown in~\cite{Zhang:2018fkk}, production of the first generation scalar leptoquarks at LHeC can have a much higher cross section than at the LHC. The authors also show that a sensitivity to the Yukawa coupling, for the LQs called  $R_2^{5/3}\sim$({\textbf 3},{\textbf 2},7/6) and $\tilde R_2^{2/3}\sim$({\textbf 3},{\textbf 2},1/6), better than the electromagnetic strength ($\sim 0.3)$ of 5$\sigma$ can be reached up to a mass of 1.2\,TeV. 

For the $S_1$ scalar leptoquark ({\boldmath $\bar 3$},{\textbf 1},1/3), an estimate of the sensitivity of the LHeC and the FCC-eh as a function of the LQ mass and LL Yukawa coupling is shown in Fig.~\ref{fig:LQ},  assuming 1\,ab$^{-1}$ of integrated luminosity. Here, the signal was generated at leading order using MadGraph with the model files from Ref.~\cite{Dorsner:2018ynv}, with hadronisation performed by Herwig7~\cite{Bahr:2008pv,Bellm:2015jjp} and detector simulation with Delphes~\cite{Ovyn:2009tx}. The SM background $e^- p \to e^- j$ was also generated at leading order. A simple set of cuts on the $p_T$ of the leading electron and jet and a window on the invariant mass of the $e$-jet system was applied.


The $\tilde R_2^{2/3}$ scalar LQ
allows for coupling to right-handed neutrinos, providing interesting search channels. Its signatures at $ep$ colliders have been investigated recently~\cite{Mandal:2018qpg,Padhan:2019dcp}. In the lepton + jet final state, it is found that LHeC can probe up to 1.2\,TeV at $3\sigma$ significance with an $e^-$ beam, and at $5\sigma$ discovery with an $e^+$ beam and 1\,ab$^{-1}$ of integrated luminosity. At FCC-eh, a $5\sigma$ discovery can be reached with an $e^-$ beam up to $\sim 2.3$\,TeV and 1\,ab$^{-1}$ of integrated luminosity.


\subsection{Z' mediated charged lepton flavour violation}
\begin{figure}
\centering
\includegraphics[width=0.33\textwidth]{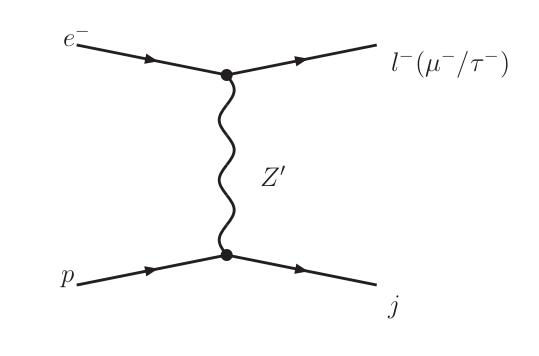}
\includegraphics[width=0.63\textwidth]{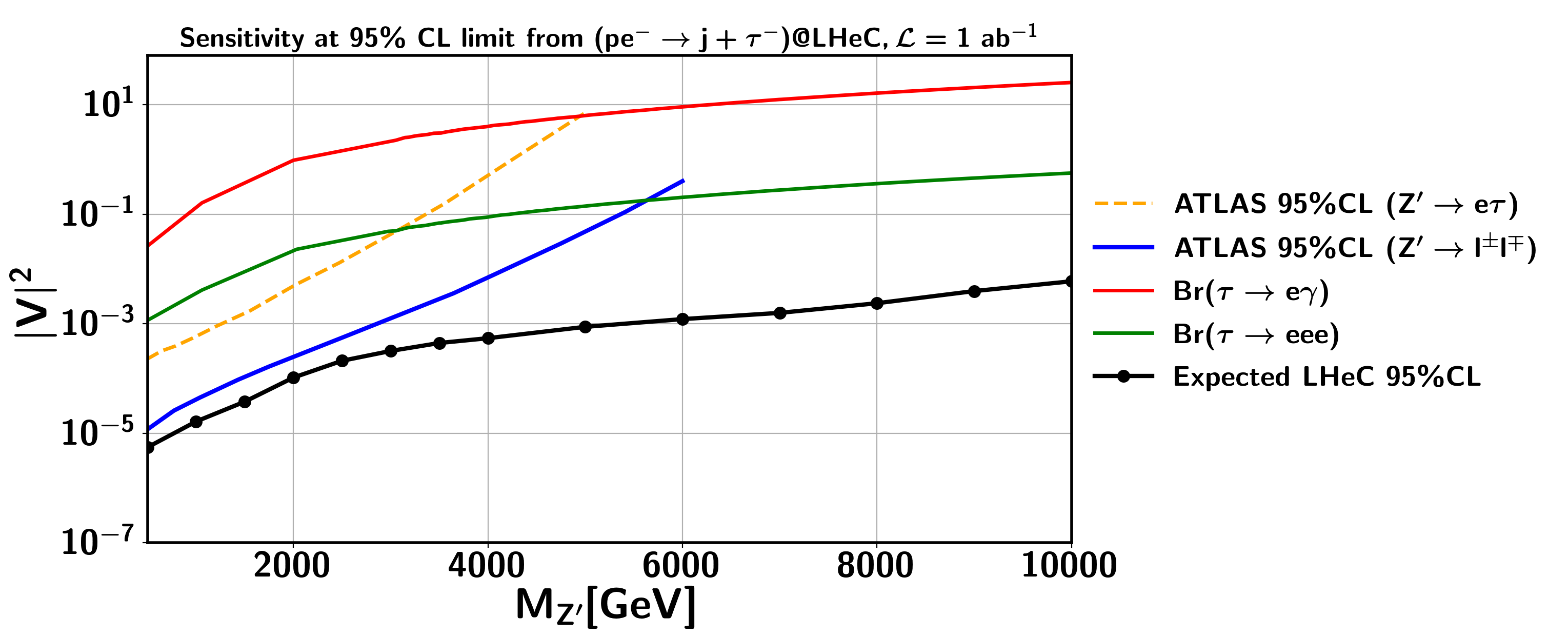}
\caption{Left: Feynman diagram for the $e$-$\tau$ (and $e$-$\mu$) conversion processes $p  e^- \to \tau^- + j$ (and $p  e^- \to \mu^- + j$) mediated by a $Z'$ with flavour-violating couplings to charged leptons at the LHeC~\cite{Antusch:2020fyz}.
Right: Limits on the coupling parameter $|V|^2$ for signal hypothesis compared with the existing limits from experimental constraints on the relevant flavour conserving and flavour violating processes. The black line is the LHeC sensitivity for the process $pe^-\to\tau j$. For the other limits, see text.}
\label{fig:LFV}
\end{figure}

Charged lepton flavour violating signatures are well tested involving electrons and muons, but less so when they involve tau leptons. Interestingly, in many extensions of the SM lepton flavour is much more strongly violated in the tau sector whilst weaker experimental constraints at low energy exist. In Ref.~\cite{Antusch:2020fyz} the $Z'$ mediated $e-\tau$ (and $e-\mu$) conversion processes are studied at the LHeC, considering the lepton flavour violating processes $pe^- \to \tau^- j$ (and $pe^- \to \mu^- j$). 

For this LHeC study, a 60-GeV electron beam with up to 80\% polarization is considered, to achieve a centre-of-mass energy close to 1.3 TeV with a total of 1 ab$^{-1}$ integrated luminosity.  Several backgrounds featuring tau leptons are considered, a parameterised reconstruction efficiency and mis-identification for tau jets is included in the analysis. To distinguish between the signal events and all relevant backgrounds, 31 kinematic variables (at the reconstruction level after the detector simulation) are used as input to a tool for Multi-Variate Analysis (TMVA). A BDT algorithm is used to separate the signal events from the background events. Systematic uncertainties are evaluated and are found to be around $2\%$. 

Asuming equal couplings $|V^{ij}_R|=|V^{ij}_L| \equiv |V|$ of the $Z'$ to quark-quark or lepton-lepton flavours $i,j$,  
the LHeC is found to be sensitive to $Z'$ masses up to $\mathcal{O}(10)$ TeV, as depicted in Fig.~\ref{fig:LFV} by the black line.
Included in the Figure are also the existing limits from ATLAS searches for $Z^\prime$ decays into $e  \tau$~\cite{Aaboud:2018jff} and the search for same flavour final states~\cite{Aad:2019fac}. The experimental limits based on the branching ratio $\mathrm{BR}(\tau \to e \gamma)$~\cite{Aubert:2009ag} and $\mathrm{BR}(\tau \to eee)$~\cite{Hayasaka:2010np} are also reported.

Overall, lepton flavour violation in the tau sector can be tested extremely well at the LHeC, surpassing the sensitivity of the LHC and low energy experiments in the whole considered mass range by more than two orders of magnitude.
This is particularly interesting for very heavy Z' that are not accessible for direct production, where the LHeC provides an exciting new discovery channel for this kind of lepton flavour violating processes.

\subsection{Vector-like quarks}
In composite Higgs models, new vector-like quarks are introduced. The third generation is favored, in particular the top-partner ($T$) with charge 2/3. 
The prospects of detecting $T$ at the LHeC are discussed in Ref.~\cite{Zhang:2017nsn}.
For this search a simplified model is considered where $T$ is produced from positron proton scattering via intergenerational mixing and decays  as $T\to t Z$, with the final state $\nu_e \ell^+\ell^- b j j'$, considering $E_e=140$\,GeV.
The authors find that for $L=1$/ab masses for the top partner $T$ around 800\,GeV can be tested when the model-related coupling constants are ${\mathcal O}(0.1)$ and that mixing between $T$ and the first generation quarks can significantly enhance the LHeC sensitivity.

Another search strategy for singly produced top partners is given by their decays $T \to W b$ and $T \to th$, which is presented in Ref.~\cite{Liu:2017rjw}.
The analysis is based on a simplified model where the top partner is an $SU_L(2)$ singlet and interacts only with the third generation of quarks. 
It considers collisions of positrons and protons with $E_e=140$\,GeV. The analysis, carried out at the parton level, investigates the kinematic distributions of the final states.
Useful kinematic variables for the $bW$ final state were found to be the transverse momentum of the lepton, $b$-jet missing energy, while for the $th$ final state the most useful observable is the transverse hadronic energy. 
For masses of ${\mathcal O}(1)$\,TeV the LHeC is found to be sensitive to the new interactions when they are ${\mathcal O}(0.1)$ for $L=1$/ab, in agreement with \cite{Zhang:2017nsn}.
A very similar analysis was performed for the $T\to Wb$ signal channel with comparable results \cite{Han:2017cvu}.

\subsection[Excited fermions ($\nu^*, e^*, u^*$)]{\boldmath Excited fermions ($\nu^*, e^*, u^*$)}
The potential of searches for excited spin-1/2 and spin-3/2 neutrinos are discussed in Ref.~\cite{Ozansoy:2016ivj}.
For the analysis the authors consider effective currents that describe the interactions between excited fermions, gauge bosons, and SM leptons. 
For the signature, the production of the excited electron neutrino $\nu^*$ and its subsequent decay $\nu^* \to W e$ with $W \to jj$ was chosen. 
The analysis, carried out at the parton level, considers $E_e=60$\,GeV, and consists in a study of the kinematic distributions of the final states.
It is concluded that the signature can be well distinguished from backgrounds, and that other lepton-hadron colliders would be required to test the excited neutrinos of different flavours.

Analyses in similar models, considering electron-proton collisions at energies of the FCC-eh and beyond, were carried out for excited electron neutrinos and are presented in Ref.~\cite{Caliskan:2017fts}.
An analysis for the reach for testing excited electrons is discussed in Ref.~\cite{Caliskan:2018vsk}, and for excited quarks in a composite model framework in Ref.~\cite{Gunaydin:2017enp}.

\subsection{Colour octet leptons} 
Unresolved issues of the SM, like family replication and quark-lepton symmetry, can be addressed by composite models, where quarks, leptons, and gauge
bosons are composite particles made up of more basic constituents. 
One general class of particles, predicted in most composite models, are colour octet leptons, which are bound states of a heavy fermion and a heavy scalar particle that is assumed to be colour-charged.
In this scenario each SM lepton is accompanied by a colour octet lepton, which may have spin 1/2 or 3/2. Since they are unobserved, the compositeness scale is expected to be at least ${\mathcal O}(1)$\,TeV. 

At the LHeC, the colour octet partner of the electron $e_8$ can be produced through the process $e^- p \to e_8 g + X$ and studied via its decays products.
An analysis including the study of kinematic distributions that were obtained at the parton level is presented in Ref.~\cite{Sahin:2013vha}. 
It was shown that discovery prospects exist for masses of ${\mathcal O}(\TeV)$. 
A similar analysis is performed for the FCC-eh at much higher energies in Ref.~\cite{Acar:2016rsw}.

%
\subsection{Quark substructure and Contact interactions}

Several long-standing questions arise in the SM, such as those enumerated in Section~1.1. Perhaps most seriously, the SM does not appear to provide a clear, dynamical raison d'\^{e}tre for the existence of quarks. Leptons and quarks appear in  the Standard Model in a symmetric way, sharing electromagnetic interaction with the same charge quantization and with a cancellation of anomaly in the family structure.  This strongly suggests that they may be composed of the same fundamental constituents, or that they form a representation of an extended gauge symmetry group of a Grand Unified Theory.

Assuming that the electron is a point-like particle, the quark substructure can be investigated by introducing a form factor $f_q(Q^2)$ to describe deviations of the $ep$ scattering cross section:
\begin{eqnarray}
   \frac{d\sigma}{dQ^2} = \frac{d\sigma^{SM}}{dQ^2} f_q^2(Q^2)\\
    f_q^2(Q^2) \simeq 1 - \frac{R^2}{6}Q^2
\end{eqnarray}
Here, $R$ is the rms electric charge quark radius. The present limit from HERA is $4.3 \times 10^{-19}$ m~\cite{Abramowicz:2016xzf} while it is estimated that LHeC will be sensitive up to $\sim 10^{-19}$ m~\cite{Zarnecki:2008cp}.

\begin{figure}
\centering
\includegraphics[width=0.64\textwidth]{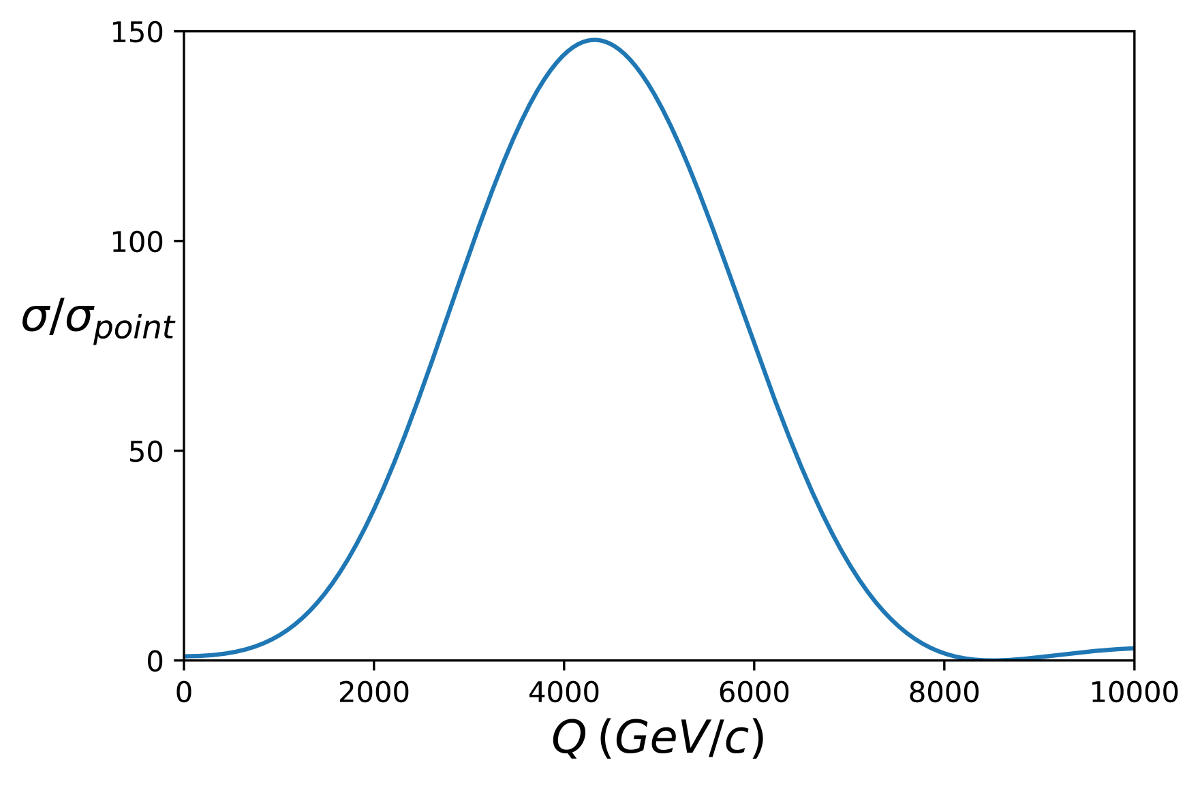}
\caption{Form factor effect in the e-p interaction produced by substructure according to Model II of Hofstadter \cite{Hofstadter:1956qs} with the model parameters given in the text.}
\label{fig:FormFactor}
\end{figure}

An electric precursor to QCD was formulated in 1969 that assumed that hadron constituents are highly electrically charged and where the strong attraction between positive and negative constituents was assumed to bind them together \cite{Yock:1969yv}. Neither the electric model nor Schwinger's comparable model of monopoles \cite{Schwinger:1969ib} reproduce the observed particle spectrum of hadrons, or the observed pattern of weak interactions. 
The ATLAS Collaboration has recently reported searches for free magnetic monopoles and free highly electrically charged particles produced in $pp$ collisions at 13 TeV \cite{Aad:2019pfm}. No candidates were detected with one or two Dirac magnetic charges or with electric charges $20 e < |z| < 100 e$. This extends the results of previous searches made at lower energies and in cosmic rays or bulk matter. 
A simple picture of what might emerge with highly electrically charged constituents is obtained by modeling the proton's substructure by a charge of (say) $21|e|$ smeared uniformly over a region of radius $10^{-19}$ m, and two charges of $-10|e|$ smeared over a larger region of radius $2 \times 10^{-19}$ m. The model II by Hofstadter \cite{Hofstadter:1956qs} predicts the form factor results shown in Fig.~\ref{fig:FormFactor}, consistent the HERA upper limit. 

More generally~\cite{Zarnecki:1999je}, contact interactions can be parameterized in the Lagrangian by coupling coefficients $\eta_{ij}^q$ where the indices $i,j$ indicate the left-handed or right-handed fermion helicities and $q$ the quark flavor. The interaction can be of a scalar, vector or tensor nature and the interference with SM currents can be constructive or destructive. It has been estimated that the LHeC can be sensitive to a scale of contact interaction of $\sim 40-60$ TeV with 100 fb$^{-1}$ of integrated luminosity~\cite{Zarnecki:2008cp} while the present LHC limits are between 20 and 40 TeV, depending on the sign of the interference~\cite{Aaboud:2017buh,Sirunyan:2018ipj}.

\section{Summary and conclusion}
The lack of new physics at the LHC to date forces the community to develop new theoretical ideas as well as to explore the complementarities of $pp$ machines with other possible future facilities. In the context of $ep$ colliders, several studies are being carried out to understand the potential to search for new physics, considering that many interactions can be tested at high precision that are otherwise not easily accessible.

At $ep$ colliders, most BSM physics is accessed via vector-boson fusion, which suppresses the production cross section quickly with increasing mass.
Nonetheless, scalar extensions of the SM as well as neutrino-mass related BSM physics can be well tested at $ep$ due to the smallness and reducibility of the SM backgrounds.
The absence of pile up and complicated triggering makes searches for soft-momenta final state particles feasible, so that results for BSM theories for example characterised by the presence of non-prompt, long-lived particles are complementary to those at the LHC. Additionally, the excellent angular acceptance and resolution of the detector also renders the LHeC a very suitable environment for displaced vertex searches. An increase in the centre-of-mass energy as high as the one foreseen at the FCC would naturally boost the reach in most scenarios considerably. 

Finally, it is worth noting that the LHeC can offer different or indirect ways to search for new physics. 
It was shown recently that Lorentz invariance violation in the weak vector-boson sector can be studied in electron-proton scattering~\cite{Michel:2019tti} via a Fourier-analysis of the parity violating asymmetry in deep inelastic scattering.
Moreover, New Physics could be related to nucleon, nuclear, and top structure functions as discussed in Refs.~\cite{Boroun:2015fwa,Boroun:2015yea,Boroun:2016mke}. 
Investigating of the $B_c^{(*)}$ meson and doubly heavy baryon also was shown to have discovery potential for New Physics~\cite{Bi:2016vbt,Kai:2017cba,Huan-Yu:2017emk}.

%% file: LHeC_and_HLLHC/LHeC_and_HLLHC.tex
\linenumbers
\lhectitlepage
\lhecinstructions
\subfilestableofcontents

\chapter{Influence of the LHeC on Physics at the HL-LHC \ourauthor{Maarten Boonekamp}}   
\label{chap:HLLHC}
After almost 10 years of scientific exploitation of the LHC and about 175\,fb$^{-1}$ of proton-proton collision data delivered to each of the ATLAS and CMS experiments, the sensitivity of a significant fraction of leading measurements and searches becomes limited by systematic uncertainties. Uncertainties induced by the strong interaction, in particular related to the proton structure, play a prominent role and tend to saturate the physics reach of the experiments. This context will only become more evident when the LHC enters its high-luminosity era.

With high precision PDFs measured independently from the other LHC experiments, the LHeC project can resolve this situation. It allows a clean study of the pure QCD effects it aims at measuring, resolving the ambiguity between new physics effects at high mass and PDF uncertainties that intrinsically affects the interpretation of proton-proton data alone. At the weak scale, improved PDFs provide a significant boost to the achievable precision of measurements of the Higgs boson properties and of fundamental electroweak parameters. The LHeC is thus a perfect companion machine for the HL-LHC, allowing a full exploitation of the data and significantly extending its reach.

The present chapter illustrates this with a few selected examples in the domain of precision measurements of the $W$-, $Z$- and Higgs boson properties. The impact of precise PDFs on searches for TeV-scale new physics is also illustrated. Besides, the complementarity of PDF studies at the LHeC and the HL-LHC and the impact of new QCD dynamics at small $x$ on measurements at hadronic colliders, as well as the impact of electron-nucleus scattering data on heavy-ion physics at the LHC, are presented.

\section{Precision Electroweak Measurements at the HL-LHC \ourauthor{Maarten Boonekamp}}

\subsection{The effective weak mixing angle}

Prospective studies for the measurement of the effective weak mixing angle using the forward-backward asymmetry, $A_\text{FB}$, in Drell-Yan di-lepton events at the HL-LHC were performed at ATLAS~\cite{ATL-PHYS-PUB-2018-037}, CMS~\cite{CMS-PAS-FTR-17-001} and LHCb~\cite{Barter:2647836} and reported in the CERN report on Standard Model physics at the HL-LHC~\cite{Azzi:2019yne}. A brief summary is given here, focusing on the impact of the LHeC on this measurement.

At leading order, lepton pairs are produced through the annihilation of a quark and antiquark via the exchange of a Z boson or a virtual photon. The definition of $A_\text{FB}$ is based on the angle $\theta^*$ between the initial- and final-state fermions:
\begin{equation}
A_\text{FB} = \frac{\sigma_\text{F} - \sigma_\text{B}}{\sigma_\text{F} + \sigma_\text{B}}\,,
\end{equation}
where $\sigma_\text{F}$ and $\sigma_\text{B}$ are the cross sections in the forward ($\cos\theta^*>0$) and backward ($\cos\theta^*<0$) hemispheres, respectively.

A non-zero $A_\text{FB}$ in dilepton events arises from the vector and axial-vector couplings of electroweak bosons to fermions. At tree level, the vector and axial-vector couplings of the $Z$ boson to a fermion $f$ are
\begin{equation}
g_V^f = T_3^f - 2 Q_f \sinsqtheta, \,\,\,\,\,\,\,\,\,\,\,\, g_A^f = T_3^f.
\label{eq:gvga}
\end{equation}
The coupling ratio, $g_V^f/g_A^f = 1 - 4|Q_f|\sinsqtheta$, generates the asymmetry. Defining
\begin{equation}
\mathcal{A}_f = 2\frac{g_V^f/g_A^f}{1+(g_V^f/g_A^f)^2}
\end{equation}
one finds, for a given sub-process $q\bar{q} \to Z \to \ell^+ \ell^-$,
\begin{equation}
A_\text{FB} = \frac{3}{4} \mathcal{A}_q \mathcal{A}_\ell.
\end{equation}
As discussed in Sects.~\ref{sec:EW} and~\ref{sec:ewfit} below, Eq.~\eqref{eq:gvga} is subject to radiative corrections introducing the effective weak mixing angle \sinleff in replacement of the leading order observable \sinsqtheta. The asymmetry definitions downstream are however unchanged.

The angle $\theta^*$ is uniquely defined in $e^+e^-$ collisions, where the directions of the $e^+$ and $e^-$ beams are known. In proton-antiproton collisions, at the Tevatron, the incoming quarks and anti-quarks also have preferred directions, and a non-zero asymmetry exists for all lepton-pair rapidities. At the LHC the beams are symmetric, and a non-zero asymmetry only appears for high-rapidity events, as the direction of the longitudinal boost reflects, on average, the direction of the incoming valence quark. While the expected $Z$-boson statistics are very large, with $\mathcal{O}(3\times 10^9)$ events expected in ATLAS and CMS, the measuremend is thus highly affected by PDF uncertainties, and in particular by the $u$ and $d$ valence and sea distributions.

Prospective studies were performed by ATLAS, CMS and LHCb, including a discussion of expected PDF uncertainties. The impact of LHeC PDFs was evaluated by ATLAS and is discussed further. Tab.\,\ref{tab:sinatlas} compares the published ATLAS result~\cite{ATLAS:2018gqq} with the prospects for 3~ab$^{-1}$, for a variety of PDF sets. The statistical uncertainty is at the level of $3\times 10^{-5}$ with this sample, and the experimental systematic uncertainties are improved by $10-25$\,\% depending on the PDF scenario considered. While MMHT2014~\cite{Harland-Lang:2014zoa} and CT14~\cite{Dulat:2015mca} claim comparable PDF uncertainties, the size of the PDF uncertainty is reduced at the HL-LHC thanks to the increased sample size, which helps constraining this component $in$ $situ$. The HL-LHC PDF set~\cite{Khalek:2018mdn}, which incorporates the expected constraints from present and future LHC data, further decreases the associated uncertainty by about 20\%. The LHeC projection~\cite{Klein:1564929} results from a QCD fit to 1\,ab$^{-1}$ of $ep$ scattering pseudodata, with $E_e=60$\,GeV and $E_p=7$\,TeV; in this case, the PDF uncertainty is subleading compared to the experimental systematics.

\begin{table}[ht]
  \centering
  \small
  \begin{tabular}{lccccc}
    \toprule
    Parameter       &  Unit &  ATLAS\,\footnotesize{(Ref.\,\cite{ATLAS:2018gqq})} & \multicolumn{3}{c}{HL-LHC projection} \\
    \cmidrule(lr){3-3} \cmidrule(lr){4-6} 
        &  & MMHT2014 & CT14  & HL-LHC\,PDF & LHeC\,PDF \\ 
    \midrule
    Centre-of-mass energy, $\sqrt{s}$ & TeV         & 8 & 14 & 14 & 14 \\
    Int. luminosity, $\mathcal{L}$ &  $\text{fb}^{-1}$ & 20  & 3000   & 3000 & 3000  \\ 
    \addlinespace
    Experimental uncert.  & $10^{-5}$ & $\pm~23$ & $\pm~9$    &  $\pm~7$     &  $\pm~7$        \\  
    PDF  uncert.   & $10^{-5}$      &$ \pm~24 $ &$ \pm~16 $  & $ \pm~13 $   & $ \pm~3 $      \\ 
    Other syst. uncert. & $10^{-5}$ & $\pm~13$  & --          & --            & --                \\  
    \addlinespace
    Total uncert.,  $\Delta \sinsqtheta$  & $10^{-5}$&  $\pm~36 $ & $ \pm~18 $ & $ \pm~15 $ & $ \pm~8 $  \\ 
    \bottomrule
  \end{tabular}
    \caption{
The  breakdown of uncertainties of \sinsqtheta\ from the ATLAS preliminary results at $\sqrt{s} = 8$\,TeV with 20\,fb$^{-1}$~\cite{ATLAS:2018gqq} is compared to the projected measurements with 3000\,fb$^{-1}$ of data at $\sqrt{s} = 14$\,TeV for two PDF sets considered in this note. All uncertainties are given in units of  $10^{-5} $. Other sources of systematic uncertainties, such as the impact of the MC statistical uncertainty, evaluated in Ref.~\cite{ATLAS:2018gqq} are not considered in the HL-LHC prospect analysis.
    }
    \label{tab:sinatlas}
\end{table}

\begin{figure}[!ht]
  \centering
    \includegraphics[width=0.95\textwidth,trim={0 0 60 0},clip]{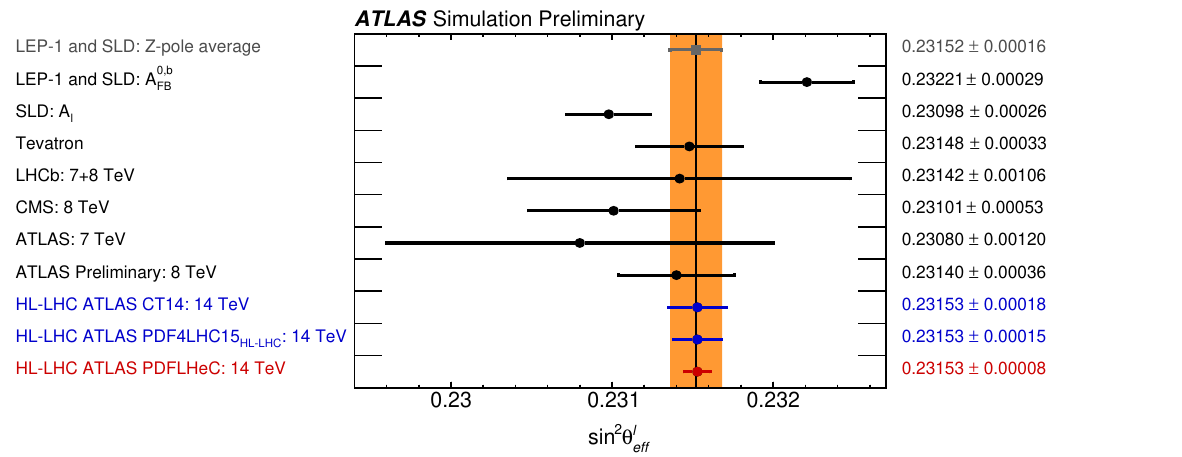}
  \caption{Comparison of measurements or combinations of \sinleff with the world average value (orange band) and the projected uncertainties of measurements at the HL-LHC from Ref.~\cite{ATL-PHYS-PUB-2018-037}.
    For the HL-LHC the central values are set to the world average value and uncertainties are displayed for different assumptions of the available PDF sets, similar to Tab.\,\ref{tab:sinatlas}.}
  \label{fig:sinatlas}
\end{figure}

Fig.\,\ref{fig:sinatlas} compares the ATLAS sensitivity studies of \sinleff to previous measurements from the LHC experiments~\cite{ATLAS:2018gqq,Sirunyan:2018swq,Aaij:2015lka,Aad:2015uau}, and to the legacy measurements by the experiments at LEP and SLC~\cite{ALEPH:2005ab} and the Tevatron~\cite{Aaltonen:2018dxj}. The precision of the measurement of the weak mixing angle in $Z$-boson events, using 3000~fb$^{-1}$ of $pp$ collision data at $\sqrt{s} = 14$ TeV, exceeds the precision achieved in all previous single-experiments  to date. The LHeC is thus essential in exploiting the full potential of the HL-LHC data for this measurement.

\subsection[The $W$-boson mass]{\boldmath The $W$-boson mass}
\label{subsec:Wbosonmass}

This section summarises a
study describing prospects for the measurement of $m_W$ with the upgraded ATLAS detector, using low pile-up data collected during the HL-LHC period~\cite{ATL-PHYS-PUB-2018-026}. Similar features and performance are expected for CMS.

Proton-proton collision data at low pile-up are of large interest for $W$ boson physics, as the low detector occupancy allows an optimal reconstruction of missing transverse momentum, and the $W$ production cross section is large enough to achieve small statistical uncertainties in a moderate running time. At $\sqrt{s}=14$\,TeV and for an instantaneous luminosity 
of $\mathcal{L}\sim 5\times 10^{32}$\,$\text{cm}^{-2} \text{s}^{-1}$, corresponding to two collisions per bunch crossing on average at the LHC, about $\times 10^7$ W boson events can be collected in one month. Such a sample provides a statistical sensitivity at the permille level for cross section measurements, at the percent level for measurements of the $W$ boson transverse momentum distribution, and below 4\,MeV for a measurement of $m_W$.

Additional potential is provided by the upgraded tracking detector, the ITk, which extends the coverage in pseudorapidity beyond $|\eta|<2.5$ to $|\eta|<4$. The increased acceptance allows $W$-boson measurements to probe a new region in Bjorken $x$ at $Q^2\sim m_W^2$. This will in turn allow further constraints on the parton density functions (PDFs) from cross section
measurements, and reduce PDF uncertainties in the measurement of $m_W$. A possible increase of the LHC centre-of-mass energy, such as the HE-LHC program with $\sqrt{s}=27$\,TeV~\cite{Zimmermann:2017bbr}, could play a similar role on a longer timescale.

Leptonic $W$ boson decays are characterised by an energetic, isolated electron or muon, and significant missing transverse momentum reflecting the decay neutrino. The hadronic recoil, \uT, is defined
from the vector sum of the transverse momenta of all reconstructed particles in the event excluding the charged lepton, and provides a measure of the $W$ boson transverse momentum. Lepton transverse momentum, \pTl, missing
transverse momentum, \met, and the hadronic recoil are related through
$\vecmet = -(\vecpTl + \vec{\uT})$. The \pTl and \met distributions have sharp peaks at $\pTl\sim\met\sim m_W/2$. The transverse mass \mT, defined as $\mT=\sqrt{2 \pTl \met \cos(\phil-\phimiss)}$, peaks at $\mT\sim m_W$.

Events are selected applying the following cuts to the object kinematics, after resolution corrections:
\begin{itemize}
\item $\pTl>25$\,GeV, $\met>25$\,GeV, $\mT>50$\,GeV and $\uT<15$\,GeV;
\item $|\etal|<2.4$ or $2.4<|\etal|<4$.
\end{itemize}
The first set of cuts select the range of the kinematic peaks of the $W$ boson decay products, restricting to the region of small \pTW to maximise the sensitivity of the
distributions to $m_W$. Two pseudorapidity ranges are considered, corresponding to the central region accessible with the current ATLAS detector, and to the forward region accessible in the electron channel with the ITk.

The $W$-boson mass is determined comparing the final state kinematic peaks in the simulation to those observed in the data, and adjusting the value of $m_W$ assumed in the former to optimise the agreement. The shift in the measured value of $m_W$ resulting from a change in the assumed PDF set is estimated using a set of template distributions obtained for different values of $m_W$ and a given reference PDF set, and ``pseudo-data'' distributions obtained for an alternate set representing, for example, uncertainty variations with respect to the reference set. The PDF uncertainty for a given set is calculated by summing the shifts obtained for all uncertainty variations in quadrature.

The PDF uncertainty is calculated for the CT14~\cite{Dulat:2015mca}, MMHT2014~\cite{Harland-Lang:2014zoa}, HL-LHC~\cite{Khalek:2018mdn} and LHeC~\cite{Klein:1564929} PDF sets and their associated uncertainties. Compared to current sets such as CT14 and MMHT2014, the HL-LHC set incorporates the expected constraints from present and future LHC data; it starts from the PDF4LHC convention~\cite{Butterworth:2015oua} and comes in three scenarios corresponding to more or less optimistic projections of the experimental uncertainties. 

\begin{figure}[!th]
  \centering
    \includegraphics[width=0.6\textwidth]{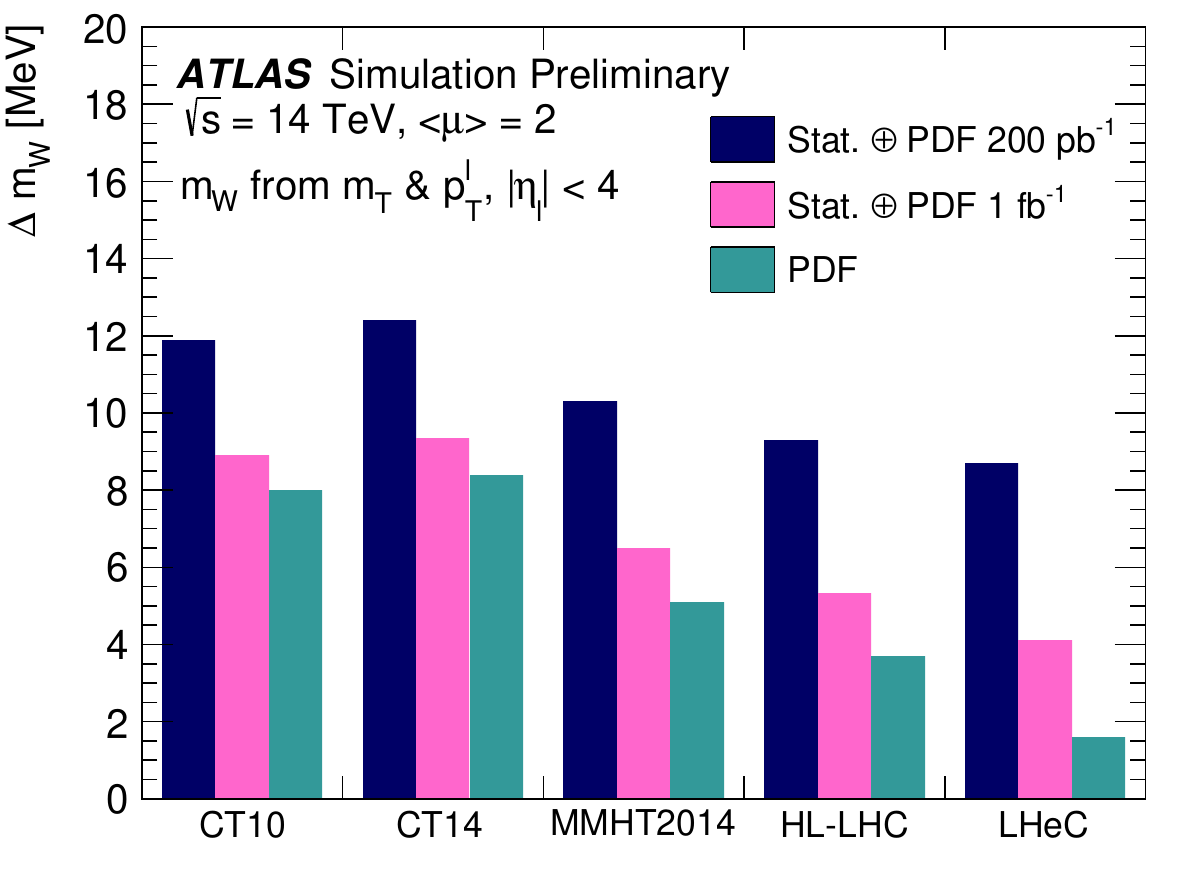}
  \caption{Measurement uncertainty of $m_W$ at the HL-LHC with 200\,\ipb (dark blue) and 1\,\ifb (pink) of collected low pile-up data for different present and future PDF sets from Ref.~\cite{ATL-PHYS-PUB-2018-026}.
    The green area indicates the PDF uncertainty from those sets alone.
    The projections are obtained from a combined fit to the simulated \pTl and \mT distributions in the acceptance $|\eta|<4$.}
  \label{fig:pdfcomp}.
\end{figure}

The expected statistical and PDF uncertainties are illustrated in Tab.\,\ref{tab:pdfcomp} and Fig.\,\ref{fig:pdfcomp}. The CT10 and CT14 sets yield comparable uncertainties. The MMHT2014 uncertainties are about 30\,\% lower. The three projected HL-LHC PDF sets give very similar uncertainties; scenario 2 is the most conservative and shown here. Compared to CT10 and CT14, a reduction in PDF uncertainty of about a factor of two is obtained in this case.

The LHeC sample can be collected in about three years, synchronously with the HL-LHC operation. In this configuration, the neutral- and charged-current DIS samples are sufficient to disentangle the first ($d,u$) and second ($s,c$) generation parton densities without ambiguity, and reduce the PDF uncertainty below 2\,MeV, a factor 5--6 compared to present knowledge. Also in this case the $m_W$ measurement will benefit from the large $W$ boson samples collected at the LHC, and from the combination of the central and forward categories. In this context, PDF uncertainties would be sub-leading even with 1\,fb$^{-1}$ of low pile-up LHC data. 

\begin{table}[ht]
  \centering
  \small
  \begin{tabular}{lcccccc}
    \toprule
    Parameter       &  Unit &  ATLAS\,\footnotesize{(Ref.\,\cite{Aaboud:2017svj})} & \multicolumn{4}{c}{HL-LHC projection} \\
    \cmidrule(lr){3-3} \cmidrule(lr){4-7} 
         &      & CT10 & CT14  & HL-LHC & LHeC & LHeC \\ 
    \midrule
    Centre-of-mass energy, $\sqrt{s}$ & TeV &       7  & 14 & 14 & 14 & 14 \\
    Int. luminosity, $\mathcal{L}$ &  $\text{fb}^{-1}$ & 5  & 1 & 1 & 1 & 1  \\
Acceptance  &  & $|\eta|<2.4$ & $|\eta|<2.4$ & $|\eta|<2.4$ & $|\eta|<2.4$ & $|\eta|<4$ \\ 
\addlinespace
Statistical uncert.    &  MeV& $\pm~7$    & $\pm~5$    &  $\pm~4.5$     &  $\pm~4.5$      &  $\pm~3.7$ \\
      PDF uncert.         &  MeV & $ \pm~9 $  &$ \pm~12 $  & $ \pm~5.8 $   &  $\pm~2.2$ &  $\pm~1.6$    \\ 
      Other syst. uncert.    &  MeV& $\pm~13$   & -          & -            & -                \\  
\addlinespace
        Total uncert. $\Delta m_W$  &  MeV     & $\pm~19 $ & 13 & 7.3 & 5.0 & 4.1 \\ 
      \bottomrule
    \end{tabular}
  \caption{Measurement uncertainty of the $W$-boson mass at the HL-LHC for different PDF sets (CT14, HL-LHC PDF and LHeC PDF) and lepton acceptance regions in comparison with a measurement by ATLAS\,\cite{Aaboud:2017svj}. The HL-LHC projections are obtained from a combined fit to the simulated \pTl and \mT distributions.}
  \label{tab:pdfcomp}
\end{table}

\subsection{Impact on electroweak precision tests \label{sec:ewfit}}

The theoretical expressions for the electroweak parameters discussed above are functions of the other fundamental constants of the theory. In the Standard Model, an approximate expression for \MW, valid at one loop for $\MH>\MW$, is~\cite{ALEPH:2005ab}

\begin{eqnarray}
\label{eqn:wmassho}
\MW^2 &=& \frac{\MZ^2}{2} \left(1+\sqrt{1-\frac{\sqrt{8} \pi  \aem}{\gf  \MZ^2}\frac{1}{1-\Delta r}}\right), \\
\Delta r &=& \Delta\aem -\frac{\cossqtheta}{\sinsqtheta} \Delta\rho, \\
\Delta\rho &=& \frac{3 \gf  \MW^2}{8\sqrt{2}\pi^2}\left[\frac{m_\text{top}^2}{\MW^2} -\frac{\sinsqtheta}{\cossqtheta}\left( \ln{\frac{\MH^2}{\MW^2} - \frac{5}{6}}\right) + \cdots \right].
\end{eqnarray}
$\Delta r$ includes all radiative corrections to \MW, $\Delta\aem$ is the difference between the electromagnetic coupling constant evaluated at $q^2=0$ and $q^2=\MZ^2$, and $\Delta\rho$ is the quantum correction to the tree-level relation $\rho \equiv \MW^2/(\MZ^2\cossqtheta) = 1$, defined as $\rho=1+\Delta\rho$.

Similarly, approximate one-loop expressions for the vector and axial-vector couplings between the $Z$ boson and the fermions, $g_V$ and $g_A$, are
\begin{eqnarray}
\label{eqn:sinthho}
g_V &=& \sqrt{1+\Delta\rho} \,\, \left(T_3 - 2 Q (1+\Delta\kappa)\sinsqtheta\right),\\
g_A &=& \sqrt{1+\Delta\rho} \,\, T_3\,,
\end{eqnarray}
where 
\begin{equation}
\Delta\kappa = \frac{3 \gf  \MW^2}{8\sqrt{2}\pi^2}\left[\frac{\cossqtheta}{\sinsqtheta}\frac{m_\text{top}^2}{\MW^2} -\frac{10}{9}\left( \ln{\frac{\MH^2}{\MW^2} - \frac{5}{6}}\right) + \cdots \right].
\label{eqn:kappa}
\end{equation}
At two loops, also the strong coupling constant enters.

A large class of theories beyond the SM predict particles that contribute to the $W$- and $Z$-boson self-energies, modifying the above expressions. From the point of view of on-shell observables of the $W$ and $Z$, these modifications are usually parameterized using the so-called $oblique$ parameters, called $S$, $T$ and $U$~\cite{Peskin:1991sw}. Their values are by definition 0 in the SM and, for example, a significant violation of the relation between \MW, \MH and \mt would translate into non-zero values for $S$ and $T$.

A typical application of this formalism consists in using the measured properties of the $W$ and $Z$ bosons, the top quark mass, and the values of coupling constants, to derive an indirect determination of the Higgs boson mass in the SM and compare the latter to the measured value. Beyond the SM, the measured values can be used to derive allowed contours in the ($S$, $T$) plane. 


Present and future measurement uncertainties for the most relevant electroweak parameters are summarised in Tab.\,\ref{tab:ewfit}, and are used to evaluate the impact of the improved measurements on electroweak precision tests. Specifically, we consider the effect of improved measurements of $m_W$ and \sinleff discussed in this chapter, and of the improved precision of \as\ discussed in Chapter~\ref{sec:QCD}. In addition, we consider an ultimate precision of 300\,MeV for the top quark mass measured at the LHC.

\begin{table}[ht]
  \centering
  \small
\begin{tabular}{lcccc}
\toprule
Parameter & Unit & Value & \multicolumn{2}{c}{Uncertainty}\\
\cmidrule(lr){4-5}
 & &       & Present & Expected \\
\midrule
$m_Z$              & MeV          &      91187.6      & 2.1 & 2.1 \\
$m_W$              & MeV          &      80385      & 15 & 5 \\
\sinleff           &              & 0.23152         & 0.00016 & 0.00008        \\
$m_\text{top}$     & GeV          & 173.1 & 0.7      & 0.3           \\
$\asmz$            &              & 0.1179 & 0.0010 & 0.0001 \\
\bottomrule
\end{tabular}
\caption{Present uncertainties for the relevant EW precision observables~\cite{ALEPH:2005ab,Tanabashi:2018oca,Tanabashi:2019}, and their expected precision in the LHeC and HL-LHC era.}
\label{tab:ewfit}
\end{table}

\begin{figure}[ht!]
  \centering
    \includegraphics[width=0.6\textwidth]{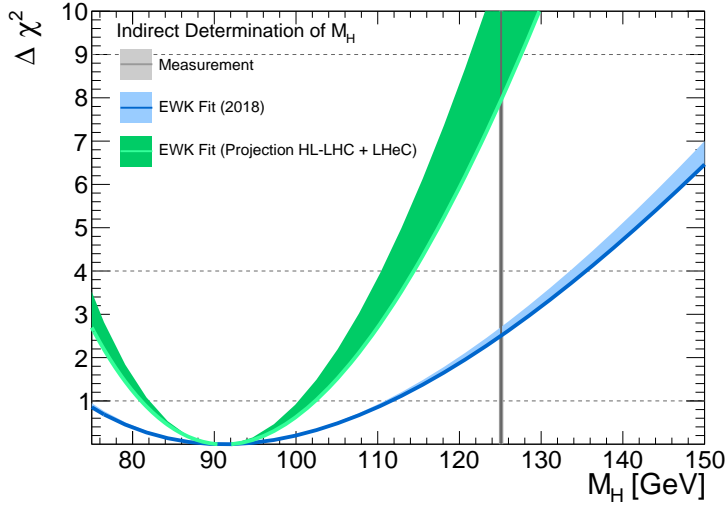}
  \caption{Comparisons of $\chi^2$ distributions for different Higgs boson mass values, using present and future experimental uncertainties. The theoretical uncertainties are indicated by the filled areas. The Gfitter program~\cite{Haller:2018nnx} was used for this analysis. }\label{fig:mhplot}
\end{figure}
\begin{figure}[ht!]
  \centering
     \includegraphics[width=0.46\textwidth]{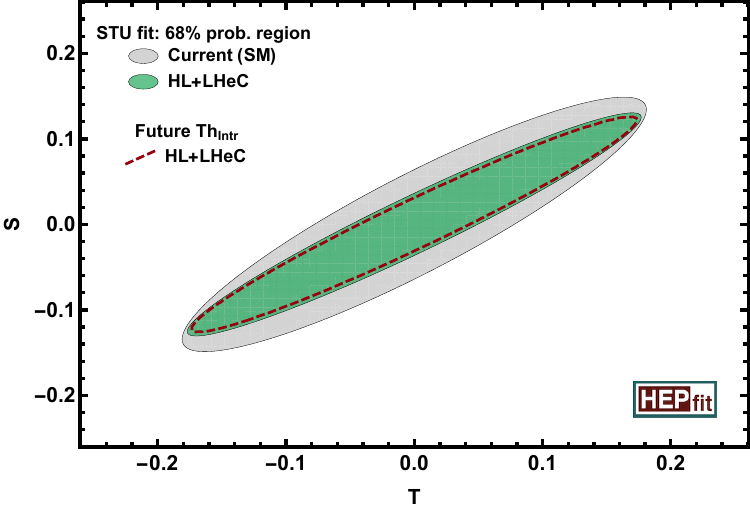}
     \hspace{0.04\textwidth}
     \includegraphics[width=0.46\textwidth]{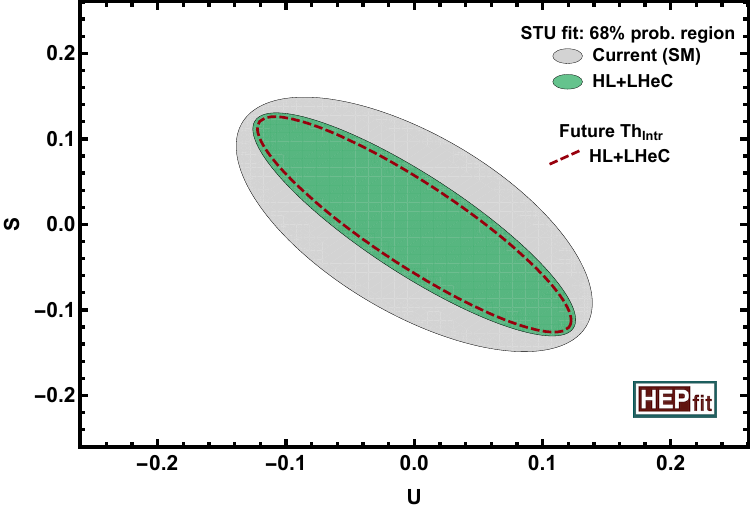}
     \includegraphics[width=0.46\textwidth]{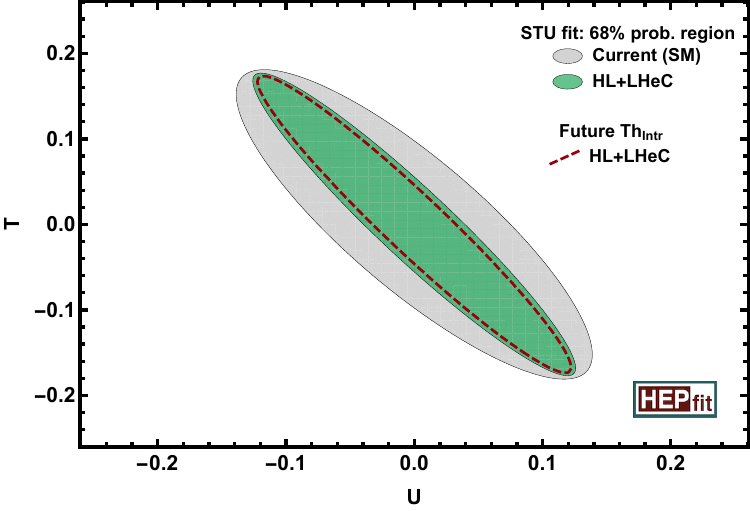}
  \caption{Allowed regions in the ($S$,$T$,$U$) plane showing all three combinations: $S$ vs $T$ (top-left), $S$ vs $U$ (top-right), $T$ vs $U$ (bottom). The grey and green areas indicate the currently allowed region and the LHeC projection, respectively. The dashed line indicates the effect of expected theoretical improvements. The HEPfit program~\cite{deBlas:2019okz} was used for this analysis. }\label{fig:stplot}
\end{figure}

The results are illustrated in Figs.\,\ref{fig:mhplot} and~\ref{fig:stplot}. The former results from a fit performed using the GFitter framework~\cite{Haller:2018nnx}, and compares the indirect determinations of the Higgs boson mass for the present and expected measurement precisions. The indirect uncertainty in \MH reduces from about 20\,\% to 10\,\%.

Fig.\,\ref{fig:stplot} was performed using HEPfit~\cite{deBlas:2019okz}, and compares allowed contours for the $S$ and $T$ parameters. Here also, the allowed region is reduced by a factor of about two from the improved measurements of \MW, \sinleff, $m_\text{top}$ and \as. Improved theoretical calculations in the SM will provide an additional reduction of 10-15\,\%.


In summary, the LHeC data promises significant improvements in the measurement precision of fundamental electroweak parameters such as \MW and \sinleff. The improved measurements enhance the sensitivity of electroweak tests by a factor of two or more.

\section{Higgs Physics}

\subsection{Impact of LHeC data on Higgs cross section predictions at the LHC}

A detailed analysis of Higgs boson production cross sections was given in the report on Higgs Physics at the HL-LHC and HE-LHC~\cite{Cepeda:2019klc}. Central values at $\sqrt{s}=14$\,TeV and the corresponding uncertainties are reported in Tab.\,\ref{tab:hxs}. Perturbative uncertainties (labelled $\Delta\sigma_{\textrm{scales}}$ in Tab.\,\ref{tab:hxs}) generally dominate compared to the contributions of $\as$ and the PDFs. This is especially true for gluon fusion, where the residual theoretical uncertainties correspond to missing corrections beyond N$^3$LO in QCD, and for $t\bar{t}H$ production which is known to NLO QCD+EW accuracy. The weak boson fusion, $WH$ and $ZH$ cross sections are known to NNLO QCD + NLO EW accuracy; residual theoretical uncertainties are smaller for these weak interaction processes.

In Ref.~\cite{Cepeda:2019klc}, $\as$-related uncertainties are propagated assuming $\as=0.118 \pm 0.0015$, and the assumed PDF uncertainties reflect the HL-LHC prospects~\cite{Khalek:2018mdn}. They are in excess of 3\,\% for gluon fusion and $t\bar{t}H$, below 2\,\% for $WH$ and $ZH$, and 0.4\% for weak boson fusion. The LHeC uncertainties in Tab.\,\ref{tab:hxs} are calculated using MCFM~\cite{Campbell:2019dru}, interfaced to PDFs determined from LHeC pseudodata as described in Chapter~\ref{chapter:pdf}. Assuming the prospects for $\as$ and PDFs described in Chapters~\ref{chapter:pdf} and~\ref{sec:QCD}, and with the exception of weak-boson fusion production, the corresponding uncertainties decrease by a factor 5 to 10. 

\begin{table}[ht]
  \centering
  \small
\begin{tabular}{lcccc}
\toprule
Process         & $\sigma_H$ [pb] & $\Delta\sigma_{\textrm{scales}}$ & \multicolumn{2}{c}{$\Delta\sigma_{\textrm{PDF}+\as}$}\\ 
\cmidrule(lr){4-5}
&  &  & HL-LHC\,PDF & LHeC\,PDF  \\ 
\midrule
Gluon-fusion               & 54.7 & 5.4\,\% & 3.1\,\% & 0.4\,\% \\
Vector-boson-fusion        &  4.3 & 2.1\,\% & 0.4\,\% & 0.3\,\% \\
$pp\rightarrow WH$         &  1.5 & 0.5\,\% & 1.4\,\% & 0.2\,\% \\
$pp\rightarrow ZH$         &  1.0 & 3.5\,\% & 1.9\,\% & 0.3\,\% \\
$pp\rightarrow t\bar{t}H$  &  0.6 & 7.5\,\% & 3.5\,\% & 0.4\,\% \\
\midrule
\end{tabular}
\caption{Predictions for Higgs boson production cross sections at the HL-LHC at $\sqrt{s}=14$\,TeV and its associated
  relative uncertainties from scale variations and two PDF projections, HL-LHC and LHeC PDFs, $\Delta\sigma$.
  The PDF uncertainties include uncertainties of \as.
}
\label{tab:hxs}
\end{table}

\begin{figure}[!th]
\centering
\includegraphics[width=.7\textwidth]{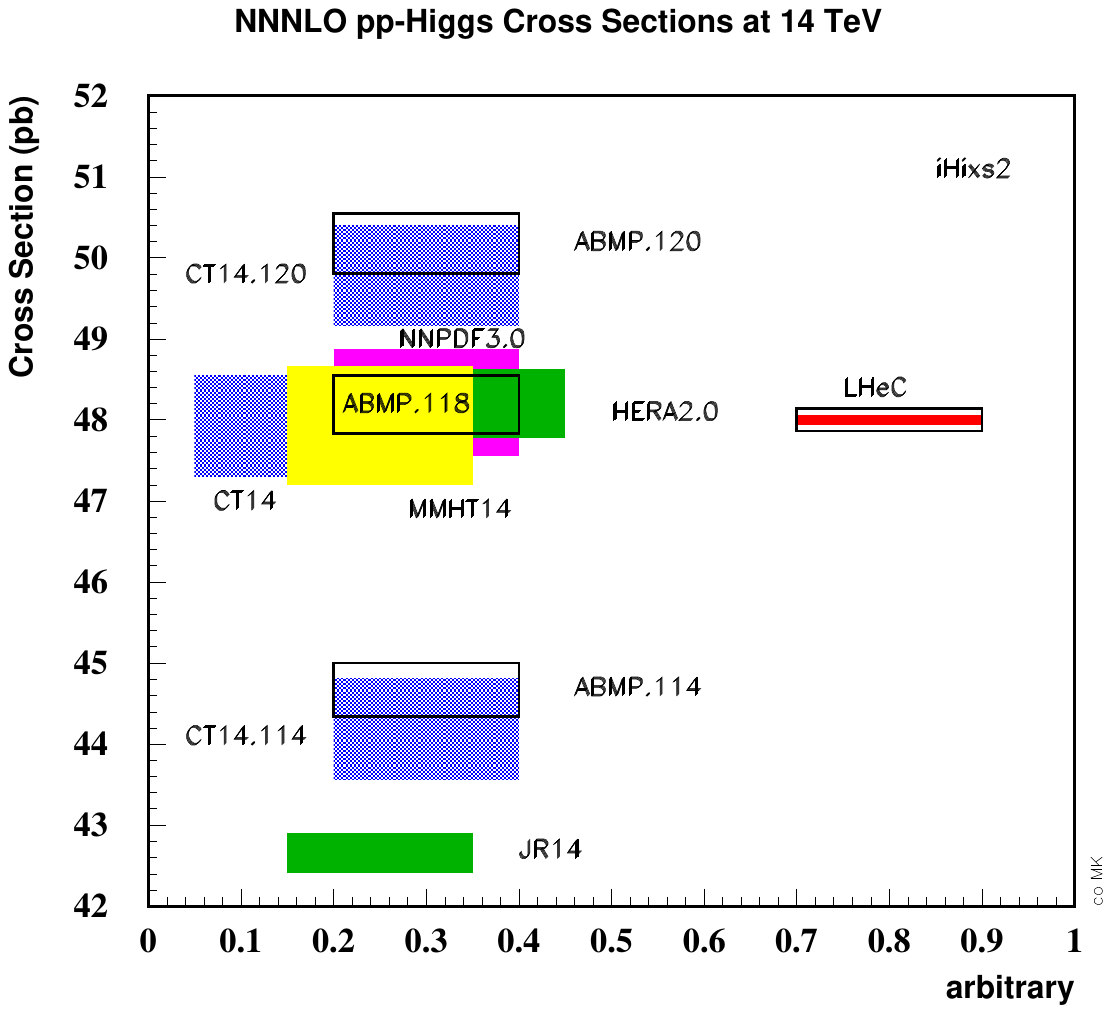}
\vspace{-3.cm}
\caption{\footnotesize{Cross sections of Higgs production calculated to
N$^3$LO using
the iHix program~\cite{Dulat:2018rbf} for existing
PDF parameterisation sets (left side) and for the LHeC
PDFs (right side).  The widths of the areas correspond to the
uncertainties as quoted by the various sets, having rescaled the CT14
uncertainties from 90 to 68\,\% C.L. Results (left) are included also for
different values of the strong coupling constant $\as(M_Z^2)$, from
0.114 to 0.120.
The inner LHeC uncertainty band (red) includes the expected systematic
uncertainty due to the PDFs while the outer box illustrates the expected
uncertainty resulting from the determination of $\as$ with the LHeC.}
}
\label{fig:ihix}
\end{figure}

The important, beneficial role of $ep$ PDF information for LHC Higgs
physics can also be illustrated using the predictions for the total cross section, $pp \to H X$ at the LHC. This has recently
been calculated~\cite{Mistlberger:2018etf} to N$^3$LO pQCD. In
Fig.\,\ref{fig:ihix} calculations of this cross section are
shown for several recent sets of parton distributions, calculated with the
iHix code~\cite{Dulat:2018rbf}, including the LHeC set.

The effect of these improvements on Higgs boson coupling determination at the HL-LHC is at present modest, due to the combined effect of still significant perturbative uncertainties and of the expected experimental systematic uncertainties. The influence of the LHeC on these measurements is further discussed in the next section.

\subsection{Higgs Couplings from a simultaneous analysis of \emph{pp} and \emph{ep} collision data \ourauthor{Jorge De Blas}} 

The LHC data collected during the Runs I and II have provided a first exploration of the properties
of the Higgs boson. The so-called $\kappa$ framework~\cite{LHCHiggsCrossSectionWorkingGroup:2012nn}
-- which allows modifications of the SM-like couplings of the Higgs boson to each SM particle $i$,
parameterised by coupling modifiers $\kappa_i$ -- has been widely used for the interpretation of these measurements.
With current data, the $\kappa$ parameters associated to the main couplings of the Higgs can be determined to a precision
of roughly $10$-$20\,\%$, see e.g.\ \cite{Aad:2019mbh}.\footnote{Note that at the LHC one can only determine coupling ratios.}
This knowledge will be further improved at the high-luminosity phase of the LHC, reaching a precision
in many cases well below the 10\,$\%$ level~\cite{Cepeda:2019klc}. Even at the HL-LHC it will  however, be difficult
to obtain sensible measurements of certain Higgs interactions, especially the coupling to charm quarks.
Such gap would be covered by the precise measurements of that channel at the LHeC, as described in Sect.~\ref{sec:smhiggsinep}. Channels measured to  a few percent accuracy at both HL-LHC and LHeC
would provide important cross checks and additional physics information because 
of the different dominant Higgs production mechanisms,  $gg \to H$ in $pp$ and
$WW \to H$ in $ep$. There follows a remarkable
complementarity between the measurements from  both machines and a joint precision
that is comparable to that at ILC or CLIC~\cite{deBlas:2019rxi}, which yet have the important possibility
to determine the total cross section through the $e^+e^- \to Z^* \to ZH$ reaction.
Furthermore, as also explained in Sect.~\ref{sec:smhiggsinep}, 
the LHeC environment allows very precise determinations of certain interactions,
well beyond of what will be possible at the high-luminosity $pp$ collider. In this subsection we briefly
describe the complementarity between the Higgs measurements at the $pp$ and $ep$ colliders,
illustrated via a combined fit to the HL-LHC and LHeC projections in the $\kappa$ framework.

For a detailed description of the Higgs physics program at the LHeC we refer to Chapter~\ref{chapter:higgs}. The only
information not included in the fit presented in this section 
is that of the determination of the top Yukawa coupling, since projections
from that study are performed assuming any coupling other than $\kappa_t$ to be SM like. Comments in this regard will be made,
when necessary, below. 

For the HL-LHC inputs of the combined fit we rely on the projections presented in Ref.~\cite{Cepeda:2019klc},
as used in the comparative study in Ref.~\cite{deBlas:2019rxi}. These HL-LHC inputs include projections for the total rates in the main
production (ggF, VBF, $VH$ and $ttH$) and decay channels ($H\to bb,~\tau\tau,~\mu\mu,~ZZ^*,~WW^*,~\gamma\gamma,~Z\gamma$).
They are available both for ATLAS and CMS. Regarding the theory systematics in these projections, we assume the scenario S2
described in~\cite{Cepeda:2019klc}, where the SM theory uncertainties are reduced by roughly a factor of two with respect to their
current values, a reduction to which LHeC would contribute by eliminating the PDF and $\alpha_s$
parts of the uncertainty, see Fig.\,\ref{fig:ihix}.
Theory systematics are assumed to be fully correlated between ATLAS and CMS. These projections are combined with
LHeC ones, where, as in Ref.~\cite{deBlas:2019rxi}, we use the future projections for the SM theory uncertainties in the different production
cross sections and decay widths. In the $\kappa$ fit performed here we assume: (1) no Higgs decays into particles other than the SM ones;
(2) heavy particles are allowed to modify the SM loops, so we use effective $\kappa$ parameters to describe the SM loop-induced processes,
i.e.\ we use $\kappa_g,~\kappa_\gamma, \kappa_{Z\gamma}$ as free parameters. The total list of free parameters considered for this
combined HL-LHC+LHeC $\kappa$ fit is, therefore,
\begin{equation}
\left\{\kappa_b,~\kappa_t,~\kappa_\tau,~\kappa_c,~\kappa_\mu,~,\kappa_Z,~\kappa_W,~\kappa_g,~\kappa_\gamma,~\kappa_{Z\gamma}\right\},
\end{equation}
for a total of 10 degrees of freedom. Coupling modifiers associated to any other SM particles are assumed to be SM-like, $\kappa_i=1$.

\begin{table}[hbt]
  \centering
  \small
\begin{tabular}{lcccc}
\toprule
Parameter   & \multicolumn{3}{c}{Uncertainty}  \\
 \cmidrule(lr){2-4}
&HL-LHC   &LHeC   & HL-LHC+LHeC   \\
\midrule
$\kappa_{W}$  & 1.7 & 0.75  & 0.50  \\
$\kappa_{Z}$  & 1.5 & 1.2  & 0.82  \\
$\kappa_{g}$  & 2.3 & 3.6 & 1.6  \\
$\kappa_{\gamma}$  & 1.9 & 7.6 & 1.4 \\
$\kappa_{Z\gamma}$  & 10 & -- & 10  \\
$\kappa_{c}$  & -- & 4.1 & 3.6  \\
$\kappa_{t}$  & 3.3 & -- & 3.1 \\
$\kappa_{b}$  & 3.6 & 2.1  & 1.1 \\
$\kappa_{\mu}$  & 4.6 & -- & 4.4  \\
$\kappa_{\tau}$  & 1.9 & 3.3 & 1.3 \\
\bottomrule
\end{tabular}
\caption{
Results of the combined HL-LHC + LHeC $\kappa$ fit. The output of the fit is compared with the results of the HL-LHC and LHeC stand-alone fits. The uncertainties of the $\kappa$ values are given in per cent.
}
\label{tab:Kappa_HLLHC_LHeC}
\end{table}

\begin{figure}[!th]
\centering{\includegraphics[width=0.9\textwidth]{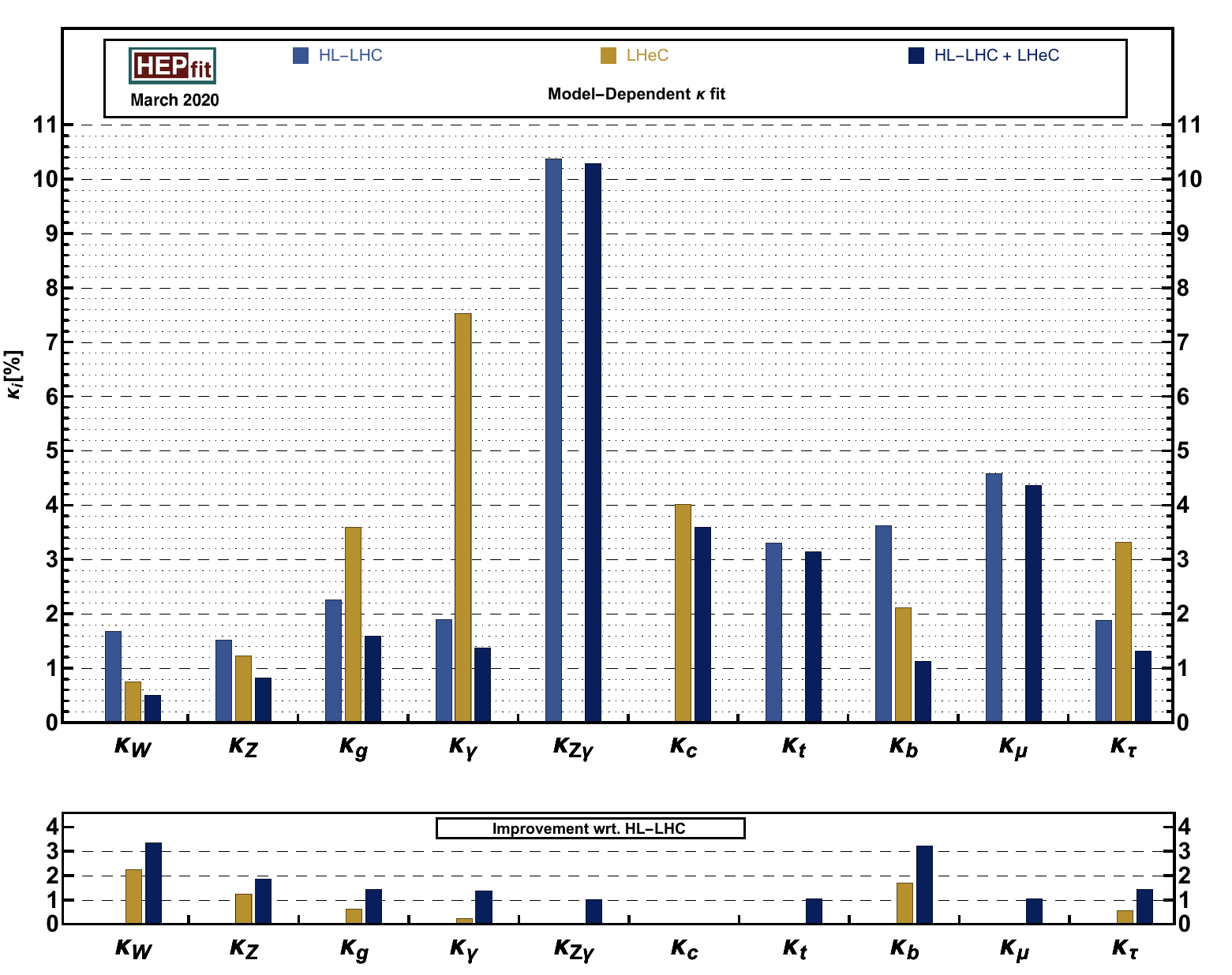}} \\
\caption{Top: Uncertainty of the determination of the scale factor
$\kappa$ in the determination of the Higgs couplings, in per cent. Results are given of the combined HL-LHC + LHeC $\kappa$ fit (dark blue) and of the HL-LHC (blue) and LHeC (gold) stand-alone fits. There is no accurate measurement expected of  $\kappa_{c}$ at the LHC. Likewise the precision of the
rare channels $Z\gamma$, $t\bar t$ and $\mu \mu$ will be very limited
at the LHeC. Bottom: Improvement of the $\kappa$ determinations 
through the addition of the $ep$ information (gold)
and by the combined $ep+pp$ analysis (dark blue), calculated with respect to the
HL-LHC prospects. Strong improvements are seen for the $W$, $Z$ and $b$ 
couplings, while that for charm cannot be illustrated here as $\kappa_c$ is
considered to be not measurable at the HL-LHC.}
\label{fig:Kappa_HLLHC_LHeC}
\end{figure}

\begin{figure}[!th]
\centering{\includegraphics[width=0.8\textwidth]{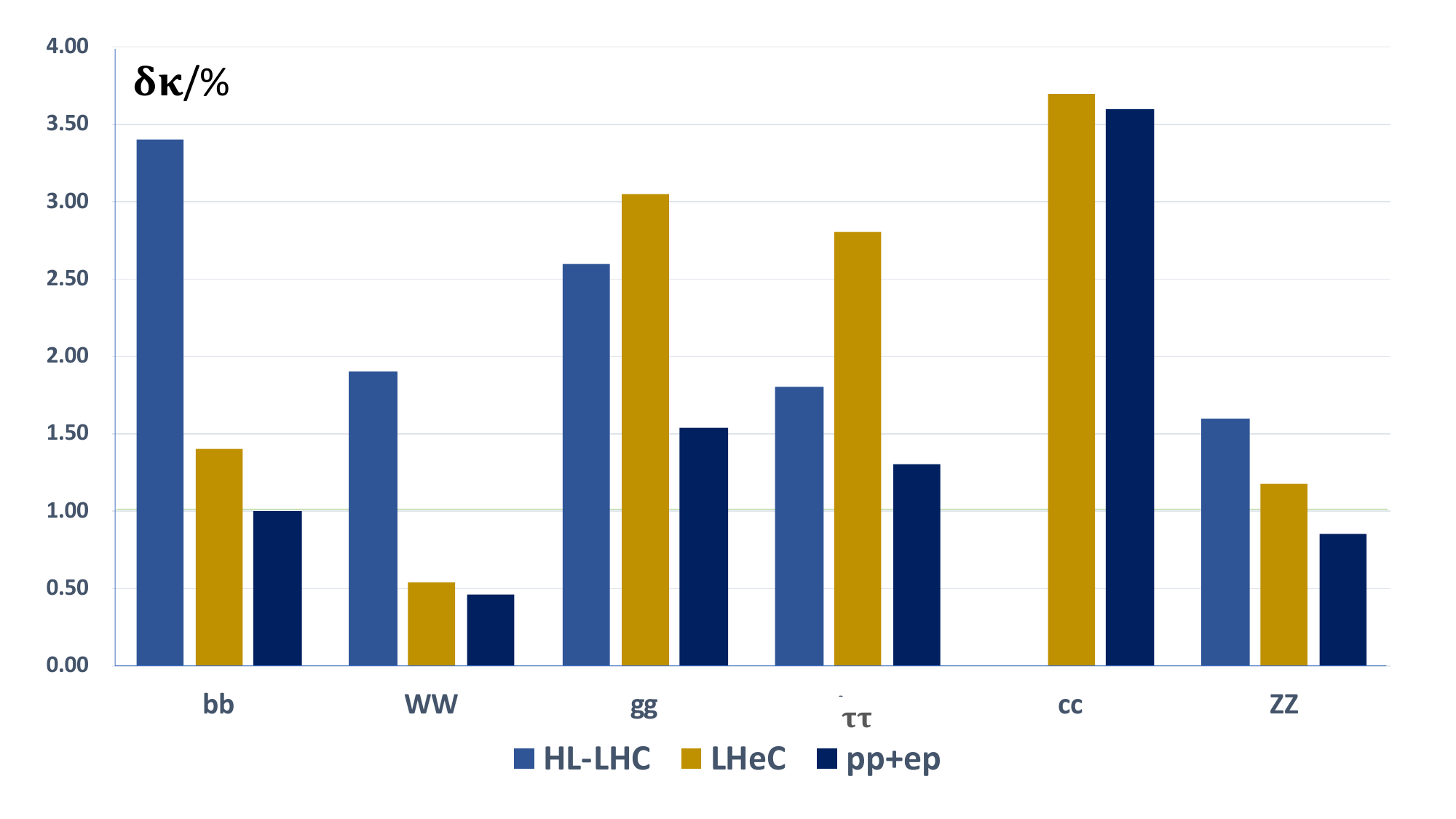}}
\caption{Uncertainty of the determination of the scale factor
$\kappa$ in the determination of the Higgs couplings, in per cent.
Zoom of  Fig.\,\ref{fig:Kappa_HLLHC_LHeC} into the six most frequent H decay channels.
 Results are given of the combined HL-LHC + LHeC $\kappa$ fit (dark blue) and of the HL-LHC (blue) and LHeC (gold) stand-alone fits. There is no accurate measurement expected of  $\kappa_{c}$ at the LHC. }
\label{fig:Kappa_HLLHC_LHeCzoom}
\end{figure}
The results of the HL-LHC+LHeC fit, which has been performed using the HEPfit code~\cite{deBlas:2019okz}, are shown in
Tab.\,\ref{tab:Kappa_HLLHC_LHeC}, and 
Fig.\,\ref{fig:Kappa_HLLHC_LHeC} and its zoomed version Fig.\,\ref{fig:Kappa_HLLHC_LHeCzoom}~\footnote{The $\kappa$ analysis of the LHeC
has been performed independently using a MINUIT based fit program leading to
perfect agreement with the HEPfit result.}.
%
The increment in constraining power after adding the
LHeC measurements is especially apparent for the couplings to $W$ bosons and $b$ quarks, bringing an improvement with respect to the HL-LHC
result of a factor $\simeq 3$. As explained at the beginning of this section, the LHeC measurements also bring the possibility of setting sensible
constraints on the Higgs interactions with charm quarks, with a precision of roughly 4$\%$. The HL-LHC measurements, in turn, fill some of the
\emph{gaps} in the fit at the LHeC, where there is little sensitivity to the couplings involved in rare Higgs decays, e.g.\  $H\to \mu\mu$ and $H\to Z\gamma$.
This makes apparent the complementarity between the measurements at $ep$ and $pp$ machines, with the former leading in terms of precision
in the largest Higgs couplings, while the high-luminosity of the latter brings sensitivity to the smaller interactions. 

Finally, as mentioned at the beginning, we did not include in this combined $ep$+$pp$ fit the projections for top Yukawa interactions
at the LHeC from Section~\ref{sec:topHinep}, 
as these were not derived in a global setup, but rather setting all other interactions involved in $\bar{t}H\nu_e$ product
to their SM values. However, the main uncertainty from the other 
$\kappa$ parameters is expected to come from the $W$ and $b$ couplings, $\kappa_W$
and $\kappa_b$, which are determined with an overall precision of $\sim 0.8\%$ and $2\%$. Therefore one expects the LHeC result, $\delta \kappa_t \sim 17\%$
for $L=1$\,ab$^{-1}$, to be minimally affected. This number is to be compared
with the HL-LHC projection of $\sim 4\%$, which is
expected to dominate in a combined result.
%
%

\section{Further precision SM measurements at the HL-LHC \ourauthor{Daniel, Maarten}}
        
The LHeC measurements and the results from their phenomenological interpretations will have an important impact on many areas of the HL-LHC physics programme.
This goes far beyond the precision electroweak and the Higgs physics, as discussed at hand of dedicated analyses in the previous sections, and BSM or $e$A physics as discussed in the subsequent sections.
In this section a few further selected topics of the Standard Model (SM) physics programme at the LHC and HL-LHC are discussed, where substantial improvements due to the LHeC can be expected.

In general, two distinct aspects can be considered for any SM measurement in that respect\,\footnote{In some cases, a model- or physics parameter is directly extracted from the experiment data and the two applications are merged into a single analysis workflow, for instance in many LHC top-quark mass analyses.
Additionally to these two aspects, of course, the complementarity of the physics case of $ep$ collisions enhances our understanding of the fundamental laws of physics.}:
\begin{compactitem}
\item improvements of the analysis of the recorded event data, and
\item improvements of the phenomenological interpretation of the measurements.
\end{compactitem}
In order to assess the impact of the LHeC for the first bullet, one must recollect that an essential key ingredient of the analysis of any hadron collider data is the utilisation of phenomenological models, and commonly QCD inspired Monte Carlo (MC) event generators are employed.
These are used for calibration, corrections of limited acceptance and resolution effects (\emph{unfolding}), training of machine learning algorithms for event or object classification, extrapolations from the \emph{fiducial} to the \emph{full} phase space, estimates of different background sources and also signal extraction.
Although the implemented models are derived from more fundamental equations like the QCD Lagrangian, a number of model parameters remain poorly known and have to be \emph{tuned} with data.
Also, since most models involve approximations and may be numerically limited, any model needs to be validated, or invalidated, with independent measurements prior to its usage, of course. 
With more and more data being recorded at the (HL-)LHC, statistical uncertainties become very small and systematic uncertainties are reduced due to improved calibration and analysis algorithms, so that uncertainties associated to the MC event models become important and are limiting the accuracy of the HL-LHC measurements.
It must be noted, that the MC parameters should be tuned with data from another experiment in order to avoid a potential bias of the actual measurement due to experimental correlations.

For the second bullet, the phenomenological interpretation of hadron collider measurements, like for instance tests of pQCD or the determination of SM parameters (e.g.~\asmz, \sinleff, $m_W$, the $\kappa$ parameters, \ldots), the proton PDFs and SM parameter which are input to the prediction must be known with high accuracy, most noteworthy the value of \asmz.

The most important inputs of the LHeC to the HL-LHC measurements are of course the precise determination of the PDFs and~\asmz, see Chapter~\ref{chapter:pdf}. These will improve both, the data analysis and its interpretation.
Beyond that, the measurements of charged particle spectra, jet shape and jet substructure observables, jet cross sections and event shape observables or heavy flavor cross sections will help to improve MC models further, for instance with the determination of charm and bottom-quark masses, heavy quark ($c$, $b$) fragmentation functions and fragmentation fractions, finding optimal choices for all scales involved in a MC model, or determining the optimal parameters for the parton shower. 
Such measurements can be performed with high precision at the LHeC, since DIS represents a superior QCD laboratory.
This is because in the final state there is always a lepton, which is used for trigger and vertexing, and simultaneously a hadronic system which is then subject of interest.
In addition, the overconstrained kinematic system allows for the precise calibration of hadronic final state objects, and furthermore limiting effects like minimum bias definition or pile-up are absent.



In the following, a few selected subjects are discussed at hand of LHC analyses performed with Run-I data at $\sqrt{s}=8$ or 13\,\TeV, and thus giving a tangible indication about challenges at future HL-LHC measurements:
\begin{itemize}
  %
\item
    \begin{figure}[!tb]
    \centering
    \includegraphics[width=0.46\textwidth]{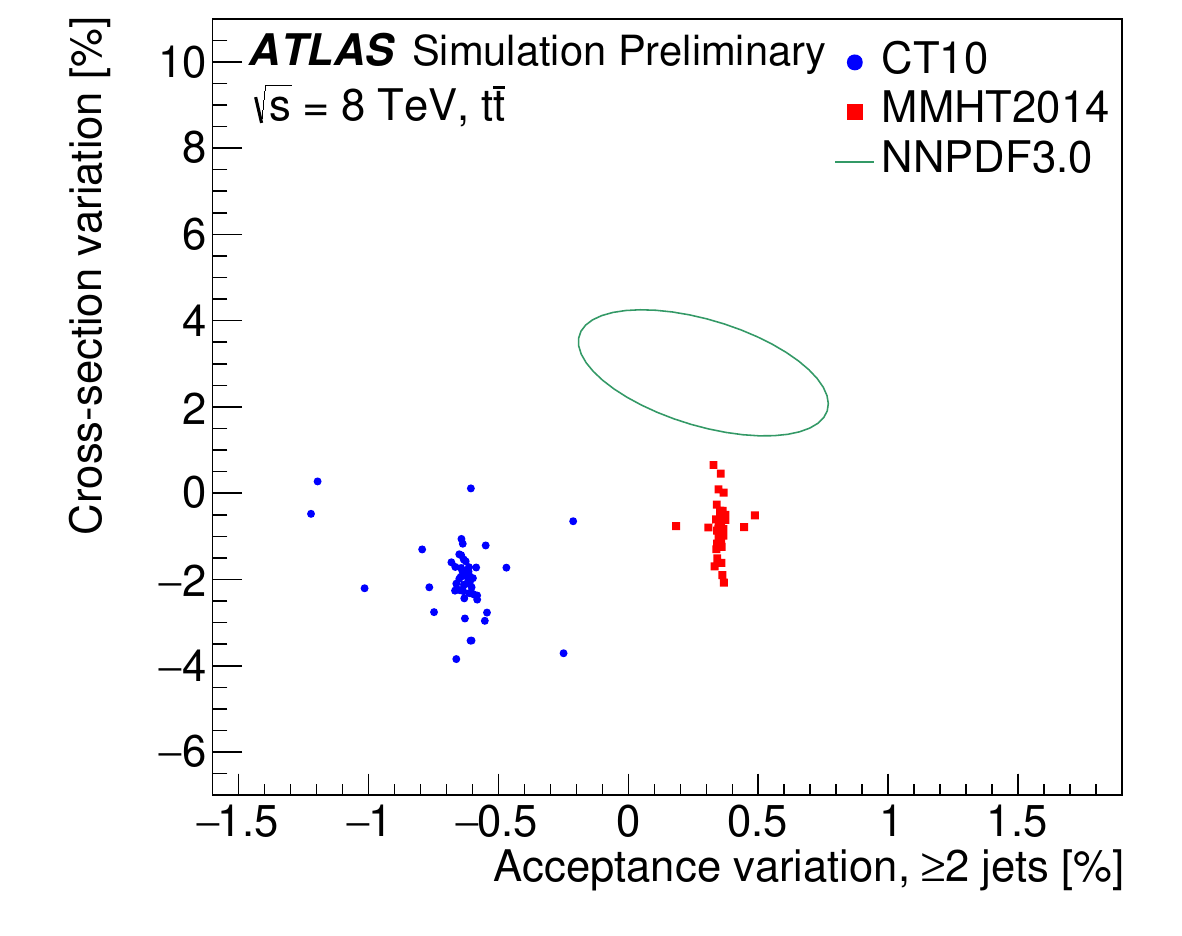}
    \hspace{0.02\textwidth}
    \includegraphics[width=0.46\textwidth]{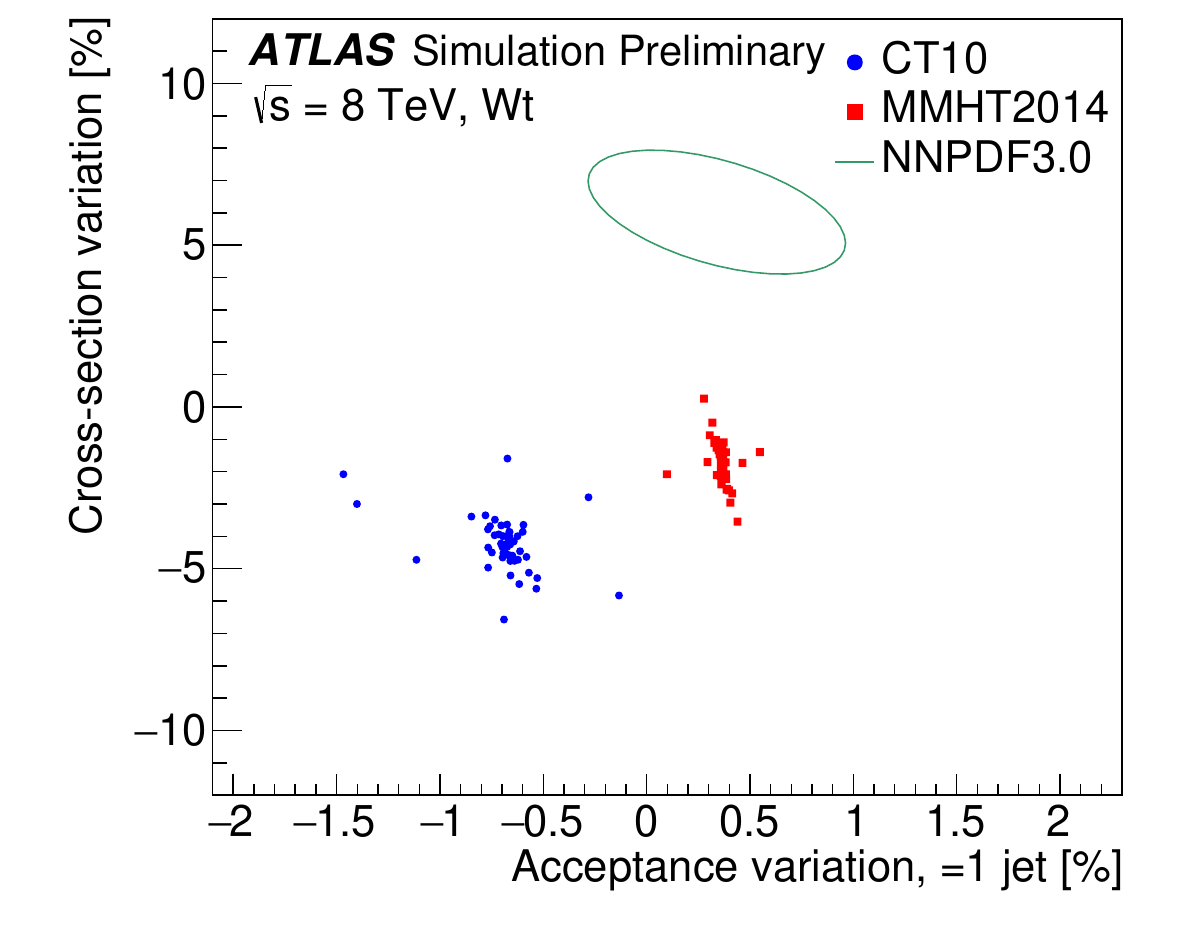}
    \caption{
      Impact of PDF uncertainty from CT10 and MMHT2014 eigenvectors or NNPDF3.0 replicas, on the cross section and the acceptance correction for top pair production $t\bar{t}$  (left) and single top production $Wt$ (right) (taken from Ref.~\cite{ATL-PHYS-PUB-2015-010}).
      Events are selected with at least two jets or with exactly one jet, respectively.
      Depending on the PDF set and eigenvector employed, the cross sections varies by up to 5--7\,\% for top-pair and more than 10\,\% for single-top production. Also the acceptance correction varies by about 0.5--1\,\% for different PDF sets, and can become as large as 2.5\,\% for different PDF sets and eigenvectors. Since the acceptance correction has to be imposed for the measurement, the limited knowledge of the PDFs introduces a sizeable modelling uncertainty on the measurement.
    }
    \label{fig:top_acceptance}
  \end{figure}
  The measurement of the integrated top-quark pair cross section represents an outstanding benchmark quantity for the entire field of top-quark physics.
  Its measurement for top-transverse momenta $p_\text{T}^t>400\,\GeV$ in the \emph{lepton+jets} decay channel yields a high experimental precision with both, small statistical and systematic uncertainties. However, its measurement precision is limited by theoretical uncertainties (also called \emph{modelling} uncertainties), and the largest individual source stems from the PDFs~\cite{Khachatryan:2016gxp,Tanabashi:2019}.
  A related study of PDF effects on the acceptance correction for the integrated top-pair production cross section and single-top production $Wt$ is displayed in Fig.~\ref{fig:top_acceptance}.
  The acceptance correction changes by up to  0.5--1\,\% for different PDF sets, and can become as large as 2.5\,\% for different PDF sets and eigenvectors.
  Another very important uncertainty for top-quark measurements is from the modelling of the parton shower.
  Both uncertainties from the PDFs and from parton shower modelling, are expected to be significantly reduced with LHeC data.
\item The determination of the top-quark mass $m_t$ from LHC data requires the precise modelling of the physics and all background processes with suitable MC models.
  Today, the value of $m_t$ is determined most precisely from a combination of such individual analyses, and uncertainties of 0.4--0.8\,\GeV are reported~\cite{TevatronElectroweakWorkingGroup:2016lid,Khachatryan:2015hba,Aaboud:2016igd,Sirunyan:2017huu,Sirunyan:2018gqx,Tanabashi:2019}.
  Any of these individual precision determinations are limited by model uncertainties, and therefore improvements at the HL-LHC cannot be obtained with more data, but only with improved models.
  Some of the model uncertainties, e.g.\ PDF, parton shower, hadronisation or fragmentation related uncertainties can expected to be reduced with LHeC data. 
\item At the HL-LHC also rare decay channels can be exploited for precision measurements. For example, the top-quark mass can be determined from top-quark pair production with a subsequent decay, where one $b$-quark hadronises into a $B$-hadron which then decays through a $J/\psi$-meson into a pair of muons, $t\bar{t}\rightarrow W^+bW^-b \rightarrow \ell\nu_\ell J/\psi(\rightarrow \mu^+\mu^-)X qq^\prime b $~\cite{ATL-PHYS-PUB-2018-042}.
  Such a measurement requires the precise knowledge of $b$-quark fragmentation, which can be well measured at the LHeC, and will thus improve the HL-LHC measurement.
\item The value of the strong coupling constant \asmz is one of the least known fundamental parameters in physics and an improved determination with new measurement constitutes a real challenge for LHC an HL-LHC experiments.
  A large number of observables at the LHC are \emph{per-se} sensitive to \asmz, and its value was determined in the past from various definitions of jet cross section observables (see e.g.~\cite{Khachatryan:2016mlc,Britzger:2017maj,Aaboud:2018hie,Rabbertz:2017ssq}) or transverse energy-energy correlations~\cite{Aaboud:2017fml}, $Z$+jet cross sections~\cite{Johnson:2017ttl}, integrated~\cite{Klijnsma:2017eqp} or differential top-quark cross sections~\cite{Sirunyan:2019zvx}, inclusive $W$ or $Z$ production~\cite{Sirunyan:2019crt,dEnterria:2019aat}, prompt photon data~\cite{Bouzid:2017uak}, and many other observables (see Ref.~\cite{Tanabashi:2019} for a review).
  Although the harsh environment in high-luminosity hadron-hadron collisions requires sophisticated analysis techniques and dedicated measurements, small experimental uncertainties for \asmz\ could be achieved.
  Hence, \as\ determinations are nowadays limited due to theoretical uncertainties and
  the dominant uncertainties are most commonly PDF related~\cite{Johnson:2017ttl,Klijnsma:2017eqp,Sirunyan:2019crt,dEnterria:2019aat} (only for observables where NNLO predictions are not yet applicable, the scale uncertainties may overshoot the PDF uncertainties).
  Therefore, already today the knowledge of the PDFs represent the limiting factor, and a significant reduction of the total uncertainty for \asmz\ can (only) be achieved with PDFs determined at the LHeC.
  \begin{figure}[!tb]
\centering{\includegraphics[width=0.55\textwidth]{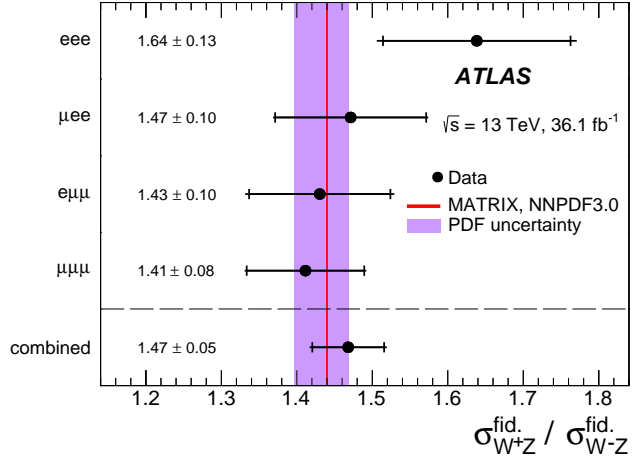}}
\caption{Measurement of the ratio of di-boson $\sigma(W^+Z)/\sigma(W^-Z)$ integrated cross sections in a fiducial phase space for four different decay channels and their combination at $\sqrt{s}=13\,\TeV$
  in comparison with NNLO predictions~\cite{Grazzini:2016swo,Grazzini:2017mhc} (taken from Ref.~\cite{Aaboud:2019gxl}).
  The total uncertainties of the data points are dominated by statistical uncertainties and will be reduced in the future.
  The shaded violet band indicates the size of the PDF uncertainties that limits the overall interpretation of the measurement.
  }
  \label{fig:WZ_ratio}
\end{figure}
\item The production of $W^\pm Z$ pairs in $pp$ collisions provides a crucial test of the electroweak sector of the SM, since di-boson production is sensitive to the gauge-boson self-interactions.
  Already small deviations in the observed distributions could provide indications for new physics.
  The process can be well measured in a high-pile up environment and can be well separated from its huge QCD background. However, due to the relatively small $W^\pm Z$ cross sections high statistical precision can only be achieved with high luminosity.
  Recent measurement of  $W^\pm Z$ pairs at $\sqrt{s}=13\,\TeV$ based on 36\,\ifb of integrated luminosity have been performed by ATLAS and CMS~\cite{Sirunyan:2019bez,Aaboud:2019gxl}.
  In Fig.~\ref{fig:WZ_ratio} the ratio of fiducial cross sections $\sigma_{W^+ Z}/\sigma_{W^-Z}$ is displayed.
  The  largest individual uncertainty is the statistical uncertainty and therefore future measurements at the LHC and HL-LHC are of great importance in order to reach higher precision.
  Nontheless, already today, the overall phenomenological interpretation is limited by PDF uncertainties, as visible from Fig.~\ref{fig:WZ_ratio}, and these can be improved best with PDFs from LHeC.
\end{itemize}

In the situation of the absence of indications for new physics, an important goal of the future LHC and HL-LHC physics programme has to be devoted to precision measurements.
From the examples discussed above ($W$-boson mass and Higgs measurements are discussed in previous sections), it becomes obvious that limiting factors of such measurements arise from the signal and MC modelling, where PDF uncertainties constitute a limiting factor, and also improved understandings of parton shower, hadronisation and fragmentation processes are of importance.
These aspects can all be ameliorated with independent precision measurements at the LHeC.

Similarly, the phenomenological interpretation of many processes is already today limited by PDF uncertainties, and as outlined, \as determinations, di-boson processes, top-mass or top-cross section measurements,  and many other topics, require a higher precision for PDFs.
In the HL-LHC era, where data and predictions are more precise, the detailed knowledge of the PDFs will become of even greater importance.

\section{High Mass Searches at the LHC \ourauthor{Uta Klein}}

\subsection{Strongly-produced supersymmetric particles}

The potential of the HL- and HE-LHC to discover supersymmetry was extensively discussed in Ref.~\cite{CidVidal:2018eel}. Here we focus on searches for gluinos within MSSM scenarios. Gluino pairs are produced through the strong interaction, and their production cross section is relatively large; naturalness considerations indicate that gluino masses should not exceed a few TeV and lie not too far above the EW scale. Hence, they are certainly among the first particles that could be discovered at the HL-LHC.

In the following we assume that a simplified topology dominates the gluino decay chain, culminating in jets plus missing energy originating from a massless LSP, $\tilde{\chi}_0$. Ref.~\cite{CidVidal:2018eel} evaluated the sensitivity of the HL- and HE-LHC to gluino pair production with gluinos decaying exclusively to $q\bar{q}\tilde{\chi}_0$, through off-shell first and second generation squarks, using a standard search for events with jets and missing transverse energy. Currently, the reach for this simplified model with 36\,fb$^{-1}$ of 13\,TeV data is roughly 2\,TeV gluinos, for a massless LSP~\cite{Aaboud:2017vwy,Sirunyan:2018vjp}. Extrapolating to 3\,ab$^{-1}$ at 14\,TeV, the limit grows to 3.2\,TeV. For 15\,ab$^{-1}$ at 27\,TeV, a limit of 5.7\,TeV was found.

When deriving limits, an overall systematic uncertainty of 20\% was assumed on the SM background contributions, and a generic 10\,\% uncertainty was assumed on the signal normalisation, not taking into account PDF-related uncertainties which are as large as 50\% for gluinos around 3\,TeV. The effect of this additional source of uncertainty was found to induce a variation in the mass limit by $\pm$200\,GeV at the HL-LHC, and as much as $\pm$500\,GeV at the HE-LHC.

We can revert this argument, and claim that with present PDF knowledge, mass limits could be as low as 3.0\,TeV and 5.3\,TeV at the HL- and HE-LHC, respectively. Data from the LHeC would make this contribution negligible compared to other sources of uncertainty.
Compared to the most conservative scenario, the increase in sensitivity would correspond to an increase in centre-of-mass energy by approximately 5 to 10\,\%.


\subsection{Contact interactions}

New, high-mass gauge bosons are most often searched for in resonant final states. Peaks in the invariant-mass distributions of electron, muon or jet pairs directly reflect the presence of such new particles; the accessible mass range is limited by the available centre-of-mass energy.

Particles with a mass beyond the kinematic limit generally interfere with the $Z$ boson and the photon, generating non-resonant deviations in the invariant mass distributions. Such models can be parameterised as contact interactions (CI) between two initial-state quarks and two final-state leptons of given chirality:
\begin{equation}
\mathcal{L}_\text{CI} = \frac{g^2}{\Lambda^2} \eta_{ij} (\bar{q}_i \gamma_\mu q_i) (\bar{\ell}_i \gamma^\mu \ell_i),
\end{equation}
where $i,j = $ L or R (for left- or right-handed chirality), $g$ is a coupling constant set to be $4\pi$ by convention, and $\Lambda$ is the CI scale. The sign of $\eta_{ij}$ determines whether the interference between the SM Drell–Yan (DY) process, $q\bar{q}\rightarrow Z/\gamma^*\rightarrow \ell^+\ell^-$, is constructive or destructive. 

The size and sign of the observed deviation with respect to the SM probes the scale and interference pattern of the interaction. The sensitivity of the search is limited by experimental uncertainties (finite statistics and experimental systematic uncertainties) and by uncertainties in the theoretical modelling of the DY background.

The most recent results of the ATLAS and CMS Collaborations~\cite{Aaboud:2017buh,Sirunyan:2018ipj} are based on $e^+e^-$ and $\mu^+\mu^-$ final states in 36\,fb$^{-1}$ of data, and probe CI's up to a typical scale of 25\,TeV, depending on the chirality and sign of the interaction coupling parameter. The limits derived by ATLAS, summarised in Tab.\,\ref{tab:cilimits}, accounted for theoretical uncertainties induced by the PDFs and by \as. The dominant PDF uncertainty was estimated from the 90\% CL uncertainty in the CT14nnlo PDF set, adding an envelope from the comparison of the CT14nnlo, MMHT2014 and NNPDF3.0~\cite{Ball:2014uwa} central sets. The strong coupling constant uncertainty was propagated assuming $\as=0.118\pm0.003$, with a subleading effect.

The present study evaluates the sensitivity of this search at the HL-LHC. The increase in sensitivity is estimated using samples of Standard-Model like pseudo data, corresponding to the integrated luminosity of 3\,ab$^{-1}$. In a first step, both the experimental and theoretical systematic uncertainties are kept in the publication. In this regime, the extrapolated statistical uncertainty is typically a factor 5 to 10 smaller than the theoretical uncertainty. Improvements from the LHeC in \as\ and in the proton PDFs are incorporated in a second step. Assuming the prospects described in Chapter~\ref{chapter:pdf}, \as and PDF uncertainties are smaller than the statistical fluctuations and can be neglected in a first approximation.

The results are summarised in Tab.\,\ref{tab:cilimits}. Everything else equal, increasing the sample size from 36\,fb$^{-1}$ to 3\,ab$^{-1}$ enhances the CI reach by a typical factor of two. Accounting for the improvement in the theoretical modelling of the DY process brought by the LHeC brings another factor of 1.5--1.8 in the limits. In the last case, the limits reach well into range directly accessible with proton-proton collisions at $\sqrt{s}=100$\,TeV, as envisioned at the FCC-hh.

\begin{table}[ht]
  \centering
  \small
\begin{tabular}{lcccccc}
\toprule
Model         & ATLAS\,\footnotesize{(Ref.\,\cite{Aaboud:2017buh})}    &  \multicolumn{2}{c}{HL-LHC} \\
\cmidrule(lr){2-2 }\cmidrule(lr){3-4}
& $\mathcal{L} = 36\,\text{fb}^{-1}$ (CT14nnlo) & $\mathcal{L} = 3\,\text{ab}^{-1}$ (CT14nnlo) & $\mathcal{L} = 3\,\text{ab}^{-1}$ (LHeC) \\
\midrule
LL (constr.) & 28\,TeV & 58\,TeV & 96\,TeV \\
LL (destr.) & 21\,TeV & 49\,TeV & 77\,TeV \\
RR (constr.) & 26\,TeV & 58\,TeV & 84\,TeV \\
RR (destr.) & 22\,TeV & 61\,TeV & 75\,TeV \\
LR (constr.) & 26\,TeV & 49\,TeV & 81\,TeV \\
LR (destr.) & 22\,TeV & 45\,TeV & 62\,TeV \\
\bottomrule
\end{tabular}
\caption{Contact interaction limits from ATLAS based on 36\,fb$^{-1}$ of data~\cite{Aaboud:2017buh}, and extrapolated to the full HL-LHC dataset (3\,ab$^{-1}$). The extrapolation is performed assuming the same PDF and $\as$ uncertainties as in Ref.~\cite{Aaboud:2017buh}, and assuming the improved uncertainties as obtained from the LHeC.\label{tab:cilimits}}
\end{table}

\section{PDFs and the HL-LHC and the LHeC}
\label{sec:HLLHCPDFs}
      

As discussed in the previous Sections, a precise determination of PDFs is an essential ingredient for the success of the HL-LHC. Conversely, the HL-LHC itself offers a significant opportunity to
improve our understanding of proton structure. In this Section we will discuss the possibilities that the combination of HL-LHC ad LHeC measurements offer for the determination of PDFs in the proton.

\subsection{PDF Prospects with the HL-LHC and the LHeC}
In~Ref.\,\cite{Khalek:2018mdn} the HL-LHC potential to constrain PDFs was analysed in detail, focussing on SM processes that are expected to have the most impact at higher $x$. In particular, projections for the production of top quark pairs, inclusive jets, forward $W$ + charm quark and direct photons, as well as forward and high--mass Drell-Yan and the $Z$ boson $p_\perp$ distribution were included. It was found that PDF uncertainties on LHC processes can be reduced by a factor between two and five,
depending on the specific flavour combination and on the optimistic assumptions about the reduction of the (experimental) systematic uncertainties. 

It is of interest to compare these constraints with those expected to come from the LHeC itself, as well as potential improvements from a combined PDF fit to the HL-LHC and LHeC datasets; this was studied in~\cite{AbdulKhalek:2019mps}. The basic procedure consists in generating HL-LHC and LHeC pseudodata with the  PDF4LHC15 set~\cite{Butterworth:2015oua} and then applying Hessian PDF profiling~\cite{Paukkunen:2014zia,Schmidt:2018hvu}, in other words a simplified version of a full refit, to this baseline to assess the expected impact of the data. While the HL-LHC datasets are described above,  the LHeC pseudodata correspond to the most recent publicly available official LHeC projections, see Section~\ref{sect:DISdata}, for electron and positron neutral-current (NC) and charged-current (CC) scattering.  As well as inclusive data at different beam energies ($E_p=1,7$\,TeV), charm and bottom heavy quark NC and charm production in $e^- p$ CC scattering are included.

\begin{figure}[th!b]
  \centering
  \includegraphics[width=0.45\textwidth]{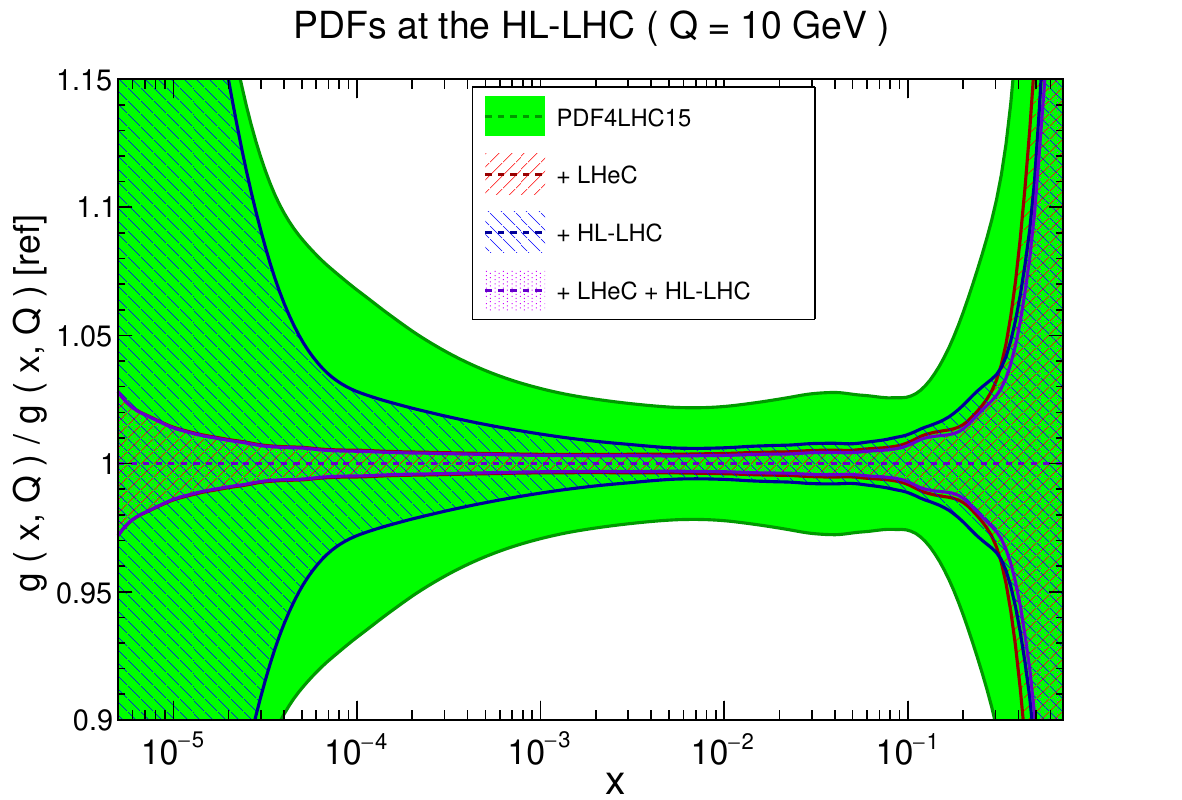}
  \includegraphics[width=0.45\textwidth]{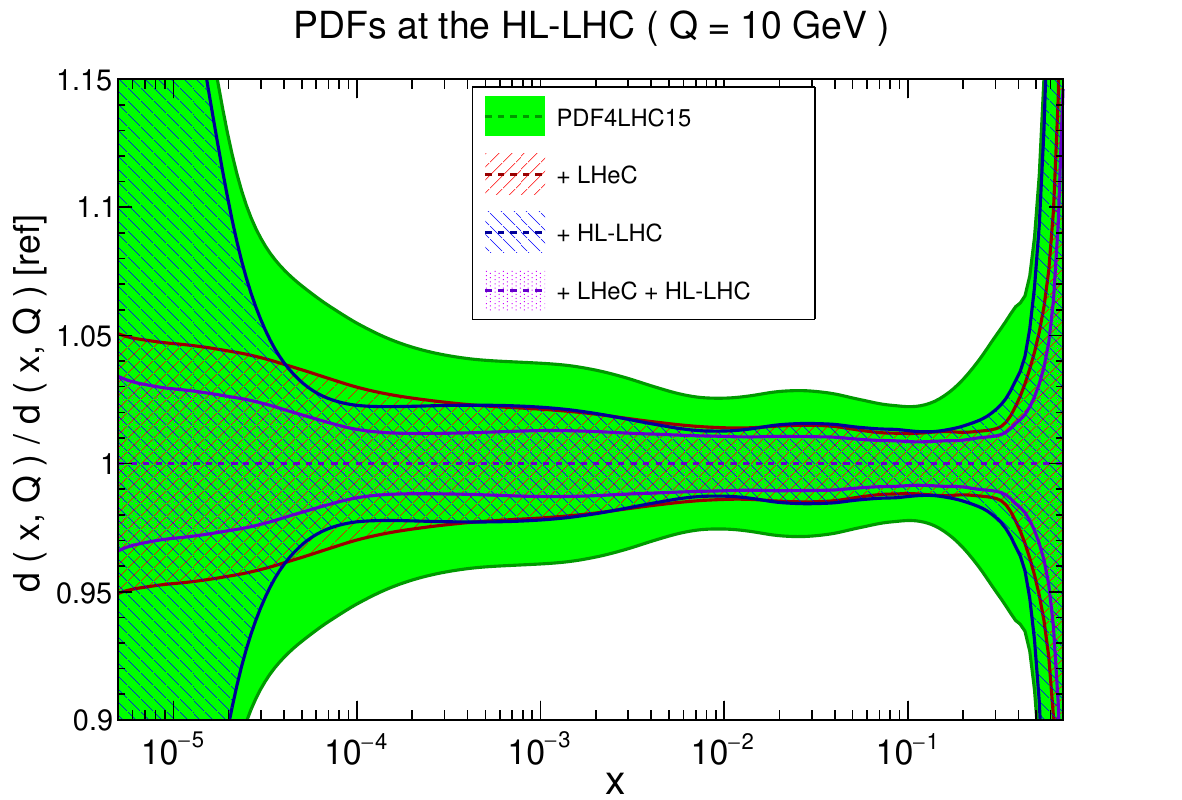}
  \includegraphics[width=0.45\textwidth]{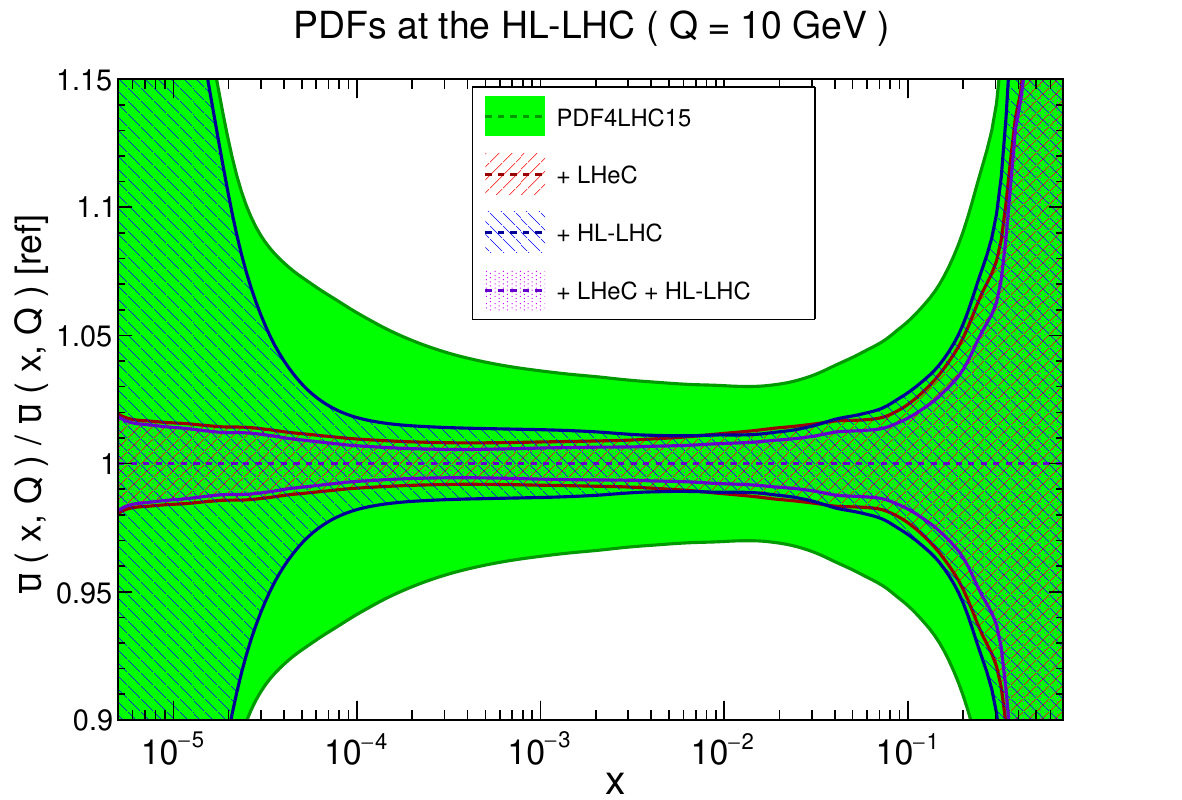}
  \includegraphics[width=0.45\textwidth]{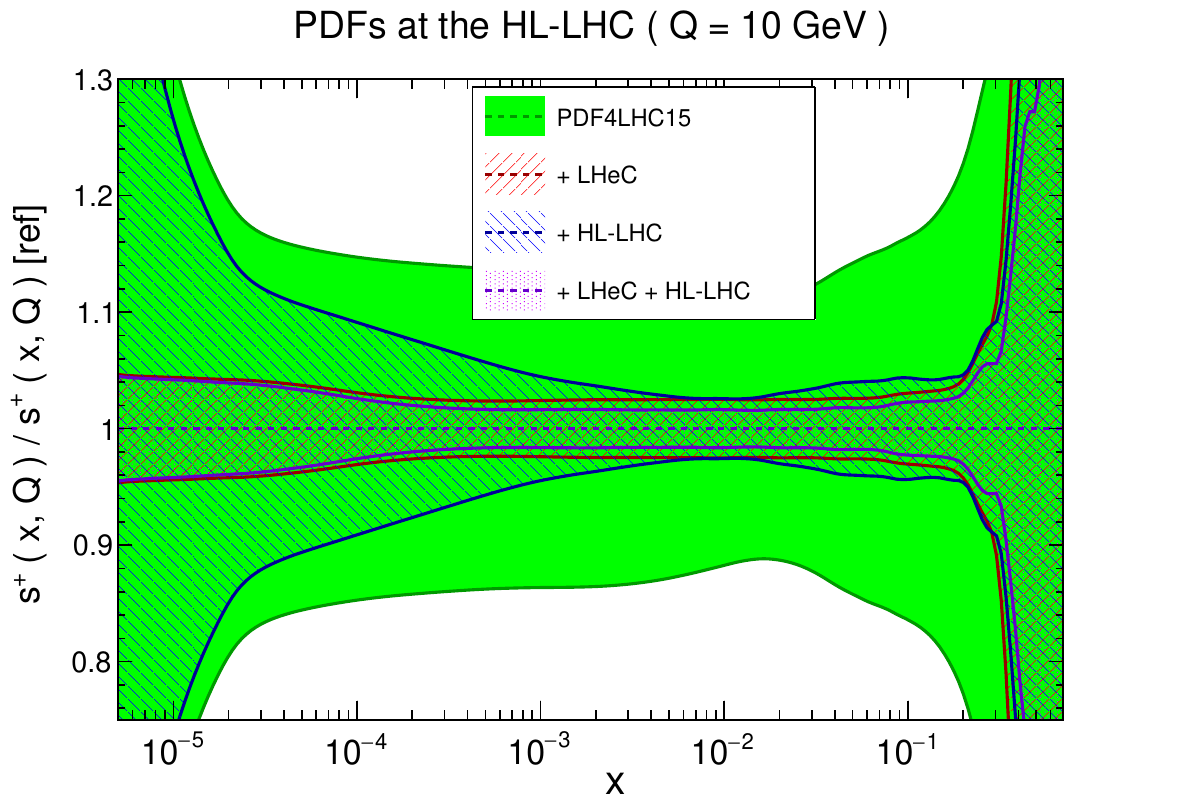}
  \caption{Impact of LHeC on the 1-$\sigma$ relative PDF uncertainties of the gluon, down quark, anti--up quark and strangeness distributions, with respect to the PDF4LHC15 baseline set (green band). Results for the LHeC (red), the HL-LHC (blue) and their combination (violet) are shown.}
  \label{fig:pdf_lhec-hlllhc-abs}
\end{figure}

The expected impact of the HL-LHC, LHeC and their combination on the PDF uncertainties of the gluon, down quark, anti--up quark and strangeness distributions are shown in Fig.~\ref{fig:pdf_lhec-hlllhc-abs}.
One observes that at low $x$ the LHeC data place in general by far the strongest constraint, in particular for the gluon, as expected from its greatly extended coverage at small $x$.
At intermediate $x$ the impact of the HL-LHC and LHeC are more comparable in size, but nonetheless the LHeC is generally expected to have a larger impact. 
At higher $x$ the constraints are again comparable in size, with the HL-LHC resulting in a somewhat larger reduction in the gluon and strangeness uncertainty, while the LHeC has a somewhat larger impact for the down and anti--up quark distributions.
Thus, the combination of both HL-LHC and LHeC pseudodata
nicely illustrate a clear and significant reduction in PDF uncertainties over a very wide range of $x$, improving upon the constraints from the individual datasets in a non-negligible way.

\subsection{Parton luminosities at the HL-LHC}
\begin{figure}
\centering
\includegraphics[width=0.45\textwidth]{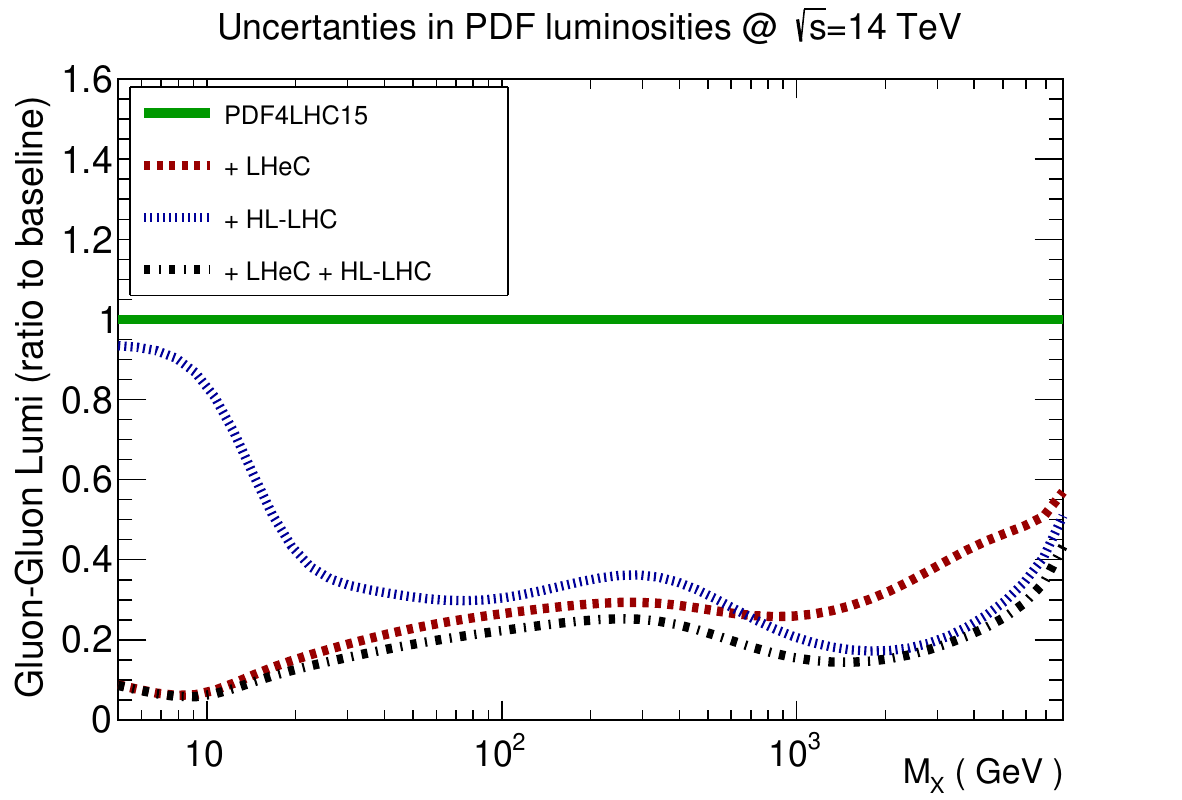}
\includegraphics[width=0.45\textwidth]{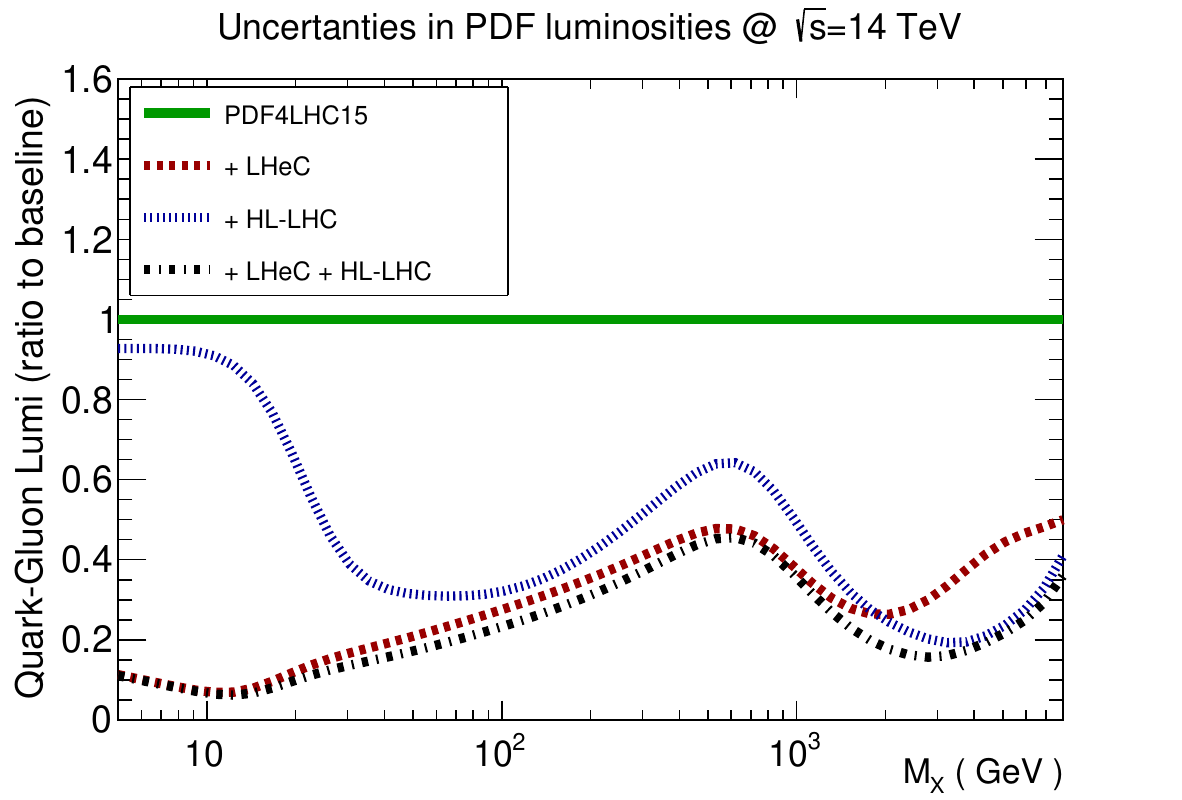}
\includegraphics[width=0.45\textwidth]{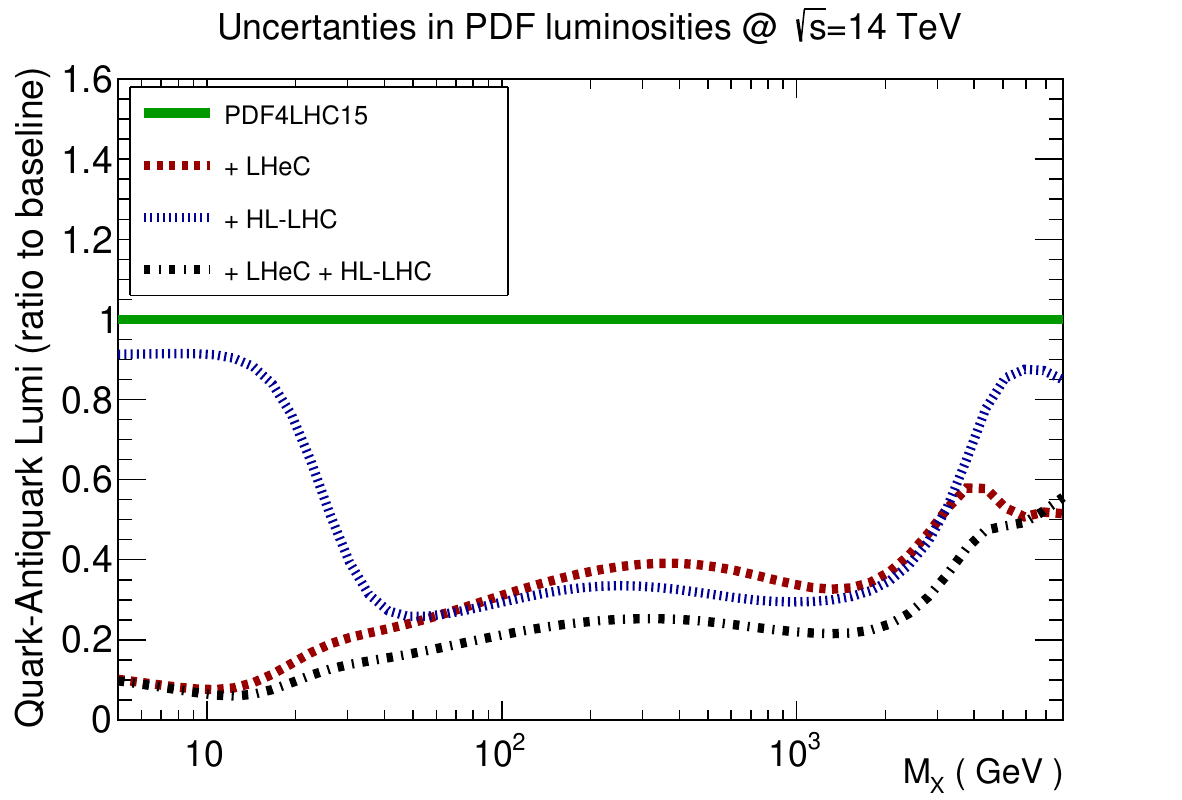}
\includegraphics[width=0.45\textwidth]{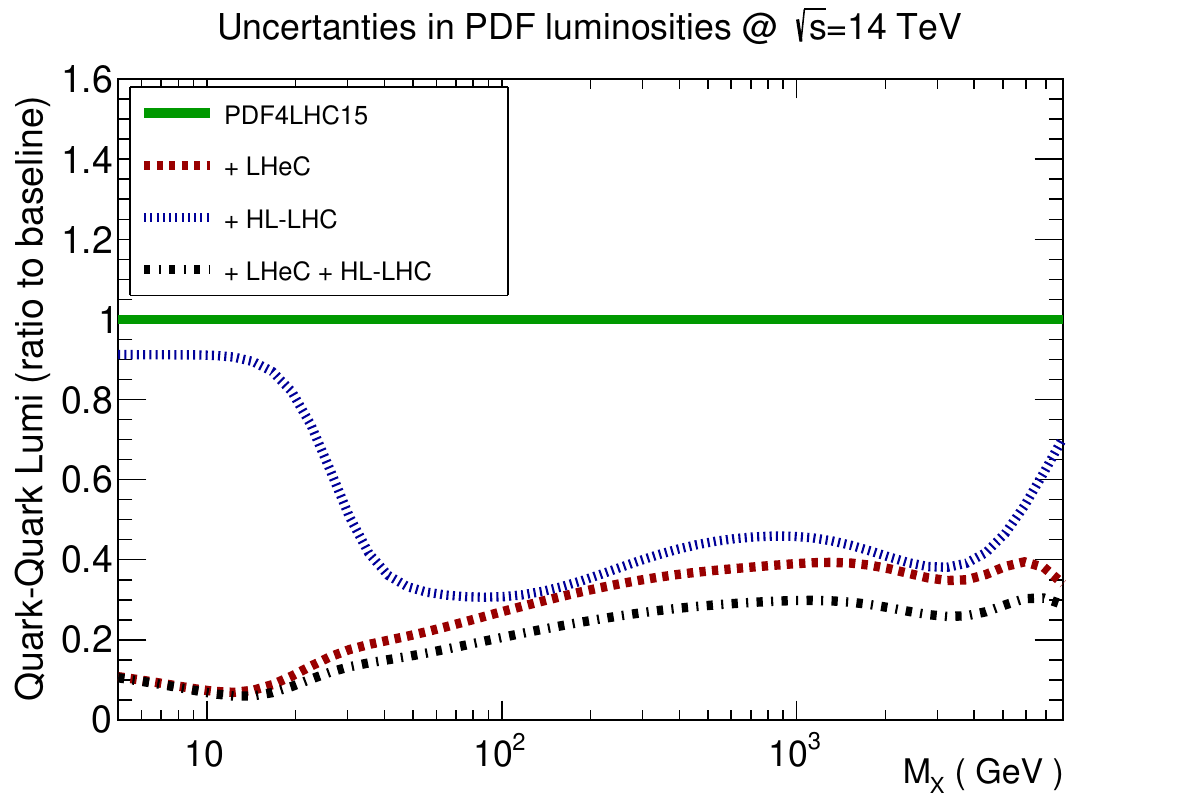}
\caption{Impact of LHeC, HL-LHC and combined LHeC + HL-LHC pseudodata on the uncertainties of the gluon-gluon, quark-gluon, quark-antiquark and quark-quark luminosities, with respect to the PDF4LHC15 baseline set.
  In this comparison we display the relative reduction of the PDF uncertainty in the luminosities compared to the baseline.
}
\label{fig:lumi_lhec-hlllhc}
\end{figure}
In Fig.~\ref{fig:lumi_lhec-hlllhc} we show the impact on the gluon-gluon, quark-gluon, quark-antiquark and quark-quark partonic luminosities for a centre-of-mass energy $\sqrt{s}=14\,\TeV$.
Some clear trends are evident from this comparison, consistent with the results from the individual PDFs. 
We can in particular observe that at low mass the LHeC places the dominant constraint, while at intermediate masses the LHeC and HL-LHC constraints are comparable in size, and at high mass the stronger constraint on the gluon-gluon and quark-gluon luminosities comes from the HL-LHC, with the LHeC dominating for the quark-quark and quark-antiquark luminosities.
As in the case of the PDFs, for the partonic luminosities the combination of the HL-LHC and LHeC constraints
leads to a  clear reduction in the PDF uncertainties in comparison
to the individual cases, by up to an order of magnitude over a wide range of invariant masses, $M_X$, of the produced final state.

In summary, these results demonstrate that while the HL-LHC alone is expected to have a sizeable impact on PDF constraints, the LHeC can improve our current precision on PDFs significantly in comparison to this, in particular at low to intermediate $x$. Moreover, the combination of both the LHeC and HL-LHC pseudodata leads
to a significantly superior PDF error reduction in comparison to the two facilities individually. Further details, including LHeC-only studies as well as an investigation of the impact of the PDF baseline on the uncertainty projections, can be found in Ref.~\cite{AbdulKhalek:2019mps}.

\subsection{PDF Sensitivity: Comparing HL-LHC and LHeC}
While the experimental reach of each facility in the $\{x,Q^2\}$ kinematic plane
provides a useful comparison, 
there are more factors to consider -- especially
when we are striving for ultra-high precision measurements. 
One measure that provides a dimension beyond  the $\{x,Q^2\}$ plane is the \textit{sensitivity}; 
this is a combination of the correlation coefficient times a scaled residual~\cite{Wang:2018heo,Hobbs:2019sut}.
This provides an extra dimension of information in comparison to a 
simple $\{x,Q^2\}$ map and represents a measure of the impact of the data.

\begin{figure}[!th]
  \centering
  \includegraphics[width=0.90\textwidth]{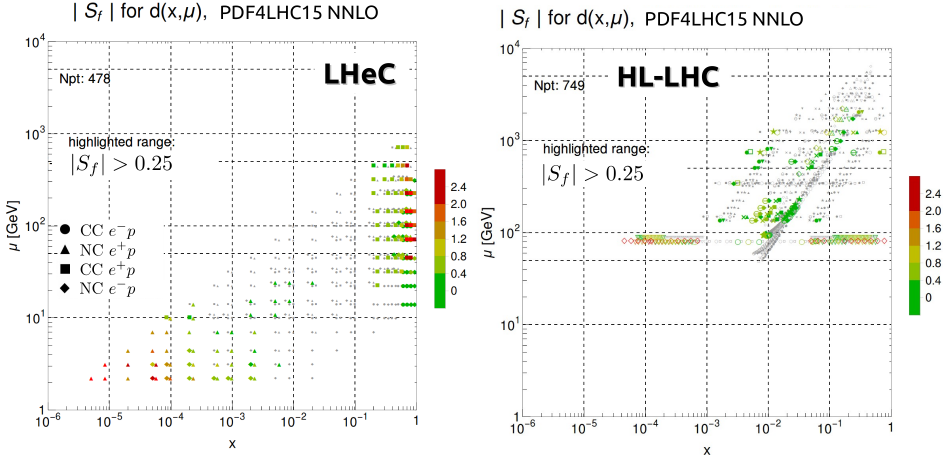}
  \caption{Sensitivity for a sample flavour $\{d(x,Q)\}$ in the  $\{x,Q^2\}$ kinematic plane
    for the LHeC (left) and the HL-LHC (right) calculated with pseudodata~\cite{Hobbs:2019sut}.
    We observe the LHeC is particularly sensitive in both the high and low $x$ regions,
    and the HL-LHC covers the intermediate $x$ region out to large $Q$ scales.     
  }
    \label{fig:pdfs_sensSx}
\end{figure}
%
In Fig.~\ref{fig:pdfs_sensSx} this PDF sensitivity for a sample PDF flavour is displayed for the LHeC and the HL-LHC pseudo-data.
In particular, one observe that the LHeC provides strong sensitivity in  the high-$x$ region, which is of great importance for BSM searches,
and also in the low-$x$ region, which is relevant for QCD phenomena such as saturation.
The HL-LHC provides constraints coming from $W$/$Z$ production ($Q\sim M_{W/Z}$) as well as
from jets at high scales.
The combination of these measurements will provide strongest constraints on
the various PDF flavours across the broad  $\{x,Q^2\}$ kinematic plane.

%


\section{Impact of New Small-$x$ Dynamics on Hadron Collider Physics}
\label{sec:smallx_impact}

As discussed in Subsections~\ref{sec:PSM_Disc_smallx} and \ref{sec:FL}, the presence of new dynamics at small $x$ as claimed in Refs.~\cite{Bonvini:2016wki,Ball:2017otu,Abdolmaleki:2018jln} will have impact on hadronic observables. The impact is stronger for larger energies, therefore more important for the FCC-hh than for the LHC. But it may compete with other uncertainties and thus become crucial for precision studies even at LHC energies. Studies on the impact of non-linear dynamics at hadron colliders have been devoted mainly to photoproduction in UPCs, see e.g.~\cite{BrennerMariotto:2012zz,Coelho:2020lyd,Coelho:2020qeh} and Refs. therein for the case of gauge boson production. In this section we focus on the effect of resummation at small $x$.

While hadronic data like jet, Drell-Yan or top production at existing energies do not have much constraining  power at low $x$~\cite{Bonvini:2016wki} and thus need not be included in the extraction of PDFs using resummed theoretical predictions, this fact does not automatically mean that the impact of resummation is not visible at large scales for large energies.
Indeed the PDFs obtained with small-$x$ resummation may change at low energies in the region of $x$ relevant for hadronic data, thereby giving an effect also at higher energies after evolving to those scales.
A consistent inclusion of resummation effects on hadronic observables is thus crucial for achieving precision.
The difficulty for implementing resummation on different observables lies in the fact that not only evolution equations should include it but also the computation of the relevant matrix elements for the observable must be performed with matching accuracy. 

Until present, the only observable that has been examined in detail is Higgs production cross section through gluon fusion~\cite{Bonvini:2018iwt}. 
Other observables like Drell-Yan~\cite{BGTMDY} or heavy quark~\cite{BSHQ} production are under study and they will become available in the near future.

\begin{figure}[!th]
    \centering
\includegraphics[width=0.7\textwidth]{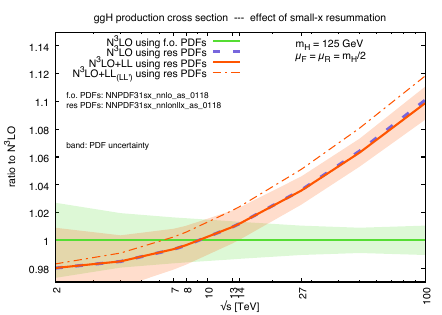}
\caption{Ratio of the N$^3$LO Higgs cross section with and without resummation to the N$^3$LO fixed-order cross section,
    as a function of the collider centre-of-mass energy. ``f.o.'' denotes fixed order, ``res'' denotes resummed and ``LL$^\prime$'' a different anomalous dimension matching at leading logarithmic accuracy, see the legend on the plot and Ref.~\cite{Bonvini:2018iwt} for details.
    The PDFs used are from the global dataset of Ref.~\cite{Ball:2017otu}. Figure taken from Ref.~\cite{Bonvini:2018iwt}.}
\label{fig:resum_Higgs}
\end{figure}

\begin{figure}[!th]
    \centering
    \includegraphics[width=0.42\textwidth]{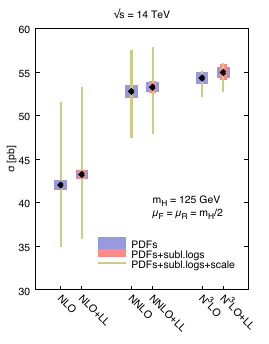}
    \hspace{0.05\textwidth}
    \includegraphics[width=0.42\textwidth]{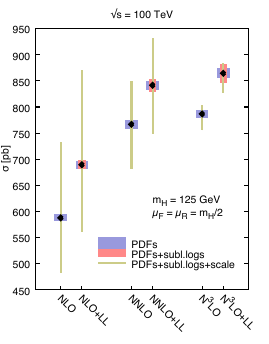}
\caption{Perturbative progression of the Higgs cross section for two collider energies $\sqrt{s}=\left\{14,100\right\}$~TeV.
    In each plot the NLO, NLO+LL, NNLO, NNLO+LL, N$^3$LO and N$^3$LO+LL results are shown.
    The results are supplemented by uncertainty bands from PDF, subleading logarithms and scale uncertainties. Figure taken from Ref.~\cite{Bonvini:2018iwt}.}
\label{fig:resum_Higgs2}
\end{figure}

For $gg\to H$, the LL resummation of the matrix elements matched to fixed order at N$^3$LO was done in Refs.~\cite{Bonvini:2018ixe, Bonvini:2018iwt} and the results are shown in Figs.~\ref{fig:resum_Higgs} and \ref{fig:resum_Higgs2}. Fig.~\ref{fig:resum_Higgs} shows the increasing impact of resummation on the cross section with increasing energy. It also illustrates the fact that the main effect of resummation comes through the modification of the extraction of parton densities and their extrapolation, not through the modification of the matrix elements or the details of the matching.
 
Fig.~\ref{fig:resum_Higgs2} indicates the size of the different uncertainties on the absolute values of the cross section with increasing accuracy of the perturbative expansion, at HL-LHC and FCC-hh energies. For N$^3$LO(+LL) it can be seen that while at the HL-LHC, the effect of resummation is of the same order as other uncertainties like those coming scale variations, PDFs and subleading logarithms, this is not the case for the FCC where it can be clearly seen that it will be the dominant one. Resummation should also strongly affect the rapidity distributions, a key need for extrapolation of observed to total cross sections.
In particular, rapidity distributions are more directly sensitive to PDFs at given values of momentum fraction $x$, and therefore in regions where this momentum fraction is small (large rapidities) the effect of resummation may be sizeable also at lower collider energies.
These facts underline the need of understanding the dynamics at small $x$ for any kind of precision physics measurements at future hadronic colliders, with increasing importance for increasing energies.

Finally, it should be mentioned that a different kind of factorisation, called transverse momentum (TMD) factorisation~\cite{Collins:1984kg,Collins:1989gx,Collins:2011zzd,Angeles-Martinez:2015sea,Diehl:2015uka,Rogers:2015sqa}, may have an effect on large scale observables in hadronic colliders. The extension of the TMD evolution equations towards small $x$~\cite{Balitsky:2016dgz} and the relation of such factorisation with new dynamics at small $x$, either through high-energy factorisation~\cite{Catani:1990xk,Catani:1990eg,Collins:1991ty,Levin:1991ry} or with the CGC~\cite{Gelis:2010nm,Kovchegov:2012mbw}, is under development~\cite{Altinoluk:2019wyu}.

\section{Heavy Ion Physics with \emph{e}A Input  \ourauthor{Nestor Armesto}}
\label{sec:LHeConHLLHC_heavyions}

\begin{figure}[!hbt]
\centering{\includegraphics[width=1.0\textwidth]{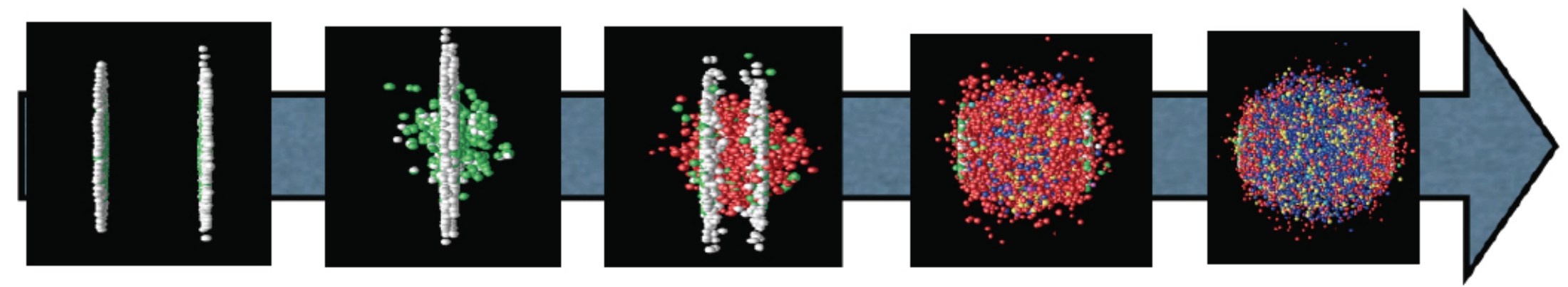}}
\caption{Sketch of a heavy ion collision with time running left to right, going from the approach of two ultrarelativistic Lorentz-contracted nuclei, the collision and parton creation in the central rapidity region, the beginning of expansion and formation of the QGP, the expansion of the QGP until hadronisation, and, finally, the expansion of the hadronic gas.}
\label{fig:NPP_LHeConHLLHC_heavyions_fig1}
\end{figure}

The study of hadronic collisions at RHIC and the LHC, proton-proton, proton-nucleus and nucleus-nucleus, has produced several observations of crucial importance for our understanding of QCD in complex systems where a large number of partons is involved~\cite{Armesto:2015ioy,Busza:2018rrf}. The different stages of a heavy ion collision, as we presently picture it, are schematically drawn in Fig\,\ref{fig:NPP_LHeConHLLHC_heavyions_fig1}.

First, the hot and dense partonic medium created in heavy ion collisions, the quark-gluon plasma (QGP), experiences a collective behaviour of which azimuthal asymmetries and transverse spectra with a specific ordering in particle masses are the most prominent observables. This collectivity can be very well described by relativistic hydrodynamics~\cite{Romatschke:2017ejr}. For this description, the system has to undergo some dynamics leading to rough isotropisation in a short time, $\lesssim 1$ fm/c, for which both strong and weak coupling explanations have been proposed~\cite{Romatschke:2016hle}.

Second, collisions between smaller systems, $pp$ and $p$A, show many of the features~\cite{Schlichting:2016sqo,Loizides:2016tew,Schenke:2017bog} that in heavy ion collisions are taken as indicative of the production of a dense hot partonic medium. The most celebrated of such features, the long rapidity range particle correlations collimated in azimuth, named the ridge (see Sect.~\ref{sect:ridge}), has been found in all collisions systems. The dynamics underlying this phenomena, either the formation of QGP and the existence of strong final state interactions, or some initial state dynamics that leaves imprint on the final observables, is under discussion~\cite{Romatschke:2016hle}.

Finally, the QGP is extremely opaque to both highly energetic partons~\cite{Mehtar-Tani:2013pia} and quarkonia~\cite{Andronic:2015wma} traversing it. These observables, whose production in $pp$ can be addressed through perturbative methods, are called hard probes~\cite{Proceedings:2018mei}. The quantification of the properties of the QGP extracted through hard probes is done by a comparison with  predictions based on assuming a nuclear collision to be a superposition of collisions among free nucleons. Such predictions contain uncertainties coming  both from nuclear effects other than those in QGP (named cold nuclear matter effects), and from uncertainties in the dynamics determining the interaction between the energetic parton or bound state and the medium. In the case of partons, this has motivated the development of sophisticated jet studies in heavy ion collisions~\cite{Andrews:2018jcm}.

$e$A collisions studied in the energy range relevant for the corresponding hadronic accelerator -- the LHeC for the LHC -- would substantially improve our knowledge on all these aspects and, indeed, on all stages of a heavy ion collisions depicted in Fig.\,\ref{fig:NPP_LHeConHLLHC_heavyions_fig1}. Besides, they can reduce sizeably the uncertainties in the extracted QGP parameters, the central goal of the heavy program for the understanding of the different phases of QCD. Here we provide three examples of such synergies:

\begin{itemize}

\item Nuclear parton densities: ~The large lack of precision presently existing in the determination of parton densities induce large uncertainties in the understanding of several signatures of the QGP. For example, for J/$\psi$ suppression, its magnitude at midrapidity at the LHC is compatible with the sole effect of nuclear shadowing on nPDFs~\cite{Andronic:2015wma}, see Fig.\,\ref{fig:NPP_LHeConHLLHC_heavyions_fig2}. While from data at lower energies and at forward and backward rapidities it is clear that this is not the only effect at work, only a reduction on the nPDF uncertainty as feasible at the LHeC , see Sect.~\ref{sec:NPP_nPDFs}, will make possible a precise quantification of the different mechanisms producing either suppression (screening, gluon dissociation, energy loss) or enhancement (recombination or coalescence), that play a role for this observable.

\begin{figure}[!hbt]
\centering{\includegraphics[width=0.65\textwidth]{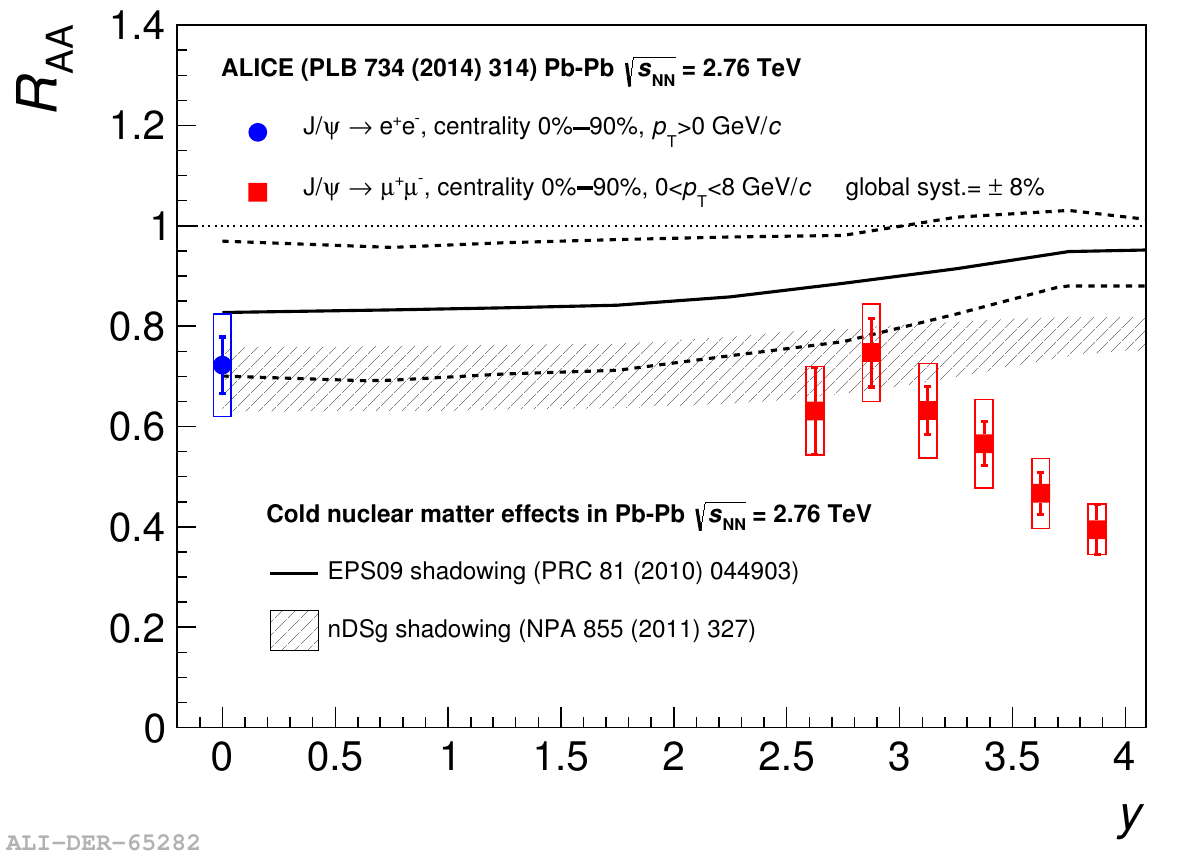}}
\caption{ALICE inclusive J/$\psi$ nuclear modification factor versus rapidity~\cite{Abelev:2013ila}, 
  compared to nPDF calculations. Taken from~\cite{Andronic:2015wma}.}
\label{fig:NPP_LHeConHLLHC_heavyions_fig2}
\end{figure}

\item Initial conditions for the collective expansion and the small system problem: ~At present, the largest uncertainty in the determination of the transport coefficients of the partonic matter created in heavy ion collisions~\cite{Song:2010mg,Niemi:2015qia} (see Fig.\,\ref{fig:NPP_LHeConHLLHC_heavyions_fig3}), required in hydrodynamic calculations, and in our understanding of the speed of the approach to isotropisation and of the dynamics prior to it~\cite{Liu:2015nwa}, comes from our lack of knowledge of the nuclear wave function and of the mechanism of particle production at small to moderate scales -- i.e.\ the soft and semihard regimes. Both aspects determine the initial conditions for the application of relativistic hydrodynamics. This is even more crucial in the discussion of small systems, where details of the transverse structure of protons are key~\cite{Schenke:2012wb} not only to provide such initial conditions but also to establish the relative role of initial versus final state dynamics. For example, the description of azimuthal asymmetries in $pp$ and $p$Pb collisions at the LHC demands that the proton is modelled as a collection of constituent quarks or hot spots~\cite{Romatschke:2017ejr,Schenke:2012wb}. $ep$ and $e$A collisions at the LHeC can constrain both aspects in the pertinent kinematic region, see Sects.~\ref{sec:PSM_Disc_3D} and \ref{sec:NPP_nonconventional}. Besides, they can clarify the mechanisms of particle production and the possible relevance of initial state correlations on the final state observables as suggested e.g.\ by CGC calculations, see Sects.~\ref{sec:PSM_Disc_smallx} and \ref{sec:NPP_smallx}, whose importance for LHC energies can be established at the LHeC.

\begin{figure}[!hbt]
\centering{\includegraphics[width=0.9\textwidth,trim={80 160 80 160 },clip]{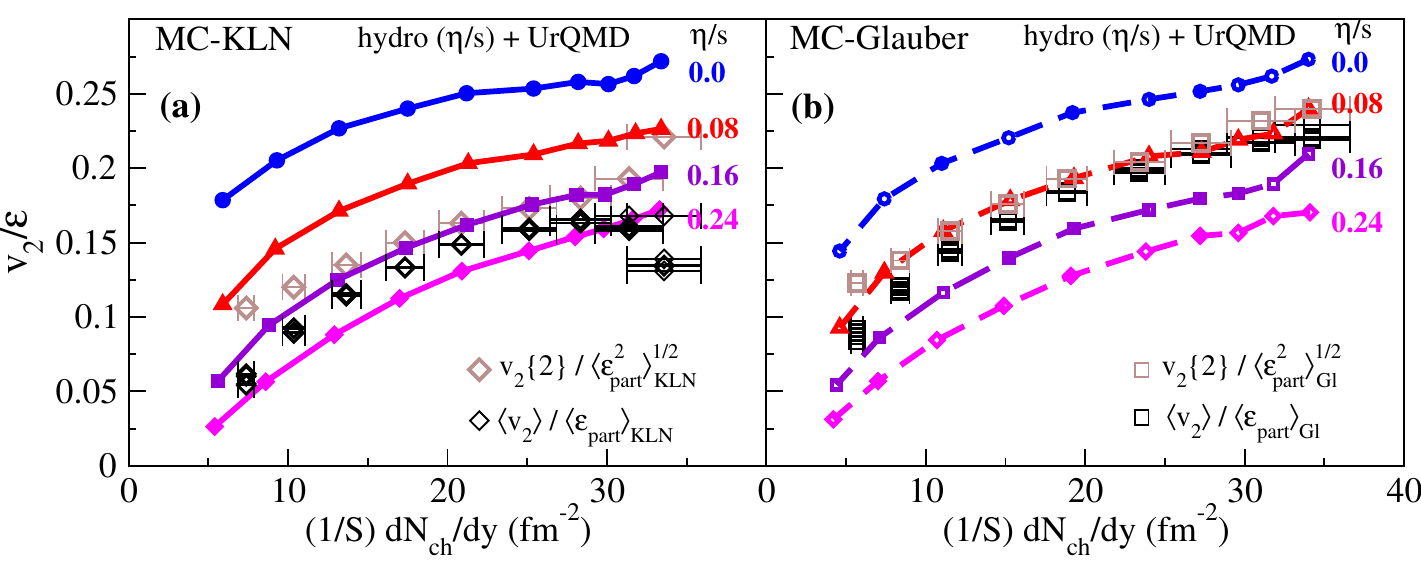}}
\caption{Comparison of the universal
$v_2(\eta/s)/\varepsilon$ vs. $(1/S)(dN_\mathrm{ch}/dy)$ curves
with experimental data for $\langle v_2\rangle$
\protect\cite{Ollitrault:2009ie}, $v_2\{2\}$ \protect\cite{Adams:2004bi},
and $dN_\mathrm{ch}/dy$ \protect\cite{Abelev:2008ab} from the STAR Collaboration.
The experimental data used in (a) and (b) are identical, but the
normalisation factors $\langle \varepsilon_\mathrm{part}\rangle$ and $S$ used
on the vertical and horizontal axes, as well as the factor
$\langle \varepsilon_\mathrm{part}^2 \rangle^{1/2}$ used to normalize
the $v_2\{2\}$ data, are taken from the MC-KLN model in (a) and from
the MC-Glauber model in (b). Theoretical curves are from simulations
with MC-KLN initial conditions in (a) and with MC-Glauber initial
conditions in (b). Taken from~\cite{Song:2010mg}.}
\label{fig:NPP_LHeConHLLHC_heavyions_fig3}
\end{figure}

\item Impact on hard probes: ~Besides the improvement in the determination of nPDFs that affects the quantification of hard probes, commented above, $e$A collisions can help to understand the dynamics of the probes by analysing the effects of the nuclear medium on them. As two examples, the abundant yields of jets and large transverse momentum particles at the LHeC~\cite{AbelleiraFernandez:2012cc} will allow precise studies of the nuclear effects on jet observables and of hadronisation inside the nuclear medium. These two aspects are of capital importance not only in heavy ion collisions but also in small systems where the lack of jet modification is the only QGP-like characteristics not observed in $p$Pb. On the other hand, measurements of exclusive quarkonium production at the LHeC~\cite{AbelleiraFernandez:2012cc} will provide a better understanding of the cold nuclear matter effects on this probe, on top of which the effects of the QGP will provide a  quantitative characterisation of this new form of QCD matter.

\end{itemize}

As discussed in Sect.~\ref{sec:NPP_nPDFs}, $p$Pb and PbPb collisions at the LHC offer possibilities for constraning nPDFs, through the measurement of EW vector boson production~\cite{Sirunyan:2019dox}, dijets~\cite{Eskola:2019dui}, D mesons at forward rapidities~\cite{Eskola:2019bgf} and exclusive charmonium and dijet photoproduction in ultraperipheral collisions~\cite{Guzey:2013qza,Contreras:2016pkc,Guzey:2019kik}. Specifically, dijets in UPCs could constrain nPDFs in the region $10^{-3}\lesssim x\lesssim 0.7$ and $200 \lesssim Q^2\lesssim  10^4$\,GeV$^2$. $e$A collisions would provide more precise nPDFs, whose compatibility with these mentioned observables would clearly establish the validity of collinear factorisation and the mechanisms of particle production  in collisions involving nuclei.

Furthermore, $e$A  offers another system where photon-photon collisions, recently measured in UPCs at the LHC~\cite{Aaboud:2018eph}, can be studied. For example, the observed acoplanarity of the produced muon pairs can be analysed in $e$A in order to clarify its possible origin and constrain the parton densities in the photon.

Finally,  the possible existence of a new non-linear regime of QCD - saturation - at small $x$ is also under study at the LHC, for example using dijets in the forward rapidity region in $p$Pb collisions~\cite{Aaboud:2019oop}. As discussed in Sect.~\ref{sect:ridge}, the ridge phenomenon (two particle correlations peaked at zero and $\pi$ azimuthal angles and stretched along the full rapidity of the detector) observed in all collision systems, $pp$, $p$Pb and PbPb at the LHC, has  been measured in photoproduction on Pb in UPCs at the LHC~\cite{ATLAS:2019gsn}. For the time being, its existence in smaller systems like $e^+e^-$~\cite{Badea:2019vey} at LEP and $ep$ at HERA~\cite{ZEUS:2019jya} has been scrutinised but the results are not conclusive. These studies are fully complementary to those in $ep$ and $e$A, where its search  at the smallest possible values of $x$ at the LHeC would be most interesting. For example, the collision of the virtual photon with the proton at the  LHeC can be
considered as a high energy collision of two jets or  ``flux tubes".

\begin{figure}[!hbt]
\centering{\includegraphics[width=0.75\textwidth]{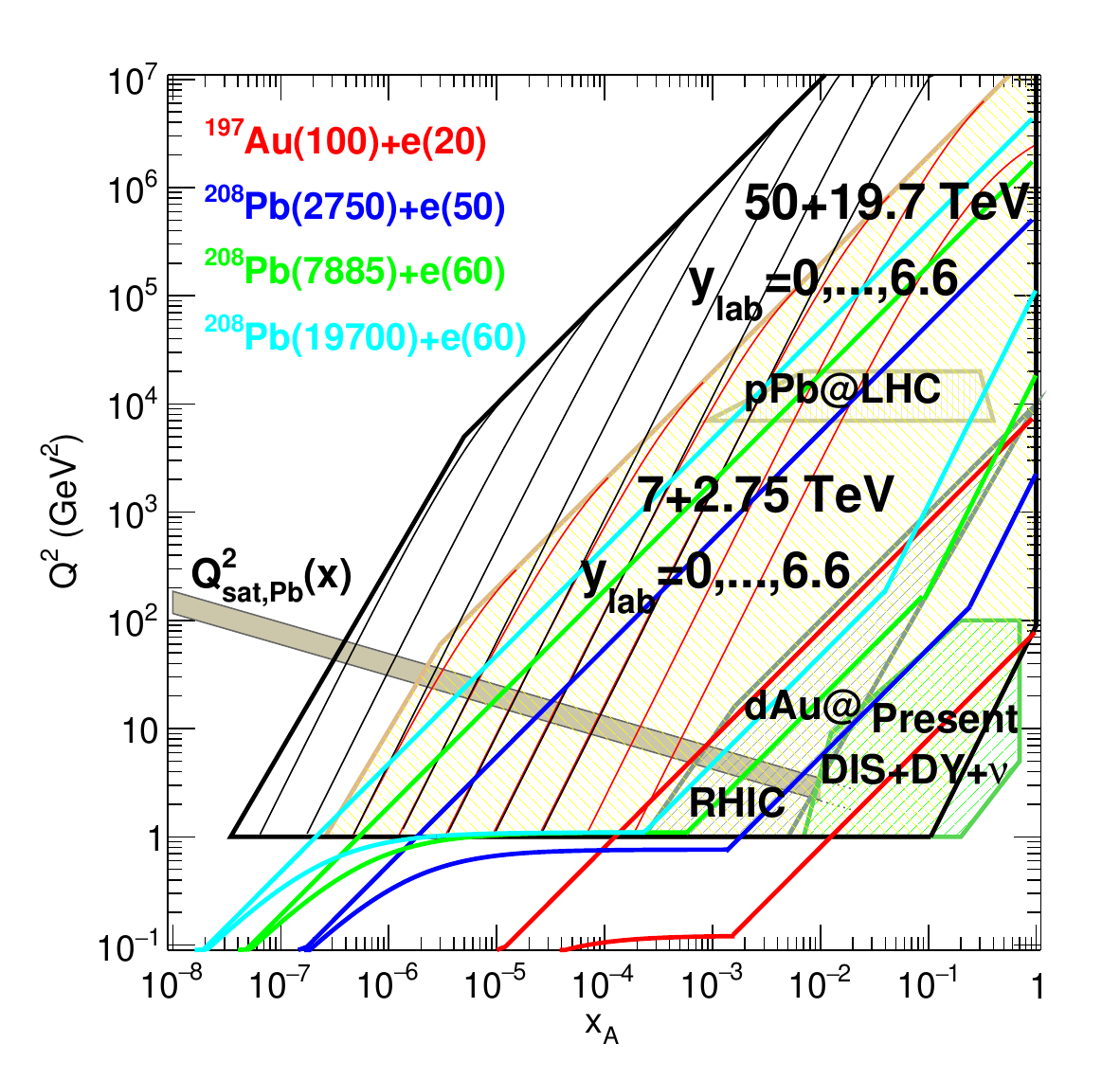}}
\caption{Kinematic regions in the $x-Q^2$ plane explored by data sets (charged lepton and neutrino DIS, DY, $d$Au at RHIC and $p$Pb at the LHC) used in present nPDF analyses~\cite{Eskola:2016oht}, compared to the ones achievable at the EIC (red), the LHeC (ERL against the HL-LHC beams, dark blue) and two FCC-eh versions (with Pb beams corresponding to proton energies of 20\,TeV - green and 50\,TeV - light blue). Acceptance is taken to be $1^\circ < \theta < 179 ^\circ$, and $0.01 (0.001)<y < 1$ for the EIC (all other colliders). The areas delimitated by thick brown and black  lines show the regions accessible in pPb collisions at the LHC and the FCC-hh (50\,TeV) respectively, while the thin lines represent constant rapidities from 0 (right) to 6.6 (left) for each case. The saturation scale $Q_{sat}$ shown here for indicative purposes only, see also~\cite{Salgado:2011wc}, has been drawn for a Pb nucleus considering an uncertainty $\sim 2$ and a behaviour with energy following the model in~\cite{GolecBiernat:1998js}. Note that it only indicates a region where saturation effects are expected to be important but there is no sharp transition between the linear and non-linear regimes.}
\label{fig:NPP_LHeConHLLHC_heavyions_fig4}
\end{figure}

In conclusion, $ep$ and $e$A collisions as studied at the LHeC will have a large impact on the heavy ion programme, as the comparison of the kinematic reach of DIS and hadronic machines shown in Fig.\,\ref{fig:NPP_LHeConHLLHC_heavyions_fig4} makes evident. It should be noted that there exist proposals for extending such programme into Run 5 and 6 of the LHC~\cite{Citron:2018lsq}, by running lighter ions and with detector upgrades in ATLAS and CMS (starting in Run 4) and LHCb (Upgrade II~\cite{lhcbupgradeII}).

%

\biblio
\end{document}

%% file: accelerator/accelerator.chapter.tex
\linenumbers
\lhectitlepage
\lhecinstructions
\subfilestableofcontents
\input{\main/accelerator/accelerator.tex}
\biblio

%% file: accelerator/accelerator.tex
\chapter{The Electron Energy Recovery Linac \ourauthor{Erk Jensen,  Gianluigi Arduini, Rogelio Tomas}}
We studied different options for the electron accelerator for LHeC in Ref.~\cite{AbelleiraFernandez:2012cc}, of which the Energy Recovery Linac (ERL) option is retained in this update of the CDR. This is due to the higher achievable luminosity of the Linac-Ring option, as compared to the Ring-Ring option, as well as the interference of the installation of an electron ring in the LHC tunnel with its operation~\cite{bib:Bruening:LHeCWorkshop2015}.
The clear advantage of the ERL compared to its contenders in 2012 is the possibility to keep the overall energy consumption at bay, albeit, in its baseline configuration and size of the return arcs,  
operation 
is still limited to lepton energies below \SI{70}{\giga\electronvolt} to avoid excessive synchrotron radiation losses. Since there is no fundamental beam loading in an ERL by its principle, higher average currents and thus higher luminosities would not lead to larger power consumption. 
\section{Introduction -- Design Goals \ourauthor{Gianluigi Arduini, Erk Jensen, Rogelio Tomas }}
The main guidelines for the design of the Electron ERL and the Interaction Region (IR) with the LHC are:

\begin{itemize}
    \item electron-hadron operation in parallel with high luminosity hadron-hadron collisions in LHC/HL-LHC;
    \item centre-of-mass collision energy in the TeV scale; 
    \item power consumption of the electron accelerator smaller than \SI{100}{MW};
    \item peak luminosity approaching $10^{34}\,\si{cm^{-2}s^{-1}}$; 
    \item integrated luminosity exceeding by at least two orders of magnitude that achieved by HERA at DESY.
\end{itemize}

The electron energy $E_e$ chosen in the previous version of the CDR~\cite{AbelleiraFernandez:2012cc} was \SI{60}{GeV}. This could be achieved with an ERL circumference of 1/3 of that of the LHC. Cost considerations and machine--detector performance aspects, in particular the amount of synchrotron radiation losses in the IR, have led to define a new reference configuration with $E_e = \SI{49.2}{GeV}$ and a circumference of $\approx\SI{5.4}{km}$, 1/5 of that of the LHC.

\begin{figure}[tbh]
  \centering
  \includegraphics[width=0.75\textwidth]{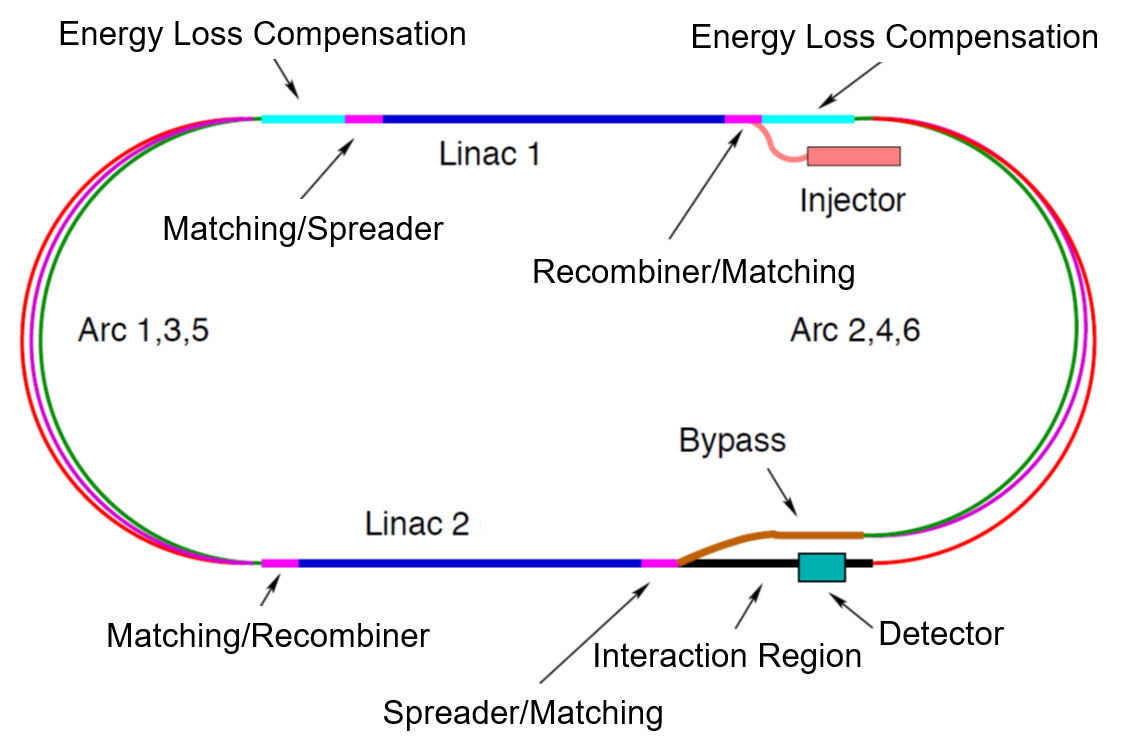}
  \caption{Schematic layout of the LHeC design based on an Energy Recovery Linac.}
  \label{fig:ERL_sketch}
\end{figure}

The ERL consists of two superconducting (SC) linacs operated in CW connected by at least three pairs of arcs to allow three accelerating and three decelerating passes (see Fig.~\ref{fig:ERL_sketch}). The length of the high energy return arc following the interaction point should be such as to provide a half RF period wavelength shift to allow the deceleration of the beam in the linac structures in three passes down to the injection energy and its safe disposal.  SC Cavities with an unloaded quality factor $Q_0$ exceeding $10^{10}$ are required to minimise the requirements on the cryogenic cooling power and to allow an efficient ERL operation. The choice of having three accelerating and three decelerating passes implies that the circulating current in the linacs is six times the current colliding at the Interaction Point (IP) with the hadron beam.

The choice of an Energy Recovery Linac offers the advantage of a high brightness beam and it avoids performance limitations due to the beam-beam effect seen by the electron beam~\cite{Brandt:2000xk}, which was a major performance limitation in many circular lepton colliders (e.g.\ LEP) and for the LHeC Ring-Ring option. The current of the ERL is limited by its source and an operational goal of $I_e = \SI{20}{mA}$ has been set, corresponding to a bunch charge of \SI{500}{pC} at a bunch frequency of \SI{40}{MHz}. This implies operating the SRF cavities with the very high current of \SI{120}{mA} for a virtual beam power (product of the beam current at the IP times the maximum beam energy) of \SI{1}{GW}. The validation of such performance in terms of source brightness and ERL 3-turn stable and efficient operation in the PERLE facility~\cite{Klein:2652336} is a key milestone for the LHeC design.

A small beam size at the IP is required to maximize luminosity and approach peak luminosities of $10^{34}\,\si{cm^{-2}s^{-1}}$ and integrated luminosities of \SI{1}{\per\atto\barn} in the LHeC lifetime. In particular $\beta^* < \SI{10}{cm}$ needs to be achieved for the colliding proton beam compatibly with the optics constraints imposed by the operation in parallel to proton-proton physics in the other IPs during the HL-LHC era~\cite{ApollinariG.:2017ojx}. The peak luminosity values quoted above exceed those at HERA by 2-3 orders of magnitude. The operation of HERA in its first, extended running period 1992--2000, provided and integrated luminosity of about \SI{0.1}{\per\femto\barn} for the H1 and ZEUS experiments, corresponding to the expected integrated luminosity collected over 1~day of LHeC operation.

%
%

\section{The ERL Configuration of the LHeC \ourauthor{Alex Bogacz}}
The main parameters of the LHeC ERL are listed in Tab.~\ref{tab:ERLparameters}; their choices and optimisation criteria will be discussed in the following sections.
\begin{table}[!ht]
  \centering
  \small
  \begin{tabular}{lcc} 
    \toprule
    Parameter & Unit & Value  \\
    \midrule
    Injector energy & \si{GeV} & 0.5 \\
    Total number of linacs & & 2 \\
    Number of acceleration passes & & 3 \\
    Maximum electron energy & \si{GeV} & 49.19 \\
    Bunch charge & \si{pC} & $499$ \\
    Bunch spacing & \si{ns}	& 24.95 \\
    Electron current & \si{mA}	& 20 \\
    Transverse normalized emittance & \si{\micro\meter} & 30 \\
    Total energy gain per linac & \si{GeV} & 8.114\\
    Frequency & \si{MHz} & 801.58  \\
    Acceleration gradient & \si{MV/m} & 19.73 \\
    Cavity iris diameter & \si{\milli\meter} & 130 \\
    Number of cells per cavity & & 5 \\
    Cavity length (active/real estate) & \si{\meter} & 0.918/1.5 \\
    Cavities per cryomodule &  & 4  \\
    Cryomodule length & \si{m} & 7 \\
    Length of 4-CM unit & \si{m} & 29.6 \\
    Acceleration per cryomodule (4-CM unit) & \si{MeV} & 289.8 \\
    Total number of cryomodules (4-CM units) per linac & & 112 (28) \\
    Total linac length (with with spr/rec matching) & \si{m} & 828.8 (980.8) \\
    Return arc radius (length) & \si{m} & 536.4 (1685.1) \\
    Total ERL length & \si{km} & 5.332 \\
    \bottomrule
  \end{tabular}
  \caption{Parameters of LHeC Energy Recovery Linac (ERL).}
  \label{tab:ERLparameters}
\end{table}

\subsection{Baseline Design -- Lattice Architecture   \ourauthor{Alex Bogacz}}
The ERL, as sketched in Fig.~\ref{fig:ERL_sketch}, is arranged in a racetrack configuration; hosting two superconducting linacs in the parallel straights and three recirculating arcs on each side. The linacs are \SI{828.8}{m} long and the arcs have \SI{536.4}{m} radius, additional space of \SI{76}{m} is taken up by utilities like Spreader/Recombiner (Spr/Rec), matching and energy loss compensating sections adjacent to both ends of each linac (total of 4 sections)~\cite{Pellegrini:2015rdx}. The total length of the racetrack is \SI{5.332}{km}: 1/5 of the LHC circumference $2 \cdot (828.8+2\cdot 76 +  536.4\pi)~\si{m}$. Each of the two linacs provides 8.114 GV accelerating voltage, therefore a \SI{49.19}{GeV} energy is achieved in three turns. After the collision with the protons in the LHC, the beam is decelerated in the three subsequent turns. The injection and dump energy has been chosen at \SI{0.5}{GeV}.

Injection into the first linac is done through a fixed field injection chicane, with its last magnet (closing the chicane) being placed at the beginning of the linac.
It closes the orbit \emph{bump} at the lowest energy, injection pass, but the magnet (physically located in the linac) will deflect the beam on all subsequent linac passes. In order to close the resulting higher pass \emph{bumps}, the so-called re-injection chicane is instrumented, by placing two additional opposing bends in front of the last chicane magnet.
The chosen arrangement is such that, the re-injection chicane magnets are only \emph{visible} by the higher pass beams.
The second linac in the racetrack is configured exactly as a mirror image of the first one, with a replica of the re-injection chicane at its end, which facilitates a fixed-field extraction of energy recovered beam to the dump.

\begin{figure}[t]
  \centering
  \includegraphics[width=14cm]{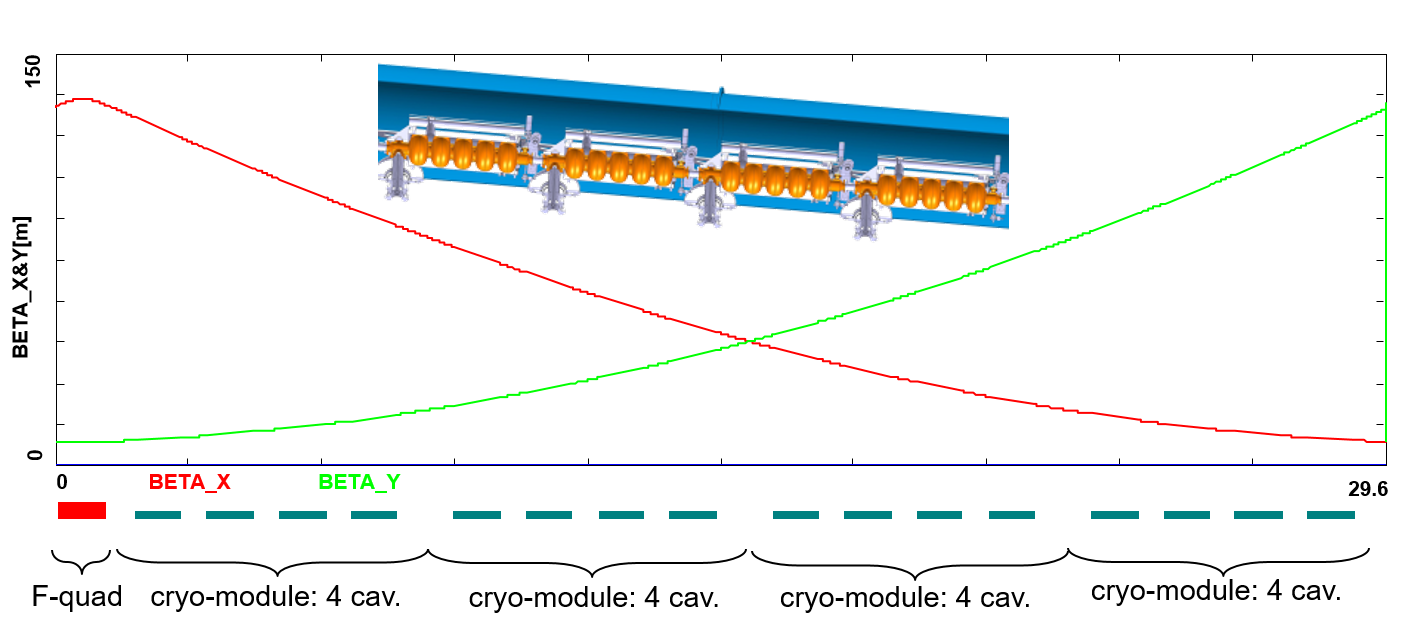}
  \caption{Layout of a half-cell composed out of four cryomodules (each hosting four, 5-cell cavities: top insert) and a focusing quad. Beta functions reflect \SI{130}{\degree} FODO optics.}
  \label{fig:Half_cell}
\end{figure}

\subsubsection{Linac Configuration and Multi-pass Optics}
Appropriate choice of the linac optics is of paramount importance for the transverse beam dynamics in a multi-pass ERL. The focusing profile along the linac (quadrupole gradients) need to be set (and they stay constant), so that multiple pass beams within a vast energy range may be transported efficiently. The chosen arrangement is such that adequate transverse focusing is provided for a given linac aperture. The linac optics is configured as a strongly focusing, \SI{130}{\degree} FODO. In a basic FODO cell a quadrupole is placed every four cryomodules, so that the full cell contains two groups of 16 RF cavities and a pair of quads (F, D) as illustrated in Fig.~\ref{fig:Half_cell}. The entire linac is built out of 14 such cells.
Energy recovery in a racetrack topology explicitly requires that both the
accelerating and decelerating beams share the individual return arcs~\cite{ICFA_Bogacz}. This in turn, imposes specific requirements for TWISS function at the linacs ends: TWISS functions have to be identical for both the accelerating and decelerating linac passes converging to the same energy and therefore entering the same arc. There is an alternative scheme, proposed by Peter Williams~\cite{erl.15}, who has argued that it would be beneficial to separate the accelerating and decelerating arcs. This would simplify energy compensation systems and linac-to-arc matching, but at an higher cost of the magnetic system of the arcs. However, doubling number of arcs is a very costly proposition. On the other hand, C-BETA experiment is pioneering a multi-pass arcs to transport a vast energy range through the same beam-line and it still intends to use them for energy recovery. Our approach, based on proven, CEBAF-like, RLA technology~\cite{CEBAF_12} is somewhere in the 'middle'. 

\begin{figure}[th]
  \centering
  \includegraphics[width=0.8\textwidth]{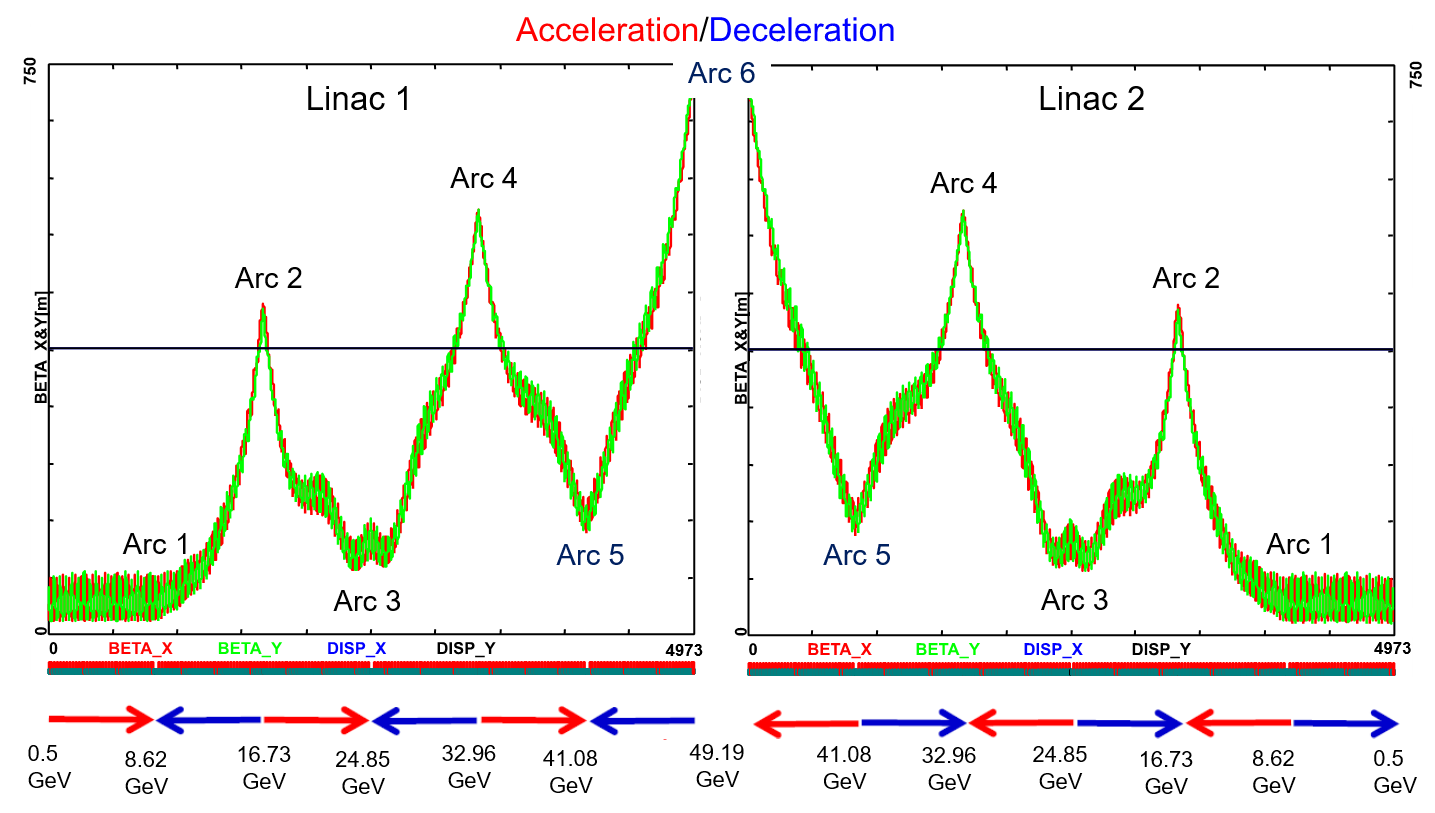}
  \caption{Beta function in the optimised multi-pass linacs (3 accelerating passes and 3 decelerating passes in each of two linacs). The matching conditions are automatically built into the resulting multi-pass linac beamline.}
  \label{fig:Multi_pass}
\end{figure}

To visualize beta functions for multiple accelerating and decelerating passes through a given linac, it is convenient to reverse the linac direction for all decelerating passes and string them together with the interleaved accelerating passes, as illustrated in Fig.~\ref{fig:Multi_pass}.
This way, the corresponding accelerating and decelerating passes are joined together at the arc's entrance/exit.
Therefore, the matching conditions are automatically built into the resulting multi-pass linac beamline. One can see that both linacs uniquely define the TWISS functions for the arcs: Linac 1 fixes input to all odd arcs and output to all even arcs, while Linac 2 fixes input to all even arcs and output to all odd arcs.
The optics of the two linacs are mirror-symmetric; They were optimised so that, Linac 1 is periodic for the first accelerating pass  and Linac 2 has this feature for  last decelerating one.
In order to maximize the BBU threshold current~\cite{Hoffstaetter:2004qy}, the optics is tuned so that the integral of $\beta/E$ along the linac is minimised. The resulting phase advance per cell is close to \SI{130}{\degree}.
Non-linear strength profiles and more refined merit functions were tested, but they only brought negligible improvements.

\subsubsection{Recirculating Arcs -- Emittance Preserving Optics}
Synchrotron radiation effects on beam dynamics, such as the transverse emittance dilution induced by quantum excitations have a paramount impact on the collider luminosity. All six horizontal arcs are accommodated in a tunnel of \SI{536.4}{m} radius.  
The transverse emittance dilution accrued through a given arc is proportional to the emittance dispersion function, $H$, averaged over all arc's bends~\cite{Schwinger:1996mc}: 
\begin{equation}
  \Delta \epsilon = \frac{2 \pi}{3} C_q r_0 <H> \frac{\gamma^5}{\rho^2}\,,
  \label{eq:Emit_dil_0}
\end{equation}
where
\begin{equation}
  C_q = \frac{55}{32 \sqrt{3}} \frac{\hbar}{m c}
  \label{eq:C_q}
\end{equation}
and $r_0$ is the classical electron radius and $\gamma$ is the Lorentz boost.
Here,  $H = (1+\alpha^2)/\beta \cdot D^2 + 2 \alpha \ D D' + \beta \cdot D'^2$ where $D, D'$ are the bending plane dispersion and its derivative, with $<...>~=~\frac{1}{\pi}\int_\text{bends}...~\text{d}\theta$.

Therefore, emittance dilution can be mitigated through appropriate choice of arc optics (values of $\alpha, \beta, D, D'$ at the bends). In the presented design, the arcs are configured with a FMC (Flexible Momentum Compaction) optics to ease individual adjustment of, $<H>$, in various energy arcs. 

\begin{figure}[th]
  \centering
  \includegraphics[width=0.8\textwidth]{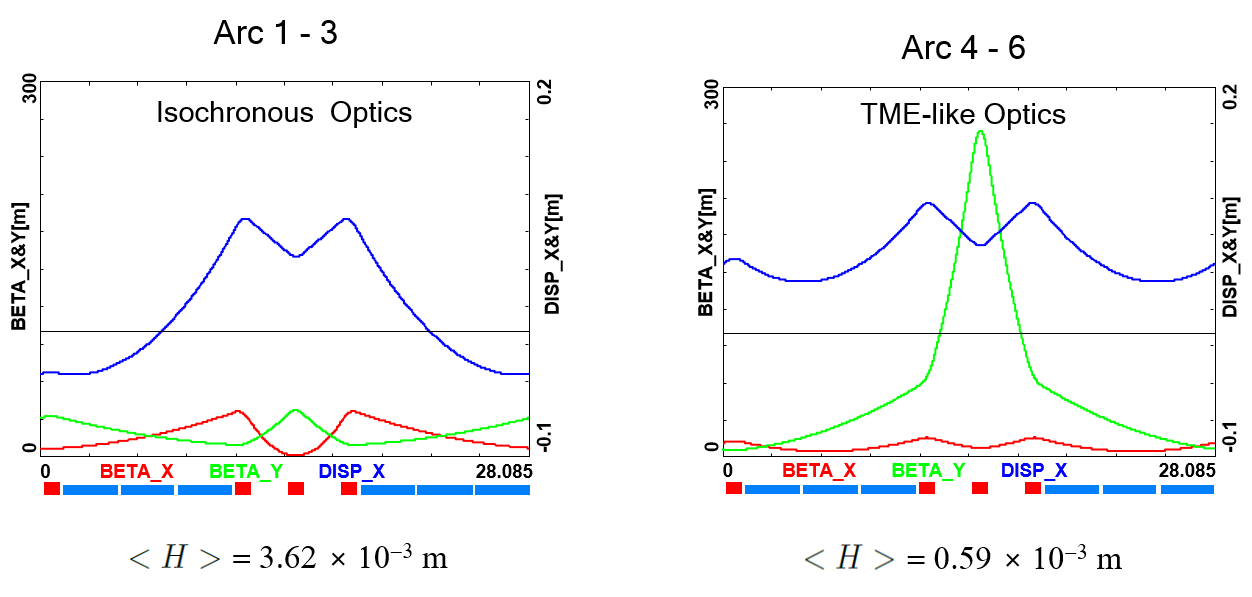}
  \caption{Two styles of FMC cells appropriate for different energy ranges. Left: lower energy arcs (Arc 1--3) configured with \emph{Isochronous} cells, Right: higher energy arcs configured with \emph{TME-like} cells. Corresponding values of the emittance dispersion averages, $<H>$, are listed for both style cells.}
  \label{fig:Arc_cells}
\end{figure}

Optics design of each arc takes into account the impact of synchrotron radiation at different energies. At the highest energy, it is crucial to minimise the emittance dilution due to quantum excitations; therefore, the cells are tuned to minimise the emittance dispersion, $H$, in the bending sections, as in the TME (Theoretical Minimum Emittance) lattice. On the other hand, at the lowest energy, it is beneficial to compensate for the bunch elongation with isochronous optics. The higher energy arcs (4,5 and 6) configured with the TME cells are still quasi-isochronous. To fully compensate remnant bunch elongation one could set higher pass linacs slightly off-crest to compress the bunches, since one has full control of gang-phases for individual linac passes. 
All styles of FMC lattice cells, as illustrated in Fig.~\ref{fig:Arc_cells}, share the same footprint for each arc. This allows us to stack magnets on top of each other or to combine them in a single design. Here, we use substantially shorter then in the 60 GeV design, \SI{28.1}{m}, FMC cell configured with six \SI{3}{m} bends, in groups of flanked by a quadrupole singlet and a triplet, as illustrated in Fig.~\ref{fig:Arc_cells}.
The dipole filling factor of each cell is \SI{63}{\percent}; therefore, the effective bending radius $\rho$ is \SI{336.1}{m}.
Each arc is followed by a matching section and a recombiner (mirror symmetric to spreader and matching section). Since the linacs are mirror-symmetric, the matching conditions described in the previous section, impose mirror-symmetric arc optics (identical betas and sign reversed alphas at the arc ends).

Path-length adjusting chicanes were also foreseen to tune the beam time of flight in order to hit the proper phase at each linac injection. Later investigations proved them to be effective only with lower energy beams, as these chicanes trigger unbearable energy losses, if applied to the highest energy beams. A possible solution may consist in distributing the perturbation along the whole arc with small orbit excitations. This issue will be fully addressed in a subsequent section on 'Synchrotron Radiation Effects - Emittance Dilution'.

\subsubsection{Spreaders and Recombiners}
\begin{figure}
  \centering
  \includegraphics[width=0.7\textwidth]{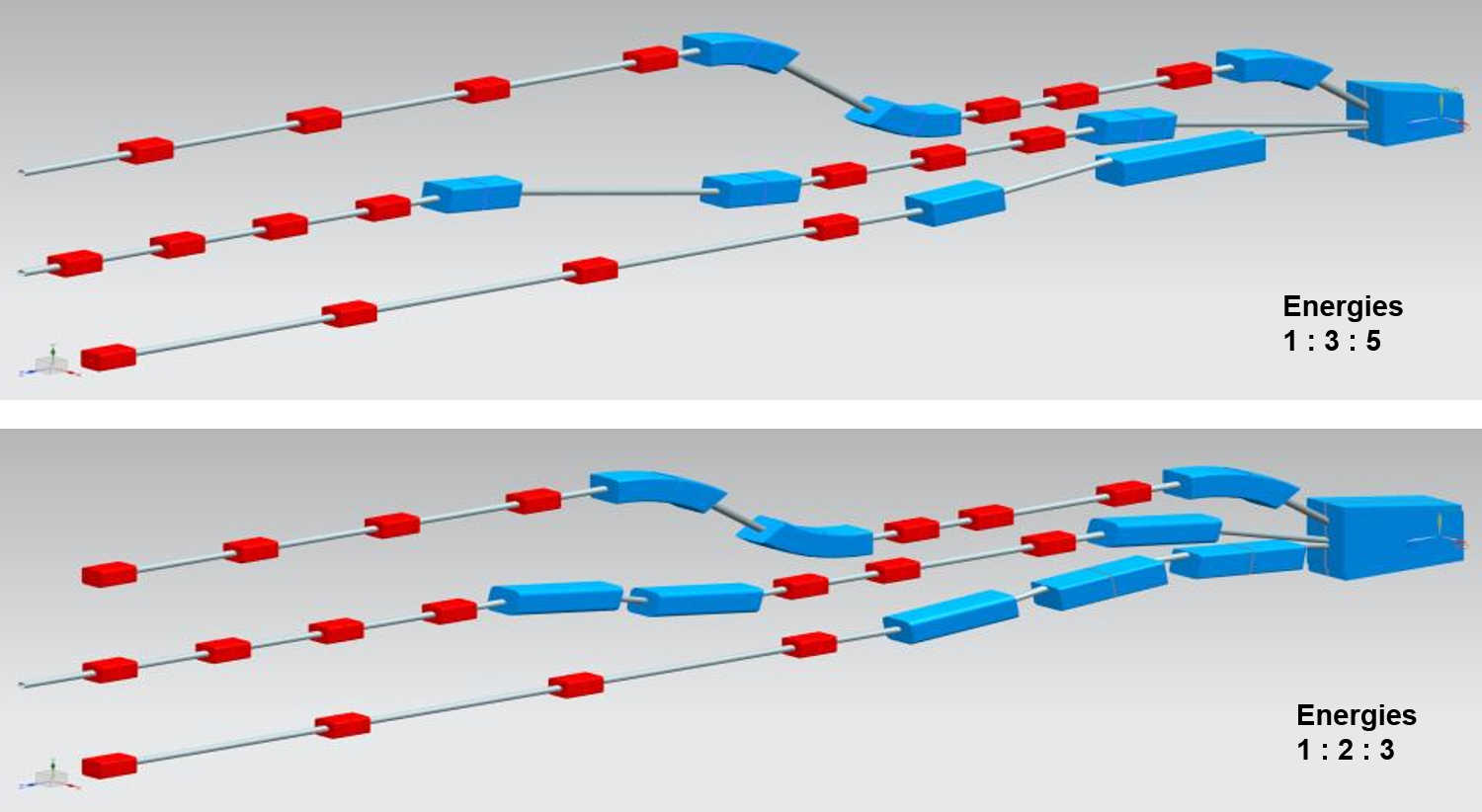}
  \caption{Layout of a three-beam switch-yard for different energy ratios: 1\,:\,3\,:\,5 and  1\,:\,2\,:\,3 corresponding to specific switch-yard geometries implemented on both sides of the racetrack}
  \label{fig:Switchyard}
\end{figure}

The spreaders are placed directly after each linac to separate beams of different energies and to route them to the corresponding arcs. The recombiners facilitate just the opposite: merging the beams of different energies into the same trajectory before entering the next linac.
As illustrated in Fig.~\ref{fig:Switchyard}, each spreader starts with a vertical bending magnet, common for all three beams, that initiates the separation. The highest energy, at the bottom, is brought back to the horizontal plane with a chicane.
The lower energies are captured with a two-step vertical bending adapted from the CEBAF design~\cite{CEBAF_12}.

\begin{figure}
  \centering
  \includegraphics[width=0.6\textwidth]{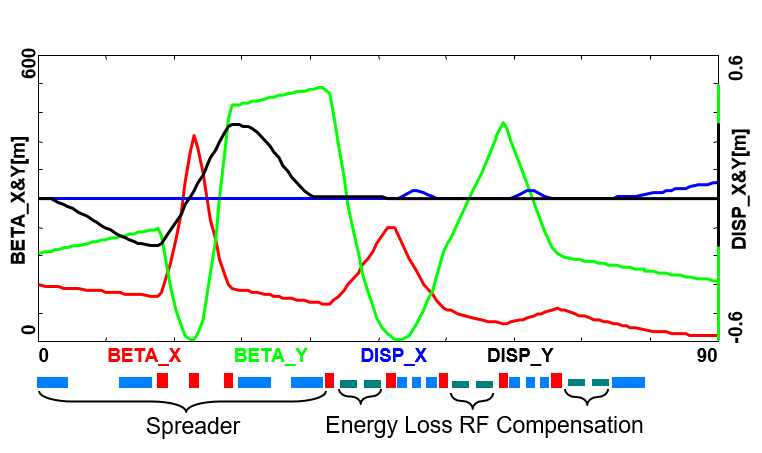}
  \caption{Spreader 3 (\SI{24.8}{GeV}) optics; featuring a vertical achromat with three dispersion suppressing quads in-between the two steps, a pair of path-length adjusting dogleg chicanes and four betatron matching quads, interleaved with three energy loss compensating sections (2-nd harmonic RF cavities marked in green).}
  \label{fig:Spreader}
\end{figure}

\begin{figure}
  \centering
  \includegraphics[width=0.7\linewidth]{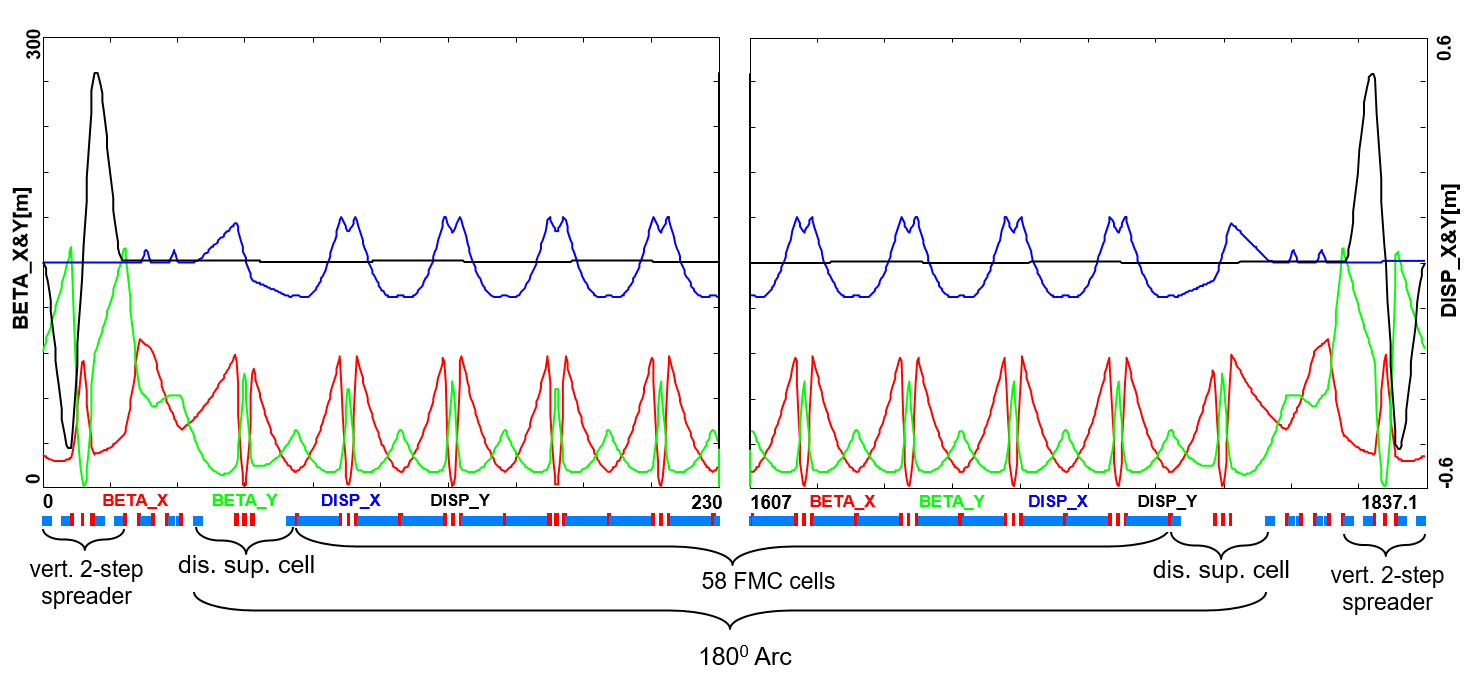}
  \caption{Complete Optics for Arc~3 (including switch-yard); featuring: low emittance \SI{180}{\degree} arc based on isochronous cells (30 cells flanked by dispersion suppression cell with missing dipoles on each side), spreaders and recombiners with matching sections and doglegs symmetrically placed on each side of the arc proper.}
  \label{fig:Arc}
\end{figure}

Functional modularity of the lattice requires spreaders and recombiners to be achromats (both in the horizontal and vertical plane). To facilitate that, the vertical dispersion is suppressed by a pair of quadrupoles located in-between vertical steps; they naturally introduce strong vertical focusing, which needs to be compensated by the middle horizontally focusing quad. The overall spreader optics is illustrated in Fig.~\ref{fig:Spreader}. 
Complete layout of two styles of switch-yard with different energy ratios is depicted in Fig.~\ref{fig:Switchyard}. Following the spreader, there are four matching quads to \emph{bridge} the Twiss function between the spreader and the following \SI{180}{\degree} arc (two betas and two alphas). 
Combined spreader-arc-recombiner optics, features a high degree of modular functionality to facilitate momentum compaction management, as well as orthogonal tunability for both the beta functions and dispersion, as illustrated in Fig.~\ref{fig:Arc}. 

\begin{figure}
  \centering
  \includegraphics[width=0.8\linewidth]{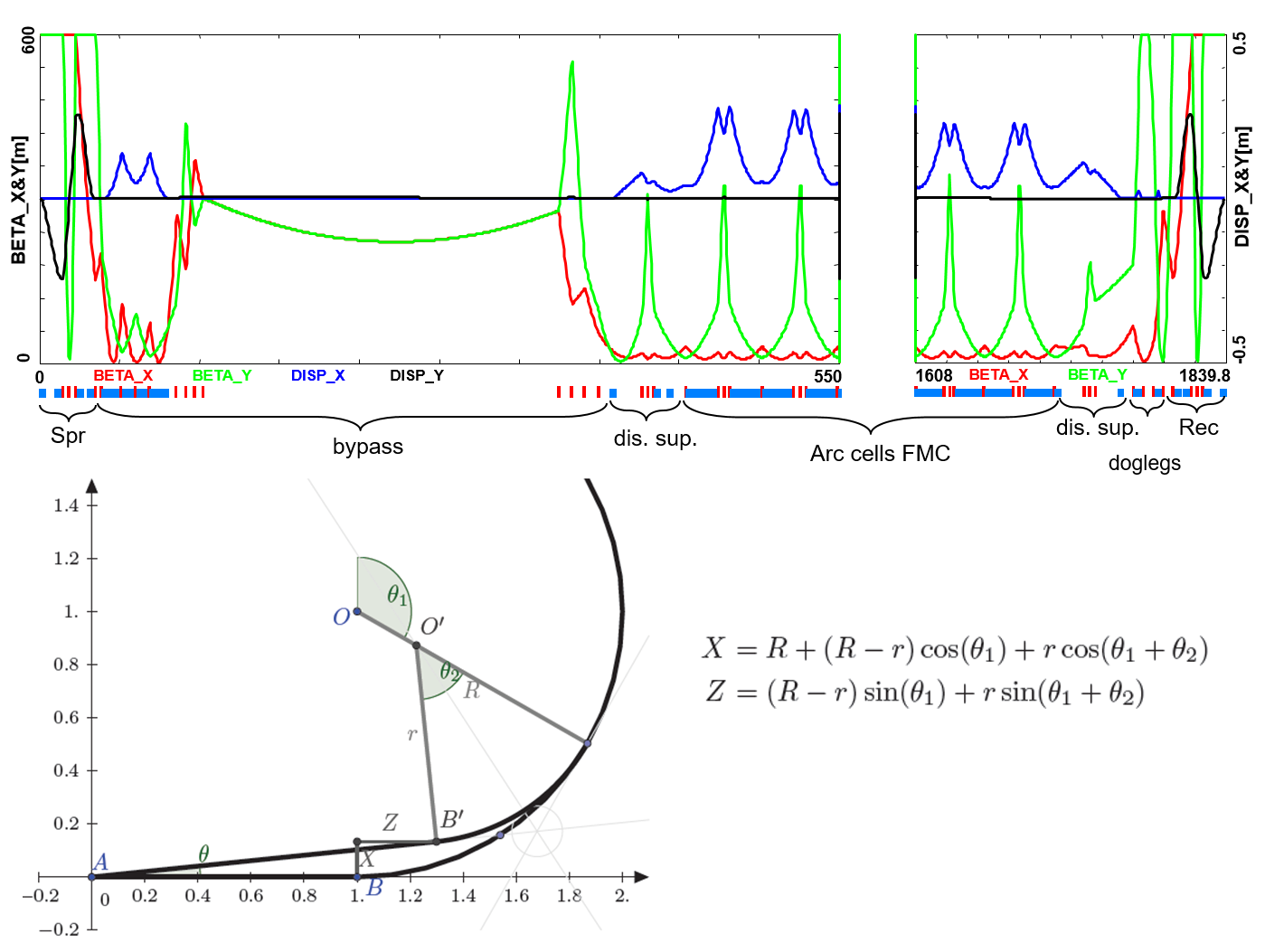}
  \caption{Optics and layout of Arc~4 including the detector bypass. The lattice (top insert) features a vertical spreader, an initial horizontal bending, a straight section, a modified dispersion suppressor, seven junction cells, and four regular cells. The bypass geometry (bottom insert), features a long IP line, AB,  which for visual reasons has been purposely stretched, being actually about $1/5$ of the arc radius. All geometric dependencies of the bypass parameters are summarized in the inserted formulae.
  }
  \label{fig:Bypass}
\end{figure}
\subsubsection{Alternative design of the spreader/recombiner}
\begin{figure}[hbtp!]
    \centering
    \includegraphics[width=0.75\textwidth]{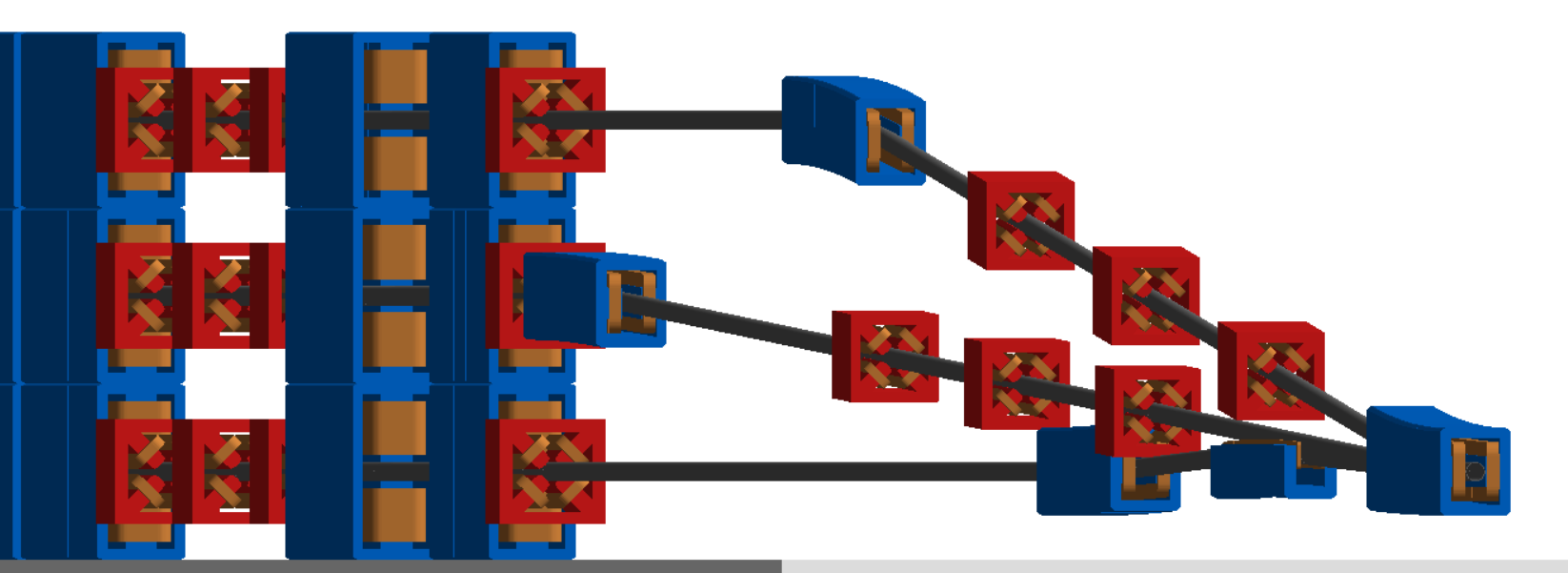}
    \caption{ \label{fig:spreader135arc} 3D visualisation of the spreader 1,3,5 inserted between the end of the linac and routing the different beamlines into their respective dispersion suppresors.}
\end{figure}

\begin{figure}[hbtp!]
    \centering
    \includegraphics[width=0.75\textwidth]{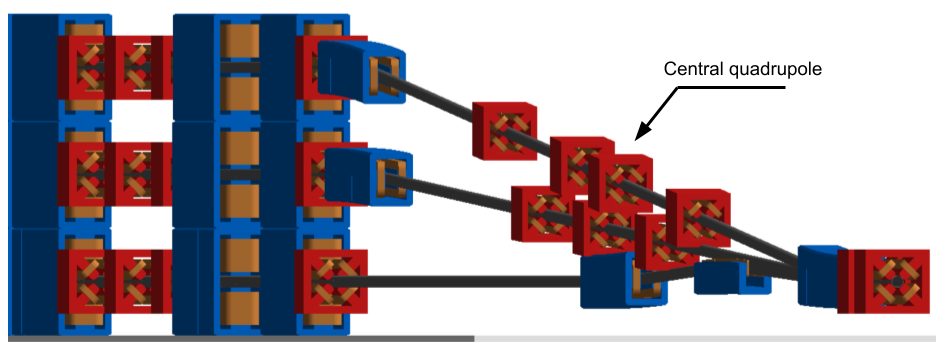}
    \caption{ \label{fig:spreader246arc} 3D visualisation of the spreader 2,4,6 inserted between the end of the linac and routing the different beamlines into their respective dispersion suppressors.}
\end{figure}

The desire to reduce the number of elements included in the spreader led to the reduction of the number of steps required to separate vertically the different beams and route them into their specific arcs. In particular, this alternative spreader design uses a single vertical step instead of two. Although the concept has been briefly discussed in \cite{AbelleiraFernandez:2012cc} it was not retained due to the superconducting technology needed for the quadrupoles that must be avoided in this highly radiative section. Nevertheless, recent studies have been pursuing a one step spreader version, based on normal conducting magnet technology. It assumes a pole tip field of less than \SI{1}{T} for an aperture radius of \SI{30}{mm}, allowing the use of thin quadrupoles and thus minimise potential overlap with the other beamlines. With respect to the previous study, the use of normal conductors was made possible by increasing the overall spreader length and reducing the number of quadrupoles. In particular, the focusing magnets are limited to two outer quadrupoles for the achromatic function and one quadrupole in the middle, where the dispersion is zero, to control the beta function in the defocusing plane. Two visualisations are given Fig \ref{fig:spreader135arc} and \ref{fig:spreader246arc}.

Both spreader types start with a first dipole that separates vertically the different beamlines in a 1-3-5 ratio for the odd number spreaders and in a 1-2-3 ratio for the even number spreaders. Theses ratios are defined by  the beam energies of the corresponding turn. Therefore by fixing the length of the longest beamline for each spreaders (odd and even numbers) one obtains the required angle to get a \SI{50}{cm} vertical offset between each beamlines. The equations below represent the required bending angle in the dipole and beamline lengths in order to meet the requirements,
\begin{align}
   \theta_3 &= \cfrac{0.5}{l_3 - L}   &l_1 = \cfrac{2 E_1}{E_3}(l_3 - L) + L \\
   \theta_2 &= \cfrac{1}{l_2 - L}  &l_4 = \cfrac{1}{2} \cfrac{E_4}{E_2}(l_2 - L) + L
\end{align}
where the index i corresponds to the beamline number associated to an energy $E_i$. L is the dipole length and l is the whole spreader beamline length.

For the beam line 2 and 4 we obtain $l_4 \approx l_2$ according to the energy ratio of 1.97. On the other hand, $l_3$ will be longer than $l_1$ because the energy ratio is 2.88. One can therefore tune the angle of the even number spreaders by defining $l_2$ as it will be the longest. However the angle of the odd number spreaders will be determined by the length of $l_3$. Regarding the chicane used for the highest energy only a minimum separation between the highest energy and the intermediate energy allows the introduction of the opposite bending dipole. It constraints in return the placement of the quadrupoles of the intermediate energy. The location of the magnets is the main limitation towards a minimization of the spreader length: the shorter the lattice gets, the stronger the quadrupoles need to be in order to preserve the achromatic function.

As a result, a one step spreader halves the number of dipoles present in the lattice and relaxes the constraint on magnets interference and overlap favorable for compactness. Dividing by two the number of dipoles has a noticeable effect on the synchrotron power radiated in the spreader which is, in addition, beneficial for the emittance growth. The dipole fields required, for a maximal length of \SI{50}{m}, are \SI{226}{mT} for the odd number spreaders and \SI{326}{mT} for the even number spreaders.
The multipass linac optics Fig. \ref{fig:Multi_pass} shows that the even number spreaders ,\textit{i.e.} 2, 4 and 6, have the highest beta functions at their entrance which is detrimental from the perspective of minimizing the emittance growth within the spreader lattice. A solution to solve this issue is to insert a doublet of quadrupoles at the exit of the linac. All three energies will go through the doublet and, therefore, a compromise has to be found for the gradients. Finally, the reduction of the $\mathcal{H}$ function over the length of the spreader and specifically in the dipoles contributes to a further reduction of the emittance growth.

The energy loss for spreader 1 is low due to the low beam energy; spreader 2 and 3 have similar values that are acceptable as well as the one of spreader 4. Spreader 5 and 6 have the highest beam energies and therefore the largest energy loss. In addition, the dipoles used to produce the chicane need double the field strength  compared to the other dipoles for the same length, \textit{i.e.} half the bending radius, in order to save space for the other elements in the other beamlines. The vertical emittance growth is well controlled just as in the even number spreaders. Only spreader 6 has an order of magnitude higher contribution but one has to keep in mind that the even number spreaders will only act as recombiner since there will be the horizontal bypass doing the separation with the detector and a vertical separation will only occur for arc 2 and 4. Consequently spreader 6 should not be taken into account for the emittance growth contribution until the interaction point.

\begin{figure}[hbtp!]
    \centering
    \includegraphics[width=0.47\textwidth]{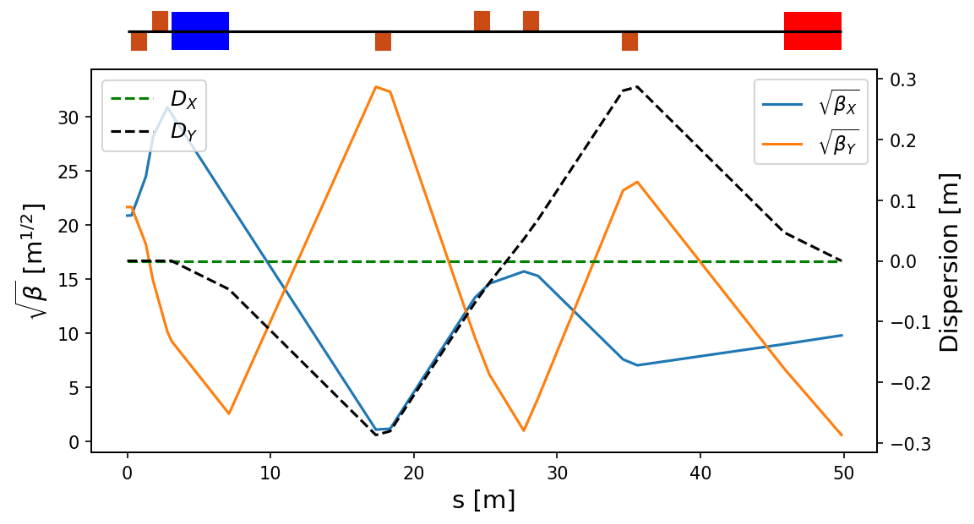}
    \includegraphics[width=0.47\textwidth]{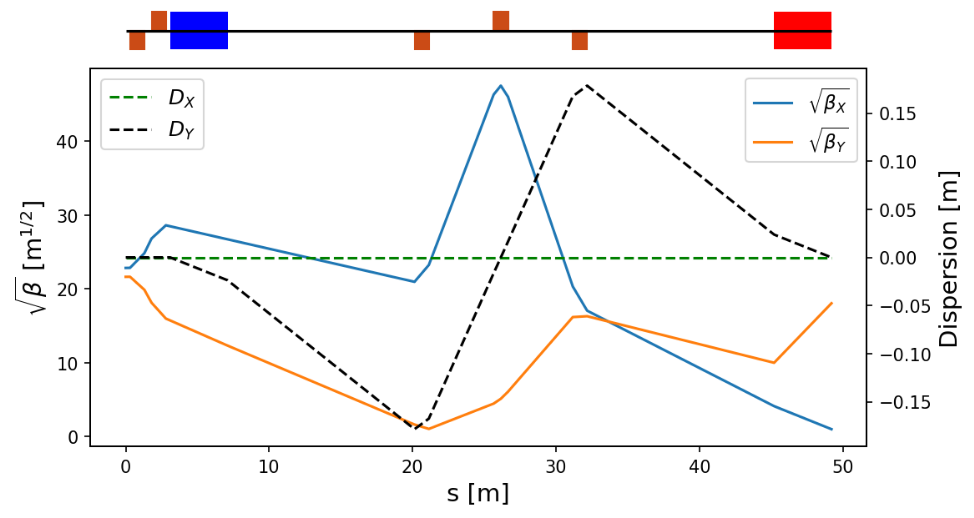}
    \caption{ \label{fig:spreaderoptics24} Left : Spreader/Recombiner optics of the arc 2 for the \SI{16.73}{GeV} electron beam. Right : Spreader/Recombiner optics of the arc 4 for the \SI{32.96}{GeV} electron beam.}
\end{figure}

The optics for the Spreader/Recombiner of arc 2 and 4 are presented Fig. \ref{fig:spreaderoptics24}, it shows the achromatic function supported by the outer quadrupoles while the control of the horizontal beta function is provided by the "middle" quadrupole. One sees that the lattice of the arc 2 requires to split the "middle" quadrupole in two, in order to avoid overlap with the other beamline (arc 4), see Fig.\ref{fig:spreader246arc}. These two optics are the most challenging as they have high beta functions at the entrance of their lattices, due to the multi pass linac optics as previously explained.

\subsubsection {IR Bypasses}
After the last spreader the \SI{49.19}{GeV} beam goes straight to the interaction region. However the lower energy beams; at \SI{16.7} and \SI{33.0}{GeV}, need to be further separated horizontally in order to avoid interference with the detector. Different design options for the bypass section were explored~\cite{Thesis_Pellegrini} and the one that minimises the extra bending has been chosen and implemented in the lattice.

Ten arc-like dipoles are placed very close to the spreader, to provide an initial bending, $\theta$, which results in $X=\SI{10}{m}$ separation from the detector located \SI{120}{m} downstream. The straight section of the bypass is approximately \SI{240}{m} long. After the bypass, in order to reconnect to the footprint of Arc~6, 7 of 30 standard cells in Arc~2 and Arc~4 are replaced with 7 higher field, junction cells. The number of junction cells is a compromise between the field strength increase and the length of additional bypass tunnel, as can be inferred from the scheme summarised in Fig.~\ref{fig:Bypass}.
The stronger bending in the junction cells creates a small mismatch, which is corrected by adjusting the strengths of the quadrupoles in the last junction cell and in the first regular cell.
%

\subsubsection{Synchrotron Radiation Effects -- Emittance Dilution}
ERL efficiency as a source of multi-GeV electrons for a high luminosity collider is limited by the incoherent synchrotron radiation effects on beam dynamics; namely the transverse emittance dilution and the longitudinal momentum spread (induced by quantum excitations).
The first effect, the transverse emittance increase, will have a paramount impact on the collider luminosity, due to stringent limits on the allowed emittance increase.
The second one, accrued momentum spread, governs asymmetries of accelerated and decelerated beam profiles. These asymmetries substantially complicate multi-pass energy recovery and matching, and ultimately they limit the energy reach of the ERLs due to recirculating arc momentum acceptance.

Arc optics was designed to ease individual adjustment of momentum compaction (needed for the longitudinal phase-space control, essential for operation with energy recovery) and the horizontal emittance  dispersion, $H$, in each arc.
Tab.~\ref{tab:SREmittance} lists arc-by-arc dilution of the transverse, $\Delta \epsilon$, and longitudinal, $\Delta \sigma_{\frac{\Delta E}{E}}$, emittance due to quantum excitations calculated using analytic formulas, Eqs.~\eqref{eq:Emit_dil_1},~\eqref{eq:Emit_dil_2} and~\eqref{eq:Emit_dil_3}, introduced by M.~Sands~\cite{Schwinger:1996mc}:  
\begin{equation}
  \Delta E = \frac{2 \pi}{3} r_0 ~mc^2~ \frac{\gamma^4}{\rho}\,
  \label{eq:Emit_dil_1}
\end{equation}
\begin{equation}
  \Delta \epsilon_N = \frac{2 \pi}{3} C_q r_0 <H> \frac{\gamma^6}{\rho^2}\,,
  \label{eq:Emit_dil_2}
\end{equation}
\begin{equation}
  \frac{\Delta \epsilon_E^2}{E^2} = \frac{2 \pi}{3} C_q r_0~ \frac{\gamma^5}{\rho^2}\,,
  \label{eq:Emit_dil_3}
\end{equation}
where $C_q$ is given by Eq.~\eqref{eq:C_q}.
Here, $\Delta \epsilon^2_E$ is an increment of energy square variance, $r_0$ is the classical electron radius, $\gamma$ is the Lorentz boost and $C_q \approx 3.832 \cdot 10^{-13}\,\text{m}$ for electrons (or positrons).

\begin{table}[!ht]
  \centering
  \small
  \begin{tabular}{lcccc} 
  \toprule
  Beamline & Beam energy & $\Delta E$  & $\Delta \epsilon^x_N$ & $\Delta \sigma_{\frac{\Delta E}{E}}$  \\
    & [\si{GeV}] &  [\si{MeV}]&  [\si{mm~mrad}] &  [\si{\percent}]\\
  \midrule
  arc 1 & 8.62 & 0.7 & 0.0016 & 0.0005\\
  arc 2 & 16.73 & 10 & 0.085 & 0.0027\\
  arc 3 & 24.85 & 49 & 0.91 & 0.0072\\
  arc 4 & 32.96 & 152 & 0.81 & 0.015\\
  arc 5 & 41.08 & 368 & 3.03 & 0.026\\
  arc 6 & 49.19 & 758 & 8.93 & 0.040\\
  \bottomrule
  \end{tabular}
  \caption{Energy loss and emittance dilution (horizontal and longitudinal) due to synchroton radiation generated by all six \SI{180}{\degree} arcs (not including Spreaders, Recombiners and Doglegs). Here, $\Delta \sigma_{\frac{\Delta E}{E}} = \sqrt{\tfrac{\Delta \epsilon_E^2}{E^2}}$}.
  \label{tab:SREmittance}
\end{table}

\begin{table}[!ht]
  \centering
  \small
  \begin{tabular}{lcccc} 
    \toprule
    Beamline & Beam energy & $\Delta E$ & $\Delta \epsilon^y_N$& $\Delta \sigma_{\frac{\Delta E}{E}}$\\
     &  [\si{GeV}] & [\si{MeV}]&  [\si{mm~mrad}] & [\si{\percent}]\\
    \midrule
    Spr/Rec 1 & 8.62 & 0.2 & 0.035 & 0.0008\\
    Spr/Rec 2 & 16.73 & 3.0 & 0.540 & 0.0044\\
    Spr/Rec 3 & 24.85 & 6.0 & 0.871 & 0.0066\\
    Spr/Rec 4 & 32.96 & 21.6 & 5.549 & 0.0143\\
    Spr/Rec 5 & 41.08 & 7.1 & 0.402 & 0.0062\\
    Spr/Rec 6 & 49.19 & 39.2 & 3.92 & 0.0205\\
    \bottomrule
  \end{tabular}
  \caption{Energy loss and  emittance dilution (vertical and longitudinal) due to synchroton radiation generated by the two step Spreader, or Recombiner design of a given arc. Here, $\Delta \sigma_{\frac{\Delta E}{E}} = \sqrt{\frac{\Delta \epsilon_E^2}{E^2}}$}.
  \label{tab:SprRecEmittance}
\end{table}

Apart from the horizontal \SI{180}{\degree} arcs, there are other sources of emittance dilution due to synchrotron radiation, namely vertical Spreaders and Recombiners, as well as horizontal 'Doglegs' used to compensate seasonal variation of path-length. To minimise their contribution to the vertical emittance dilution, special optics with small vertical $<H>$ has been introduced in Spr/Rec sections. The effects on vertical emittance dilution coming from these beamlines (Spr/Rec) are summarized in Tab.~\ref{tab:SprRecEmittance} for the two-step spreaders and in  Tab.~\ref{tab:onestep} for the alternative version of a one-step spreader.

\begin{table}[!ht]
  \centering
  {\tabcolsep=0pt\def\arraystretch{1.0}
  \begin{tabularx}{340pt}{c *4{>{\Centering}X}}\toprule
  & Beam Energy &  $\Delta$E & $\Delta\epsilon^y_N$  & $\Delta\sigma_{\frac{\Delta E}{E}}$ \tabularnewline
  &  [GeV] & [MeV] &  [mm.mrad] &  [\%] \tabularnewline \midrule
  Spreader 1 & 8.62 & 0.04 & 0.004 & 0.0002 \tabularnewline 
  Spreader 2 & 16.73 & 0.31 & 0.004 & 0.0007 \tabularnewline 
  Spreader 3 & 24.85 & 0.32 & 0.012 & 0.0006 \tabularnewline 
  Spreader 4 & 32.96 & 1.18 & 0.112 & 0.0013 \tabularnewline 
  Spreader 5 & 41.08 & 2.64 & 0.083 & 0.0019 \tabularnewline 
  Spreader 6 & 49.19 & 7.92 & 1.060 & 0.0040\tabularnewline \bottomrule
  \end{tabularx} }
  \caption{ \label{tab:onestep} Energy loss and  emittance dilution (vertical and longitudinal) due to synchroton radiation generated by a one-step Spreader, or Recombiner design of a given arc. Here, $\Delta \sigma_{\frac{\Delta E}{E}} = \sqrt{\frac{\Delta \epsilon_E^2}{E^2}}$}
\end{table}

Similarly, the horizontal emittance dilution induced by the Doglegs (four dogleg chicanes per arc) in various arcs is summarized in Tab.~\ref{tab:DogEmittance}. Each dogleg chicane is configured with four 1 meter bends (1 Tesla each), so that they bend the lowest energy beam at \SI{8.6}{GeV} by 2 degrees. The corresponding path-lengths gained in the Doglegs of different arcs are also indicated.

\begin{table}[!ht]
  \centering
  \small
  \begin{tabular}{lccccc} 
    \toprule
    Beamline & Beam energy  & $\Delta E$ & $\Delta \epsilon^x_N$  & $\Delta \sigma_{\frac{\Delta E}{E}}$  & path-length\\
     & [\si{GeV}] & [\si{MeV}]& [\si{mm~mrad}] & [\si{\percent}] & [\si{mm}] \\
    \midrule
    Doglegs 1 & 8.62 & 2 & 0.201 & 0.007 & 7.32\\
    Doglegs 2 & 16.73 & 9 & 0.667 & 0.009 & 1.96\\
    Doglegs 3 & 24.85 & 19 & 5.476 & 0.014 & 0.84\\
    Doglegs 4 & 32.96 & 33 & 5.067 & 0.014 & 0.52\\
    Doglegs 5 & 41.08 & 52 & 12.067 & 0.028 & 0.36\\
    Doglegs 6 & 49.19 & 74 & 2.836 & 0.011 & 0.28\\\bottomrule
  \end{tabular}
  \caption{Energy loss and  emittance dilution (horizontal and longitudinal) due to synchroton radiation generated by the Doglegs (four dogleg chicanes) of a given arc. Here, $\Delta \sigma_{\frac{\Delta E}{E}} = \sqrt{\frac{\Delta \epsilon_E^2}{E^2}}$}.
  \label{tab:DogEmittance}
\end{table}

As indicated in Tab.~\ref{tab:DogEmittance}, the Doglegs in the highest energy arcs, Arc 5 and Arc 6, provide only sub mm path-length gain with large synchrotron radiation effects. They are not very effective and generate strong, undesired emittance dilution. Therefore, it is reasonable to eliminate them from both Arc 5 and 6. Instead,  one could resort to an alternative path-length control via appropriate orbit steering with both horizontal and vertical correctors present at every girder and distributed evenly throughout the arc.

Combining all three contributions: (\SI{180}{\degree} arc, Spreader, Recombiner and Doglegs (no Doglegs in Arcs 5 and 6), the net cumulative emittance dilution is summarized in Tab.~\ref{tab:CummEmittance} for the case of the two-step spreader.

\begin{table}[!ht]
  \centering
  \small
  \begin{tabular}{lccccc} 
    \toprule
        Beamline & Beam energy& $\Delta E$ & $\Delta^\text{cum} \epsilon^x_N$ & $\Delta^\text{cum} \epsilon^y_N$  & $\Delta^\text{cum} \sigma_{\frac{\Delta E}{E}}$ \\
       & [\si{GeV}] & [\si{MeV}]& [\si{mm~mrad}] & [\si{mm~mrad}] & [\si{\percent}]\\
    \midrule
    Arc 1 &  8.62 & 3 & 0.2 & 0.1 & 0.01\\
    Arc 2 & 16.73 & 25 & 1.0 & 1.2 & 0.03\\
    Arc 3 & 24.85 & 80 & 7.3 & 2.9 & 0.06\\
    Arc 4 & 32.96 & 229 & 13.2 & 14.0 & 0.12\\
   Arc 5 & 41.08 & 383 & 16.2 & 14.8 & 0.16\\
 \textbf{IR} & 49.19 & 39 & \textbf{16.2} & \textbf{18.7} & \textbf{0.18}\\
    Arc 6 & 49.19 & 797 & 25.2 & 22.6 & 0.24\\
    Arc 5 & 41.08 & 383 & 28.2 & 23.4 & 0.28\\
    Arc 4 & 32.96 & 229 & 34.1 & 34.5 & 0.33\\
    Arc 3 & 24.85 & 80 & 40.5 & 36.3 & 0.37\\
    Arc 2 & 16.73 & 25 & 41.2 & 37.4 & 0.39\\
    Arc 1 &  8.62 & 3 & 41.4 & 37.4 & 0.40\\
    Dump  &  0.5  & & 41.4 & 37.4 & 0.40\\
    \bottomrule
  \end{tabular}
  \caption{Energy loss and cumulative emittance dilution (transverse and longitudinal) due to synchroton radiation at the end of a given beam-line (complete Arc including: \SI{180}{\degree} arc, Spreader, Recombiner and Doglegs in arcs 1-4). The table covers the entire ER cycle: 3 passes 'up' + 3 passes 'down'. Cumulative emittance dilution values just before the IP (past Arc 5 and Spr 6), which are critical for the luminosity consideration are highlighted in 'bold'. That row accounts for contributions from Spr 6 (the last bending section before the IR) to energy loss, as well as the vertical and longitudinal emittance dilutions. Here, $\Delta \sigma_{\frac{\Delta E}{E}} = \sqrt{\frac{\Delta \epsilon_E^2}{E^2}}$}.
  \label{tab:CummEmittance}
\end{table}

Tab.~\ref{tab:CummEmittance} shows, the LHeC luminosity requirement of total transverse emittance dilution in either plane (normalized) at the IP (at the end of Arc 5), not to exceed \SI{20}{mm~mrad} (hor: \SI{16.2}{mm~mrad} and ver: \SI{18.7}{mm~mrad}) is met by-design, employing presented low emittance lattices in both the arcs and switch-yards. 
In the case of the optimised one-step spreader design, another reduction - mainly of the vertical emittance budget -  is obtained, providing a comfortable safety margin of the design.

Finally, one can see from Eqs.~\eqref{eq:Emit_dil_2} and~\eqref{eq:Emit_dil_3} an underlying universal scaling of the transverse (unnormalized) and longitudinal emittance dilution with energy and arc radius; they are both proportional to $\gamma^5/\rho^2$.
This in turn, has a profound impact on arc size scalability with energy; namely the arc radius should scale as $\gamma^{5/2}$ in order to preserve both the transverse and longitudinal emittance dilutions, which is a figure of merit for a synchrotron radiation dominated ERL.

\subsection{30\,GeV~ERL Options \ourauthor{Alex Bogacz}}
One may think of an upgrade path from 30~to~\SI{50}{GeV} ERL, using the same  $1/5$ of the LHC circumference (\SI{5.4}{km}), footprint.  In this scenario, each linac straight (front end) would initially be loaded  with $18$ cryomodules, forming two \SI{5.21}{GV} linacs. One would also need to decrease the injector energy by a factor of $5.21/8.11$.
The top ERL energy, after three passes, would reach \SI{31.3}{GeV}.
Then for the upgrade to \SI{50}{GeV}, one would fill the remaining space in the linacs with additional \SI{10} cryomodules each; \SI{2.9}{GV} worth of RF in each linac.  This way the energy ratios would be preserved for both 30 and \SI{50}{GeV} ERL options, so that the same switch-yard geometry could be used. Finally, one would scale up the entire lattice; all magnets (dipoles and quads) by $8.11/5.21$ ratio.

If one wanted to stop at the \SI{30}{GeV} option with no upgrade path, then a $1/14$ of the LHC circumference (\SI{1.9}{km}) would be a viable footprint for the racetrack, featuring: two linacs, \SI{503}{m} each, (17~cryomodules) and arcs of \SI{94.5}{m} radius. Again, assuming \SI{0.32}{GeV} injection energy, the top ERL energy would reach \SI{30.2}{GeV}. Such a configuration may become of interest
if time and funds may permit a small version of the LHeC or none.
This version of the LHeC would have a reduced Higgs, top
and BSM physics potential. Yet,
owing to the high proton beam energy, this configuration
would still have a TeV in the centre-of-mass such that the core
QCD, PDF and electroweak programme would still be striking.

\subsection{Component Summary\ourauthor{Alex Bogacz}}
This closing section will summarise active accelerator components: magnets (bends and quads) and RF cavities for the \SI{50}{GeV} baseline ERL. The bends (both horizontal and vertical) are captured in Tab.~\ref{tab:DipolesComponents}, while the quadrupole magnets and RF cavities are collected in Tab.~\ref{tab:QuadRFComponents}.

One would like to use a combined aperture (3-in-one) arc magnet design with 50\,cm vertical separation between the three apertures, proposed by Attilo Milanese~\cite{AM14}. That would reduce net arc bend count from 2112 to 704. As far as the Spr/Rec vertical bends are concerned, the design was optimised to include an additional common bend separating the two highest passes. So, there are a total of 8 trapezoid B-com magnets, with second face tilted by \SI{3}{\degree} and large 10\,cm vertical aperture, the rest are simple rectangular bends with specs from the summary Tab.~\ref{tab:DipolesComponents}.

\begin{table}[!ht]
  \centering
  \small
  \begin{tabular}{l@{\hspace{2em}}c@{\hspace{0.8em}}c@{\hspace{0.8em}}c@{\hspace{0.8em}}c@{\hspace{0.8em}}c@{\hspace{0.8em}}c@{\hspace{0.8em}}c@{\hspace{0.8em}}c@{\hspace{0.8em}}c@{\hspace{0.8em}}c@{\hspace{0.8em}}c@{\hspace{0.8em}}c@{\hspace{0.8em}}c@{\hspace{0.8em}}c} 
    \toprule
    & \multicolumn{4}{c}{Arc dipoles (horiz.)} & & \multicolumn{4}{c}{Spr/Rec dipoles (vert.)} & & \multicolumn{4}{c}{\emph{Dogleg} dipoles (horiz.)} \\
    \cmidrule{2-5}  \cmidrule{7-10}  \cmidrule{12-15} 
    Section & $N$ & $B [\text{T}]$ & $g/2[\text{cm}]$ & $L [\text{m}]$ &
    & $N$ & $B [\text{T}]$ & $g/2[\text{cm}]$ & $L [\text{m}]$ &
    & $N$ & $B [\text{T}]$ & $g/2[\text{cm}]$ & $L [\text{m}]$ \\
    \midrule
    Arc 1 & 352 & 0.087 & 1.5 & 3 & & 8 & 0.678 & 2 & 3 & & 16 & 1  & 1.5 & 1\\
    Arc 2 & 352 & 0.174 & 1.5 & 3 & & 8 & 0.989 & 2 & 3 & & 16 & 1 & 1.5 & 1\\
    Arc 3 & 352 & 0.261 & 1.5 & 3 & & 6 & 1.222 & 2 & 3 & & 16 & 1  & 1.5 & 1\\
    Arc 4 & 352 & 0.348 & 1.5 & 3 & & 6 & 1.633 & 2 & 3 & & 16 & 1 & 1.5 & 1\\
    Arc 5 & 352 & 0.435 & 1.5 & 3 & & 4 & 1.022 & 2 & 3 & &  &  &  & \\
    Arc 6 & 352 & 0.522 & 1.5 & 3 & & 4 & 1.389 & 2 & 3 & &  &  &  & \\
    \midrule
    Total & 2112 &  &  &  &  & 36 &  &  &  &  & 64 &  &  &\\
    \bottomrule
  \end{tabular}
  \caption{\SI{50}{GeV} ERL -- Dipole magnet count along with basic magnet parameters: Magnetic field $(B)$, Half-Gap $(g/2)$, and Magnetic length $(L)$.}
  \label{tab:DipolesComponents}
\end{table}

\begin{table}[!ht]
\centering
  \small
  \begin{tabular}{l@{\hspace{2.5em}}c@{\hspace{0.8em}}c@{\hspace{0.8em}}c@{\hspace{0.8em}}c@{\hspace{0.8em}}c@{\hspace{0.8em}}c@{\hspace{0.8em}}c@{\hspace{0.8em}}c@{\hspace{0.8em}}c} 
    \toprule
    & \multicolumn{4}{c}{Quadrupoles} &  & \multicolumn{4}{c}{RF cavities} \\
    \cmidrule{2-5}  \cmidrule{7-10} 
    Section & $N$ & $G [\text{T/m}]$ & $a[\text{cm}]$ & $L [\text{m}]$ & &  $N$ & $f [\text{MHz}]$ & cell & $G_\text{RF}[\text{T/m}]$ \\
    \midrule
    Linac 1 &  29 &  7.7 & 3   & 0.25 & & 448 & 802 & 5 & 20 \\
    Linac 2 &  29 &  7.7 & 3   & 0.25 & & 448 & 802 & 5 & 20 \\
    Arc 1   & 255 &  9.25 & 2.5 & 1 & &   &   &   &   \\
    Arc 2   & 255 & 17.67 & 2.5 & 1 & &  &  &  &  \\
    Arc 3   & 255 & 24.25 & 2.5 & 1 & & 6 & 1604 & 9 & 30 \\
    Arc 4   & 255 & 27.17 & 2.5 & 1 & & 12 & 1604 & 9 & 30 \\
    Arc 5   & 249 & 33.92 & 2.5 & 1 & & 18 & 1604 & 9 & 30 \\
    Arc 6   & 249 & 40.75 & 2.5 & 1 & & 36 & 1604 & 9 & 30 \\
        \midrule
    Total & 1576 &  &  &  & & 968 &  &  & \\
    \bottomrule
  \end{tabular}
  \caption{\SI{50}{GeV} ERL -- Quadrupole magnet and RF cavities count along with basic magnet/RF parameters: Magnetic field gradient $(G)$, Aperture radius $(a)$, Magnetic length $(L)$, Frequency $(f)$, Number of cells in RF cavity (cell), and RF Gradient $(G_\text{RF})$.}
  \label{tab:QuadRFComponents}
\end{table}

\section{Electron-Ion Collisions \label{sec:eAoperation}\ourauthor{John Jowett}} 
Besides colliding proton beams, the LHC also provides collisions of nuclear (fully-stripped ion) beams with each other (AA collisions) or with protons ($p$A).  
Either of these operating modes offers the possibility of electron-ion ($e$A) collisions in the LHeC configuration\footnote{
In $p$A operation of the LHC the beams may be reversed (A$p$) for some part of the operating time.
Only one direction (ions in Beam~2) would provide $e$A collisions while the other would provide ep collisions at significantly reduced luminosity compared to the $pp$ mode, since there would be fewer proton bunches of lower intensity.
}

Here we summarise the considerations leading to the  luminosity estimates given in Tab.\,\ref{tab:lumeA} for collisions of electrons with \Pb\ nuclei, the nominal heavy ion species collided in the LHC.   
Other, lighter, nuclei are under consideration for future LHC operation~\cite{Citron:2018lsq} and could also be considered for electron-ion collisions. 

The  heavy ion beams that the CERN injector complex 
can provide   to  the LHC, the HE-LHC and the FCC provide a unique basis
for high energy, high luminosity deep inelastic electron-ion scattering physics.
Since HERA was restricted to protons only, the LHeC or FCC-eh  would  extend the
kinematic range in $Q^2$ and $1/x$ by 4 or 5 orders of magnitude.
This 
is a huge increase in coverage and would be set to radically change the understanding of
parton dynamics in nuclei and of the formation of the quark gluon plasma. 

 An initial set of
 parameters  in the maximum energy configurations was given in~\cite{LHeClumi}.  
 The Pb beam parameters are essentially those foreseen for operation of the LHC (or HL-LHC) in Run~3 and Run~4 (planned for the 2020s).  These parameters  have already been largely demonstrated~\cite{Jowett:2019jni} except for the major remaining step of implementing slip-stacking injection in the SPS which would reduce the basic bunch spacing 
 from 100 to 50~ns~\cite{Argyropoulos:2019dnj}.  
 With respect to the proton spacing of 25~ns, this allows the electron bunch intensity to be doubled while still respecting the limit on total electron current. 
In fact, without the slip-stacking in the SPS, the initial luminosity would be the same with a 100~ns Pb spacing (and quadrupled electron bunch intensity). 
However one must remember that the evolution of the Pb beam intensity will be dominated by luminosity burn-off by the concurrent PbPb collisions at the other interaction points and integrated luminosity for both PbPb and ePb collisions will be higher with the higher total Pb intensity.  
The details of this will depend on the operating scenarios, number of active experiments, etc, and are not considered further here. 
The time-evolution of eA luminosity will be determined by that of  PbPb and pPb collisions, as discussed, for example, in Ref.~\cite{Schaumann:2015fsa,Citron:2018lsq,Benedikt:2018csr}.

Combining these assumptions with  the default $50$\,GeV electron ERL for LHeC and 60~GeV for FCC-eh, yields the updated parameter sets and initial luminosities given in 
Tab.\,\ref{tab:lumeA}, earlier in the present report.

Radiation damping of Pb beams in the hadron rings is about twice as fast as for protons and can be fully exploited since it takes longer to approach the beam-beam limit at the PbPb collisions points. 
For the case of the FCC-hh~\cite{Benedikt:2018csr}, one can expect the emittance values in Tab.~\ref{tab:lumeA} to be reduced  during  fills~\cite{Schaumann:2015fsa,Citron:2018lsq,Benedikt:2018csr}. 

The Pb beam will be affected by ultraperipheral collision effects, mainly bound-free pair production and Coulomb dissociation of the nuclei, induced by the electromagnetic fields of the electrons, seen as pulses of virtual photons.   
The relevant cross-sections will be similar to those in pPb collisions which are down by a factor of $Z^2$  compared to those in PbPb collisions and can be neglected in practice.

\section{Beam-Beam Interactions \ourauthor{Kevin Andre, Andrea Latina, Daniel Schulte}}

In the framework of the Large Hadron electron Collider, the concept of an Energy Recirculating Linac (ERL) allows to overcome the beam-beam limit that one would face in a storage ring. The electron beam can be heavily disturbed by the beam collision process, while the large acceptance of the ERL will still allow for a successful energy recovery during the deceleration of the beam so that the power consumption is minimised. In order to compare the relevant beam-beam parameters and put them into the context of other colliders, two tables are shown highlighting, on the one hand, the parameters from LEP and LHC runs in  Tab.~\ref{tab:LEPLHCparams}, and on the other hand, the parameters planned for LHeC at HL-LHC in Tab.~\ref{tab:LHeCparams}.

\begin{table}[!htb]
    \centering
    \small
    \begin{tabular}{lccc}
      \toprule
         Parameter & Unit & LEP & LHC \\
      \midrule
         Beam sizes $\sigma_x$ / $\sigma_y$ & \si{\micro m} & 180 / 7 & 16.6 / 16.6 \\
         Intensity  & $10^{11}$ particles/bunch & 4.00 & 1.15 \\
         Energy & GeV & 100 & 7000 \\
         $\beta^*_x/\beta^*_y$ & cm & 125/5 & 55/55 \\
         Crossing angle & \si{\micro rad} & 0 & 0/285 \\
         Beam-beam tune shift $\Delta Q_x / \Delta Q_y$ &  & 0.0400/0.0400 & 0.0037/0.0034 \\
         Beam-beam parameter $\xi$ & & 0.0700 & 0.0037 \\
    \bottomrule
    \end{tabular}
    \caption{Comparison of parameters for the LEP collider and LHC. Taken from CDR 2012, p.286.}
      \label{tab:LEPLHCparams}
\end{table}

\begin{table}[!htb]
  \centering
  \small
      \begin{tabular}{lccc}
         \toprule
         Beam parameter & Unit & \multicolumn{2}{c}{LHeC at HL-LHC} \\
         \cmidrule{3-4}
         & & Proton beam & Electron beam  \\
         \midrule
         Energy & GeV & 7000 & 49.19 \\
         Normalized emittance & mm$\cdot$mrad & 2.5 & 50\\
         Beam sizes $\sigma_{x,y}$ & \si{\micro m} & 5.8 & 5.8 \\
         Intensity & $10^{9}$ particles/bunch & 220.00 & 3.1 \\
         Bunch length \textrm{$\sigma_s$} & mm & 75.5 & 0.6 \\
         $\beta^*_{x,y}$ & cm & 10.00 & 6.45 \\
         Disruption factor & & $1.2 \times 10^{-5}$ & 14.5  \\
         Beam-beam parameter $\xi$ & & $1.52 \times 10^{-4}$ & 0.99 \\
         \bottomrule
      \end{tabular}
      \caption{Comparison of parameters for the LHeC at HL-LHC. The parameters presented correspond to the default design.}
      \label{tab:LHeCparams}
\end{table}

In the case of LHeC, the $\beta$-functions at the interaction point are chosen such that the transverse beam sizes of the $e$- and $p$- beams are equal in both transverse planes. Although the proton and electron emittances are different, the beta functions at the interaction point are set accordingly so that the two beams conserve
$\sigma_x^e = \sigma_x^p$ and $\sigma_y^e = \sigma_y^p$.

\subsection{Effect on the electron beam}

\begin{figure}[htb]
    \centering
    \includegraphics[width=0.45\textwidth]{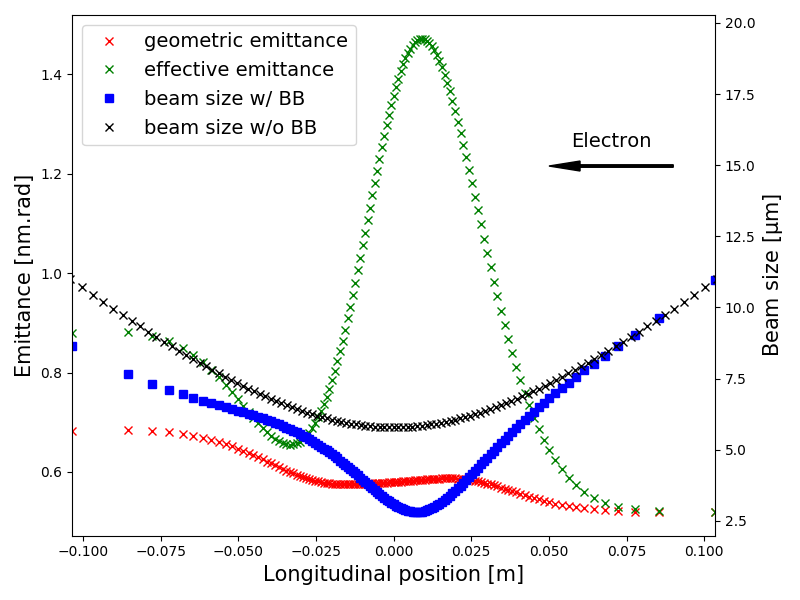}
    \hspace{0.02\textwidth}
    \includegraphics[width=0.45\textwidth]{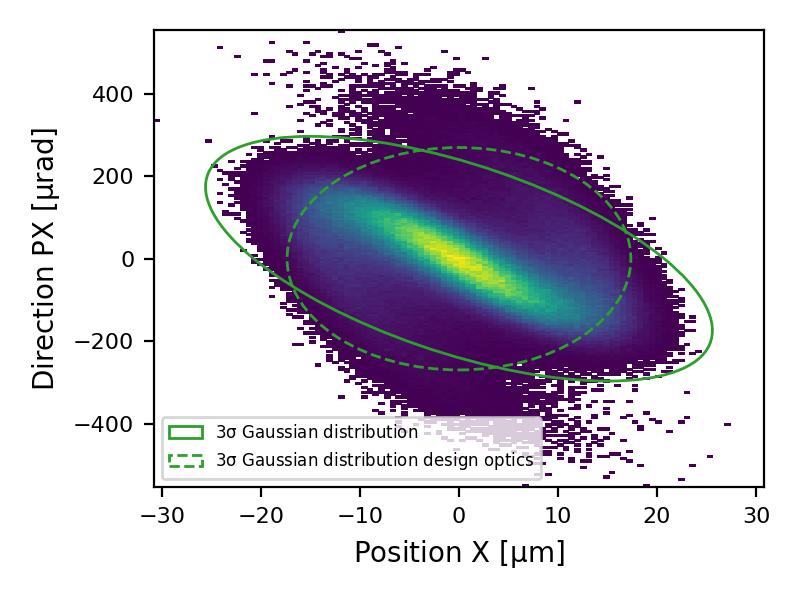}
    \caption{Left: Electron beam sizes with (blue) and without (black) the beam-beam forces exerted on the electron beam. The geometric emittance is represented in red and the effective emittance that takes into account the mismatch from the original optics is illustrated in green. Right: The horizontal phase space of the spent electron distribution backtracked to the interaction point. $3\sigma$ Gaussian distribution are highlighted for the post-collided distribution (solid line) and the design optics (dashed line).}
    \label{fig:spentelectron}
\end{figure}

The disruption parameter for the electron beam is of the order of 14.5 which corresponds, in linear approximation, to almost 2 oscillations of the beam envelope within the proton bunch.
The non linearity of the interaction creates a distortion of the phase space and a mismatch from the design optics (see Fig.~\ref{fig:spentelectron}). The mismatch and distortion can be minimized by tuning the Twiss parameters ($\alpha^*, \beta^*$) at the interaction point.

\begin{figure}[htb]
    \centering
    \includegraphics[width=0.47\textwidth]{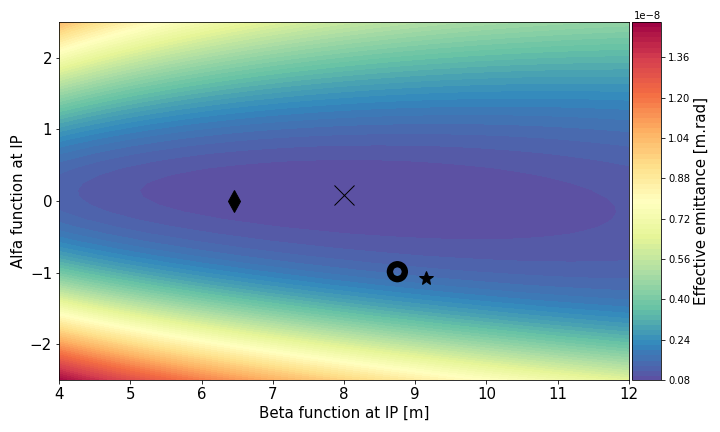}
    \includegraphics[width=0.47\textwidth]{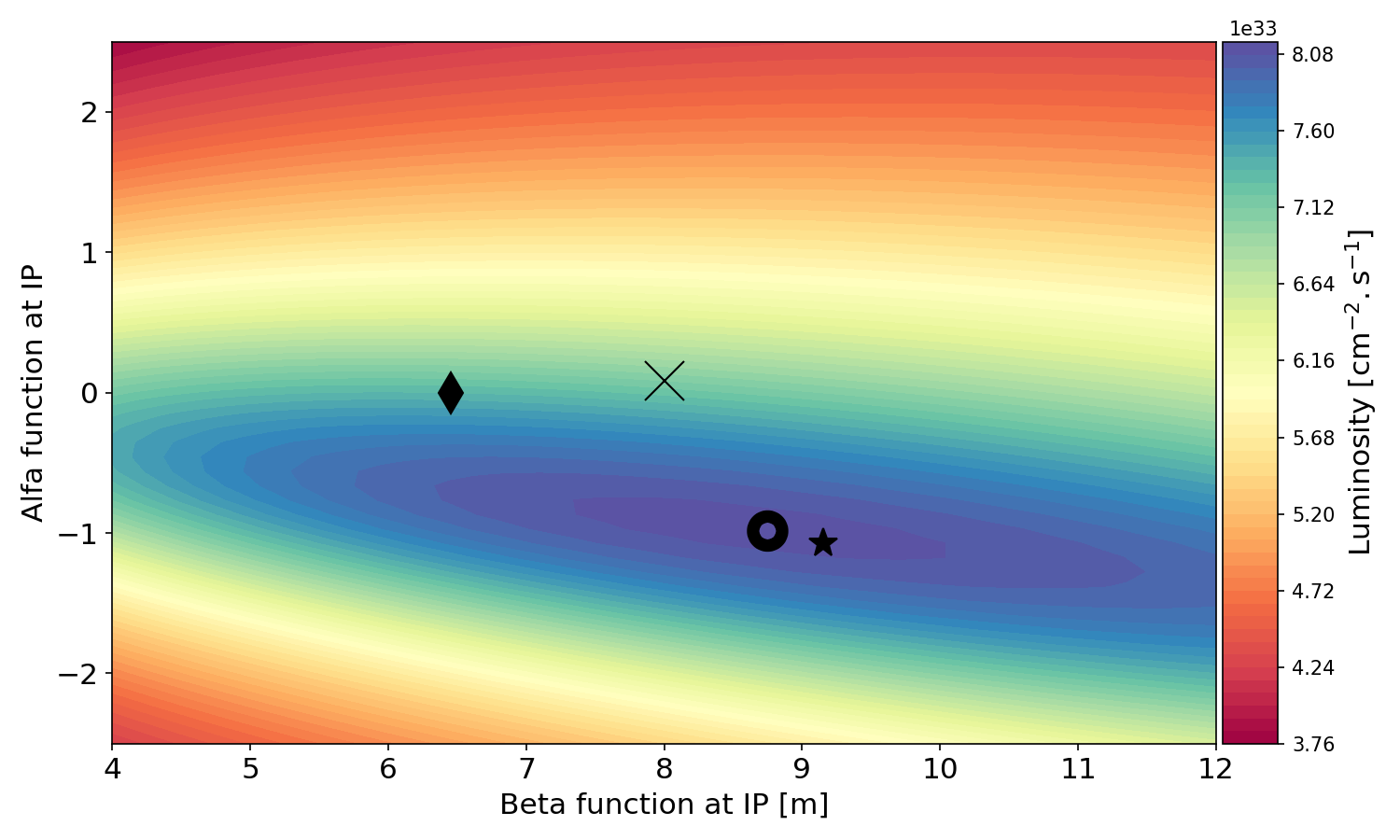}
    \caption{Left : Contour plot describing the effective emittance post collision as a function of the alfa and beta functions at IP. Right : Contour plot describing the luminosity as a function of the alfa and beta functions at the IP. The diamond marker represents the initial Twiss parameters, the circle shows the luminosity optimum, the cross symbolizes the smallest mismatch from the original optics and the star illustrates the minimal geometric emittance growth.}
    \label{fig:EmittanceScan}
\end{figure}

In a series of studies  the optics parameters of the electron beam were tracked back to the interaction point in presence of the beam-beam forces in order to show the impact of the beam-beam effect for different values of the electron Twiss parameters at the IP. In addition, the influence of a waist shift from the IP (proportional to $\alpha^*$), similar to changing the foci of the interacting beams, has been studied and allows to keep the electron beam for a longer time within the proton bunch, thus  optimizing the luminosity. The modification of the electron beta function ($\beta^*$) leads to more freedom and gives access, among all the possibilities, to two different optima regarding the luminosity and the mismatch from the design optics. The results are summarized in the contour plots of Fig.~\ref{fig:EmittanceScan}.

As a consequence, the Twiss parameters at the interaction point can be set in a way, to minimize the mismatch of the optics (i.e.\ the effective emittance)  or to maximize the luminosity. In case the optimization of the luminosity is chosen (see the circle marker in Fig.~\ref{fig:EmittanceScan}), a modified capture optics in the beam transfer to the arc structure will be needed to re-match the modified Twiss functions perturbed by the non-linear beam-beam effects. 

\begin{figure}[htb]
    \centering
    \includegraphics[width=0.5\textwidth]{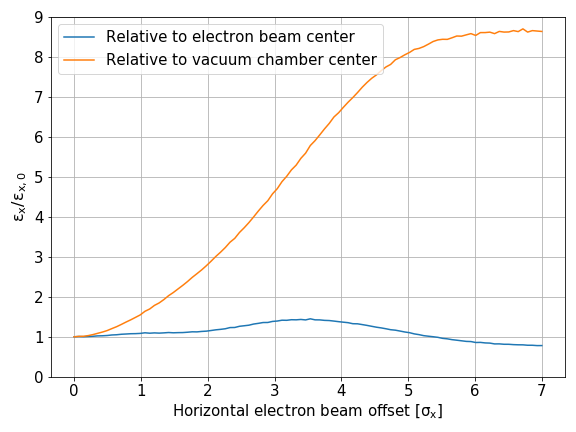}
    \caption{Electron bean emittance relative change with respect to its centroid (blue) and with respect to the vacuum chamber center (orange).}
    \label{fig:EmittanceOffset}
\end{figure}

The effect of possible offsets between the two colliding beams has been characterized in previous beam-beam studies~\cite{BeamDynamicsNewsletter_n68}, and -- if uncorrected -- might lead to an electron beam emittance growth.  The parameters for these studies have been updated and the results are presented in Fig.~\ref{fig:EmittanceOffset}. As any offset between the two beams is amplified, it results in a larger increase of the beam envelope. As a solution, a fast feed-forward system is proposed, across the Arc 6, which would aim at damping the transverse motion so that the beam emittance can be recovered. Using two sets of kickers placed at the center and at the end of the arc, an offset of $0.16 \sigma$ can be damped. A single set cutting across the whole arc can correct a $1 \sigma$ offset with approximately \SI{4.4}{kV}.

Additionally, the coupling of the beam-beam effect with long range wakefields has been addressed~\cite{BeamDynamicsNewsletter_n68}. Assuming a misaligned bunch injected among a train of nominal bunches, the coupling of the beam-beam effect with the wakefields leads to a reduction of the damping of the excitation created by the misaligned bunch. Nevertheless it can be shown that the beam stability is conserved and the total amplification remains acceptable with respect to the study that was not considering the coupling.

\subsection{Effect on the proton beam}
The beam-beam interaction between the electron and proton beams is asymmetric in terms of beam rigidities. Although the less energetic \SI{49.19}{GeV} electron beam is heavily distorted by the strong \SI{7}{TeV} proton beam, the proton beam will suffer from an emittance growth adding up turn by turn~\cite{BeamDynamicsNewsletter_n68} due to the build up of the tiny disruption created by the offset between the beams. In fact, the previous studies gave a growth rate of around \SI{0.01}{\%/s} for a jitter of 0.2~ $\sigma_x$. As long as an adequate control of the bunches is preserved, this effect should lie in the shadow of other effects leading to emittance blow-up in the LHC (e.g. IBS). Since the electron beam energy decreased from \SI{60}{GeV} to \SI{49.19}{GeV} this study needs to be updated and the results should remain in agreement with the previous statement.

\section{Arc Magnets \ourauthor{Pierre Thonet, Cynthia Vallerand}}
In this section, a conceptual design of the main magnets needed for the Linac-Ring (LR) accelerator at \SI{50}{GeV} is described. The number and types of magnets is listed in Tabs.~\ref{tab:DipolesComponents} and~\ref{tab:QuadRFComponents}.

\subsection{Dipole magnets}
The bending magnets are used in the arcs of the recirculator. Each of the six arcs needs 352 horizontal bending dipoles. Additional dipoles are needed in the straight sections: 36 vertical bending dipoles in the spreader/recombiner and 64 horizontal bending dipoles for the “dogleg”. These magnets are not considered at the moment.

In the CDR issued in 2012 for a \SI{60}{GeV} lepton ring (LR) , a design based on three independent dipoles stacked on top of each other was proposed. A post-CDR design with three apertures dipoles was introduced in 2014~\cite{AM14}. This solution allows reducing the Ampere-turns and the production cost of the dipoles. For a \SI{50}{GeV} LR, the three apertures dipole design is adapted to fulfil new magnetic field requirements. 

The 352 horizontal bending dipoles needed for each arc, combined in three apertures dipoles result in a total of 704 units. These magnets are \SI{3}{m} long and provide a field in the \SI{30}{mm} aperture ranging from \SI{0.087}{T} to \SI{0.522}{T} depending on the arc energy, from \SI{8.62}{GeV} to \SI{49.19}{GeV}.

\begin{table}[!htb]
  \centering
  \small
      \begin{tabular}{lcc}
        \toprule
        Parameter & Unit & Value \\
        \midrule
         Beam energy & GeV & 8.62 to 49.19 \\
         Magnetic field & T & 0.087 to 0.522 \\
         Magnetic length & m & 3 \\
         Vertical aperture & mm & 30 \\
         Pole width & mm & 90 \\
         Number of apertures &  & 3 \\
         Distance between apertures & mm & 500 \\
         Mass & 8000 & kg \\
         Number of magnets & & 704 \\
         Current & A & 4250 \\
         Number of turns per magnet & & 4 \\
         Current density & A/mm$^2$ & 1 \\
         Conductor material &  & aluminum \\
         Magnet resistance & m$\Omega$ & 0.17 \\
         Power & kW & 3 \\
         Total power consumption six arcs & MW & 2.1 \\
         Cooling & & air \\
         \bottomrule
      \end{tabular}
      \caption{\SI{50}{GeV} ERL -- Main parameters of the three apertures bending magnets.}
      \label{tab:LHeCArcDipole}
\end{table}

\begin{figure}[!htb]
    \centering
    \includegraphics[width=0.45\textwidth]{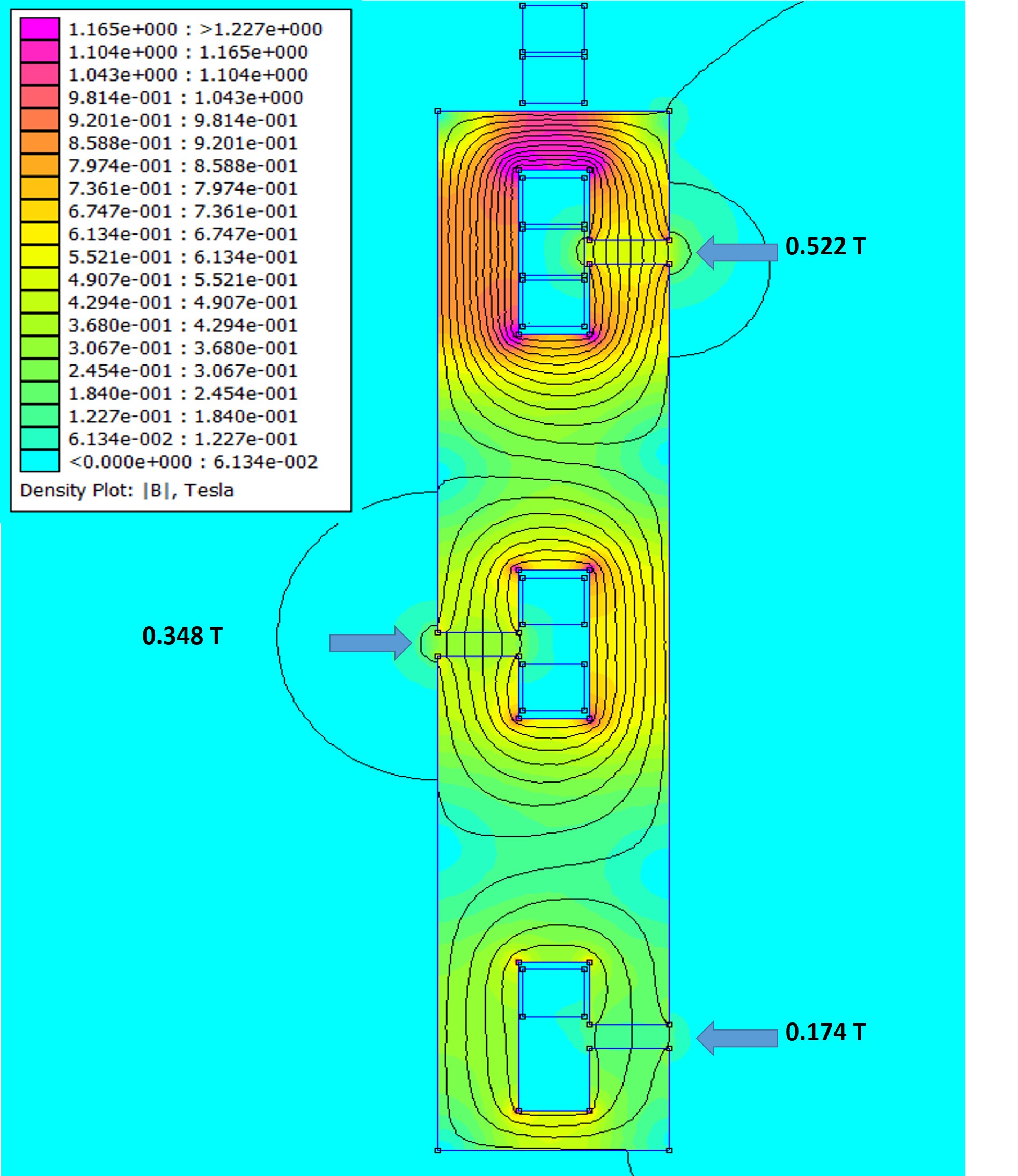}
    \caption{\label{fig:ArcDipole}\SI{50}{GeV} ERL - Cross section of the three apertures bending magnet, arc 2, 4 and 6 with \SI{500}{mm} between consecutive arcs - Finite Element Method (FEM).}
\end{figure}

In the proposed design, the three apertures are stacked vertically but offset transversely. This allows recycling the Ampere-turns from one aperture to the other. The coils are centrally located on the yoke and are made of simple aluminium bus-bars all powered in series. A current density of \SI{1}{A/mm^2} in the coils is sufficiently low to not have water-cooling but in order to limit the temperature in the tunnel it may be required.
Trim coils can be added on two of the apertures to provide some tuning. Alternatively, each stage could be powered separately. The dipole yokes are made of low carbon steel plates.
The relevant parameters are summarised in Tab.~\ref{tab:LHeCArcDipole} and the cross section is illustrated in Fig.~\ref{fig:ArcDipole} for \SI{500}{mm} between consecutive arcs.

\subsection{Quadrupole magnets}
\subsubsection{Quadrupoles for recirculator arcs}
\begin{figure}[!htb]
    \centering
    \includegraphics[width=0.45\textwidth]{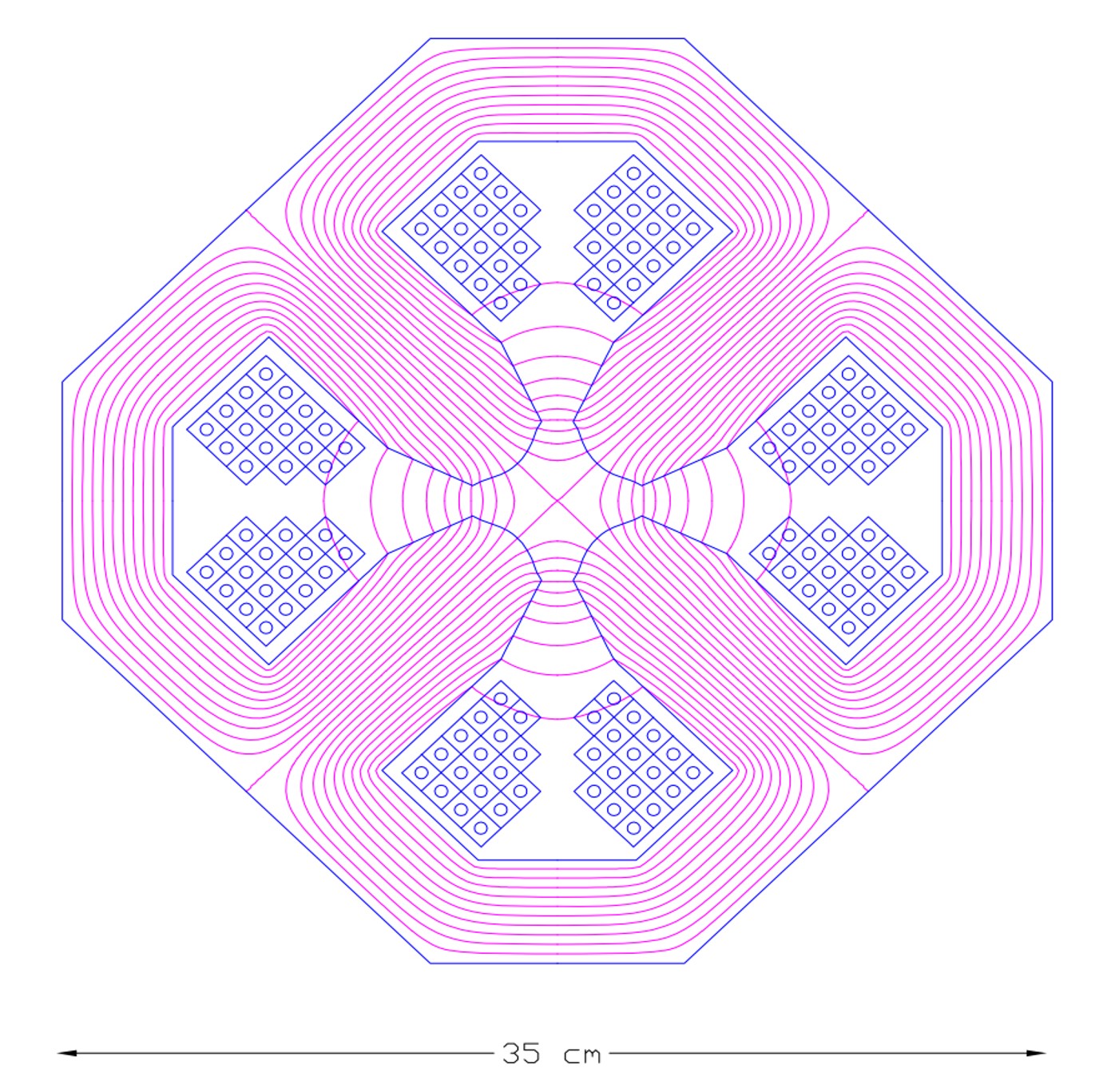}
    \hspace{0.03\textwidth}
    \includegraphics[width=0.45\textwidth]{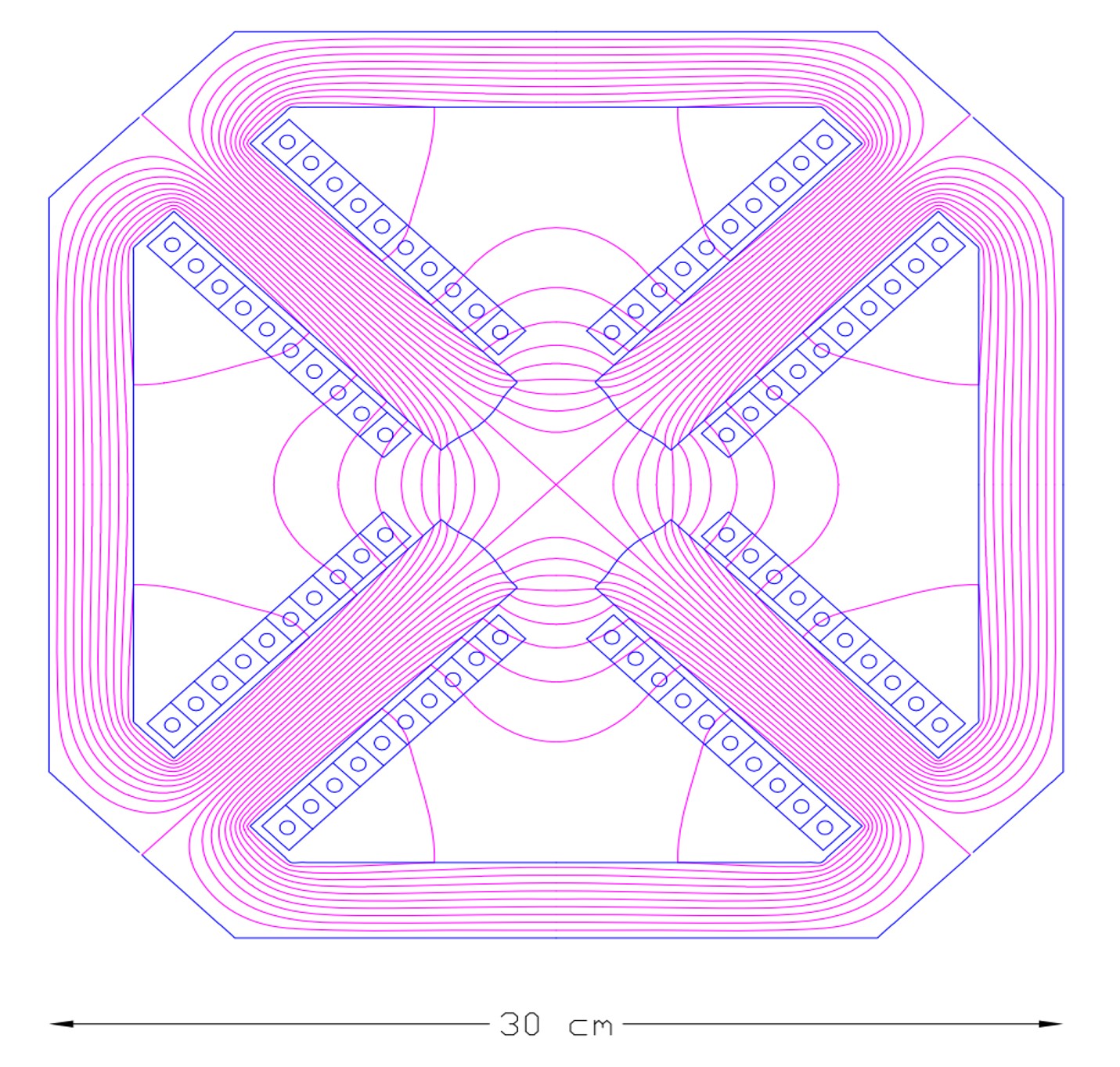}
    \caption{\SI{50}{GeV} ERL. Left: Cross section of the arc quadrupole magnets.
      Right: Cross section of the linac quadrupole magnets.
    }
    \label{fig:ArcQuadLinacQuad}
\end{figure}

In total 1518 quadrupoles are needed for the recirculator arcs: 255 for each of the arcs one to four and 249 for each of the arcs five and six. The required integrated gradients, comprised between \SI{9.25}{T} and \SI{40.75}{T}, can be achieved using one type of quadrupole one meter long. However, instead of operating the magnets at low current for lower arcs energy, it can be considered to have a shorter model 0.6 meter long for arcs one to three. These quadrupoles require water-cooling for the coils. The relevant parameters are summarised in Tab.~\ref{tab:LHeCArcQuad} and the cross section is illustrated in Fig.~\ref{fig:ArcQuadLinacQuad} (left).

In order to reduce the power consumption, it could be envisaged to use a hybrid configuration for the quadrupoles, with most of the excitation given by permanent magnets. The gradient strength could be varied by trim coils or by mechanical methods.

\begin{table}[!htb]
  \centering
  \small
      \begin{tabular}{lcc}
        \toprule
        Parameter  &  Unit  & Value \\
        \midrule
         Beam energy & GeV & 8.62 to 49.19 \\
         Field gradient & T/m &  9.25 to 40.75 \\
         Magnetic length & m & 1 \\
         Aperture radius & mm & 25 \\
         Mass & kg & 550 \\
         Number of magnets & & 1518 \\
         Current at \SI{40.75}{T/m} & A & 560 \\
         Number of turns per pole & & 17 \\
         Current density at 40.75 T/m & A/mm$^2$ & 6.7 \\
         Conductor material & & copper \\
         Magnet resistance & m$\Omega$ & 33 \\
         Power at 8.62\,GeV  & kW & 0.5 \\
         Power at 16.73\,GeV & kW & 1.9 \\
         Power at 24.85\,GeV & kW & 3.7 \\
         Power at 32.96\,GeV & kW & 4.6 \\
         Power at 41.08\,GeV & kW & 7.2 \\
         Power at 49.19\,GeV & kW & 10.3 \\
         Total power consumption six arcs & MW & 7.1 \\
         Cooling &  & water \\
         \bottomrule
      \end{tabular}
      \caption{\label{tab:LHeCArcQuad}\SI{50}{GeV} ERL -- Main parameters of the arc quadrupoles.}
\end{table}

\subsubsection{Quadrupoles for the two 8.1 GeV linacs}

In the two 8.1\,GeV linacs, 29 + 29 quadrupoles, each providing \SI{1.93}{T} integrated strength are required. The present design solution considers \SI{30}{mm} aperture radius magnets. The relevant parameters are summarised in Tab.~\ref{tab:LHeCLinacQuad} and the cross section is illustrated in Fig.~\ref{fig:ArcQuadLinacQuad} (right).

\begin{table}[!ht]
  \centering
  \small
      \begin{tabular}{lcc}
        \toprule
        Parameter & Unit & Value \\
        \midrule
         Beam energy & GeV & 8.62 to 49.19 \\
         Field gradient & T/m & 7.7 \\
         Magnetic length & m & 0.25 \\
         Aperture radius & mm & 30 \\
         Mass & kg & 110 \\
         Number of magnets &  & 56 \\
         Current at \SI{7.7}{T/m} & A & 285 \\
         Number of turns per pole & & 10 \\
         Current density at \SI{7.7}{T/m} & A/mm$^2$ & 3 \\
         Conductor material &  & copper \\
         Magnet resistance & m$\Omega$ & 6 \\
         Power at 8.1\,GeV & kW & 0.5 \\
         Total power consumption 2 linacs & MW & 0.03 \\
         Cooling & & water \\
         \bottomrule
      \end{tabular}
      \caption{\SI{50}{GeV} ERL -- Main parameters of the linac quadrupoles.}
      \label{tab:LHeCLinacQuad}
\end{table}


%
\section{LINAC and SRF \ourauthor{Erk Jensen}} \label{LandSRF}
Each of the two main linacs has an overall length of \SI{828.8}{m} and provides an acceleration of \SI{8.114}{GV}.
Each linac consists of 112 cryomodules, arranged in 28 units of 4 cryomodules with their focussing elements -- each cryomodule contains four 5-cell cavities, optimised to operate  with large beam current (up to \SI{120}{mA} at the High Order Mode -- HOM -- frequencies).
The operating temperature is \SI{2}{K}; the cavities are based on modern SRF technology and are fabricated from bulk Nb sheets; they are described in detail in section~\ref{cavprot} below. The nominal acceleration gradient is \SI{19.73}{MV/m}.

In addition to the main linacs, the synchrotron losses in the arcs will make additional linacs necessary, referred to here as the \emph{loss compensation linacs}.
These will have to provide different accelerations in the different arcs, depending on the energy of the beams as shown in Tab.~\ref{tab:SRlosses}.
The quoted beam energies are at entry into the arc. Their natural placement would be at the end of the arcs just before the combiner, where the different energy beams are still separate. The largest of these linacs would have to compensate the SR losses at the highest energy, requiring a total acceleration of about \SI{700}{MV}.
The loss compensation linacs will be detailed in section~\ref{SRcomp} below.
\begin{table}[!ht]
  \centering
  \small
  \begin{tabular}{lcc} 
    \toprule
    Section & Beam energy [\si{GeV}] & $\Delta E$ [\si{MeV}]  \\
    \midrule
    Arc 1 & 8.62 & 3 \\
    Arc 2 & 16.73 & 25 \\
    Arc 3 & 24.85 & 80 \\
    Arc 4 & 32.96 & 229 \\
    Arc 5 & 41.08 & 383 \\
    Arc 6 & 49.19 & 836 \\
    \bottomrule
  \end{tabular}
  \caption{Synchroton radiation losses for the different arc energies}
  \label{tab:SRlosses}
\end{table}

Through all arcs but Arc 6, the beam passes twice, once while accelerated and once while decelerated.
It is planned to operate these additional \emph{loss compensation linacs} at \SI{1603.2}{MHz}, which allows energy compensation of both the accelerated and the decelerated beam simultaneously.
This subject will be discussed in detail in a subsequent section~\ref{SRcomp}.

\subsection{Choice of Frequency  \ourauthor{Frank Marhauser}} \label{choffreq}
The RF frequency choice primarily takes into account the constraints of the LHC bunch repetition frequency, $f_0$, of \SI{40.079}{\mega\hertz}, while allowing for a sufficiently high harmonic, $h$, for a flexible system.
For an ERL with $n_{pass}=3$ recirculating passes and in order to enable equal bunch spacing for the 3 bunches -- though not mandatory -- it was originally considered to suppress all harmonics that are not a multiple of $n_{pass} \cdot f_0= \SI{120.237}{\mega\hertz}$.
Initial choices for instance were \SI{721.42}{\mega\hertz} ($h = 18$) and \SI{1322.61}{\mega\hertz} ($h = 33$) in consideration of the proximity to the frequencies used for state-of-the-art SRF system developments worldwide~\cite{Calaga:2013fpa}.
In synergy with other RF system developments at CERN though, the final choice was \SI{801.58}{\mega\hertz} ($h = 20$), where the bunching between the 3 recirculating bunches can be made similar but not exactly equal.
Note that this frequency is also very close to the \SI{805}{\mega\hertz} SRF proton cavities operating at the Spallation Neutron Source (SNS) at ORNL, so that one could leverage from the experience in regard to cryomodule and component design at this frequency.
  
Furthermore, in the frame of an independent study for a \SI{1}{\giga\electronvolt} CW proton linac, a capital plus operational cost optimisation was conducted~\cite{Marhauser:2014fya}.
This optimisation took into account the expenditures for cavities, cryomodules, the linac tunnel as well as the helium refrigerator expenses as a function of frequency and thus component sizes. Labor costs were included based on the existing SNS linac facility work breakdown structure.
It was shown that capital plus operating costs could be minimised with a cavity frequency between \SI{800}{\mega\hertz} and \SI{850}{\mega\hertz}, depending also on the choice of the operating He bath temperature (\SI{1.8}{\kelvin} to \SI{2.1}{\kelvin}).
Clear benefit of operating in this frequency regime are the comparably small dynamic RF losses per installation length due to a relatively small BCS surface resistance as well as low residual resistance of the niobium at the operating temperature.
This could be principally verified as part of the prototyping effort detailed in the next sub-section.
Note that the cost optimum also favors cavities operating at rather moderate field levels ($<\SI{20}{MV/m}$).
This comes as a benefit in concern of field emission and associated potential performance degradations. 

\subsection{Cavity Prototype   \ourauthor{Frank Marhauser}} \label{cavprot}
\begin{figure}[!ht]
  \centering
  \includegraphics[width=0.8\textwidth]{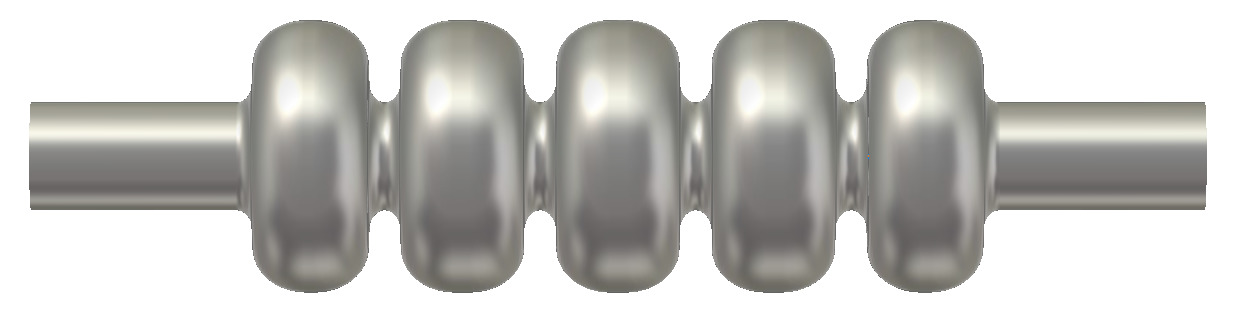}
  \caption{Bare \SI{802}{\mega\hertz} five-cell cavity design (RF vacuum) with a \SI{130}{\milli\meter} iris and beam tube aperture.}
  \label{FIG:BARECAVITY}
\end{figure}

Given the RF frequency of \SI{801.58}{\mega\hertz}, JLab has collaborated with CERN, and consequently proposed a five-cell cavity design that was accepted for prototyping, see Fig.~\ref{FIG:BARECAVITY}. The cavity shape has also been adopted for PERLE. 
Tab.~\ref{tab:rfparameters} summarises the relevant cavity parameters.  

\begin{table}[!ht]
  \centering
  \small
  \begin{tabular}{lcc} 
    \toprule
    Paramater & Unit & Value  \\
    \midrule
    Frequency & \si{\mega\hertz} & 801.58\\
    Number of cells & & 5  \\
    active length $l_{act}$ & \si{\milli\meter} & $917.9$ \\
    loss factor & \si{\volt\per\pico\coulomb}	& 2.742 \\
    $R/Q$ (linac convention)	& \si{\ohm}	& 523.9 \\
    $R/Q\cdot G$ per cell & \si{\ohm^2} & 28788 \\
    Cavity equator diameter & \si{\milli\meter} & 327.95 \\
    Cavity iris diameter & \si{\milli\meter} & 130 \\
    Beam tube inner diameter & \si{\milli\meter} & 130 \\
    diameter ratio equator/iris & & 2.52  \\
    $E_{peak}/E_{acc}$ & & 2.26  \\
    $B_{peak}/E_{acc}$ & \si{\milli\tesla/(\mega\volt\per\meter)}& 4.2  \\
    cell-to-cell coupling factor $k_{cc}$ & \% & 3.21 \\
    TE$_{11}$ cutoff frequency & \si{\giga\hertz} & 1.35 \\
    TM$_{01}$ cutoff frequency & \si{\giga\hertz} & 1.77 \\
    \bottomrule
  \end{tabular}
  \caption{Parameter table of the 802\,MHz prototype five-cell cavity.}
  \label{tab:rfparameters}
\end{table}

The cavity exhibits a rather large iris and beam tube aperture (\SI{130}{\milli\meter}) to consider beam-dynamical aspects such as HOM-driven multi-bunch instabilities. Despite the comparably large aperture, the ratio of the peak surface electric field, $E_{pk}$, respectively the peak surface magnetic field, $B_{pk}$, and the accelerating field, $E_{acc}$, are reasonably low, while the factor $R/Q\cdot G$ is kept reasonably high, concurrently to limit cryogenic losses. This is considered as a generically well \emph{balanced} cavity design~\cite{bib:marhauserAMST}. The cavity cell shape also avoids that crucial HOMs will coincide with the main spectral lines (multiples of \SI{801.58}{\mega\hertz}), while the specific HOM coupler development is pending. 

\begin{figure}[!th]
\centering
\includegraphics[width=0.7\textwidth]{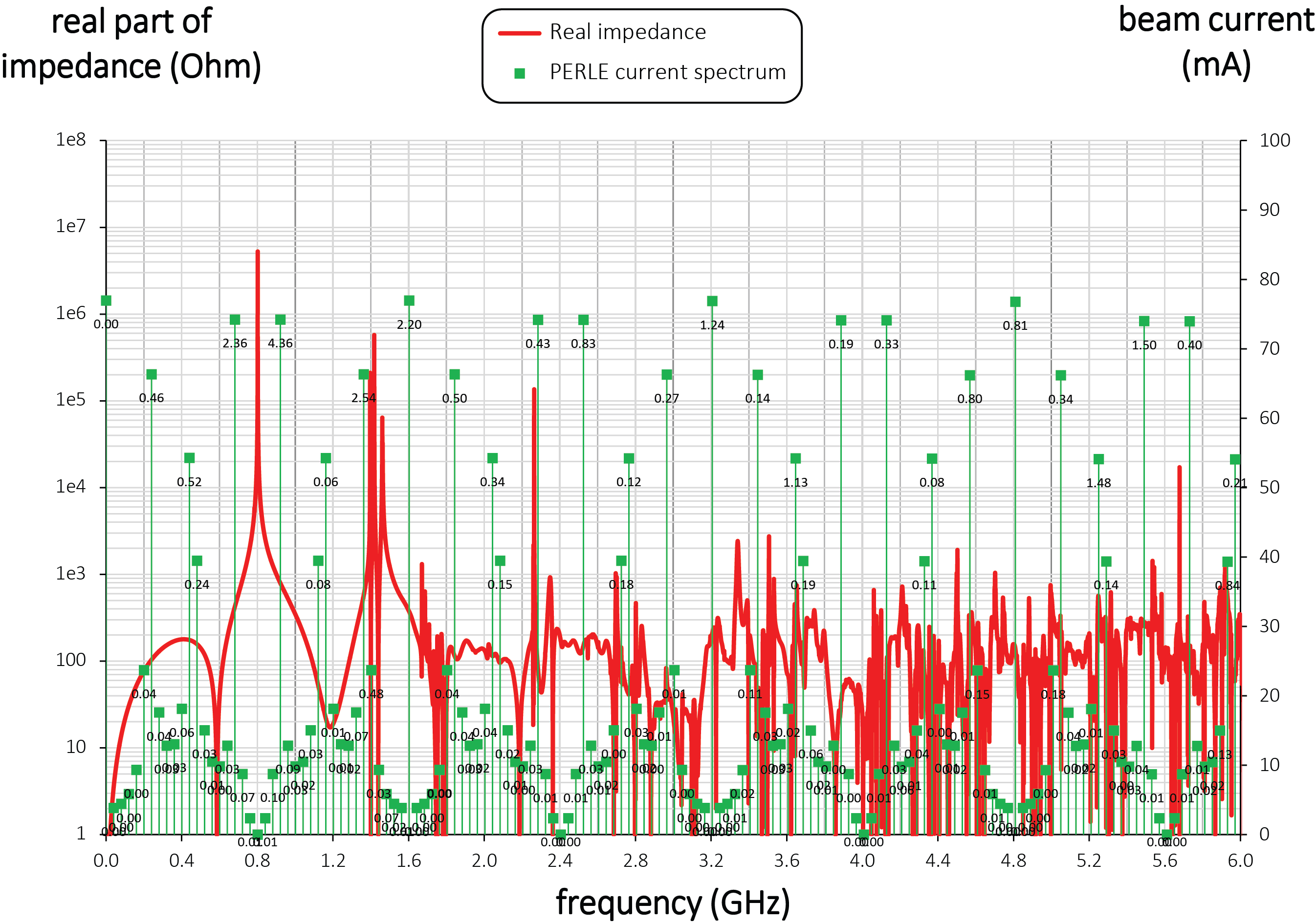}
\caption{Real monopole impedance spectrum of the five-cell \SI{802}{\mega\hertz} cavity prototype (red) together with the considered beam current lines (green) for the 3-pass PERLE machine (\SI{25}{\milli\ampere} injected current). The numbers associated with the spectral lines denote the power dissipation (in Watt).}
\label{FIG:monopolespectrum}
\end{figure}
Furthermore, as shown in Fig.~\ref{FIG:monopolespectrum} for the case of the bunch recombination pattern considered for PERLE originally, the much denser intermediate beam current lines (green) are not coinciding with cavity HOMs. Here the figure plots the real part of the beam-excited cavity monopole impedance spectrum up to \SI{6}{\giga\hertz}, and denotes the power deposited at each spectral line (in Watt) for an injected beam current of \SI{25}{\milli\ampere}. For instance, the summation of the power in this spectral range results in a moderate \SI{30}{W}.
This covers the monopole modes with the highest impedances residing below the beam tube cutoff frequency.
The HOM-induced heat has to be extracted from the cavity and shared among the HOM couplers attached to the cavity beam tubes.
The fraction of the power escaping through the beam tubes above cutoff can be intercepted by beam line absorbers. 

Note that for Fig.~\ref{FIG:monopolespectrum} a single HOM-coupler end-group consisting of three scaled TESLA-type coaxial couplers was assumed to provide damping. Instead of coaxial couplers, waveguide couplers could be utilized, which for instance have been developed at JLab in the past for high current machines.
These are naturally broadband and designed for high power capability, though some penalty is introduced as this will increase the complexity of the cryomodule.
Ultimately, the aim is to efficiently damp the most parasitic longitudinal and transverse modes (each polarization).
The evaluation of the total power deposition is important for LHeC to decide which HOM coupler technology is most appropriate to cope with the dissipated heat and whether active cooling of the couplers is a requirement. 

\begin{figure}[!thb]
  \centering
  \includegraphics[width=0.8\textwidth]{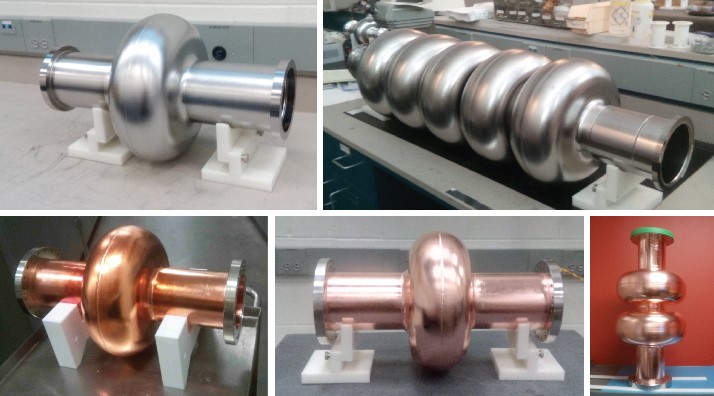}
  \caption{Ensemble of \SI{802}{\mega\hertz} cavities designed and built at JLab for CERN. The Nb cavities have been tested vertically at 2 Kelvin in JLab’s vertical test area.}
  \label{FIG:jlabcavities}
\end{figure}

Though the prototype efforts focused on the five-cell cavity development, JLab also produced single-cell cavities, i.e.\ one further Nb cavity and two OFE copper cavities.
The former has been shipped to FNAL for N-doping/infusion studies, whereas the latter were delivered to CERN for Nb thin-film coating as a possible alternative to bulk Nb cavities.
In addition, a copper cavity was built for low power bench measurements, for which multiple half-cells can be mechanically clamped together.
Presently, a mock-up can be created with up to two full cells.
This cavity has been produced in support of the pending HOM coupler development.
The ensemble of manufactured cavities resonating at \SI{802}{MHz} is shwon in Fig.~\ref{FIG:jlabcavities}.

\begin{figure}[th]
  \centering
  \includegraphics[width=0.6\textwidth]{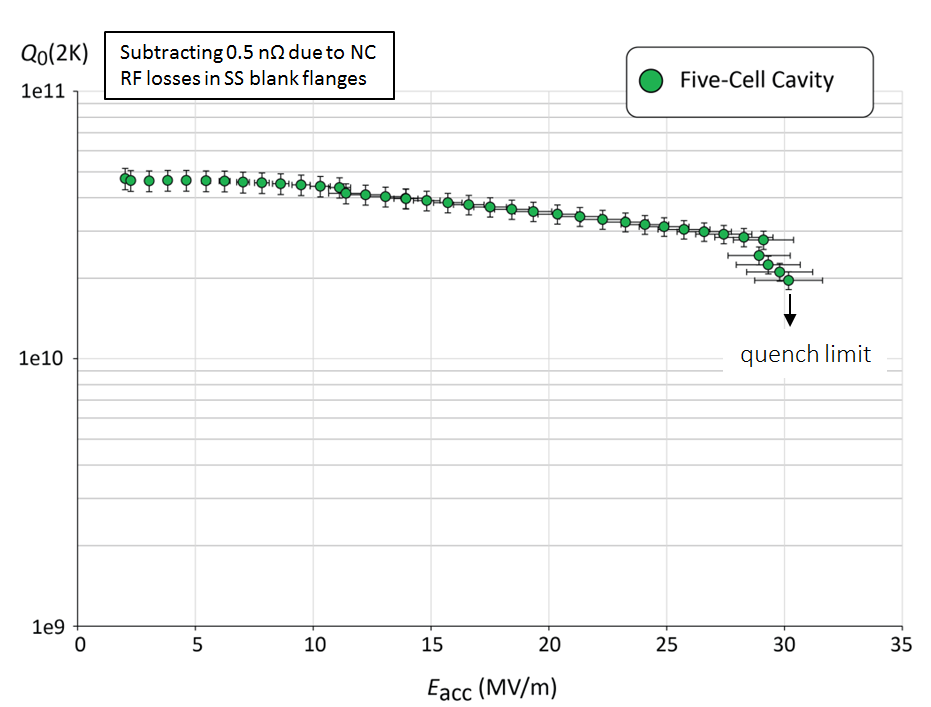}
  \caption{Vertical test result of the five-cell \SI{802}{\mega\hertz} niobium cavity prototype.}
  \label{FIG:qvse802}
\end{figure}

Results for the Nb cavities - made from fine grain high-RRR Nb - were encouraging since both cavities reached accelerating fields, $E_\text{acc}$, slightly above \SI{30}{MV/m} ultimately limited by thermal breakdown (quench).
Moreover, the RF losses were rather small as a benefit of the relatively low RF frequency as anticipated. The residual resistance extracted from the measurement data upon cooldown of the cavity was \SI{3.2}{\ohm} $\pm$ \SI{0.8}{\ohm}.
This resulted in unloaded quality factors, $Q_0$, well above \SI{4e10}{} at \SI{2}{K} at low field levels, while $Q_0$-values beyond \SI{3e10}{} could be maintained for the five-cell cavity up to \SI{\sim 27}{MV/m} (see Fig.~\ref{FIG:qvse802}). Only standard interior surface post-processing methods were applied including bulk buffered chemical polishing, high temperature vacuum annealing, light electropolishing, ultrapure high-pressure water rinsing, and a low temperature bake-out.  While the vertical test results indicate generous headroom for a potential performance reduction once a cavity is equipped with all the ancillary components and installed in a cryomodule, clean cavity assembly procedure protocols must be established for the cryomodules to minimise the chance of introducing field-emitting particulates.

\subsection{Cavity-Cryomodule \ourauthor{Sebastien Bousson}}

The ERL cryomodules hosting the superconducting RF cavities are a key component of the accelerator. They should provide the proper mechanical, vacuum and cryogenic environment to the SRF cavities equipped with their ancillaries systems: helium tank, power coupler and HOM couplers. Each cryomodule is containing 4 superconducting \SI{801.58}{MHz} 5-cells elliptical cavities described in the previous chapters.

Recently, several projects worldwide have designed cryomodules for elliptical cavities with a cavity configuration (number, length and diameter) quite close to the one required by LHeC ERL:
\begin{itemize}
\item SNS\,\cite{schneider2001design}: two different sized cryomodules host either 4 elliptical 6-cells 805 MHz cavities of $\beta$ = 0.81 or 4 elliptical 6-cells \SI{805}{MHz} of $\beta$ = 0.61;
\item SPL\,\cite{parma2011conceptual}: the cryomodule is designed to integrate 4 elliptical 5-cells \SI{704}{MHz} cavities of $\beta$ = 1;
\item ESS\,\cite{olivier2013ess}: two cryomodules of the same length can host either 4 elliptical 6-cells \SI{704}{MHz} cavities of $\beta$ = 0.67 or 4 elliptical 5-cells \SI{704}{MHz} cavities of $\beta$ = 0.85.
\end{itemize}

These three cryomodule designs are based on two completely different concepts for the cavity string support structure. 
SNS and ESS cryomodules are based on an intermediate support system, called the spaceframe, which is horizontally translated inside the cryomodule vacuum vessel. The low pressure cryogenic line is located above the cavities string and connected to the cryogenic transfer line by a double angled connection, the jumper. RF waveguides are connected underneath the cryomodule, using door-knob transition to the couplers. All the hanging and alignment operations of the cavities string and shielding are implemented outside the vacuum tank, using the spaceframe. In the ESS case, each cavity is hanged by 2 sets of 4 cross rods. The thermal shield is also hanged to these rods by the mean of an aluminium “elastic boxes” that allow the thermal shrinkage while maintaining the transverse stability. The thermal shield is made of \SI{2.5}{mm} thick aluminium and wrapped with multi-layer insulation. It is fastened directly to the support rods of the cavities string.

\begin{figure}[!htb]
    \centering
    \includegraphics[width=0.8\textwidth]{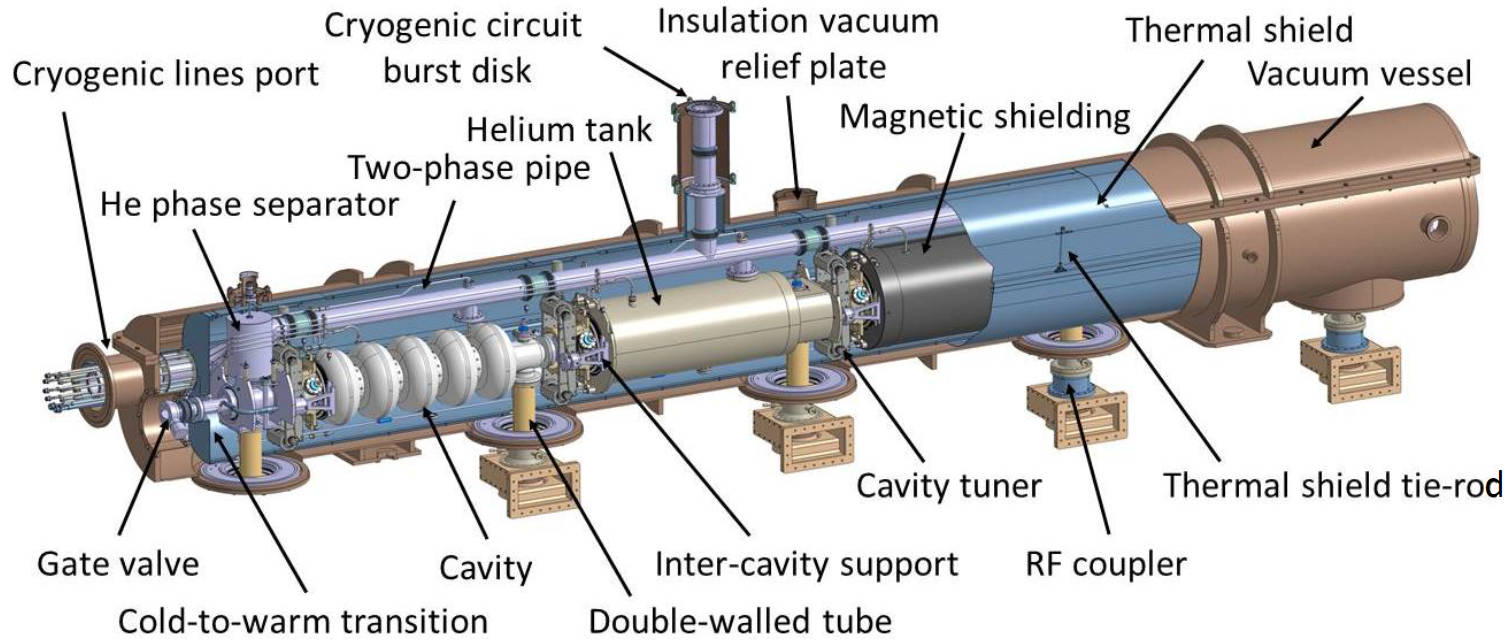}
    \caption{SPL cryomodule general assembly view}
    \label{fig:splCryomoduleGeneralAssembly}
\end{figure}
In the SPL cryomodule, the cavity string is directly supported by the power coupler and with dedicated inter-cavity support features. Moreover, the SPL cryomodule integrates a full length demountable top lid, enabling the cavity string assembly from the cryomodule top (Fig.\,\ref{fig:splCryomoduleGeneralAssembly}). The thermal shield is made of rolled aluminium sheets, and is composed of four main parts assembled before the vertical insertion of the string of cavities. The shield, wrapped with multi-layer insulation, is suspended to the vacuum vessel via adjustable tie rods in titanium alloy which also cope, by angular movements, with its thermal contractions. The cavity stainless steel helium tanks are connected by a 100-mm-diameter two-phase pipe placed above the cavities. This pipe ensures liquid feeding to the cavities by gravity, and is also used as a pumping line for gaseous helium.

\begin{figure}[!htb]
    \centering
    \includegraphics[width=0.65\textwidth]{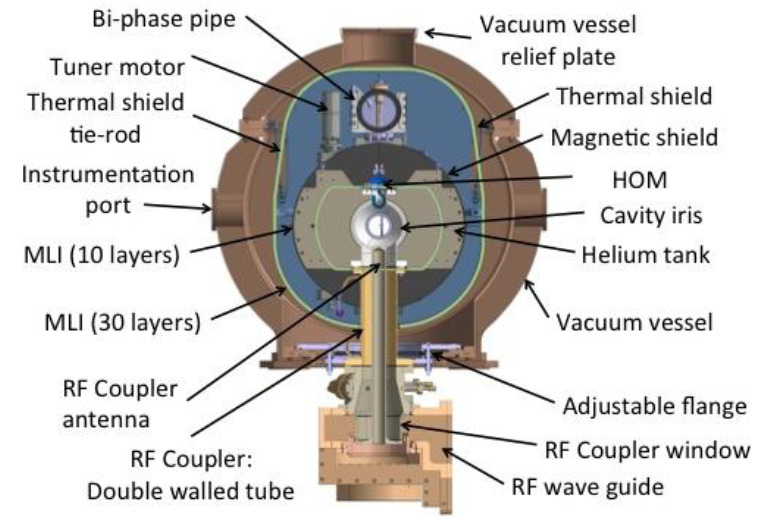}
    \caption{Cross-view of the SPL cryomodule}
    \label{fig:splCryomoduleCrossveiw}
\end{figure} 
With the aim of minimizing static heat loads from room temperature to \SI{2}{K} by solid thermal conduction, the number of mechanical elements between the two extreme temperatures is reduced to the strict minimum: the cavities are supported directly via the external conductor of the RF coupler (Fig.\,\ref{fig:splCryomoduleCrossveiw}), the double-walled tube (DWT). The latter is made out of a stainless steel tube with an internal diameter of 100 mm, which is actively cooled by gaseous helium circulating inside a double-walled envelope in order to improve its thermal efficiency.

An additional supporting point to keep cavity straightness and alignment stability within requirements is obtained by supporting each cavity on the adjacent one via the inter-cavity support, which is composed of a stem sliding inside a spherical bearing. As a result, a pure vertical supporting force is exchanged by adjacent cavities whereas all other degrees of freedom remain unrestrained allowing thermal contraction movements to occur unhindered. The thermo-mechanical behaviour of this supporting system has been extensively studied on a dedicated test bench at CERN, proving its efficiency and reliability.

There are some specific additional constraints or requirements for a cryomodule to be used in an ERL, and some of them are quite challenging, 
The first set of constraints is linked to the CW operation of the cryomodules (contrary to SNS, SPL and ESS which are pulsed accelerators), where dynamic heat loads are much larger than the static ones. Thus, reaching high $Q_0$ (low cryogenic losses) is a main objective in these machines and beside specific optimization on cavity design and preparation (such as N-doping), magnetic shielding should be carefully studied: material, operating temperature, numbers of layers, active and/or passive shielding.
Another important constraint is linked to relative high power to be extracted by the HOM couplers: thermal analysis should be carefully performed to have an optimized evacuation of the HOM thermal load not to degrade the cryogenic performances of the cryomodule.

\begin{figure}[!htb]
    \centering
    \includegraphics[width=0.95\textwidth]{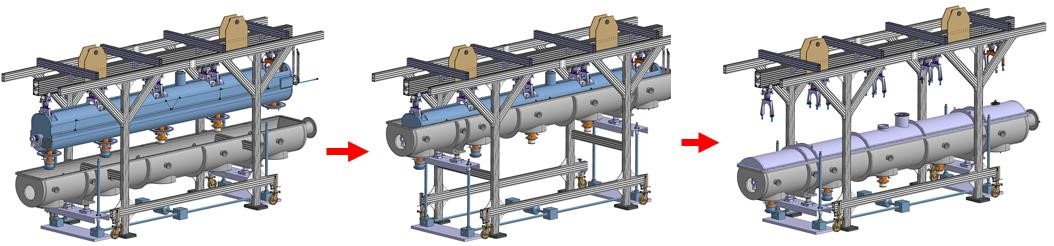}
    \caption{Cryomodule assembly procedure main steps}
    \label{fig:splCryomoduleAssembly}
\end{figure}

We recently decided to push further away the analysis to use the SPL cryomodule for the LHeC ERL, thanks to its geometrical compatibility with the LHeC ERL superconducting cavities, but also because it fits quite well the overall ERL requirements. One of the clear advantages of the SPL configuration is a much simplified assembly procedure (Fig.\,\ref{fig:splCryomoduleAssembly}), with its top-lid configuration which also allows an easier maintenance.

The first study performed was to analyse the possibility to integrate the ERL cavities instead of the SPL ones. The \SI{802}{MHz} cavities are a little bit shorter than the SPL ones and the cells are also smaller in diameter. The beam port internal diameter is about the same, as well as the power coupler port. As a result, the SPL cryomodule is well fitted to the ERL 802 MHz superconducting cavities from the geometrical point of view, and they could be easily integrated providing minor mechanical features adaptations.

\begin{figure}[!htb]
    \centering
    \includegraphics[width=0.65\textwidth]{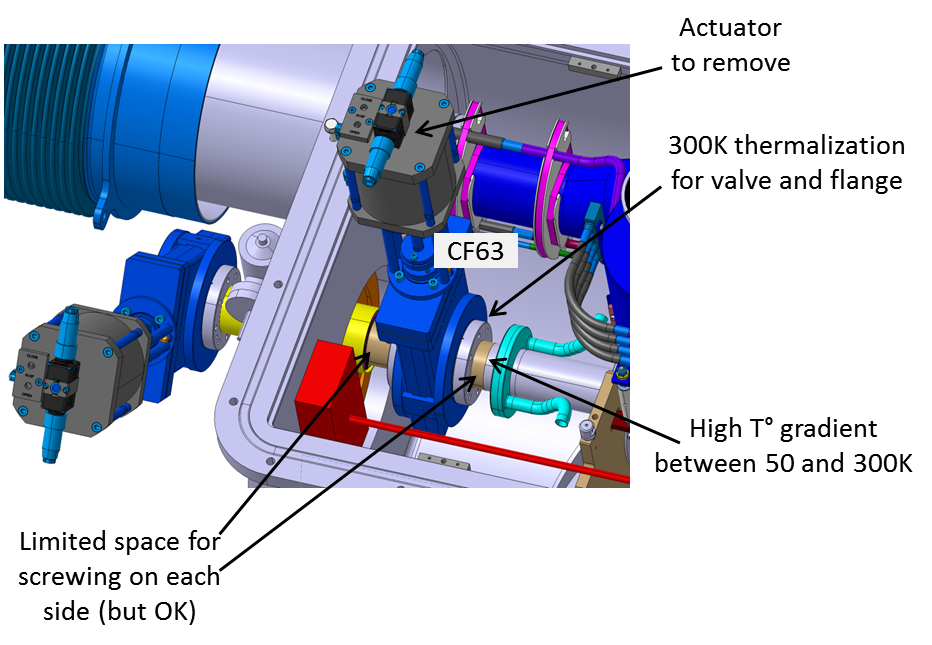}
    \caption{The two stages vacuum valve solution for adapting the SPL cryomodule prototype to the 802 MHz cavities of the LHeC ERL.}
    \label{fig:splTwoStageValve}
\end{figure}

The second analysed point is the beam vacuum. As the SPL cryomodule existing design was done for a prototype, intended for RF and cryogenic test only, without beam, the vacuum valve is a VAT CF63 “vatterfly” valve with viton seal and manual actuator, which is not adapted for a real operating cryomodule. Integration of an all-metal gate valve instead is not an issue and we also designed a specific solution based on a two stages valves (Fig.\,\ref{fig:splTwoStageValve}) to adapt the already fabricated SPL prototype cryomodule in order to be able to integrate the 802 MHz cavities.

The third study performed is the compatibility of the SPL cryogenic features with the ERL requirements. SPL was designed to operate 702 MHz cavities at 25 MV/m with a $Q_0$ of $5\times 10^9$ with a 8.2~\% duty cycle. The LHeC ERL will operate SRF cavities in CW regime, but at a lower field (20 MV/m) and with a higher expected $Q_0$ at the nominal gradient (about $1.5\times 10^{10}$). As a result, and despite the different duty cycle, the dynamic cryogenic losses are estimated to be only about 30~\% more in the ERL case. The overall cryogenic dimensioning is then fully compatible, providing some unavoidable adaptation of a few internal cryogenic piping. The main issue still to address is the need and consequences of the HOM coupler cooling. Even if the present engineering analysis showed that this point will not be a showstopper, it might have an impact on some cryogenic piping and cooling circuit.

Detailed engineering studies are being pursued to transform the SPL cryomodule prototype into an ERL LHeC cryomodule prototype. We are taking benefit of all the design and fabrication work previously performed on the SPL, and also on the fact some parts, such as the thermal and magnetic shielding, are not yet fabricated and could be exactly adapted to the ERL requirements. This will give the possibility to have an earlier full prototype cryomodule RF and cryogenic test as compared to a standard experimental plan where the complete study and fabrication is starting from scratch.

\subsection{Electron sources and injectors  \ourauthor{Boris Militsyn, Ben Hounsell, Matt Poelker}}

\subsubsection{Specification of electron sources}
Operation of the LHeC with an electron beam, delivered by a full energy ERL imposes specific requirements on the electron source.
It should deliver a beam with the charge and temporal structure required at the Interaction Point.
Additionally as during acceleration in a high energy ERL both longitudinal and transverse emittances of the beam are increased due to Synchrotron Radiation (SR), the 6D emittance of the beam delivered by electron source should be small enough to mitigate this effect.
The general specification of the electron source are shown in Tab.~\ref{tab:injector_specification}.
  Some parameters in this table such as RMS bunch length, uncorrelated energy spread and normalised transverse emittance are given on the basis of the requirements for the acceleration in ERL and to pre-compensate the effects of SR.
The most difficult of the parameters to specify is injector energy.
It should be as low as possible to reduce the unrecoverable power used to accelerate the beam before injection into the ERL while still being high enough to deliver short electron bunches with high peak current.
Another constraint on the injection energy is the average energy and energy spread of the returned beam.
The average energy cannot be less than the energy of electron source, but the maximum energy in the spectrum should not exceed 10~MeV the neutron activation threshold.
An injection energy of 7\,MeV is a reasonable compromise to meet this constraint.
\begin{table}[!ht]
   \centering
   \small
   \begin{tabular}{lcc}
     \toprule
     Parameter &  Unit & Value \\
     \midrule
     Booster energy & MeV & 7* \\
     Bunch repetition rate & MHz & 40.1 \\
     Average beam current& mA & 20 \\
     Bunch charge & pC & 500 \\
     RMS bunch length & mm & 3 \\
     Normalised transverse emittance &  $\pi\cdot$mm$\cdot$mrad & \textless 6 \\
     Uncorrelated energy spread & keV & 10 \\
     Beam polarisation & &  Unpolarised/Polarised \\
     \bottomrule
   \end{tabular}
   \caption{General specification of the LHeC ERL electron source.}
   \label{tab:injector_specification}
\end{table}

The required temporal structure of the beam and the stringent requirements for beam emittance do not allow the use of conventional thermionic electron sources for the LHeC ERL without using a bunching process involving beam losses. While this option cannot completely be excluded as a source of unpolarised electrons. The additional requirement to deliver polarised beam can only be met with photoemission based electron sources.

There are now four possible designs of electron sources for delivering unpolarised beams and (potentially) three for delivering polarised beams:
\begin{enumerate}
\item A thermionic electron source with RF modulated grid or gate electrode with following (multi)stage compression and acceleration. The electron source could be either a DC electron gun or an RF electron source in this case. Although these sources are widely used in the injectors of Infra-Red FELs~\cite{Bluem:ERL11-WG1010} their emittance is not good enough to meet the specification of the LHeC injector. Moreover, thermionic sources cannot deliver polarised electrons.
\item A VHF photoemission source. This is a type of normal conducting RF source which operates in the frequency range 160\,MHz -- 200\,MHz. The relatively low frequency of these sources means that they are large enough that sufficient cooling should be provided to permit CW operation. This type of source has been developed for the new generation of CW FELs such as LCLS-II~\cite{Sannibale:16}, SHINE~\cite{Wang:IPAC2019-TUPRB053} and a back-up option of the European XFEL upgrade~\cite{Shu:IPAC2019-TUPRB010}, but they have not yet demonstrated the average current required for the LHeC injector. The possibility of generating polarised electrons with this type of source has not investigated yet. 
\item A superconducting RF photoemission source. This type of sources are under development for different applications such as CW FEL’s (ELBE FEL~\cite{TEICHERT2014114}, SRF option of LCLS-II injector~\cite{Bisognano:NAPAC13-TUPMA19}, European XFEL upgrade~\cite{Vogel:413970}), as a basis of injectors for ERL’s (bERLinPro~\cite{Neumann:ERL19-THCOZBS02}) and for electron cooling (BNL~\cite{Belomestnykh:SRF15-THPB058}). Though this type of sources has already demonstrated the possibility of delivering the average current, required for the LHeC with unpolarised beams (BNL), and has the potential for operation with GaAs type photocathodes (HZDR) which are required for delivery of polarised beams, the current technology of SRF photoelectron source cannot be considered as mature enough for use in the LHeC.
\item A DC photoemission source. In this type of source the electrons are accelerated immediately after emission by a potential difference between the source cathode and anode. This type of source is the most common for use in ERL injectors. It has been used in the projects which are already completed (JLAB~\cite{Hern:ERL09-WG107}, DL~\cite{Jones_2011}), is being used for ongoing projects (KEK~\cite{Kato:FEL19-THA03}, Cornell/CBeta~\cite{Hoffstaetter:LINAC16-TUOP02}) and is planned to be used in new projects such as the LHeC prototype PERLE~\cite{Hounsell:2019PERLEgun}. The technology of DC photoemission sources is well-developed and has demonstrated the average current and beam emittance required for the LHeC ERL (Cornell). Another advantage of the photoelectron source with DC acceleration is the possibility of operation with GaAs based photocathodes for delivering of polarised beam. Currently it’s the only source, which can deliver highly polarised electron beams with the current of several mA's which is already in the range of LHeC specifications (JLab~\cite{doi:10.1063/1.5040226}).
\end{enumerate}

Based on this analysis at CDR stage we consider the use of DC photoemission source as a basic option, keeping in mind that in the course of the injector development other types of electron sources may be considered, especially for providing of unpolarised beam. 

\subsubsection{The LHeC unpolarised injector}
\begin{figure}[ht]
    \centering
    \includegraphics[width=.8\textwidth]{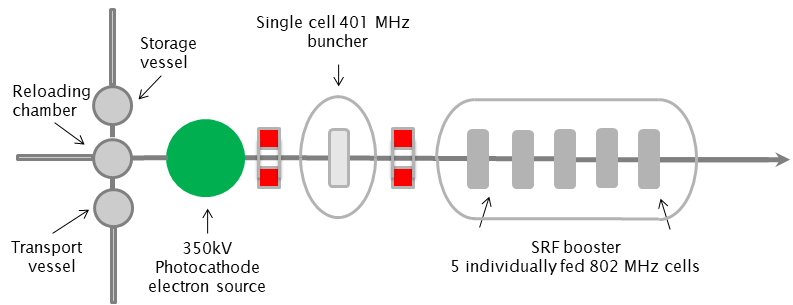}
    \caption{The layout of the unpolarised injector.}
    \label{fig:1_6_5_unpolarisedInjector}
\end{figure}

The injector layout follows the scheme depicted in Fig.~\ref{fig:1_6_5_unpolarisedInjector}.
Its design will be similar to the unpolarised variant of the PERLE injector~\cite{Hounsell:2019PERLEgun}.
The electron source with DC acceleration delivers a CW beam with the required bunch charge and temporal structure.
Immediately after the source is a focusing and bunching section consisting of two solenoids with a normal conducting buncher placed between them. The solenoids have two purposes.
Firstly to control the transverse size of the space charge dominated beam which will otherwise rapidly expand transversely. This ensures that the beam will fit through all of the apertures in the injector beamline. Secondly the solenoids are used for emittance compensation to counter the space charge induced growth in the projected emittance.
This is then followed by a superconducting booster linac. This accelerates the beam up to its injection energy, provides further longitudinal bunch compression and continues the emittance compensation process.

The DC electron source will have an accelerating voltage of 350\,kV using a high quantum efficiency antimonide based photocathode such as Cs$_2$KSb. The photoinjector laser required for this cathode type will be a 532\,nm green laser. There will be a load lock system to allow photocathodes to be replaced without breaking the source vacuum. This significantly reduces the down time required for each replacement which is a major advantage in a user facility such as the LHeC where maximising uptime is very important.
The cathode electrode will be mounted from above similar to the Cornell~\cite{Sinclair:PAC07-TUPMS021} and KEK~\cite{doi:10.1063/1.4811158} sources.
This electrode geometry makes the addition of a photocathode exchange mechanism much easier as the photocathode can be exchanged through the back of the cathode electrode. In addition the cathode electrode will be shaped to provide beam focusing.
The operational voltage of 350\,kV for the source was chosen as practical estimate of what is achievable. A higher voltage would produce better performance but would be challenging to achieve in practice. The highest operational voltage successfully achieved is 500\,kV by the DC electron source that is used for the cERL injector~\cite{PhysRevAccelBeams.22.053402}.
However 350\,kV is sufficient to achieve the required beam quality~\cite{Hounsell:2019PERLEgun}.

\subsubsection{Polarised electron source for ERL}
Providing polarised electrons has always been a challenging process, especially at relatively high average current as required for the LHeC.
The only practically usable production mechanism of polarised electrons is the illumination of activated to Negative Electron Affinity (NEA) state GaAs based photocathodes with circularly polarised laser light.
The vacuum requirements for these cathodes mean that this must be done in a DC electron source only.
In the course of the last 30 years significant progress has been achieved in improving the performance of polarised electron sources.
The maximum achievable polarisation has reached 90\,\% and the maximum Quantum Efficiency (QE) of the photocathode at the laser wavelength of maximum polarisation has reached 6\,\%.
Meanwhile the implementation of a polarised electron source into the LHeC remains a challenge as the practical operational charge lifetime of the GaAs based photocathode does not exceed hundreds Coulombs~(JLAB~\cite{doi:10.1063/1.4972180}) at an operational current in mA range.

\begin{figure}[!htb]
    \centering
    \includegraphics[width=.95\textwidth]{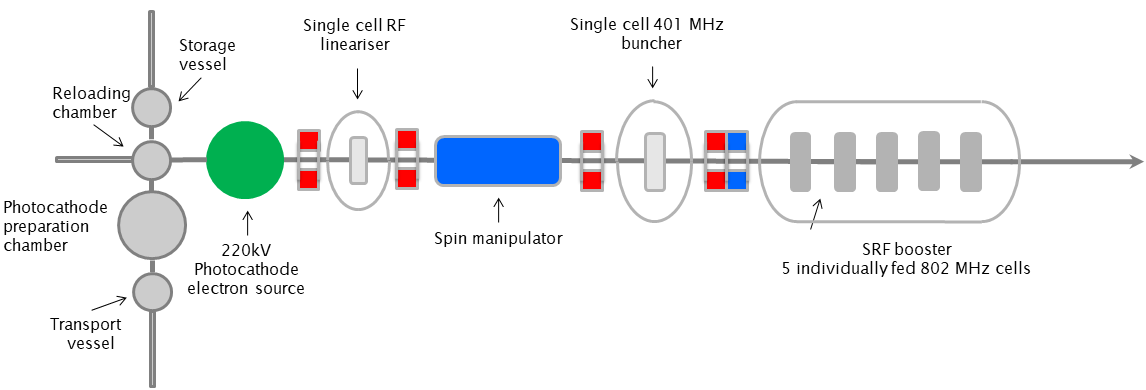}
    \caption{The layout of the polarised injector.}
    \label{fig:1_6_5_polarisedInjector}
\end{figure}

In Fig.~\ref{fig:1_6_5_polarisedInjector} a preliminary design of the LHeC polarised injector is shown.
In general, the design of the polarised electrons injector is close to that of the unpolarised injector and is based on a DC electron source where a photocathode is illuminated by a pulsed laser beam.
The choice of a DC source is dictated by the necessity of achieving extra high vacuum, with a pressure at a level of 10$^{-12}$\,mbar, in the photocathode area. This level of vacuum is neccesary for providing long lifetime of the photocathode.
In order to reduce photocathode degradation caused by electron stimulated gas desorption, the accelerating voltage in the source is reduced to 220\,kV.
The main differences with unpolarised injector are the presence of a photocathode preparation system, permanently attached to the source, and a Wien filter based spin manipulator between the source and the buncher.
In order to reduce depolarisation of the beam in the spin manipulator, caused by the space charge induced energy spread of the beam, an RF dechirper is installed between the source and the spin manipulator.
The injector is also equipped with a Mott polarimeter to characterise the polarisation of the beam delivered by the source.

An important consideration of the operation with interchangeable photocathodes is minimisation of the down time required for the photocathode exchange.
It typically takes few hours to replace the photocathode and to characterise polarisation of the beam. For large facility like LHeC this is unacceptable.
A practical solution could be operation with 2 or more electron sources which operate in rotation similar to the way which was proposed at BNL~\cite{doi:10.1063/1.5040227}.
Another motivation for using multi-source injector is the nonlinear dependence of photocathode charge lifetime on average beam current (JLAB~\cite{doi:10.1063/1.5040226}), which reduces with increasing of the average current.
For example in case of 3 electron sources 2 of them can be operated with half operation frequency 20.05\,MHz in opposite phase delivering average current of 10\,mA each, while the third is in stand by regime with freshly activated photocathode.
The only time which is necessary to switch it on is the time required for rising the high voltage.
Another advantage of using a multi-source scheme is the reduction of the average laser power deposited on the photocathode and as result relaxing requirements for the photocathode cooling.
In order to implement the multi-source polarised electron injector, development of a deflection system which is able to merge the beams from different sources before the spin rotator is required.

\subsubsection{Lasers for electron sources}
In the proposed design of the LHeC injection system at least 2 lasers must be used. In the unpolarised electron injector, which is going to operate with antimionide-based photocathode, a laser with a wavelength of 532\,nm is required.
Typical initial QE of these photocathodes is 10\% and for practical application reduction of QE up to 1\,\% may be expected.
For polarised electron source typical QE varies from 1\,\% down to 0.1\,\% and laser with a wavelength of 780\,nm is required.
The optimised parameters of the required lasers are summarised in Tab.~\ref{tab:1_6_5_laser}.
Laser temporal profile and spot size on the photocathode are given on the basis of source optimisation for operation at 350\,kV for unpolarised regime and 220\,kV for polarised.
\begin{table}[!ht]
  \centering
  \small
   \begin{tabular}{lccc}
       \toprule
       Laser beam parameter & Unit & Unpolarised & Polarised\\
                            &      & mode & mode\\
       \midrule
	Laser wavelength & nm & 532 & 780  \\
	Laser pulse repetition rate & MHz & 40.1 & 40.1 \\
	Energy in the single pulse at photocathode QE=1\,\% & $\mu$J & 0.12 & \\	
	Average laser power at photocathode QE=1\,\% & W	& 4.7* & \\
	Energy in the single pulse at photocathode QE=0.1\,\% & $\mu$J & & 0.79 \\
	Average laser power at photocathode Qe=0.1\,\% & W & & 32* \\
	Laser pulse duration & ps FWHM & 118 & 80 \\
	Laser pulse rise time & ps & 3.2 & 3.2 \\
	Laser pulse fall time & ps & 3.2 & 3.2 \\
	Spot diameter on the photocathode surface & mm & 6.4 & 8 \\
	Laser spot shape on the photocathode surface &  & Flat top \\
       \bottomrule
   \end{tabular}
   \caption{Parameters of the electron source drive laser.}
   \label{tab:1_6_5_laser}
\end{table}

\subsection{Positrons}
\subsubsection{Physics and Intensity Considerations}
Variation of the beam conditions (energies, lepton charge and polarisation, hadron types) provides a considerable extension of the physics programme of the LHeC.
The LHC permits a variation of the proton beam energy between about 1 and 7\,TeV. It is a proton collider with options for heavy ions, primarily Pb and possibly lighter ones.
The electron beam energy may be varied between about 10 and the maximum of 
50 (or eventually 60)\,GeV. Highly intense polarised electron beam sources
are under development which shall allow detailed investigations of weak interactions
and searches for new physics to be performed through variations of the electron
beam helicity, P. It is an advantage of the linac to achieve very high values of P,
as compared to a ring electron accelerator where the polarisation build up due to
the Sokolov-Ternov effect~\cite{Sokolov:1963zn} runs into serious difficulties at higher energies. The electron beam polarisation at HERA was limited to about 40\,\%. 
Positrons are the genuine challenge for the LHeC and as well for future $e^+e^-$
linear colliders. The reason is the difficulty to generate intense beams as was 
discussed for the LHeC in quite some detail already in the 2012 
CDR~\cite{AbelleiraFernandez:2012cc}. 

The physics reasons for positrons at the
LHeC are somehow obvious: positrons permit to establish, exploit and
question the existence of charge symmetry which may lead to discovery. For example,
the charm tagging process in electron initiated charged current scattering measures
the anti-strange quark density $x\bar{s}(x,Q^2)$ in the proton. There are expectations that the difference $x(s-\bar{s})$ may be different from zero, i.e. that there existed
a strange-quark valence component.  That requires a precision measurement of
also  $xs(x,Q^2)$ for which one needs about $1$\,fb$^{-1}$ of integrated 
$e^+p$ luminosity, desirably of course more. 
Further reasons, presented
in the CDR~\cite{AbelleiraFernandez:2012cc} regard the nature of excited leptons,
the origin of contact interactions, the spectroscopy of lepto-quarks, the understanding 
of DIS, as for the measurement of $F_L$ where the signal is charge sensitive but
background at high inelasticity dominantly charge symmetric, the thorough 
resolution of the parton contents of hadrons etc. Thus, yes, 
one has many reasons to operate LHeC as a positron-hadron collider also.

However, from today's perspective, one has to be realistic in one's assumptions
about what intensity may be realistically achievable and required 
in the positron linac - proton ring configuration. The current luminosity goal of the LHeC had been set with the
observation that the Higgs production cross section in $ep$ is about $200$\,fb, 
comparable to
that of the $e^+e^-$ colliders, and the LHeC could
become a Higgs factory~\cite{Zimmermann:2013aga}.
 Higgs production at the LHeC
 is dominantly due to $e^-p \to \nu H X$ scattering, i.e.
the LHeC has a competitive Higgs  physics potential which is complementary
to $e^+e^-$ and $pp$ as the dominant production mechanisms are $WW-H$,
$Z^*-HZ$ and $gg-H$, respectively. The electron-proton CC Higgs cross section 
is much larger than the one in positron-proton scattering which is related
to the dominance of up quarks as compared to down quarks in the proton.
Much of the running optimisation used in this paper has targeted to maximise
the number of Higgs events and preferred electron over positron operation.

The target electron current to achieve $10^{34}$\,cm$^{-2}$s$^{-1}$ luminosity 
has been set to $20$\,mA. This origins from a $500$\,pC gun which for $40$\,MHz LHC operation frequency leads to a charge of $3 \cdot 10^9$ electrons per bunch
corresponding to $1.2 \cdot 10^{17}$\,$e^-$/s. Given the current and near future
status of positron intensity requirements, one may set an LHeC target 
of order $10^{15-16}$\,$e^+$/s. Note that the normalised transverse
 emittance of of the electron beam is 50 mm mrad and the longitudinal 
 emittance 5 MeV mm.

The intensity above would potentially provide a luminosity of order
of $1-10$\,fb$^{-1}$ within one year. With a drastic difference in the 
electron and positron intensities, later operation would favour $e^-p$ over $e^+p$
running to maximise the statistics. We thus assumed $e^+p$ would operate
for about one year, somehow comparable to the heavy ion operation of the LHC.
 In the physics studies, as on PDF and electroweak measurements, we have used values of integrated $e^+p$ luminosities corresponding to these assumptions. It was also assumed the positrons were not polarised.
The linac-ring $ep$ configuration thus has a highly polarised, intense electron beam and a less intense unpolarised positron beam. The ring-ring configuration, a still
possible back-up for the LHeC at HE-LHC, has intense electron and positron beams but 
with rather lower polarisations.

\subsubsection{Positron Sources}
One can compare the LHeC $e^+$ intensity target with the goals
for CLIC and ILC as listed  in 
Tab.\,\ref{tab:Positron_collider}. One finds that the chosen LHeC value is
more demanding than that of CLIC and ILC.
\begin{table}[!htb]
    \centering\small
    \begin{tabular}{lccc}
      \toprule
         Parameter & CLIC & ILC  & LHeC\\
        \midrule
          Energy  [GeV] & 1500 & 250 &  50 \\
          $e^+$/bunch  [$10^9$] & 3.7 & 20 & 2 \\
          Norm. emittance [mm.mrad] & 0.66 (H) & 10 (H) & 50 \\
          & 0.02 (V) & 0.04 (V) & 50\\
          Norm. emittance [eV.m] & 5000 & 60000 & 5000\\
          Repetition rate [Hz] & 50 & 10 & CW \\
          Bunches / s & 15600 & 26250 & $2\cdot10^7$\\ 
          $e^+$ flux [$10^{15}$ $e^+$/s]  & 0.1 & 0.4 & 1-10 \\
      \bottomrule
    \end{tabular}
    \caption{Characteristics of positron beams for CLIC, ILC and LHeC. Note that
    the muon collider target value in the LEMMA scheme is  about $4\cdot 10^{16}$\,$e^+$/s.}
    \label{tab:Positron_collider}
\end{table}
Recently the interest in very intense $e^+$ production has been renewed with the
revival of the muon collider studies and the so-called 
LEMMA proposal~\cite{Antonelli:2015nla} to generate
muons from $e^+e^-$ pairs, i.e.
 not from pion decays to achieve small emittance beams. This requires
to generate an intense, $45$\,GeV energy positron beam annihilating with electrons
from a target for muon pair production near threshold. 
In a study following the LEMMA idea, a target
positron intensity of $3.9~10^{16}$\,$e^+$/s was set~\cite{Boscolo:2018tlu} which
requires considerable R+D efforts.

%
%
%

A conventional positron source uses only a single amorphous target. An
electron beam hits the target where Bremstrahlung and pair-production take
place. Downstream the target, particular devices (Quarter Wave Transformer
QWT or Adiabatic Matching Device AMD) allow capturing as many positrons as possible, with a large emittance. The 
CLIC $e^+$ source~\cite{Aicheler:2012bya} takes advantage of a hybrid target design. A thin crystal target allows reducing the peak power deposition and enhances photon production via a channelling process. An amorphous target converting the photons into positrons follows it. In between, a magnet sweeps out charged particles. 

The ILC $e^+$ source~\cite{Adolphsen:2013kya} 
takes advantage of a long helical undulator using the high-energy electron beam of the collider. The electron beam passing through the undulator produces polarised photons by impinging on a moving target the design of which is still to be finalised. 
This target converts photons into positrons. 
The ILC-type positron source is not an option for the LHeC since it requires an electron beam energy
above 100\,GeV

One option considered  for the initial LHeC $e^+$ source~\cite{AbelleiraFernandez:2012cc}
was using ten hybrid targets in parallel.
Bunch intensity and density could be enhanced by a tri-ring transformer system converting from CW to pulsed mode for accumulation, and again back to CW.

To evaluate the performance of $e^+$ sources, one defines a \emph{positron yield} parameter. This parameter is the number of positrons, at a given place along the production channel, per electron impinging onto the target. 
It is crucial to improve the positron yield while keeping the peak energy density deposition PEDD
and the shockwave inside the target within acceptable limits.
The target lifetime suffers from the cyclic thermal loads and stresses due to the beam pulses. The evacuation of the average power (kW to MW) from the target is challenging and should be investigated for the reliability of the target. Heat dissipation in the amorphous target may be improved by replacing it with a granular target (experiment at KEK). The capture and accelerating sections should also be optimised. Peak magnetic field and its shape, aperture and accelerating gradient of the RF structures are important parameters. Given the large emittance of the $e^+$ beam, a damping ring is mandatory. Due to the high requested $e^+$ flux, an accumulation process should be considered. The $e^+$ flux is 
\begin{equation}
    \cfrac{dN^+}{dt} = a \cdot y \cdot N^-  \cdot f\,,
\end{equation}
where $a$ is the accumulation efficiency and is a function of the damping time, $y$ is the yield as defined above depending on the
electron beam energy $E^-$, further
$N^-$ is the number of electron impinging on the target,
and $f$ the linac repetition rate. 
This accumulation could be realised by means of a tri-ring system
as presented in the CDR.

%


\subsubsection{Approaches towards LHeC Positrons}

It is to be mentioned that positrons have not been in the focus
of our recent LHeC design activity such that basic discussions as presented
in the CDR~\cite{AbelleiraFernandez:2012cc} still hold. 

%
%
The CLIC positron source was studied in great detail and many pertinent simulations were performed. Based
on the expected flux of the CLIC $e^+$ source, we have identified three possibilities for the LHeC:
\begin{itemize}
    \item Option 1: Keep the CW mode and the bunch spacing of 25 ns. This implies a bunch charge of $2.5\times10^6$\,$e^+$/bunch and a current of 16\,$\mu$A;
    \item Option 2: 
    Keep the CW mode with a bunch charge of $2.5\times10^{9}$\,$e^+$/bunch. This implies a bunch spacing of 25\,$\mu$s  and a current of 16\,$\mu$A;
    \item Option 3: Keep the bunch spacing of 25\,ns with a bunch charge of $1\times10^9$\,$e^+$/bunch. This implies a pulsed mode with a repetition rate of 50\,Hz. The beam current is now 6.4\,mA.
\end{itemize}
The CLIC source, however, just provides O($10^{14}$)\,$e^+/s$ which would provide maximally $100$\,pb$^{-1}$ $ep$ luminosity in an efficient year. 

One notes that the ILC luminosity upgrade foresees a positron rate up to eight times higher than the CLIC rate. A recent ILC status report cites novel concepts for
high intense and polarised positron beams, obtained in an electron beam driven
configuration~\cite{bambade:2019fyw}.

Two alternative options, not yet studied in greater detail, promise to deliver 
a still much higher positron rate, indeed close to that of the electrons, providing
 1000 times more positron per second than a CLIC-based source: One possibly
 may  i) convert high-energy photons from the LHC-based gamma 
 factory~\cite{Krasny:2015ffb} for
  producing a positron rate of up 
  to $10^{17}$ $e^+$/s~\cite{Zimmermann:2018wfu}
   or 
ii) using the photons from an LHeC based FEL~\cite{Zimmermann:2019tru} to
 generate a similarly high rate of positrons, which
in both cases would already be at the correct bunch spacing.  These two options rely on either the LHC hadron or the LHeC lepton-beam infrastructure, and thus do not need other, possibly additional investments. 

Depending on how challenging the parameter requirements are,  a more or less
radical change of paradigm is necessary. There is no easy path even
to $10^{14}$\,$e^+$/s. As to  LHeC, it may profit from recent and forthcoming
developments for lepton colliders for which higher intense
positron sources and beams are a matter of existence. When approved, however,
serious R+D efforts will be inevitable also for LHeC, and later FCC-he, for which 
positron beams may even be more important, if, for example, lepto-quarks or SUSY
particles in the few TeV range were found and to be examined in $e^{\pm}p$ 
scattering at the FCC.

\subsection{Compensation of Synchrotron Radiation Losses \ourauthor{Alex Bogacz}} \label{SRcomp}
\begin{figure}[th]
  \centering
  \includegraphics[width=0.7\textwidth]{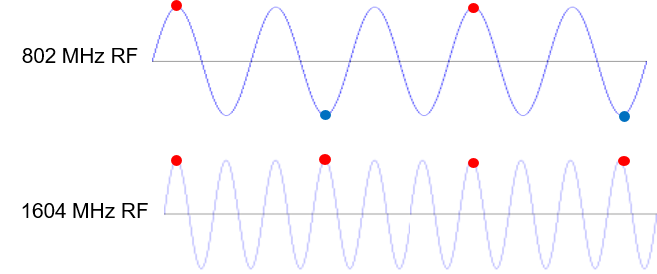}
  \caption{The second-harmonic RF restores the energy loss in both
    the accelerating and decelerating passes.}
  \label{fig:2_Harm}
\end{figure}

Depending on energy, each arc exhibits fractional energy loss  due to the synchrotron radiation, which scales as $\gamma^4/\rho$ (see Eq.~\eqref{eq:Emit_dil_1}).
Arc-by-arc energy loss was previously summarised in Tab.~\ref{tab:SRlosses}. 
That energy loss has to be replenished back to the beam, so that at the entrance of each arc the accelerated and decelerated beams have the same energy, unless separate arcs are used for the accelerated and decelerated beams.
Before or after each arc, a matching section adjusts the optics from and to the linac.
Adjacent to these, additional cells are placed, hosting the RF compensating sections.
The compensation makes use of a second harmonic RF at \SI{1603.2}{MHz} to replenish the energy loss for both the accelerated and the decelerated beams, therefore allowing them to have the same energy at the entrance of each arc, as shown in Fig.~\ref{fig:2_Harm}.

Parameters of the RF compensation cryomodules,  
shown in Table ~\ref{tab:Harm__cryos}, have been extrapolated from the ILC cavity design, expecting that the higher frequency and lower gradient would support continuous operation. 
\begin{table}[!ht]
  \centering
  \small
  \begin{tabular}{lcc} 
    \toprule
    Parameter & Unit & Value \\
    \midrule
    Frequency & MHz & \SI{1603.2}{}\\
    Gradient & MV/m & \SI{30}{}\\
    Design &  & Nine cells\\
    Cells length  & mm & \SI{841}{}\\
    Structure length & m &  \SI{1}{}\\
    Cavity per cryomodule &  & {6}\\
    Cryomodule length & m & \SI{6}{}\\
    Cryomodule voltage & MV &\SI{150}{}\\
    \bottomrule
  \end{tabular}
  \caption{ A tentative list of parameter for the compensating RF cryomodules extrapolated from the ILC design.}
  \label{tab:Harm__cryos}
\end{table}

As illustrated schematically in  Fig.~\ref{fig:2_Harm}, there are two beams in each arcs (with exception of Arc 6) one needs to replenish energy loss for: the accelerated and the decelerated beams. Assuming nominal beam current of \SI{20}{\milli\ampere}, the net current for two beams doubles. Therefore, \SI{40}{\milli\ampere} current in Arcs 1-5, was used to evaluated power required to compensate energy loss by 2-nd harmonic RF system, as summarized in Table  ~\ref{tab:Comp_cryos}.

\begin{table}[!ht]
  \centering
  \small
  \begin{tabular}{lccc} 
    \toprule
    Section &  $\Delta E$ [\si{MeV}] & $P$ [\si{MW}] & Cryomodules \\
    \midrule
    Arc 1 & 3 & 0.12 & 0\\
    Arc 2 & 25 & 1.0 & 0\\
    Arc 3 & 80 & 3.2 & 1\\
    Arc 4 & 229 & 9.16 & 2\\
    Arc 5 &  383 & 15.32 & 3\\
    Arc 6 & 836 & 16.7 & 6\\
    \bottomrule
  \end{tabular}
  \caption{Arc-by-arc synchrotron radiated power for both the accelerated and decelerated beams (only one beam in Arc 6) along with a number of 2-nd harmonic RF cryomodules required to compensate energy loss.}
  \label{tab:Comp_cryos}
\end{table}

The compensating cryomodules are placed into Linac 1 side of the racetrack, before the bending section of Arc~1, Arc~3, and Arc~5 and after the bending section of Arc~2, Arc~4, and Arc~6.
This saves space on Linac 2 side to better fit the IP line and the bypasses.
Note that with the current vertical separation of 0.5\,m it will not be possible to stack the cryomodules on top of each other; therefore, they will occupy 36\,m on the Arc~4 and Arc~6 side and 18 m on the Arc~3 and Arc~5 side of the racetrack.
Each of the compensating cavities in Arc~5 needs to transfer up to \SI{1}{MW} to the beam.
Although a \SI{1}{MW} continuous wave klystron are available~\cite{Zaltsman}, the cryomodule integration and protection system will require a careful design.
Tab.~\ref{tab:Comp_cryos} shows the energy loss for each arc and the corresponding synchrotron radiated power, along with number of cryomodules at \SI{1603.2}{MHz} RF frequency required to replenish the energy loss. 
\subsection{LINAC Configuration and Infrastructure \ourauthor{Erk Jensen}}\label{LCaIS}
Since the power supplied to the beam in the main linacs will be recovered, the average RF power requirements at \SI{802}{MHz} are relatively small and determined by the needs to handle transients and microphonics. 

The RF power required for the second-harmonic RF system however is substantial -- it can be estimated from Tab.~\ref{tab:SRlosses} with the nominal current of \SI{20}{mA}. Tab.~\ref{tab:Comp_cryos} above summarizes the estimated power lost in each arc depending on beam energy; these power values must be supplied by the 6 2-nd harmonic RF systems. 

The RF infrastructure required at 802 MHz

\section{Interaction Region \ourauthor{Emilia Cruz Alaniz, Kevin Andre', Bernhard Holzer,  Roman Martin,  Rogelio Tomas}}
The design of the LHeC Interaction region has been revised with respect to the LHeC CDR to take into account the reduction of the electron energy from \SI{60}{GeV} to \SI{50}{GeV} and the latest design of the HL-LHC optics and it has been optimized to minimize synchrotron radiation power and critical energy at the IP.

\subsection{Layout  \ourauthor{Emilia Cruz Alaniz, Roman Martin, Rogelio Tomas}}
\begin{figure}[tbh]
  \centering
  \includegraphics[width=0.65\textwidth]{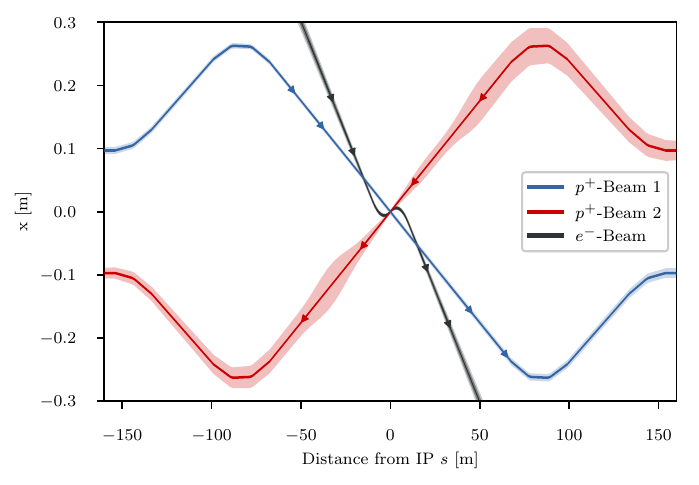}
  \caption{Geometry of the interaction region with 10\,$\sigma$ envelopes. The electron beam is colliding with the focussed anti-clockwise rotating LHC beam (Beam~2) while the clockwise rotating LHC beam is unfocussed and passes the Interaction Region without interacting with the other two beams}
  \label{fig:IR_geometry}
\end{figure}

The basic principle of the Linac-Ring IR design remains unchanged and it is shown in Fig.~\ref{fig:IR_geometry}: the two proton beams are brought onto intersecting orbits by strong separation and recombination dipoles. A collision of the proton beams at the IP is avoided by selecting appropriately its location, i.e. by displacing it longitudinally with respect to the point where the two counter-rotating proton beams would collide. The large crossing angle keeps the long range beam-beam effect small and separates the beams enough to allow septum quadrupoles to focus only the colliding beam (the anti-clockwise rotating LHC beam -- Beam 2).
The non-colliding beam (the clockwise rotating LHC beam -- Beam 1) is unfocused and passes the septum quadrupoles in a field free aperture. The electron beam is brought in with an even larger angle, partly sharing the field free aperture of the septum quadrupoles with the non-colliding beam. A weak dipole in the detector region bends the electron beam into head-on collisions with the colliding proton beam. The two proton beams are also exposed to the dipole field but, due to the large beam rigidity, they are barely affected. After the interaction point a dipole with opposite polarity separates the orbits of the electron and proton beam.

The high electron current (cf.\ Tab.~\ref{tab:ERLparameters}) required to approach the goal peak luminosity of $10^{34}\si{cm^{-2}s^{-1}}$ poses a potential problem for the interaction region (IR) as it increases the already high synchrotron radiation.

The ERL parameters are not the only major change the new IR design has to account for. The first design of the quadrupole septa featured a separation of \SI{68}{mm} for the two proton beams.
However, this design focused strongly on providing a field free region for the non-colliding beam.
Unfortunately, this lead to a poor field quality for the strongly focused colliding beam.
The first quadrupole Q1 was a half quadrupole design effectively acting as a combined function magnet with a dipole component of \SI{4.45}{T}~\cite{bib:BParker:LHeCWorkshop17}. The sextupole field component was also prohibitively high.
Consequently, a new design approach focusing on the field quality in the quadrupole aperture was necessary.
The parameters relevant for the interaction region design are summarised in Tab.~\ref{tab:quad_septa_params}.
\begin{table}
  \centering
  \small
  \begin{tabular}{lccc}
	\toprule
	Magnet & Gradient [T/m] & Length [m] &  Free aperture radius [mm]\\
        \midrule
	Q1A & 252 &	3.5 & 20 \\ 
	Q1B & 164 & 3.0 & 32\\ 
	Q2 type & 186 & 3.7 & 40 \\ 
	Q3 type & 175 & 3.5 & 45 \\
        \bottomrule
  \end{tabular}
  \caption{Parameters of the final focus quadrupole septa. The parameters of Q1A/B and Q2 are compatible with the Nb$_3$Sn based designs from~\cite{bib:BParker:LHeCWorkshop18} assuming the inner protective layer of Q2 can be reduced to \SI{5}{mm} thickness.}
  \label{tab:quad_septa_params}
\end{table}

It is noteworthy that the minimum separation of the two beams at the entrance of the first quadrupole Q1A increased from \SI{68}{mm} to \SI{106}{mm} requiring a stronger bending of the electron beam.
This would increase the already high synchrotron radiation in the detector region even more.
In order to compensate this increase, it was decided to increase $\lstar$ (i.e. the distance from the IP to the first superconducting septum quadrupole focussing Beam 2) to \SI{15}{m}, an approach that was shown to have a strong leverage on the emitted power~\cite{bib:ECruz:LHeC:prab}.

The increased separation of the two proton beams, the longer $\lstar$ and the overall longer final focus triplet make longer and stronger separation and recombination dipoles necessary.
The dipoles differ from the arc dipoles in that the magnetic field in both apertures has the same direction.
Consequently the cross talk between both apertures is significant and the maximum reachable field is lower. The new geometry keeps the required field below \SI{5.6}{T}.
The required lengths and strength of these dipoles are listed in Tab.~\ref{tab:ir_dipole_params}. It should be noted that the inter--beam distance is different for each of the five magnets per side, so each magnet will likely require an individual design.
The design of the D1 dipoles is further complicated by the fact that an escape line for neutral collision debris traveling down the beam pipe will be necessary~\cite{AbelleiraFernandez:2012cc}, as well as a small angle electron tagger.
These issues have not been addressed so far, further studies will require detailed dipole designs.
\begin{table}
  \centering
  \small
  \begin{tabular}{lcccc}
	\toprule
	Magnet & Field strength [T] &  Interbeam distance [mm] & Length [m] & Number \\
        \midrule
	D1 & 5.6 & $\geq \SI{496}{mm}$ & 9.45 & 6 \\ 
	D2 & 4.0 & $\geq \SI{194}{mm}$ & 9.45 & 4 \\ 
	IP Dipole & 0.21 & - & 10 & - \\
	\bottomrule
  \end{tabular}
  \caption{Parameters of the separation and recombination dipoles. The respective interbeam distances are given for the magnet with the lowest value.}
  \label{tab:ir_dipole_params}
\end{table}

The first design of the LHeC interaction region featured detector dipoles occupying almost the entire drift space between the interaction point and first quadrupole.
The approach was to have the softest synchrotron radiation possible to minimise the power.
However, since the purpose of the dipoles is to create a spacial separation at the entrance of the first quadrupole, it is possible to make use of a short drift between dipole and quadrupole to increase the separation without increasing the synchrotron radiation power.
A dipole length of $\frac{2}{3} \lstar$ is the optimum in terms of synchrotron radiation power~\cite{bib:RMartin:dipole_optimization}.
Compared to the full length dipole it reduces the power by \SI{15.6}{\%} at the cost of a \SI{12.5}{\%} higher critical energy.
With an $\lstar$ of \SI{15}{m} the optimum length of the detector dipoles is \SI{10}{m}.
A magnetic field of \SI{0.21}{T} is sufficient to separate the electron and proton beams by \SI{106}{mm} at the entrance of the first quadrupole.
With these dipoles and an electron beam current of \SI{20}{mA} at \SI{49.19}{GeV} the total synchrotron radiation power is \SI{38}{kW} with a critical energy of \SI{283}{keV} to be compared with a power of \SI{83}{kW} and a critical energy of \SI{513}{keV} for the electron beam energy of \SI{60}{GeV}.
More detailed studies on the synchrotron radiation for different options and including a beam envelope for the electron beam are summarised in Tab.~\ref{tab:all_optimum} below.

\begin{figure}[th]
  \centering
  \includegraphics[width=0.8\textwidth]{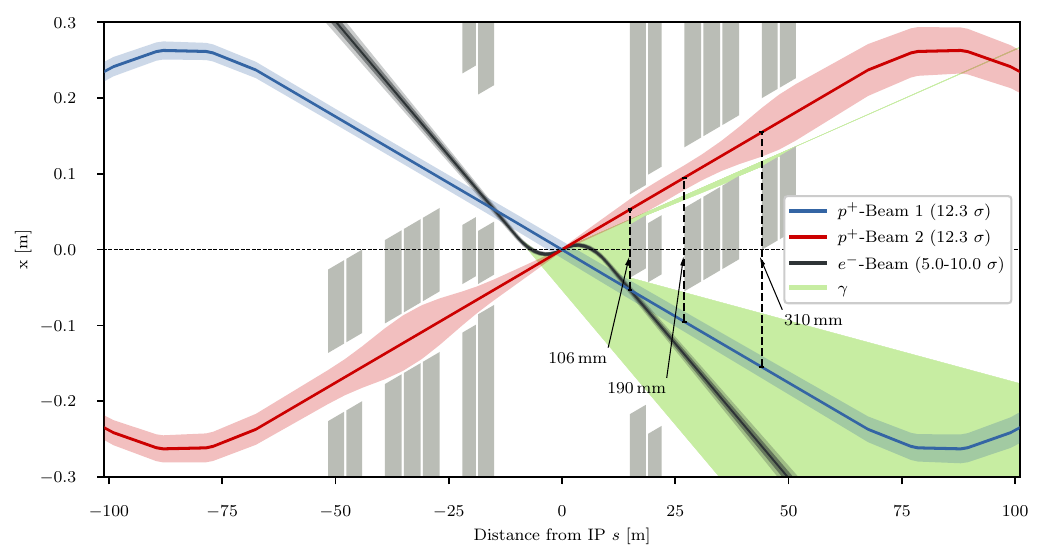}
  \caption{Schematic layout of the LHeC interaction region. The colliding proton beam and the electron beam are shown at collision energy while the non-colliding beam is shown at injection energy when its emittance is the largest.}
  \label{fig:IR_schematic}
\end{figure}

A schematic layout of the LHeC interaction region with the dipoles discussed above is shown in Fig.~\ref{fig:IR_schematic}.
The corresponding beam optics will be discussed in the following sections.

\subsection{Proton Optics  \ourauthor{Emilia Cruz Alaniz}}
As discussed above, the $\lstar$ was increased to \SI{15}{m} in order to compensate the increased synchrotron radiation due to the larger separation.
The final focus system is a triplet consisting of the quadrupoles Q1A and Q1B (see Tab.~\ref{tab:quad_septa_params}), three elements of the Q2 type and two of the Q3 type.
Between the elements a drift space of \SI{0.5}{m} was left to account for the magnet interconnects in a single cryostat.
Between Q1 and Q2 as well as Q2 and Q3 a longer drift of \SI{5}{m} is left for cold-warm transitions, Beam Position Monitors (BPMs) and vacuum equipment.
Behind Q3, but before the first element of the recombination dipole D1, another \SI{16}{m} of drift space are left to allow for the installation of non-linear correctors in case the need arises, as well as a local protection of the triplet magnets from asynchronous beam dumps caused by failures of the beam dump kickers (MKD) as discussed below.

As the recombination dipoles D1 and D2 for the LHeC interaction region require more space than the current ALICE interaction region, the quadrupoles Q4 and Q5 had to be moved further away from the IP.
The position of Q6 is mostly unchanged but due to a need for more focusing the length was increased by replacing it with two elements of the MQM magnet class of LHC.

\begin{figure}[th]
  \centering
  \includegraphics[width=0.7\textwidth]{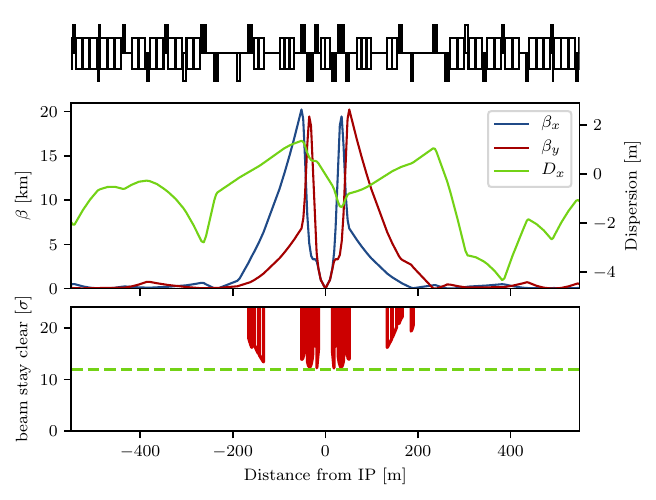}
  \caption{Optics (top) and beam stay clear (bottom) of the colliding beam with $\beta^* = \SI{10}{cm}$.
  }
  \label{fig:IR_colliding_beam_optics}
\end{figure}

\begin{figure}[th]
  \centering
  \includegraphics[width=0.7\textwidth]{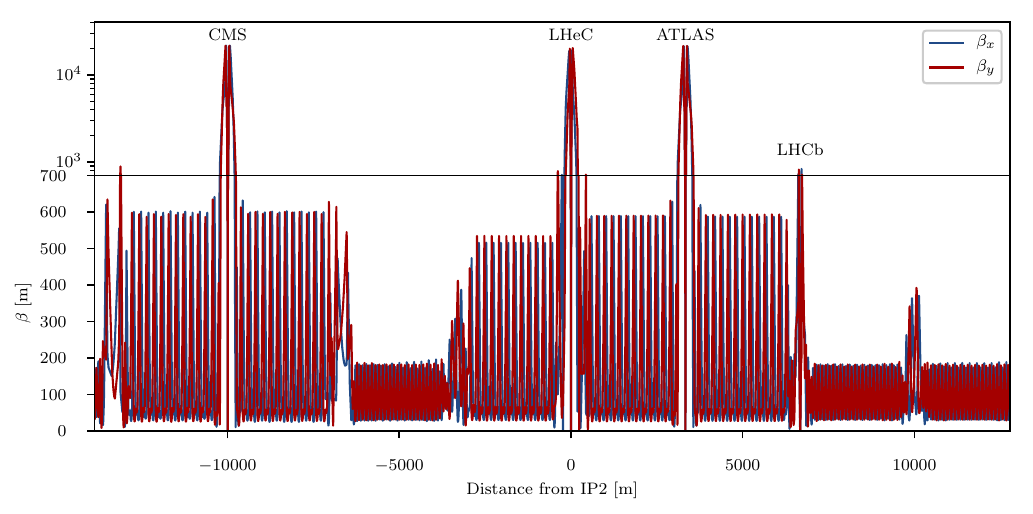}
  \caption{Optics of full ring of the colliding LHC proton beam (Beam 2).}
  \label{fig:full_ring}
\end{figure}

With the triplet quadrupole parameters provided in Tab.~\ref{tab:quad_septa_params} we were able to match optics with a minimum $\beta^*$ of \SI{10}{cm}.
The corresponding optics are shown in Fig.~\ref{fig:IR_colliding_beam_optics} and feature maximum $\beta$ functions in the triplet in the order of \SI{20}{km}.
With these large $\beta$ functions, the free apertures of the quadrupoles leave just enough space for a beam stay clear of \SI{12.3}{\sigma}, the specification of the LHC.
This is illustrated in Fig~\ref{fig:IR_colliding_beam_optics}.
However, since the LHeC is supposed to be incorporated in the HL-LHC lattice, this minimum beam stay clear requires specific phase advances from the MKD kicker to the protected aperture as detailed later.
The large $\beta$ functions not only drive the aperture need in the final focus system, but also the required chromaticity correction in the adjacent arcs.
To increase the leverage of the arc sextupoles, the Achromatic Telescopic Squeezing scheme (ATS) developed for HL-LHC~\cite{bib:Fartoukh:ATS} was extended to the arc upstream of IP2 for the colliding beam (Beam 2) (see Fig.~\ref{fig:full_ring}).
This limited the optical flexibility in the matching sections of IR2, specifically of the phase advances between arc and IP2.
As a consequence, the optical solution that has been found (Fig.~\ref{fig:IR_colliding_beam_optics}) still has a residual dispersion of \SI{15}{cm} at the IP and the polarities of the quadrupoles Q4 and Q5 on the left side of the IP break up the usual sequence of focusing and defocusing magnets. It needs to be studied whether this is compatible with the injection optics.
The latest optics designs can be found at the webpage~\cite{bib:repo}.

The free apertures given in Tab.~\ref{tab:quad_septa_params} include a \SI{10}{mm} thick shielding layer in Q1 and \SI{5}{mm} in Q2 and Q3.
This is necessary to protect the superconducting coils from synchrotron radiation entering the magnets as can be seen in Fig.~\ref{fig:IR_schematic}. The absorber must also protect the magnets from collision debris.
Simulations of both synchrotron radiation and collision debris are yet to be conducted in order to confirm the feasibility of this design.

A separation between the two proton beams in time is currently foreseen, i.e.\ while the orbits of the two proton beams do cross, the bunches do not pass through the IP at the same time.
This approach is complicated by the fact that the timing of the bunches in the other three interaction points should not be affected.
The easiest way to accomplish this is by shifting the interaction point of LHeC by a quarter of a bunch separation, i.e.\ $\SI{6.25}{\nano s} \times c \approx \SI{1.87}{m}$ upstream or downstream of the current ALICE IP, similar to what has been done for the LHCb detector in Point 8 of the LHC.
This will of course have an impact in the integration of the detector in the underground cavern~\cite{bib:AGaddi:LHeCWorkshop17}, however it seems feasible~\cite{bib:AGaddi:PrivateCom}.

The LHC protected aperture in the event of an asynchronous beam dump significantly depends on the phase advance between the MKD kicker and the local aperture protection~\cite{bib:Bruce:HLLHC_aperture_calculation}.
This is due to the oscillation trajectory of bunches deflected during the kicker rise time. With a phase advance of $0\degree$ or $180\degree$  
from the kicker to the protected aperture, a direct hit should be unlikely, so aperture bottlenecks should be close to that. For a beam stay clear of \SI{12.3}{\sigma} a phase advance of less than \SI{30}{\degree} from either $0 \degree$ or $180 \degree$ 
was calculated to be acceptable~\cite{bib:Bruce:HLLHC_aperture_calculation}.
The major complication comes from the fact that not only the final focus system of LHeC, but also of the two main experiments ATLAS and CMS need to have to correct phase advances and since the phase advances between IP2 (LHeC) and IP1 (ATLAS) are locked in the achromatic telescopic squeezing scheme there are few degrees of freedom to make adaptations.

The Achromatic Telescopic Squeezing (ATS) scheme~\cite{bib:Fartoukh:ATS} is a novel optical solution proposed for the HL-LHC to strongly reduce the $\beta^*$ while controlling the chromatic aberrations induced, among other benefits.

The principles of the ATS as implemented for the HL-LHC are as follows: first, in the presqueeze stage, a standard matching procedure is performed in the interaction regions to obtain a value of $\beta^*$ which is achievable in terms of quadrupole strengths and chromaticity correction efficiency, in the case of HL-LHC this corresponds to IR1 and IR5.
A further constraint at this point is to match the arc cell phase advance on the regions adjacent to the low $\beta^*$ interaction regions to exactly $\pi/2$.
Later, at the collision stage, the low $\beta^*$ insertions remain unchanged and instead the adjacent interaction regions contribute to the reduction of $\beta^*$, that is IR8 and IR2 for IR1, and IR4 and IR6 for IR5.
The $\pi/2$ phase advance allows the propagation of $\beta$-waves in the arc.
If phased correctly with the IP, these $\beta$-waves will reach their maximum at every other sextupoles, increasing the $\beta$ function at their location at the same rate that the decrease in $\beta^*$. The increase of the $\beta$ function at the location of the sextupoles will result in an increase of their efficiency, allowing the system to correct the high chromaticity produced by the high-$\beta$ function in the inner triplet.
This way, the ATS allows a further reduction of the $\beta^*$ at the same time that correcting the chromaticity aberrations produced in the low $\beta$ insertions. 

Following the experience for HL-LHC, the ATS scheme was proposed for the LHeC project to overcome some of the challenges of this design in terms of limits in the quadrupole strengths of the interaction region and in the chromaticity correction.

A first integration of the LHeC IR into the HL-LHC lattice using the ATS scheme for the previous nominal case with $\beta^* = \SI{10}{cm}$ and $L^*=\SI{10}{m}$ was presented by extending the $\beta$ wave into the arc 23~\cite{bib:ECruz:LHeC:prab}. The flexibility of this design was later explored to study the feasibility of minimising $\beta^*$, to increase the luminosity, and increasing $L^*$, to minimise the synchrotron radiation. It was found that increasing $L^*$ to \SI{15}{m} provided a good compromise but keeping the $\beta^*$ to \SI{10}{cm}.

The changes made to the HLLHCV1.3 lattice~\cite{bib:hllhc13rep} to obtain the LHeC lattice and the detailed matching procedure are described in Ref.~\cite{bib:ECruzTomas:LHeCoptics}.
At the end of this process a lattice for the required collision optics in all IRs ($\beta^*$=15~cm for IR1 and IR5 and $\beta^*$=10~cm for IR2) has been obtained, with the appropriate corrections (crossing, dispersion, tune and chromaticity).
The phases between the MKD kicker in IR6 and the different low $\beta^*$ triplets were also checked, resulting in 15$\degree$ from the horizontal for IR1, $22\degree$ for IR2 and $26\degree$ for IR5, therefore fulfilling the $<30\degree$ requirement for all three IRs.

Similarly the chromaticity correction for the LHeC lattice further develops from the HL-LHC chromaticity correction scheme~\cite{bib:ECruzTomas:LHeCoptics} allowing to correct the chromaticity for the case with $\beta^* = \SI{10}{cm}$ in IP2 within the available main sextupole strength.
Lattices with $\beta^*$= 7, 8 and \SI{9}{cm} and $L^*=15$~m were also successfully matched in terms of both the $\beta^*$ and the chromaticity correction. It must be noted however that these cases require a larger aperture in the inner triplet. 

Dynamic aperture (DA) studies were performed to analyze the stability of the lattice designs using SixTrack~\cite{bib:SixTrack} on a thin-lens version of the LHeC lattice at collision ($\beta^*= \SI{0.15}{m}$ in IP1 and IP5, $\beta^* = \SI{10}{cm}$ in IP2) over $10^5$ turns with crossing angles on, 30 particles pairs per amplitude step of \SI{2}{\sigma}, 5 angles in the transverse plane and a momentum offset of \num{2.7e-4}.
The energy was set to \SI{7}{TeV} and the normalised emittance of the proton beam to $\epsilon = \SI{2.5}{\micro\metre}$.
No beam-beam effects were included in this study. 

Previous DA studies had been performed for an earlier version of the LHeC lattice~\cite{bib:ECruz:LHeC:prab}. These studies did not include triplet errors of either of the low-$\beta$ interaction regions, as these errors were not available at that stage.
These studies were updated for the newer version of the LHeC lattice described in the previous sections and included errors on the triplets of IR1 and IR5.
For the case of IR2 errors tables for the new triplet are not yet available but it was estimated that the same field quality than the triplets for the HL-LHC IR can be achieved for these magnets, and therefore the same field errors were applied but adjusted to the LHeC triplet apertures. 

\begin{figure}[th]
  \centering
  \includegraphics[width=0.5\textwidth]{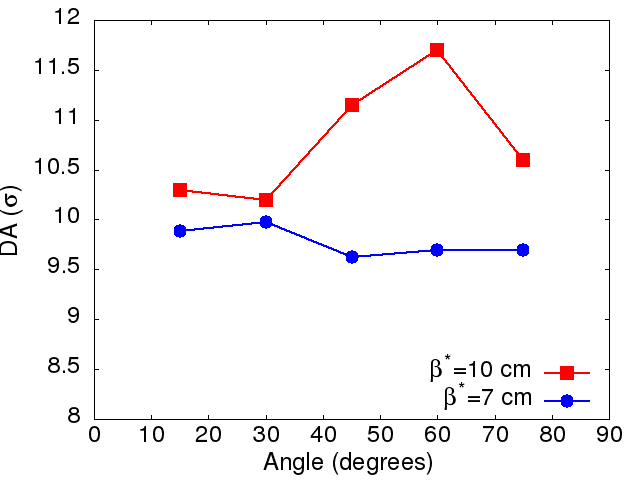}
  \caption{Dynamic aperture vs angle for 60 seeds for the LHeC lattice at collision for the cases $\beta^*=10$~cm (red) and $\beta^*=5$~cm in IP2. }
  \label{fig:LHeC_DA} 
\end{figure}

The initial DA resulted in \SI{7}{\sigma} but following the example of HL-LHC and FCC studies~\cite{bib:ECruz:FCC} two further corrections were implemented: the use of non-linear correctors to compensate for the non linear errors in the LHeC IR, and the optimisation of the phase advance between IP1 and IP5. With these corrections the DA was increased to \SI{10.2}{\sigma}, above the target of \SI{10}{\sigma}.
The case for lower $\beta^*$, particularly for the case of interest with $\beta^* = \SI{7}{cm}$ proved to be more challenging, as expected, when adding errors on the LHeC IR; however with the use of the latest corrections a DA of \SI{9.6}{\sigma} was achieved, that is not far off from the target.
The DA versus angle for both these cases are shown in Fig.~\ref{fig:LHeC_DA}.
It is important to point out that the challenge for the $\beta^*$=\SI{7}{cm} case comes instead from the quadrupole aperture and gradient requirements, particularly in the first magnet. 

\begin{figure}[th]
  \centering
  \includegraphics[width=0.7\textwidth]{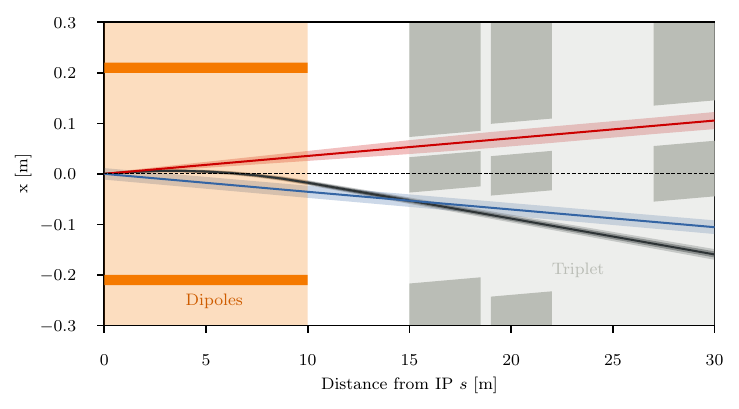}
  \caption{Empty space between the detector dipole and the superconducting quadrupoles of the final focus triplet.}
  \label{fig:IP2_empty_space}
\end{figure}

$\beta^*$ values lower than \SI{10}{cm} require a completely different final focus system as the lower $\beta^*$ means the beam size in the triplet will become larger. Larger apertures are required and consequently the gradients in the quadrupoles will decrease.
However similar integrated focusing strengths will be required so the overall length of the triplet will increase.
As this will in turn increase the $\beta$ functions in the triplet further it is imperative to optimise the use of the available space.
An example of available space is the drift between the detector region dipoles and the triplet magnets as shown in Fig.~\ref{fig:IP2_empty_space}.
The optimum dipole lengths in terms of synchrotron radiation power was determined to be $2/3 \cdot \lstar$ so a drift of \SI{5}{m} is left.
Now it is immediately clear that this region cannot be occupied by a superconducting  quadrupole septum as that would effectively decrease $\lstar$ and thus increase the synchrotron radiation power as a stronger separation is necessary.
Instead it is thinkable that a normal conducting quadrupole septum can be built that either does not require a yoke or similar structure between the beams or has a very thin yoke, or a septum that has a very limited and controlled field in the region of the electron beam trajectory.
In the later case it might even be used as part of the final focus system of the electron beam.
Either way, it is clear that such a normal conducting septum must have a pole tip field way below the saturation limit of iron.
The section on electron optics shows that a normal quadrupole of this kind can also have benefits in terms of synchrotron radiation, but studies remained to be done to make sure the parameters work for both cases.
For our calculation a pole tip field of \SI{1}{\tesla} was assumed.
For $\beta^* = \SI{5}{cm}$ an aperture radius of \SI{20}{mm} is required at a distance of \SI{14}{m} from the IP, resulting in a pole tip field of \SI{50}{T/m} for the normal conducting septum called Q0.
Possible ratios of apertures and gradients for the remaining triplet magnets were approximately based on the quadrupole parameters shown in Tab.~\ref{tab:quad_septa_params}, however these parameters would require a magnet design for confirmation.
With the quadrupole parameters shown in Tab.~\ref{tab:quad_septa_params_5cm} we were able to obtain triplet optics that can accommodate a beam with a minimum $\beta^*$ of \SI{5}{cm}.
\begin{table}[!ht]
  \centering
  \small
  \begin{tabular}{lccc}
	\toprule
	Magnet & Gradient  $[\si{T/m}]$ & Length $[\si{\metre}]$ &  Aperture radius  $[\si{mm}]$\\
        \midrule
	Q0 (nc) & 50 & 3.0 & 20 \\ 
	Q1A & 110 & 3.5 & 27 \\ 
	Q1B & 162 & 5.0 & 37\\ 
	Q2 & 123 & 5.0 & 62 \\ 
	Q3 & 123 & 4.5 & 62 \\
	\bottomrule
  \end{tabular}
  \caption{Parameters of the final focus quadrupole septa required to accommodate a $\beta^*$ of \SI{5}{cm}. The normal conducting quadrupole is called Q0 although it has the same polarity as Q1A/B.}
  \label{tab:quad_septa_params_5cm}
\end{table}

\begin{figure}[th]
  \centering
  \includegraphics[width=0.7\textwidth]{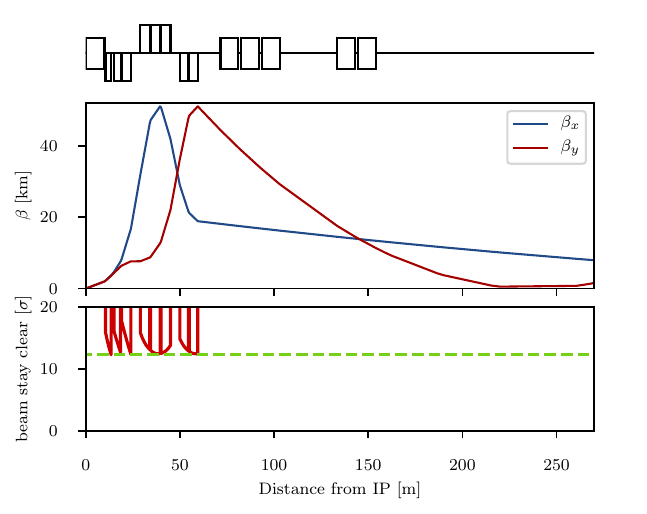}
  \caption{Optics (top) and beam stay clear (bottom) in the triplet region of colliding beam with $\bstar = \SI{5}{cm}$.}
  \label{fig:LHeC_aperture_beam_5cm}
\end{figure}

The corresponding optics are shown in Fig.~\ref{fig:LHeC_aperture_beam_5cm}.
So from the triplet point of view it appears possible to reach lower $\beta^*$, however many assumptions need verification: First the magnetic design for the normal conducting quadrupole septum must be shown to be possible. If there is a residual field in the space of the electron beam trajectory, the impact on the electron beam and the synchrotron radiation power must be evaluated.
The parameters of the modified superconducting triplet quadrupole septa, although scaled conservatively, must be confirmed.
Furthermore the larger aperture radius of Q1 might require a larger separation at the entrance of Q1, increasing the synchrotron power that is already critical. Thus a full design of such magnets is required. 
Lastly, the interaction region must be integrated into the full ring to verify that chromaticity correction is possible.
Studies in Ref.~\cite{bib:ECruzTomas:LHeCoptics} that were conducted on the normal triplet without regard for aperture constraints suggest that a chromaticity correction is only possible for a $\beta^*$ down to around~\SI{7}{cm}.

So far, the optics of the final focus system featured asymmetrically powered triplets on the two sides of the IP. This is inherited from the ALICE final focus system where the aperture is shared and the antisymmetry guarantees the same optics for both beams and similar chromaticities in both horizontal and vertical planes. In the LHeC final focus system however, the apertures of the quadrupoles are not shared between both beams, so the antisymmetry is not strictly necessary, although it eases the integration in the full ring. An alternative approach that is worth studying is a symmetric doublet. Doublets feature a large $\beta$ function in one plane and a relatively low one in the other plane for equal $\beta$ functions at the IP. Since the non-colliding proton beam is of no concern for LHeC it makes sense to create doublets on each side of the IP that have the peak $\beta$ function in the horizontal plane as the chromaticity correction was limited in the vertical plane. Furthermore, in a doublet the integrated focusing strength needed is lower as fewer quadrupoles act against each other. This further reduces the chromaticity and should also reduce the overall length of the final focus system. With the space saved by the doublet it is possible to either shift the recombination dipoles D1 and D2 closer to the IP, reducing the needed integrated strengths, or even to increase $\lstar$ to further reduce the synchrotron radiation power and critical energy. In order to make best use of the available doublet quadrupole aperture, it is also thinkable to collide with flat beams. The main disadvantage of symmetric doublets is the breaking of the sequence of focusing and defocusing quadrupoles. As no changes should be made to the arcs, the left-right symmetry needs to be broken up again in one of the matching sections, either by introducing another quadrupole on one side of the IP, or by overfocusing the beam.

At collision energy the non-colliding beam has no optics specification within the straight section. Consequently the optics should transfer the beam from the left arc to the right arc without hitting the aperture and at a specific phase advance. The same is true at injection energy, but with a larger emittance, making the satisfaction of the aperture constraint more difficult. Thus it is sufficient to find working injection optics, as no squeeze will be required for this beam. This approach of course will require some tuning as at least one arc will apply the ATS scheme at collision, but as the aperture constraint is less tight at higher energy there should be enough degrees of freedom available.

\begin{figure}[th]
  \centering
  \includegraphics[width=0.7\textwidth]{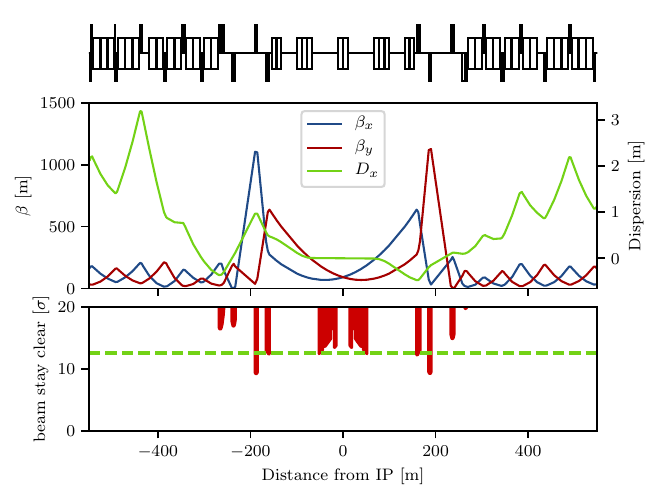}
  \caption{Optics (top) and beam stay clear of the non-colliding beam at injection energy. The Q5 quadrupole magnets on either side of the IP currently are aperture bottlenecks. It should be possible to mitigate this problem by replacing the magnets with longer, larger aperture magnets.}
  \label{fig:IR_noncolliding_beam_optics_inj}	
\end{figure}

Finding injection optics appears trivial at first but is complicated by the fact that the distance between the IP and the first quadrupole magnet Q4 is larger than \SI{159}{m}. A total distance of \SI{318}{m} needs to be bridged without any focusing available. A solution has been  found with $\beta^* = \SI{92}{m}$ and $\alpha^{*} = \pm 0.57$ with the required beam size in the quadrupole septa and Q4~\cite{bib:ECruzTomas:LHeCoptics}. The corresponding optics are shown in Fig.~\ref{fig:IR_noncolliding_beam_optics_inj}.
For the magnets Q4 and Q5 LHC quadrupoles of the large aperture MQY type with \SI{70}{mm} aperture diameter and a \SI{160}{T/m} gradient were assumed. As can be seen in the aperture plot, the triplet quadrupole septa and Q4 are just below the minimum beam stay clear at injection of \SI{12.6}{\sigma} but it is expected that nominal aperture can be achieved With some minor optimisation. However the Q5 magnets only have a beam stay clear of about \SI{9.2}{\sigma} with little chance of decreasing the beam size without increasing it both in Q4 and in the quadrupole septa. Consequently it will be necessary to use quadrupoles with apertures larger than \SI{106}{mm} and make up for the lower gradient by increasing the length or by using Nb$_3$Sn technology. At injection energy the remaining magnets in the IR have strengths according to the HL-LHC specification and thus do not pose any problems. However the injection optics shown in Fig.~\ref{fig:IR_noncolliding_beam_optics_inj} will require some changes during the ramp as Q4, Q5 and Q6 would become too strong at collision energy. This is not considered a problem though, as the emittance shrinking will ease the aperture requirements.

The non-colliding proton beam does not need to be focused and consequently passes the quadrupole septa of the colliding beam in the field free region. 

The large angle of \SI{7200}{\micro rad} between the two beams (compared to \SI{590}{\micro rad} in the high luminosity IPs) should suffice to mitigate long range beam-beam effects, considering that the shared aperture is only \SI{30}{m} long as opposed to the main experiments where the shared aperture exceeds a length of \SI{70}{m}.

\subsection{Electron Optics  \ourauthor{Kevin André, Bernhard Holzer}}
First ideas of a possible layout and design of the LHeC IR have already been presented in Ref.~\cite{AbelleiraFernandez:2012cc}.
Based on the principles explained there, a further optimisation of the beam separation scheme  has been established, with the ultimate goal of lowest synchrotron radiation power and critical energy in the direct environment of the particle detector.
Depending on the requests from the actual detector geometry and shielding, the flexibility of the new IR layout allows to optimise for either side. 

The basic principle is -- as before -- based on the large ratio (approximately 140) of the proton to electron beam momentum (or beam rigidity, $B\rho=p/e$)  that makes a magnetic field based separation scheme the straightforward solution to the problem, using effective dipole fields.

Boundary conditions are set however due to the limited longitudinal space, resulting from the distance of the first focusing elements of the proton lattice, located at $L*=15$\,m, and the need for sufficient transverse separation, defined by the technical design of this first proton quadrupole. The size of the two beams and -- clear enough -- the  power of the emitted synchrotron radiation $P_\text{syn}$ and the critical energy $E_\text{crit}$ have to be taken into account in addition.
The well known dependencies of these two parameters on the beam energy $E_e=m_e c^2 \gamma$ and bending radius $\rho$ are given by
\begin{equation}
  P_\text{syn}=\frac{e^2 c}{6 \pi \varepsilon_0} \frac{\gamma^4}{ \rho^2} ~~~~{\text{and}}~~~
  E_\text{crit}=\frac{3}{2} \frac{ \hbar c \gamma^3}{ \rho}\,.
  \label{eq:Syn_Pow}
\end{equation}

\begin{figure}[th]
	\centering
	\includegraphics[width=0.8\textwidth]{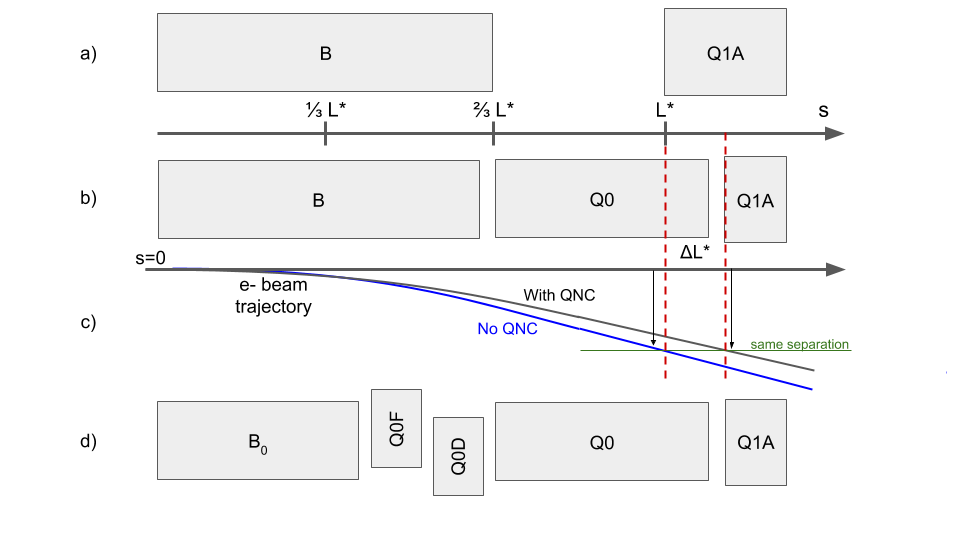}
	\caption{Separation scheme based on a long dipole magnet B (a) and improved layout using Q0, a normal conducting half-quadrupole as first focusing element of the proton beam (b). The last design features a doublet of off-centered quadrupoles to minimise the electron beam size at the entrance of Q1A (d).}
        \label{fig:sep_schem_alt}
\end{figure}

The schematic layout of the original design of the electron interaction region shown in Fig.~\ref{fig:IR_schematic} is reproduced in Fig.~\ref{fig:sep_schem_alt}~(a).
The long dipole magnet B, used to deflect the electron beam, is embedded inside the detector structure which is ranging from \SI{-6}{m} to \SI{4}{m} around the interaction point, extended by $\pm \SI{1.65}{m}$ of muon chamber. Basic interaction region designs with and without chromaticity correction were presented~\cite{bib:Zimmermann:IPAC2010-TUPEB037,bib:Abelleira:IPAC2012-TUPPR023} but were not fully integrated in the ERL.
The electron final quadrupoles were placed at \SI{30}{m} from the IP~\cite{bib:pos}, compatible with the proton layout described above. While this approach is straightforward, the only parameter that can be used to minimise the power of the emitted synchrotron radiaton is the length of the separator-dipole field~\cite{bib:RMartin:dipole_optimization}.
In addition, the installation of the first focusing elements of the electron beam downstream of the triplet focussing the colliding proton beam leads to a considerable increase of the electron beam size in the separation plane.

Lattices including chromaticity correction had a significant length of \SI{150}{m}.
However, the whole straight section between Linac and arc is only \SI{290}{m} long~\cite{AbelleiraFernandez:2012cc} and the IR design did not include a matching and splitting section or a focus system for the spent, outgoing electron beam.
Without chromaticity correction in the electron final focus, aberrations at the IP decrease luminosity by about 20\%~\cite{bib:tomasbonn}.

Investigations have been launched  to minimise critical energy and emitted synchrotron radiation power by reducing the separation in two main steps:
\begin{itemize}
\item introduce a compact mirror-plate half quadrupole (QNC) in front of Q1A (on the IP side)  to focus the colliding proton beam and provide a field free region for the electron and non-interacting proton beam.
  This reduces the required bending field of the separation dipole B for the same separation at Q1A.
  In addition, the normal conducting magnet QNC will act as shielding of the superconducting triplet magnets that would otherwise be subject to direct synchrotron radiation.
  Additional shielding is foreseen, to protect the SC magnets and avoid as much as possible backshining to the detector.
  In addition, sufficient space will be provided to correct the vertical orbit and coupling of the electrons coming from the solenoid. 
\item reduce the beam size of the electron beam by a very early focusing of the beam.
  As positive side effect this leads to a considerable reduction of the chromaticity of the electron lattice. \end{itemize}

The first step is sketched in Fig.~\ref{fig:sep_schem_alt}~(b) and the corresponding electron beam trajectory is shown in Fig.~\ref{fig:sep_schem_alt}~(c).

The introduction of the mirror plate half quadrupole QNC allows to reduce the length of the Q1A quadrupole while conserving the total integrated gradient, therefore leaving the overall focusing properties of the proton lattice quasi untouched.
The entry of Q1A is therefore moved away from the IP to relax the separation fields. 

\begin{figure}[tbh]
	\centering
	\includegraphics[width=0.9\textwidth]{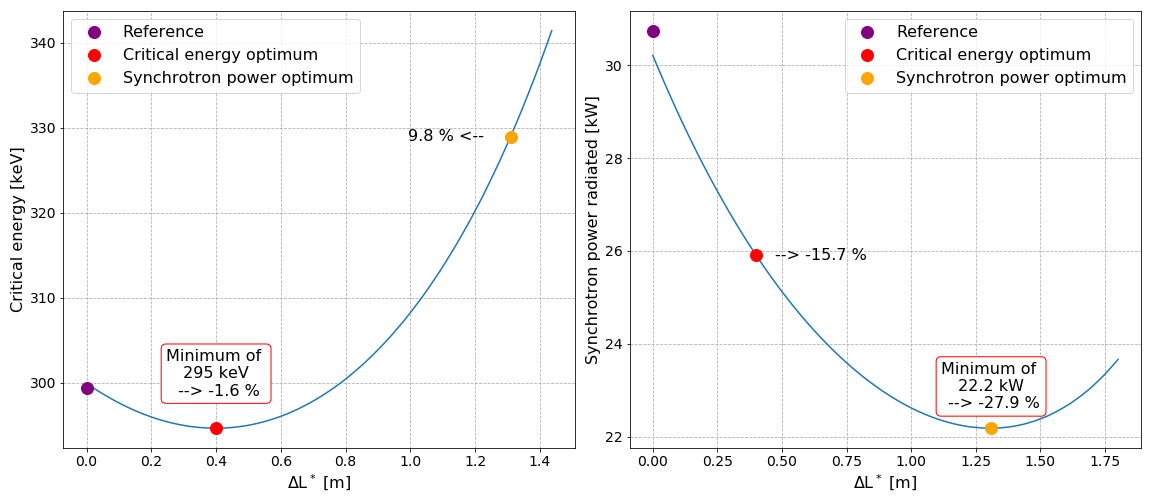}
	\caption{Improved critical energy and power of the synchrotron radiation for the half quadrupole based proton lattice. Left side: critical energy, right side: synchrotron radiation power. The horizontal axis refers to the shift $\Delta L^*$ of the position of the first proton superconducting magnet Q1A.}
        \label{fig:half_quad_lattice_ecrit}
\end{figure}

Scanning the Q1A entry position leads to either an optimum of the critical energy or to a minimum of the emitted synchrotron power.
Both cases are shown in Fig.~\ref{fig:half_quad_lattice_ecrit} and for each of them the new Q1A entry position  has been determined. The power of the emitted radiation is reduced by up to 28\,\%.
The colliding proton beam, passing through this half quadrupole with a certain offset to guarantee sufficient beam stay clear, will receive a deflecting kick in the horizontal plane of about \SI{90}{\micro rad}.
It supports the dipole based beam separation, provided by the so-called D1 / D2 magnets in LHC, and will be integral part of the LHC design orbit.

The resulting beam optics of the protons differs only marginally from the original version and only a slight re-match is needed. However by carefully choosing the gradient of the new magnet the parameters of the superconducting proton quadrupoles are untouched and the phase advance at the end of the interaction region lattice is conserved in both planes.

\subsubsection{Improved Electron lattice}
A further improvement of the emitted synchrotron power and critical energy is obtained by introducing an early focusing scheme of the electrons, which leads to a reduced electron beam size and thus to softer separation requirements. 

The reduction of the electron beam size is obtained by installing a quadrupole  doublet in the electron lattice between the separation dipole and the QNC (half-) quadrupole. A carefully matched focusing strength of this doublet will minimise the $\beta$ function of the electrons at the location of Q1A. At the same time an effective dipole field, that is needed to maintain the separation of proton and electron beams, is provided by  shifting the magnet centres of the doublet lenses off axis. The horizontal offset of these quadrupoles has been chosen to provide the same bending radius as the separation dipole, thus leading in first order to the same critical energy of the emitted light in all separation fields. A detailed calculation of the divergence of the photons, the geometry of the radiation fan and the position of the absorbers and collimators will be one of the essential next steps within the so-called machine-detector-interface considerations.

Fig.~\ref{fig:sep_schem_alt}~(d) shows the new layout -- compared to the previous version. The doublet providing the early focusing of the electron beam is embedded in the separator dipole, i.e. it is positioned at $ s=\SI{6.3}{m}$  and acts in combination with the separation dipole. The quadrupole gradients have been chosen for optimum matching conditions of the electron beam and the transverse shift of the field centres provide the same separation dipole effect as used in the long dipole. 

\begin{figure}[th]
  \centering
  \includegraphics[width=0.47\textwidth]{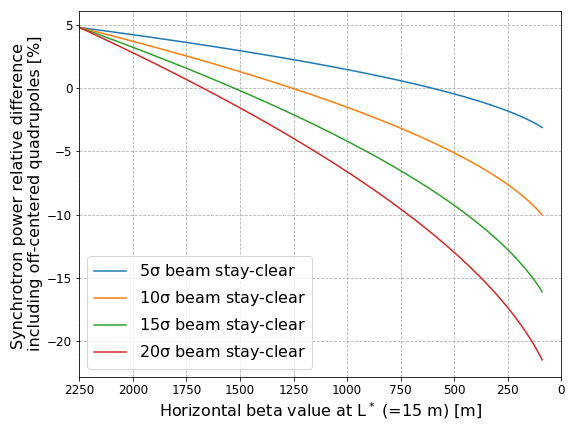}
  \includegraphics[width=0.47\textwidth]{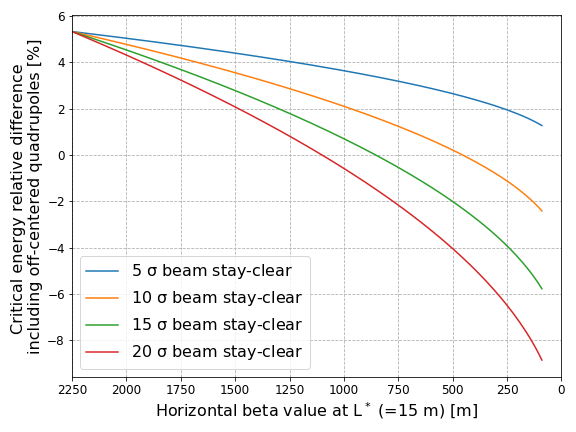}
  \caption{Relative difference with respect to the single dipole separation scheme for different values of the required beam stay-clear expressed in $\sigma$. Left : for the power of the emitted radiation, as function of the $\beta$-function of the electron beam at position s=15m. Left: for the critical energy of the emitted radiation, as function of the $\beta$-function of the electron beam at position s=15m. The early focusing of the electron beam allows for a much reduced separation field and thus to a reduced critical energy and power of the emitted radiation. The initial beta value is \SI{2250}{m}.}
  \label{fig:syrad_improvement_equads_crit}
\end{figure}

The early focusing of the electron beam allows for a softer separation of the beams, and leads therefore directly to a reduced critical energy $E_\text{crit}$ and power $P_\text{syn}$ of the emitted radiation.
Fig.~\ref{fig:syrad_improvement_equads_crit} shows the dependence of $E_\text{crit}$ and $P_\text{syn}$ on the $\beta$-function at $s=L^*$ for the electron optics for different values of the required electron beam stay-clear expressed in units of the electron beam size $\sigma$.
The beam separation has been re-calculated and the critical energy and radiation power are plotted. The graphs include different assumptions for the beam size considered. Including orbit tolerances, a beam stay-clear of 20 $\sigma$ is considered as the most relevant case, which refers to the red curve in the graph.

In order to provide a complete study with the lattice featuring the off-centered quadrupoles, 
the new interaction region has been embedded in between the high energy end of the acceleration part of the linac and the \emph{Arc~6}
of the ERL, which marks the start of the energy recovery lattice. An optimum has been found for a beam optics with  a  beta function in the plane of the beam separation (i.e. horizontal) of $\beta_x = \SI{90}{m}$ at $L^* \approx \SI{15}{m}$

\begin{figure}[th]
  \centering
  \includegraphics[width=0.8\textwidth]{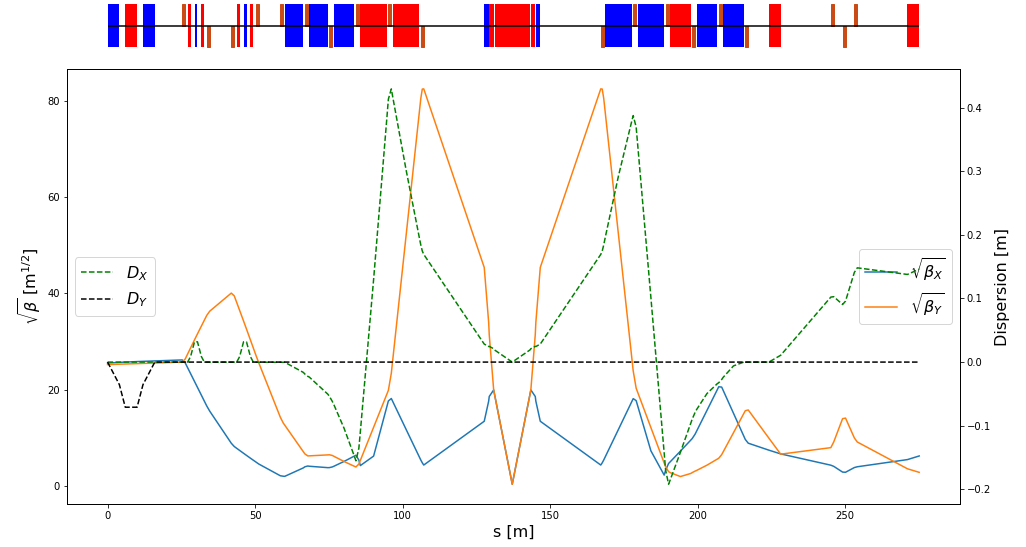}
  \caption{Electron beam optics for the new lattice including the early focusing scheme. The offset of the new doublet quadrupoles are chosen to provide the same separation field as in the dipole. The new optics is matched on the left side of the plot to the end of the acceleration linac. The right hand side is connected to Arc~6, the beginning of the decelerating ERL part. At the position of the first superconducting proton magnet the $\beta$-function in the (horizontal) separation plane of the electron beam is reduced to \SI{90}{m} for lowest possible synchrotron radiation load.}
  \label{fig:e_optics}
\end{figure}

An improvement of about 9\,\% for the critical energy and close to 25\,\% of the radiated power is obtained, if an electron beam optics with $\beta_x=\SI{90}{m}$ at the entrance of Q1A is used. 
For this most promising case the matched beam optics is shown in Fig.~\ref{fig:e_optics}. 

The lower $\beta$-function of the electron beam at the focusing elements has the additional positive feature of reducing considerably the chromaticity of the new lattice, which is a crucial parameter for the performance of the energy recovery process (details are described below in the chapter on tracking calculations).
Compared to the dipole based separation and a late focusing, $Q'$ is reduced to a level of 13\,\% horizontally and to a level of 11\,\% in the vertical plane. 
The details are listed in Tab.~\ref{tab:chroma_summary}. Further studies will investigate the orbit correction scheme of the new IR, and an eventual interplay of the solenoid fringe field and the quadrupoles. 
\begin{table}[!ht]
  \centering
  \small
  \begin{tabular}{lcc}
    \toprule
    & Dipole based separation &  Early focusing scheme \\
    \midrule
    $\xi_x$ & -116 & -15 \\
    $\xi_y$ & -294 & -32 \\
    \bottomrule
  \end{tabular}
  \caption{Chromaticity of the dipole based separation scheme and the new lattice based on early focusing, off-axis quadrupole lenses.}
  \label{tab:chroma_summary}
\end{table}

The influence of the electron doublet magnets on the proton optics is marginal -- as can be expected due to the large difference in beam rigidity: If uncorrected, the electron doublet creates a distortion (a so-called \emph{beta-beat}) of the proton optics of roughly 1\,\%.
Still it has been calculated and taken into account in the context of a re-match of the proton beam optics.

\begin{figure}[tbh]
	\centering
	\includegraphics[width=0.7\textwidth]{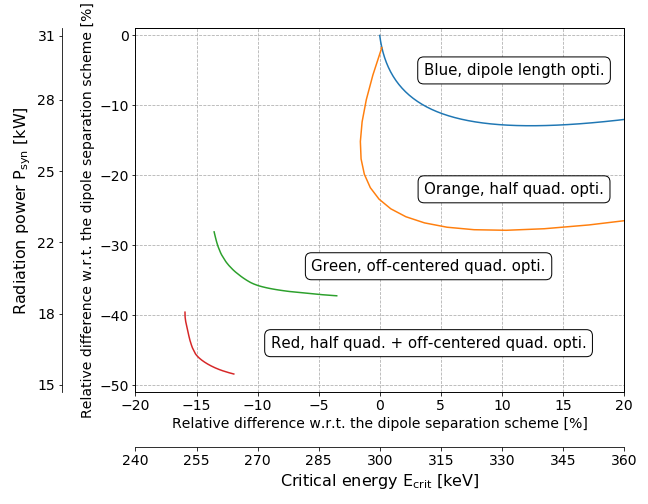}
	\caption{Relative differences with respect to the original single dipole separation scheme. The synchrotron radiated power is plotted as a function of the critical energy for different optimisation results: only optimising the dipole length (blue), only using a mirror quadrupole (orange), only using off-centered quadrupoles (green) and combining the mirror quadrupole with an earlier focusing (red).}
        \label{fig:allinall_optimum_power} 
\end{figure}

Combining the two improvement factors, namely the effective lengthening of $L^*$ due to the use of a half quadrupole in front of the superconducting triplet, and the early focusing scheme in the lattice of the electrons, leads to an overall improvement of the interaction region with respect to synchrotron radiation power and critical energy that is shown in Fig.~\ref{fig:allinall_optimum_power}. 
The overall improvement factor is plotted with reference to the baseline dipole separation design with originally $\beta=\SI{2250}{m}$ at the separation point $s=L^*$. Using a normal conducting half quadrupole in combination with the early focusing scheme, the power of the emitted synchrotron radiation is reduced by 48\,\% for an electron beam stay-clear of 20\,$\sigma$.

The estimated synchrotron radiation power and critical energy for the different optimisations are plotted in Fig.~\ref{fig:allinall_optimum_power} and the results are summarised in Tab.~\ref{tab:all_optimum}.
Referring to a beam energy of 49.19\,GeV and the design current of \SI{20}{mA} an overall power of 16.2\,kW is emitted within one half of the interaction region.

\begin{table}[!ht]
  \centering
  \small
  \begin{tabular}{lccccc}
    \toprule
    Optimised scheme  & \multicolumn{2}{c}{Synchrotron radiation} & & \multicolumn {2}{c}{Critical energy} \\
    \cmidrule{2-3}    \cmidrule{5-6} 
    & Radiation  &  Critical     & & Radiation  &  Critical  \\
    & power [kW] &  energy [keV] & & power [kW] &  energy [keV]  \\
    \midrule
    Reference design                      & 30.8 & 300 & & 30.8 & 300 \\
    Dipole length optimum                 & 26.8 & 336 & & 30.8 & 300  \\
    Half quadrupole optimum               & 22.2 & 331 & & 26.1 & 295  \\
    Off-centered quadrupoles opti.        & 19.3 & 290 & & 22.1 & 259  \\
    Half quad. + Off-centered quad. opti. & 16.2 & 265 & & 17.4 & 255  \\
    \bottomrule
  \end{tabular}
  \caption{Synchrotron radiation power and critical energy for the different optimised separation schemes.}
  \label{tab:all_optimum}
\end{table}

Depending on the boundary conditions imposed by the integration of the particle detector, one of the two optimum layouts can be chosen --  or a combination of both, i.e.\ an overall minimum defined by critical energy and radiated power.

The basic main parameters of the proton mirror plate half quadrupole are summarised in Tab.~\ref{tab:half_quad} for the two optimum scenarios explained above: the optimum found for smallest synchrotron radiation power and the optimum for smallest critical energy of the emitted radiation. The values result from the optics studies of the previous sections. The presented gradients lead to a pole tip field of $B_p \approx \SI{1.3}{T}$.

\begin{table}[!ht]
  \centering
  \small
  \begin{tabular}{lccc}
    \toprule
    Half quadrupole & Unit & Minimum synchrotron & Minimum critical \\
    parameter       &      & radiation power     & energy \\
    \midrule
    $\gamma \varepsilon_p$ & mm$\cdot$mrad & 2.50 & 2.50 \\
    Gradient & T/m & 48.2 & 50.7 \\
    Aperture radius & mm & 27.0 & 25.6 \\
    Length & m & 6.84 & 2.08 \\
    \bottomrule
  \end{tabular}
  \caption{Magnet gradient of the proposed half quadrupole for lowest synchrotron radiaton power and lowest critical energy. An aperture of $15\,\sigma$ + 20\,\% beta-beating + \SI{2}{mm} orbit tolerances has been assumed.}
  \label{tab:half_quad}
\end{table}

\begin{figure}[tbh]
	\centering
	\includegraphics[width=0.8\textwidth]{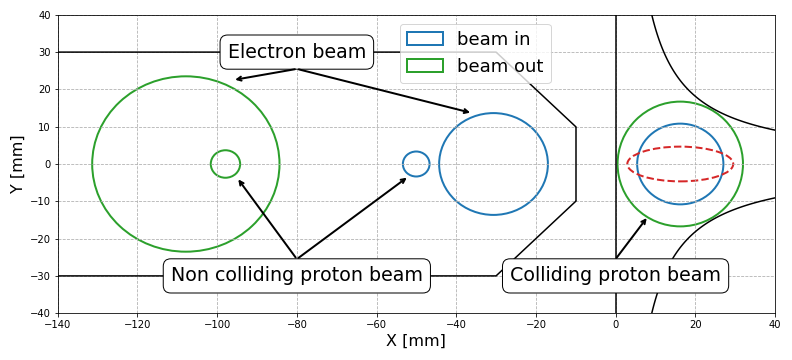}
	\caption{The position of the three beams at the entrance (blue) and exit (green) of the half quadrupole. The colliding proton beam is centered inside the main magnet aperture, while the second proton beam and the electrons are located in the field free region. The dashed red line represents the injection proton beam at the output of the half quadrupole.}
        \label{fig:half_quad_threebeams} 
\end{figure}

In both cases, the proton aperture radius has been chosen to include an orbit tolerance of 2\,mm, a 10\,\% tolerance on the beam size due to optics imperfections (beta-beating) and a beam size that corresponds to $n=15\,\sigma$ for a proton beam normalised emittance $ \varepsilon_p = 2.50 \mu m$.
A value that is comfortably larger than the requirements of the HL-LHC standard lattice.
The injection proton optics has been taken into account and although it features a larger emittance it clearly fit in the aperture, see the red dashed line in Fig.~\ref{fig:half_quad_threebeams}.
The electron beam and the non-colliding proton beam will pass through the field free region delimited by the mirror plate.

The aperture requirements inside the half quadrupole are determined on one side by the colliding proton beam optics in the main aperture of the magnet.
The beam separation scheme and optics of electron and non-colliding proton beam on the other side have to fit into the field free region beyond the mid plane of the mirror plate. As described below, a crossing angle of 7\,mrad is assumed for the non-colliding protons.
These requirements are illustrated in Fig.~\ref{fig:half_quad_threebeams}.
For the case of smallest synchrotron radiation power, the three beams are plotted at the entrance and exit of the quadrupole lens.
For both proton beams the beam size shown in the graph corresponds to 15 sigma plus \SI{2}{mm} orbit tolerance and 10\,\% beam size beating.
Due to the mini-beta optics the colliding proton beam fills nearly the given aperture of the magnet.
The non-colliding proton beam follows a relaxed optics with very limited aperture need. The envelope of the electron beam is shown for 20\, $\sigma$ beam size in both transverse planes. 

In contrast to the proton half quadrupole, the doublet magnets of the early focusing scheme will house the three beams in one single aperture.
In addition to the beam envelopes, the offset that has been chosen to provide the beam separation effect has to be taken into account and included in the aperture considerations. 

\begin{figure}[th]
  \centering
  \includegraphics[width=0.9\textwidth]{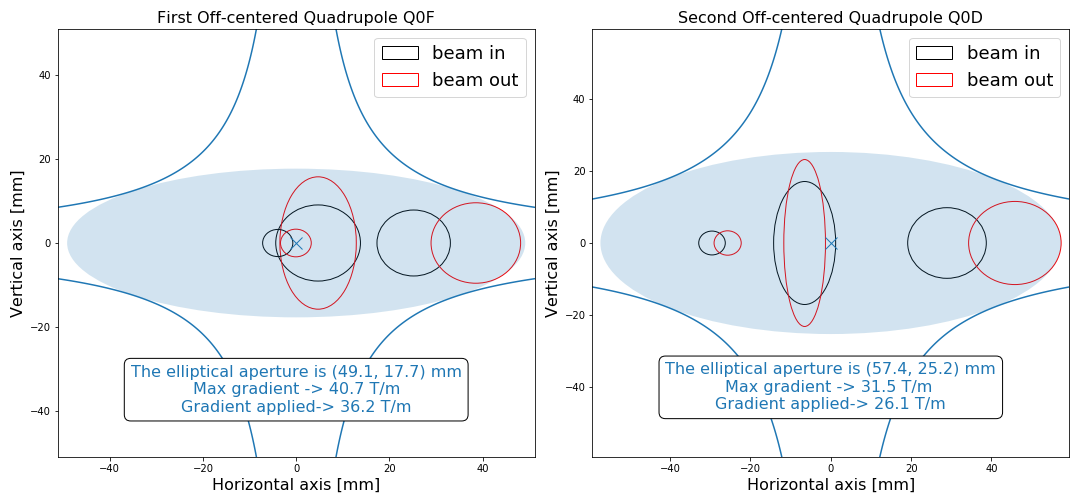} 
  \caption{The position of the three beams at the entrance (black) and exit (red) of the electron doublet magnets. Following the internal convention, 15\,$\sigma$ plus 20\,\% beta beating plus \SI{2}{mm} orbit tolerances beam envelopes are chosen for the proton beams. The beam size of the electrons refer to 20\,$\sigma$. From left to right the three beams are respectively the non colliding proton beam (tiny circles), electron beam (squeezed ellipses) and the colliding proton beam.}
  \label{fig:doublet}
\end{figure}

In Fig.~\ref{fig:doublet} the situation is visualised.
On the left side the first off-center quadrupole  (powered as focusing lens) is presented. Following the field direction, the electron beam is offset towards the outer side of the ring (right side of the plot) as defined by the proton beam closed orbit. The right part of the figure  shows the second quadrupole (powered as defocusing lens) with the electron beam offset shifted to the other direction. In order to provide sufficient aperture for the three beams, an elliptical shape has been chosen for the vacuum chamber. It defines enough space for the beam envelopes and the off-centre design trajectories. The black ellipses correspond to the beams at the entrance of the magnet while the red  shapes represent the beams at the exit. From left to right the three beams are respectively the non colliding proton beam (tiny circles), electron beam (squeezed ellipses) and the colliding proton beam. As defined before we refer to a beam size of 20\,$\sigma$ in case of the electrons and 15 sigma plus beta-beating plus 2 mm orbit tolerance for the colliding and non-colliding proton beam.

In this context it should be pointed out that the non-colliding proton beam, travelling in the same direction as the electrons, is shifted in time by half the bunch spacing. While the projected beam envelopes in Figs.~\ref{fig:doublet} and~\ref{fig:half_quad_threebeams} seem to overlap in the transverse plane, they are well separated by 12.5\,ns, corresponding to 3.75\,m, in the longitudinal direction.

The minimum required gradients and pole tip radius of the quadrupoles of the doublet are listed in Tab.~\ref{tab:doublet_specs}.
\begin{table}[!ht]
  \centering
  \small
  \begin{tabular}{lccc}
    \toprule
    Parameter & Unit & Q0F & Q0D \\
    \midrule
    $\gamma \varepsilon_e$  & mm$\cdot$mrad & 50 & 50 \\
    $\gamma \varepsilon_p$ & mm$\cdot$mrad & 2.50 & 2.50\\
    Gradient & T/m & 36.2 & 26.1 \\
    Min. pole-tip radius & mm & 28.9 & 38.1 \\
    Length & m & 1.86 & 1.86 \\
    \bottomrule
  \end{tabular}
  \caption{Magnet gradient and pole tip aperture of the quadrupoles of the doublet for the synchrotron power optimum.}
  \label{tab:doublet_specs}
\end{table}
Following the increasing beam size after the IP, the two quadrupoles are optimised for sufficient free aperture for the collidng beams and their design orbits. Accordingly a different layout has been chosen for the magnets, to provide the best conditions for the radiation power and critical energy. An alternative approach has been studied, based on a single quadrupole design for both lenses of the doublet. While an optics solution still is possible, it does however not allow for minimum radiation power and sets more stringent requirements on the shielding and absorption of the synchrotron light fan.

\begin{figure}[th]
    \centering
    \includegraphics[width=0.8\textwidth]{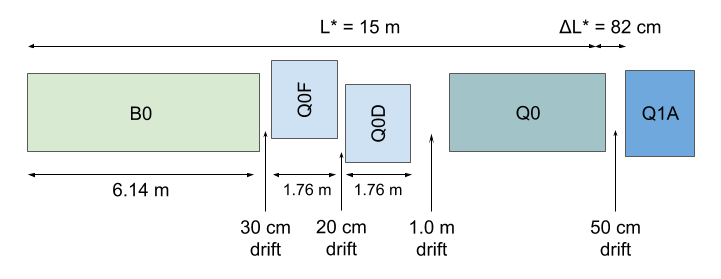}
    \caption{Possible optimised design featuring a 1.0\,m drift between the off-centered quadrupoles and the half quadrupole in order to leave space for shielding material.
    }
    \label{fig:my_label}
\end{figure}

\begin{figure}[th]
	\centering
	\includegraphics[width=0.44\textwidth]{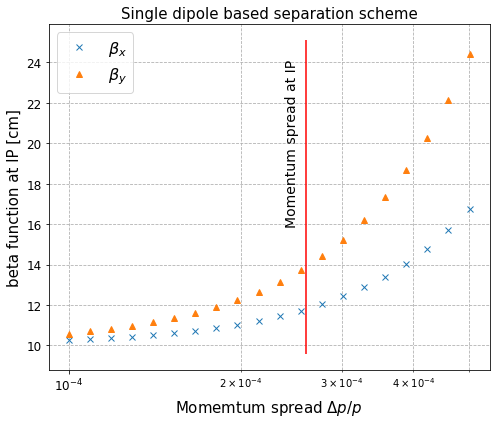} 
        \hspace{0.02\textwidth}
	\includegraphics[width=0.44\textwidth]{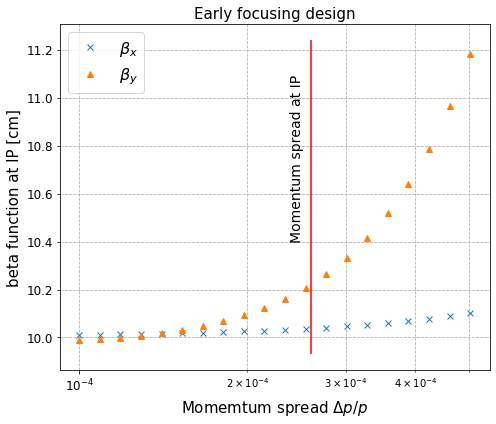} 
	\caption{Beta function at the IP as a function of the momentum spread. Left : Situation for the single dipole based separation scheme. Right : With the design featuring an earlier focusing. The graphs show the increase of $\beta^*$ due to the chromaticity of the lattice.}
        \label{fig:beamsize_momemtum_acceptance}
\end{figure}

\begin{figure}[tbh]
	\centering
	\includegraphics[width=0.5\textwidth]{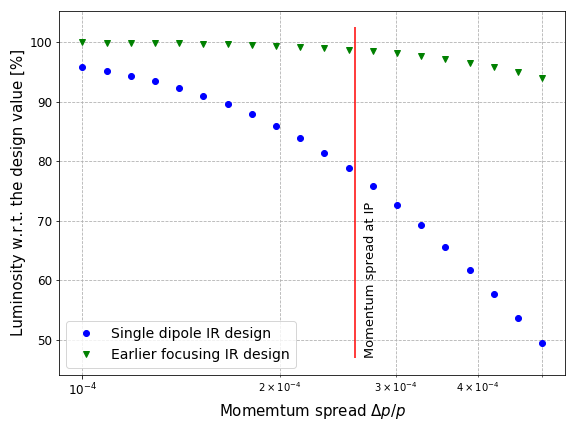} 
	\caption{Luminosity as a function of the momentum spread for the single dipole based separation scheme (blue circles) and the design featuring an earlier focusing (green triangles).}
        \label{fig:lumi_momemtum_acceptance}
\end{figure}

The chromatic effect of the two lattice versions as a function of the momentum spread
is shown in Fig. \ref{fig:beamsize_momemtum_acceptance}.
The lattice based on a single dipole magnet and late focusing of the electron beam shows an increase of the $\beta$ function of up to 40\% in the vertical plane for particles with a momentum deviation up to the design value of $\frac{\Delta p}{p} = 2.6 \cdot 10^{-4} $ (vertical cursor line in the graph) and a corresponding luminosity loss of 20\% for those particles (see Fig. \ref{fig:lumi_momemtum_acceptance}). The optimised design, based on the early focusing scheme, shows a much reduced chromatic effect and the resulting off-momentum beta-beating at the IP is limited to a few percent. As direct consequence the luminosity loss is well below the 1.5\% level. A special local chromaticity correction scheme, therefore, dealing with the aberrations at IP, is thus not considered as necessary. Further studies will include the recirculation of the beam post-collision and the energy recovery performance and might nevertheless highlight the need of explicit sextupoles to mitigate the growing momentum spread through the deceleration process and to avoid beam losses.

\subsection{Interaction Region Magnet Design \ourauthor{Stephan Russenschuck, Kevin Andre', Bernhard Holzer}}
%
\subsubsection{Triplet Magnet Design}

\begin{figure}[th]
  \centering
    \includegraphics [width=0.45\textwidth]{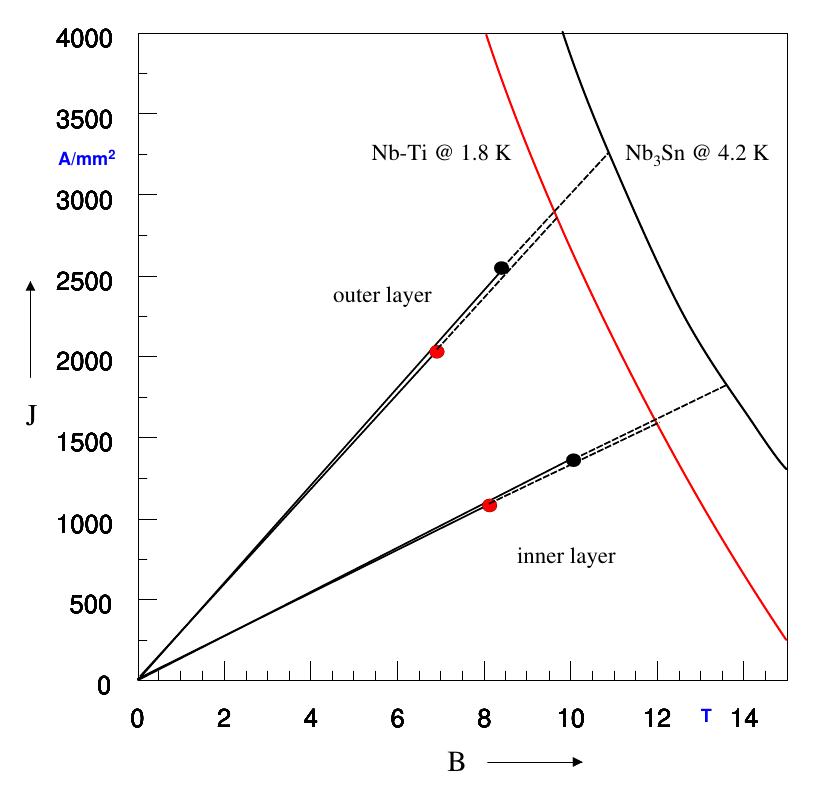}
  \caption{Working points on the load-line for both Nb-Ti and Nb$_3$Sn variants of Q1A.}
  \label{figll} 
\end{figure}

While the Q1 magnets remain in the range achievable with the well proven Nb-Ti superconductors, operated at 1.8\,K, the Q2 magnets require Nb$_3$Sn technology at an operation temperature of 4.2\,K.
The working points on the load-line are given for both superconducting technologies in Fig.~\ref{figll}.

The thickness of a coil layer is limited by the flexural rigidity of the cable, which will make the coil-end design difficult.
Therefore multi-layer coils must be considered.
However, a thicker, multi-layer coil will increase the beam separation between the proton and the electron beams.
The results of the field computation are given in Tab.~\ref{tab:mag_data}.

\begin{table}[!ht]
  \centering
  \small
  \begin{tabular}{lcccccc}
    \toprule
    Magnet parameter  & Unit & \multicolumn{4}{c}{Magnet type} \\
    \cmidrule{3-6}
    &  &     Q1A    &  Q1B      &  Q2 type  &   Q3 type       \\
    \midrule
    Superconductor type              &       &    Nb-Ti   &  Nb-Ti   &  Nb$_3$Sn   &   Nb$_3$Sn  \\
    Coil aperture radius $R$             &  mm   &     20      & 32     &  40   &  45    \\ 
    Nominal current $I_\text{nom}$  &  A    &   7080     &   6260      &  7890  &  9260   \\
    Nominal gradient $g$             &  T/m  &   252   &  164      &  186   & 175     \\
    Percentage on the load line              &   \%  &  78      &  64        & 71   &  75    \\
    Beam separation distance $S_\text{beam}$ &  mm   &  106-143     &  148-180      &  233-272  &   414-452   \\
    \bottomrule
    \end{tabular}
  \caption{Main triplet magnet parameters
  }
  \label{tab:mag_data}
\end{table}
Unlike with the design proposed in the CDR of 2012~\cite{AbelleiraFernandez:2012cc}, the increased beam separation distance between the colliding proton beam and the electron beam makes it possible to neglect the fringe fields in the electron beam pipe.
For the Q2 and Q3 magnets, the electron beam is outside of the quadrupole cold-mass and consequently, an HL-LHC inner-triplet magnet design can be adapted.

For the Nb$_3$Sn material we assume composite wire produced with the internal Sn process (Nb rod extrusions)~\cite{Parrell:2004vgc}.
The non-Cu critical current density is 2900\,A/mm$^2$ at 12\,T and 4.2\,K.
The filament size of 46 $\mu$m in Nb$_3$Sn strands give rise to higher persistent current effects in the magnet.
The choice of Nb$_3$Sn would impose a considerable R\&D and engineering design effort, which is however, not more challenging than other accelerator magnet projects, such as the HL-LHC.

\begin{figure}[th]
  \centering
  \includegraphics [width=0.35\textwidth]{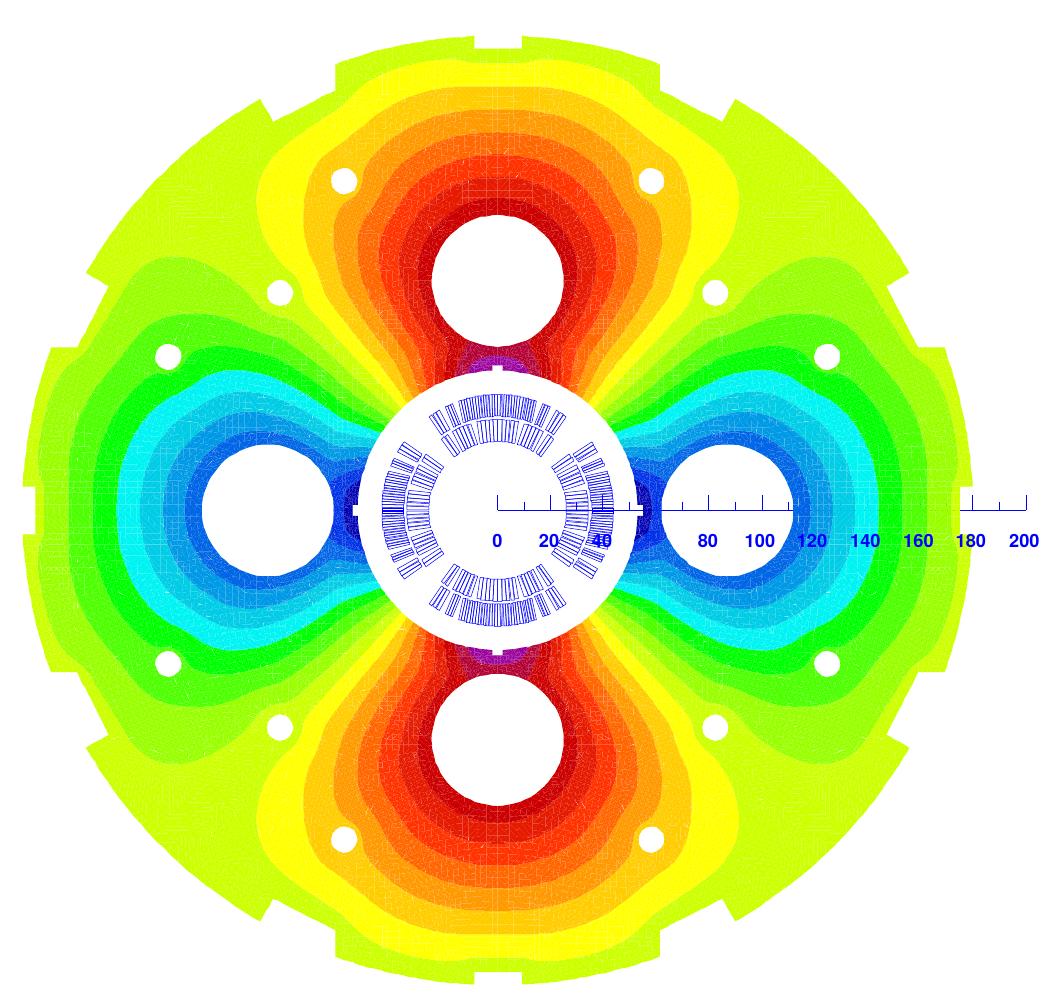}
  \hspace{0.02\textwidth}
  \includegraphics [width=0.35\textwidth]{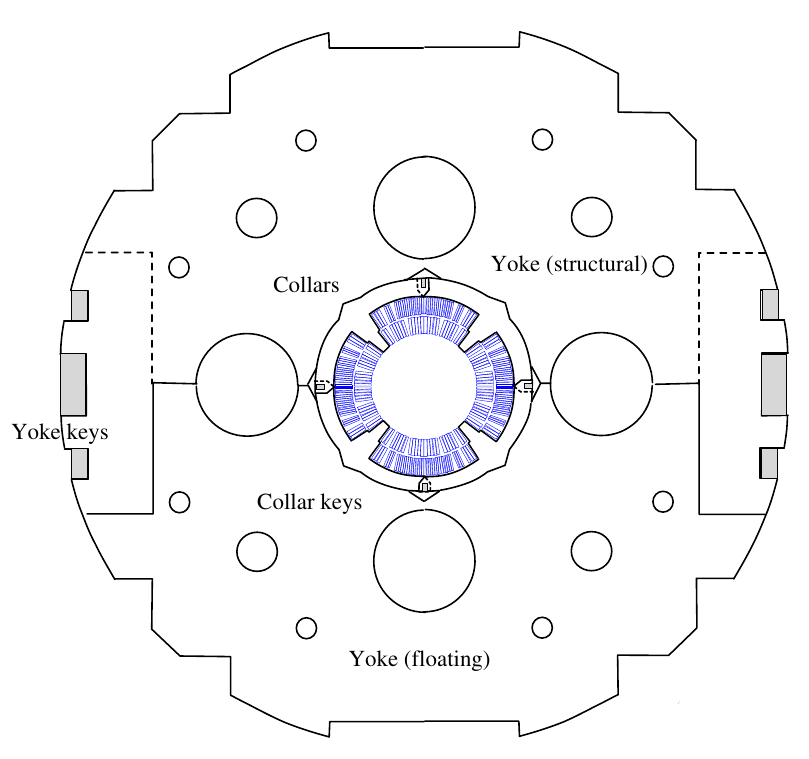}
  \caption{Conceptual design of the final focus septa Q1. Left: Magnetic vector potential (field lines). Right: Sketch of the mechanical structure.}
  \label{fig:elec_fig5}
\end{figure}

The conceptual design of the mechanical structure of the Q1 magnets is shown in Fig.~\ref{fig:elec_fig5} (right).
The necessary prestress in the coil-collar structure, which must be high enough to avoid unloading at full excitation, cannot be exerted with the stainless-steel collars alone.
Two interleaved sets of yoke laminations (a large one comprising the area of the yoke keys and a smaller, floating lamination with no structural function) provide the necessary mechanical
stability of the magnet during cooldown and excitation.
Preassembled yoke packs are mounted around the collars and put under a hydraulic press, so that the keys can be inserted.
The sizing of these keys and the amount of prestress before the cooldown will have to be calculated using mechanical FEM programs.
This also depends on the elastic modulus of the coil, which has to be measured with a short-model equipped with pressure gauges.
Special care must be taken to avoid nonallowed multipole harmonics because the four-fold symmetry of the quadrupole will not entirely be maintained.

For the Q2 and Q3 magnets, a HL-LHC inner triplet desing using a bladder and key mechanical structure can be adapted.

\subsubsection{Normal-Conducting Magnet Design}
\begin{figure}[tbh]
	\centering
	\includegraphics [width=0.45\textwidth]{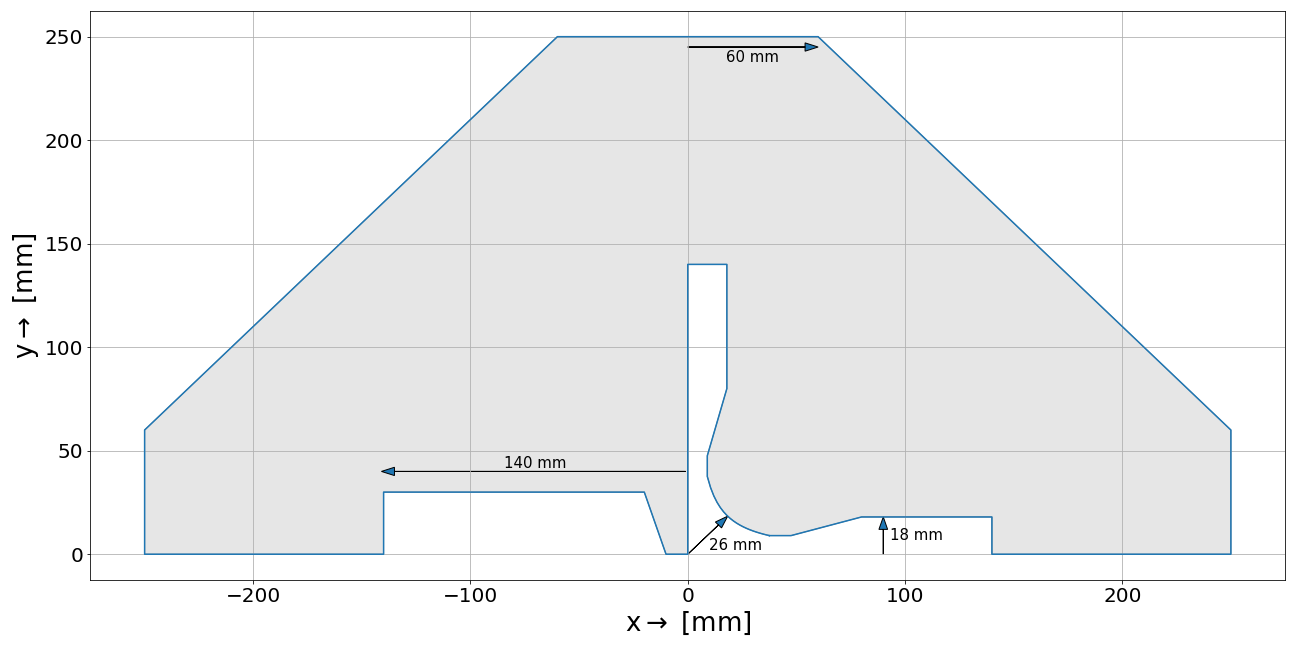}
	\includegraphics[width=0.45\textwidth]{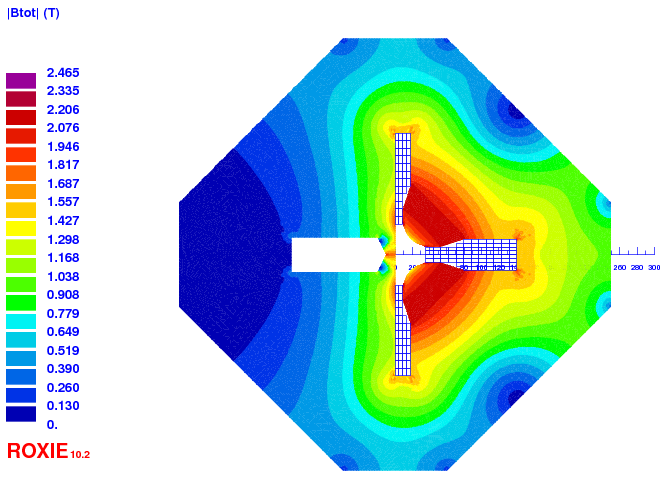}
	\caption{\label{fig:half_quad_geo} Left: Mechanical layout of the new half quadrupole for the proton beam. Right : Field distribution in the half quadrupole for the proton beam.}
\end{figure}

The proposed mini-beta doublet of the electron lattice, providing an early focusing of the beam, and the normal conducting proton-half quadrupole are new magnet concepts. These have been studied conceptually to determine their technical feasibility.  The geometry of the QNC magnet is shwon in Fig.~\ref{fig:half_quad_geo}~(left). 
Left of the mirror plate, the field free region will provide space for the electron beam and the non-colliding proton beam. 
The thickness of the mirror plate at the magnet mid-plane is \SI{20}{mm}, allowing for sufficient mechanical stability at the minimal beam separation between the electron and proton beams.

Field calculations, using the magnet design code ROXIE~\cite{bib:Russenschuck} are presented in Fig.~\ref{fig:half_quad_geo}~(right).
The achieved field gradient is \SI{50}{T/m} for a current of \SI{400}{A}, assuming a current density of \SI{21.14}{A/mm^2}.
This is in line with conductor geometries used for normal conducting magnets installed in the CERN injector complex, for example, ID: PXMQNDD8WC, which is rated at \SI{860}{A} corresponding to \SI{45.45}{A/mm^2}.
A more comprehensive design study must also include a further reduction of the multipole field components.

The geometry of the Q0F and Q0D quadrupoles are given in Fig.~\ref{fig:doublet} and the main specifications are provided in Tab.~\ref{tab:doublet_specs}.
A maximum magnetic field of \SI{1.2}{T} at the pole tip is well within reach for a normal conducting quadrupole. 

\section{Civil Engineering \ourauthor{Alexandra Tudora, John Osborne}}
\begin{figure}[th]
    \centering
    \includegraphics[width=0.8\textwidth]{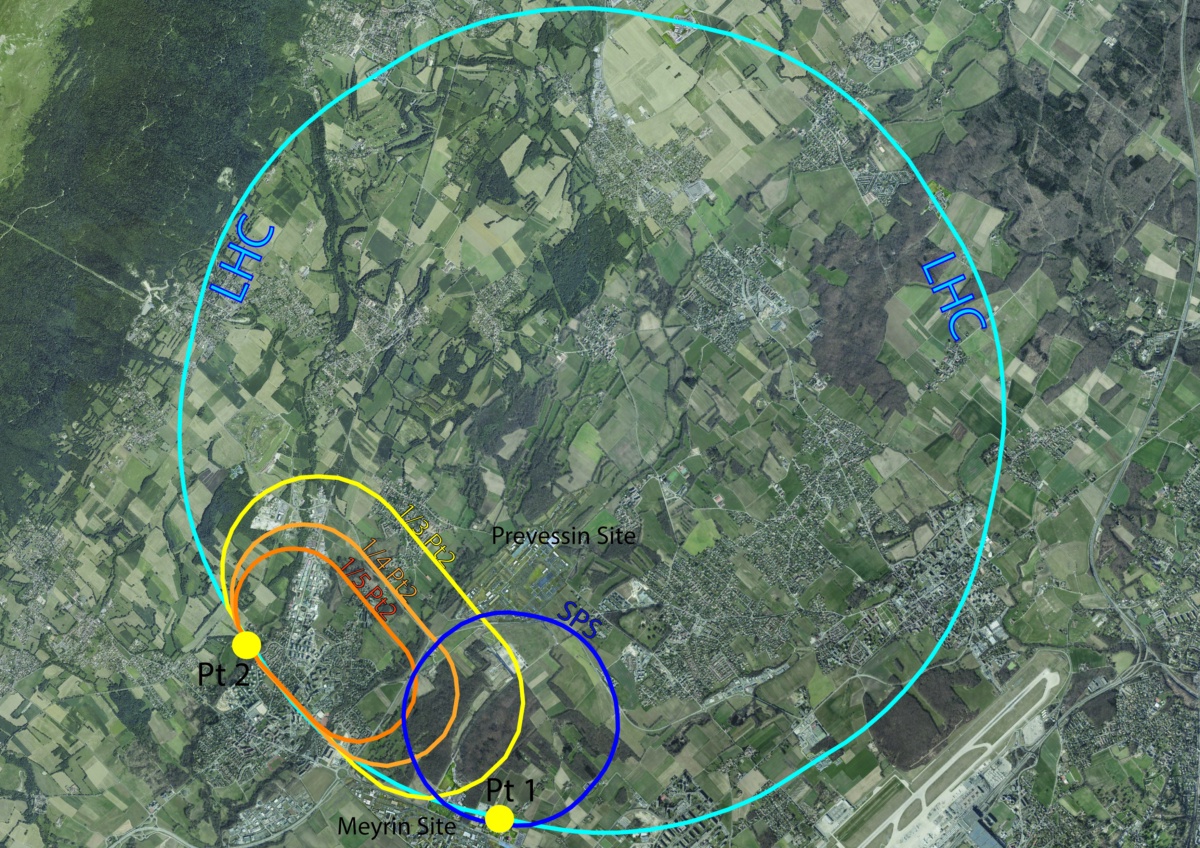}
    \caption{Racetrack options proposed for LHeC at Point 2 of the LHC.
    The color coding illustrated different options with 1/3, 1/4 and 1/5 of the LHC circumference, resulting in different electron beam energies.}
    \label{fig:LHeC options}
\end{figure}

Since the beginning of the LHeC study which proposes a electron-hadron collider, various shapes and sizes of the $eh$ collider were studied around CERN region. Two main options were initially considered, namely the Ring-Ring and the Linac-Ring. For civil engineering, these options were studied taking into account geology, construction risks, land features as well as technical constraints and operations of the LHC. The Linac-Ring configuration was selected, favouring a higher achievable luminosity. This chapter describes the civil engineering infrastructure required for an Energy Recovery Linac (ERL) injecting into the ALICE cavern at Point 2 LHC.
Fig.~\ref{fig:LHeC options} shows three options for the ERL of different sizes, represented as fractions of the LHC circumference, respectively 1/3, 1/4 and 1/5 of the LHC circumference.

\subsection{Placement and Geology}
The proposed siting for the LHeC is in the North-Western part of the Geneva region at the existing CERN laboratory. The proposed Interaction Region is fully located within existing CERN land at LHC Point~2, close to the village of St.~Genis, in France. The CERN area is extremely well suited to housing such a large project, with well understood ground conditions having several particle accelerators in the region for over 50 years. Extensive geological records exist from previous projects such as LEP and LHC and more recently, further ground investigations have been undertaken for the High-Luminosity LHC project. Any new underground structures will be constructed in the stable molasse rock at a depth of 100--150\,m in an area with low seismic activity. 

\begin{figure}
    \centering
    \includegraphics[width=0.8\textwidth]{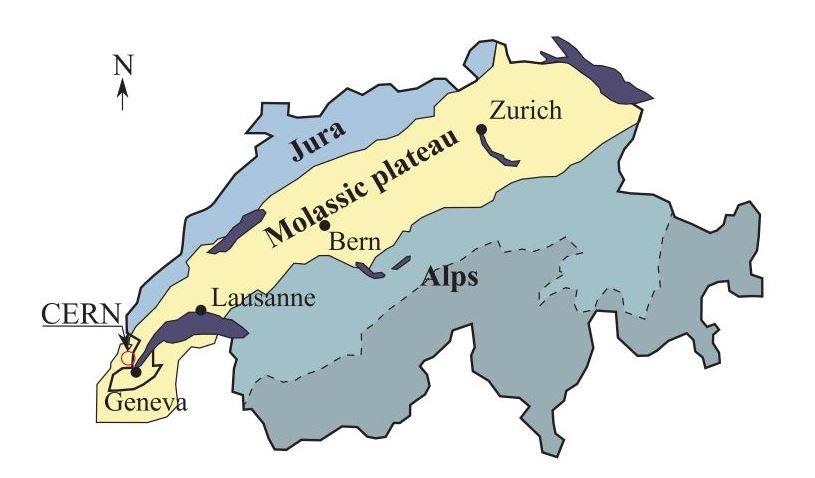}
    \caption{Simplified map of Swiss geology.}
    \label{fig:swiss-geology}
\end{figure}

\begin{figure}
    \centering
    \includegraphics[width=0.8\textwidth]{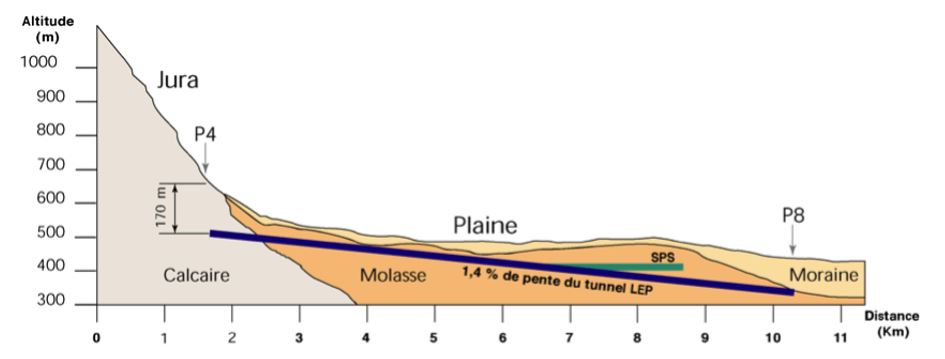}
    \caption{Geological profile of the LHC tunnel.}
    \label{fig:LHC profile}
\end{figure}

The LHeC is situated within the Geneva basin, a sub-basin of the large molassic plateau (Fig.~\ref{fig:swiss-geology}).
The molasse is a weak sedimentary rock which formed from the erosion of the Alps. It comprises of alternating layers of marls and sandstones (and formations of intermediate compositions), which show a high variety of strength parameters\cite{molasse}. The molasse is overlaid by the Quaternary glacial moraines.
A simplified geological profile of the LHC is shown in Fig.~\ref{fig:LHC profile}.
Although placed mainly within the molasse plateau, one sector of the LHC is situated in the Jura limestone.

The physical positioning of the LHeC has been developed based on the assumption that the maximum underground volume should be placed within the molasse rock and should avoid as much as possible any known geological faults or environmentally sensitive areas. Stable and dry, the molasse is considered a suitable rock type for Tunnel Boring Machines (TBM) excavation.  In comparison, CERN has experienced significant issues with the underground construction of sector 3-4 in the Jura limestone. There were major issues with water ingress at and behind the tunnel face\cite{karstic}. Another challenging factor for limestone is the presence of karsts. These are formed by chemical weathering of the rock and often they are filled with water and sediment, which can lead to water infiltration and instability of the excavation. 

\begin{figure}
    \centering
    \includegraphics[width=0.8\textwidth]{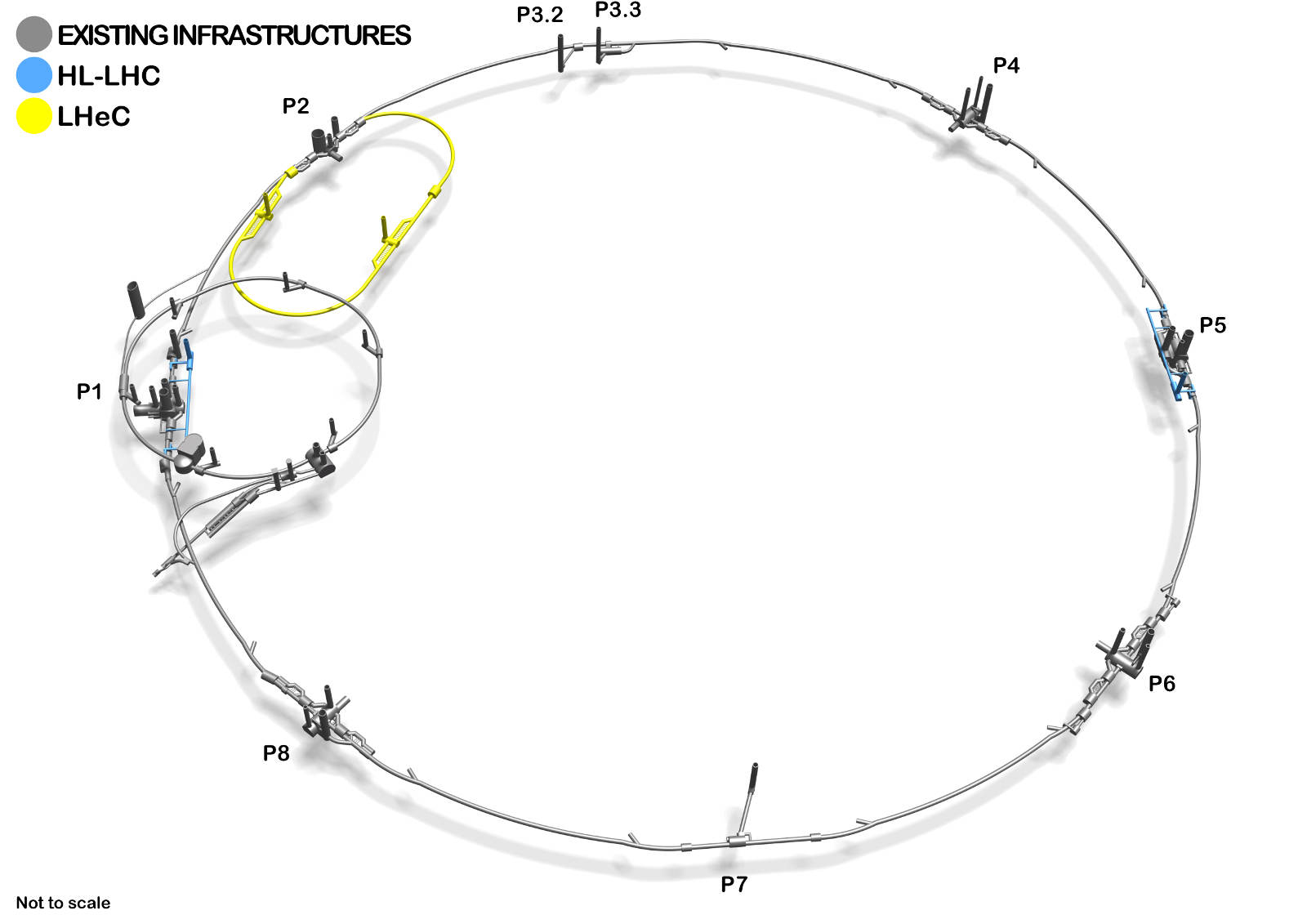}
    \caption{3D Schematic showing proposed underground structures of LHeC (shwon in yellow). The HL-LHC structures are highlighted in blue.}
    \label{fig:3D schematic}
\end{figure}

The ERL will be positioned inside the LHC layout, in order to ensure that new surface facilities are located on existing CERN land. The proposed underground structures for the LHeC with an electron beam energy of 60 GeV are shown in Fig.~\ref{fig:3D schematic}. The LHeC tunnel will be tilted similarly to the LHC at a slope of 1.4\% to follow a suitable layer of molasse rock.

\subsection{Underground infrastructure}
The underground structures proposed for LHeC option 1/3 LHC require a 9\,km long tunnel including two LINACs. The internal diameter of the tunnel is 5.5\,m. 
Parallel to the LINACs, at 10m distance apart, there are the RF galleries, each 1070\,m long.
Waveguides of 1\,m diameter and four connection tunnels are connecting the RF galleries and LINACs.
These structures are listed in Tab.~\ref{tab:List of structures}.
\begin{table}[!bth]
  \centering
  \small
  \begin{tabular}{lccccc}
    \toprule
    Structure & Quantities & Span [m] & 1/3 LHC & & 1/5 LHC \\
    \cmidrule{4-4} \cmidrule{6-6}
     & & & Length [m] & & Length [m] \\
        \midrule
        Machine tunnels       &  - & 5.5  & 9000 & & 5400 \\
        Service caverns       &  2 & 25   &   50 & &   50 \\
        Service shafts        &  2 & 9    &   80 & &   80 \\
        Injection caverns     &  1 & 25   &   50 & &   50 \\
        Dump cavern           &  1 & 16.8 &   90 & &   90 \\
        Junction caverns      &  3 & 16.8 &   20 & &   20 \\
        RF galleries          &  2 & 5.5  & 1070 & &  830 \\
        Waveguide connections & 50 & 1    &   10 & &   10 \\
        Connection tunnels    &  4 & 3    &   10 & &   10 \\
        \bottomrule
    \end{tabular}
    \caption{List of underground structures for LHeC for two different options with 1/3 or 1/5 of the LHC circumference.}
    \label{tab:List of structures}
\end{table}
Two additional caverns, 25\,m wide and 50\,m long are required for cryogenics and technical services.
These are connected to the surface via two \SI{9}{m} diameter shafts, provided with lifts to allow access for equipment and personnel. 
Additional caverns are needed to house injection facilities and a beam dump.
As shown in Tab. \ref{tab:List of structures}, the underground structures proposed for LHeC options 1/5 LHC and 1/3 LHC are similar with the exception of the main tunnel and the RF galleries which have different lengths. 

Shaft locations were chosen such that the surface facilities are located on CERN land.
The scope of work for surface sites is still to be defined.
New facilities are envisaged for housing technical services such as cooling and ventilation, cryogenics and electrical distribution.

\begin{figure}
    \centering
    \includegraphics[width=0.7\textwidth]{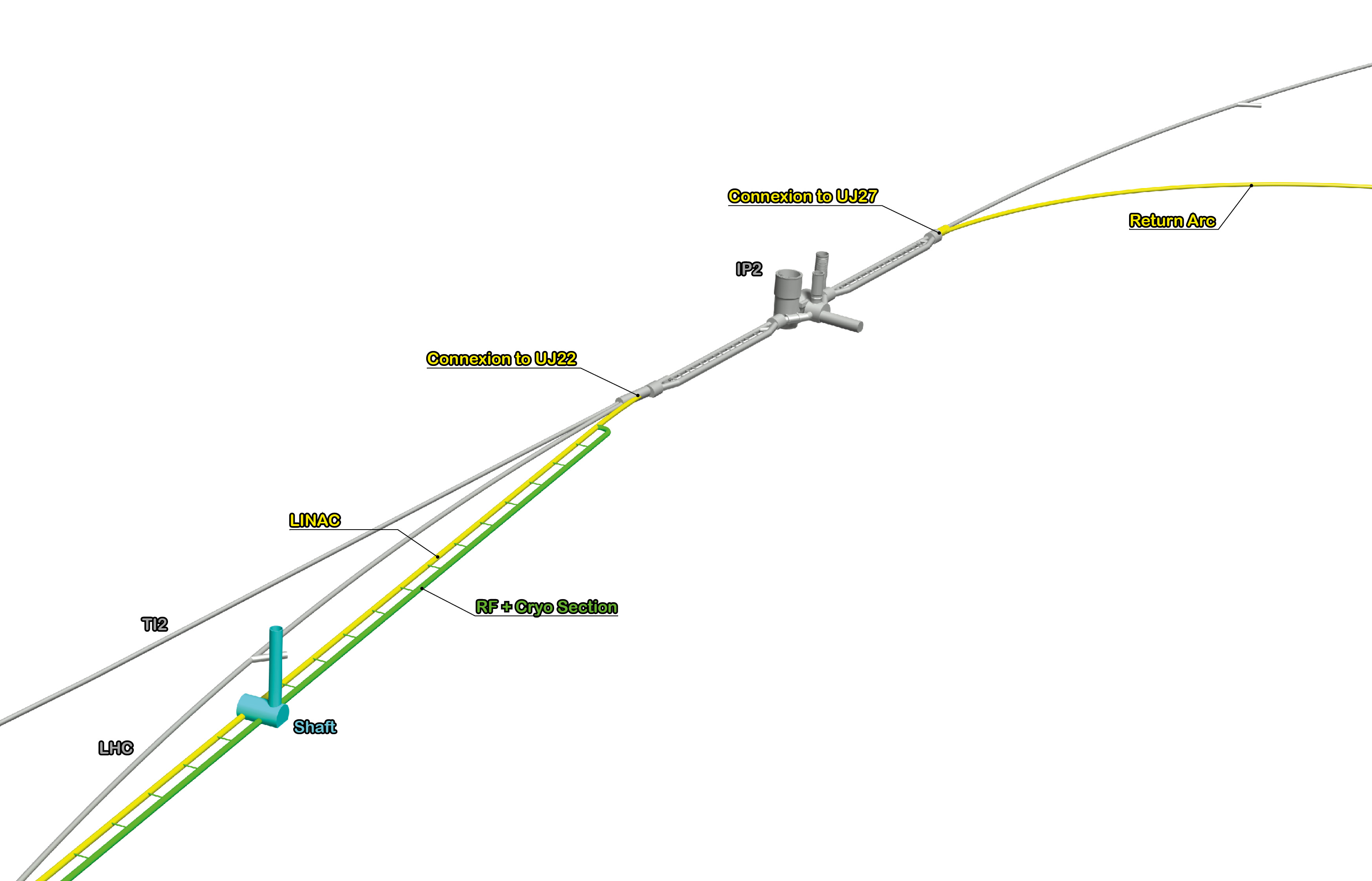}
    \caption{ERL injection area into IP2 and junction cavern}
    \label{fig:LHeC P2 legend}
\end{figure}

In addition to the new structures, the existing LHC infrastructure requires some modifications. To ensure connection between LHC and LHeC tunnels, the junction caverns UJ22 and UJ27 need to be enlarged. Fig.~\ref{fig:LHeC P2 legend} shows the location of these caverns.
Localised parts of the cavern and tunnel lining will be broken out to facilitate the excavation of the new spaces and the new connections, requiring temporary support.

Infrastructure works for LEP were completed in 1989, for which a design lifespan of 50 years was specified. If the LHC infrastructure is to be re-used, refurbishment and maintenance works are needed. 

\subsection{Construction Methods}
A TBM would be utilised for the excavation of the main tunnel to achieve the fastest construction. When ground conditions are good and the geology is consistent, TBMs can be two to four times faster than conventional methods. A double shield TBM could be employed, installing pre-cast segments as primary lining, and injecting grouting behind the lining.

\begin{figure}
    \centering
    \includegraphics[height=0.29\textwidth]{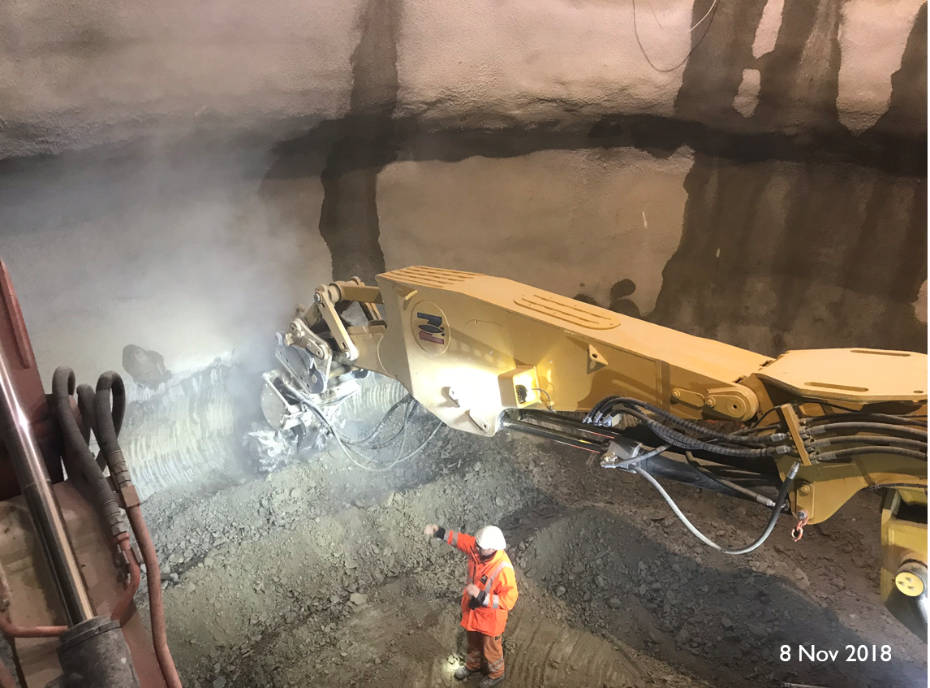}
    \hspace{0.02\textwidth}
    \includegraphics[height=0.29\textwidth]{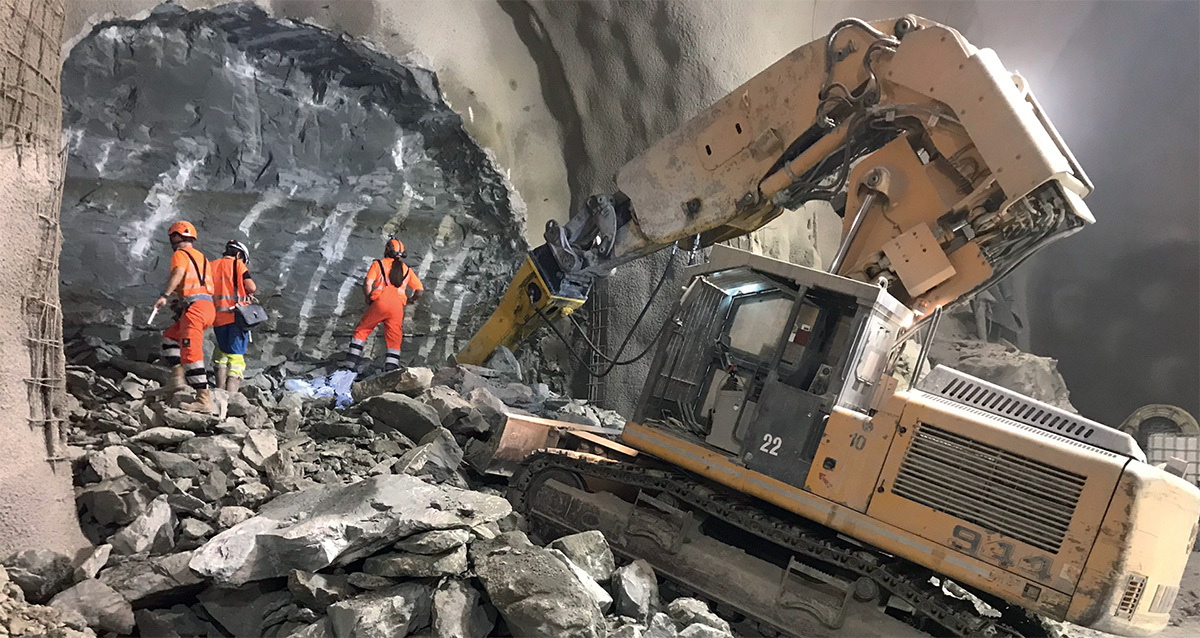}
    \caption{Left: Roadheader being used for shaft excavation at HL-LHC Point 1. Right: Rockbreaker used for new service tunnels excavation at HL-LHC Point~5 (Credit: Z. Arenas).}
    \label{fig:HL-LHC roadheader}
\end{figure}

For the excavation of the shafts, caverns and connection tunnels, typical conventional techniques could be used.
Similar construction methods used during HL-LHC construction can be adopted for LHeC, for example using roadheaders and rockbreakers.
This machinery is illustrated in Fig.~\ref{fig:HL-LHC roadheader}, showing the excavation works at Point~1.
One main constraint that dictated what equipment to be used for the HL-LHC excavation, was the vibration limit.
Considering the sensitivity of the beamline, diesel excavators have been modified and equipped with an electric motor in order to reduce vibrations that could disrupt LHC operation.
Similar equipment could be required for LHeC, if construction works are carried out during operation of the LHC.

Existing boreholes data around IP2 shows that the moraines layer is approximately 25--35\,m deep before reaching the molasse.
Temporary support of the excavation, for example using diaphragm walls is recommended.
Once reaching a stable ground in dry conditions, common excavation methods can be adopted.
The shaft lining will consist of a primary layer of shortcrete with rockbolts and an in-situ reinforced concrete secondary lining, with a waterproofing membrane in between the two linings. 

\subsection{Civil Engineering for FCC-eh}
\begin{figure}
    \centering
    \includegraphics[width=0.7\textwidth]{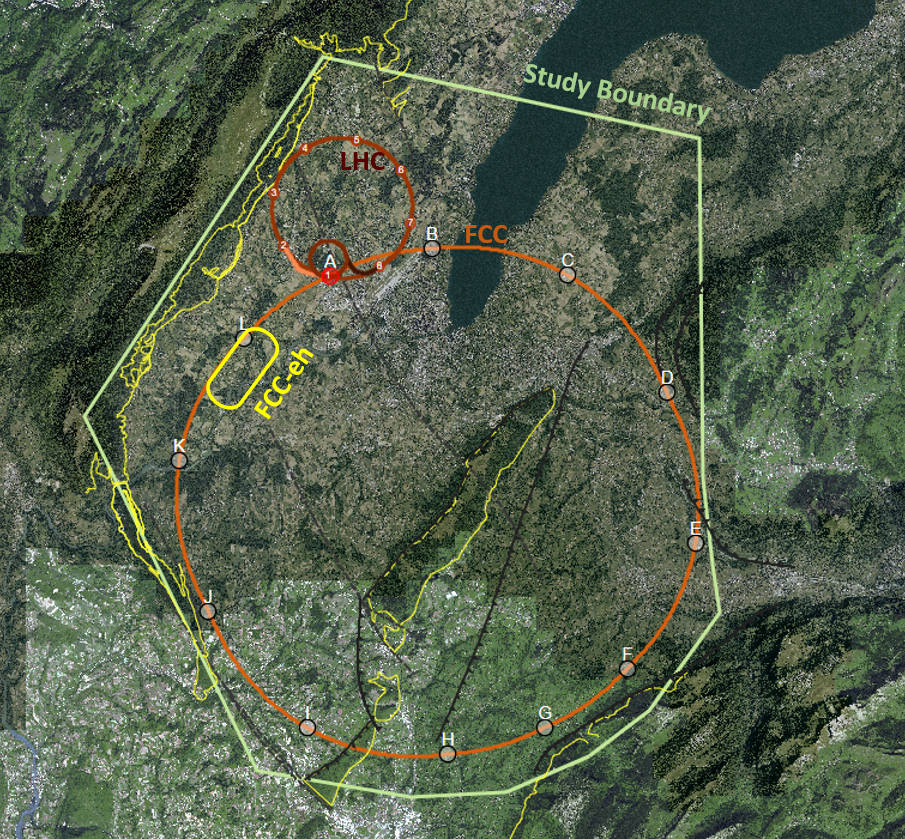}
    \caption{Baseline position and layout for FCC. The lepton ring location is shown at Point L. }
    \label{fig:FCCeh plan view}
\end{figure}

A facility allowing collisions between protons and electrons was considered in the study for the Future Circular Collider (FCC). Fig.~\ref{fig:FCCeh plan view} shows the baseline position for FCC and the lepron ring located at Point L.

\begin{figure}
    \centering
    \includegraphics[width=0.9\textwidth]{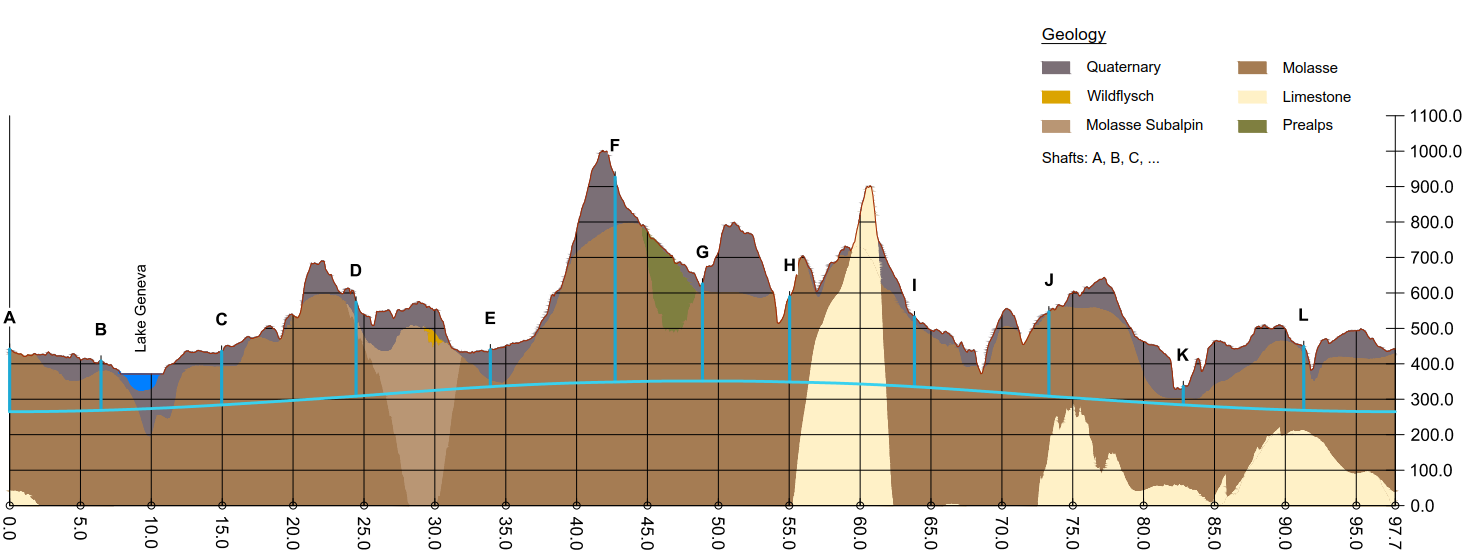}
    \caption{Geological profile along FCC tunnel circumference}
    \label{fig:FCC geological profile}
\end{figure}

During FCC feasibility stage, a bespoke GIS based tool (the Tunnel Optimisation Tool -- TOT) was used to optimise the placement and layout of the FCC ring. The current baseline location was chosen such that the FCC tunnel is placed in preferable geology (90\,\% of the tunnel is in molasse), the depth of the shafts and the overburden is minimised and tunnel under the Geneva Lake goes through the lake bed, passing though reasonably stable ground. More investigations are needed to determine the feasibility of tunnelling under the Geneva Lake. The baseline position also allows connections to the LHC. Fig.~\ref{fig:FCC geological profile} shows the geological profile of the tunnel in baseline position. TOT was used to evaluate different layouts and positions for the FCC ring and assess the impact on the location of the lepton ring. The candidate locations for the $eh$ IR were the experimental points A, B, G and L. Point L was selected because it provides good geological conditions, being fully housed in the molasse layer at a depth of around 180\,m. In comparison, Point G is much deeper, Point A is challenging due to proximity of the LHC and Point B is located in a congested urban area. Similarly to LHeC, the lepton ring will be located inside the FCC ring, in this instance to avoid the Jura limestone. The entire FCC-eh infrastructure is located in the molasse. 

The geological data captured within the TOT tool was collected from various sources including previous underground projects at CERN, the French Bureau de Recherches Géologiques et Minières (BRGM), and existing geological maps and boreholes for geothermal and petroleum exploration. The data was processed to produce rock-head maps and to create the geological layers. No ground investigations have been conducted specifically for the FCC project~\cite{Benedikt:2018csr}. In order to validate its baseline alignment and determine the geotechnical parameters required for the detailed design, site investigation campaigns will need to be carried out. Some boreholes exist in the region where the tunnel for the lepton ring will be built, reducing the uncertainty of the ground conditions. However, further ground investigations are needed in order to verify the boundary between geological layers. The geological features of interest in this region are the Allondon Fault and possible zones of poor rock and level of limestone, which should be avoided. 

\begin{figure}
    \centering
    \includegraphics[width=0.7\textwidth]{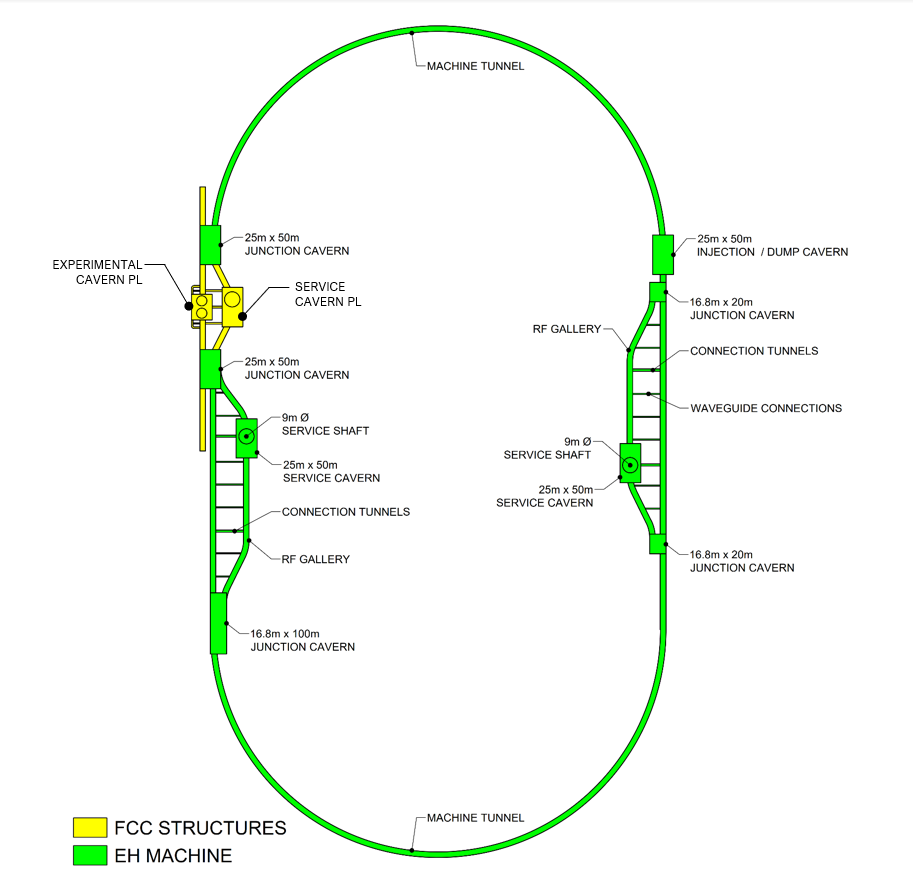}
    \caption{Schematic layout showing the proposed underground structures for FCC-eh}
    \label{fig:FCCeh schematic}
\end{figure}

The IP will be in the experimental cavern at point L, defined as an experimental point for FCC-hh. The layout of the ERL and the underground infrastructure for the FCC-eh is similar to LHeC (see Table ~\ref{tab:List of structures}), with the exception of the shafts which are 180\,m deep. The schematic layout and proposed civil engineering structures are shown in Fig.~\ref{fig:FCCeh schematic}.

The upper excavation for each shaft will be through the moraines. Based on available geological data, the moraines layer should be approximately 30\,m deep. Similar construction methods as described in Section 8.8.3 could be used. For FCC, the alternative technology that has been considered for deep shafts is using a Vertical Shaft Sinking Machine. 
The junction caverns connecting the ERL tunnel with the FCC tunnel must be designed such that they fit the requirements for the new collider and the lepton machine. The junction caverns near Point L will connect three tunnels, the FCC main tunnel, the ERL tunnel and the RF galleries. These caverns will have a 25\,m span and 50\,m length.

For the FCC TBM excavations, different lining designs have been developed corresponding to conditions of the rock~\cite{Benedikt:2018csr}. Good ground conditions have been assumed based on available geological information in the area where the ERL tunnels are positioned and a single-pass pre-cast lining is proposed.

\subsection{Cost estimates}
The cost for underground civil engineering for FCC-eh facility was estimated to be approximately 430\,MCHF. 
The construction programme for the lepton accelerator tunnels, caverns and shafts is currently integrated into the overall FCC construction schedule.

A detailed cost estimate was prepared for a 9\,km ERL located at Point~2 of LHC, using the same unit prices as for FCC. More recently for LHeC, the cost figures were adapted to fit the smaller version, the 5.4\,km racetrack at Point~2 (option 1/5~LHC). The civil engineering costs amount to about 25\,\% of the total project costs. For the 9\,km ERL (1/3~LHC option) the civil engineering was estimated to 386\,MCHF and for a 5.4\,km configuration (1/5~LHC) the costs would be 289\,MCHF. These costs do not include surface structures. Where possible, existing surface infrastructure will be re-used. 

The cost estimates include the fees for preliminary design, approvals and tender documents (12\,\%), site investigations (2\,\%) and contractor’s profit (3\,\%). The accuracy range of the cost estimates at feasibility stage is $\pm$ 30\,\%.

\subsection{Spoil management}
As with all construction projects, environmental aspects play an important role. A detailed study is being conducted at CERN to find a potential re-use for of the spoil that will be generated from the FCC underground excavations. The total amount of spoil calculated is approximately 10 million cubic meters, of which 778,000 cubic metres of spoil would be generated from the lepton ring tunnel construction. 

%% file: erl/erl.chapter.tex
\linenumbers
\lhectitlepage
\lhecinstructions
\subfilestableofcontents

\input{\main/erl/erl.tex}

\biblio

%% file: erl/erl.tex
\chapter{Technology of ERL and PERLE \ourauthor{Alex Bogacz, Walid Kaabi}}
Energy recovery has been proposed in 1965~\cite{Tigner:1965wf}
as a means for efficient colliding beam interactions. It has been 
demonstrated to indeed work at a number of laboratories, at BINP 
Novosibirsk, at Daresbury, at Jefferson Lab and very recently 
at Cornell. The striking technology developments of
high quality superconducting cavities of the last decades 
and the need for 
high collider intensities at economic use of power have now lead
to a wider recognition of  ERL applications as one of the most promising and fundamental developments of energy frontier
accelerator physics. For the LHeC, it had become clear
already with and before the CDR, that ERL was the only way to
achieve high luminosity in $ep$ within the given power limit of
$100$\,MW wall plug for the linac-ring $ep$ 
collider configuration.  For FCC-ee it has been promoted as
an alternative to conventional synchrotron technology
for extending the energy range and increasing the luminosity
in the WW and top mass ranges~\cite{Litvinenko:2019txu}.
A high current electron ERL is designed to reach high lumosities
with proton beam cooling at the EIC~\cite{Litvinenko:2019qlc}.
High energy particle and nuclear physics colliders do await
high current ERLs to become available.

Following the LHeC CDR it became increasingly
clear, much emphasised by the IAC of the LHeC,
 that the basic concept, of high current, multi-turn ERL,
needed a smaller size facility for gaining experience and 
developing the technology. This lead to the development of 
the PERLE concept as was described in a Conceptual Design Report
published in 2017~\cite{Angal-Kalinin:2017iup}. PERLE imported
the main characteristics of the LHeC, the $802$\,MHz frequency and the 3-turn racetrack configuration of two oppositely positioned
linacs. With the $20$\,mA current goal and a $500$\,MeV
beam it represents a first ERL facility in the $10$\,MW power range. 
Its intensity, exceeding that of ELI by 2-3 orders of magnitude,
is the base for novel low energy experiments which are
envisaged to follow the first years of dedicated accelerator 
design study and technology development. PERLE therefore has 
a physics and technical programme which reaches beyond
supporting the LHeC design and possible future operation.
 
An international collaboration has recently been established 
with the aim to realise PERLE in a few stages at IJC Laboratory at Orsay near 
Paris within the next years. The following chapter has two parts, the first describing
challenges and status of ERL developments and the second one 
briefly summarising PERLE. It has to be noted that crucial parts
of PERLE are described above in the LHeC Linac part, such as
the choice of frequency, the electron source, and the  
successful design, construction and test of the first 5-cell
SC Niobium cavity, because  all these characteristics are shared between the LHeC and its development facility.
  
\section{Energy Recovery Linac Technology - Status and Prospects \ourauthor{Chris Tennant}}
%
In instances where high beam power is required, the concept of energy recovery presents an attractive solution. Energy recovering linacs (ERLs) are a class of novel accelerators which are uniquely qualified to meet the demands for a wide variety of applications by borrowing features from traditional architectures to generate linac quality beams with near storage ring efficiency~\cite{tennant.1}. After acceleration through a linac section, the electrons in an ERL are returned 180$^\circ$ out of phase with respect to the radio frequency (RF) accelerating field for energy recovery. The beam deposits energy into cavity fields, which can then accelerate newly injected bunches, thereby effectively canceling the beam loading effects of the accelerated beam. Therefore ERLs can accelerate very high average currents with only modest amounts of RF power. Because the beam is constantly being renewed, it never reaches an equilibrium state. Consequently this provides flexibility to manipulate the phase space and tailor the beam properties for a specific application. Further, since the energy of the decelerated beam is approximately equal to the injection energy, the dump design becomes considerably easier.

\subsection{ERL Applications}
Historically, nearly all ERLs built and operated were used to drive a free-electron laser (FEL). The requirement for high peak current bunches necessitated bunch compression and handling the attendant beam dynamical challenges. In recent years, ERLs have turned from being drivers of light sources toward applications for nuclear physics experiments, Compton backscattering sources and strong electron cooling. Unlike an FEL, these latter use cases require long, high charge bunches with small energy spread. Where once a short bunch length was the key performance metric, now there is a premium on maintaining a small correlated energy spread (with a commensurately long bunch).

\subsection{Challenges}
Energy recovery linacs are not without their own set of challenges. In the following sections a brief survey of some of the most relevant are given. These include collective effects, such as space charge, the multipass beam breakup (BBU) instability, coherent synchrotron radiation (CSR) and the microbunching instability ($\mu$BI), beam dynamic issues such as halo, the interaction of the beam with the RF system and other environmental impedances as well as issues related to common transport lines. 

\subsubsection{Space Charge}
The role of space charge forces (both transverse and longitudinal) often dictate many operational aspects of the machine. Maintaining beam brightness during the low energy injection stage is vitally important. In addition to the low energy, ERL injectors must also preserve beam quality through the merger system that directs the beam to the linac axis. Once injected into the linac, the beam energy at the front end is often still low enough that space charge forces cannot be neglected. Just as important is the longitudinal space charge (LSC) force which manifests itself by an energy spread asymmetry about the linac on-crest phase~\cite{Tennant:2009zz}. The LSC wake acts to accelerate the head of the bunch while decelerating the tail. Operating on the rising part of the waveform leads to a decrease in the correlated energy spread, while accelerating on the falling side leads to an increase. These observations inform where acceleration, and how the longitudinal match, is performed.

\subsubsection{Beam Breakup Instability}
The beam breakup instability is initiated when a beam bunch passes through an RF cavity off-axis, thereby exciting dipole higher-order modes (HOMs). The magnetic field of an excited mode deflects following bunches traveling through the cavity. Depending on the details of the machine optics, the deflection produced by the mode can translate into a transverse displacement at the cavity after recirculation. The recirculated beam induces, in turn, an HOM voltage which depends on the magnitude and direction of the beam displacement. Thus, the recirculated beam completes a feedback loop which can become unstable if the average beam current exceeds the threshold for stability~\cite{PhysRevSTAB.7.054401}. Beam breakup is of particular concern in the design of high average current ERLs utilizing superconducting RF (SRF) technology. If not sufficiently damped by the HOM couplers, dipole modes with quality factors several orders of magnitude higher than in normal conducting cavities can exist, providing a threat for BBU to develop. For single pass ERLs, beam optical suppression techniques – namely, interchanging the horizontal and vertical phase spaces to break the feedback loop between the beam and the offending HOM – are effective at mitigating BBU~\cite{PhysRevSTAB.9.064403}.

\subsubsection{Coherent Synchrotron Radiation}
Coherent synchrotron radiation poses a significant challenge for accelerators utilizing high brightness beams. When a bunch travels along a curved orbit, fields radiated from the tail of the bunch can overtake and interact with the head. Rather than the more conventional class of head-tail instabilities where the tail is affected by the actions of the head, CSR is a tail-head instability. The net result is that the tail loses energy while the head gains energy leading to an undesirable redistribution of particles in the bunch. Because the interaction takes place in a region of dispersion, the energy redistribution is correlated with the transverse positions in the bend plane and can lead to projected emittance growth. While there has been much progress in recent years to undo the effects of CSR in the bend plane with an appropriate choice of beam optics~\cite{PhysRevLett.110.014801}, it is more difficult to undo the gross longitudinal distortion caused by the CSR wake. This is particularly true in applications where the intrinsic energy spread is small and/or where the effect can accumulate over multiple recirculations. One possible mitigation is shielding the CSR wake using an appropriately sized beam pipe~\cite{Fedurin:2011ak}. 

\subsubsection{Microbunching Instability}
Microbunching develops when an initial density modulation, either from shot noise or from the drive laser, is converted to energy modulations through short-range wakefields such as space charge and CSR. The energy modulations are then transformed back to density modulations through the momentum compaction of the lattice. Danger arises when a positive feedback is formed and the initial modulations are enhanced. This phenomenon has been studied extensively, both theoretically and experimentally, in bunch compressor chicanes~\cite{PhysRevSTAB.5.064401,PhysRevSTAB.5.074401}. Only recently has there been a concerted effort to study the microbunching instability in recirculating arcs~\cite{pub.1064226665,PhysRevAccelBeams.19.114401,PhysRevAccelBeams.20.024401}. Because the beam is subject to space charge and/or CSR throughout an ERL, density modulations can be converted to energy modulations. And because of the native momentum compaction of the lattice (in arcs, spreaders/recombiners, chicanes, etc.) those energy modulations may be converted back to density modulations. Therefore, ERLs offer potentially favorable conditions for seeding the microbunching instability, which requires careful attention in the early design stages. 

\subsubsection{Halo}
Halo is defined as the relatively diffuse and potentially irregularly distributed components of beam phase space that can reach large amplitudes. It is of concern because ERL beams are manifestly non-Gaussian and can have beam components of significant intensity beyond the beam core~\cite{erl.9}. Though sampling large amplitudes, halo responds to the external focusing of the accelerator transport system in a predictable manner. It is therefore not always at large spatial amplitude, but will at some locations instead be small in size but strongly divergent. Halo can therefore present itself as \emph{hot spots} in a beam distribution, and thus may be thought of as a lower-intensity, co-propagating beam that is mismatched to the core beam focusing, timing, and energy. Beam loss due to halo scraping is perhaps the major operational challenge for higher-power ERLs. Megawatt-class systems must control losses at unshielded locations to better than 100 parts-per-million to stay within facility radiation envelopes. Scaling to 100\,MW suggests that control must be at the part-per-million level. This has been demonstrated – but only at specific locations within an ERL~\cite{PhysRevLett.111.164801}.

\subsubsection{RF Transients}
Dynamic loading due to incomplete energy recovery is an issue for all ERLs~\cite{erl.11}. In some machines it is due to unintentional errors imposed on the energy recovered beam; for instance, path length errors in large-scale systems. In other machines, such as high power ERL-based FEL drivers, it is done intentionally. In cases where there is the potential for rapid changes in the relative phase of the energy recovered beam, dynamic loading would be difficult to completely control using fast tuners. In such cases adequate headroom in the RF power will have to be designed into the system. These transient beam-loading phenomena are widely unrecognized and/or neglected, however studies have been exploring these issues and the dependence on factors such as the bunch injection pattern~\cite{Setiniyaz:2020hrx}. RF drive requirements for an ERL are often viewed as \emph{minimal}, because in steady-state operation the recovered beam notionally provides RF power for acceleration. It has however been operationally established that RF drive requirements for ERLs are defined not by the steady-state, but rather by beam transients and environmental/design factors such as microphonics~\cite{erl.12}. As a result, the RF power required for stable ERL operation can differ dramatically from naïve expectations.

\subsubsection{Wakefields and Interaction of Beam with Environment}
As with other system architectures intended to handle high-brightness beams, ERLs can be performance-limited by wakefield effects. Not only can beam quality be compromised by interaction of the beam with environmental impedances, there is also significant potential for localized power deposition in beamline components. Resistive wall and RF heating have proven problematic during ERL operation in the past~\cite{Beard:2007zz}. Extrapolation of this experience to higher bunch charges and beam powers leads to serious concern regarding heating effects. Careful analysis and management of system component impedances is required.

\subsubsection{Multi-turn, Common Transport}
Future systems must evolve to utilize multiple turns; it is a natural cost optimization method~\cite{erl.14} and multi-turn systems can in principle provide performance equal to that of 1-pass up/down ERLs at significantly lower cost. In addition to the use of multiple turns, cost control motivates use of extended lengths of common transport, in which both accelerated and recovered passes are handled simultaneously using the same beam lines. This presents unique challenges for high energy ERLs, like LHeC in particular, where energy loss due to synchrotron radiation cannot be ignored and causes an energy mismatch for common transport lines. But addressing these challenges will open up exciting new opportunities for ERLs. In addition to PERLE and LHeC, a multi-turn ERL design from Daresbury illustrates the manner in which the cost/complexity optimum lies toward shorter linacs, more turns, and multiple beams in fewer beam lines~\cite{erl.15}. This also drives the use of multiple turns in stacking rings for hadron cooling; the more turns the cooling beam can be utilized, the lower the current required from the driver ERL, which mitigates challenges associated with source lifetime~\cite{Benson:2018jzt}.

\subsection{ERL Landscape}
\begin{figure}[htb]
\centering
    \includegraphics[width=0.85\textwidth]{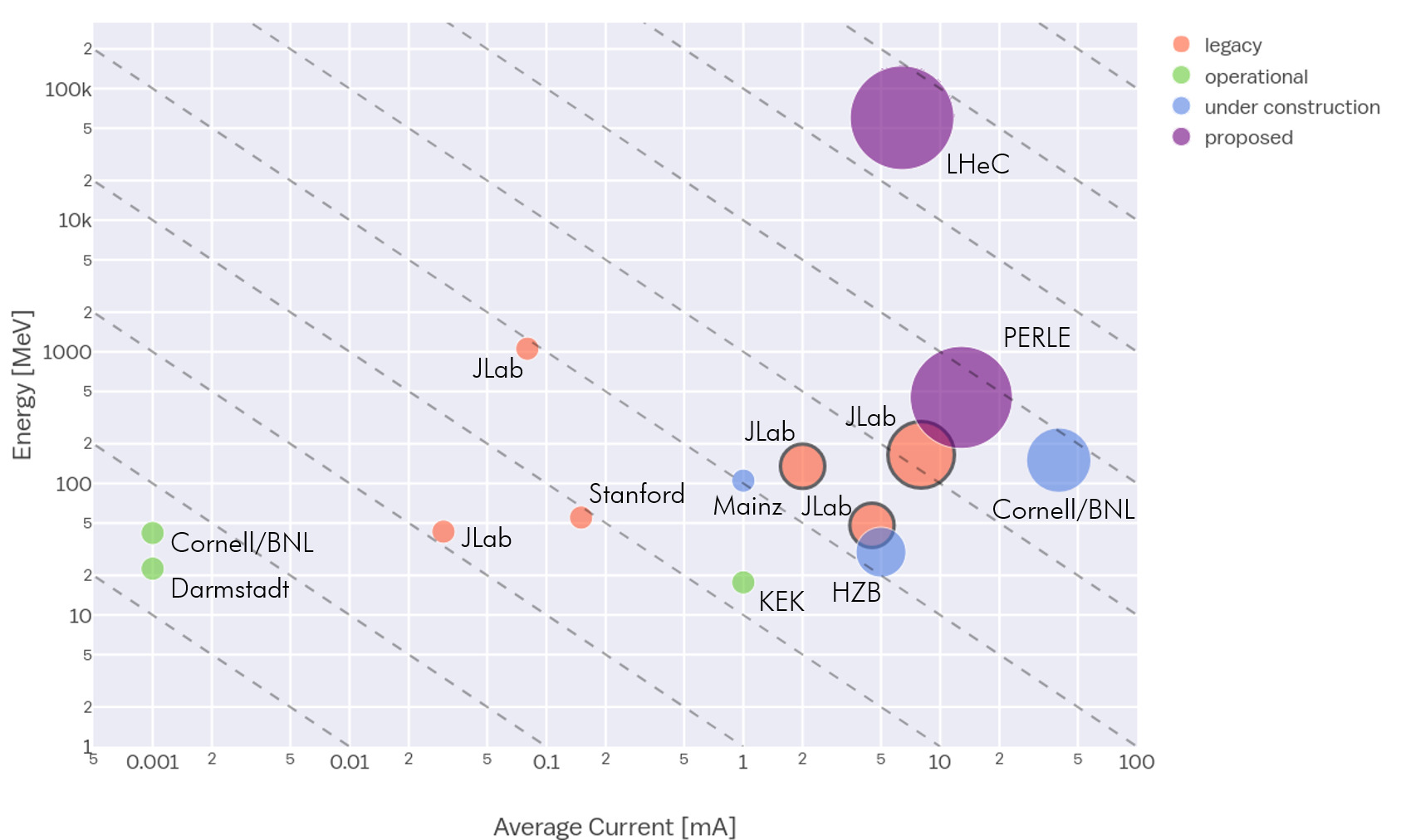}
    \caption{The \emph{ERL landscape}, where data points are restricted to CW, SRF-based ERLs. The dashed lines represent lines of constant beam power – starting from 10\,W in the lower left and going to 10\,GW in the upper right. Note that both axes use a log scale.}
\label{fig:ERL_Development}
\end{figure}

One way to view the current state of ERLs globally is the so-called \emph{ERL landscape} shown in Fig.~\ref{fig:ERL_Development}~\cite{erl.17}. Every data point represents a machine that demonstrated energy recovery and is positioned in (maximum) energy and (average) current parameter space. For clarity, the plot is restricted to continuous-wave (CW), SRF-based ERLs only and includes legacy machines, those under construction and currently in operation as well as the LHeC and PERLE (proposed). The size of the marker is indicative of the charge per bunch while a black line around the marker indicates it was/is a \emph{true ERL}. That is, where the beam power exceeds the installed RF power (they are represented in the plot by the three FEL drivers that were designed, built, commissioned and operated at Jefferson Laboratory).

A cursory look at Fig.~\ref{fig:ERL_Development} illustrates several of the challenges facing the next generation of ERLs. While getting from the current state-of-the-art to the LHeC requires only a modest increase in average current, it requires a significant increase in bunch charge and addressing the consequent collective effects~\cite{erl.18}. Most significantly, however, is the leap in energy from systems that have operated in the 100 MeV range to several tens of GeV. Note that PERLE is strategically positioned to address incremental changes in both average current, bunch charge and energy. As such, it provides a convenient test bed facility to address the issues described previously~\cite{erl.19}. Several ERLs are still in the nascent stages and as they ramp up beam power, will also be valuable in advancing the state-of-the-art. For instance, though it uses a Fixed Field Alternating Gradient (FFAG) arc, the Cornell/Brookhaven ERL Test Accelerator (CBETA) will address multi-turn energy recovery for the first time in an SRF system~\cite{Hoffstaetter:2017jei}. Note that with only minor modifications Jefferson Laboratory’s Continuous Electron Beam Accelerator Facility (CEBAF) could be operated with multi-pass energy recovery at several GeV using common transport with the same topology as LHeC (i.e.\ bisected linacs of equal energy gain with arcs vertically separated by energy using spreaders and recombiners)~\cite{erl.21}.

\section{The ERL Facility PERLE}
%
PERLE is a compact three-pass ERL based on SRF technology,  a new generation machine uniquely covering the $10$\,MW power regime of beam current and energy. 
Its Conceptual Design Report appeared recently~\cite{Angal-Kalinin:2017iup}.
Apart from low energy experiments it could host, thanks to its beam characteristics, PERLE will serve as a hub for the validation of a broad range of accelerator phenomena and the development of ERL technology for future colliders as
introduced above. Particularly, the basic 3-turn configuration, design challenges and beam parameters (see Tab.\,\ref{tab:perleBeamParameters})
 are chosen to enable PERLE as a testbed for the injection line and SRF technology development, as well as multi-turn and high current ERL operation techniques for the Large Hadron electron Collider. While the concept and promise of ERL's has been kick-started by demonstration machines based on existing accelerator technology, PERLE will be the first machine designed from the ground up to use fully optimised ERL-specific designs and hardware. 
 
 The PERLE collaboration involves today CERN, Jefferson Laboratory, STFC-Daresbury (AsTEC with Cockcroft), University of Liverpool, BINP-Novosibirsk and the newly formed Ir\`ene Joliot-Curie Lab (IJCLab) at Orsay. Four of these international partners have been pioneering the development of ERL technology, the other are leading laboratories on SRF technology and accelerator physics. The Orsay Lab, 
 belonging to CNRS and IN2P3, is leading the effort to develop and host PERLE at the Orsay campus
 in close collaboration with the LHeC coordination.
 
 The following PERLE summary focuses on the power challenge, the lattice, site and time schedule. PERLE uses a cryo-module with four 5-cell cavities like the LHeC. The prototype
cavity production and test as well as the design status of the cryo-module are described
in the LHeC linac chapter. Above one also finds a section on the source and injector and as well arc magnets, dipoles of a
3-in-1 design and quadrupoles, which are foreseen to also be used, $mutatis~mutandis$, for PERLE. 
 
\subsection{Configuration} 
 
In the final PERLE configuration, a high current electron beam ($20$\,mA) is accelerated through three passes to the maximum energy ($500$\,MeV) in the superconducting RF CW linear accelerator cryo-cavity units.
The 3-passes up in energy do increase the energy spread and emittance while the major part of the beam power remains. The beam is then sent back through the accelerators again only this time roughly $180$\,degrees off the accelerating RF phase such that the beam is decelerated through the same number of passes and may be sent to a beam dump at the injection energy. Several benefits arise from this configuration: the required RF power (and its capital cost and required electricity) is significantly reduced to that required to establish the cavity field;  the beam power that must be dissipated in the dump is reduced by a large factor, and often the electron beam dump energy can be reduced below the photo-neutron threshold so that activation of the dump region can be reduced or eliminated. 
\subsection{Importance of PERLE towards the LHeC}
PERLE is an important and necessary step accompanying the LHeC realisation. Together with other ERL facilities, CBETA, hopefully
bERLin-Pro and possibly others, it will bridge the gap of power level between the currently reached maximum (CEBAF-ER at $1$\,MW) and the targeted performances of LHeC ($1$\,GW) by exploring a next higher operational power regime of around $10$\,MW. Moreover, sharing the same conceptual design with the LHeC, a racetrack configuration with 3 acceleration and 3 deceleration passes, identical injection line and the same SRF system, as well as the same beam current in the SRF cavities will allow to acquire with PERLE an enormous insight on multiple pass operation and common transport from full energy, before and possibly during LHeC operation.

Up to date, existing SRF systems have demonstrated stability at only a modest fraction ($ \leq 20$\,\%) of the current envisaged for the LHeC. Though threshold currents have been indirectly measured at higher values, there is no direct evidence that multi-pass systems will be sufficiently resistant to BBU at the higher current, nor has the sensitivity of the instability threshold to linac length, dynamic range, and number of passes been directly or systematically measured as yet. PERLE will provide a single datum on linac length, and can directly measure the dependence on the number of passes and the turn-to-turn transfer matrix. 

The dynamic range (which is the ratio of injected/extracted energy to full energy) is a critical design parameter, in as much as it defines the sensitivity of the overall system to magnetic field errors. Errors at full energy drive phase/energy errors that are magnified by adiabatic anti-damping during recovery,  can exceed the dump acceptance should the errors be too large. Thus, the field quality needed is inversely proportional to the ratio of full energy to dump energy: that is, a very high energy machine (or one with very low dump energy) needs very high-quality magnets. For PERLE, the dynamic range is $70:1$ ($7$\,MeV injected and  $490$\,MeV full energy). This implies a need of $\Delta B/B_{dipole} \simeq 0.001$\,\% field flatness (extrapolated from JLAB ERL needs) to recover cleanly enough. This implies a tight constraint on magnet performances and also affects their cost, even when it is the SRF which drives the overall cost of the facility, for LHeC. PERLE has a very large dynamic range and a transport system with considerable symmetry and flexibility. It is therefore a suitable tool to explore this issue and evaluate the cost implications for larger scale systems.

Existing systems have operated at maximum $1$\,MW full beam power.  This is too low 
for a precise understanding and control of beam halo. Extrapolation to $10$\,MW will demand suppression of localised losses to, or below, parts per million. Higher power requires a lower fractional loss. It is not yet well understood how to do this - in particular, collimation systems require a more optimised control of CW losses at rates observed in linacs. PERLE will provide a platform on which the next step in understanding can be taken. Other halo effects may become visible at only the higher CW powers under consideration in PERLE (including Touschek and intra-beam scattering, beam-gas scattering, and ion trapping). These lead to scattering events that adiabatically anti-damp and result in intolerable loss in the back end of the machine, limiting dynamic range. There is no experience with these phenomena, although theoretical studies suggest they are problematic. PERLE will be the first system capable of directly exploring these issues. 

There are many collective effects that have  proven challenging at lower beam powers - including RF heating, resistive wall heating, THz emission heating... - that will have greater impact at both higher power and higher energy. There are at present no operating ERL systems that can study these. PERLE is the only system proposed or under construction that combines sufficient beam power with sufficient operational flexibility to study and test mitigation algorithms and methods. Without PERLE, higher energy/power machines will have very little insight regarding these problems and lack the ability to test solutions. 

Beam quality preservation in the presence of collective effects is a significant challenge for modern machines. In particular, Longitudinal Space Charge (LCS), Coherent Synchrotron Radiation (CSR), and the micro-bunching instability have serious deleterious impact on performance, and can prevent a machine from producing beam consistent with user requirements - or, worse, from being able to operate at significant powers. PERLE probes the regions of parameter space where these effects are observable, and offers an opportunity to benchmark models and explore mitigation methods.

\subsection{PERLE Layout and Beam Parameters}
The PERLE accelerator complex is arranged in a racetrack configuration hosting two cryo-modules (containing 4 five-cell cavities operating at $801.6$\,MHz
frequency), each located in one of two parallel straights completed with a vertical stack of three recirculating arcs on each side. The straights are $10$\,m long and the $180^{\circ}$ arcs are $5.5$\,m across. Additional space is taken by $4$\,m long spreaders/recombiners, including matching sections. 
\begin{figure}
    \centering
    \includegraphics[width=0.75\textwidth]{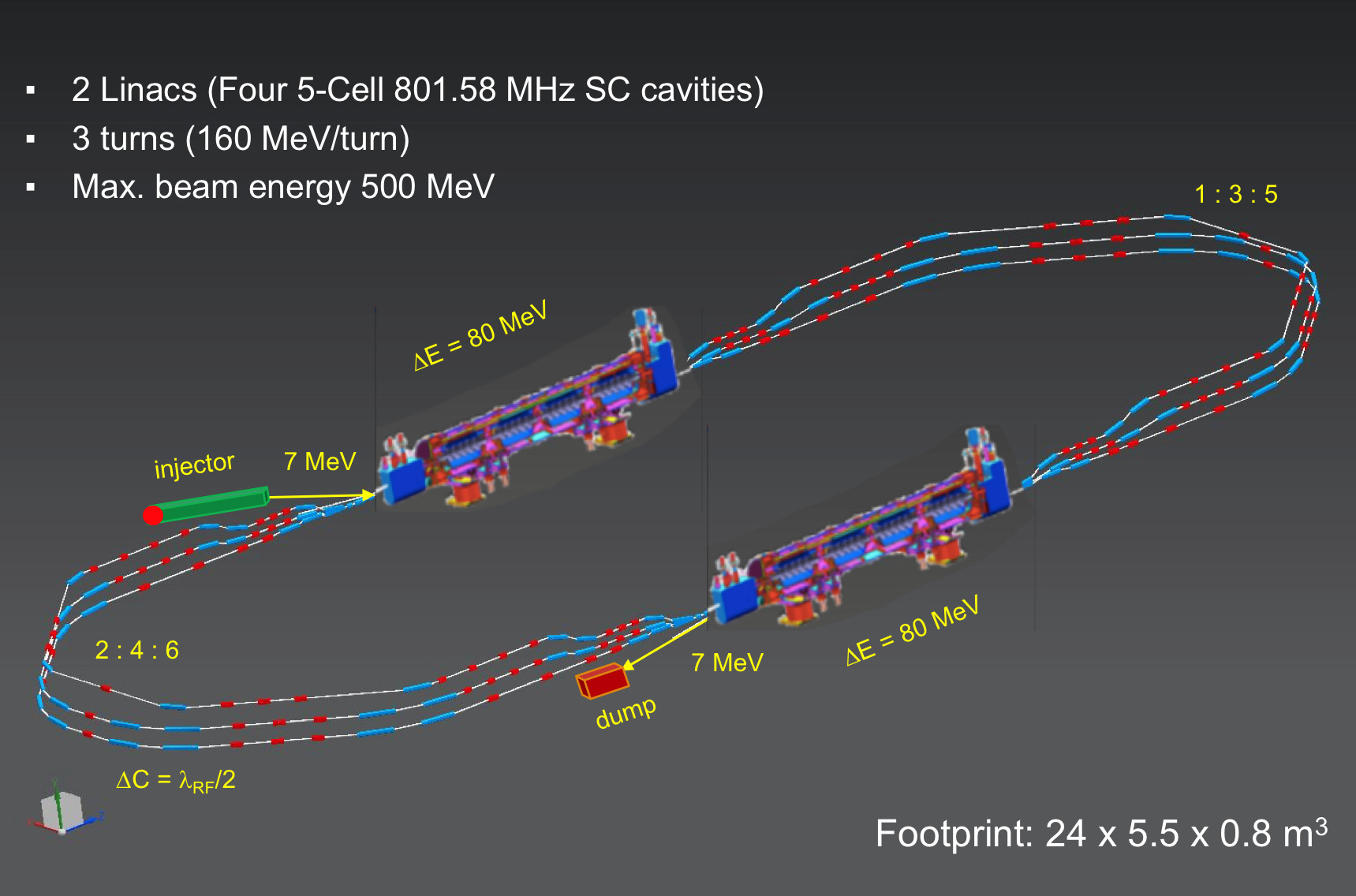}
    \caption{PERLE facility layout featuring two parallel linacs each hosting a cryomodule housing 4 five-cell SC cavities, achieving 500 MeV in three passes, see text.}
    \label{fig:perleLattice}
\end{figure}
As illustrated in Fig.\,\ref{fig:perleLattice}, the  PERLE footprint, excluding shielding and experiments,  is: 
$ 24 \times 5.5 \times 0.8 $\,m$^{3}$, accounting for $40$\,cm vertical separation between arcs. Each of the two cryo-modules provides up to $82$\,MeV energy boost
per path. Therefore, in three turns, a $492$\,MeV energy beam is generated. Adding the initial injection energy of $7$\,MeV yields the total energy of approximately $500$\,MeV. The main beam parameters of PERLE facility are summarised in 
Tab.\,\ref{tab:perleBeamParameters} 
\begin{table}[ht]
  \centering
  \small
  \begin{tabular}{lcc} 
  \toprule
    Target parameter & Unit & Value  \\
   \midrule
Injection energy & MeV & 7 \\
Electron beam energy & MeV & 500 \\
Norm. emittance $\gamma \epsilon_{x,y}$ & mm$\cdot$mrad & 6 \\
Average beam current & mA & 20 \\
Bunch charge & pC & 500 \\
Bunch length	& mm & 3 \\
Bunch spacing & ns & 25 \\
RF frequency & MHz & 801.6 \\
Duty factor & & CW \\
 \bottomrule
  \end{tabular}
   \caption{Summary of main PERLE beam parameters.}
  \label{tab:perleBeamParameters}
\end{table}

As mentioned in the introduction, the essential PERLE parameters are the same as the LHeC. The frequency choice, emittance, beam current and the time structure are chosen regarding the requirements of the electron-proton collisions in the LHeC.
%
\subsection{PERLE Lattice}
Multi-pass energy recovery in a racetrack topology explicitly requires that both the accelerating and decelerating beams share the individual return arcs 
(Fig.\,\ref{fig:perleLattice}). Therefore, the TWISS functions at the linac ends have to be identical, for both the accelerating and decelerating linac passes converging to the same energy and therefore entering the same arc. 

Injection at $7$\,MeV into the first linac is done through a fixed field injection chicane, with its last magnet (closing the chicane) being placed at the beginning of the linac. It closes the orbit bump at the lowest energy, injection pass, but the magnet (physically located in the linac) will deflect the beam on all subsequent linac passes. In order to close the resulting higher pass bumps, the so-called re-injection chicane is instrumented, by placing two additional bends in front of the last chicane magnet. This way, the re-injection chicane magnets are only visible by the higher pass beams. The spreaders are placed directly after each linac to separate beams of different energies and to route them to the corresponding arcs. The recombiners facilitate just the opposite: merging the beams of different energies into the same trajectory before entering the next linac. 
\begin{figure}
    \centering
    \includegraphics[width=0.75\textwidth]{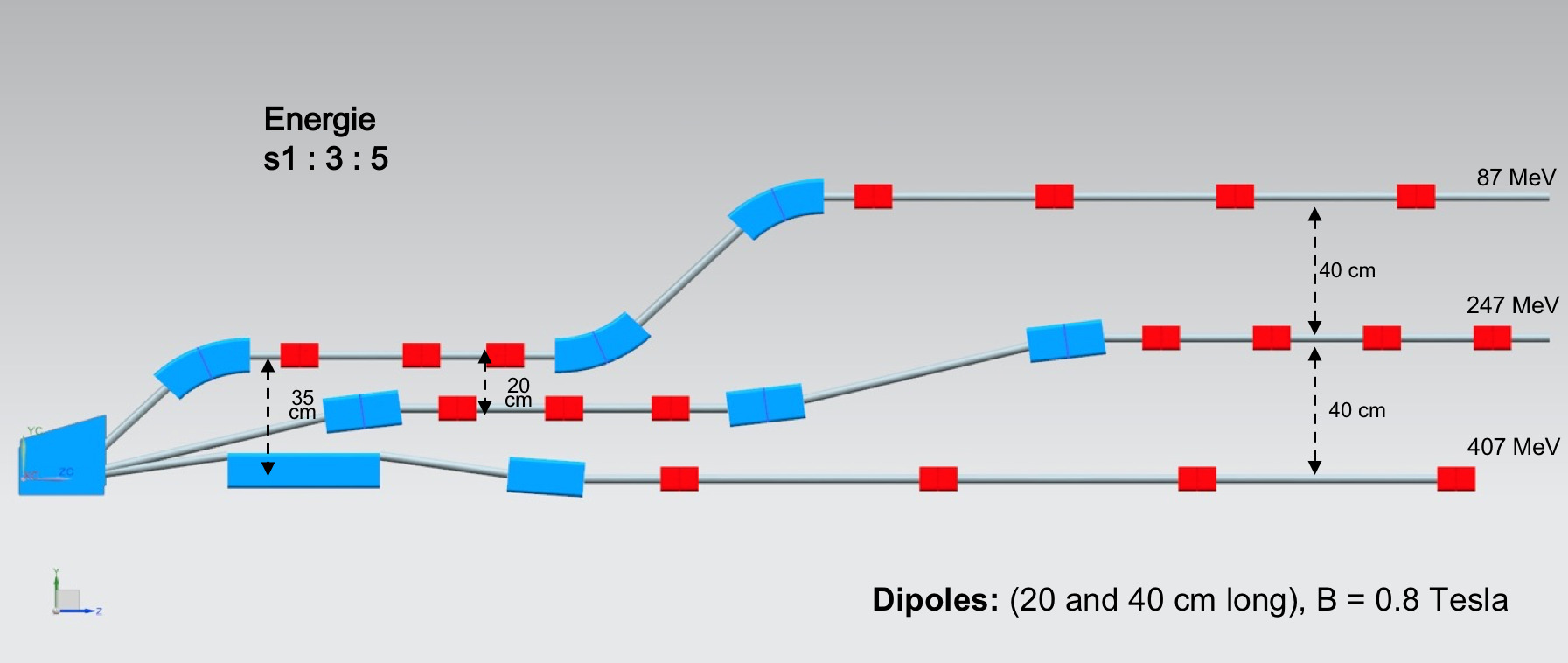}
    \caption{PERLE spreader design and matching to three circulating arcs.}
    \label{fig:perleSpreader}
\end{figure}
The current spreader design (Fig. \ref{fig:perleSpreader}) consists of a vertical bending magnet, common for all three beams, that initiates the separation. The highest energy, at the bottom, is brought back to the horizontal plane with a chicane. The lower energies are captured with a two-step vertical bending. The vertical dispersion introduced by the first step bends is suppressed by the three quadrupoles located appropriately between the two steps. The lowest energy spreader is configured with three curved bends following the common magnet, because of a large bending angle (45$^{\circ}$) the spreader is configured with. This minimises adverse effects of strong edge focusing on dispersion suppression in the spreader. Following the spreader there are four matching quads to bridge the TWISS function between the spreader and the following 180$^{\circ}$ arc (two betas and two alphas). All six, 180$^{\circ}$ horizontal arcs are configured with Flexible Momentum Compaction (FMC) optics to ease individual adjustment of M56 in each arc (needed for the longitudinal phase-space reshaping, essential for operation with energy recovery). The lower energy arcs (1, 2, 3) are composed of four 45.6\,cm long curved $45^{\circ}$ bends and of a series of quadrupoles (two triplets and one singlet), while the higher arcs (4, 5, 6) use double length, 91.2 cm long, curved bends. The usage of curved bends is dictated by a large bending angle (45$^{\circ}$). If rectangular bends were used, their edge focusing would have caused significant imbalance of focusing, which in turn, would have had adverse effect on the overall arc optics. Another reason for using curved bends is to eliminate the problem of magnet sagitta, which would be especially significant for longer, 91.2 cm, bends. Each arc is followed by a matching section and a recombiner (both mirror symmetric to previously described spreader and matching segments). As required in case of identical linacs, the resulting arc features a mirror symmetric optics (identical betas and sign reversed alphas at the arc ends). 

The presented arc optics with modular functionality facilitates momentum compaction management (isochronicity), as well as orthogonal tunability for both beta functions and dispersion. The path-length of each arc is chosen to be an integer number of RF wavelengths except for the highest energy pass, arc 6, whose length is longer by half of the RF wavelength to shift the RF phase from accelerating to decelerating, switching to the energy recovery mode. 
\subsection{The Site}
The IJCLab Orsay intends to host PERLE. The footprint of this facility occupies a rectangle of $24 \times 5.5$\,m$^{2}$. This area should be enclosed by shielding at a sufficient distance to allow passage and maintenance operations. We estimate the required passage and half thickness of the accelerator component to $2$\,m. A concrete shielding is assumed here to stop photons and neutrons produced by halo electrons. A more detailed study of the radiation generated by the impinging electron will be necessary at a following stage. An increase of the shielding required could be alleviated by the use of denser materials. 

The PERLE operation at the design beam parameters 
(Tab.\,\ref{tab:perleBeamParameters})
 required an in-depth study of the machine failure scenario to estimate the power left in the machine during operation after beam losses and how to handle and control it. The study aimed at looking if the PERLE facility will be classified as INB (Infrastructure Nucleaire de Base) or not, with respect to the French radioprotection and nuclear safety rules. This conclusion is crucial for the decision of hosting PERLE at Orsay as such INB facilities require heavy regulation procedures and a very high investment to fulfil the requirements and ensure the safety provisions to be implemented. The outcome of the study had concluded  that PERLE shall not be considered as INB, even if the beam parameters are quite demanding, because for several failure scenarios the energy of the beam is brought back to the injection energy and safely dumped, in a few tenths of micro seconds thanks to the recovery mode. For other scenarios,  hard interlocks and the machine safety system are fast enough to manage the situations. The complete report of this study has been delivered by the IRSD team at Orsay. 

Besides the central area required for machine implementation, space needs to be allocated for the auxiliary systems (power converters for magnets, septa and kickers, RF power, Water cooling, Cryogenics, Electron source, Dump). One has also to consider sufficient space for experiments that may use the PERLE beam. These have been
sketched in the PERLE CDR~\cite{Angal-Kalinin:2017iup}. 
As a rough estimate one would need to triple the area of the accelerator itself to accommodate all services, with shielding included. The building that is foreseen to host this version of PERLE is a former experimental hall (Super ACO). 
It is equipped with cranes and electricity. The ground of the building is made of concrete slabs with variable ground resistance. More than half of the hall area has a sufficient resistance to allow the installation PERLE. Being next to the tunnel of the old Orsay Linac and close to the \emph{Iglooo}, where new accelerators are being installed currently, the building is partially shielded and some equipment (water-cooling circuits, electrical transformer) can be shared with the other machines. The building gives the possibility to install the RF source and the power supplies at a different level than the accelerator. An existing control room that overlooks the experimental hall may be used for PERLE. Since all the accelerators installed nearby are based on warm technology, a cryogenic plant will be built. All the needed support for infrastructure could be assured by the CPER program. Altogether, this appears to be a well suitable place which has the great advantage to be available.
\subsection{Building PERLE in Stages}
\begin{figure}[th]
    \centering
    \includegraphics[width=0.55\textwidth]{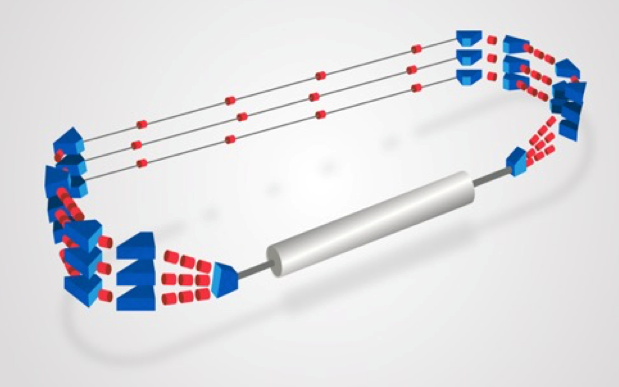}
    \caption{PERLE-Phase 1 layout featuring a single Linac in the first straight and beam line in the second straight, achieving 250 MeV in three passes.}
    \label{fig:perlePhaseOneLattice}
\end{figure}

The PERLE realisation starts with a design and prototyping phase that ends with the PERLE TDR. This phase will include the design, the simulations and the test of the main component prototypes allowing to define the technical choices and needs prior to the construction phases. The PERLE configuration (cf. Fig. \ref{fig:perleLattice}) entails the possibility to construct PERLE in subsequent stages. Three phases of construction, commissioning and exploitation are foreseen to achieve the final configuration. Briefly these are characterised as follows:
\begin{itemize}
\item	$\bf{Phase~0}$: Installation of the injection line with a beam dump at its end:
The injection line includes the DC gun, the load lock photocathode system, solenoids, buncher, booster, merger and required beam instrumentations to qualify the generated beam. 
The commissioning of the injection line will require the installation of cryogenics, RF power source, power supplies for the optics, photocathode laser, beam dump, control-command, vacuum systems, site shielding, safety control system, fluids, etc. Many of these installations must be already sized according to the final configuration of PERLE. 
\item	$\bf{Phase~1}$: 250 MeV Version of PERLE, 
see~Fig.\,\ref{fig:perlePhaseOneLattice}: 
Installation of a single linac in the first straight and installation of beam pipe and complete return arcs. The switchyards have to be chosen according to the beam energy at each end (energy acceptance ratio: 1:2:3 for the spreader and combiner). This version of the race track is connected to the injection line built in phase 0, via the merger.
This particular staging is determined by the existence of the SPL cryomodule at CERN (see the discussion in Chapter\,10) which will permit a rather rapid realisation of a 250 MeV machine likely still using the ALICE gun which resides already at IJCLab. This will permit
first tests with beam of the various SRF components, to explore the multi-turn ERL operation and to gain essential operation experience.  
\item $\bf{Phase~2}$: 500 MeV version of PERLE: 
This phase is for the realisation of PERLE at its design parameters, as a  10\,MW power machine which requires the nominal electron current, i.e. the upgraded electron gun and the completion of the production of a dedicated further cryomodule. Also, a second spreader and recombiner at the required acceptance ratio need to be installed on both sides of the second cryomodule. 
\end{itemize}
The PERLE collaboration is currently developing a detailed time schedule for the project at its different phases, which depends on
the suitability to use the SPL cryomodule adapted for PERLE and the availability of further essential components. In relation to
the LHeC it may be noticed that PERLE offers to gain the
experience one would have to gain with LHeC, years before it starts. LHeC cannot begin before early thirties while
PERLE will operate in the twenties. 

%
%
%
%
\subsection{Concluding Remark}
Currently the focus of the planning for PERLE is on the development of ERL as a means for high
power, large energy accelerator design, technology and realisation. PERLE has a considerable 
potential for low energy particle and nuclear physics too. Its intensity is orders of magnitude
higher than that of ELI. This opens a huge field of physics and industrial applications
for a user facility once the machine has been understood and operates close to 
its design in a reliable manner. With recent increased interest in energy recovery
technology applications at LHeC, but also FCC and EIC, PERLE may become an
important cornerstone for future high energy and nuclear physics. The re-use of 
power is a \emph{per~se} green technology which is an example as to how science
may react to the low power requirements of our time.
%

%% file: detector/detector.tex
\linenumbers
\lhectitlepage
\lhecinstructions
\subfilestableofcontents

\chapter{Experimentation at the LHeC  \ourauthor{ Paul Newman, Peter Kostka, Max Klein}}
\label{chap:detector}

\section{Introduction \ourauthor{Alessandro Polini,  Paul Newman}}

The LHeC Conceptual Design Report~\cite{AbelleiraFernandez:2012cc}
contained a very detailed description of a core detector concept for the
LHeC. At the time of writing, the target luminosity was of order
$10^{33}$\,cm$^{-2}$s$^{-1}$ and, whilst evidence was building,
the Higgs boson had yet to be discovered. 
A detector design based on established technologies either in use by the 
LHC General Purpose Detectors, ATLAS and CMS, or being 
developed for their upgrades was 
found to be adequate to realise the physics priorities of the project at 
the time and could comply with the $ep$ machine constraints at an 
affordable cost, provided the angular acceptance was sufficient 
(nominally to within $1^\circ$ of the beamline). A salient feature of 
experimentation at the LHeC, as compared to the LHC,  is the complete
absence of pile-up which can be estimated~\footnote{
The pile-up is given as the number of events per bunch crossing, 
which is obtained from the instantaneous luminosity, $L=10^{34}$\,cm$^{-2}$s$^{-1}$,
the total cross section, $\sigma_{tot} \simeq \sigma(\gamma p) \cdot \Phi_{\gamma}$,
and the bunch distance of $25$\,ns. The total 
photo-production cross section,
with a minimum $Q^2 = (M_e \cdot y)^2/(1-y)$,
is estimated as $220~(260)$\,$\mu$b at the LHeC (FCC-he)
using the parameterisation given in Ref.\,\cite{Donnachie:1992ny}.
Here  $y$ is the inelasticity variable and $M_e$ the electron mass.
 The photon flux factor
in the Weizs\"{a}cker-Williams approximation is calculated as $\Phi_{\gamma} =1.03~(1.25)$ for
$W=\sqrt{ys} > 1$\,GeV. The hadronic final state at
very small scattering angles, $\theta_h \leq 0.7^{\circ}$ or $|\eta| \geq 5$ , is not reaching the central detector acceptance such that
at the LHeC   $W_{min}$ would be larger, about $10$\,GeV, which reduces the
flux to about $0.6$. A conservative estimate is to use $W > 1$\,GeV.
This translates to an estimated pile-up of 0.06 at the LHeC and 0.09 at the FCC-eh which  favourably compares with 
an estimated pile-up of
150 at HL-LHC.}
 to be around $0.1$ in $ep$ at the LHeC
as compared to $\simeq 150$ in $pp$ at HL-LHC. Similarly, there is a
reduced level of radiation in $ep$, by orders of magnitude lower than in $pp$,
 which enables to also consider novel technologies that are
less radiation hard than conventional ones, with 
HV CMOS Silicon detectors as an example.

This chapter provides a short overview of a partially revised LHeC detector design,
with more detail on those aspects which have developed significantly
since the 2012 version (notably the central tracking). 
To a large extent, the considerations in the CDR are still valid and are taken
forward here. However, this update also  profits from the evolution of
the design in the subsequent years, 
the updated and
long term physics priorities with the higher achievable luminosities.
It also introduces new technologies where they are becoming available.
In more detail, the major considerations which motivate an update of the
detector with respect to the 2012 baseline are:

\begin{itemize}


\item The increased luminosity and the confirmation of a Higgs boson
discovery at a mass of around 125\,GeV opens the
opportunity for the LHeC to provide a set of precision measurements of
the Higgs properties, in particular, percent-level measurements of several of
its couplings. The possibility of obtaining world-leading measurements
of couplings to beauty and charm place a heavy emphasis on the inner
tracking and vertexing. The tracking region has therefore been extended
radially with an also increased segmentation. 
The requirement to maximise the acceptance for Higgs decays
places an even heavier requirement on angular coverage than was the case
in 2012, with forward tracking and vertexing being of particular
importance.

\item The fast development of detector technologies and related
infrastructure in some areas necessitates a fresh look at the optimum
choices. Most notably, silicon detector technologies have advanced
rapidly in response to both commercial and particle physics
requirements. The low material budget, potential high granularity,
and cost-effectiveness offered by monolithic active pixel sensor (MAPS)
solutions such as HV-CMOS are particularly attractive and can reasonably
be assumed to be in wide use in future particle physics collider
detector contexts.


\item The long term, high energy hadron collider physics
program, including FCC and possibilities in Asia, as well as the
ultimate use of the LHC 
for two more decades, require the precise, independent, and comprehensive 
measurements to determine PDFs, over a wider
range of $x$ and $Q^2$ than has previously been possible. The
implication for the LHeC is a need to further improve and extend the
detector acceptance and overall performance.

\item Options in 
which the $ep$ centre-of-mass energy is increased, at HE LHC or FCC-eh,
require a further reinforcement of the detector design in the forward
(outgoing proton) direction, increasing the overall size of the
detector. In particular, the calorimeter depth scales logarithmically with $E_p$
so as to fully contain particles from very high energy forward-going hadronic showers
and to allow for precise measurements of actual and missing energy. Using such
scaling considerations, the LHeC design has been applied also to the post LHC
hadron beam configurations.

\end{itemize}

The design described in the following addresses the points above.
The updated detector requirements point in 
the tracking region to the need for higher
spatial resolution, improved precision in momentum
measurements and enhanced primary and secondary vertexing capabilities. 
The most significant change compared with 2012 is 
therefore a more ambitious tracking detector design.
The detector must also provide accurate measurements of hadronic jets
and missing transverse energy, as well as isolated electrons and photons.
As an option compared to the CDR, the 
liquid argon (LAr) choice for the main electromagnetic barrel calorimeter sampling material 
is here changed to a scintillator-based solution. Both options are subsequently compared, and as
expected the long term stability and resolution performance favour a LAr calorimeter
while the modularity and installation aspects are easier solved with a warm crystal calorimeter. 

Both the overall event kinematics (much larger proton than electron beam
energy) and the specific acceptance requirements for the key Higgs
production process imply an asymmetric design with enhanced hadronic
final state detection capabilities in the forward  direction where the deposited hadronic and electromagnetic
energies are much higher than in the backward (the electron beam) direction, see
Fig.\,\ref{fig:kinisomax} in Sect.\,\ref{sect:DISdata}.

A dipole magnet bends the electron beam
into head-on collision with the colliding proton beam and after the
interaction point a further dipole with opposite polarity separates the
orbits of the electron and proton beam. These weak bending dipoles are
placed outside of the tracker and electromagnetic calorimeter regions. The total length 
is $10$\,m or $2/3 L*$ as explained in the IR section.
The resulting synchrotron radiation fan has to be given free space and
the beam pipe geometry is designed specifically to accommodate it. The
residual synchrotron radiation background poses a constraint to 
the inner detector components.

The 2012 and 2020 versions of the LHeC 
detector design are both realisable in terms of technology readiness.
It has been a goal of this conceptual design to study the feasibility, performance
and integration of the detector, which will eventually be designed by a future  
$ep/eA$ experiment collaboration. The two designs, albeit being still similar,
 can be considered as two example solutions to the LHeC 
requirements with differences in where the emphasis is placed in terms of 
performance and cost. The current design 
is performed using the DD4hep~\cite{frank_markus_2018_1464634} software framework.

 \section{Overview of Main Detector Elements}

\begin{figure}[!htbp]
  \centering
\includegraphics[width=0.95\textwidth]{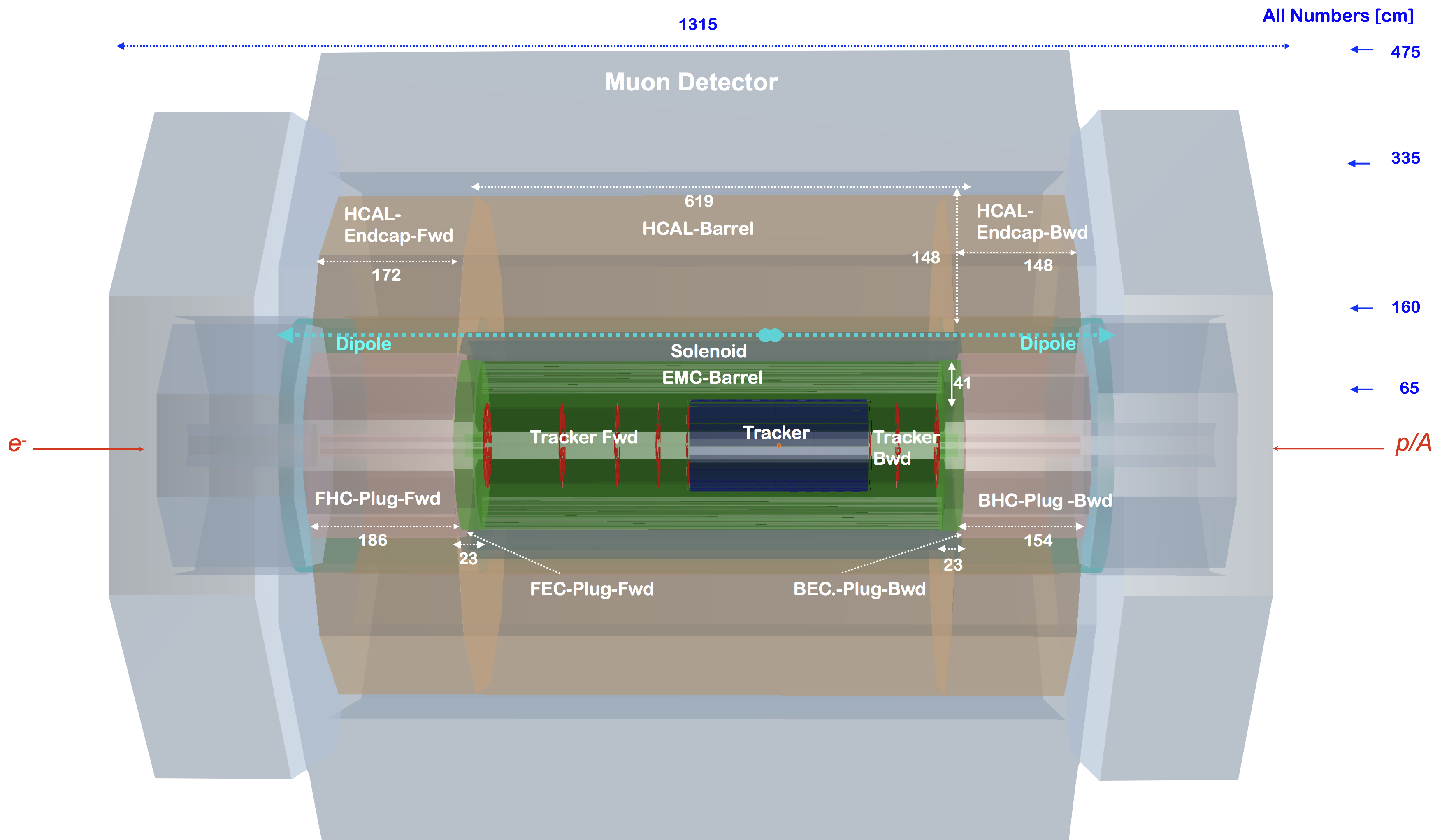}
\caption{Side view of the updated baseline LHeC detector concept, providing
an overview of the main detector components and their locations. The detector dimensions
are about $13$\,m length and $9$\,m diameter. 
The
central detector is complemented with forward ($p,~n$) and backward ($e,~\gamma$)
spectrometers mainly for diffractive physics and for photo-production and luminosity
measurements, respectively.  See text
for details.}
\label{FIG:LHeC-maindetector2}
\end{figure}

A side projection overview of the current detector design is shown in
Fig.~\ref{FIG:LHeC-maindetector2}, illustrating the main detector
components. The overall size remains compact by recent standards, with 
overall dimensions of approximately $13 \,{\mathrm m}$ in length and $9
\,{\mathrm m}$ in diameter, small compared with ATLAS ($45 \times 25 \,
{\mathrm m}^2$) and even CMS ($21 \times 15 \, {\mathrm m}^2$). The inner
silicon tracker contains a central barrel component (`Tracker'), with
additional disks in the forward and backward directions (`Tracker Fwd'
and `Tracker Bwd', respectively). It is surrounded at larger radii by
the Electromagnetic Barrel (`EMC-Barrel') and in the forward and
backward directions by the electromagnetic forward and backward plug
calorimeters (`FEC-Plug-Fwd' and `BEC-Plug-Bwd', respectively). The
solenoid magnet is placed at radii immediately outside the EMC-Barrel,
and is housed in a cryostat, which it shares with the weak dipole
magnet that ensures head-on collisions.  It is the dipole and cost considerations
which suggest to place the solenoid there instead of surrounding the 
HCAL which in terms of performance surely would have been preferable.

The Hadronic-Barrel calorimeter
(HCAL-Barrel) is located at radii beyond the solenoid and dipole, whilst
the forward and backward hadronic plug detectors (FHC-Plug-Fwd and
BHC-Plug-Bwd, respectively) lie beyond their electromagnetic
counterparts in the longitudinal coordinate.
The Muon Detector forms a near-hermetic envelope around all other parts
of the main detector. 
It uses similar technologies to those employed
by ATLAS, at much smaller surface, see below.  
 
\begin{figure}[!ht]
  \centering
  \includegraphics[width=0.82\textwidth]{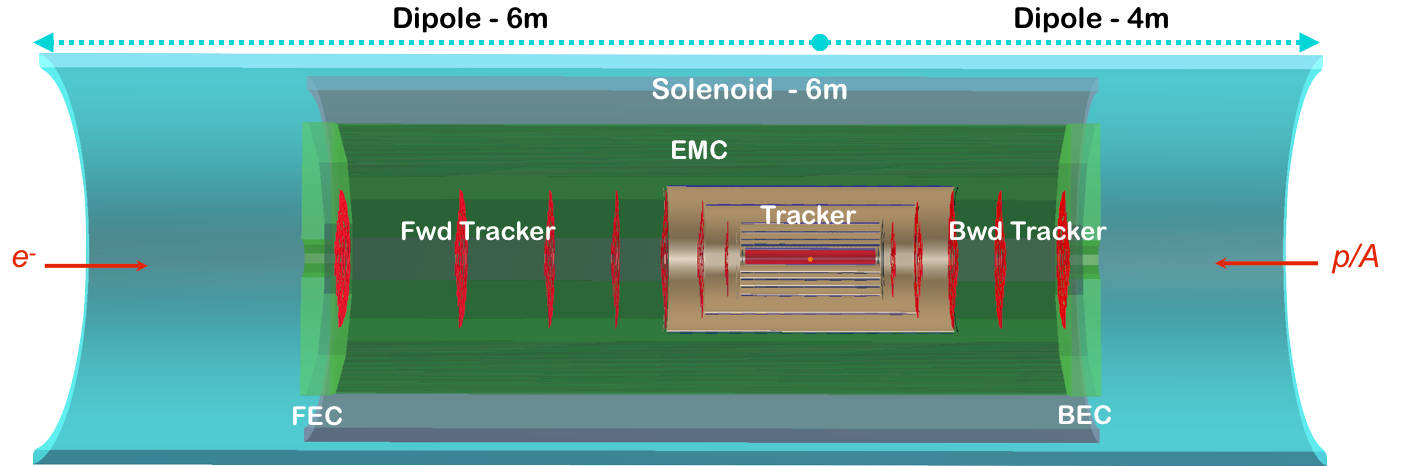}
\caption{Side projection of the central part of the LHeC
detector, illustrating also the solenoid and electron-beam-steering 
dipoles. See text for further details.}
\label{FIG:dipoles_solenoid_tracker_ECAL_cut}
\end{figure}

A magnified view of the inner part of the detector, including the magnet
elements, is shown in
Fig.~\ref{FIG:dipoles_solenoid_tracker_ECAL_cut}. The solenoid and
steering dipoles enclose the electromagnetic calorimeters and the
tracker setup completely, the steering dipoles extending over the full
$10 \, {\mathrm m}$ length of the inner detector and forward and backward
plugs. If liquid argon is chosen for the sensitive material in the EMC
as in the 2012 design, the EMC will be mounted inside the cryostat,
alongside the solenoid and dipoles. The hadronic calorimeter components
remain outside the cryostat and magnet elements in all circumstances.

%

\begin{figure}[!htbp]
  \centering
  \includegraphics[width=0.7\textwidth]{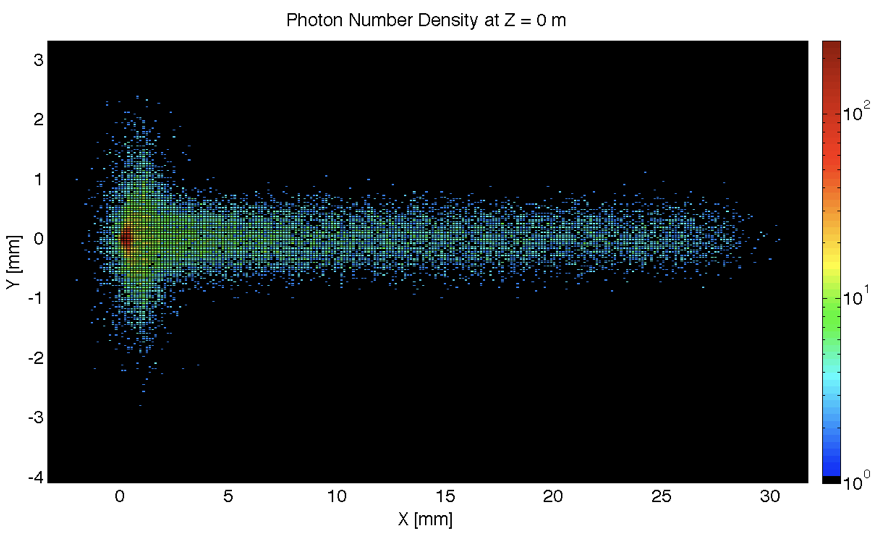}
\caption{The simulation of synchrotron radiation profile at the IP using 
GEANT4~\cite{Agostinelli:2002hh,AbelleiraFernandez:2012cc}.  }
\label{FIG:Synchr-Rad-at-IP}
\end{figure}
\begin{figure}[!htbp]
  \centering
  \includegraphics[width=0.7\textwidth]{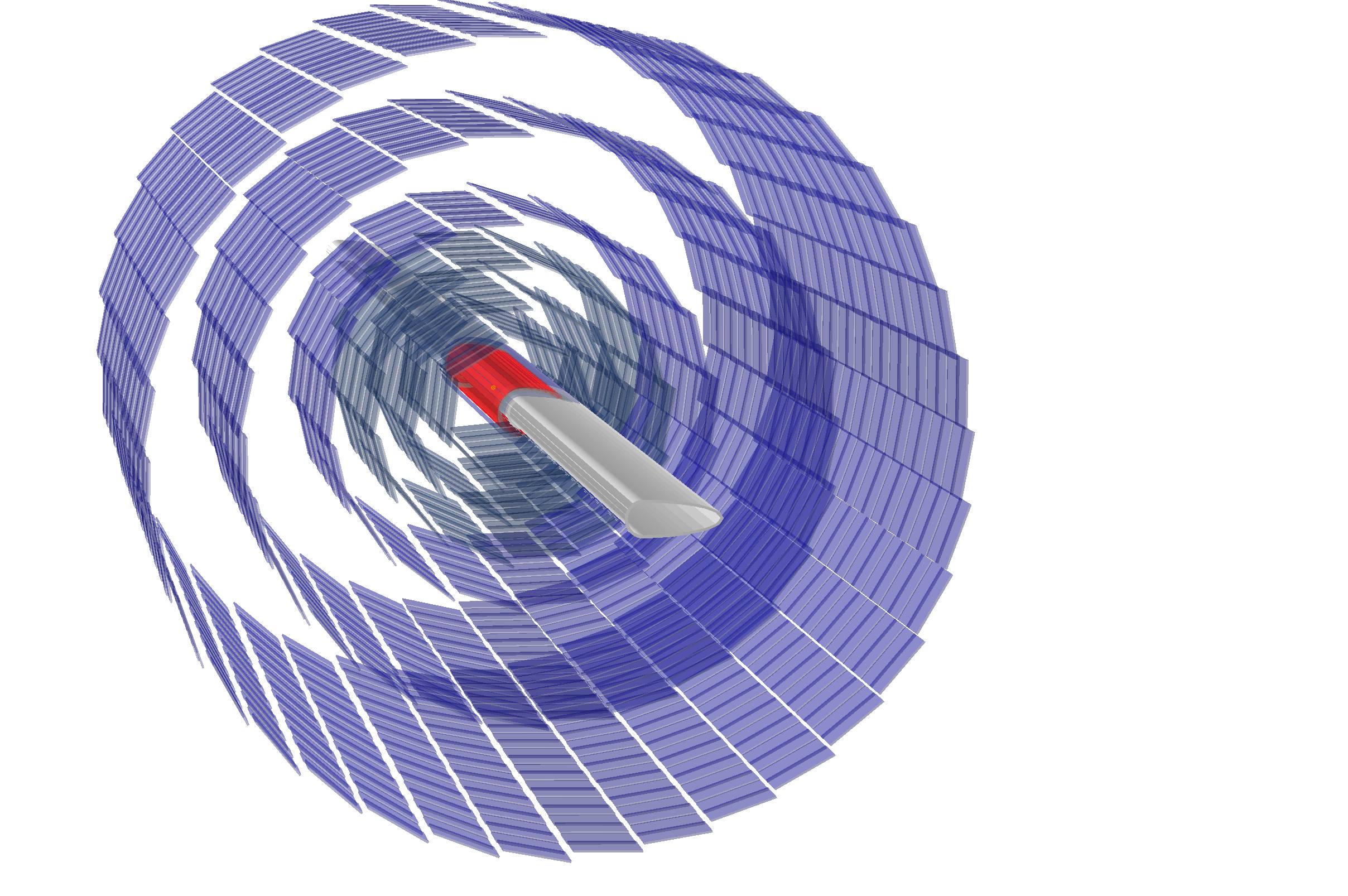}
\caption{End-on view of the arrangement of the inner barrel tracker layers
around the beam pipe. }
\label{FIG:Barrel_Tracker_ell-BP}
\end{figure}

Exploiting the current state of the art, the beam pipe is constructed of
berylium of $2.5$--$3$\,mm thickness. As in the 2012 CDR, the
beam pipe has an asymmetric shape in order to accommodate the
synchrotron radiation fan from the dipole magnets. 
It is thus 2.2\,cm
distant from the interaction region, comparable to the HL-LHC beam pipes of the GPDs,
except in the  direction of the synchrotron fan, where it is increased to 
10.0\,cm, giving rise to an overall 
circular-elliptical profile ( illustration of the profile at IP in 
Fig.~\ref{FIG:Synchr-Rad-at-IP} ). The beam pipe shape has
implications for the design of the inner detector components, as
illustrated in Fig.~\ref{FIG:Barrel_Tracker_ell-BP}. The first layer
of the barrel tracker follows the circular-elliptical beam pipe shape as
closely as possible, with the profiles of subsequent layers reverting to
a circular geometry. 

\section{Inner Tracking \ourauthor{ Peter Kostka}}
\subsection{Overview and Performance}
\begin{figure}[th]
  \centering
 \includegraphics[width=0.99\textwidth]{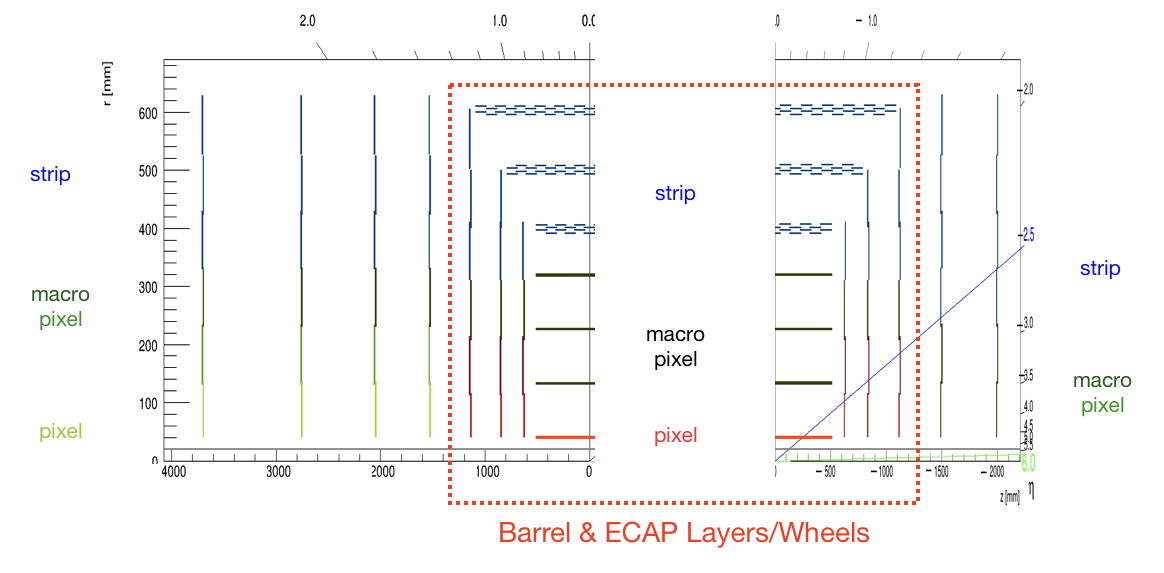}
\caption{Schematic side-view
of the tracker, subdivided into forward and backward parts and including
disks as well as barrel components. The layers/wheels forming the barrel
part are enclosed by the red-dotted box. The innermost pixel layers are
coloured red, the macro-pixel layers are shown in black and the strip
detectors in blue.  For the forward and backward wheels, possibly formed with
separate rings, (outside the
dashed red box), the pixels, macro-pixels and strip detectors are shown
in light green, dark green and blue, respectively.}
\label{FIG:LHeC_tracker_barrel_fwd_bwd_sensors}
\end{figure}

A schematic view of the updated tracking region is shown in
Fig.~\ref{FIG:LHeC_tracker_barrel_fwd_bwd_sensors}. The layouts in the
central, forward and backward directions have been separately optimised
using the tkLayout performance estimation tool for silicon
trackers~\cite{Bianchi:2014mpa}. The result is seven concentric barrel
layers with the innermost layer approximately 3\,cm from
the beam line at its closest distance and with 
approximately equal radial spacing thereafter. The tracker barrel is
supplemented by seven forward wheels and five backward wheels of which
three in each direction comprise the central tracker end-cap and,
respectively, four and two, respectively, are mounted beyond the central tracker
enclosure.

For reasons described in Sect.\,\ref{sec:cmos}, 
HV-CMOS MAPS sensors can be employed, restricting material associated
with the pixel sensors to just 0.1\,mm per layer. The strip detector sensors
have a larger thickness of 0.2\,mm. The preferred active silicon
solutions vary with radial distance from the interaction point,
so as to provide the highest spatial resolution in the layers closest to
the the interaction point. The barrel is formed from one layer of
pixel-wafers, with three layers of macro-pixels between 
10\,cm and 30\,cm radius and a further three layers of
strip-sensors beyond 30\,cm. 
The end-cap wheels and the forward tracker also contain combinations of the three types of sensor, 
whilst the backward tracker consists of macro-pixels and strips only.

\input{\main/detector/tables/LHeC_Tracker_main-properties.tex}

\begin{figure}[th]
  \centering
  \includegraphics[width=0.49\textwidth]{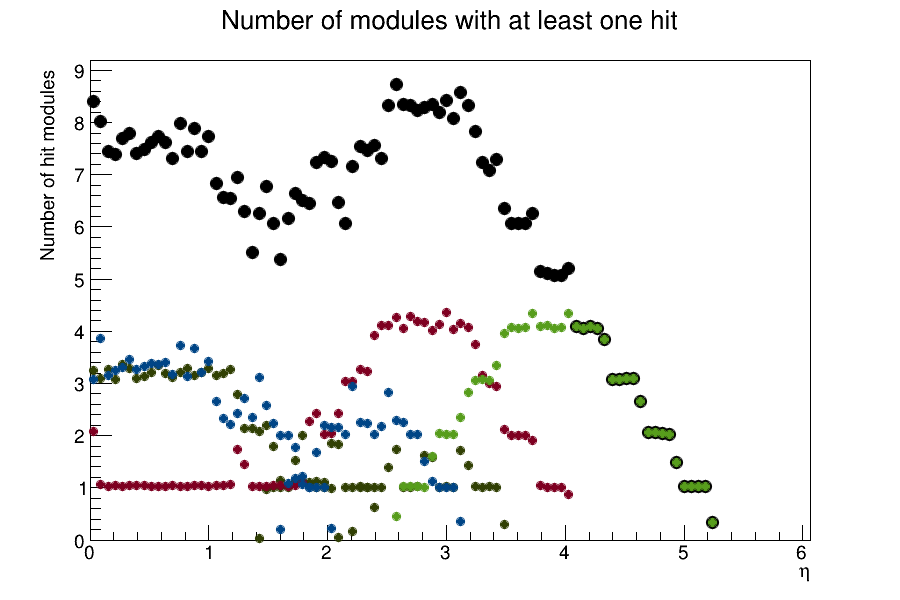}
  \includegraphics[width=0.49\textwidth]{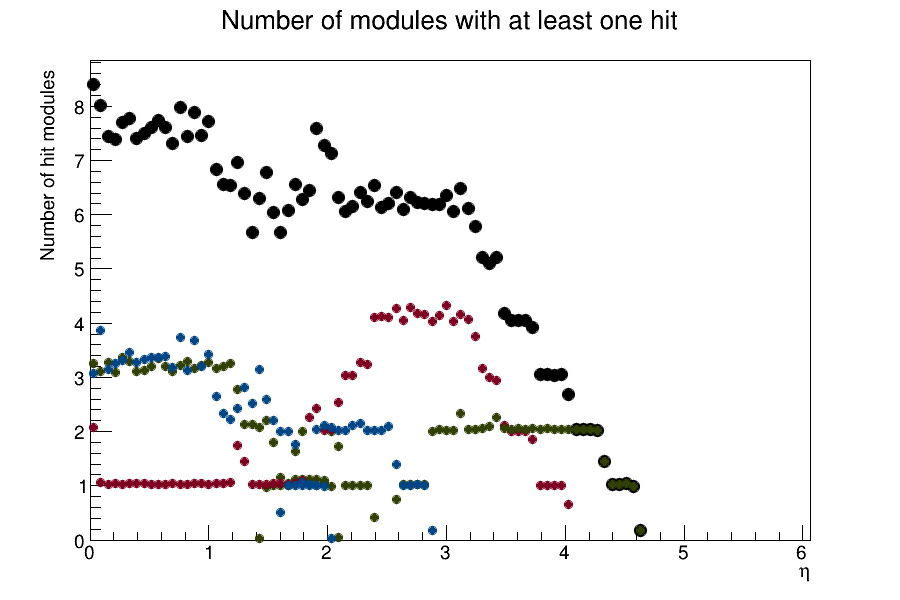}
\caption{Numbers of silicon layers that provide acceptance for charged
particles as a function of absolute value of pseudorapidity in the
forward (left) and backward (right) directions, summed across the
central, forward and backward trackers. The distributions are broken
down according to sensor type, with colour coding of red for pixels,
light or dark green for macro-pixels, blue for strips and black for the
sum.}
\label{FIG:Nof_modules_atleast_1hit}
\end{figure}

Tabs.~\ref{tab:LHeC_Tracker_main-properties_1} and \ref{tab:LHeC_Tracker_main-properties_2} 
summarise the overall
basic properties of the tracker modules, including total numbers of
channels and total area of silicon coverage, as well as spatial
resolutions and material budgets. The inner barrel has a pseudorapidity
coverage $|\eta| < 3.3$ for hits in at least one layer, increasing to
$|\eta| < 4.1$ when the endcaps are also taken into account. The
additional disks beyond the central tracker enclosure extend the
coverage to $\eta = 5.3$ and $\eta = -4.6$ in the forward and backward
directions, respectively. Fig.~\ref{FIG:Nof_modules_atleast_1hit}
illustrates the coverage in more detail, displaying the numbers of layers
that provide acceptance as a function of pseudorapidity in both the
forward and backward directions, also broken down into different sensor
types. Charged particles are sampled in between 5 and 8 layers
throughout the entire range $-3.5 < \eta < 4$, with sampling in at least
two layers provided for $-4.2 < \eta < 5$.

\begin{figure}[th]
  \centering
  \includegraphics[width=0.6\textwidth]{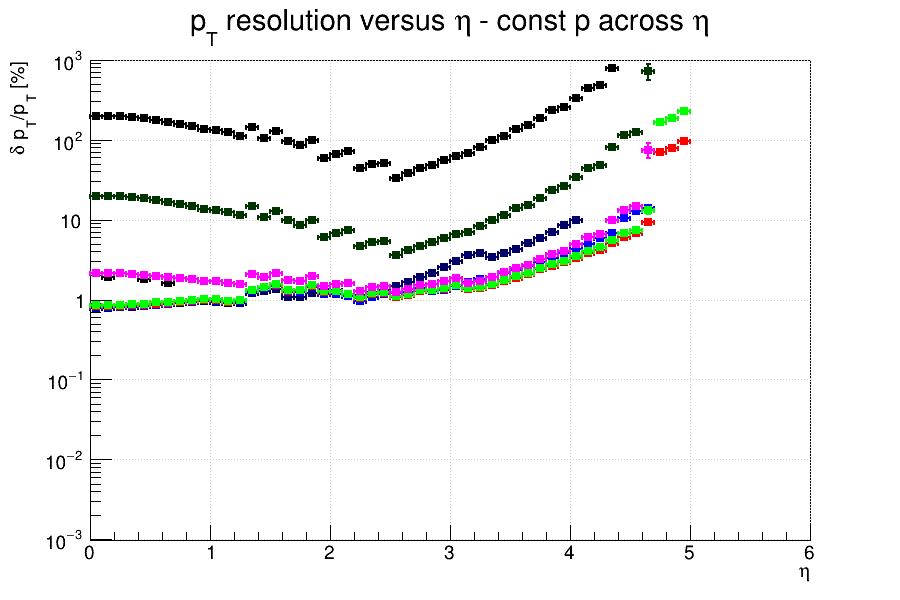}
\caption{Simulated transverse momentum track resolution using all modules in the 
revised LHeC tracking system.
Results are shown in terms of fractional $p_T$ resolution as a
function of pseudorapidity for several constant momenta,
$p = 100$\,MeV (Black, bottom, obscured), $1$\,GeV (Dark Blue, obscured),
$2$\,GeV (Light Blue, obscured),	$5 $\,GeV (Red), $10$\,GeV
(Light Green), $100$\,GeV (Magenta),	$1$\,TeV
(Dark Green) and $10$\,TeV (Black, top).}
\label{FIG:pt-resolution_tracker}
\end{figure}

Spatial
resolutions in the $r-{\phi}$ plane, driven by the sensor pitches, reach
${7.5 \,\mathrm{\mu{m}} }$ for the pixel layers.
The resolutions are
propagated using tkLayout to produce simulated charged particle
transverse momentum resolutions, as shown in
Fig.~\ref{FIG:pt-resolution_tracker}. Both active and passive material
contributions are included, with a ${2.5}$\,mm Be beam pipe
thickness. An excellent resolution ($\delta p_T / p_T$) at the level of
$1 -2 \% $ is achieved over a wide range of pseudorapidity and momentum.  
The precision degrades slowly in the forward
direction, remaining at the sub $10\,{\%}$ level up to very forward
pseudorapidities ${\eta \sim 4.5}$. Central tracks with transverse momenta
up to ${1 \,\mathrm{TeV} }$ are measured with ${10 -20 \,{\%}}$ precision.
Similar results are achieved in the (negative $\eta$) backward direction (not shown). 

\begin{figure}[thbp]
  \centering
  \includegraphics[width=0.47\textwidth]{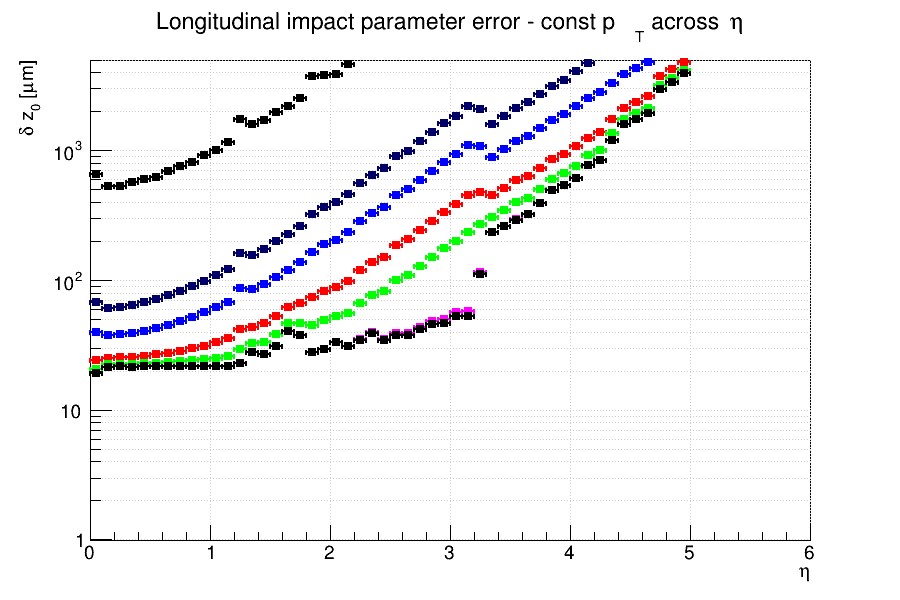}
  \includegraphics[width=0.47\textwidth]{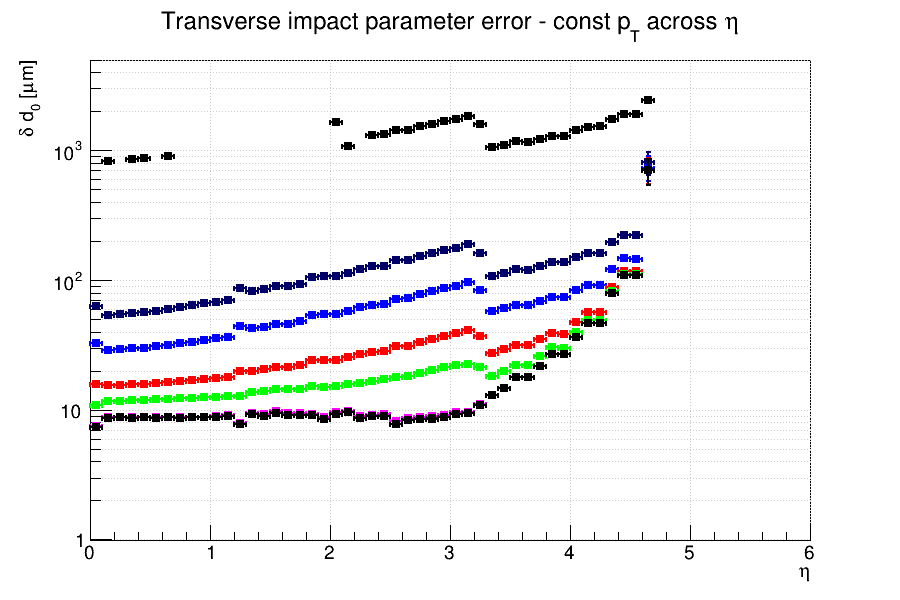}
\caption{Simulated longitudinal (left) and transverse (right) 
impact parameter resolutions using all modules in the 
revised LHeC tracking system. 
Results are shown as a
function of pseudorapidity for several constant momenta, 
$p = 100 \ \mathrm{MeV}$ (Black, top), $1 \ \mathrm{GeV}$ (Dark Blue), 
$2 \ \mathrm{GeV}$ (Light Blue),	$5 \ \mathrm{GeV}$ (Red), $10 \ \mathrm{GeV}$
(Light Green), $100 \ \mathrm{GeV}$ (Magenta, obscureed),	$1 \ \mathrm{TeV}$
(Dark Green, obscured) and $10 \ \mathrm{TeV}$ (Black, bottom).
}
\label{FIG:impact-parameter-resolution_at_const-pT-fwd-tracker} 
\end{figure}

A major requirement of the tracking detectors will be the precise determination of vertex coordinates and 
track impact parameters relative to the primary vertex in order to give the best possible sensitivity 
to secondary vertices from heavy flavour decays, for example for the study of the Higgs in its 
dominant $b \bar{b}$ decay mode. The simulated results for longitudinal and transverse track 
impact parameter resolutions using the full new tracking layout are shown 
in Fig.~\ref{FIG:impact-parameter-resolution_at_const-pT-fwd-tracker}. 
The transverse spatial resolutions are at the level of $10 - 50 \,\mathrm{\mu{m}}$ over a 
wide range of transverse momentum and pseudorapidity, extending well into the forward direction.

\begin{figure}[thbp] 
  \centering 
  \includegraphics[width=0.42\textwidth,trim={20 20 650 10},clip]{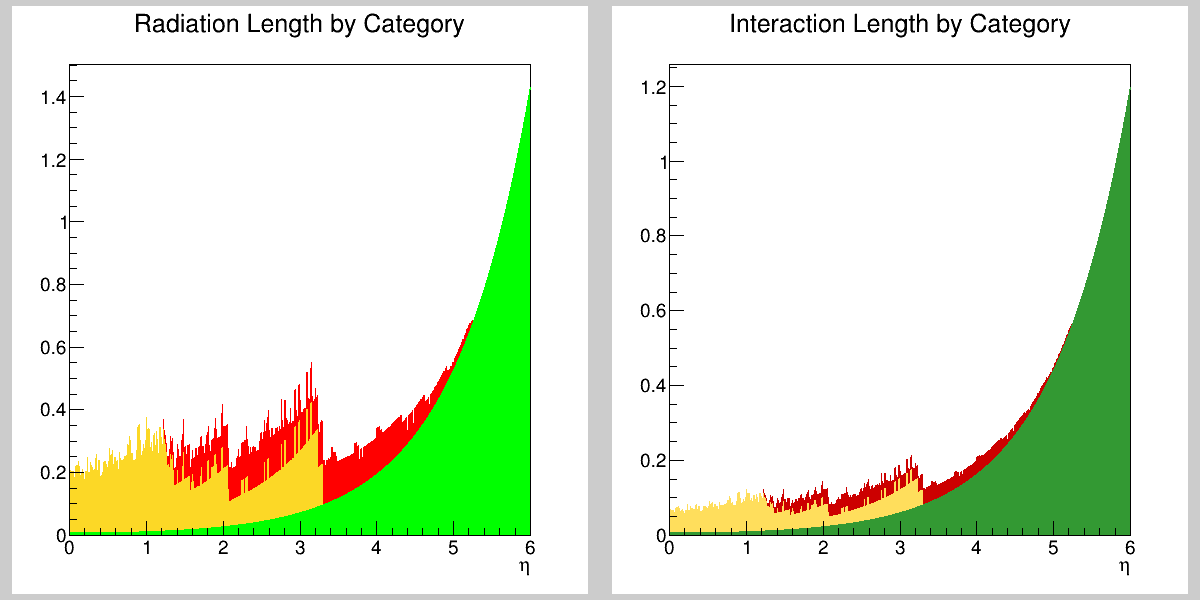}
  \hspace{0.03\textwidth}
  \includegraphics[width=0.42\textwidth,trim={650 20 20 10},clip]{figures/fwd_riDistrCateg002.png}
\caption{Material contributions from the tracking modules as a function
of pseudorapidity. Results are given in terms of radiation lengths
(left) and hadronic interaction lengths (right). The results are broken
down into contributions from barrel modules (yellow) and endcap /
additional disk modules (red) and are compared with the contribution
from the $2.5$\,mm  beam pipe (green).}
\label{FIG:material_tracker} 
\end{figure}

The material budget contributions from the sensors summed across
all layers are given in
Tabs.~\ref{tab:LHeC_Tracker_main-properties_1} and
This is largest for the inner barrel, where it amounts to $7.2\,\%$ of a
radiation length. The sensors in the central tracker endcap and the
forward and backward tracking rings contribute $2.2\, \%$, $6.7\, \%$ and $6.1\, \%$ of a radiation length, respectively.
The material budget simulations, propagated for the full system and including passive contributions, 
are shown in Fig.~\ref{FIG:material_tracker}. The use of thin sensors
keeps the total material to the level of $0.2 - 0.4 X_0$ throughout the
entire tracking region up to $\eta \sim 4.5$. 
At the most forward (and backward) pseudorapidities, particles travel through a 
large effective thickness of material as they pass through the beam pipe; this becomes 
the dominant contribution for $\eta > 3.5$.

\subsection{Silicon Technology Choice}
\label{sec:cmos}
\begin{figure}[!th]
\centering
\includegraphics[width=0.8\textwidth]{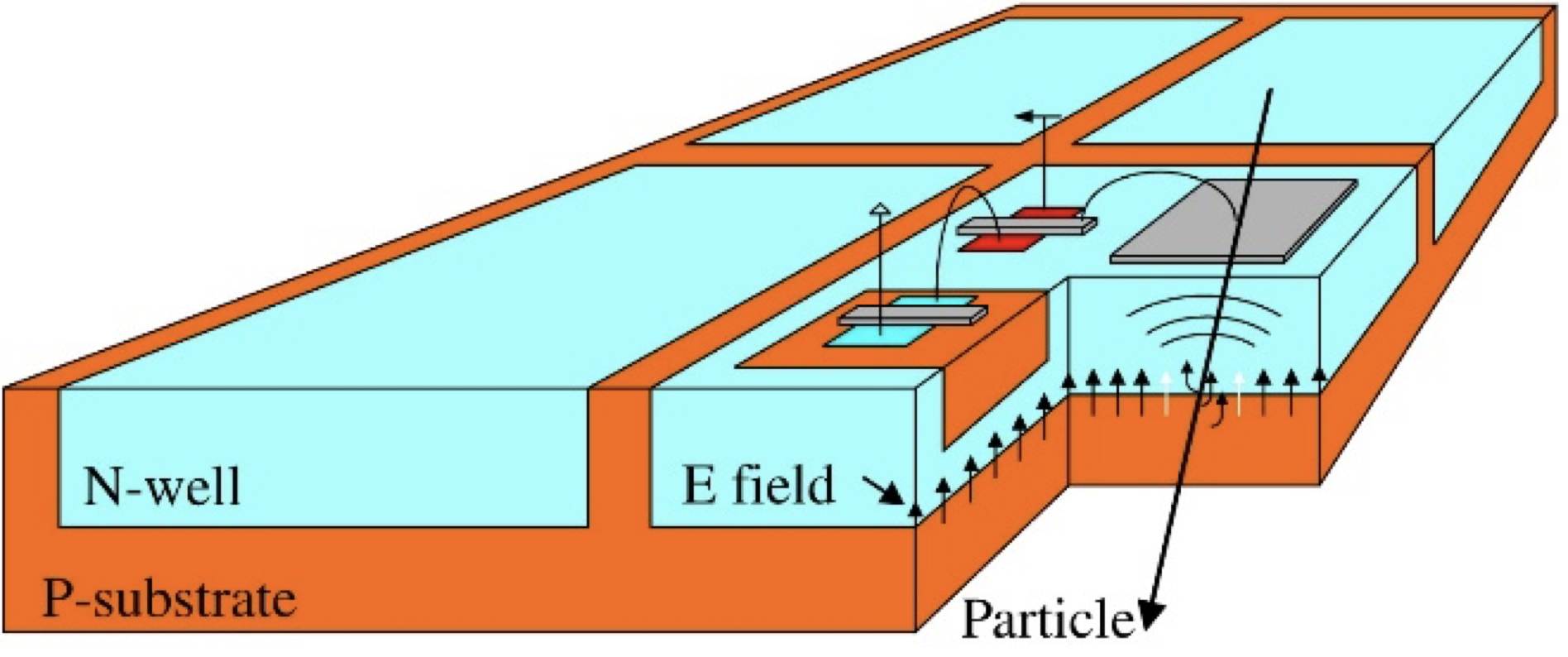}
\caption{Typical sensor cross-section of a DMAPS detector in a HV-CMOS process~\cite{Peric:2007zz}. 
}
\label{FIG:DMAPS_HV-CMOS}
\end{figure}
%
Being developed for several High Luminosity-LHC (HL-LHC) upgrades and the proposed CLIC high-energy 
linear collider we envisage depleted CMOS sensor technology, also known as Depleted Monolithic 
Active Pixel Sensors (DMAPS), to be used as position sensitive detectors in industry standard CMOS processes or 
High Voltage-CMOS (HV-CMOS) processes~\cite{Peric:2007zz}. These sensors are extremely attractive for 
experiments in particle physics as they integrate the sensing element and the readout 
ASIC in a single layer of silicon, which removes the need for interconnection with complex 
and expensive solder bump technology. Depleted CMOS sensors also benefit from faster 
turnaround times and lower production costs when compared to hybrid silicon sensors.
To achieve fast charge collection and high radiation tolerance, DMAPS can be implemented 
following two different approaches known as low fill-factor and large fill-factor. 
Low fill-factor DMAPS benefit from High Resistivity (HR) substrates and thick epitaxial 
layers accessible from large-scale CMOS imaging processes, while large fill-factor 
DMAPS exploit the High Voltage (HV) option developed by commercial CMOS foundries 
for power electronics. Recently, HR wafers have become available in the production line of 
foundries that manufacture HV-CMOS processes, thus DMAPS in HR/HV-CMOS are also possible to 
further improve the performance of the sensor. Today’s most performant DMAPS detectors are 
50$\mathrm{\mu{m}}$ thin and have 50$\mathrm{\mu{m}}$~x~50$\mathrm{\mu{m}}$ cell size 
with integrated mixed analogue and digital readout electronics, 6$ns$ time resolution and 
2x10$^{15} 1\mathrm{MeV}$~neq/cm$^2$ radiation tolerance.
The typical cross-section of a large fill-factor DMAPS in a HV-CMOS process is shown in 
Fig.~\ref{FIG:DMAPS_HV-CMOS}. 

DMAPS in HR/HV-CMOS have been adopted as a world first as the sensor technology of choice 
for the \textbf{Mu3e} experiment at the Paul Scherrer Institute (PSI) in Switzerland~\cite{Blondel:2013ia}. 
 MuPix, the DMAPS detector for Mu3e, implements active pixels that amplify the collected charge in the collecting electrode and peripheral readout electronics that discriminate and process the amplified signals. MuPix10, the first reticle size detector for Mu3e ($\simeq 2 \times 2$\,cm$^2$) and currently in production, features $250 \times 256$ pixels with an $80 \times 80 \mu$m$^2$ pixel size, 11-bit time-stamp, 6-bit Time-over-Threshold (ToT) and continuous readout. Its peripheral readout electronics include readout buffers, a state machine, a Phase Locked Loop (PLL) and Voltage-Controlled Oscillator (VCO), 8/10-bit encoders and 3 serialisers for data transmission with a rate of up to 1.6 Gbit/s. Previous MuPix 
 prototypes have been thinned to $50$\,$\mu$m successfully and tested to achieve a 6\,ns time resolution after time-walk correction~\cite{Schoning:2020zed}.
  ATLASPix, the DMAPS development in HR/HV-CMOS that was originally aimed at providing an alternative sensor technology for the outermost pixel layer of the new ATLAS Inner Tracker (ITk) upgrade, has been tested to have an approximate 150 mW/cm$^2$ power consumption and be radiation tolerant up to $2 \cdot 10^{15}$ 1\,MeV neq/cm$^2$
 fluences\cite{Prathapan:2020mel} 
 DMAPS in HR-CMOS, such as the MALTA development originally aimed at the new ATLAS ITk upgrade as well, have achieved full efficiency after $1 \cdot 10^{15}$ 
 1\,MeV neq/cm$^2$ fluences\cite{Dyndal:2019nxt}. However, further research is still needed to demonstrate reticle size DMAPS in HR-CMOS. Research to further develop DMAPS to meet the extreme requirements of future experiments in particle physics is on-going.

The here presented tracker design of the LHeC utilises pixel detectors for high
resolution tracking in the inner barrel and as well the barrel endcaps and the
forward tracker. The number of readout channels is close to $10^9$, with a high
transverse and longitudinal segmentation provided by a pitch of $25 \times
50$\,$\mu m^2$. One can expect that such a fine segmentation is in reach for a
detector which would be built in a decade hence. The radiation level in
electron-proton scattering is by orders of magnitude lower than in
proton-proton interactions at the LHC and is indeed in a range of $10^15$\,MeV
n$_{eq}$/cm$^2$ for which radiation hardness has been proven as indicated
above. The monolithic CMOS detector technology leads to a significant
simplification of the production of these detectors and a considerably reduced
cost. We thus conclude that the LHeC pixel tracker represents a particularly
suitable device for a large scale implementation of HV CMOS Silicon in a
forthcoming collider detector.

\section{Calorimetry \ourauthor{ Peter Kostka}}

\begin{figure}[!htbp]
  \centering
  \includegraphics[width=0.99\textwidth]{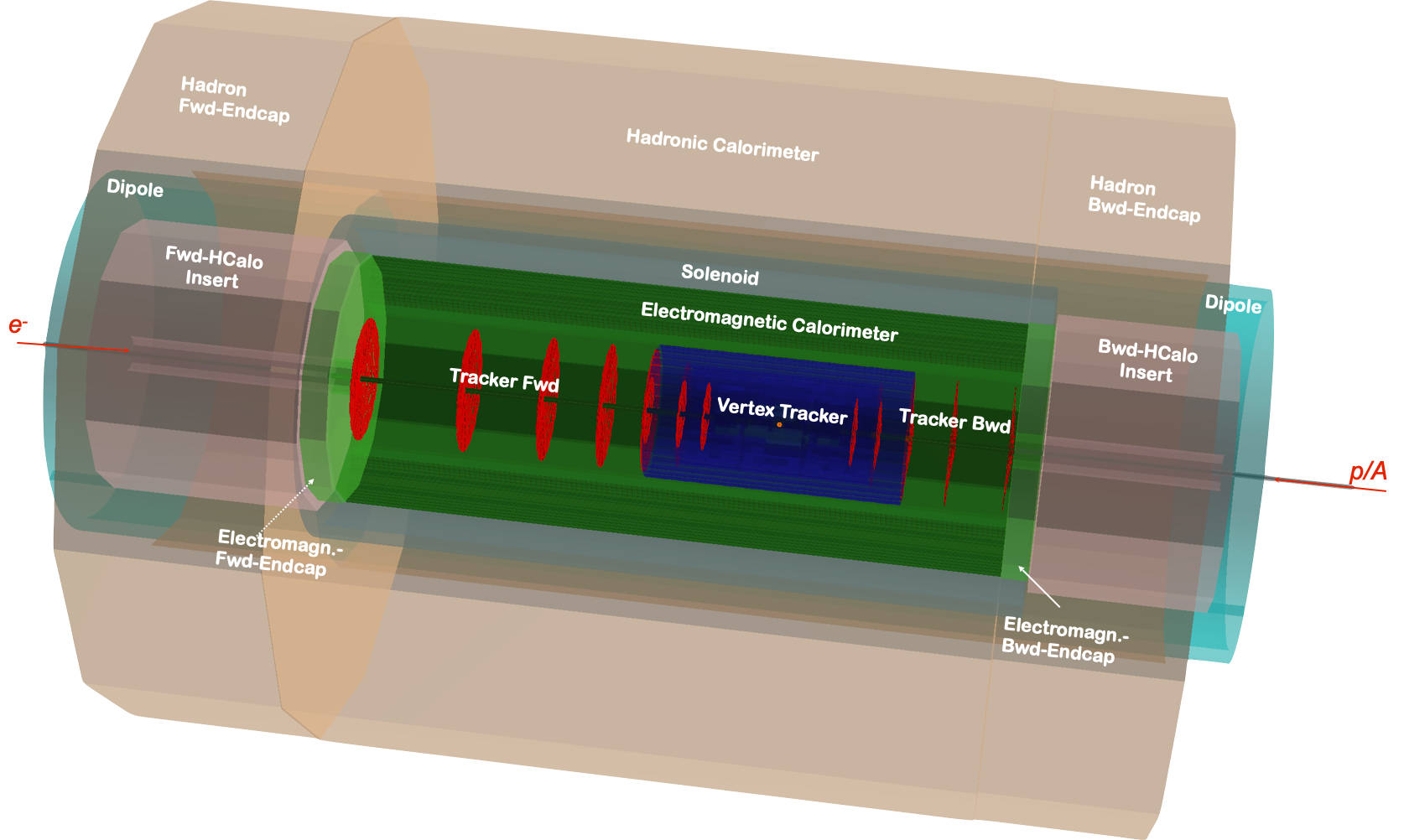}
\caption{Three-dimensional view of the arrangement of Hadronic-Calorimeter, experimental magnets (solenoid and dipoles), 
the electromagnetic calorimeter and tracking detector layers. }
\label{FIG:lhec-HCAL-magnets-EMC-tracker}
\end{figure}
\begin{figure}[!htbp]
  \centering
  \includegraphics[width=0.97\textwidth]{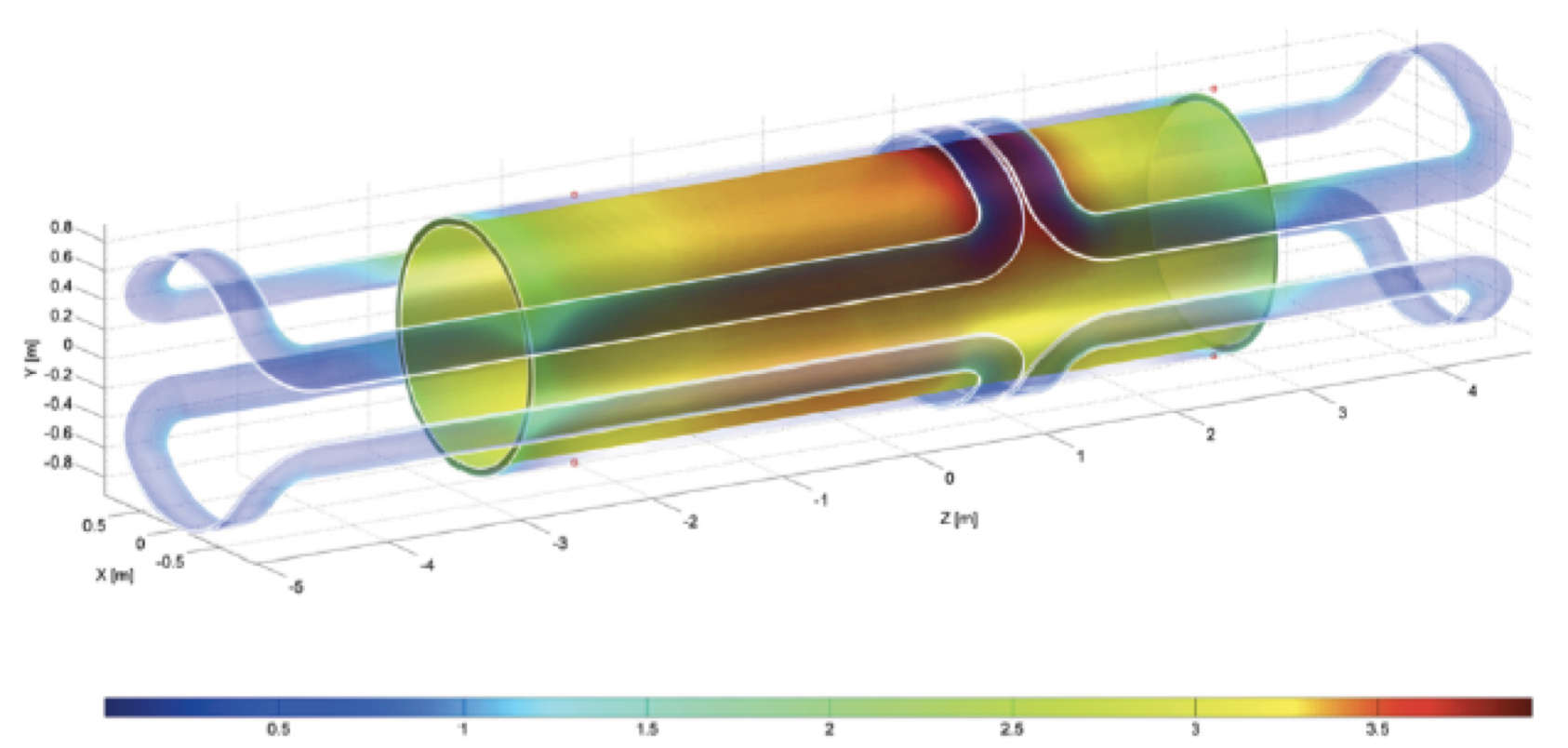}
\caption{The coil arrangement of the solenoid and dipoles system housing in a common cryostat. }
\label{FIG:LHeC-solenoid-dipole-sytem}
\end{figure}

The 2012 CDR detector design leaned on  technologies
employed by ATLAS for calorimetry in the barrel region, adopting a lead
/ liquid argon sampling electromagnetic calorimeter with an accordion
geometry and a steel / scintillating tile sampling hadronic component.
For the version of the LHeC detector described here, an alternative solution of a
lead / scintillator electromagnetic calorimeter has been investigated.
This has the advantage of removing the need for cryogenics, whilst
maintaining an acceptable performance level. 
Comparing the lead-scintintillator designs for the electromagnetic 
barrel calorimeter for the 2012 CDR with the updated setup, the 
\textbf{a}-term for shower fluctuations and transverse leakages and 
the \textbf{b}-term describing the back-leakages of the calorimeter 
the resolution performance of the updated design is better  
(\textbf{a}$=20\%$ and \textbf{b}$=0.14\%$ in the 2012 CDR and 
\textbf{a}$=12.4\%$ and \textbf{b}$=1.9\%$ in the new design).
Although it is not discussed here, the liquid argon solution 
very much remains the favorable option due to its high level of performance 
and  stability / radiation hardness.
The fit-results in CDR 2012 the LAr calorimeter option show a slightly 
better resolution performance than the lead-scintillator variant. 
Due to the accordion shaped absorber it forces more energy deposit in the calorimeter volume. 
The CDR values for comparison: \textbf{a}$=8.47\%$ and \textbf{b}$= 0.318\%$.
The hadronic calorimeter
retains the steel and scintillating tile design, similar to ATLAS. As in
the 2012 CDR, plug sampling calorimeters are also incorporated at large
$|\eta|$, the forward and backward components using tungsten and lead
absorber material, respectively, with both using silicon based sensitive
readout layers. The steel structures in the central and plug calorimetry
close the outer field of the central solenoid. 
The solenoid and the dipoles are placed between the Electromagnetic-Barrel and the Hadronic-Calorimeter.
The HCAL-Barrel sampling calorimeter using steel and scintillating tiles as absorber and active material, respectively,
provides the mechanical stability for the Magnet/Dipole cryostat and the tracking system 
Fig.~\ref{FIG:lhec-HCAL-magnets-EMC-tracker}. How the solenoid/dipoles-system would look like has been discussed
in more detail in ~\cite{AbelleiraFernandez:2012cc} and is illustrated by Fig.~\ref{FIG:LHeC-solenoid-dipole-sytem}. 
(and the LAr cryostat in a cold EMC version) along with the iron required for the return flux of the solenoidal field.
The main features of the new calorimeter layout are summarised in
Tab.~\ref{tab:LHeC_Calo_main-properties_1} and \ref{tab:LHeC_Calo_main-properties_2}. 
The pseudorapidity coverage
of the electromagnetic barrel is $-1.4 < \eta < 2.4$, whilst the
hadronic barrel and its end cap cover $-1.5 < \eta < 1.9$. Also
including the forward and backward plug modules, the total coverage is
very close to hermetic, spanning $-5.0 < \eta < 5.5$. The total depth of
the electromagnetic section is 30 radiation lengths in the barrel and
backward regions, increasing to almost $50 X_0$ in the forward direction
where particle and energy densities are highest. The hadronic
calorimeter has a depth of between 7.1 and 9.6 interaction lengths, with
the largest values in the forward plug region.

\input{\main/detector/tables/LHeC_Calo_main-properties}

\begin{figure}[!th]
  \centering
  \makebox[\textwidth]{
    \includegraphics[width=0.44\textwidth,trim={ 0 0 10 0 },clip]{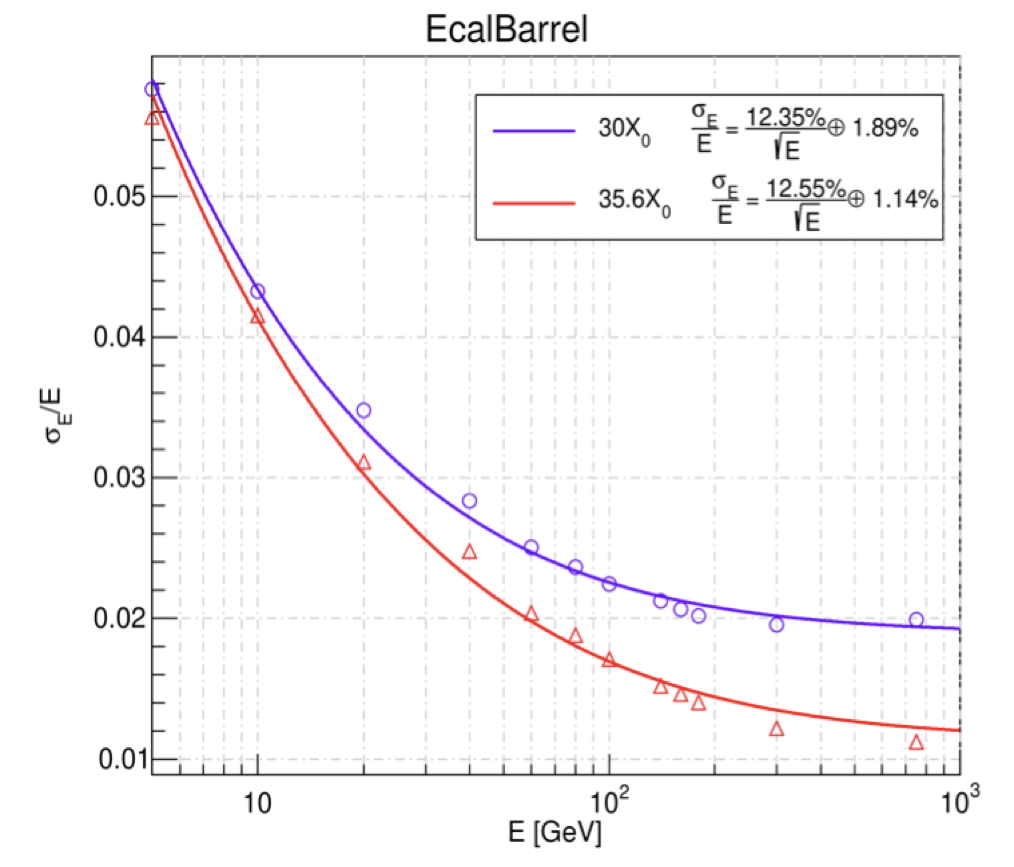}
    \hspace{0.03\textwidth}
    \includegraphics[width=0.52\textwidth,trim={ 10 5 5 5 },clip]{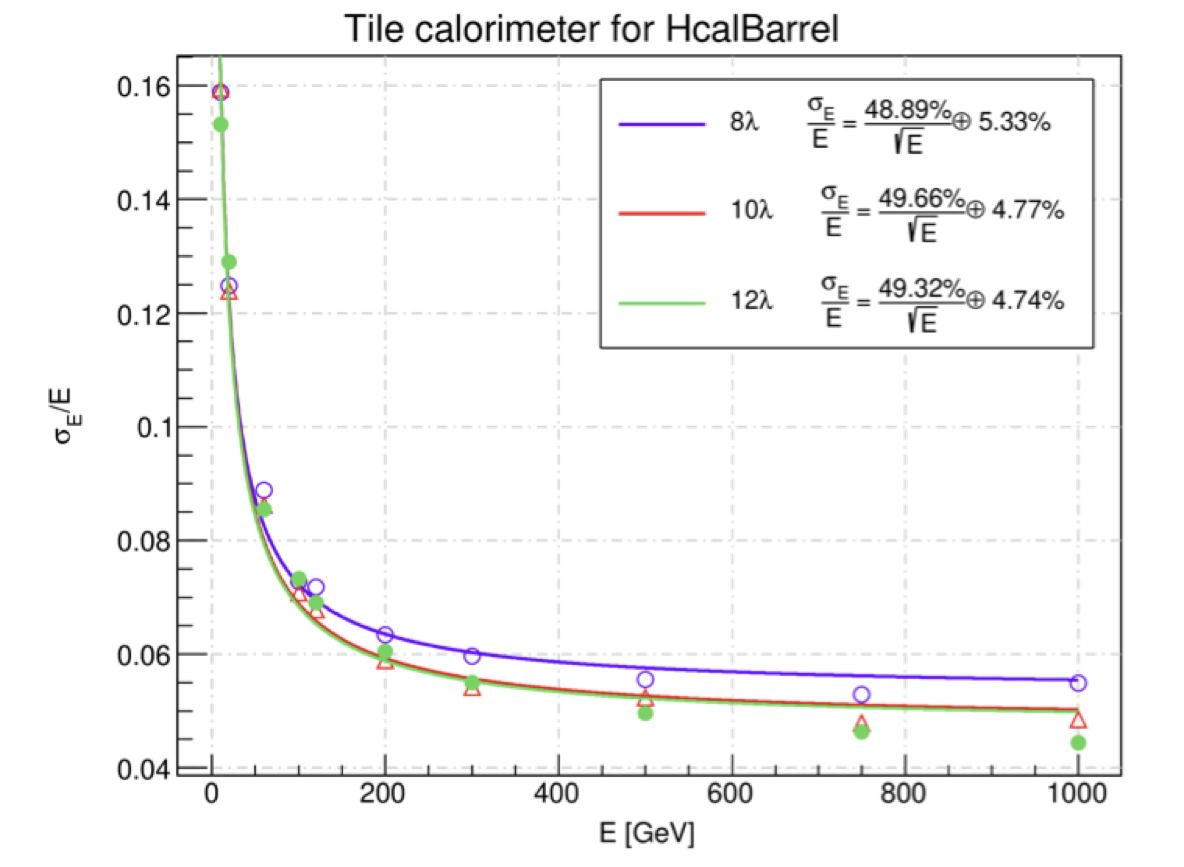}
  }
\caption{Crystal Ball fitted energy dependent resolution for the barrel electromagnetic (left) 
and barrel hadronic (right) calorimeters EMC and HCAL, respectively. 
The first ($a$) term includes shower fluctuations and transverse leakages and the second ($b$) term 
includes leakages from the calorimeter volume longitudinally. 
}
\label{FIG:lhec-barrel-energy-resolution} 
\end{figure}

\begin{figure}[!th]
  \centering
    \includegraphics[width=0.51\textwidth]{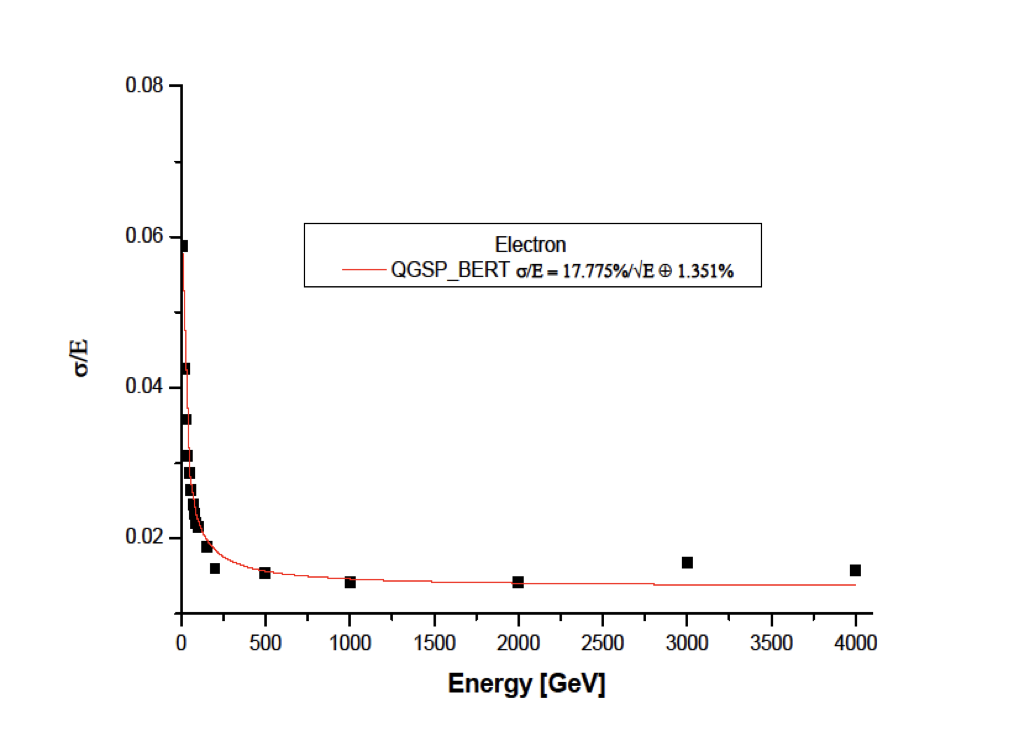}
    \includegraphics[width=0.47\textwidth]{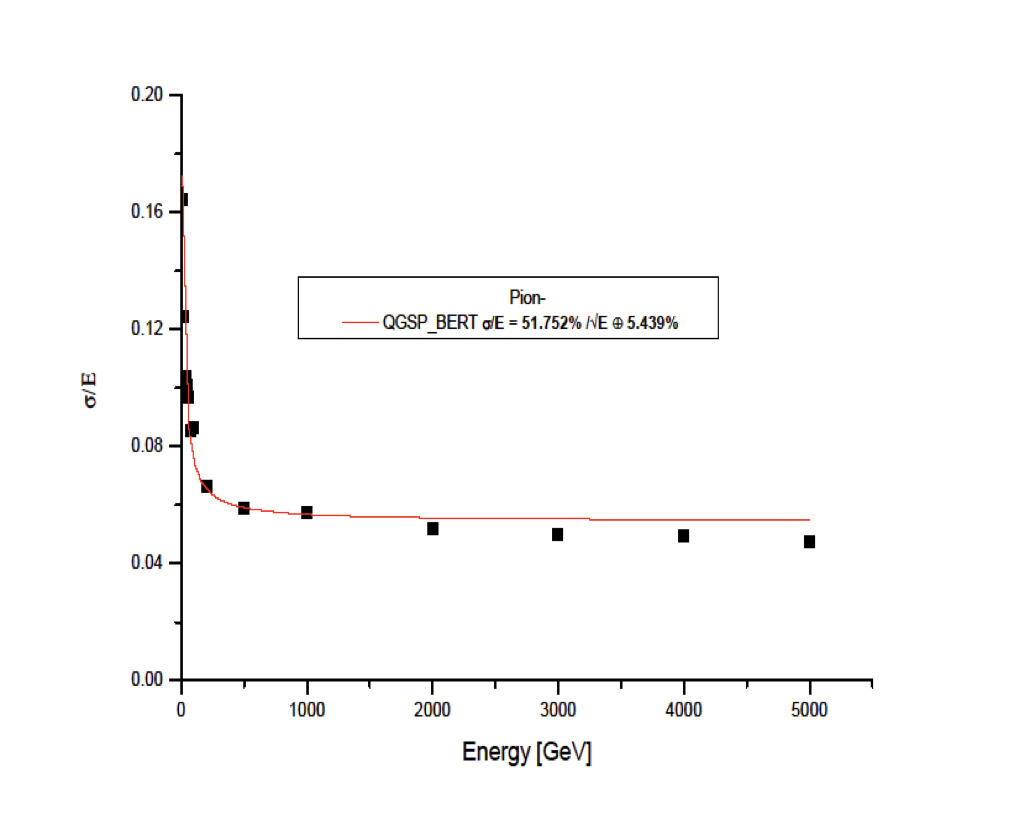}
\caption{Crystal Ball fitted energy dependent resolution for the forward electromagnetic (left) 
and forward hadronic (right) plug calorimeters FEC and FHC, respectively. 
The first ($a$) term includes shower fluctuations and transverse leakages and the second ($b$) term 
includes leakages from the calorimeter volume longitudinally. 
}
\label{FIG:lhec-plug-energy-resolution} 
\end{figure}

The performance of the new calorimeter layout has been simulated by
evaluating the mean simulated response to electromagnetic (electron) and hadronic (pion)
objects with various specific energies using
GEANT4~\cite{Agostinelli:2002hh} and interpreting the results as a
function of energy in terms of sampling $(a)$ and material / leakage $(b)$ terms
in the usual form $\sigma_E / E = a / \surd E \oplus b$. 
Example results from fits are shown for the barrel electromagnetic
and hadronic calorimeters in Fig.~\ref{FIG:lhec-barrel-energy-resolution} and for the
forward plug electromagnetic and hadronic calorimeters in 
Fig.~\ref{FIG:lhec-plug-energy-resolution}.
The results for the $a$ and $b$ parameters are
summarised in Tabs.~\ref{tab:LHeC_Calo_main-properties_1} and \ref{tab:LHeC_Calo_main-properties_2}. 
The response of
the barrel electromagnetic calorimeter to electrons in terms of both
sampling ($a = 12.4\, \%$) and material ($b = 1.9\, \%$) terms is only
slightly worse than that achieved with liquid argon sampling in the 2012
CDR. The resolutions of the forward and backward electromagnetic plug
calorimeters are comparable to those achieved in the 2012 design. A
similar pattern holds for the hadronic response, with sampling terms at
the sub-$50\, \%$ level and material terms of typically $5\, \%$ throughout
the barrel end-caps and forward and backward plugs.

\section{Muon Detector}

Muon identification is an important aspect for any general purpose HEP experiment.
At the LHeC the muon detector can widen the scope and the spectrum of many measurements, of which only a few are listed here:
\begin{itemize}
\item Higgs decay,
\item Semi-leptonic decays of heavy flavoured hadrons,
\item Vector meson production,
\item Direct $W$ and $Z$ production,
\item Di-muon production,
\item Leptoquarks, lepton flavour violation, and other BSM phenomena.
\end{itemize}

%
The primary target of the muon detector at the LHeC is to provide a
reliable muon tag signature which can be uniquely used in conjunction
with the central detector for muon identification, triggering and
precision measurements. This 
specification is appropriate to 
the constraints of
limited space~\footnote{As in the 2012 CDR, the baseline LHeC detector
including the muon system and all of the services and supports is
expected to fit into the octagonal shape envelope of the L3 magnet (11.6\,m minimum diameter).} and the lack of a dedicated magnetic field as in
the baseline design. The muon chambers surround
the central detector and cover the maximum possible
solid angle. They have a compact multi-layer structure,
providing a pointing trigger
and a precise timing measurement which is used to separate muons coming
from the interaction point from cosmics, beam halo 
and non prompt
particles. This tagging feature does not include the muon momentum
measurement which is performed only in conjunction with the central
detector. 
A trigger candidate in the muon detector is
characterised by a
time coincidence over a majority of the layers in a
range of $\eta$ and $\phi$, compatible with an $ep$ interaction of
interest in the main detector. The muon candidates are combined with the
trigger information coming from the central detector (mainly the
calorimetry at Level 1 trigger) to reduce the fake rate or more complex
event topologies.

In terms of technology choices, 
the options in use in ATLAS and CMS
and their planned upgrades are adequate for LHeC.  
Generally, muon and background rates in LHeC are expected
to be lower than in $pp$.
The option
of an LHeC muon detector composed by layers of Resistive Plate Chambers
(RPC), providing the Level 1 trigger and a two coordinate ($\eta$,
$\phi$) measurement possibly aided by Monitored Drift Tubes (MDT) for
additional precision measurements,
as chosen for the 2012 CDR, is still valid. Recent
developments as presented in the 
LHC Phase 2 Upgrade Technical Design
Reports \cite{atlas-phase2-upgrade-tdr, cms-phase2-upgrade-tdr}
further strengthen this choice. 
A new thin-RPC (1\,mm gas gap) operated
with lower HV, provides 
a sharper time response (few ns), a higher rate
capability (tens of kHz/cm$^2$), 
and extends the already good aging
perspective. Advances in low-noise,
high-bandwidth front-end electronics can improve the
performance of older detectors. Similar arguments also hold for smaller
tube MDTs (15\,mm diameter) which provide lower occupancy and 
higher
rate capability.

\begin{figure}[!th]
\centering
\includegraphics[width=0.95\textwidth]{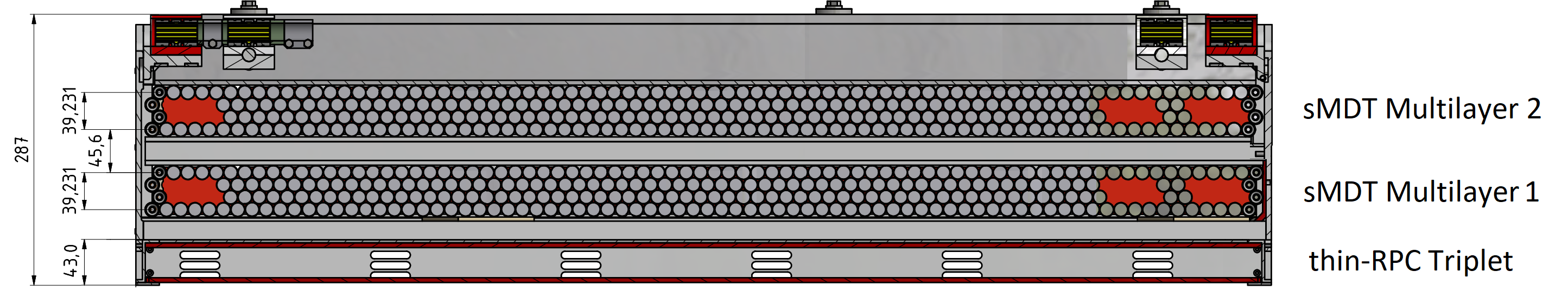}
\caption{A transverse view of a RPC-MDT assembly as adapted from a drawing of the ATLAS Phase-1 
muon upgrade\cite{atlas-phase2-upgrade-tdr}.  In this case a station is composed of an RPC triplet 
for trigger and tw0-coordinate readout and two MDT superlayers for precise track measurements. 
}
\label{fig:det:muon}
\end{figure}

Fig.~\ref{fig:det:muon} shows an adaptation for LHeC of an RPC-MDT assembly 
as will be implemented for the inner muon layer of ATLAS already during the Phase-1 
upgrade as a pilot for Phase-2. A triplet of thin gap RPCs,
each with 2 coordinate measurement, is combined with two superlayers of
small MDTs. It is also important to note the reduced volume of this
structure, in particular the RPC part which would provide
the muon tag. For the LHeC a baseline would be 
to have one or two such stations forming
a near-hermetic envelope around the central detector.

Finally, as already presented in the 2012 CDR, detector extensions, with
a dedicated magnetic field in the muon detector, be this a second
solenoid around the whole detector or extra dipole or toroid in the
forward region are, at this stage, left open as possible developments
only for upgrade scenarios.

%
%
%
\section{Forward and Backward Detectors \ourauthor{ Paul Newman}}
\label{sec:forbackdet}
In the 2012 CDR, initial plans for beamline instrumentation were
provided for the LHeC. In the backward direction, low angle electron and
photon calorimeters were included with the primary intention of measuring
luminosity via the Bethe-Heitler process $ep \rightarrow eXp$,
also offering an electron tagger to identify photoproduction
($\gamma p \rightarrow X$) processes at intermediate $y$ values. The
current design carries forward the 2012 version of this backward
instrumentation.

In the forward direction, Roman pot detectors were included in the
region of $z \sim 420 \,\mathrm{m}$, capable of detecting scattered
protons over a range of fractional energy loss $10^{-3} < \xi < 3 \times
10^{-2}$ and wide transverse momentum acceptance, based on extensive
previous work in the LHC context by the FP420 
group \cite{Albrow:2008pn}. This also
forms the basis of forward proton tagging in the revised design.
However, as is the case at ATLAS and CMS / TOTEM, further Roman pot
detectors in the region of $200 \, \mathrm{m}$ and (with HL-LHC optics)
perhaps around $320 \, \mathrm{m}$ would extend the acceptance towards
higher $\xi$ values up to around $0.2$ allowing the study of diffractive
processes $ep \rightarrow eXp$ where the dissociation system $X$ has a
mass extending into the TeV regime. It is worth noting that Roman pot
technologies have come of age at the LHC, with the TOTEM collaboration
operating 14 separate stations at its high point. Silicon sensor
designs borrowed from the innermost regions of the ATLAS and CMS
vertexing detectors have been used, providing high spatial resolution
and radiation hardness well beyond the needs of LHeC. Very precise
timing detectors based on fast silicon or Cherenkov radiation signals
from traversing protons in quartz or diamond have also been deployed. It
is natural that these advances and the lessons from their deployment at
the LHC will be used to inform the next iteration of the LHeC design.

The forward beamline design also incorporates a zero angle calorimeter,
designed primarily to detect high energy leading neutrons from
semi-inclusive processes in $ep$ scattering and to determine whether
nuclei break up in $e$A events. This component of the detector was not
considered in detail in 2012 and is therefore discussed here.

 
 \subsection{Zero-Degree (Neutron) Calorimeter}

The Zero-Degree Calorimeter (ZDC) measures final state neutral particles
produced at angles near the incoming hadron beam direction. They
typically have large longitudinal momentum ($x_F \gg 10^{-2}$), but with
transverse momentum of order of $\Lambda_\mathrm{QCD}$. Such a
calorimeter has been instrumented in experiments for $ep$ collisions (H1
and ZEUS) and for $pp$, $p$A and AA collisions at RHIC (STAR and
PHENIX) and at the LHC (ATLAS, CMS, ALICE and LHCf at the ATLAS IP). The
detector's main focus is to study the soft-hard interplay in the QCD
description of $ep$ and $e$A collisions by studying the dependence of
forward-going particles with small transverse momentum on 
variables such as  $Q^2$ and $x$ that 
describe the hard scattering. The detector also allows the tagging of
spectator neutrons to detect nuclear breakup in $e$A collisions
and
enables the precise study of the EMC effect by using neutron-tagged DIS
on small systems, such as $e\,^3He \rightarrow ed + n \rightarrow eX +
n$. For heavier ions, several tens of neutrons may enter within the
aperture of the ZDC. Inclusive $\pi^0$ production 
has been measured
by the LHCf experiments for $pp$ collisions. It is of great interest to
compare with DIS measurements at the same proton energies. Precise
understanding of the inclusive spectrum of the forward-going particles
is a key ingredient in simulating air showers from ultra-high energy
cosmic rays.

\subsubsection[Physics requirement for forward neutron and $\pi^0$
production measurement]{\boldmath Physics requirement for forward neutron and $\pi^0$
production measurement}

It is known from various HERA measurements that the slope parameter $b$
is about $8$\,GeV$^{-2}$ in the exponential parameterisation $e^{bt}$ of
the $t$ distribution of leading neutrons. In order to precisely
determine the slope parameter it is necessary to measure the transverse
momentum of the neutrons up to or beyond $1$\,GeV. The aperture for
forward neutral particles does not have to be very large, thanks to the
large energy of the proton and heavy ion beam. For example, collisions
with $E_p = 7$\,TeV need 0.14\,mrad for $p_T = 1$\,GeV neutrons at
$E_\mathrm{particle}/E_\mathrm{beam} \equiv x_F = 1.0$, or 0.56\,mrad
for $x_F = 0.25$.

The energy or $x_F$ resolution for neutrons will not be a dominant
factor thanks to the high energy of the produced particles. The energy
resolution of a neutron with $x_F = 0.1$ is about 2\% for cutting-edge
hadron calorimeters with $\sigma_E/E = 50\,\%/\sqrt{E}$, where $E$ is in
GeV. Such a resolution can be achieved if non-unity $e/h$ can be
compensated either by construction of the calorimeter or by software
weighting, and if the size of the calorimeter is large enough so that
shower leakage is small.

On the other hand, the resolution requirement on the transverse momentum
is rather stringent. For example, 1\,mm resolution on hadronic showers
from the neutron measured at $100$\,m downstream from the interaction
point corresponds to 0.01\,mrad or 70\,MeV, which is rather moderate
($\leq 10\,\%$ resolution for large $p_T$ hadrons with $p_T > 700$\,MeV).
For smaller $p_T$ it is more appropriate to evaluate the resolution in
terms of $t \simeq -(1-x_F)p_T^2$ i.e.\ $\Delta t \simeq 2(\Delta
p_T)p_T$ at $x_F = 1$. At $t = 0.1$\,GeV$^2$ or $p_T \simeq 300$\,MeV,
$\Delta t$ is about 50\,\%. A shower measurement with significantly better
than $1$\,mm position resolution, therefore, would improve the
$t$-distribution measurement significantly.

According to the current LHC operation conditions with $\beta^\ast =
5$\,cm, the beam spread is $8\times 10^{-5}$\,rad or 0.56\,GeV. This is
much larger than the required resolution in $p_T$. It is 
therefore
neither possible to measure the particle flow nor to control the acceptance
of the forward aperture. For precision measurement of forward particles,
it is necessary to have runs with $\beta^{\ast} \geq 1$\,m,
corresponding to $\sigma(p_T) < 70$\,MeV.

The calorimeter should be able to measure more than 30 neutrons of
5\,TeV to tag spectator neutrons from heavy-ion collisions. The
dynamic range of the calorimeter should exceed 100\,TeV with good
linearity.

As for $\pi^0$ measurements, the LHCf experiment has demonstrated that a
position resolution of $200\,\mu$m on electromagnetic showers provides
good performance for the inclusive photon spectrum
measurements\cite{Adriani:2017jys}. This also calls for fine
segmentation sampling layers.

\subsubsection{ZDC location}

\begin{figure}[!ht]
\centering
\includegraphics[width=0.7\textwidth]{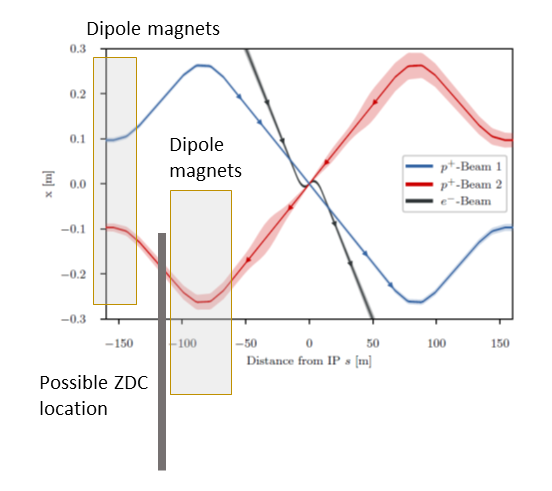}
\caption{
Possible location for a ZDC for the linac--ring IR
design of LHeC as shown in Fig.~\ref{fig:IR_geometry}.
The solid rectangle represents the potential location and magnets overlaid.
}
\label{fig:zdcloc}
\end{figure}

According to the IP design, a possible location for the ZDC is after the
first bending of the outgoing colliding proton beam at around $Z =
110$\,m, where no beam magnet is placed (see Fig.\,\ref{fig:zdcloc}). It
is anyhow planned to place a neutral particle dump around this location 
in order to protect
accelerator components. A ZDC could
serve as the first absorbing layer at zero degrees.

The aperture to the ZDC would be determined by the last
quadruple magnet at around $z = 50$\,m. Assuming a typical
aperture for the LHC magnets of 35\,mm, the aperture could
be as large as 0.7\,mrad. The horizontal aperture of the dipole
magnets between 75 and 100\,m would be larger, since otherwise
the magnets receive significant radiation from neutral particles
produced from the collisions at the IP. Even if the aperture is limited by
the vertical aperture of the last
dipole at $z = 100$\,m, the aperture is 0.35\,mrad,
corresponding to $2.4$\,GeV in $p_T$ for 7\,GeV particles.
This fulfills the physics requirement.

The space for the ZDC location in the transverse direction should be
at least $\pm 2\lambda_I$ to avoid large leakage of hadronic showers.
This can be achieved if the proton beam passes inside the calorimeter,
about 20\,cm from the centre of the calorimeter. The total
size of the calorimeter could then be $60\times 60\times 200$\,cm$^3$ or
larger according to the current layout of the beam and accelerator components.
This would provide
about $\pm 3\lambda_I$ in the transverse direction and about
$10\lambda_I$ in depth.

\subsubsection{Radiation requirement for the ZDC}

It can safely be assumed that the energy spectrum of the 
forward neutral
particles produced in $ep$ and $pp$ events are very similar.
According to the LHCf simulation, their tungsten--scintillator
sandwich calorimeter receives about 30\,Gy/nb$^{-1}$ or
$10^8$\,events/nb$^{-1}$ assuming $\sigma^\mathrm{tot}_{pp} = 100$\,mb,
i.e.\ $3\times 10^{-7}$\,Joule/event.
This means that about 1/4 of the total proton beam energy
($7\,\mathrm{TeV}\simeq 1.12\times 10^{-6}$\,Joule/event) is deposited in 1\,kg
material in $pp$ collisions. The $ep$ total cross section is
expected to be approximately
68\,$\mu$b or 680\,kHz at $10^{34}$\,cm$^2$s$^{-1}$. A 7\,TeV
beam or $1.12\times 10^{-6}$\,Joule/event 
corresponds to 0.76\,Joule/s at this instantaneous luminosity.
A quarter of the total dose is then about 0.2\,Gy/sec
or 0.02\,Gy/nb.
The contribution from beam-gas interactions is estimated to be
much smaller (${\mathcal O}(100\,\mathrm{kHz})$).

Assuming that the ZDC is always operational during LHeC running, one year of $ep$
operation amounts to $2.5$\,MGy/year assuming $10^7$\,sec operation,
or ${\mathcal O}(10$\,MGy$)$ throughout the lifetime of the LHeC operation.
This approximately corresponds to $10^{14} - 10^{15}$ 1\,MeV
neutron equivalent.

\subsubsection{Possible calorimeter design}

The high dose of ${\mathcal O}(10\,\mathrm{MGy})$ requires calorimeters
based on modern crystals (e.g.\  LYSO) or silicon as sampling
layers, at least for the central part of the calorimeter
where the dose is concentrated.
Since we also need very fine segmentation for
photons, it is desirable to use finely segmented silicon pads
of order of 1\,mm. As for the absorbers,
tungsten should be used for good position resolution
of photons and the initial part of hadronic showers.

In the area outside the core of the shower i.e.\ well outside
the aperture, the dose may be much smaller and small scintillator
tiles could be used for absorbers, which allows
measurements with good $e/h$ ratio. If we choose a uniform
design using silicon across the detector, the segmentation
of the outer towers could be order of a few cm, which still
makes it possible to use software compensation technology, as 
developed for example for the calorimeters in the ILC design. 
It may also be possible to use lead
instead of tungsten for outer towers to reduce
the cost.

%
%

\section{Detector Installation and Infrastructure}
\begin{figure}[!ht]
  \centering
  \includegraphics[width=0.8\textwidth]{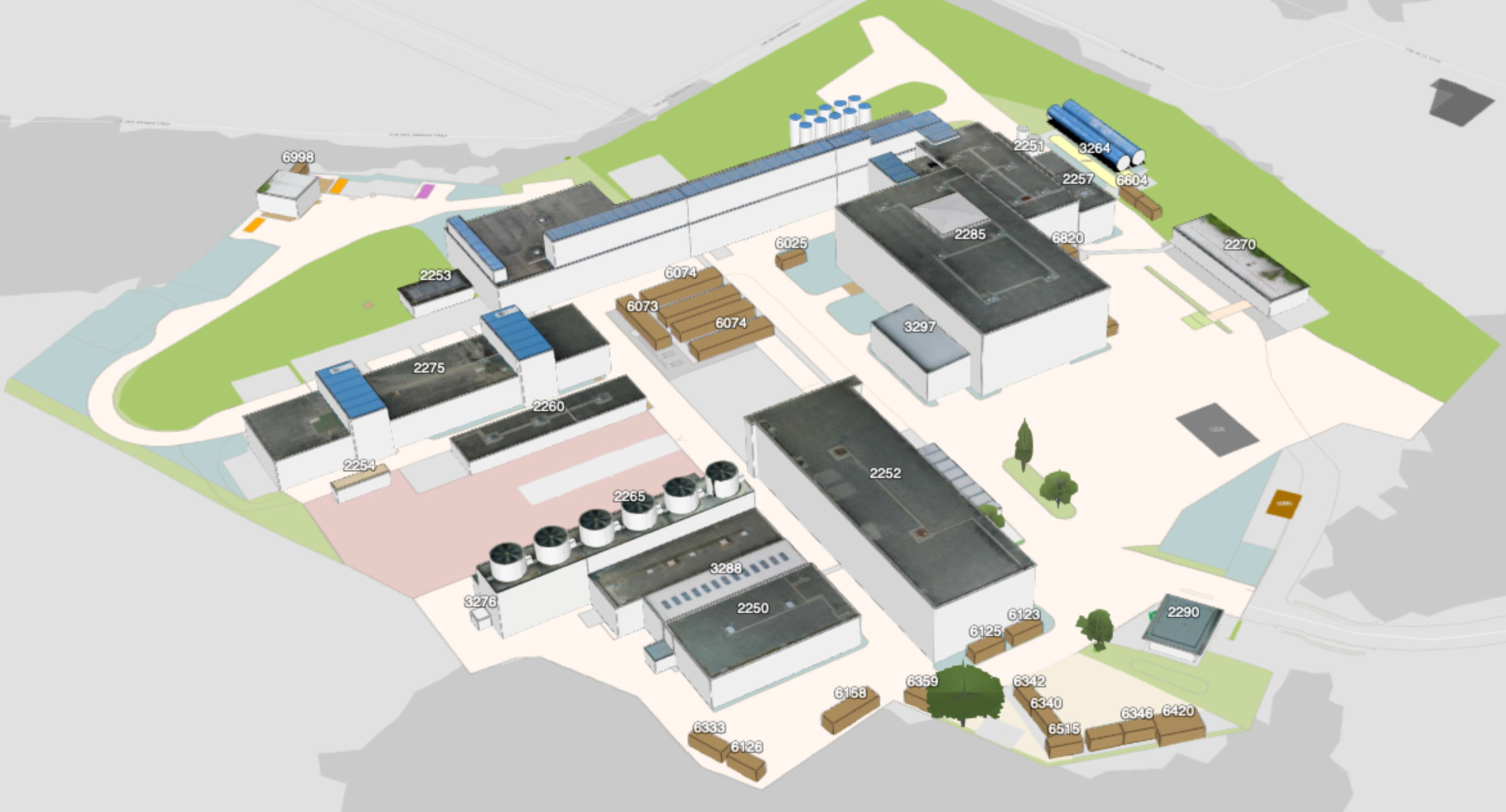}
  \caption{
    View of the surface infrastructure at Point 2, near St. Genis~\cite{mapscern}.
  }
  \label{fig:gaddi1}
\end{figure}

The usual constraints that apply to HEP detector integration and assembly studies also apply to the LHeC. 
In places, they are even tighter since the detector has to be installed 
in a relatively short time, as given by the duration of an LHC machine shutdown, 
which is typically two years. 
For the purposes of this report, it is assumed that
the LHeC detector will be installed at 
IP2, see Fig.\,\ref{fig:gaddi1}. The magnet formerly used by
L3 and now in use by ALICE is already present at IP2 and its support structure will be used once again my LHeC.
However, the time needed to remove the remainder of the existing detector and its services has to be included to the overall schedule.
Thus the only realistic possibility to accomplish the timely dismantling of the 
old detector and the installation of the new one is to complete as much as possible of
the assembly and testing of the LHeC detector on 
the surface, where the construction 
can proceed without impacting on the LHC physics runs. 
The condition for doing this is 
the availability of equipped free space at the 
LHC-P2 surface, namely a large assembly hall 
with one or two cranes. To save time, most of the detector components have been designed 
to match the handling means available on site, i.e.\ 
a bridge crane in the surface hall and 
experiment cavern. Nevertheless, a heavy lifting facility (about 300\,tons capacity) 
will be rented for the time needed to lower the heaviest detector components, such as the
 HCal barrel and plug modules. Large experience with this will be acquired during LHC 
Long Shutdown 3, when a significant part of 
the ATLAS and CMS detectors will be replaced by 
new elements. At CMS, for instance, a new Endcap Calorimeter weighting about 220\,tons will 
be lowered into the experiment cavern, a scenario very close to what is envisaged for 
the LHeC detector assembly.

The detector has been split into the following main parts for assembly purposes:
\begin{itemize}
\item	Coil cryostat, including the superconducting coil, the two integrated dipoles and eventually the EMCal.
\item	Five HCal tile calorimeter barrel modules, fully instrumented and cabled (5).
\item	Two HCal plugs modules, forward and backward (2).
\item	Two EMCal plugs, forward and backward (2).
\item	Innner Tracking detector (1).
\item	Beam-pipe (1).
\item	Central Muon detector (1 or 2).
\item	Endcaps Muon detector (2).
\end{itemize}
The full detector, including the Muon chambers, fits inside the former L3 detector Magnet Yoke,
once the four large doors are taken away. The goal is to prevent losing time in dismantling 
the L3 Magnet barrel yoke and to make use of its sturdy structure to hold the detector central 
part on a platform supported by the magnet crown, whilst the Muon chambers are inserted 
into lightweight structures (space-frames) attached to the inner surface of the octagonal L3 magnet.

\begin{figure}[!ht]
  \centering
  \includegraphics[width=0.7\textwidth]{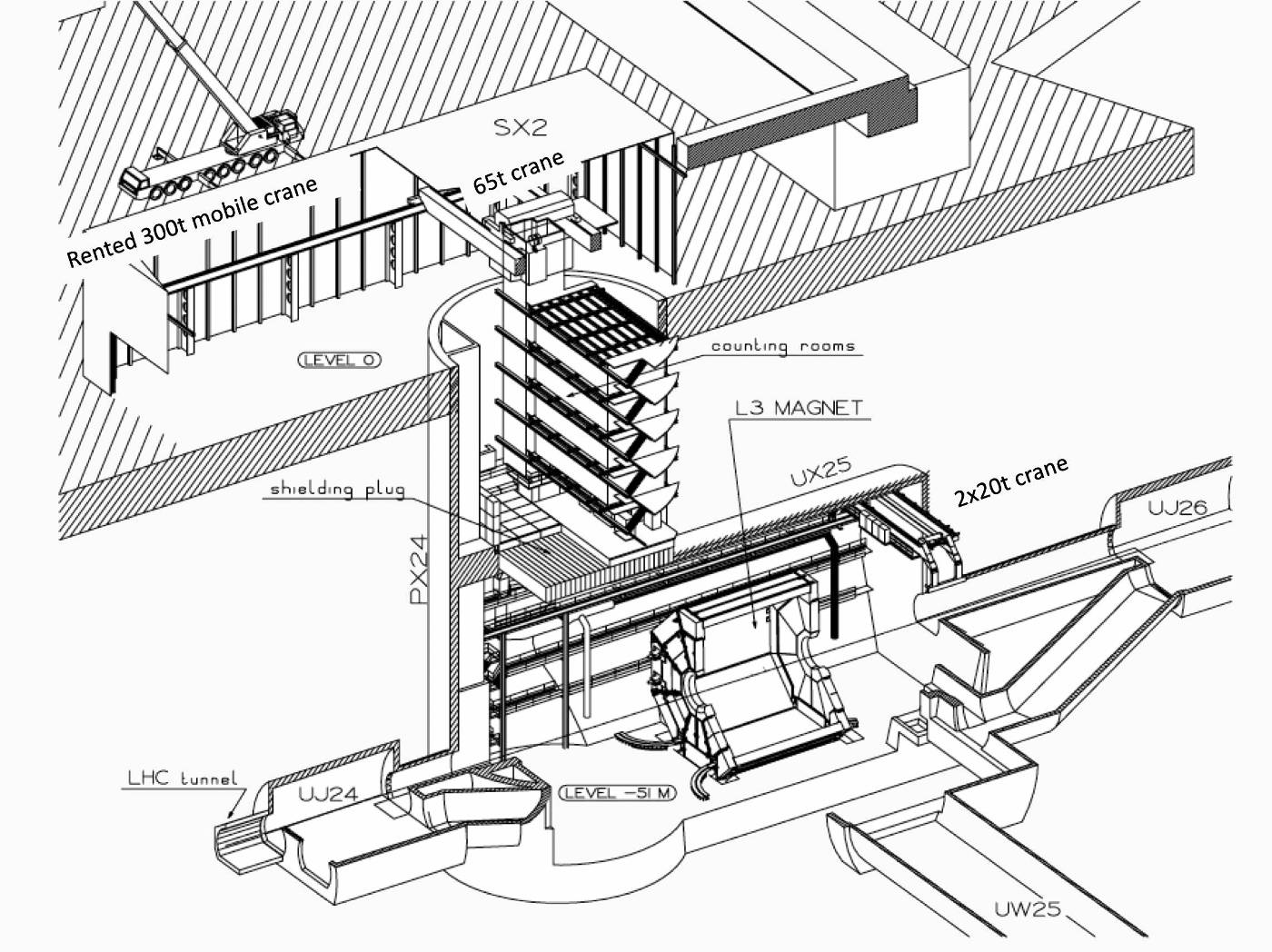}
  \caption{
    View of the cavern infrastructure at Point 2~\cite{Dellacasa:2000kh}.
    The support structure of the magnet of the L3 experiment (at the centre) will  
    house and support the LHeC detector.
  }
  \label{fig:gaddi2}
\end{figure}

\begin{figure}[!ht]
  \centering
  \includegraphics[width=0.7\textwidth]{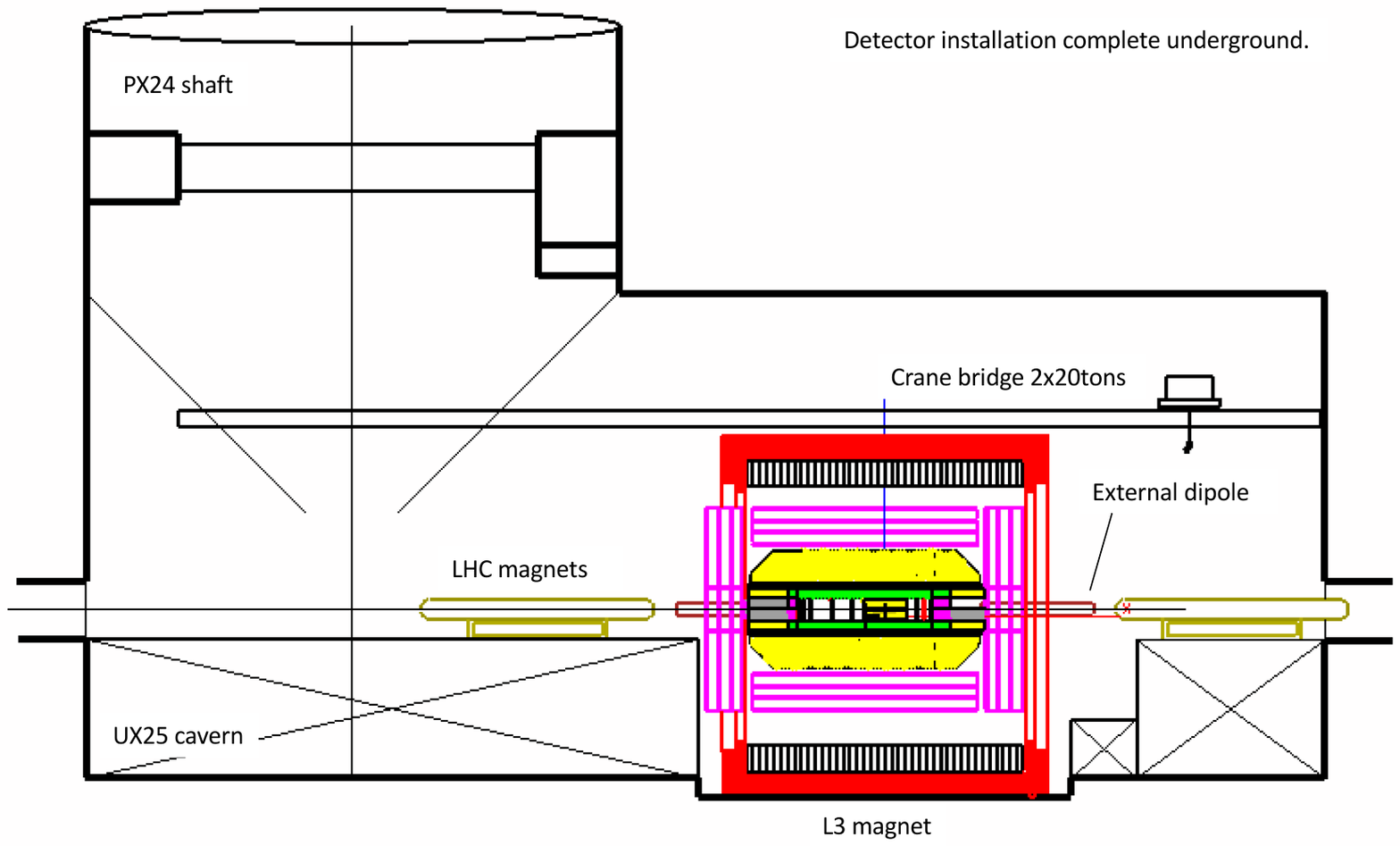}
  \caption{
    View of the LHeC detector, housed in the L3 magnet support structure, after installation at the interaction point.
  }
  \label{fig:gaddi3}
\end{figure}

The assembly of the main detector elements
on the surface can start at any time,
without sensible impact on the LHC run,
providing that the surface facilities are available.
The Coil system commissioning on site is estimated to require three months and preparation for lowering 
a further three months, including some contingency. In the same time window, the L3 Magnet will be freed 
up and prepared for the new detector~\footnote{
The actual delay depends on the level of activation and the procedure adopted for dismantling 
the existing detector. Here again the experience acquired during the long shutdown LS2 
with the upgrades of ALICE and LHCb and later 
with the ATLAS and CMS upgrades during LS3 
will provide important insight for defining procedures and optimising the schedule.}.
Lowering of the main detector components into the cavern, illustrated in Fig.\,\ref{fig:gaddi2},
 is expected to take one week per piece (15 pieces in total). Underground 
integration of the central detector elements inside the L3 Magnet would 
require about 6 months, cabling and connection to services some 8 to 10 months, 
in parallel with the installation of the Muon chambers, the Tracker and 
the Calorimeter Plugs. Fig.\,\ref{fig:gaddi3} shows the installed complete 
detector housed in the L3 magnet support.

\begin{figure}[!ht]
  \centering
  \includegraphics[width=0.8\textwidth]{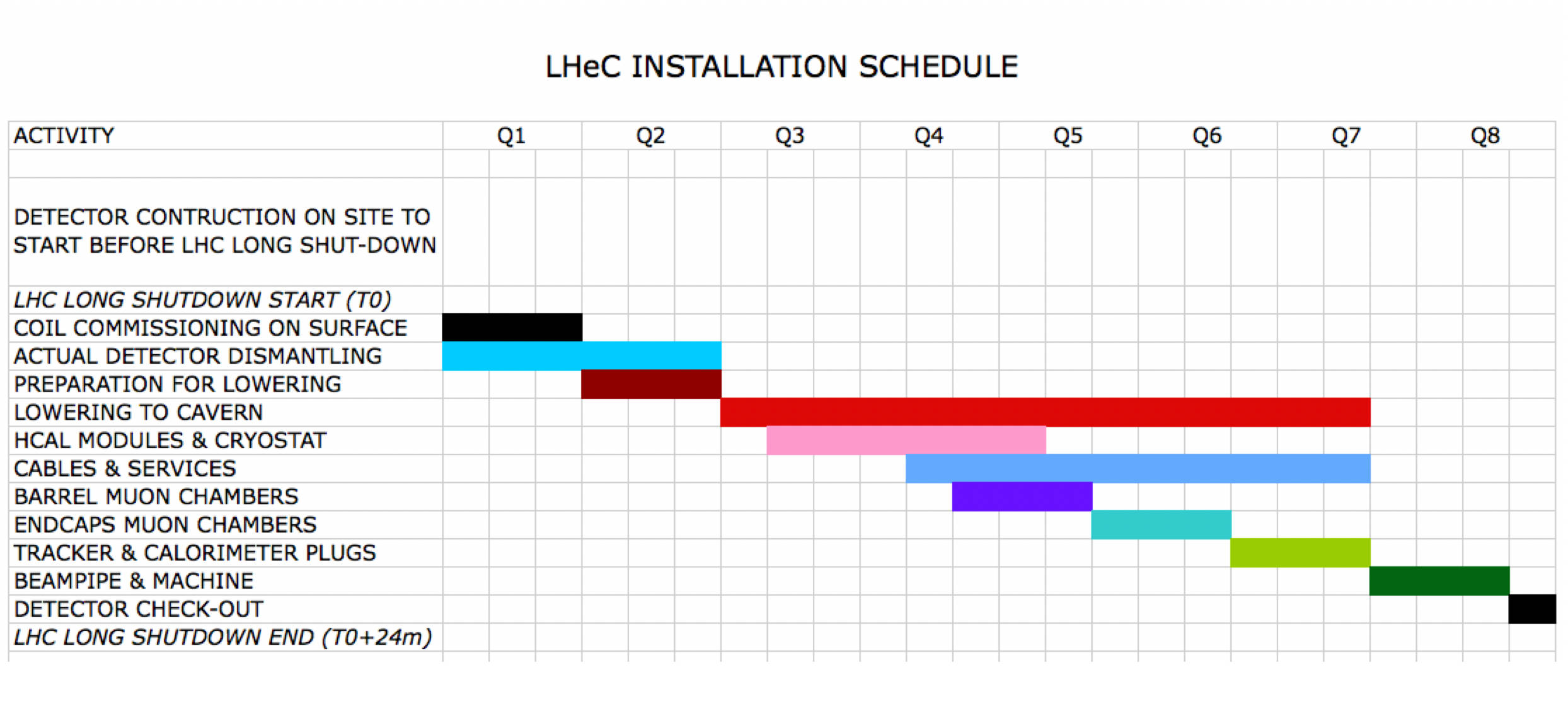}
  \caption{
    Time schedule of the sequential installation of the LHeC detector at point 2, as described in the text.
  }
\label{fig:gaddi4}
\end{figure}

The total estimated time, from the starting of the testing of the Coil system on 
surface to the commissioning of the detector underground is thus 20 months. The 
beam-pipe bake out and vacuum pumping could take another 3 months and the final 
detector check-out one additional month. Some contingency (2--3 months in total) 
is foreseen at the beginning and the end of the installation period. A sketch of 
the installation schedule is provided in Fig.\,\ref{fig:gaddi4}

Concerning the detector infrastructures, not much can be said at this stage. 
The LHeC detector superconducting coil will need cryogenic services and a 
choice has to be made between purchasing a dedicated liquid helium refrigeration 
plant or profiting from the existing LHC cryogenic infrastructure to feed the detector
 magnet. 
 The electrical and water-cooling 
 networks present at LHC-P2 are already well 
sized for the new detector and only minor interventions are expected there.

\clearpage
\section{Detector Design for a Low Energy FCC-eh}

Although not the primary focus of this report, 
a full detector design has also been carried out for an $ep$ facility based on an FCC tunnel with proton-ring 
magnet strengths limited such that the proton energy is 20\,TeV.
For ease of comparison, the basic layout and the technology choices are currently similar to 
those of the LHeC detector. Similar or improved performance is obtained compared with the LHeC, 
provided that additional disks are included in the forward and backward trackers and the calorimeter 
depths are scaled logarithmically with the beam energies.

\begin{figure}[!htbp]
  \centering
\includegraphics[width=0.95\textwidth]{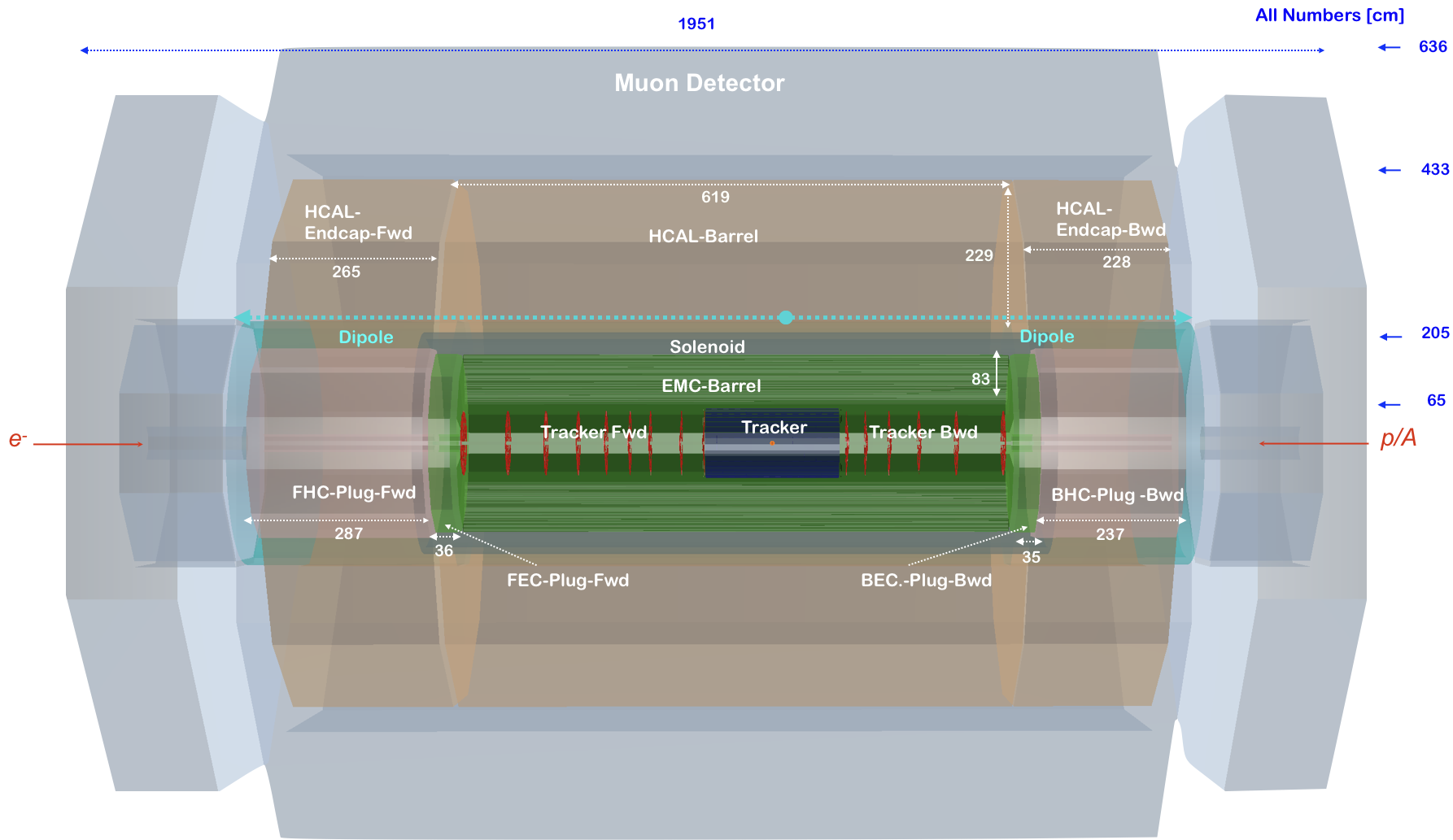}
\caption{Side view of a low energy FCCeh ($E_{p}=20$\,TeV) concept detector, designed using the 
DD4hep framework~\cite{frank_markus_2018_1464634}, showing the essential features.
The solenoid is again placed
between the ECAL-Barrel and Hadronic-Barrel calorimeters 
and is housed in 
a cryostat in common with the beam steering dipoles extending over the full length of 
the barrel and plug hadronic calorimeters. 
The sizes have been chosen such that 
the solenoid/dipoles and ECAL-Barrel systems as well as the whole tracker are also suitable to operate 
after an upgrade of the beam energy to $E_{p}=50$\,TeV.}
\label{FIG:lowE-FCCeh-maindetector2}
\end{figure}

\begin{figure}[th]
  \centering
    \includegraphics[width=0.65\textwidth]{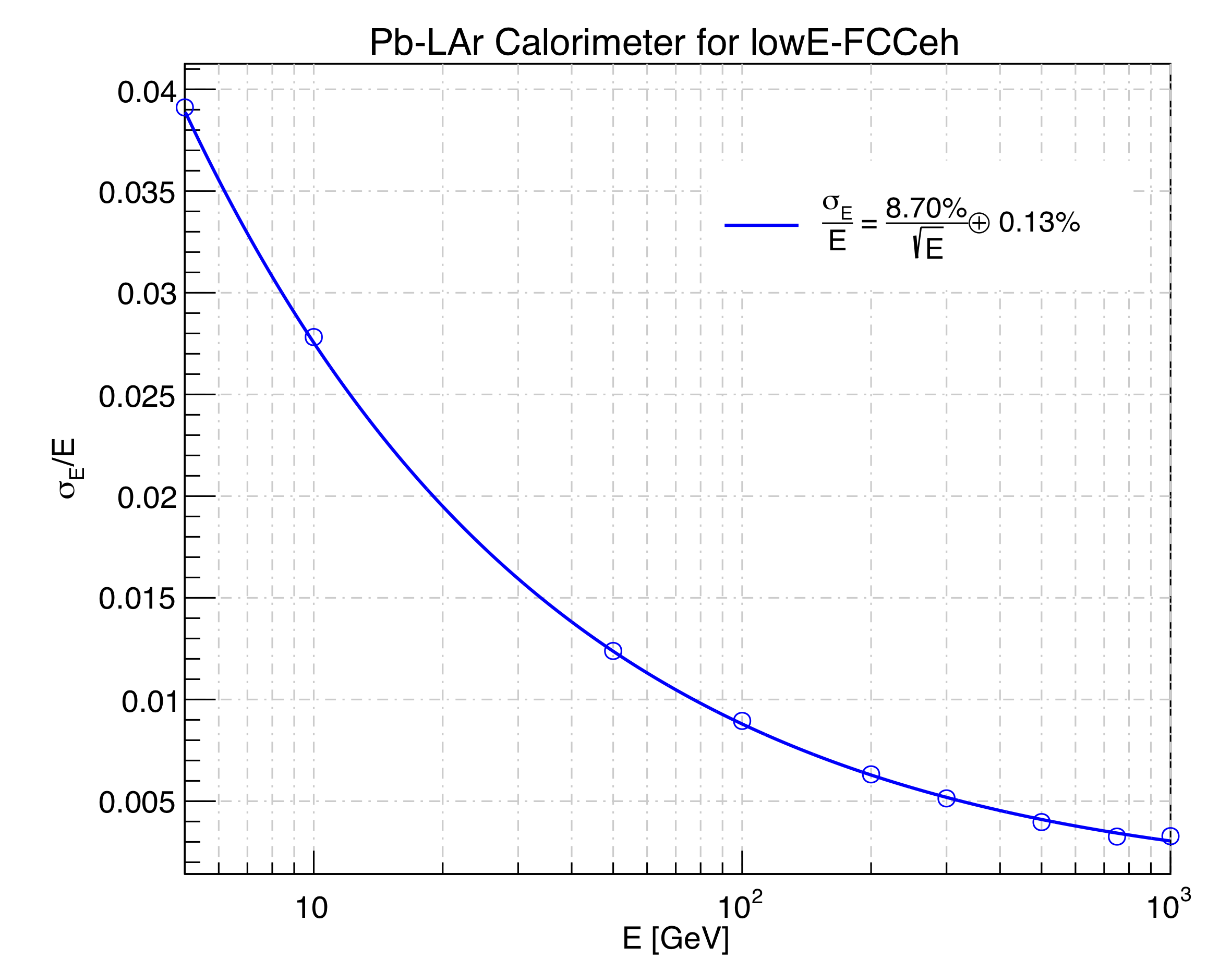}
\caption{For comparison the achievable resolution of a cold version of an EM-calorimeter stack is shown. 
The sampling calorimeter setup (ATLAS type) is characterised by
lead as absorber $2.2$\,mm thick and $3.8$\,mm gaps filled with liquid argon as detecting medium, 
a cartesian accordion geometry and stack folds having a length of $40.1$\,mm and an 
inclination angle of $\pm{45}$\textdegree to each other.
The radiation length for the setup described is estimated from geantino scans using \textbf{GEANT4}~\cite{Agostinelli:2002hh}.
The simulated calorimeter stack has a depth of $83.7$\,cm (approximately $58$\,X$_0$).
The fits have been performed as for Fig.~\ref{FIG:lhec-barrel-energy-resolution}.
}
\label{FIG:Pb-LAr-cal-resolution-lowE-FCCeh} 
\end{figure}

The basic layout is shown in Fig.~\ref{FIG:lowE-FCCeh-maindetector2}. The barrel and end-caps of 
the central tracker are identical to those of the LHeC design, as given in table~\ref{tab:LHeC_Tracker_main-properties_1}.
The design parameters for the FCC-eh versions of the forward and backward trackers, the 
barrel calorimeters and the plug calorimeters are given in 
tables~\ref{FCC-tracker},~\ref{tab:lowE-FCCeh_Calo_main-properties_1} and~\ref{tab:lowE-FCCeh_Calo_main-properties_2}, 
respectively. 
Comparing the performance of "warm" solution (Pb-Sctillator) with the "cold" variant (Pb-LAr) for the 
barrel electromagnetic calorimeter (EMC) the superior performance of the "cold" calorimeter setup again favorises the 
Pb-LAr option for the lowE-FCCeh detector (see figure~\ref{FIG:Pb-LAr-cal-resolution-lowE-FCCeh} and table~\ref{tab:lowE-FCCeh_Calo_main-properties_1}).

\input{\main/detector/tables/lowE-FCCeh_fwd-bwd_Tracker_main-properties.tex}

\input{\main/detector/tables/lowE-FCCeh_Calo_main-properties}

\biblio

%% file: detector/tables/LHeC_Tracker_main-properties.tex

\begin{table}[htbp]
    \centering
    \small
    \begin{tabular}{lcccccc}
      \toprule
      Tracker (LHeC) &    \multicolumn{3}{c}{Inner Barrel}  & \multicolumn{3}{c}{ECAP} \\ 
      \cmidrule(lr){2-4} \cmidrule(lr){5-7}
      &      pix &      pix$_\text{macro}$ &  strip &  pix & pix$_\text{macro}$ & strip \\
      \midrule
      $\eta{_\text{max}}$,$\eta{_\text{min}}$  &  
      $3.3,-3.3 $  & 
      $2.1,-2.1 $   &  
      $1.4,-1.4 $   & 
      $\pm[4.1,1.8] $   & 
      $\pm[2.4,1.5] $   &  
      $\pm[2.0,0.9]   $   \\

Layers (Barrel) &
1 &
3 & 
3 &
 
  & & \\
  
Wheels (ECAP) & & & &
2 &
1 & 
1-3  \\

Modules/Sensors  & 
320  & 
4420  & 
3352  & 

192  & 
192  & 
552  \\

Total Si area  \hfill [m$^2$]  &  
0.3  & 
4.6  & 
17.6  & 

0.8  & 
5.6  & 
3.3  \\

Read-out-Channels  \hfill [$10^6$]   & 
224.5  & 
1738  & 
20.6  & 

322.4  & 
73.3  & 
17.0   \\

pitch${^{r-\phi}}$ \hfill [$\mu$m] & 
{25}  & 
{100}   & 
{100}  &  

{25} & 
{100} & 
{100}   \\

pitch${^{z}}$ \hfill [$\mu$m] & 
{50}   &    
{400}  & 
50k\,$^{2)}$ & 

{50} & 
{400} & 
10k\,$^{1)}$   \\

\cmidrule(lr){2-4} \cmidrule(lr){5-7}
Average\,$X{_0}$/$\Lambda{_I}$ \hfill [${\%}$] & 
\multicolumn{3}{c} {7.2\,/\,2.2} & 
\multicolumn{3}{c} {2.2\,/\,0.7} \\
\bottomrule
\multicolumn{6}{l}{ $^{1)}$ Reaching pitch${^{r-\phi}}$ when using two wafer layers rotated by 20\,mrad is achievable.} 
\end{tabular}
\caption{Summary of the main properties of the Barrel and Endcap tracker modules 
based on calculations performed using 
tkLayout~\cite{Bianchi:2014mpa}.
For each module, the rows correspond to the pseudorapidity coverage, numbers of barrel and 
disk layers, numbers of sensors, total area covered by silicon sensors, numbers of 
readout channels, the hardware pitches affecting the ($r - \phi$) and the $z$ resolution, respectively, and
the average material budget in terms of radiation lengths and interaction lengths. Where appropriate, 
the numbers are broken down into separate contributions from pixels, macro-pixels and strips.
See Tab.\,\ref{tab:LHeC_Tracker_main-properties_2} for a sum of all tracker components.
}  \label{tab:LHeC_Tracker_main-properties_1} 

\end{table}

\begin{table}[htbp]
  \centering
  \small
  \begin{tabular}{@{\hskip0.3em}lc@{\hskip0.3em}c@{\hskip0.3em}cc@{\hskip0.3em}c@{\hskip0.3em}cc@{\hskip0.3em}}
    \toprule
    Tracker (LHeC) & \multicolumn{3}{c}{Fwd\,Tracker}  & \multicolumn{2}{c}{Bwd\,Tracker}  &   Total\\ 
    \cmidrule(lr){2-4} \cmidrule(lr){5-6} \cmidrule(lr){7-7}

    &
    pix &
    pix$_\text{macro}$ &
    strip & 

    pix$_\text{macro}$ &
    strip  &
    \footnotesize{(incl.\,Tab.\,\ref{tab:LHeC_Tracker_main-properties_1})} \\
    \midrule
    $\eta_\text{max}$,$\eta_\text{min}$  &  

5.3,2.6   & 
3.5,2.2   &  
3.1,1.6   & 
    
$-4.6,-2.5 $ &  
$-2.9,-1.6 $ & 

5.3,$-4.6$     \\
  
Wheels &
2 &
1 & 
3 &

2 & 
4 &   \\

Modules/Sensors  & 
180 & 
180  & 
860  & 
 
72  & 
416  & 

10736 \\

Total Si area  \hfill [m$^2$]  &  
0.8  & 
0.9  & 
4.6  & 
 
0.4  & 
1.8  & 

40.7 \\

Read-out-Channels \hfill [$10^6$]   & 
404.9  & 
68.9 & 
26.4  & 
 
27.6  & 
10.6  & 

2934.2  \\ 

pitch${^{r-\phi}}$ \hfill [$\mu$m] & 
{25}  & 
{100}   & 
{100}  &  
 
{100} & 
{100}   \\

pitch${^{z}}$ \hfill [$\mu$m] & 
{50}   &    
{400}  & 
50k\,$^{2)}$ & 

{400} & 
10k\,$^{1)}$   \\

\cmidrule(lr){2-4} \cmidrule(lr){5-6} \cmidrule(lr){7-7}
Average\,$X{_0}$/$\Lambda{_I}$ \hfill [${\%}$] & 
\multicolumn{3}{c}{6.7\,/\,2.1} &
\multicolumn{2}{c}{6.1\,/\,1.9} &  \\
~~~~incl. beam pipe\hfill[\%] &
\multicolumn{3}{c}{} &
\multicolumn{2}{c}{} & 
40\,/\,25 \\
\bottomrule
\multicolumn{6}{l}{ $^{1)}$ Reaching pitch${^{r-\phi}}$ when using two wafer layers rotated by 20\,mrad is achievable.} 
\end{tabular}
\caption{Summary of the main properties of the forward and backward tracker modules in the revised LHeC detector 
configuration based on calculations performed 
using tkLayout~\cite{Bianchi:2014mpa}.
For each module, the rows correspond to the pseudorapidity coverage, numbers of 
disk layers, numbers of sensors, total area covered by silicon sensors, numbers of 
readout channels, the hardware pitches affecting the ($r - \phi$) and the $z$ resolution, respectively, and
the average material budget in terms of radiation lengths and interaction lengths. The polar angle dependence and decomposition
of $X_0$ and $\Lambda_I$ are shown in Fig.\,\ref{FIG:material_tracker}. Where appropriate, 
the numbers are broken down into separate contributions from pixels, macro-pixels and strips.
The column \emph{Total} contains the sum of corresponding values in tables~\ref{tab:LHeC_Tracker_main-properties_1} 
and~\ref{tab:LHeC_Tracker_main-properties_2}.
}  \label{tab:LHeC_Tracker_main-properties_2} 
\end{table}

%% file: detector/tables/LHeC_Calo_main-properties.tex

\begin{table}[htbp] 
  \centering
  \small
  \begin{tabular}{lcccc}
    \toprule
        Calo (LHeC) &
        EMC&
        \multicolumn{3}{c}{HCAL}
        \\
        \cmidrule(lr){2-2} \cmidrule(lr){3-5}
        &
        Barrel&
        Ecap {Fwd} & Barrel   & Ecap {Bwd}  
        \\
        \midrule
        Readout, Absorber   &
        Sci,Pb &
        Sci,Fe & Sci,Fe & Sci,Fe 
        \\
        Layers &
        38 &
        58 &
        45 &   
        50  \\
        Integral Absorber Thickness [cm] &
        16.7 &
        134.0 &
        119.0 &
        115.5 \\
        $\eta{_\text{max}}$, $\eta_\text{min}$  \hfill     &
        $2.4  $, $ -1.9 $ &
        $1.9  $, $ 1.0  $ &
        $1.6  $, $ -1.1 $  &
        $-1.5 $, $ -0.6 $   \\
        $\sigma{_E}/{E}=a/\sqrt{E}\oplus b$\hfill[\%]  &
        12.4/1.9&
        46.5/3.8&
        48.23/5.6&
        51.7/4.3  \\
        \small{${\Lambda{_{I}}}\,/\,{X{_0}}$} \hfill  &
        $X{_0}={30.2}$&
        $\Lambda{_I}={8.2}$&
        $\Lambda{_I}={8.3}$&
        $\Lambda{_I}={7.1}$\\
        \small{Total area Sci}  \hfill \small{[m$^2$]}  &
        1174  &
        1403 &
        3853 &
        1209   \\
        \bottomrule
  \end{tabular}
\caption{
Basic properties and simulated resolutions of barrel calorimeter modules in the new LHeC detector configuration. 
For each of the modules, the rows indicate the absorber and sensitive materials, the number of layers and total 
absorber thickness, the pseudorapidity coverage, 
the contributions to the simulated resolution from the sampling ($a$) and material ($b$) terms in the form $a/b$, 
the depth in terms or radiation or interaction lengths and the total area covered by the sensitive material.
GEANT4~\cite{Agostinelli:2002hh} simulation based fits using crystal ball
function~\cite{Oreglia:1980cs,Gaiser:1982yw,Skwarnicki:1986xj}.  }
\label{tab:LHeC_Calo_main-properties_1}
\end{table}

\begin{table}[htbp] 
  \centering
  \small
  \begin{tabular}{lcccc}
    \toprule
        Calo (LHeC) & 
        FHC &
        FEC &
        BEC&
        BHC \\
        &
        Plug {Fwd} &
        Plug {Fwd} &
        Plug {Bwd} &   
        Plug {Bwd}   \\
        \midrule
        Readout, Absorber   &
        Si,W &
        Si,W &
        Si,Pb &   
        Si,Cu   \\
        Layers &
        300 &
        49 &
        49 &   
        165  \\  
        Integral Absorber Thickness [cm] &
        156.0 &
        17.0 &
        17.1 &
        137.5 \\      
        $\eta{_\text{max}}$, $\eta_\text{min}$ \hfill     &
        $5.5  $, $ 1.9  $ &
        $5.1  $, $ 2.0  $ &
        $-1.4 $, $ -4.5 $    &
        $-1.4 $, $ -5.0 $   \\
        $\sigma{_E}/{E}=a/\sqrt{E}\oplus b$\hfill[\%]  &
        51.8/5.4&
        17.8/1.4 &
        14.4/2.8&
        49.5/7.9   \\
        \small{${\Lambda{_{I}}}\,/\,{X{_0}}$} \hfill  &
        $\Lambda{_I}={9.6}$&
        $X{_0}={48.8}$&
        $X{_0}={30.9}$ &
        $\Lambda{_I}={9.2}$  \\
        \small{Total area Si}  \hfill \small{[m$^2$]}  &
        1354  &
        187 &
        187  &
        745      \\
        \bottomrule
  \end{tabular}
\caption{
Basic properties and simulated resolutions of forward and backward plug calorimeter modules in the new LHeC detector configuration. 
For each of the modules, the rows indicate the absorber and sensitive materials, the number of layers and total absorber thickness, the pseudorapidity coverage, 
the contributions to the simulated resolution from the sampling ($a$) and material ($b$) terms in the form $a/b$, 
the depth in terms or radiation or interaction lengths and the total area covered by the sensitive material. 
GEANT4~\cite{Agostinelli:2002hh} simulation based fits using crystal ball
function~\cite{Oreglia:1980cs,Gaiser:1982yw,Skwarnicki:1986xj}.  }
\label{tab:LHeC_Calo_main-properties_2}
\end{table}

%% file: detector/tables/lowE-FCCeh_fwd-bwd_Tracker_main-properties.tex

\begin{table}[htbp]
  \centering
  \small
    \begin{tabular}{lcccccc}
    \toprule
    Tracker (lowE-FCCeh)\,$^{1)}$ & \multicolumn{3}{c}{Fwd\,Tracker}  & \multicolumn{2}{c}{Bwd\,Tracker}  &   Total\\ 
    \cmidrule(lr){2-4} \cmidrule(lr){5-6} \cmidrule(lr){7-7}

    &
    pix &
    pix$_\text{macro}$ &
    strip & 

    pix$_\text{macro}$ &
    strip  &
    \footnotesize{(incl.\,Tab.\,\ref{tab:LHeC_Tracker_main-properties_1})} \\
    \midrule
    $\eta_\text{max}$,$\eta_\text{min}$  &  

5.6,2.6   & 
3.8,2.2   &  
3.5,1.6   & 

$-4.6,-2.6 $ &  
$-2.8,-1.6 $ & 

5.3,$-4.6 $     \\
  
Wheels &
2 &
1 & 
3 &

3 & 
3 &   \\

Modules/Sensors  & 
288  & 
288  & 
1376  & 

216  & 
1248  & 

 12444 \\

Total Si area  \hfill [m$^2$]  &  
1.35  & 
1.45  & 
7.35  & 

1.0  & 
6.5  & 

49.85 \\

Read-out-Channels \hfill [$10^6$]   & 
647.9  & 
110.2 & 
42.3  & 

82.7 & 
38.3  & 

3317.2  \\ 

pitch${^{r-\phi}}$ \hfill [$\mu$m] & 
{25}  & 
{100}   & 
{100}  &  
 
{100} & 
{100}   \\

pitch${^{z}}$ \hfill [$\mu$m] & 
{50}   &    
{400}  & 
50k\,$^{2)}$ & 

{400} & 
10k\,$^{2)}$   \\

\cmidrule(lr){2-4} \cmidrule(lr){5-6} \cmidrule(lr){7-7}
Average\,$X{_0}$/$\Lambda{_I}$ \hfill [${\%}$] & 
\multicolumn{3}{c}{6.7\,/\,2.1} &
\multicolumn{2}{c}{6.1\,/\,1.9} &  \\
~~~~incl. beam pipe\hfill[\%] &
\multicolumn{3}{c}{} &
\multicolumn{2}{c}{} & 
40\,/\,25 \\
\bottomrule
\multicolumn{6}{l}{ $^{1)}$ Based on tklayout calculations~\cite{Bianchi:2014mpa} } \\
\multicolumn{6}{l}{ $^{2)}$ Reaching pitch${^{r-\phi}}$ when using two wafer layers rotated by 20\,mrad is achievable.} 
\end{tabular}
\caption{Summary of the main properties of the forward and backward tracker modules in the low energy FCC-eh 
detector 
configuration, based on calculations performed using tkLayout.
For each module, the rows correspond to the pseudorapidity coverage, the numbers of disk layers and 
of sensors, the total area covered by silicon sensors, the numbers of 
readout channels, the hardware pitches affecting the ($r - \phi$) and the $z$ resolution, and
the average material budget in terms of radiation lengths and interaction lengths. 
The numbers are broken down into separate contributions from pixels, macro-pixels and strips.
The column \emph{Total} contains the sum of corresponding values of barrel tracker modules
(identical to the LHeC barrel layout,  table~\ref{tab:LHeC_Tracker_main-properties_1})and
the forward and backward trackers in this 
table,~\ref{FCC-tracker}.
}  \label{FCC-tracker} 
\end{table}

%% file: detector/tables/lowE-FCCeh_Calo_main-properties.tex

\begin{table}[htbp] 
  \centering
  \small
  \begin{tabular}{lcccc}
    \toprule
        Calo (lowE-FCCeh) &
        EMC&
        \multicolumn{3}{c}{HCAL}
        \\
        \cmidrule(lr){2-2} \cmidrule(lr){3-5}
        &
        Barrel&
        Ecap {Fwd} & Barrel   & Ecap {Bwd}  
        \\
        \midrule
        Readout, Absorber   &
        Sci,Pb &
        Sci,Fe & Sci,Fe & Sci,Fe 
        \\
        Layers &
        49 &
        91 &
        68 &   
        78  \\
        Integral Absorber Thickness [cm] &
        36.6 &
        206.0 &
        184.0 &
        178.0 \\        
        $\eta{_\text{max}}$, $\eta_\text{min}$  \hfill     &
        $2.8  $, $ -2.5 $ &
        $2.0  $, $  0.8 $ &
        $1.6  $, $ -1.4 $  &
        $-0.7 $, $ -1.8 $   \\
        $\sigma{_E}/{E}=a/\sqrt{E}\oplus b$\hfill[\%]  &
        12.6/1.1 &
        38.9/3.3 &
        42.4/4.2 &
        40.6/3.5  \\
        \small{${\Lambda{_{I}}}\,/\,{X{_0}}$} \hfill  &
        $X{_0}={66.2}$&
        $\Lambda{_I}={12.7}$&
        $\Lambda{_I}={11.3}$&
        $\Lambda{_I}={11.0}$\\
        \small{Total area Sci}  \hfill \small{[m$^2$]}  &
        2915  &
        4554 &
        12298 &
        3903   \\
        \bottomrule
  \end{tabular}
\caption{
Basic properties and simulated resolutions of barrel calorimeter modules in a scaled configuration, 
suitable for a low energy FCC detector.
For each of the modules, the rows indicate the absorber and sensitive materials, 
the numbers of layers and the total absorber thickness,
the pseudorapidity coverage, 
the contributions to the simulated resolution from the sampling ($a$) and material ($b$) terms in the form $a/b$, 
the depth in terms or radiation or interaction lengths and the total area covered by the sensitive material. 
The resolutions are obtained from a 
GEANT4~\cite{Agostinelli:2002hh} simulation, with fits using a crystal ball function~\cite{Oreglia:1980cs,Gaiser:1982yw,Skwarnicki:1986xj}.  }
\label{tab:lowE-FCCeh_Calo_main-properties_1}
\end{table}

\begin{table}[htbp] 
  \centering
  \small
  \begin{tabular}{lcccc}
    \toprule
        Calo (lowE-FCCeh)  &
        FHC &
        FEC &
        BEC&
        BHC \\
        &
        Plug {Fwd} &
        Plug {Fwd} &
        Plug {Bwd} &   
        Plug {Bwd}   \\
        \midrule
        Readout, Absorber   &
        Si,W &
        Si,W &
        Si,Pb &   
        Si,Cu   \\
        Layers &
        296 &
        49 &
        59 &   
        238  \\  
        Integral Absorber Thickness [cm] &
        256.9 &
        29.6 &
        27.9 &
        220.8 \\      
        $\eta{_\text{max}}$, $\eta_\text{min}$ \hfill     &
        $5.8  $, $ 1.8  $ &
        $5.4  $, $ 1.8  $ &
        $-1.5 $, $ -5.2 $    &
        $-1.5 $, $ -5.6 $   \\
        $\sigma{_E}/{E}=a/\sqrt{E}\oplus b$\hfill[\%]  &
        61.9/0.5 &
        26.5/0.4 &
        24.7/0.4 &
        46.7/4.4   \\
        \small{${\Lambda{_{I}}}\,/\,{X{_0}}$} \hfill  &
        $\Lambda{_I}={15.5}$&
        $X{_0}={84.7}$&
        $X{_0}={50.2}$ &
        $\Lambda{_I}={14.7}$  \\
        \small{Total area Si}  \hfill \small{[m$^2$]}  &
        2479  &
        364 &
        438  &
        1994      \\
        \bottomrule
  \end{tabular}
\caption{
Basic properties and simulated resolutions of forward and backward plug calorimeter modules in a scaled 
configuration, suitablle for a low energy FCC detector. 
For each of the modules, the rows indicate the absorber and sensitive materials, 
the numbers of layers and the total absorber thickness,
the pseudorapidity coverage, 
the contributions to the simulated resolution from the sampling ($a$) and material ($b$) terms in the form $a/b$, 
the depth in terms or radiation or interaction lengths and the total area covered by the sensitive material. 
The resolutions are obtained from a 
GEANT4~\cite{Agostinelli:2002hh} simulation, with fits using a crystal ball function~\cite{Oreglia:1980cs,Gaiser:1982yw,Skwarnicki:1986xj}.  }
\label{tab:lowE-FCCeh_Calo_main-properties_2}
\end{table}

%% file: conclusion/conclusion.tex
\linenumbers
\lhectitlepage
\lhecinstructions
\subfilestableofcontents

\chapter{Conclusion}
%
The Large Hadron Collider determines the energy frontier of experimental
collider physics for the next two decades.  Following the current luminosity
upgrade, the LHC can be further upgraded with a high energy,
intense electron beam such that it becomes a twin-collider facility, in which
 $ep$ collisions are registered  concurrently with $pp$. 
 A joint ECFA, CERN and NuPECC initiative led to a 
detailed conceptual design report (CDR)~\cite{AbelleiraFernandez:2012cc} for the 
 Large Hadron Electron Collider published in 2012.
The present paper represents an update of the original CDR in view
of new physics and technology developments.

The LHeC uses a novel, energy recovery linear electron accelerator  which
enables TeV energy  electron-proton collisions at high luminosity,  
of $\mathcal{O}(10^{34})$\,cm$^{-2}$s$^{-1}$, 
exceeding that of HERA by nearly three orders of magnitude.   
The discovery  of the Higgs boson and the surprising absence of BSM physics 
at the LHC demand to  extend the experimental base of particle physics suitable
to explore the energy frontier, beyond $pp$ collisions at the LHC.
The LHC infrastructure is the largest single investment the European and
global particle physics community ever assembled, and the addition 
of an electron accelerator a seminal opportunity way to build on it, and
to sustain the HL-LHC programme by adding necessary elements
which are provided by high energy deep inelastic scattering. As has been
shown in this paper, the external DIS input transforms the LHC to a much 
more powerful facility, with a new level of resolving matter substructure,
a more precise Higgs programme, challenging and complementing 
that of a next $e^+e^-$ collider, and with a hugely extended potential
to discover physics beyond the Standard Model.

The very high luminosity and the substantial extension of the kinematic range
in deep inelastic scattering, compared to HERA, make the LHeC on its own
a uniquely powerful TeV energy  collider.
Realising the  $\it{Electrons~for~LHC}$ programme developed with the 
previous and the present ``white" papers, will create the
cleanest, high resolution microscope accessible to the world, which one may term the
 ``CERN Hubble Telescope for the Micro-Universe". 
 It is directed to unravel
the substructure of matter encoded in 
the complex dynamics of the strong interaction, and to provide the necessary input 
for  precision and discovery physics at 
the HL-LHC and for future hadron colliders. 

This programme, as has been described
in this paper, comprises the 
complete resolution of the partonic densities in an unexplored range of
kinematics, the foundations for new, generalised views on proton structure
and the long awaited clarification of the QCD dynamics at high densities, as
are observed at small Bjorken $x$. New high precision measurements on
diffraction and vector mesons will shed new light on the puzzle of confinement.
As a complement to the LHC and a possible future e$^+$e$^-$ machine, 
the LHeC would scrutinise the Standard Model deeper than 
ever before, and possibly discover new physics in the electroweak 
and chromodynamic sectors as is outlined in the paper. 

Through the extension of the kinematic range 
by about three orders of magnitude in lepton-nucleus ($e$A) scattering, 
the LHeC is the most powerful electron-ion research facility one can 
build in the next decades, 
for clarifying the partonic substructure 
and dynamics inside nuclei for the first time and elucidating the chromodynamic origin
of the Quark-Gluon-Plasma. 

The Higgs programme at the LHeC  is
 astonishing in its precision. It relies on CC and NC precision measurements for
which an inverse atobarn of luminosity is desirable to achieve.  The prospective 
results on the Higgs couplings
from the HL-LHC, when combined with those here presented from the LHeC, will determine
the couplings in the most frequent six Higgs decay channels to one per cent
level accuracy. This is as precise as one expects measurements from linear $e^+e^-$
colliders but obtained dominantly from $gg$ and $WW$ fusion respectively, as compared
to Higgs-strahlung in electron-positron scattering which has the advantage of 
providing a Higgs width determination too.  The combined $pp$+$ep$ LHC facility 
at CERN may then be expected to remain the centre of Higgs physics for two more decades.

Searches for BSM physics at the LHeC offer great complementarity to similar searches at the HL-LHC.
The core advantage of the LHeC is the clean, QCD-background and pileup-free environment of an
electron-proton collider with a cms energy exceeding a TeV.
This enables discoveries of signatures that could be lost in the hadronic noise at $pp$ 
or possibly unaccessible due to the limited com energy at $ee$.
Prominent examples for discovery enabled with $ep$
are heavy neutral leptons (or sterile neutrinos) that mix with the electron flavour, 
dark photons below the di-muon threshold,
which are notoriously difficult to detect in other experiments,
 long-lived new particles in general or new
physics scenarios with a compressed mass spectrum, such as SUSY electrowikinos and heavy scalar resonances with masses around and below 500\,GeV, which may exist but  literally be buried in di-top backgrounds at the LHC.


The LHeC physics programme reaches far beyond any  specialised 
goal which underlines the unique opportunity for particle physics
to build a novel laboratory for accelerator based energy frontier
research at CERN. The project is fundable within the CERN budget, 
and not preventing much  more massive investments into the further future. 
It offers the possibility for the current generation of accelerator
physicists to build a new collider using and developing novel
technology while preparations proceed for the next grand step in particle
physics for generations ahead.

The main technical innovation through the LHeC
is the first ever high energy application of energy recovery technology, based on
high quality superconducting RF developments, a major contribution
to the development of $green$ collider technology which is an appropriate
 response to demands of our time. The ERL technique is more and more seen to
have major further applications, beyond ep at HE-LHC and FCC-eh, such
 as  for FCC-ee, as a $\gamma \gamma$ Higgs facility or, 
 beyond particle physics, as the highest energy XFEL
of hugely increased brightness.

The paper describes the plans and configuration of PERLE, the
first $10$\,MW power ERL facility, which is being prepared 
in international collaboration for built
at Ir\`ene Joliot-Curie Laboratory at Orsay. PERLE has adopted the
3-pass configuration, cavity and cryomodule technology, source 
and injector layout, frequency and electron current parameters
from the LHeC. This qualifies it to be the ideal machine to 
accompany the development of the LHeC. However, through 
its technology innovation  and its challenging 
parameters, such as an intensity exceeding that of ELI by orders
of magnitude, PERLE has an independent, far reaching low energy
nuclear and particle physics programme with new and particularly 
precise measurements. It also has a possible program on industrial
applications, which has not been discussed in the present paper.

The LHeC provides an opportunity for building a novel
collider detector which is sought for as the design of the HL-LHC
detector upgrades is approaching completion.
A novel $ep$ experiment enables modern detection technology, such as 
HV CMOS Silicon tracking, to be further developed
and exploited in a new generation, $4\pi$ acceptance, no pile-up, 
high precision collider detector in the decade(s) hence. This
paper presented an update of the 2012 detector design, 
in response to developments of physics,
 especially Higgs and BSM, and of technology in detectors and 
analysis. The LHeC requires to be installed at IP2 at the LHC for there is no other
interaction region available while the heavy ion programme at the LHC is
presently limited to the time until LS4. In the coming years it  will have to be
decided whether this or alternative proposals for using IP2 during the 
final years of LHC operation are considered attractive enough and realistic
to be realised.

The next steps in this development are rather clear: the emphasis on ERL, beyond LHeC,
requires the PERLE development to rapidly
proceed. Limited funds are to be found  for essential components with the
challenging IR quadrupole as the main example.  ECFA is about to establish a
detector and physics series of workshops, including possible future Higgs facilities, and $ep$, 
which is a stimulus for further developing the organisational base of the LHeC towards
a detector proto-Collaboration. These developments shall include preparations for FCC-he
and provide a necessary basis when in a few years time, as recommended by the
IAC, a decision on building the LHeC at CERN may be taken. 

The recent history teaches a lesson about the complementarity required
for energy frontier particle physics. In the seventies and eighties, 
CERN hosted the $p \bar{p}$ energy frontier, with UA1 and UA2,
and the most powerful DIS experiments with muons (EMC, BCDMS, NMC)
and neutrinos (CDHSW, CHARM), while $e^+e^-$ physics was pursued
at PEP, PETRA and also TRISTAN. Following this, the Fermi scale could be
explored with the Tevatron, HERA and LEP. The here advertised
next logical step is to complement the HL-LHC by a most powerful DIS facility,
the LHeC, while preparations will take shape for  a next $e^+e^-$ 
collider, currently at CERN and in Asia.  Hardly a decision on LHeC may be taken
independently of how the grand future unfolds. Still, this scenario would give a realistic
and yet exciting base for completing the exploration of TeV
scale physics which may not be achieved with solely the LHC. 

The ERL concept and technology here presented has the
potential to accompany the FCC for realising the FCC-eh machine
when the time comes for the next, higher energy hadron collider, and
the search for new physics at the $\mathcal{O}(10)$\,TeV scale.

\vspace{1cm}
\begin{Large}
$\bf{Acknowledgement}$ \\
\end{Large}
\vspace{0.4cm}

\noindent
The analyses and developments here presented would not have been possible
without the CERN Directorate and other laboratories, universities
and groups supporting this
study. We admire the skills of the technicians who successfully built
the first $802$\,MHz SC cavity. We 
thank many of our colleagues for their interest in this work and a supportive
attitude when time constraints could have caused  lesser understanding.
 Special thanks are also due to the members and Chair of the International
Advisory Committee for their attention and guidance to the project.
From the beginning of the LHeC study, it has been supported by ECFA
and its chairs, which was a great help and stimulus for undertaking this
study performed outside our usual duties. During the time,
a number of students, in Master and PhD courses, have made very essential
contributions to this project for which we are especially grateful. This
also extends to colleagues with whom we have been working closely
but who meanwhile left this development, perhaps temporarily,
or work at non-particle physics institutions while wishing LHeC success. The current
situation of particle physics reminds us on the potential we have
when resources and prospects are combined, for which this study
is considered to be a contribution.

The authors would like to thank the Institute of Particle Physics Phenomenology (IPPP) at Durham
for the award of an IPPP Associateship to support this work, and gratefully acknowledges support
from the state of Baden-W\"urttemberg through bwHPC and the German Research Foundation (DFG) through grant no INST 39/963-1 FUGG.
Financial support by Ministerio de Ciencia e Innovaci\'on
of Spain under projects FPA2017-83814-P and Unidad de
Excelencia Mar\'{\i}a de Maetzu under project MDM-2016-0692, and
by Xunta de Galicia (project ED431C 2017/07 and Centro singular de investigación de Galicia accreditation 2019-2022) and the European Union (European Regional Development Fund – ERDF), is gratefully acknowledged.
It has been performed in the framework of COST Action CA
15213 “Theory of hot matter and relativistic heavy-ion
collisions” (THOR), MSCA RISE 823947 “Heavy ion collisions: collectivity and precision in saturation physics” (HIEIC) and has received funding from the European Union’s Horizon 2020 research and innovation programme
under grant agreement No. 824093.

\biblio

%% file: main/appendix.tex
\linenumbers
\lhectitlepage
\lhecinstructions
\subfilestableofcontents

\newpage
\chapter{Statement of the International Advisory Committee}
%
End of 2014, the CERN Directorate appointed an International
Advisory Committee (IAC) for advice on the direction of
energy frontier electron-hadron scattering at CERN, for their mandate see below. The committee and its chair, em. DG of CERN Herwig Schopper, was reconfirmed when a new DG 
had been appointed. The IAC held regular sessions at the annual
LHeC workshops in which reports were heard by the co-coordinators
of the project, Oliver Br{\"u}ning and Max Klein. Its work and 
opinion shaped the project development considerably and it was pivotal for the foundation of the PERLE project.  The committee
was in close contact and advised especially on the documents,
on the LHeC~\cite{Bruning:2652313,Bruning:2652335} and PERLE~\cite{Klein:2652336},  submitted end of 2018 to the update of the European strategy on particle physics. In line with the present updated LHeC design
report and the strategy process,
the IAC formulated a brief report 
to the CERN DG, in which its observations and recommendations 
have been summarised. This report was also sent to the members of the European particle physics strategy group. It is reproduced here.

\vspace{0.7cm}
\section*{Report by the IAC on the LHeC to the DG of CERN}
The development of the LHeC project was initiated by CERN and ECFA, in cooperation with NuPECC. It culminated in the publication of the Conceptual Design Report (CDR), arXiv:1206.2913 in 2012, which received by now about 500 citations. In 2014, the CERN Directorate invited our committee to advise the CERN Directorate, and the Coordination Group, on the directions of future energy frontier electron-hadron scattering as are enabled with the LHC and the future FCC (for the mandate see below). In 2016, Council endorsed the HL-LHC, which offers a higher LHC performance and strengthened the interest in exploring the Higgs phenomenon. In view of the imminent final discussions for the European Road Map for particle physics, a short summary report is here presented.

\subsection*{Main Developments 2014--2019}

A series of annual workshops on the LHeC and FCC-eh was held, and this report is given following the latest workshop https://indico.cern.ch/event/835947 , October 24/25, 2019. 

Based on recent developments concerning the development of the LHC accelerator and physics, and the progress in technology, a new default configuration of the LHeC and FCC-eh has been worked out with a tenfold increased peak luminosity goal, of 
$10^{34}$\,cm$^{-2}$s$^{-1}$, as compared to the CDR. A comprehensive paper, “The LHeC at the HL-LHC”, is being finalised for publication this year. 

Within this work, it has been shown that the LHeC represents the cleanest, high resolution microscope the world can currently build, a seminal opportunity to develop and explore QCD, to study high precision Higgs and electroweak physics and to substantially extend the range and prospects for accessing BSM physics, on its own and in combination of pp with ep. The LHeC, in eA scattering mode, has a unique discovery potential on nuclear structure, dynamics and QGP physics.

Intense eh collisions with LHeC and FCC-eh are enabled through a special electron-beam racetrack arrangement with energy recovery linac (ERL) technology. If LHeC were to be considered either on its own merits, or as a bridge project to FCC-eh, it seemed important to find a configuration, which could be realised within the existing CERN budget. Several options were studied and found.  

Before a decision on such a project can be taken, the ERL technology has to be further developed. Considerable progress has been made in the USA, and a major effort is now necessary to develop it further in Europe. An international collaboration (ASTeC, BINP, CERN, Jefferson Lab, Liverpool, Orsay) has been formed to realise the first multi-turn 10 MW ERL facility, PERLE at Orsay, with its main parameters set by the LHeC and producing the first encouraging results on $802$\,MHz cavity technology, for the CDR see arXiv:1705.08783.

This radically new accelerator technology, ERL, has an outstanding technical (SRF), physics (nuclear physics) and industrial (lithography, transmutations, ..) impact, and offers possible applications beyond ep (such as a racetrack injector or ERL layout for FCC-ee, a high energy FEL or $\gamma \gamma$  collider).

\subsection*{In conclusion it may be stated}
\begin{compactitem}
\item  The installation and operation of the LHeC has been demonstrated to be commensurate with the currently projected HL-LHC program, while the FCC-eh has been integrated into the FCC vision; 
\item  The feasibility of the project as far as accelerator issues and detectors are concerned has been shown. It can only be realised at CERN and would fully exploit the massive LHC and HL-LHC investments; 
\item  The sensitivity for discoveries of new physics is comparable, and in some cases superior, to the other projects envisaged; 
\item  The addition of an ep/A experiment to the LHC substantially reinforces the physics program of the facility, especially in the areas of QCD, precision Higgs and electroweak as well as heavy ion physics; 
\item  The operation of LHeC and FCC-eh is compatible with simultaneous pp operation; for LHeC the interaction point\,2 would be the appropriate choice, which is currently used by ALICE; 
\item  The development of the ERL technology needs to be intensified in Europe, in national laboratories but with the collaboration of CERN; 
\item  A preparatory phase is still necessary to work out some time-sensitive key elements, especially the  high power ERL technology (PERLE) and the prototyping of Intersection Region magnets. 
\end{compactitem}
\subsection*{Recommendations}

i) It is recommended to further develop the ERL based ep/A scattering plans, both at LHC and FCC, as attractive options for the mid and long term programme of CERN, resp. Before a decision on such a project can be taken, further development work is necessary, and should be supported, possibly within existing CERN frameworks (e.g. development of SC cavities and high field IR magnets). \\

\noindent
ii) The development of the promising high-power beam-recovery technology ERL should be intensified in Europe. This could be done mainly in national laboratories, in particular with the PERLE project at Orsay. To facilitate such a collaboration, CERN should express its interest and continue to take part. \\

\noindent
iii) It is recommended to keep the LHeC option open until further decisions have been taken. An investigation should be started on the compatibility between the LHeC and a new heavy ion experiment in Interaction Point 2, which is currently under discussion. \\

\noindent
After the final results of the European Strategy Process will be made known, the IAC considers its task to be completed. A new decision will then have to be taken for how to continue these activities. \\

\noindent
Herwig Schopper,  Chair of the Committee, ~~~~~~~~\hfill Geneva, November 4, 2019


\section*{Mandate of the International Advisory Committee}

Advice to the LHeC Coordination Group and the CERN directorate by following the development of options of an ep/eA collider at the LHC and at FCC, especially with: Provision of scientific and technical direction for the physics potential of the ep/eA collider, both at LHC and at FCC, as a function of the machine parameters and of a realistic detector design, as well as for the design and possible approval of an ERL test facility at CERN. Assistance in building the international case for the accelerator and detector developments as well as guidance to the resource, infrastructure and science policy aspects of the ep/eA collider. (December 2014)

\section*{Members of the Committee}
%
                                                                    
  \begin{tabular}{@{}l@{\hspace{0.05\textwidth}}l}
Sergio Bertolucci    (U Bologna)                 &    Max Klein            (U Liverpool, coordinator) \\     
Nichola Bianchi      (INFN, now Singapore)     &    Shin-Ichi Kurokawa   (KEK) \\                        
Frederick Bordy      (CERN)                    &    Victor Matveev       (JINR Dubna) \\                 
Stan Brodsky         (SLAC)                    &    Aleandro Nisati      (Rome I) \\                     
Oliver Br{\"u}ning      (CERN, coordinator)    &    Leonid Rivkin        (PSI Villigen) \\               
Hesheng Chen         (IHEP Beijing)                 &    Herwig Schopper      (CERN, em.DG, Chair) \\         
Eckhard Elsen        (CERN)                    &    J{\"u}rgen Schukraft    (CERN) \\                       
Stefano Forte        (U Milano)                  &    Achille Stocchi      (IJCLab Orsay) \\                      
Andrew Hutton        (Jefferson Lab)           &    John Womersley       (ESS Lund)  \\
Young-Kee Kim        (U Chicago) \\
\end{tabular}   
%
%

\vskip -1cm
\chapter{Membership of Coordination}
%
\vspace{-1.2cm}
{\Large \bf Coordinating Group}
\vspace{0.1cm}
\\ \noindent
Gianluigi Arduini (CERN) \\
N\'estor Armesto (University of Santiago de Compostela) \\
Oliver Br\"{u}ning (CERN) – Co-Chair \\
Andrea Gaddi (CERN) \\
Erk Jensen (CERN) \\
Walid Kaabi (IJCLab Orsay) \\
Max Klein (University of Liverpool) – Co-Chair \\
Peter Kostka (University of Liverpool) \\
Bruce Mellado (University of Witwatersrand) \\
Paul R. Newman (University of Birmingham) \\
Daniel Schulte (CERN) \\
Frank Zimmermann (CERN)

\vspace{0.2cm}
{\Large \bf Physics Convenors}
\vspace{0.1cm}
\\ \noindent
$\bf{Parton~Distributions~and~QCD}$ \\
Claire Gwenlan (University of Oxford) \\
Fred Olness (Texas University, Dallas) \\
$\bf{Physics~at~Small~x}$ \\
Paul R. Newman (University of Birmingham)\\
Anna M. Stasto (Pennsylvania State University) \\
$\bf{Top~and~Electroweak~Physics}$ \\
Olaf Behnke (DESY Hamburg)\\
Daniel Britzger (MPI Munich) \\
Christian Schwanenberger (DESY Hamburg) \\
$\bf{Electron-Ion~Physics}$ \\
N\'estor Armesto (University of Santiago de Compostela)\\
$\bf{Higgs~Physics}$ \\
Uta Klein (University of Liverpool)\\
Masahiro Kuze (Institute of Technology Tokyo) \\
$\bf{BSM~Physics}$ \\
Georges Azuelos (University of Montreal)\\
Monica D'Onofrio (University of Liverpool)\\
Oliver Fischer (MPIK Heidelberg) \\
$\bf{Detector~Design}$ \\
Peter Kostka (University of Liverpool)\\
Alessandro Polini (INFN Bologna)
\biblio